%% file: main.tex
\documentclass[10pt,twoside,openany]{scrbook}

\usepackage[
  paperwidth=439.42bp,
  paperheight=683.15bp,
  inner=0.74in,
  outer=0.58in,
  top=0.56in,
  bottom=0.66in,
  headheight=13pt,
  headsep=0.16in
]{geometry}
\usepackage[T1]{fontenc}
\usepackage[utf8]{inputenc}
\usepackage{newtxtext,newtxmath}
\usepackage{microtype}
\usepackage{amsmath,mathtools}
\usepackage{array,booktabs,longtable}
\usepackage{graphicx}
\usepackage{xcolor}
\usepackage{enumitem}
\usepackage[most]{tcolorbox}
\usepackage{tikz}
\usepackage{prodint}
\usepackage{scrlayer-scrpage}
\usepackage{xr-hyper}
\usepackage[hyperindex=true]{hyperref}
\usepackage{bookmark}
\usepackage[quiet]{imakeidx}
\usepackage[
  backend=biber,
  style=trad-alpha,
  sorting=nyt,
  maxbibnames=99,
  maxalphanames=4,
  giveninits=true,
  natbib=true,
  indexing=cite,
  doi=false,
  url=false,
  isbn=false
]{biblatex}
\makeindex[name=person,title={Name Index},columns=2,noautomatic]
\makeindex[name=subject,title={Subject Index},columns=2,noautomatic]
\indexsetup{level=\section*,noclearpage}
\newcommand{\bookpersonbibindex}[1]{\index[person]{#1|hyperpage}}
\DeclareIndexNameFormat{default}{%
  \usebibmacro{index:name}
    {\bookpersonbibindex}
    {\namepartfamily}
    {\namepartgiven}
    {\namepartprefix}
    {\namepartsuffix}}
\renewbibmacro*{citeindex}{%
  \ifciteindex
    {\indexnames{labelname}}
    {}}
\usetikzlibrary{arrows.meta,calc,positioning}

\definecolor{bookblue}{RGB}{31,47,160}
\definecolor{bookred}{RGB}{228,48,39}
\definecolor{bookgreen}{RGB}{0,176,34}
\definecolor{bookteal}{RGB}{0,148,166}
\definecolor{bookgold}{RGB}{204,176,30}
\definecolor{chaptercream}{RGB}{253,251,246}
\definecolor{theoremback}{RGB}{255,254,246}
\definecolor{definitionback}{RGB}{255,255,255}
\definecolor{exampleback}{RGB}{255,255,255}
\definecolor{noteback}{RGB}{255,255,247}

\providecommand{\BookTitle}{The Statistical Compass}
\providecommand{\BookSubtitle}{From Data to Decision, from Measure to Meaning}
\providecommand{\BookAuthor}{Elvis Han Cui, PhD}
\providecommand{\BookAffiliation}{Department of Biostatistics, UCLA Fielding School of Public Health}
\providecommand{\BookPDFTitle}{The Statistical Compass}
\ifdefined\CHAPTERONE
\renewcommand{\BookPDFTitle}{Introduction: Statistics as Translation}
\fi
\ifdefined\CHAPTERTWO
\renewcommand{\BookPDFTitle}{Randomness in Data: Examples That Force Theory}
\fi
\ifdefined\CHAPTERTHREE
\renewcommand{\BookPDFTitle}{Probability and Measure: Events, Laws, and Integration}
\fi
\ifdefined\CHAPTERFOUR
\renewcommand{\BookPDFTitle}{Design Before Data: Randomization, Sampling, and Collection}
\fi
\ifdefined\CHAPTERFIVE
\renewcommand{\BookPDFTitle}{Observation, Conditioning, and Likelihood}
\fi
\ifdefined\CHAPTERSIX
\renewcommand{\BookPDFTitle}{How Data Objects Are Built: Products, Kernels, and Processes}
\fi
\ifdefined\CHAPTERSEVEN
\renewcommand{\BookPDFTitle}{Modern Data Objects: What Kind of Object Was Observed?}
\fi
\ifdefined\CHAPTEREIGHT
\renewcommand{\BookPDFTitle}{Statistical Targets: What Claim Is Being Made?}
\fi
\ifdefined\CHAPTERNINE
\renewcommand{\BookPDFTitle}{Laws of Large Numbers and Concentration: Stability of Empirical Summaries}
\fi
\ifdefined\CHAPTERTEN
\renewcommand{\BookPDFTitle}{Weak Convergence of Random Objects}
\fi
\ifdefined\CHAPTERELEVEN
\renewcommand{\BookPDFTitle}{Uniform Laws and Empirical Processes: Controlling Indexed Noise}
\fi
\ifdefined\CHAPTERTWELVE
\renewcommand{\BookPDFTitle}{Karhunen--Loeve Expansions and Functional Data}
\fi
\ifdefined\CHAPTERSEVENTEEN
\renewcommand{\BookPDFTitle}{From Inference to Use: Prediction, Risk, Feedback, and Deployment}
\fi
\ifdefined\CHAPTERSIXTEEN
\renewcommand{\BookPDFTitle}{Continuous-Time Processes, Event Histories, and Martingales}
\fi
\ifdefined\CHAPTERFIFTEEN
\renewcommand{\BookPDFTitle}{Delta Method, Influence Functions, and Local Uncertainty}
\fi
\ifdefined\CHAPTERTHIRTEEN
\renewcommand{\BookPDFTitle}{M- and Z-Estimation: Consistency and Asymptotic Normality}
\fi
\ifdefined\CHAPTERFOURTEEN
\renewcommand{\BookPDFTitle}{Testing, Hellinger Geometry, and Local Alternatives}
\fi
\ifdefined\APPENDICES
\renewcommand{\BookPDFTitle}{Appendices: Mathematical and Biomedical Toolkits}
\fi

\hypersetup{
  colorlinks=true,
  linktoc=page,
  linkcolor=bookblue,
  urlcolor=bookblue,
  citecolor=bookblue,
  pdftitle={\BookPDFTitle},
  pdfauthor={Elvis Han Cui, PhD}
}
\makeatletter
\let\book@orig@hyperpage\hyperpage
\renewcommand*{\hyperpage}[1]{%
  \begingroup
  \hypersetup{linkcolor=bookred!82!black}%
  \book@orig@hyperpage{#1}%
  \endgroup}
\makeatother
\bookmarksetup{numbered,open,openlevel=1,depth=2}

\newif\ifappendixrefsneeded
\ifdefined\CHAPTERONE\appendixrefsneededtrue\fi
\ifdefined\CHAPTERTHREE\appendixrefsneededtrue\fi
\ifdefined\CHAPTERFIVE\appendixrefsneededtrue\fi
\ifdefined\CHAPTERSIX\appendixrefsneededtrue\fi
\ifdefined\CHAPTERTWELVE\appendixrefsneededtrue\fi
\ifdefined\CHAPTERFIFTEEN\appendixrefsneededtrue\fi
\ifdefined\CHAPTERSIXTEEN\appendixrefsneededtrue\fi
\ifappendixrefsneeded
\fi
\ifdefined\CHAPTERTWELVE
  \IfFileExists{build/main_ch10.aux}{%
    \externaldocument[][nocite]{build/main_ch10}[main_ch10.pdf]%
  }{}%
\fi
\ifdefined\CHAPTERTHREE
  \IfFileExists{build/main_ch06.aux}{%
    \externaldocument[][nocite]{build/main_ch06}[main_ch06.pdf]%
  }{}%
\fi
\ifdefined\CHAPTERSIXTEEN
  \IfFileExists{build/main_ch11.aux}{%
    \externaldocument[][nocite]{build/main_ch11}[main_ch11.pdf]%
  }{}%
\fi
\ifdefined\APPENDICES
  \IfFileExists{build/main_ch10.aux}{%
    \externaldocument[][nocite]{build/main_ch10}[main_ch10.pdf]%
  }{}%
  \IfFileExists{build/main_ch11.aux}{%
    \externaldocument[][nocite]{build/main_ch11}[main_ch11.pdf]%
  }{}%
\fi
\newcommand{\Appref}[1]{\hyperref[#1]{Appendix~\ref*{#1}}}

\KOMAoptions{headings=small}
\setkomafont{disposition}{\normalfont\bfseries}
\setkomafont{chapter}{\normalfont\huge\bfseries\color{bookblue}}
\setkomafont{section}{\normalfont\large\bfseries\color{bookred}}
\setkomafont{subsection}{\normalfont\normalsize\bfseries}

\RedeclareSectionCommand[
  beforeskip=-1.7\baselineskip,
  afterskip=1.05\baselineskip
]{chapter}
\RedeclareSectionCommand[
  beforeskip=1.02\baselineskip,
  afterskip=0.46\baselineskip
]{section}

\clearpairofpagestyles
\automark[chapter]{chapter}
\ohead{\small\normalfont\bfseries\pagemark}
\setlist[enumerate]{leftmargin=2.0em,itemsep=0.28\baselineskip,topsep=0.32\baselineskip}
\setlist[enumerate,1]{label=(\roman*)}
\newcommand{\bookdescriptionlabelfont}{\normalfont\scshape\color{bookblue!85!black}}
\setlist[description]{
  leftmargin=0pt,
  labelsep=0.65em,
  style=unboxed,
  font=\bookdescriptionlabelfont,
  itemsep=0.28\baselineskip,
  topsep=0.32\baselineskip
}
\numberwithin{equation}{chapter}
\DeclareTOCStyleEntry[numwidth=3.1em]{tocline}{section}
\DeclareTOCStyleEntry[indent=2.1em,numwidth=4.0em]{tocline}{subsection}
\makeatletter
\BeforeClosingMainAux{\let\scr@dte@donumwidth\@empty}
\makeatother

\newcounter{definition}[section]
\renewcommand{\thedefinition}{\thesection.\arabic{definition}}
\NewDocumentEnvironment{definition}{o}{%
  \refstepcounter{definition}%
  \begin{tcolorbox}[
    enhanced,
    breakable,
    colback=white,
    colframe=bookgreen!92!black,
    boxrule=0.72pt,
    arc=5pt,
    boxsep=1pt,
    left=1.0em,
    right=0.92em,
    top=0.66em,
    bottom=0.66em,
    before skip=0.74\baselineskip,
    after skip=0.78\baselineskip
  ]%
  \noindent\mbox{\color{bookred}\scshape Definition~\thedefinition%
  \IfNoValueF{#1}{~(#1)}.}\quad
}{%
  \end{tcolorbox}%
}

\newcounter{example}[section]
\renewcommand{\theexample}{\thesection.\arabic{example}}
\NewDocumentEnvironment{example}{o}{%
  \refstepcounter{example}%
  \par\addvspace{0.34\baselineskip}%
  \noindent{\color{bookred}\scshape Example~\theexample%
  \IfNoValueF{#1}{~(#1)}.}\quad
}{%
  \par\vspace{0.16\baselineskip}%
  \noindent{\color{black!72}\rule{\linewidth}{0.42pt}}%
  \par\addvspace{0.36\baselineskip}%
}

\newcounter{exercise}[section]
\renewcommand{\theexercise}{\thesection-E\arabic{exercise}}
\NewDocumentEnvironment{exercise}{o}{%
  \refstepcounter{exercise}%
  \par\medskip
  \noindent{\color{bookred!82!black}\scshape Exercise~\theexercise%
  \IfNoValueF{#1}{~(#1)}.}\quad
}{%
  \par\medskip
}

\NewDocumentEnvironment{realdatacapsule}{m}{%
  \refstepcounter{realdatacapsule}%
  \addcontentsline{rdc}{realdatacapsule}{\protect\numberline{\therealdatacapsule}#1}%
  \begin{tcolorbox}[
    enhanced,
    breakable,
    colback=noteback,
    colframe=bookteal!85!black,
    boxrule=0.62pt,
    arc=4pt,
    boxsep=1pt,
    left=0.96em,
    right=0.9em,
    top=0.62em,
    bottom=0.62em,
    before skip=0.68\baselineskip,
    after skip=0.72\baselineskip
  ]%
  \noindent{\color{bookteal!72!black}\bfseries Real-data capsule~\therealdatacapsule: #1}\par\smallskip
  \begin{description}[
    leftmargin=0pt,
    labelsep=0.55em,
    style=unboxed,
    font=\bookdescriptionlabelfont,
    itemsep=0.16\baselineskip,
    topsep=0.12\baselineskip
  ]%
}{%
  \end{description}%
  \end{tcolorbox}%
}

\newcounter{realdatacapsule}[chapter]
\renewcommand{\therealdatacapsule}{\thechapter.\arabic{realdatacapsule}}

\makeatletter
\newcommand{\listofrealdatacapsules}{%
  \chapter*{Real-Data Capsules and Case Studies}%
  \markboth{Real-Data Capsules and Case Studies}{Real-Data Capsules and Case Studies}%
  \begingroup
  \small
  \setlength{\parskip}{0pt}%
  \@starttoc{rdc}%
  \endgroup
}
\newcommand*{\l@realdatacapsule}{\@dottedtocline{1}{0em}{4.8em}}
\makeatother

\newcounter{theoremlike}[section]
\renewcommand{\thetheoremlike}{\thesection.\arabic{theoremlike}}

\NewDocumentCommand{\bookresultentry}{m m}{}

\NewDocumentEnvironment{proposition}{o}{%
  \refstepcounter{theoremlike}%
  \bookresultentry{Proposition}{#1}%
  \begin{tcolorbox}[
    enhanced,
    breakable,
    colback=theoremback,
    colframe=bookblue!92!black,
    boxrule=0.74pt,
    arc=5pt,
    boxsep=1pt,
    left=1.0em,
    right=0.92em,
    top=0.68em,
    bottom=0.68em,
    before skip=0.76\baselineskip,
    after skip=0.82\baselineskip
  ]%
  \noindent{\color{bookblue}\bfseries Proposition~\thetheoremlike%
  \IfNoValueF{#1}{~(#1)}.}\quad\itshape
}{%
  \end{tcolorbox}%
}

\NewDocumentEnvironment{theorem}{o}{%
  \refstepcounter{theoremlike}%
  \bookresultentry{Theorem}{#1}%
  \begin{tcolorbox}[
    enhanced,
    breakable,
    colback=theoremback,
    colframe=bookblue!92!black,
    boxrule=0.78pt,
    arc=5pt,
    boxsep=1pt,
    left=1.04em,
    right=0.96em,
    top=0.70em,
    bottom=0.70em,
    before skip=0.80\baselineskip,
    after skip=0.86\baselineskip
  ]%
  \noindent{\color{bookblue}\bfseries Theorem~\thetheoremlike%
  \IfNoValueF{#1}{~(#1)}.}\quad\itshape
}{%
  \end{tcolorbox}%
}

\NewDocumentEnvironment{lemma}{o}{%
  \refstepcounter{theoremlike}%
  \bookresultentry{Lemma}{#1}%
  \begin{tcolorbox}[
    enhanced,
    breakable,
    colback=theoremback,
    colframe=bookblue!88!black,
    boxrule=0.68pt,
    arc=5pt,
    boxsep=1pt,
    left=0.96em,
    right=0.9em,
    top=0.62em,
    bottom=0.62em,
    before skip=0.70\baselineskip,
    after skip=0.76\baselineskip
  ]%
  \noindent{\color{bookblue}\bfseries Lemma~\thetheoremlike%
  \IfNoValueF{#1}{~(#1)}.}\quad\itshape
}{%
  \end{tcolorbox}%
}

\NewDocumentEnvironment{corollary}{o}{%
  \refstepcounter{theoremlike}%
  \bookresultentry{Corollary}{#1}%
  \begin{tcolorbox}[
    enhanced,
    breakable,
    colback=theoremback,
    colframe=bookblue!88!black,
    boxrule=0.68pt,
    arc=5pt,
    boxsep=1pt,
    left=0.96em,
    right=0.9em,
    top=0.62em,
    bottom=0.62em,
    before skip=0.70\baselineskip,
    after skip=0.76\baselineskip
  ]%
  \noindent{\color{bookblue}\bfseries Corollary~\thetheoremlike%
  \IfNoValueF{#1}{~(#1)}.}\quad\itshape
}{%
  \end{tcolorbox}%
}

\newcommand{\Pow}{\operatorname{Pow}}
\newcommand{\fieldF}{\mathcal{F}}
\newcommand{\classS}{\mathcal{S}}
\newcommand{\classG}{\mathcal{G}}
\newcommand{\classR}{\mathcal{R}}
\newcommand{\classC}{\mathcal{C}}
\providecommand{\Borel}{\mathcal{B}}
\providecommand{\Nat}{\mathbb{N}}
\providecommand{\Int}{\mathbb{Z}}
\providecommand{\Rat}{\mathbb{Q}}
\providecommand{\Real}{\mathbb{R}}
\providecommand{\Hilbert}{\mathbb{H}}
\providecommand{\R}{\mathbb{R}}
\providecommand{\Rplus}{\mathbb{R}_{+}}
\providecommand{\Prob}{\mathbb{P}}
\providecommand{\Expect}{\mathbb{E}}
\providecommand{\Var}{\operatorname{Var}}
\providecommand{\Cov}{\operatorname{Cov}}
\providecommand{\Corr}{\operatorname{Corr}}
\providecommand{\Normal}{\mathcal{N}}
\providecommand{\Poisson}{\operatorname{Poisson}}
\providecommand{\Bernoulli}{\operatorname{Bernoulli}}
\providecommand{\Binomial}{\operatorname{Binomial}}
\providecommand{\NegBin}{\operatorname{NegBin}}
\providecommand{\Unif}{\operatorname{Unif}}
\providecommand{\Dirichlet}{\operatorname{Dirichlet}}
\providecommand{\DP}{\operatorname{DP}}
\providecommand{\Law}{\operatorname{Law}}
\providecommand{\argmin}{\operatorname*{arg\,min}}
\providecommand{\argmax}{\operatorname*{arg\,max}}
\providecommand{\graph}{\operatorname{graph}}

\providecommand{\diam}{\operatorname{diam}}
\providecommand{\Beta}{\operatorname{Beta}}
\providecommand{\outerProb}{\Prob^{*}}
\providecommand{\innerProb}{\Prob_{*}}
\providecommand{\outerExpect}{\Expect^{*}}
\providecommand{\innerExpect}{\Expect_{*}}

\providecommand{\sign}{\operatorname{sign}}
\providecommand{\diag}{\operatorname{diag}}
\providecommand{\tr}{\operatorname{tr}}
\providecommand{\matA}{\mathbf{A}}

\providecommand{\matI}{\mathbf{I}}
\providecommand{\matP}{\mathbf{P}}
\providecommand{\matQ}{\mathbf{Q}}

\providecommand{\logit}{\operatorname{logit}}
\providecommand{\IF}{\operatorname{IF}}

\newcommand{\toP}{\xrightarrow{\Prob}}
\newcommand{\toPstar}{\xrightarrow{\Prob^{*}}}
\providecommand{\opstar}{o_{\Prob}^{*}}

\newcommand{\weakto}{\Rightarrow}
\providecommand{\norm}[1]{\lVert#1\rVert}
\providecommand{\indep}{\mathrel{\perp\!\!\!\perp}}
\providecommand{\ind}[1]{\mathbf{1}\{#1\}}
\providecommand{\indset}[1]{\mathbf{1}_{#1}}
\newcommand{\qedmark}{\ifmmode\quad\square\else\hfill$\square$\fi}

\newif\ifprintbibliography
\printbibliographyfalse
\newif\ifbookbackmatter
\bookbackmatterfalse
\newif\ifprintindexes
\printindexesfalse

\newcommand{\conceptindex}[1]{\index[subject]{#1|hyperpage}}
\newcommand{\personindex}[1]{\index[person]{#1|hyperpage}}
\ExplSyntaxOn
\NewDocumentCommand{\conceptindexes}{m}{%
  \clist_map_inline:nn {#1} {\conceptindex{##1}}%
}
\NewDocumentCommand{\personindexes}{m}{%
  \clist_map_inline:nn {#1} {\personindex{##1}}%
}
\ExplSyntaxOff
\newcommand{\term}[2][]{%
  \if\relax\detokenize{#1}\relax
    #2\conceptindex{#2}%
  \else
    #1\conceptindex{#2}%
  \fi
}
\newcommand{\person}[2][]{%
  \if\relax\detokenize{#1}\relax
    #2\personindex{#2}%
  \else
    #1\personindex{#2}%
  \fi
}
\newcommand{\bookmanualfigure}[2]{%
  \refstepcounter{figure}\label{#1}%
  \addcontentsline{lof}{figure}{\protect\numberline{\thefigure}#2}%
}

\newcommand{\bookfrontquote}{%
  \cleardoublepage
  \thispagestyle{empty}%
  \vspace*{0.31\textheight}%
  \begin{flushright}
  {\large\itshape ``'Tis the good reader that makes the good book.''\par}
  \vspace{0.85\baselineskip}
  {\normalfont ---Ralph Waldo Emerson, \emph{Society and Solitude}\par}
  \end{flushright}
  \cleardoublepage
}

\newcommand{\bookcoverpage}{%
  \cleardoublepage
  \thispagestyle{empty}%
  \phantomsection
  \pdfbookmark[0]{Cover}{cover}%
  \begin{tikzpicture}[remember picture,overlay]
    \node[inner sep=0pt] at (current page.center)
      {\includegraphics[width=\paperwidth,height=\paperheight]{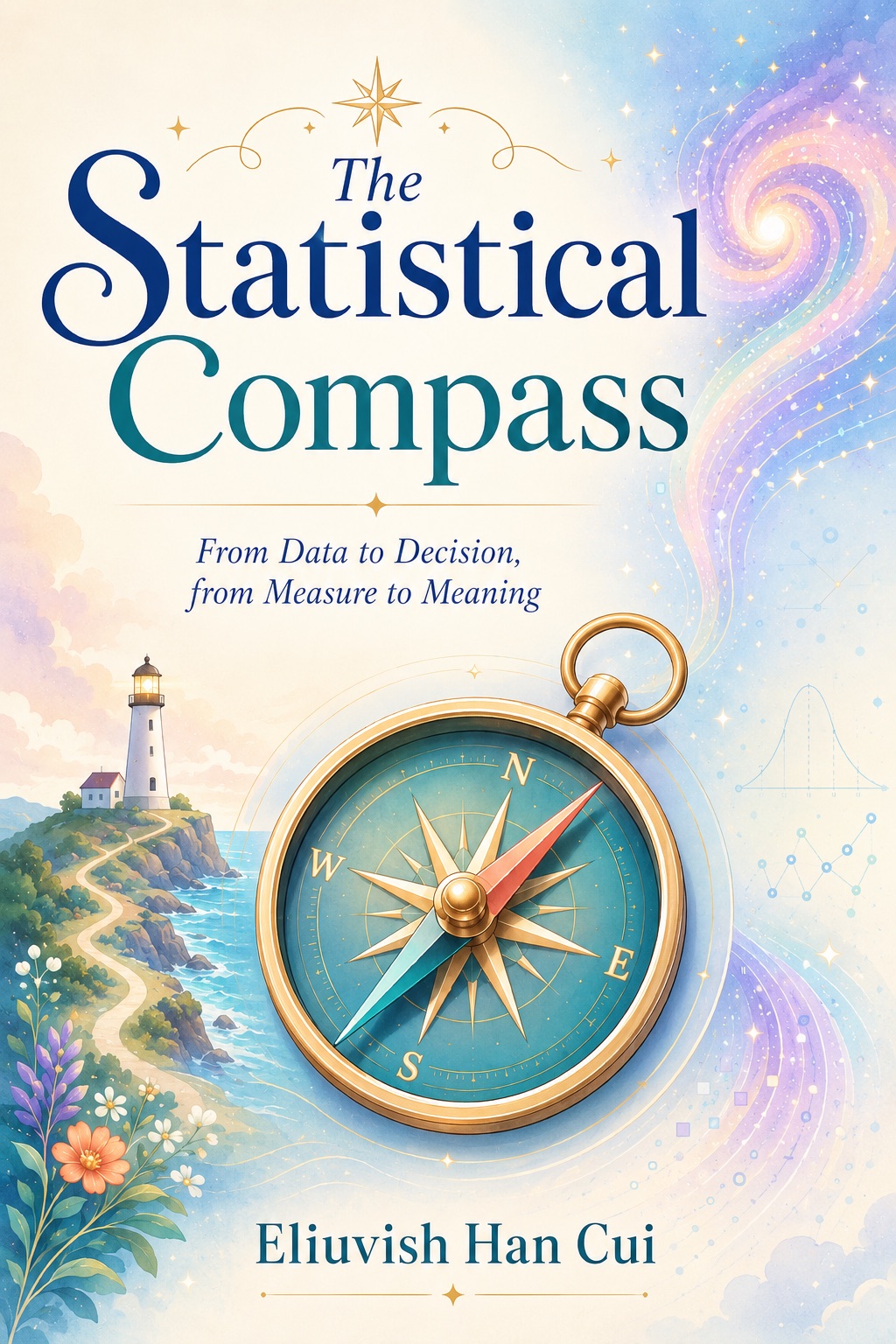}};
  \end{tikzpicture}%
  \null
  \cleardoublepage
}

\newcommand{\booktitlepage}{%
  \cleardoublepage
  \thispagestyle{empty}%
  \phantomsection
  \pdfbookmark[0]{Title Page}{titlepage}%
  \vspace*{0.18\textheight}%
  \begin{center}
    {\color{bookblue}\fontsize{30}{36}\selectfont\bfseries \BookTitle\par}%
    \vspace{0.7\baselineskip}%
    {\color{bookteal!72!black}\Large\itshape \BookSubtitle\par}%
    \vspace{4.2\baselineskip}%
    {\Large \BookAuthor\par}%
    \vspace{0.55\baselineskip}%
    {\normalsize \BookAffiliation\par}%
  \end{center}%
  \cleardoublepage
}

\newcommand{\bookpart}[3]{%
  \cleardoublepage
  \phantomsection
  \refstepcounter{part}%
  \addcontentsline{toc}{part}{Part~\thepart.\ #1}%
  \markboth{Part~\thepart.\ #1}{Part~\thepart.\ #1}%
  \thispagestyle{empty}%
  \vspace*{0.16\textheight}%
  {\noindent\color{bookred}\Large\bfseries Part~\thepart\par}%
  \vspace{0.75\baselineskip}%
  {\noindent\color{bookblue}\Huge\bfseries #1\par}%
  \vspace{0.75\baselineskip}%
  {\noindent\large\itshape #2\par}%
  \vspace{1.35\baselineskip}%
  \begin{tcolorbox}[
    enhanced,
    breakable,
    colback=chaptercream,
    colframe=bookblue!80!black,
    boxrule=0.7pt,
    arc=5pt,
    boxsep=1pt,
    left=1.05em,
    right=1.0em,
    top=0.82em,
    bottom=0.84em
  ]
  \noindent\textbf{Route through this part.}\quad #3
  \end{tcolorbox}%
  \cleardoublepage
}

\newcommand{\bookappendices}{%
  \cleardoublepage
  \phantomsection
  \addcontentsline{toc}{part}{Appendices. Mathematical and Biomedical Toolkits}%
  \markboth{Appendices}{Appendices}%
  \thispagestyle{empty}%
  \vspace*{0.18\textheight}%
  {\noindent\color{bookred}\Large\bfseries Appendices\par}%
  \vspace{0.75\baselineskip}%
  {\noindent\color{bookblue}\Huge\bfseries Mathematical and Biomedical Toolkits\par}%
  \vspace{0.75\baselineskip}%
  {\noindent\large\itshape Set language, measure construction, mathematical infrastructure, and computational translation. \par}%
  \vspace{1.35\baselineskip}%
  \begin{tcolorbox}[
    enhanced,
    breakable,
    colback=chaptercream,
    colframe=bookblue!80!black,
    boxrule=0.7pt,
    arc=5pt,
    boxsep=1pt,
    left=1.05em,
    right=1.0em,
    top=0.82em,
    bottom=0.84em
  ]
  \noindent\textbf{How to use these appendices.}\quad These chapters are not a sixth part of the main route.  They are reusable toolkits for set-theoretic grammar, measure-theoretic construction, mathematical backbone results, and computational or biomedical translation.
  \end{tcolorbox}%
  \cleardoublepage
}

\newcommand{\bookbackmatter}{%
  \cleardoublepage
  \backmatter
  \bookmarksetup{startatroot}%
}

\newcommand{\bookbibliography}{%
  \cleardoublepage
  \printbibliography[heading=bibintoc,title={Bibliography}]%
}

\newcommand{\bookindexes}{%
  \ifprintindexes
    \cleardoublepage
    \phantomsection
    \chapter*{Index}%
    \addcontentsline{toc}{chapter}{Index}%
    \markboth{Index}{Index}%
    \printindex[person]%
    \clearpage
    \printindex[subject]%
  \fi
}

\begin{document}

\ifdefined\CHAPTERONE
\printbibliographytrue
\setcounter{chapter}{0}
\input{chapters/ch01_introduction}
\else
\ifdefined\CHAPTERTWO
\printbibliographytrue
\setcounter{chapter}{1}
\input{chapters/ch02_randomness}
\else
\ifdefined\CHAPTERFOUR
\printbibliographytrue
\setcounter{chapter}{3}
\input{chapters/ch04_design_data_collection}
\else
\ifdefined\CHAPTERFIVE
\printbibliographytrue
\setcounter{chapter}{4}
\input{chapters/ch05_observation_likelihood}
\else
\ifdefined\CHAPTERTEN
\printbibliographytrue
\setcounter{chapter}{9}
\input{chapters/ch10_weak_convergence_random_objects}
\else
\ifdefined\CHAPTERELEVEN
\printbibliographytrue
\setcounter{chapter}{10}
\input{chapters/ch11_uniform_laws_empirical_processes}
\else
\ifdefined\CHAPTERTWELVE
\printbibliographytrue
\setcounter{chapter}{11}
\input{chapters/ch12_karhunen_loeve_functional_data}
\else
\ifdefined\CHAPTERFOURTEEN
\printbibliographytrue
\setcounter{chapter}{13}
\input{chapters/ch14_testing_hellinger_local_asymptotics}
\else
\ifdefined\CHAPTERFIFTEEN
\printbibliographytrue
\setcounter{chapter}{14}
  \input{chapters/ch15_local_approximation_influence}
\else
\ifdefined\CHAPTERSIXTEEN
\printbibliographytrue
\setcounter{chapter}{15}
  \input{chapters/ch16_processes_continuous_time}
\else
\ifdefined\CHAPTERSEVENTEEN
\printbibliographytrue
\setcounter{chapter}{16}
  \input{chapters/ch17_future_statistics}
\else
\ifdefined\CHAPTERTHIRTEEN
\printbibliographytrue
\setcounter{chapter}{12}
\input{chapters/ch13_m_z_estimation}
\else
\ifdefined\CHAPTERTHREE
\printbibliographytrue
\setcounter{chapter}{2}
\input{chapters/ch03_probability_measure}
\else
\ifdefined\CHAPTERSIX
\printbibliographytrue
\setcounter{chapter}{5}
\input{chapters/ch06_product_spaces_processes}
\else
\ifdefined\CHAPTERSEVEN
\printbibliographytrue
\setcounter{chapter}{6}
\input{chapters/ch08_common_grammar_modern_data_structures}
\else
\ifdefined\CHAPTEREIGHT
\printbibliographytrue
\setcounter{chapter}{7}
\input{chapters/ch09_data_structures_to_targets}
\else
\ifdefined\CHAPTERNINE
\printbibliographytrue
\setcounter{chapter}{8}
\input{chapters/ch07_laws_concentration}
\else
\ifdefined\APPENDICES
\printbibliographytrue
\bookappendices
\appendix
\input{appendices/set_theory_compass}
\input{appendices/measure_theoretic_toolkit}
\input{appendices/mathematical_backbone}
\input{appendices/computational_biomedical_translation}
\else
\printbibliographytrue
\bookbackmattertrue
\printindexestrue
\bookcoverpage
\booktitlepage
\bookfrontquote
\input{frontmatter/preface}
\pdfbookmark[0]{Contents}{toc}
\tableofcontents
\cleardoublepage
\pdfbookmark[0]{List of Figures}{lof}
\listoffigures
\cleardoublepage
\pdfbookmark[0]{List of Tables}{lot}
\listoftables
\cleardoublepage
\pdfbookmark[0]{Real-Data Capsules and Case Studies}{rdc}
\listofrealdatacapsules
\cleardoublepage
\input{frontmatter/notation}
\bookpart{Examples Before Models}
  {Chapters~1--2: begin with translation, examples, and the questions that force formal language.}
  {The book starts before technique. Chapter~1 states the compass; Chapter~2
  turns examples into the problem bank that makes later formal language
  necessary.}
\setcounter{chapter}{0}
\input{chapters/ch01_introduction}
\cleardoublepage
\setcounter{chapter}{1}
\input{chapters/ch02_randomness}
\bookpart{Observation and Probability Objects}
  {Chapters~3--6: turn observed records into measurable probability objects.}
  {Events say what can be seen; design decides what records can exist;
  likelihood reads observed records; products and kernels show how tables,
  histories, fields, and processes are built as laws.}
\setcounter{chapter}{2}
\input{chapters/ch03_probability_measure}
\cleardoublepage
\setcounter{chapter}{3}
\input{chapters/ch04_design_data_collection}
\cleardoublepage
\setcounter{chapter}{4}
\input{chapters/ch05_observation_likelihood}
\cleardoublepage
\setcounter{chapter}{5}
\input{chapters/ch06_product_spaces_processes}
\bookpart{Data Objects and Targets}
  {Chapters~7--8: turn observed structures into named statistical claims.}
  {With observation grammar in place, the book pauses before asymptotics:
  first name the empirical object, then name the target it is meant to support.}
\cleardoublepage
\setcounter{chapter}{6}
\input{chapters/ch08_common_grammar_modern_data_structures}
\cleardoublepage
\setcounter{chapter}{7}
\input{chapters/ch09_data_structures_to_targets}
\bookpart{Stability, Limits, and Representations}
  {Chapters~9--12: move from named targets to stable empirical and random objects.}
  {Laws of large numbers and concentration ask when fixed empirical summaries
  can be trusted; weak convergence turns whole errors into random objects;
  empirical processes control searched fields; functional representation turns
  curves into coordinates.}
\cleardoublepage
\setcounter{chapter}{8}
\input{chapters/ch07_laws_concentration}
\cleardoublepage
\setcounter{chapter}{9}
\input{chapters/ch10_weak_convergence_random_objects}
\cleardoublepage
\setcounter{chapter}{10}
\input{chapters/ch11_uniform_laws_empirical_processes}
\cleardoublepage
\setcounter{chapter}{11}
\input{chapters/ch12_karhunen_loeve_functional_data}
\bookpart{Inference, Local Geometry, and Event Time}
  {Chapters~13--16: turn stable empirical objects into estimates, tests, uncertainty, and time-indexed event histories.}
  {M/Z-estimation locates peaks and roots; testing reads nearby laws through
  local geometry; influence functions and delta methods translate residual
  noise into uncertainty; continuous-time processes make information flow,
  stopping, and event histories explicit.}
\cleardoublepage
\setcounter{chapter}{12}
\input{chapters/ch13_m_z_estimation}
\cleardoublepage
\setcounter{chapter}{13}
\input{chapters/ch14_testing_hellinger_local_asymptotics}
\cleardoublepage
\setcounter{chapter}{14}
\input{chapters/ch15_local_approximation_influence}
\cleardoublepage
\setcounter{chapter}{15}
\input{chapters/ch16_processes_continuous_time}
\bookpart{From Inference to Use}
  {Chapter~17: return statistical claims to prediction, risk, feedback, and deployment.}
  {The final chapter is a closing part rather than another technical block:
  it asks how targets, uncertainty, loss, policy, and feedback behave when
  statistical procedures act inside the systems they measure.}
\cleardoublepage
\setcounter{chapter}{16}
\input{chapters/ch17_future_statistics}
\cleardoublepage
\bookappendices
\appendix

\input{appendices/set_theory_compass}
\input{appendices/measure_theoretic_toolkit}
\input{appendices/mathematical_backbone}
\input{appendices/computational_biomedical_translation}
\fi
\fi
\fi
\fi
\fi
\fi
\fi
\fi
\fi
\fi
\fi
\fi
\fi
\fi
\fi
\fi
\fi
\fi

\ifprintbibliography
\ifbookbackmatter
\bookbackmatter
\fi
\bookbibliography
\ifbookbackmatter
\bookindexes
\fi
\fi

\end{document}

%% file: chapters/ch01_introduction.tex
\chapter{Introduction: Statistics as Translation}
\label{chap:introduction}
\conceptindexes{statistical compass, statistical translation, recurring ledger, measurement, data structures, assumptions, targets, procedures, uncertainty, use}

\begin{flushright}
\small\itshape
A journey of a thousand miles begins beneath one's feet.\\
\normalfont ---Laozi, \emph{Dao De Jing}, Chapter~64 \citep{laozi1963tao}
\end{flushright}

\begin{tcolorbox}[
  enhanced,
  breakable,
  colback=chaptercream,
  colframe=bookblue!88!black,
  boxrule=0.72pt,
  arc=5pt,
  boxsep=1pt,
  left=1.0em,
  right=0.95em,
  top=0.82em,
  bottom=0.82em,
  before skip=0.55\baselineskip,
  after skip=1.0\baselineskip
]
\noindent\textbf{Book charter.}
This opening chapter states the book's route before the technical machinery
arrives. It introduces the statistical compass and the recurring ledger:
measurement, data structure, assumptions, target, procedure, uncertainty, and
use. Measure theory, products, stochastic processes, asymptotics, and
continuous-time martingales are introduced later as the grammar needed to keep
that ledger honest.
\end{tcolorbox}

As Kai Lai Chung memorably insisted in the preface to
\emph{A Course in Probability Theory}, a course is ``not a stockpile of raw
materials nor a random selection of vignettes''; it should guide the reader
through a field with a sustained route and a point of view
\citep[preface]{chung1974course}.
This book accepts that charge. It is not arranged as a shelf of methods; its
route begins by asking how questions become measurements, how measurements
become data structures, and how models carry claims back to the world.

Statistics is often introduced as a cabinet of techniques. A student first
meets means, variances, confidence intervals, regression lines, likelihoods,
tests, classifiers, and prediction rules. Later come measure theory,
asymptotics, stochastic processes, Bayesian computation, causal diagrams,
high-dimensional models, and machine learning. The curriculum is powerful, but
it can feel like entering a workshop after all the tools have been arranged in
drawers. The tools are here; what is missing is the story of the work.

This book begins with that story. Before there is a model, there is a data
structure. Before there is a data structure, there is a reason why some part of
the world was observed, measured, recorded, filtered, stored, and transformed.
A row in a biomedical table may have begun as a patient visit, a biopsy, a cell
captured in a droplet, an image taken under a protocol, or a molecular read
that survived laboratory and computational choices. By the time the row reaches
the statistician, it looks quiet. It is not quiet. It is already a processed
record.

An old Western warning says that ``the map is not the territory''
\citep{korzybski1933science}. Statistics lives inside that warning. A model is
a map, but a good map is not a lie. It tells a traveler where the mountains
are likely to be, which paths are impossible, where the fog is thick, and what
would have to be true for the journey to continue. A statistical analysis is
therefore not just a calculation. It is an argument about which features of a
world may be preserved in a simplified representation.

The same idea can be drawn as a compass. Statistical work moves from a
question in the world to measurement, from measurement to data structure, from
data structure to model, and from model to action. Then the action sends the analysis
back to the world with sharper questions. The compass is simple, but almost
the entire technical architecture of the book is a refinement of one of its
arrows.

\input{figures/ch01_statistical_compass}

The story behind a dataset has at least three layers. First, how were the data
generated? Chemistry data arise from assays, compounds, instruments, batches,
reactions, and measurement limits. Tumor efficacy data arise from patients,
lesions, imaging schedules, censoring, treatment protocols, and clinical
endpoints. Single-cell RNA sequencing data arise from tissues, dissociation
protocols, cells, reads, genes, batch effects, and biological heterogeneity.
ImageNet data arise from photographs, labels, taxonomies, human annotators,
and visual variation. Videos, human resource records, recommender-system logs,
and spatial records of wartime bomb impacts each carry their own grammar of
observation.

Landmark datasets make the point concrete. The Netflix Prize turned movie
ratings into a problem about sparse user--item matrices, privacy, and
recommendation under missingness \citep{bennett2007netflix}. ImageNet and
Microsoft Common Objects in Context (MS COCO) turned photographs into
taxonomies, bounding boxes, segmentations, and crowd-labeled visual context
\citep{deng2009imagenet,lin2014microsoft}. The Alzheimer's Disease
Neuroimaging Initiative (ADNI) and the Anti-Amyloid Treatment in Asymptomatic
Alzheimer's Disease (A4) solanezumab trial turned aging and Alzheimer's disease
into linked clinical, imaging, biomarker, and longitudinal cognitive records
\citep{petersen2010adni,sperling2023solanezumab}. Real-world studies and
pragmatic clinical trials turned routine care into a contested source of
evidence, where eligibility, treatment choice, adherence, follow-up, and
endpoint capture must be made explicit before a clinical comparison can be
trusted \citep{schwartz1967explanatory,zwarenstein2008consort,
loudon2015precis2,hernan2016targettrial,fda2018rweframework}. RiskMetrics and
stock-return series turned market risk into time-indexed covariances, tail
events, and portfolio decisions
\citep{jpmorgan1996riskmetrics,fama1965behavior}. Political polls and votes
turn opinions, turnout, aggregation, and time into a moving inferential target
\citep{gelman1993polls}. These examples are not impressive because they are
large. They are instructive because each one forces a different answer to the
question: what, exactly, is the statistical object?

Second, why were the data collected? A biologist may ask which gene affects a
phenotype. A clinician may ask whether a therapy delays progression. A company
may ask which movie should be recommended next. A city may ask whether
accidents cluster near a particular intersection. A historian or military
analyst may ask whether impacts in a city are compatible with spatial
randomness. The same mathematical object, such as a table, an image, a graph,
or a point pattern, can mean different things depending on the question that
called it into existence.

Third, who needs the data and what must be done with them? Some data support
explanation, some support prediction, some support decision making, and some
support exploration. Sometimes the goal is to estimate a parameter. Sometimes
it is to discover a latent structure, compare interventions, rank alternatives,
compress information, detect anomalies, or produce a calibrated uncertainty
statement. The task determines what counts as a successful analysis. A
beautiful estimator may be useless for a decision problem; a highly predictive
black-box model may be unsatisfactory for a scientific question that demands a
mechanism. This is the first habit the book tries to teach: never ask what
method to use before asking what kind of object the data are and what kind of
claim the analysis must support.

\section{The Core View: Statistics as Translation}
\conceptindexes{statistical translation, questions, measurement, data structures, models, action, statistical compass}

The central act of statistics is translation, but translation is not
transcription. We translate a concrete problem into mathematical language, work
inside that language, and translate the result back into the original context.
Something is always carried across, and something is always left behind. The
art is to know which loss is harmless, which loss is dangerous, and which loss
is the price of making a question answerable at all.

The word ``assumption'' is the hinge of this process. Assumptions are sometimes
treated as technical decorations attached to theorems. In practice, they are
the places where the world enters mathematics. Independence, exchangeability,
linearity, smoothness, sparsity, stationarity, positivity, missing at random,
ignorability, proportional hazards, Markovian dynamics, and low-dimensional
structure are not merely conditions in a proof. They are claims, sometimes
bold and sometimes modest, about how a phenomenon can be simplified without
losing the feature that matters for the problem at hand.

This is why statistics sits between mathematics and reality in a distinctive
way. Mathematics asks what follows from precise premises. Science asks how the
world behaves. Statistics asks which premises are useful, defensible, and
informative when the world has been observed through noisy, partial, and often
biased data. The beauty of statistics is not only in elegant formulas. It is in
the disciplined movement from a messy question to a clear representation, and
then back again to a statement that someone can interpret, criticize, and use.

\section{Data Structures Before Models}
\conceptindexes{data structures, models!before data structures, observation processes, ImageNet, MS COCO, ADNI, RiskMetrics, real-world evidence, recommender systems}

The first discipline is to look at the data before reaching for a model. This
does not mean drawing a few plots as a ritual warm-up. It means asking what
kind of object has been produced and what claim that object can reasonably
support. Tukey's program of data analysis was radical precisely because it
treated looking, describing, and diagnosing as intellectual work, not as
clerical preparation for ``real'' inference
\citep{tukey1962future,tukey1977exploratory}. Box's warning that scientific
models are useful approximations gives the same lesson from the modeling side:
the first question is not whether a model is elegant, but what it is an
approximation to \citep{box1976science}.

Consider a familiar phrase: ``patient data.'' It sounds like one object, but
it is many. A flat table of baseline covariates asks about units, columns,
missingness, confounding, and clustering. A path through diagnosis, treatment,
response, toxicity, and relapse asks about order and transition. A radiology
image asks about spatial locality, scale, annotation, and acquisition
protocol. A gene-expression matrix asks about high dimension, sparsity,
multiplicity, batch effects, and biological interpretation. A follow-up time
asks about censoring, truncation, and risk sets. An electronic health record
extract asks a different question again: which parts of illness became care,
which parts of care became codes, and which codes became analyzable variables?
The model has not yet been chosen, but the inferential question has already
changed.

The following list is therefore not a data-type encyclopedia. It is a set of
constitutional examples. Each structure illustrates the same rule: the form of
the data suggests certain assumptions, makes other assumptions dangerous, and
changes what a model is allowed to mean.

Common structures therefore have statistical personalities:
\begin{description}[leftmargin=0pt,labelsep=0.55em,style=unboxed,font=\bookdescriptionlabelfont,itemsep=0.28\baselineskip]
\item[Tables.]
Rows and columns make units and covariates visible, but they also hide
clustering, repeated measurement, missingness, and the process by which rows
entered the file. Missing-data assumptions are therefore part of the data
structure, not a later nuisance \citep{rubin1976inference}.

\item[Clinical records.]
Clinical datasets are often tables, but they are not just tables. Electronic
health records and claims data are traces of encounters, documentation habits,
billing incentives, coding systems, medication orders, laboratory calendars,
and linkage rules. Their difficulty is not only censoring in time; it is that
clinical facts reach the dataset through care-seeking and record-making
processes \citep{weiskopf2013methods,goldstein2017opportunities}.

\item[Real-world and pragmatic clinical evidence.]
Real-world studies (RWS) and pragmatic clinical trials (PCTs) sit at the
boundary between designed evidence and recorded evidence. A PCT embeds a
randomized comparison into ordinary care; an RWS asks whether routine records
can be disciplined enough to answer a trial-like question. Their statistical
personality is therefore not ``messy clinical data'' in general, but the
alignment among eligibility, treatment strategy, time zero, follow-up, outcome,
and estimand \citep{schwartz1967explanatory,hernan2016targettrial}.

\item[Paths and sequences.]
Order carries information. Once transitions matter, independence of rows is no
longer the natural starting point; filtrations, Markov structure, stopping
rules, and feedback may become part of the statistical object.

\item[Images.]
Nearby pixels or voxels are usually related, but locality is shaped by optics,
scanner protocol, preprocessing, segmentation, and annotation. The spatial
regularity that helps modeling can also carry acquisition bias
\citep{besag1974spatial}.

\item[Text.]
Text must be tokenized, counted, embedded, or otherwise encoded before it can
be analyzed. Each encoding preserves some structure and discards some context;
Shannon's communication theory is a useful reminder that representation and
information are already statistical choices \citep{shannon1948mathematical}.

\item[Networks.]
Dependence is not noise around the observation; it is often the observation.
Edges, paths, communities, and exposure through neighbors define the object of
study, so independent-unit reasoning can be badly misleading
\citep{wasserman1994social}.

\item[Historical records and regions.]
Archives, gazetteers, prices, administrative reports, and local chronologies
are not raw windows into the past.  They are institutional traces.  A
historical ``region'' may be an administrative unit, a market basin, a transport
network, or a latent economic system; changing that definition changes the
statistical object.  Skinner's macroregional view of China is therefore useful
in this book not as background color, but as an example of how geography,
markets, records, and spatial dependence jointly create data structure
\citep{skinner1964marketing,skinner1977regional}.

\item[Omics.]
Modern molecular assays produce high-dimensional, sparse, and often
zero-inflated measurements with heavy multiplicity and batch effects. The
statistical question is inseparable from the assay and preprocessing pipeline,
as the RNA-seq and single-cell literature repeatedly shows
\citep{li2011sparse,li2015statistics,li2018scimpute}.

\item[Distributional objects.]
Sometimes the observation is not a number or vector but a distribution: a
histogram of patient-level measurements, a cell-type composition, a density of
waiting times, or an empirical distribution attached to a hospital, city, or
experimental condition. Then the geometry of the sample space becomes part of
the assumption. Distribution-to-distribution and Wasserstein regression belong
to this family of random-object problems
\citep{petersen2019frechet,chen2023wasserstein}.

\item[Event-time data.]
An observed time is rarely simply a true time. Delayed entry, right censoring,
competing events, inspection schedules, and treatment switching determine who
is at risk and when comparisons are valid
\citep{cox1972regression,andersen1982cox}.
\end{description}

Data are also made, not found. A dataset has a provenance. Someone or something
entered it; some variable definitions were chosen; some measurements were
rounded, discretized, coded, or transformed; some records were filtered away;
some information was compressed by storage formats, privacy rules, laboratory
pipelines, or business logic. This is why selection bias in large data can
overwhelm sample size \citep{meng2018statistical}, why modern datasets benefit
from explicit documentation \citep{gebru2021datasheets}, and why philosophers
of data-intensive biology emphasize the material and institutional work that
turns biological traces into usable data \citep{leonelli2016data}.

\begin{example}[Rao's mournful numbers]
C.~R. Rao uses a small public-risk table to make this point almost painfully
clear.  Reproducing figures from Cohen and Lee's catalog of risks, he lists
estimated losses or gains in life expectancy, measured in days, for familiar
conditions and actions \citep[Ch.~6, pp.~166--167]{rao1997statisticsTruth};
the original catalog is \citet{cohenLee1979catalog}.  A few entries already
show the trap: being unmarried
for males is listed as \(3500\) days, being left-handed as \(3285\) days,
male cigarette smoking as \(2250\) days, being \(30\%\) overweight as \(1300\)
days, a smoke alarm in the house as \(-10\) days, and mobile coronary-care
units as \(-125\) days.  In the original table a negative number means an
increase in life expectancy.

The statistical question is not whether these numbers are interesting.  They
are.  The question is what kind of translation they support.  The entries are
contrasts among groups, built from records in which sex, marital status,
handedness, exposure, and age at death have already been encoded.  They do not
say that marriage adds ten years to every man, that left-handedness is a
biological death sentence, or that a single person can read his future directly
from the table.  Rao's interpretation is exactly the habit this book wants:
an average can summarize a population comparison, but before it becomes advice
for an individual it must pass through subgroup structure, selection,
confounding, mechanism, and purpose.
\end{example}

\begin{example}[Rounded numbers and royal records]
Rao gives an older example in the same spirit: a royal inscription from ancient
Egypt reports the capture of \(120{,}000\) prisoners of war, \(400{,}000\)
oxen, and \(1{,}422{,}000\) goats \citep[Ch.~2]{rao1997statisticsTruth}.  The
figures look authoritative because they are numerical, but the statistical
object is ambiguous.  Were they literal tallies, rounded administrative
records, ceremonial exaggerations, or a mixture of counting and display?  The
lesson is not cynicism about old sources.  It is provenance.  A number carved
into a public record is still a measurement produced by people, incentives,
rounding conventions, memory, and political purpose.  Long before we choose a
probability model, one has to ask what work the number was doing in its own
world.
\end{example}

\begingroup
\small
\setlength{\LTpre}{0.45\baselineskip}
\setlength{\LTpost}{0.45\baselineskip}
\setlength{\tabcolsep}{0.35em}
\renewcommand{\arraystretch}{1.08}
\begin{longtable}{@{}>{\raggedright\arraybackslash}p{0.24\linewidth}
                  >{\raggedright\arraybackslash}p{0.30\linewidth}
                  >{\raggedright\arraybackslash}p{0.36\linewidth}@{}}
\caption{Data structures and the assumptions they tempt}
\label{tab:ch1-data-structures-assumptions}\\
\toprule
\textbf{Data structure} & \textbf{Tempting assumption} & \textbf{Risk} \\
\midrule
\endfirsthead
\caption[]{Data structures and the assumptions they tempt (continued)}\\
\toprule
\textbf{Data structure} & \textbf{Tempting assumption} & \textbf{Risk} \\
\midrule
\endhead
\bottomrule
\endlastfoot
Table & Rows are independent units & Clustered, repeated, or selected rows \\
Public risk summaries & Group averages are individual prescriptions &
Heterogeneity, selection, confounding, and causal overinterpretation \\
Historical counts & Recorded numbers are exact measurements &
Rounding, institutional purpose, display, and survival of records \\
Clinical/EHR records & Codes are direct clinical facts & Care-seeking,
billing, documentation, and linkage mechanisms \\
RWS/PCT evidence & Routine care directly answers the clinical question &
Eligibility, treatment choice, time zero, adherence, and endpoint capture \\
Image & Nearby pixels share signal & Artifacts, scanner bias, annotation error \\
Text & Word counts represent meaning & Context loss, ambiguity, changing usage \\
Network & Nodes are ordinary samples & Edges create dependence and interference \\
Omics & More features mean more discovery & Sparsity, batch effects, multiplicity \\
Distributional object & A mean or vector summary is enough & Geometry,
alignment, support, and within-object sample size \\
Event-time data & Observed time is true time & Censoring, truncation, changing risk sets \\
\end{longtable}
\endgroup

The useful chain is therefore
\[
\text{data structure}\;\longrightarrow\;\text{assumptions}\;\longrightarrow\;\text{model}\;\longrightarrow\;\text{inference task}.
\]
The arrows matter. Likelihoods, priors, loss functions, sampling designs, and
algorithms encode assumptions about generation, regularity, error, comparison,
and approximation. Before modeling, a student should be able to ask six plain
questions:
\begin{enumerate}[itemsep=0.04\baselineskip,topsep=0.12\baselineskip,parsep=0pt,partopsep=0pt]
\item What is the unit: subject, visit, encounter, lesion, cell, pixel, word,
edge, distribution, or event?
\item Where is dependence: cluster, time, space, graph, family, batch, or
feedback loop?
\item How were non-observations, missing values, and filtered records created?
\item Does time or order change the meaning of the observation?
\item Where could measurement error, annotation error, or acquisition bias
enter?
\item Is the target prediction, explanation, comparison, decision, or
exploration?
\end{enumerate}
Only after these questions are visible does the choice of model become a
statistical choice rather than a reflex.

\section{Historical Pressure: Modeling Cultures and Modern Data}
\conceptindexes{two cultures, algorithmic modeling, data modeling, statistical practice, statistical culture, statistical genomics, uniform design}

A second thread of the book takes a historical view of how statisticians handle
data. The shift is from a clean contrast between modeling cultures to the
messier reality of modern scientific data. Breiman famously contrasted two cultures of
statistical modeling: one centered on stochastic data models and one centered
on algorithmic prediction \citep{breiman2001two_cultures}. That contrast
remains useful, but today's students inherit a world in which the boundary is
no longer clean. A modern statistician may use a randomized trial
\citep[Sec.~9]{neyman1923agricultural}, a survival model
\citep{cox1972regression}, a neural network, a Bayesian nonparametric prior
\citep{ferguson1973bayesian}, a causal graph \citep{pearl1995causal}, and a
scalable optimization routine in the same project.

The informal name ``Jessica'' stands here for the contemporary researcher, student, or analyst
who faces real data before having a perfect theory. She may work on
genome-wide association studies, spatial transcriptomics, medical imaging,
clinical trial data, online experiments, or music perception. She does not
merely choose between classical inference and machine learning. She has to
decide what the data mean, which comparison is valid, which uncertainty is
relevant, and which mathematical simplification is faithful enough to the
problem.

The name can also be read more literally as a pointer to Jingyi Jessica Li and
to the contemporary field of statistical genomics. Li's work makes the abstract
``Jessica'' concrete: the data are produced by modern molecular assays, the
scientific questions are biological, and the statistical issues are
foundational. Her papers use sparse linear modeling to address identifiability
in RNA-seq isoform discovery \citep{li2011sparse}, argue for statistical
quantification of the central dogma \citep{li2015statistics}, implement
Neyman--Pearson error control in modern classification
\citep{tong2018neyman}, and develop methods for single-cell and spatial omics:
dropout-aware imputation \citep{li2018scimpute}, calibrated pseudotime
differential expression \citep{song2021pseudotimede}, p-value-free FDR control
\citep{ge2021clipper}, and transparent simulation for benchmarking
\citep{sun2021scdesign2,song2024scdesign3}, as well as reliability diagnostics
for two-dimensional embeddings \citep{xia2024scdeed}. In that sense, Jessica is
not only a fictional modern analyst; she is also a reminder that current
statistics is often built at the junction where data generation, computation,
biological interpretation, and error control all meet.

One focused single-cell thread will return later in the book.  scImpute asks
what an observed zero means, scDesign3 asks what generative mechanism could
produce a realistic assay, and scGTM asks how a structured biological map is
estimated from the whole empirical field.  The point is not to collect method
names.  It is to show how modern statistical genomics turns measurement,
simulation, and inference into one connected probability story.

The historical citations here are meant as representative anchors rather than
as exhaustive priority claims. One possible arc starts with Bayes's essay on
inverse probability and Laplace's analytic theory of probability
\citep{bayes1763essay,laplace1812theorie}, passes through Gauss's
least-squares astronomy \citep{gauss1809theoria}, Galton's regression and
Pearson's chi-square criterion
\citep{galton1886regression,pearson1900criterion}, and then reaches Student's
small-sample theory, Fisher's likelihood program, and Neyman--Pearson testing
\citep{student1908probable,fisher1922foundations,neyman1933tests}.
Kolmogorov's axiomatization then gave probability a measure-theoretic language
in which random variables, stochastic processes, and convergence could be
handled with modern precision \citep{kolmogorov1933grundbegriffe}.

There is also a Chinese line inside this international arc.  Pao-Lu Hsu
(Xu Baolu) helped give probability and mathematical statistics in China a
modern foundation; his work ranged from multivariate statistical theory and
determinantal equations to the Hsu--Robbins theorem on complete convergence
\citep{hsuRobbins1947complete,chenOlkin2012hsu}.  Kai Lai Chung carried
probability theory, Markov chains, and mathematical
exposition into a form that generations of students could actually use
\citep{chung1960markov,chung1974course}.  Xiru Chen represents the later
discipline of mathematical statistics in China: asymptotic theory,
nonparametric thinking, regression, and rigorous education as a national
infrastructure for statistical work \citep{fan2008master,chen2008rigorous}.
Kaitai Fang connects that theoretical tradition to design and computation:
uniform design, developed with Wang Yuan and later collaborators, made
space-filling experimental plans a practical language for industrial,
scientific, and biomedical experimentation
\citep{fangLinWinkerZhang2000uniform,fangLiuQinZhou2018uniform}.  For this
book, these names are not a nationalist sidebar.  They show the same compass
from another direction: probability, statistical inference, computation, and
scientific design have always had to grow together.

This historical view is not a parade of names. It is a constitutional
casebook: each name marks a moment when a new form of data pressure exposed
what the older language could not yet say. It is a way to understand why
statistical ideas were invented. Fisher's missing-species problem
\citep{fisher1943species}, and its later Shakespeare-vocabulary form
\citep{efronThisted1976unseen}, are not only curiosities in sampling; they are
early lessons in unobserved heterogeneity. The problem of musical
temperament is not only an art problem; it is a problem of approximation, loss,
and representation. Genome-wide
association studies are not only large regressions; they are case studies in
multiplicity, dependence, population structure, and scientific interpretation.
Each example shows how a practical problem forces a mathematical refinement.

The book will return repeatedly to this historical pattern. A method becomes
important when it solves a tension that earlier language could not express
well. Likelihood theory clarified the connection between data and parameters
\citep{fisher1922foundations}. The Neyman--Pearson line clarified testing,
design, and long-run operating behavior
\citep{neyman1933tests,neyman1923agricultural}. Measure theory clarified what
probability statements mean \citep{kolmogorov1933grundbegriffe}. Survival
analysis, resampling, and exploratory data analysis each reshaped what counted
as a statistical object
\citep{cox1972regression,efron1979bootstrap,tukey1962future}. Bayesian
nonparametrics clarified how to put distributions on distributions
\citep{ferguson1973bayesian,ferguson1974prior}. Machine learning clarified
prediction at scale \citep{breiman2001two_cultures}. None of these
developments is isolated; each is a response to a new kind of data pressure.
Figure~\ref{fig:historical-terrain} draws that pressure as a terrain rather
than as a straight timeline: the names mark moments when new questions raised
the landscape, not a closed canon of heroes.

\input{figures/ch01_historical_terrain}

A second historical thread will appear in a more concrete form.  Historical
China supplies a compact data system: basic economic regions ask what a region
is, famine relief asks how reports become decisions, and the Zhu/Chu Ko-chen
climate curve asks how proxy traces become evidence about a latent
environmental process.  Later the same climate thread will also be read as a
collection of small empirical distributions, one per century, so that a
Wasserstein regression is not an imported trick but a new question asked of an
old example.  The point is the same as in modern single-cell genomics: data are
not the world itself, but filtered traces from which a model tries to recover
structure.

\section{Modern Practice: Fine Structure and Examples}
\conceptindexes{fine structure, recurrent examples, biomedical data, clinical data, industrial data, platform experiments, single-cell data}

A third thread of the book studies modern statistics through representative
examples. The phrase ``fine structure'' is deliberate. From far away,
statistics may look like a few large territories: inference, prediction,
computation, and decision. Up close, each territory has a rich internal
geometry.

Consider regression. A classical linear model, a generalized linear model, a
survival regression, a kernel regression, a sparse high-dimensional regression,
a Wasserstein regression, and a deep representation model all share the broad
idea of relating an input to an output. But they differ in the kind of object
being regressed, the geometry of the sample space, the role of noise, the
meaning of coefficients, the computational method, and the scientific claim
that can be made after fitting. A mathematician who wants to ``apply'' a new
geometric idea may discover that the hard part is not writing down a distance,
but identifying the data structure and inferential target for which that
distance is meaningful.

This is why examples matter. Examples are not decorations placed after theory.
They test interpretation. They reveal which parts of a theory are structural
and which parts are conveniences. They expose hidden assumptions. They show
where computation changes the question. They also prevent the common mistake of
thinking that a method is understood merely because its formula is known.

Throughout the book, examples will be chosen from both academic and industrial
settings. Academic examples often emphasize explanation, theory, and
uncertainty. Industrial examples often emphasize scale, prediction, deployment,
and feedback. Both are necessary. A recommender system without statistical
thinking can optimize the wrong objective. A scientific model without
computational discipline can remain beautiful but unusable. Modern statistics
lives in the conversation between these demands.

The industrial examples are included with a narrow purpose.  The book is not
trying to survey companies or advertise systems.  It uses company-scale
settings when they expose a statistical translation that small textbook
examples can hide: a platform team must decide which metric is a target and
which metrics are guardrails; an oncology real-world-evidence group must turn
routine-care traces into an endpoint; a molecular-prediction system must
separate benchmark accuracy from biochemical mechanism; a precision-agriculture
system must translate image classification into field action.  These examples
will be treated with the same discipline as the classical examples: observed
record, hidden or target object, assumptions, procedure, uncertainty, and use.

\section{Reading the Book: Questions and Roadmap}
\conceptindexes{roadmap, reading protocol, mathematical spine, real-data capsules, recurrent examples}

The intended audience is broad but not casual.  The primary readers are
graduate students and early researchers in statistics, biostatistics, data
science, machine learning, bioinformatics, clinical research, and related
quantitative fields.  A second group is made of applied scientists,
clinicians, data scientists, and industry researchers who need to understand
what statistical reasoning tries to preserve when a scientific or operational
question becomes a model.  A third group is the expert reader who wants the
familiar probability-and-inference spine, but wants it connected to modern data
objects, targets, and feedback systems.  Advanced undergraduates with strong
preparation can read for orientation, but the book is not calibrated as a
first mathematical statistics course.

The book should be read at two speeds. At the first speed, read for the
statistical question: What is being observed? What is random? What comparison
or prediction is desired? What uncertainty must be reported? At the second
speed, read for the mathematical mechanism: Which sigma-field makes the object
measurable? Which product construction builds the joint law? Which convergence
theorem justifies the approximation? Which stochastic clock or martingale turns
an event history into an inferential object?

Some chapters are mathematical, some historical, and some example-driven. They
are not meant to be read at exactly the same speed.  The advanced chapters are
grammar and bridges, not substitutes for monographs.  Their job is to show
which mathematical language a data structure and target require.

A useful way to read the book is to keep four questions nearby:
\begin{enumerate}
\item What is the data structure?
\item What task is being performed?
\item What assumptions make the task mathematically possible?
\item What is gained, lost, or hidden by those assumptions?
\end{enumerate}
These questions are simple, but they are not elementary. They are the questions
that separate routine analysis from statistical thinking.

\subsection*{Roadmap of the Book}

The book follows the compass in Figure~\ref{fig:statistical-compass}.  The
roadmap below is intentionally short; the rest of the chapter carries the
examples.  A dataset becomes a random object only after we specify what counts
as observable.  This is why probability and measure appear early.  They are
not formal ornaments.  They are the grammar that makes it possible to say what data are.
Figure~\ref{fig:chapter-dependencies} gives the corresponding dependency map:
Chapter~2 turns examples into demands on language, Chapters~3--6 build the
grammar of observation, Chapters~7--8 move from data objects to targets,
Chapters~9--12 build stability, limit, and representation tools,
Chapters~13--16 turn those tools into local and time-indexed inference, and
Chapter~17 returns statistical claims to use, feedback, and deployment.  The
map should not be read as a separation between theory and model building.  It is
a braid: classical anchors such as likelihood, standard errors, regression,
testing, confidence intervals, and event-time models remain visible, while the
same chapters also prepare the reader for modern data objects, selection,
representation, feedback, and deployment.

\input{figures/ch01_chapter_dependencies}

The design chapter sits inside this first movement, at the compass step marked
measurement and collection. It asks what happens before \(P_n\) exists: who is
eligible, what is randomized, which dose is assigned, where observations are
placed, when outcomes are read, and what information a future model is allowed
to use. In that sense, design is the first statistical model, because it
determines what traces the world will be allowed to leave.

The observation and likelihood chapter then asks what survives into the data
object actually analyzed. It separates the scientific world, designed full
data, and observed data; it uses conditioning to describe what has been learned
from the observation; and it treats likelihood as a way of reading the same
observed object under different model maps. This chapter is also the caution
sign before formal inference: if the observation mechanism or nuisance
structure changes the meaning of a parameter, the likelihood may be precise
about the wrong target.

The second movement is from one random object to many. Product spaces explain
how vectors, panels, experiments, trajectories, and sequential observations
acquire joint laws. Kernels express conditional generation. Kolmogorov-type
extensions explain how local finite-dimensional specifications become process
laws. This is where the book begins to connect probability with the data
structures that appear in regression, Markov models, missing-data problems,
Bayesian hierarchies, and empirical processes.

The third movement names what the book is trying to protect before
asymptotics begin. The data-object chapter identifies the empirical object,
and the target chapter asks what claim that object is meant to support:
estimand, risk, causal contrast, prediction target, or scientific summary.
Only after those names are in place do stability, weak convergence, empirical
processes, and functional representation enter as the fourth movement.

The fifth movement is inference in local and event-time form.  Estimators,
tests, risk functions, likelihoods, estimating equations, and optimization
criteria are all ways of turning data into claims.  An $M$-estimator is not
just an argmax; it is a choice of target, criterion, geometry, approximation,
and error.  A $Z$-estimator is not just an equation; it is a way of encoding
balance, score, calibration, or moment restriction.  Filtrations then describe
what has been seen; stopping times describe decisions made without future
knowledge; predictable processes describe quantities fixed just before the next
instant; compensators turn irregular jumps into martingales.  The final
movement is from inference to use: prediction, risk, feedback, and deployment
ask what happens when statistical procedures act inside the systems they
measure.

The appendices support these movements without turning the main text into a
technical warehouse.  \Appref{app:set-theory-compass} gives the set
language needed to read sigma-fields.  \Appref{app:measure-theoretic-toolkit}
collects the extension, conditioning, and measure-theoretic tools used by the
chapters.  \Appref{app:computational-biomedical-translation} gives a
compact translation guide for algorithmic, bioinformatics, biochemical, and
clinical vocabulary.

\subsection*{Model and design atlas}
\conceptindexes{model atlas, design taxonomy, model taxonomy, data objects, target machinery}

Readers often ask for a classification of statistical models: which model is
for which kind of data, and which theory supports it?  The useful answer is not
a flat list of method names.  A linear model, a Cox model, a random forest, a
mixed model, a state-space model, and a causal estimator do not live at the
same level.  Some describe a law, some describe an observation mechanism, some
define a target, and some are procedures for optimizing a loss or solving an
estimating equation.  This book therefore classifies models by the work they
do in the translation from record to claim.

\input{figures/ch01_model_design_atlas}

The atlas in Figure~\ref{fig:ch1-model-design-atlas} gives the rule.  First
ask how the record was made: sampling, randomization, measurement, censoring,
linkage, follow-up, or an adaptive policy.  Then ask what kind of object was
observed: vector, table, curve, event history, image, graph, point pattern,
text, count matrix, or log.  Only then ask what probability grammar is
appropriate: product laws, conditional kernels, likelihoods, missing-data
maps, stochastic processes, empirical fields, random measures, or feedback
systems.  The target and the theory come after those choices.

\begingroup
\footnotesize
\setlength{\LTpre}{0.45\baselineskip}
\setlength{\LTpost}{0.45\baselineskip}
\setlength{\tabcolsep}{0.42em}
\renewcommand{\arraystretch}{1.22}
\begin{longtable}{@{}>{\raggedright\arraybackslash\bfseries}p{0.17\linewidth}
                  >{\raggedright\arraybackslash}p{0.25\linewidth}
                  >{\raggedright\arraybackslash}p{0.25\linewidth}
                  >{\raggedright\arraybackslash}p{0.23\linewidth}@{}}
\caption{Statistical model and design families used in the book\label{tab:ch1-model-design-atlas}}\\
\toprule
\textbf{Design/data pressure} & \textbf{Model language} & \textbf{Typical target} & \textbf{Main theory in this book} \\
\midrule
\endfirsthead
\caption[]{Statistical model and design families used in the book (continued)}\\
\toprule
\textbf{Design/data pressure} & \textbf{Model language} & \textbf{Typical target} & \textbf{Main theory in this book} \\
\midrule
\endhead
\multicolumn{4}{@{}l}{\textsc{Classical records and designed comparisons}}\\
\midrule
Random samples and tables &
Product laws, regression functions, parametric and semiparametric families &
Means, contrasts, risks, regression curves, interpretable parameters &
Chs.~3, 6, 9, 13, 15: random elements, products, LLNs, M/Z-estimation, local
uncertainty. \\
\addlinespace[0.22em]

Designed experiments &
Assignment kernels, randomization laws, blocking, dose or exposure designs &
Causal contrasts, estimands, design-based comparisons, guardrail summaries &
Chs.~4, 5, 8, 13--14: design before data, conditioning, targets, estimators,
tests. \\
\addlinespace[0.22em]

Coarsened, missing, or selected records &
Full-data law plus observation map; observed-data likelihood; nuisance
structure &
Full-data or target-population claims under stated identification assumptions &
Chs.~5, 8, 13, 15: likelihood, conditioning, identification, estimating
equations, influence functions. \\
\addlinespace[0.34em]

\multicolumn{4}{@{}l}{\textsc{Structured random objects}}\\
\midrule

Longitudinal and event-time data &
Stochastic processes, filtrations, hazards, risk sets, counting processes &
Survival, cumulative hazard, treatment over time, dynamic risk &
Chs.~6, 10, 15, 16: process laws, weak convergence, local linearization,
martingales. \\
\addlinespace[0.22em]

Curves, spatial fields, and geometric objects &
Random elements in function, measure, metric, or manifold spaces &
Mean curves, covariance modes, spatial intensity, barycenters, geometric
summaries &
Chs.~6, 10--12, 15: random-object grammar, weak convergence, empirical fields,
KL expansion, delta methods. \\
\addlinespace[0.34em]

\multicolumn{4}{@{}l}{\textsc{High-dimensional, latent, and deployed systems}}\\
\midrule

High-dimensional screening and learning &
Indexed empirical fields, regularized criteria, margins, selected models &
Selected features, classifiers, heterogeneous relationships, prediction risk &
Chs.~9, 11, 13--14, 17: concentration, uniform laws, M-estimation, testing,
deployment risk. \\
\addlinespace[0.22em]

Latent-structure scientific data &
Mixtures, hidden states, hierarchical laws, factor or trajectory models,
simulation laws &
Cell states, trajectories, dependence patterns, biological or physical maps &
Chs.~5, 6, 8, 11--13: observation mechanisms, kernels, targets, empirical
processes, M/Z-estimation. \\
\addlinespace[0.22em]

Adaptive and deployed systems &
Sequential kernels, policies, logs, stopping rules, feedback mechanisms &
Prediction risk, policy value, monitoring, regret-like loss, safety claims &
Chs.~4, 10, 16--17: adaptive design, histories, stopping, martingales,
feedback and use. \\
\bottomrule
\end{longtable}
\endgroup

This taxonomy is deliberately not a menu of algorithms.  The same procedure
can play different roles in different rows.  A logistic regression may be a
conditional mean model, a propensity model, a classifier, a nuisance estimate,
or a component of a policy.  A neural network may be a prediction rule, a
feature map, a simulator, or a nuisance function inside a semiparametric
estimator.  A survival model may be a likelihood for event times, a causal
working model, or a deployment risk monitor.  The classification that matters
for this book is therefore structural: what was observed, what was designed,
what target was named, and what theory makes the resulting claim stable.

\subsection*{Recurrent examples matrix}

The examples are meant to recur with different mathematical roles.
Table~\ref{tab:ch1-recurrent-examples-matrix} is a compact map of those roles:
observed object, hidden or target object, statistical pressure, and the chapters
where the example does work.
The new industrial rows are source-backed rather than decorative: online
experimentation draws on platform experimentation accounts
\citep{kohavi2020trustworthy,xu2015infrastructure,kaufman2017democratizing,
uber2018experimentation}; real-world oncology draws on EHR-derived endpoint and
estimand work \citep{griffith2019tumorBurden,fda2018rweframework,ich2021e9r1};
molecular prediction draws on AlphaFold and AlphaFold~3
\citep{jumper2021alphafold,abramson2024alphafold3}; and precision agriculture
uses John Deere's See \& Spray documentation \citep{johnDeere2026seeSpray}.

\begingroup
\footnotesize
\setlength{\LTpre}{0.45\baselineskip}
\setlength{\LTpost}{0.45\baselineskip}
\setlength{\tabcolsep}{0.42em}
\renewcommand{\arraystretch}{1.22}
\begin{longtable}{@{}>{\raggedright\arraybackslash\bfseries}p{0.15\linewidth}
                  >{\raggedright\arraybackslash}p{0.23\linewidth}
                  >{\raggedright\arraybackslash}p{0.22\linewidth}
                  >{\raggedright\arraybackslash}p{0.30\linewidth}@{}}
\caption{Recurrent examples and their mathematical roles\label{tab:ch1-recurrent-examples-matrix}}\\
\toprule
\textbf{Example thread} & \textbf{Observed object} & \textbf{Hidden or target object} & \textbf{Where it returns} \\
\midrule
\endfirsthead
\caption[]{Recurrent examples and their mathematical roles (continued)}\\
\toprule
\textbf{Example thread} & \textbf{Observed object} & \textbf{Hidden or target object} & \textbf{Where it returns} \\
\midrule
\endhead
\multicolumn{4}{@{}l}{\textsc{Historical and public records}}\\
\midrule
London bombing &
Impact locations or cell counts on a map &
Spatial randomness versus targeting &
Ch.~2: visible randomness. Ch.~3: observable events. Chs.~6, 10, 11:
point processes, random measures, and indexed summaries. \\
\addlinespace[0.25em]

Missing species and Shakespeare vocabulary &
Repeated labels summarized by frequency-of-frequency counts &
Unseen types and missing probability mass &
Ch.~2: absence as information. Ch.~5: unseen mass. Chs.~7, 9--11:
data grammar, empirical laws, random-object limits, and indexed summaries.
Ch.~15: rare-type influence functions. \\
\addlinespace[0.25em]

Rao's public numbers &
Life-expectancy risk summaries and rounded royal counts &
Population averages, individual interpretation, and provenance of recorded
numbers &
Ch.~1: numerical authority versus statistical translation. Chs.~4, 5:
record-making and target claims. Chs.~9, 14: averages, heterogeneity, and
local interpretation. \\
\addlinespace[0.25em]

Zhu/Chu Ko-chen climate curve &
Historical climate proxies, chronicles, phenology, disasters, and regional
records &
Latent climate signal under measurement and recording bias &
Ch.~2: proxy traces. Ch.~9: smoothing and bias--variance. Chs.~10, 11:
curves, distribution-valued summaries, and empirical fields. Ch.~15:
Wasserstein coordinates and local uncertainty. \\
\addlinespace[0.25em]

Chi/Skinner basic economic regions &
Water-control, transport, grain, market, population, and administrative traces &
Latent spatial-economic regions, partitions, graphs, and fields &
Ch.~1: historical regions as data structures. Ch.~2: traces. Ch.~6:
regions as random elements beyond ordinary vectors. \\
\addlinespace[0.25em]

Chinese famine relief &
Reports, grain prices, granary stocks, transport delays, migration, and relief
actions &
An information and decision system under crisis &
Ch.~4: design before analysis. Ch.~5: the archive as observed data. Ch.~16:
drought, price spikes, migration, and intervention as event histories. \\
\addlinespace[0.25em]

\emph{Dream of the Red Chamber} &
Chapter-level textual features, function words, n-grams, sentence patterns, and
verse/prose structure &
Latent style distribution and authorship evidence &
Ch.~2: textual traces. Ch.~11: feature classes as empirical-process
contrasts. Later inference chapters: classification and model comparison. \\
\addlinespace[0.45em]

\multicolumn{4}{@{}l}{\textsc{Biomedicine and scientific measurement}}\\
\midrule
Single-cell genomics &
Cells by genes, modalities, pseudotime, batches, donors, perturbations, and
spatial locations &
Observation mechanism, generative law, and structured biological map &
Ch.~4: experimental budget and simulation design. Ch.~5: dropout, batching,
and processed counts. Ch.~7: count matrices, latent states, simulations, and
targets. Chs.~11, 13: empirical processes and M/Z-estimation. \\
\addlinespace[0.25em]

Clinical longitudinal and event-time data &
Visits, biomarkers, imaging, treatments, endpoints, censoring, and risk sets &
Disease process, treatment effect, and event-time mechanism &
Ch.~4: trial design. Ch.~5: censoring and observed-data likelihood.
Ch.~15: survival influence functions. Ch.~16: hazards, counting processes,
compensators, and martingales. \\
\addlinespace[0.25em]

Real-world studies and pragmatic trials &
EHR, claims, registries, routine-care randomization, adherence, follow-up, and
endpoint capture &
Trial-like estimand under routine care, with confounding and observation
mechanisms made visible &
Ch.~4: randomization inside care. Ch.~5: target-trial emulation and
observed-data maps. Chs.~13--16: estimating equations, time-varying treatment,
censoring, and event histories. \\
\addlinespace[0.25em]

Real-world oncology endpoints &
EHR-derived visits, imaging reports, chart abstraction, treatment lines,
progression labels, and mortality links &
Real-world survival, progression, treatment duration, and external-control
estimands &
Chs.~5, 8: routine-care observation versus trial-like endpoints. Chs.~14, 15:
survival contrasts, censoring, and event-history uncertainty. \\
\addlinespace[0.25em]

Molecular prediction and scientific AI &
Sequences, structural databases, templates, alignments, ligands, biomolecular
contexts, and benchmark splits &
Structure prediction, confidence, interaction modeling, laboratory decisions,
and physical mechanism &
Chs.~7, 8: structured data objects and benchmark risk. Ch.~17: scientific AI
as a deployed design-and-feedback problem. \\
\addlinespace[0.45em]

\multicolumn{4}{@{}l}{\textsc{Platforms, robotics, and engineered systems}}\\
\midrule
Online experimentation platforms &
Assignments, exposure logs, user histories, outcome metrics, guardrails, and
marketplace context &
Short-run metric effects, long-run product value, welfare, interference, and
feedback &
Ch.~4: exposure as design. Ch.~8: metric lift versus target value.
Ch.~11: searched metrics and model classes. Ch.~17: deployed feedback systems. \\
\addlinespace[0.25em]

Netflix, ImageNet, and large benchmark datasets &
Sparse ratings, labeled images, captions, boxes, masks, and human annotations &
Prediction target shaped by missingness, labels, and evaluation design &
Ch.~7: data-structure grammar. Ch.~8: benchmark target versus real target.
Later learning chapters: risk, representation, feedback, and evaluation. \\
\addlinespace[0.25em]

Precision agriculture and field robotics &
Field images, GPS traces, crop and weed detections, nozzle actions, machine
logs, and as-applied maps &
Action policies balancing weed control, crop injury, chemical use, drift,
timing, and cost &
Ch.~7: sensor-and-action records. Ch.~8: policy loss. Ch.~17: deployment,
feedback, and changing field conditions. \\
\addlinespace[0.25em]

RiskMetrics and stock returns &
Returns, volatilities, covariances, and portfolio losses over time &
Time-varying risk under dependence and tail behavior &
Chs.~9, 15: concentration, temporal dependence, and process risk for
decision-oriented financial records. \\
\addlinespace[0.25em]

Autonomous laboratories &
Time-stamped interventions, robot logs, sensor curves, yields, failures, and
acquisition decisions &
Closed-loop experimental learning under an adapted information set &
Ch.~2: adaptive traces. Ch.~4: the next experiment as design action.
Ch.~10: action--observation histories. Ch.~16: filtrations and stopping rules. \\
\bottomrule
\end{longtable}
\endgroup

\subsection*{How to read a returning example}

The matrix above is not a promise to mention the same examples by name.  It is
a promise to make them do work.  When an example returns in a later chapter, it
should answer four application questions:
\[
\begin{gathered}
\text{theory: what mathematical object is being named?}\\
\text{data: what observed record or empirical object is being used?}\\
\text{computation: what fitting, simulation, or diagnostic action is taken?}\\
\text{return claim: what uncertainty, interpretation, or decision goes back to the world?}
\end{gathered}
\]
A single-cell example, for instance, begins as a count matrix with cells,
genes, donors, batches, modalities, and biological annotations.  In the design
chapter it becomes a budgeting problem: how many donors, cells, batches,
conditions, and reads should be bought, and how could scDesign3-like simulation
stress-test that choice before sequencing?  In the observation chapter the same
matrix becomes a likelihood question: are zeros, overdispersion, and batch
effects biological signals, technical artifacts, or both?  In the empirical
process and estimation chapters, the matrix becomes a field of gene-, cell-,
trajectory-, and spatial summaries; scDesign3 is read as a fitted generative
law checked against the summaries downstream analysis will use, while scGTM is
read as a constrained likelihood procedure for an interpretable pseudotime
trend.  The point is not to turn the book into a software manual.  The point is
to teach the reader how a theory chapter, a data object, and a computation
should meet in one defensible statistical translation.

The same rule applies beyond single-cell data.  A returning example should
eventually do at least one real analysis: estimate a treatment contrast, draw a
survival curve, decompose a curve sample, inspect a count matrix, backtest a
risk rule, or audit a deployed target.  That analysis should then feed back into
the theory by exposing the assumption that mattered: independent units,
unconfounded assignment, time zero, censoring, missingness, dependence,
preprocessing, metric choice, or feedback.  Otherwise the book would only use
real examples as scenery.  The goal is for real data to make the mathematical
language necessary.

\section{From Viewpoint to Formal Language}
\conceptindexes{grammar, statistical grammar, measure-theoretic language, stochastic processes, asymptotics}

Statistics has repeatedly changed when new data regimes made old habits
insufficient. Tukey made data analysis an intellectual activity
\citep{tukey1962future}; Donoho tied computation to reproducible data science
\citep{donoho2017fifty}; Breiman forced a confrontation between stochastic
modeling and algorithmic prediction \citep{breiman2001two_cultures}. The future
will not replace inference by prediction or modeling by algorithms. It will ask
for a deeper integration of data structure, computation, uncertainty, and
decision.

The movement after this charter is therefore from story to grammar, then from
grammar to named objects and targets. Chapter~2 trains the eye to see
randomness before probability is formalized. Probability and measure then name
what can be observed and integrated; product spaces and processes build joint
and time-indexed laws; the data-object and target chapters say what must be
stabilized before asymptotics begin; estimation theory turns stable random
objects into claims; continuous-time process theory supplies the clock,
information flow, and martingale structure needed for event histories. The
chapters are technical because the questions are delicate.

\section{Orientation Exercises}
\conceptindexes{orientation exercises, translation protocol}

These orientation exercises are not proof drills. They are meant to slow the
reader down before the formal machinery begins and to make the compass in
Figure~\ref{fig:statistical-compass} operational.

\begin{exercise}[The translation protocol]
Choose one concrete statistical problem and write a seven-line protocol:
\[
\begin{aligned}
\text{world question}&\longrightarrow\text{measurement}
\longrightarrow\text{data structure}\\
&\longrightarrow\text{assumptions}\longrightarrow\text{model}
\longrightarrow\text{uncertainty}\longrightarrow\text{decision}.
\end{aligned}
\]
For each arrow, write one sentence explaining what is preserved and one
sentence explaining what might be lost. Which arrow is the most fragile, and
what evidence would make it more defensible?
\end{exercise}

\begin{exercise}[Data provenance]
Choose a dataset you know: a clinical registry, an electronic health record
extract, a genomics matrix, an image archive, a recommender log, or a survival
dataset. Write a one-page provenance account for the dataset. Who or what entered it?
How were the main variables measured? What was filtered out? What information
was compressed, rounded, coded, anonymized, or lost before analysis began?
\end{exercise}

\begin{exercise}[One phrase, many structures]
Take the phrase ``patient data'' and describe three different data structures
it might refer to, for example a baseline table, a treatment path, an image, an
omics profile, an event-time record, or an EHR extract. For each structure,
state one natural inferential question and one assumption that would be
dangerous if made automatically.
\end{exercise}

\begin{exercise}[From structure to assumptions]
Pick one row of the data-structure table above. Replace the tempting assumption
with a more careful one. Explain what extra design information, metadata,
diagnostic plot, sensitivity analysis, or scientific knowledge would make that
assumption more defensible.
\end{exercise}

\begin{exercise}[A method is a translation]
Choose a familiar method, such as linear regression, logistic regression,
Kaplan--Meier estimation, a random forest, a clustering algorithm, or a neural
network. Describe the chain
\[
\text{data structure}\longrightarrow\text{assumptions}
\longrightarrow\text{model}\longrightarrow\text{inference task}.
\]
What is gained by the method, and what is hidden?
\end{exercise}

\begin{exercise}[Bridge to probability and measure]
For one of your examples above, identify a sample space, a few events that
should be observable, and a random variable or random object that would appear
in the analysis. The goal is not formal perfection, but to see why the next
mathematical step is to define what counts as measurable.
\end{exercise}

\begin{exercise}[Bridge to stochastic processes]
Find an example in which time matters. Specify the clock, the information
available at each time, the event or transition of interest, and one way in
which censoring, delayed entry, feedback, or observation schedule could change
the analysis.
\end{exercise}

\section*{Sources and Further Reading}
\addcontentsline{toc}{section}{Sources and Further Reading}

Historical references are used selectively. The goal is not to settle priority
disputes, but to give the reader concrete papers and books that mark turning
points in the statistical imagination: inverse probability,
least-squares estimation, regression, goodness-of-fit testing, small-sample
inference, likelihood, hypothesis testing, axiomatic probability, survival
modeling, resampling, exploratory data analysis, Bayesian nonparametrics,
causal diagrams, algorithmic prediction, and data science.

\begin{description}[leftmargin=0pt,labelsep=0.65em,style=unboxed,font=\normalfont,itemsep=0.45\baselineskip]
\item[\textsc{Classical probability and inverse probability.}]
The essay by \citet{bayes1763essay} and the analytic theory of
\citet{laplace1812theorie} are useful historical anchors for the
inverse-probability tradition. They also show why probability entered
statistics as a language for reasoning backward from observations to uncertain
causes.

\item[\textsc{Least squares, regression, and early sampling distributions.}]
The astronomical work of \citet{gauss1809theoria} is a classical source for
least squares, \citet{galton1886regression} introduced regression as a
statistical phenomenon, \citet{pearson1900criterion} gave the chi-square
goodness-of-fit criterion, and \citet{student1908probable} made small-sample
uncertainty visible in a way that still shapes introductory statistics.

\item[\textsc{Likelihood, tests, and design.}]
\citet{fisher1922foundations} is one of the central sources for likelihood,
sufficiency, efficiency, and the organization of theoretical statistics.
\citet{neyman1933tests} anchor the testing tradition, while
\citet{neyman1923agricultural} is a representative source for
randomization-based design and causal comparison. Fisher's later experimental
design work \citep{fisher1935design} is one historical anchor for the book's
design principle that data collection is already statistical.

\item[\textsc{Probability as measure.}]
\citet{kolmogorov1933grundbegriffe} gave probability its modern axiomatic
language. The probability-and-measure chapter develops the measure-theoretic
version of this viewpoint in detail.

\item[\textsc{Chinese probability and statistics.}]
The short Chinese line in the historical section is necessarily selective.
For a broad recent account of the development of statistics in China, see the
two-volume history edited by \citet{yuan2025zhongguoTongjiXueshi}.
For Pao-Lu Hsu, useful anchors are the technical Hsu--Robbins theorem on
complete convergence \citep{hsuRobbins1947complete}, Hsu's work on roots of
determinantal equations \citep{hsu1939roots}, and later historical accounts by
\citet{andersonChungLehmann1979hsu} and \citet{chenOlkin2012hsu}.  For Chung,
the opening passage cites his probability text \citep{chung1974course}, while
\citet{chung1960markov} represents his role in Markov-chain theory.  Chen's
role as a rigorous statistician and educator is summarized in the Statistica
Sinica memorial pieces of \citet{fan2008master} and
\citet{chen2008rigorous}.  Fang's uniform-design line is represented by
\citet{fangLinWinkerZhang2000uniform} and the later monograph
\citet{fangLiuQinZhou2018uniform}.  These references are included because
they connect the book's themes--probability, asymptotic statistics, design,
computation, and scientific application--to a tradition that is also part of
the author's intellectual inheritance.

\item[\textsc{Readers, maps, and journeys.}]
The opening passage takes its cue from \citet{chung1974course}, whose preface
argues that a mathematical course should be more than a collection of raw
materials. That idea fits the present book: the aim is to give a coherent route
through probability, stochastic
processes, inference, and modern data analysis. The full-book frontispiece line
from \citet{emerson1870society} frames reading as an active statistical habit:
the book becomes useful only when the reader tests each idea against a problem.
The opening proverb from
\citet[Ch.~64]{laozi1963tao} is used as a reminder that statistical work
begins locally, with the first trace under one's feet.
\citet{korzybski1933science} supplies the map-and-territory warning that helps
frame models as useful simplifications rather than copies of reality.

\item[\textsc{Data structures and data biographies.}]
The emphasis on looking before modeling follows the tradition of exploratory
data analysis in \citet{tukey1962future,tukey1977exploratory} and the
model-building caution in \citet{box1976science}. Specific data-structure
warnings are anchored by \citet{rubin1976inference} on missingness,
\citet{weiskopf2013methods,goldstein2017opportunities} on clinical record
quality, \citet{besag1974spatial} on spatial dependence,
\citet{wasserman1994social} on networks, \citet{petersen2019frechet} and
\citet{chen2023wasserstein} on distributional random objects, and
\citet{cox1972regression,andersen1982cox} on event-time data.
The phrase ``data provenance'' is used here for the generated, measured,
filtered, stored, and documented history of a dataset; it is closest in spirit
to \citet{leonelli2016data}, \citet{meng2018statistical}, and
\citet{gebru2021datasheets}.

\item[\textsc{Rao's public understanding examples.}]
The two examples from \citet{rao1997statisticsTruth} are included because they
make numerical translation visible before any formal model is introduced.  The
``mournful numbers'' table in Rao's Chapter~6 reproduces selected
life-expectancy losses and gains from \citet{cohenLee1979catalog}; it is used
here as a warning that a population average is not automatically an individual
or causal prescription.  Rao's ancient royal-count example is used as a compact
provenance warning: a historical number may be a tally, a rounded
administrative record, a public display, or a mixture of these.

\item[\textsc{Real-world and pragmatic clinical evidence.}]
\citet{schwartz1967explanatory} introduced the explanatory--pragmatic
distinction for therapeutic trials. The pragmatic-trial reporting and design
literature is represented here by the CONSORT extension of
\citet{zwarenstein2008consort} and the PRECIS-2 tool of
\citet{loudon2015precis2}. Real-world evidence is represented by the FDA
framework \citep{fda2018rweframework}; target-trial emulation follows
\citet{hernan2016targettrial}. Together these references make RWS/PCT a
full-book example of the boundary between designed and recorded evidence.

\item[\textsc{Modern data pressure.}]
The regression model of \citet{cox1972regression}, the bootstrap of
\citet{efron1979bootstrap}, the data-analysis manifesto of
\citet{tukey1962future}, the Dirichlet process of
\citet{ferguson1973bayesian,ferguson1974prior}, the causal diagrams of
\citet{pearl1995causal}, the two cultures of
\citet{breiman2001two_cultures}, and the data-science essay of
\citet{donoho2017fifty} are included because each marks a pressure that made
older statistical language insufficient on its own.

\item[\textsc{Jingyi Jessica Li and statistical genomics.}]
Li's papers are included to make the book's contemporary ``Jessica'' concrete.
They show how statistical questions in identifiability, uncertainty
calibration, asymmetric error control, FDR control, and simulation arise
directly from RNA-seq, single-cell RNA-seq, and spatial omics assays. The
sequence from \citet{li2011sparse} through
\citet{sun2021scdesign2} and scDesign3 to \citet{xia2024scdeed} is especially useful
for seeing how statistical modeling becomes a language for designing,
validating, and interpreting modern biological data analyses.
\end{description}

%% file: figures/ch01_statistical_compass.tex
\begin{center}
\includegraphics[width=\linewidth]{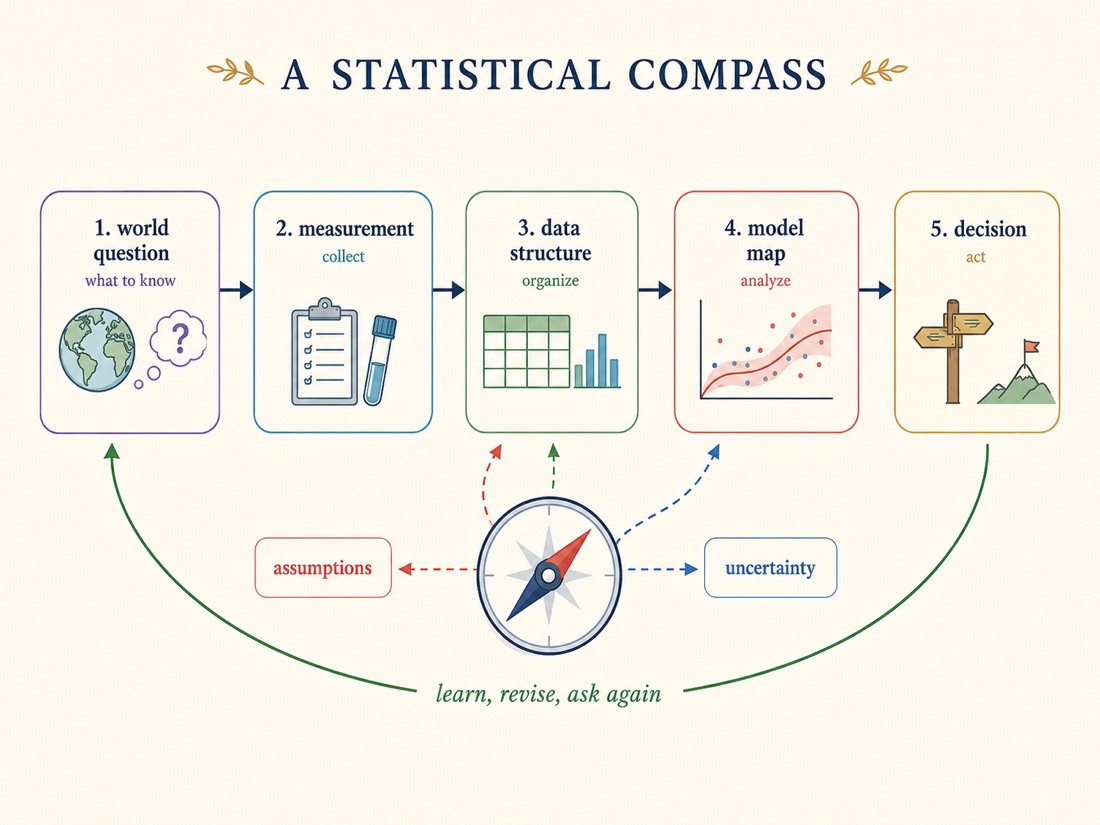}
\bookmanualfigure{fig:statistical-compass}{A statistical compass}
\par\smallskip
\small Figure~\thefigure: A statistical compass. A question becomes
measurements, measurements become data, data motivate models, and models guide
decisions. Assumptions and uncertainty calibrate each translation.
\end{center}

%% file: figures/ch01_historical_terrain.tex
\begin{center}
\includegraphics[width=\linewidth]{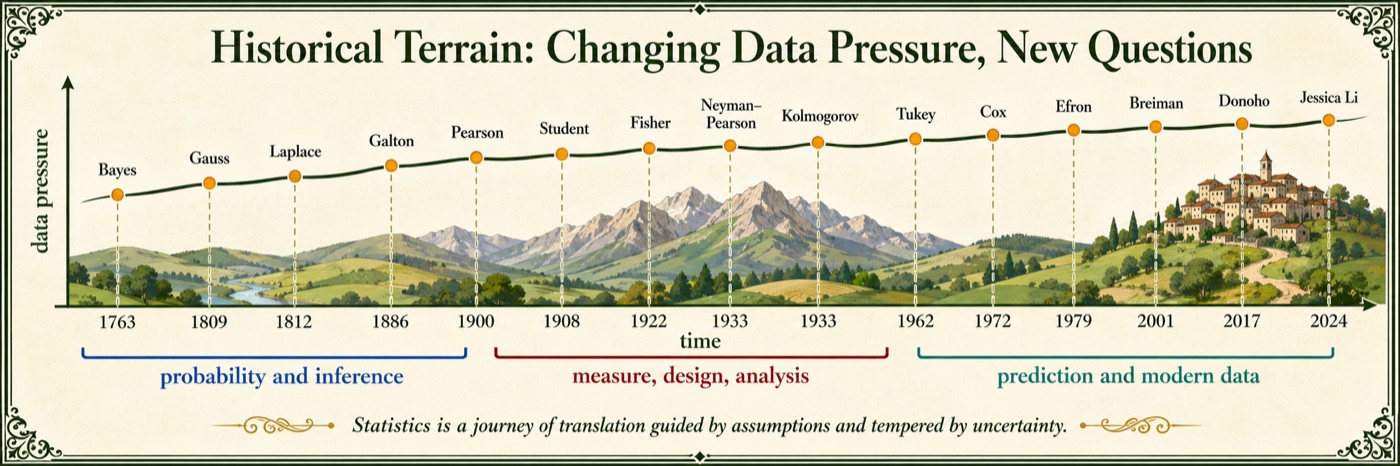}
\bookmanualfigure{fig:historical-terrain}{A historical terrain for Chapter 1}
\par\smallskip
\small Figure~\thefigure: A historical terrain for Chapter~1. The figure uses
representative anchors, not exhaustive priority claims, to show how changing
data pressure repeatedly asks statistics for new languages: probability,
inference, design, measure, processes, computation, prediction, and modern
biological data analysis.
\end{center}

%% file: figures/ch01_chapter_dependencies.tex
\begin{center}
\resizebox{\linewidth}{!}{%
\begin{tikzpicture}[
  >=Latex,
  band/.style={
    rounded corners=4pt,
    draw=bookblue!18,
    fill=bookblue!3,
    line width=0.35pt
  },
  finalband/.style={
    rounded corners=4pt,
    draw=bookred!24,
    fill=bookred!5,
    line width=0.35pt
  },
  partlabel/.style={
    anchor=west,
    align=left,
    text width=2.24cm,
    text=bookblue!88!black,
    font=\sffamily\bfseries\scriptsize
  },
  nodebase/.style={
    draw=bookblue!76!black,
    fill=white,
    rounded corners=3pt,
    minimum width=2.00cm,
    minimum height=0.63cm,
    inner xsep=3pt,
    inner ysep=3pt,
    align=center,
    font=\sffamily\scriptsize
  },
  hinge/.style={
    nodebase,
    draw=bookred!82!black,
    fill=chaptercream
  },
  finish/.style={
    nodebase,
    draw=bookred!82!black,
    fill=bookred!6
  },
  supportbox/.style={
    draw=bookteal!82!black,
    fill=white,
    rounded corners=2pt,
    minimum width=2.16cm,
    minimum height=0.50cm,
    inner xsep=2pt,
    inner ysep=2pt,
    align=center,
    font=\sffamily\tiny
  },
  bridgebox/.style={
    draw=bookred!70!black,
    fill=bookred!4,
    rounded corners=4pt,
    line width=0.42pt,
    inner xsep=5pt,
    inner ysep=5pt,
    align=left,
    font=\sffamily\tiny,
    text width=5.05cm
  },
  classicbox/.style={
    draw=bookgold!78!black,
    fill=noteback,
    rounded corners=4pt,
    line width=0.42pt,
    inner xsep=5pt,
    inner ysep=5pt,
    align=left,
    font=\sffamily\tiny,
    text width=6.62cm
  },
  arrow/.style={->, line width=0.78pt, draw=bookblue!78!black},
  downarrow/.style={->, line width=0.70pt, draw=bookblue!74!black},
  support/.style={->, dashed, line width=0.55pt, draw=bookteal!78!black},
  braid/.style={->, dashed, line width=0.62pt, draw=bookred!74!black},
  note/.style={font=\sffamily\tiny, text=black!72}
]

\foreach \y/\style in {0/band,-1.06/band,-2.12/band,-3.18/band,-4.24/band,-5.30/finalband} {
  \path[\style] (-0.05,\y+0.43) rectangle (13.25,\y-0.43);
}

\draw[braid] (7.71,-1.48) .. controls (8.34,-2.28) and (2.18,-3.38) .. (2.34,-3.86);
\draw[braid] (6.92,-2.46) .. controls (8.36,-2.92) and (6.20,-3.48) .. (6.84,-3.86);
\draw[braid] (5.38,-4.60) .. controls (6.10,-5.02) and (8.22,-5.00) .. (8.98,-5.28);

\node[partlabel] at (0.16,0) {Part I\\Examples first};
\node[partlabel] at (0.16,-1.06) {Part II\\Observation};
\node[partlabel] at (0.16,-2.12) {Part III\\Objects/targets};
\node[partlabel] at (0.16,-3.18) {Part IV\\Stability/limits};
\node[partlabel] at (0.16,-4.24) {Part V\\Inference\\event time};
\node[partlabel] at (0.16,-5.30) {Part VI\\Use};

\node[nodebase] (c1) at (3.05,0) {Ch 1\\Compass};
\node[nodebase] (c2) at (5.38,0) {Ch 2\\Problem bank};

\node[nodebase] (c3) at (3.05,-1.06) {Ch 3\\Probability};
\node[nodebase] (c4) at (5.38,-1.06) {Ch 4\\Design};
\node[nodebase] (c5) at (7.71,-1.06) {Ch 5\\Observation};
\node[nodebase] (c6) at (10.04,-1.06) {Ch 6\\Kernels/processes};

\node[hinge] (c7) at (4.22,-2.12) {Ch 7\\Data objects};
\node[hinge] (c8) at (6.55,-2.12) {Ch 8\\Targets};

\node[nodebase] (c9) at (3.05,-3.18) {Ch 9\\Stability};
\node[nodebase] (c10) at (5.38,-3.18) {Ch 10\\Weak conv.};
\node[nodebase] (c11) at (7.71,-3.18) {Ch 11\\Emp. fields};
\node[nodebase] (c12) at (10.04,-3.18) {Ch 12\\Functional data};

\node[nodebase] (c13) at (3.05,-4.24) {Ch 13\\M/Z};
\node[nodebase] (c14) at (5.38,-4.24) {Ch 14\\Testing};
\node[nodebase] (c15) at (7.71,-4.24) {Ch 15\\Delta/IF};
\node[nodebase] (c16) at (10.04,-4.24) {Ch 16\\Event time};

\node[finish] (c17) at (10.04,-5.30) {Ch 17\\Prediction/use};

\node[supportbox] (appAB) at (12.15,-1.06) {App A--B\\foundations};
\node[supportbox] (appC) at (12.15,-3.18) {App C\\limit toolkit};
\node[supportbox] (appD) at (12.15,-5.30) {App D\\translation};

\draw[arrow] (c1) -- (c2);
\draw[downarrow] (c1.south) -- ++(0,-0.20) -| (c3.north);
\draw[arrow] (c3) -- (c4);
\draw[arrow] (c4) -- (c5);
\draw[arrow] (c5) -- (c6);
\draw[downarrow] (c5.south) -- ++(0,-0.20) -| (c7.north);
\draw[arrow] (c7) -- (c8);
\draw[downarrow] (c8.south) -- ++(0,-0.20) -| (c9.north);
\draw[arrow] (c9) -- (c10);
\draw[arrow] (c10) -- (c11);
\draw[arrow] (c10) -- (c12);
\draw[downarrow] (c8.south) -- ++(0,-0.96) -| (c13.north);
\draw[arrow] (c13) -- (c14);
\draw[arrow] (c14) -- (c15);
\draw[arrow] (c15) -- (c16);
\draw[downarrow] (c16) -- (c17);

\draw[support] (appAB.west) -- (c6.east);
\draw[support] (appC.west) -- (c12.east);
\draw[support] (appD.west) -- (c17.east);

\node[bridgebox,anchor=north west] at (0.16,-6.03) {
  \textbf{\color{bookred!80!black}Theory--model braid}\\[-1pt]
  Ch 2 supplies modeling pressure; Ch 4--6 make records and laws.\\
  Ch 7--8 name the object and target before any procedure.\\
  Ch 9--16 alternate between theory tools and model-building debts.\\
  Ch 17 asks whether the fitted claim survives use and feedback.
};

\node[classicbox,anchor=north west] at (5.62,-6.03) {
  \textbf{\color{bookgold!62!black}Classical anchors remain visible}\\[-1pt]
  Ch 5: likelihood, nuisance structure, exponential-family logic.\\
  Ch 9 and Ch 15: means, variances, standard errors, confidence intervals.\\
  Ch 13--14: MLE, method of moments, GLM, \(t\)/Wald/score/LR tests.\\
  Ch 16: Cox/counting-process models as event-history statistics.
};
\end{tikzpicture}%
}
\bookmanualfigure{fig:chapter-dependencies}{Chapter dependencies for the book}
\par\smallskip
\small Figure~\thefigure: Chapter dependencies for the book. Solid arrows show
the main local reading path; dashed red arrows mark places where model-building
questions return to the theory.  The lower boxes make the intended reading
explicit: classical statistics remains as an anchor layer, while modern examples
force the same objects--targets--theory discipline at larger scale.
\end{center}

%% file: figures/ch01_model_design_atlas.tex
\begin{center}
\resizebox{\linewidth}{!}{%
\begin{tikzpicture}[
  >=Latex,
  stage/.style={
    draw=bookblue!78!black,
    fill=white,
    rounded corners=3pt,
    minimum width=2.05cm,
    minimum height=0.72cm,
    inner xsep=4pt,
    inner ysep=3pt,
    align=center,
    font=\sffamily\scriptsize
  },
  hinge/.style={
    stage,
    draw=bookred!82!black,
    fill=chaptercream
  },
  usebox/.style={
    stage,
    draw=bookred!82!black,
    fill=bookred!5
  },
  tag/.style={
    draw=bookteal!75!black,
    fill=bookteal!5,
    rounded corners=2pt,
    align=center,
    font=\sffamily\tiny,
    inner xsep=3pt,
    inner ysep=2pt,
    text width=2.02cm
  },
  question/.style={
    draw=bookgold!75!black,
    fill=noteback,
    rounded corners=4pt,
    align=left,
    font=\sffamily\tiny,
    inner xsep=5pt,
    inner ysep=5pt,
    text width=4.55cm
  },
  arrow/.style={->, line width=0.75pt, draw=bookblue!80!black},
  feedback/.style={->, dashed, line width=0.65pt, draw=bookred!80!black}
]

\node[stage] (design) at (0,0) {Design\\before data};
\node[stage] (object) at (2.50,0) {Observed\\data object};
\node[hinge] (model) at (5.00,0) {Model\\grammar};
\node[hinge] (target) at (7.50,0) {Target\\or task};
\node[stage] (theory) at (10.00,0) {Theory\\and uncertainty};
\node[usebox] (use) at (12.50,0) {Use\\and feedback};

\draw[arrow] (design) -- (object);
\draw[arrow] (object) -- (model);
\draw[arrow] (model) -- (target);
\draw[arrow] (target) -- (theory);
\draw[arrow] (theory) -- (use);
\draw[feedback] (use.south) .. controls +(0,-1.05) and +(0,-1.05) .. (design.south);

\node[tag] at (0,1.05) {sampling\\randomization\\measurement\\follow-up};
\node[tag] at (2.50,-1.05) {table\\curve\\event time\\image/graph/log};
\node[tag] at (5.00,1.05) {product laws\\kernels\\likelihoods\\processes\\empirical fields};
\node[tag] at (7.50,-1.05) {estimand\\risk\\test\\selected feature\\policy value};
\node[tag] at (10.00,1.05) {LLN/CLT\\weak limits\\M/Z\\IF\\martingales};
\node[tag] at (12.50,-1.05) {decision\\deployment\\audit\\next design};

\node[question,anchor=north west] at (0,-2.08) {
  \textbf{\color{bookred!82!black}Classification question}\\[-1pt]
  What record was allowed to exist?\\
  What law or mechanism describes it?\\
  What claim is the model protecting?\\
  Which theorem controls the error?
};
\node[question,anchor=north west,text width=7.55cm] at (4.95,-2.08) {
  \textbf{\color{bookred!82!black}Reading rule}\\[-1pt]
  A statistical model is not just a named regression or software routine.  In
  this book it is the map from a designed or observed random object to a
  target, together with the assumptions and theory that make the target
  estimable, testable, predictable, or usable.
};

\end{tikzpicture}%
}
\bookmanualfigure{fig:ch1-model-design-atlas}{Model and design atlas}
\par\smallskip
\small Figure~\thefigure: A model-and-design atlas for the book.  Design
decisions determine the record; the record determines the probability grammar;
the target determines which theory is needed; use can feed back into the next
design.
\end{center}

%% file: chapters/ch02_randomness.tex
\chapter{Randomness in Data: Examples That Force Theory}
\label{chap:randomness}
\conceptindexes{randomness, clusters, gaps, point patterns, missing species, unseen words, latent objects, data streams, adaptive laboratories}

\begin{tcolorbox}[
  enhanced,
  breakable,
  colback=chaptercream,
  colframe=bookblue!88!black,
  boxrule=0.72pt,
  arc=5pt,
  boxsep=1pt,
  left=1.0em,
  right=0.95em,
  top=0.82em,
  bottom=0.82em,
  before skip=0.55\baselineskip,
  after skip=1.0\baselineskip
]
\noindent\textbf{Chapter overview.}
This chapter is the book's problem bank.  It trains the statistical eye before
formal probability arrives by letting examples make demands on language.
Random does not mean evenly spread; visible pattern does not by itself prove a
mechanism; absence from the sample is not absence from the world. Bomb impacts,
missing species, unseen words, climate proxies, historical regions, text,
multi-omics profiles, adaptive laboratories, and clinical records do more than
decorate the theory.  They force the analysis to invent the mathematical objects that
later chapters will formalize: events, random elements, product spaces,
histories, targets, empirical processes, estimands, and process laws.
\end{tcolorbox}

The first mistake many students make about randomness is also the most human:
they expect it to be fair to the eye. If points are thrown at random in a
square, they imagine a polite scatter, with roughly equal spacing and no
awkward empty region. But independent random points do not know that they are
supposed to reassure the eye. Some land close together. Some leave holes. Clusters
and gaps are not failures of randomness; they are among its most recognizable
signatures.

This chapter uses a small set of stories to make that lesson durable. It is not
a gallery of curiosities.  It is a sequence of examples before formulas:
examples appear first because they force the abstractions that the rest of the
book will need.  The
London bombing example asks whether a spatial pattern of impacts is evidence
of targeting or compatible with chance \citep{clarke1946poisson,feller1968introduction}.
The missing-species story asks how a sample can carry information about
biological types, or word types, that were not observed at all
\citep{fisher1943species,good1953population,efronThisted1976unseen}. Historical
climate reconstruction asks how frost dates, flowering records, harvest reports,
and chronicles can speak about a latent temperature curve
\citep{chu1973climatic,ge2013temperature}. Chi Ch'ao-ting's basic economic
areas ask how water-control works, transport, grain movement, and political
economy can speak about latent spatial regions \citep{chi1936key}. Stylometric
authorship asks how chapter-level word patterns can speak about a latent
writing regime \citep{hu2014redchamber,tu2013redchamber,zhu2021redchamber}.
Multi-omics asks what happens when a dataset is no longer a single table but a
layered, partly matched record \citep{argelaguet2018mofa}.
Autonomous laboratories add time-stamped robot logs, continuous monitoring, and
policy-dependent experimentation
\citep{burger2020mobile,abolhasani2023rise,xtalpi2025autonomousLab}. One
story warns the analyst not to overread visible clusters. The others teach how to read
the invisible through the visible. The modern examples add one more warning:
data can be created by the very model or policy that later analyzes them.
Together they turn the book's opening compass into a concrete demand: before
calculation can begin, the observed record, random summaries, information
available at each decision, claimed target, and events measured by a probability
model must be named.

The chapter therefore repeats one small protocol:
\[
  \text{data structure}
  \;\longrightarrow\;
  \text{assumptions}
  \;\longrightarrow\;
  \text{working model}
  \;\longrightarrow\;
  \text{observable question}.
\]
The protocol is deliberately modest. It does not yet require measure theory.
It asks the reader to pause before formalism and decide what kind of thing the
data are: points in space, counts in boxes, a list of types, a frequency table,
or something else. Only then can one ask whether a cluster is surprising, whether
a zero is real absence or sampling silence, and which summary of the data a
future probability model is supposed to judge.

The chapter is easiest to read as a training sequence.  Each section removes
one tempting but wrong shortcut:
\begin{description}[leftmargin=0pt,labelsep=0.55em,style=unboxed,font=\bookdescriptionlabelfont,itemsep=0.18\baselineskip]
\item[Point patterns.]
Section~\ref{sec:ch2-texture} teaches the eye that random variation is allowed
to look uneven.  This is the warm-up: before asking whether a pattern is
meaningful, the reader must learn what ordinary randomness can look like.

\item[Null models.]
Section~\ref{sec:ch2-visible-pattern} turns that habit into a testable
question through the London bombing example.  A cluster becomes evidence only
after a null model says what clusters chance itself would commonly create.

\item[Unobserved types.]
Section~\ref{sec:ch2-unseen} reverses the mistake.  It shows that what is
unseen may still leave information in low-frequency counts.  The sample is not
the world; it is a trace of the world.

\item[Traces.]
Section~\ref{sec:ch2-historical-traces} moves from simple samples to historical
and textual evidence.  Proxy records, public works, and chapters must first be
encoded as features before they can speak about latent climate, regions, or
authorship.

\item[Histories.]
Section~\ref{sec:ch2-modern-data-streams} brings the same logic into modern
data systems.  Multi-omics and robot laboratories show that data can be layered,
time-stamped, and generated by earlier decisions.

\item[Formal objects.]
Section~\ref{sec:ch2-two-stories} gathers the examples and points forward:
Chapter~3 supplies events and measurable summaries; later chapters supply
design, product spaces, empirical fields, estimation, perturbation, and dynamic
process language.
\end{description}

The same protocol will organize less toy-like examples later in the book. The
following cases are cameo examples, not a syllabus of datasets.  Each cameo has
one job: to make a later mathematical pressure visible before the formal
machinery arrives.  Figure~\ref{fig:ch2-data-forms-atlas} gives the
corresponding visual atlas: each panel asks what kind of random object the
observations can become before any formula is chosen.

\begin{center}
\centering
\includegraphics[width=0.96\linewidth]{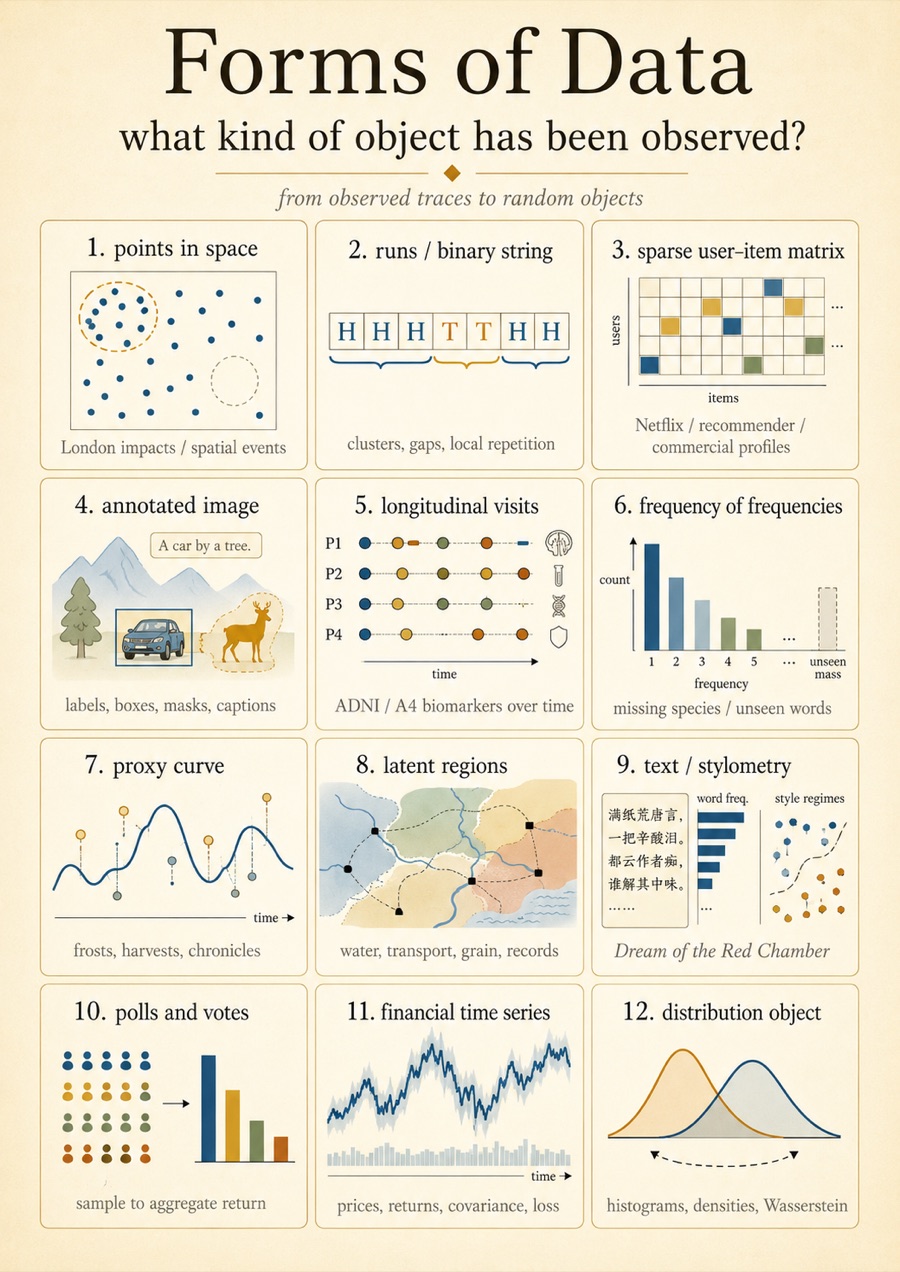}
\bookmanualfigure{fig:ch2-data-forms-atlas}{Data-form atlas as statistical pressure}
\par\smallskip
\parbox{0.92\linewidth}{\small\textbf{Figure~\thefigure.} A schematic atlas of
data forms used as recurring examples in this chapter and later chapters.  The
panels are synthetic visual mnemonics rather than reproductions of the datasets
themselves: spatial points, binary runs, sparse matrices, annotated images,
longitudinal records, frequency-of-frequency summaries, proxy curves, latent
regions, stylometric traces, layered omics, adaptive laboratory histories,
polls, financial time series, and distributional objects each ask for a
different path from observation to random object.}
\end{center}

Table~\ref{tab:ch2-data-form-atlas} keeps the same atlas in textual form.  Its
purpose is not to multiply examples, but to make explicit what each example
forces the book to name later: a data structure, an observation law, a target,
or a model object.
\begingroup
\footnotesize
\setlength{\LTpre}{0.45\baselineskip}
\setlength{\LTpost}{0.45\baselineskip}
\setlength{\tabcolsep}{0.35em}
\renewcommand{\arraystretch}{0.96}
\begin{longtable}{@{}>{\raggedright\arraybackslash}p{0.25\linewidth}>{\raggedright\arraybackslash}p{0.31\linewidth}>{\raggedright\arraybackslash}p{0.35\linewidth}@{}}
\caption{Data-form atlas as statistical pressure}
\label{tab:ch2-data-form-atlas}\\
\toprule
\textbf{Example} & \textbf{Data structure} & \textbf{Statistical pressure} \\
\midrule
\endfirsthead
\caption[]{Data-form atlas as statistical pressure (continued)}\\
\toprule
\textbf{Example} & \textbf{Data structure} & \textbf{Statistical pressure} \\
\midrule
\endhead
\bottomrule
\endlastfoot
Netflix Prize & Sparse user--movie rating matrix & Missingness, prediction,
privacy, and feedback in recommender systems \citep{bennett2007netflix} \\
ImageNet and MS COCO & Images with labels, boxes, masks, and captions &
Annotation, hierarchy, spatial locality, and benchmark design
\citep{deng2009imagenet,lin2014microsoft} \\
ADNI and A4 & Clinical visits, MRI/PET, biomarkers, and longitudinal cognitive
endpoints & Measurement protocol, disease staging, selection, and endpoint
definition \citep{petersen2010adni,sperling2023solanezumab} \\
Multi-omics and spatial omics & Matched or partially matched molecular layers:
RNA, chromatin, protein, metabolites, spatial coordinates & Alignment,
modality-specific missingness, batch effects, latent cell states, and
simulation-based checking \citep{argelaguet2018mofa} \\
Autonomous wet labs & Time-stamped interventions, robot logs, sensor traces,
reaction outcomes, and decisions & Time order, adaptive design, closed-loop
experimentation, and policy-dependent data
\citep{burger2020mobile,abolhasani2023rise,xtalpi2025autonomousLab} \\
Zhu/Chu Ko-chen curve & Historical climate proxies over time and place &
Proxy measurement, alignment, smoothing, and latent process reconstruction
\citep{chu1973climatic,ge2013temperature} \\
Polls and votes & Individual survey responses and aggregate election returns &
Sampling, nonresponse, turnout, aggregation, and temporal updating
\citep{gelman1993polls} \\
RiskMetrics and stock returns & Time series of prices, returns, covariances,
and losses & Dependence, heavy tails, volatility, and decision-oriented risk
\citep{jpmorgan1996riskmetrics,fama1965behavior} \\
Wasserstein examples & Distributions, histograms, densities, simulated random
objects, and century-level climate anomaly distributions & Geometry, alignment,
support, and distribution-to-distribution regression
\citep{petersen2019frechet,chen2023wasserstein} \\
\end{longtable}
\endgroup

\begin{realdatacapsule}{London bombing grid}
\item[Data object.] Clarke's South London impact map, usually read as counts in
a fixed grid of cells \citep{clarke1946poisson}.
\item[Observation mechanism.] Wartime impact locations are recorded, then
coarsened by a chosen map window and grid resolution.
\item[Target.] A baseline question about spatial randomness: is the observed
clustering unusual under a homogeneous chance allocation?
\item[Model.] A Poisson point process, or its finite-cell multinomial/Poisson
shadow, supplies the null distribution for cell counts.
\item[Uncertainty.] Tail probabilities and goodness-of-fit summaries compare
the observed frequency of empty, singly hit, and multiply hit cells with the
null.
\item[Limitation.] The capsule tests one coarse null; it does not reconstruct
military targeting, reporting error, blast physics, or spatial heterogeneity.
\end{realdatacapsule}

\begin{realdatacapsule}{PBMC3k single-cell counts}
\item[Data object.] A public 10x Genomics PBMC3k cell-by-gene count matrix
\citep{tenx2016pbmc3k}.
\item[Observation mechanism.] Cells are captured, lysed, barcoded, sequenced,
aligned, filtered, and summarized into molecular counts with zeros from both
biology and technology.
\item[Target.] Early chapters use the matrix to ask what a high-dimensional
biological record is before it becomes a trajectory, cluster, or fitted
generative law.
\item[Model.] Count models with overdispersion, zero structure, donor/batch
effects, and later simulation checks give possible readings of the matrix.
\item[Uncertainty.] Gene-level and cell-level summaries, bootstrap or
simulation variation, and sensitivity to preprocessing quantify which patterns
are stable.
\item[Limitation.] The fallback matrix has no real pseudotime or perturbation
design, so it supports count-matrix diagnostics but not a biological trajectory
claim.
\end{realdatacapsule}

\begin{realdatacapsule}{Autonomous laboratory log}
\item[Data object.] Time-stamped actions, sensor curves, robot logs, quality
flags, yields, failed runs, and acquisition decisions in a self-driving
laboratory \citep{burger2020mobile,abolhasani2023rise}.
\item[Observation mechanism.] The system observes only experiments the policy
chooses to run, with instrument failures, operator interventions, and
post-run quality filters.
\item[Target.] Learn a scientific response surface or policy value while
choosing future experiments adaptively.
\item[Model.] Bayesian optimization, active learning, or sequential
generative models map the current history to the next action.
\item[Uncertainty.] Posterior uncertainty, predictive intervals, exploration
diagnostics, and repeated-run variability decide whether action is justified.
\item[Limitation.] The data law depends on the policy that created the log, so
ordinary iid validation can overstate generality.
\end{realdatacapsule}

\section{Random Point Patterns and Uneven Spacing}
\label{sec:ch2-texture}
\conceptindexes{randomness!texture, random points, spatial randomness, clustering, gaps, point patterns}

This section is the reader's first sensory exercise.  It does not ask for a
formal probability space yet.  It asks for a change in expectation: randomness
is not a visual synonym for evenness.  Once that is learned, the rest of the
chapter can ask sharper questions about which features are genuinely rare.

\subsection{Random points are not evenly spaced}
\conceptindexes{random points, even spacing, nearest-neighbor distances, homogeneous Poisson process}
\label{subsec:ch2-random-points}

Imagine placing \(n\) points independently and uniformly in a square. The
model says that each small region receives points in proportion to its area, on
average. It does not say that each small region receives exactly the same
number of points. It certainly does not say that nearest-neighbor distances
will be nearly equal. A regular grid has equal spacing because it has been
designed. A random point pattern has accidental crowds and accidental silences.

\begin{example}[Clusters in a flat world]
Divide the unit square into \(m\) equal boxes \(A_1,\ldots,A_m\), and let
\[
  N_j=\sum_{i=1}^n\ind{X_i\in A_j}
\]
be the number of random points falling in box \(A_j\). If
\(X_1,\ldots,X_n\) are independent and uniform, then each \(N_j\) has the same
mean \(n/m\). Equal means, however, do not produce equal counts. Some boxes
will be empty, some will have several points, and two adjacent boxes can look
strikingly different. The eye sees a local story; the model asks whether that
story is more surprising than chance itself.
\end{example}

\begin{example}[Runs in a coin string]
Let \(X_1,\ldots,X_n\) be independent fair coin tosses, coded as \(0\) and \(1\).
A common human expectation is that a random string should alternate often. But
if
\[
  C_n=\sum_{i=2}^n \ind{X_i\ne X_{i-1}},
\]
then \(\Expect C_n=(n-1)/2\), not \(n-1\). Long local repetitions are also not
surprising. The expected number of length-\(r\) all-heads blocks is
\[
  (n-r+1)2^{-r}.
\]
For \(n=100\) and \(r=6\), this expectation is \(95/64\), about \(1.48\). Thus a
run such as HHHHHH is not, by itself, evidence that the coin was unfair. It is
one of the ordinary textures of independence.  Figure~\ref{fig:ch2-long-runs}
turns this point into the kind of visual feature a reader is tempted to
overinterpret.
\end{example}

\begin{center}
\centering
\includegraphics[width=0.96\linewidth]{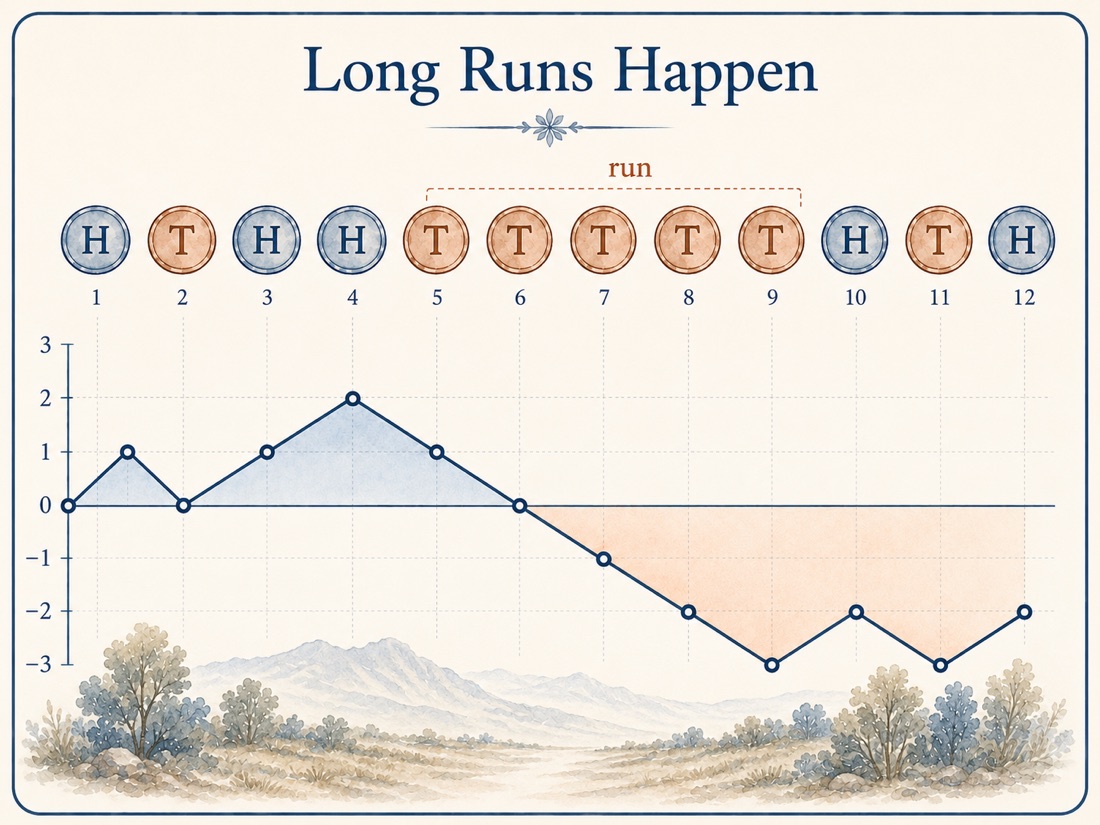}
\bookmanualfigure{fig:ch2-long-runs}{Long runs in a coin sequence}
\par\smallskip
\parbox{0.92\linewidth}{\small\textbf{Figure~\thefigure.} A schematic display
of long runs in a coin sequence.  The visual lesson is that repeated symbols and
large local excursions can be ordinary features of independence; a run is a
statistic to be evaluated under a model, not visual proof of bias.}
\end{center}

This distinction is the first statistical habit of the chapter. We should not
ask whether a display looks random. We should ask what random mechanism would
produce displays of this kind, and what features of the observed display would
be rare under that mechanism.

\section{Visible Patterns and Null Models}
\label{sec:ch2-visible-pattern}
\conceptindexes{visible pattern, mechanism, null model, spatial point process, London bombing}

This section turns the sensory lesson into a statistical question.  Once the
reader accepts that random points can cluster, the next temptation is to treat
every cluster on a meaningful map as a cause.  The London bombing story is
placed here to slow that impulse down.  A map with clusters feels intentional;
statistical thinking asks what kinds of clusters chance itself would commonly
produce.

\subsection{London bombing as a point-pattern lesson}
\conceptindexes{London bombing, point patterns, quadrat counts, Poisson approximation, spatial clustering}
\label{subsec:ch2-london-bombing}

The classic London bombing example is often retold because it is compact and
unsettling. The observed data are locations: points of impact on a map. The
question is not whether the map looks smooth. It is whether the point pattern
is closer to random falling or to deliberate targeting. Clarke's well-known
analysis divided a region of South London into \(24\times24=576\) equal
squares and recorded \(537\) flying-bomb impacts
\citep{clarke1946poisson}. The corresponding homogeneous Poisson idealization
sets
\[
  N_j\sim\Poisson(\lambda),
  \qquad
  \lambda=\frac{537}{576}\approx0.932,
\]
for the number of impacts in square \(j\). Hence the expected number of squares
with exactly \(k\) impacts is
\[
  E_k=576e^{-\lambda}\frac{\lambda^k}{k!},\qquad k=0,1,2,\ldots .
\]
Using the grouped last category \(k\ge5\), Clarke's observed frequencies and
the fitted Poisson frequencies are:
\begin{center}
\small
\textbf{Clarke cell-count diagnostic.}\par\smallskip
\setlength{\tabcolsep}{0.45em}
\begin{tabular}{@{}lrrrrrr@{}}
\toprule
Impacts in a square & \(0\) & \(1\) & \(2\) & \(3\) & \(4\) & \(\ge5\) \\
\midrule
Observed squares & 229 & 211 & 93 & 35 & 7 & 1 \\
Poisson fit & 226.74 & 211.39 & 98.54 & 30.62 & 7.14 & 1.57 \\
\bottomrule
\end{tabular}
\end{center}
As a rough diagnostic,
\[
  X^2=\sum_k\frac{(O_k-E_k)^2}{E_k}\approx1.17,
\]
with one fitted parameter and six grouped cells. This is not large on the
usual \(\chi^2_4\) scale. Feller later used the same data as a textbook example
of spatial Poisson behavior \citep[Ch.~VI, Sec.~7, pp.~160--161]{feller1968introduction}.

For this book, the historical details are less important than the statistical
eye trained by the example:
\begin{description}[leftmargin=0pt,labelsep=0.55em,style=unboxed,font=\bookdescriptionlabelfont,itemsep=0.25\baselineskip]
\item[Data structure.]
A spatial point pattern, or equivalently counts of points in selected regions.

\item[Perceptual trap.]
Nearby impacts feel like evidence of deliberate targeting, because the eye
treats a cluster as a story.

\item[Working assumption.]
As a null idealization, the local intensity is roughly constant over the region
and the cell counts behave approximately as independent Poisson fluctuations.

\item[Model map.]
Map of impacts \(\rightarrow\) cell counts \(N_j\) \(\rightarrow\) fitted
Poisson law \(\rightarrow\) comparison of observed and expected frequencies.

\item[Observable question.]
Is the observed amount of clustering, or a statistic such as \(X^2\), rare under
this spatial null?

\item[Statistical lesson.]
Clustering can be produced by randomness itself. A cluster is evidence only
after the typical clusters generated by the null mechanism have been described.
\end{description}

The Poisson point process will later give a formal language for this idea. At
this stage it is enough to keep the intuition: if points fall independently in
space with nearly constant intensity, disjoint regions have counts that
fluctuate, and those fluctuations make clusters and gaps. The surprise is not
that randomness creates pattern. The surprise is that visual intuition so often expects it not
to.

\section{Unobserved Types and Missing Mass}
\label{sec:ch2-unseen}
\conceptindexes{absence, unseen classes, missing species, singleton counts, sampling silence}

This section reverses the direction of attention.  The London example protects
the reader from overreading what is visible.  Missing species protects the
reader from underreading what is nearly invisible.  Sometimes the important
feature of the data is not an obvious pattern, but a structured absence. What is
barely visible can show something about what has not been seen at all.

\subsection{Fisher's missing species and Shakespeare's unused words}
\conceptindexes{missing species, unseen words, vocabulary problem, species sampling, Good--Turing idea}
\label{subsec:ch2-missing-species}

In the bombing example, the visible pattern tempts the analyst to infer too much
mechanism. In Fisher's missing-species problem, the danger is the opposite: the
sample shows only observed species, but the scientific question concerns the
population beyond the sample. How many species remain unseen? How much
probability mass belongs to types that did not appear?

Let \(f_r\) denote the number of species observed exactly \(r\) times in a
sample. The singletons \(f_1\), doubletons \(f_2\), and other low-frequency
counts are not statistical leftovers. They are the edge of the visible world.
Fisher, Corbet, and Williams used species-abundance data to show how these
rare counts contain information about the unseen part of an animal population
\citep{fisher1943species}. Good's later work made the same theme central in
frequency estimation: the number of things seen once is evidence about the
probability of seeing something new next \citep{good1953population}.

The same abstraction becomes especially vivid when species are replaced by
words. Efron and Thisted treated Shakespeare's known writings as a sample and
word types as species. Shakespeare's known works contained \(31{,}534\)
different words, including \(14{,}376\) words appearing once and \(4{,}343\)
appearing twice; the question was how many words he knew but did not use. Their
analysis concluded that the models were consistent with at least \(35{,}000\)
additional words beyond those observed
\citep{efronThisted1976unseen}. Efron and Hastie revisit this missing-species
problem as an empirical Bayes example in Chapter~6, Section~6.2 of
\emph{Computer Age Statistical Inference}
\citep[Ch.~6, Sec.~6.2]{efronHastie2016empiricalBayes}.

Here the same protocol has a different shape:
\begin{description}[leftmargin=0pt,labelsep=0.55em,style=unboxed,font=\bookdescriptionlabelfont,itemsep=0.25\baselineskip]
\item[Data structure.]
A sample of repeated labels, summarized by the frequency-of-frequency counts
\(f_1,f_2,\ldots\).

\item[Perceptual trap.]
The observed catalog looks like the world itself, so unseen types are easily
mistaken for nonexistent types.

\item[Working assumption.]
The sample is treated as repeated draws from a population of possible types,
with unknown type probabilities.

\item[Model map.]
Observed labels \(\rightarrow\) low-frequency counts \(f_r\) \(\rightarrow\)
estimated missing mass or unseen vocabulary.

\item[Observable question.]
How much probability, or how many types, may lie outside the observed sample?
\end{description}

A toy calculation explains why this problem is unavoidable. Suppose for a
moment that there are \(K\) equally likely types and that sampling is repeated \(n\) times
with replacement. For any fixed type,
\[
  \Prob\{\text{the type is unseen after \(n\) draws}\}
  =\left(1-\frac1K\right)^n\approx e^{-n/K}.
\]
Therefore
\[
  \Expect\{\text{observed types}\}
  =K\left\{1-\left(1-\frac1K\right)^n\right\},
  \qquad
  \Expect\{\text{unseen types}\}
  =K\left(1-\frac1K\right)^n .
\]
The discovery curve rises quickly at first and then slows, but slowing does not
mean exhaustion. In real species and vocabulary problems the type probabilities
are far from equal, which is exactly why the low-frequency counts
\(f_1,f_2,\ldots\) become informative.

\begin{center}
\itshape
The unseen leaves evidence in the seen.
\end{center}

That sentence is the chapter's second statistical habit. Data are not only a
list of observed cases. They are a pattern of frequencies, absences, repeats,
near-repeats, and rare events. A species never seen cannot be counted directly,
but the abundance of species barely seen can still speak on its behalf.

\section{Historical Traces and Latent Objects}
\label{sec:ch2-historical-traces}
\conceptindexes{historical traces, latent objects, proxy records, climate reconstruction, latent regions, stylometry}

This section raises the stakes.  The first two stories used data structures
that could be stated cleanly: spatial points and repeated labels.  Historical
and textual problems are less tidy.  The object of interest cannot be observed
again. Past climate cannot be remeasured by placing modern instruments in the
Han, Tang, Ming, or Qing dynasties.  The formation of historical economic
regions cannot be replayed under controlled conditions.  The authorship
history of a classical novel cannot be rerun as an experiment.  Yet these
questions leave traces: frost dates, lake freezing, flowering times, harvest
failures, water-control works, gazetteer entries, grain prices, famine reports,
function words, chapter lengths, phrase patterns, and verse/prose style.  The
statistical question is not whether these traces are the hidden object.  They
are not.  The question is how traces become evidence.

\subsection{The Zhu/Chu Ko-chen curve}
\conceptindexes{Zhu/Chu Ko-chen curve, climate proxies, latent temperature curve, historical climate reconstruction}
\label{subsec:ch2-zhu-climate}

Zhu Kezhen, often cited in English as Chu Ko-chen, assembled historical and
natural records into a long view of climate fluctuations in China
\citep{chu1973climatic}.  His curve is important here because it is not an
instrumental thermometer series in the modern sense.  It is a disciplined
translation from proxy records to an unobserved environmental process.  Modern
reconstructions of Chinese temperature over the past two millennia use richer
proxy archives and calibration methods, but they preserve the same statistical
logic: a latent climate signal is inferred from records whose creation,
survival, and interpretation are themselves part of the data-generating story
\citep{ge2013temperature,ge2016recent}.  Figure~\ref{fig:ch2-zhu-ke-zhen-curve}
is used to keep that proxy-to-latent-process map visible while introducing the
notation.

\begin{center}
\centering
\includegraphics[width=0.82\linewidth]{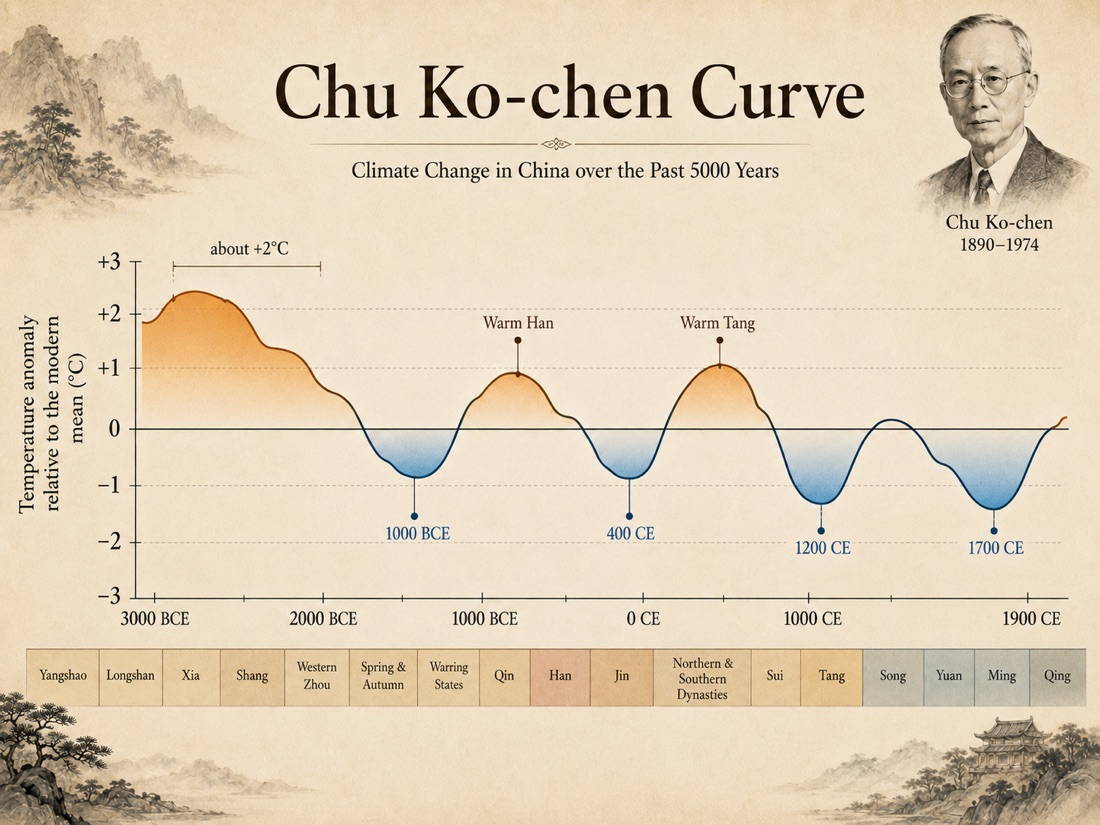}
\bookmanualfigure{fig:ch2-zhu-ke-zhen-curve}{The Zhu/Chu Ko-chen climate curve as proxy evidence}
\par\smallskip
\parbox{0.88\linewidth}{\small\textbf{Figure~\thefigure.} An illustrative
schematic of the Zhu/Chu Ko-chen climate curve.  The
image should be read as a pedagogical rendering of the proxy-to-latent-process
idea, not as a reproduction of the original historical data display.}
\end{center}

A compact formalization is
\[
  Y_{r,t}
  =
  a_r+b_r m(t)+\varepsilon_{r,t},
\]
where \(m(t)\) is the latent climate curve, \(Y_{r,t}\) is a proxy record of
type \(r\) at time \(t\), \(a_r\) and \(b_r\) encode proxy-specific level and
sensitivity, and \(\varepsilon_{r,t}\) contains measurement error, dating
error, local weather, and recording bias.  A flowering date, a frozen river, a
snow disaster, or a grain-price spike is therefore not climate itself.  It is
a filtered coordinate of a historical data system.

This example also links statistics to environmental history.  Environmental
historians of China and of the modern world study land use, forests, animals,
water control, population pressure, energy, state capacity, and ecological
change as interacting systems \citep{elvin2004retreat,marks2012china,mcneill2000something}.
For a statistician, their work is a warning against treating historical proxies
as neutral measurements.  A dynasty that recorded disasters more intensely, a
region with denser local gazetteers, or a state with stronger reporting
institutions can change the observation process.  The model must therefore ask
not only what the climate did, but also which parts of climate were allowed to
leave a legible trace.

\subsection{Basic economic areas as latent regions}
\conceptindexes{basic economic areas, latent regions, spatial regions, political economy}
\label{subsec:ch2-basic-economic-areas}

Chi Ch'ao-ting's \emph{Key Economic Areas in Chinese History} gives the chapter
a spatial-economic version of the same problem.  The object of interest is not
a temperature curve or an authorial signature, but a region: an economic center
whose importance is revealed through water-control works, river systems,
canals, transport costs, grain movement, fiscal capacity, military pressure,
and state investment \citep{chi1936key}.  The statistical lesson is that a
region need not be identical to an administrative boundary.  It can be a latent
spatial-economic structure inferred from many partial traces.

One compact encoding is
\[
  \{W_t,C_t,G_t,R_t\}
  \longrightarrow
  H_t
  \longrightarrow
  \Pi_t ,
\]
where \(W_t\) records water-control works, \(C_t\) transport and canal
connections, \(G_t\) grain, population, price, or fiscal traces, \(R_t\)
historical reports, \(H_t\) a spatial-economic field or network, and \(\Pi_t\)
a partition into basic economic areas at time \(t\).  The visible traces do not
directly announce the partition.  They make some partitions more plausible than
others.  Figure~\ref{fig:ch2-china-basic-economic-areas} renders the same
idea as a spatial picture: traces are not regions, but evidence from which
regional structure is inferred.

\begin{center}
\centering
\includegraphics[width=0.96\linewidth]{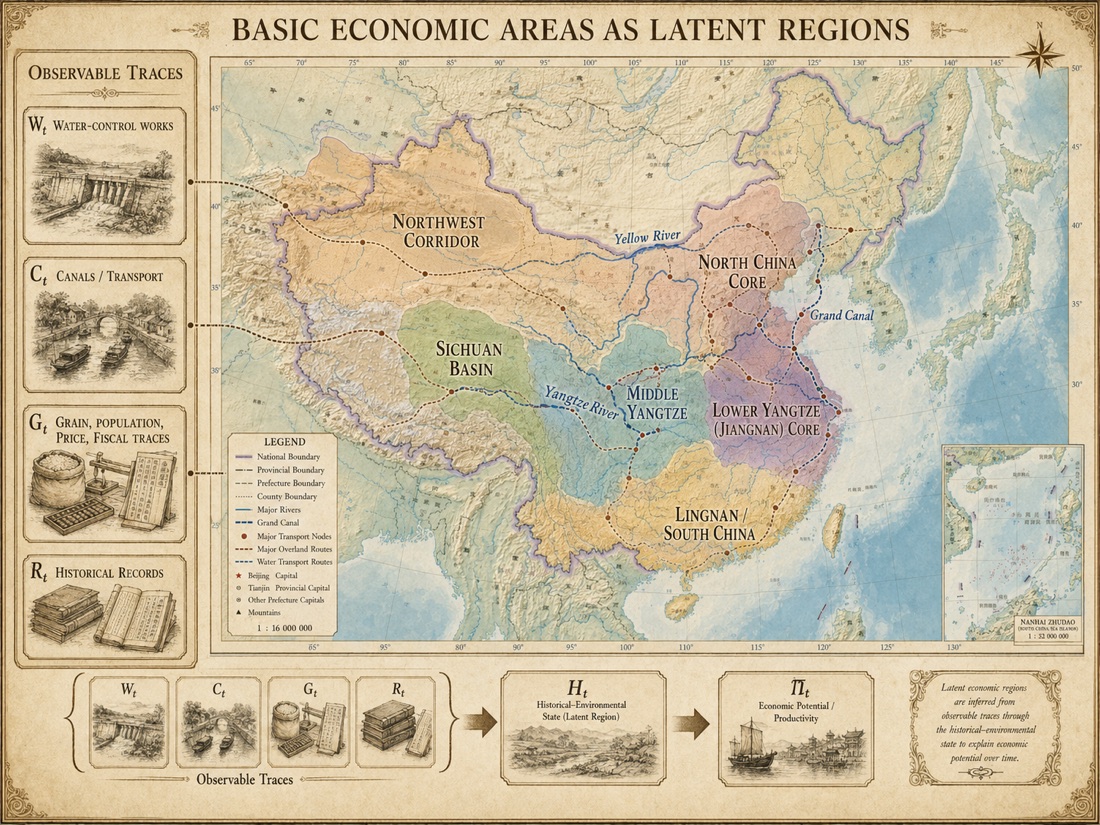}
\bookmanualfigure{fig:ch2-china-basic-economic-areas}{Basic economic areas as latent regions}
\par\smallskip
\parbox{0.92\linewidth}{\small\textbf{Figure~\thefigure.} A schematic rendering
of basic economic areas as latent regions.  Observable traces such as
water-control works, transport links, fiscal and grain records, and historical
reports are first read as a spatial-economic field \(H_t\), then summarized as a
partition \(\Pi_t\).  The picture is a schematic model rather than a
literal reconstruction of historical boundaries.}
\end{center}

This example is useful precisely because it resists a simple vector encoding.
It points toward graphs, spatial fields, partitions, and time-varying systems.
Skinner's later macroregional work can be read in the same statistical spirit:
regions are objects produced by markets, geography, transport, and records, not
labels that arrive fully formed before analysis begins
\citep{skinner1964marketing,skinner1977regional}.

\subsection{\emph{Dream of the Red Chamber} and stylometric traces}
\conceptindexes{Dream of the Red Chamber, stylometry, textual features, authorship attribution}
\label{subsec:ch2-red-chamber}

The authorship question around \emph{Dream of the Red Chamber} has a parallel
statistical form.  The hidden object is not temperature but authorship or style
regime.  The traces are chapter-level features: word frequencies, function
words, bigrams, character usage, sentence structure, prose/verse patterns, and
edition-dependent textual variants.  Quantitative studies do not turn this
literary problem into a mechanical verdict.  They ask a more careful
statistical question: do the first eighty chapters and the last forty chapters
look like samples from the same stylistic distribution, after one has chosen a
feature representation and a comparison rule?

One way to write the comparison is
\[
  \sup_{f\in\mathcal F}
  \left|
    P_{1:80}f-P_{81:120}f
  \right|,
\]
where \(P_{1:80}\) and \(P_{81:120}\) are empirical measures over chapters and
\(\mathcal F\) is a class of textual features.  Hu, Wang, and Wu use
quantitative feature analysis to study multiple-author evidence in the novel;
Tu and Hsiang frame the problem through text mining; and Zhu, Lei, and Craig
give a stylometric analysis that separates prose, verse, and authorship
evidence \citep{hu2014redchamber,tu2013redchamber,zhu2021redchamber}.  The
statistical lesson is not that a classifier settles Cao Xueqin versus Gao E.
The lesson is that authorship becomes a probability question only after text
has been encoded into observable features, and that every encoding leaves
something literary outside the model.

The climate, regional, and authorship examples are deliberately grouped.  Each
asks how traces become evidence:
\[
  \begin{gathered}
  \text{records, works, or chapters}
  \longrightarrow
  \text{features} \\
  \longrightarrow
  \text{latent climate, region, style, or authorship}.
  \end{gathered}
\]
Proxy records, public works, and chapter features are not the hidden object
itself.  They are the parts of the hidden object that survived measurement,
recording, editing, transmission, and representation.

\section{Modern Data Streams and Designed Observation}
\label{sec:ch2-modern-data-streams}
\conceptindexes{modern data streams, designed observation, multi-omics, autonomous laboratories, policy-dependent data}

This section is the final step before formal probability.  It shows that the
same statistical eye is needed in settings that look nothing like old maps or
historical archives.  Modern datasets are often wider, faster, and more
engineered: a cell is measured through several molecular technologies; a lab
robot chooses the next experimental condition after reading the previous
result; a platform logs not only outcomes but also the policy that produced
them.  These examples are not inserted to make the chapter fashionable.  They
make Chapter~1's constitution harder to evade.  Before probability begins, it is
necessary to know not only what was observed, but also how the observation was made and
which earlier observations were allowed to influence the next one.

\subsection{Multi-omics as aligned partial views}
\conceptindexes{multi-omics, aligned partial views, latent factors, omics layers}
\label{subsec:ch2-multiomics}

Multi-omics data turn the missing-species lesson into a high-dimensional
biological problem.  A sample may have RNA counts, chromatin accessibility,
protein abundance, metabolites, spatial coordinates, image features, donor
labels, treatment assignments, and time points.  These layers are not merely
extra columns.  Each layer has its own measurement scale, noise, censoring,
batch effects, missingness pattern, and biological resolution.  The scientific
target is often a latent state: a cell type, developmental trajectory,
disease state, pathway activity, or treatment response that is not equal to
any single assay.

A compact encoding is
\[
  X_i^{(m)} = g_m(Z_i) + \varepsilon_i^{(m)},\qquad
  R_i^{(m)}=\ind{\text{modality \(m\) observed}},
\]
where \(Z_i\) is a latent biological state, \(X_i^{(m)}\) is the observation in
modality \(m\), \(g_m\) is the modality-specific measurement map, and
\(R_i^{(m)}\) records whether that layer was observed for unit \(i\).  The
formal object is therefore not
\[
  \text{omics table} \longrightarrow \text{answer}.
\]
It is closer to
\[
  \begin{gathered}
  \{X_i^{(m)},R_i^{(m)}:m=1,\ldots,M\}
  \longrightarrow
  \text{aligned features}\\
  \longrightarrow
  Z_i\ \text{or a population distribution of }Z.
  \end{gathered}
\]
Methods such as multi-omics factor analysis make this latent-view logic
explicit by separating shared factors from modality-specific variation
\citep{argelaguet2018mofa}.  Synthetic-data tools such as scDesign3 push the
same lesson into design and validation: a simulator for multimodal single-cell
and spatial-omics data is useful only if it reproduces the summaries that a
downstream scientific question will actually use.

The perceptual trap is the opposite of the old tidy spreadsheet.  Because the
data are rich, one is tempted to believe that the hidden biology must already
be visible somewhere inside them.  But more modalities can also mean more
ways to be misaligned.  The statistical question is not ``Is there enough
data?''  It is: which coordinates are measurements, which coordinates are
design choices, which absences are technical, which absences are biological,
and which summaries are stable across donors, batches, modalities, and
analysis pipelines?

\subsection{Robot laboratories as adaptive data streams}
\conceptindexes{robot laboratories, adaptive data streams, self-driving laboratories, automated experimentation}
\label{subsec:ch2-robot-labs}

Autonomous laboratories make the time order visible.  In a classical dataset,
one may pretend that the observations arrived and then the analysis began.  A
robot laboratory refuses that fiction.  A model proposes a reaction condition
or compound, a robot executes it, sensors monitor the run, software checks
quality, and the next action is chosen from the growing record.  XtalPi's
public description of its autonomous laboratory platform places AI prediction,
robotic execution, data acquisition, and closed-loop experimental decision
making in one industrial workflow \citep{xtalpi2025autonomousLab}.  This is a
company-specific instance of a broader self-driving-lab movement in chemistry
and materials science \citep{burger2020mobile,abolhasani2023rise}.

The probability object is a growing history.  At this point, ``history'' means
an ordered record, not yet a sigma-field.  At experimental round \(t\), write
\[
  H_t
  =
  \{(A_s,Y_s(\cdot),L_s,Q_s):1\le s\le t\},
\]
where \(A_s\) is the chosen action or experimental condition, \(Y_s(\cdot)\)
is a sensor curve or continuously monitored readout, \(L_s\) records robot and
laboratory logs, and \(Q_s\) records quality-control flags, failed runs, or
human interventions.  The next action is not drawn from nowhere.  It is chosen
by a policy
\[
  A_{t+1}=\pi_t(H_t),
\]
possibly with randomization or exploration.  A final scalar response, if one
is reported, may be only a compression
\[
  Z_{t+1}=h\{Y_{t+1}(\cdot),L_{t+1},Q_{t+1}\}
\]
of a much richer observed path.

This example is valuable because it makes several later chapters feel
necessary before they arrive.  Chapter~4 will treat the action \(A_{t+1}\) as
design, not as a passive covariate.  Chapter~6 will need product spaces,
kernels, and process laws to describe histories.  Chapter~7 will give a common
grammar for modern data structures, Chapter~8 will ask which target a
closed-loop record is meant to support, Chapter~9 will ask which single
summaries remain stable, Chapter~10 will explain what it means for histories
and empirical objects to converge weakly, and Chapter~11 will ask which
summaries remain stable when a model searches over many candidates.
Chapters~13--15 will turn closed-loop objectives into estimating criteria,
local tests, and perturbation questions.  Chapter~16 gives filtrations,
stopping times, intensities, and martingales their natural home, and
Chapter~17 returns the loop to use, deployment, and feedback.  The laboratory
becomes a stochastic process; the statistical eye asks what was observable
just before each experimental choice.

\section{From Examples to Formal Objects}
\label{sec:ch2-two-stories}
\conceptindexes{formal objects, observable questions, stress tests, recurring examples}

This final section tells the reader what to carry forward.  The examples do
different work, but they teach the same discipline: an example is useful when
it exposes the question hidden inside a data structure.  The point is not to
remember every story.  The point is to know which kind of formal language each
story is asking for.  Table~\ref{tab:ch2-examples-demands} makes this contract
explicit: Chapter~2 is not a collection of anecdotes, but the problem bank that
generates the book.

\begingroup
\footnotesize
\setlength{\LTpre}{0.45\baselineskip}
\setlength{\LTpost}{0.45\baselineskip}
\setlength{\tabcolsep}{0.35em}
\renewcommand{\arraystretch}{1.04}
\begin{longtable}{@{}>{\raggedright\arraybackslash}p{0.22\linewidth}
                >{\raggedright\arraybackslash}p{0.34\linewidth}
                >{\raggedright\arraybackslash}p{0.34\linewidth}@{}}
\caption{Examples as demands on statistical language}
\label{tab:ch2-examples-demands}\\
\toprule
\textbf{Example} & \textbf{Problem it forces} & \textbf{Later language} \\
\midrule
\endfirsthead
\caption[]{Examples as demands on statistical language (continued)}\\
\toprule
\textbf{Example} & \textbf{Problem it forces} & \textbf{Later language} \\
\midrule
\endhead
London bombs / point pattern &
Random points are not evenly spaced; visible clustering is not automatically a
mechanism. &
Events, spatial null models, Poisson counts, point processes, and spatial
randomness. \\
Missing species and unseen words &
Absence is information; the unobserved can leave a trace in low-frequency
counts. &
Frequency-of-frequency summaries, missing mass, sampling laws, empirical
distributions, and unseen-species estimation. \\
Zhu/Chu Ko-chen curve &
Proxy records are not the latent temperature curve; traces must be aligned,
smoothed, and calibrated. &
Random curves, latent processes, smoothing, inverse problems, and functional
data. \\
Chi's basic economic areas &
Administrative borders need not be economic regions; a region may be a latent
spatial-economic object. &
Spatial fields, networks, partitions, graphical structure, and historical
measurement. \\
\emph{Dream of the Red Chamber} stylometry &
Text becomes data only after features are chosen; a classifier is not by itself
a literary verdict. &
High-dimensional features, empirical processes, text-as-data objects, and
model-based evidence. \\
Multi-omics and spatial omics &
Aligned partial views are not one clean table; missing layers may be technical,
biological, or design-induced. &
Random elements, product spaces, latent states, batch effects, simulation, and
high-dimensional empirical objects. \\
Autonomous laboratory / robot lab &
The observed record is adaptive; later data are partly produced by earlier
decisions. &
Growing histories \(H_t\), kernels, sequential design, policies, feedback, and
later filtrations. \\
Clinical and RWE records &
The observation law is not the target law; endpoints, censoring, eligibility,
and treatment choice shape the claim. &
Design, observation mechanisms, estimands, target trials, censoring, survival
objects, and sensitivity analysis. \\
\bottomrule
\end{longtable}
\endgroup

This table is also a contract with the rest of the book.  A returning example
should not merely decorate a theorem.  It should move through a repeatable
application protocol:
\[
\begin{aligned}
  \text{early story}
  &\longrightarrow
  \text{formal object}
  \longrightarrow
  \text{empirical summary}\\
  &\longrightarrow
  \text{computational action}
  \longrightarrow
  \text{substantive return}.
\end{aligned}
\]
For the multi-omics row, this means that later chapters should tell the reader
how to move from a layered assay to a product space, from a product space to a
target such as a latent cell-state distribution or a pseudotime trend, from the
target to a fitting or simulation procedure such as scDesign3 or scGTM, and from
the fitted object back to a biological statement with uncertainty.  The same
rule applies to historical climate curves, text authorship, bombing maps, and
adaptive laboratories: each example has to earn its return by showing a new
piece of the translation.

The first story says: do not mistake a visible cluster for a mechanism until
you have imagined randomness carefully enough. The second says: do not mistake
absence from the sample for absence from the world. The historical examples
add a third habit: do not mistake a surviving trace for the hidden process
itself. The modern data-stream examples add a fourth: do not forget that data
may be shaped by design, software, and previous decisions. Together they
prepare the reader for the measure-theoretic language of the next chapter.
Probability is not introduced there merely to decorate familiar calculations.
It gives a precise way to say which events are observable, which random
objects are being measured, which information histories are available, and
which features of data should count as evidence.

\subsection{What these examples force the analysis to define}
\conceptindexes{events, measurable summaries, random object, working model}
\label{subsec:ch2-formal-pressure}

The formal pressure sits just beneath the table. The London example cannot be
studied seriously until clustering has been described as an event and its
probability under a spatial null. The missing-species and Shakespeare examples
cannot be studied seriously until low-frequency counts are treated as random
summaries carrying information about unseen probability mass. The climate and
regional examples cannot be studied seriously until proxy records, public
works, flows, and partitions are treated as summaries of latent processes. The
authorship example cannot be studied seriously until text features are treated
as random summaries of a latent style regime. Multi-omics cannot be studied
seriously until modality-specific measurements, missing layers, and latent
states are separated. Autonomous labs cannot be studied seriously until the
available information before each action is named. Each story therefore forces
a piece of formal language into view:
\begin{description}[leftmargin=0pt,labelsep=0.55em,style=unboxed,font=\bookdescriptionlabelfont,itemsep=0.25\baselineskip]
\item[Possible worlds.]
What are the possible maps, samples, or datasets that could have been observed?

\item[Observable events.]
Which features of those worlds can be decided from the recorded data, such as
``at least this much clustering'' or ``exactly \(f_1\) types appeared once''?

\item[Random summaries.]
Which functions of the data are being studied: cell counts, nearest-neighbor
distances, low-frequency species or word counts, or estimates of missing mass?

\item[Assumptions.]
Which simplification makes the data structure analyzable without requiring the analyst to
pretend that the simplification is literally the world?

\item[Probability laws.]
Which mechanism assigns probabilities to those events and summaries, and what
does it make rare or common?

\item[Limits and refinements.]
What happens when the partition gets finer, the sample grows, or the observed
object becomes a path, process, or high-dimensional array?

\item[Information histories.]
If decisions are made sequentially, what information was available before each
decision, and which later data were generated because of that earlier
information?
\end{description}

Chapter~3 begins where this list becomes grammar. Its first act is not to
define probability, but to decide which features of possible worlds are allowed
to count as events. That is why sigma-fields appear before probability
measures. A sample space lists possible worlds. A sigma-field records the
features treated as observable. Random variables are measurable summaries of
those worlds. Probability measures assign mass to observable events. The
formalism is not ceremony; it is what keeps the examples from becoming vague
intuition.

\subsection{Examples as stress tests}
\conceptindexes{stress tests, assumptions, data structure, model checking}
\label{subsec:ch2-examples-stress-tests}

Examples in this book are not decorations. A good example should reveal one
statistical problem sharply enough that the abstraction becomes necessary.
London bombing reveals the problem of reading spatial randomness.
Missing species and Shakespeare's unused words reveal the problem of learning
about the unobserved from the pattern of the observed. The Zhu/Chu Ko-chen
curve, Chi's basic economic areas, and \emph{Dream of the Red Chamber} add
historical, regional, and textual versions of the same problem: surviving
traces must be encoded before they can support a probabilistic claim. Later
multi-omics examples will do the same for layered biological measurements,
simulation, and high-dimensional empirical fields. Robot laboratories will do
it for filtrations, adaptive design, stopping times, likelihoods, empirical
processes, hazards, martingales, and high-dimensional prediction. The point is
not to make the mathematics lively after the fact. The point is to make the
mathematics answer to something.

The short cameo examples in the opening atlas have the same discipline but a
faster rhythm.  A recommender log makes absence behavioral: the movie not rated
is part of the observation law.  An image benchmark turns pixels into a
negotiated object: taxonomies, boxes, masks, captions, annotators, and
evaluation metrics decide what the machine is allowed to see.  ADNI and A4 make
disease progression a protocol of visits, eligibility rules, imaging schedules,
biomarkers, endpoints, and censoring rather than a single medical fact.  Polls
separate noisy samples from comparatively stable vote totals.  RiskMetrics
turns market history into time-indexed covariance, tail loss, and decision
rules.  Wasserstein examples make the observed unit an entire distribution, not
a scalar.  These cameos are useful precisely because they do not require a new
chapter here.  They teach the reader to ask, quickly and repeatedly: what was
observed, what was absent by design, what target is being claimed, and which
formal object must be invented next?

\subsection{How these examples return}
\conceptindexes{recurring examples, chapter threads, object-level inference}
\label{subsec:ch2-examples-return}

The examples in this chapter are also planted so that later technical chapters
can be read as answers to concrete pressures rather than as formal
interruptions. Each example returns when a piece of mathematical language
becomes unavoidable.

\begin{description}[leftmargin=0pt,labelsep=0.55em,style=unboxed,font=\bookdescriptionlabelfont,itemsep=0.25\baselineskip]
\item[Events and measurable summaries.]
The London example asks whether ``this much clustering'' is rare; the
missing-species example asks what can be inferred from ``exactly \(f_1\)
singletons.'' Chapter~3 turns such phrases into events, measurable random
variables, probability laws, null sets, and limits.

\item[Observation before inference.]
The climate, regional, authorship, multi-omics, and robot-lab examples show
that data do not arrive as pure mathematical objects. They are collected,
recorded, filtered, edited, encoded, and sometimes generated by an adaptive
policy. Chapter~4 returns to this point through design, randomization,
sequential rules, and information.

\item[Products, kernels, and process laws.]
Random points, repeated word draws, proxy records over time, spatial-economic
networks, partitions, chapter-level features, multi-omics layers, and robot
histories all require probability laws for many coordinates at once. Chapter~6
develops the product-space and stochastic-process architecture that lets these
many-coordinate objects be built and transformed.

\item[Objects, targets, stability, and empirical fields.]
The missing-species, stylometric, and multi-omics examples all ask what kind of
empirical object has been observed, what target it is meant to support, and
when its summaries are stable enough to carry a general claim. Chapter~7
organizes data structures by object, observation mechanism, empirical object,
structure, target functional, and decision. Chapter~8 names the target layer.
Chapter~9 studies stability for averages, tails, quantiles, and concentration.
Chapter~10 supplies the language of weak convergence for random objects;
Chapter~11 extends the question from one statistic to whole indexed fields of
features.

\item[Criteria, local noise, and time.]
Once features have been chosen, many later procedures become optimization or
estimating-equation problems, the domain of Chapter~13. Chapter~14 then asks
how nearby probability laws are separated by tests, and Chapter~15 asks how
small perturbations in the data move those estimates. Robot laboratories add
the problem of adaptive time order. The London point pattern and the
robot-lab history both return in Chapter~16, where events in time, marks,
filtrations, intensities, martingales, and compensators give dynamic data
their grammar.
\end{description}

The reader should therefore keep a memory of the examples while moving forward.
When a later chapter introduces a sigma-field, a kernel, an empirical process,
an \(M\)-estimator, an influence function, or a compensator, the question is
not merely ``what is the definition?'' but ``which data structure needed this
language in the first place?''

\section{Exercises}
\conceptindexes{exercises, randomness exercises, point-pattern exercises, missing-species exercises}
\label{sec:ch2-exercises}

\begin{exercise}[Clusters and gaps]
Simulate, sketch, or mentally construct \(100\) independent uniform points in
the unit square. Divide the square into \(25\) equal boxes. Which features
would make the display look nonrandom to a casual viewer? Which of those
features might be common under independent sampling?
\end{exercise}

\begin{exercise}[A point-pattern question]
Suppose accident locations are recorded along a city street network. State a
null model under which clustering can occur by chance, and name one feature
that would make you suspect a real spatial mechanism.
\end{exercise}

\begin{exercise}[Fingerprints of the unseen]
In a sample of \(n\) organisms or words, let \(f_1\) be the number of types
observed once and \(f_2\) the number observed twice. Explain why these counts
may be more informative about unseen types than the counts of very common
types.
\end{exercise}

\begin{exercise}[A simple discovery curve]
Suppose \(K\) equally likely types are sampled independently \(n\) times. Find
the expected number of observed types and the expected number of unseen types.
For what approximate value of \(n\) is half of the population still unseen?
\end{exercise}

\begin{exercise}[Proxy traces]
Choose either a historical climate proxy, a water-control or transport trace
for an economic region, or a textual feature from a novel. Write a model of the
form
\[
  \text{observed trace}
  =
  \text{latent signal}
  +
  \text{measurement or recording error}.
\]
Name one source of bias that would not disappear merely by increasing the
number of traces.
\end{exercise}

\begin{exercise}[Multi-omics as partial views]
Suppose each biological unit may have RNA, protein, and spatial measurements,
but not all modalities are observed for all units. Write down \(X_i^{(m)}\),
\(R_i^{(m)}\), and one latent state \(Z_i\) you might want to learn. Name one
missingness mechanism that would make naive integration misleading.
\end{exercise}

\begin{exercise}[Robot-lab history]
In a closed-loop experiment, let \(A_t\) be the condition chosen at round \(t\)
and \(Y_t(\cdot)\) the sensor path recorded during the run.  Define a
reasonable history \(H_t\).  Which objects belong to the history before the
choice of \(A_{t+1}\), and which should only enter after round \(t+1\)?
\end{exercise}

\begin{exercise}[From example to formal language]
Choose either the bombing example or the missing-species example. Name the data
structure, one working assumption, one possible sample space, one observable
event, one random summary, and one probability law that could be used to study
the example.
\end{exercise}

\begin{exercise}[Examples as stress tests]
Choose one example from your own field. Describe the observed data structure,
the hidden question, and the abstraction that the example forces you to learn.
\end{exercise}

\begin{exercise}[A cameo without a detour]
Choose one cameo example from the opening atlas, such as Netflix, ImageNet,
ADNI/A4, polls and votes, RiskMetrics, or a Wasserstein distributional example.
Write three short lines: the data object, the tempting but wrong shortcut, and
the later formal language that the example demands.
\end{exercise}

\section*{Sources and Further Reading}
\addcontentsline{toc}{section}{Sources and Further Reading}

Clarke's flying-bomb analysis is the classical reference for the London
point-pattern story \citep{clarke1946poisson}. Feller's presentation in
Chapter VI, Section 7 of the third edition of Volume I helped make the example
a standard probability-textbook illustration of spatial Poisson variation
\citep[pp.~160--161]{feller1968introduction}. It should be read here as an
entry point into spatial randomness, not as a complete historical account of
wartime London. The species-sampling discussion begins with
\citet{fisher1943species}; \citet{good1953population} is the standard anchor
for the Good--Turing view of frequency estimation and missing mass.
\citet{efronThisted1976unseen} translated the missing-species question into
Shakespeare's vocabulary, and \citet[Ch.~6, Sec.~6.2]{efronHastie2016empiricalBayes}
place the same example inside the empirical Bayes development of
\emph{Computer Age Statistical Inference}. Zhu/Chu Ko-chen's climate
reconstruction is cited here as a proxy-data example rather than as an
instrumental series \citep{chu1973climatic}; modern reconstructions and reviews
include \citet{ge2013temperature} and \citet{ge2016recent}. The environmental
history references \citet{elvin2004retreat}, \citet{marks2012china}, and
\citet{mcneill2000something} are used to emphasize that proxy archives are
created by ecological and institutional systems. Chi Ch'ao-ting's
\emph{Key Economic Areas in Chinese History} is used here as an early
historical example of region as a spatial-economic object inferred from water
control, transport, and political economy \citep{chi1936key}; a convenient
Chinese translation is \citet{chi2014basicChinese}. Skinner's
macroregional studies provide a later related reference point
\citep{skinner1964marketing,skinner1977regional}. For \emph{Dream of the Red
Chamber}, the statistical references are \citet{hu2014redchamber},
\citet{tu2013redchamber}, and \citet{zhu2021redchamber}. The multi-omics
discussion uses \citet{argelaguet2018mofa} for the latent-factor view of
integrated omics layers and scDesign3 for simulation of multimodal single-cell
and spatial-omics data. The autonomous-laboratory
example is anchored by \citet{burger2020mobile} and
\citet{abolhasani2023rise} on self-driving laboratories, with
\citet{xtalpi2025autonomousLab} used as a contemporary industrial example of
an AI-and-robotics laboratory platform. The opening cameo atlas cites the
Netflix Prize \citep{bennett2007netflix}, ImageNet and MS COCO
\citep{deng2009imagenet,lin2014microsoft}, ADNI and the A4 solanezumab trial
\citep{petersen2010adni,sperling2023solanezumab}, presidential polling
\citep{gelman1993polls}, RiskMetrics and classical stock-return behavior
\citep{jpmorgan1996riskmetrics,fama1965behavior}, and distribution-valued
regression examples in Wasserstein or Frechet geometry
\citep{petersen2019frechet,chen2023wasserstein}. This chapter treats these
examples as
stress tests for the next chapter's formal language: sample spaces, observable
events, random variables, probability measures, information histories, and
limits.

%% file: chapters/ch04_design_data_collection.tex
\chapter{Before the Data Arrive: Design, Randomization, and Information}
\label{chap:design-data-collection}
\conceptindexes{design, randomization, information, data collection, design-indexed law, optimal design, adaptive design}

\begin{tcolorbox}[
  enhanced,
  breakable,
  colback=chaptercream,
  colframe=bookblue!88!black,
  boxrule=0.72pt,
  arc=5pt,
  boxsep=1pt,
  left=1.0em,
  right=0.95em,
  top=0.82em,
  bottom=0.82em,
  before skip=0.55\baselineskip,
  after skip=1.0\baselineskip
]
\noindent\textbf{Chapter overview.}
This is the data-collection chapter: it asks how a question is turned into
records that later probability models are allowed to read. A design assigns
interventions, chooses observation locations, fixes labels and endpoints, or
decides the next action from accumulated information. Across clinical trials,
pharmacology, AIDD \citep{vamathevan2019applications}, computer-vision
datasets, single-cell genomics, simulation, and biomedical measurement, the
point is the same: design decides what random object the later analysis can
read.
\end{tcolorbox}

A dataset can look innocent after it arrives.  It has rows, columns, time
stamps, treatments, doses, endpoints, visits, inclusion rules, and missing
values.  But those features were not created by probability alone.  Someone
decided which population could enter, when measurements would be made, what
counted as toxicity, how a dose could be escalated, which comparison deserved
randomization, which labels or categories were allowed to exist, and what
information was too costly or too unethical to collect.
The observed empirical measure \(P_n\) is therefore never merely a pile of
points.  It is a record of choices.

This chapter gives those choices a statistical language.  If \(d\) denotes a
design, then the law of the data is more honestly written as
\[
  X\sim P_{\theta,d}.
\]
The parameter \(\theta\) describes the scientific mechanism, but the design
\(d\) describes the window through which the mechanism is observed.  A bad
window can make a simple world look confusing.  A good window can make a
complicated world answer a focused question.

The lesson is especially sharp in clinical trials.  In a randomized comparison,
the design creates the comparison before a regression model is fitted.  In a
phase I dose-finding study, the design is an ethical policy for learning under
danger.  In optimal design, the design is a geometry problem: choose the
locations and weights of observations so that the information matrix has the
right shape.  In pharmacology, it can decide when blood samples are taken; in
AIDD, which molecule is worth synthesizing next; in computer vision, which
ontology, labeling pipeline, benchmark split, and metric turn images into a
learning problem; in single-cell work, how many donors, time points, batches,
modalities, and cells are enough to separate a biological signal from technical
noise.  These cases look different, but they are all versions of the same act.
From Neyman's finite-population view of randomized experiments to
Kiefer--Wolfowitz information geometry and Wang--Fang uniform design, design
turns a question about the world into data that can bear the weight of a model
\citep{neyman1923agricultural,kiefer1960equivalence,wangFang1979uniform}.

\section{Design as the First Model}
\label{sec:ch04-design-first-model}
\conceptindexes{design!as first model, design-indexed law, observed-data map, information before data}

Suppose the statistician starts with a world question: Does a treatment help?
Which dose is tolerable?  Where should sensors be placed?  Which biomarker
trajectory should be measured densely?  A design answers the practical
question that comes before estimation:
\[
  \text{What data will be allowed to exist?}
\]

\begin{definition}[Design-indexed law]
Let \(\mathcal D\) be a class of possible designs.  For each
\(d\in\mathcal D\) and parameter \(\theta\in\Theta\), let
\[
  P_{\theta,d}
\]
denote the probability law of the observed data under parameter \(\theta\) and
design \(d\).  A design-indexed model is the family
\[
  \mathcal P_{\mathcal D}
  =
  \{P_{\theta,d}:\theta\in\Theta,\ d\in\mathcal D\}.
\]
\end{definition}

In this chapter the design index \(d\) has three recurring faces:
\[
  d=(d_{\mathrm{assign}},d_{\mathrm{observe}},d_{\mathrm{adapt}}).
\]
The assignment part determines who can receive which intervention and with what
randomization law.  The observation part determines visits, sensors, sampling
locations, ontologies, annotation protocols, endpoint maps, quality-control
filters, and missingness rules.  The adaptive part determines
how future actions or stopping decisions depend on past information.  Not every
study uses all three pieces, but every example below is meant to be read through
this same design-indexed law \(P_{\theta,d}\).

The notation is simple, but it prevents a common mistake.  If the observed data
come from a trial in which only high-risk patients were eligible, then the law
of the data is not the same as the law in a general clinic.  If an adaptive
trial assigns later doses using earlier toxicity outcomes, then the joint law
of the data is not the same as iid sampling from a fixed dose.  If a spatial
study samples only near roads, then the sampling design is part of the
probability law.  The design is not an administrative detail; it is part of the
mathematics.  Fisher's experimental tradition made this point operational:
randomization, blocking, and replication are not afterthoughts but devices that
make a scientific comparison interpretable \citep{fisher1935design}.

\begin{example}[A design changes the estimand]
Consider a binary treatment and potential outcomes \(Y(1)\) and \(Y(0)\).
If a randomized trial enrolls patients from a narrow eligibility region
\(\mathcal E\), a natural estimand is
\[
  \tau_{\mathcal E}
  =
  \Expect\{Y(1)-Y(0)\mid X\in\mathcal E\}.
\]
The same treatment in a registry with broad inclusion may point toward
\[
  \tau_{\mathrm{pop}}
  =
  \Expect\{Y(1)-Y(0)\}.
\]
The estimator may still be a difference in means, but the design has changed
what the difference is trying to estimate.
\end{example}

\begin{example}[One design question in several laboratories]
The question ``What data will be allowed to exist?'' changes its surface form
across scientific settings.
\begin{description}[leftmargin=0pt,labelsep=0.65em,style=unboxed,font=\bookdescriptionlabelfont,itemsep=0.25\baselineskip]
\item[Medicine and biotech.]
An adaptive clinical trial may drop an inferior arm, re-estimate sample size, or
change allocation probabilities at an interim analysis, but only if those
changes are built into the design before the trial begins
\citep{bhatt2016adaptive}.  In an early-development biotech program, the same
logic appears as a go/no-go design: the protocol must say which biomarker,
endpoint, interim evidence threshold, and safety signal are strong enough to
justify moving a program forward.

\item[Toxicology.]
A concentration-response experiment must choose doses that are not all too low
and not all too high.  The modern toxicology agenda explicitly links testing
strategies to value of information, chemical prioritization, and reduced animal
use \citep{nationalresearchcouncil2007toxicity}.

\item[Pharmacology.]
A population pharmacokinetic study often cannot draw a full blood-sampling
curve from every subject.  Sparse sampling is therefore a design problem:
choose visit times and subject groups so that clearance, exposure, and
variability remain estimable \citep{mentre1995sparse,mentre1997optimal}.

\item[AIDD.]
In artificial intelligence-driven drug discovery, an active-learning screen
treats the next assayed or synthesized compound as a design action.  The model
is useful only if its uncertainty is allowed to steer future experiments, not
merely rank compounds after the budget is gone
\citep{vamathevan2019applications,reker2017active,shields2021bayesian}.

\item[Computer vision.]
ImageNet made visual recognition trainable at scale by designing a data
infrastructure: a label ontology, an image-search and collection pipeline,
human-verification rules, benchmark splits, and evaluation tasks
\citep{deng2009imagenet}.  A later neural network did not see ``the visual
world'' directly.  It saw the world through this designed benchmark, whose
classes and labels shaped what counted as progress.

\item[Single-cell biology.]
Before sequencing, the design has already allocated budget among donors,
batches, perturbations, time points, spatial locations, modalities, and cells.
The scDesign series is useful here because it lets analysts stress-test these
choices on realistic in silico data before spending the real experiment budget
in the lab \citep{sun2021scdesign2,song2024scdesign3}.
\end{description}
The fields differ, but the statistical grammar is the same: define the target,
decide what can be observed, and write the rule that turns uncertainty into the
next measurement or action.
\end{example}

\begin{example}[What changed in the dataset?]
Imagine two studies of the same antihypertensive drug.  Study A enrolls adults
from primary-care clinics and measures systolic blood pressure after eight
weeks.  Study B enrolls only patients with chronic kidney disease, excludes
people with unstable medication histories, and measures the same endpoint after
four weeks.  Both datasets may contain a column called ``change in systolic
blood pressure.''  They do not answer the same question.  Eligibility, timing,
and endpoint definition have already moved the target before any model sees the
rows.
\end{example}

\begin{example}[CGM endpoints as measurement design]
Continuous glucose monitoring (CGM) is a useful example because the design does
not only choose who receives which treatment; it also chooses the observed
endpoint object.  Let \(G_i(t)\) be the latent glucose trajectory for subject
\(i\), and let \(d\) specify the device, wear window \(W_d\), sampling grid,
valid-reading rule, and minimum coverage requirement.  The observed CGM object is
not the whole path but
\[
  O_i(d)
  =
  \{G_i(t):t\in W_d,\ R_i(t;d)=1\},
  \qquad
  N_i(d)=|O_i(d)|,
\]
where \(R_i(t;d)\) records whether the reading at time \(t\) is usable under the
protocol.  A scalar endpoint applies a map \(\phi\) to this object, for example
\(\phi(O_i)=N_i(d)^{-1}\sum_{g\in O_i(d)}g\) for mean glucose, or a
time-in-range proportion.

Glucodensity changes the endpoint map.  Instead of reducing \(O_i(d)\) to one
number, it estimates the subject-level glucose distribution
\[
  \widehat F_i(g)
  =
  \frac{1}{N_i(d)}
  \sum_{u\in O_i(d)}\mathbf 1\{u\le g\},
  \qquad
  \widehat f_i(g)
  =
  \int K_h(g-u)\,d\widehat F_i(u),
\]
or an equivalent smoothed density object \citep{cui2023glucodensity}.  The
estimand is therefore no longer merely a contrast of scalar means; it may be a
contrast of distribution-valued summaries, such as
\[
  \Delta_\Psi
  =
  \Expect\{\Psi(F_i^{(1)})-\Psi(F_i^{(0)})\},
\]
for a clinically chosen functional \(\Psi\), or a distance between glucose
distributions.  For example, \(\Psi\) might extract time below range, tail mass,
a variability index, or a Wasserstein distance from a target profile.  This
example belongs here not as a randomized-treatment-design result, but as
endpoint construction: the design \(d\) determines which random object the later
clustering, comparison, or model is allowed to read.
\end{example}

\begin{example}[Famine relief as a designed information system]
Historical famine relief in China is a design problem before it is an analysis
problem.  A relief decision depended on what local officials reported, when
reports arrived, how harvest failure, grain prices, migration, mortality, and
granary stocks were recorded, and which signals triggered opening granaries,
tax remission, transport, or direct aid.  The state did not observe ``famine''
as a clean variable.  It observed a filtered stream of administrative traces.
Studies of Qing famine bureaucracy and the civilian granary system make clear
that reporting channels, incentives, logistics, and storage institutions were
part of the relief data system itself
\citep{will1990bureaucracy,will1991nourish,li2007fighting}.

In the notation of this chapter, a relief design \(d\) includes the reporting
calendar, the administrative levels through which information passes, the
price and granary variables monitored, and the decision rules for action.  The
law \(P_{\theta,d}\) therefore includes both the environmental shock
\(\theta\) and the observation system that turns drought, crop failure, and
prices into official evidence.  This is why historical relief belongs in a
statistics book: it shows that data collection, delay, missingness, and
decision are one system.
\end{example}

The rest of the chapter keeps returning to the same three mathematical objects.
The symbol \(d\) denotes the concrete design rule.  In optimal-design sections,
an approximate design is written as a probability measure
\[
  \xi=\sum_{j=1}^k w_j\delta_{x_j}
\]
on a design space \(\mathcal X\).  In adaptive sections,
\(\mathcal F_k\) denotes the information available after stage \(k\).  Thus the
same principle appears as a law \(P_{\theta,d}\), an information geometry
\(M(\xi)\), or an adapted rule with \(A_{k+1}\) \(\mathcal F_k\)-measurable.

\section{Randomization and Design-Based Guarantees}
\label{sec:ch04-randomization}
\conceptindexes{randomization, finite-population view, treatment assignment, causal comparison, potential outcomes}

Randomization is often introduced as a way to remove bias.  That is true, but
too thin.  Randomization is a promise about what is allowed to depend on what.
It says that the treatment assignment is produced by the design, not by a
clinician's hunch, a patient's prognosis, or an analyst's preference.  The
promise is mathematical: the assignment mechanism is known, and the comparison
can be evaluated under that mechanism.  This is the force of the
Neyman--Fisher tradition: the probability calculation can be made with the
potential outcomes held fixed and the random assignment as the only source of
chance \citep{neyman1923agricultural,fisher1935design}.

The cleanest version is finite-population randomization.  There are \(N\)
units.  Unit \(i\) has two fixed potential outcomes \(Y_i(1)\) and \(Y_i(0)\).
Only one can be observed, because the unit receives either treatment or
control.  Let \(Z_i=1\) if unit \(i\) is assigned to treatment and \(Z_i=0\)
otherwise.  Under complete randomization, exactly \(n_1\) of the \(N\) units
are assigned to treatment, and every subset of size \(n_1\) is equally likely.
Write \(n_0=N-n_1\).

\begin{proposition}[Randomization unbiasedness; \citealp{neyman1923agricultural,fisher1935design}]
Under complete randomization,
\[
  \hat\tau
  =
  \frac1{n_1}\sum_{i=1}^N Z_iY_i(1)
  -
  \frac1{n_0}\sum_{i=1}^N (1-Z_i)Y_i(0)
\]
is unbiased, over the randomization distribution, for the finite-population
average treatment effect
\[
  \tau_N
  =
  \frac1N\sum_{i=1}^N\{Y_i(1)-Y_i(0)\}.
\]
\end{proposition}

\emph{Proof.}
The potential outcomes are fixed in this calculation; only the assignment
vector \(Z=(Z_1,\ldots,Z_N)\) is random.  By symmetry of complete
randomization,
\[
  \Expect_d Z_i=\frac{n_1}{N},
  \qquad
  \Expect_d(1-Z_i)=\frac{n_0}{N},
\]
where \(\Expect_d\) denotes expectation over the design randomization.  Hence
\[
  \Expect_d\left\{
  \frac1{n_1}\sum_{i=1}^N Z_iY_i(1)
  \right\}
  =
  \frac1{n_1}\sum_{i=1}^N
  \frac{n_1}{N}Y_i(1)
  =
  \frac1N\sum_{i=1}^N Y_i(1),
\]
and similarly
\[
  \Expect_d\left\{
  \frac1{n_0}\sum_{i=1}^N (1-Z_i)Y_i(0)
  \right\}
  =
  \frac1N\sum_{i=1}^N Y_i(0).
\]
Subtracting the two displays gives \(\Expect_d\hat\tau=\tau_N\).
\qedmark

\begin{proposition}[Randomization variance; \citealp{neyman1923agricultural,fisher1935design}]
Write \(\bar Y(z)=N^{-1}\sum_iY_i(z)\),
\(\tau_i=Y_i(1)-Y_i(0)\), and
\[
  S_z^2=\frac{1}{N-1}\sum_{i=1}^N\{Y_i(z)-\bar Y(z)\}^2,
  \qquad
  S_\tau^2=\frac{1}{N-1}\sum_{i=1}^N(\tau_i-\tau_N)^2 .
\]
Under complete randomization,
\[
  \Var_d(\hat\tau)
  =
  \frac{S_1^2}{n_1}
  +
  \frac{S_0^2}{n_0}
  -
  \frac{S_\tau^2}{N}.
\]
\end{proposition}

The proof of unbiasedness is short because the design has done the hard work.
No regression model, normal approximation, or large-sample limit is needed to
justify the target of the comparison.  The variance formula makes the same point
for uncertainty: the randomization law, sample allocation, and heterogeneity of
potential outcomes determine the design-based noise.  Since \(S_\tau^2\) depends
on unobserved unit-level treatment effects, the usual conservative variance
estimate drops the negative term.  Later modeling may improve precision or
handle complications, but the comparison is already legible because the
assignment mechanism was designed.

\begin{example}[Stratified randomization]
Suppose a trial stratifies by disease stage.  Within each stratum \(s\), it
randomizes \(n_{1s}\) of \(N_s\) patients to treatment.  The same proof applied
inside each stratum gives an unbiased estimator of the stratum-specific finite
population effect.  A weighted average of stratum estimators then estimates the
weighted target
\[
  \sum_s \frac{N_s}{N}\tau_s.
\]
The design has made the comparison more stable by forcing treatment and control
to be balanced on a variable known before assignment.
\end{example}

\begin{example}[A trial list before the trial begins]
Suppose an oncology trial will enroll \(60\) patients, with \(30\) in early
stage disease and \(30\) in late stage disease.  A statistician can generate two
separate randomization lists before the first patient arrives:
\[
\begin{array}{c|c|c|c}
\text{stratum} & N_s & n_{1s} & n_{0s}\\
\hline
\text{early stage} & 30 & 15 & 15\\
\text{late stage} & 30 & 15 & 15
\end{array}
\]
When the last patient is enrolled, treatment and control are guaranteed to be
balanced on disease stage.  A regression model may still adjust for additional
covariates, but the most clinically obvious imbalance was handled by design
rather than by hope.
\end{example}

Randomization also clarifies what regression adjustment can and cannot do.
Adjustment can improve precision or correct chance imbalance, but it should not
be asked to create the comparison from scratch.  The design creates the first
comparison; the model refines the reading.

\section{Dose Finding as a Sequential Design}
\label{sec:ch04-dose-finding}
\conceptindexes{dose finding, sequential design, phase I trials, toxicity target, BOIN, CRM}

Dose finding makes the design problem visceral.  We want to learn which dose is
acceptable, but learning requires exposing people to doses whose risks are not
fully known.  A phase I oncology design is therefore both a statistical rule
and an ethical policy: it decides who is treated next while the evidence is
still incomplete.  The historical \(3+3\) rule, the continual reassessment
method, and BOIN are useful to compare because they make different promises
about model structure, transparency, and operating characteristics
\citep{storer1989phasei,oquigley1990crm,liu2015boin}.

\subsection{The 3+3 Rule: A Rule Without a Model}
\label{sec:ch04-three-plus-three}
\conceptindexes{3+3 rule, rule-based design, oncology dose finding}

The traditional \(3+3\) rule is simple.  It treats patients in cohorts of
three.  At a current dose, let \(Y_1\) be the number of dose-limiting toxicities
among the first three patients.
\[
\begin{array}{ccl}
Y_1=0 &\Longrightarrow& \text{escalate to the next dose},\\
Y_1=1 &\Longrightarrow& \text{treat three more patients at the same dose},\\
Y_1\ge2 &\Longrightarrow& \text{de-escalate or stop escalation}.
\end{array}
\]
If three more patients are treated at the same dose, let \(Y_2\) be the number
of toxicities in the second cohort.  The usual rule escalates if
\(Y_1+Y_2\le1\) and de-escalates if \(Y_1+Y_2\ge2\).

\begin{proposition}[Operating probabilities of a single 3+3 step]
At a dose with true toxicity probability \(p\), suppose toxicity outcomes are
independent Bernoulli\((p)\) within cohorts.  The probability that the \(3+3\)
rule escalates from this dose is
\[
  \pi_E(p)
  =
  (1-p)^3
  +
  3p(1-p)^2(1-p)^3.
\]
The probability that it de-escalates or stops from this dose is
\[
  \pi_D(p)
  =
  \{3p^2(1-p)+p^3\}
  +
  3p(1-p)^2\{1-(1-p)^3\}.
\]
\end{proposition}

\emph{Proof.}
The rule escalates in two disjoint cases.  First, the first cohort has no
toxicity, an event with probability \((1-p)^3\).  Second, the first cohort has
exactly one toxicity and the second cohort has none.  These events have
probabilities \(3p(1-p)^2\) and \((1-p)^3\), respectively, and the cohorts are
independent.  This proves the formula for \(\pi_E(p)\).

The rule de-escalates or stops in two disjoint cases.  First, the first cohort
has at least two toxicities, with probability \(3p^2(1-p)+p^3\).  Second, the
first cohort has exactly one toxicity and the second cohort has at least one
toxicity, with probability \(3p(1-p)^2\{1-(1-p)^3\}\).  Adding these
probabilities gives \(\pi_D(p)\).
\qedmark

The formula tells a story that the verbal rule hides.  The \(3+3\) design has
operating characteristics, but they are implicit.  It does not begin with a
target toxicity probability \(\phi\), a loss function, or a model for the dose
toxicity curve.  It behaves like a design, but it does not say exactly what
statistical question it is optimizing.

\begin{example}[A target can be hidden]
If \(p=0.30\), then
\[
  \pi_E(0.30)
  =
  0.70^3+3(0.30)(0.70)^2(0.70)^3
  \approx 0.494.
\]
Thus, at a dose with a \(30\%\) toxicity probability, the one-step probability
of escalation is still close to one half.  This is not automatically wrong:
some trials may accept such behavior.  The issue is that the target and the
tradeoff were not declared before the rule was used.
\end{example}

\begin{example}[A six-patient story]
Suppose the current dose is the second of five planned dose levels.  The first
three patients have outcomes \(0,0,1\), where \(1\) denotes a dose-limiting
toxicity.  The \(3+3\) rule pauses escalation and treats three more patients at
the same dose.  If the next outcomes are \(0,0,0\), the design escalates.  If
they are \(1,0,0\), it de-escalates or stops escalation.  One extra toxicity
among six patients flips the action.  That sharp threshold is the practical
face of an unstated loss function.
\end{example}

The \(3+3\) rule is worth teaching because it reveals the minimum facts a
serious design should expose.  What is the target?  What is the action space?
How does the rule behave if the current dose is too safe, near target, or too
dangerous?  What information is carried from earlier patients to later
patients?  Modern model-based and model-assisted designs answer these questions
more explicitly.

\subsection{BOIN Dose-Finding Boundaries}
\label{sec:ch04-boin}
\conceptindexes{BOIN, escalation boundaries, de-escalation boundaries, ethics and learning}

The Bayesian optimal interval design, or BOIN, is a model-assisted phase I
dose-finding design \citep{liu2015boin}.  It keeps the bedside action simple:
escalate, stay, or de-escalate.  But unlike \(3+3\), it declares a target
toxicity probability and derives decision boundaries from likelihood
comparisons.

Let \(\phi\) be the target toxicity probability.  Choose two reference
probabilities
\[
  \phi_1<\phi<\phi_2.
\]
The value \(\phi_1\) represents a dose that is meaningfully below target and
therefore too safe to stay at if higher doses exist; \(\phi_2\) represents a
dose that is meaningfully above target and therefore too toxic.  At the current
dose, suppose \(n\) patients have been treated and \(x\) have experienced a
dose-limiting toxicity.  Write \(\hat p=x/n\).

\begin{theorem}[BOIN escalation and de-escalation boundaries; \citealp{liu2015boin}]
Assume \(0<\phi_1<\phi<\phi_2<1\).  Compare the binomial likelihood at
\(\phi_1\) with the likelihood at \(\phi\), and compare the likelihood at
\(\phi_2\) with the likelihood at \(\phi\), using equal prior weights and
equal action losses for the local comparisons.  Define
\[
  \lambda_e
  =
  \frac{\log\{(1-\phi_1)/(1-\phi)\}}
       {\log\{\phi(1-\phi_1)/[\phi_1(1-\phi)]\}},
\]
and
\[
  \lambda_d
  =
  \frac{\log\{(1-\phi)/(1-\phi_2)\}}
       {\log\{\phi_2(1-\phi)/[\phi(1-\phi_2)]\}}.
\]
Then the local likelihood rule chooses the low-toxicity action when
\(\hat p<\lambda_e\), chooses the high-toxicity action when
\(\hat p>\lambda_d\), and stays near the target region when
\(\lambda_e\le \hat p\le \lambda_d\).
\end{theorem}

\emph{Proof.}
The binomial likelihood at toxicity probability \(q\), up to the common
binomial coefficient, is
\[
  L(q)=q^x(1-q)^{n-x}.
\]
First compare the low-toxicity reference \(\phi_1\) with the target \(\phi\).
The inequality \(L(\phi_1)>L(\phi)\) is equivalent to
\[
  x\log\frac{\phi_1}{\phi}
  +(n-x)\log\frac{1-\phi_1}{1-\phi}
  >0.
\]
Divide by \(n\) and write \(\hat p=x/n\).  Let
\[
  A=\log\frac{1-\phi_1}{1-\phi}>0.
\]
Because \(\phi_1<\phi\),
\[
  \log\frac{\phi_1}{\phi}-\log\frac{1-\phi_1}{1-\phi}
  =
  -\log\frac{\phi(1-\phi_1)}{\phi_1(1-\phi)}
  <0.
\]
Thus the likelihood comparison becomes
\[
  A
  -
  \hat p
  \log\frac{\phi(1-\phi_1)}{\phi_1(1-\phi)}
  >0,
\]
or
\[
  \hat p
  <
  \frac{\log\{(1-\phi_1)/(1-\phi)\}}
       {\log\{\phi(1-\phi_1)/[\phi_1(1-\phi)]\}}
  =
  \lambda_e.
\]
When the data look more like \(\phi_1\) than \(\phi\), the current dose is
locally too low, so escalation is the corresponding action.

Now compare the high-toxicity reference \(\phi_2\) with the target \(\phi\).
The inequality \(L(\phi_2)>L(\phi)\) is equivalent to
\[
  x\log\frac{\phi_2}{\phi}
  +(n-x)\log\frac{1-\phi_2}{1-\phi}
  >0.
\]
After division by \(n\), this is
\[
  \hat p
  \log\frac{\phi_2(1-\phi)}{\phi(1-\phi_2)}
  -
  \log\frac{1-\phi}{1-\phi_2}
  >0.
\]
Since \(\phi_2>\phi\), both logarithms in the preceding display are positive.
Therefore
\[
  \hat p
  >
  \frac{\log\{(1-\phi)/(1-\phi_2)\}}
       {\log\{\phi_2(1-\phi)/[\phi(1-\phi_2)]\}}
  =
  \lambda_d.
\]
When the data look more like \(\phi_2\) than \(\phi\), the current dose is
locally too toxic, so de-escalation is the corresponding action.  If neither
comparison wins, the current dose remains in the local target interval.
\qedmark

At exact equality, the two local likelihoods tie.  Operational BOIN tables use
a fixed tie convention and integer rounding; below, the common convention is used:
``escalate when \(\hat p\le\lambda_e\), de-escalate when
\(\hat p\ge\lambda_d\), and otherwise stay.''

\begin{example}[A BOIN table]
Let the target toxicity probability be \(\phi=0.30\), and choose
\(\phi_1=0.18\), \(\phi_2=0.42\).  Then
\[
  \lambda_e\approx0.236,
  \qquad
  \lambda_d\approx0.359.
\]
The rule escalates if \(x/n\le0.236\), de-escalates if
\(x/n\ge0.359\), and otherwise stays.  For \(n=3\), this gives:
\[
\begin{array}{c|c|c}
x & x/3 & \text{action}\\
\hline
0 & 0 & \text{escalate}\\
1 & 0.333 & \text{stay}\\
2,3 & \ge0.667 & \text{de-escalate}
\end{array}
\]
For \(n=6\), the rule escalates at \(x=0,1\), stays at \(x=2\), and
de-escalates at \(x\ge3\).  The resulting table is as easy to use as \(3+3\),
but its boundaries are tied to a declared target and reference toxicities.
\end{example}

\begin{example}[Why a table can be more honest than a black box]
With \(\phi=0.30\), a current dose with \(1/6\) toxicities looks below the BOIN
escalation boundary, so escalation is allowed.  With \(2/6\), the design stays.
With \(3/6\), it de-escalates.  These decisions are simple enough to print on a
trial card, yet the card has a derivation and can be stress-tested by simulation
before the study opens.  The attraction of BOIN is not that it removes judgment;
it puts the judgment into visible quantities \(\phi,\phi_1,\phi_2\) and the
resulting operating characteristics.
\end{example}

BOIN illustrates a recurring design principle.  A good design should be simple
at the point of action and explicit at the point of justification.  The
clinician needs a rule that can be executed; the statistician needs operating
characteristics that can be simulated, criticized, and revised before the trial
opens.

\section{Optimal Design and Information Geometry}
\label{sec:ch04-optimal-design}
\conceptindexes{optimal design, information geometry, design measure, Fisher information, equivalence theorem, Riemannian manifold design, functional approach to optimal design}

Randomization makes comparisons legible.  Dose-finding design makes learning
ethical under uncertainty.  Optimal design asks a third question: if the
scientist can choose where observations are taken, where should they be placed?
The classical literature turns this into an optimization problem over
information matrices, with broad treatments in
\citet{fedorov1972optimal}, \citet{pukelsheim2006optimal}, and
\citet{atkinson2007optimum}.

Consider a regression experiment.  At design point \(x\in\mathcal X\), the
observation satisfies
\[
  Y=\mathbf f(x)^T\beta+\varepsilon,
  \qquad
  \Expect(\varepsilon)=0,\qquad
  \Var(\varepsilon)=\sigma^2.
\]
Here \(\mathbf f(x)\in\R^p\) is the vector of regression functions.  An approximate
design is a probability measure
\[
  \xi=\sum_{j=1}^k w_j\delta_{x_j},
  \qquad
  w_j>0,\quad \sum_{j=1}^k w_j=1.
\]
Its information matrix is
\[
  M(\xi)
  =
  \int_{\mathcal X} \mathbf f(x)\mathbf f(x)^T\,d\xi(x)
  =
  \sum_{j=1}^k w_j \mathbf f(x_j)\mathbf f(x_j)^T.
\]
For a fixed total sample size \(n\), one may allocate approximately \(nw_j\)
observations to \(x_j\).  The approximate design idealizes integer allocation
so that the geometry can be seen clearly.

\begin{definition}[Common optimality criteria]
Assume \(M(\xi)\) is nonsingular.
\begin{enumerate}
\item \(D\)-optimality maximizes \(\log\det M(\xi)\), equivalently the
information volume.
\item \(A\)-optimality minimizes \(\tr\{M(\xi)^{-1}\}\), the
average variance of the least-squares coordinates after scaling by
\(\sigma^2/n\).
\item \(c\)-optimality minimizes \(\mathbf c^TM(\xi)^{-1}\mathbf c\), the variance of the
estimated contrast \(\mathbf c^T\beta\).
\item \(G\)-optimality minimizes
\[
  \sup_{x\in\mathcal X} \mathbf f(x)^TM(\xi)^{-1}\mathbf f(x),
\]
the worst prediction variance factor over the design space.
\end{enumerate}
\end{definition}

The criteria are different ways of turning a scientific goal into a geometry.
If all coordinates of \(\beta\) matter jointly, \(D\)-optimality is natural.  If
a particular contrast matters, \(c\)-optimality is more honest.  If prediction
over the whole region matters, \(G\)-optimality is the right language.  The
criterion is part of the design, because it declares what information is
valuable.

\paragraph{Riemannian-manifold optimal design.}
The design space need not be a Euclidean interval or box.  If \(x\) lives on a Riemannian manifold
\(\mathcal M\), an approximate design is still a probability measure
\(\xi\) on the design space, and a familiar information matrix may still have
the form
\[
  M(\xi)=\int_{\mathcal M} \mathbf f(x)\mathbf f(x)^T\,d\xi(x).
\]
What changes is the geometry behind the candidate set, local neighborhoods,
prediction targets, and algorithms.  A design point is no longer moved by
ordinary vector addition; it is moved along charts, tangent spaces, or geodesic
steps.  The Riemannian optimal-design treatment of
\citet{liDelCastillo2024riemannianDesign} develops this idea for
\(D\)- and \(G\)-optimality, including equivalence-theorem language on
manifold-valued design regions.

\paragraph{The functional approach to optimal design.}
``Functional optimal design'' has a specific meaning in part of the
optimal-design literature.  In \citet{melas2006functional}, the word
``functional'' refers to a functional-analytic way of studying how optimal
support points and weights change with model parameters, design intervals, and
implicit optimality equations.  That is different from functional data
analysis, where the observations themselves are curves.  The Melas monograph
therefore belongs to optimal-design theory rather than to FDA proper.

\begin{example}[Uniform design as military-engineering design]
Uniform design should not appear in this chapter merely as another item in a
catalogue of optimality criteria.  It is one of the clearest examples of design
as cross-disciplinary engineering.  Wang Yuan and Kai-Tai Fang's late-1970s
work, publicly associated with their 1979 paper on uniform distribution and
experimental design, grew out of a defense and aerospace engineering problem:
how to build accurate approximate models for missile command-instrument
simulations when the number of runs was severely limited
\citep{wangFang1979uniform,cas2009uniformDesign,ckcest2024wangUniformDesign}.

The mathematical move is different from model-specific \(D\)-optimality.  When
the response surface is unknown, nonlinear, possibly multi-modal, and expensive
to query, the safest first promise is often coverage.  After rescaling each
factor to \([0,1]\), a uniform design chooses points
\[
  U_n=\{u_1,\ldots,u_n\}\subset[0,1]^s
\]
so that they spread evenly through the \(s\)-dimensional design region.  One
idealized way to express the goal is to make a discrepancy small:
\[
  D(U_n)
  =
  \sup_{B}
  \left|
    \frac1n\sum_{i=1}^n \mathbf 1\{u_i\in B\}
    -
    \lambda(B)
  \right|,
\]
where \(B\) ranges over a chosen class of rectangular boxes and \(\lambda(B)\)
is volume.  The operational idea is simple: if only a small number of computer
or physical experiments can be run, do not waste points by clustering them in
one corner of a high-dimensional space.

This is why uniform design belongs near the center of this chapter.  It was not
invented as a classroom embellishment to analysis of variance.  It was a design
technology for military-industrial computation: limited runs, many factors,
restricted information, and a need for an approximate model good enough over the
whole operating region.  In this book's notation, the design \(d\) is the
space-filling table itself, and \(P_{\theta,d}\) is the law induced after a
complex engineering system is only observed through those selected runs.
\end{example}

\begin{example}[The same clinic, different information promises]
A dose-response study might care about the whole response curve because the
future label will include several dose levels.  That points toward a
\(D\)- or \(G\)-type criterion.  A different study might care mainly about the
dose at which response crosses a clinically meaningful threshold; then a
contrast or nonlinear function of the parameters is the object worth protecting.
The patients, clinics, and assay may be the same, but the design criterion
changes because the scientific promise changed.
\end{example}

\begin{example}[Straight-line regression on an interval]
Let \(\mathcal X=[-1,1]\) and \(\mathbf f(x)=(1,x)^T\).  The design that puts weight
\(1/2\) at \(-1\) and weight \(1/2\) at \(1\) has
\[
  M(\xi)
  =
  \frac12
  \begin{pmatrix}1\\-1\end{pmatrix}
  \begin{pmatrix}1&-1\end{pmatrix}
  +
  \frac12
  \begin{pmatrix}1\\1\end{pmatrix}
  \begin{pmatrix}1&1\end{pmatrix}
  =
  \begin{pmatrix}1&0\\0&1\end{pmatrix}.
\]
The prediction variance factor is
\[
  \mathbf f(x)^TM(\xi)^{-1}\mathbf f(x)=1+x^2\le2.
\]
The maximum is attained at the two support points.  The equivalence theorem
below shows that this design is \(D\)-optimal.
\end{example}

\subsection{Computational Design in Biomedical Examples}
\label{sec:ch04-computational-design}
\conceptindexes{computational design, biomedical design, sparse sampling, pharmacokinetics, sea urchin toxicology}

The preceding formulas can make optimal design look like a closed-form
exercise.  In modern biomedical work it is often a constrained computation.
The design space may include dose ranges, visit windows, assay budgets,
toxicity constraints, recruitment limits, and nonlinear mean functions.
A useful bridge from classical equivalence theorems to modern biomedical
design comes from work by Wong and collaborators.  This line of work treats
optimal design not only as information geometry, but also as a computational
workflow: generate candidate designs, evaluate operating characteristics, and
revise the design under scientific, ethical, and logistical constraints
\citep{sverdlov2020review}.

Binary dose-response and quantal-response studies give a concrete version of
this problem.  In a two-parameter binary regression model
\[
  \Prob(Y=1\mid x)=F(\beta_0+\beta_1x),
\]
the information supplied by dose \(x\) is small when the response probability is
near \(0\) or \(1\), because almost everyone has the same outcome.  The most
informative doses often sit near the transition region of the curve, not at
every equally spaced point.  A recent arXiv treatment of \(D\)-optimal binary
regression works out many such two-point designs and checks them with
sensitivity plots \citep{cui2022doptimal}.

\begin{definition}[\(D\)-optimal sensitivity function for binary regression]
Let \(\mathbf z(x)=(1,x)^T\), let \(\eta(x)=\beta_0+\beta_1x\), and let
\(\pi(x)=F\{\eta(x)\}\).  Write \(f=F'\).  For a candidate design \(\xi\),
define
\[
  M_\beta(\xi)
  =
  \int_{\mathcal X}
  \omega_\beta(u)\mathbf z(u)\mathbf z(u)^T\,d\xi(u),
  \qquad
  \omega_\beta(u)
  =
  \frac{f\{\eta(u)\}^2}{F\{\eta(u)\}[1-F\{\eta(u)\}]}.
\]
The \(D\)-optimal sensitivity function is
\[
  \psi_D(x;\xi,\beta)
  =
  \omega_\beta(x)\mathbf z(x)^TM_\beta(\xi)^{-1}\mathbf z(x)-p,
\]
where \(p\) is the number of regression parameters, so \(p=2\) here.  The
equivalence theorem says that a locally \(D\)-optimal design \(\xi^\star\)
satisfies
\[
  \sup_{x\in\mathcal X}\psi_D(x;\xi^\star,\beta)\le0,
\]
with equality at support points with positive design weight.  For the logit
link, \(\omega_\beta(x)=\pi(x)\{1-\pi(x)\}\).
\end{definition}

\begin{example}[Two-point logit design]
For the logit link, \(F(\eta)=\{1+\exp(-\eta)\}^{-1}\), the locally
\(D\)-optimal two-point design for the simple linear predictor
\(\eta=\beta_0+\beta_1x\) places equal weight at the two points satisfying
\[
  \eta=\pm 1.5434.
\]
Thus
\[
  x_1=\frac{-1.5434-\beta_0}{\beta_1},
  \qquad
  x_2=\frac{ 1.5434-\beta_0}{\beta_1}.
\]
This rule is more memorable than it first appears.  It says that, for a
logistic dose-response curve, the design should not spend all its observations
where the curve is flat.  It should place them on the two shoulders of the
curve, where a small movement in dose changes the response probability.  At
\(\eta=-1.5434\) and \(1.5434\), the response probabilities are about
\(0.176\) and \(0.824\).  The two-point design is therefore not saying ``put
everything at the median effective dose.''  It is saying ``bracket the steep
part of the curve.''
\end{example}

\begin{example}[Why equally spaced doses can be expensive]
Suppose a lab has enough animals or embryos for \(80\) observations and is
tempted to use eight equally spaced concentrations with \(10\) observations at
each.  If the lowest two concentrations almost never produce the event and the
highest two almost always do, half of the concentration grid is mostly confirming
what the investigator already knows.  A locally \(D\)-optimal follow-up design
might instead put most or all observations near the two shoulders of the fitted
curve.  The scientific gain is not just a smaller standard error; it is a more
honest use of scarce experimental units.
\end{example}

\begin{example}[Sparse pharmacokinetic sampling]
In an early pharmacology study, the endpoint may be a concentration-time curve
rather than a binary response.  A full profile might require samples at
\[
  0.25,\ 0.5,\ 1,\ 2,\ 4,\ 8,\ 12,\ 24
  \quad\text{hours after dosing,}
\]
but hospitalized patients, children, or oncology patients may tolerate only two
or three draws.  A pharmacometric design then chooses sampling windows, not
just sample size.  For a one-compartment model, early samples identify
absorption, middle samples locate peak exposure, and late samples protect the
elimination slope.  Population optimal-design work formalizes this problem with
Fisher information for nonlinear mixed-effects models
\citep{mentre1997optimal,nyberg2012poped}.  The design question is concrete:
should the next blood draw buy more information about \(C_{\max}\), area under
the curve, clearance, or between-subject variability?
\end{example}

Modern toxicology amplifies the same issue.  A high-throughput screen can test
many chemicals, cell types, and concentrations, but a grid with too many
uninformative concentrations wastes biological material and follow-up effort.
The National Research Council's vision for twenty-first-century toxicity
testing pushed the field toward pathway-based, information-rich designs; binary
and quantal-response optimal design is one mathematical expression of that
agenda \citep{nationalresearchcouncil2007toxicity,cui2022doptimal}.

\begin{example}[Sea urchin toxicology]
In a developmental toxicology study using sea urchin embryos
\citep{collins2022seaurchin}, the concentration-response data can be modeled
with either a logit link or a complementary log-log link.  After rescaling
concentration by \(1000\), one fitted logit model has approximately
\[
  \beta_0=-4.5,\qquad \beta_1=20.
\]
Plugging these values into the two-point logit rule gives
\[
  x_1=\frac{-1.5434+4.5}{20}\approx0.1478,
  \qquad
  x_2=\frac{ 1.5434+4.5}{20}\approx0.3022.
\]
On the original scale, the suggested concentration levels are about
\(147.8\) and \(302.2\,\mu\mathrm{M}\), with half the experimental units at
each level.  Using a complementary log-log link moves the two support points to
about \(168.7\) and \(334.3\,\mu\mathrm{M}\).  The application is a useful
warning: the optimal design is local to a model and plausible parameter values,
so the design calculation should be paired with sensitivity analysis.
\end{example}

\begin{figure}[htbp]
\centering
\includegraphics[
  width=0.98\linewidth,
  trim=0 196bp 0 0,
  clip
]{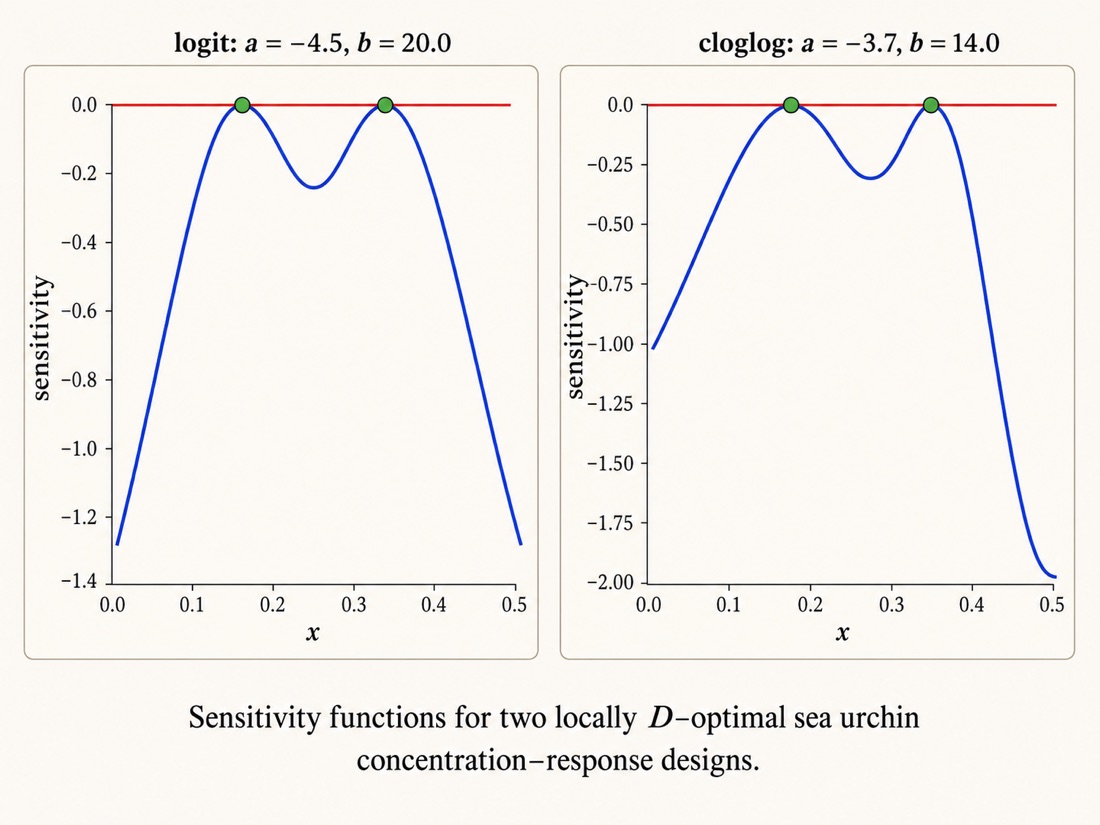}
\caption[Sea urchin \(D\)-optimal design sensitivity functions]{Sensitivity functions for two locally \(D\)-optimal sea urchin
concentration-response designs.  The curve stays below zero and touches zero at
the support points, which is the graphical certificate supplied by the
equivalence theorem.  The left panel uses a logit link; the right panel uses a
complementary log-log link.}
\label{fig:ch04-sea-urchin-sensitivity}
\end{figure}

\begin{example}[How to read a sensitivity plot]
Figure~\ref{fig:ch04-sea-urchin-sensitivity} plots
\(\psi_D(x;\xi^\star,\beta)\) over the allowed concentration range.  If the
plot rises above zero somewhere, the design has left information on the table:
moving mass toward that \(x\) would improve the determinant.  If the plot stays
below zero and touches zero at the proposed support points, the picture is not
decoration.  It is a certificate of optimality.  In the sea urchin logit panel,
the two touches occur near \(147.8\) and \(302.2\,\mu\mathrm{M}\), exactly the
two concentrations recommended by the local \(D\)-optimal calculation.
\end{example}

\clearpage

\begin{example}[ImageNet as dataset design]
ImageNet is a useful reminder that modern ``data collection'' can be an
industrial design problem even when no treatment is assigned.  The project did
not begin with iid photographs falling from the sky.  It began with a semantic
design: choose object categories from a lexical hierarchy, collect candidate
images for each category, verify labels with human annotation, and define
benchmark tasks and splits \citep{deng2009imagenet}.  A convenient abstraction is
\[
  d_{\mathrm{ImageNet}}
  =
  (\mathcal C,\ q_c,\ A,\ Q,\ S,\ m),
\]
where \(\mathcal C\) is the class ontology, \(q_c\) are search queries or
collection routes for class \(c\), \(A\) is the annotation protocol, \(Q\) is the
quality-control rule, \(S\) is the train--validation--test split, and \(m\) is
the evaluation metric.  The observed object is therefore
\[
  O(d_{\mathrm{ImageNet}})
  =
  \{(I_i,Y_i): I_i \text{ passes } Q,\ Y_i\in\mathcal C
  \text{ is assigned by } A\}.
\]

This notation is not decorative.  Changing \(\mathcal C\) changes the question:
fine-grained dog breeds, household objects, and natural scenes produce different
learning problems.  Changing \(A\) changes label noise.  Changing \(S\) changes
what can be claimed about generalization.  Changing \(m\) changes what kinds of
mistakes count as progress.  The 2012 deep-learning breakthrough on ImageNet
was therefore not only a model story \citep{krizhevsky2012imagenet}; it was also
a design story.  A large, hierarchical, human-verified benchmark made a certain
kind of visual intelligence measurable at industrial scale.
\end{example}

\begin{example}[AIDD as a concrete active-learning batch]
Consider a kinase hit-to-lead project with \(8{,}000\) purchasable analogs and
a weekly budget for \(24\) biochemical assays.  At round \(t\), let
\(\mathcal F_t\) contain the molecular structures already tested, observed
\(\mathrm{pIC}_{50}\) values, solubility flags, cytotoxicity flags, failed
syntheses, prices, and delivery times.  A model trained on \(\mathcal F_t\)
predicts, for each untested molecule \(x\), a mean activity
\(\widehat\mu_t^{\mathrm{pIC}_{50}}(x)\), an uncertainty
\(\widehat\sigma_t(x)\), a toxicity risk \(\widehat r_t^{\mathrm{tox}}(x)\),
and a practical cost \(\mathrm{cost}(x)\).  One possible acquisition score is
\[
  a_t(x)
  =
  \widehat\mu_t^{\mathrm{pIC}_{50}}(x)
  +
  \kappa\,\widehat\sigma_t(x)
  -
  \lambda_1\widehat r_t^{\mathrm{tox}}(x)
  -
  \lambda_2\mathrm{cost}(x).
\]
Now the design is the batch
\[
  B_{t+1}
  =
  \{x:\text{\(x\) is selected among the next \(24\) molecules}\}.
\]
A project team might deliberately reserve \(16\) positions for high-scoring
near-neighbor analogs of known hits and \(8\) positions for chemically diverse
molecules with large \(\widehat\sigma_t(x)\).  The first group tries to improve
potency quickly; the second group prevents the model from becoming confident
only in one familiar chemical neighborhood.  Molecules predicted to be potent
but insoluble, cytotoxic, unavailable, or too expensive are down-weighted before
the assay is run.  After the \(24\) results return, \(\mathcal F_{t+1}\) is
formed and the next batch is chosen.  This is optimal design in a different
accent: the design points are molecules, the response is an assay readout, and
the information geometry is learned sequentially
\citep{vamathevan2019applications,reker2017active,frazier2018tutorial,shields2021bayesian}.
\end{example}

\begin{example}[Single-cell design before sequencing]
In a single-cell perturbation experiment, it is tempting to say that the design
variable is the number of cells.  Usually that is too narrow.  A treatment
effect seen in \(50{,}000\) cells from one donor is not the same evidence as a
smaller number of cells from many donors, batches, and capture lanes.  A useful
design record is closer to
\[
  d_{\mathrm{scRNA}}
  =
  (D,\ T,\ B,\ L,\ C,\ R,\ M,\ Q),
\]
where \(D\) is the donor or biological-replicate plan, \(T\) the perturbation and
time-point grid, \(B\) the batch assignment, \(L\) the library or capture
chemistry, \(C\) the cells targeted per unit, \(R\) the sequencing depth, \(M\)
the modality plan, and \(Q\) the quality-control and doublet-filtering rule.  The
observed object is not a generic matrix.  It is a count array plus metadata,
\[
  O(d_{\mathrm{scRNA}})
  =
  \{Y_{gci},\ \text{donor}_i,\ \text{batch}_i,\ \text{time}_i,\ \text{condition}_i\},
\]
after cell capture, library preparation, sequencing, alignment, and filtering.

This is an industrial and biological design problem before it is a generalized
linear model problem.  Dissociation can destroy fragile cell types; multiplexing
can reduce batch confounding; sequencing depth trades off against the number of
biological replicates; spatial or multiome assays change the random object from
a gene-count vector to a coupled measurement across modalities.  The scDesign
series, from scDesign2 to scDesign3, is useful in this chapter's language
because it generates realistic in silico single-cell and spatial-omics data; one can
simulate candidate designs, run the planned analysis, and ask which designs
actually recover the cell-type composition, differential expression, trajectory,
or perturbation signal of interest
\citep{sun2021scdesign2,song2024scdesign3}.
\end{example}

\subsection{The Equivalence Theorem for Optimal Design}
\label{sec:ch04-equivalence-theorem}
\conceptindexes{equivalence theorem, D-optimality, sensitivity function, design compass}

The most elegant result in classical optimal design says that two different
goals agree.  Maximizing the determinant of the information matrix is
equivalent, in the linear model, to minimizing the worst prediction variance.
This is the Kiefer--Wolfowitz equivalence theorem
\citep{kiefer1960equivalence}.  It is also a perfect example of the Chapter 1
compass: a design criterion, a model class, and a decision rule become one
mathematical object.

\begin{theorem}[Kiefer--Wolfowitz equivalence theorem, linear-model form; \citealp{kiefer1960equivalence}]
Let \(\mathcal X\) be compact and let \(\mathbf f:\mathcal X\to\R^p\) be continuous.
Assume at least one design has nonsingular information matrix.  For a design
\(\xi^\star\) with nonsingular \(M(\xi^\star)\), the following are equivalent:
\[
  \xi^\star \text{ maximizes } \log\det M(\xi)
\]
over all approximate designs \(\xi\), and
\[
  \sup_{x\in\mathcal X}
  \mathbf f(x)^TM(\xi^\star)^{-1}\mathbf f(x)
  \le p.
\]
Moreover, equality holds at every support point of \(\xi^\star\) with positive
weight.
\end{theorem}

\emph{Proof.}
Write
\[
  \Phi(\xi)=\log\det M(\xi),
  \qquad
  d(x,\xi)=\mathbf f(x)^TM(\xi)^{-1}\mathbf f(x).
\]
The map \(M\mapsto\log\det M\) is concave on the cone of positive definite
matrices, and \(M(\xi)\) is linear in \(\xi\).  Hence \(\Phi\) is concave on
the set of designs with nonsingular information matrix.

Fix a nonsingular design \(\xi\), and perturb it toward a point mass at \(x\):
\[
  \xi_\alpha=(1-\alpha)\xi+\alpha\delta_x,
  \qquad 0\le\alpha\le1.
\]
Then
\[
  M(\xi_\alpha)
  =
  M(\xi)+\alpha\{\mathbf f(x)\mathbf f(x)^T-M(\xi)\}.
\]
The derivative of \(\log\det A\) in direction \(H\) is
\(\tr(A^{-1}H)\).  Therefore
\[
  \left.\frac{d}{d\alpha}\Phi(\xi_\alpha)\right|_{\alpha=0}
  =
  \tr
  \left[
    M(\xi)^{-1}\{\mathbf f(x)\mathbf f(x)^T-M(\xi)\}
  \right].
\]
Using \(\tr\{M^{-1}\mathbf f\mathbf f^T\}=\mathbf f^TM^{-1}\mathbf f\) and
\(\tr\{M^{-1}M\}=p\), the derivative equals
\[
  d(x,\xi)-p.
\]

Suppose \(\xi^\star\) is \(D\)-optimal.  Moving an infinitesimal amount of mass
from \(\xi^\star\) toward any point \(x\) cannot increase \(\Phi\), so
\[
  d(x,\xi^\star)-p\le0
  \qquad\text{for every }x\in\mathcal X.
\]
Thus \(\sup_x d(x,\xi^\star)\le p\).

Conversely, suppose \(\sup_x d(x,\xi^\star)\le p\).  For any design \(\eta\),
consider the direction \(\eta-\xi^\star\).  The directional derivative at
\(\xi^\star\) is
\[
  \int_{\mathcal X} d(x,\xi^\star)\,d\eta(x)-p.
\]
The assumed bound gives
\[
  \int_{\mathcal X} d(x,\xi^\star)\,d\eta(x)-p
  \le
  p-p
  =
  0.
\]
Since \(\Phi\) is concave, a nonpositive directional derivative in every
feasible direction implies
\[
  \Phi(\eta)-\Phi(\xi^\star)\le0
\]
for every design \(\eta\).  Hence \(\xi^\star\) is \(D\)-optimal.

It remains to prove equality on the support.  Since
\[
  \int_{\mathcal X} d(x,\xi^\star)\,d\xi^\star(x)
  =
  \tr
  \left[
    M(\xi^\star)^{-1}
    \int \mathbf f(x)\mathbf f(x)^T\,d\xi^\star(x)
  \right]
  =
  \tr(I_p)
  =
  p,
\]
and \(d(x,\xi^\star)\le p\) everywhere, the function
\(d(x,\xi^\star)\) must equal \(p\) at \(\xi^\star\)-almost every support
point.  For a finite design with positive weights, this means equality at each
support point.  By continuity, the usual support statement follows for compact
designs as well.
\qedmark

The theorem is a design compass.  To check a proposed \(D\)-optimal design, one
does not need to compare it with every possible design directly.  Compute the
variance function \(d(x,\xi)\).  If it never rises above \(p\), and if it
touches \(p\) at the design's support points, the design has solved the
global problem.  The geometry supplies both a certificate and a diagnostic.

\begin{example}[Constrained and compound design]
Real studies often ask for two design promises at once.  A pharmacokinetic
study may want precise clearance estimation while keeping the number of late
blood draws small.  A toxicology study may want information about the slope of
a concentration-response curve while maintaining enough support near a
clinically important benchmark concentration.  Write the two design utilities
as \(\phi_1\{M(\xi)\}\) and \(\phi_2\{M(\xi)\}\).

A constrained design treats the first promise as a hard requirement:
\[
  \max_{\xi}\ \phi_2\{M(\xi)\}
  \qquad
  \text{subject to}\qquad
  \phi_1\{M(\xi)\}\ge c .
\]
A compound design turns the two promises into one weighted criterion:
\[
  \max_{\xi}\ 
  \lambda \phi_1\{M(\xi)\}
  +(1-\lambda)\phi_2\{M(\xi)\},
  \qquad 0\le\lambda\le1 .
\]
The first formulation sounds regulatory: ``protect at least this much
efficiency for objective 1.''  The second sounds operational: ``choose a
trade-off weight and optimize.''  Cook and Wong's equivalence theorem says
that, under the usual regularity conditions for these information criteria,
the efficient constrained designs are recovered by compound criteria as the
weight \(\lambda\) varies \citep{cookWong1994equivalence}.  Thus a constraint
and a weighted compromise are not two unrelated design philosophies.  They are
two coordinate systems on the same boundary of attainable information.

This is useful in industrial design work because teams often negotiate in
constraints but compute in weighted objectives.  A clinical pharmacology team
may insist on a minimum precision for exposure; a modeling team may then search
over \(\lambda\) to find the compound design that lands on that constraint.
The Chapter 1 compass is visible again: a scientific promise becomes a
criterion, the criterion becomes a geometric boundary, and the design is chosen
on that boundary before the data exist.
\end{example}

\section{Adaptive Rules and Practical Design}
\label{sec:ch04-adaptive-practical-design}
\conceptindexes{adaptive design, filtrations, interim monitoring, pragmatic trials, real-world study design}

\subsection{Adaptive Design and Filtrations}
\label{sec:ch04-adaptive-filtrations}
\conceptindexes{adaptive design, filtration, information flow, stopping rules}

Adaptive design sounds dangerous if the word ``adaptive'' is heard as
``improvised.''  In a statistical design, adaptation is not improvisation.  It
is a rule that says how future actions depend on past information.
This distinction matters in medicine: a platform trial, group-sequential trial,
or response-adaptive trial may change what happens next, but the possible
changes, timing, and decision rules must be part of the design that regulators,
clinicians, and patients can inspect before the study starts
\citep{bhatt2016adaptive}.

Let \(\mathcal F_k\) be the information available after the first \(k\) stages
of a study.  A sequential design action \(A_{k+1}\), such as the next dose, the
next allocation probability, or the next sampling location, is validly adapted
when
\[
  A_{k+1}\ \text{is}\ \mathcal F_k\text{-measurable}.
\]
This condition says that the next action may use what has already happened but
may not use future outcomes.  It is the design version of the no-looking-ahead
principle that will reappear in martingales and counting processes.  If fresh
randomization is used at stage \(k+1\), the action can be written as
\[
  A_{k+1}=a_k(H_k,U_{k+1}),
\]
where \(H_k\) is the recorded history generating \(\mathcal F_k\) and
\(U_{k+1}\) is a new random number generated before the next action.  The rule
\(a_k\) belongs to the design.

\begin{example}[BOIN as an adapted rule]
In a dose-finding trial, let \(\mathcal F_k\) contain all doses assigned and
toxicity outcomes observed through cohort \(k\).  The BOIN action for cohort
\(k+1\) is a function of the current dose, the accumulated number \(n\) treated
at that dose, and the accumulated toxicity count \(x\).  Since these are all
\(\mathcal F_k\)-measurable, the next dose is \(\mathcal F_k\)-measurable.
The rule adapts, but it does not look ahead.
\end{example}

This filtration language also distinguishes two kinds of design changes.  A
pre-specified adaptive rule belongs to the design.  A change made after seeing
unblinded interim results but not encoded in the design may change the law of
the data and therefore the meaning of the later analysis.  The same numerical
dataset can carry different inferential meaning depending on whether its
adaptation was part of \(P_{\theta,d}\).

\begin{example}[Adaptation that is planned, not improvised]
In a response-adaptive trial, the next allocation probability may favor an arm
that has performed well so far.  The design remains statistically coherent if
the update rule was written before the trial and uses only
\(\mathcal F_k\), the information available after stage \(k\).  By contrast,
quietly changing the randomization ratio after an unplanned look at unblinded
data creates a different data-generating law.  The problem is not adaptation
itself; the problem is adaptation without a design.
\end{example}

\begin{example}[A platform trial is still a design]
Suppose a hospital network studies three treatments for the same disease, with
an option to drop an arm for futility after \(100\) outcome observations and an
option to add a new arm after an external safety review.  Those features do not
make the trial vague if they are written into \(d\).  The randomization ratios,
futility rule, information times, safety rule, and final estimator are all part
of \(P_{\theta,d}\).  The design can be simulated under optimistic, null, and
harmful scenarios before the first patient is enrolled.  The adaptive trial is
therefore not a trial that changes its mind; it is a trial whose possible
changes were designed.
\end{example}

\begin{example}[Group-sequential monitoring in the wild]
A group-sequential trial is a trial whose design includes planned looks at the
accumulating data.  If \(T_k\) is a test statistic at information time
\(t_k=I_k/I_K\), the stopping rule is the random time
\[
  T
  =
  \inf\{k:T_k\ge b_k\},
\]
with \(T=\infty\) if no boundary is crossed.  The boundary sequence
\((b_1,\ldots,b_K)\) is part of \(d_{\mathrm{adapt}}\), and it is chosen so that
\[
  P_{H_0,d}\{T\le K\}
  =
  P_{H_0,d}\{\exists k\le K:T_k\ge b_k\}
  \le \alpha .
\]
Pocock-type boundaries spend type-I error more evenly across looks;
O'Brien--Fleming-type boundaries are conservative early and closer to the usual
fixed-sample cutoff near the final look.  Lan--DeMets spending functions express
the same idea as
\[
  P_{H_0,d}\{T\le k\}\le A_\alpha(t_k),
  \qquad A_\alpha(1)=\alpha,
\]
so the boundary can be calibrated to the actual information times
\citep{pocock1977group,obrien1979multiple,lan1983discrete}.

The Women's Health Initiative estrogen-plus-progestin trial is a real clinical
example of why this is a design, not a post-hoc reaction.  The trial had
prospective safety monitoring, and it was stopped early after the monitoring
process showed an unfavorable risk-benefit profile, including increased breast
cancer and cardiovascular risks \citep{rossouw2002whi,anderson2007whiMonitoring}.
The important statistical point is not merely that the trial stopped.  It is
that stopping rules, information looks, endpoint priorities, and data-monitoring
responsibilities were part of the law \(P_{\theta,d}\) that generated the final
dataset.
\end{example}

\begin{example}[I-SPY 2 as a biotech clinical-design example]
I-SPY 2, a neoadjuvant breast-cancer platform trial, makes the biotech
go/no-go idea concrete.  Experimental regimens enter a common platform, patients
are characterized by biomarker signatures, randomization probabilities are
updated as response information accumulates, and regimens can graduate or stop
for futility according to pre-specified Bayesian predictive rules
\citep{barker2009ispy2,park2016neratinib}.  A stylized graduation rule has the
form
\[
  \Prob\{\text{success in a future confirmatory trial}\mid\mathcal F_k\}
  >
  \gamma_G,
\]
with a lower threshold \(\gamma_F\) for futility.  The constants
\(\gamma_G,\gamma_F\), biomarker strata, endpoint definition, adaptive
randomization rule, and simulation plan are all design objects.  From the
viewpoint of this chapter, the design is not just ``randomize patients.''  It is
the whole learning policy that turns accumulated evidence into the next action.
\end{example}

\subsection{Pragmatic Trials and Real-World Study Design}
\label{sec:ch04-pragmatic-rws}
\conceptindexes{pragmatic trials, real-world studies, target-trial emulation, estimands, PRECIS-2}

The same design logic applies when the study is deliberately close to routine
care.  \citet{schwartz1967explanatory} distinguished trials designed to ask
whether an intervention can work under ideal conditions from trials designed
to ask whether it does work in ordinary practice.  A pragmatic clinical trial
(PCT) is not a trial with less design.  It is a trial whose design deliberately
allows ordinary patients, clinicians, sites, adherence patterns, and endpoint
capture to enter the evidence.

The design object \(d\) must therefore include more than a randomization list.
It includes the care setting, eligibility rule, treatment strategy, follow-up
schedule, outcome source, adherence policy, and analysis population.  Tools
such as PRECIS-2 ask whether these choices make the trial more explanatory or
more pragmatic along domains such as eligibility, recruitment, setting,
organization, flexibility of delivery, follow-up, primary outcome, and primary
analysis \citep{loudon2015precis2}.  The CONSORT pragmatic-trials extension
makes the same point for reporting: the trial must say which real-world
conditions were part of the design rather than accidental background
\citep{zwarenstein2008consort}.

Dynamic treatment regimes are the multi-stage version of the same promise.  A
regime is not a single treatment label but a sequence of rules that maps a
patient's observed history to the next clinical action; recent work on
preference-aware Q-learning makes this explicit by treating patient-specific
tradeoffs among multiple outcomes as part of the policy target
\citep{zitovsky2023patientPreferencesQLearning}.

Real-world studies (RWS) sit on the other side of the boundary.  Instead of
embedding randomization in care, they read care as it was recorded in EHRs,
claims, registries, devices, or linked administrative systems.  Their design
work is retrospective or protocolized before analysis: define time zero,
eligibility, treatment strategies, follow-up, outcome, censoring, and the
causal contrast before looking for an estimator.  This is why target-trial
emulation is a design idea before it is a modeling idea
\citep{hernan2016targettrial}.  The FDA framework for real-world evidence uses
the same discipline: real-world data can support regulatory evidence only when
the data source and study design are fit for the question being asked
\citep{fda2018rweframework}.

\begin{center}
\small
\textbf{Evidence modes and design risks.}\par\smallskip
\begin{tabular}{@{}>{\raggedright\arraybackslash}p{0.24\linewidth}>{\raggedright\arraybackslash}p{0.32\linewidth}>{\raggedright\arraybackslash}p{0.34\linewidth}@{}}
\toprule
\textbf{Evidence mode} & \textbf{Design promise} & \textbf{Main risk} \\
\midrule
Explanatory trial & Control conditions to isolate biological efficacy &
Narrow eligibility and artificial care conditions \\
Pragmatic clinical trial & Randomize while preserving routine-care context &
Outcome capture, adherence, and site heterogeneity \\
Real-world study & Emulate a trial using already recorded care & Confounding,
time-zero errors, selection, and missing observation mechanisms \\
\bottomrule
\end{tabular}
\end{center}

The practical lesson is simple but unforgiving.  PCT and RWS evidence becomes
statistical only when the clinical question is translated into a design:
\[
  \text{population},\quad
  \text{treatment strategies},\quad
  \text{time zero},\quad
  \text{follow-up},\quad
  \text{endpoint},\quad
  \text{estimand}.
\]
Only after those pieces are named does adjustment, weighting, likelihood, or a
machine-learning nuisance model have a target to protect.

\subsection{A Compact Design SOP}
\label{sec:ch04-design-sop}
\conceptindexes{design SOP, design audit, operating characteristics, estimand alignment}

The practical work of design is never only algebra.  It is the act of making
scientific commitments early enough that the data can answer the question
later.  The following checklist is deliberately short.

\begin{description}[leftmargin=0pt,labelsep=0.65em,style=unboxed,font=\bookdescriptionlabelfont,itemsep=0.45\baselineskip]
\item[Name the design lever.]
Say whether \(d\) acts mainly through assignment, observation, adaptation, or a
combination of them.  The later probability law cannot be understood until this
part is named.

\item[Estimand before estimator.]
State what population quantity or decision target the study is meant to learn.
An estimator without an estimand is a calculation looking for a purpose.

\item[Endpoint before model.]
Define the observed outcome as a random object: scalar, vector, curve,
distribution, event time, or process.  Then define the assessment window,
censoring rule, and missingness handling before choosing the model that will
read it.

\item[Randomization before adjustment.]
Use design to make the primary comparison legible.  Use regression adjustment
to improve precision or handle planned structure, not to rescue an avoidable
confounding problem.

\item[Operating characteristics before launch.]
For dose-finding and adaptive trials, simulate the design under plausible dose
toxicity curves and response scenarios.  Track selection probability, overdose
risk, futility stopping, power, type-I error, expected sample size, and expected
patient allocation as appropriate.  A rule is not understood until its behavior
has been studied under worlds that might actually occur.

\item[Interim looks before peeking.]
If a study may stop early for efficacy, futility, or harm, define the
information times, boundaries, endpoint hierarchy, and monitoring roles before
outcomes accumulate.

\item[Information criterion before optimality.]
Optimal design is only meaningful after the criterion is chosen.  \(D\)-,
\(A\)-, \(c\)-, and \(G\)-optimal designs answer different scientific promises.

\item[Adaptation requires a filtration.]
Every adaptive action should be a function of information already available.
This is the design-level version of the martingale discipline developed later.

\item[Pragmatism is a design choice.]
If the study is meant to answer a routine-care question, say which ordinary
features are deliberately preserved and which are controlled.  If the study is
real-world rather than randomized, write the target trial before writing the
model.

\item[Ethics enters through constraints.]
In clinical studies, safety constraints, stopping rules, maximum sample sizes,
and exclusion rules are not decorations added after the mathematics.  They are
part of the design space \(\mathcal D\).
\end{description}

The message of the chapter is not that every study needs a sophisticated
design.  Sometimes simple randomization, a fixed sample size, and a transparent
endpoint are exactly right.  The message is that simplicity should be chosen,
not inherited by habit.  The design should make clear what the future dataset
will be capable of saying.

\section{Exercises}
\label{sec:ch04-exercises}
\conceptindexes{design exercises, randomization exercises, optimal-design exercises}

\begin{exercise}[Randomization with unequal allocation]
In the finite-population setup of Section~\ref{sec:ch04-randomization}, suppose
complete randomization assigns \(n_1\) units to treatment and \(n_0=N-n_1\) to
control.  Prove directly that the treated sample mean is unbiased for
\(N^{-1}\sum_iY_i(1)\) and the control sample mean is unbiased for
\(N^{-1}\sum_iY_i(0)\).  Explain why the result does not require iid outcomes.
\end{exercise}

\begin{exercise}[One-step behavior of 3+3]
Using the formula for \(\pi_E(p)\), compute the one-step escalation
probabilities at \(p=0.10,0.20,0.30,0.40\).  Plot or tabulate the values.
Where does the design become conservative, and where does it still escalate
often?
\end{exercise}

\begin{exercise}[BOIN boundary algebra]
Show that, when \(0<\phi_1<\phi<\phi_2<1\), the BOIN boundaries satisfy
\[
  \phi_1<\lambda_e<\phi<\lambda_d<\phi_2.
\]
Hint: use the strict concavity of \(q\mapsto x\log q+(1-x)\log(1-q)\) as a
function of \(q\) for fixed \(x\in(0,1)\).
\end{exercise}

\begin{exercise}[A D-optimal straight-line design]
For \(f(x)=(1,x)^T\) on \([-1,1]\), verify from the equivalence theorem that
the design placing weight \(1/2\) at \(-1\) and \(1/2\) at \(1\) is
\(D\)-optimal.
\end{exercise}

\begin{exercise}[A two-point logit design]
For a logit model with \(\eta=\beta_0+\beta_1x\), use the rule
\(\eta=\pm1.5434\) to compute the two locally \(D\)-optimal support points when
\((\beta_0,\beta_1)=(-2,4)\).  Which support points would be infeasible if the
allowable dose interval were restricted to \([0,1]\), and how should the design
problem be modified?
\end{exercise}

\begin{exercise}[Reading a sensitivity plot]
Explain why a proposed \(D\)-optimal design is suspect if its sensitivity
function is positive at a point outside the proposed support.  In the sea
urchin example in Figure~\ref{fig:ch04-sea-urchin-sensitivity}, what do the
two zero crossings say about the selected concentration levels?
\end{exercise}

\begin{exercise}[Uniform design as an engineering promise]
Suppose an aerospace simulation has \(s=6\) controllable factors, each scaled to
\([0,1]\), but the budget allows only \(n=20\) runs.  Explain why a full
factorial design is infeasible.  What does a uniform design promise that a
one-factor-at-a-time design does not?  In words, relate this promise to
discrepancy over rectangular regions of \([0,1]^6\).
\end{exercise}

\begin{exercise}[ImageNet as an observation design]
For the ImageNet example, list four components of \(d_{\mathrm{ImageNet}}\).
How would the learning problem change if the ontology \(\mathcal C\), the
annotation protocol \(A\), or the train--test split \(S\) were changed?  Why is
this a design question rather than merely a data-cleaning question?
\end{exercise}

\begin{exercise}[AIDD acquisition as a design rule]
In the AIDD example, suppose \(\widehat\mu_t(x)\) estimates activity,
\(\widehat\sigma_t(x)\) estimates predictive uncertainty, and
\(\mathrm{cost}(x)\) measures synthetic difficulty.  For
\[
  a_t(x)=\widehat\mu_t(x)+\kappa\widehat\sigma_t(x)
  -\lambda\mathrm{cost}(x),
\]
explain how increasing \(\kappa\) or \(\lambda\) changes the next batch of
compounds.  Define a filtration \(\mathcal F_t\) under which this acquisition
rule is adapted.
\end{exercise}

\begin{exercise}[Single-cell budget allocation]
A lab can afford either \(4\) donors with \(10{,}000\) cells per donor or
\(12\) donors with \(3{,}000\) cells per donor.  Suppose the main estimand is a
treatment effect that should generalize across donors.  Which design is more
convincing, and what assumptions would make the answer change?  Describe how a
single-cell simulator could be used to compare the two designs before
sequencing.
\end{exercise}

\begin{exercise}[Filtration check]
Consider an adaptive trial in which the next allocation probability is a
function of all previously observed outcomes and a fresh random number
generated before the next patient is assigned.  Define a filtration under which
the assignment is adapted.  What would go wrong if the rule also used an
outcome that had not yet been observed at the time of assignment?
\end{exercise}

\begin{exercise}[A real endpoint as a design object]
For a CGM-based diabetes study, compare two endpoint designs: mean glucose over a
valid wear window and the full glucodensity distribution over the same window.
What extra design choices are required for the distributional endpoint?  Which
scientific questions become easier or harder to answer?
\end{exercise}

\begin{exercise}[Pragmatic or explanatory?]
Take a trial question such as whether a diabetes, oncology, or cardiovascular
treatment improves a clinically meaningful outcome.  Write two designs: one
explanatory and one pragmatic.  Compare eligibility, setting, treatment
delivery, adherence, follow-up, outcome source, and primary analysis.  Which
version is closer to routine care, and which version gives a cleaner
biological contrast?
\end{exercise}

\section*{Sources and Further Reading}
\addcontentsline{toc}{section}{Sources and Further Reading}

The randomization argument in this chapter is in the spirit of Neyman's
finite-population view of experiments \citep{neyman1923agricultural} and the
classical design tradition associated with Fisher's experimental work
\citep{fisher1935design}.  The \(3+3\) rule is historically common in oncology
phase I trials; \citet{storer1989phasei} gives an influential early discussion
of phase I design.  The continual reassessment method of
\citet{oquigley1990crm} is a landmark model-based alternative.  BOIN was
introduced by \citet{liu2015boin} as a model-assisted design with simple
interval decisions and strong practical operating characteristics.

Optimal design has a large literature.  The equivalence theorem used here goes
back to \citet{kiefer1960equivalence}; broader treatments include
\citet{fedorov1972optimal}, \citet{pukelsheim2006optimal}, and
\citet{atkinson2007optimum}.  Riemannian-manifold design regions are treated
explicitly by \citet{liDelCastillo2024riemannianDesign}.  The
functional-analytic treatment of how optimal supports and weights vary is
developed in \citet{melas2006functional}.  Work by
Wong and collaborators has been
especially influential in making optimal design computationally usable for
biomedical and nonlinear-model applications; see, for example,
\citet{sverdlov2020review} for a modern review perspective.  Uniform design,
developed by Wang Yuan and Kai-Tai Fang in the late 1970s and published in
their 1979 number-theoretic design paper, is treated here as a central
industrial-design example rather than an aside: public accounts connect its
origin to a defense and aerospace simulation problem with very few affordable
runs
\citep{wangFang1979uniform,cas2009uniformDesign,ckcest2024wangUniformDesign}.
Later treatments include \citet{fangLinWinkerZhang2000uniform} and
\citet{fangLiuQinZhou2018uniform}.  The binary
regression and sea urchin examples draw on \citet{cui2022doptimal} and the
developmental toxicology application of \citet{collins2022seaurchin}.
The broader toxicology framing is influenced by
\citet{nationalresearchcouncil2007toxicity}.  Sparse pharmacokinetic and
pharmacometrics examples are connected to
\citet{mentre1995sparse}, \citet{mentre1997optimal}, and
\citet{nyberg2012poped}.  The AIDD example is motivated by the broader
machine-learning-in-drug-discovery review of \citet{vamathevan2019applications}
and uses the active-learning and Bayesian-optimization viewpoint of
\citet{reker2017active}, \citet{frazier2018tutorial}, and
\citet{shields2021bayesian}.  The ImageNet dataset-design example follows the
large-scale hierarchical benchmark construction of \citet{deng2009imagenet} and
the later deep-learning benchmark result of \citet{krizhevsky2012imagenet}.  The
glucodensity example uses
\citet{cui2023glucodensity} as an endpoint-construction and CGM measurement
design example, not as a randomized-treatment-design example.  The
group-sequential discussion points to the classical monitoring ideas of
\citet{pocock1977group}, \citet{obrien1979multiple}, and
\citet{lan1983discrete}; the Women's Health Initiative example follows
\citet{rossouw2002whi} and the monitoring account of
\citet{anderson2007whiMonitoring}.  The I-SPY 2 platform example follows the
adaptive breast-cancer trial design of \citet{barker2009ispy2} and a published
I-SPY 2 regimen report \citep{park2016neratinib}.  The single-cell design
example points to
the scDesign simulation series
\citep{sun2021scdesign2,song2024scdesign3}.  The pragmatic-trials discussion follows the
explanatory--pragmatic distinction of \citet{schwartz1967explanatory}, the
CONSORT extension for pragmatic trials \citep{zwarenstein2008consort}, and the
PRECIS-2 design tool \citep{loudon2015precis2}.  The real-world-study framing
uses the FDA real-world evidence framework \citep{fda2018rweframework} and the
target-trial-emulation principle of \citet{hernan2016targettrial}.  The brief
dynamic-treatment-regime comment points to preference-aware Q-learning in
\citet{zitovsky2023patientPreferencesQLearning}.  The
famine-relief example uses historical work on Qing famine bureaucracy, state
granaries, markets, and environmental decline
\citep{will1990bureaucracy,will1991nourish,li2007fighting} to emphasize that
administrative records are designed information systems.

%% file: chapters/ch05_observation_likelihood.tex
\chapter{Observation, Conditioning, and Likelihood}
\label{chap:observation-likelihood}
\conceptindexes{observation, conditioning, likelihood, observed-data map, conditional laws, missingness, censoring, coarsening}

\begin{tcolorbox}[
  enhanced,
  breakable,
  colback=chaptercream,
  colframe=bookblue!88!black,
  boxrule=0.72pt,
  arc=5pt,
  boxsep=1pt,
  left=1.0em,
  right=0.95em,
  top=0.82em,
  bottom=0.82em,
  before skip=0.55\baselineskip,
  after skip=1.0\baselineskip
]
\noindent\textbf{Chapter overview.}
This chapter names the observed object: what survives from the designed or
measured world into the record that likelihood and conditioning can read. A
study may filter a richer scientific object through eligibility, treatment,
visits, censoring, missingness, instruments, labels, and preprocessing.
Conditioning names what is known once the observed object \(O\) is seen;
likelihood compares model readings of that same \(O\). Do not trust a
likelihood until the observed object, its information, the conditioning
operation, and the target have been named.
\end{tcolorbox}

A dataset is not the world in miniature.  It is a survived record: a trace of
the world after design, classification, measurement, storage, and processing
have decided what can be seen.  This is why the chapter belongs here.  The
design chapter made the law of the full data visible.  But even the full-data
record \(W\) may still be too generous.  The product-space chapter will soon
build joint laws from product spaces and transition kernels.  Before doing
that, the object whose law is being built must be named.  What the analyst
receives is often
\[
  O=\phi(W,U),
\]
where \(\phi\) is an observation map and \(U\) records measurement noise,
sampling indicators, censoring times, label decisions, library preparation,
platform logging, or other auxiliary randomness.  The statistical question is
therefore not only ``What is the law of the world \(W\)?'' but also ``What has
survived into \(O\), and under which model is \(O\) being read?''  This chapter
keeps the notation light: \(O\) is the observed object, \(P_\theta\) is usually
a model for \(O\), and design labels such as \(d\) are shown only when the
design mechanism itself is part of the point.

This question is older and broader than likelihood theory.  Tukey's program
for data analysis insisted that data analysis begins with the character of the
data, not with a predetermined formalism \citep{tukey1962future}.  Box's view
of models as useful approximations warns against mistaking a fitted model for
the system itself \citep{box1976science}.  Shannon's communication theory made
the channel part of the message; in statistics, the observation mechanism plays
a similar role \citep{shannon1948mathematical}.  Rubin's missing-data framework
made non-observation part of the probability law \citep{rubin1976inference}.
Cox's survival model showed that risk sets and censoring rules can be the
natural language of a clinical endpoint \citep{cox1972regression}.  Outside
statistics, Porter and Bowker--Star remind the analyst that measurement, classification,
and objectivity are social and technical infrastructures, not neutral
background conditions \citep{porter1995trust,bowker1999sorting}.

The same lesson appears in modern data industries.  A Netflix rating matrix is
not a map of all preferences; it is a platform log shaped by exposure,
selection, and missing ratings \citep{bennett2007netflix}.  ImageNet and COCO
are not vision itself; they are labeled visual worlds created by category
systems, image sources, and annotation protocols
\citep{deng2009imagenet,lin2014microsoft}.  ADNI is not Alzheimer's disease in
the abstract; it is a harmonized neuroimaging and clinical protocol
\citep{petersen2010adni}.  RiskMetrics is not financial risk itself; it is a
convention for reading market histories into portfolio risk
\citep{jpmorgan1996riskmetrics}.  Clinical-trial estimands make the same point
regulatory: the endpoint, population, intercurrent-event strategy, and summary
measure must be named before an analysis can be interpreted
\citep{ich2021e9r1}.

\begin{realdatacapsule}{CARMELINA and real-world oncology endpoints}
\item[Data object.] A randomized cardiovascular outcomes trial such as
CARMELINA \citep{rosenstock2019linagliptin}, paired conceptually with
EHR-derived oncology endpoints from routine care
\citep{griffith2019tumorBurden,fda2018rweframework}.
\item[Observation mechanism.] Trial endpoints are protocol-scheduled and
adjudicated; routine-care oncology records are produced by visits, imaging
frequency, chart abstraction, treatment-line coding, and mortality linkage.
\item[Target.] A trial estimand or real-world endpoint such as survival,
progression, treatment duration, or a treatment-strategy contrast.
\item[Model.] Observed-data likelihoods, censoring models, target-trial
emulation, and time-to-event regression connect the recorded endpoint to the
target law.
\item[Uncertainty.] Standard errors, survival bands, sensitivity analyses for
censoring and endpoint misclassification, and protocol deviations carry the
main uncertainty.
\item[Limitation.] A routine-care endpoint is not automatically a trial
endpoint; visit schedules, documentation, and confounding can change the
meaning of the target.
\end{realdatacapsule}

\begin{realdatacapsule}{RiskMetrics and daily returns}
\item[Data object.] A frozen series of asset returns, portfolio weights, and
losses, read through the RiskMetrics volatility convention
\citep{jpmorgan1996riskmetrics,fama1965behavior}.
\item[Observation mechanism.] Prices are sampled at trading times, cleaned,
adjusted for corporate actions, and converted into returns under market
microstructure and calendar choices.
\item[Target.] A risk functional such as next-day volatility, value at risk,
expected shortfall, or exceedance probability.
\item[Model.] Exponentially weighted covariance, heavy-tailed return models,
or conditional-volatility models summarize dependence and scale.
\item[Uncertainty.] Backtesting exceedances, block bootstrap variation, tail
sensitivity, and stress periods diagnose whether the risk rule is calibrated.
\item[Limitation.] A short return history can miss regime changes, liquidity
constraints, dependence in extremes, and policy feedback from the risk rule
itself.
\end{realdatacapsule}

In a clinical trial, the scientific object might include potential outcomes,
toxicity times, biomarker trajectories, adherence histories, and clinical
decisions.  The observed data may include only eligibility variables, assigned
treatment, visit-time measurements, adverse-event indicators, and a censored
endpoint.  In a single-cell study, the biological object includes cells,
lineages, cell states, donors, spatial neighborhoods, and molecular counts
before technical loss.  The observed matrix records a processed version of that
object after sampling, library preparation, sequencing, alignment, and
normalization.  In a historical archive, the observed record may be a report
written under administrative pressure rather than a direct measurement of
famine, migration, or local prices.

Nancy Reid, at the University of Toronto, gave a UCLA lecture called ``When
likelihood goes wrong'' that is a useful warning sign for this chapter
\citep{reid2024whenlikelihood}.  Likelihood is one of statistics' most
powerful languages because it compares parameter values through a model for the
observed data.  But that strength is also its vulnerability.  If the observed
object has been misnamed, if the model is a poor representation of its law, if
the parameter of interest changes meaning when nuisance structure changes, or
if an asymptotic approximation is asked to do more than it can, likelihood can
become confident about the wrong object.  Recent work by \citet{battey2024role}
sharpens this point: parameterization and orthogonality can decide whether a
parameter of interest remains consistently recoverable when a nuisance
component is misspecified.

\section{Observed Data as Functions on a Full Data Space}
\label{sec:ch05-observed-data-map}
\conceptindexes{observed data, observed-data map, full data, measurement, coarsening, data biographies}

The point is simple before it is formal.  The analyst usually does not see
every scientific quantity that matters.  The analyst sees a record made by a
protocol, an instrument, a platform, or an archive.  A survival study may care
about the event time \(T\), but the case report form records a censored pair.
A single-cell study may care about biological abundance, but the dataset
records counts after capture and sequencing.  A historical study may care about
hunger or migration, but the archive records reports that an administration
could collect and transmit.

The cleanest way to keep this distinction honest is to separate three objects:
\[
  \text{scientific world} \quad
  \longrightarrow \quad
  \text{designed full data} \quad
  \longrightarrow \quad
  \text{observed data}.
\]
The first arrow is substantive and often unmodeled.  The second arrow is the
statistical object available for formalization.  Before writing the formal map, it helps
to see what such a map does.

\input{figures/ch05_observed_object_pipeline}

\begin{example}[Censored event time]
Let \(T\) be a failure time and \(C\) a censoring time.  The full data may be
\((T,C)\), but a standard right-censored observation is
\[
  O=(Z,\delta)
  =
  \{\min(T,C),\ind{T\le C}\}.
\]
The observation forgets the exact failure time whenever \(\delta=0\).  A
likelihood for survival data is therefore not automatically a likelihood for
the full pair \((T,C)\) unless the censoring law is included.  Under independent
censoring, parts of the censoring law may be irrelevant for inference about the
failure-time distribution; without such a condition, the same observed
\((Z,\delta)\) can be compatible with different scientific stories.
\end{example}

\begin{example}[Single-cell observation]
Let \(X_{ig}\) be the true molecular abundance for cell \(i\) and gene \(g\),
and let \(R_{ig}\in\{0,1\}\) indicate whether the molecule is captured and
reported after technical processing.  A simplified observed count may be
\[
  O_{ig}=R_{ig}X_{ig}.
\]
If \(R_{ig}\) depends on cell type, batch, transcript abundance, or sequencing
depth, then the observed zero is not automatically a biological zero.  The
observation mechanism has created a data feature that the model must interpret.
\end{example}

\begin{example}[Historical reports as observed data]
A famine archive may record county reports, relief requests, grain transfers,
local prices, and migration records.  The full historical object includes
household hunger, local storage, transport bottlenecks, local political
incentives, and unreported suffering.  The observed archive is not a neutral
sample from that object.  It is the output of an administrative observation
mechanism.  This is why the design discussion from the preceding chapter
returns here as a likelihood problem: a model for the archive is a model for
what the state could see, record, and transmit.
\end{example}

These examples share the same mathematical shape.  The designed full data live
in one space, the observed data live in another, and the observation mechanism
is the rule that carries the first object into the second.

\begin{definition}[Observed-data map]
Let \(W\) denote the designed full-data object and let \(O\) denote the record
actually analyzed.  An observed-data map is a measurable function
\[
  \phi:\mathcal W\longrightarrow\mathcal Y
\]
with \(O=\phi(W)\).  If \(W\sim P\), the observed law is the push-forward
\[
  Q
  =
  P\circ \phi^{-1}.
\]
When the full law is \(P_\theta\), write \(Q_\theta=P_\theta\circ\phi^{-1}\).
\end{definition}

The letter \(Q\) is just a temporary reminder that this is the law after
observation.  Once the observed object has been named, the notation usually returns to the
lighter notation \(P_\theta\) for a model of \(O\).  If important pieces of
\(W\) are hidden, coarsened, or sampled with unequal probability, then a model
for \(O\) must say how that happened or deliberately choose a target that does
not require it.

\begin{proposition}[The observed law is a probability law]
If \(P\) is a probability measure on \((\mathcal W,\mathcal A)\) and
\(\phi:\mathcal W\to\mathcal Y\) is measurable, then
\(Q=P\circ\phi^{-1}\) is a probability measure on the observed-data space.
\end{proposition}

\noindent\textit{Proof.}
For a measurable observed-data event \(B\), define
\[
  Q(B)=P\{\phi(W)\in B\}.
\]
Measurability of \(\phi\) ensures that \(\phi^{-1}(B)\in\mathcal A\).
Countable additivity follows because inverse images preserve countable
disjoint unions.  Finally,
\[
  Q(\mathcal Y)
  =
  P\{\phi(W)\in\mathcal Y\}
  =
  P(\mathcal W)=1.
\qedmark
\]

The proof is elementary, but it is one of the most important habits in applied
probability.  Do not write a density for the data until it is clear which data
space the density lives on.

\conceptindexes{quantum statistical inference, quantum measurement, density matrix, Born rule}
\begin{example}[Quantum measurements as observation laws]
Quantum statistical inference is an extreme version of the same warning
\citep{holevo1982probabilistic,barndorffNielsenGillJupp2003quantum,petz2008quantum}.
The unknown object is not first a classical distribution on the reported data;
it may be a density matrix \(\rho_\theta\) on a Hilbert space.  A measurement
choice is part of the observation mechanism.  For a finite-outcome measurement
given by positive operators \(E_b\) with \(\sum_b E_b=\matI\), the reported
classical outcome \(B\) has law
\[
  P_{\theta,E}(B=b)
  =
  \operatorname{tr}(\rho_\theta E_b).
\]
Once \(E\) is fixed, the observed outcomes can be analyzed by ordinary
likelihood methods.  What is new is upstream of that likelihood: different
measurements can produce different classical experiments from the same
\(\rho_\theta\), and noncommuting observables should not be silently treated
as if they were jointly observed classical coordinates.  Thus the statistical
object is the pair consisting of the quantum state model and the measurement
law that turns it into data.
\end{example}

\section{Information and Conditional Laws}
\label{sec:ch05-conditioning}
\conceptindexes{conditioning, conditional expectation, conditional law, conditional independence, information}

Once the observed object \(O\) has been defined, the next question is what can
be learned from it.  Mathematically, information is represented by a
\(\sigma\)-field.  If the full experiment lives on
\((\Omega,\mathcal F,\Prob)\), then the information carried by \(O\) is
\[
  \sigma(O)=\{O^{-1}(G):G\in\mathcal G\}.
\]
A statistic \(T=T(O)\) is a coarser reading of the same data; it carries the
smaller information set \(\sigma(T)\subseteq\sigma(O)\).  This inclusion is
the formal version of a familiar practical fact: summarizing data can make the
analysis clearer, but it can also discard evidence.

For this chapter, conditioning has three connected meanings.
First, conditional expectation is a projection: it is the part of a random
quantity that can be read using a chosen information set.  Second, a
conditional law is the object that will later be formalized as a kernel: it
describes how the remaining randomness is distributed after some variables,
history, or design features are known.  Third, conditional likelihood is a
modeling decision: it treats some part of
the observed record as fixed in order to focus inference on a parameter of
interest.  These are not three unrelated techniques.  They are three faces of
the same question: what is being regarded as already known when the probability
calculation is made?

\begin{definition}[Conditional expectation as projection]
Let \(Y\) be integrable and let \(\mathcal H\subseteq\mathcal F\) be a
\(\sigma\)-field.  A conditional expectation of \(Y\) given \(\mathcal H\) is
an \(\mathcal H\)-measurable random variable, denoted
\(\Expect(Y\mid\mathcal H)\), such that
\[
  \int_H \Expect(Y\mid\mathcal H)\,d\Prob
  =
  \int_H Y\,d\Prob,
  \qquad H\in\mathcal H.
\]
If \(\mathcal H=\sigma(X)\), also write \(\Expect(Y\mid X)\).
\end{definition}

\noindent\textit{Remark (conditioning on a random variable).}
Conditional expectation is defined with respect to a \(\sigma\)-field because
conditioning first means choosing what information is available, not choosing a
numerical value.  A random variable \(Y\) carries exactly the information
\(\sigma(Y)\), so \(\Expect(X\mid Y)\) means
\(\Expect(X\mid\sigma(Y))\); this also avoids pretending that \(\{Y=y\}\) has
positive probability in continuous problems.  The pointwise notation used in
statistics comes from a second step: by the Doob--Dynkin factorization principle, a
\(\sigma(Y)\)-measurable version can be written as \(m(Y)\), with \(m\) unique
only \(P_Y\)-almost surely.  Thus
\(\Expect(X\mid Y)(\omega)=m(Y(\omega))\) for the chosen version, and
\(\Expect(X\mid Y=y)\) denotes the same version evaluated at \(y\).  When a
regular conditional law \(K(y,dx)=\Prob(X\in dx\mid Y=y)\) exists, the
disintegration theorem in \Appref{cor:appB-disintegration-random-variable} gives
\(m(y)=\int x\,K(y,dx)\) for \(P_Y\)-almost every \(y\).  If a joint density exists, one familiar version is
\[
  m(y)
  =
  \int x\,p_{X\mid Y}(x\mid y)\,dx
  =
  \frac{\int x\,p_{X,Y}(x,y)\,dx}{p_Y(y)},
  \qquad p_Y(y)>0.
\]
On \(P_Y\)-null sets, or outside the support of \(Y\), the pointwise value is a
version choice rather than an identifiable feature of the law.

The definition says that conditional expectation is the best version of \(Y\)
that can be expressed using the information in \(\mathcal H\).  In regression,
\(m(x)=\Expect(Y\mid X=x)\) is the part of the response readable from \(X\).

\begin{proposition}[\(L^2\) projection reading]
If \(Y\in L^2(\Prob)\), then \(\Expect(Y\mid\mathcal H)\) is the orthogonal
projection of \(Y\) onto the closed subspace of \(L^2(\Prob)\) consisting of
\(\mathcal H\)-measurable square-integrable random variables.  Equivalently, it
is the unique \(Z\) in that subspace such that
\[
  \Expect\{(Y-Z)W\}=0
  \qquad
  \text{for every } \mathcal H\text{-measurable }W\in L^2(\Prob),
\]
and it minimizes \(\Expect(Y-W)^2\) over all such \(W\).
\end{proposition}

\noindent\textit{Proof.}
Let \(Z=\Expect(Y\mid\mathcal H)\).  The defining identity gives
\(\Expect\{(Y-Z)\indset H\}=0\) for every \(H\in\mathcal H\).  By linearity,
bounded approximation, and \(L^2\) truncation, the same identity holds with
\(\indset H\) replaced by every \(\mathcal H\)-measurable \(W\in L^2(\Prob)\).
Thus \(Y-Z\) is orthogonal to the \(\mathcal H\)-measurable subspace.  The
Pythagorean identity
\[
  \Expect(Y-W)^2
  =
  \Expect(Y-Z)^2+\Expect(Z-W)^2
\]
then proves the minimizing property and uniqueness.
\qedmark

\input{figures/ch05_conditioning_projection}

\begin{proposition}[The tower property]
If \(\mathcal H_0\subseteq\mathcal H_1\subseteq\mathcal F\) and \(Y\) is
integrable, then
\[
  \Expect\{\Expect(Y\mid\mathcal H_1)\mid\mathcal H_0\}
  =
  \Expect(Y\mid\mathcal H_0)
  \quad\text{a.s.}
\]
\end{proposition}

\noindent\textit{Proof.}
The left side is \(\mathcal H_0\)-measurable.  For every \(H\in\mathcal H_0\),
the defining property of conditional expectation gives
\[
  \int_H \Expect\{\Expect(Y\mid\mathcal H_1)\mid\mathcal H_0\}\,d\Prob
  =
  \int_H \Expect(Y\mid\mathcal H_1)\,d\Prob
  =
  \int_H Y\,d\Prob,
\]
because \(H\in\mathcal H_1\).  Uniqueness up to almost sure equality gives the
claim.
\qedmark

The tower property is the quiet engine behind many statistical calculations.
It is why residuals have mean zero after conditioning on covariates, why
sequential prediction can be updated one information set at a time, and why an
adaptive design can remain coherent if each action depends only on the past.

\begin{proposition}[Pulling out known information]
Let \(Y\) be integrable, let \(\mathcal H\subseteq\mathcal F\), and let \(Z\)
be bounded and \(\mathcal H\)-measurable.  Then
\[
  \Expect(ZY\mid\mathcal H)
  =
  Z\,\Expect(Y\mid\mathcal H)
  \quad\text{a.s.}
\]
\end{proposition}

\noindent\textit{Proof.}
The right side is \(\mathcal H\)-measurable.  For \(H\in\mathcal H\),
\[
  \int_H Z\,\Expect(Y\mid\mathcal H)\,d\Prob
  =
  \int_H ZY\,d\Prob,
\]
first for indicator \(Z=\indset A\), \(A\in\mathcal H\), then for bounded
simple \(Z\), and finally for bounded \(\mathcal H\)-measurable \(Z\) by
bounded approximation.  The defining property of conditional expectation gives
the claim.
\qedmark

This rule is the small algebra behind a large amount of applied statistics.
Once covariates, histories, or risk-set membership are treated as known, they
can be moved outside the conditional averaging operation.  That is why
conditional mean-zero residuals become estimating equations:
\[
  \Expect\{Y-m(X)\mid X\}=0
  \quad\Longrightarrow\quad
  \Expect\{h(X)(Y-m(X))\}=0
\]
for every bounded measurable instrument \(h\).

\begin{definition}[Conditional law]
Let \(Y\) take values in a measurable space \((\mathcal Y,\mathcal B)\), and
let \(\mathcal H\subseteq\mathcal F\).  A conditional law of \(Y\) given
\(\mathcal H\) is a family of conditional probability measures, written in the
kernel notation
\[
  K(\omega,B),\qquad B\in\mathcal B,
\]
such that \(K(\cdot,B)\) is a version of
\(\Prob(Y\in B\mid\mathcal H)\) for every \(B\in\mathcal B\).  When
\(\mathcal H=\sigma(X)\), write \(K(x,B)\) informally as
\(\Prob(Y\in B\mid X=x)\).
\end{definition}

Regular conditional laws require some care on general measurable spaces.  For
standard Borel spaces, the versions needed in most statistical models exist.
Chapter~6 gives the formal definition of a transition kernel and turns this
preview notation into machinery.  For now, the
important point is conceptual: a conditional law is the mathematical form of a
data-generating instruction.
\Appref{sec:appB-conditioning-toolkit} records the Radon--Nikodym and
regular-conditional-law facts behind this statement; the present chapter uses
them as modeling language.

\begin{example}[Bayesian conditioning]
Bayesian statistics is a particularly direct use of the same conditioning
language.  Its philosophical claim is not that an unknown constant must be
physically random.  Rather, uncertainty about that value may be represented by a
coherent probability law and then updated by the information actually observed.
The historical and axiomatic lines run from Bayes and Laplace through
de Finetti's coherence view and Savage's decision-theoretic foundations
\citep{bayes1763essay,laplace1812theorie,definetti1937prevision,savage1954foundations};
modern hierarchical modeling uses the same calculus at larger scale
\citep{gelman2013bayesian}.

Let \(\Theta\) be the unknown parameter, taking values in
\((\mathsf T,\mathcal T)\) with prior law \(\Pi\), and suppose that, given
\(\Theta=\theta\) and design \(d\), the observed object \(O\) has law
\(P^O_{\theta,d}\) on \((\mathcal O,\mathcal G)\).  The Bayesian joint law is
\[
  M_d(d\theta,do)=\Pi(d\theta)\,P^O_{\theta,d}(do).
\]
A posterior distribution is a regular conditional law of \(\Theta\) given the
observed information \(\sigma(O)\):
\[
  \Pi_d(B\mid O)=M_d(\Theta\in B\mid \sigma(O)),
  \qquad B\in\mathcal T.
\]
In standard Borel parameter and observation spaces,
\Appref{sec:appB-regular-conditional-laws} guarantees that such a version can be
chosen as a probability kernel \(o\mapsto\Pi_d(\cdot\mid o)\); it also records
the measure-theoretic disintegration argument behind the usual Bayes formula.
This is the same distinction as above.  The mathematically primitive object is
conditioning on the information \(\sigma(O)\).  Once a kernel version is chosen,
the posterior random probability \(\Pi_d(B\mid O)\) can be written as
\(\Pi_d(B\mid O(\omega))\) on each outcome \(\omega\), and after observing
\(O=o_{\mathrm{obs}}\) the analyst reports
\(\Pi_d(B\mid o_{\mathrm{obs}})\).  For continuous
data this is not conditioning on a positive-probability event
\(\{O=o_{\mathrm{obs}}\}\); it is evaluating the disintegrated conditional law
at the observed record.

When the observed laws are dominated by a common measure \(\nu\), with density
\(p_{\theta,d}(o)=dP^O_{\theta,d}/d\nu(o)\), this conditional law becomes the
usual Bayes formula:
\[
  \Pi_d(B\mid o)
  =
  \frac{\int_B p_{\theta,d}(o)\,\Pi(d\theta)}
       {\int_{\mathsf T} p_{\theta,d}(o)\,\Pi(d\theta)}
\]
whenever the denominator is finite and positive.  Thus the posterior is not
built from the full-data law unless the full data are observed; it is built from
the observed-data law determined by the observation map.  Chapter~6 uses the
same conditioning grammar when the unknown object is itself a probability
measure \(G\in\mathcal P(S)\), as in Ferguson's Dirichlet-process prior
\citep{ferguson1973bayesian,ferguson1974prior}.
\end{example}

\begin{definition}[Conditional independence]
Let \(\mathcal F_1,\mathcal F_2,\mathcal H\subseteq\mathcal F\).  The
\(\sigma\)-fields \(\mathcal F_1\) and \(\mathcal F_2\) are conditionally
independent given \(\mathcal H\) if
\[
  \Prob(A\cap B\mid\mathcal H)
  =
  \Prob(A\mid\mathcal H)\Prob(B\mid\mathcal H)
  \quad\text{a.s.}
\]
for all \(A\in\mathcal F_1\) and \(B\in\mathcal F_2\).
\end{definition}

Conditional independence is the grammar of many statistical models.  In a
trial, treatment assignment may be independent of future potential outcomes
given baseline information because the design enforces randomization.  In a
longitudinal model, the future may be conditionally independent of the remote
past given the current state, which is the Markov reading of a patient history.
In a missing-data analysis, whether nonresponse is ignorable is a conditional
independence question about the response indicator and the unseen value after
the observed record is known.  The notation is compact, but the scientific
claim is large.

\begin{example}[Treatment assignment and outcome]
In a randomized trial, let \(A\in\{0,1\}\) be treatment assignment and \(Y\) the
outcome.  The design specifies
\[
  P_d(A=1\mid X)=\pi(X),
\]
where \(X\) denotes baseline information.  The outcome model specifies
\[
  P_\theta(Y\in dy\mid A,X).
\]
The joint law can be read as
\[
  P_{\theta,d}(dx,da,dy)
  =
  P_X(dx)\,P_d(da\mid x)\,P_\theta(dy\mid a,x).
\]
This display is already a kernel product.  The design kernel and the outcome
kernel have different scientific roles, and confusing them is a common source
of causal mistakes.
\end{example}

\begin{example}[Prediction after partial observation]
Suppose \(Y\) is a future biomarker and \(O_t\) is the patient record available
through visit \(t\).  The prediction
\[
  \Expect(Y\mid O_t)
\]
is not a fixed property of the patient alone.  It is a property of the patient
as observed through a particular schedule, assay, and clinical record.  Two
studies can target the same biological quantity but have different predictions
because their observed \(\sigma\)-fields are different.
\end{example}

\section{Likelihood Under Observation Models}
\label{sec:ch05-likelihood}
\conceptindexes{likelihood, likelihood ratio, Radon--Nikodym derivative, reference measure, observed-data law}

At its simplest, a likelihood is an ordinary probability formula read after
the data have been observed.  Suppose \(Y\in\{0,1\}\) and the model says
\(P_\theta(Y=1)=\theta\).  Before observing \(Y\), the model assigns
\[
  P_\theta(Y=y)=\theta^y(1-\theta)^{1-y},
  \qquad y\in\{0,1\}.
\]
After observing \(Y=y\), the same expression is read as a function of
\(\theta\):
\[
  L(\theta;y)\propto \theta^y(1-\theta)^{1-y}.
\]
For independent binary observations \(y_1,\ldots,y_n\), with
\(s=\sum_i y_i\), this becomes
\[
  L(\theta;y_1,\ldots,y_n)
  \propto
  \theta^s(1-\theta)^{n-s}.
\]
If the data are recorded only through the success count \(s\), a binomial
coefficient \(\binom ns\) appears, but it does not depend on \(\theta\).  This
is the basic reading rule: likelihood is not a probability distribution over
\(\theta\); it is a way to compare parameter values by the probability or
density they assign to the data actually seen.

\input{figures/ch05_likelihood_reading}

The formal version uses the Radon--Nikodym derivative.  Likelihood begins once
a model assigns probabilities to the observed data.  Let
\(\{P_\theta:\theta\in\Theta\}\) be a family of laws for the observed object
\(O\).  If these laws are dominated by a common measure \(\nu\), write
\[
  p_\theta(o)=\frac{dP_\theta}{d\nu}(o).
\]
After observing \(O=o\), the likelihood is
\[
  L(\theta;o)\propto p_\theta(o).
\]
Thus the familiar probability mass function and probability density function
are both special cases of the same Radon--Nikodym object.  The common
dominating measure \(\nu\) is bookkeeping: changing \(\nu\) can multiply
\(L(\theta;o)\) by a factor depending on \(o\), but not on \(\theta\), so the
likelihood comparison is unchanged.
The proportionality sign is important: likelihood compares parameter values
for the same observed data.  It is not a probability distribution over
\(\theta\) unless a prior distribution is supplied.

\begin{example}[Zero-inflated observations need a mixed reference measure]
Some observed-data laws have mass on pieces of different dimension.  Suppose
an outcome \(O=(X,Y)\) lives on \([0,\infty)^2\), but zeros can be structural:
both components may be zero, exactly one component may be positive, or both may
be positive.  Let \(\lambda_+\) denote Lebesgue measure on \((0,\infty)\).  A
natural dominating measure is
\[
  \nu
  =
  \delta_{(0,0)}
  +\lambda_+\otimes\delta_0
  +\delta_0\otimes\lambda_+
  +\lambda_+\otimes\lambda_+ .
\]
A model can then have density, relative to \(\nu\),
\[
q_\theta(x,y)=
\begin{cases}
  p_{00,\theta}, & (x,y)=(0,0),\\
  p_{10,\theta}f_{10,\theta}(x), & x>0,\ y=0,\\
  p_{01,\theta}f_{01,\theta}(y), & x=0,\ y>0,\\
  p_{11,\theta}f_{11,\theta}(x,y), & x>0,\ y>0,
\end{cases}
\]
where the four mixing probabilities sum to one and the displayed component
densities integrate to one on their supports.  An observation on an axis
therefore contributes a one-dimensional density term, while an observation in
the interior contributes a two-dimensional density term.  Writing only a
density with respect to \(dx\,dy\) would erase the structural-zero events.
\end{example}

\begin{definition}[Likelihood ratio]
For two parameter values \(\theta_0,\theta_1\), the likelihood ratio based on
the observed data is
\[
  \Lambda(\theta_1,\theta_0;O)
  =
  \frac{p_{\theta_1}(O)}{p_{\theta_0}(O)}
  =
  \frac{dP_{\theta_1}}{dP_{\theta_0}}(O),
\]
when \(P_{\theta_1}\) is absolutely continuous with respect to
\(P_{\theta_0}\) on the relevant support.
\end{definition}

This is the Radon--Nikodym theorem in statistical clothing.  A likelihood ratio
does not merely say which parameter is more plausible; it says which observed
events have been reweighted when one model is changed into another.  That is
why likelihood is so useful for tests, confidence intervals, Bayesian updating,
and asymptotic approximation.

\begin{example}[Iid sampling as a special observation mechanism]
If \(O=(Y_1,\ldots,Y_n)\) and the model says that the observations are iid with
density \(f_\theta\), then
\[
  L(\theta;O)
  =
  \prod_{i=1}^n f_\theta(Y_i).
\]
The familiar product form is not a definition of likelihood.  It is a
consequence of a strong observation model: the \(n\) coordinates are conditionally
independent and have the same distribution.  A cluster sample, repeated-measures
study, adaptive trial, or spatial lattice usually needs a different
factorization.
\end{example}

\begin{example}[Randomization likelihood and model likelihood]
In a randomized experiment, the assignment mechanism has a known law
\(P_d(A\mid X)\).  For a sharp null hypothesis, the missing potential outcomes
can be filled in from the observed outcomes, and a randomization test can use
the distribution of a statistic over possible assignments.  This is likelihood
or probability under the design, not under a superpopulation outcome model.
For a parametric outcome model, one instead writes a density for the observed
outcomes given the assignments.  Both calculations are legitimate, but they
condition on and randomize over different parts of the experiment.
\end{example}

\subsection{Real datasets, different likelihoods}
\label{sec:ch05-real-likelihoods}
\conceptindexes{dataset likelihoods, ImageNet, London bombing, Dream of the Red Chamber, missing species, climate proxies, clinical endpoints, single-cell data}

The phrase ``the likelihood of the data'' hides a plural.  Different observed
objects from the same scientific area lead to different likelihoods, even when
the substantive question sounds similar.  We first return to several data
types from Chapters~1 and~2, then move to clinical and single-cell examples.

\begin{example}[Image annotations]
An image benchmark is not only a collection of pictures.  It is a collection
of pictures read through labels, boxes, masks, and annotation rules
\citep{deng2009imagenet,lin2014microsoft}.  Let \(I_i\) be image \(i\) and
let \(A_i\in\{1,\ldots,K\}\) be its observed class label.  A discriminative
classification analysis uses the conditional likelihood
\[
  L_{\mathrm{cls}}(\theta)
  =
  \prod_{i=1}^n \pi_\theta(A_i\mid I_i),
  \qquad
  \sum_{k=1}^K \pi_\theta(k\mid I_i)=1 .
\]
This is a likelihood for labels given images, not a generative likelihood for
photographs.  The conditioning is substantive: scanner artifacts, image
sources, cropping, and category construction are being held fixed as part of
the input object.

If the observed annotation is noisy, introduce a latent semantic class
\(Y_i\) and an annotation matrix
\(\Gamma_{ay}=\Prob(A_i=a\mid Y_i=y)\).  The observed-label contribution is
\[
  L_i(\theta,\Gamma)
  =
  \sum_{y=1}^K
  \Gamma_{A_i y}\,\pi_\theta(y\mid I_i).
\]
For segmentation, the observed object may instead be a pixel mask
\(M_i=(M_{iv}:v\in\mathcal V_i)\).  A conditionally independent pixel reading
uses
\[
  L_i^{\mathrm{pix}}(\theta)
  =
  \prod_{v\in\mathcal V_i}
  \pi_{\theta,v,M_{iv}}(I_i),
\]
whereas a spatially coupled reading might use a conditional random field
\[
  p_\theta(m\mid I_i)
  =
  \frac{1}{Z_\theta(I_i)}
  \exp\left\{
    \sum_{v\in\mathcal V_i}h_{\theta v}(I_i,m_v)
    +
    \rho\sum_{u\sim v}\ind{m_u=m_v}
  \right\}.
\]
The same image archive therefore supports different likelihoods depending on
whether the analyst treats the object as a classified image, a noisy human
annotation, or a spatially structured mask.
\end{example}

\begin{example}[Spatial point patterns]
The London bombing example in Chapter~2 began with impact locations on a map
\citep{clarke1946poisson}.  If the observed record is the unordered point
pattern \(S=\{s_1,\ldots,s_n\}\) in a window \(W\), an inhomogeneous Poisson
point-process reading writes the likelihood, relative to a unit-rate Poisson
reference measure, as
\[
  L_{\mathrm{pp}}(\theta)
  =
  \exp\left\{-\int_W \lambda_\theta(s)\,ds\right\}
  \prod_{i=1}^n \lambda_\theta(s_i).
\]
The integral is exposure over space; the product is the contribution of the
observed impact locations.

If the same map is first divided into cells \(A_1,\ldots,A_J\), the observed
object becomes counts \(N_j=\#(S\cap A_j)\).  With
\(\mu_j(\theta)=\int_{A_j}\lambda_\theta(s)\,ds\), the binned likelihood is
\[
  L_{\mathrm{cell}}(\theta)
  =
  \prod_{j=1}^J
  \exp\{-\mu_j(\theta)\}
  \frac{\mu_j(\theta)^{N_j}}{N_j!}.
\]
If the analyst conditions on the total number of impacts \(n=\sum_jN_j\), the
same counts lead instead to a multinomial likelihood
\[
  L_{\mathrm{cond}}(\theta)
  =
  \frac{n!}{\prod_jN_j!}
  \prod_{j=1}^J
  p_j(\theta)^{N_j},
  \qquad
  p_j(\theta)=
  \frac{\mu_j(\theta)}{\sum_{\ell=1}^J\mu_\ell(\theta)}.
\]
Thus a map, a set of points, and a table of cell counts are related, but they
do not carry identical likelihoods.  The conditioning choice decides whether
the total event rate is part of the evidence.
\end{example}

\begin{example}[Text features and authorship]
Text becomes data only after an encoding has been chosen.  For a stylometric
reading of \emph{Dream of the Red Chamber}, let \(N_{jv}\) be the count of
feature \(v\) in chapter \(j\), let \(m_j=\sum_vN_{jv}\), and let \(a_j\) be
the hypothesized author or style regime for that chapter
\citep{hu2014redchamber,tu2013redchamber,zhu2021redchamber}.  A simple
multinomial style likelihood is
\[
  L_{\mathrm{text}}(\theta,a)
  =
  \prod_j
  \frac{m_j!}{\prod_v N_{jv}!}
  \prod_v
  \theta_{a_jv}^{N_{jv}},
  \qquad
  \sum_v\theta_{av}=1 .
\]
Here the feature map is part of the observation model: function words,
character names, punctuation, bigrams, sentence lengths, and verse/prose
features would give different count arrays.

Chapter-to-chapter heterogeneity can be made explicit by letting each chapter
have its own feature probability vector
\(\Theta_j\sim\Dirichlet(\alpha_{a_j})\).  Integrating
\(\Theta_j\) out gives the Dirichlet--multinomial contribution
\[
  L_j(\alpha_{a_j})
  =
  \frac{m_j!}{\prod_vN_{jv}!}
  \frac{\Gamma(\alpha_{a_j+})}{\Gamma(\alpha_{a_j+}+m_j)}
  \prod_v
  \frac{\Gamma(\alpha_{a_jv}+N_{jv})}{\Gamma(\alpha_{a_jv})},
  \qquad
  \alpha_{a+}=\sum_v\alpha_{av}.
\]
If the question is instead Shakespeare's unseen vocabulary, the same token
stream is summarized by low-frequency counts \(f_1,f_2,\ldots\)
\citep{efronThisted1976unseen}.  Then the likelihood must include a model for
unseen types or missing mass.  Authorship and unseen-vocabulary problems both
start from text, but they read different summaries through different
likelihoods.
\end{example}

\begin{example}[Historical proxy records]
The Zhu/Chu Ko-chen climate curve in Chapter~2 is a proxy-data example rather
than a direct thermometer series \citep{chu1973climatic,ge2013temperature,
ge2016recent}.  Let \(Y_{rt}\) be proxy record \(r\) at time \(t\), such as a
phenological record, frozen-river report, disaster chronicle, or local
temperature reconstruction.  Let \(m_t\) be a latent climate signal.  A
Gaussian calibration likelihood for the observed records is
\[
  Y_{rt}\mid m_t,a_r,b_r,\sigma_r^2
  \sim
  \Normal(a_r+b_rm_t,\sigma_r^2),
  \qquad (r,t)\in\mathcal I ,
\]
where \(\mathcal I\) is the set of proxy-time pairs that survived into the
archive.  If \(p_\eta(m)\) is a smoothing or process model for the latent
curve \(m=(m_t)\), the observed likelihood is
\[
  L(\theta)
  =
  \int
  p_\eta(m)
  \prod_{(r,t)\in\mathcal I}
  \varphi\{y_{rt};a_r+b_rm_t,\sigma_r^2\}
  \,dm .
\]
If record survival is itself informative, with \(R_{rt}=\ind{(r,t)\in\mathcal
O}\), the likelihood must be enlarged to include a reporting model
\[
  \prod_{r,t}
  q_\gamma(R_{rt}\mid m_t,r,t).
\]
The statistical object is therefore not simply a curve.  It is a curve read
through proxy sensitivity, smoothing assumptions, and the survival of records.
\end{example}

\begin{example}[Cardiovascular trial endpoints]
The CARMELINA randomized clinical trial compared linagliptin with placebo in
adults with type 2 diabetes and high cardiovascular and renal risk
\citep{rosenstock2019linagliptin}.  For the primary cardiovascular endpoint,
let \(T_i\) be the event time, \(C_i\) the censoring time,
\[
  Z_i=T_i\wedge C_i,\qquad \delta_i=\ind{T_i\le C_i}.
\]
Let \(A_i\) denote treatment and \(\mathbf X_i\) baseline
covariates.  Under independent censoring and a proportional hazards model,
\[
  \lambda_i(t\mid A_i,\mathbf X_i)
  =
  \lambda_0(t)\exp(\psi A_i+\beta^T\mathbf X_i),
  \qquad
  \Lambda_i(t)=\int_0^t\lambda_i(u\mid A_i,\mathbf X_i)\,du .
\]
The observed event-time likelihood for
\(\theta=(\psi,\beta,\lambda_0)\), ignoring factors involving only the
censoring law, is
\[
  L_T(\theta)
  =
  \prod_{i=1}^n
  \{\lambda_i(Z_i\mid A_i,\mathbf X_i)\}^{\delta_i}
  \exp\{-\Lambda_i(Z_i)\}.
\]
If the baseline hazard is treated as a nuisance component, the Cox partial
likelihood conditions on the ordered event times and their risk sets:
\[
  L_{\mathrm{partial}}(\psi,\beta)
  =
  \prod_{j:\delta_j=1}
  \frac{\exp(\psi A_j+\beta^T\mathbf X_j)}
  {\sum_{k\in R(Z_j)}\exp(\psi A_k+\beta^T\mathbf X_k)},
\]
with the usual separate choices needed for ties.

The same trial also contains longitudinal clinical measurements, such as
hemoglobin A1c.  Now the observed object is
\[
  O_i^{L}=(V_i,Y_{i,V_i},A_i,\mathbf X_i),
\]
where \(V_i\) is the set of visits actually observed.  A Gaussian
repeated-measures model starts from a full visit vector
\[
  Y_i\mid A_i,\mathbf X_i
  \sim
  \Normal_m\{\mu_i(A_i,\mathbf X_i;\theta),\Sigma_i(\theta)\}.
\]
Under a missing-at-random reading that permits the visit process to be ignored
for inference on \(\theta\), the contribution becomes the marginal normal
density of the observed coordinates,
\[
  L_i^L(\theta)
  =
  \varphi_{|V_i|}
  \{y_{i,V_i};
  \mu_{i,V_i}(A_i,\mathbf X_i;\theta),
  \Sigma_{i,V_i,V_i}(\theta)\}.
\]
If visit attendance depends on unobserved outcomes, the likelihood must include
the visit mechanism instead:
\[
  L_i^L(\theta,\gamma)
  =
  \int
  \varphi_m\{y_i;\mu_i(\theta),\Sigma_i(\theta)\}
  q_\gamma(V_i\mid y_i,A_i,\mathbf X_i)\,dy_{i,V_i^c}.
\]
The patients are the same, but survival, repeated measurement, and visit
processes generate different likelihoods.
\end{example}

\begin{example}[Time to clinical improvement]
The Lancet trial of remdesivir in adults with severe COVID-19 used time to
clinical improvement as a central endpoint \citep{wang2020remdesivir}.  The
statistical form looks like a survival analysis, but the event is improvement
on an ordinal clinical scale, not death.  Let \(A_i\) be treatment, \(\mathbf X_i\)
baseline covariates, \(T_i^I\) time to clinical improvement, \(T_i^D\) time to
death before improvement, and \(C_i\) administrative or loss-to-follow-up
censoring.  The observed record can be written as
\[
  Z_i=\min(T_i^I,T_i^D,C_i),
  \qquad
  J_i\in\{0,1,2\},
\]
where \(J_i=1\) means improvement, \(J_i=2\) means death before improvement,
and \(J_i=0\) means censoring.  A competing-risk event-time reading uses
cause-specific hazards
\[
  \lambda_{ri}(t\mid A_i,\mathbf X_i)
  =
  \lambda_{r0}(t)\exp(\psi_rA_i+\beta_r^T\mathbf X_i),
  \qquad r=1,2,
\]
and gives the observed-data contribution
\[
  L_i(\theta)
  =
  \{\lambda_{1i}(Z_i)\}^{\ind{J_i=1}}
  \{\lambda_{2i}(Z_i)\}^{\ind{J_i=2}}
  \exp\left\{-\sum_{r=1}^2\Lambda_{ri}(Z_i)\right\}.
\]
A simpler ``time to improvement'' analysis that treats death as censoring uses
only the \(r=1\) part and requires a much stronger censoring interpretation.
If the daily ordinal clinical score is modeled directly instead, an ordinal
likelihood would use
\[
  \Prob(S_{it}\le k\mid A_i,\mathbf X_i)
  =
  \logit^{-1}\{\alpha_k-\psi A_i-\beta^T\mathbf X_i\},
  \qquad
  L_i^{\mathrm{ord}}(\theta)
  =
  \prod_{t\in V_i}p_{it,S_{it}}(\theta).
\]
Thus the same clinical record can be read as an improvement-time likelihood, a
competing-risk likelihood, or a repeated ordinal likelihood.  The likelihood is
not determined by the word ``COVID trial''; it is determined by the endpoint
map and the conditioning assumptions.
\end{example}

\begin{example}[Process records]
Process likelihoods can appear in this chapter if the notation is read
modestly.  At this point, a process is a time-indexed observed record; the
formal probability law on path space comes later.  Suppose subject \(i\) is
followed until \(\tau\), has recurrent event times
\[
  0<T_{i1}<\cdots<T_{im_i}\le \tau,
\]
and is under observation or at risk according to \(Y_i(t)\in\{0,1\}\).  Define
the counting record
\[
  N_i(t)=\sum_{j=1}^{m_i}\ind{T_{ij}\le t}.
\]
If the conditional event rate while observed is \(Y_i(t)\lambda_i(t;\theta)\),
then, ignoring factors from an independent censoring or visit process, the
observed event-time likelihood is
\[
  L_i^N(\theta)
  =
  \left\{\prod_{j=1}^{m_i}\lambda_i(T_{ij};\theta)\right\}
  \exp\left\{-\int_0^\tau Y_i(u)\lambda_i(u;\theta)\,du\right\}.
\]
The same likelihood is often written in counting-process shorthand as
\[
  \ell_i^N(\theta)
  =
  \int_0^\tau \log\lambda_i(u;\theta)\,dN_i(u)
  -
  \int_0^\tau Y_i(u)\lambda_i(u;\theta)\,du .
\]
This display is not asking the reader to have the full theory of stochastic
processes yet.  It is a compact way to say which jump times were seen and
which exposure time was available for seeing them.

If events also have types, such as hospitalization, adverse-event grade,
purchase category, or state transition, the observed record is marked:
\[
  (T_{ij},M_{ij}),\qquad M_{ij}\in\{1,\ldots,K\}.
\]
With type-specific intensities \(Y_{ik}(t)\lambda_{ik}(t;\theta)\), the
marked-event likelihood is
\[
  L_i^M(\theta)
  =
  \left\{\prod_{j=1}^{m_i}
  \lambda_{iM_{ij}}(T_{ij};\theta)\right\}
  \exp\left\{-\int_0^\tau
  \sum_{k=1}^K Y_{ik}(u)\lambda_{ik}(u;\theta)\,du\right\}.
\]
Here conditioning on the internal history decides whether the next-event rate
depends only on current state, on accumulated past events, or on a richer
patient or user history.

Other stochastic-process likelihoods can be introduced the same way.  If a
smooth biological, imaging, or environmental trajectory is observed only at
times \(V_i\), a Gaussian trajectory reading might use the finite-dimensional
density
\[
  Y_{i,V_i}
  \sim
  \Normal_{|V_i|}
  \{m_{V_i}(\theta),
  K_{V_i,V_i}(\theta)+\sigma^2I_{|V_i|}\}.
\]
The likelihood contribution is this multivariate normal density evaluated at
the observed coordinates.  The product-space chapter will explain when such
finite-dimensional laws determine a genuine stochastic process; the
continuous-time-process chapter will return to counting-process intensities
carefully.  The role of the present chapter is only to make the observed path
fragment, the exposure process, and the conditioning assumptions visible.
\end{example}

\begin{example}[Single-cell count data: from matrix to likelihood]
A single-cell assay gives the analyst a count matrix
\[
  X=(x_{ig}),\qquad i=1,\ldots,n,\quad g=1,\ldots,G,
\]
together with cell-level information such as size factors \(s_i\), batch,
condition, cell type, spatial position, or pseudotime \(t_i\).  The scientific
object is not the matrix alone.  It also includes latent abundance, capture
efficiency, technical loss, biological state, and the downstream question.  A
likelihood appears only after the matrix is read through one of these
observation stories.  The displays below are simplified versions meant to expose
that map from data to likelihood, not to reproduce every computational detail of
the software.  Figure~\ref{fig:ch05-scrna-matrix-likelihood} keeps the same
map visible before the formulas begin.

\input{figures/ch05_scrna_matrix_likelihood}

In a dropout-aware observation reading inspired by scImpute
\citep{li2018scimpute}, an observed zero can mean either biological silence or
technical non-observation.  One transparent likelihood introduces a dropout
indicator \(D_{ig}\) and a latent count \(Y^*_{ig}\):
\[
  D_{ig}\mid c_i\sim\Bernoulli(\pi_{ig}),
  \qquad
  Y^*_{ig}\mid c_i\sim \mathrm{NB}(s_i\mu_{ig},\kappa_g),
  \qquad
  X_{ig}=
  \begin{cases}
  0, & D_{ig}=1,\\
  Y^*_{ig}, & D_{ig}=0.
  \end{cases}
\]
Here \(c_i\) denotes the cell-level covariates being conditioned on, \(\mu_{ig}\)
is the expected expression level after size-factor adjustment, \(\kappa_g\) is a
gene-specific dispersion parameter, and \(\pi_{ig}\) is the probability that the
measurement process hides an expressed molecule count.  Integrating out
\(D_{ig}\) gives the observed-count contribution
\[
  p_{\mathrm{drop}}(x_{ig}\mid c_i;\eta)
  =
  \begin{cases}
  \pi_{ig}+(1-\pi_{ig})\,\mathrm{NB}(0;s_i\mu_{ig},\kappa_g),
  & x_{ig}=0,\\
  (1-\pi_{ig})\,\mathrm{NB}(x_{ig};s_i\mu_{ig},\kappa_g),
  & x_{ig}>0.
  \end{cases}
\]
Thus a stylized dropout likelihood is
\[
  L_{\mathrm{drop}}(\eta)
  =
  \prod_{i,g}p_{\mathrm{drop}}(x_{ig}\mid c_i;\eta),
\]
or the same product over a neighborhood of similar cells when the method borrows
local information.  The data-to-likelihood decision is clear: zeros are not
automatically expression values; some are modeled as the image of an unobserved
positive count under the observation map.

scDesign3 \citep{song2024scdesign3} uses the same kind of matrix for a different
task: fitting a generative law that can produce realistic synthetic single-cell
or spatial-omics data.  For each feature \(g\), the first layer is a conditional
marginal law
\[
  X_{ig}\mid c_i \sim f_g(\cdot\mid c_i;\theta_g),
  \qquad
  F_g(\cdot\mid c_i;\theta_g)\ \text{its distribution function},
\]
where \(c_i\) may include cell type, pseudotime, spatial location, batch, or
condition.  The second layer ties genes or features together through dependence
among the probability-scale variables
\[
  U_{ig}=F_g(X_{ig}\mid c_i;\theta_g).
\]
In compact continuous notation, the resulting cell-level likelihood has the
form
\[
  L_{\mathrm{gen}}(\theta,R)
  =
  \prod_{i=1}^n
  c_R\{u_{i1},\ldots,u_{iG}\}
  \prod_{g=1}^G f_g(x_{ig}\mid c_i;\theta_g),
\]
where \(c_R\) is a copula density or other dependence model.  A vine copula is
one high-dimensional way to implement this \(c_R\): it replaces one large
dependence object by a product of pair-copula factors, often arranged over
trees.  For three features, a simplified vine with feature \(2\) in the middle
would use
\[
  c_R(u_1,u_2,u_3)
  =
  c_{12}(u_1,u_2)c_{23}(u_2,u_3)
  c_{13\mid 2}\{u_{1\mid 2},u_{3\mid 2}\},
\]
where \(u_{1\mid 2}=F_{1\mid 2}(x_1\mid x_2)\) and
\(u_{3\mid 2}=F_{3\mid 2}(x_3\mid x_2)\).  Without the simplifying assumption,
the last factor may also depend on the conditioning value \(x_2\).  For
discrete counts, the same idea is implemented with the appropriate count
probabilities, finite differences, distributional transforms, or transformed
residuals, because the probability integral transform is no longer a smooth
one-to-one map.  The likelihood is not mainly an imputation likelihood.  It is
a generative likelihood whose fitted law is judged by whether simulated cells
preserve the marginal distributions, zero patterns, correlations, trajectories,
spatial structure, and multimodal relationships needed for downstream analysis.

scGTM \citep{cui2022scgtm} gives a third reading, aimed at interpretable
gene-trend inference along pseudotime.  Now the observed object for one gene is
\[
  (x_{1g},\ldots,x_{ng};\, t_1,\ldots,t_n;\, s_1,\ldots,s_n),
\]
and the conditioning decision is to treat the pseudotimes and size factors as
given.  A finite catalogue \(\mathcal M\) of interpretable trend shapes, such as
increasing, decreasing, transient, hill, or valley patterns, supplies candidate
mean curves.  For trend class \(m\in\mathcal M\),
\[
  X_{ig}\mid t_i,m
  \sim
  \mathrm{NB}\{s_i\mu_{gm}(t_i;\vartheta_{gm}),\kappa_g\}.
\]
The gene-level likelihood is therefore
\[
  L_g(m,\vartheta_{gm},\kappa_g)
  =
  \prod_{i=1}^n
  \mathrm{NB}\{x_{ig};s_i\mu_{gm}(t_i;\vartheta_{gm}),\kappa_g\}.
\]
The trend label is selected or weighted by profiling or integrating over the
shape-specific parameters:
\[
  \ell_g(m)
  =
  \sup_{\vartheta_{gm},\kappa_g}
  \sum_{i=1}^n
  \log \mathrm{NB}\{x_{ig};s_i\mu_{gm}(t_i;\vartheta_{gm}),\kappa_g\},
  \qquad
  \hat m_g\in\argmax_{m\in\mathcal M}\ell_g(m).
\]
Here the matrix is read neither as a dropout problem nor as a simulator by
itself.  It becomes evidence for a constrained biological trend class.  The same
assay matrix is therefore not one likelihood waiting to be found; it becomes an
observation likelihood, a generative likelihood, or a trend-class likelihood
after the observed object, conditioning variables, and scientific target have
been named.
\end{example}

\subsection{Scores, information, and nuisance parameters}
\label{sec:ch05-scores-nuisance}
\conceptindexes{score, Fisher information, nuisance parameters, information orthogonality}

When \(p_\theta\) is smooth and \(\theta\in\Real^p\), write
\[
  \ell(\theta)=\log L(\theta;O).
\]
The score and observed information are
\[
  \dot\ell(\theta)=\nabla_\theta\ell(\theta),
  \qquad
  j(\theta)=-\nabla_\theta^2\ell(\theta).
\]
Under regular correctly specified models, the maximum likelihood estimator
\(\hat\theta\) solves \(\dot\ell(\hat\theta)=0\) and is approximately normal,
with variance controlled by Fisher information.  These are the approximations
that make likelihood a practical language for inference.

Most scientific parameters do not live alone.  Write
\[
  \theta=(\psi,\lambda),
\]
where \(\psi\) is the parameter of interest and \(\lambda\) is nuisance
structure.  In a clinical model, \(\psi\) might be a treatment effect and
\(\lambda\) a baseline hazard, site effect, overdispersion, random-effect
distribution, missingness model, or covariance structure.  The profile
likelihood
\[
  \ell_p(\psi)=\ell\{\psi,\hat\lambda_\psi\},
  \qquad
  \hat\lambda_\psi\in\argmax_\lambda \ell(\psi,\lambda),
\]
tries to compare \(\psi\) values after letting the nuisance component choose
its best explanation.  This is often effective, but it is not magic.  If
\(\lambda\) can absorb or distort the scientific contrast, the profile
likelihood may make \(\psi\) look stable for the wrong reason.

\begin{definition}[Information orthogonality]
In a regular parametric model with \(\theta=(\psi,\lambda)\), the parameters
\(\psi\) and \(\lambda\) are information-orthogonal at \(\theta\) if the
off-diagonal block of expected information vanishes:
\[
  I_{\psi\lambda}(\theta)
  =
  \Expect_\theta
  \left\{-\frac{\partial^2\ell(\theta)}
  {\partial\psi\,\partial\lambda^T}\right\}
  =
  0.
\]
\end{definition}

Orthogonality says that, locally, changing \(\lambda\) does not tilt the score
for \(\psi\).  It is not merely algebraic neatness.  It can make inference for
\(\psi\) less sensitive to nuisance estimation.  The modern literature on
profile likelihood, semiparametric inference, and double/debiased machine
learning all returns to this idea in different languages.

\section{Missingness, Censoring, and Coarsening}
\label{sec:ch05-missingness-censoring}
\conceptindexes{missingness, censoring, coarsening, missing at random, observed zeroes}

Observation mechanisms matter most when they hide data in a structured way.
Missingness, censoring, truncation, subsampling, batching, and coarsening are
not defects outside probability theory.  They are maps from a richer object to
the data actually analyzed.

\begin{definition}[A missing-data notation]
Let \(X=(X_{\mathrm{obs}},X_{\mathrm{mis}})\) be a full data vector and let
\(R\) denote the response pattern that determines which components are
observed.  The observed object is
\[
  O=(R,X_{\mathrm{obs}}).
\]
The missingness mechanism is the conditional law
\[
  \Prob(R=r\mid X_{\mathrm{obs}},X_{\mathrm{mis}}).
\]
It is missing at random, relative to the chosen full-data model, if this
conditional law depends on \(X\) only through \(X_{\mathrm{obs}}\):
\[
  \Prob(R=r\mid X_{\mathrm{obs}},X_{\mathrm{mis}})
  =
  \Prob(R=r\mid X_{\mathrm{obs}}).
\]
\end{definition}

Rubin's missing-data framework \citep{rubin1976inference} is often remembered
through labels such as MCAR, MAR, and MNAR.  The deeper point is structural:
missingness is not one event.  It is part of the observed-data law.  Whether it
can be ignored depends on the parameter, the model, the sampling design, and
the likelihood factorization.

\begin{example}[Ignorable and nonignorable missingness]
Suppose \(Y\) is a continuous outcome and \(R\) indicates whether it is
observed.  If
\[
  \Prob(R=1\mid Y,X)=\Prob(R=1\mid X),
\]
then, after conditioning on \(X\), missingness does not directly select on the
unseen value of \(Y\).  A likelihood for the regression of \(Y\) on \(X\) may
be built from the observed cases under appropriate conditions.  If instead
\[
  \Prob(R=1\mid Y,X)
\]
depends on \(Y\) even after \(X\) is known, then the observed response
distribution is tilted.  Complete-case regression is then reading a selected
population unless the estimand is changed.
\end{example}

\begin{example}[Right censoring likelihood]
If \(T\) has density \(f_T(t;\theta)\) and survival function \(S_T(t;\theta)\),
and if censoring is independent of \(T\), then the part of the likelihood for
\(\theta\) contributed by \(O=(Z,\delta)\) is
\[
  \{f_T(Z;\theta)\}^{\delta}
  \{S_T(Z;\theta)\}^{1-\delta}.
\]
The expression is a compact summary of the observation map.  When
\(\delta=1\), the event time is known; when \(\delta=0\), the data say only
that \(T>Z\).  Cox's proportional hazards model builds a partial likelihood by
conditioning on risk sets so that a baseline hazard nuisance component drops
out of the treatment-effect comparison \citep{cox1972regression}.
\end{example}

\begin{example}[Benchmark data and label mechanisms]
Large benchmark datasets such as image collections or product-rating matrices
also have observation mechanisms.  A missing rating may mean a user never saw
the movie, disliked it enough not to rate it, watched it outside the platform,
or was never offered it.  A class label may reflect annotator instructions,
ambiguous categories, or a benchmark's decision about what counts as ground
truth.  Prediction models trained on such data inherit the observation map.
\end{example}

\subsection{Real-World Studies and Target-Trial Emulation}
\label{sec:ch05-rws-target-trial}
\conceptindexes{real-world studies, target-trial emulation, clinical estimands, routine-care data}

Real-world studies (RWS) make the observation map unavoidable.  Their data are
not collected by a protocol that fixes every visit, test, treatment, and
endpoint in advance.  They are assembled from routine care, claims, registries,
devices, or administrative systems.  The scientific object may be a treatment
effect under a clinical strategy, but the observed object is a care trace:
\[
  O_i
  =
  (L_{i0}, A_i(t), L_i(t), R_i(t), C_i(t), Y_i(t):0\le t\le \tau),
\]
where \(L\) denotes covariate and clinical history, \(A\) treatment, \(R\)
measurement or visit processes, \(C\) censoring, and \(Y\) outcomes.  The
notation is schematic, but it records the main warning: treatment, observation,
and outcome are all time-indexed and institutionally produced.

Target-trial emulation is a way to discipline this situation
\citep{hernan2016targettrial}.  Before fitting a model, write the randomized
trial that one would have preferred to run:
\[
\begin{array}{ll}
\text{eligibility} & \text{who could enter the comparison?}\\
\text{time zero} & \text{when does follow-up begin?}\\
\text{strategies} & \text{which treatment rules are being compared?}\\
\text{outcome} & \text{what event or measurement defines success?}\\
\text{follow-up} & \text{when does observation stop, and why?}\\
\text{estimand} & \text{which contrast is the target?}
\end{array}
\]
Then ask whether the real-world data contain enough information to emulate
those components.  This is not a cosmetic protocol.  It decides which records
enter the risk set, which treatment histories violate a strategy, which
censoring events need adjustment, and which covariates must be measured before
treatment rather than after it.

\begin{example}[Time zero is an observed-data problem]
Suppose an analyst compares treated and untreated patients in an EHR database.
If treated patients enter follow-up at treatment initiation but untreated
patients enter at diagnosis, the comparison can create immortal-time bias: the
treated group had to survive long enough to receive treatment.  A target-trial
emulation forces both strategies to share a time zero, such as eligibility or
clinical decision time, before any model is fitted.  The likelihood or
estimating equation can then be read as a model for a better-defined observed
law.
\end{example}

Regulatory real-world evidence uses the same statistical discipline.  The FDA
framework emphasizes that real-world data become evidence only after the data
source, study design, endpoint, and analysis are judged fit for the question
\citep{fda2018rweframework}.  In the notation of this chapter, that means the
observation rule and the target parameter must be named together.
Otherwise a highly precise analysis may only estimate the behavior of a
recording system.

\section{Model Misspecification and Likelihood Failure}
\label{sec:ch05-when-likelihood-goes-wrong}
\conceptindexes{likelihood!failure modes, misspecification, pseudo-true parameter, target-stable score}

Likelihood can fail in several different ways.  Reid's UCLA lecture grouped
many of them under a practical warning: standard likelihood approximations can
be poor, parameter spaces can be irregular or high-dimensional, computation can
force approximate likelihoods, and the fitted model can be misspecified
\citep{reid2024whenlikelihood}.  For this chapter, the most important failure
mode is misspecification of the observed-data law.

Suppose the true observed-data density is \(m(o)\), but the analyst fits
\[
  \{p_\theta(o):\theta\in\Theta\}.
\]
If no \(\theta\) satisfies \(p_\theta=m\), maximum likelihood no longer
converges to a true parameter in the literal sense.  Under suitable conditions
it converges to the pseudo-true value
\[
  \theta_m
  =
  \argmin_{\theta\in\Theta}
  \int m(o)\log\frac{m(o)}{p_\theta(o)}\,d\nu(o),
\]
the member of the fitted model closest to the truth in Kullback--Leibler
distance \citep{huber1967behavior,white1982maximum}.  This can still be useful
if \(\theta_m\) has a scientific interpretation.  It can be dangerous if the
scientific parameter was something else.

\begin{definition}[Pseudo-true parameter]
Let \(m\) be the true observed-data density and let
\(\{p_\theta:\theta\in\Theta\}\) be a dominated fitted model.  A pseudo-true
parameter is any minimizer
\[
  \theta_m
  \in
  \argmin_{\theta\in\Theta}
  \Expect_m
  \left\{\log\frac{m(O)}{p_\theta(O)}\right\}.
\]
Equivalently, it maximizes \(\Expect_m\log p_\theta(O)\).
\end{definition}

The pseudo-true parameter is not a consolation prize; it is the actual target
of the fitting procedure under misspecification.  The question is whether that
target is acceptable.

Information theory gives the same point in a compact form.  The fitted
log-likelihood estimates the cross-entropy \( -\Expect_m\log p_\theta(O) \);
subtracting the entropy of the true law \(m\) leaves
\[
  D_{\mathrm{KL}}(m\|p_\theta)
  =
  \Expect_m
  \left\{\log\frac{m(O)}{p_\theta(O)}\right\}.
\]
Thus a better likelihood fit is a better code for the observed data under the
candidate model family.  In a genomic classifier, for example, a smaller
cross-entropy loss means the fitted conditional probabilities assign more
mass to the observed labels.  It does not by itself say that the chosen genes
are causal, that the calibration transfers to a new hospital, or that the
scientific estimand is the one the study needs.

\begin{example}[Robust standard errors do not repair the estimand]
In a misspecified regression, a sandwich covariance estimate can give a more
honest large-sample variance for the estimator's limiting target.  But the
sandwich does not decide whether the limiting target is the treatment effect,
prediction contrast, or scientific parameter the analyst wanted.  Variance
robustness is not target robustness.
\end{example}

\begin{example}[Random effects as nuisance structure]
In a clustered clinical study, a treatment effect \(\psi\) may be modeled with
random site or subject effects \(\lambda\).  If the random-effects distribution
is wrong, the maximum likelihood estimator of \(\psi\) may or may not remain
consistent for the intended effect.  The answer depends on how the parameter
enters the model, what moments or symmetries are preserved, and whether the
nuisance component can distort the score for \(\psi\).  This is the setting in
which \citet{battey2024role} emphasize parameterization and generalized
orthogonality.
\end{example}

\subsection{Stable parameters under misspecified nuisance}
\label{sec:ch05-stable-parameters}
\conceptindexes{stable parameters, misspecified nuisance, generalized Bayesian updating, parameterization}

Write the fitted density as \(p_{\psi,\lambda}\), where \(\psi\) is the
scientific parameter and \(\lambda\) is nuisance structure.  The true law is
\(m\).  Even if no \(\lambda\) makes \(p_{\psi,\lambda}=m\), it can still be useful to ask whether
the likelihood score for \(\psi\) still points to the right value \(\psi^*\).
A sufficient condition has the form
\[
  \Expect_m
  \left\{
    \frac{\partial}{\partial\psi}
    \log p_{\psi^*,\lambda}(O)
  \right\}
  =
  0
  \quad\text{for relevant }\lambda.
\]
This says that, under the true law, the score for the parameter of interest is
unbiased at the target value even when the nuisance component is not correctly
specified.

Battey and Reid's contribution is to study when such score stability is forced
by the structure of the parameterization itself \citep{battey2024role}.  Their
matched-comparison and two-group examples show that symmetry, orthogonality,
and the way nuisance parameters enter the model can protect a parameter of
interest from some forms of nuisance misspecification.  The message for this
book is not that every model should be made symmetric.  It is that a parameter
is not only a symbol in a density.  It is a reading of the observed-data law,
and its meaning can change when the nuisance part of the model changes.

\begin{definition}[Target-stable score, informal]
Let \(m\) be the true observed law and \(p_{\psi,\lambda}\) a fitted model.  A
score for \(\psi\) is target-stable at \(\psi^*\) over a nuisance class
\(\Lambda_0\) if
\[
  \Expect_m\{\dot\ell_\psi(\psi^*,\lambda;O)\}=0,
  \qquad \lambda\in\Lambda_0.
\]
When this condition holds with enough curvature and empirical-process control,
estimating equations based on \(\dot\ell_\psi\) can target \(\psi^*\) even if
the nuisance model is wrong.
\end{definition}

This definition is deliberately informal because the full theory belongs later
in the book.  The M- and Z-estimation chapter studies estimators as roots and
maximizers; the local approximation chapter studies local
linearization, influence functions, and orthogonal scores.  Here the idea is
used as a reading rule: before trusting a likelihood, ask whether the score is
aimed at the scientific target or merely at the best-fitting member of the
assumed model.

\section{Partial Likelihood, Composite Likelihood, and Estimating Functions}
\label{sec:ch05-partial-composite}
\conceptindexes{partial likelihood, composite likelihood, estimating functions, pseudo-likelihood}

When the full observed-data likelihood is unavailable, inconvenient, or too
fragile, statisticians often use a different inference function.  This is not
an admission that probability has failed.  It is a choice to preserve a
meaningful part of the observed law.

\begin{description}[leftmargin=0pt,labelsep=0.65em,style=unboxed,font=\bookdescriptionlabelfont,itemsep=0.45\baselineskip]
\item[Partial likelihood.]
Condition on enough information that a nuisance component drops out.  Cox's
partial likelihood for proportional hazards models compares failures within
risk sets and avoids specifying the baseline hazard in full
\citep{cox1972regression}.

\item[Conditional likelihood.]
Condition on a statistic whose distribution carries the nuisance parameter,
leaving a likelihood for the parameter of interest.  In matched or stratified
problems, this can turn a nuisance-heavy model into a cleaner comparison.

\item[Composite likelihood.]
Multiply lower-dimensional marginal or conditional likelihood pieces when the
full joint likelihood is too complex.  Pairwise likelihoods and spatial
pseudo-likelihoods are examples; Besag's lattice pseudo-likelihood is a classic
case \citep{besag1974spatial}.

\item[Quasi-likelihood and estimating equations.]
Model a mean and variance relation, or a moment restriction, without claiming a
complete density.  The resulting estimator targets the solution of an
estimating equation rather than the maximizer of a full likelihood.
\end{description}

Each device answers a slightly different question.  The practical discipline
is to state which part of the observed law is being trusted and which part is
being left unspecified.

\begin{example}[Pairwise likelihood for dependent measurements]
Suppose \(O_i=(O_{i1},\ldots,O_{id})\) is a high-dimensional vector and the
joint density is hard to specify.  A pairwise likelihood uses bivariate
densities:
\[
  L_{\mathrm{pair}}(\theta)
  =
  \prod_i\prod_{s<t} p_{\theta,st}(O_{is},O_{it}).
\]
The estimator is not usually fully efficient if the full model is correct, but
it may be more stable and computationally usable.  Its variance must be read
through an estimating-function, or Godambe, information rather than ordinary
Fisher information.
\end{example}

\begin{example}[Estimating a treatment contrast without a full outcome law]
Let \(T\in\{0,1\}\), covariates \(X\), and outcome \(Y\).  A working model may
postulate
\[
  \Expect(Y\mid T,X)=\psi T+\omega(X),
\]
while leaving the full conditional distribution of \(Y\) unspecified.  If the
scientific target is a weighted average treatment contrast, an estimating
equation can sometimes be designed so that misspecification of \(\omega\) has
limited first-order effect.  This is the same instinct as orthogonal scoring:
protect the target from nuisance error.
\end{example}

\section{Sequential Observation and Transition Kernels}
\label{sec:ch05-sequential-bridge}
\conceptindexes{sequential observation, kernels, filtration, transition kernel}

The final bridge to the product-space chapter is sequential
observation.  This section is only an intuitive bridge, not a formal definition
of transition kernels.  A study rarely produces one monolithic object.  It
produces baseline variables, assignments, visits, interim decisions, endpoints,
and analysis summaries in time.

Let
\[
  O=(O_1,O_2,\ldots,O_n)
\]
and let \(\mathcal F_k^O=\sigma(O_1,\ldots,O_k)\).  A coherent sequential
model specifies
\[
  \mu_1(do_1),\quad
  K_2(o_1,do_2),\quad
  K_3(o_1,o_2,do_3),\quad \ldots
\]
where each \(K_k\) should be read informally for now as the conditional law of
the next observation given the past.  Chapter~6 will make these objects precise
as measurable transition kernels.  With that warning in place, the intended
joint law is
\[
  \mu_1(do_1)
  K_2(o_1,do_2)
  K_3(o_1,o_2,do_3)
  \cdots
  K_n(o_1,\ldots,o_{n-1},do_n).
\]
This is the informal version of the kernel product theorem and Ionescu--Tulcea
construction developed in the next chapter.

\begin{example}[Adaptive dose finding]
Let \(A_k\) be the dose assigned to patient \(k\), and let \(Y_k\) be the
toxicity indicator.  A model for the trial can be written sequentially:
\[
  A_k\sim Q_k(\cdot\mid \mathcal F_{k-1}),
  \qquad
  Y_k\sim P_\theta(\cdot\mid A_k,\mathcal F_{k-1}),
\]
where \(\mathcal F_{k-1}\) contains the previously assigned doses and observed
toxicities.  The design kernel \(Q_k\) must be adapted to the past.  It may be
deterministic, randomized, or model-assisted, but it may not depend on future
outcomes.  This is exactly the filtration discipline introduced in the design
chapter.
\end{example}

\begin{example}[Autonomous laboratory]
In a closed-loop experiment, the next compound or perturbation \(A_{k+1}\) is
chosen after observing earlier assay results.  The next readout \(Y_{k+1}\)
then updates the model.  The observed law alternates design kernels and
scientific kernels:
\[
  P(dA_1)\,P_\theta(dY_1\mid A_1)\,
  Q_2(dA_2\mid A_1,Y_1)\,
  P_\theta(dY_2\mid A_2,A_1,Y_1)\cdots.
\]
Ignoring this alternation can make the data look iid when they are actually
the product of a learning system.
\end{example}

\subsection{A Compact Likelihood Checklist}
\label{sec:ch05-checklist}
\conceptindexes{reading checklist, likelihood checklist, observation checklist}

Before writing down a likelihood or an estimating equation, ask:

\begin{description}[leftmargin=0pt,labelsep=0.65em,style=unboxed,font=\bookdescriptionlabelfont,itemsep=0.45\baselineskip]
\item[What is the observed object?]
Name the measurable object \(O\), not just the scientific ideal \(X\).

\item[What was filtered?]
Identify eligibility, sampling, censoring, missingness, measurement, and
processing maps.

\item[What is conditioned on?]
State which covariates, risk sets, histories, or design features are treated as
given.

\item[What is random under the likelihood?]
Distinguish design randomness, scientific randomness, measurement randomness,
and posterior uncertainty.

\item[What is the parameter reading?]
Explain what \(\psi\) means under the fitted model and whether that meaning
survives plausible nuisance misspecification.

\item[What approximation is being trusted?]
Check whether normal, chi-square, sandwich, bootstrap, or profile-likelihood
approximations match the sample size, dimension, and regularity of the problem.
\end{description}

The transition to the product-space chapter is now natural.
Once the observed data are many-coordinate or sequential, the mathematics must
explain how conditional laws multiply, how integrals can be rearranged, and how
finite-dimensional specifications become process laws.  That is the job of
product spaces and kernels.

\section{Exercises}
\label{sec:ch05-exercises}
\conceptindexes{likelihood exercises, conditioning exercises, missing-data exercises}

\begin{exercise}[Observed-data push-forward]
Let \(W\sim P_\theta\), let \(O=\phi(W)\), and set
\(Q_\theta=P_\theta\circ\phi^{-1}\).  Prove that \(Q_\theta\) is the unique
probability law satisfying
\[
  \Expect_\theta h(O)
  =
  \int h(o)\,Q_\theta(do)
\]
for every bounded measurable \(h\).
\end{exercise}

\begin{exercise}[Censoring map]
Let \(T\) and \(C\) be nonnegative event and censoring times, and define
\(Z=\min(T,C)\), \(\delta=\ind{T\le C}\).  Show explicitly that the event
\(\{Z\le t,\delta=1\}\) is measurable with respect to \(\sigma(T,C)\).  If
\(T\) and \(C\) are independent with densities, derive the observed-data
density for \((Z,\delta)\).
\end{exercise}

\begin{exercise}[Paired censoring and a survival surface]
\label{ex:ch05-paired-censoring-survival-surface}
The bivariate Kaplan--Meier problem begins from paired right-censored
observations rather than from a fully observed survival surface; see
\citet{dabrowska1988kaplan} and the counting-process background in
\citet{dabrowskaStochasticProcessesCommunication}.  Let
\[
  T=(T_1,T_2),\qquad C=(C_1,C_2),
\]
and suppose the analyst observes
\[
  O=(Z_1,Z_2,\delta_1,\delta_2),\qquad
  Z_j=T_j\wedge C_j,\quad
  \delta_j=\ind{T_j\le C_j}.
\]
\begin{enumerate}
\item Show that \(O\) is a measurable function of \((T,C)\).
\item For fixed \(s,t\), write each of the four probabilities
\[
\begin{aligned}
  H_{00}(s,t)&=\Prob(Z_1>s,Z_2>t),\\
  H_{10}(s,t)&=\Prob(Z_1>s,Z_2>t,\delta_1=1),\\
  H_{01}(s,t)&=\Prob(Z_1>s,Z_2>t,\delta_2=1),\\
  H_{11}(s,t)&=\Prob(Z_1>s,Z_2>t,\delta_1=\delta_2=1)
\end{aligned}
\]
as an expectation of a bounded function of \(O\).
\item Explain why estimating the joint survival surface
\(S(s,t)=\Prob(T_1>s,T_2>t)\) cannot be reduced to two separate univariate
Kaplan--Meier estimators unless the dependence structure between \(T_1\) and
\(T_2\) is supplied by an additional assumption.
\end{enumerate}
This exercise is only about the observed-data objects.  Later chapters return to
the product-integral estimator, uniform convergence, and confidence bands.
\end{exercise}

\begin{exercise}[Tower property in regression]
Suppose \(X=(X_1,X_2)\) and \(Y\) is integrable.  Use the tower property to
show that
\[
  \Expect\{\Expect(Y\mid X_1,X_2)\mid X_1\}
  =
  \Expect(Y\mid X_1).
\]
Explain in words why a prediction using both covariates can be averaged down to
a prediction using only \(X_1\).
\end{exercise}

\begin{exercise}[Pulling out what is known]
Let \(Y\) be integrable and let \(Z\) be bounded and \(\sigma(X)\)-measurable.
Prove that
\[
  \Expect(ZY\mid X)=Z\,\Expect(Y\mid X).
\]
Use this to show that if \(\Expect(Y-m(X)\mid X)=0\), then
\[
  \Expect\{h(X)(Y-m(X))\}=0
\]
for every bounded measurable function \(h\).
\end{exercise}

\begin{exercise}[Conditioning changes the information set]
Let \(X_1,X_2\) be iid with mean \(0\), and let \(S=X_1+X_2\).  Find
\(\Expect(X_1\mid X_1)\), \(\Expect(X_1\mid S)\), and
\(\Expect(X_1\mid X_1,X_2)\).  Explain why these are different readings of
the same random variable under different information sets.
\end{exercise}

\begin{exercise}[A conditional-independence check]
Let \(X\), \(Y\), and \(Z\) be random variables.  Suppose that for every
bounded measurable \(h\),
\[
  \Expect\{h(Y)\mid X,Z\}
  =
  \Expect\{h(Y)\mid Z\}.
\]
Explain why this is a conditional-independence statement.  In a longitudinal
study, what scientific claim is being made if \(X\) is the remote past, \(Z\)
is the current state, and \(Y\) is the future?
\end{exercise}

\begin{exercise}[Likelihood under a coarsening map]
Let \(X\) have density \(f_\theta\) on \(\Real\), but suppose the observation is
only \(O=\ind{X>c}\).  Write the likelihood for \(\theta\) after observing
\(O=1\) and after observing \(O=0\).  How much information about \(\theta\) has
the observation map discarded?
\end{exercise}

\begin{exercise}[Missing at random check]
Let \(Y\) be an outcome, \(\mathbf X\) covariates, and \(R\) the indicator that \(Y\) is
observed.  For each mechanism below, decide whether missingness is missing at
random relative to \(\mathbf X\):
\[
  \Prob(R=1\mid Y,\mathbf X)=\logit^{-1}(\alpha+\beta^T\mathbf X),
\]
\[
  \Prob(R=1\mid Y,\mathbf X)=\logit^{-1}(\alpha+\gamma Y),
\]
and
\[
  \Prob(R=1\mid Y,\mathbf X)=\logit^{-1}(\alpha+\beta^T\mathbf X+\gamma Y).
\]
\end{exercise}

\begin{exercise}[Sketch a target-trial emulation]
Choose a real-world clinical question using EHR, claims, registry, or device
data.  Write the trial you would like to emulate: eligibility, treatment
strategies, assignment time, follow-up, outcome, censoring rule, and estimand.
Then list three variables or events that must be observed before treatment for
the emulation to be credible.
\end{exercise}

\begin{exercise}[Pseudo-true normal mean]
Suppose the true distribution of \(Y\) has mean \(\mu\) and finite variance but
is not normal.  The analyst fits the normal model \(N(\theta,\sigma^2)\) with
\(\sigma^2\) known.  Show that the pseudo-true value of \(\theta\) is \(\mu\).
What changes if both \(\theta\) and \(\sigma^2\) are fitted?
\end{exercise}

\begin{exercise}[Sandwich target]
In a misspecified model, explain why a sandwich variance estimate can correct
the large-sample variance around the pseudo-true parameter but cannot by itself
make the pseudo-true parameter scientifically meaningful.
\end{exercise}

\begin{exercise}[Orthogonal score algebra]
Let \(\theta=(\psi,\lambda)\) and suppose the expected information matrix is
partitioned into blocks \(I_{\psi\psi},I_{\psi\lambda},I_{\lambda\psi}\), and
\(I_{\lambda\lambda}\).  Show that the decorrelated score
\[
  \dot\ell_\psi
  -
  I_{\psi\lambda}I_{\lambda\lambda}^{-1}\dot\ell_\lambda
\]
has zero covariance with \(\dot\ell_\lambda\) under the correctly specified
model.  What part of this calculation might fail under misspecification?
\end{exercise}

\begin{exercise}[Sequential factorization]
Let \(O=(O_1,O_2,O_3)\).  Suppose \(O_1\sim\mu\),
\(O_2\mid O_1=o_1\sim K_2(o_1,\cdot)\), and
\(O_3\mid(O_1,O_2)=(o_1,o_2)\sim K_3(o_1,o_2,\cdot)\).  Write the joint
probability of a rectangle \(A_1\times A_2\times A_3\).  Identify where
the product-space machinery is needed to make this construction
fully rigorous.
\end{exercise}

\section*{Sources and Further Reading}
\addcontentsline{toc}{section}{Sources and Further Reading}

This chapter is a bridge chapter, so its references point in several
directions.  Its opening situates the observed-data viewpoint between Tukey's
data analysis program \citep{tukey1962future}, Box's model-building caution
\citep{box1976science}, Shannon's channel view of information
\citep{shannon1948mathematical}, and social studies of quantification and
classification \citep{porter1995trust,bowker1999sorting}.  The applied
examples from platform recommendation, computer vision, neuroimaging,
financial risk, and clinical regulation are included to stress that
``observed data'' is an industry-specific object, not a universal raw material
\citep{bennett2007netflix,deng2009imagenet,lin2014microsoft,petersen2010adni,jpmorgan1996riskmetrics,ich2021e9r1}.
The observed-data viewpoint is connected to the push-forward and
Radon--Nikodym material in the probability-measure chapter; the missing-data
discussion follows the language introduced by \citet{rubin1976inference}.  The
route through conditioning, kernels, conditional expectation, and
Radon--Nikodym derivatives is also informed by
\citet{dabrowskaAdvancedProbabilityCommunication}.  The
image, spatial, text, and historical-proxy examples in the real-data section
return to the data structures introduced in Chapters~1 and~2: annotated image
benchmarks \citep{deng2009imagenet,lin2014microsoft}, London's spatial
point-pattern story \citep{clarke1946poisson,feller1968introduction},
stylometric and unseen-vocabulary problems
\citep{efronThisted1976unseen,hu2014redchamber,tu2013redchamber,
zhu2021redchamber}, and Chinese climate proxy reconstructions
\citep{chu1973climatic,ge2013temperature,ge2016recent}.  The
real-world-study discussion uses the target-trial-emulation viewpoint of
\citet{hernan2016targettrial} and the FDA real-world evidence framework
\citep{fda2018rweframework}.  The survival example uses the standard
right-censoring likelihood and points toward Cox's proportional hazards model
\citep{cox1972regression}.  The clinical examples use CARMELINA from
\emph{JAMA} \citep{rosenstock2019linagliptin} and the Lancet remdesivir trial
\citep{wang2020remdesivir} to emphasize that
endpoint definitions, censoring rules, visit schedules, and missingness
assumptions are part of the observed-data likelihood rather than clerical
details.  The process-record example is deliberately only a preview: it uses
counting-process likelihood notation in the observed-data sense, while the
formal construction of stochastic processes is deferred to the product-space
chapter and the intensity and compensator theory is deferred to the
continuous-time-process chapter \citep{andersen1982cox,andersen1993statistical,
fleming1991counting}.
The paired-censoring exercise points to the bivariate Kaplan--Meier estimator
on the plane studied by \citet{dabrowska1988kaplan}; related
counting-process background is treated in
\citet{dabrowskaStochasticProcessesCommunication}.  This chapter uses the example only
to identify the observed-data pieces, while the consistency and
confidence-band arguments appear later with empirical-process and
local-approximation tools.

The conditioning section is deliberately reader-facing.  Conditional
expectation is presented as projection onto an information set; conditional
laws are presented as kernels; conditional likelihood is presented as a choice
about which information is held fixed.  \Appref{sec:appB-conditioning-toolkit}
gives the supporting details: Radon--Nikodym construction of conditional expectation,
the tower and pull-out rules, conditional Jensen, regular conditional laws, and
conditional independence.

The likelihood discussion is influenced by Fisher's likelihood program
\citep{fisher1922foundations}, by modern likelihood texts such as
\citet{davison2003statistical} and \citet{cox2011principles}, and by Nancy
Reid's UCLA talk on ways likelihood can fail in practice
\citep{reid2024whenlikelihood}.  The misspecification discussion uses the
classical large-sample view of \citet{huber1967behavior} and
\citet{white1982maximum}: under misspecification, maximum likelihood targets a
Kullback--Leibler projection and requires sandwich-type variance calculations.
The parameterization and nuisance-orthogonality theme follows
\citet{battey2024role}, now published in \emph{Proceedings of the National
Academy of Sciences}, who study when a parameter of interest remains
recoverable under misspecification of nuisance components.  Composite and
pseudo-likelihood ideas are represented here by Besag's spatial
pseudo-likelihood \citep{besag1974spatial}; later chapters return to estimating
equations, empirical processes, and influence functions with more formal tools.
The single-cell examples link this observation chapter to the book's later
thread on scImpute \citep{li2018scimpute}, scDesign3
and scGTM \citep{cui2022scgtm}: one assay matrix
can be read as an observation problem, a generative simulation problem, or a
structured trend-inference problem, depending on the likelihood being trusted.

%% file: figures/ch05_observed_object_pipeline.tex
\begin{center}
\begingroup
\includegraphics[width=0.88\linewidth]{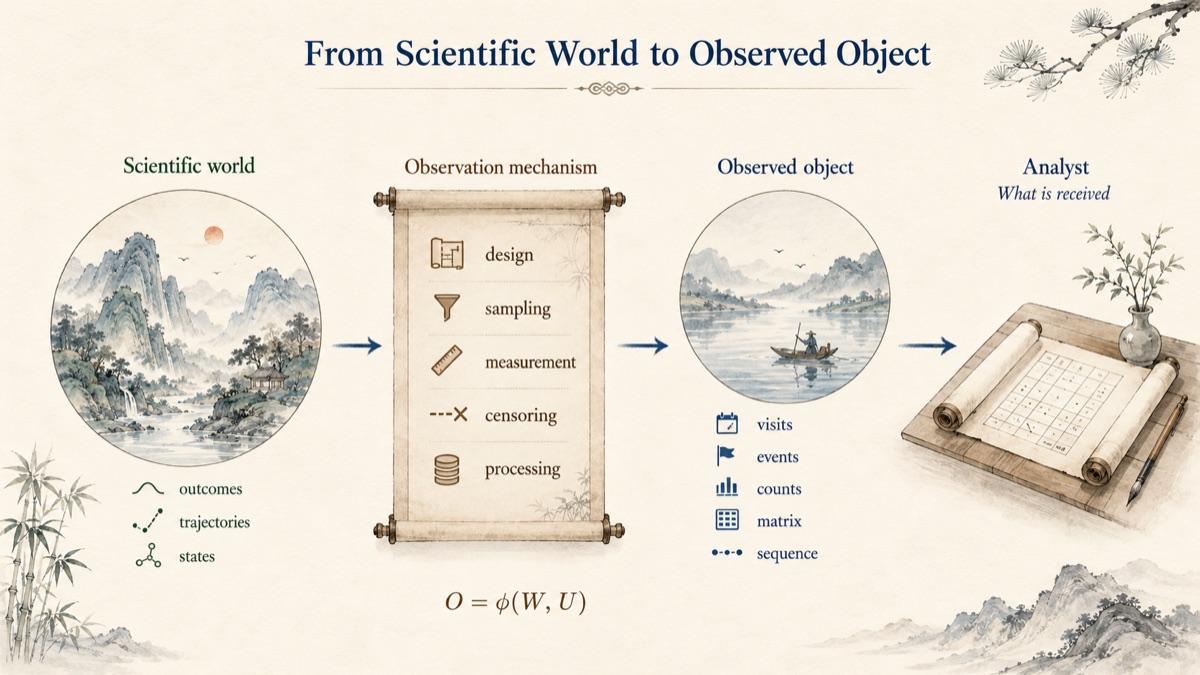}
\par\smallskip
\bookmanualfigure{fig:ch05-observed-object-pipeline}{The observed object pipeline}
\parbox{0.94\linewidth}{\small\raggedright
\textbf{Figure~\thefigure.} The observed object is a survived record, not the
scientific world itself.  A design, sampling rule, measurement device,
censoring mechanism, and processing pipeline turn a richer world \(W\), together
with auxiliary randomness \(U\), into \(O=\phi(W,U)\), the object delivered to
the analyst.\par}
\endgroup
\end{center}

%% file: figures/ch05_conditioning_projection.tex
\begin{center}
\begingroup
\includegraphics[width=0.72\linewidth]{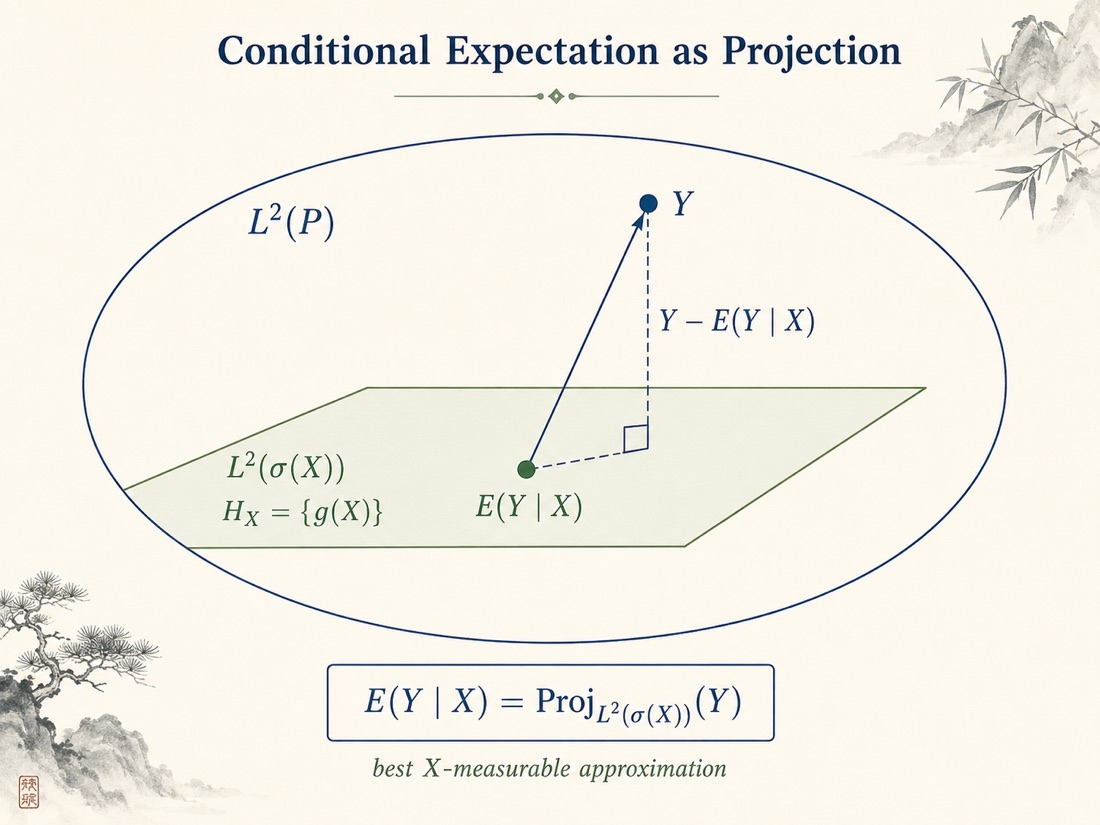}
\par\smallskip
\bookmanualfigure{fig:ch05-conditioning-projection}{Conditional expectation as projection}
\parbox{0.88\linewidth}{\small\raggedright
\textbf{Figure~\thefigure.} Conditional expectation can be read geometrically as
projection.  In \(L^2(\Prob)\), \(\Expect(Y\mid X)\) is the closest
square-integrable function of the observed information \(X\), and the residual
\(Y-\Expect(Y\mid X)\) is orthogonal to that information subspace.\par}
\endgroup
\end{center}

%% file: figures/ch05_likelihood_reading.tex
\begin{center}
\begingroup
\includegraphics[width=0.88\linewidth]{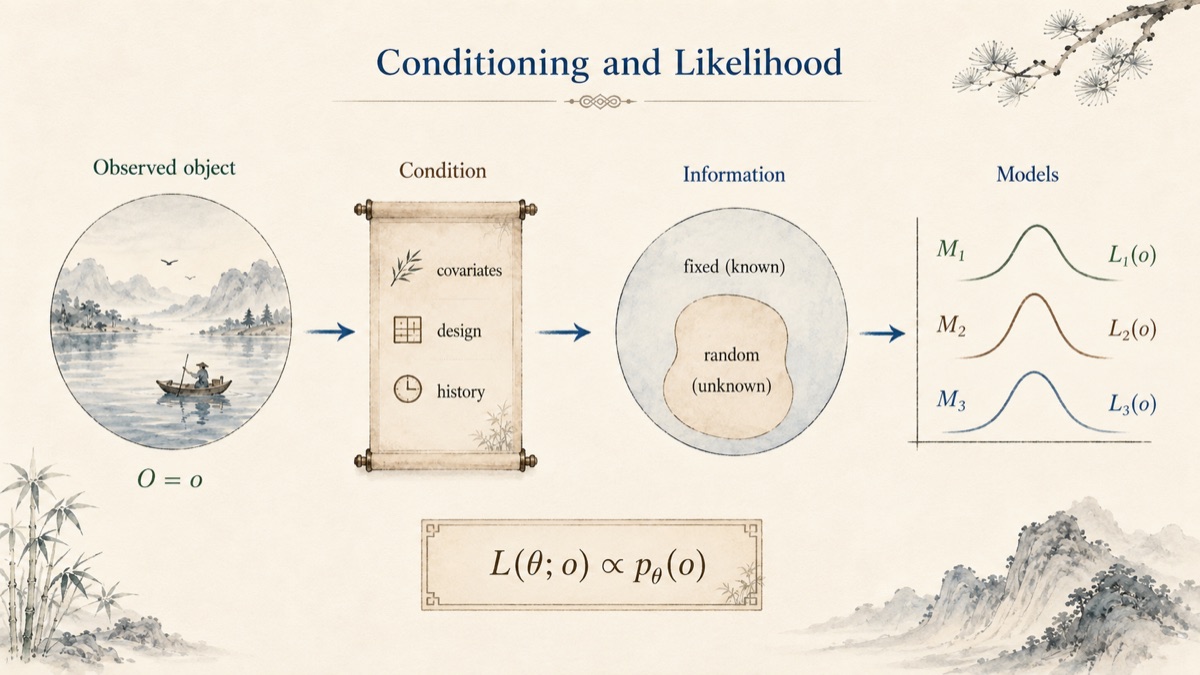}
\par\smallskip
\bookmanualfigure{fig:ch05-likelihood-reading}{Likelihood as a change of reading}
\parbox{0.94\linewidth}{\small\raggedright
\textbf{Figure~\thefigure.} Likelihood is a change of reading after the observed
object is fixed.  Conditioning choices decide which information is treated as
known; competing models then assign densities or probabilities to the same
observed \(o\), producing likelihood comparisons such as
\(L(\theta;o)\propto p_\theta(o)\).\par}
\endgroup
\end{center}

%% file: figures/ch05_scrna_matrix_likelihood.tex
\begin{center}
\begingroup
\includegraphics[width=0.84\linewidth]{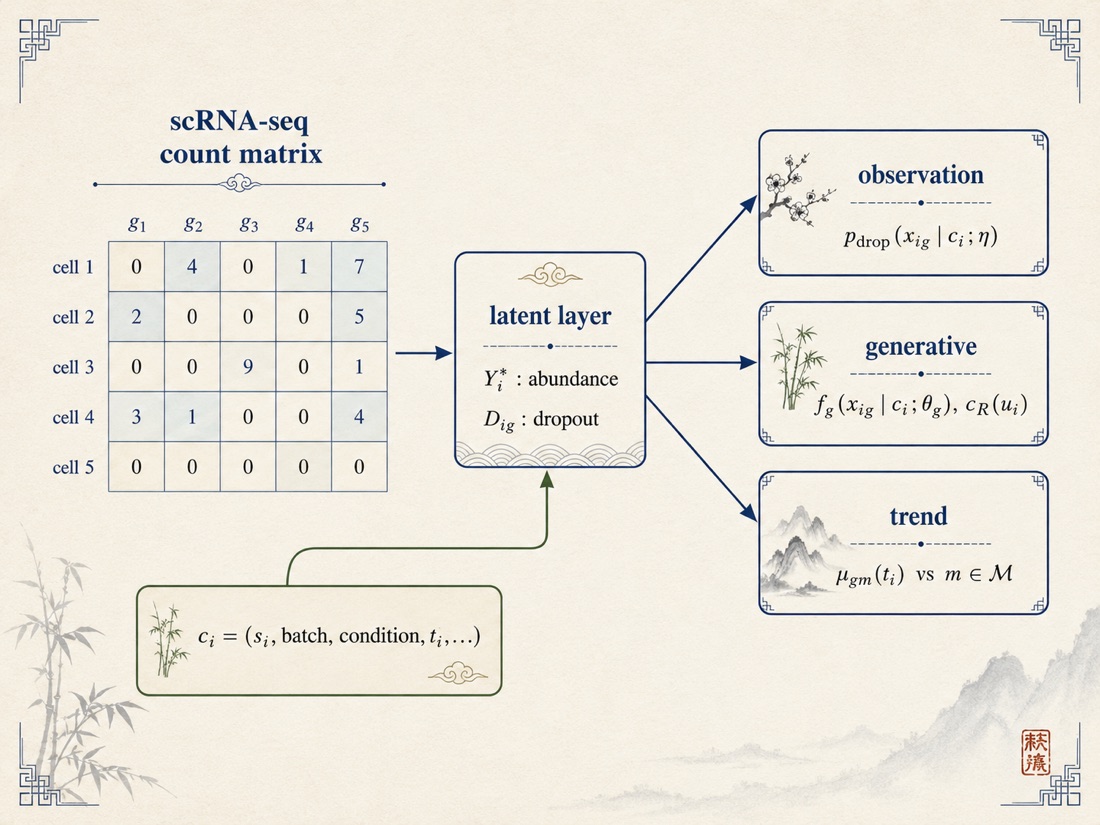}
\par\smallskip
\bookmanualfigure{fig:ch05-scrna-matrix-likelihood}{A single-cell count matrix as observed object}
\parbox{0.94\linewidth}{\small\raggedright
\textbf{Figure~\thefigure.} A single-cell count matrix is an observed object,
not a likelihood by itself.  The same sparse matrix can be read as a dropout
observation problem, a generative simulation problem, or a trend-class
inference problem once the conditioning variables and scientific target are
specified.\par}
\endgroup
\end{center}

%% file: chapters/ch10_weak_convergence_random_objects.tex
\chapter{Weak Convergence of Random Objects}
\label{chap:weak-convergence-random-objects}
\conceptindexes{weak convergence, random elements, random objects, metric spaces, portmanteau theorem, tightness, continuous mapping theorem}

\begin{tcolorbox}[
  enhanced,
  breakable,
  colback=chaptercream,
  colframe=bookblue!88!black,
  boxrule=0.72pt,
  arc=5pt,
  boxsep=1pt,
  left=1.0em,
  right=0.95em,
  top=0.82em,
  bottom=0.82em,
  before skip=0.55\baselineskip,
  after skip=1.0\baselineskip
]
\noindent\textbf{Chapter overview.}
This chapter begins the inference spine by asking how entire random objects,
not only numbers, have limits. Empirical distribution functions converge as
random curves, partial-sum processes as random paths, empirical processes as
random functions, and counting-process estimators as event-history objects. Weak
convergence, Portmanteau, tightness, finite-dimensional convergence, and
continuous mapping supply the bridge language.
\end{tcolorbox}

Chapter~3 introduced probability measures on measurable spaces.  Chapter~6
then enlarged the kinds of random things we can build: product-coordinate
objects, kernels, paths, random measures, point processes, and fields.  The
present chapter asks what it means for such random things to have a limit.

\begin{tcolorbox}[
  enhanced,
  breakable,
  colback=noteback,
  colframe=bookgold!75!black,
  boxrule=0.55pt,
  arc=4pt,
  boxsep=1pt,
  left=0.95em,
  right=0.9em,
  top=0.7em,
  bottom=0.7em,
  before skip=0.9\baselineskip,
  after skip=1.0\baselineskip
]
\noindent\textbf{How the early examples return.}
The recurrent examples from Chapters~1 and~2 are not decorative here; they
decide which state spaces are natural.  The London bombing example becomes a
point measure or a vector of cell counts, so weak convergence must say which
spatial events have negligible boundaries.  Missing species and Shakespeare
vocabulary become empirical measures and frequency-count processes, so a
limit must control more than one observed count.  The Zhu/Chu Ko-chen climate
thread becomes a proxy curve, clinical endpoints become event-history paths,
single-cell assays become empirical laws over cells, and a robot laboratory
becomes an adapted path of actions, sensor traces, and decisions.  Thus this
chapter does not repeat the scalar laws of Chapter~9 or the entropy arguments
of Chapter~11.  It names the spaces and maps that make those later technical
statements meaningful.
\end{tcolorbox}

The question is unavoidable.  A central limit theorem for one number says
\[
  \sqrt n(\bar X_n-\mu)\weakto \Normal(0,\sigma^2).
\]
But an empirical distribution function is not one number.  It is the whole
curve \(t\mapsto F_n(t)\).  A sequential experiment is not one outcome.  It is
a path of actions, observations, and filtrations.  A survival analysis is not
only an estimated hazard at one time.  It is often a counting process, a
compensator, or a martingale residual observed over a time interval.  Once the
statistic is a curve, a measure, or a field, convergence in distribution must
become convergence of random objects.

This chapter is deliberately a bridge, not a replacement for a full treatise on
weak convergence.  Its job is to make later chapters readable.  Chapter~7
supplied the common grammar that shows which empirical object is being studied,
Chapter~8 supplied the target language that shows why that object is being
studied, and Chapter~9 gave laws and concentration for fixed empirical
summaries.
Chapter~11 studies indexed empirical fields in spaces such as
\(\ell^\infty(\mathcal F)\).  Chapter~12 turns Gaussian process limits and
random curves into covariance modes through the Karhunen--Loeve expansion.
Chapter~13 uses continuous mapping, argmax, and root-finding logic for
estimators.  Chapter~15 uses the functional delta method.  Chapter~16 returns
to path spaces and counting-process limits.  All of those arguments share the
same quiet grammar: identify the random object, identify its topology, prove
weak convergence, and then apply a map.

\section{Random Elements and Their Laws}
\label{sec:ch10-random-elements}
\conceptindexes{random elements, laws on metric spaces, weak convergence, bounded continuous functions}

Let \((S,d)\) be a metric space and let \(\Borel(S)\) be its Borel
\(\sigma\)-field.  A random element in \(S\) is a measurable map
\[
  X:(\Omega,\mathcal A,\Prob)\to(S,\Borel(S)).
\]
Its law is the probability measure \(\Law(X)\) on \(S\) defined by
\[
  \Law(X)(B)=\Prob\{X\in B\},\qquad B\in\Borel(S).
\]
Nothing in this definition requires \(S=\R\).  The space \(S\) might be
\(\R^d\), \(C[0,1]\), \(D[0,T]\), \(\ell^\infty(\mathcal F)\), a space of
probability measures, or a space of locally finite counting measures.

\begin{definition}[Weak convergence in a metric space]
Let \(X_n\) and \(X\) be random elements in a metric space \(S\), with laws
\(\mu_n=\Law(X_n)\) and \(\mu=\Law(X)\).  We say that
\[
  X_n\weakto X
\]
weakly, or in distribution, if
\[
  \int_S h\,d\mu_n\to \int_S h\,d\mu
\]
for every bounded continuous function \(h:S\to\R\).  Equivalently,
\[
  \Expect h(X_n)\to\Expect h(X)
\]
for every bounded continuous \(h\).
\end{definition}

This definition is the right one because a probability law on a metric space
is probed by continuous tests.  If every bounded continuous question has the
same limiting answer under \(X_n\) as under \(X\), then the laws of \(X_n\)
approach the law of \(X\) in the weak sense.

\begin{example}[Real-valued convergence as a special case]
When \(S=\R\), the bounded-continuous definition is equivalent to the familiar
distribution-function criterion
\[
  F_n(t)\to F(t)
\]
at every continuity point \(t\) of \(F\).  The continuity-point restriction is
not a technical irritation.  It is the real-line version of the same principle:
weak convergence cannot be tested reliably at a set whose boundary receives
positive limiting probability.

\noindent\textit{Verification.}
Apply the Portmanteau theorem below to the sets \((-\infty,t]\).  Their
boundaries are the singletons \(\{t\}\), so their probabilities converge
whenever \(\Prob\{X=t\}=0\), exactly the continuity-point condition for \(F\).
Conversely, if the distribution functions converge at every continuity point
of \(F\), then probabilities converge on intervals whose endpoints are
continuity points.  Finite disjoint unions of such intervals form an algebra
that approximates every continuity set of the limiting law from inside and
outside.  Portmanteau then gives convergence against every bounded continuous
test function. \qedmark
\end{example}

\begin{example}[The empirical distribution function as a random object]
For iid real-valued observations, the empirical distribution function
\[
  F_n(t)=P_n\ind{X\le t}
\]
can be viewed as a random element of a function space.  Pointwise convergence
at a fixed \(t\) is only a one-dimensional statement.  Uniform convergence of
\(F_n\) and process convergence of \(\sqrt n(F_n-F)\) are statements about the
whole curve.  This is why the empirical-process chapter needs weak convergence
beyond \(\R\).

\noindent\textit{Details.}
For each sample point \(\omega\), the function \(t\mapsto F_n(t,\omega)\) is
nondecreasing, right-continuous, and bounded between \(0\) and \(1\).  Thus it
can be treated as an element of \(D(\R)\) on compact time windows, or as a
bounded function in \(\ell^\infty(\R)\) when the sup norm is the statistic of
interest.  At a fixed \(t\),
\[
  F_n(t)=\frac1n\sum_{i=1}^n\ind{X_i\le t}
\]
is only a Bernoulli average.  The whole-curve object is the map
\[
  \omega\longmapsto
  \left\{t\mapsto \frac1n\sum_{i=1}^n\ind{X_i(\omega)\le t}\right\}.
\]
Once this map is placed in a function space, statistics such as
\(\sup_t|F_n(t)-F(t)|\) become ordinary functionals of a random element.
\qedmark
\end{example}

\begin{example}[Wigner's semicircle law as weak convergence of random laws]
\label{ex:ch10-wigner-semicircle-pointer}
Weak convergence also appears when the random object is itself a probability
measure.  Let \(W_n\) be a normalized real symmetric Wigner matrix, and let
\(\lambda_1(W_n),\ldots,\lambda_n(W_n)\) be its eigenvalues.  The empirical
spectral distribution
\[
  \mu_n=\frac1n\sum_{j=1}^n\delta_{\lambda_j(W_n)}
\]
is a random element of the space of probability measures on \(\R\), equipped
with the topology of weak convergence.  Wigner's semicircle law says that
\[
  \mu_n\toP \mu_{\mathrm{sc}},
  \qquad
  \mu_{\mathrm{sc}}(dx)=
  \frac{1}{2\pi}\sqrt{4-x^2}\,\ind{|x|\le2}\,dx .
\]
Equivalently, for every bounded continuous test function \(h:\R\to\R\),
\[
  \int h\,d\mu_n
  \toP
  \int h\,d\mu_{\mathrm{sc}} .
\]
This is not a statement about one eigenvalue or one coordinate.  It says that
the whole random eigenvalue cloud, read as an empirical probability law, has a
deterministic weak limit.  The complete moment-method proof, including the
closed-walk count and Catalan moments, is given in
\Appref{sec:appC-wigner-semicircle}.
\qedmark
\end{example}

\begin{tcolorbox}[
  enhanced,
  breakable,
  colback=noteback,
  colframe=bookgold!75!black,
  boxrule=0.55pt,
  arc=4pt,
  boxsep=1pt,
  left=0.95em,
  right=0.9em,
  top=0.7em,
  bottom=0.7em,
  before skip=0.9\baselineskip,
  after skip=1.0\baselineskip
]
\noindent\textbf{The practical rule.}
Do not say that a random function, path, or measure converges until the space
and topology have been named.  The same algebraic object can have different
limits under different topologies, because different topologies declare
different features to be continuous.
\end{tcolorbox}

\section{Portmanteau Theorem and Equivalent Tests}
\label{sec:ch10-portmanteau}
\conceptindexes{portmanteau theorem, closed-set test, open-set test, convergence in distribution}

The bounded-continuous definition is compact, but applications often ask about
events.  Does the probability of a closed set have the right upper limit?  Does
the probability of an open set have the right lower limit?  Do probabilities
of continuity sets converge?  The Portmanteau theorem says that these are all
faces of the same weak-convergence statement.

\begin{theorem}[Portmanteau theorem]
Let \(\mu_n\) and \(\mu\) be probability measures on a metric space \(S\).
The following statements are equivalent:
\begin{enumerate}
\item \(\int h\,d\mu_n\to\int h\,d\mu\) for every bounded continuous
\(h:S\to\R\).
\item \(\limsup_n \mu_n(F)\le \mu(F)\) for every closed set \(F\subset S\).
\item \(\liminf_n \mu_n(G)\ge \mu(G)\) for every open set \(G\subset S\).
\item \(\mu_n(A)\to\mu(A)\) for every Borel set \(A\) whose boundary satisfies
\(\mu(\partial A)=0\).
\end{enumerate}
\end{theorem}

\noindent\textit{Proof.}
Write (i)--(iv) for the four displayed statements.  First assume (i).  If
\(F\subset S\) is closed, define
\[
  h_m(x)=\max\{1-md(x,F),0\},\qquad m=1,2,\ldots .
\]
Then \(h_m\) is bounded and continuous, \(1_F\le h_m\le1\), and
\(h_m(x)\downarrow1_F(x)\).  Hence
\[
  \limsup_n\mu_n(F)
  \le
  \lim_n\int h_m\,d\mu_n
  =
  \int h_m\,d\mu
\]
for each fixed \(m\).  Letting \(m\to\infty\) and using bounded convergence
gives \(\limsup_n\mu_n(F)\le\mu(F)\), so (ii) holds.

Statement (iii) follows from (ii) by applying the closed-set bound to
\(G^c\):
\[
  \liminf_n\mu_n(G)
  =
  1-\limsup_n\mu_n(G^c)
  \ge
  1-\mu(G^c)
  =
  \mu(G).
\]
If \(A\) is a \(\mu\)-continuity set, then
\[
  A^\circ\subseteq A\subseteq \overline A
  \quad\text{and}\quad
  \mu(A^\circ)=\mu(A)=\mu(\overline A),
\]
because \(\overline A\setminus A^\circ\subseteq\partial A\) has \(\mu\)-mass
zero.  Applying (ii) to \(\overline A\) and (iii) to \(A^\circ\) gives
\[
  \mu(A)
  \le
  \liminf_n\mu_n(A)
  \le
  \limsup_n\mu_n(A)
  \le
  \mu(A),
\]
which proves (iv).

It remains to show that (iv) implies (i).  Let \(h:S\to\R\) be bounded and
continuous.  After replacing \(h\) by \(a h+b\), it is enough to handle the
case \(0\le h\le1\).  Let \(\nu_n=\mu_n\circ h^{-1}\) and
\(\nu=\mu\circ h^{-1}\) be the laws of \(h\) under \(\mu_n\) and \(\mu\).
For \(t\in[0,1]\),
\[
  \partial\{x:h(x)\le t\}\subseteq\{x:h(x)=t\}.
\]
The set of \(t\)'s for which \(\mu\{h=t\}>0\) is at most countable, since the
positive masses of disjoint level sets must sum to at most one.  Thus (iv)
implies
\[
  \nu_n((-\infty,t])\to\nu((-\infty,t])
\]
at every continuity point of \(\nu\).  This is ordinary real-line weak
convergence of \(\nu_n\) to \(\nu\).  Because the variables are bounded in
\([0,1]\), convergence in distribution implies convergence of expectations;
for instance,
\[
  \int_0^1 y\,d\nu_n(y)
  =
  \int_0^1 \nu_n((t,1])\,dt
  \to
  \int_0^1 \nu((t,1])\,dt
  =
  \int_0^1 y\,d\nu(y)
\]
by dominated convergence, since the tail probabilities converge for all but
countably many \(t\).  Therefore
\(\int h\,d\mu_n\to\int h\,d\mu\), proving (i). \qedmark

This formulation is the standard metric-space version of Portmanteau; see
\citet[Chapter~1]{billingsley1999convergence} and
\citet[Chapter~II]{pollard1984convergence}.

The fourth form is the one statisticians use most often without naming it.  In
\(\R\), the event \((-\infty,t]\) has boundary \(\{t\}\), so the convergence
of distribution functions is guaranteed exactly at continuity points of the
limit distribution.  In a path space, an event such as
\[
  \left\{x:\sup_{0\le t\le T}|x(t)|\le c\right\}
\]
has a boundary determined by paths whose supremum is exactly \(c\).  If the
limiting process hits that boundary with probability zero, then weak
convergence of the process implies convergence of the event probabilities.

\begin{example}[Why boundaries matter]
Let \(X_n=1/n\) deterministically and \(X=0\).  Then \(X_n\weakto X\) in
\(\R\).  For the closed set \(F=\{0\}\), \(\Prob\{X_n\in F\}=0\) but
\(\Prob\{X\in F\}=1\), which is allowed by the Portmanteau upper-bound
inequality.  For the open set \(G=(-1,1)\), the probabilities converge to
one.  Weak convergence does not require every event probability to converge;
it requires convergence at events whose limiting boundary is invisible.

\noindent\textit{Verification.}
For every bounded continuous \(h:\R\to\R\), \(h(X_n)=h(1/n)\to h(0)=h(X)\);
boundedness is irrelevant here because the convergence is deterministic.
Hence \(X_n\weakto X\).  The boundary of \(F=\{0\}\) is again \(\{0\}\), and
the limiting law assigns it mass one.  Therefore \(F\) is not a continuity set,
so Portmanteau does not promise convergence of \(\Prob\{X_n\in F\}\).
\qedmark
\end{example}

\begin{example}[Continuity sets in statistical inference]
Suppose \(Z_n\weakto Z\) in a metric space and a test rejects when
\(Z_n\in A\).  If \(\Prob\{Z\in\partial A\}=0\), then
\[
  \Prob\{Z_n\in A\}\to\Prob\{Z\in A\}.
\]
This is the hidden step behind many asymptotic level calculations.  The limit
law supplies a usable approximation only when the rejection boundary is not
hit with positive limiting probability.

\noindent\textit{Proof.}
Let \(\mu_n=\Law(Z_n)\) and \(\mu=\Law(Z)\).  Since
\(\mu(\partial A)=0\), the set \(A\) is a \(\mu\)-continuity set.  Part (iv)
of the Portmanteau theorem gives \(\mu_n(A)\to\mu(A)\), which is exactly the
displayed probability convergence. \qedmark
\end{example}

\section{Tightness and Relative Compactness}
\label{sec:ch10-tightness}
\conceptindexes{tightness, compact containment, Prokhorov theorem, finite-dimensional convergence, Donsker theorem}

Weak convergence is a statement about laws.  Before laws can converge, their
mass must stay in controllable parts of the space.  On the real line this is
familiar: a sequence of random variables cannot have a proper weak limit if
more and more mass drifts to infinity.  In a function space, mass can escape
in more subtle ways: paths can oscillate faster and faster, develop jumps
under a topology that does not allow jumps, or place variation at smaller and
smaller scales.

\begin{definition}[Tightness]
A probability measure \(\mu\) on a metric space \(S\) is tight if for every
\(\varepsilon>0\) there exists a compact set \(K_\varepsilon\subset S\) such
that
\[
  \mu(K_\varepsilon)>1-\varepsilon.
\]
A family of probability measures \(\{\mu_\alpha:\alpha\in A\}\) is tight if
the same compact set can be chosen for all members of the family:
\[
  \inf_{\alpha\in A}\mu_\alpha(K_\varepsilon)>1-\varepsilon.
\]
A sequence of random elements \(X_n\) is tight if the sequence of laws
\(\Law(X_n)\) is tight.
\end{definition}

In \(\R^d\), compact sets are closed and bounded, so tightness mostly says
that the random vectors do not run off to infinity.  In \(C[0,1]\), compact
sets are controlled not only by boundedness but by equicontinuity.  That is
why process-level tightness often appears as a modulus-of-continuity bound.

\begin{tcolorbox}[
  enhanced,
  breakable,
  colback=noteback,
  colframe=bookgold!75!black,
  boxrule=0.55pt,
  arc=4pt,
  boxsep=1pt,
  left=0.95em,
  right=0.9em,
  top=0.7em,
  bottom=0.7em,
  before skip=0.9\baselineskip,
  after skip=1.0\baselineskip
]
\noindent\textbf{Prokhorov principle.}
On a Polish space, tightness is equivalent to relative compactness for weak
convergence: every tight sequence of probability laws has a weakly convergent
subsequence, and every relatively compact family is tight.  This is usually
cited as Prokhorov's theorem; modern treatments include
\citet[Chapter~1]{billingsley1999convergence} and
\citet[Chapter~III]{pollard1984convergence}.  The book uses it as a
background principle rather than a theorem proved here: tightness is the
condition that makes subsequential weak limits exist.  Identifying the
possible subsequential limit then turns tightness into convergence.
\end{tcolorbox}

\begin{example}[Tightness in \(\R\)]
If \(X_n\) are real-valued and
\[
  \sup_n \Expect |X_n|^p<\infty
\]
for some \(p>0\), then the laws of \(X_n\) are tight.  Markov's inequality
gives
\[
  \sup_n\Prob\{|X_n|>M\}
  \le
  M^{-p}\sup_n\Expect |X_n|^p,
\]
which can be made arbitrarily small by choosing \(M\) large.
\qedmark
\end{example}

\begin{example}[What tightness means for paths]
For random elements in \(C[0,1]\), tightness requires two kinds of control:
the paths must be bounded with high probability, and their oscillations over
small time intervals must be small with high probability.  A typical condition
has the form
\[
  \lim_{\delta\downarrow0}\limsup_n
  \Prob\left\{\sup_{|s-t|<\delta}|X_n(s)-X_n(t)|>\eta\right\}=0
\]
for every \(\eta>0\), together with tightness of \(X_n(0)\).  This is the
path-space version of saying that mass cannot escape through wild oscillation.

\noindent\textit{Verification.}
Here is the compactness mechanism.  Fix \(\varepsilon>0\).  Tightness of
\(X_n(0)\) gives \(M_0\) with
\[
  \inf_n\Prob\{|X_n(0)|\le M_0\}>1-\varepsilon/2.
\]
The modulus condition makes it possible to choose a decreasing sequence
\(\delta_k\downarrow0\) such that
\[
  \sup_n\Prob\{w(X_n,\delta_k)>1/k\}<\varepsilon 2^{-k-1},
  \qquad
  w(x,\delta)=\sup_{|s-t|<\delta}|x(s)-x(t)|.
\]
With probability at least \(1-\varepsilon\), a path \(X_n\) lies in the set
\[
  K=\{x\in C[0,1]: |x(0)|\le M_0,\ w(x,\delta_k)\le 1/k
      \text{ for all }k\ge1\}.
\]
The set \(K\) is uniformly bounded because each path can move from \(0\) to
any \(t\) through at most \(\lceil1/\delta_1\rceil\) intervals of length
less than \(\delta_1\), and it is equicontinuous because the bounds
\(w(x,\delta_k)\le1/k\) hold uniformly over \(x\in K\).  By the
Arzela--Ascoli theorem, \(K\) has compact closure in \(C[0,1]\).  Thus the
laws are tight. \qedmark
\end{example}

\subsection{Finite-Dimensional Limits Are Not Enough}
\label{sec:ch10-fidi-tightness}
\conceptindexes{finite-dimensional convergence, tightness, process convergence, moving spike example}

A stochastic process \(X_n=\{X_n(t):t\in T\}\) has many finite-dimensional
shadows.  For \(t_1,\ldots,t_k\in T\), the vector
\[
  (X_n(t_1),\ldots,X_n(t_k))
\]
may converge in distribution.  This is necessary for process convergence, but
not sufficient.  The shadows do not control what happens between the chosen
time points.

\begin{definition}[Finite-dimensional convergence]
Processes \(X_n=\{X_n(t):t\in T\}\) converge in finite-dimensional
distributions to \(X=\{X(t):t\in T\}\) if, for every finite collection
\(t_1,\ldots,t_k\in T\),
\[
  (X_n(t_1),\ldots,X_n(t_k))
  \weakto
  (X(t_1),\ldots,X(t_k))
\]
in the corresponding finite-dimensional Euclidean space.
\end{definition}

Finite-dimensional convergence identifies the candidate limit.  Tightness
decides whether the whole random object is forced to stay near that candidate
between the observed coordinates.

\begin{example}[A moving-spike counterexample]
\label{ex:ch10-moving-spike-fidi-not-weak}
Let the state space be \(C[0,1]\) with the sup-norm topology.  Define the
deterministic continuous path
\[
  x_n(t)=\max\{1-n|t-1/n|,0\},\qquad 0\le t\le1,
\]
and let \(X_n=x_n\).  The graph of \(x_n\) is a triangular spike of height
one, centered at \(1/n\), with support contained in \([0,2/n]\).  Let
\(X\equiv0\) be the zero process.

For every fixed \(t\in[0,1]\), \(x_n(t)\to0\).  Indeed, \(x_n(0)=0\) for all
\(n\), and if \(t>0\), then \(t>2/n\) for all sufficiently large \(n\), so
\(x_n(t)=0\) eventually.  Therefore, for any fixed
\(t_1,\ldots,t_k\),
\[
  (X_n(t_1),\ldots,X_n(t_k))
  \to
  (0,\ldots,0)
\]
deterministically.  Hence \(X_n\) converges to \(X\) in finite-dimensional
distributions.

But \(X_n\) does not converge weakly to \(X\) as a \(C[0,1]\)-valued random
element.  The bounded continuous functional
\[
  \Phi(x)=\min\{\|x\|_\infty,1\}
\]
satisfies \(\Phi(x_n)=1\) for every \(n\), while \(\Phi(0)=0\).  If
\(X_n\weakto X\) in \(C[0,1]\), then the definition of weak convergence would
give
\[
  \Expect\Phi(X_n)\to\Expect\Phi(X),
\]
that is, \(1\to0\), a contradiction.

The lesson is that fixed coordinates can miss moving local features.  The
spike escapes every preassigned finite set of time points, but it never
escapes the sup-norm view of the whole curve.  Tightness rules out exactly
this kind of uncontrolled motion between coordinates. \qedmark
\end{example}

\begin{tcolorbox}[
  enhanced,
  breakable,
  colback=chaptercream,
  colframe=bookblue!82!black,
  boxrule=0.58pt,
  arc=4pt,
  boxsep=1pt,
  left=0.95em,
  right=0.9em,
  top=0.72em,
  bottom=0.72em,
  before skip=0.9\baselineskip,
  after skip=1.0\baselineskip
]
\noindent\textbf{Billingsley's \(D[0,1]\) tightness criterion.}
Let \(X_n\) and \(X\) be \(D[0,1]\)-valued processes.  Suppose the
finite-dimensional distributions of \(X_n\) converge to those of \(X\).
Suppose also that there are constants \(\gamma>0\), \(\alpha>1/2\), and
\(C<\infty\), and bounded nondecreasing right-continuous functions \(H_n\),
such that for all \(0\le t_1<t<t_2\le1\),
\[
\begin{aligned}
  &\Expect\left[
    |X_n(t)-X_n(t_1)|^\gamma
    |X_n(t_2)-X_n(t)|^\gamma
  \right]  \\
  &\qquad\le
  C\{H_n(t)-H_n(t_1)\}^{\alpha}
   \{H_n(t_2)-H_n(t)\}^{\alpha}.
\end{aligned}
\]
If \(H_n\) converges pointwise at the continuity points of a bounded
nondecreasing right-continuous limit \(H\), then the laws of \(X_n\) are
tight in \(D[0,1]\).  Consequently, together with the finite-dimensional
convergence, \(X_n\weakto X\) in \(D[0,1]\).

\smallskip
\noindent This is a compact form of Billingsley's criterion
\citep[Chapter~13]{billingsley1999convergence}.  We quote it as a standard
path-space compactness theorem; the two examples below verify its moment
condition explicitly.
\end{tcolorbox}

\begin{tcolorbox}[
  enhanced,
  breakable,
  colback=noteback,
  colframe=bookgold!75!black,
  boxrule=0.55pt,
  arc=4pt,
  boxsep=1pt,
  left=0.95em,
  right=0.9em,
  top=0.7em,
  bottom=0.7em,
  before skip=0.9\baselineskip,
  after skip=1.0\baselineskip
]
\noindent\textbf{Process convergence recipe.}
For many process limits, the proof has two jobs:
\[
  \begin{gathered}
  \text{finite-dimensional convergence}+\text{tightness}\\
  \Longrightarrow\\
  \text{weak convergence of random objects}.
  \end{gathered}
\]
The first job is often a multivariate central limit theorem.  The second job
is usually the hard one.
\end{tcolorbox}

\begin{example}[Partial-sum process: weak convergence in \(D[0,1]\)]
\label{ex:ch10-partial-sum-donsker}
Let \(X_1,X_2,\ldots\) be iid with mean \(0\) and variance \(\sigma^2>0\), and
define
\[
  W_n(t)
  =
  \frac{1}{\sigma\sqrt n}
  \sum_{i=1}^{\lfloor nt\rfloor}X_i,\qquad 0\le t\le1.
\]
The finite-dimensional distributions converge to those of Brownian motion
because increments over disjoint time intervals are asymptotically normal and
asymptotically independent.  Tightness supplies control of the path between
the grid points.  The finite-dimensional calculation is already enough to
show why the coin-string examples of Chapter~2 become path-valued objects:
partial sums remember the order of the random signs, not only their final
average.

\noindent\textit{Finite-dimensional proof.}
Fix \(0\le t_1<\cdots<t_k\le1\).  It is enough, by the Cramer--Wold device, to
show convergence of every linear combination
\[
  \sum_{j=1}^k a_j W_n(t_j).
\]
Rewrite this linear combination in terms of disjoint increments.  Put
\(t_0=0\) and
\[
  b_\ell=\sum_{j=\ell}^k a_j,\qquad \ell=1,\ldots,k.
\]
Then
\[
  \sum_{j=1}^k a_j W_n(t_j)
  =
  \sum_{\ell=1}^k
  b_\ell\{W_n(t_\ell)-W_n(t_{\ell-1})\}.
\]
The \(\ell\)th increment is a normalized sum of
\(\lfloor nt_\ell\rfloor-\lfloor nt_{\ell-1}\rfloor\) iid variables, over a
block disjoint from the other increments.  The ordinary central limit theorem
therefore gives
\[
  W_n(t_\ell)-W_n(t_{\ell-1})
  \weakto
  \Normal(0,t_\ell-t_{\ell-1}),
\]
and the block sums are independent.  Hence the linear combination converges
to a centered normal variable with variance
\[
  \sum_{\ell=1}^k b_\ell^2(t_\ell-t_{\ell-1}),
\]
which is exactly the variance of
\(\sum_{j=1}^k a_j W(t_j)\) for standard Brownian motion \(W\).  Thus the
finite-dimensional distributions converge.

\noindent\textit{Tightness and weak convergence.}
Use the Billingsley criterion above.  Put
\[
  H_n(t)=\frac{\lfloor nt\rfloor}{n}.
\]
Then \(H_n\) is bounded, nondecreasing, right-continuous, and
\(H_n(t)\to t\) for every \(t\in[0,1]\).  For \(t_1<t<t_2\), define
\[
  A_n=W_n(t)-W_n(t_1),
  \qquad
  B_n=W_n(t_2)-W_n(t).
\]
The sums \(A_n\) and \(B_n\) use disjoint blocks of observations, hence they
are independent and centered.  Their variances are
\[
  \Var(A_n)=H_n(t)-H_n(t_1),
  \qquad
  \Var(B_n)=H_n(t_2)-H_n(t),
\]
because the normalization divides the block variances by \(n\sigma^2\).
Therefore
\[
\begin{aligned}
  \Expect(A_n^2B_n^2)
  &=
  \Expect A_n^2\,\Expect B_n^2  \\
  &=
  \{H_n(t)-H_n(t_1)\}
  \{H_n(t_2)-H_n(t)\}.
\end{aligned}
\]
Billingsley's criterion applies with \(\gamma=2\), \(\alpha=1\), and
\(C=1\).  Hence \(W_n\) is tight in \(D[0,1]\).  Combined with the
finite-dimensional convergence above,
\[
  W_n\weakto W
  \qquad\text{in }D[0,1],
\]
where \(W\) is standard Brownian motion. \qedmark
\end{example}

\begin{example}[Empirical CDF fluctuations: weak convergence on the \(F\)-scale]
\label{ex:ch10-empirical-cdf-donsker}
For iid observations with continuous distribution function \(F\), the process
\[
  \alpha_n(t)=\sqrt n\{F_n(t)-F(t)\},\qquad t\in\R,
\]
can be proved on the uniform scale.  Let \(U_i=F(X_i)\), so
\(U_i\sim\Unif(0,1)\), let \(G_n\) be the empirical cdf of
\(U_1,\ldots,U_n\), and define
\[
  \beta_n(u)=\sqrt n\{G_n(u)-u\},\qquad 0\le u\le1.
\]
The original process satisfies \(\alpha_n(t)=\beta_n(F(t))\).

\noindent\textit{Finite-dimensional proof.}
Fix \(u_1,\ldots,u_k\in[0,1]\) and \(a_1,\ldots,a_k\).  Define
\[
  V_i=\sum_{j=1}^k a_j\{\ind{U_i\le u_j}-u_j\}.
\]
Then
\[
  \sum_{j=1}^k a_j\beta_n(u_j)
  =
  \frac1{\sqrt n}\sum_{i=1}^n V_i.
\]
The \(V_i\)'s are iid, mean zero, and bounded, so the ordinary central limit
theorem applies.  The limiting variance is
\[
  \Var(V_1)
  =
  \sum_{j=1}^k\sum_{\ell=1}^k
  a_ja_\ell\{u_j\wedge u_\ell-u_ju_\ell\}.
\]
This is the covariance form of a standard Brownian bridge \(B\), since
\[
  \Cov\{B(u),B(v)\}
  =
  u\wedge v-uv.
\]
By Cramer--Wold, the vector
\((\beta_n(u_1),\ldots,\beta_n(u_k))\) converges to the corresponding
Brownian-bridge vector.

\noindent\textit{Tightness and weak convergence.}
We again use the Billingsley criterion above.  For
\(u_1<u<u_2\), put
\[
  p=u-u_1,\qquad q=u_2-u,
\]
and
\[
  Z_i=\ind{u_1<U_i\le u}-p,
  \qquad
  R_i=\ind{u<U_i\le u_2}-q.
\]
Then
\[
  \beta_n(u)-\beta_n(u_1)=n^{-1/2}\sum_{i=1}^n Z_i,
  \qquad
  \beta_n(u_2)-\beta_n(u)=n^{-1/2}\sum_{i=1}^n R_i.
\]
Because the two intervals are disjoint,
\[
  \Expect Z_1^2R_1^2\le2pq,\qquad
  \Expect Z_1^2\le p,\qquad
  \Expect R_1^2\le q,\qquad
  |\Expect Z_1R_1|=pq.
\]
Expanding the fourth mixed moment gives
\[
\begin{aligned}
  &\Expect\left[
    \left(\sum_{i=1}^n Z_i\right)^2
    \left(\sum_{i=1}^n R_i\right)^2
  \right] \\
  &\quad =
  n\Expect Z_1^2R_1^2
  +n(n-1)\Expect Z_1^2\,\Expect R_1^2
  +2n(n-1)(\Expect Z_1R_1)^2 .
\end{aligned}
\]
After division by \(n^2\),
\[
\begin{aligned}
  &\Expect\{|\beta_n(u)-\beta_n(u_1)|^2
  |\beta_n(u_2)-\beta_n(u)|^2\}  \\
  &\qquad\le
  \frac{2pq}{n}
  +\left(1-\frac1n\right)pq
  +2\left(1-\frac1n\right)p^2q^2
  \le 5pq .
\end{aligned}
\]
The criterion applies with \(\gamma=2\), \(\alpha=1\), \(C=5\), and
\(H_n(u)=u\).  Thus
\[
  \beta_n\weakto B
  \qquad\text{in }D[0,1].
\]
This is Donsker's theorem on the uniform scale.  The passage back to the
original \(F\)-scale is a mapping step, taken up after the continuous mapping
theorem in Example~\ref{ex:ch10-empirical-cdf-f-scale}.  The classical
Kolmogorov--Smirnov thread runs through
\citet{doob1949heuristic}, \citet{donsker1952justification}, and
\citet{pollard1982beyond}. \qedmark
\end{example}

\section{Mapping Random Objects Into Statistics}
\label{sec:ch10-continuous-mapping}
\conceptindexes{continuous mapping theorem, Skorokhod--Dudley representation, Slutsky theorem, random elements}

Weak convergence becomes useful when a statistic is a function of a random
object.  A supremum, an integral, an estimated quantile, an argmax, a root, a
projection, and a confidence band are all maps applied to an underlying random
object.  The main question is whether the map is continuous at the limiting
object.

\begin{theorem}[Skorokhod--Dudley representation theorem; \citealp{skorokhod1956limit,dudley1968distances}]
\label{thm:ch10-skorokhod-dudley}
Let \(S\) be a Polish space with metric \(d\).  If \(X_n\weakto X\) as
\(S\)-valued random elements, then there exist \(S\)-valued random elements
\(\widetilde X_n,\widetilde X\), defined on a common probability space, such
that
\[
  \Law(\widetilde X_n)=\Law(X_n),\qquad
  \Law(\widetilde X)=\Law(X),
\]
and
\[
  d(\widetilde X_n,\widetilde X)\to0
  \qquad\text{almost surely.}
\]
\end{theorem}

\noindent\textit{Proof.}
Write \(\mu_n=\Law(X_n)\) and \(\mu=\Law(X)\).  Since \(S\) is Polish, it is
separable; fix a countable dense set \(\{s_1,s_2,\ldots\}\).  We first build a
nested sequence of countable Borel partitions
\[
  \mathcal A_m=\{A_\alpha:|\alpha|=m\},\qquad m=1,2,\ldots,
\]
where \(\alpha\) ranges over finite words of positive integers, such that
\(\mathcal A_{m+1}\) refines \(\mathcal A_m\),
\[
  \diam(A_\alpha)<2^{-m},\qquad \mu(\partial A_\alpha)=0
  \quad (|\alpha|=m).
\]
Here is the construction.  Start with \(A_{\varnothing}=S\).  Given a cell
\(C\) at level \(m-1\), choose radii
\[
  r_{m,j}\in(2^{-(m+2)},2^{-(m+1)}),\qquad j=1,2,\ldots,
\]
so that \(\mu\{\partial B(s_j,r_{m,j})\}=0\).  This is possible because, for
fixed \(s_j\), the spheres \(\partial B(s_j,r)\) are disjoint as \(r\) varies,
and a probability measure can assign positive mass to at most countably many
disjoint spheres.  The balls \(B(s_j,r_{m,j})\) cover \(S\), because the centers
are dense and every radius is larger than \(2^{-(m+2)}\).  Define the children
of \(C\) by
\[
  C_j
  =
  C\cap B(s_j,r_{m,j})\setminus\bigcup_{\ell<j}B(s_\ell,r_{m,\ell}),
  \qquad j=1,2,\ldots,
\]
and discard the empty children.  These children are disjoint, cover \(C\), have
diameter less than \(2^{-m}\), and have \(\mu\)-null boundary because their
boundaries are contained in \(\partial C\) together with countably many
\(\mu\)-null sphere boundaries.  Induction gives the desired partitions.

For a finite word \(\alpha\), the set \(A_\alpha\) is a \(\mu\)-continuity set.
Since \(\mu_n\weakto\mu\), Portmanteau gives
\[
  \mu_n(A_\alpha)\to\mu(A_\alpha)
  \qquad\text{for every finite word }\alpha.
\]
Let \(E=\mathbb N^{\mathbb N}\) be the code space, and let
\(\phi:S\to E\) record the nested cells containing a point: the first \(m\)
coordinates of \(\phi(x)\) are the word \(\alpha\) for which
\(x\in A_\alpha\).  The map \(\phi\) is one-to-one, because two distinct points
cannot remain in the same cells once the cell diameters are smaller than their
distance.  Let
\[
  \nu_n=\mu_n\circ\phi^{-1},
  \qquad
  \nu=\mu\circ\phi^{-1}.
\]
For a cylinder \([\alpha]=\{z\in E:z_1,\ldots,z_m=\alpha\}\),
\[
  \nu_n([\alpha])=\mu_n(A_\alpha)\to\mu(A_\alpha)=\nu([\alpha]).
\]

We now realize all these code laws on the same probability space
\((0,1),\mathcal B(0,1),\lambda)\).  For \(\nu\), split \((0,1)\) into
consecutive half-open intervals \(I_\alpha\), nested by cylinders, so that
\(\lambda(I_\alpha)=\nu([\alpha])\).  More precisely, the children of
\(I_\alpha\) are placed from left to right with lengths
\(\nu([\alpha 1]),\nu([\alpha 2]),\ldots\).  Construct intervals
\(I_{n,\alpha}\) in the same way from the cylinder probabilities of \(\nu_n\).
Because each fixed cylinder probability converges, the two endpoints of
\(I_{n,\alpha}\) converge to the corresponding endpoints of \(I_\alpha\) for
every fixed finite word \(\alpha\).

Define \(Z(u)\in E\) to be the unique code whose level-\(m\) prefix
\(\alpha\) satisfies \(u\in I_\alpha\) for every \(m\); define \(Z_n(u)\) from
the intervals \(I_{n,\alpha}\) in the same way.  Then \(Z\sim\nu\) and
\(Z_n\sim\nu_n\), since cylinder probabilities are exactly the interval
lengths.  Let \(D\) be the countable set of all endpoints of the limiting
intervals \(I_\alpha\).  If \(u\notin D\), then for every fixed level \(m\), the
level-\(m\) prefix of \(Z_n(u)\) equals the level-\(m\) prefix of \(Z(u)\) for
all sufficiently large \(n\), because the relevant interval endpoints converge
and \(u\) lies in the interior of its limiting level-\(m\) interval.

It remains only to decode.  For each nonempty cell \(A_\alpha\), choose one
representative point \(a_\alpha\in A_\alpha\).  For \(z\in E\), put
\[
  \psi_m(z)=a_{z_1,\ldots,z_m}
\]
when the corresponding cell is nonempty, and use a fixed point of \(S\)
otherwise.  If \(z=\phi(x)\), then \(\psi_m(z)\in A_{z_1,\ldots,z_m}\) and
\(x\in A_{z_1,\ldots,z_m}\), so
\[
  d\{\psi_m(z),x\}\le \diam(A_{z_1,\ldots,z_m})<2^{-m}.
\]
Thus \(\psi_m(\phi(x))\to x\).  Define \(\psi(z)=\lim_m\psi_m(z)\) on the set
where the limit exists, and define it arbitrarily elsewhere.  This \(\psi\) is
Borel measurable as a pointwise limit on its convergence set, and
\(\psi\circ\phi\) is the identity on \(S\).

Set
\[
  \widetilde X_n(u)=\psi\{Z_n(u)\},
  \qquad
  \widetilde X(u)=\psi\{Z(u)\}.
\]
Because \(Z_n\sim\mu_n\circ\phi^{-1}\), \(Z\sim\mu\circ\phi^{-1}\), and
\(\psi\circ\phi=\mathrm{id}_S\), the decoded random elements satisfy
\[
  \Law(\widetilde X_n)=\mu_n=\Law(X_n),
  \qquad
  \Law(\widetilde X)=\mu=\Law(X).
\]
Since each \(\nu_n\) and \(\nu\) is supported on \(\phi(S)\), after removing one
more null set we may also assume \(Z(u),Z_1(u),Z_2(u),\ldots\in\phi(S)\).  For
such a \(u\notin D\), fix \(m\).  For all large \(n\), the first \(m\)
coordinates of \(Z_n(u)\) and \(Z(u)\) agree.  The decoded points
\(\widetilde X_n(u)\) and \(\widetilde X(u)\) therefore belong to the same
level-\(m\) cell, so
\[
  d\{\widetilde X_n(u),\widetilde X(u)\}\le 2^{-m}
\]
for all sufficiently large \(n\).  Since \(m\) is arbitrary and
\(\lambda(D)=0\), \(d(\widetilde X_n,\widetilde X)\to0\) almost surely.
\qedmark

\noindent This theorem is often called Skorokhod's representation theorem; in
this Polish-space form it is also called the Skorokhod--Dudley representation
theorem \citep{skorokhod1956limit,dudley1968distances,billingsley1999convergence}.
It does not say that the original variables \(X_n\) converge almost surely on
their original probability space.  It says that their laws can be represented
by copies on another probability space where almost-sure convergence holds.

\begin{example}[Why the representation is not original almost-sure convergence]
\label{ex:ch10-weak-not-original-as}
Let \(X_1,X_2,\ldots\) be iid Bernoulli\((1/2)\) random variables on one
probability space, and let \(X\) be another Bernoulli\((1/2)\) random variable.
Since all \(X_n\)'s have the same law as \(X\),
\[
  X_n\weakto X.
\]
But the original sequence \(X_n\) does not converge almost surely.

\noindent\textit{Proof.}
For a \(0\)-\(1\) sequence to converge, it must eventually be constant.  Hence
\[
  \{X_n\text{ converges}\}
  =
  \left(\bigcup_{N=1}^\infty\bigcap_{n\ge N}\{X_n=0\}\right)
  \cup
  \left(\bigcup_{N=1}^\infty\bigcap_{n\ge N}\{X_n=1\}\right).
\]
For each fixed \(N\),
\[
  \Prob\left(\bigcap_{n\ge N}\{X_n=0\}\right)
  =
  \lim_{m\to\infty}\Prob(X_N=\cdots=X_m=0)
  =
  \lim_{m\to\infty}2^{-(m-N+1)}
  =
  0,
\]
and the same calculation holds for eventual ones.  A countable union of null
sets is null, so \(\Prob\{X_n\text{ converges}\}=0\).  Thus weak convergence
does not force almost-sure convergence of the original sequence.

The representation theorem says something different: because the laws are all
the same, one may take \(\widetilde X_n=\widetilde X\) for every \(n\) on a new
probability space with \(\widetilde X\sim\mathrm{Bernoulli}(1/2)\).  Then
\(\widetilde X_n\to\widetilde X\) almost surely.  The laws agree with the
original laws, but the pathwise behavior has been changed by the coupling.
\qedmark
\end{example}

\begin{theorem}[Continuous mapping theorem]
\label{thm:ch10-continuous-mapping}
Let \(X_n\weakto X\) in a metric space \(S\), and let \(g:S\to T\) be a
measurable map into another metric space \(T\).  If \(g\) is continuous at
\(X\) with probability one, then
\[
  g(X_n)\weakto g(X).
\]
\end{theorem}

\noindent\textit{Proof.}
Let \(h:T\to\R\) be bounded and continuous.  We must show
\[
  \Expect h\{g(X_n)\}\to \Expect h\{g(X)\}.
\]
Put \(f=h\circ g\).  The function \(f\) is bounded and measurable, and its
discontinuity set is contained in the discontinuity set of \(g\).  Hence
\(\Prob\{X\in D_f\}=0\).

We use the following consequence of Portmanteau: if \(X_n\weakto X\) and
\(f\) is bounded measurable with \(\Prob\{X\in D_f\}=0\), then
\(\Expect f(X_n)\to\Expect f(X)\).  To prove this consequence, first reduce
to \(0\le f\le1\).  For \(t\in[0,1]\), set
\[
  A_t=\{x:f(x)\le t\}.
\]
The boundary of \(A_t\) is contained in \(D_f\cup\{x:f(x)=t\}\).  The first
set has zero \(X\)-probability, and the second has positive probability for
at most countably many \(t\).  Therefore \(A_t\) is a continuity set for all
but countably many \(t\), so Portmanteau gives
\[
  \Prob\{f(X_n)\le t\}\to \Prob\{f(X)\le t\}
\]
at all continuity points of the law of \(f(X)\).  The bounded real random
variables \(f(X_n)\) therefore converge in distribution to \(f(X)\), and their
expectations converge by the tail-integral identity on \([0,1]\).  Applying
this consequence to \(f=h\circ g\) proves the theorem. \qedmark

The continuous mapping theorem is one of the basic transfer devices of weak
convergence; see \citet[Chapter~1]{billingsley1999convergence} and
\citet[Section~1.3]{vaart2023weak}.

\begin{example}[Empirical CDF fluctuations: returning to the \(F\)-scale]
\label{ex:ch10-empirical-cdf-f-scale}
Continue Example~\ref{ex:ch10-empirical-cdf-donsker}.  The uniform-scale
process satisfies \(\beta_n\weakto B\) in \(D[0,1]\), where \(B\) is a
Brownian bridge.  For continuous \(F\), define
\[
  g(x)(t)=x\{F(t)\},\qquad t\in\R.
\]
At every continuous path \(x\),
\[
  \sup_{t\in\R}|g(x_m)(t)-g(x)(t)|
  \le
  \sup_{0\le u\le1}|x_m(u)-x(u)|.
\]
Since \(B\) has continuous paths almost surely, \(g\) is continuous at \(B\).
The continuous mapping theorem gives
\[
  \alpha_n(\cdot)=\beta_n\{F(\cdot)\}
  \weakto
  B\{F(\cdot)\}.
\]
This is Donsker's theorem for the empirical cdf in the continuous case.
\qedmark
\end{example}

\begin{example}[Skorokhod representation as a proof engine]
\label{ex:ch10-skorokhod-cmt-engine}
Assume \(S\) is Polish, \(X_n\weakto X\) in \(S\), and \(g:S\to T\) is
continuous at \(X\) with probability one.  The representation theorem gives
copies \(\widetilde X_n,\widetilde X\) with
\(\widetilde X_n\to\widetilde X\) almost surely.  On the event where
\(\widetilde X\) is a continuity point of \(g\),
\[
  g(\widetilde X_n)\to g(\widetilde X).
\]
Therefore \(g(\widetilde X_n)\weakto g(\widetilde X)\).  Since
\[
  \Law\{g(\widetilde X_n)\}=\Law\{g(X_n)\},
  \qquad
  \Law\{g(\widetilde X)\}=\Law\{g(X)\},
\]
one recovers \(g(X_n)\weakto g(X)\).

\noindent\textit{Application.}
For the empirical-cdf process above, the theorem makes it possible to work on a
representation where
\(\widetilde\beta_n\to B\) almost surely in \(D[0,1]\).  Since the Brownian
bridge \(B\) is continuous, this convergence is uniform after the Skorokhod
time change, and continuous functionals such as \(x\mapsto\sup_u|x(u)|\) may
be applied path by path.  Translating back to laws gives the usual
Kolmogorov--Smirnov weak limit. \qedmark
\end{example}

\begin{example}[A supremum statistic]
If \(X_n\weakto X\) in \(\ell^\infty(T)\) with the sup-norm topology, then
the map
\[
  x\mapsto \sup_{t\in T}|x(t)|
\]
is Lipschitz.  Therefore
\[
  \sup_{t\in T}|X_n(t)|
  \weakto
  \sup_{t\in T}|X(t)|.
\]
Kolmogorov--Smirnov limits and confidence bands are built from this simple
mapping idea.

\noindent\textit{Proof.}
For \(x,y\in\ell^\infty(T)\),
\[
  \left|\sup_{t\in T}|x(t)|-\sup_{t\in T}|y(t)|\right|
  \le
  \sup_{t\in T}|x(t)-y(t)|.
\]
Thus the supremum map is Lipschitz with constant one under the sup norm.  It
is therefore continuous everywhere, and the continuous mapping theorem gives
the displayed weak convergence. \qedmark
\end{example}

\begin{theorem}[Slutsky theorem for random elements; \citealp{slutsky1925asymptoten}]
Let \(X_n\weakto X\) in a separable metric space \(S\), and let \(Y_n\toP c\)
in a separable metric space \(T\), where \(c\) is a
constant.  Then
\[
  (X_n,Y_n)\weakto (X,c)
\]
in \(S\times T\).  Consequently, if \(g:S\times T\to U\) is continuous at
\((X,c)\) with probability one, then
\[
  g(X_n,Y_n)\weakto g(X,c).
\]
\end{theorem}

\noindent\textit{Proof.}
Bounded Lipschitz test functions determine weak convergence on separable
metric spaces.  Let \(h:S\times T\to\R\) be bounded by \(B\) and Lipschitz
with constant \(L\) under the product metric.  Then, for every \(\eta>0\),
\[
\begin{aligned}
  \left|
  \Expect h(X_n,Y_n)-\Expect h(X_n,c)
  \right|
  &\le
  L\eta+
  2B\,\Prob\{d_T(Y_n,c)>\eta\}.
\end{aligned}
\]
The second term tends to zero because \(Y_n\toP c\).  Letting \(\eta\downarrow0\)
shows that
\[
  \Expect h(X_n,Y_n)-\Expect h(X_n,c)\to0.
\]
The map \(x\mapsto h(x,c)\) is bounded and continuous on \(S\), so
\[
  \Expect h(X_n,c)\to \Expect h(X,c)
\]
by \(X_n\weakto X\).  Thus \((X_n,Y_n)\weakto(X,c)\).  The final statement is
the continuous mapping theorem applied to the pair \((X_n,Y_n)\). \qedmark

This random-element version is the usual Slutsky theorem written on product
spaces; see \citet[Chapter~1]{billingsley1999convergence}.

This is the form that appears whenever an estimated nuisance quantity is
plugged into an asymptotic approximation.  If the nuisance estimator converges
in probability and the main fluctuation converges weakly, the joint pair can
often be treated as if the nuisance component had settled at its limit.

\begin{example}[Argmax as a map]
Let \(M_n\) be a random criterion function and \(M\) a deterministic limit.
If \(M_n\) converges to \(M\) in a topology strong enough to preserve the
location of a well-separated maximum, and if \(M\) has a unique maximizer
\(\theta_0\), then the maximizer of \(M_n\) converges to \(\theta_0\).  This is
the continuous-mapping idea behind consistency of \(M\)-estimators.  When the
local process
\[
  h\mapsto r_n^2\{M_n(\theta_0+h/r_n)-M_n(\theta_0)\}
\]
has a weak limit, the same idea can produce the limiting distribution of
\(r_n(\hat\theta_n-\theta_0)\).  Chapter~13 develops this statistical version
of object mapping.

\noindent\textit{Consistency proof on a compact parameter space.}
Assume \(\Theta\) is compact, \(M\) is continuous, \(M\) has unique maximizer
\(\theta_0\), and
\[
  \sup_{\theta\in\Theta}|M_n(\theta)-M(\theta)|\toP0.
\]
Let \(\hat\theta_n\) satisfy
\[
  M_n(\hat\theta_n)\ge \sup_{\theta\in\Theta}M_n(\theta)-o_{\Prob}(1).
\]
Fix \(\varepsilon>0\).  The closed set
\(\{\theta:d(\theta,\theta_0)\ge\varepsilon\}\) is compact and does not contain
\(\theta_0\).  Therefore
\[
  \Delta_\varepsilon
  =
  M(\theta_0)-
  \sup_{d(\theta,\theta_0)\ge\varepsilon}M(\theta)
  >
  0.
\]
On the event
\(\sup_\theta|M_n(\theta)-M(\theta)|<\Delta_\varepsilon/3\) and with the
approximation error smaller than \(\Delta_\varepsilon/3\), every
\(\theta\) outside the \(\varepsilon\)-ball has criterion value below
\(M_n(\theta_0)\) by at least \(\Delta_\varepsilon/3\).  Hence
\(d(\hat\theta_n,\theta_0)<\varepsilon\).  The event has probability tending
to one, so \(\hat\theta_n\toP\theta_0\). \qedmark
\end{example}

\subsection{Function Spaces Used Later}
\label{sec:ch10-spaces-return}
\conceptindexes{function spaces, Skorokhod space, Hilbert space, probability-measure spaces, empirical process spaces}

Weak convergence is not a single technique attached to one space.  It is a
habit for treating statistics as random elements.  The following spaces will
return repeatedly.

\begin{center}
\small
\textbf{Common weak-convergence state spaces.}\par\smallskip
\setlength{\tabcolsep}{0.42em}
\renewcommand{\arraystretch}{1.16}
\begin{tabular}{@{}>{\raggedright\arraybackslash}p{0.23\linewidth}
                >{\raggedright\arraybackslash}p{0.31\linewidth}
                >{\raggedright\arraybackslash}p{0.34\linewidth}@{}}
\toprule
Space & Random object & Later role \\
\midrule
\(\R^d\) &
Vectors of means, scores, regression coefficients &
Multivariate CLTs, Slutsky arguments, sandwich limits \\
\(C[0,1]\) &
Continuous paths and Gaussian limits &
Brownian motion, partial-sum limits, smooth process approximations \\
\(D[0,T]\) &
Cadlag paths with jumps &
Counting processes, survival paths, event histories, martingales \\
\(\ell^\infty(\mathcal F)\) &
Bounded functions indexed by a class \(\mathcal F\) &
Empirical processes, confidence bands, stochastic equicontinuity \\
\(\mathcal P(S)\) &
Probability measures on a state space &
Empirical measures, bootstrap laws, distributional data objects \\
Counting-measure spaces &
Integer-valued random measures &
Point patterns, marked point processes, compensators \\
\bottomrule
\end{tabular}
\end{center}

The common pattern is this:
\[
  \text{data}
  \longmapsto
  \text{random object}
  \longmapsto
  \text{weak limit}
  \longmapsto
  \text{statistical functional}.
\]
The topology controls which functionals are continuous and therefore which
statistical summaries can be transferred through the limit.

\begin{example}[London bombing as a counting-measure functional]
Return to the London bombing example from Chapter~2.  Let \(W\subset\R^2\) be
the observation window, for instance the part of London being studied, and let
\[
  N=\sum_{i=1}^M\delta_{Z_i}
\]
be the random counting measure of impact locations.  This notation only means
that each impact point \(Z_i\) contributes one unit of mass at its location.
If \(A\subset W\) is one grid cell, then
\[
  N(A)=\sum_{i=1}^M\ind{Z_i\in A}
\]
is just the number of impacts falling inside that cell.

Here is the concrete use.  Suppose the map is split into grid cells and we want
to study whether one cell \(A\) has an unusually large count.  The data analyst
does three things:
\[
  \text{impact coordinates}
  \longmapsto
  \text{counting measure }N
  \longmapsto
  \text{cell count }N(A).
\]
For example, if three impacts fall at \(z_1,z_2,z_3\), with \(z_1,z_2\in A\)
and \(z_3\notin A\), then \(N(A)=2\).  If a fitted point-process model or a
large-sample approximation moves these points by a very small amount, the
count remains \(2\), provided none of the points lies exactly on the boundary
line of the cell.  If \(z_2\in\partial A\), however, an arbitrarily small move
can push it inside or outside, changing the count from \(1\) to \(2\).  This is
why the boundary condition appears.

Under the vague topology on locally finite counting measures, the cell-count
map \(N\mapsto N(A)\) is continuous at counting measures that put no point on
\(\partial A\).  Thus weak convergence of point-pattern laws transfers to
convergence of grid-cell count probabilities whenever the limiting process has
no boundary point in the chosen cell.  In statistical language, the asymptotic
approximation is stable for ordinary cells, but it cannot decide a case where
the limiting process puts positive probability on the cell boundary.

\noindent\textit{Proof of the continuity statement.}
Suppose \(n_m\to n\) vaguely and \(n(\partial A)=0\), with \(A\) relatively
compact.  By the Portmanteau theorem for finite measures on \(W\),
\[
  \limsup_m n_m(\overline A)\le n(\overline A),
  \qquad
  \liminf_m n_m(A^\circ)\ge n(A^\circ).
\]
Since \(n(\partial A)=0\),
\[
  n(A^\circ)=n(A)=n(\overline A).
\]
Also \(n_m(A^\circ)\le n_m(A)\le n_m(\overline A)\).  The squeeze theorem
gives \(n_m(A)\to n(A)\).  Therefore the grid-cell count is a continuous
functional at all configurations with no boundary impact.  The Chapter~2
lesson now has a topological form: a cluster statistic can be transferred
through a weak limit only after its boundary cases have been named. \qedmark
\end{example}

\begin{example}[Why \(\ell^\infty(\mathcal F)\) appears]
The empirical process
\[
  \mathbb G_n f=\sqrt n(P_n-P)f,\qquad f\in\mathcal F,
\]
is a random function of the index \(f\).  Placing it in
\(\ell^\infty(\mathcal F)\) allows one to discuss the whole field at once.
Finite-dimensional convergence checks finitely many functions
\((f_1,\ldots,f_k)\).  Tightness, often proved through entropy and stochastic
equicontinuity, prevents the field from oscillating too much over nearby
functions.  This is exactly the architecture of Chapter~11.

\noindent\textit{Finite-dimensional verification.}
For any fixed \(f_1,\ldots,f_k\in\mathcal F\) and constants \(a_1,\ldots,a_k\),
\[
  \sum_{j=1}^k a_j\mathbb G_n f_j
  =
  \frac1{\sqrt n}\sum_{i=1}^n
  \sum_{j=1}^k a_j\{f_j(X_i)-Pf_j\}.
\]
If the displayed summand has finite variance, the ordinary central limit
theorem gives a normal limit with variance
\[
  \Var\left\{\sum_{j=1}^k a_j f_j(X)\right\}
  =
  \sum_{j=1}^k\sum_{\ell=1}^k
  a_ja_\ell\,\Cov\{f_j(X),f_\ell(X)\}.
\]
By Cramer--Wold, the finite vector
\((\mathbb G_nf_1,\ldots,\mathbb G_nf_k)\) has a Gaussian limit.  What remains
for Chapter~11 is the genuinely indexed part: proving tightness over the whole
class \(\mathcal F\). \qedmark
\end{example}

\begin{example}[Why \(D[0,T]\) appears]
Counting processes jump.  A topology designed only for continuous paths would
reject the natural sample paths of event-history data.  The Skorokhod space
\(D[0,T]\) keeps right-continuous paths with left limits and gives a topology
under which small timing errors in jumps can be harmless.  This is why
event-time asymptotics in Chapter~16 are naturally formulated in a path space
rather than one time point at a time.

\noindent\textit{Details.}
For a one-event time \(T_0\), the process
\[
  N(t)=\ind{T_0\le t},\qquad 0\le t\le T,
\]
is right-continuous and has left limits.  It is not continuous when the event
occurs inside \([0,T]\), because the path jumps from \(0\) to \(1\).  Hence
\(N\notin C[0,T]\) on that event, but \(N\in D[0,T]\).  A robot-lab log or a
clinical event history from Chapter~2 has the same structure: actions,
failures, enrollments, and alarms arrive as jumps in an information path.
The state space must allow those jumps before weak convergence of such paths
can even be stated. \qedmark
\end{example}

\subsection{Wasserstein Regression as Object-Level Inference}
\label{sec:ch10-wasserstein-regression}
\conceptindexes{Wasserstein regression, object-level inference, distributional data, conditional Wasserstein means, Wasserstein measurability}

A distribution-valued response is a natural test of this chapter's grammar.
The observed unit is not a number but a probability law: a patient's biomarker
profile, a plant's quality distribution, a store's demand distribution, a
single-cell population, or a regional climate histogram.  Wasserstein
regression treats such responses as random elements in a metric space of
probability measures, then studies regression as a map from covariates to
distributional objects \citep{petersen2019frechet,chen2023wasserstein}.

\begin{proposition}[Measurability checkpoint for Wasserstein responses]
\label{prop:ch10-wasserstein-measurability}
Let \(D\subseteq\R\) be a compact interval, and let \(\mathcal W_2(D)\) be
equipped with the \(L^2\)-Wasserstein metric \(d_W\).  Then
\(\mathcal W_2(D)\) is a Polish metric space.  The distance map
\((\mu,\nu)\mapsto d_W(\mu,\nu)\) is Borel measurable, indeed continuous.  If
\(Y_1,\ldots,Y_m\) are \(D\)-valued random variables, then
\[
  \hat\nu=\frac1m\sum_{r=1}^m\delta_{Y_r}
\]
is an \(\mathcal W_2(D)\)-valued random element.
\end{proposition}

\noindent\textit{Proof.}
For a law \(\mu\in\mathcal W_2(D)\), let \(Q_\mu\) be its quantile function.
In one dimension,
\[
  d_W^2(\mu,\nu)
  =
  \int_0^1\{Q_\mu(u)-Q_\nu(u)\}^2\,du .
\]
Thus the map \(J:\mu\mapsto Q_\mu\) embeds \(\mathcal W_2(D)\) isometrically
into \(L^2(0,1)\).  Its image is the set of \(L^2\)-classes that have a
nondecreasing representative taking values in \(D\).  This set
is closed in \(L^2(0,1)\): if \(q_k\) are nondecreasing and
\(q_k\to q\) in \(L^2\), a subsequence converges almost everywhere, and the
almost-everywhere limit has a nondecreasing version.  Since \(L^2(0,1)\) is
separable and complete, the closed image is Polish, and the isometry transfers
that Polish structure to \(\mathcal W_2(D)\).  The distance \(d_W\) is the
metric of this Polish space, so \((\mu,\nu)\mapsto d_W(\mu,\nu)\) is
continuous.

It remains to check the empirical-law map.  Define
\[
  T(y_1,\ldots,y_m)=\frac1m\sum_{r=1}^m\delta_{y_r},\qquad y_r\in D.
\]
If \(y^{(k)}=(y_1^{(k)},\ldots,y_m^{(k)})\to y=(y_1,\ldots,y_m)\), then the
coupling that pairs the \(r\)th atom of \(T(y^{(k)})\) with the \(r\)th atom of
\(T(y)\) gives
\[
  d_W^2\{T(y^{(k)}),T(y)\}
  \le
  \frac1m\sum_{r=1}^m |y_r^{(k)}-y_r|^2
  \to0 .
\]
Hence \(T:D^m\to\mathcal W_2(D)\) is continuous.  Since
\((Y_1,\ldots,Y_m)\) is a measurable \(D^m\)-valued map, the composition
\(\hat\nu=T(Y_1,\ldots,Y_m)\) is a measurable \(\mathcal W_2(D)\)-valued
random element. \qedmark

\noindent The same conclusion holds for general Polish \(D\) with finite
second moments \citep{villani2009optimal,panaretos2019statistical}; the
compact-interval proof above is the version needed for the
examples in this chapter.  For a Wasserstein regression functional, the
additional issue is not the metric but the argmin.  If a criterion
\(M(x,\nu)\) is measurable in \(x\), continuous or lower semicontinuous in
\(\nu\), and has a unique minimizer, then \(x\mapsto\argmin_\nu M(x,\nu)\) is
Borel measurable by the measurable argmin theorem; without uniqueness one
uses a measurable selector under the usual nonempty closed-valued argmin
conditions.  This is the measurable-selection layer behind Frechet regression
for random objects \citep{himmelberg1975measurable,petersen2019frechet}.

\begin{theorem}[Conditional Wasserstein means on the line; \citealp{petersen2019frechet,chen2023wasserstein}]
Let \(D\subseteq\R\) be an interval and let \(\mathcal W_2(D)\) be the space of
probability measures on \(D\) with finite second moment, equipped with the
\(L^2\)-Wasserstein metric.  Let \(Y\) be an \(\mathcal W_2(D)\)-valued random
element with quantile process \(Q_Y\), and let \(X\) be a Euclidean covariate.
If
\[
  \Expect\int_0^1 Q_Y(u)^2\,du<\infty,
\]
then, for each \(x\) for which the conditional expectation is defined, a
conditional Frechet mean
\[
  m(x)\in
  \argmin_{\nu\in\mathcal W_2(D)}
  \Expect\{d_W^2(Y,\nu)\mid X=x\}
\]
has quantile function
\[
  Q_{m(x)}(u)=\Expect\{Q_Y(u)\mid X=x\},\qquad 0<u<1,
\]
up to the usual almost-everywhere identification of quantile functions.
\end{theorem}

\noindent\textit{Proof.}
In one dimension,
\[
  d_W^2(\mu,\nu)=\int_0^1\{Q_\mu(u)-Q_\nu(u)\}^2\,du .
\]
Let \(q_x(u)=\Expect\{Q_Y(u)\mid X=x\}\).  Since each \(Q_Y\) is nondecreasing,
the conditional mean \(q_x\) has a nondecreasing version and therefore
represents a probability measure in \(\mathcal W_2(D)\).  For any candidate
quantile function \(q\),
\[
\begin{aligned}
  \Expect\left[\int_0^1\{Q_Y(u)-q(u)\}^2\,du\,\middle|\,X=x\right]
  &=
  \Expect\left[\int_0^1\{Q_Y(u)-q_x(u)\}^2\,du\,\middle|\,X=x\right]  \\
  &\quad + \int_0^1\{q_x(u)-q(u)\}^2\,du ,
\end{aligned}
\]
so the minimum is attained at \(q=q_x\).  This is the \(L^2\) projection
identity written in quantile coordinates. \qedmark

The one-dimensional quantile identity is standard in optimal transport
\citep{villani2009optimal,panaretos2019statistical}.  Its statistical use as a
conditional Frechet mean belongs to the random-object regression framework of
\citet{petersen2019frechet}; tangent-space Wasserstein regression is developed
by \citet{chen2023wasserstein}.

\begin{example}[A worked bridge: recovery time, Wasserstein regression, and tangent space]
\label{ex:ch10-worked-clinical-wasserstein}
A Chapter~2 clinical endpoint such as recovery by day \(t_0\) can be read in
two ways.  The scalar reading reports one number, for example a day-\(t_0\)
recovery probability.  The object-level reading keeps the whole response
distribution.  This is the difference between asking whether a treatment moves
one reported endpoint and asking how it moves the distribution of patient
trajectories.  Trials such as the CARMELINA cardiovascular outcomes trial in
\emph{JAMA} and the remdesivir trial in \emph{The Lancet} motivate exactly
this kind of time-to-event representation
\citep{rosenstock2019linagliptin,wang2020remdesivir}.

To keep the bridge visible, first suppose the event or recovery time \(T\) is
observed on \(D=[0,\tau]\).  The censored version replaces \(T\) by the
event-history paths \((N,R)\) used later in this chapter and in Chapter~16.
For treatment arm \(a\), let
\[
  \nu_a=\Law(T\mid A=a),
  \qquad
  \hat\nu_{n,a}
  =
  \frac1{n_a}\sum_{i:A_i=a}\delta_{T_i},
  \qquad
  n_a=\sum_{i=1}^n\ind{A_i=a}.
\]
Then \(\hat\nu_{n,a}\) is a random element of \(\mathcal W_2(D)\).  The usual
day-\(t_0\) endpoint is the functional
\[
  \hat p_{n,a}(t_0)=\hat\nu_{n,a}([0,t_0]),
  \qquad
  p_a(t_0)=\nu_a([0,t_0]).
\]
The distributional endpoint keeps \(\nu_a\) itself.  If units are centers,
subgroups, donors, or repeated platform cohorts, each unit can carry a
distribution-valued response \(Y\in\mathcal W_2(D)\), and Wasserstein
regression studies the conditional distributional mean \(m(x)\) rather than
only \(x\mapsto \Expect(T\mid X=x)\).

\noindent\textit{How the chapter is used.}
The workflow is
\[
\begin{gathered}
  \text{data}\longmapsto
  \hat\nu\in\mathcal W_2(D),\\
  \hat\nu\longmapsto
  \text{weak/Wasserstein limit}
  \longmapsto
  \text{reported functional}.
\end{gathered}
\]
Portmanteau checks boundary-sensitive functionals such as
\(\nu([0,t_0])\).  Continuous mapping checks stable summaries such as means
or smooth transforms.  Wasserstein regression keeps the whole distribution in
the analysis.  Chapter~15 then asks the complementary question: how does the
reported functional change under a small perturbation of the observation law?

\noindent\textit{Details and proof.}
Assume \(\hat\nu_{n,a}\to\nu_a\) in \(d_W\) and that
\(\nu_a(\{t_0\})=0\).  Since \(D\) is compact, \(d_W\)-convergence implies
weak convergence.  The map \(t\mapsto t\) is bounded and continuous on \(D\),
so
\[
  \int_D t\,d\hat\nu_{n,a}(t)
  \to
  \int_D t\,d\nu_a(t).
\]
For the day-\(t_0\) endpoint, the set \([0,t_0]\) has boundary
\(\{t_0\}\).  The no-atom condition \(\nu_a(\{t_0\})=0\) makes
\([0,t_0]\) a \(\nu_a\)-continuity set, so Portmanteau gives
\[
  \hat p_{n,a}(t_0)
  =
  \hat\nu_{n,a}([0,t_0])
  \to
  \nu_a([0,t_0])
  =
  p_a(t_0).
\]
This is the precise mathematical reason that an endpoint reported at a fixed
day is well behaved only away from a point mass at that day.

Now let \(Y\) be a \(\mathcal W_2(D)\)-valued response.  The theorem above
shows that the conditional Wasserstein mean satisfies
\[
  Q_{m(x)}(u)=\Expect\{Q_Y(u)\mid X=x\}.
\]
If \(\mu_\ast\) is an atomless reference law on \(D\), the Wasserstein tangent
coordinate obeys
\[
\begin{aligned}
  \operatorname{Log}_{\mu_\ast}(m(x))(s)
  &=
  Q_{m(x)}(F_\ast(s))-s                                      \\
  &=
  \Expect\{Q_Y(F_\ast(s))-s\mid X=x\}                         \\
  &=
  \Expect\{\operatorname{Log}_{\mu_\ast}(Y)(s)\mid X=x\}.
\end{aligned}
\]
Thus a nonlinear regression over probability laws becomes an ordinary
conditional-mean problem in the Hilbert space \(L^2(\mu_\ast)\), as long as
one remembers that the coordinate represents a distributional object.

Finally, Chapter~15's Bickel tangent-space idea enters through the observation
law.  For the complete-data endpoint
\[
  p_a(t_0)=\Prob(T\le t_0\mid A=a),
  \qquad
  \pi_a=\Prob(A=a)>0,
\]
consider a regular parametric submodel \(P_\varepsilon\) with score
\(s\in L_0^2(P)\).  Writing
\(\eta_a=\Expect\{\ind{A=a}\ind{T\le t_0}\}\), one has
\(p_a=\eta_a/\pi_a\).  Differentiating the ratio gives
\[
  \left.\frac{d}{d\varepsilon}p_{a,\varepsilon}(t_0)
  \right|_{\varepsilon=0}
  =
  \Expect\left[
    \frac{\ind{A=a}}{\pi_a}
    \{\ind{T\le t_0}-p_a(t_0)\}s(O)
  \right],
\]
where \(O=(T,A,X)\).  Hence the influence function is
\[
  \phi_{a,t_0}(O)
  =
  \frac{\ind{A=a}}{\pi_a}\{\ind{T\le t_0}-p_a(t_0)\}.
\]
The two tangent spaces therefore do different jobs in the same example:
\(\operatorname{Log}_{\mu_\ast}\) linearizes the response geometry inside
\(\mathcal W_2(D)\), while Bickel's \(L_0^2(P)\) tangent space linearizes
perturbations of the law that generated the observations. \qedmark
\end{example}

\begin{example}[Wasserstein regression in practice]
Suppose \(Y_i\) is a histogram, density estimate, or empirical distribution
attached to unit \(i\), and \(X_i\) records covariates.  In manufacturing,
\(Y_i\) might be the distribution of particle sizes or defect depths for one
batch.  In biomedicine, it might be a distribution of cell-level expression,
tumor-image intensities, or patient-level biomarker trajectories.  In climate,
transportation, finance, or retail operations, it might be a distribution of
rainfall, travel times, losses, or lead times.  The estimand is not a scalar
mean response but a conditional law \(x\mapsto m(x)\).

The weak-convergence viewpoint says what must be checked.  First, the
distribution-valued observations must be random elements of
\(\mathcal W_2(D)\).  Second, the empirical criterion or tangent-coordinate
process must converge in a topology strong enough to control the fitted
conditional law.  Third, the map that sends the limiting coordinate back to a
probability distribution must be continuous at the relevant limit.  This is
exactly the same chain used elsewhere in the chapter:
\[
\begin{aligned}
  \text{distribution-valued data}
  &\longmapsto
  \text{random object in } \mathcal W_2(D) \\
  &\longmapsto
  \text{weak or } L^2 \text{ limit}
  \longmapsto
  \text{regression functional}.
\end{aligned}
\]
In the tangent-coordinate formulation of \citet{chen2023wasserstein}, an
atomless reference law \(\mu_\ast\) gives
\[
  \operatorname{Log}_{\mu_\ast}(\mu)
  =
  Q_\mu\circ F_\ast-\operatorname{id}
  \in L^2(\mu_\ast),
\]
so a nonlinear regression problem over distributions can be studied through a
Hilbert-space coordinate and then mapped back to a probability measure.
Chapter~15 returns to this same coordinate from Bickel's semiparametric
tangent-space viewpoint: the Wasserstein coordinate linearizes the response
geometry, while the influence function linearizes perturbations of the
observation law.

\noindent\textit{Details for an empirical distribution response.}
Suppose unit \(i\) contributes observations \(Y_{i1},\ldots,Y_{im_i}\) in a
bounded interval \(D\), and let
\[
  \hat\nu_i=\frac1{m_i}\sum_{r=1}^{m_i}\delta_{Y_{ir}}.
\]
Then \(\hat\nu_i\in\mathcal W_2(D)\) automatically, because bounded support
implies a finite second moment.  Its quantile function is the empirical
quantile step function.  In one dimension,
\[
  d_W^2(\hat\nu_i,\nu)
  =
  \int_0^1\{Q_{\hat\nu_i}(u)-Q_\nu(u)\}^2\,du,
\]
so regression on distribution-valued outcomes can be written as regression on
quantile functions with the monotonicity constraint preserved.  This is why
the single-cell and climate-histogram examples from Chapters~1--2 fit the
same grammar: each observed unit can carry a whole empirical law, and the
target is a covariate-indexed population law rather than a scalar mean.
The companion Ge et al. capsule uses this grammar directly: each century is
represented by the empirical law of its ten decadal anomaly values, and the
fitted coefficient is a quantile-indexed slope rather than one scalar trend.
\qedmark
\end{example}

\subsection{Scientific and Industrial Random Objects}
\label{sec:ch10-scientific-industrial-random-objects}
\conceptindexes{scientific random objects, industrial random objects, curves, images, graphs, distributions}

The preceding examples explain the grammar.  This section makes the book's
application promise explicit.  A modern scientific or industrial dataset often
arrives as a path, a random measure, an empirical law, or a closed-loop history
before anyone reduces it to a scalar endpoint.  Weak convergence is the
language that lets the reduction be checked instead of guessed: name the state
space, name the topology, and identify which reported functional is continuous
away from its boundary cases.

\begin{realdatacapsule}{Clinical trial as an event-history path}
\item[Data object.] Patient-level treatment assignment, baseline covariates,
follow-up time, event indicator, and censoring path from a trial such as
CARMELINA or remdesivir \citep{rosenstock2019linagliptin,wang2020remdesivir}.
\item[Observation mechanism.] Protocol visits, endpoint adjudication,
censoring, loss to follow-up, and intercurrent events determine the observed
path.
\item[Target.] Survival probability, cumulative incidence, restricted mean,
hazard contrast, or treatment-strategy estimand over a stated time window.
\item[Model.] Cadlag counting-process paths in \(D[0,\tau]\), Kaplan--Meier or
Cox functionals, and later martingale/intensity models represent the endpoint.
\item[Uncertainty.] Weak convergence of path-valued empirical laws, Greenwood
or sandwich variance, log-rank fluctuation, and bootstrap paths quantify error.
\item[Limitation.] The endpoint path is only interpretable after time zero,
risk sets, censoring, and intercurrent-event rules are fixed.
\end{realdatacapsule}

\begin{example}[Clinical trials as event-history random objects]
The CARMELINA cardiovascular outcomes trial in \emph{JAMA} and the remdesivir
trial in \emph{The Lancet} both report clinically meaningful time-to-event
quantities, although the medical endpoints differ
\citep{rosenstock2019linagliptin,wang2020remdesivir}.  For patient \(i\), let
\[
  \tilde T_i=T_i\wedge C_i,
  \qquad
  \Delta_i=\ind{T_i\le C_i},
\]
where \(T_i\) is the event time of interest and \(C_i\) is the censoring time.
The observed endpoint can be written as the pair of cadlag paths
\[
  N_i(t)=\ind{\tilde T_i\le t,\Delta_i=1},
  \qquad
  R_i(t)=\ind{\tilde T_i>t},
  \qquad 0\le t\le\tau .
\]
Thus the patient-level random object is not only
\((\tilde T_i,\Delta_i)\).  It is naturally an element of
\[
  D[0,\tau]\times D[0,\tau],
\]
possibly enlarged by baseline covariates and treatment assignment.  A
Kaplan--Meier curve, a cumulative incidence curve, a log-rank statistic, or a
Cox score is then a functional of the empirical law of these paths.

\noindent\textit{Details.}
The event count \(N_i\) is right-continuous and has left limits.  It jumps
from \(0\) to \(1\) at the observed event time and is constant otherwise, so
\(N_i\in D[0,\tau]\).  The right-continuous risk-indicator path \(R_i\) is
also cadlag; the predictable at-risk process used in counting-process
intensities is the corresponding left-continuous convention.  The
evaluation map \(x\mapsto x(t_0)\) on \(D[0,\tau]\) is continuous at every
path \(x\) that is continuous at \(t_0\).  To see this, suppose
\(x_m\to x\) in the Skorokhod \(J_1\) topology.  Then there are increasing
homeomorphisms \(\lambda_m\) with
\[
  \sup_t|\lambda_m(t)-t|\to0,
  \qquad
  \sup_t|x_m(\lambda_m(t))-x(t)|\to0 .
\]
Put \(s_m=\lambda_m^{-1}(t_0)\).  Then \(s_m\to t_0\), and
\[
  |x_m(t_0)-x(t_0)|
  \le
  |x_m(\lambda_m(s_m))-x(s_m)|
  +
  |x(s_m)-x(t_0)|
  \to0
  \]
whenever \(x\) is continuous at \(t_0\).  For a one-jump event path, the only
discontinuity is the event time.  Therefore fixed-time event probabilities,
risk-set probabilities, and rejection events defined by continuous path
functionals transfer through weak limits as long as the limiting law assigns no
mass to the relevant boundary time or boundary value. \qedmark
\end{example}

\begin{example}[Single-cell assays as empirical laws over cells]
Single-cell and spatial-omics platforms do not observe one scalar per
biological unit.  A donor, tissue region, perturbation, or time point may
produce a population of cells, each with expression, protein, chromatin,
spatial, or pseudotime coordinates.  Methods such as scImpute and scDesign3
make this population-level view explicit: the scientific object is often a
cell-state distribution, not a single average
\citep{li2018scimpute}.  For a scalar marker, pseudotime
coordinate, image-derived score, or other one-dimensional summary
\(G_{ir}\in D\subset\R\), write the empirical law for unit \(i\) as
\[
  \hat\nu_i
  =
  \frac1{m_i}\sum_{r=1}^{m_i}\delta_{G_{ir}} .
\]
If \(D\) is bounded, then \(\hat\nu_i\in\mathcal W_2(D)\).  A treatment,
disease, batch, or donor covariate can therefore be related to the whole
distribution \(Y_i=\hat\nu_i\), not merely to \(m_i^{-1}\sum_rG_{ir}\).

\noindent\textit{Details.}
Boundedness of \(D\) gives
\[
  \int_D z^2\,d\hat\nu_i(z)
  =
  \frac1{m_i}\sum_{r=1}^{m_i}G_{ir}^2
  <\infty,
\]
so the empirical cell distribution is an element of \(\mathcal W_2(D)\).  The
ordinary mean is a continuous functional under the \(W_2\) topology.  Indeed,
for any coupling \((U,V)\) of two laws \(\mu,\nu\in\mathcal W_2(D)\),
\[
  \left|\int z\,d\mu(z)-\int z\,d\nu(z)\right|
  =
  |\Expect(U-V)|
  \le
  \Expect|U-V|
  \le
  \{\Expect(U-V)^2\}^{1/2}.
\]
Taking the infimum over all couplings gives
\[
  \left|\int z\,d\mu(z)-\int z\,d\nu(z)\right|
  \le d_W(\mu,\nu).
\]
Thus a weak or \(W_2\) limit for cell-population laws automatically transfers
to the reported population mean.  In contrast, a threshold functional
\(\nu\mapsto\nu((c,\infty))\) is continuous only at laws with no atom at
\(c\), by the Portmanteau continuity-set condition.  This is the Chapter~2
lesson in modern biological form: the whole empirical distribution shows
which summaries are stable and which summaries sit on a boundary. \qedmark
\end{example}

\begin{example}[Autonomous laboratories as closed-loop histories]
Self-driving laboratories and Bayesian reaction-optimization systems make
adaptation part of the data object.  In Burger et al.'s mobile robotic chemist
\citep{burger2020mobile} and Shields et al.'s Bayesian reaction-optimization
workflow \citep{shields2021bayesian}, actions, sensor outputs, failed runs,
quality flags, and decisions are accumulated over time.  For a finite horizon
\(T\), write
\[
  H_T
  =
  \{(A_t,Y_t(\cdot),L_t,Q_t):1\le t\le T\},
\]
where \(A_t\) is the chosen condition, \(Y_t(\cdot)\in D[0,\tau]\) is a sensor
or reaction trace, \(L_t\) is the laboratory log, and \(Q_t\) records quality
or intervention flags.  The next action is chosen by a policy
\[
  A_{t+1}=\pi_t(H_t),
\]
possibly with randomized exploration.  The natural state space is a finite
product of action spaces, path spaces, log spaces, and quality-flag spaces,
with the product topology.

\noindent\textit{Details.}
Let \(Z_t=h\{Y_t(\cdot),L_t,Q_t\}\) be a continuous scalar performance summary,
such as yield or purity after a prespecified preprocessing rule.  The terminal
best value
\[
  M_T(H_T)=\max_{1\le t\le T} Z_t
\]
is continuous whenever \(h\) is continuous in the chosen product topology,
because coordinatewise convergence gives \(Z_{tm}\to Z_t\) for each fixed
\(t\), and the maximum over a finite set is a continuous map:
\[
  \left|\max_t Z_{tm}-\max_t Z_t\right|
  \le
  \max_t |Z_{tm}-Z_t|.
\]
A stopping or success functional
\[
  \tau_b(H_T)=\inf\{t\le T:Z_t\ge b\}
\]
is more delicate.  It is continuous at histories for which no \(Z_t\) equals
the threshold \(b\).  If every \(Z_t\ne b\), then
\(\eta=\min_t|Z_t-b|>0\), and all sufficiently close histories have the same
indicators \(\ind{Z_t\ge b}\).  Hence the first crossing time is eventually
unchanged.  If some \(Z_t=b\), arbitrarily small perturbations can create or
remove a crossing.  This is exactly why adaptive industrial systems need the
same boundary discipline as clinical event histories and point patterns.
\qedmark
\end{example}

\section{Measurability and Outer Probability}
\label{sec:ch10-measurability-outer}
\conceptindexes{measurability, outer probability, outer expectation, asymptotic measurability, measurable maps}

One technical warning belongs in the bridge.  In separable metric spaces,
random elements and their suprema usually behave measurably.  In empirical
process theory, however, the index class \(\mathcal F\) can be uncountable and
the map
\[
  \omega\mapsto \sup_{f\in\mathcal F}|\mathbb G_n f(\omega)|
\]
need not be measurable without extra assumptions.  Chapter~11 therefore uses
outer probability and outer expectation in some statements.  This is not a new
form of probability; it is a bookkeeping device that lets asymptotic arguments
proceed before all separability details have been settled.

\noindent\textbf{Notation.}
The notation follows the convention introduced in Chapter~6's discussion of
supremum measurability repairs.  For any \(A\subseteq\Omega\),
\[
  \outerProb(A)=\Prob^{*}(A)
  =
  \inf\{\Prob(B):A\subseteq B,\ B\in\mathcal F\},
\]
and \(\outerExpect Z=\Expect^{*}Z\) denotes the corresponding outer
expectation, the infimum of \(\Expect Y\) over measurable \(Y\ge Z\).  We write
\[
  Z_n\toPstar 0
  \qquad\text{or equivalently}\qquad
  Z_n\to0\ \text{in }\outerProb
\]
when \(\outerProb(|Z_n|>\varepsilon)\to0\) for every \(\varepsilon>0\).  The
notation \(R_n=\opstar(a_n)\) means \(R_n/a_n\toPstar 0\).  If the
quantities are measurable, the star disappears: \(\outerProb=\Prob\) on the
events involved, and outer convergence is ordinary convergence in probability.

For the main flow of the book, the moral is simple.  If the random object lives
in a well-behaved Polish space, ordinary weak convergence is usually enough.
If the object lives in a large function space indexed by an arbitrary class,
one must either impose separability or use outer-probability language.  The
statistical idea remains the same: finite-dimensional limits identify the
candidate Gaussian field, and tightness or stochastic equicontinuity makes the
field a genuine object-level limit.  Chapter~11 uses this notation in earnest
for \(\ell^\infty(\mathcal F)\)-valued empirical processes; Chapter~15 returns
to the same convention for functional delta-method remainders.

\section{Exercises}
\label{sec:ch10-exercises}
\conceptindexes{weak-convergence exercises, tightness exercises, random-object exercises}

\begin{exercise}[Name the space]
For each statistic below, name a natural state space and a topology: an
empirical distribution function \(F_n\), a partial-sum path \(W_n\), a counting
process \(N(t)\), an empirical process \(\mathbb G_n\), and a bootstrap
distribution estimate.  State one functional of the object that would be
continuous in the chosen topology and one that might fail to be continuous.
\end{exercise}

\begin{exercise}[Portmanteau boundaries]
Let \(Z_n\weakto Z\) in \(\R\).  Explain why
\(\Prob\{Z_n\le c\}\to\Prob\{Z\le c\}\) need not hold for every \(c\).  Give a
condition on \(c\) under which it does hold, and interpret that condition as a
boundary condition.
\end{exercise}

\begin{exercise}[Tightness intuition]
Construct, or describe, a sequence of random continuous functions on
\([0,1]\) whose finite-dimensional distributions converge to zero at every
fixed time but whose paths have increasingly narrow spikes.  Explain why
finite-dimensional convergence alone misses the problem.
\end{exercise}

\begin{exercise}[Continuous mapping]
Suppose \(X_n\weakto X\) in \(\ell^\infty[0,1]\).  Show that
\[
  \sup_{0\le t\le1}|X_n(t)|
  \weakto
  \sup_{0\le t\le1}|X(t)|.
\]
Which property of the supremum map is being used?
\end{exercise}

\begin{exercise}[A bridge to estimation]
Let \(M_n\) be random continuous functions on a compact parameter space
\(\Theta\), and suppose \(\sup_{\theta\in\Theta}|M_n(\theta)-M(\theta)|\toP0\).
If \(M\) has a unique maximizer \(\theta_0\), explain why any approximate
maximizer of \(M_n\) should be consistent for \(\theta_0\).  Identify this as
a continuous-mapping or argmax principle.
\end{exercise}

\begin{exercise}[A Wasserstein regression object]
Let \(Y_i\in\mathcal W_2(\R)\) have quantile function \(Q_i\), and suppose a
scalar covariate \(X_i\) is observed.  Propose a state space for the random
pair \((X_i,Y_i)\), write the conditional Frechet mean in quantile
coordinates, and name one step in which weak convergence or a continuous
mapping argument would enter an asymptotic analysis.
\end{exercise}

\begin{exercise}[Embedding-valued summaries]
Suppose each document, image, or molecule is mapped by a fixed representation
model to an embedding \(Z_i\in\R^d\), and the analyst studies the empirical
mean and covariance of the embeddings.  Name the random object before and
after the embedding map.  Which conclusions depend only on continuity of the
map, and which conclusions depend on the embedding model being scientifically
meaningful?
\end{exercise}

\begin{exercise}[Graph-valued convergence]
For each \(n\), let \(G_n\) be a random network observed from a platform,
cell-cell interaction system, or supply chain.  Propose one topology or metric
for comparing \(G_n\) to a limiting object.  Then name one statistic that
should be continuous in that topology and one statistic that may be unstable
because of boundary or threshold effects.
\end{exercise}

\begin{exercise}[Distributional treatment effect]
In a wearable-device study, the response for subject \(i\) is a distribution
of daily glucose values rather than a scalar.  Define a treatment effect as a
difference between two distribution-valued means.  State the state space and
one continuous mapping step that would turn convergence of empirical
quantile functions into convergence of the treatment-effect estimate.
\end{exercise}

\begin{exercise}[Bootstrap after representation learning]
An analyst first trains a representation on a large corpus and then uses the
learned representation to estimate a downstream target on a smaller study.
Describe two asymptotic regimes: one in which the representation is treated as
fixed and one in which its training uncertainty is part of the random object.
What would the bootstrap need to resample in each regime?
\end{exercise}

\section*{Sources and Further Reading}
\addcontentsline{toc}{section}{Sources and Further Reading}

The bounded-continuous definition, Portmanteau theorem, tightness, Prokhorov
theory, representation theorems, and process convergence criteria are standard
topics in weak convergence; see \citet{billingsley1999convergence} and
\citet{pollard1984convergence}.  The local route through weak convergence,
Portmanteau, tightness, Cramer--Wold, Slutsky, representation, and empirical
process examples is also informed by
\citet{dabrowskaAdvancedProbabilityCommunication} and
\citet{dabrowskaStochasticProcessesCommunication}.  Skorokhod's path-space topology and
representation theorem originate in \citet{skorokhod1956limit}; Dudley's
metric-space representation and coupling viewpoint is in
\citet{dudley1968distances}.  Donsker-type empirical-process limits are linked
historically to \citet{doob1949heuristic}, \citet{donsker1952justification},
and \citet{pollard1982beyond}.  The empirical-process version of this
chapter's bridge language is developed in \citet{pollard1990empirical} and
\citet{vaart2023weak}.  For Wasserstein geometry and statistics, see
\citet{villani2009optimal}, \citet{panaretos2019statistical},
\citet{petersen2019frechet}, and \citet{chen2023wasserstein}.  The clinical
event-history examples are anchored by the CARMELINA cardiovascular outcomes
trial in \emph{JAMA} \citep{rosenstock2019linagliptin} and the remdesivir
trial in \emph{The Lancet} \citep{wang2020remdesivir}.  The single-cell
random-law example points to scImpute \citep{li2018scimpute} and scDesign3
as recurring examples.  The autonomous-laboratory example follows Burger
et al.'s mobile robotic chemist \citep{burger2020mobile} and Shields et al.'s
Bayesian reaction-optimization workflow \citep{shields2021bayesian}.  The present
chapter proves the transfer tools it uses directly, while treating
Prokhorov's theorem and full Donsker-type tightness proofs as cited background
results.  Its purpose is connective: it supplies the language needed to pass
from random objects in Chapter~6 to laws, empirical processes, estimators,
delta-method functionals, and continuous-time event histories in the chapters
that follow.

%% file: chapters/ch11_uniform_laws_empirical_processes.tex
\chapter{Uniform Laws and Empirical Processes: Controlling Indexed Noise}
\label{chap:uniform-laws-empirical-processes}
\conceptindexes{empirical processes, uniform laws, indexed fields, Glivenko--Cantelli classes, Donsker classes, entropy, stochastic equicontinuity}

\begin{tcolorbox}[
  enhanced,
  breakable,
  colback=chaptercream,
  colframe=bookblue!88!black,
  boxrule=0.72pt,
  arc=5pt,
  boxsep=1pt,
  left=1.0em,
  right=0.95em,
  top=0.82em,
  bottom=0.82em,
  before skip=0.55\baselineskip,
  after skip=1.0\baselineskip
]
\noindent\textbf{Chapter overview.}
In Part~IV, this chapter controls searched empirical fields, the technical
step that keeps flexible procedures from mistaking random roughness for signal.
A law of large numbers controls one translation at a time.  Statistical
procedures rarely stop at one translation: a likelihood scans a parameter
space, a quantile scans a distribution function, a classifier scans a class of
decision rules, and a residual diagnostic scans many possible cutpoints.  This
chapter studies the random field
\[
  \{(P_n-P)f:f\in\mathcal F\}.
\]
The empirical distribution function is the first example; Glivenko--Cantelli
and Donsker theory are the two central forms of stability; entropy, brackets,
outer probability, Skorohod path spaces, and stochastic equicontinuity are the
tools that keep the field mathematically honest.  The goal is not to reproduce
a full empirical process monograph, but to give the structural language needed
for M- and Z-estimation, semiparametric inference, and event-time applications.
\end{tcolorbox}

Chapter~7 asked what kind of empirical object a modern data structure creates
in the first place, Chapter~8 asked which target the object is meant to
support, and Chapter~9 made one empirical translation stable.  That is already
powerful: a mean estimates a mean, a proportion estimates a probability, a
fixed loss estimates a risk.  But a statistical procedure
usually chooses after looking.  Least squares searches a landscape of empirical
losses.  Maximum likelihood searches a landscape of empirical log-likelihoods.
A quantile searches for a crossing point of the empirical distribution
function.  A classifier searches through a class of possible decision rules.
The danger is not that any one ruler is broken; it is that a sufficiently
flexible collection of rulers can be moved around until random roughness looks
like structure.

This is where the opening examples become a technical requirement.  The
mathematical object is no longer a single average but an indexed field, and
the danger is search.  Empirical-process theory asks whether a pattern found
after scanning many summaries is still larger than the random texture expected
from the sampling law, and whether the whole observed field is stable enough to
support the population feature it is meant to represent.

The difficulty is subtle.  For each fixed $f$, the difference
$P_nf-Pf$ may be small.  But if many $f$'s are inspected, random errors may align
somewhere in the class.  The question becomes
\[
  \sup_{f\in\mathcal F}|P_nf-Pf|.
\]
This is the mathematical form of scanning a whole indexed empirical field.  In
Chapter 2, the unseen left traces in the seen.  Here the
population law is read through every function in $\mathcal F$, and the
statistician asks whether the whole indexed field is stable.

In the book's recurring ledger, this chapter lives between target and
procedure.  Chapters~7 and~8 insist that the object and the target be named
before inference begins.  This chapter adds the next question: after the
target has been named, did the procedure search so widely that the empirical
record can no longer stand in for the population law?  That is why the same
mathematics appears in cdf bands, survival surfaces, single-cell pipelines,
market screens, and learning algorithms.  The surface changes; the ledger
entry is the same.

\section{From One Statistic to an Indexed Field}
\label{sec:ch11-indexed-field}
\conceptindexes{indexed fields, empirical measure, empirical process, function classes, uniform convergence}

Let $X_1,\ldots,X_n$ be iid with law $P$ on a measurable space
$(\mathcal X,\mathcal A)$.  The empirical measure is
\[
  P_n=\frac1n\sum_{i=1}^n\delta_{X_i}.
\]
For a class $\mathcal F$ of measurable real functions on $\mathcal X$, define
\[
  \|P_n-P\|_{\mathcal F}
  =
  \sup_{f\in\mathcal F}|P_nf-Pf|.
\]
If this quantity converges to zero in probability or almost surely, then
$\mathcal F$ is a class over which the empirical law is uniformly stable.

\begin{definition}[Glivenko--Cantelli class]
A class $\mathcal F$ is called $P$-Glivenko--Cantelli if
\[
  \|P_n-P\|_{\mathcal F}\to0
\]
in outer probability, or almost surely when the supremum is measurable and the
strong version is intended.
\end{definition}

\begin{definition}[Empirical process]
The empirical process indexed by $\mathcal F$ is
\[
  \mathbb G_n f=\sqrt n(P_n-P)f,\qquad f\in\mathcal F.
\]
When $\mathbb G_n$ converges weakly in $\ell^\infty(\mathcal F)$ to a tight
mean-zero Gaussian process $\mathbb G$, the class $\mathcal F$ is called
$P$-Donsker.
\end{definition}

\begin{definition}[Random criterion field]
Let \(\Theta\) be a parameter or model class, and let \(m_\theta\) be the
criterion contribution of one observation: a log-likelihood, loss,
pseudo-likelihood, estimating contrast, or diagnostic score.  The empirical
criterion
\[
  M_n(\theta)=P_n m_\theta
\]
is random for each fixed \(\theta\).  The collection
\[
  \{M_n(\theta)-M(\theta):\theta\in\Theta\}
  =
  \{(P_n-P)m_\theta:\theta\in\Theta\},
  \qquad M(\theta)=P m_\theta,
\]
is the \emph{random criterion field}.  It is simply the empirical-process
field indexed by the functions \(m_\theta\).  The word ``criterion'' reminds
us that an estimator will search this field, usually by maximizing or
minimizing \(M_n(\theta)\).
\end{definition}

The two definitions separate two levels of stability.  Glivenko--Cantelli says
the whole empirical field converges to its population field.  Donsker says the
remaining fluctuation, after multiplying by $\sqrt n$, has a process-level
Gaussian limit.  Consistency of estimators usually needs the first.  Standard
errors, confidence bands, and local asymptotic approximations often need the
second.

\begin{tcolorbox}[
  enhanced,
  breakable,
  colback=noteback,
  colframe=bookgold!75!black,
  boxrule=0.55pt,
  arc=4pt,
  boxsep=1pt,
  left=0.95em,
  right=0.9em,
  top=0.7em,
  bottom=0.7em,
  before skip=0.9\baselineskip,
  after skip=1.0\baselineskip
]
\noindent\textbf{Four jobs of the chapter.}
Uniform laws keep empirical criteria from choosing false locations.  Empirical
process limits describe the residual fluctuation of the whole field.
Measurability and path-space topology say when the field is a genuine random
object.  Stochastic equicontinuity lets random plug-in indices be replaced by
their population limits.  These four jobs are distinct, but in applications
they usually appear together.
\end{tcolorbox}

\begin{tcolorbox}[
  enhanced,
  breakable,
  colback=chaptercream,
  colframe=bookblue!75!black,
  boxrule=0.55pt,
  arc=4pt,
  boxsep=1pt,
  left=0.95em,
  right=0.9em,
  top=0.7em,
  bottom=0.7em,
  before skip=0.9\baselineskip,
  after skip=1.0\baselineskip
]
\noindent\textbf{Historical map: two doors, one theory.}
There are two classical ways into empirical processes.  The first is the
probabilist's entrance: the empirical distribution function is treated as a
random element of a function space, and Donsker's theorem becomes weak
convergence of probability measures on spaces such as \(C[0,1]\) or
\(D[0,1]\).  This line runs from the Kolmogorov--Smirnov problem through
Doob's heuristic argument, Donsker's correction, Skorokhod's topology, and
Billingsley's systematic weak-convergence framework
\citep{doob1949heuristic,donsker1952justification,skorokhod1956limit,billingsley1999convergence}.

The second is the statistician's entrance: replace half-lines by a general
class \(\mathcal F\), view
\(\mathbb G_n=\{\sqrt n(P_n-P)f:f\in\mathcal F\}\) as an indexed random field,
and ask which classes are Glivenko--Cantelli or Donsker.  This line brings in
VC combinatorics, entropy, symmetrization, chaining, bracketing, and outer
probability.  It is associated with Vapnik--Chervonenkis, Dudley's Gaussian
process and empirical-measure work, Pollard's statistical synthesis, and the
van der Vaart--Wellner treatment
\citep{vapnik1971uniform,dudley1967sizes,dudley1978central,pollard1984convergence,pollard1990empirical,vaart2023weak}.

These are not competing theories.  Billingsley supplies the clean language for
weak convergence of random elements; the Dudley--Pollard--van der
Vaart--Wellner line answers the extra question that statistics forces on us:
how large, irregular, or nonmeasurable may the indexing class be before the
empirical field stops behaving like a stable object?
\end{tcolorbox}

\subsection{The Empirical Distribution Function}
\label{sec:ch11-edf}
\conceptindexes{empirical distribution function, empirical process, Brownian bridge}

The empirical distribution function is the canonical example.  Let
\[
  \mathcal F_{\mathrm{cdf}}
  =
  \{\ind{x\le t}:t\in\R\}.
\]
Then
\[
  F_n(t)-F(t)=(P_n-P)\ind{x\le t}.
\]
The Glivenko--Cantelli theorem says
\[
  \sup_{t\in\R}|F_n(t)-F(t)|\to0
  \quad\text{almost surely}.
\]
\citep{glivenko1933sulla,cantelli1933sulla}
The Dvoretzky--Kiefer--Wolfowitz inequality gives finite-sample control:
\[
  \Prob\left(
    \sqrt n\sup_t|F_n(t)-F(t)|>u
  \right)
  \le 2e^{-2u^2}.
\]
\citep{dvoretzky1956asymptotic,massart1990tight}

\begin{tcolorbox}[
  enhanced,
  breakable,
  colback=noteback,
  colframe=bookgold!75!black,
  boxrule=0.55pt,
  arc=4pt,
  boxsep=1pt,
  left=0.95em,
  right=0.9em,
  top=0.7em,
  bottom=0.7em,
  before skip=0.9\baselineskip,
  after skip=1.0\baselineskip
]
\noindent\textbf{Why the classical real-line theorem is ordinary probability.}
The outer-probability convention in the definition of a general
Glivenko--Cantelli class is not needed for the classical empirical cdf on
\(\R\).  Although \(\R\) is uncountable, both \(F_n\) and \(F\) are
right-continuous, so path by path
\[
  \sup_{t\in\R}|F_n(t)-F(t)|
  =
  \sup_{q\in\Rat}|F_n(q)-F(q)|.
\]
For each fixed rational \(q\), \(F_n(q)\) is measurable; the right side is a
countable supremum of measurable random variables.  Hence the
Kolmogorov--Smirnov statistic and the classical Glivenko--Cantelli theorem are
ordinary measurable statements.  The star notation becomes useful only after
the half-line class \(\{\ind{x\le t}:t\in\R\}\) is replaced by a general
possibly nonseparable class \(\mathcal F\); see the empirical-process
treatments of \citet{pollard1984convergence}, \citet{pollard1990empirical},
and \citet{vaart2023weak}.
\end{tcolorbox}

The central limit version is Donsker's theorem.  If $F$ is continuous, then
\[
  \{\sqrt n(F_n(t)-F(t)):t\in\R\}
  \weakto
  \{B(F(t)):t\in\R\},
\]
where $B$ is a Brownian bridge.  The covariance is
\[
  \Cov(B(F(s)),B(F(t)))
  =
  F(s\wedge t)-F(s)F(t).
\]

\begin{example}[A confidence band as a process statement]
The statistic
\[
  \sqrt n\sup_t|F_n(t)-F(t)|
\]
does not ask whether one cutpoint is well estimated.  It asks whether the whole
distribution curve is close.  A Kolmogorov--Smirnov band is therefore a
process-level confidence statement, not merely a collection of pointwise
intervals.
\end{example}

\begin{example}[Quantiles]
The quantile map sends a distribution function $F$ to
$Q(p)=\inf\{x:F(x)\ge p\}$.  Uniform control of $F_n-F$ gives consistency of
$Q_n(p)$, and process-level control gives asymptotic approximations for
quantile processes.  The same object that was a finite-sample concentration
example in Chapter 9 becomes a functional of an
empirical process here.
\end{example}

\subsection{When a Process Is Really a Random Element}
\label{sec:ch11-process-random-element}
\conceptindexes{random element, process convergence, ell-infinity space, Donsker theorem}

Donsker's theorem is sometimes stated as if it were only a central limit
theorem at many time points:
\[
  \sqrt n\{F_n(t)-F(t)\}\weakto B(F(t)).
\]
That sentence is incomplete unless it says where the convergence takes place.
Finite-dimensional convergence follows from the multivariate central limit
theorem, but a process limit also needs a measurable function space and
tightness.  This is the lesson behind Donsker's correction of Doob's heuristic
approach to the Kolmogorov--Smirnov theorem
\citep{doob1949heuristic,donsker1952justification,pollard1982beyond}.

The first tempting choice is the space \(D[0,1]\) of right-continuous functions
with left limits, equipped with the uniform metric
\citep[Chapter~12]{billingsley1999convergence}
\[
  w(x,y)=\sup_{0\le t\le1}|x(t)-y(t)|.
\]
But this metric makes \(D[0,1]\) nonseparable.  For each \(s\in[0,1]\), let
\[
  x_s(t)=\ind{s\le t}.
\]
If \(s\ne r\), then \(w(x_s,x_r)=1\).  Thus an uncountable family of points is
separated by distance one.

Nonseparability is not just cosmetic.  Let \(H\subset[0,1]\) be a non-Borel
set and put
\[
  A=\bigcup_{s\in H} B_w(x_s,1/2),
\]
where \(B_w(x_s,1/2)\) is the open \(w\)-ball around \(x_s\).  The set \(A\) is
open in the uniform topology on \(D[0,1]\), hence Borel in that topology.  Now
define \(T:[0,1]\to D[0,1]\) by \(T(s)=x_s\).  Since the balls above are
disjoint around the points \(x_s\),
\[
  T^{-1}(A)=H.
\]
Therefore \(T\) is not Borel measurable as a map from
\(([0,1],\mathcal B[0,1])\) to \(D[0,1]\) with the uniform Borel sigma-field.
This is already the \(n=1\) empirical cdf path \(t\mapsto\ind{U\le t}\) when
\(U\sim\Unif(0,1)\).
This small construction is the concrete version of a recurring warning: once
statistics becomes a statement about a whole path or a whole indexed field, the
ambient measurable space matters.  Chapter 6 introduced this issue abstractly
through product sigma-fields, path regularity, and nonmeasurable suprema; here
it appears inside the empirical-distribution process itself.

The usual repair is the Skorohod topology.  Let \(\Lambda\) be the strictly
increasing continuous maps of \([0,1]\) onto itself, and define
\[
  \|\lambda\|_0
  =
  \sup_{s<t}
  \left|
    \log\frac{\lambda(t)-\lambda(s)}{t-s}
  \right|.
\]
One version of the Skorohod metric is
\[
  d_0(x,y)
  =
  \inf_{\lambda\in\Lambda}
  \left\{
    \|\lambda\|_0\vee\|x-y\circ\lambda\|_\infty
  \right\}.
\]
It allows a small wiggle in time as well as in vertical height.  Under this
topology \(D[0,1]\) becomes a separable path space suitable for weak
convergence \citep{skorokhod1956limit,billingsley1999convergence}.  When the
limit process is continuous, as the Brownian bridge is, convergence in
Skorohod space is especially friendly: the time wiggle disappears in the
limit, and continuous functionals such as the supremum norm can be applied.

Chapter~10 already proves the path-space step that this example needs.  In
the empirical-cdf example there, Billingsley's \(D[0,1]\) tightness criterion
\citep[Chapter~13]{billingsley1999convergence} is applied to the uniform
empirical cdf and yields, for continuous \(F\),
\[
  \sqrt n\{F_n(\cdot)-F(\cdot)\}
  \weakto
  B\{F(\cdot)\}.
\]
The Kolmogorov--Smirnov limit is then just the continuous mapping theorem:
\[
  \sqrt n\sup_t|F_n(t)-F(t)|
  \weakto
  \sup_{0\le u\le1}|B(u)|.
\]

\begin{tcolorbox}[
  enhanced,
  breakable,
  colback=noteback,
  colframe=bookgold!75!black,
  boxrule=0.55pt,
  arc=4pt,
  boxsep=1pt,
  left=0.95em,
  right=0.9em,
  top=0.7em,
  bottom=0.7em,
  before skip=0.9\baselineskip,
  after skip=1.0\baselineskip
]
\noindent\textbf{What changes in this chapter.}
The empirical cdf uses the ordered class
\(\{\ind{x\le t}:t\in\R\}\).  Chapter~10 explains how this class becomes a
genuine path-valued random element and how Billingsley's criterion turns
finite-dimensional convergence into weak convergence.  The present chapter
starts from that solved case and asks the larger empirical-process question:
what if the index set is a general class \(\mathcal F\), chosen by a model, a
subgroup search, a classifier, a residual diagnostic, or a scientific feature
map?  The new work is not another proof of the cdf case; it is controlling the
size, measurability, and continuity of the whole searched field.
\end{tcolorbox}

\subsection{Why Pointwise Laws Are Not Enough}
\label{sec:ch11-pointwise-not-enough}
\conceptindexes{pointwise convergence, uniform convergence, stochastic process limits}

For every fixed $f\in\mathcal F$, the law of large numbers may give
$P_nf\to Pf$.  But the supremum over $\mathcal F$ can still fail to converge.
The reason is the same reason overfitting exists: if the class is too flexible,
it can chase noise.

\begin{example}[A class that memorizes the sample]
Let $\mathcal X=[0,1]$ and let $\mathcal F$ be the class of all measurable
indicators.  If $X_1,\ldots,X_n$ are iid continuous, the random set
$A_n=\{X_1,\ldots,X_n\}$ is measurable and satisfies
\[
  P_nA_n=1,\qquad P A_n=0.
\]
Thus $\sup_{A}|P_nA-PA|=1$ for every $n$.  Pointwise laws hold for each fixed
set $A$, but uniform convergence over all measurable sets is impossible.
\end{example}

This example is intentionally extreme.  It shows that uniform laws are not
automatic consequences of pointwise laws.  A useful class must be rich enough
to express the scientific question and restrained enough not to turn random
noise into apparent signal.

\section{Brackets, Covers, and Entropy}
\label{sec:ch11-entropy}
\conceptindexes{bracketing, covers, entropy, envelope functions, Glivenko--Cantelli criterion}

Empirical-process theory measures the size of a function class by asking how
many simple pieces are needed to approximate it.  One useful device is
bracketing.

\begin{definition}[Envelope and brackets]
An envelope for $\mathcal F$ is a measurable function $F$ such that
$|f(x)|\le F(x)$ for every $f\in\mathcal F$.  A bracket $[l,u]$ is a pair of
measurable functions with $l\le u$.  It contains all $f$ satisfying
$l\le f\le u$.  Its $L_r(P)$ size is
\[
  \|u-l\|_{P,r}=(P|u-l|^r)^{1/r}.
\]
The bracketing number $N_{[]}(\varepsilon,\mathcal F,L_r(P))$ is the smallest
number of brackets of size at most $\varepsilon$ needed to cover
$\mathcal F$.
\end{definition}

\begin{theorem}[A bracketing Glivenko--Cantelli criterion; \citealp{dudley1978central,pollard1984convergence,kosorok2008empirical}]
If $\mathcal F$ has an integrable envelope and
\[
  N_{[]}(\varepsilon,\mathcal F,L_1(P))<\infty
  \qquad\text{for every }\varepsilon>0,
\]
then $\mathcal F$ is $P$-Glivenko--Cantelli.
\end{theorem}

\noindent\textit{Proof.}
Fix \(\varepsilon>0\), and choose finitely many \(L_1(P)\)-brackets
\([l_j,u_j]\), \(j=1,\ldots,J\), such that every \(f\in\mathcal F\) belongs to
one bracket and \(P(u_j-l_j)\le\varepsilon\).  The endpoints are integrable
because they are dominated by an integrable envelope up to an \(L_1(P)\)
bracket error.  If \(f\in[l_j,u_j]\), then
\[
  P_nf-Pf
  \le
  P_nu_j-Pl_j
  =
  (P_n-P)u_j+P(u_j-l_j),
\]
and similarly
\[
  Pf-P_nf
  \le
  Pu_j-P_nl_j
  =
  (P-P_n)l_j+P(u_j-l_j).
\]
Therefore
\[
  \sup_{f\in\mathcal F}|P_nf-Pf|
  \le
  \max_{1\le j\le J}|(P_n-P)u_j|
  \vee
  \max_{1\le j\le J}|(P_n-P)l_j|
  +\varepsilon .
\]
The ordinary strong law sends each finitely many endpoint deviation to zero
almost surely.  Hence the limsup of the left side is at most \(\varepsilon\)
almost surely.  Since \(\varepsilon\) is arbitrary, the supremum converges to
zero almost surely, and therefore in probability.
\qedmark

The theorem explains why monotone one-dimensional classes are friendly.  The
class $\{\ind{x\le t}:t\in\R\}$ can be bracketed by a finite partition of the
line into intervals of small $P$-mass.  Event-time classes such as
$\{\ind{T\le t,\Delta=1}:t\le\tau\}$ have the same monotone structure.  A
small entropy condition is the formal version of a simple idea: the class has
many functions, but only finitely many distinguishable shapes at a given
resolution.

For Donsker theory the entropy condition is stronger.  One common sufficient
condition is the finiteness of a bracketing integral,
\[
  \int_0^\delta
  \sqrt{\log N_{[]}(\varepsilon,\mathcal F,L_2(P))}\,d\varepsilon
  <\infty
\]
for some $\delta>0$, together with a square-integrable envelope.  This
condition prevents the class from having too many nearly distinguishable
directions for Gaussian fluctuation to remain tight.

Dudley's entropy integral is the parallel signpost on the Gaussian-process
side \citep{dudley1967sizes}.  If \(G=\{G_t:t\in T\}\) is a centered Gaussian
process with canonical semimetric
\[
  d_G(s,t)^2=\Expect\{G_s-G_t\}^2,
\]
then chaining bounds the size of the Gaussian field by a covering integral of
the form
\[
  \Expect\sup_{t\in T}|G_t-G_{t_0}|
  \le
  C\int_0^{\operatorname{diam}(T)}
  \sqrt{\log N(\varepsilon,T,d_G)}\,d\varepsilon .
\]
For this chapter, the message is more important than the proof: entropy
integrals translate geometric size of an index set into tightness and
stochastic equicontinuity.  The bracketing integral above is the empirical-
process sibling used here as a sufficient condition rather than as a
new theorem to prove here.

\subsection{Outer Probability and Measurability}
\label{sec:ch11-outer-probability}
\conceptindexes{outer probability, measurability, empirical-process measurability}

There is a small technical door in this subject, and it should not be hidden.
The map
\[
  \omega\mapsto\sup_{f\in\mathcal F}|P_nf(\omega)-Pf|
\]
need not be measurable for an arbitrary uncountable class $\mathcal F$.  This
is not philosophical fussing.  Supremum, projection, and argmax operations are
exactly the operations that modern statistics uses, and measurability can fail
outside separable settings.

The standard repair is outer probability:
\[
  \outerProb(A)=\inf\{\Prob(B): A\subseteq B,\ B\in\fieldF\},
\]
with outer expectation
\[
  \outerExpect Z
  =
  \inf\{\Expect Y:Y\text{ is measurable and }Y\ge Z\}.
\]
This is the notation announced in Chapter~10.  Convergence in outer probability is
written
\[
  Z_n\to0\quad\text{in }\outerProb
\]
when $\outerProb(|Z_n|>\varepsilon)\to0$ for every $\varepsilon>0$.  If the
quantities are measurable, outer convergence reduces to ordinary convergence.
Equivalently, this chapter writes \(Z_n\toPstar 0\); the notation
\(R_n=\opstar(a_n)\) means \(R_n/a_n\toPstar 0\).

\begin{tcolorbox}[
  enhanced,
  breakable,
  colback=noteback,
  colframe=bookgold!75!black,
  boxrule=0.55pt,
  arc=4pt,
  boxsep=1pt,
  left=0.95em,
  right=0.9em,
  top=0.7em,
  bottom=0.7em,
  before skip=0.9\baselineskip,
  after skip=1.0\baselineskip
]
\noindent\textbf{When ordinary measurability is enough, and when outer
probability is needed.}
Use ordinary probability once the displayed statistic has been proved to be a
random variable.  This is automatic for finite or countable index sets, for
countable suprema of measurable variables, for Borel functionals of a random
element in a separable path space, and for integrals of jointly measurable
processes.  It is also automatic after a separability argument reduces an
uncountable supremum to a countable dense subclass, for example
\[
  \sup_{t\in T}X_t
  =
  \sup_{t\in T_0}X_t
\]
with \(T_0\) countable.

Use \(\outerProb\) or \(\outerExpect\) when the argument scans an uncountable
class and no such reduction has been verified.  The typical warning signs are
\[
  \sup_{f\in\mathcal F},\qquad
  \inf_{\theta\in\Theta},\qquad
  \argmax_{\theta\in\Theta},\qquad
  \{\exists f\in\mathcal F:\cdots\},
\]
with \(\mathcal F\) or \(\Theta\) not known to be separable in the relevant
semimetric.  In that case the outer notation is not a change of theorem; it is
a promise not to pretend that the displayed object is measurable before the
proof has earned it.

\medskip
\noindent\textbf{Operational map.}
The references below mark where these cases appear in the book.
\begin{description}[leftmargin=1.4em,style=nextline]
\item[Classical empirical cdf on \(\R\).]
Use ordinary probability.  Right-continuity reduces
\(\sup_{t\in\R}|F_n(t)-F(t)|\) to the rational cutpoints.
\emph{Book locations:} the real-line Glivenko--Cantelli/DKW bridge appears in
Section~\ref{sec:ch07-edf-bridge}, the empirical-process reading appears in
Section~\ref{sec:ch11-edf}, and the empirical-cdf Donsker examples appear in
Examples~\ref{ex:ch10-empirical-cdf-donsker} and
\ref{ex:ch10-empirical-cdf-f-scale}.
\item[Continuous-time processes.]
Use ordinary probability when the process has continuous, cadlag, jointly
measurable, progressive, optional, or predictable paths/process structure.
Use outer probability only if one scans an uncountable time set before such a
path-space or separability statement has been proved.
\emph{Book locations:} process suprema and path-space measurability are set up
in Sections~\ref{sec:supremum-measurability-repairs} and
\ref{sec:ch10-measurability-outer}, with the cadlag-supremum reduction in
Example~\ref{ex:ch06-cadlag-sup-measurable}; filtrations,
progressive/optional measurability, predictable processes, and event-history
examples are developed in Sections~\ref{sec:ch16-stochastic-basis},
\ref{sec:ch16-progressive-measurability},
\ref{sec:ch16-predictable-processes}, and
\ref{sec:ch16-marked-point-processes-real-time}.
\item[General empirical-process classes.]
Use ordinary probability for finite, countable, pointwise measurable, or
separable/admissible classes.  Use outer probability for a general uncountable
\(\mathcal F\) until this measurability has been verified.
\emph{Book locations:} bracketing and entropy are introduced in
Section~\ref{sec:ch11-entropy}; stochastic equicontinuity in
Section~\ref{sec:ch11-equicontinuity}; and the single-cell and learning
examples in Sections~\ref{sec:ch11-single-cell-thread} and
\ref{sec:ch11-modern-learning-examples}.
\item[Optimization and Z-estimation.]
Use ordinary probability once the estimator or root is a measurable map, often
by explicit construction, compact-continuous criteria, or a measurable
selection theorem.  Use outer probability for approximate argmax/root
statements before that step.
\emph{Book locations:} Section~\ref{sec:ch11-criterion-processes} introduces
criterion and equation processes; Chapter~\ref{chap:m-z-estimation} uses them
in Examples~\ref{ex:ch13-glm}, \ref{ex:ch13-median-regression-consistency},
and \ref{ex:ch13-logistic-separation}.
\item[U-process and higher-order fields.]
Use ordinary probability for a fixed kernel or a measurable/separable kernel
class.  Use outer probability for uniform statements over a large kernel class
when the supremum or the degenerate remainder has not yet been shown
measurable.
\emph{Book locations:} U-process survival examples appear in
Examples~\ref{ex:ch11-panel-count-empirical-process},
\ref{ex:ch11-dabrowska-u-process}, and
\ref{ex:ch11-dabrowska-semi-markov-application}; the same event-history models
return in Section~\ref{sec:ch16-semi-markov-transformation}.  Spatial and
multiparameter random fields, including the brain-voxel example, appear in
Section~\ref{sec:ch06-random-objects} and
Example~\ref{ex:ch06-brain-voxels-random-field}.
\end{description}

Before reporting an estimator, a confidence band, or a continuous-mapping
limit as an ordinary probabilistic statement, return to measurability.  Either
exhibit a measurable version, verify pointwise measurability or separability,
work in a Polish path space where the needed functional is Borel, invoke a
measurable-selection or analytic-set repair, or keep the conclusion explicitly
in outer probability.  Once measurability is established, the outer notation
collapses back to the usual one.
\end{tcolorbox}

The message is not that every reader must become a descriptive set theorist.
The message is that the product-space warnings from Chapter 6 are now
operational.  Once an uncountable class is scanned, the statistic is not
automatically a random variable.
Separability, analytic sets, measurable selection, and outer probability are the
devices that let the statistical argument proceed without pretending this issue
is absent.

\section{Stochastic Equicontinuity and Tightness}
\label{sec:ch11-equicontinuity}
\conceptindexes{stochastic equicontinuity, tightness, asymptotic equicontinuity, criterion processes}

Uniform laws control the size of the empirical field.  Weak convergence of the
field needs a second idea: nearby indices should have nearby fluctuations.  Let
$\rho$ be a semimetric on $\mathcal F$.  The empirical process is
asymptotically equicontinuous if for every $\eta>0$,
\[
  \lim_{\delta\downarrow0}
  \limsup_n
  \outerProb\left(
    \sup_{\rho(f,g)<\delta}
    |\mathbb G_nf-\mathbb G_ng|>\eta
  \right)
  =0.
\]
This condition says that the process cannot oscillate wildly over small
neighborhoods of the index set.

In $\ell^\infty(\mathcal F)$, weak convergence is built from two ingredients:
finite-dimensional convergence and tightness.  Finite-dimensional convergence
usually comes from the multivariate central limit theorem applied to
\[
  (f_1(X),\ldots,f_k(X)).
\]
Tightness is the hard part; stochastic equicontinuity is the usable form of
that hard part.

\begin{example}[Lipschitz parametric classes]
Suppose
\[
  \mathcal F=\{f_\theta:\theta\in\Theta\subset\R^d\}
\]
and
\[
  |f_\theta(x)-f_{\theta'}(x)|
  \le
  \|\theta-\theta'\|H(x)
\]
with $PH^2<\infty$ and bounded $\Theta$.  Then the class can be covered by
Euclidean balls in $\Theta$, and its entropy grows polynomially in
$1/\varepsilon$.  This is why smooth finite-dimensional parametric models
usually have manageable empirical processes.
\end{example}

\begin{example}[Monotone event-time classes]
For right-censored data, classes such as
\[
  \{\ind{T\le t,\Delta=1}:0\le t\le\tau\}
\]
and
\[
  \{\ind{T\ge t}:0\le t\le\tau\}
\]
are controlled by monotonicity.  Their brackets are created by time partitions
whose expected increments are small.  This is the empirical-process skeleton
behind uniform consistency of Nelson--Aalen and Kaplan--Meier type estimators.
For example, the numerator and denominator processes in a cumulative-hazard
estimate are built from averages of
\[
  N_i(t)=\ind{T_i\le t,\Delta_i=1},
  \qquad
  Y_i(t)=\ind{T_i\ge t}.
\]
Uniform stability in $t$ is what lets a whole estimated hazard curve behave
like its population counterpart, rather than merely matching it at one selected
time.
\end{example}

\subsection{Event-History Processes as Indexed Fields}
\label{sec:ch11-event-history-dictionary}
\conceptindexes{event-history indexed fields, counting processes, marked point processes, compensators, martingales, predictable processes}

The survival examples below use a little continuous-time language before the
book gives the full treatment.  Only a small dictionary is needed here.  A
subject's event history may be encoded by a counting process
\[
  N_i(t)=\sum_k \ind{T_{ik}\le t},
\]
or, when event types matter, by a marked point process
\[
  N_i(t,B)
  =
  \sum_k \ind{T_{ik}\le t,\ Z_{ik}\in B},
  \qquad B\subseteq\mathcal Z .
\]
Equivalently, \(N_i(dt,dz)\) is a random measure that puts mass at observed
event times and marks.  In a clinical trial, the mark \(z\) might be adverse
event type, transition destination, cause of failure, or recurrent-event label.
The at-risk process \(Y_i(t)\) records whether subject \(i\) is still able to
experience the event just before time \(t\).

The information available just before time \(t\) is represented by a
filtration \(\mathbb F_i=\{\mathcal F_{i,t}:t\ge0\}\).  A process is
predictable if its value at time \(t\) is determined from the history just
before \(t\).  The compensator is the predictable cumulative rate of the
counting process.  Thus \(\Lambda_i(t,B)\) is chosen so that
\[
  M_i(t,B)=N_i(t,B)-\Lambda_i(t,B)
\]
is a martingale with respect to \(\mathbb F_i\).  If an intensity exists, this
often has the schematic form
\[
  \Lambda_i(dt,dz)
  =
  Y_i(t)\lambda(t,z\mid\mathcal F_{i,t-})\,dt\,\nu(dz).
\]
For a one-event survival endpoint,
\[
  N_i(t)=\ind{T_i\le t,\Delta_i=1},
  \qquad
  \Lambda_i(t)
  =
  \int_0^t Y_i(u)\alpha(u\mid Z_i)\,du,
\]
and \(M_i=N_i-\Lambda_i\) is the residual event noise after subtracting the
predictable hazard clock.

There are two complementary large-sample routes.  The martingale route uses
the identity \(N=\Lambda+M\), predictable variation, and martingale central
limit theorems.  The empirical-process route treats the observed event history
as the observation \(O_i\) and controls indexed averages such as
\[
  \left\{
    P_n f_t:
    f_t(O)=N(t),\ Y(t),\ \int_0^t H(u)\,N(du),
    \ 0\le t\le\tau
  \right\}.
\]
Panel-count likelihoods, Kaplan--Meier surfaces, U-process survival
regressions, and safety dashboards usually use both grammars: martingales
explain the predictable event clock, while empirical processes control the
searched field over times, marks, transitions, parameters, and subgroups.

This subsection is only a bridge.  The detailed construction of marked point
processes, predictable processes, compensators, martingales, and transition
intensities belongs to the continuous-time chapter.  The classical survival
references are \citet{aalen1978nonparametric}, \citet{andersen1982cox},
\citet{andersen1993statistical}, and \citet{fleming1991counting}.

\begin{example}[Panel counts as an empirical criterion field]
\label{ex:ch11-panel-count-empirical-process}
Survival analysis often looks like martingale theory from the outside, but its
large-sample arguments also have a very empirical-process face.  In panel count
data, subject \(i\) has an unobserved recurrent-event counting process
\(N_i(t)\), but the analyst sees it only at random inspection times
\[
  0<T_{i1}<\cdots<T_{iK_i}.
\]
The target in \citet{wellnerZhang2000panel} is the mean function
\[
  \Lambda_0(t)=\Expect\{N_i(t)\}.
\]
Under a non-homogeneous Poisson working model, the increment
\(\Delta N_{ij}=N_i(T_{ij})-N_i(T_{i,j-1})\) has mean
\[
  \Delta\Lambda(T_{ij})
  =
  \Lambda(T_{ij})-\Lambda(T_{i,j-1}).
\]
Thus one pseudo-likelihood criterion has the schematic form
\[
  M_n(\Lambda)
  =
  P_n m_\Lambda
  =
  \frac1n\sum_{i=1}^n
  \sum_{j=1}^{K_i}
  \left\{
    \Delta N_{ij}\log \Delta\Lambda(T_{ij})
    -\Delta\Lambda(T_{ij})
  \right\},
\]
maximized over nondecreasing mean functions \(\Lambda\).  The empirical-process
question is not whether \(M_n(\Lambda_0)\) is close to \(M(\Lambda_0)\) at one
function.  It is whether the random criterion field
\[
  \{(P_n-P)m_\Lambda:\Lambda\in\mathcal L\}
\]
is uniformly controlled on a monotone function class \(\mathcal L\).  Here
\(\mathcal L\) is not a ``monotone class'' in the measure-theoretic sense of
the monotone-class theorem.  It is a statistical model class, for example
\[
  \mathcal L
  =
  \{\Lambda:[0,\tau]\to[0,\infty):
    \Lambda \text{ is nondecreasing, right-continuous, and bounded}\}.
\]
The monotonicity restriction is what makes a very large curve-indexed search
manageable: brackets can be built from time partitions and increments, rather
than from arbitrary oscillating functions.  Once this random criterion field is
stable uniformly over \(\mathcal L\), argmax arguments can identify the
population target and support consistency of the pseudo-likelihood and
likelihood estimators.  Chapter~16 returns to the same data as an observation
scheme for counting processes; here the lesson is that the likelihood itself
is an empirical field indexed by a curve.
\end{example}

\begin{example}[Dabrowska's bivariate survival surface]
\label{ex:ch11-dabrowska-bivariate-km}
Dabrowska's bivariate Kaplan--Meier estimator is not introduced here as
another survival-analysis formula.  It is the same paired-censoring problem
that appeared as an observed-data exercise in Chapter~5, as a uniform-law
stress test in Chapter~9, and later as a bootstrap-band example in Chapter~15.
The role of the present chapter is narrower: it explains where the empirical
process enters.

Let \(T=(T_1,T_2)\) be paired event times and \(C=(C_1,C_2)\) paired
censoring times.  We observe
\[
  O=(Y_1,Y_2,\Delta_1,\Delta_2),
  \qquad
  Y_j=T_j\wedge C_j,\quad \Delta_j=\ind{T_j\le C_j},
\]
but the target is the unobserved survival surface
\[
  S(s,t)=\Prob(T_1>s,T_2>t).
\]
Dabrowska's notation starts from four observable marked survival surfaces.
For example,
\[
  H_{00,n}(s,t)=P_n\ind{Y_1>s,Y_2>t}
\]
is the empirical risk surface.  The companion surfaces attach the censoring
marks:
\[
\begin{aligned}
  H_{10,n}(s,t)&=P_n\ind{Y_1>s,Y_2>t,\Delta_1=1},\\
  H_{01,n}(s,t)&=P_n\ind{Y_1>s,Y_2>t,\Delta_2=1},\\
  H_{11,n}(s,t)&=P_n\ind{Y_1>s,Y_2>t,\Delta_1=\Delta_2=1}.
\end{aligned}
\]
The subscript \(10\), for instance, does not mean the time point
\((1,0)\); it means that the first coordinate carries the failure mark and
the second does not.  These four functions are ordinary empirical averages
indexed by \((s,t)\), but the estimator itself is not a pointwise average.  It
is a product-limit functional of the vector surface
\(H_n=(H_{00,n},H_{10,n},H_{01,n},H_{11,n})\):
\[
  \widehat S_n=\Psi(H_n).
\]
Dabrowska's 1988 paper builds this product-integral estimator and proves
uniform strong consistency.  Her 1989 paper is the empirical-process part:
it represents \(\widehat S_n-S\) as a sum of mean-zero iid surface-valued
processes plus a smaller remainder, then proves weak convergence, an LIL
statement, and bootstrap confidence sets
\citep{dabrowska1988kaplan,dabrowska1989kaplan}.  Thus the modern shorthand
is
\[
  \sqrt n(\widehat S_n-S)
  =
  \frac1{\sqrt n}\sum_{i=1}^n \phi_P(O_i)
  +o_{\Prob}(1)
  \quad\text{in a two-parameter function space,}
\]
for a mean-zero random surface \(\phi_P\).  Chapter~15 returns to this display
when discussing simultaneous bootstrap bands; Chapter~16 returns to related
survival data through the counting-process and compensator language.
\end{example}

\begin{proposition}[U-processes are projected empirical processes; \citealp{hoeffding1948class,delapenaGine1999decoupling}]
\label{prop:ch11-u-process-projection}
Let \(O_1,\ldots,O_n\) be iid and let \(\mathcal H\) be a class of symmetric
square-integrable kernels on \(m\)-tuples.  Define
\[
  U_n h
  =
  \binom{n}{m}^{-1}
  \sum_{1\le i_1<\cdots<i_m\le n}
  h(O_{i_1},\ldots,O_{i_m}),
  \qquad h\in\mathcal H.
\]
For \(\theta_h=P^m h\), set
\[
  h_1(o)
  =
  \Expect\{h(o,O_2,\ldots,O_m)\}-\theta_h .
\]
Let \(\pi_rh\), \(r=1,\ldots,m\), denote the canonical Hoeffding projections
defined in the proof below, so \(\pi_1h=h_1\), and let \(U_{n,r}\) be the
order-\(r\) U-statistic.  If the first-projection class
\(\mathcal H_1=\{h_1:h\in\mathcal H\}\) is Donsker and
\[
  \sup_{h\in\mathcal H}
  \left|
    \sqrt n\sum_{r=2}^m \binom{m}{r}U_{n,r}(\pi_rh)
  \right|
  \toPstar 0,
\]
then in \(\ell^\infty(\mathcal H)\),
\[
  \sqrt n\{U_n h-\theta_h\}
  =
  m\,\mathbb G_n h_1
  +\opstar(1).
\]
\end{proposition}

\noindent\textit{Proof.}
For \(0\le r\le m\), write \(P^{m-r}h(o_1,\ldots,o_r)\) for the result of
integrating the last \(m-r\) coordinates of \(h\) with respect to \(P\); for
\(r=0\) this is \(\theta_h=P^mh\).  Define the canonical projections
recursively by
\[
\begin{aligned}
  \pi_0h=\theta_h,
  \qquad
  \pi_rh(o_1,\ldots,o_r)
  &=
  P^{m-r}h(o_1,\ldots,o_r)\\
  &\quad-
  \sum_{\substack{A\subsetneq\{1,\ldots,r\}}}
  \pi_{|A|}h(o_A),
  \qquad 1\le r\le m,
\end{aligned}
\]
where \(o_A=(o_j:j\in A)\).  Equivalently,
\[
  P^{m-r}h(o_1,\ldots,o_r)
  =
  \sum_{A\subseteq\{1,\ldots,r\}}\pi_{|A|}h(o_A).
  \tag{11.1}
\]
For \(r=1\), this gives
\[
  \pi_1h(o)=P^{m-1}h(o)-P^mh=h_1(o),
\]
so \(P\pi_1h=0\).  More generally, each \(\pi_rh\) is degenerate: integrating
it with respect to any one of its arguments gives zero.  This follows from
the recursive definition by integrating both sides of (11.1) in one displayed
coordinate and canceling the lower-order terms.

Now fix \(h\).  Apply (11.1) with \(r=m\) to every ordered \(m\)-tuple
\((O_{i_1},\ldots,O_{i_m})\).  Since \(h\) is symmetric, averaging over
unordered \(m\)-subsets gives
\[
\begin{aligned}
  U_nh-\theta_h
  &=
  \binom{n}{m}^{-1}
  \sum_{|I|=m}
  \sum_{\emptyset\ne A\subseteq I}
  \pi_{|A|}h(O_A) .
\end{aligned}
\]
Group the terms by \(r=|A|\).  A fixed \(r\)-subset \(J\) is contained in
\(\binom{n-r}{m-r}\) different \(m\)-subsets \(I\), hence
\[
\begin{aligned}
  U_nh-\theta_h
  &=
  \sum_{r=1}^m
  \binom{n}{m}^{-1}\binom{n-r}{m-r}
  \sum_{|J|=r}\pi_rh(O_J)  \\
  &=
  \sum_{r=1}^m
  \binom{m}{r}U_{n,r}(\pi_rh),
\end{aligned}
\]
because
\[
  \binom{n}{m}^{-1}\binom{n-r}{m-r}\binom{n}{r}
  =
  \binom{m}{r}.
\]
This is Hoeffding's decomposition.  The \(r=1\) term is
\[
  \binom{m}{1}U_{n,1}(\pi_1h)
  =
  mP_nh_1
  =
  m(P_n-P)h_1,
\]
because \(Ph_1=0\).  Multiplying by \(\sqrt n\) gives
\[
  \sqrt n\{U_nh-\theta_h\}
  =
  m\,\mathbb G_nh_1
  +
  \sqrt n\sum_{r=2}^m\binom{m}{r}U_{n,r}(\pi_rh).
\]
Taking the supremum over \(h\in\mathcal H\), the assumed uniform
negligibility of the higher-order degenerate terms makes the second term
\(\opstar(1)\) in \(\ell^\infty(\mathcal H)\).  Therefore
\[
  \sqrt n\{U_nh-\theta_h\}
  =
  m\,\mathbb G_nh_1+\opstar(1)
\]
uniformly in \(h\).  The Donsker assumption on \(\mathcal H_1\) supplies the
process-level tight Gaussian limit for the leading empirical-process term.
Sufficient entropy and moment conditions for the displayed negligibility are
standard U-process results; see \citet[Chapter~5]{vaart2023weak}, building on
\citet{hoeffding1948class}. \qedmark

\begin{example}[Two-stage renewal regression: where the U-process enters]
\label{ex:ch11-dabrowska-u-process}
Dabrowska's two-stage renewal regression model
\citep{dabrowska2009twoStageRenewal} is a good place to see why ordinary
counting-process martingale arguments are sometimes not the right large-sample
language.  The motivating data are matched-pair experiments on two organs of
one subject, for instance eyes or kidneys.  One organ receives an experimental
treatment, the other a conventional treatment.  Starting from the randomization
state \(R\), the process may move to failure of the experimental organ \(E\),
failure of the conventional organ \(C\), or simultaneous failure \(E+C\); after
a first single-organ failure, the second organ may fail later.  Censoring adds
a loss-to-follow-up state.

Let \(h\) index the possible one-step transitions.  For an observed subject,
write \(N_{hi}(t)\) for the transition counting process and \(Y_{hi}(t)\) for
the corresponding at-risk indicator.  The model assumes transition intensities
of the schematic form
\[
  \lambda_h(u\mid Z_i)
  =
  Y_{hi}(u)\,
  q_h(Z_{hi},u,\theta)\exp\{r(Z_i,u,\beta)\}\alpha(u),
\]
where \(\alpha\) is an unknown baseline hazard, \(r\) carries common covariate
effects, and the \(q_h\)'s describe transition-specific effects, with
\(\sum_{h\in\mathcal H_0}q_h=1\) for the first-stage transitions.  The
Euclidean parameter is \(\xi=(\beta,\theta)\).

After profiling out \(\alpha\) by a weighted Nelson--Aalen estimator, the
estimating equation for \(\xi\) has the Cox-like form
\[
  \Phi_n(\xi)
  =
  \frac1n\sum_{i=1}^n\sum_h
  \int_0^\tau
  \left\{
    \varphi_h(Z_{hi},u,\xi)
    -
    \frac{S^{(1)}(u,\xi)}{S^{(0)}(u,\xi)}
  \right\}N_{hi}(du),
\]
where \(S^{(p)}(u,\xi)=n^{-1}\sum_i\sum_h S_{hi}^{(p)}(u,\xi)\) are empirical
risk-set averages.  The ratio \(S^{(1)}/S^{(0)}\) makes the score nonlinear in
the empirical law.  When it is expanded around its population counterpart, the
score is no longer just a sum of iid terms.  Schematically,
\[
  \Phi_n(\xi_0)
  =
  U_{n,2}g^{(2)}
  +
  U_{n,3}g^{(3)}
  +
  V_{n,4}g^{(4)}
  +\text{smaller terms}.
\]
Hoeffding projection then isolates the first-order empirical-process part:
\[
  U_{n,2}g^{(2)}
  =
  P_n\psi
  +
  U_{n,2}\pi_2g^{(2)}.
\]
The first term gives the Gaussian limit; the canonical degenerate U-statistic
and higher-order V-process terms are shown to be \(o_{\Prob}(n^{-1/2})\)
uniformly over the local parameter set.  This is exactly the mechanism in
Proposition~\ref{prop:ch11-u-process-projection}, with the extra survival
work needed for censoring and the renewal clock.

The statistical payoff is concrete.  The score equation
\(\Phi_n(\hat\xi)=o_{\Prob}(n^{-1/2})\) has a local root with probability
tending to one, and
\[
  \sqrt n(\hat\xi-\xi_0)
  \weakto
  N\{0,\Sigma(\xi_0)^{-1}\}.
\]
The profiled baseline estimator \(\widehat A(x,\hat\xi)\) has a process limit,
after the usual correction for estimating \(\xi\), given by a time-transformed
Brownian motion, and it is asymptotically independent of
\(\sqrt n(\hat\xi-\xi_0)\).  The supplement verifies the information bound,
so the root is asymptotically efficient under the model.

The 2009 paper is mostly a methodology paper rather than a worked data
analysis.  Its landing place is nevertheless clear: in a paired-organ trial,
the analyst would report the common covariate effects \(\hat\beta\), the
transition-specific effects \(\hat\theta\), and the profiled baseline
transition law \(\widehat A\), with covariance estimates justified by the
U-process expansion rather than by a common-time martingale argument that the
renewal scale does not support.
\end{example}

\begin{example}[Semi-Markov transformation models: from U-processes to transplant curves]
\label{ex:ch11-dabrowska-semi-markov-application}
The later semi-Markov transformation paper
\citep{dabrowska2012semiMarkovTransformation} takes the same issue into a
full multistate survival analysis.  A subject moves through a finite state
space; for a transition \(j=(j_1,j_2)\), the compensator is modeled on elapsed
sojourn time, not merely calendar time:
\[
\begin{aligned}
  \Lambda_j(t)
  &=
  \Lambda_j(T_m)
  +
  \int_0^{t-T_m}
  \ind{J_m=j_1}\,
  \alpha_j\{\Gamma_{(j_1,\cdot)}(u),\theta,Z_{j_1m}\}
  \,\Gamma_j(du),\\
  &\hspace{8.5em} T_m<t\le T_{m+1}.
\end{aligned}
\]
Here \(\theta\) is finite-dimensional, while
\(\Gamma=(\Gamma_j)\) is a vector of unknown increasing transformation
functions.  The model includes proportional-hazard modulated renewal processes
as a special case, but also allows proportional-odds-type transformation
behavior, where covariate effects may dissipate or change strength over time.
Read as continuous-time model grammar, this display defines the event-history
law.  The later continuous-time chapter keeps that model grammar and its
observation/target interpretation; here it becomes the input to the
empirical-process argument.

The estimator solves two coupled empirical problems.  For a fixed \(\theta\),
\(\Gamma_{n\theta}\) is obtained from nonlinear Volterra-type equations built
from transition counts and risk sets.  Then \(\theta\) solves an estimating
equation
\[
  U_n^{\varphi_n}(\theta)
  =
  \frac1n\sum_i\sum_j\sum_m
  \int_0^\tau
  \widehat b_{jmi}\{\Gamma_{n\theta}(u),\theta,u\}\,
  N_{jmi}(du)
  =
  \opstar(n^{-1/2}).
\]
Because the relevant time scale is the time since entry into the current
state, the usual multiplicative-intensity machinery is not enough by itself.
The proof uses Hoeffding projections, empirical-process Donsker arguments for
Euclidean classes, stochastic equicontinuity in
\(\ell^\infty([0,\tau]\times\mathcal J)\), and outer probability to handle the
measurability of the estimated score field.

The main limit theorem is an \(M\)- and empirical-process statement at once:
\[
  \left[
    \sqrt n(\hat\theta-\theta_0),
    \sqrt n\left\{
      \Gamma_{n\hat\theta}-\Gamma_{\theta_0}
      -(\hat\theta-\theta_0)^T\dot\Gamma_{n\hat\theta}
    \right\}
  \right]
  \weakto
  (\Xi,W_0),
\]
where the limit is tight Gaussian in
\(\R^d\times\ell^\infty([0,\tau]\times\mathcal J)\).  A Gaussian multiplier
process consistently approximates this joint limit.  By the functional delta
method, plug-in one-step and multistep transition probabilities inherit
Gaussian limits and simultaneous confidence bands.

The data analysis uses CIBMTR transplant data: HLA-identical sibling
transplants from 1995--2004 for AML or ALL patients in first remission,
comparing bone marrow transplant (BMT) with peripheral blood stem cell
transplant (PBSCT).  The fitted five-state recovery model starts at transplant,
then allows acute GVHD, chronic GVHD, relapse, death in remission, and
censoring.  The empirical object is not a single hazard ratio.  It is a family
of path probabilities, for example transplant \(\to\) relapse directly versus
transplant \(\to\) acute GVHD \(\to\) relapse, each evaluated under covariate
profiles such as age group, disease type, donor--recipient sex match, and graft
source.

That is where the empirical-process machinery changes the interpretation.  The
paper compares whole transition-probability curves with pointwise and
simultaneous multiplier bands.  Younger age is associated with lower endpoint
probabilities, but a pointwise signal for one-step relapse weakens under the
simultaneous band, while multistep death probabilities remain clearly lower.
Older age has a discordant pattern: lower relapse probabilities but higher
death probabilities, especially through chronic GVHD paths.  For young AML
patients, PBSCT has lower one-step death probability, but after intermediate
GVHD states are included, the multistep comparison can reverse direction.  The
model check is also process-based: Kolmogorov--Smirnov residual statistics over
transition-specific residual curves are calibrated by 5000 Gaussian-multiplier
resamples and give an acceptable fit, with caveats about center and calendar
effects.  In applied terms, the method turns a multistate transplant history
into interpretable path-specific risk curves, with uncertainty bands attached
to the curves rather than to isolated coefficients.
\end{example}

\subsection{Criterion and Equation Processes}
\label{sec:ch11-criterion-processes}
\conceptindexes{criterion processes, equation processes, M-estimation, Z-estimation}

The reason this chapter sits before M- and Z-estimation is that estimators are
often features of empirical fields.  An M-estimator may minimize
\[
  M_n(\theta)=P_nm_\theta
\]
over $\theta\in\Theta$, while the population target minimizes
\[
  M(\theta)=Pm_\theta.
\]
Uniform convergence
\[
  \sup_{\theta\in\Theta}|M_n(\theta)-M(\theta)|\to0
\]
is the condition that prevents the empirical criterion from inventing a false
peak far from the population peak.

A Z-estimator solves
\[
  \Psi_n(\theta)=P_n\psi_\theta=0,
\]
where the target solves
\[
  \Psi(\theta)=P\psi_\theta=0.
\]
Uniform convergence of $\Psi_n$ to $\Psi$ on neighborhoods of the target is
the corresponding stability condition for roots.

\begin{tcolorbox}[
  enhanced,
  breakable,
  colback=noteback,
  colframe=bookgold!75!black,
  boxrule=0.55pt,
  arc=4pt,
  boxsep=1pt,
  left=0.95em,
  right=0.9em,
  top=0.7em,
  bottom=0.7em,
  before skip=0.9\baselineskip,
  after skip=1.0\baselineskip
]
\noindent\textbf{Connection to curve representations and M/Z-estimation.}
Chapter 9 handled fixed summaries; this chapter handles indexed fields.  The
next chapter pauses on one especially important indexed limit, the Brownian
bridge, and asks how a random curve can be decomposed into covariance modes.
The estimation chapter after that uses empirical-field stability to choose:
peaks of $M_n$ should be near peaks of $M$, and roots of $\Psi_n$ should be
near roots of $\Psi$.  The technical chain is field, representation or
criterion, uniform law, selected feature.
\end{tcolorbox}

\section{A Single-Cell Thread: Observation, Generation, and Inference}
\label{sec:ch11-single-cell-thread}
\conceptindexes{single-cell thread, observation mechanism, generative model, inference layer, scImpute, scDesign3, scGTM, generalized Pearson correlation squares, K-lines clustering}

Single-cell genomics is a compact example of why empirical-process thinking is
more than technical decoration.  A modern assay does not give one statistic.
It gives an empirical field: cells by genes, cells by coordinates, cells by
pseudotime, cells by batches, and cells by experimental conditions.  The
scientific question is usually not whether one fixed average is stable.  It is
whether a whole fitted structure remains stable while the analyst scans across
genes, neighborhoods, latent states, smooth trends, and simulation scenarios.

It helps to read the same object in two languages.  In the biological language,
the data are a noisy assay matrix with metadata: a row may be a cell, a column
may be a gene or feature, and the annotations may include batch, modality,
spatial coordinate, condition, or pseudotime.  In the mathematical language,
one cell-level record is a random element
\[
  X_i=(Y_i,C_i),
\]
where \(Y_i\) is the measured expression or feature vector and \(C_i\) stores
the covariates and annotations.  A biological summary is a function
\(f(X_i)\).  A marker mean, a zero rate, a neighborhood contrast, a smooth
pseudotime trend, or a spatial statistic are all examples of such functions.
The empirical field is the collection
\[
  \{P_nf:f\in\mathcal F\},
\]
where \(\mathcal F\) is the set of summaries the analysis is willing to scan.
Uniform stability asks whether this whole collection is close to its
population counterpart, not merely whether one chosen summary looks stable.

\begin{center}
\small
\textbf{Biological questions as empirical-process objects.}\par\smallskip
\begin{tabular}{@{}p{0.46\linewidth}p{0.47\linewidth}@{}}
\toprule
\textbf{Biological question} & \textbf{Mathematical object} \\
\midrule
Is an observed zero true silence or technical dropout? &
An observation model for events such as \(\ind{Y_{ig}=0}\), indexed by cell
and gene. \\
Can synthetic cells reproduce the real assay well enough to test methods? &
A uniform distance between observed and synthetic empirical fields. \\
What structured trend or map is learned from the full field? &
An \(M\)- or \(Z\)-estimator obtained from indexed criteria or estimating
equations. \\
Which gene pairs hide subtype-specific dependence? &
A selected empirical functional
\(\hat T_{gh}^{(K)}=\Phi_K(P_n;Y_g,Y_h)\), indexed by gene pairs, line
memberships, and the number of linear components. \\
\bottomrule
\end{tabular}
\end{center}

The thread is easiest to see in three connected methods.  The first is
observation.  \citet{li2018scimpute} start from the fact that, in single-cell
RNA-seq, an observed zero may record true absence of expression, or it may be a
dropout created by capture and sequencing.  Biologically, scImpute asks whether
a zero should be treated as silence or as missing signal.  Mathematically, the
zero is not just the single event \(X=0\) for one gene.  It is part of a
cell-by-gene family of events and conditional probabilities, for example
\[
  \pi_g(c)=\Prob(Y_g=0\mid C=c),
  \qquad
  \mu_g(c)=E(Y_g\mid C=c),
\]
estimated while borrowing information from similar cells and genes.  The
uniform question is whether the whole dropout-correction rule remains stable
over the genes, cell states, and neighborhoods used by the procedure.  This is
what prevents random sparsity from being converted into biological signal.

The second step is generation.  \citet{song2024scdesign3} move the same logic
from observed zeros to synthetic assays.  Biologically, scDesign3 asks for
simulated cells that preserve the features that make a real experiment hard:
marginal expression, zero rates, gene dependence, multimodal measurements,
cell-state structure, and spatial or temporal covariates.  Mathematically, a
simulator is not checked by one goodness-of-fit number.  It is checked over a
class of functions:
\[
  \begin{gathered}
  f\in\mathcal F\ \text{may encode means, zero rates, correlations,}\\
  \text{marker contrasts, trajectory trends, or spatial summaries.}
  \end{gathered}
\]
The useful target is therefore
\[
  \sup_{f\in\mathcal F}
  |P_n^{\mathrm{sim}}f-P_n^{\mathrm{obs}}f|
  \quad\text{small for the summaries that matter.}
\]
In biological terms, the synthetic assay should preserve the signals a
downstream analyst will use.  In mathematical terms, this is a uniform
approximation problem between two empirical fields.

The companion PBMC3k capsule makes this sentence concrete.  On the public
fallback matrix, the analysis keeps 2,700 cells and 100 genes; the median
cell-level zero rate is 0.03 and the median gene-level zero rate is 0.0174.
A small scDesign3 run on 30 genes and 500 cells gives a mean absolute
zero-rate difference of 0.00773 between reference and synthetic genes.  This
number is not a universal success certificate.  It is an empirical-process
diagnostic: one selected class of summaries, zero-rate functions indexed by
genes, was checked before the fitted simulator was allowed to stand in for the
assay.

The third step is inference.  scGTM \citep{cui2022scgtm} belongs at this end of
the thread because its target is a structured biological map learned from the
full single-cell field.  Biologically, the analyst wants an interpretable
expression trend along pseudotime or another ordered cellular coordinate, not
just a list of significant genes.  Mathematically, the object is a parameter or
function \(\theta\) chosen by optimizing or solving a collection of empirical
criteria across cells, genes, and latent structure.  The relevant probability
questions are therefore the questions of this chapter and the next two
chapters:
\[
  \sup_{\theta\in\Theta}|M_n(\theta)-M(\theta)|,
  \qquad
  \sup_{\theta\in\Theta}\|\Psi_n(\theta)-\Psi(\theta)\|,
  \qquad
  \sqrt n\{\hat\theta-\theta_0\}.
\]
Uniform stability says that the learned map is not merely a pattern found by
searching; M- and Z-estimation describe the criterion or equation that locates
it; local approximation explains how the remaining uncertainty should be read.
The same capsule also records a negative result: the PBMC3k fallback data have
no real pseudotime column, so the scGTM input audit is complete but no
biological trajectory target is claimed.  That limitation is exactly the
target discipline of Chapter~8, now enforced by the data rather than by a
definition on the page.

\begin{example}[Gene-expression heterogeneity as a selected empirical field]
\citet{liZhouBickelTong2024geneHeterogeneity} study what changes when the
biological target is not a marginal mean, a zero rate, or one smooth trend, but
heterogeneity in the relation between two genes.  Ordinary Pearson correlation
summarizes one linear dependence.  In expression data, the same pair of genes
may have one linear relation in one subtype and a different, even oppositely
signed, relation in another.  If the subtype label \(S\) is observed, the
field already has an extra index:
\[
  \{\rho_{gh}(s)=\operatorname{Corr}(Y_g,Y_h\mid S=s):
    (g,h)\in\mathcal G^2,\ s\in\mathcal S\}.
\]
If the subtype label is not observed, the line membership is also estimated.
This is where the \(K\)-lines and generalized-correlation language enters.

A useful abstraction is the following.  For a gene pair \((g,h)\), let
\(Z_{gh}=(Y_g,Y_h)\).  For a candidate number of components \(K\), let
\(\mathcal L_K\) be a class of \(K\)-line descriptions, including both the
linear pieces and the rule assigning a point \(Z_{gh}\) to a piece.  Write
\[
  q_{\ell,gh}(z)=d^2(z,\ell),
  \qquad \ell\in\mathcal L_K,
\]
for the squared residual from the assigned line.  The population and empirical
\(K\)-line risks are
\[
  Q_{gh}^{(K)}=\inf_{\ell\in\mathcal L_K}Pq_{\ell,gh},
  \qquad
  \hat Q_{gh}^{(K)}=\inf_{\ell\in\mathcal L_K}P_nq_{\ell,gh}.
\]
A generalized Pearson-type score can then be read schematically as a normalized
reduction in residual variation,
\[
  T_{gh}^{(K)}
    =1-\frac{Q_{gh}^{(K)}}{Q_{gh}^{(0)}},
  \qquad
  \hat T_{gh}^{(K)}
    =1-\frac{\hat Q_{gh}^{(K)}}{\hat Q_{gh}^{(0)}},
\]
where \(Q_{gh}^{(0)}\) and \(\hat Q_{gh}^{(0)}\) denote the corresponding
null or total-variation baselines.  The precise normalization may vary across
versions of the method.  The important point for this chapter is the empirical
field, not the name of one final score.

For fixed \((g,h,K)\), ordinary asymptotics ask whether
\(\hat T_{gh}^{(K)}\) is close to \(T_{gh}^{(K)}\).  A screening analysis asks a
larger question because it searches across gene pairs, component counts, and
line assignments.  A sufficient uniform target has the form
\[
  \sup_{(g,h)\in\mathcal G^2}
  \sup_{K\in\mathcal K}
  \sup_{\ell\in\mathcal L_K}
  |(P_n-P)q_{\ell,gh}|
  \quad\hbox{small}.
\]
This bound controls the empirical risks uniformly:
\[
  \sup_{(g,h),K}
  |\hat Q_{gh}^{(K)}-Q_{gh}^{(K)}|
  \le
  \sup_{(g,h),K,\ell}|(P_n-P)q_{\ell,gh}|.
\]
If the normalizing risks \(Q_{gh}^{(0)}\) stay away from zero, the same control
transfers to
\[
  \sup_{(g,h),K}
  |\hat T_{gh}^{(K)}-T_{gh}^{(K)}|.
\]
Thus the selected heterogeneous gene-pair is credible only after the scanned
field has been named and controlled.  The detailed behavior of the fitted
\(K\)-line estimator belongs to the later M/Z-estimation and local-asymptotic
chapters; the role of this example here is to show why a high-throughput
biological discovery is an empirical-process problem before it is a single
correlation calculation.
\end{example}

For a working analyst, the translation is concrete.  Before trusting a
single-cell pipeline, name the field being scanned.  Are the indices genes,
cells, neighborhoods, pseudotime locations, batches, simulation settings, or
model parameters, or candidate line memberships?  Then name the functions
being compared.  Are they zero indicators, expression means, marker contrasts,
correlations, heterogeneous dependence scores, smooth trends, or losses?
Finally, ask whether the conclusion depends on one selected function or on the
whole class \(\mathcal F\).  The first question is an ordinary average.  The
second is an empirical-process problem.

The same advice can be turned into a working single-cell protocol.  The table
below is intentionally schematic: the software names are examples of the
statistical roles, not commands the reader must memorize.
\begin{description}[leftmargin=0pt,labelsep=0.55em,style=unboxed,font=\bookdescriptionlabelfont,itemsep=0.22\baselineskip]
\item[Name the assay field.]
The theoretical object is \(\{P_nf:f\in\mathcal F\}\).  Choose the summaries
the analysis will scan--means, zero rates, marker contrasts, correlations,
pseudotime trends, spatial neighborhoods--so that the field is visible before
the search begins.

\item[Use simulation as design.]
Fit a scDesign3-type generative law, simulate candidate donor, batch, depth,
condition, or modality designs, and check whether
\[
  \sup_{f\in\mathcal F}|P_n^{\mathrm{sim}}f-P_n^{\mathrm{obs}}f|
\]
is small for the summaries needed by the planned analysis.

\item[Estimate a biological trend.]
Fit a scGTM-type constrained count model along pseudotime or another ordered
cell coordinate.  The theoretical object is an \(\argmax_\theta M_n(\theta)\)
or a score root; the selected trend shape is a target of a stated criterion,
not a visual afterthought.

\item[Screen heterogeneous dependence.]
Use a generalized-correlation or \(K\)-lines-type analysis when a gene pair may
follow different linear relations in different latent states.  The theoretical
object is a selected functional \(\Phi_K(P_n;Y_g,Y_h)\), and the scientific
claim should say how the gene-pair and line-membership search was disciplined.

\item[Stress-test the conclusion.]
Repeat the key summaries under alternative normalizations, pseudotime
constructions, batches, and gene sets.  The biological statement is stronger
when it survives this reasonable indexed class of analyses.
\end{description}

The three methods therefore form one statistical sentence:
\[
\begin{gathered}
\text{observation: what was recorded and what was missed}\\
\longrightarrow\ \text{generation: what law could reproduce the assay}\\
\longrightarrow\ \text{inference: what structured biological target is learned}.
\end{gathered}
\]
scImpute, scDesign3, and scGTM are not scattered examples.  They are a focused
case study in the same theme that organizes the inference chapters: empirical
stability is what lets a high-dimensional biological pattern become an
inferential object.

\subsection{Historical Traces as Empirical Fields}
\label{sec:ch11-historical-traces}
\conceptindexes{historical traces, empirical fields, climate reconstruction, stylometric fields}

The historical examples from Chapter~2 fit the same mathematical pattern.  In
the Zhu/Chu Ko-chen climate thread, the index is not just time.  It may include
proxy type, region, season, archival source, and calibration model.  A class of
summaries \(\mathcal F\) may contain smoothed temperature contrasts, regional
averages, extreme-event indicators, or comparisons between warm and cold
periods.  Uniform stability asks whether conclusions are stable across the
whole class of proxy summaries, not merely at one chosen year.

The \emph{Dream of the Red Chamber} authorship problem is even closer to the
language of empirical processes.  Treat chapters as observations and let
\(\mathcal F\) contain function-word frequencies, character n-grams, sentence
length summaries, verse/prose indicators, or classifier scores.  Then a
stylometric contrast has the form
\[
  \Delta_n(\mathcal F)
  =
  \sup_{f\in\mathcal F}
  \left|
    P_{1:80}f-P_{81:120}f
  \right|.
\]
The statistical question is not whether this number can be made large by
searching over enough features.  It is whether a chosen feature class gives
stable evidence that two chapter groups behave like different samples from a
style distribution.  This is the cautious reading of quantitative work on the
novel: stylometry can organize evidence about Cao Xueqin, Gao E, editions, and
chapter groups, but it cannot avoid the modeling choices that turn literary
text into features \citep{hu2014redchamber,tu2013redchamber,zhu2021redchamber}.

The same warning applies to climate proxies.  Uniform laws control random
variation over a class of summaries.  They do not by themselves certify that
the summaries are historically unbiased.  Representation remains part of the
model.

\section{Application Map for Empirical Processes}
\label{sec:ch11-application-map}
\conceptindexes{application map, empirical-process grammar, multi-field applications}

The examples in this chapter differ in surface language, but they share the
same grammar.  An analyst chooses a class of fingerprints \(\mathcal F\); the
data produce the empirical field \(P_nf\); the scientific or operational claim
depends on whether the whole field is stable enough to survive scanning.  This
section is a map, not a second theory.  Each example names the observation, the
searched index set, the target, and the practical decision that would be
misleading if only the final selected statistic were analyzed.

\begin{example}[Online experiments as subgroup-time fields]
In an online experiment, the observed record may be
\[
  O_i=(A_i,Y_i,X_i,T_i),
\]
where \(A_i\in\{0,1\}\) is treatment assignment, \(Y_i\) is an outcome or
guardrail metric, \(X_i\) contains pre-treatment user features, and \(T_i\) is
time since exposure.  For a subgroup \(G\), horizon \(t\), and metric \(m\),
an inverse-probability contrast can be written as
\[
  f_{G,t,m}(O_i)
  =
  \ind{X_i\in G,T_i\le t}
  \left\{
    \frac{A_iY_{im}}{e}
    -
    \frac{(1-A_i)Y_{im}}{1-e}
  \right\},
  \qquad e=\Prob(A_i=1).
\]
The dashboard is the empirical field
\[
  \{\hat\tau_{G,t,m}=P_nf_{G,t,m}:(G,t,m)\in\mathcal I\}.
\]
If the launch decision uses the largest positive subgroup lift, the relevant
noise level is not the standard error of one prespecified cell.  It is
\[
  \sup_{(G,t,m)\in\mathcal I}
  |(P_n-P)f_{G,t,m}|.
\]
In practice this means that product teams should predefine the metric,
subgroup, and horizon index set; use simultaneous bands, resampling, or
multiplicity rules for the displayed field; and keep guardrail metrics inside
the same inferential object rather than checking them after a winner is found.
\end{example}

\begin{example}[Machine learning and model selection]
For supervised learning, let
\[
  R(f)=P\ell_f,\qquad R_n(f)=P_n\ell_f,
  \qquad
  \hat f\in\argmin_{f\in\mathcal F}R_n(f).
\]
The basic excess-risk inequality is
\[
  R(\hat f)-\inf_{f\in\mathcal F}R(f)
  \le
  2\sup_{f\in\mathcal F}|(P_n-P)\ell_f|,
\]
up to optimization error.  Thus the familiar generalization gap
\[
  \sup_{f\in\mathcal F}|P_n\ell_f-P\ell_f|
\]
is not decorative theory; it is what lets an empirical minimizer be read as a
population-risk minimizer.  A Rademacher version of the same control is
\[
  \Expect\sup_{f\in\mathcal F}|(P_n-P)\ell_f|
  \le
  2\Expect_\varepsilon
  \sup_{f\in\mathcal F}
  \left|
    \frac1n\sum_{i=1}^n\varepsilon_i\ell_f(O_i)
  \right|,
\]
where \(\varepsilon_i\)'s are iid signs.  VC dimension, entropy, margins, and
norm controls are different ways to make the right-hand side small enough
\citep{vapnik1971uniform,bartlett2017spectrally}.

The applied rule is plain: if a validation set is used to choose architectures,
thresholds, feature sets, or recommender policies, then the searched class is
the whole tuning path, not only the final model.  Off-policy recommendation
evaluation makes the same point with inverse-propensity weighted losses
\citep{schnabel2016recommendations}: the estimator is useful only if the policy
class being compared has been controlled as a class.
\end{example}

\begin{example}[Clinical safety as a hazard-curve field]
In a trial or pharmacovigilance system, a safety endpoint is often a marked
counting process.  Let \(N_{ia}(t)\) count adverse event type \(a\) for subject
\(i\), \(Y_i(t)\) be the at-risk indicator, and \(G_i\) a subgroup label.  A
subgroup-specific Nelson--Aalen-type curve is
\[
  \widehat\Lambda_{a,g}(t)
  =
  \int_0^t
  \frac{P_n\{\ind{G_i=g}N_{ia}(du)\}}
       {P_n\{\ind{G_i=g}Y_i(u)\}}.
\]
The safety review does not inspect one curve.  It scans
\[
  \left\{
    \widehat\Lambda_{a,g}(t)-\Lambda_{a,g}(t):
    a\in\mathcal A,\ g\in\mathcal G,\ 0\le t\le\tau
  \right\}.
\]
Uniform weak convergence of the numerator and denominator processes is what
supports simultaneous bands for the hazard or survival curves, while
multiplicity control decides which event-subgroup pairs may be escalated.

The practical effect is the difference between a noisy safety table and an
auditable signal-detection workflow.  Before a data-monitoring committee sees
the dashboard, the event dictionary, subgroup set, time window, and escalation
rule should be part of the indexed field.  For recurrent events or multistate
outcomes, the same logic leads to the panel-count, U-process, and
semi-Markov examples earlier in this chapter.
\end{example}

\begin{example}[Finance: value-at-risk as a quantile field]
Let \(R_i\in\R^d\) be daily asset returns and \(w\) a portfolio weight vector.
For loss \(L_i(w)=-w^TR_i\), the population value-at-risk is
\[
  q_\alpha(w)=\inf\{x:\Prob(L(w)\le x)\ge\alpha\}.
\]
The empirical version \(\hat q_\alpha(w)\) is a quantile process indexed by
both \(\alpha\) and \(w\).  Uniform control can be phrased through the halfspace
class
\[
  \mathcal F
  =
  \{\ind{-w^TR\le x}:w\in\mathcal W,\ x\in\R\},
\]
because \(\hat q_\alpha(w)\) is found by inverting
\[
  F_{n,w}(x)=P_n\ind{-w^TR\le x}.
\]
A risk desk that searches portfolios, tail levels, and stress windows is
therefore using a whole empirical quantile field, not one scalar VaR number
\citep{jpmorgan1996riskmetrics}.

The operational consequence is familiar: backtesting one reported VaR level is
not enough if the number was chosen after scanning many portfolios or horizons.
The portfolio class \(\mathcal W\), the tail levels, and the stress-window
rules define the statistical object.  Confidence bands or stress limits should
be calibrated for that object, especially when the selected portfolio is the
worst-looking one.
\end{example}

\begin{example}[Quant trading as an empirical-process problem]
The public literature around quantitative trading gives a concrete version of
the same lesson.  Citadel, Morgan Stanley, Renaissance Technologies, and Jane
Street make the institutional setting recognizable, but the manuscript should
not pretend to know proprietary systems.  The mathematical object we can study
is public: a large class of data-dependent trading rules whose performance is
estimated after searching over signals, assets, horizons, transaction-cost
assumptions, and risk limits.

At the daily or intraday return-prediction scale, let
\[
  O_t=\{(R_{t+1,j},X_{t,j},C_{t,j}):j\in\mathcal A_t\},
  \qquad t=1,\ldots,n,
\]
where \(R_{t+1,j}\) is the next-period return of asset \(j\),
\(X_{t,j}\) contains characteristics or signals observable at time \(t\), and
\(C_{t,j}\) records tradability, costs, borrow constraints, or liquidity.  A
model \(f_\theta\) turns signals into scores
\[
  s_{\theta,t,j}=f_\theta(X_{t,j}),
\]
and a portfolio rule maps the score vector into predictable weights
\[
  w_{\theta,t}
  =
  \Pi_{\mathcal C_t}\{A_t s_{\theta,t}\},
\]
where \(A_t\) may be a covariance or risk-scaling matrix and
\(\Pi_{\mathcal C_t}\) projects onto leverage, neutrality, turnover, and
capacity constraints.  The next-period net performance is not raw return but
\[
  G_{t+1}(\theta)
  =
  w_{\theta,t}^{T}R_{t+1}
  -
  \kappa_t\{w_{\theta,t}-w_{\theta,t-1}\}
  -
  \lambda\, w_{\theta,t}^{T}\widehat\Sigma_t w_{\theta,t},
\]
with transaction-cost functional \(\kappa_t\) and an inventory or volatility
penalty.  The statistical target for a fixed rule is
\[
  \psi(\theta)=\Expect G_{t+1}(\theta),
  \qquad
  \hat\psi_n(\theta)=\frac1n\sum_{t=1}^n G_{t+1}(\theta).
\]
If \(\theta\) ranges over neural networks, trees, shrinkage regressions,
portfolio constraints, and rebalancing horizons, the selected backtest
\[
  \sup_{\theta\in\Theta}\hat\psi_n(\theta)
\]
is an extreme value of an empirical field, not an estimate for one prespecified
strategy.  The relevant uncertainty object is therefore
\[
  \left\{
    \sqrt n\bigl(\hat\psi_n(\theta)-\psi(\theta)\bigr):
    \theta\in\Theta
  \right\},
\]
with time dependence handled by blocking, sample splitting by calendar period,
or other dependence-aware resampling.  This is the precise sense in which
data-snooping bias becomes an empirical-process problem
\citep{loMacKinlay1990dataSnooping}.  The machine-learning asset-pricing
literature studies a public version of this score-building problem, where the
signals include momentum, liquidity, volatility, valuation, and other
cross-sectional predictors \citep{guKellyXiu2020mlAssetPricing}.

Pairs trading gives a smaller, fully visible rule.  In a formation window,
choose a pair \((a,b)\) and hedge ratio \(\hat\beta\), then monitor the spread
\[
  S_t=\log P_{t,a}-\hat\beta\log P_{t,b},
  \qquad
  Z_t=\frac{S_t-\hat\mu}{\hat\sigma}.
\]
A threshold rule with parameter \(\tau\) enters a dollar-neutral position when
\(|Z_t|>\tau\), exits near zero, and reports net out-of-sample P\&L after
transaction costs.  The target is not simply whether \(S_t\) looks
mean-reverting in the same window used to choose the pair.  It is the
out-of-sample net value of a selected rule over the pair universe and threshold
grid, as in the relative-value study of
\citet{gatevGoetzmannRouwenhorst2006pairs}.  Again the statistic is a maximum
over a searched class.

Market making makes the observation mechanism still sharper.  In a simplified
limit-order-book model, let \(S_t\) be the midprice, \(q_t\) inventory, and let
buy and sell fills have intensities
\[
  \lambda^\pm(\delta)=A\exp(-k\delta),
\]
where \(\delta\) is the distance of the quote from the midprice.  The
Avellaneda--Stoikov model chooses reservation price and spread of the form
\[
  r_t=S_t-q_t\gamma\sigma^2(T-t),
  \qquad
  \Delta_t=\gamma\sigma^2(T-t)+\frac{2}{\gamma}\log(1+\gamma/k),
\]
and quotes \(r_t-\Delta_t/2\) and \(r_t+\Delta_t/2\) under a stylized utility
criterion \citep{avellanedaStoikov2008hft}.  The data object is now order-book
states, quotes, cancellations, fills, inventory, and realized cash; the target
is a fill-adjusted expected utility or risk-adjusted P\&L; and the limitation
is that the act of quoting changes the future observation process.  This is
why a high-frequency backtest cannot be read as iid prediction accuracy.

Panageas's finance work supplies the complementary target lesson.  A trading
desk or hedge fund is not only a forecasting machine; it is embedded in
contracts and equilibrium.  With high-water-mark compensation, the manager's
state includes fund value \(X_t\) and high-water mark
\[
  H_t=\sup_{s\le t}X_s,
\]
and the objective is a dynamic portfolio-choice problem for incentive-fee
payments, not merely \(\max_\theta\Expect G_{t+1}(\theta)\).  Panageas and
Westerfield show that this contract can lead to disciplined long-horizon
portfolio choice rather than unbounded risk taking
\citep{panageasWesterfield2009highWater}.  In asset-pricing models with
heterogeneous agents or technological adoption, the state also includes a
distribution of investors, constraints, and growth shocks
\citep{panageas2020heterogeneity,garleanuPanageasYu2012technology}.  For this
book, the message is structural: Chapter~5 names the observation mechanism,
Chapter~8 names the target, and this chapter controls the searched field.  The
statistical deliverable is therefore not a story about famous firms; it is a
reproducible target, a documented search class, and uncertainty for the selected
procedure under market frictions.
\end{example}

\begin{example}[Text and single-cell biology: selected feature fields]
Stylometry begins by turning a chapter, author segment, or edition into a
feature vector.  If \(X_c\) is a chapter and \(\phi_j(X_c)\) is the frequency
of word, character \(n\)-gram, or syntactic feature \(j\), then a two-group
contrast has the form
\[
  T_j
  =
  P_n\left[
    \frac{\ind{G=A}}{\hat p_A}\phi_j(X)
    -
    \frac{\ind{G=B}}{\hat p_B}\phi_j(X)
  \right].
\]
The statistical claim is rarely about a prespecified word.  It is about a
selected feature, score, or classifier drawn from a large textual field.  A
permutation analysis or holdout classifier should therefore be calibrated over
the feature set actually searched, not only over the final displayed contrast
\citep{hu2014redchamber,tu2013redchamber,zhu2021redchamber}.

Single-cell biology has the same shape with different features.  The
generalized Pearson correlation squares and \(K\)-lines clustering of
\citet{liZhouBickelTong2024geneHeterogeneity} can be summarized as an indexed
field
\[
  \{\hat T_{gh}^{(K)}:(g,h)\in\mathcal G^2,\ K\in\mathcal K\},
\]
where \(\hat T_{gh}^{(K)}\) measures whether genes \(g\) and \(h\) show a
mixture of linear dependences.  In a real analysis, the output is not simply
``this gene pair is interesting.''  It is a selected gene-pair and component
structure after searching many pairs and several \(K\)'s.  The landing rule is
the same as in stylometry: state the searched feature field, reserve data or
resampling for validation, and attach uncertainty to the selected pattern
rather than pretending it was fixed in advance.
\end{example}

\begin{tcolorbox}[
  enhanced,
  breakable,
  colback=noteback,
  colframe=bookgold!75!black,
  boxrule=0.55pt,
  arc=4pt,
  boxsep=1pt,
  left=0.95em,
  right=0.9em,
  top=0.7em,
  bottom=0.7em,
  before skip=0.9\baselineskip,
  after skip=1.0\baselineskip
]
\noindent\textbf{How the examples line up.}
The empirical cdf uses cutpoints as fingerprints.  The nonmeasurability example
shows why the fingerprint field must live in a legitimate measurable space.
The Chapter~10 Billingsley proof explains how finite-dimensional convergence
becomes a process limit in the cdf case.  The memorizing class shows what
happens when the field is too large.  Event-time, single-cell, climate, text,
industrial, boosting, and neural-network examples then differ mainly in what
their index set represents: time, threshold, feature, model, subgroup, state,
or policy.  This is the chapter's main claim: empirical-process theory is not
an extra layer placed on top of applications; it is the grammar that says when
a searched empirical pattern can still be interpreted as evidence about a
population feature.
\end{tcolorbox}

\section{Modern Learning Examples}
\label{sec:ch11-modern-learning-examples}
\conceptindexes{Vapnik--Chervonenkis theory, support vector machines, structural risk minimization, boosting, neural networks, deep learning theory, ReLU networks, margins, linear regions}

The application map included learning as one field among many.  This final
section zooms in because modern learning is where the chapter's vocabulary is
especially visible: a training algorithm searches a loss class, and the
statistical question is whether empirical risk, margin, or validation
performance remains meaningful after that search.  Vapnik's route gives the
classical uniform-law entry point; boosting and deep networks show how the
same question survives in more flexible algorithmic forms.

\subsection{Vapnik's margin machine: from uniform laws to support vector machines}
\conceptindexes{support vector machines, structural risk minimization, margin bounds, hinge loss, kernel machines}

Vapnik's statistical-learning program is one of the cleanest places where
empirical-process theory becomes an algorithm.  The starting point is empirical
risk minimization.  For \(O=(X,Y)\), \(Y\in\{-1,1\}\), and a classifier \(g\),
the population and empirical misclassification risks are
\[
  R(g)=P\ind{Yg(X)\le0},
  \qquad
  R_n(g)=P_n\ind{Yg(X)\le0}.
\]
If a learning rule searches a class \(\mathcal G\), the basic Vapnik--
Chervonenkis question is exactly the question of this chapter:
\[
  \sup_{g\in\mathcal G}|R_n(g)-R(g)|
  =
  \sup_{g\in\mathcal G}
  |(P_n-P)\ind{Yg(X)\le0}|.
\]
VC dimension, growth functions, and later Rademacher averages give ways to
control this supremum \citep{vapnik1971uniform,vapnik1998statistical}.  The
lesson is not merely that a classifier should fit the training sample.  The
class being searched must be small enough, or structured enough, that training
error still says something about prediction error.

Support vector machines (SVMs) turn that principle into a concrete
optimization problem \citep{boser1992training,cortes1995support}.  Let
\(\Phi(x)\) be a feature map into a Hilbert space \(\mathcal H\) with kernel
\(K(x,x')=\langle\Phi(x),\Phi(x')\rangle\), and consider affine scores
\[
  f_{w,b}(x)=\langle w,\Phi(x)\rangle+b.
\]
For separable data, the hard-margin SVM solves
\[
  \min_{w,b}\frac12\|w\|_{\mathcal H}^2
  \qquad\text{subject to}\qquad
  Y_i f_{w,b}(X_i)\ge1,\quad i=1,\ldots,n.
\]
The constraint forces zero training error, but the objective chooses the
largest geometric margin \(1/\|w\|_{\mathcal H}\).  This is structural risk
minimization in geometric form: among separating classifiers, choose the one
whose effective class is controlled by a norm or margin scale.

With nonseparable data, the soft-margin SVM minimizes the regularized hinge
criterion
\[
  P_n(1-Yf_{w,b}(X))_+ + \lambda \|w\|_{\mathcal H}^2,
  \qquad (u)_+=\max(u,0).
\]
For a fixed radius \(B\), define the searched loss class
\[
  \mathcal F_B
  =
  \{(x,y)\mapsto (1-yf_{w,b}(x))_+:
    \|w\|_{\mathcal H}\le B,\ |b|\le B_0\}.
\]
If \(K(x,x)\le \kappa^2\), this class has Rademacher complexity of order
\[
  \frac{B\kappa}{\sqrt n}
\]
up to constants and the intercept term.  Therefore the generalization gap is
controlled by the radius of the selected function class, not by the ambient
dimension of the feature representation.  The representer theorem then explains
why the infinite-dimensional problem is computationally finite
\citep{scholkopf2001generalized}:
\[
  \hat w=\sum_{i=1}^n \alpha_iY_i\Phi(X_i),
\]
so the fitted score depends only on kernel evaluations at the training records.

This is why SVMs belong in an empirical-process chapter.  The visible
algorithm is a quadratic program; the statistical engine is uniform control of
an indexed loss field.  The practical message is also plain.  Kernel choice,
\(\lambda\), margin scale, and the tuning path are part of the searched class.
When cross-validation chooses among kernels and penalty values, the object is
not one trained SVM but the empirical field generated by the whole tuning
family.

\subsection{Boosting as empirical-process search}
\conceptindexes{boosting, AdaBoost, margin distribution, exponential loss}

Boosting is a good modern example because it looks algorithmic before it looks
statistical.  Let \(Y\in\{-1,1\}\), let \(\mathcal H\) be a class of base
classifiers \(h:\mathcal X\to\{-1,1\}\), and consider additive voting rules
\[
  F_a(x)=\sum_{m=1}^M a_m h_m(x),
  \qquad h_m\in\mathcal H,\quad a_m\ge0,\quad \sum_m a_m\le B .
\]
The class of possible losses is
\[
  \mathcal F_B
  =
  \{(x,y)\mapsto \phi(yF_a(x)):F_a\text{ as above}\},
\]
where \(\phi(u)=e^{-u}\) gives the exponential-loss view of AdaBoost and
\(\phi(u)=\log(1+e^{-u})\) gives the logistic-loss view.  Training decreases
\[
  P_n\phi(YF(X)),
\]
but the statistical question is the uniform one:
\[
  \sup_{F_a}
  \left|
    P_n\phi(YF_a(X))-P\phi(YF_a(X))
  \right|.
\]
Thus boosting is an empirical-process problem over the \(\ell_1\)-hull of a
base class, not merely a clever reweighting trick.

The more interesting object is the margin
\[
  \rho_i=\frac{Y_iF_a(X_i)}{\sum_m a_m}.
\]
After the training error reaches zero, the algorithm can still change the
margin distribution.  The classical margin explanation says that test error is
controlled not only by whether the training labels are correct, but by how
strongly the vote separates them from the decision boundary
\citep{schapire1998boosting}.  Friedman, Hastie, and Tibshirani's statistical
view then reads boosting as stagewise additive modeling in function space
\citep{friedman2000additive}.  In the language of this chapter, shrinkage,
early stopping, and base-class restrictions are ways of keeping the searched
empirical field from becoming a memorizing class.

\subsection{Deep ReLU networks as empirical fields with geometry}
\conceptindexes{ReLU networks, linear regions, Montufar, neural tangent kernel, norm bounds}

For a neural network, the indexed class is no longer a small parametric family
in any practical sense.  A fixed architecture and a parameter constraint
produce
\[
  \mathcal F_{\mathrm{net}}
  =
  \{(x,y)\mapsto \ell(y,f_\theta(x)):\theta\in\Theta_B\},
\]
and training minimizes
\[
  P_n\ell(Y,f_\theta(X))+\lambda_n J(\theta).
\]
The empirical-process question remains familiar:
\[
  \sup_{\theta\in\Theta_B}
  \left|
    P_n\ell(Y,f_\theta(X))-P\ell(Y,f_\theta(X))
  \right|,
\]
but the geometry of \(\mathcal F_{\mathrm{net}}\) is very different from the
geometry of linear regression.

For ReLU networks, \(f_\theta\) is piecewise affine in the input.  Activation
patterns divide \(\R^d\) into linear regions, and depth can multiply the number
of such regions.  If a network has input dimension \(d\) and hidden-layer
widths \(n_1,\ldots,n_L\), one lower bound from
\citet{montufar2014number} has the form
\[
  \left\{
    \prod_{\ell=1}^{L-1}
    \left\lfloor\frac{n_\ell}{d}\right\rfloor^d
  \right\}
  \sum_{j=0}^d \binom{n_L}{j},
\]
under the usual width assumptions.  This is why the Montufar line of work is a
natural example here: a deep network is not just a large vector of parameters;
it is a learned arrangement of many affine charts on the input space.
Expressivity is geometric before it is asymptotic.

The statistical warning is equally important.  Large expressivity alone does
not explain generalization; deep networks can fit random labels
\citep{zhang2017understanding}.  Modern theory therefore studies additional
structure: norm and margin controls \citep{bartlett2017spectrally}, kernel or
linearized regimes such as the neural tangent kernel
\citep{jacot2018neural}, implicit bias of optimization, and architecture-driven
geometry.  The empirical-process lens does not solve all of deep learning.  It
does give a clean diagnostic question: which constrained field is being
searched, and which part of the theory keeps that search from confusing
memorization with signal?

\begin{tcolorbox}[
  enhanced,
  breakable,
  colback=chaptercream,
  colframe=bookblue!75!black,
  boxrule=0.55pt,
  arc=4pt,
  boxsep=1pt,
  left=0.95em,
  right=0.9em,
  top=0.7em,
  bottom=0.7em,
  before skip=0.9\baselineskip,
  after skip=1.0\baselineskip
]
\noindent\textbf{Where this chapter leaves the reader.}
The chapter has done one job for the book's larger argument.  It has taken the
ordinary empirical average \(P_nf\) and asked what happens when \(f\) is chosen
from a class after seeing data.  Glivenko--Cantelli theory controls the whole
field at law-of-large-numbers scale; Donsker theory and stochastic
equicontinuity control its fluctuation; measurability and path topology keep
the field honest as a random object.  Chapter~12 studies one important random
curve more geometrically, Chapter~13 uses the same uniform control to locate
peaks and roots, Chapter~15 turns local expansions into uncertainty, and
Chapter~16 develops the continuous-time machinery that the survival examples
only previewed.
\end{tcolorbox}

\section{Exercises}
\label{sec:ch11-exercises}
\conceptindexes{empirical-process exercises, entropy exercises, stochastic-equicontinuity exercises}

\begin{exercise}[A finite class]
Let $\mathcal F=\{f_1,\ldots,f_k\}$ with $P|f_j|<\infty$ for each $j$.  Show
that $\mathcal F$ is $P$-Glivenko--Cantelli.
\end{exercise}

\begin{exercise}[Indicators of half-lines]
Use the bracketing proof to show that
$\{\ind{x\le t}:t\in\R\}$ is $P$-Glivenko--Cantelli for every probability
distribution $P$ on $\R$.
\end{exercise}

\begin{exercise}[Lipschitz class entropy]
Suppose $\Theta\subset\R^d$ is bounded and
$|f_\theta-f_{\theta'}|\le\|\theta-\theta'\|H$ with $PH^r<\infty$.  Derive a
polynomial upper bound for
$N(\varepsilon\|H\|_{P,r},\mathcal F,L_r(P))$.
\end{exercise}

\begin{exercise}[SVM margin and empirical-process control]
Let \(Y\in\{-1,1\}\), let \(K(x,x)\le\kappa^2\), and consider kernel scores
\(f_w(x)=\langle w,\Phi(x)\rangle\) with \(\|w\|_{\mathcal H}\le B\).  Explain
why the hinge-loss class
\[
  \{(x,y)\mapsto (1-yf_w(x))_+:\|w\|_{\mathcal H}\le B\}
\]
is the object whose uniform deviation controls the soft-margin SVM's
generalization gap.  Then use Cauchy--Schwarz to show that
\(|f_w(x)|\le B\kappa\).  Why does this bound depend on the margin/norm scale
rather than directly on the dimension of the feature map?
\end{exercise}

\begin{exercise}[Boosting margins]
For additive classifiers \(F_a=\sum_m a_mh_m\) with \(a_m\ge0\) and
\(\sum_m a_m\le B\), explain why bounding \(B\) restricts the searched field.
Then compare two fitted rules with the same training error but different
empirical margin distributions.  Which one should be more stable, and why?
\end{exercise}

\begin{exercise}[ReLU regions and statistical control]
For a one-hidden-layer ReLU network on \(\R\), draw the linear regions created
by three hidden units.  Then explain why counting regions is an expressivity
statement, while a uniform law for the loss class is a generalization
statement.  Why are these related but not identical?
\end{exercise}

\begin{exercise}[Uniform consistency of criteria]
Let $M_n(\theta)=P_nm_\theta$ and $M(\theta)=Pm_\theta$.  Suppose
$\{m_\theta:\theta\in\Theta\}$ is $P$-Glivenko--Cantelli and $M$ has a unique
well-separated minimizer $\theta_0$.  Show that every approximate minimizer of
$M_n$ is consistent for $\theta_0$.
\end{exercise}

\begin{exercise}[Outer probability]
Give an example of a class $\mathcal F$ for which
$\sup_{f\in\mathcal F}|P_nf-Pf|$ is trivially measurable, and explain why a
countable dense subclass often repairs measurability.
\end{exercise}

\begin{exercise}[Single-cell simulation and trend checking]
Suppose a pilot single-cell experiment has metadata for donor, batch,
condition, and pseudotime.  Choose a feature class \(\mathcal F\) containing at
least one zero-rate summary, one marker-mean contrast, one correlation summary,
and one smooth pseudotime trend.  Describe how you would use a fitted
scDesign3-type simulator to compare two candidate follow-up designs, and then
explain how a scGTM-type trend fit would be checked against the same
\(\mathcal F\).  Which parts of your conclusion are statements about
observation, generation, and inference?
\end{exercise}

\begin{exercise}[Heterogeneous gene-pair screening]
Suppose an assay measures \(p\) genes and the analyst wants to screen pairs for
a mixture of \(K\) linear dependences, as in generalized Pearson correlation
squares and \(K\)-lines clustering.  Write a feature or functional class whose
indices include \((g,h)\), \(K\), and the line-assignment rule.  Explain what
would be proved by a fixed-pair asymptotic normal approximation, and what
additional question is raised by selecting the largest score over many gene
pairs.
\end{exercise}

\section*{Sources and Further Reading}
\addcontentsline{toc}{section}{Sources and Further Reading}

The chapter is a compact bridge from empirical distribution functions and weak
convergence to the empirical-process tools used later in the book.  The main
monographs are
\citet{pollard1984convergence}, \citet{pollard1990empirical}, and
\citet{vaart2023weak}.  The chapter's route through Glivenko--Cantelli and
Donsker classes, outer probability, bracketing, stochastic equicontinuity, and
survival-process examples is also informed by
\citet{dabrowskaAdvancedProbabilityCommunication} and
\citet{dabrowskaStochasticProcessesCommunication}.  The panel-count example is
based on \citet{wellnerZhang2000panel}; the bivariate survival-surface example
uses \citet{dabrowska1988kaplan} and \citet{dabrowska1989kaplan}.  The
U-process remarks use Hoeffding's projection idea \citep{hoeffding1948class},
Dabrowska's renewal and semi-Markov regression examples
\citep{dabrowska2009twoStageRenewal,dabrowska2012semiMarkovTransformation},
and the empirical-process treatment in \citet{vaart2023weak}.  Dudley's entropy integral
\citep{dudley1967sizes} is included only as a signpost toward chaining:
covering numbers control the size and continuity of Gaussian limits, while
bracketing entropy gives a convenient empirical-process sufficient condition.
The path-space discussion follows
\citet{skorokhod1956limit} and \citet{billingsley1999convergence}; the
nonmeasurability warning is a concrete reason that the topology on
\(D[0,1]\) cannot be chosen casually.  Donsker's theorem originates with
\citet{donsker1952justification}, correcting the heuristic finite-dimensional
argument in \citet{doob1949heuristic}; see also
\citet{pollard1982beyond}.  The Glivenko--Cantelli theorem begins with
\citet{glivenko1933sulla} and \citet{cantelli1933sulla}.  The Vapnik/SVM
example uses the VC uniform-convergence paper
\citep{vapnik1971uniform}, Vapnik's statistical-learning synthesis
\citep{vapnik1998statistical}, the optimal-margin and support-vector papers
\citep{boser1992training,cortes1995support}, and the representer-theorem
formulation of \citet{scholkopf2001generalized}.  The boosting example
uses the margin explanation of \citet{schapire1998boosting} and the additive
model interpretation of \citet{friedman2000additive}.  The deep-network example
uses the piecewise-linear expressivity theory of \citet{montufar2014number},
the random-label warning of \citet{zhang2017understanding}, the margin-bound
view of \citet{bartlett2017spectrally}, and the neural-tangent-kernel view of
\citet{jacot2018neural}.  For efficient and semiparametric uses of empirical
processes, see \citet{bickel1993efficient}.  The single-cell thread uses scImpute
\citep{li2018scimpute}, scDesign3, scGTM
\citep{cui2022scgtm}, and the generalized Pearson correlation squares and
\(K\)-lines clustering of \citet{liZhouBickelTong2024geneHeterogeneity} as
modern examples of the same structure: an observed assay, a generative
mechanism, and an inferred biological map or heterogeneous dependence pattern
must be compared through indexed families of summaries rather than through one
fixed statistic.  The historical-traces
section uses Zhu/Chu Ko-chen's climate reconstruction and later climate
reconstruction work \citep{chu1973climatic,ge2013temperature,ge2016recent},
and the \emph{Dream of the Red Chamber} stylometric studies
\citep{hu2014redchamber,tu2013redchamber,zhu2021redchamber}, as examples in
which the feature class is part of the statistical claim.

%% file: chapters/ch12_karhunen_loeve_functional_data.tex
\chapter{Karhunen--Loeve Expansions and Functional Data}
\label{chap:karhunen-loeve-functional-data}
\conceptindexes{functional data, random curves, covariance operators, Karhunen--Loeve expansion, Brownian motion, Brownian bridge, functional principal components, maximal correlation}

\begin{tcolorbox}[
  enhanced,
  breakable,
  colback=chaptercream,
  colframe=bookblue!88!black,
  boxrule=0.72pt,
  arc=5pt,
  boxsep=1pt,
  left=1.0em,
  right=0.95em,
  top=0.82em,
  bottom=0.82em,
  before skip=0.55\baselineskip,
  after skip=1.0\baselineskip
]
\noindent\textbf{Chapter overview.}
In Part~IV, this chapter has one job: representation.  Chapters~10 and~11 made
random objects and empirical fields converge; some of their most important
limits are curves rather than scalars.  Before an estimator chooses a point,
and before uncertainty is attached to that point, the curve itself has to be
read without being flattened too early.  The Karhunen--Loeve expansion
decomposes a square-integrable random function into orthogonal modes determined
by its covariance operator.  For Brownian motion and the Brownian bridge the
modes can be written down exactly; in functional data analysis they become
estimated principal component curves and subject-level scores.  The examples
serve this single bridge:
\[
  \hbox{random curve}
  \longrightarrow
  \hbox{covariance modes}
  \longrightarrow
  \hbox{scores or interpretable coordinates}.
\]
Everything below serves that bridge.  The abstract theorem gives the operator
language, the Brownian examples show what happens to empirical-process limits,
the path-geometry interlude marks what covariance representation does not
claim, and the functional-data examples show how the representation is
estimated and used.
\end{tcolorbox}

The opening chapters insisted on a sequence:
\[
  \text{data structure}
  \longrightarrow
  \text{assumptions}
  \longrightarrow
  \text{model}
  \longrightarrow
  \text{inference task}.
\]
This chapter is one of the cleanest places to see that sequence.  A glucose
monitor, a wearable accelerometer, a tumor-burden profile, a drug-concentration
curve, or an empirical distribution function is not naturally a single number.
It is a function of time, dose, space, or probability level.  Once the data
object is a curve, the model must say what kinds of curve variation are
scientifically meaningful and what kinds are measurement noise, timing error,
or nuisance structure.

Karhunen--Loeve theory gives one answer.  It says that a second-order random
curve can be decomposed into covariance modes.  Functional data analysis uses
the same idea statistically: estimate the mean curve, estimate the covariance,
find its leading eigenfunctions, and represent each subject by a few scores.
The compact-operator facts used for this decomposition are collected in
\Appref{sec:appC-compact-spectral}; this chapter uses only that reading
version, not a full treatment of functional analysis.
The trick is powerful because it does not flatten the curve too early.  It
lets the curve remain a curve until the last responsible moment.
That is the chapter's scope: not every theorem about curves, and not every
method for functional data, but the translation from a random function to a
coordinate system that later inference can inherit.

\begin{tcolorbox}[
  enhanced,
  breakable,
  colback=noteback,
  colframe=bookgold!75!black,
  boxrule=0.55pt,
  arc=4pt,
  boxsep=1pt,
  left=0.95em,
  right=0.9em,
  top=0.7em,
  bottom=0.7em,
  before skip=0.9\baselineskip,
  after skip=1.0\baselineskip
]
\noindent\textbf{The chapter's input and output.}
The input is a random curve, either because an empirical process has a curve as
its weak limit or because the observed data are themselves trajectories,
profiles, spectra, or longitudinal records.  The output is not yet a final
scientific conclusion.  It is a representation:
\[
  X(t)
  =
  \mu(t)+\sum_{k\ge1}\xi_k\phi_k(t),
\]
with a mean function, covariance operator, eigenfunctions, and subject-level
scores.  Later chapters may optimize, test, regress, predict, or attach
uncertainty to those scores or to functionals of the original curve.  This
chapter explains the representation step that makes those later moves
meaningful.
\end{tcolorbox}

\section{Random Curves and Covariance Operators}
\label{sec:ch12-random-curves-covariance}
\conceptindexes{random curves, covariance operator, eigenfunctions, functional principal component scores, maximal correlation, conditional expectation operator}

Let \(\mathcal T=[0,1]\) for simplicity, and let \(X\) be a random curve with
\[
  \Expect\int_0^1 X(t)^2\,dt<\infty .
\]
Write
\[
  \mu(t)=\Expect X(t),
  \qquad
  K(s,t)=\Cov\{X(s),X(t)\}.
\]
The covariance kernel defines an operator on \(L^2[0,1]\):
\[
  (\mathcal K f)(s)=\int_0^1 K(s,t)f(t)\,dt .
\]
Under the square-integrability condition above, \(\mathcal K\) is a compact,
self-adjoint, positive operator.  This is the functional version of a covariance
matrix, in the precise sense summarized in
\Appref{sec:appC-compact-spectral}.  The finite-dimensional matrix formula
\[
  \Sigma v_k=\lambda_k v_k
\]
has become
\[
  \mathcal K\phi_k=\lambda_k\phi_k.
\]
The eigenvectors are now eigenfunctions.

\begin{definition}[Functional principal component scores]
Let \((\lambda_k,\phi_k)\) be the nonzero eigenvalue-eigenfunction pairs of
\(\mathcal K\), ordered so that
\[
  \lambda_1\ge\lambda_2\ge\cdots>0,
  \qquad
  \|\phi_k\|_{L^2}=1.
\]
The \(k\)th functional principal component score is
\[
  \xi_k=\int_0^1 \{X(t)-\mu(t)\}\phi_k(t)\,dt.
\]
\end{definition}

These scores are random variables.  The functions \(\phi_k\) are deterministic
directions of variation.  The curve is therefore split into two roles: common
shape directions and subject-specific coordinates along those directions.  That
split is the chapter's main translation move: a path remains a path, but it
also becomes a coordinate system that later inference can use.

\begin{theorem}[Karhunen--Loeve expansion]
\label{thm:ch12-kl}
Let \(X\) be a square-integrable random element of \(L^2[0,1]\), with mean
\(\mu\) and covariance operator \(\mathcal K\).  Let
\((\lambda_k,\phi_k)\) be the nonzero eigenvalue-eigenfunction pairs of
\(\mathcal K\).  Define
\[
  \xi_k=\langle X-\mu,\phi_k\rangle
  =
  \int_0^1\{X(t)-\mu(t)\}\phi_k(t)\,dt .
\]
Then
\[
  \Expect \xi_k=0,
  \qquad
  \Cov(\xi_j,\xi_k)=\lambda_k\ind{j=k},
\]
and
\[
  X(t)=\mu(t)+\sum_{k\ge1}\xi_k\phi_k(t)
\]
with convergence in \(L^2(\Omega;L^2[0,1])\).  Moreover,
\[
  \Expect\left\|
    X-\mu-\sum_{k=1}^m\xi_k\phi_k
  \right\|_{L^2}^2
  =
  \sum_{k>m}\lambda_k .
\]
If \(X\) is Gaussian, the scores \(\xi_k\) are independent normal random
variables with \(\xi_k\sim \Normal(0,\lambda_k)\).
\end{theorem}

\noindent\textit{Proof.}
The covariance operator is compact, self-adjoint, and positive, so the spectral
theorem gives an orthonormal eigenbasis for the closed range generated by
\(X-\mu\).  For the scores,
\[
  \Expect\xi_k
  =
  \left\langle \Expect(X-\mu),\phi_k\right\rangle
  =
  0.
\]
Also,
\[
\begin{aligned}
  \Cov(\xi_j,\xi_k)
  &=
  \Expect\{\langle X-\mu,\phi_j\rangle
          \langle X-\mu,\phi_k\rangle\}  \\
  &=
  \langle \mathcal K\phi_j,\phi_k\rangle
   =
  \lambda_j\langle\phi_j,\phi_k\rangle
   =
  \lambda_j\ind{j=k}.
\end{aligned}
\]
This is the displayed covariance formula, since the value is zero unless
\(j=k\).  Parseval's identity then gives
\[
  \Expect\|X-\mu\|_{L^2}^2=\sum_{k\ge1}\lambda_k
\]
and, after removing the first \(m\) orthogonal components,
\[
  \Expect\left\|
    X-\mu-\sum_{k=1}^m\xi_k\phi_k
  \right\|_{L^2}^2
  =
  \sum_{k>m}\lambda_k .
\]
If \(X\) is Gaussian, every finite vector \((\xi_1,\ldots,\xi_m)\) is
multivariate normal.  Its covariance matrix is diagonal, so its coordinates
are independent. \qedmark

The theorem is usually named after \citet{karhunen1947lineare} and
\citet{loeve1978probability}.  The operator language is the modern way to
teach it; see \citet{hsingEubank2015fda}.  When the covariance kernel is
continuous, Mercer's theorem gives a corresponding kernel expansion
\citep{mercer1909functions},
\[
  K(s,t)=\sum_{k\ge1}\lambda_k\phi_k(s)\phi_k(t),
\]
with convergence in the appropriate continuous-kernel sense.  This identity is
often the more visible form in applications, because the covariance surface is
what we estimate from data.

\begin{proposition}[Why the first modes are the best low-dimensional summary]
\label{prop:ch12-best-basis}
Among all \(m\)-dimensional linear summaries of a centered random curve in
\(L^2[0,1]\), the span of \(\phi_1,\ldots,\phi_m\) minimizes mean squared
reconstruction error.  The minimum error is
\[
  \sum_{k>m}\lambda_k .
\]
\end{proposition}

\noindent\textit{Proof.}
Let \(\psi_1,\ldots,\psi_m\) be any orthonormal system and let \(P_\psi\) be
projection onto its span.  For \(Y=X-\mu\),
\[
  \Expect\|Y-P_\psi Y\|^2
  =
  \Expect\|Y\|^2-\sum_{\ell=1}^m\Expect\langle Y,\psi_\ell\rangle^2
  =
  \tr(\mathcal K)
  -
  \sum_{\ell=1}^m\langle\mathcal K\psi_\ell,\psi_\ell\rangle .
\]
Thus minimizing the error is the same as maximizing the second sum.  The
Rayleigh--Ritz principle for compact self-adjoint operators says that this
maximum is \(\lambda_1+\cdots+\lambda_m\), achieved by
\(\psi_\ell=\phi_\ell\).  Since
\(\tr(\mathcal K)=\sum_{k\ge1}\lambda_k\), the minimized error is
\(\sum_{k>m}\lambda_k\). \qedmark

This proposition is the reason functional principal components are not just a
plotting device.  They are the optimal linear compression of a random curve
under squared \(L^2\) loss.  Whether that compression answers the scientific
question is a separate question.  The first eigenfunction may capture a boring
batch effect, a device-calibration shift, or a timing artifact.  The theorem
shows what the covariance geometry is doing; Chapter~1 still shows us to
ask whether that geometry is the right translation of the problem.

\begin{example}[R{\'e}nyi's maximal correlation]
Pearson correlation fixes the two summaries before optimizing: it correlates
\(X\) with \(Y\).  R{\'e}nyi's maximal correlation asks a more functional
question: among all square-integrable transformations of \(X\) and \(Y\), how
large can the correlation become \citep{renyi1959measures}?  For a joint law
\(P_{XY}\), define
\[
  \rho_{\max}(X,Y)
  =
  \sup
  \Expect\{f(X)g(Y)\},
\]
where the supremum is over measurable functions satisfying
\[
  \Expect f(X)=\Expect g(Y)=0,
  \qquad
  \Expect f(X)^2=\Expect g(Y)^2=1.
\]
The data object is therefore not a pair of numbers alone, but the joint
distribution \(P_{XY}\).  The target is a nonlinear dependence coefficient,
together with the transformations \(f\) and \(g\) that reveal it.

The connection to this chapter is the conditional-expectation operator.  Let
\[
  T:L_0^2(P_X)\longrightarrow L_0^2(P_Y),
  \qquad
  (Tf)(y)=\Expect\{f(X)\mid Y=y\}.
\]
Here \(L_0^2(P_X)\) and \(L_0^2(P_Y)\) denote the mean-zero square-integrable
function spaces under the two marginal laws.
Then
\[
  \Expect\{f(X)g(Y)\}
  =
  \langle Tf,g\rangle_{L^2(P_Y)}.
\]
Hence
\[
  \rho_{\max}(X,Y)=\|T\|_{\mathrm{op}}.
\]
When \(T\) is compact, the maximal-correlation directions are leading singular
functions of \(T\).  This is the same operator grammar as KL, but with a
cross-operator between two \(L^2\) spaces rather than one covariance operator
on a single curve space.

This example is useful because linear correlation can be silent about strong
dependence.  If \(X\) is symmetric about zero and \(Y=X^2\), then
\(\Corr(X,Y)=0\), but \(\rho_{\max}(X,Y)=1\), because a nonconstant function
of \(Y\) is also a function of \(X\).  In a clinical or financial data set,
the same issue appears when a biomarker or return has a U-shaped relationship
with risk: the raw variables may look nearly uncorrelated, while transformed
features carry the dependence.

The statistical warning is as important as the definition.  With unrestricted
functions, the empirical maximal correlation can degenerate: if the observed
\(X_i\)'s and \(Y_i\)'s are all distinct, one can often choose functions on the
observed support that make the sample correlation equal to one.  A usable
estimator therefore restricts or regularizes the function classes, for example
by splines, kernels, orthogonal series, neural-network classes with validation,
or low-rank operator smoothing.  Uncertainty is then attached to the regularized
target and to the learned transformations, not to an unconstrained empirical
supremum.
\end{example}

\section{Brownian Motion and Brownian Bridge Examples}
\label{sec:ch12-brownian-kl}
\conceptindexes{Brownian motion, Brownian bridge, Karhunen--Loeve expansion, sample-path regularity, zero set, local time, occupation time, threshold functionals}

The Brownian bridge from Chapter~11 is the most useful first example because it
shows why this chapter sits here.  Donsker's theorem gives a curve-valued
limit; the KL expansion asks how that limit is organized.  This is not a
detour into a special Gaussian identity.  It is the first instance of the
book's representation step.  Let \(B\) be the standard
Brownian bridge on \([0,1]\), with covariance
\[
  K_B(s,t)=s\wedge t-st.
\]
This is the covariance of the empirical-process limit
\[
  \sqrt n\{F_n(t)-t\}\weakto B(t)
\]
for uniform data.  The next proposition turns that limit into a spectral
object.

\begin{proposition}[The Brownian bridge expansion]
\label{prop:ch12-brownian-bridge-kl}
For the Brownian bridge covariance
\[
  K_B(s,t)=s\wedge t-st,
\]
the eigenfunctions and eigenvalues are
\[
  \phi_k(t)=\sqrt2\sin(k\pi t),
  \qquad
  \lambda_k=\frac{1}{k^2\pi^2},
  \qquad k=1,2,\ldots .
\]
Thus
\[
  B(t)
  =
  \sqrt2\sum_{k=1}^{\infty}
  Z_k\frac{\sin(k\pi t)}{k\pi},
  \qquad
  Z_k\stackrel{\mathrm{iid}}{\sim}\Normal(0,1),
\]
with convergence in \(L^2(\Omega;L^2[0,1])\).
\end{proposition}

\noindent\textit{Proof.}
Let
\[
  g(t)=\int_0^1 K_B(s,t)\phi(s)\,ds.
\]
Because \(K_B(0,s)=K_B(1,s)=0\), any eigenfunction with nonzero eigenvalue
satisfies \(g(0)=g(1)=0\).  The Brownian bridge kernel is the Green's function
for the operator \(-d^2/dt^2\) with zero boundary conditions, so
\[
  g''(t)=-\phi(t).
\]
If \(g=\lambda\phi\), then
\[
  \lambda\phi''(t)=-\phi(t),
  \qquad
  \phi(0)=\phi(1)=0.
\]
The nonzero solutions are \(\phi_k(t)=\sqrt2\sin(k\pi t)\), and the
corresponding eigenvalues are \(\lambda_k=(k^2\pi^2)^{-1}\).  Since the
Brownian bridge is Gaussian, Theorem~\ref{thm:ch12-kl} gives independent
normal scores and the displayed expansion. \qedmark

The formula is more than a neat identity.  It says that the rough fluctuation
of an empirical cdf can be read as a random mixture of smooth sine waves.  The
first mode is a broad up-versus-down bending of the distribution curve.  Higher
modes are increasingly local oscillations, but their variances decay like
\(1/k^2\).  Randomness is still random, but it is not featureless.

Brownian motion has a closely related expansion.  If \(W\) is standard
Brownian motion with covariance \(K_W(s,t)=s\wedge t\), then
\[
  \phi_k(t)=\sqrt2\sin\{(k-1/2)\pi t\},
  \qquad
  \lambda_k=\frac{1}{(k-1/2)^2\pi^2},
\]
and
\[
  W(t)
  =
  \sqrt2\sum_{k=1}^{\infty}
  Z_k
  \frac{\sin\{(k-1/2)\pi t\}}{(k-1/2)\pi}.
\]
The boundary condition has changed.  A Brownian bridge is pinned at both ends,
so its modes vanish at \(0\) and \(1\).  Brownian motion is pinned at \(0\) but
free at \(1\), so its eigenfunctions satisfy a zero-slope condition at the
right endpoint.  The covariance remembers the scientific constraint.

Brownian motion can be read through several grammars.  The Markov and strong
Markov properties read it through filtrations and stopping times; reflection
principles read it through crossings and hitting times; local time reads it
through occupation density.  The KL expansion reads the same process through
covariance geometry.  This chapter uses that last reading because its task is
representation: it asks which orthogonal coordinates carry the path variation,
not which path property proves a hitting-time identity.
The next few results are included for one reason: they keep the representation
honest.  A useful coordinate system is not the same thing as a complete
description of the path.

\begin{tcolorbox}[
  enhanced,
  breakable,
  colback=noteback,
  colframe=bookgold!75!black,
  boxrule=0.55pt,
  arc=4pt,
  boxsep=1pt,
  left=0.95em,
  right=0.9em,
  top=0.7em,
  bottom=0.7em,
  before skip=0.9\baselineskip,
  after skip=1.0\baselineskip
]
\noindent\textbf{Path geometry, not covariance geometry.}
The KL expansion should not make Brownian motion look smooth.  Each sine basis
function is smooth, but the random path is not.  The point is not a paradox.
The KL expansion describes covariance geometry; differentiability, zero sets,
and local time describe sample-path geometry.  Both descriptions are true, but
they answer different questions.
\end{tcolorbox}

\begin{lemma}[Canonical path-space warning]
\label{lem:ch12-continuity-not-cylinder-event}
Let \(\Omega=\mathbb R^{[0,\infty)}\), and let
\[
  \mathcal H=\sigma\{X_t:t\ge0\},
  \qquad
  X_t(\omega)=\omega(t),
\]
be the cylinder, equivalently product Borel, \(\sigma\)-algebra on \(\Omega\).
Then
\[
  C=\{\omega:t\mapsto\omega(t)\text{ is continuous}\}
\]
is not an element of \(\mathcal H\).
\end{lemma}

\noindent\textit{Proof.}
Call a set \(A\subset\Omega\) countably determined if there is a countable
set \(J\subset[0,\infty)\) such that
\[
  \omega|_J=\eta|_J
  \quad\Longrightarrow\quad
  \{\omega\in A\Longleftrightarrow \eta\in A\}.
\]
The collection of countably determined sets is a \(\sigma\)-algebra: complements
use the same \(J\), and a countable union uses the union of the countably many
corresponding \(J\)'s.  Every finite-coordinate cylinder is countably
determined, so every set in \(\mathcal H\) is countably determined.

The continuity set \(C\) is not countably determined.  If a countable
\(J\subset[0,\infty)\) were sufficient to decide continuity, choose
\(t_0\notin J\).  Let \(f(t)\equiv0\), and let \(g\) agree with \(f\) except
that \(g(t_0)=1\).  Then \(f|_J=g|_J\), but \(f\in C\) and \(g\notin C\), a
contradiction.  Hence \(C\notin\mathcal H\). \qedmark

This is not a technical nuisance about notation.  If Brownian motion is placed
on the enormous product space \(\mathbb R^{[0,\infty)}\) with only the
cylinder \(\sigma\)-algebra, continuity is not even an event.  The usual
statement that Brownian motion has continuous paths means that we work on
\(C[0,\infty)\) with its Borel \(\sigma\)-algebra, or on a filtered probability
space carrying a continuous modification.  Sample-path geometry begins with
that choice of measurable space.

\begin{proposition}[Smooth KL terms do not make Brownian paths differentiable]
\label{prop:ch12-brownian-not-smooth}
For each \(t\in[0,1)\), Brownian motion has no derivative in quadratic mean at
\(t\).  Moreover, if the truncated KL expansion is differentiated term by term,
the derivative variances diverge at every interior point \(t\in(0,1)\).
\end{proposition}

\noindent\textit{Proof.}
For \(h\ne0\) such that \(t+h\in[0,1]\),
\[
  \Expect\left[
    \left\{
      \frac{W(t+h)-W(t)}{h}
    \right\}^2
  \right]
  =
  \frac{\Expect\{W(t+h)-W(t)\}^2}{h^2}
  =
  \frac{|h|}{h^2}
  =
  \frac{1}{|h|}.
\]
This diverges as \(h\to0\), so the difference quotients cannot converge in
\(L^2\).  Thus Brownian motion has no quadratic-mean derivative at \(t\).

The same warning is visible directly from the KL series.  Write the \(m\)-term
Brownian expansion as
\[
  W_m(t)
  =
  \sqrt2\sum_{k=1}^{m}
  Z_k\frac{\sin\{(k-1/2)\pi t\}}{(k-1/2)\pi}.
\]
Term-by-term differentiation gives
\[
  W_m'(t)
  =
  \sqrt2\sum_{k=1}^{m}Z_k\cos\{(k-1/2)\pi t\}.
\]
For \(t\in(0,1)\), independence and unit variance of the \(Z_k\)'s give
\[
\begin{aligned}
  \Expect\{W_m'(t)^2\}
  &=
  2\sum_{k=1}^{m}\cos^2\{(k-1/2)\pi t\}  \\
  &=
  m+\sum_{k=1}^{m}\cos\{(2k-1)\pi t\}  \\
  &=
  m+\frac{\sin(2m\pi t)}{2\sin(\pi t)} .
\end{aligned}
\]
The second term is bounded in \(m\), while the first term diverges.  Hence the
formal derivative series has exploding variance and is not an \(L^2\) random
function. \qedmark

\begin{theorem}[Paley--Wiener--Zygmund nondifferentiability; \citealp{paleyWienerZygmund1933notes}]
\label{thm:ch12-brownian-nowhere-differentiable}
With probability one, a Brownian path is nowhere differentiable on \([0,1]\).
\end{theorem}

\noindent\textit{Proof.}
We first rule out differentiability at interior points.  Let
\[
  D=\{\omega: t\mapsto W(t,\omega)\text{ is differentiable at some }
  s\in(0,1)\}.
\]
For integers \(m\ge1\) and \(n\ge3\), define
\[
  D_{mn}
  =
  \bigcup_{k=1}^{n-2}
  \bigcap_{j=k}^{k+2}
  \left\{
    \left|
      W\!\left(\frac{j}{n}\right)
      -
      W\!\left(\frac{j-1}{n}\right)
    \right|
    \le
    \frac{3m}{n}
  \right\}.
\]
Set
\[
  \Gamma
  =
  \bigcup_{m=1}^{\infty}\liminf_{n\to\infty}D_{mn},
  \qquad
  \liminf_{n\to\infty}D_{mn}
  =
  \bigcup_{\ell=1}^{\infty}\bigcap_{n\ge\ell}D_{mn}.
\]
We claim that \(D\subset\Gamma\).  If a continuous function \(f\) is
differentiable at some \(s\in(0,1)\), then for some integer \(m\) and some
\(\delta>0\),
\[
  |f(u)-f(s)|\le m|u-s|
  \qquad\text{whenever } |u-s|\le\delta .
\]
For all sufficiently large \(n\), choose \(k=k(n)\) so that
\[
  \frac{k}{n}\le s<\frac{k+1}{n}.
\]
Then \(1\le k\le n-2\), and for \(j=k,k+1,k+2\) the two endpoints
\((j-1)/n\) and \(j/n\) are within the \(\delta\)-neighborhood of \(s\).  More
precisely, their distances from \(s\) add to at most \(3/n\).  Hence
\[
  \left|
    f\!\left(\frac{j}{n}\right)
    -
    f\!\left(\frac{j-1}{n}\right)
  \right|
  \le
  \frac{3m}{n}
  \qquad j=k,k+1,k+2.
\]
Thus \(D_{mn}\) occurs for all sufficiently large \(n\), proving
\(D\subset\Gamma\).

For fixed \(m\), the increments in each three-increment block are independent
and distributed as \(\Normal(0,1/n)\).  Hence
\[
\begin{aligned}
  \Prob(D_{mn})
  &\le
  n
  \left[
    \Prob\left\{
      |\Normal(0,1/n)|\le \frac{3m}{n}
    \right\}
  \right]^3  \\
  &=
  n
  \left[
    \Prob\left\{
      |\Normal(0,1)|\le \frac{3m}{\sqrt n}
    \right\}
  \right]^3  \\
  &\le
  n
  \left(
    \frac{6m}{\sqrt{2\pi n}}
  \right)^3
  =
  O(n^{-1/2}).
\end{aligned}
\]
Thus \(\liminf_n\Prob(D_{mn})=0\).  Since
\(\ind{\liminf_{n\to\infty}D_{mn}}
 \le\liminf_{n\to\infty}\ind{D_{mn}}\), Fatou's lemma gives
\[
  \Prob\left(\liminf_{n\to\infty}D_{mn}\right)=0 .
\]
Taking the countable union over \(m\) gives \(\Prob(\Gamma)=0\), and the
inclusion \(D\subset\Gamma\) rules out differentiability anywhere in
\((0,1)\) with probability one.

It remains only to rule out one-sided derivatives at the endpoints.  At \(0\),
define
\[
  E^0_{mn}
  =
  \bigcap_{j=1}^{2}
  \left\{
    \left|
      W\!\left(\frac{j}{n}\right)
      -
      W\!\left(\frac{j-1}{n}\right)
    \right|
    \le
    \frac{3m}{n}
  \right\}.
\]
If the right derivative at \(0\) exists, then for some \(m\) the event
\(E^0_{mn}\) occurs for all sufficiently large \(n\).  Since the two increments
are independent \(\Normal(0,1/n)\) variables,
\[
  \Prob(E^0_{mn})
  \le
  \left(
    \frac{6m}{\sqrt{2\pi n}}
  \right)^2
  =
  O(n^{-1}),
\]
so Fatou's lemma gives
\[
  \Prob\left(\liminf_{n\to\infty}E^0_{mn}\right)=0
\]
for each fixed \(m\).  A countable union over \(m\) rules out a right
derivative at \(0\).  The endpoint \(1\) is identical, using the last two
increments
\[
  E^1_{mn}
  =
  \bigcap_{j=n-1}^{n}
  \left\{
    \left|
      W\!\left(\frac{j}{n}\right)
      -
      W\!\left(\frac{j-1}{n}\right)
    \right|
    \le
    \frac{3m}{n}
  \right\}.
\]
Thus, with probability one, there is no differentiability point in
\([0,1]\). \qedmark

\subsection{Zero Sets Are Statistical Targets}
\label{subsec:ch12-zero-sets-targets}

The zero set and local time encode a different part of the same sample-path
geometry, but they are not only path curiosities.  They are prototypes of
statistical targets built from a whole curve.  For a deterministic curve \(x\)
and a level \(a\), define the level set, excursion set, and occupation time
\[
\begin{gathered}
  Z_a(x)=\{t:x(t)=a\},
  \qquad
  E_a^+(x)=\{t:x(t)>a\},\\
  O_A(x;T)=\int_0^T\ind{x(t)\in A}\,dt .
\end{gathered}
\]
The exact set \(Z_a(x)\) is usually too fragile to be the final data-analysis
target when observations are noisy or discrete.  The nearby targets are more
statistical: first passage time, number of crossings at a recorded resolution,
duration of excursions above a threshold, or occupation time in a band.  These
are the curve analogues of the book's general question from Chapter~8: what
claim is being made from the observed object?

This reading also connects back to Chapters~1 and~2.  A proxy climate curve
can be summarized by warm-period crossings relative to a baseline.  A robot-lab
sensor trace can be summarized by how long a run stays near a quality-control
boundary.  A single-cell pseudotime curve can be summarized by the activation
interval of a gene or pathway.  In each case, the target is not the whole curve
and not merely its first covariance mode; it is a path functional whose
definition depends on the scientific threshold and the observation mechanism.

\begin{theorem}[Zero set of Brownian motion]
\label{thm:ch12-brownian-zero-set}
For Brownian motion on \([0,\infty)\), define
\[
  Z=\{t\ge0:W(t)=0\}.
\]
With probability one, \(Z\) is perfect, hence closed and uncountable.  Its
one-dimensional Lebesgue measure satisfies \(\lambda(Z)=0\).
\end{theorem}

\noindent\textit{Proof.}
Continuity of the path makes \(Z=W^{-1}(\{0\})\) closed.  For the measure
claim, write \(\lambda=\lambda^1\) for Lebesgue measure on \(\mathbb R\), and
fix \(T<\infty\).  By Tonelli's theorem,
\[
\begin{aligned}
  \Expect\{\lambda(Z\cap[0,T])\}
  &=
  \Expect\int_0^T \ind{\{W(t)=0\}}\,dt        \\
  &=
  \int_0^T \Prob\{W(t)=0\}\,dt
  =
  0,
\end{aligned}
\]
because \(W(t)\sim\Normal(0,t)\) has no atom at \(0\) for every \(t>0\), and
the point \(t=0\) has \(\lambda\)-measure zero.  Hence
\(\lambda(Z\cap[0,T])=0\) almost surely.  Taking \(T=1,2,\ldots\) gives
\(\lambda(Z)=0\) almost surely on \([0,\infty)\).

It remains to rule out isolated zeros.  First observe that Brownian motion
started at \(0\) has zeros arbitrarily close to \(0\) on the right.  Indeed,
for each \(\epsilon>0\), the reflection principle gives
\[
  \Prob\left\{\sup_{0\le u\le\epsilon}W(u)\le0\right\}=0,
  \qquad
  \Prob\left\{\inf_{0\le u\le\epsilon}W(u)\ge0\right\}=0 .
\]
Thus, with probability one, on every interval \((0,\epsilon)\) with rational
\(\epsilon>0\), the path takes both positive and negative values.  By
continuity, there is then a zero in \((0,\epsilon)\).

Now fix rational \(0\le a<b\), and set
\[
  T_a=\inf\{t\ge a:W(t)=0\}.
\]
By continuity, \(T_a\) is a stopping time.  On the event \(\{T_a<b\}\), the
strong Markov property at \(T_a\) says that
\[
  \{W(T_a+u)-W(T_a):u\ge0\}
\]
is a Brownian motion started at \(0\), conditionally on the information up to
\(T_a\).  The immediate-return result above therefore implies that, almost
surely on \(\{T_a<b\}\), zeros occur in every interval
\((T_a,T_a+\epsilon)\).  Hence \(T_a\), whenever it is finite, is not isolated
from the right.

The point \(0\) itself is not isolated by the immediate-return result.  If some
zero \(t>0\) were isolated, there would be an \(\epsilon>0\) such that
\(Z\cap(t-\epsilon,t+\epsilon)=\{t\}\).  Choose rationals \(a,b\) with
\[
  t-\epsilon<a<t<b<t+\epsilon .
\]
Then \(T_a=t<b\), but \(t\) is isolated from the right, contradicting the
previous paragraph.  A countable union over rational \(a,b\) rules out all
isolated zeros.  Therefore \(Z\) is closed and has no isolated points; that is,
it is perfect.  Since \(0\in Z\), a nonempty perfect subset of the real line is
uncountable. \qedmark

\subsection{Local Time as Threshold Sensitivity}
\label{subsec:ch12-local-time-statistics}

Local time makes recurrence quantitative.  For fixed \(t\), the occupation
measure of the path is
\[
  \nu_t(A)=\int_0^t\ind{W(s)\in A}\,ds .
\]
Brownian local time is the density of this random measure: there is a jointly
continuous process \(L_t(a)\) such that, for bounded measurable \(f\),
\[
  \int_0^t f\{W(s)\}\,ds
  =
  \int_{\mathbb R} f(a)L_t(a)\,da .
\]
Thus \(Z_a(W)\) records where the path returns to level \(a\), while
\(L_t(a)\) records how much occupation time accumulates near that level.
Equivalently,
\[
  \int_0^t\ind{a\le W(s)\le b}\,ds
  =
  \int_a^b L_t(u)\,du .
\]
If the upper threshold \(b\) is moved by a small amount \(h>0\), then
\[
  \int_0^t\ind{b<W(s)\le b+h}\,ds
  =
  \int_b^{b+h}L_t(u)\,du
  \approx
  hL_t(b).
\]
This is the statistical reading of local time: it is the first-order
sensitivity of a time-in-band summary to a threshold or calibration shift.
When local time near a decision boundary is large, tiny changes in the boundary
can change many records.  When it is small, the same boundary perturbation may
have little practical effect.

Neither \(Z_a(W)\) nor \(L_t(a)\) is visible from the covariance operator alone.
For the purpose of this chapter, this is the boundary line.  KL modes organize
second-order variation; they do not replace sample-path analysis.  Once that
boundary is clear, the same representation grammar can be used safely for
empirical-process limits and for observed functional data.

\begin{realdatacapsule}{Zhu/Chu Ko-chen climate curve as a crossing target}
\item[Data object.] Historical climate proxies, chronicles, phenology,
disaster reports, and regional records summarized as evidence about a latent
temperature anomaly curve \(m(t)\) \citep{chu1973climatic,ge2013temperature}.
\item[Observation mechanism.] Proxy records are created, preserved, dated,
selected, and calibrated with proxy-specific bias and temporal resolution.
\item[Target.] Crossings and excursions relative to a baseline:
\(\{t:m(t)=0\}\), warm-period duration \(\int\ind{m(t)>0}\,dt\), or time spent
near the baseline \(\int\ind{|m(t)|\le\epsilon}\,dt\).
\item[Model.] A latent curve model such as
\(Y_{r,t}=a_r+b_rm(t)+\varepsilon_{r,t}\), with smoothing or basis
representation for \(m\).
\item[Uncertainty.] Bands for \(m(t)\), bootstrap or posterior uncertainty for
crossing times, and sensitivity of occupation summaries to baseline choice.
\item[Limitation.] Exact zero sets are unstable under dating error,
smoothing, and proxy calibration.  The defensible target is usually a band,
excursion duration, or threshold-crossing statement with explicit resolution.
\end{realdatacapsule}

\begin{center}
\footnotesize
\textbf{Returning examples through zero-set language.}\par\smallskip
\setlength{\tabcolsep}{0.32em}
\renewcommand{\arraystretch}{1.14}
\begin{tabular}{p{0.23\linewidth}p{0.25\linewidth}p{0.39\linewidth}}
\toprule
\textbf{Returning example} & \textbf{Curve or field} & \textbf{Zero-set or local-time reading} \\
\midrule
Zhu/Chu Ko-chen climate thread &
Latent temperature anomaly \(m(t)\) &
Baseline crossings, warm/cold excursion lengths, and time near a historical
threshold; local-time language explains why borderline periods are sensitive
to calibration. \\
Single-cell trajectory thread &
Smoothed gene or pathway expression \(g(t)\) over pseudotime &
Activation boundaries \(Z_a(g)\), duration above an expression threshold, and
the warning that count zeros are an observation issue before they become
trajectory-level zero sets. \\
Autonomous laboratory thread &
Sensor curve \(Y_s(u)\) or quality trace during a run &
Time near a failure, yield, pressure, or purity boundary; high occupation near
the boundary means the pass/fail decision is fragile to sensor noise and
calibration. \\
Empirical distribution thread &
\(\sqrt n\{F_n(t)-F_0(t)\}\) as an empirical-process curve &
Near-zero occupation marks where the fitted null is locally close; supremum
statistics, spectral modes, and occupation summaries answer different
goodness-of-fit questions. \\
\bottomrule
\end{tabular}
\end{center}

\begin{example}[Empirical-process zeroes and closeness bands]
A Kolmogorov--Smirnov statistic looks at
\[
  \sup_{0\le t\le1}|B(t)|.
\]
The KL expansion instead asks which spectral components contribute to the
bridge.  A third question asks where the bridge is close to a level.  For the
finite-sample empirical process
\[
  G_n(t)=\sqrt n\{F_n(t)-F_0(t)\},
\]
one may inspect a closeness-band functional
\[
  O_{\epsilon,n}
  =
  \int_0^1\ind{|G_n(t)|\le\epsilon}\,w(t)\,dt ,
\]
where \(w\) emphasizes the center, tails, or a scientifically important
quantile region.  Under the Brownian-bridge approximation, this behaves like
an occupation functional of \(B\).  If \(w\equiv1\) and \(\epsilon\) is small,
the occupation-density idea says heuristically that
\[
  \int_0^1\ind{|B(t)|\le\epsilon}\,dt
  \approx
  2\epsilon L_1^B(0),
\]
where \(L_1^B(0)\) is the bridge local time at level zero.  The supremum norm
asks for the largest vertical discrepancy.  The expansion asks whether the
discrepancy is low-frequency drift, mid-frequency shape change, or
high-frequency texture.  The occupation view asks where the discrepancy is
near a decision boundary.  In goodness-of-fit, quality monitoring, and
distribution-shift diagnostics, those are three different statistical
questions.

The measurability bookkeeping is exactly the one from
Chapters~6 and~11.  If
\(F_n(t)=n^{-1}\sum_{i=1}^n\ind{Y_i\le t}\), then
\((\omega,t)\mapsto\ind{Y_i(\omega)\le t}\) is product-measurable, so
\((\omega,t)\mapsto G_n(t,\omega)\) is product-measurable and the occupation
integral \(O_{\epsilon,n}\) is a measurable random variable by Tonelli whenever
\(w\) is measurable and integrable.  For the bridge limit we use the continuous
modification \(B\in C[0,1]\).  On \(C[0,1]\), \(x\mapsto\|x\|_\infty\) is
continuous; equivalently,
\[
  \sup_{0\le t\le1}|x(t)|
  =
  \sup_{q\in\Rat\cap[0,1]}|x(q)|.
\]
Thus the Kolmogorov--Smirnov functional is measurable, and the bridge
occupation functional is measurable because the evaluation map
\((x,t)\mapsto x(t)\) is Borel on \(C[0,1]\times[0,1]\).
\end{example}

\section{Functional Data Analysis with KL Scores}
\label{sec:ch12-fda}
\conceptindexes{functional data analysis, sparse functional data, PACE, curve registration, glucodensity, Wasserstein regression, distribution-valued data, tangent-space regression}

Functional data analysis is the statistical version of the same representation
problem.  Instead of receiving the covariance kernel of a Gaussian limit from a
theorem, we estimate the mean and covariance from observed curves.  The goal in
this chapter is not to survey the field, but to show how the theoretical
decomposition becomes a working representation.  The Brownian section used a
known covariance to understand a limiting curve; this section uses data to
estimate the covariance coordinates themselves.  Functional data analysis
begins when we observe several curves
\[
  X_1,\ldots,X_n,
\]
or noisy, partial versions of them.  A common observation model is
\[
  Y_{ij}=X_i(t_{ij})+\varepsilon_{ij},
  \qquad
  i=1,\ldots,n,\quad j=1,\ldots,m_i.
\]
The times \(t_{ij}\) may be dense and regular, as in some sensor streams; or
sparse and irregular, as in many clinical visits.  The measurement errors
\(\varepsilon_{ij}\) add a diagonal noise component to the covariance.  The
underlying curve \(X_i\) carries the biological, behavioral, financial, or
physical signal.

The functional principal component workflow is:
\begin{enumerate}
\item estimate the mean curve \(\mu(t)\);
\item estimate the covariance surface \(K(s,t)\), separating smooth covariance
      from measurement-error noise when needed;
\item compute eigenfunctions \(\hat\phi_k\) and eigenvalues \(\hat\lambda_k\);
\item estimate subject scores \(\hat\xi_{ik}\);
\item use those scores, curves, or reconstructions in regression,
      classification, clustering, monitoring, or prediction.
\end{enumerate}

For densely observed curves, the score estimate is close to a numerical
integral:
\[
  \hat\xi_{ik}
  \approx
  \int_0^1\{Y_i(t)-\hat\mu(t)\}\hat\phi_k(t)\,dt .
\]
For sparse longitudinal data, one usually estimates scores by conditional
expectation under a working Gaussian model, using the observed time points and
the estimated covariance surface.  This is the PACE idea of
\citet{yaoMullerWang2005pace}.  More explicitly, write
\[
  Y_i=(Y_{i1},\ldots,Y_{im_i})^\top,
  \qquad
  \hat{\boldsymbol\mu}_i
  =
  \{\hat\mu(t_{i1}),\ldots,\hat\mu(t_{im_i})\}^\top,
\]
and
\[
  \hat{\boldsymbol\phi}_{ik}
  =
  \{\hat\phi_k(t_{i1}),\ldots,\hat\phi_k(t_{im_i})\}^\top .
\]
If \(\hat\Sigma_i\) is the estimated covariance matrix of the observed vector,
with entries
\[
  (\hat\Sigma_i)_{j\ell}
  =
  \hat K(t_{ij},t_{i\ell})+\hat\sigma^2\ind{j=\ell},
\]
then the conditional-expectation score predictor is
\[
  \hat\xi_{ik}
  =
  \hat\lambda_k\,
  \hat{\boldsymbol\phi}_{ik}^{\top}
  \hat\Sigma_i^{-1}
  (Y_i-\hat{\boldsymbol\mu}_i).
\]
The formula is a reminder that sparse FPCA is not just numerical integration
with missing grid points.  It borrows strength through the estimated covariance
surface and the observation-time pattern.

\subsection{Scores Are Inputs to Statistical Targets}
\label{subsec:ch12-scores-to-targets}

The scores are usually not the final scientific answer.  They are coordinates
used to estimate or explain a target.  Some targets are linear functionals of
the curve, such as an area under a concentration curve.  Some are extrema, such
as a peak value.  Others are threshold or path-function targets:
\[
  \inf\{t:X_i(t)>a\},
  \qquad
  \int_0^1\ind{a\le X_i(t)\le b}\,dt,
  \qquad
  \{t:X_i(t)=a\}.
\]
The first is a crossing time, the second is an occupation time, and the third
is a level set.  A KL or FPCA representation can make these targets estimable
or modelable, but it does not make them linear.  This is where the local-time
reading above becomes statistical: if many curves spend time near a threshold,
then time-in-range, pass/fail, toxicity, or activation summaries can be highly
sensitive to small measurement and calibration changes.

This is also the promised return of the Chapter~1 and Chapter~2 examples.  A
historical climate curve asks about excursions from a baseline, not only its
first smooth mode.  A robot-lab trace asks how long a run lives near a failure
boundary, not only its mean sensor pattern.  A single-cell pseudotime curve
asks where a pathway turns on, but observed molecular zeros must first be read
through the observation mechanism.  The representation is therefore a middle
step:
\[
  \text{curve}
  \longrightarrow
  \text{coordinates}
  \longrightarrow
  \text{path target}
  \longrightarrow
  \text{uncertainty or decision}.
\]

\begin{example}[Continuous glucose monitoring]
A continuous glucose monitor may record a patient's glucose every few minutes.
Let \(G_i(t)\) be the glucose curve for subject or day \(i\), scaled to
\([0,1]\), and suppose the recorded values satisfy
\[
  Y_{ij}=G_i(t_{ij})+\varepsilon_{ij}.
\]
Two familiar scalar summaries are
\[
  \bar G_i=\int_0^1G_i(t)\,dt,
  \qquad
  \mathrm{TIR}_i=\int_0^1\ind{a\le G_i(t)\le b}\,dt,
\]
where \([a,b]\) is the target range.  The mean glucose \(\bar G_i\) summarizes
average glycemic exposure over the day; it is easy to compare across subjects,
but a day with steady moderate glucose and a day with alternating lows and
highs can have the same mean.  Time-in-range \(\mathrm{TIR}_i\) is the
fraction of the day spent in clinically acceptable control, so it is more
directly tied to daily management and trial endpoints, but it still compresses
all values below \(a\), all values inside \([a,b]\), and all values above \(b\)
into a few bins.

The local-time reading says what makes this endpoint stable or unstable.  A
smoothed empirical version of time spent near the upper boundary is
\[
  L_{i,\epsilon}(b)
  =
  \frac{1}{2\epsilon}
  \int_0^1\ind{|G_i(t)-b|\le\epsilon}\,dt .
\]
If \(L_{i,\epsilon}(b)\) is large, a small calibration shift of the glucose
sensor or a small change in the clinical boundary can move a visible amount of
time into or out of range.  Thus time-in-range is not just a scalar summary; it
is a threshold functional whose uncertainty depends on how the whole curve
sits near the boundary.

A related distributional representation is the glucodensity of
\citet{matabuena2021glucodensities}; see also the exploratory type~1 diabetes
analysis of \citet{cui2023glucodensity}.  Define the time-spent distribution
of the glucose values by
\[
  F_i(x)=\int_0^1\ind{G_i(t)\le x}\,dt,
\]
and, when a smoothed density representation is used, write
\[
  f_i(x)=\frac{d}{dx}F_i(x).
\]
Then
\[
  \mathrm{TIR}_i=F_i(b)-F_i(a-),
\]
so time-in-range is one interval probability under this glucose-value
distribution.  The glucodensity keeps the whole distribution of glucose
concentrations, not just one target interval.  It can distinguish whether poor
control comes from frequent mild hyperglycemia, rare extreme excursions, or a
bimodal mixture of lows and highs.  What it gives up is clock time: two days
with the same histogram of glucose values but different meal timing can have
the same glucodensity.  Thus the time curve \(G_i(t)\) and the glucodensity
\(f_i(x)\) answer complementary clinical questions.

Returning to the time-domain curve, these summaries are useful, but they
collapse timing.  If
\[
  G_i(t)=\mu(t)+\sum_{k\ge1}\xi_{ik}\phi_k(t),
\]
then the mean glucose is already a linear functional of the scores:
\[
  \bar G_i
  =
  \int_0^1\mu(t)\,dt
  +
  \sum_{k\ge1}\xi_{ik}\int_0^1\phi_k(t)\,dt .
\]
Time-in-range is nonlinear, because of the indicator, but it is still a
functional of the same curve.  A score representation keeps more information
available before that nonlinear collapse.

Let \(R_i\) denote a downstream clinical endpoint or risk variable attached to
the same subject or day.  It might be a binary indicator of a future severe
hypoglycemic episode, a continuous marker such as next-period HbA1c, or a
numeric risk score used for monitoring.  Conditioning on \(G_i\) means that
the entire observed glucose trajectory is treated as the predictor, not only
its mean or time-in-range.  Thus \(\Expect(R_i\mid G_i)\) is the conditional
mean of that endpoint given the glucose curve; when \(R_i\) is binary, it is
the conditional probability of the event.  A simple score model is
\[
  \Expect(R_i\mid G_i)
  =
  \alpha+\sum_{k=1}^m\beta_k\xi_{ik},
\]
which is the truncated version of the functional linear model
\[
  \Expect(R_i\mid G_i)
  =
  \alpha+\int_0^1\beta(t)\{G_i(t)-\mu(t)\}\,dt,
  \qquad
  \beta(t)\approx\sum_{k=1}^m\beta_k\phi_k(t).
\]
The first component may represent overall glycemic load; another may represent
a post-meal excursion; another may separate morning instability from evening
instability.  The statistical point is not that FPCA is automatically the right
endpoint.  The point is that the data structure was a curve, so reducing it to
one scalar should be a decision, not a reflex.
\end{example}

\begin{example}[Wasserstein tangent-space FDA]
The glucodensity in the previous example is already a distribution-valued
functional datum.  A similar object appears if a climate study records, for
each region and decade, the empirical distribution of daily temperature
anomalies, or if a single-cell study records, for each donor, the empirical
distribution of an expression score across cells.  In all three cases the
data object attached to unit \(i\) is
\[
  Y_i\in\mathcal W_2(D),
\]
a probability distribution on a bounded interval \(D\), rather than a scalar
response.  The observation mechanism is usually an empirical or smoothed
distribution built from finite measurements: glucose readings within a day,
daily climate observations within a region-period cell, or cell-level
molecular measurements within a donor.

Choose an atomless reference distribution \(\mu_\ast\) on \(D\), with
distribution function \(F_\ast\).  If \(Q_i\) is the quantile function of
\(Y_i\), the one-dimensional Wasserstein log map is
\[
  V_i(s)
  =
  \operatorname{Log}_{\mu_\ast}(Y_i)(s)
  =
  Q_i\{F_\ast(s)\}-s,
  \qquad s\in D .
\]
Thus each distribution becomes a function \(V_i\in L^2(\mu_\ast)\).  This is
where Wasserstein regression and FDA meet: after the geometry-respecting
coordinate change, the statistical model can be written as a functional
regression or FPCA model in the tangent space,
\[
  V_i(s)=\alpha(s)+\mathbf x_i^\top\beta(s)+U_i(s),
  \qquad
  U_i(s)=\sum_{k\ge1}\eta_{ik}\psi_k(s).
\]
\hyperref[sec:ch10-wasserstein-regression]{Section~\ref*{sec:ch10-wasserstein-regression}}
introduced this coordinate as a way to do regression over probability laws.
Here the same coordinate is read as ordinary functional data analysis after
mapping the response to \(L^2(\mu_\ast)\).

The statistical target is a coefficient function, covariance operator, or
conditional mean distribution, not only a scalar endpoint.  For a CGM study,
\(\beta(s)\) says which quantiles of the glucose distribution move with
treatment, diet, or device behavior.  A positive coefficient in the upper
tail means the covariate shifts high glucose values upward, even if mean
glucose changes little.  For the Chapter~1 climate example, the same
coefficient function separates lower-tail cooling, median shifts, and
upper-tail heat extremes.  For the Chapter~2 single-cell example, it can
describe how a treatment changes the whole donor-level distribution of an
activation score, not merely the donor mean.

After fitting in tangent coordinates, the fitted distribution is obtained by
the Wasserstein exponential map,
\[
  \widehat T_i(s)=s+\widehat V_i(s),
  \qquad
  \widehat Y_i
  =
  \operatorname{Exp}_{\mu_\ast}(\widehat V_i)
  =
  \widehat T_i{}_\#\mu_\ast .
\]
Finite-sample fitted maps may need monotone rearrangement or projection before
they define valid distributions.  Uncertainty is also functional: one can use
standard errors or bootstrap bands for \(\beta(s)\), covariance estimation for
the residual process \(U_i\), and FPCA-score uncertainty for the subject-level
tangent functions.  The main limitation is locality.  The tangent
linearization is most faithful near the reference distribution
\(\mu_\ast\); large support changes, atoms, or strong multimodal deformation
can make the tangent model easier to fit than to interpret
\citep{petersen2019frechet,chen2023wasserstein,panaretos2019statistical}.
\end{example}

\begin{example}[Pharmacokinetic and pharmacodynamic curves]
In drug development, concentration and response profiles are functions of time:
absorption, distribution, metabolism, and elimination leave different curve
shapes.  Sparse pharmacokinetic sampling and nonlinear mixed-effects design
make this a pharmacometric problem as well as a representation problem
\citep{mentre1995sparse,mentre1997optimal,nyberg2012poped}.  Let \(C_i(t)\)
be a subject-level concentration curve.  The usual summaries include
\[
  \mathrm{AUC}_i=\int_0^\tau C_i(t)\,dt,
  \qquad
  C_{\max,i}=\sup_{0\le t\le\tau} C_i(t),
\]
and, if a response curve is observed, a pharmacodynamic object \(E_i(t)\).
Under the representation
\[
  C_i(t)=\mu_C(t)+\sum_{k\ge1}\xi_{ik}\phi_k(t),
\]
the exposure summary is a linear functional:
\[
  \mathrm{AUC}_i
  =
  \int_0^\tau\mu_C(t)\,dt
  +
  \sum_{k\ge1}\xi_{ik}\int_0^\tau\phi_k(t)\,dt .
\]
The peak \(C_{\max,i}\) is not linear, but the modes can still say whether a
large peak comes from high overall exposure, delayed absorption, or prolonged
elimination.  For example, a downstream safety or efficacy endpoint might be
modeled as
\[
  \Expect(R_i\mid C_i)
  =
  g^{-1}\left(\alpha+\sum_{k=1}^m\beta_k\xi_{ik}\right),
\]
with \(g\) a link function.  If both concentration and response curves are
available, one can model a response surface by
\[
  \Expect\{E_i(s)\mid C_i\}
  =
  m(s)+\int_0^\tau H(s,t)\{C_i(t)-\mu_C(t)\}\,dt,
\]
and then approximate the integral through the concentration scores.  This is
still not a full pharmacometric model; it is the representation step that
decides which curve features enter dose-response modeling, safety monitoring,
or adaptive design.
\end{example}

\begin{example}[Registered biological trajectories]
Curve registration, discussed in Chapter~7 through the work of
\citet{telescaInoue2008curve}, is a warning that covariance modes are not
always biological modes.  If two cells, patients, or experiments follow the
same shape at different speeds, ordinary FPCA may spend its first components
on timing variation.  A simple mathematical picture is
\[
  X_i(t)=Z_i\{h_i^{-1}(t)\}+\varepsilon_i(t),
\]
where \(Z_i\) is an aligned biological trajectory and \(h_i\) is an increasing
time-warping map with \(h_i(0)=0\) and \(h_i(1)=1\).  The function \(h_i\)
describes phase variation; the aligned curve \(Z_i\) carries amplitude
variation.

The danger can be seen from a one-parameter shift model.  If
\[
  X_i(t)=f(t-\tau_i),
\]
then for small \(\tau_i\),
\[
  X_i(t)\approx f(t)-\tau_i f'(t).
\]
Consequently,
\[
  \Cov\{X_i(s),X_i(t)\}
  \approx
  \Var(\tau_i) f'(s)f'(t),
\]
so the leading empirical eigenfunction may be proportional to \(f'\).  That
component records timing mismatch, not a new biological amplitude pattern.

Registration tries to estimate \(h_i\) and then decompose
\[
  \widetilde X_i(u)=X_i\{h_i(u)\}
  =
  \mu(u)+\sum_{k\ge1}\zeta_{ik}\phi_k(u).
\]
In practice, \(h_i\) is chosen by minimizing an alignment criterion such as
\[
  \sum_i\int_0^1\{X_i(h_i(u))-\mu(u)\}^2\,du
  +
  \lambda\sum_i J(h_i),
\]
where \(J\) penalizes overly rough or implausible warpings.  In single-cell
pseudotime, development, speech, gait, and longitudinal biomarker studies, this
separation can be the difference between discovering a pathway and discovering
that subjects were not aligned.  This is why representation is already a
statistical claim, not a neutral preprocessing step.
\end{example}

\begin{tcolorbox}[
  enhanced,
  breakable,
  colback=noteback,
  colframe=bookgold!75!black,
  boxrule=0.55pt,
  arc=4pt,
  boxsep=1pt,
  left=0.95em,
  right=0.9em,
  top=0.7em,
  bottom=0.7em,
  before skip=0.9\baselineskip,
  after skip=1.0\baselineskip
]
\noindent\textbf{How to read an eigenfunction.}
An eigenfunction is a direction of variation in the observed population.  It is
not automatically a mechanism.  Before naming it biologically, ask whether it
could be driven by measurement noise, time warping, batch, recruitment
patterns, calendar effects, missingness, or device behavior.  The expansion is
a microscope.  It does not decide what the specimen is.
\end{tcolorbox}

Three practical caveats belong next to every FPCA display.  First, the sign of
an eigenfunction is arbitrary: \(\phi_k\) and \(-\phi_k\) define the same
one-dimensional direction if the score changes sign as well.  Second, if an
eigenvalue has multiplicity greater than one, the individual eigenfunctions
inside that eigenspace are not uniquely identified; only the subspace is.
Third, the truncation point \(m\) is a modeling decision.  A large fraction of
variance explained is useful, but it does not by itself prove that the retained
modes answer the scientific question.

\section{Representation Language Revisited}
\label{sec:ch12-grammar-revisited}
\conceptindexes{representation language, random curves, coordinates, covariance modes, functional targets}

The KL expansion fits the book because it is not merely a technique.  It is the
representation bridge for one common data structure.

\begin{center}
\textbf{Functional-data version of the book grammar.}\par\smallskip
\begin{tabular}{p{0.25\linewidth}p{0.64\linewidth}}
\toprule
\textbf{Book grammar} & \textbf{Functional version} \\
\midrule
Data structure &
Curves, paths, profiles, trajectories, empirical cdf processes. \\
Representation question &
Which modes of variation should be kept visible before the curve is reduced to
a target, score, prediction, or decision variable? \\
Assumptions &
Square integrability, smooth covariance, meaningful alignment, noise model,
and often approximate Gaussianity for score prediction. \\
Model object &
Mean function \(\mu\), covariance operator \(\mathcal K\), eigenfunctions
\(\phi_k\), scores \(\xi_{ik}\). \\
Inference use &
Compression, prediction, monitoring, regression, clustering, testing, or
scientific interpretation of dominant modes. \\
\bottomrule
\end{tabular}
\end{center}

This is why the chapter belongs between empirical processes and estimation.
Chapter~11 produced random curves and fields as stable empirical objects.  This
chapter asks what to do when that stable object is a curve: build coordinates,
name the variation they preserve, and remember what they do not preserve.
Chapter~13 then asks how stable empirical objects, often after such a
representation step, select estimators as peaks, valleys, or roots.

The bridge matters because many modern procedures begin by changing the
coordinate system: a curve becomes scores, a distribution becomes a transport
map, an image becomes an embedding, and a network becomes a latent position.
Each representation is a statistical claim about structure, and every later
estimate inherits that claim.  That is the full chapter in one sentence:
before inference can act on a complex object, statistics often has to decide
which coordinates make the object legible.

\section{Exercises}
\label{sec:ch12-exercises}
\conceptindexes{functional-data exercises, Karhunen--Loeve exercises, Brownian-motion exercises}

\begin{exercise}[Brownian motion boundary conditions]
For \(K_W(s,t)=s\wedge t\), show that \(g(t)=\int_0^1K_W(s,t)\phi(s)\,ds\)
satisfies \(g(0)=0\), \(g'(1)=0\), and \(g''(t)=-\phi(t)\).  Derive the
eigenfunctions and eigenvalues displayed in Section~\ref{sec:ch12-brownian-kl}.
\end{exercise}

\begin{exercise}[Best two-dimensional approximation]
Assume \(\lambda_1\ge\lambda_2\ge\cdots\).  Use
Proposition~\ref{prop:ch12-best-basis} to show that the fraction of curve
variation explained by the first \(m\) functional principal components is
\[
  \frac{\lambda_1+\cdots+\lambda_m}{\sum_{k\ge1}\lambda_k}.
\]
Explain why this number can be large even when the first modes are not
scientifically interpretable.
\end{exercise}

\begin{exercise}[Measurement error]
Suppose \(Y(t_j)=X(t_j)+\varepsilon_j\), where
\(\varepsilon_j\) are independent with variance \(\sigma^2\).  What happens to
the covariance matrix of the observed vector
\((Y(t_1),\ldots,Y(t_m))\)?  Explain why estimating the smooth covariance
surface requires special care near the diagonal.
\end{exercise}

\begin{exercise}[From FPCA to prediction]
Suppose each subject has scores \(\xi_{i1},\ldots,\xi_{im}\) and outcome
\(R_i\).  Write a regression model using the scores as predictors.  What
scientific interpretation is lost if the eigenfunctions are not examined?
\end{exercise}

\begin{exercise}[Threshold occupation as a target]
For a continuous observed curve \(x\), define
\[
  O_{[a,b]}(x)=\int_0^1\ind{a\le x(t)\le b}\,dt .
\]
Explain why \(O_{[a,b]}\) is a natural target for a glucose curve, a climate
anomaly curve, or a robot-lab sensor trace.  Then discuss why estimating the
exact level set \(\{t:x(t)=b\}\) is less stable than estimating
\(\int_0^1\ind{|x(t)-b|\le\epsilon}\,dt\) from noisy discrete observations.
\end{exercise}

\begin{exercise}[Wearable curves on irregular grids]
A wearable device records heart rate or glucose at irregular times, with gaps
caused by device removal and battery failure.  Write an observation model for
the latent curve, the observation times, and the missingness process.  Which
part of the problem is functional data analysis, and which part is an
observation-mechanism problem?
\end{exercise}

\begin{exercise}[ICU alarm trajectory]
In an ICU, a patient has a multivariate sensor trajectory before an alarm:
heart rate, blood pressure, oxygen saturation, medication changes, and lab
times.  Describe how one could form functional principal component scores from
the continuous sensor channels.  Which variables should instead be treated as
events, marks, or time-dependent covariates?
\end{exercise}

\begin{exercise}[Robot-lab sensor paths]
An autonomous laboratory records a reaction-temperature curve, a pressure
curve, image-derived color summaries, and final yield for each experiment.
Propose an FPCA-based representation for the curves and a regression target
for yield.  What scientific interpretation would be lost if only the first two
scores were used without inspecting the eigenfunctions?
\end{exercise}

\begin{exercise}[Climate curves and time alignment]
Two regions have reconstructed temperature-anomaly curves over centuries, but
the timing of historical shocks may be uncertain.  Explain why ordinary
\(L^2\) comparison may confuse amplitude differences with time-alignment
differences.  Propose one preprocessing or modeling step that separates curve
shape from time warping.
\end{exercise}

\section*{Sources and Further Reading}
\addcontentsline{toc}{section}{Sources and Further Reading}

The Karhunen--Loeve expansion is associated with
\citet{karhunen1947lineare} and \citet{loeve1978probability}; modern
functional-analytic treatments can be found in \citet{hsingEubank2015fda}.
Mercer's theorem \citep{mercer1909functions} explains the kernel expansion for
continuous positive-definite covariance kernels.  Functional data analysis as
a statistical field is represented here by \citet{ramsaySilverman2005fda};
sparse longitudinal FPCA and conditional score prediction follow
\citet{yaoMullerWang2005pace}.  The Brownian bridge connection returns to
Donsker's theorem from Chapter~11 and to the weak-convergence framework of
\citet{billingsley1999convergence} and \citet{vaart2023weak}.  For a
Brownian-motion-centered route through Markov properties, reflection, local
time, KL expansions, and Donsker's theorem, see
\citet{cui2024brownianTutorial}; the Brownian and Karhunen--Loeve discussion
is also informed by \citet{dabrowskaStochasticProcessesCommunication}.  The
curve registration example points back
to \citet{telescaInoue2008curve}, because alignment is often the practical step
that decides whether a functional principal component is scientifically
meaningful or only a timing artifact.

%% file: chapters/ch14_testing_hellinger_local_asymptotics.tex
\chapter{Testing, Hellinger Geometry, and Local Alternatives}
\label{chap:testing-hellinger-local}
\conceptindexes{hypothesis testing, Hellinger geometry, local alternatives, contiguity, local asymptotic normality, likelihood-ratio tests}

\begin{tcolorbox}[
  enhanced,
  breakable,
  colback=chaptercream,
  colframe=bookblue!88!black,
  boxrule=0.72pt,
  arc=5pt,
  boxsep=1pt,
  left=1.0em,
  right=0.95em,
  top=0.82em,
  bottom=0.82em,
  before skip=0.55\baselineskip,
  after skip=1.0\baselineskip
]
\noindent\textbf{Chapter overview.}
A hypothesis is a set of possible laws, and a test is a decision rule read
under those laws.  This chapter asks when two data-generating laws can be
separated, and at what scale.  Total variation gives the operational distance
for simple tests; Hellinger, KL, and \(\alpha\)-divergences give useful geometries;
contiguity and local likelihood expansions explain local power.  The applied
examples widen the map: goodness-of-fit, discriminant analysis, generalized
\(R^2\), liquid association, randomization and permutation tests in modern
omics, group-sequential monitoring, model misspecification, multiple-endpoint
FWER control, gatekeeping, and false discovery rate control.  The point is not
to turn testing into a recipe list, but to show how many practical tests are
readings of the same probability law.
\end{tcolorbox}

\section{Hypotheses, Likelihood Ratios, and Testing Geometry}
\label{sec:ch14-hypotheses-laws}
\conceptindexes{hypotheses, sets of laws, null hypothesis, alternative hypothesis, test size, power}

A statistical hypothesis is not a sentence about a parameter in isolation.  It
is a restriction on the law that could have generated the observation.  If the
sample space is \((\mathcal X,\mathcal A)\), a statistical model is a family
\(\mathcal P\) of probability measures on that space.  Throughout this
chapter, \(P\) and \(Q\) denote probability laws; when an event
probability is written as \(P(A)\), the same symbol is being used for the
law and its evaluation on \(A\).  A null hypothesis and an alternative are subfamilies
\[
  H_0:\ P\in\mathcal P_0,
  \qquad
  H_1:\ P\in\mathcal P_1,
  \qquad
  \mathcal P_0,\mathcal P_1\subseteq\mathcal P .
\]
In a parametric model \(\{P_\theta:\theta\in\Theta\}\), this often means
\(\theta\in\Theta_0\) versus \(\theta\in\Theta_1\).  The law-based wording is
more portable: it covers simple hypotheses, composite hypotheses, nuisance
parameters, nonparametric restrictions, and sequential data structures.

It also keeps the examples from becoming a list of tests by data format.  A
single-cell matrix, randomized trial, event-history record, classifier score,
or functional observation enters this chapter only after
Chapters~\ref{chap:grammar-modern-data-structures}
and~\ref{chap:data-structures-to-targets} have turned it into an observed
object, a target, and a family of laws.  Testing theory then asks how those
laws can be separated, how the null reading is calibrated, and which
dependence, censoring, randomization, or multiplicity structure the data
object forces the calibration to respect.

\begin{definition}[Test, size, and power]
A test is a measurable function \(\phi:\mathcal X\to[0,1]\).  When
\(\phi(x)=1\) the null is rejected; when \(\phi(x)=0\) it is not rejected; and
intermediate values represent randomization.  The size of \(\phi\) is
\[
  \alpha(\phi)=\sup_{P\in\mathcal P_0} P\phi ,
\]
and its power at \(Q\in\mathcal P_1\) is
\[
  \beta_\phi(Q)=Q\phi .
\]
The type-I error is \(P\phi\) under \(P\in\mathcal P_0\), and the type-II
error at \(Q\) is \(Q(1-\phi)\).
\end{definition}

The notation \(P\phi=\int\phi\,dP\) emphasizes that a test is a statistic read
by different probability laws.  This is the same habit used throughout the
book: one observed object, many possible readings.  Under \(P\in\mathcal P_0\)
the statistic is calibrated as a false-positive probability.  Under
\(Q\in\mathcal P_1\) it is calibrated as power.

The simplest case already contains the central geometry.  Suppose
\(\mathcal P_0=\{P\}\) and \(\mathcal P_1=\{Q\}\).  A test that rejects often
under \(Q\) and rarely under \(P\) can exist only if the two laws place
different mass on some measurable sets.  Testing is therefore a way of asking:
how far apart are \(P\) and \(Q\) as probability measures?

\subsection{Simple Tests and Likelihood Ratios}
\label{sec:ch14-simple-tests}
\conceptindexes{simple tests, likelihood ratio, Neyman--Pearson lemma, most powerful tests}

Assume \(P\) and \(Q\) are dominated by a common \(\sigma\)-finite measure
\(\mu\), with densities \(p\) and \(q\).  The likelihood ratio is
\[
  \Lambda(x)=\frac{q(x)}{p(x)}
\]
on the set where \(p>0\).  Under the null, large values of \(\Lambda\) are
evidence in favor of \(Q\) because the observation is more likely under the
alternative density than under the null density.

\begin{theorem}[Neyman--Pearson lemma; \citealp{neyman1933tests}]
For testing the simple null \(P\) against the simple alternative \(Q\), among
all tests with \(P\phi\le\alpha\), a most powerful level-\(\alpha\) test
rejects for large values of the likelihood ratio \(q/p\), with randomization
on the boundary if needed.
\end{theorem}

\noindent\textit{Proof.}
Let \(\phi\) be any level-\(\alpha\) test and let \(\phi^*\) reject when
\(\Lambda>c\), not reject when \(\Lambda<c\), and randomize on
\(\Lambda=c\) so that \(P\phi^*=\alpha\).  The sign of
\((\phi^*-\phi)(q-cp)\) is nonnegative pointwise.  Integrating gives
\[
  Q\phi^*-Q\phi
  =
  \int(\phi^*-\phi)q\,d\mu
  \ge
  c\int(\phi^*-\phi)p\,d\mu
  =
  c(P\phi^*-P\phi)\ge0 .
\]
Thus no other level-\(\alpha\) test has larger power. \qedmark

\begin{example}[Normal mean]
Let \(X_1,\ldots,X_n\) be iid \(N(\theta,\sigma^2)\), with \(\sigma^2\) known.
To test \(H_0:\theta=\theta_0\) against \(H_1:\theta=\theta_1>\theta_0\), the
log-likelihood ratio is
\[
  \log\Lambda_n
  =
  \frac{\theta_1-\theta_0}{\sigma^2}\sum_{i=1}^n
  \left\{X_i-\frac{\theta_0+\theta_1}{2}\right\}.
\]
It is increasing in \(\bar X_n\), so the most powerful test rejects for large
\(\bar X_n\).  The familiar one-sided \(z\)-test is the likelihood-ratio test
in this simple-vs-simple problem.
\end{example}

For composite hypotheses the likelihood-ratio idea remains, but a single
ratio \(q/p\) is replaced by a comparison between the best fit under the
alternative and the best fit under the null.  The asymptotic theory of those
comparisons is postponed until Section~\ref{sec:ch14-wald-score-lr}; first we
need a language for saying when alternatives are separated or local.

\begin{example}[Two proportions and the denominator]
Return to the risk tables from Chapter~1.  Suppose group \(a=0,1\) contributes
\(X_a\) events among \(m_a\) people at risk, with independent
\[
  X_a\sim\Binomial(m_a,p_a).
\]
The null of equal risk is \(H_0:p_0=p_1\).  The alternative lets the two
probabilities differ.  The binomial log-likelihood is
\[
  \ell(p_0,p_1)
  =
  \sum_{a=0}^1
  \{X_a\log p_a+(m_a-X_a)\log(1-p_a)\}
  +
  \hbox{constant}.
\]
Under the unrestricted model the fitted risks are
\(\hat p_a=X_a/m_a\).  Under the null the fitted common risk is
\[
  \hat p=\frac{X_0+X_1}{m_0+m_1}.
\]
The likelihood-ratio statistic compares these two fits:
\[
  2\{\ell(\hat p_0,\hat p_1)-\ell(\hat p,\hat p)\}.
\]
This example is deliberately simple, because it shows the basic discipline of
testing.  A difference between \(7/100\) and \(9/100\) is not the same
statistical object as a difference between \(70/1000\) and \(90/1000\), even
though the displayed percentages match.  The denominators are part of the
law, hence part of the null and the evidence against it.
\end{example}

\begin{example}[PBMC3k cell-type discrimination as likelihood-ratio testing]
Discriminant analysis is easiest to remember when it is attached to a modern
data object.  Return to the PBMC3k single-cell count matrix from
Chapter~2 \citep{tenx2016pbmc3k}.  The raw object is not a ready-made vector
for a textbook classifier.  It is a sparse cell-by-gene matrix produced by cell
capture, barcoding, sequencing, alignment, filtering, and normalization.  In
the local real-data capsule used by this book, the fallback matrix has
2,700 cells; among the 100 retained high-expression genes, the median
cell-level zero rate is \(0.03\), and the median gene-level zero rate is
\(0.0174\).  Those numbers belong in the example because they remind us that
the feature vector is an observed-data construction, not the biological cell
itself.

Suppose an analyst wants to distinguish two annotated immune-cell classes,
say B cells and T cells, using a training subset.  Let \(Y=1\) denote the
class of interest, and let \(X\in\R^d\) contain log-normalized marker
expression and quality-aware features: for instance MS4A1 and CD79A for the
B-cell side, CD3D and CD3E for the T-cell side, plus library-size or detection
summaries if they are needed to stabilize the comparison.  Conditional on
\(Y=y\), let \(X\) have density \(f_y\), and let
\(\pi_y=\Prob(Y=y)\).  With false-positive cost \(C_{10}\) and false-negative
cost \(C_{01}\), the Bayes rule calls a new cell class \(1\) when
\[
  \log\frac{f_1(X)}{f_0(X)}
  >
  \log\frac{\pi_0}{\pi_1}
  +
  \log\frac{C_{10}}{C_{01}} .
\]
Thus a cell-type classifier is a likelihood-ratio test with a threshold moved
by prevalence and scientific loss.  If false B-cell calls would contaminate a
downstream marker analysis, the threshold moves right; if missed rare cells
are more costly, it moves left.  This is exactly the statistical-compass
translation from Chapters~1--2: data structure, assumptions, target,
procedure, and use all enter the rule.

In the equal-covariance Gaussian working model,
\[
  X\mid Y=y\sim \Normal(\mu_y,\Sigma),
  \qquad y=0,1,
\]
the log-likelihood ratio is linear in \(X\):
\[
  (\mu_1-\mu_0)^T\Sigma^{-1}X
  -
  \frac12(\mu_1+\mu_0)^T\Sigma^{-1}(\mu_1-\mu_0).
\]
Fisher's linear discriminant \citep{fisher1936discriminant} is therefore the
direction that compresses a high-dimensional cell profile into a
one-dimensional evidence score.  In PBMC3k or other omics data, the classical
formula survives only after modern repairs: regularize \(\Sigma\), screen or
shrink marker features, account for batch and depth, and validate the score on
held-out cells or by sample-level resampling.

This example uses the theory above rather than merely borrowing its notation.
The likelihood ratio is the Radon--Nikodym derivative that powers the
Neyman--Pearson rule.  Total variation controls the best possible
classification error between \(P_0=\Law(X\mid Y=0)\) and
\(P_1=\Law(X\mid Y=1)\).  Hellinger or KL separation says whether more cells,
cleaner features, or a better observation protocol can make the two class laws
distinguishable.  Model misspecification is also visible: a Gaussian LDA score
may be useful as a calibrated low-dimensional summary even when the true
single-cell law is zero-inflated, overdispersed, and batch-dependent.
\end{example}

\subsection{Distances and Divergences Between Laws}
\label{sec:ch14-hellinger}
\conceptindexes{distances between laws, divergences, total variation, Hellinger distance, KL divergence, chi-square divergence, Rao's distance, Fisher metric}

Total variation gives the operational distance for simple testing.  Hellinger
distance gives a smoother geometry for products and local likelihood
expansions.  KL, chi-square, and \(\alpha\)-divergences measure expected
log-likelihood or likelihood-ratio discrepancy.  These quantities are not all
metrics, but each answers a recurring statistical question.

\begin{definition}[Total variation]
For probability measures \(P\) and \(Q\) on \((\mathcal X,\mathcal A)\),
\[
  \|P-Q\|_{\mathrm{TV}}
  =
  \sup_{A\in\mathcal A}|P(A)-Q(A)|.
\]
If \(P\) and \(Q\) have densities \(p,q\) with respect to \(\mu\), then
\[
  \|P-Q\|_{\mathrm{TV}}
  =
  \frac12\int |p-q|\,d\mu .
\]
\end{definition}

\begin{proposition}[Best total error]
For simple testing \(P\) versus \(Q\),
\[
  \inf_{\phi}\{P\phi+Q(1-\phi)\}
  =
  1-\|P-Q\|_{\mathrm{TV}} .
\]
Thus total variation is exactly the amount by which the best rule improves on
the trivial total error \(1\).
\end{proposition}

\noindent\textit{Verification.}
If \(\phi=\ind{q>p}\), then
\[
  P\phi+Q(1-\phi)
  =
  \int_{\{q>p\}}p\,d\mu+\int_{\{p\ge q\}}q\,d\mu
  =
  \int \min(p,q)\,d\mu .
\]
Since \(\min(p,q)=\{p+q-|p-q|\}/2\), the value is
\[
  1-\frac12\int |p-q|\,d\mu
  =
  1-\|P-Q\|_{\mathrm{TV}}.
\]
The same pointwise comparison shows that no other test has smaller total
error. \qedmark

\begin{definition}[Hellinger distance and affinity]
If \(P\) and \(Q\) have densities \(p,q\) with respect to a common dominating
measure \(\mu\), the squared Hellinger distance is
\[
  H^2(P,Q)
  =
  \frac12\int(\sqrt p-\sqrt q)^2\,d\mu .
\]
The Hellinger affinity is
\[
  \rho(P,Q)
  =
  \int\sqrt{pq}\,d\mu
  =
  1-H^2(P,Q).
\]
\end{definition}

The definition does not depend on the chosen dominating measure.  The
important feature is that Hellinger distance is an \(L^2\)-distance between
square-root densities.  This turns likelihood problems into geometry in a
Hilbert space.

\begin{proposition}[Comparison with total variation]
For all probability measures \(P,Q\),
\[
  H^2(P,Q)
  \le
  \|P-Q\|_{\mathrm{TV}}
  \le
  H(P,Q)\sqrt{2-H^2(P,Q)}
  \le
  \sqrt2\,H(P,Q).
\]
\end{proposition}

Only some of these quantities are distances in the metric sense.  Total
variation and Hellinger distance are symmetric and satisfy the triangle
inequality.  KL and chi-square divergences are usually asymmetric; they are
better read as losses attached to likelihood ratios.

\begin{definition}[KL, chi-square, and \(\alpha\)-divergence]
Assume first that \(P\ll Q\).  The Kullback--Leibler divergence and
chi-square divergence are
\[
  D_{\mathrm{KL}}(P\|Q)
  =
  \int \log\!\left(\frac{dP}{dQ}\right)\,dP,
  \qquad
  \chi^2(P\|Q)
  =
  \int\left(\frac{dP}{dQ}-1\right)^2\,dQ.
\]
For densities \(p,q\) with respect to a common dominating measure \(\mu\), the
information-geometric \(\alpha\)-divergence is
\[
  D_\alpha(P,Q)
  =
  \frac{4}{1-\alpha^2}
  \left\{
    1-\int p^{(1-\alpha)/2}q^{(1+\alpha)/2}\,d\mu
  \right\},
  \qquad -1<\alpha<1.
\]
With this convention,
\begin{align*}
  D_0(P,Q)&=4H^2(P,Q),\\
  D_{-1}(P,Q)&=D_{\mathrm{KL}}(P\|Q),\\
  D_1(P,Q)&=D_{\mathrm{KL}}(Q\|P),
\end{align*}
where the last two are understood as limiting cases.
\end{definition}

This is the small amount of information geometry needed here
\citep{amari2016information}.  A parametric model is a surface inside the
space of probability laws.  KL divergence is not a symmetric distance, but its
second-order expansion creates Fisher information as the local quadratic form
on that surface.  Hellinger distance gives a closely related
square-root-density embedding.  The chapter uses this geometry only when it
buys an operational statement about tests: separation, local power, or the
effect of nuisance directions.

\begin{definition}[Rao's distance]
Let \(\mathcal M=\{P_\theta:\theta\in\Theta\}\) be a regular parametric model
with Fisher information matrix \(I(\theta)\).  A smooth path
\(\gamma:[0,1]\to\Theta\) has Fisher length
\[
  L(\gamma)
  =
  \int_0^1
  \{\dot\gamma(t)^T I(\gamma(t))\dot\gamma(t)\}^{1/2}\,dt.
\]
The induced geodesic distance between two model points is
\[
  d_R(\theta_0,\theta_1)
  =
  \inf_\gamma L(\gamma),
\]
where the infimum is over smooth paths joining \(\theta_0\) to \(\theta_1\)
inside the model.  This is Rao's distance
\citep{rao1945information,rao1949distance}.  It is intrinsic: changing
coordinates changes the matrix representation of \(I(\theta)\), but not the
length of the path.  Locally,
\[
  d_R(\theta_0,\theta_0+h)^2
  =
  h^T I(\theta_0)h+o(\|h\|^2).
\]
For \(n\) iid observations the Fisher information is \(nI(\theta)\), so Rao
distance in the product experiment is multiplied by \(\sqrt n\).
\end{definition}

The related Renyi divergence is
\[
  D_\gamma^{\mathrm R}(P\|Q)
  =
  \frac{1}{\gamma-1}
  \log\int p^\gamma q^{1-\gamma}\,d\mu,
  \qquad \gamma>0,\quad \gamma\neq 1.
\]
It is another way to record the same likelihood-ratio moments.  For
\(\gamma=1/2\), the integral inside the logarithm is the Hellinger affinity
\(\rho(P,Q)\).

\begin{example}[Rao distance for Bernoulli and multinomial laws]
For \(P_p=\Bernoulli(p)\), \(0<p<1\), the Fisher information is
\[
  I(p)=\frac{1}{p(1-p)}.
\]
Therefore
\[
  d_R(p,q)
  =
  \left|\int_p^q \frac{du}{\sqrt{u(1-u)}}\right|
  =
  2\left|\arcsin\sqrt q-\arcsin\sqrt p\right|.
\]
Thus the risk change from \(p=0.20\) to \(q=0.23\) has intrinsic one-record
length about \(0.073\), close to the local Wald scale
\[
  \frac{|q-p|}{\sqrt{p(1-p)}}=0.075.
\]
With \(n\) independent binary records the product-experiment distance is
\(\sqrt n\,d_R(p,q)\), which is why local power calculations see
\(n(q-p)^2/\{p(1-p)\}\) at second order.

For the full finite-alphabet model,
\[
  \Delta_{K-1}^{\circ}
  =
  \{p:p_j>0,\ \sum_j p_j=1\},
\]
the Fisher inner product on tangent vectors with \(\sum_j u_j=\sum_j v_j=0\)
is
\[
  \langle u,v\rangle_p
  =
  \sum_{j=1}^K\frac{u_jv_j}{p_j}.
\]
The square-root map \(p\mapsto 2(\sqrt{p_1},\ldots,\sqrt{p_K})\) sends this
metric to the ordinary Euclidean metric on the radius-\(2\) sphere.  Hence the
full-simplex Rao distance has the closed form
\[
  d_R(p,q)
  =
  2\arccos\left\{\sum_{j=1}^K\sqrt{p_jq_j}\right\}
  =
  2\arccos\{\rho(p,q)\}.
\]
Hellinger distance is the chord length in the square-root embedding; Rao
distance is the corresponding spherical arc length.  This gives a
coordinate-free effect size for compositional endpoints such as cell-type
proportions or multinomial response profiles.  If the statistical model is a
curved submanifold of the simplex, the geodesic must stay inside that
submanifold, so the intrinsic distance can be larger than the full-simplex
angular distance \citep{atkinsonMitchell1981rao,amari2016information}.
\end{example}

\begin{example}[Two Bernoulli risks]
Return to the Chapter~1 two-group risk comparison.  For one binary record let
\[
  P=\Bernoulli(p),\qquad Q=\Bernoulli(q).
\]
Then
\[
  \|P-Q\|_{\mathrm{TV}}=|p-q|,
  \qquad
  H^2(P,Q)=1-\sqrt{pq}-\sqrt{(1-p)(1-q)},
\]
and
\[
  D_{\mathrm{KL}}(P\|Q)
  =
  p\log\frac{p}{q}
  +(1-p)\log\frac{1-p}{1-q},
  \qquad
  \chi^2(P\|Q)=\frac{(p-q)^2}{q(1-q)}.
\]
If \(p=0.20\) and \(q=0.23\), the per-record values are
\[
  \|P-Q\|_{\mathrm{TV}}=0.03,\qquad
  H^2(P,Q)\approx 0.00067,\qquad
  D_{\mathrm{KL}}(P\|Q)\approx0.00262.
\]
The single record barely separates the laws.  With \(1000\) independent records
from the changed group, the KL divergence adds to about \(2.62\), while the
product Hellinger distance becomes
\[
  H^2(P^{\otimes 1000},Q^{\otimes 1000})
  =
  1-\{1-H^2(P,Q)\}^{1000}
  \approx 0.49.
\]
This is the same lesson as the denominator example: the visible risk
difference is only half the information; the other half is the amount of
independent evidence behind it.
\end{example}

\begin{example}[Poisson exposure]
For the Chapter~2 count/exposure examples, compare
\[
  P=\Poisson(E\lambda_0),
  \qquad
  Q=\Poisson(E\lambda_1).
\]
The KL divergence and Hellinger distance are explicit:
\[
  D_{\mathrm{KL}}(P\|Q)
  =
  E\left\{
    \lambda_0\log\frac{\lambda_0}{\lambda_1}
    +\lambda_1-\lambda_0
  \right\},
\]
and
\[
  H^2(P,Q)
  =
  1-\exp\left\{
    -\frac{E}{2}(\sqrt{\lambda_1}-\sqrt{\lambda_0})^2
  \right\}.
\]
If \(\lambda_0=0.5\), \(\lambda_1=0.6\), and \(E=100\), then
\[
  D_{\mathrm{KL}}(P\|Q)\approx0.88,
  \qquad
  H^2(P,Q)\approx0.20.
\]
The rate difference is numerically small, but the exposure turns it into a
meaningful separation between probability laws.  This is why a zero count, a
small count, or a modest excess cannot be interpreted without its search time,
area, sequencing depth, or text length.
\end{example}

Some comparisons use the geometry of the sample space itself rather than only
likelihood ratios.  If \((\mathcal X,d)\) is a metric space, the
Wasserstein--1 distance is
\[
  W_1(P,Q)
  =
  \inf\{\Expect\,d(X,Y): X\sim P,\ Y\sim Q\}
  =
  \sup_{\mathrm{Lip}(f)\le1}|Pf-Qf|,
\]
in the usual settings where the dual formula holds.  The right side is an
integral probability metric: it compares expectations over a chosen class of
test functions.  For point masses, \(P=\delta_x\) and \(Q=\delta_y\), total
variation and Hellinger only report that the laws are disjoint when \(x\neq y\);
\(W_1(P,Q)=d(x,y)\) also records how far apart the two outcomes are.

\begin{example}[Goodness-of-fit and the Kolmogorov--Smirnov statistic]
Goodness-of-fit testing asks whether the whole distributional shape is
compatible with a proposed law, not only whether a mean or coefficient differs
from zero.  For iid real-valued observations with empirical cdf \(F_n\), the
Kolmogorov--Smirnov statistic for the null cdf \(F_0\) is
\[
  D_n=\sup_t |F_n(t)-F_0(t)|.
\]
The statistic is an integral-probability-metric idea with the test-function
class reduced to indicators of half-lines.  The null is distribution-free when
\(F_0\) is continuous, because \(F_0(X_i)\) are uniform under the null.  Its
weak limit belongs to Chapters~10 and~11; its testing interpretation belongs
here.

This example is useful precisely because it is not a coefficient test.  A
single-cell simulator, a residual model, or a clinical risk model may match the
mean while failing in tails, zero rates, or mixture shape.  A KS statistic, a
Cramer--von Mises statistic, or a collection of calibrated marginal and
conditional checks asks whether the proposed law can survive comparison as a
distribution, not just as a fitted average.
\end{example}

The inequalities say that Hellinger convergence and total-variation
convergence are equivalent at the level of going to zero.  Hellinger distance,
however, has a product formula:
\[
  \rho\!\left(\bigotimes_{i=1}^n P_i,\bigotimes_{i=1}^n Q_i\right)
  =
  \prod_{i=1}^n \rho(P_i,Q_i).
\]
Equivalently,
\[
  H^2\!\left(\bigotimes_{i=1}^n P_i,\bigotimes_{i=1}^n Q_i\right)
  =
  1-\prod_{i=1}^n\{1-H^2(P_i,Q_i)\}
  \le
  \sum_{i=1}^n H^2(P_i,Q_i).
\]
This multiplicativity is the reason Hellinger distance is so natural in
large-sample likelihood theory.

\begin{example}[Fixed and local alternatives]
Let \(P\neq Q\), and compare \(P^{\otimes n}\) with \(Q^{\otimes n}\).  If
\(\rho(P,Q)<1\), then
\[
  \rho(P^{\otimes n},Q^{\otimes n})=\rho(P,Q)^n\to0.
\]
Hence \(H^2(P^{\otimes n},Q^{\otimes n})\to1\), and the two product laws
become asymptotically separated.  This is why fixed alternatives are usually
easy in large samples.

Local alternatives must move toward the null as \(n\) grows.  In regular
parametric models the meaningful scale is typically
\(\theta_n=\theta_0+h/\sqrt n\).  At that scale the accumulated Hellinger
distance stays of constant order rather than exploding to perfect separation.
\end{example}

\begin{example}[Poisson exposure and the meaning of zero counts]
The count examples from Chapter~2 have the same product geometry.  Suppose
row \(i\) records a count with exposure \(e_i\):
\[
  X_i\sim\Poisson(e_i\lambda).
\]
Under rate \(\lambda_0\) the law is \(P_i\), and under rate \(\lambda_1\) it
is \(Q_i\).  The Hellinger affinity between two Poisson laws is
\[
  \rho\{\Poisson(a),\Poisson(b)\}
  =
  \exp\left\{-\frac12(\sqrt a-\sqrt b)^2\right\}.
\]
Therefore
\[
  \rho\!\left(\bigotimes_i P_i,\bigotimes_i Q_i\right)
  =
  \exp\left[
    -\frac12(\sqrt{\lambda_1}-\sqrt{\lambda_0})^2\sum_i e_i
  \right].
\]
The separation is driven by total exposure \(\sum_i e_i\), not by the row
count alone.  A zero observed after a long search area, long person-time, long
document, or deep sequencing run carries more testing information than a zero
observed after a short exposure.  This is the Hellinger version of the
``absence as information'' lesson from Chapter~2.

The local scale also becomes concrete.  If \(E_n=\sum_i e_i\), then alternatives
with
\[
  \sqrt{\lambda_n}-\sqrt{\lambda_0}=\frac{h}{\sqrt{E_n}}
\]
keep the product affinity near \(\exp(-h^2/2)\).  Fixed rate differences
separate as \(E_n\to\infty\); \(E_n^{-1/2}\)-scale rate differences remain
locally testable.
\end{example}

\section{Product Measures, Contiguity, and Local Power}
\label{sec:ch14-products-kakutani}
\conceptindexes{product measures, separation, Kakutani dichotomy, affinity}

Hellinger geometry gives a clean criterion for product experiments.  The
following statement is the form needed in this book.

\begin{theorem}[Kakutani product dichotomy; \citealp{kakutani1948equivalence}]
Let \(P=\bigotimes_{i=1}^\infty P_i\) and
\(Q=\bigotimes_{i=1}^\infty Q_i\) be infinite product measures.  Suppose
\(P_i\) and \(Q_i\) are mutually absolutely continuous for each \(i\).  Then
\[
  P\sim Q
  \quad\Longleftrightarrow\quad
  \sum_{i=1}^\infty H^2(P_i,Q_i)<\infty ,
\]
and
\[
  P\perp Q
  \quad\Longleftrightarrow\quad
  \sum_{i=1}^\infty H^2(P_i,Q_i)=\infty .
\]
\end{theorem}

The theorem has a strong testing interpretation.  If the coordinate-level
Hellinger distances are summable, the two infinite product laws remain
mutually absolutely continuous.  No event has probability one under one law
and zero under the other.  If the sum diverges, the infinite records separate:
there is an event that eventually distinguishes the two product laws with
probability one.

For finite samples the same phenomenon appears through likelihood-ratio
martingales.  Suppose \(Q_i\ll P_i\) and
\[
  \Lambda_n
  =
  \prod_{i=1}^n \frac{dQ_i}{dP_i}(X_i),
  \qquad X_i\sim P_i \hbox{ independently}.
\]
Then \((\Lambda_n)\) is a nonnegative \(P\)-martingale with mean one.  Its
square-root expectation is
\[
  \Expect_{P}\sqrt{\Lambda_n}
  =
  \prod_{i=1}^n \rho(P_i,Q_i).
\]
Thus the same Hellinger affinity that controls product distance also controls
the likelihood-ratio martingale.

This is the probabilistic origin of the local point of view.  If alternatives
do not approach the null, the likelihood ratio often runs to zero under the
null and to infinity under the alternative.  Good tests become consistent.
If alternatives approach at the right rate, the likelihood ratio has a
nondegenerate limit.  That nondegenerate limit is the domain of contiguity and
Le Cam's lemmas.

The next two lemmas are included as a compact calculator, not as a new
doctrine the reader must admire.  They say how a statistic's null limit changes
when the alternative is close enough that neither law wins automatically.
Readers who want the applied thread can read the statements and then move
directly to the examples.

\subsection{Contiguity}
\label{sec:ch14-contiguity}
\conceptindexes{contiguity, Le Cam first lemma, Le Cam third lemma, likelihood-ratio limits}

Asymptotic testing compares sequences of experiments.  Let
\((\mathcal X_n,\mathcal A_n)\) be sample spaces, and let \(P_n,Q_n\) be
probability measures on them.  The sequence \(Q_n\) is close to \(P_n\) in the
testing sense if events that are negligible under \(P_n\) are also negligible
under \(Q_n\).

\begin{definition}[Contiguity]
The sequence \(Q_n\) is contiguous to \(P_n\), written \(Q_n\triangleleft P_n\),
if
\[
  P_n(A_n)\to0
  \quad\Longrightarrow\quad
  Q_n(A_n)\to0
\]
for every sequence \(A_n\in\mathcal A_n\).  If both \(Q_n\triangleleft P_n\)
and \(P_n\triangleleft Q_n\), the sequences are mutually contiguous.
\end{definition}

Contiguity is weaker than total-variation convergence and stronger than merely
saying that two distributions are close for one statistic.  It says that no
asymptotically impossible event under \(P_n\) becomes possible under \(Q_n\).
Consequently, \(O_{P_n}(1)\) and \(o_{P_n}(1)\) statements often transfer to
\(Q_n\), once the relevant statistic is controlled.

\begin{theorem}[Le Cam's first lemma, likelihood-ratio form; \citealp{leCam1986asymptotic}]
Suppose \(Q_n\ll P_n\), and let \(\Lambda_n=dQ_n/dP_n\).  If the likelihood
ratios \(\Lambda_n\) are tight under \(P_n\), and every subsequential weak
limit has expectation one, then \(Q_n\triangleleft P_n\).  Conversely,
contiguity forces the likelihood ratios to be uniformly tight under \(P_n\).
\end{theorem}

A common sufficient condition is especially useful.  If
\[
  \log\Lambda_n\weakto N\!\left(-\frac12\tau^2,\tau^2\right)
  \quad\hbox{under } P_n,
\]
then \(\Expect\exp\{N(-\tau^2/2,\tau^2)\}=1\), and the alternatives \(Q_n\) are
contiguous to \(P_n\).  This is why log-likelihood ratios with the
\(-\tau^2/2\) mean correction occur throughout local asymptotic theory.

\begin{theorem}[Le Cam's third lemma; \citealp{leCam1986asymptotic}]
Let \(L_n=\log(dQ_n/dP_n)\).  Suppose
\[
  (T_n,L_n)\weakto (T,L)
  \quad\hbox{under } P_n,
\]
and \(Q_n\triangleleft P_n\).  Then under \(Q_n\), the limiting distribution
of \(T_n\) is the \(e^L\)-tilt of the joint limit under \(P_n\): for bounded
continuous \(f\),
\[
  \lim_n \Expect_{Q_n}f(T_n)
  =
  \Expect_{P}\{f(T)e^L\}.
\]
\end{theorem}

\begin{corollary}[Normal shift under local alternatives]
\label{cor:ch14-normal-shift}
If under \(P_n\),
\[
  \begin{pmatrix}T_n\\ L_n\end{pmatrix}
  \weakto
  N\!\left[
  \begin{pmatrix}m\\ -\tau^2/2\end{pmatrix},
  \begin{pmatrix}\Sigma & c\\ c^T & \tau^2\end{pmatrix}
  \right],
\]
then under \(Q_n\),
\[
  T_n\weakto N(m+c,\Sigma).
\]
\end{corollary}

This corollary is the workhorse for asymptotic power calculations.  A test
statistic may be centered under the null, but under a contiguous alternative
its limit shifts by its covariance with the log-likelihood ratio.  Efficient
tests are those whose statistic is aligned with that likelihood-ratio
direction.

In practice the use is usually three steps: write a local likelihood ratio,
find its joint null limit with the statistic of interest, and read the shifted
limit under the alternative.  The next examples show the calculation in the
language of the opening chapters.

\begin{example}[Two-group endpoint from Chapter~1]
\label{ex:ch14-two-group-local-power}
Suppose two groups each contribute \(m\) binary records,
\[
  X_0\sim\Binomial(m,p_0),
  \qquad
  X_1\sim\Binomial(m,p_1),
\]
independently.  This is the simplest version of a public-risk comparison,
clinical response endpoint, or online conversion metric.  The null says
\(p_0=p_1=p\).  A natural standardized difference is
\[
  Z_m
  =
  \frac{\hat p_1-\hat p_0}{\sqrt{2p(1-p)/m}},
  \qquad
  \hat p_a=\frac{X_a}{m}.
\]
Under the null, \(Z_m\weakto N(0,1)\); in practice \(p\) is replaced by the
pooled estimate.

Now take a local alternative
\[
  p_0=p,
  \qquad
  p_1=p+\frac{h}{\sqrt m}.
\]
Only the second group changes, so the log-likelihood ratio is
\[
  L_m
  =
  \log\frac{dQ_m}{dP_m}
  =
  \frac{h}{\sqrt{p(1-p)}}S_{1m}
  -
  \frac{h^2}{2p(1-p)}
  +
  o_{P_m}(1),
  \qquad
  S_{1m}=\frac{X_1-mp}{\sqrt{mp(1-p)}}.
\]
Since \(Z_m=(S_{1m}-S_{0m})/\sqrt2\), Le Cam's third lemma gives, under the
local alternative,
\[
  Z_m
  \weakto
  N\!\left(\frac{h}{\sqrt{2p(1-p)}},1\right).
\]
Thus a one-sided level-\(\alpha\) test has approximate local power
\[
  1-\Phi\!\left(
    z_{1-\alpha}
    -
    \frac{h}{\sqrt{2p(1-p)}}
  \right).
\]

For a concrete scale, take \(p=0.20\) and \(m=1000\) per group.  A risk
increase of \(0.03\) corresponds to \(h=0.03\sqrt{1000}\approx0.95\), so the
local mean shift is about \(1.68\).  A one-sided \(5\%\) test then has power
about \(1-\Phi(1.645-1.68)\approx0.51\).  With \(m=4000\) per group, the same
risk increase gives about twice the local shift and power about \(0.96\).
This is the testing version of the Chapter~1 warning: a rounded percentage
difference is not self-interpreting; its evidential meaning depends on the
law and the information scale behind it.
\end{example}

\begin{example}[Discovery counts and exposure from Chapter~2]
\label{ex:ch14-poisson-local-power}
In the missing-species, vocabulary, and sequencing examples of Chapter~2, an
observed count is tied to the amount searched.  Let \(E_n\) denote exposure:
sample size, text length, sequencing depth, or search area.  Suppose a rare
discovery count satisfies the working model
\[
  X_n\sim\Poisson(E_n\lambda).
\]
To test \(\lambda=\lambda_0\) against a local increase
\[
  \lambda_n=\lambda_0+\frac{h}{\sqrt{E_n}},
\]
use
\[
  Z_n
  =
  \frac{X_n-E_n\lambda_0}{\sqrt{E_n\lambda_0}}.
\]
Under the null, \(Z_n\weakto N(0,1)\), and the log-likelihood ratio expands as
\[
  L_n
  =
  X_n\log\frac{\lambda_n}{\lambda_0}
  -
  E_n(\lambda_n-\lambda_0)
  =
  \frac{h}{\sqrt{\lambda_0}}Z_n
  -
  \frac{h^2}{2\lambda_0}
  +
  o_{P_n}(1).
\]
This is exactly Corollary~\ref{cor:ch14-normal-shift} with
\[
  T_n=Z_n,\qquad
  \tau^2=\frac{h^2}{\lambda_0},\qquad
  c=\Cov\!\left(Z,\frac{h}{\sqrt{\lambda_0}}Z\right)
    =\frac{h}{\sqrt{\lambda_0}},
  \quad Z\sim N(0,1).
\]
The theory is doing one concrete job: it turns the null calibration of
\(Z_n\) into the local-alternative calibration by adding the covariance with
the log-likelihood ratio.  Thus Le Cam's third lemma gives
\[
  Z_n
  \weakto
  N\!\left(\frac{h}{\sqrt{\lambda_0}},1\right)
  \quad\hbox{under }\lambda_n.
\]
If exposure is measured in blocks of text and \(\lambda_0=0.5\) new rare types
per block, then changing to \(0.6\) with \(E_n=100\) gives \(h=1\) and a mean
shift \(1/\sqrt{0.5}\approx1.41\).  At \(E_n=400\), the same rate difference
has twice the shift.  A zero or a small count is therefore not interpreted by
itself; it is interpreted together with how much text, area, time, or
sequencing depth was searched.
\end{example}

\subsection{Sequential Monitoring and Local Power}
\label{sec:ch14-sequential-monitoring}
\conceptindexes{sequential testing, group-sequential testing, alpha spending, Brownian motion, local power}

Sequential testing is a different error-control problem from multiple-endpoint
testing.  In sequential monitoring, the same scientific claim is inspected at
several information times.  The danger is repeated looking: even if the null is
true, one of the interim statistics may cross a boundary by chance.  The target
is therefore the probability of crossing any monitoring boundary under the
joint null law of the monitored process.  This belongs with contiguity and
local alternatives because the Brownian-limit calculation gives both null
size and local power.

\begin{corollary}[Drifted monitoring process under local alternatives]
\label{cor:ch14-sequential-drift}
Let \(0<t_1<\cdots<t_K\le1\).  Suppose that, under the null law \(P_n\),
the monitoring statistics and a local log-likelihood ratio satisfy
\[
  \left\{
    Z_n(t_1),\ldots,Z_n(t_K),L_n
  \right\}
  \weakto
  \left\{
    \frac{B(t_1)}{\sqrt{t_1}},\ldots,\frac{B(t_K)}{\sqrt{t_K}},
    \delta B(1)-\frac12\delta^2
  \right\},
\]
where \(B\) is standard Brownian motion.  Then, under the local alternative,
\[
  Z_n(t_j)
  \weakto
  \frac{B(t_j)+\delta t_j}{\sqrt{t_j}},
  \qquad j=1,\ldots,K.
\]
\end{corollary}

\noindent\textit{Verification.}
This is the same normal-shift corollary in vector form.  Under the null limit,
\[
  \Cov\!\left\{\frac{B(t_j)}{\sqrt{t_j}},\,\delta B(1)\right\}
  =
  \delta\sqrt{t_j}.
\]
Therefore the \(j\)th monitored statistic gains mean
\(\delta\sqrt{t_j}\), which is the same as replacing \(B(t_j)\) by
\(B(t_j)+\delta t_j\) before dividing by \(\sqrt{t_j}\). \qedmark

\begin{example}[Sequential and group-sequential testing]
Sequential testing fits this chapter because the sampling rule is part of the
law.  Imagine a two-arm trial with a standardized \(Z\)-statistic monitored
after half the planned information and again at the end.  Under the null, the
two looks are not independent.  A convenient approximation writes
\[
  Z(t)=\frac{B(t)}{\sqrt t},\qquad 0<t\le1,
\]
where \(B\) is standard Brownian motion and \(t\) is information fraction.
Then
\[
  \Cov\{Z(t_j),Z(t_k)\}=\sqrt{\frac{t_j}{t_k}},
  \qquad t_j\le t_k.
\]
If the analyst repeatedly used the ordinary two-sided \(1.96\) cutoff at every
look, the probability of at least one false rejection would exceed \(0.05\).
Group-sequential boundaries choose cutoffs \(c_1,c_2,\ldots\) so that
\[
  P_0\{\hbox{cross any efficacy boundary}\}\le \alpha
\]
under the joint null law of the monitored process.

This is the first way the previous theory enters: size is not calibrated
look-by-look.  It is calibrated under the joint Gaussian law of the vector
\((Z(t_1),\ldots,Z(t_K))\), whose correlations come from the Brownian
construction above.

For example, an O'Brien--Fleming-style two-look design spends little type-I
error early: the interim boundary is high, while the final boundary remains
close to the ordinary fixed-sample boundary \citep{obrien1979multiple}.  A
Lan--DeMets alpha-spending rule turns this into a design object that can
tolerate information times that are not known exactly in advance
\citep{lan1983discrete}.  The second way the theory enters is local power:
Corollary~\ref{cor:ch14-sequential-drift} says that, under an information-scale
local effect \(\delta\), the process becomes
\[
  Z_\delta(t)=\frac{B(t)+\delta t}{\sqrt t}
  =
  Z_0(t)+\delta\sqrt t .
\]
Therefore the same boundary rule has size
\[
  P_0\{\exists j:\ |Z_0(t_j)|>c_j\}
\]
and local power
\[
  P_\delta\{\exists j:\ |Z_\delta(t_j)|>c_j\}.
\]
This is a concrete case where Le Cam's normal-shift logic, Brownian process
correlations, stopping, and the monitoring plan all live inside the testing
law, not only the final table.
\end{example}

\subsection{Hellinger Differentiability and LAN}
\label{sec:ch14-lan}
\conceptindexes{Hellinger differentiability, local asymptotic normality, score, information, quadratic mean differentiability}

The local theory becomes especially clean when square-root densities are
differentiable in \(L^2\).  This is the Hellinger form of differentiability.

\begin{definition}[Hellinger differentiability]
Let \(\{P_\theta:\theta\in\Theta\subseteq\R^d\}\) be dominated by \(\mu\), with
density \(p_\theta\).  The model is Hellinger differentiable at \(\theta_0\) if
there exists a vector \(\dot\ell_{\theta_0}\in L_0^2(P_{\theta_0})^d\) such
that, as \(h\to0\),
\[
  \int
  \left[
    \sqrt{p_{\theta_0+h}}
    -
    \sqrt{p_{\theta_0}}
    -
    \frac12 h^T\dot\ell_{\theta_0}\sqrt{p_{\theta_0}}
  \right]^2 d\mu
  =
  o(\|h\|^2).
\]
The Fisher information is
\[
  I(\theta_0)=P_{\theta_0}\{\dot\ell_{\theta_0}\dot\ell_{\theta_0}^T\}.
\]
\end{definition}

This definition is deliberately expressed through square-root densities rather
than pointwise derivatives of log densities.  It is stable under changes on
null sets and is strong enough to control likelihood ratios.  In smooth
parametric models it agrees with the usual score and Fisher information.

\begin{theorem}[Local asymptotic normality; \citealp{leCam1986asymptotic}]
Let \(X_1,\ldots,X_n\) be iid from a Hellinger differentiable model at
\(\theta_0\), with nonsingular information \(I(\theta_0)\).  For fixed
\(h\in\R^d\), put \(\theta_n=\theta_0+h/\sqrt n\).  Then, under
\(P_{\theta_0}^{\otimes n}\),
\[
  \log
  \frac{dP_{\theta_n}^{\otimes n}}{dP_{\theta_0}^{\otimes n}}
  =
  h^T\Delta_n
  -
  \frac12 h^T I(\theta_0)h
  +
  o_{P_{\theta_0}}(1),
\]
where
\[
  \Delta_n
  =
  \frac1{\sqrt n}\sum_{i=1}^n \dot\ell_{\theta_0}(X_i)
  \weakto
  N(0,I(\theta_0)).
\]
\end{theorem}

The theorem says that, at \(n^{-1/2}\)-scale, a regular statistical experiment
looks like a Gaussian shift experiment.  The parameter \(h\) enters through
the linear score \(h^T\Delta_n\), and the quadratic correction
\(\frac12 h^TIh\) keeps the likelihood ratio normalized.

\begin{example}[Normal mean again]
For \(N(\theta,\sigma^2)\) with known \(\sigma^2\),
\[
  \dot\ell_{\theta_0}(x)=\frac{x-\theta_0}{\sigma^2},
  \qquad
  I(\theta_0)=\frac1{\sigma^2}.
\]
If \(\theta_n=\theta_0+h/\sqrt n\), then
\[
  \log
  \frac{dP_{\theta_n}^{\otimes n}}{dP_{\theta_0}^{\otimes n}}
  =
  \frac{h}{\sigma}
  \left\{\frac{\sqrt n(\bar X_n-\theta_0)}{\sigma}\right\}
  -
  \frac12\frac{h^2}{\sigma^2}.
\]
The expression is exactly linear normal shift plus the normalizing quadratic
term.  No asymptotic approximation is needed in this model.
\end{example}

The same expansion explains why Fisher information is a local metric.  For
\(\theta_n=\theta_0+h/\sqrt n\),
\[
  H^2(P_{\theta_0}^{\otimes n},P_{\theta_n}^{\otimes n})
\]
stays bounded away from both automatic zero and automatic one.  The local
separation is governed by \(h^TI(\theta_0)h\).

\section{Wald, Score, and Likelihood-Ratio Tests}
\label{sec:ch14-wald-score-lr}
\conceptindexes{Wald test, score test, likelihood-ratio test, composite null, efficient score}

In regular parametric models the three classical tests are different readings
of the same local quadratic experiment.  Let
\(\ell_n(\theta)=\sum_{i=1}^n\log p_\theta(X_i)\), and suppose the model is
regular at \(\theta_0\in\R^d\).  Let
\[
  U_n(\theta_0)=\frac{\partial}{\partial\theta}\ell_n(\theta_0),
  \qquad
  I_n(\theta_0)=nI(\theta_0).
\]
Under \(H_0:\theta=\theta_0\),
\[
  I_n(\theta_0)^{-1/2}U_n(\theta_0)\weakto N(0,I_d).
\]

If \(\hat\theta_n\) is an efficient regular estimator, then
\[
  \sqrt n(\hat\theta_n-\theta_0)
  =
  I(\theta_0)^{-1}
  \frac1{\sqrt n}U_n(\theta_0)
  +
  o_{P_{\theta_0}}(1).
\]
Consequently the Wald statistic
\[
  W_n
  =
  n(\hat\theta_n-\theta_0)^T
  I(\theta_0)
  (\hat\theta_n-\theta_0)
\]
converges to \(\chi_d^2\) under the null.

The score statistic reads the same local experiment at the null value:
\[
  S_n
  =
  U_n(\theta_0)^T I_n(\theta_0)^{-1}U_n(\theta_0)
  \weakto
  \chi_d^2 .
\]
The likelihood-ratio statistic compares the unrestricted and restricted
maximized log-likelihoods:
\[
  R_n
  =
  2\{\ell_n(\hat\theta_n)-\ell_n(\theta_0)\}
  \weakto
  \chi_d^2 .
\]
The three tests can behave differently in finite samples, but their first
order null limits coincide because they are the same quadratic form written in
three coordinates: estimate, score, and likelihood height.

\begin{proposition}[Local power]
Under local alternatives \(\theta_n=\theta_0+h/\sqrt n\), the Wald, score, and
likelihood-ratio statistics converge to a noncentral chi-square law with
noncentrality
\[
  \lambda=h^TI(\theta_0)h .
\]
\end{proposition}

\noindent\textit{Reason.}
LAN and Le Cam's third lemma shift the central sequence:
\[
  \Delta_n
  =
  \frac1{\sqrt n}U_n(\theta_0)
  \weakto
  N(I(\theta_0)h,I(\theta_0))
\]
under \(P_{\theta_0+h/\sqrt n}^{\otimes n}\).  The quadratic statistic
\(\Delta_n^TI(\theta_0)^{-1}\Delta_n\) therefore has a noncentral chi-square
limit with noncentrality \(h^TI(\theta_0)h\). \qedmark

\subsection{Composite Nulls and Efficient Scores}
\conceptindexes{composite nulls, efficient scores, nuisance tangent space, projection}

Now write \(\theta=(\xi,\eta)\), where \(\xi\in\R^k\) is the parameter of
interest and \(\eta\in\R^m\) is nuisance.  We test
\[
  H_0:\xi=\xi_0
  \qquad\hbox{against}\qquad
  H_1:\xi\ne\xi_0 .
\]
Partition the score and information at \(\theta_0=(\xi_0,\eta_0)\):
\[
  \dot\ell_\theta
  =
  \begin{pmatrix}
    \dot\ell_\xi\\
    \dot\ell_\eta
  \end{pmatrix},
  \qquad
  I(\theta)
  =
  \begin{pmatrix}
    I_{\xi\xi} & I_{\xi\eta}\\
    I_{\eta\xi} & I_{\eta\eta}
  \end{pmatrix}.
\]
The efficient score for \(\xi\) removes the part of \(\dot\ell_\xi\) explained
by the nuisance score:
\[
  \dot\ell_\xi^*
  =
  \dot\ell_\xi
  -
  I_{\xi\eta}I_{\eta\eta}^{-1}\dot\ell_\eta .
\]
Its variance is the efficient information
\[
  I^*
  =
  I_{\xi\xi}
  -
  I_{\xi\eta}I_{\eta\eta}^{-1}I_{\eta\xi}.
\]

This is the finite-dimensional version of the projection geometry used in
Chapter~15.  Nuisance directions form a space of score variation that should
not be counted as information about \(\xi\).  The efficient score is the
orthogonal residual.

The classical tests again take three forms.  A Wald test uses an unrestricted
efficient estimator \(\hat\xi_n\):
\[
  n(\hat\xi_n-\xi_0)^T I^*(\hat\theta_n)(\hat\xi_n-\xi_0).
\]
A score test evaluates the efficient score at a null-constrained estimate
\(\tilde\theta_n=(\xi_0,\tilde\eta_n)\):
\[
  \left\{\frac1{\sqrt n}\sum_{i=1}^n
  \dot\ell_{\xi,\tilde\theta_n}^*(X_i)\right\}^T
  I^*(\tilde\theta_n)^{-1}
  \left\{\frac1{\sqrt n}\sum_{i=1}^n
  \dot\ell_{\xi,\tilde\theta_n}^*(X_i)\right\}.
\]
A likelihood-ratio test compares the unrestricted maximum with the maximum
under \(\xi=\xi_0\).  Under regularity, all three have \(\chi_k^2\) null
limits and noncentral \(\chi_k^2\) local limits.

\begin{example}[Testing a GLM coefficient]
\label{ex:ch14-glm-testing}
The GLM examples from Chapter~13 give concrete forms of these three tests.
Suppose \(Y_i\) is binary and
\[
  \logit\{P(Y_i=1\mid x_i,A_i)\}
  =
  \alpha+x_i^T\gamma+\tau A_i ,
\]
where \(A_i\) is a treatment, exposure, or platform-assignment indicator.  In
Chapter~1 language, this could be a clinical response endpoint, an adverse
event indicator, an online conversion metric, or a public-risk outcome.  The
null hypothesis
\[
  H_0:\tau=0
\]
says that, conditional on the covariates in the model, the binary trace has no
log-odds shift associated with \(A_i\).

Let \(\beta=(\tau,\eta)\), where \(\eta=(\alpha,\gamma^T)^T\), and let
\[
  p_i(\beta)
  =
  \frac{\exp(\alpha+x_i^T\gamma+\tau A_i)}
       {1+\exp(\alpha+x_i^T\gamma+\tau A_i)}.
\]
The unrestricted log-likelihood is
\[
  \ell_n(\beta)
  =
  \sum_i
  \{Y_i\log p_i(\beta)+(1-Y_i)\log(1-p_i(\beta))\}.
\]
The Wald test estimates \(\hat\tau\) in the unrestricted model and rejects
for large
\[
  W_n=\frac{\hat\tau^2}{\widehat{\Var}(\hat\tau)}.
\]
The likelihood-ratio test compares the unrestricted fit with the constrained
fit \(\tau=0\):
\[
  R_n
  =
  2\{\ell_n(\hat\tau,\hat\eta)-\ell_n(0,\tilde\eta)\}.
\]
The score test fits only the null model, computes residuals
\(\tilde r_i=Y_i-p_i(0,\tilde\eta)\), and asks whether the treatment column
still has residual signal after the nuisance covariates have been accounted
for.  In projection language, this is the efficient score for \(\tau\).  In
the simplest fixed-design logistic model it is the treatment column residual,
weighted by \(p_i(1-p_i)\), paired with the binary residuals \(\tilde r_i\).

The same logic applies to a Poisson rate model from Chapter~2.  If
\[
  Y_i\sim\Poisson(\mu_i),
  \qquad
  \log\mu_i=\log e_i+\alpha+x_i^T\gamma+\tau A_i,
\]
then \(H_0:\tau=0\) tests a rate ratio \(e^\tau=1\), with \(e_i\) recording
exposure time, sampled area, sequencing depth, or document length.  This is
the testing version of the ``absence as information'' motif: a zero count
under a large exposure is more informative than a zero count under a tiny
exposure.  Wald, score, and likelihood-ratio tests all read the same local
Poisson likelihood, but in estimate, residual-score, and likelihood-height
coordinates.
\qedmark
\end{example}

\section{Testing Workflows and Model Checking}
\label{sec:ch14-modern-testing-examples}
\conceptindexes{generalized R-square, liquid association, randomization tests, permutation tests, single-cell testing, conditional independence, mutual independence, DiPMInd, distance profiles, model misspecification}

The same testing grammar handles problems that look less like textbook
one-parameter hypotheses.  The null may say that a block of predictors carries
no extra information, that a gene-pair relationship does not change with cell
state, that a simulator reproduces a distributional feature, or that several
random objects are mutually independent.

\begin{example}[Generalized \(R^2\) and multivariate signal]
Classical \(R^2\) measures how much squared-error variation a linear model
explains.  In likelihood models the more portable object is the improvement
over a null model.  Let \(\ell_0\) be the maximized log-likelihood under a
baseline model and \(\ell_1\) the maximized log-likelihood after adding a block
of predictors, interactions, or nonlinear features.  The likelihood-ratio
statistic is
\[
  G=2(\ell_1-\ell_0).
\]
It tests whether the larger model improves the law enough to overcome its
extra degrees of freedom.  A Cox--Snell-style generalized coefficient of
determination records the same evidence on an effect-size scale:
\[
  R^2_{\mathrm{CS}}=1-\exp(-G/n).
\]
Nagelkerke's version rescales this quantity so that the reported number can
reach one for discrete-outcome models \citep{nagelkerke1991note}.

This belongs in the testing chapter because the statistic is not just a
prediction score.  It asks whether the conditional law of \(Y\) changes when
the whole vector \(X\) is allowed into the model.  In a linear Gaussian model
it reduces to familiar \(R^2\) language; in logistic, Poisson, survival, or
multi-omics regression it becomes a likelihood-distance summary of an
\(X\)-to-\(Y\) relationship.
\end{example}

\begin{example}[Liquid association as a score test]
Ordinary correlation asks whether two variables move together on average.
Liquid association asks whether that relationship itself changes with a third
variable \citep{li2002coexpression}.  The basic operation is deliberately
simple.  Given observations \((X_i,Y_i,Z_i)\), transform, center, and scale the
three coordinates to
\[
  x_i=\frac{X_i-\bar X}{s_X},
  \qquad
  y_i=\frac{Y_i-\bar Y}{s_Y},
  \qquad
  z_i=\frac{Z_i-\bar Z}{s_Z}.
\]
Then form the pair-product \(p_i=x_i y_i\) and compute
\[
  \widehat{\mathrm{LA}}(X,Y\mid Z)
  =
  \frac1n\sum_{i=1}^n x_i y_i z_i
  =
  \frac1n\sum_{i=1}^n p_i z_i .
\]
At the population level, with \(X,Y,Z\) already standardized,
\[
  \theta_{\mathrm{LA}}
  =
  \Expect(XYZ)
  =
  \Cov(XY,Z)
  =
  \Cov\{\Expect(XY\mid Z),Z\}.
\]
Thus liquid association is the first-order slope of the \(X\)-\(Y\)
association along the \(Z\) direction.  Equivalently, \(\widehat{\mathrm{LA}}\)
is the least-squares slope in the regression
\[
  x_i y_i=\alpha+\beta z_i+\varepsilon_i,
  \qquad \hat\beta=\widehat{\mathrm{LA}},
\]
because the \(z_i\)'s have mean zero and variance one.  A positive value means
that the \(X\)-\(Y\) product tends to be larger at high \(Z\); a negative value
means that the pair tends to move together less, or move in the opposite
direction, at high \(Z\).

In gene-expression language, \(X\) and \(Y\) may be two genes and \(Z\) may be
a regulator, cell-state score, pseudotime coordinate, perturbation intensity,
or environmental marker.  A useful diagnostic is to plot the correlation of
\((X,Y)\) within low, middle, and high quantiles of \(Z\).  The statistic above
is the formal version of that plot; the bins are only a display.

\noindent\textit{Testing.}
The sharp moment null is
\[
  H_0:\theta_{\mathrm{LA}}=0.
\]
Under iid sampling and finite sixth moments,
\[
  \sqrt n\{\widehat{\mathrm{LA}}-\theta_{\mathrm{LA}}\}
  \weakto
  \Normal(0,\sigma^2_{\mathrm{LA}}).
\]
If the centering and scaling constants are treated as fixed, then
\(\sigma^2_{\mathrm{LA}}=\Var(XYZ)\).  With estimated means and variances, a
delta-method influence function is more honest.  Writing \(x,y,z\) for the
population-standardized variables, \(\rho_{XY}=\Expect(XY)\), and similarly
for the other pairs, one convenient influence function is
\[
\begin{aligned}
  \phi(x,y,z)
  &=
  xyz-\theta_{\mathrm{LA}}
  -x\rho_{YZ}-y\rho_{XZ}-z\rho_{XY} \\
  &\qquad
  -\frac{\theta_{\mathrm{LA}}}{2}
  \{(x^2-1)+(y^2-1)+(z^2-1)\}.
\end{aligned}
\]
The plug-in sandwich estimate
\[
  \hat\sigma^2_{\mathrm{LA}}
  =
  \frac1n\sum_{i=1}^n \hat\phi_i^2
\]
gives the large-sample statistic
\[
  T_{\mathrm{LA}}
  =
  \frac{\sqrt n\,\widehat{\mathrm{LA}}}{\hat\sigma_{\mathrm{LA}}}.
\]
When the null also gives exchangeability of \(Z\) relative to the observed
\((X,Y)\) pairs, a permutation calibration is often preferable:
shuffle the \(z_i\)' labels, recompute \(\widehat{\mathrm{LA}}^{(b)}\) or its
studentized version \(T_{\mathrm{LA}}^{(b)}\), and report
\[
  p_{\mathrm{perm}}
  =
  \frac{1+\sum_{b=1}^B
  \ind{|T_{\mathrm{LA}}^{(b)}|\ge |T_{\mathrm{LA}}^{\mathrm{obs}}|}}
  {B+1}.
\]
This permutation tests a stronger null than \(\theta_{\mathrm{LA}}=0\): it
asks whether \(Z\) can be detached from the \((X,Y)\) pairs without changing
the joint law.

\noindent\textit{Why this belongs in a testing chapter.}
Liquid association is a score test for a local three-way interaction in the
law.  Start with a null law \(P_0\) for the standardized triple and tilt it by
\[
  \frac{dP_\gamma}{dP_0}(x,y,z)
  =
  \exp\{\gamma xyz-\psi(\gamma)\}.
\]
At \(\gamma=0\), the score is \(xyz-\Expect_0(XYZ)\).  Under the common null
\(\Expect_0(XYZ)=0\), the score statistic is
\[
  S_n=\frac1{\sqrt n}\sum_{i=1}^n x_i y_i z_i
  =
  \sqrt n\,\widehat{\mathrm{LA}}.
\]
Under local alternatives \(\gamma=h/\sqrt n\), the likelihood-ratio theory
above shifts the mean of \(S_n\) by
\(h\,\Var_0(XYZ)\).  So the example is not a loose analogy: it is the score
test for a concrete local alternative in which the \(X\)-\(Y\) association is
allowed to change with \(Z\).

\noindent\textit{Practical cautions.}
Liquid association is not conditional independence testing.  A zero value of
\(\Expect(XYZ)\) can miss nonlinear or nonmonotone changes in association.
Conversely, marginal effects of \(Z\) on \(X\) or \(Y\), library size, donor,
batch, spatial region, or pseudotime uncertainty can create a nonzero third
moment that is not a biological interaction.  In single-cell or multi-omics
work, one should first residualize \(X,Y,Z\) against known nuisance variables
or use blocked permutations at the donor, batch, sample, or spatial-domain
level.  Genome-wide scans then require the multiple-testing machinery at the
end of the chapter, since every gene pair and regulator produces another
hypothesis.
\end{example}

\begin{example}[Randomization and permutation tests in omics]
Permutation tests are exact when the null law makes labels exchangeable.  If
the treatment label in a randomized experiment is assigned by design, the
randomization distribution is part of the experiment \citep{fisher1935design}.
For a test statistic \(T\), a Monte Carlo permutation \(p\)-value has the form
\[
  p_{\mathrm{perm}}
  =
  \frac{1+\sum_{b=1}^B \ind{T^{(b)}\ge T_{\mathrm{obs}}}}{B+1},
\]
where \(T^{(b)}\) is computed after an allowed relabeling.

Single-cell and multi-omics analyses make the word ``allowed'' do real work.
Cells from the same donor, batch, library, tissue region, time point, or
pseudotime segment are not automatically exchangeable.  A valid permutation
scheme may need to permute donor labels, permute within blocks, preserve
library-size strata, or simulate from a fitted generative law such as
scDesign3 before checking a downstream statistic.  PseudotimeDE-style
calibration \citep{song2021pseudotimede}, Clipper's p-value-free FDR logic
\citep{ge2021clipper}, and scDesign3-style simulation
all fit this chapter's grammar: name the null law,
name the statistic, and check whether the proposed data-generating mechanism
makes the observed feature surprising.

Conditional independence is the same lesson with a sharper null:
\[
  X\indep Y\mid Z.
\]
Here and below \(\indep\) denotes stochastic independence; the symbol
\(\perp\) is reserved for geometric orthogonality or mutual singularity of
measures.  Conditional independence is harder than ordinary permutation
independence because the null law must preserve the conditional distribution
given \(Z\).
\end{example}

\begin{example}[DiPMInd and distance-profile mutual independence]
DiPMInd, the method of \citet{chenDubey2024dipmind}, is best read as a modern
random-object version of independence testing.  The observations are not
assumed to be ordinary vectors.  For subject \(i\) one may observe
\[
  O_i=(Y_i^{(1)},\ldots,Y_i^{(K)}),
  \qquad
  Y_i^{(j)}\in(\mathcal M_j,d_j),
\]
where each component can be a curve, image, graph, distribution, trajectory, or
other metric-space object.  The null is mutual independence,
\[
  H_0:\quad
  P_{Y^{(1)},\ldots,Y^{(K)}}
  =
  P_{Y^{(1)}}\otimes\cdots\otimes P_{Y^{(K)}},
  \qquad
  Y^{(1)}\indep\cdots\indep Y^{(K)},
\]
not conditional independence.

The word ``profile'' is literal.  Fix an anchor
\(a=(a_1,\ldots,a_K)\), where \(a_j\in\mathcal M_j\), and convert the random
objects into their distances from that anchor:
\[
  D_a(Y)
  =
  \bigl(d_1(Y^{(1)},a_1),\ldots,d_K(Y^{(K)},a_K)\bigr).
\]
The joint distance profile at radius \(r=(r_1,\ldots,r_K)\) is the
distribution function of this distance vector,
\[
  F_a(r)
  =
  P\{d_1(Y^{(1)},a_1)\le r_1,\ldots,d_K(Y^{(K)},a_K)\le r_K\}.
\]
The product-of-marginals profile is
\[
  F_a^\otimes(r)
  =
  \prod_{j=1}^K
  P\{d_j(Y^{(j)},a_j)\le r_j\}.
\]
Under mutual independence, \(F_a=F_a^\otimes\) for every anchor and every
radius.  Conversely, under the regularity conditions in
\citet{chenDubey2024dipmind}, the collection of such profiles characterizes
the joint law, so a systematic discrepancy between \(F_a\) and \(F_a^\otimes\)
is evidence of dependence.  This is the same testing grammar as before:
first name the null product law, then choose a statistic that is sensitive to
departures from it.

For data, a simple empirical version uses the observed objects themselves as
anchors.  With anchor \(O_i\), define
\[
  \widehat F_i(r)
  =
  \frac{1}{n}\sum_{\ell=1}^n
  \prod_{j=1}^K
  1\{d_j(Y_\ell^{(j)},Y_i^{(j)})\le r_j\},
\]
and
\[
  \widehat F_{ij}(r_j)
  =
  \frac{1}{n}\sum_{\ell=1}^n
  1\{d_j(Y_\ell^{(j)},Y_i^{(j)})\le r_j\}.
\]
The local profile residual is
\[
  \widehat\Delta_i(r)
  =
  \widehat F_i(r)-\prod_{j=1}^K\widehat F_{ij}(r_j).
\]
A DiPMInd statistic aggregates the squared residuals over anchors and radii,
for example in the schematic form
\[
  T_n
  =
  n\,\frac{1}{n}\sum_{i=1}^n
  \int \widehat\Delta_i(r)^2\,w_i(r)\,d\widehat\nu_i(r),
\]
where \(w_i\) is a weight profile and \(\widehat\nu_i\) is an empirical measure
on the observed distance radii.  Different choices of \(w_i\) and
\(\widehat\nu_i\) give different members of the DiPMInd family.  The important
point is not the particular numerical quadrature; it is that the statistic
compares the empirical joint small-ball probabilities with the product of the
empirical marginal small-ball probabilities.  This places DiPMInd in the
distance-based testing family of \citet{szekely2007distance}, but the
metric-space formulation is what makes it useful for random objects rather
than only Euclidean vectors.

Here is the same idea in a toy case where nothing infinite-dimensional is
hidden.  Let \(K=2\), let \(\mathcal M_1=\mathcal M_2=\{0,1\}\), and use the
discrete metric \(d(x,a)=1\{x\ne a\}\).  Take anchor \(a=(0,0)\) and radius
\(r=(0,0)\).  The distance ball of radius \(0\) around \(0\) is just the point
\(\{0\}\), so
\[
  F_a(0,0)=P(Y^{(1)}=0,Y^{(2)}=0),
  \qquad
  F_a^\otimes(0,0)=P(Y^{(1)}=0)P(Y^{(2)}=0).
\]
If \(Y^{(1)}\) and \(Y^{(2)}\) are independent fair bits, both quantities are
\(1/4\).  If instead \(Y^{(2)}=Y^{(1)}\) and the common bit is fair, then
\[
  F_a(0,0)=1/2,
  \qquad
  F_a^\otimes(0,0)=1/4.
\]
In this finite example, distance profiles are just cell probabilities in
disguise, and the profile residual is the familiar independence residual
\[
  \widehat p_{00}-\widehat p_{0\cdot}\widehat p_{\cdot0}.
\]
For curves, images, networks, or distributions, the same comparison is made
with metric balls instead of table cells.

The calibration is also a direct application of the testing grammar in this
section.  Under \(H_0\), the component labels can be permuted independently:
\[
  (Y_i^{(1)},Y_i^{(2)},\ldots,Y_i^{(K)})
  \longmapsto
  (Y_{\pi_1(i)}^{(1)},Y_{\pi_2(i)}^{(2)},\ldots,Y_{\pi_K(i)}^{(K)}),
\]
because the joint law factors into marginal laws.  The permutation distribution
therefore approximates the null law of the distance-profile statistic.  Under
dependence, this relabeling destroys the cross-object alignment that carries
signal.

This example belongs here for the same reason the likelihood-ratio and
permutation examples belong here: the statistic is only meaningful after the
null law is named.  DiPMInd tests
\[
  P_{Y^{(1)},\ldots,Y^{(K)}}
  =
  P_{Y^{(1)}}\otimes\cdots\otimes P_{Y^{(K)}}.
\]
It should not be advertised as a conditional-independence test
\[
  X\indep Y\mid Z
\]
unless the conditioning has been built into the construction, for example by
residualizing, stratifying, matching, or modeling the conditional laws given
\(Z\).  Without that step, the method tests mutual independence of random
objects, not conditional independence.
\end{example}

\begin{example}[Model misspecification changes the null]
The likelihood-ratio, Wald, and score tests above rely on a correctly specified
regular model for their clean \(\chi^2\) limits.  Under misspecification, the
maximum likelihood estimator converges to the best approximating parameter
\citep{white1982maximum}:
\[
  \theta^*=\argmax_\theta P_0\log p_\theta(X),
\]
not necessarily to a true data-generating parameter.  If
\[
  A=-P_0\ddot\ell_{\theta^*}(X),
  \qquad
  B=P_0\{\dot\ell_{\theta^*}(X)\dot\ell_{\theta^*}(X)^T\},
\]
then the asymptotic covariance of \(\hat\theta\) is the sandwich
\[
  A^{-1}BA^{-1}.
\]
The test target has changed: \(H_0:\xi=0\) now means that the best
approximating member of the working model has \(\xi^*=0\).  A robust Wald or
score test may still be useful, but the model-based likelihood-ratio
\(\chi^2\) calibration is no longer automatic.  This is why model checking and
misspecification are not side issues; they decide what the null means.
\end{example}

\section{Multiplicity and Error-Rate Control}
\label{sec:ch14-multiplicity-error-control}
\conceptindexes{multiplicity, multiple testing, family-wise error rate, false discovery rate, gatekeeping}

Multiplicity begins where one observed study can support several possible
claims.  This is not the same problem as sequential monitoring.  Sequential
testing controls the error made by looking repeatedly at the same claim over
information time.  Multiple testing controls the error made by selecting among
several claims, endpoints, subgroups, genes, features, or model contrasts.
Both problems calibrate a rule, not a single \(p\)-value, but the index sets
are different: information times for sequential testing; hypothesis labels for
multiple testing.

\subsection{Confirmatory Error Control: FWER and Gatekeeping}
\label{sec:ch14-confirmatory-trials-contracts}
\conceptindexes{confirmatory clinical trials, regulatory testing design, estimands, family-wise error rate, closed testing, Holm procedure, Hochberg procedure, truncated gatekeeping}

A confirmatory clinical trial can ask several public questions at once: a
primary endpoint, a key secondary endpoint, a subgroup claim, or a claim under
a particular estimand strategy.  Testing each question at level \(\alpha\)
would make the chance of at least one false confirmatory claim larger than
\(\alpha\).  The statistical problem is therefore not to calibrate a single
\(p\)-value.  It is to calibrate the whole rejection rule that turns the
observed trial into a set of claims.

Following the estimand language of ICH E9(R1), each claim first needs a target
effect.  Denote the target effect for claim \(j\) by \(\Delta_j\), with the
sign chosen so that benefit is positive.  A one-sided superiority null is
\[
  H_j:\ \Delta_j\le 0 .
\]
For \(m\) claims, the protocol specifies how to compute the \(m\) marginal
\(p\)-values
\[
  p_1,\ldots,p_m .
\]
It also specifies a multiplicity procedure: an algorithm that receives these
\(p\)-values and returns which claims may be reported as confirmatory.  The
FDA multiple-endpoints guidance and ICH E9(R1) are useful here because they
force the same discipline as the rest of this chapter: name the target effect,
name the null family, name the \(p\)-value construction, then name the
rejection rule \citep{fda2022multipleEndpoints,ich2021e9r1}.

\noindent\textit{What is being controlled.}
The target error rate is usually strong family-wise error:
\[
  \Prob\{\hbox{at least one false confirmatory rejection}\}\le \alpha .
\]
The word ``family'' says that the error event concerns the whole set of claims,
not one endpoint in isolation.  The word ``strong'' says that the guarantee
must hold no matter which subset of \(H_1,\ldots,H_m\) is actually true.  This
is stronger than checking only the global null where all \(m\) nulls are true.

\noindent\textit{How the control is implemented.}
Closed testing is the master construction.  For a subset
\(I\subseteq\{1,\ldots,m\}\), the intersection null is
\[
  H_I=\bigcap_{i\in I}H_i .
\]
This says: every null in the set \(I\) is true.  The closed-testing algorithm
rejects an elementary null \(H_j\) only after rejecting every intersection null
that contains it:
\[
  \hbox{reject }H_j
  \quad\Longleftrightarrow\quad
  \hbox{reject every }H_I\hbox{ with }j\in I .
\]
The property is the key point: if every intersection test has level
\(\alpha\), then the whole closed-testing rule controls strong FWER.  If a
false confirmatory rejection occurs, the intersection of the true nulls must
also have been falsely rejected.  That one intersection test has error
probability at most \(\alpha\) \citep{marcus1976closed}.

\noindent\textit{Holm: a Bonferroni shortcut.}
Holm implements closed testing with Bonferroni local tests
\citep{holm1979simple}.  For an intersection \(I\), Bonferroni rejects \(H_I\)
when one \(p\)-value in \(I\) is small enough:
\[
  \min_{i\in I}p_i\le\frac{\alpha}{|I|}.
\]

\begin{definition}[Holm step-down procedure]
Fix a strong-FWER target \(\alpha\).  Given \(m\) marginal \(p\)-values, order
them as
\[
  p_{(1)}\le\cdots\le p_{(m)} .
\]
Let \(H_{(j)}\) be the null hypothesis paired with \(p_{(j)}\).  Starting with
the smallest \(p\)-value, compare \(p_{(j)}\) at rank \(j\) with
\[
  \frac{\alpha}{m-j+1}.
\]
Reject in order while
\[
  p_{(j)}\le\frac{\alpha}{m-j+1}
\]
holds.  Stop at the first rank \(r\) for which the inequality fails, and reject
\(H_{(1)},\ldots,H_{(r-1)}\).  If no failure occurs, reject all \(m\) null
hypotheses.
\end{definition}

This step-down algorithm is exactly the closed-testing rule generated by
Bonferroni intersection tests.  Its main property is robustness: Holm controls
strong FWER under arbitrary dependence among valid marginal \(p\)-values.  Its
cost is conservativeness, especially when many endpoints are positively
correlated.

\begin{lemma}[Holm's strong FWER control; \citealp{holm1979simple}]
Assume every true-null \(p\)-value is valid: under its null,
\[
  \Prob\{p_i\le t\}\le t,\qquad 0\le t\le1.
\]
Then Holm's step-down procedure controls the strong family-wise error rate at
level \(\alpha\), without any dependence assumption among the \(p\)-values.
\end{lemma}

\noindent\textit{Proof.}
Let \(I_0\) be the set of true nulls and let \(m_0=|I_0|\).  If
\(m_0=0\), there is no false rejection to make.  Otherwise, suppose Holm makes
at least one false rejection.  Look at the first true null in the ordered list
\[
  p_{(1)}\le\cdots\le p_{(m)} ,
\]
and call its rank \(J\).  Before rank \(J\), only false nulls can appear, so
\[
  J\le m-m_0+1,
  \qquad\hbox{hence}\qquad
  m-J+1\ge m_0 .
\]
For this true null to be rejected, Holm must have
\[
  p_{(J)}\le \frac{\alpha}{m-J+1}\le \frac{\alpha}{m_0}.
\]
But \(p_{(J)}\) is the smallest true-null \(p\)-value.  Therefore
\[
  \{\hbox{at least one false rejection}\}
  \subseteq
  \left\{\min_{i\in I_0}p_i\le\frac{\alpha}{m_0}\right\}.
\]
By the union bound and marginal validity,
\[
  \Prob\{\hbox{at least one false rejection}\}
  \le
  \sum_{i\in I_0}\Prob\left\{p_i\le\frac{\alpha}{m_0}\right\}
  \le
  m_0\frac{\alpha}{m_0}
  =
  \alpha .
\]
Since the argument holds for every possible true-null set \(I_0\), the control
is strong. \qedmark

\noindent\textit{Hochberg: a Simes shortcut.}
Hochberg uses the same ordered \(p\)-values and the same cutoffs, but it is a
step-up rule rather than a step-down rule \citep{hochberg1988sharper}.  Find
the largest rank \(k\) such that
\[
  p_{(k)}\le\frac{\alpha}{m-k+1}.
\]
Then reject \(H_{(1)},\ldots,H_{(k)}\).  The algorithm is more aggressive than
Holm because a sufficiently small larger \(p\)-value can carry all smaller
\(p\)-values with it.  The property is therefore conditional: Hochberg gives
strong FWER control under Simes-type validity conditions, such as independence
or suitable positive dependence among the true-null \(p\)-values.  Without
such conditions, Holm is the safer default
\citep{simes1986improved,hochberg1988sharper}.

\noindent\textit{Truncated gatekeeping: encoding claim order.}
Some confirmatory programs do not want all claims to have equal standing.  A
primary endpoint may need to protect or unlock a secondary endpoint, or a
subgroup claim may be allowed only after the main-population evidence is
credible.  Gatekeeping turns this scientific priority into an error-allocation
rule.  Truncated Holm and truncated Hochberg implement it by replacing
\(\alpha\) in each intersection test by a smaller local level
\[
  e_\gamma(I)
  =
  \alpha\left\{\gamma+(1-\gamma)\frac{|I|}{m}\right\},
  \qquad 0\le\gamma\le1.
\]
Here \(I\) is the intersection being tested.  When \(\gamma=1\), the rule is
untruncated.  Smaller \(\gamma\) makes smaller intersections more
conservative.  Truncated Holm uses the Bonferroni comparison
\[
  \min_{i\in I}p_i\le\frac{e_\gamma(I)}{|I|}.
\]
Truncated Hochberg orders the \(p\)-values inside \(I\) and uses the same
Simes-style comparison with \(e_\gamma(I)\) in place of \(\alpha\).  Like
ordinary Hochberg, it needs Simes-type dependence assumptions; otherwise the
truncated Holm version is the safer default.  The property is inherited from
closed testing: once the local intersection tests are valid at their assigned
levels, the final confirmatory rejection set has the advertised strong FWER
control \citep{simes1986improved,hochberg1988sharper,dmitrienko2008gatekeeping}.

\begin{example}[FDA/ICH-style confirmatory trial as a testing design]
The industry motivation for this stylized example came from a
clinical-development discussion with Kaylor at Huadong Medicine, Hangzhou
(personal communication); the mathematical formulation below is presented in
the notation of this chapter.  The setting is a prospective confirmatory
efficacy trial, not a post-marketing signal-detection exercise.  Consider a
two-arm confirmatory trial comparing \(A=1\) with control \(A=0\).
The protocol fixes the randomization law, follow-up window, missing-data
handling, endpoint definitions, and analysis model.  Let \(P_{\theta,d}\)
denote the law of the full trial data under that fixed design.  Suppose the
protocol plans three confirmatory claims:

\begin{enumerate}[label=(\roman*),leftmargin=2em]
\item a primary-endpoint treatment effect in the all-randomized population;
\item the same primary-endpoint effect in a biomarker-positive subgroup;
\item a secondary-endpoint effect in the all-randomized population, possibly
with a hypothetical strategy for rescue therapy.
\end{enumerate}

Let \(\Delta_j(\theta)\) be the pre-specified treatment-effect summary for
claim \(j\), with the sign chosen so that benefit is positive.  The elementary
nulls are
\[
  H_j:\ \Delta_j(\theta)\le0,
  \qquad j=1,2,3.
\]
Assume the multiplicity plan uses a one-sided family-wise level
\(\alpha=0.025\) and a truncated Hochberg gatekeeping rule with
\(\gamma=1/2\).  This choice needs the same Simes/Hochberg validity conditions
as above; without them, the more conservative truncated Holm version is the
safer default.

For example, consider the pair intersection \(I=\{1,3\}\).  The local level is
\[
  e_{1/2}(\{1,3\})
  =
  0.025\left\{\frac12+\frac12\frac{2}{3}\right\}
  \approx 0.0208 .
\]
Order the two \(p\)-values as
\[
  p_{(1:\{1,3\})}\le p_{(2:\{1,3\})}.
\]
The truncated Hochberg local test rejects this pair intersection when either
both \(p\)-values are at most \(0.0208\), or the smaller \(p\)-value is at
most \(0.0104\):
\[
  p_{(2:\{1,3\})}\le0.0208
  \quad\hbox{or}\quad
  p_{(1:\{1,3\})}\le0.0104 .
\]
Finally, claim \(j\) is accepted as confirmatory only if every intersection
containing \(H_j\) is rejected.  Closed testing then gives the operating
characteristic
\[
  P_{\theta,d}\{\hbox{at least one false confirmatory claim}\}
  \le0.025
\]
for every configuration of true and false claims, provided the local
\(p\)-values have the advertised validity.  The statistical point is modest:
regulatory pre-specification means that the data object, the claim family, the
\(p\)-value calculations, and the rejection rule were fixed before looking at
the data.
\end{example}

\subsection{Discovery-Scale Multiple Testing: FDR and BH}
\label{sec:ch14-multiple-testing}
\conceptindexes{multiple testing, false discovery rate, Benjamini--Hochberg procedure, reverse martingale, p-values}

High-throughput biology, imaging, online experiments, and model-selection
pipelines rarely run one test.  They produce a selected set.  The error target
must therefore be a property of the selection rule.

The contrast with Holm is important.  Holm controls FWER: it asks for a small
probability of making even one false rejection.  BH controls FDR: it allows a
large discovery list, but asks that the expected false fraction in that list be
small.  Holm is natural for confirmatory claims; BH is natural for screening
and discovery problems where some false positives can be tolerated if the
reported list is still reliable as a list.

Think of a single-cell marker screen.  One may test thousands of genes for
differential expression between two cell populations and then hand a short
list to a biologist.  Controlling the probability of even one false marker is
often too severe: with \(m=20{,}000\) tests, a Bonferroni family-wise cutoff at
level \(0.05\) is \(2.5\times10^{-6}\).  FDR asks the question that matches the
follow-up list more closely: among the genes we call discoveries, what fraction
should we expect to be false?  If a rule returns \(R=400\) genes at FDR level
\(q=0.05\), the target is not ``a 5 percent chance of any mistake.''  It is an
expected false fraction of at most about \(5\%\), under the assumptions used to
calibrate the rule.

\begin{definition}[False discovery rate]
For \(m\) hypotheses, let \(R\) be the number rejected and \(V\) the number of
rejected true null hypotheses.  The false discovery rate is
\[
  \mathrm{FDR}
  =
  \Expect\left\{\frac{V}{R\vee1}\right\}.
\]
\end{definition}

\begin{definition}[Benjamini--Hochberg step-up procedure]
Fix an FDR target \(q\in(0,1)\).  Given \(m\) \(p\)-values, order them as
\[
  p_{(1)}\le\cdots\le p_{(m)} .
\]
The Benjamini--Hochberg procedure \citep{benjamini1995fdr} finds the largest
rank
\[
  \hat k
  =
  \max\left\{
    k:\ p_{(k)}\le \frac{qk}{m}
  \right\}.
\]
If this set is nonempty, reject every hypothesis with
\(p_i\le p_{(\hat k)}\).  If the set is empty, make no rejections.
\end{definition}

\begin{proposition}[BH control under independence; \citealp{benjamini1995fdr}]
If the \(m_0\) true-null \(p\)-values are independent uniform variables and
are independent of the false-null \(p\)-values, then the BH procedure at level
\(q\) satisfies
\[
  \mathrm{FDR}\le q\,\frac{m_0}{m}\le q.
\]
\end{proposition}

\noindent\textit{Proof.}
Let \(I_0\) be the true-null index set.  For \(0<t\le1\), define
\[
  R(t)=\sum_{i=1}^m \ind{p_i\le t},
  \qquad
  V(t)=\sum_{i\in I_0}\ind{p_i\le t},
\]
so that \(R=R(\hat t)\) and \(V=V(\hat t)\) when the BH threshold
\(\hat t=p_{(\hat k)}\) is positive.

The reverse filtration is the information available when the threshold is
moved downward from \(1\) toward \(0\):
\[
  \mathcal F_t
  =
  \sigma\!\left(
    \{\ind{p_i\le s}: i\in I_0,\ s\ge t\},
    \{p_j:j\notin I_0\}
  \right).
\]
Thus \(\mathcal F_s\supseteq\mathcal F_t\) when \(0<s<t\).  It knows all
nonnull \(p\)-values and, for each true null, it knows the exact value if that
value is above \(t\), but only knows that it lies in \([0,t]\) if it is below
\(t\).  For a true-null \(p_i\sim\mathrm{Unif}(0,1)\), independence gives, for
\(0<s<t\),
\[
  \Expect\!\left\{
    \frac{\ind{p_i\le s}}{s}\,\middle|\,\mathcal F_t
  \right\}
  =
  \frac{\ind{p_i\le t}}{t}.
\]
Indeed, if \(p_i>t\) both sides are zero; if \(p_i\le t\), the conditional law
of \(p_i\) on \([0,t]\) is uniform, so
\(\Prob(p_i\le s\mid p_i\le t)=s/t\).  Summing over \(i\in I_0\), the process
\[
  M(t)=\frac{V(t)}{t}
\]
is a nonnegative reverse martingale with \(\Expect M(1)=m_0\).

Next observe that the BH threshold is a reverse stopping time.  For fixed
\(t\), the event \(\{\hat t\ge t\}\) is determined by the nonnull \(p\)-values
and the true-null information above \(t\): it is the event that some observed
\(p\)-value at least \(t\) satisfies \(p_j\le qR(p_j)/m\).  Hence
\(\{\hat t\ge t\}\in\mathcal F_t\).  To avoid the zero-threshold case, set
\(\hat t_\epsilon=\hat t\vee\epsilon\).  For every \(\epsilon>0\),
\(\hat t_\epsilon\) is bounded away from zero, so optional stopping for the
bounded reverse martingale \(M(t)\), \(\epsilon\le t\le1\), gives
\[
  \Expect M(\hat t_\epsilon)=\Expect M(1)=m_0.
\]
On the event \(\{\hat t>0\}\), the BH step-up rule gives
\[
  \hat t\le \frac{qR(\hat t)}{m}.
\]
Therefore
\[
  \frac{V}{R\vee1}
  =
  \ind{\hat t>0}\,
  \frac{V(\hat t)}{R(\hat t)}
  \le
  \frac{q}{m}\,
  \ind{\hat t>0}\,M(\hat t).
\]
Moreover,
\[
  \Expect\{\ind{\hat t\ge\epsilon}M(\hat t)\}
  \le
  \Expect M(\hat t_\epsilon)
  =
  m_0.
\]
Letting \(\epsilon\downarrow0\) and using monotone convergence yields
\(\Expect\{\ind{\hat t>0}M(\hat t)\}\le m_0\).  Taking expectations in the
previous display gives
\[
  \mathrm{FDR}\le q\,\frac{m_0}{m}.
\]
This proof is memorable because it turns the random BH cutoff from a nuisance
into the stopping time of a simple counting martingale.  The assumptions also
matter: dependence, adaptive filtering, and data-driven hypothesis generation
change the selection law and may require modified procedures or empirical
checks.
\qedmark

\section{What Testing Adds to Statistical Claims}
\label{sec:ch14-testing-compass}
\conceptindexes{testing contributions, evidence, local alternatives, operating characteristics}

This chapter adds one missing layer to the route from probability to
inference.  Chapter~13 asks how estimators are selected.  Chapter~15 asks how
selected estimators report uncertainty.  Testing asks a different question:
which probability laws can be distinguished at all, and at what local scale?

The answer is geometric.
\[
  \begin{array}{rcl}
  \hbox{simple testing} &\longleftrightarrow& \hbox{total variation},\\[0.25em]
  \hbox{product laws} &\longleftrightarrow& \hbox{Hellinger affinity},\\[0.25em]
  \hbox{local alternatives} &\longleftrightarrow& \hbox{contiguity},\\[0.25em]
  \hbox{regular parametric models} &\longleftrightarrow& \hbox{Gaussian shift experiments}.
  \end{array}
\]
Once the experiment has become locally Gaussian, Wald, score, and
likelihood-ratio tests are no longer mysterious recipes.  They are different
coordinates for the same local quadratic form.

The practical lesson is equally important.  A small \(p\)-value is not a
measure of effect size, and failure to reject is not evidence that the null is
true.  A test is a calibrated decision rule under a specified sampling law.
Its interpretation depends on the target, the null family, the alternative
scale, and the information available in the observed data structure.  This is
why testing belongs in the book, but not as a separate cookbook.  It is one
more way to read the same statistical compass.

\section{Exercises}
\conceptindexes{testing exercises, Hellinger-distance exercises, LAN exercises}

\begin{exercise}[Simple testing and total variation]
Let \(P\) and \(Q\) have densities \(p\) and \(q\).  Show that
\[
  \inf_{\phi}\{P\phi+Q(1-\phi)\}
  =
  1-\|P-Q\|_{\mathrm{TV}}.
\]
Identify all nonrandomized tests attaining the infimum when \(P(q=p)=0\).
\end{exercise}

\begin{exercise}[Hellinger and total variation bounds]
Prove the inequalities
\[
  H^2(P,Q)\le \|P-Q\|_{\mathrm{TV}}
  \le H(P,Q)\sqrt{2-H^2(P,Q)}.
\]
\end{exercise}

\begin{exercise}[Hellinger affinity under products]
Let \(P_i,Q_i\) be probability measures with Hellinger affinities
\(\rho_i=\rho(P_i,Q_i)\).  Show that
\[
  \rho\!\left(\bigotimes_{i=1}^nP_i,\bigotimes_{i=1}^nQ_i\right)
  =
  \prod_{i=1}^n\rho_i.
\]
Deduce the corresponding formula for squared Hellinger distance.
\end{exercise}

\begin{exercise}[Contiguity for local normal shifts]
Let \(X_1,\ldots,X_n\) be iid \(N(0,1)\) under \(P_n\) and iid
\(N(h/\sqrt n,1)\) under \(Q_n\).  Compute the log-likelihood ratio and show
that \(Q_n\triangleleft P_n\).  Find the limit of
\(\sqrt n\,\bar X_n\) under \(Q_n\).
\end{exercise}

\begin{exercise}[Bernoulli LAN and local power]
Let \(X_1,\ldots,X_n\) be iid Bernoulli\((p)\), with \(p_0\in(0,1)\).  Test
\(H_0:p=p_0\) against \(p=p_0+h/\sqrt n\).
\begin{enumerate}
\item Derive the LAN expansion.
\item Identify the Fisher information.
\item Find the local asymptotic power of the score test at level \(\alpha\).
\end{enumerate}
\end{exercise}

\begin{exercise}[Equivalence of classical large-sample tests]
For a regular one-parameter model, show that the Wald, score, and
likelihood-ratio statistics are asymptotically equivalent under the null:
their pairwise differences are \(o_{P_0}(1)\).
\end{exercise}

\begin{exercise}[Efficient score as projection]
Let \(\theta=(\xi,\eta)\) and suppose the information matrix is nonsingular.
Show that
\[
  I^*=I_{\xi\xi}-I_{\xi\eta}I_{\eta\eta}^{-1}I_{\eta\xi}
\]
is the variance of the efficient score
\[
  \dot\ell_\xi^*
  =
  \dot\ell_\xi-I_{\xi\eta}I_{\eta\eta}^{-1}\dot\ell_\eta .
\]
Interpret the formula as an \(L^2(P_\theta)\) projection.
\end{exercise}

\begin{exercise}[Normal mean with unknown scale]
In the normal location-scale model \(N(\mu,\sigma^2)\), test \(H_0:\mu=0\)
with \(\sigma\) unknown.  Derive the efficient score for \(\mu\), the
efficient information, and the score test.  Compare it with the usual
one-sample \(t\)-test.
\end{exercise}

\begin{exercise}[Hellinger differentiability and local scale]
Let \(P_{\theta}\) be Hellinger differentiable at \(\theta_0\).  Show that
\[
  H^2(P_{\theta_0+h},P_{\theta_0})
  =
  \frac18 h^TI(\theta_0)h+o(\|h\|^2).
\]
Use this to explain why \(h/\sqrt n\) is the local scale for iid samples.
\end{exercise}

\begin{exercise}[Le Cam's third lemma for local power]
Let \((T_n,L_n)\) converge jointly under \(P_n\) to a bivariate normal vector
with means \((0,-\tau^2/2)\), variances \((1,\tau^2)\), and covariance
\(c\).  Use Le Cam's third lemma to find the limiting power of the test
\(\ind{T_n>z_{1-\alpha}}\) under \(Q_n\), where
\(L_n=\log(dQ_n/dP_n)\).
\end{exercise}

\begin{exercise}[Generalized \(R^2\)]
For a likelihood model with null log-likelihood \(\ell_0\), full
log-likelihood \(\ell_1\), and \(G=2(\ell_1-\ell_0)\), define
\[
  R^2_{\mathrm{CS}}=1-\exp(-G/n).
\]
Show how this quantity is related to the likelihood-ratio test.  In a Gaussian
linear model with an intercept, compare it with the usual squared multiple
correlation.
\end{exercise}

\begin{exercise}[Liquid association]
Let \(X,Y,Z\) be centered, scaled random variables.  Interpret
\(\Expect(XYZ)\) as a first-order measure of how the \(X\)-\(Y\) association
changes with \(Z\).  Propose a permutation test for the null that the
\((X,Y)\) pairs are exchangeable with respect to the observed \(Z\)'s, and
state one reason that this permutation could fail in a single-cell experiment.
\end{exercise}

\begin{exercise}[Benjamini--Hochberg by leave-one-out]
For a true null \(i\), let \(R_i^0\) be the number of BH rejections after
replacing \(p_i\) by \(0\) and leaving all other \(p\)-values fixed.  Show that
on the event that hypothesis \(i\) is rejected, \(R=R_i^0\), and that this
event is equivalent to \(p_i\le qR_i^0/m\).  Use independence of \(p_i\) from
the other \(p\)-values to prove
\[
  \Expect\left\{\frac{\ind{i\text{ is rejected}}}{R\vee1}\right\}
  \le
  \frac{q}{m}.
\]
Sum over true nulls and compare this proof with the reverse-martingale proof
in the text.
\end{exercise}

\section*{Sources and Further Reading}
\addcontentsline{toc}{section}{Sources and Further Reading}

The simple-vs-simple testing result is the Neyman--Pearson lemma
\citep{neyman1933tests}.  Fisher's discriminant analysis
\citep{fisher1936discriminant} is a classical bridge from likelihood ratios to
classification.  Hellinger distance is used here as the square-root-density
geometry behind product experiments and local differentiability.  Rao's
Fisher metric and intrinsic distance go back to
\citet{rao1945information,rao1949distance}; explicit distance calculations
and examples are developed by \citet{atkinsonMitchell1981rao}, and the modern
information-geometric background is treated broadly by
\citet{amari2016information}.  The product-measure dichotomy is due to
Kakutani \citep{kakutani1948equivalence}.  Contiguity and the three lemmas are
part of Le Cam's asymptotic theory; modern treatments appear in
\citet{vaart1998asymptotic} and \citet{bickel1993efficient}.  The
likelihood-ratio chi-square limit is associated with Wilks
\citep{wilks1938large}; the score and Wald formulations are classical large
sample tests \citep{rao1948large,wald1943tests}.
Quantum statistical inference is not developed in this chapter, but it is a
useful boundary case for the law-based language: density operators replace
ordinary probability laws until a measurement turns the problem into a
classical experiment.  For quantum testing and the associated trace-distance
and relative-entropy geometry, see
\citet{holevo1982probabilistic,barndorffNielsenGillJupp2003quantum,petz2008quantum}.

The applied examples point in several directions.  Goodness-of-fit and
Kolmogorov--Smirnov limits connect this chapter to the empirical-process
material in Chapters~10 and~11.  Generalized coefficients of determination are
represented here by Nagelkerke's likelihood-based definition
\citep{nagelkerke1991note}.  Liquid association comes from
\citet{li2002coexpression}.  Randomization tests go back to Fisher's design
program \citep{fisher1935design}; the omics examples point to PseudotimeDE
\citep{song2021pseudotimede}, Clipper \citep{ge2021clipper}, and scDesign3
as a simulation-checking example.  Distance-based independence testing is represented
by \citet{szekely2007distance} and the random-object DiPMInd framework of
\citet{chenDubey2024dipmind}.  Group-sequential monitoring follows the
Pocock, O'Brien--Fleming, and Lan--DeMets line
\citep{pocock1977group,obrien1979multiple,lan1983discrete}.  Confirmatory
clinical-trial multiplicity is connected here to the ICH E9(R1) estimand
framework and FDA multiple-endpoints guidance
\citep{ich2021e9r1,fda2022multipleEndpoints}.  Closed testing comes from
\citet{marcus1976closed}; Holm, Simes, Hochberg, and gatekeeping variants are
represented by
\citet{holm1979simple,simes1986improved,hochberg1988sharper,dmitrienko2008gatekeeping}.
FDR control is anchored by \citet{benjamini1995fdr}.  Under misspecification,
the pseudo-true-parameter and sandwich logic follows \citet{white1982maximum}.
Chapter~15 uses the same local geometry in the language of influence
functions, tangent spaces, and efficiency.

%% file: chapters/ch15_local_approximation_influence.tex
\chapter{Delta Method, Influence Functions, and Local Uncertainty}
\label{chap:local-approximation-influence}
\conceptindexes{delta method, influence functions, asymptotic linearity, functional delta method, efficiency, bootstrap, semiparametric theory}

\begin{tcolorbox}[
  enhanced,
  breakable,
  colback=chaptercream,
  colframe=bookblue!88!black,
  boxrule=0.72pt,
  arc=5pt,
  boxsep=1pt,
  left=1.0em,
  right=0.95em,
  top=0.82em,
  bottom=0.82em,
  before skip=0.55\baselineskip,
  after skip=1.0\baselineskip
]
\noindent\textbf{Chapter overview.}
In Part~V, this is the uncertainty-translation chapter.  Chapter~13 explains
how estimators are selected and localized: stable empirical criteria choose
peaks, and stable empirical equations choose roots.  Chapter~14 asks how
nearby laws are separated by tests and local likelihood ratios.  This chapter
begins after those localization steps.  Its question is what remains when the
estimator is already near the target.  The answer is local linearization:
residual empirical noise is pushed through a derivative, producing an
influence representation and therefore a variance, confidence interval,
efficiency comparison, bootstrap law, or diagnostic.

The chapter is not a survey of unrelated tools.  The classical delta method,
functional delta method, influence functions, tangent spaces, bootstrap, and
application vignettes are different languages for the same local-linear
template.  A nonlinear statistical procedure is first localized; its
first-order behavior is then read as a linear map applied to the noise left by
sampling, censoring, randomization, or event histories.

The examples are chosen to echo the book's earlier threads.  Clinical risks,
rare species, Shakespeare vocabulary, climate reconstructions, distributional
regression, robot-lab reliability, and platform policy learning all return
here in a different role.  Earlier they motivated probability models, data
structures, or estimators.  Here they ask the same inference question: once the
target has been named and localized, how does its remaining empirical noise
become uncertainty that a user of the analysis can act on?
\end{tcolorbox}

The previous estimation chapter finds the parameter.  Consistency brings
$\hat\theta_n$ close to $\theta_0$, and local empirical-process arguments
explain why the selected root or maximizer lives in a small neighborhood of
the truth.  Asymptotic normality then asks a more delicate question: once the
estimator has arrived near the right place, what is the shape of its remaining
error?

This is the moment when a statistician changes instruments.  The global
criterion was useful for finding the target; now the view narrows to a small
neighborhood around that target.  Curvature, derivative, score, and influence
function become local reading devices.  They do not show where the estimator
is headed.  They show how the last visible wobble should be converted into
standard errors, confidence intervals, and efficiency comparisons.

For a Z-estimator the picture is almost visible in one line:
\[
  0=\Psi_n(\hat\theta_n)
  \approx
  \Psi_n(\theta_0)+\dot\Psi(\theta_0)(\hat\theta_n-\theta_0).
\]
Solving the linear approximation gives
\[
  \sqrt n(\hat\theta_n-\theta_0)
  \approx
  -
  \dot\Psi(\theta_0)^{-1}\sqrt n\,\Psi_n(\theta_0).
\]
The estimator has become a linear transformation of an empirical average.  The
story of this chapter is the repeated rediscovery of that sentence in different
languages:
\[
  \begin{aligned}
  \hbox{localized object}
  &\longrightarrow
  \hbox{residual noise},\\
  \hbox{residual noise}
  &\xrightarrow{\ \hbox{linear derivative}\ }
  \hbox{reported uncertainty}.
  \end{aligned}
\]

The input to this chapter is not a raw data set and not a global optimization
problem.  It is the localized estimator or empirical object delivered by the
preceding chapters:
\[
  \sqrt n(T_n-\theta)\weakto Z
  \quad\hbox{or}\quad
  \sqrt n(T_n-\theta)
  =
  \mathbb G_n\varphi+o_{\Prob}(1).
\]
The reported target is usually not \(T_n\) itself.  It is a functional
\(\Phi(T_n)\) or \(\psi(P_n)\): a log risk, odds ratio, quantile, survival
curve, transition probability, sequential-use contrast, or geometric
regression contrast.  These are the kinds of targets that appeared earlier in
biomedical, ecological, linguistic, climate, robotic, and platform examples.
The chapter's working rule is
\[
  \hbox{localized noise}
  \xrightarrow{\ \hbox{target derivative}\ }
  \hbox{uncertainty for the reported target}.
\]
The four recurring questions are: what is the residual noise, what functional
transforms the localized object, what derivative reads the noise, and what
variance, bootstrap law, efficiency bound, or diagnostic does the linear term
produce?

This chapter therefore does not reprove the stabilization, random-object, and
empirical-process results from Chapters~9, 10, and~11, nor the localization
arguments from Chapter~13 or the local testing geometry from Chapter~14.  It
uses them after the estimator has already been localized.  Its question is:
which linear functional of the residual empirical noise is left, and what
uncertainty should be attached to the target actually reported?

The examples return in that local form.  A clinical risk becomes a log-risk or
log-odds standard error.  Missing species and Shakespeare vocabulary become
functionals of an empirical law.  The Zhu/Chu Ko-chen climate thread and the
Wasserstein-regression example become distribution-valued targets read through
tangent coordinates.  Robot-lab and event-history examples become martingale
influence representations, including transition probabilities in multi-state
systems.  The continuity is the point: the same data stories that motivated
modeling now motivate uncertainty translation.

Thus Chapters~13--15 are organized by operations rather than by raw data
formats.  Chapter~13 studies selection, Chapter~14 studies separation and
calibration, and this chapter studies uncertainty translation after
localization.  The examples return whenever a data object needs one of those
operations; the specialized domain theory is then named through the object,
target, and assumptions already built earlier in the book.

\section{The Classical Delta Method}
\label{sec:ch15-classical-delta}
\conceptindexes{classical delta method, linear approximation, residual noise, uncertainty propagation}

We begin with the finite-dimensional case because it shows the mechanism with
no extra topology.  The estimator already has a limit.  The target reported to
the reader is a smooth transformation of that estimator.  A derivative
translates the estimator's residual noise to the transformed scale.

\begin{theorem}[Classical delta method]
Suppose
\[
  \sqrt n(T_n-\theta)\weakto Z
\]
in $\R^d$, and let $\phi:\R^d\to\R^k$ be differentiable at $\theta$ with
derivative matrix $\dot\phi_\theta$.  Then
\[
  \sqrt n\{\phi(T_n)-\phi(\theta)\}
  \weakto
  \dot\phi_\theta Z.
\]
\end{theorem}

\noindent\textit{Proof.}
Differentiability at $\theta$ means
\[
  \phi(T_n)-\phi(\theta)
  =
  \dot\phi_\theta(T_n-\theta)+r_n,
\]
where $\|r_n\|/\|T_n-\theta\|\to0$ whenever $T_n\to\theta$ and
$T_n\ne\theta$.  Since
\(\sqrt n(T_n-\theta)\weakto Z\), the sequence
\(\sqrt n(T_n-\theta)\) is tight, so \(T_n-\theta=O_{\Prob}(n^{-1/2})\) and
in particular \(T_n\toP\theta\).  For every $\eta>0$, choose $\delta>0$ so
that $\|r(x)\|\le \eta\|x-\theta\|$ whenever $\|x-\theta\|\le\delta$.  Then
\[
  \Prob\{\sqrt n\|r_n\|>\eta M\}
  \le
  \Prob\{\|T_n-\theta\|>\delta\}
  +
  \Prob\{\sqrt n\|T_n-\theta\|>M\}.
\]
First choose \(M\) large using tightness, and then let \(n\to\infty\) and
\(\eta\downarrow0\).  This gives \(\sqrt n r_n=o_{\Prob}(1)\).  Hence
\[
  \sqrt n\{\phi(T_n)-\phi(\theta)\}
  =
  \dot\phi_\theta\sqrt n(T_n-\theta)+o_{\Prob}(1).
\]
The continuous mapping theorem and Slutsky's theorem give the displayed weak
limit. \qedmark

\begin{example}[Log risk]
If $\hat p_n$ estimates a risk $p\in(0,1)$ and
$\sqrt n(\hat p_n-p)\weakto \Normal(0,p(1-p))$, then
\[
  \sqrt n\{\log\hat p_n-\log p\}
  \weakto
  \Normal\left(0,\frac{1-p}{p}\right).
\]
The transformation changes the scale and the variance, but the fluctuation is
still the original empirical fluctuation read through a derivative.

\noindent\textit{Proof.}
Use the delta method with $\phi(p)=\log p$.  The derivative at the target is
\(\dot\phi(p)=1/p\).  Therefore
\[
  \sqrt n\{\log\hat p_n-\log p\}
  =
  \frac1p\sqrt n(\hat p_n-p)+o_{\Prob}(1).
\]
Multiplying a normal limit by \(1/p\) multiplies its variance by \(1/p^2\), so
\[
  \frac{p(1-p)}{p^2}=\frac{1-p}{p}.
\]
\qedmark
\end{example}

\begin{example}[Odds and log odds]
For the logit map $\phi(p)=\log\{p/(1-p)\}$,
\[
  \dot\phi(p)=\frac{1}{p(1-p)}.
\]
Thus
\[
  \sqrt n\{\phi(\hat p_n)-\phi(p)\}
  \weakto
  N\left(0,\frac{1}{p(1-p)}\right).
\]
The familiar standard error of a log odds is a delta-method statement.

\noindent\textit{Proof.}
Here
\[
  \phi'(p)
  =
  \frac{d}{dp}\{\log p-\log(1-p)\}
  =
  \frac1p+\frac1{1-p}
  =
  \frac{1}{p(1-p)}.
\]
Thus
\[
  \sqrt n\{\phi(\hat p_n)-\phi(p)\}
  =
  \frac{1}{p(1-p)}\sqrt n(\hat p_n-p)+o_{\Prob}(1).
\]
Since \(\Var\{\sqrt n(\hat p_n-p)\}\to p(1-p)\), the limiting variance is
\[
  \frac{p(1-p)}{p^2(1-p)^2}
  =
  \frac1{p(1-p)}.
\]
\qedmark
\end{example}

\subsection{Z-Estimators and Local Linearization}
\label{sec:ch15-z-linear}
\conceptindexes{Z-estimators, local linearization, estimating equations, sandwich form}

This section is the explicit hand-off from Chapters~13 and~14.  There,
empirical criteria, estimating equations, and local likelihood ratios chose
and calibrated an estimator.  Here, the same estimating equation is read only
in a shrinking neighborhood of the target.  The root-finding problem becomes a
linear equation for the remaining error.

Let
\[
  \Psi_n(\theta)=P_n\psi_\theta,
  \qquad
  \Psi(\theta)=P\psi_\theta,
\]
and suppose $\Psi(\theta_0)=0$.  If $\hat\theta_n$ solves
$\Psi_n(\hat\theta_n)=0$, then a local expansion gives
\[
  0
  =
  \Psi_n(\theta_0)
  +
  \dot\Psi_{\theta_0}(\hat\theta_n-\theta_0)
  +
  r_n.
\]
If $\dot\Psi_{\theta_0}$ is nonsingular and $r_n=o_{\Prob}(n^{-1/2})$, then
\[
  \sqrt n(\hat\theta_n-\theta_0)
  =
  -
  \dot\Psi_{\theta_0}^{-1}
  \mathbb G_n\psi_{\theta_0}
  +o_{\Prob}(1).
\]
This display is the asymptotic linear representation of the estimator.

\begin{example}[Estimating a mean]
The estimating equation for $\mu$ is
\[
  \Psi_n(\mu)=P_n(X-\mu)=0.
\]
Here $\dot\Psi_\mu=-1$, so
\[
  \sqrt n(\bar X_n-\mu)
  =
  \mathbb G_n(X-\mu).
\]
The simplest estimator is already exactly asymptotically linear.

\noindent\textit{Proof.}
Solving \(P_n(X-\mu)=0\) gives \(\hat\mu_n=\bar X_n=P_nX\).  Therefore
\[
  \sqrt n(\hat\mu_n-\mu)
  =
  \sqrt n(P_n-P)X
  =
  \sqrt n(P_n-P)(X-\mu),
\]
because \((P_n-P)\mu=0\).  This is exactly
\(\mathbb G_n(X-\mu)\), with no remainder term. \qedmark
\end{example}

\begin{example}[Smooth maximum likelihood]
Let $\ell_\theta(x)=\log f_\theta(x)$ and let
$\dot\ell_{\theta_0}$ be the score.  The score equation is
\[
  P_n\dot\ell_\theta=0.
\]
Under regularity conditions,
\[
  \dot\Psi_{\theta_0}
  =
  P\ddot\ell_{\theta_0}
  =
  -I(\theta_0),
\]
where $I(\theta_0)=P\dot\ell_{\theta_0}\dot\ell_{\theta_0}^T$ is Fisher
information.  Thus
\[
  \sqrt n(\hat\theta_n-\theta_0)
  =
  I(\theta_0)^{-1}\mathbb G_n\dot\ell_{\theta_0}
  +o_{\Prob}(1).
\]
The inverse information appears because the local curvature of likelihood
translates score fluctuation into parameter fluctuation.

\noindent\textit{Proof.}
Let \(\hat\theta_n\) solve \(P_n\dot\ell_{\hat\theta_n}=0\).  A first-order
Taylor expansion of the score equation around \(\theta_0\) gives
\[
  0
  =
  P_n\dot\ell_{\theta_0}
  +
  P_n\ddot\ell_{\theta_0}(\hat\theta_n-\theta_0)
  +
  r_n,
\]
where \(r_n=o_{\Prob}(\|\hat\theta_n-\theta_0\|)+o_{\Prob}(n^{-1/2})\) under
the usual local smoothness and consistency assumptions.  If
\(P_n\ddot\ell_{\theta_0}\toP P\ddot\ell_{\theta_0}=-I(\theta_0)\) and
\(\hat\theta_n-\theta_0=O_{\Prob}(n^{-1/2})\), then multiplying by
\(\sqrt n\) gives
\[
  0
  =
  \mathbb G_n\dot\ell_{\theta_0}
  -
  I(\theta_0)\sqrt n(\hat\theta_n-\theta_0)
  +
  o_{\Prob}(1).
\]
Solving for \(\sqrt n(\hat\theta_n-\theta_0)\) yields the displayed expansion.
The information identity
\[
  P\ddot\ell_{\theta_0}
  =
  -P\{\dot\ell_{\theta_0}\dot\ell_{\theta_0}^T\}
\]
follows by differentiating
\(\int f_\theta(x)\,dx=1\) twice under the integral sign. \qedmark
\end{example}

\subsection{Asymptotic Linearity and Influence Representations}
\label{sec:ch15-asymptotic-linearity}
\conceptindexes{asymptotic linearity, influence representation, reading rule}

The displays above suggest the common output of local estimation theory.  Once
the nonlinear selection step has been reduced to its first-order part, an
estimator is asymptotically linear if there exists a mean-zero function
$\varphi$ with $P\varphi^2<\infty$ such that
\[
  \sqrt n(\hat\theta_n-\theta_0)
  =
  \mathbb G_n\varphi+o_{\Prob}(1).
\]
For vector parameters, $\varphi$ is vector-valued and
$P\varphi\varphi^T$ is the asymptotic covariance matrix.  The function
$\varphi$ is called an influence function when it arises as the derivative of
the underlying statistical functional.

This representation is more than a convenient limit theorem.  It is a
translation rule.  The complicated estimator has been reduced, at first order,
to an average of observation-level contributions.  Once that happens, the
central limit theorem supplies the Gaussian approximation, the variance becomes
$P\varphi\varphi^T$, and bootstrap or sandwich estimators try to estimate the
same covariance from data.  Many arguments in modern asymptotic statistics are
therefore attempts to prove that a procedure admits this representation and to
identify the right $\varphi$.

\begin{tcolorbox}[
  enhanced,
  breakable,
  colback=noteback,
  colframe=bookgold!75!black,
  boxrule=0.55pt,
  arc=4pt,
  boxsep=1pt,
  left=0.95em,
  right=0.9em,
  top=0.7em,
  bottom=0.7em,
  before skip=0.9\baselineskip,
  after skip=1.0\baselineskip
]
\noindent\textbf{How to use an asymptotic-linear result.}
Suppose the analysis has produced an expansion
\[
  \hat\psi-\psi(P)
  =
  \frac1n\sum_{i=1}^n \varphi(O_i)+o_{\Prob}(n^{-1/2}),
  \qquad P\varphi=0.
\]
In practice one plugs in nuisance estimates and obtains estimated
contributions
\[
  \hat\varphi_i=\hat\varphi(O_i),\qquad i=1,\ldots,n.
\]
For a scalar target, use
\[
  \hat\sigma^2
  =
  \frac1n\sum_{i=1}^n(\hat\varphi_i-\bar{\hat\varphi})^2,
  \qquad
  \widehat{\operatorname{se}}(\hat\psi)=\frac{\hat\sigma}{\sqrt n},
\]
and report, for example,
\[
  \hat\psi\pm z_{1-\alpha/2}\widehat{\operatorname{se}}(\hat\psi).
\]
For a vector target, replace \(\hat\sigma^2\) by
\[
  \hat V
  =
  \frac1n\sum_{i=1}^n
  (\hat\varphi_i-\bar{\hat\varphi})
  (\hat\varphi_i-\bar{\hat\varphi})^T,
  \qquad
  \widehat{\Cov}(\hat\psi)=\hat V/n.
\]

The index \(i\) should be the independent unit of analysis.  In a clinical
trial it is usually the patient, in a platform experiment the user or session
cluster, in an image benchmark the image, and in event-history analysis the
whole subject history.  Treating repeated rows from the same unit as
independent gives standard errors that are too small.

The same contributions also tell the analyst how to use the result beyond a
confidence interval:
\begin{description}[leftmargin=0pt,labelsep=0.55em,style=unboxed,font=\bookdescriptionlabelfont,itemsep=0.18\baselineskip]
\item[Report.]
Give the target, estimate, standard error, interval, and the unit over which
the influence contributions were averaged.
\item[Diagnose.]
Inspect the largest \(|\hat\varphi_i|\)'s, their covariates, event histories,
sites, batches, labels, or time periods.  Large contributions identify where
the first-order uncertainty is coming from.
\item[Compare.]
Two regular estimators for the same target can be compared through the
variance of their influence functions.  Smaller \(P\varphi^2\) means smaller
first-order uncertainty for the same target.
\item[Bootstrap.]
If the full procedure is easy to recompute, bootstrap the whole estimator.  If
the expansion is trusted and recomputation is expensive, a multiplier or
resampling bootstrap applied to \(\hat\varphi_i\)'s approximates the same
first-order law.
\item[Check.]
The empirical mean of \(\hat\varphi_i\) should be small after centering, the
remainder should be negligible on the scale \(n^{-1/2}\), and nuisance
estimation should not reintroduce first-order bias.  Cross-fitting is often
used for this last purpose.
\end{description}

The output is therefore an inferential object, not only a theorem: it shows
what to report, what to bootstrap, which units drive uncertainty, and which
part of the data system deserves scrutiny.
\end{tcolorbox}

For example, in the log-risk calculation above, the observation-level
contribution is approximately
\[
  \hat\varphi_i
  =
  \frac{1}{\hat p}\{\ind{Y_i=1}-\hat p\}.
\]
A standard error for \(\log\hat p\) is obtained by taking the empirical
standard deviation of these contributions and dividing by \(\sqrt n\).  This
is the practical meaning of saying that the derivative \(1/p\) reads the
empirical noise on the log scale.

\section{Functional Delta Method}
\label{sec:ch15-functional-delta}
\conceptindexes{functional delta method, Hadamard differentiability, statistical functional, plug-in estimator}

The preceding sections treated finite-dimensional vectors and estimating
equations.  Many statistical targets are larger objects or functionals of
larger objects.  A quantile is a functional of a distribution.  A survival
probability is a functional of a hazard.  A regression curve, density, risk
curve, or ROC curve may live in an infinite-dimensional space.  The derivative
must therefore be a derivative between normed spaces.  Conceptually, however,
nothing has changed: the residual empirical process is still being pushed
through a linear derivative.

\begin{definition}[Hadamard differentiability]
Let $\Phi:\mathbb D_\Phi\subset\mathbb D\to\mathbb E$ be a map between normed
spaces.  The map $\Phi$ is Hadamard differentiable at $\theta$ tangentially to
a set $\mathbb D_0$ if there exists a continuous linear map
$\Phi'_\theta:\mathbb D_0\to\mathbb E$ such that
\[
  \frac{\Phi(\theta+t_n h_n)-\Phi(\theta)}{t_n}
  \to
  \Phi'_\theta h
\]
whenever $t_n\downarrow0$, $h_n\to h\in\mathbb D_0$, and
$\theta+t_nh_n\in\mathbb D_\Phi$.
\end{definition}

\begin{theorem}[Functional delta method]
Suppose
\[
  r_n(T_n-\theta)\weakto Z
\]
in a separable normed space $\mathbb D$, where $r_n\to\infty$ and $Z$ takes
values in the tangent set $\mathbb D_0$.  Suppose $\Phi$ is Hadamard
differentiable at $\theta$ tangentially to $\mathbb D_0$.  Then
\[
  r_n\{\Phi(T_n)-\Phi(\theta)\}
  \weakto
  \Phi'_\theta Z.
\]
\end{theorem}

\noindent\textit{Proof.}
Put \(H_n=r_n(T_n-\theta)\).  Then \(T_n=\theta+r_n^{-1}H_n\) and
\[
  r_n\{\Phi(T_n)-\Phi(\theta)\}
  =
  \Phi_n(H_n),
  \qquad
  \Phi_n(h)=r_n\{\Phi(\theta+r_n^{-1}h)-\Phi(\theta)\}.
\]
Hadamard differentiability says exactly that if \(h_n\to h\in\mathbb D_0\)
and \(\theta+r_n^{-1}h_n\in\mathbb D_\Phi\), then
\[
  \Phi_n(h_n)\to \Phi'_\theta h.
\]
Because \(H_n\weakto Z\) and the spaces used here are separable, the
Skorokhod representation theorem makes it possible to work on an equivalent probability
space where \(H_n\to Z\) almost surely.  On the event \(Z\in\mathbb D_0\), the
preceding deterministic implication gives
\[
  \Phi_n(H_n)\to \Phi'_\theta Z
  \qquad\text{almost surely}.
\]
Almost-sure convergence on the representation space implies weak convergence
of the original variables.  Hence
\[
  r_n\{\Phi(T_n)-\Phi(\theta)\}\weakto \Phi'_\theta Z.
\]
\qedmark

The theorem is the same story as the ordinary delta method, but with enough
topology to handle functions.  The derivative is no longer a matrix; it is a
continuous linear operator.  The empirical process supplies the fluctuation
$Z$; the derivative shows how the functional reads that fluctuation.

\begin{example}[The quantile functional]
Let $Q(p)=F^{-1}(p)$ and assume $F$ has a positive continuous density
$f$ at $q=Q(p)$.  If $h$ is a perturbation of the distribution function, then
the derivative of the quantile functional is
\[
  \Phi'_F h
  =
  -\frac{h(q)}{f(q)}.
\]
Since
\[
  \sqrt n(F_n-F)\weakto B\circ F,
\]
the functional delta method gives
\[
  \sqrt n\{Q_n(p)-Q(p)\}
  \weakto
  N\left(0,\frac{p(1-p)}{f(q)^2}\right).
\]
This is the process-level version of the quantile calculation begun in
Chapter 9 and made uniform in Chapter 11.

\noindent\textit{Proof.}
Let \(F_t=F+t h_t\), where \(h_t\to h\) uniformly and \(t\downarrow0\), and
let \(q_t=F_t^{-1}(p)\).  The defining equation is
\[
  F(q_t)+t h_t(q_t)=p=F(q).
\]
Because \(f(q)>0\) and \(f\) is continuous at \(q\), the inverse is locally
well behaved and \(q_t\to q\).  Taylor expansion of \(F\) at \(q\) gives
\[
  F(q_t)-F(q)=f(q)(q_t-q)+o(|q_t-q|).
\]
Also \(h_t(q_t)=h(q)+o(1)\) along perturbations that are continuous at \(q\)
or, more generally, tangentially to functions continuous at \(q\).  Therefore
\[
  f(q)(q_t-q)+t h(q)+o(|q_t-q|)+o(t)=0.
\]
Dividing by \(t\) and using the local inverse bound \(q_t-q=O(t)\) yields
\[
  \frac{q_t-q}{t}\to -\frac{h(q)}{f(q)}.
\]
Thus the Hadamard derivative of the quantile functional is
\(\Phi'_Fh=-h(q)/f(q)\).  Applying the functional delta method with
\(h=B\circ F\) gives
\[
  \sqrt n\{Q_n(p)-Q(p)\}
  \weakto
  -\frac{B(p)}{f(q)}.
\]
Since a Brownian bridge satisfies \(\Var\{B(p)\}=p(1-p)\), the stated normal
variance follows. \qedmark
\end{example}

\begin{example}[Historical climate reconstruction as a functional target]
The Zhu/Chu Ko-chen climate thread from Chapter~2 becomes concrete if we use
the public decadal temperature reconstruction of China in
\citet{ge2013temperature}.  The archived table reports, for years
\[
  t=5,15,\ldots,1995,
\]
two reconstructed temperature-anomaly curves relative to the 1851--1950
climatology: a partial least squares curve \(\hat m^{\mathrm{PLS}}(t)\), a
principal-component regression curve \(\hat m^{\mathrm{PC}}(t)\), and
pointwise 95\% uncertainty bounds.  This is not the original 1973 Zhu curve
digitized point by point.  It is a modern reconstruction built from regional
proxy evidence and instrumental calibration, and it has exactly the
statistical structure that made the Zhu example useful: proxy records are
translated into a latent climate curve.

For a reported scientific claim, the object is rarely the whole curve.  A
typical target is a contrast between two historical periods.  On the decadal
grid, let
\[
  \bar m_A=\frac1{|A|}\sum_{t\in A}m(t),
  \qquad
  \psi_{A,B}(m)=\bar m_A-\bar m_B .
\]
This target is a continuous linear functional of the curve, so its Hadamard
derivative is itself:
\[
  \psi'_{A,B}(h)
  =
  \frac1{|A|}\sum_{t\in A}h(t)
  -
  \frac1{|B|}\sum_{t\in B}h(t).
\]
Thus the functional delta method says that uncertainty in the reconstructed
curve is translated into uncertainty for the period contrast by the same
weights used to average the decades.

Using the average
\[
  \hat m(t)=\{\hat m^{\mathrm{PLS}}(t)+\hat m^{\mathrm{PC}}(t)\}/2
\]
as a simple two-method summary gives the following descriptive contrasts.

\begin{center}
\textbf{Historical temperature contrast summaries.}\par\smallskip
\begin{tabular}{llrr}
\toprule
Period & Years & Mean anomaly & Avg. half-width \\
\midrule
Early warm interval & 1--200 & \(0.159\) & \(0.289\) \\
Early cold interval & 201--350 & \(-0.119\) & \(0.338\) \\
Warm interval & 551--760 & \(0.157\) & \(0.269\) \\
Medieval warm interval & 951--1320 & \(0.266\) & \(0.324\) \\
Medieval subperiod & 981--1100 & \(0.324\) & \(0.360\) \\
Medieval subperiod & 1201--1270 & \(0.427\) & \(0.354\) \\
Broad cold interval & 1321--1920 & \(-0.118\) & \(0.285\) \\
Present warm period & 1921--2000 & \(0.306\) & \(0.220\) \\
\bottomrule
\end{tabular}
\end{center}

The present warm period is about
\[
  0.306-(-0.118)=0.424^\circ\mathrm C
\]
warmer than the broad 1321--1920 cold interval in this reconstruction.  The
medieval subperiods 981--1100 and 1201--1270 are also warm in point estimate,
with means \(0.324^\circ\mathrm C\) and \(0.427^\circ\mathrm C\).  The two
reconstruction methods agree reasonably well over the whole grid: the
correlation between the PLS and PC curves is about \(0.86\), and their mean
absolute difference is about \(0.089^\circ\mathrm C\).  Both methods identify
1995 as the warmest decade in the table and 1655 as the coldest.

The theory prevents an easy but wrong reading of the table.  Pointwise
uncertainty bands do not by themselves give the standard error of
\(\psi_{A,B}(\hat m)\).  If the reconstruction error process has covariance
kernel \(K(s,t)\), then the first-order variance of the contrast is
\[
  \Var\{\hat\psi_{A,B}-\psi_{A,B}\}
  \approx
  \sum_{s,t} w_s w_t K(s,t),
\]
where \(w_t=|A|^{-1}\) for \(t\in A\), \(w_t=-|B|^{-1}\) for \(t\in B\), and
\(w_t=0\) otherwise.  Correlated reconstruction errors can make this variance
quite different from what one would obtain by treating the pointwise error
bars as independent.  A bootstrap version should therefore resample or
perturb the proxy-calibration pipeline, not merely jitter each decade
separately.

This example also shows how an application teaches the theory back to us.
First, the target must be named before uncertainty can be computed: a curve, a
period mean, a warm-period contrast, and a tail quantile are different
functionals.  Second, nuisance structure matters.  The difference between PLS
and PC reconstructions is a visible form of method uncertainty, not a random
decade-by-decade fluctuation.  Third, historical proxy archives have an
observation process: record survival, regional coverage, dating error, and
proxy sensitivity all shape the tangent directions along which the curve can
move.  The delta method supplies the local translation, but the climate example
reminds the analyst that the correct derivative is only useful after the data-generating
and reconstruction process has been made explicit.
\qedmark
\end{example}

\subsection{Influence Functions}
\label{sec:ch15-influence-functions}
\conceptindexes{influence functions, functional derivative, contamination path, empirical measure}

The delta method linearizes a map.  Influence functions make the same
linearization observable at the level of a single data unit.  They answer the
question: if the underlying law is perturbed in the direction of one
observation, how much does the target move at first order?

Let $T(P)$ be a real or vector-valued functional.  For a point $x$, consider a
contaminated distribution
\[
  P_\varepsilon=(1-\varepsilon)P+\varepsilon\delta_x.
\]
The influence function is
\[
  \IF(x;T,P)
  =
  \left.\frac{d}{d\varepsilon}T(P_\varepsilon)\right|_{\varepsilon=0},
\]
when the derivative exists.  It measures the first-order effect of placing a
small amount of probability mass at $x$.

If the plug-in estimator $T(P_n)$ admits the expansion
\[
  T(P_n)-T(P)
  =
  (P_n-P)\IF(\cdot;T,P)+o_{\Prob}(n^{-1/2}),
\]
then
\[
  \sqrt n\{T(P_n)-T(P)\}
  =
  \mathbb G_n\IF+o_{\Prob}(1).
\]
The asymptotic variance is
\[
  P\IF\IF^T,
\]
provided the influence function has mean zero and finite second moment.

\begin{example}[Mean]
For $T(P)=\int x\,dP(x)$,
\[
  T(P_\varepsilon)
  =
  (1-\varepsilon)\mu+\varepsilon x,
\]
so
\[
  \IF(x;T,P)=x-\mu.
\]
\noindent\textit{Proof.}
Differentiate
\[
  T(P_\varepsilon)-T(P)
  =
  \{(1-\varepsilon)\mu+\varepsilon x\}-\mu
  =
  \varepsilon(x-\mu)
\]
at \(\varepsilon=0\).  The derivative is \(x-\mu\), and
\(P\{X-\mu\}=0\). \qedmark
\end{example}

\begin{example}[Quantile]
For $T(P)=Q(p)$ with positive density $f(q)$ at $q=Q(p)$,
\[
  \IF(x;T,P)
  =
  \frac{p-\ind{x\le q}}{f(q)}.
\]
Its variance is $p(1-p)/f(q)^2$, matching the functional delta method.

\noindent\textit{Proof.}
Let \(q_\varepsilon=T(P_\varepsilon)\), so
\[
  (1-\varepsilon)F(q_\varepsilon)
  +
  \varepsilon\ind{x\le q_\varepsilon}
  =
  p
\]
at continuity points of \(F\).  Put \(q_\varepsilon=q+\varepsilon a+o(\varepsilon)\).
Using \(F(q)=p\) and \(F(q_\varepsilon)=p+\varepsilon a f(q)+o(\varepsilon)\),
one obtains
\[
  p+\varepsilon a f(q)-\varepsilon p+\varepsilon\ind{x\le q}+o(\varepsilon)
  =
  p.
\]
Thus
\[
  a=\frac{p-\ind{x\le q}}{f(q)}.
\]
This is the derivative of the quantile under point-mass contamination.  Its
mean is zero because \(P(X\le q)=p\), and its variance is
\[
  \frac{\Var\{\ind{X\le q}\}}{f(q)^2}
  =
  \frac{p(1-p)}{f(q)^2}.
\]
\qedmark
\end{example}

\begin{example}[Variance]
For $T(P)=P(X-\mu)^2$, the influence function is
\[
  \IF(x;T,P)
  =
  (x-\mu)^2-T(P).
\]
The contribution from estimating $\mu$ disappears at first order because
$P(X-\mu)=0$.

\noindent\textit{Proof.}
Let \(\mu_\varepsilon=(1-\varepsilon)\mu+\varepsilon x\).  Then
\[
  T(P_\varepsilon)
  =
  (1-\varepsilon)P\{X-\mu_\varepsilon\}^2
  +
  \varepsilon(x-\mu_\varepsilon)^2.
\]
Differentiate at \(\varepsilon=0\).  The first term contributes
\[
  -P(X-\mu)^2
  +
  P\left[-2(X-\mu)\left.\frac{d\mu_\varepsilon}{d\varepsilon}\right|_0\right],
\]
and the second term contributes \((x-\mu)^2\).  Since
\(\left.d\mu_\varepsilon/d\varepsilon\right|_0=x-\mu\) and
\(P(X-\mu)=0\), the middle term is zero.  Hence
\[
  \IF(x;T,P)=(x-\mu)^2-T(P).
\]
\qedmark
\end{example}

\begin{example}[Rare-type mass in the missing-species thread]
Chapter~2 used missing species and Shakespeare vocabulary to show that absence
can be informative.  A simple local target is the probability mass of a rare
class \(A\), \(T(P)=P(A)\).  Its influence function is
\[
  \IF(x;T,P)=\ind{x\in A}-P(A).
\]
Thus rare-type uncertainty is not mysterious at first order: it is a centered
Bernoulli fluctuation.  What becomes difficult in the missing-species problem
is that the target itself may involve unobserved classes or a growing
collection of rare classes, pushing the problem from one influence function
to an empirical field.

\noindent\textit{Proof.}
Under contamination,
\[
  T(P_\varepsilon)
  =
  (1-\varepsilon)P(A)+\varepsilon\ind{x\in A}.
\]
Therefore
\[
  \left.\frac{d}{d\varepsilon}T(P_\varepsilon)\right|_0
  =
  \ind{x\in A}-P(A).
\]
The variance is \(P(A)\{1-P(A)\}\), the same Bernoulli variance controlled by
Chapter~9 and then scanned over classes in Chapter~11. \qedmark
\end{example}

Influence functions are not only algebraic conveniences.  They are diagnostic
objects.  A bounded influence function suggests local robustness to small
contamination; an unbounded one reveals sensitivity to extreme observations.
This is why influence functions sit naturally between asymptotic theory,
robust statistics, and practical diagnostics.  They show which observations
the estimator listens to most loudly.

\section{Efficiency and Information}
\label{sec:ch15-efficiency}
\conceptindexes{efficiency, Fisher information, canonical gradient, information bound}

Once an influence representation is available, variance comparison becomes a
geometric question.  Which first-order linear representation has the smallest
variance while still representing the same target derivative?  In parametric
models, Fisher information gives the answer.  In semiparametric models, the
same idea becomes a projection problem.

In a regular parametric model with score $\dot\ell_{\theta_0}$ and information
$I(\theta_0)$, the maximum likelihood estimator has influence function
\[
  I(\theta_0)^{-1}\dot\ell_{\theta_0}(X).
\]
Its asymptotic variance is $I(\theta_0)^{-1}$.  The Cram{\'e}r--Rao bound and
Le Cam--H{\'a}jek theory say that, under regularity
conditions, no regular estimator can do better uniformly in local alternatives.
Information is therefore not just curvature of a likelihood; it is the local
geometry of distinguishable probability laws.

The information-geometric form of that sentence is a second-order expansion.
For a smooth parametric model,
\[
  D_{\mathrm{KL}}(P_{\theta_0}\|P_{\theta_0+h})
  =
  \frac12 h^T I(\theta_0)h+o(\|h\|^2).
\]
Two nearby parameter values are hard to distinguish when this quadratic form
is small and easy to distinguish when it is large.  In the one-parameter
normal mean model \(N(\mu,\sigma^2)\) with known \(\sigma^2\), the information
is \(I(\mu)=1/\sigma^2\), so the local squared distance between
\(\mu\) and \(\mu+h\) is \(h^2/\sigma^2\).  A shift of one unit is large when
the noise scale is small and modest when the noise scale is large.  Fisher
information is the metric that records that local scale.

\noindent\textit{Verification.}
The score expansion from the smooth likelihood example gives
\[
  \sqrt n(\hat\theta_n-\theta_0)
  =
  I(\theta_0)^{-1}\mathbb G_n\dot\ell_{\theta_0}+o_{\Prob}(1).
\]
Thus the observation-level linear contribution is
\[
  \varphi(X)=I(\theta_0)^{-1}\dot\ell_{\theta_0}(X).
\]
It has mean zero because \(P\dot\ell_{\theta_0}=0\).  Its covariance is
\[
  P\{\varphi\varphi^T\}
  =
  I(\theta_0)^{-1}
  P\{\dot\ell_{\theta_0}\dot\ell_{\theta_0}^T\}
  I(\theta_0)^{-1}
  =
  I(\theta_0)^{-1}.
\]
\qedmark

The geometry becomes clearer if we view the score as a direction.  A
parametric submodel $P_\theta$ passing through $P_{\theta_0}$ creates a tangent
vector
\[
  \dot\ell_{\theta_0}(X).
\]
An estimator's influence function must reproduce the derivative of the target
functional along every allowed direction.  In a parametric model, the allowed
directions are spanned by the score vector.  The efficient influence function
is the one with smallest variance among all influence functions satisfying the
same derivative constraints.

\subsection{Semiparametric Tangent Spaces}
\label{sec:ch15-semiparametric}
\conceptindexes{semiparametric tangent spaces, nuisance tangent space, projection, efficient influence function}

Tangent spaces may look like a new subject, but within this chapter they serve
one purpose: they describe the directions in which residual noise is allowed
to move the statistical law.  Once those directions are known, influence
functions can be projected so that they read target-relevant noise and ignore
nuisance noise.

Semiparametric models contain both a finite-dimensional target and an
infinite-dimensional nuisance part.  A Cox model has regression coefficients
and an unspecified baseline hazard.  A missing-data model may have a treatment
effect target and nuisance functions for outcome regression and propensity.
An observational survival analysis may have censoring and event-time
mechanisms that are not parametrically specified.

\begin{tcolorbox}[
  enhanced,
  breakable,
  colback=noteback,
  colframe=bookgold!75!black,
  boxrule=0.55pt,
  arc=4pt,
  boxsep=1pt,
  left=0.95em,
  right=0.9em,
  top=0.7em,
  bottom=0.7em,
  before skip=0.9\baselineskip,
  after skip=1.0\baselineskip
]
\noindent\textbf{How to read a tangent space.}
Start with the large Hilbert space \(L_0^2(P)\) of mean-zero, square-integrable
functions of one observation.  A smooth submodel \(t\mapsto P_t\) is a path
through the statistical model, and its score \(s\) is the velocity of that path
at \(t=0\).  The tangent space \(\mathcal T\) collects all velocities that the
model permits.  A target \(\psi(P)\) is locally differentiable if each allowed
velocity \(s\) produces a first-order change \(\dot\psi(s)\).  An influence
function \(\varphi\) represents that derivative when
\[
  \dot\psi(s)=P\{\varphi(X)s(X)\},\qquad s\in\mathcal T.
\]
Nuisance directions are allowed velocities that do not move the target.  The
efficient influence function is the representative that points in the
target-relevant directions and is orthogonal to the nuisance directions.  The
picture is exactly least-squares projection, but the vectors are scores and
influence functions.
\end{tcolorbox}

\begin{definition}[Tangent and nuisance tangent spaces]
Let $\mathcal P$ be a model containing a distribution $P$.  A smooth
one-dimensional submodel $t\mapsto P_t$ through $P=P_0$ has score
\[
  s(X)
  =
  \left.\frac{d}{dt}\log p_t(X)\right|_{t=0},
  \qquad Ps=0,\quad Ps^2<\infty.
\]
The tangent space $\mathcal T$ is the closed linear span in
$L_0^2(P)=\{h:Ph=0,\ Ph^2<\infty\}$ of all scores generated by smooth submodels
inside $\mathcal P$.  If $\psi(P)$ is the target, the nuisance tangent space
$\mathcal N$ is the closed span of scores from submodels that leave
$\psi(P_t)$ unchanged to first order.
\end{definition}

This is the Hilbert-space idea behind the Bickel--Klaassen--Ritov--Wellner
framework \citep{bickel1993efficient}; see also
\citet{newey1990semiparametric}, \citet{tsiatis2006semiparametric}, and
\citet{kosorok2008empirical} for efficiency bounds, missing-data models, and
empirical-process formulations.  Scores are local directions in the
statistical law.  Nuisance scores are directions in which the data law changes
but the target does not.  The efficient influence function is the part of the
gradient that remains after irrelevant directions have been removed.

\begin{example}[Mean in the nonparametric model]
Let \(\mathcal P\) be the nonparametric model of all laws on \(\Real\) with
finite second moment, and let the target be \(\psi(P)=PX\).  In this model the
tangent space is all of \(L_0^2(P)\).  Along a smooth submodel with score
\(s\),
\[
  \left.\frac{d}{dt}P_tX\right|_{t=0}
  =
  P\{Xs(X)\}
  =
  P\{(X-\psi(P))s(X)\}.
\]
Thus the influence function is
\[
  \varphi(X)=X-\psi(P).
\]
There is no nuisance projection to perform: every mean-zero square-integrable
direction is allowed, and the sample mean is already the empirical average of
this influence function plus the target.  Semiparametric examples become more
interesting when some directions are nuisance directions that must be removed.
\qedmark
\end{example}

\begin{theorem}[Canonical gradient as a projection]
Let $\mathcal T$ be a closed tangent space in $L_0^2(P)$, and suppose a scalar
target $\psi$ is pathwise differentiable with derivative
\[
  \dot\psi(s)=\left.\frac{d}{dt}\psi(P_t)\right|_{t=0}
  =
  P\{g(X)s(X)\},
  \qquad s\in\mathcal T,
\]
for some $g\in L_0^2(P)$.  Let $\Pi_{\mathcal T}g$ be the orthogonal projection
of $g$ onto $\mathcal T$.  Then $\Pi_{\mathcal T}g$ represents the same
derivative on every tangent direction:
\[
  \dot\psi(s)=P\{(\Pi_{\mathcal T}g)s\},\qquad s\in\mathcal T,
\]
and it has the smallest $L^2(P)$ norm among all representatives of this
derivative.  If $\mathcal N\subset\mathcal T$ is a nuisance tangent space, then
the canonical gradient is orthogonal to $\mathcal N$.
\end{theorem}

\noindent\textit{Proof.}
For every $s\in\mathcal T$, the projection residual
$g-\Pi_{\mathcal T}g$ is orthogonal to $\mathcal T$, so
\[
  P\{g s\}=P\{(\Pi_{\mathcal T}g)s\}.
\]
Thus $\Pi_{\mathcal T}g$ gives the same pathwise derivative on the model.  Now
let $a\in L_0^2(P)$ be any other representative, so
\[
  P\{a s\}=P\{(\Pi_{\mathcal T}g)s\}
  \qquad\text{for all }s\in\mathcal T.
\]
Taking $s=\Pi_{\mathcal T}g$ shows that
$a-\Pi_{\mathcal T}g$ is orthogonal to $\Pi_{\mathcal T}g$.  Hence
\[
  P a^2
  =
  P(\Pi_{\mathcal T}g)^2
  +
  P(a-\Pi_{\mathcal T}g)^2
  \ge
  P(\Pi_{\mathcal T}g)^2.
\]
Finally, if \(s\in\mathcal N\), then the target is unchanged to first order
along that direction, so \(\dot\psi(s)=0\).  Therefore
\[
  P\{(\Pi_{\mathcal T}g)s\}=0,\qquad s\in\mathcal N,
\]
which means the canonical gradient lies in \(\mathcal N^\perp\). \qedmark

In a semiparametric likelihood problem this projection often appears as
residualization: subtract from the target score the part predictable from the
nuisance score space.  The variance of the resulting canonical gradient is the
efficiency bound.  The algebra is the same as ordinary least squares, but the
vectors being projected are scores and influence functions in \(L_0^2(P)\).

\begin{proposition}[Semiparametric Cram{\'e}r--Rao bound]
\label{prop:ch15-semiparametric-cramer-rao}
Let \(\psi(P)\) be a scalar pathwise differentiable target in a model with
tangent space \(\mathcal T\), and let \(\varphi_{\mathrm{eff}}\) be its canonical
gradient.  Suppose a regular asymptotically linear estimator has influence
function \(\varphi\):
\[
  \sqrt n\{\hat\psi_n-\psi(P)\}
  =
  \mathbb G_n\varphi+o_P(1).
\]
Then \(\varphi\) represents the same pathwise derivative as
\(\varphi_{\mathrm{eff}}\), and therefore
\[
  P\varphi^2\ge P\varphi_{\mathrm{eff}}^2.
\]
Consequently, whenever the estimator's asymptotic variance is given by its
influence variance,
\[
  \liminf_{n\to\infty}
  n\,\operatorname{Var}_P(\hat\psi_n)
  \ge
  P\varphi_{\mathrm{eff}}^2.
\]
The right side is the semiparametric Cram{\'e}r--Rao lower bound.
\end{proposition}

\noindent\textit{Proof.}
Regularity along every smooth submodel with score \(s\in\mathcal T\) forces the
influence function to reproduce the derivative:
\[
  \left.\frac{d}{dt}\psi(P_t)\right|_{t=0}
  =
  P\{\varphi s\}.
\]
Thus \(\varphi\) is another representative of the same derivative.  The
canonical gradient is the representative with smallest \(L^2(P)\) norm, so
\(P\varphi^2\ge P\varphi_{\mathrm{eff}}^2\).  The displayed variance bound
then follows from the asymptotic linear representation and the central limit
theorem. \qedmark

For a scalar parameter \(\xi\) in a semiparametric likelihood model, the same
bound can be read in score language.  Let \(s_\xi\) be the target score and
\(\mathcal N\) the nuisance tangent space.  The efficient score is
\[
  s_{\mathrm{eff}}
  =
  s_\xi-\Pi_{\mathcal N}s_\xi,
\]
the efficient information is
\[
  I_{\mathrm{eff}}=P s_{\mathrm{eff}}^2,
\]
and the efficient influence function is
\[
  \varphi_{\mathrm{eff}}
  =
  \frac{s_{\mathrm{eff}}}{I_{\mathrm{eff}}}.
\]
The lower bound is therefore \(I_{\mathrm{eff}}^{-1}\).  This is the exact
semiparametric analogue of the ordinary Cram{\'e}r--Rao bound \(I^{-1}\):
nuisance directions are first projected out, and only the remaining information
about the target is counted.

\begin{example}[Missing outcome mean]
Let \(O=(X,R,RY)\), where \(R=1\) means that \(Y\) is observed.  Suppose
missingness is at random:
\[
  \Prob(R=1\mid X,Y)=\Prob(R=1\mid X)=\pi(X),
  \qquad \pi(X)>0,
\]
and let the target be \(\psi=\Expect(Y)\).  Write
\[
  m(X)=\Expect(Y\mid X).
\]
The efficient influence function in the nonparametric missing-at-random model
is
\[
  \varphi(O)
  =
  m(X)-\psi
  +
  \frac{R}{\pi(X)}\{Y-m(X)\}.
\]
The first term reads the part of the mean explained by the covariate
distribution.  The second term reads the residual information in the observed
outcomes, inflated by the inverse probability of being observed.  This single
formula shows what tangent-space projection buys: the influence function uses
the target-relevant part of the observed-data law, and the estimating equation
based on it is orthogonal to first-order nuisance perturbations at the truth.
This is the geometry behind doubly robust and orthogonal-score estimators.
\qedmark
\end{example}

At this level the story sounds abstract, but it is still the same translation
pattern:
\[
  \text{estimator fluctuation}
  =
  \text{linear functional of empirical noise}
  +
  o_{\Prob}(n^{-1/2}).
\]
The difference is that the linear functional must ignore nuisance directions
that change the data law without changing the target.

This projection language is the local version of the book's opening compass.
The data structure determines which directions in the probability law are
available.  The scientific target determines which directions matter.  An
efficient procedure uses the observable directions that matter and discards, as
well as it can, the directions that only move nuisance structure.

\begin{example}[Wasserstein regression and two tangent spaces]
Chapter~10 treated one-dimensional Wasserstein regression by mapping a
distribution-valued response \(Y\) to a Hilbert coordinate.  Fix an atomless
reference law \(\mu_\ast\) on \(D\), and define
\[
  V=\operatorname{Log}_{\mu_\ast}(Y)
  =
  Q_Y\circ F_\ast-\operatorname{id}
  \in \Hilbert=L^2(\mu_\ast).
\]
Let \(Z\) be the raw covariate and let
\(\mathbf X=\phi(Z)\in\R^p\) be the design vector used in the regression, with
\(\Sigma=\Expect(\mathbf X\mathbf X^T)\) nonsingular.  A linear
tangent-coordinate Wasserstein regression target is the Hilbert-vector
\(\beta=(\beta_1,\ldots,\beta_p)\in\Hilbert^p\), where each
\(\beta_j\in\Hilbert\), solving
\[
  \Expect\{\mathbf X(V-\mathbf X^T\beta)\}=0.
\]
Here
\[
  \mathbf X^T\beta=\sum_{j=1}^p X_j\beta_j\in\Hilbert,
\]
so the fitted value and the residual are functions in the Wasserstein tangent
coordinate, not scalar responses.
Its empirical version solves
\[
  P_n\{\mathbf X(V-\mathbf X^T\hat\beta)\}=0.
\]
Then, for any fixed contrast \(\mathbf a\in\R^p\),
\[
  \sqrt n\,\mathbf a^T(\hat\beta-\beta)
  =
  \mathbb G_n\left[\mathbf a^T\Sigma^{-1}\mathbf X\{V-\mathbf X^T\beta\}\right]
  +o_{\Prob}(1)
\]
as a limit in the Hilbert space \(\Hilbert\), assuming the usual law of large
numbers for \(\mathbf X\mathbf X^T\) and finite second moments.
In the Ge et al. climate capsule, \(Z\) is century midpoint,
\(\mathbf X=\phi(Z)\) is a low-dimensional time design vector, and \(V\) is
represented by empirical quantile coordinates of the ten decadal anomaly values
inside that century.  The fitted time-slope function
\(u\mapsto\beta_{\mathrm{time}}(u)\) is therefore a readable component of this
abstract Hilbert-vector: it says how the lower, middle, and upper parts of the
century-level anomaly distribution move with historical time.
The computed capsule is intentionally modest but real: it uses 20
century-level empirical distributions, a 19-point quantile grid, and ten
decadal anomalies per century.  The fitted tangent-coordinate regression has
integrated \(R^2=0.029\).  That small value is part of the lesson rather than
a failure of the example.  The analysis gives a concrete distribution-valued
target and a reproducible coefficient function, while also warning that this
short reconstruction cannot support a strong attribution claim by itself.

\begin{realdatacapsule}{Ge et al. climate distribution regression}
\item[Data object.] The Ge et al. decadal temperature reconstruction, grouped
into century-level empirical distributions of anomaly values
\citep{ge2013temperature}.
\item[Observation mechanism.] Historical proxies, reconstruction methods,
decadal aggregation, and century grouping create the curve and distributional
records.
\item[Target.] A quantile-indexed slope or distributional contrast describing
how the century-level anomaly distribution changes with historical time.
\item[Model.] One-dimensional Wasserstein geometry represents distributions by
quantile functions; local linearization treats the fitted coefficient function
as a Hilbert-valued target.
\item[Uncertainty.] Tangent-coordinate variance, bootstrap over century units,
and sensitivity to quantile grid or period definition carry the uncertainty.
\item[Limitation.] Each century has only ten decadal points and inherited proxy
uncertainty, so the capsule is pedagogical distributional regression rather
than climate attribution.
\end{realdatacapsule}

\noindent\textit{Proof.}
Write
\[
  R=V-\mathbf X^T\beta\in\Hilbert
\]
for the population residual in Wasserstein tangent coordinates.  The empirical
normal equation gives
\[
  0
  =
  P_n\{\mathbf X(V-\mathbf X^T\hat\beta)\}.
\]
This is an equation in \(\Hilbert^p\): its \(j\)th component says that the
sample average of \(X_j\) times the Hilbert-valued residual is zero.  Expand
the residual around the population coefficient:
\[
\begin{aligned}
  0
  &=
  P_n\{\mathbf X(V-\mathbf X^T\beta)\}
  -
  P_n\{\mathbf X\mathbf X^T(\hat\beta-\beta)\}  \\
  &=
  P_n(\mathbf X R)-\hat\Sigma_n(\hat\beta-\beta),
\end{aligned}
\]
where \(\hat\Sigma_n=P_n(\mathbf X\mathbf X^T)\).  The finite matrix
\(\hat\Sigma_n\) acts on \(\Hilbert^p\) component by component.  Hence
\[
  \hat\Sigma_n(\hat\beta-\beta)
  =
  P_n(\mathbf X R).
\]
The population normal equation is \(P(\mathbf X R)=0\).  Therefore
\[
  P_n(\mathbf X R)
  =
  (P_n-P)(\mathbf X R)
  =
  n^{-1/2}\mathbb G_n(\mathbf X R).
\]
If \(\hat\Sigma_n\toP\Sigma\) and \(\Sigma\) is nonsingular, then Slutsky's
lemma, applied to the finite matrix multiplying a Hilbert-valued empirical
process term, gives
\[
  \sqrt n(\hat\beta-\beta)
  =
  \Sigma^{-1}\mathbb G_n\{\mathbf X(V-\mathbf X^T\beta)\}+o_{\Prob}(1),
\]
and multiplying by \(\mathbf a^T\) gives the displayed influence
representation.  Thus the first-order error is the average of the ordinary
design vector times a function-valued residual, premultiplied by the inverse
design covariance, exactly as in ordinary least squares.

This example deliberately uses two different meanings of ``tangent.''  The
first is geometric.  The Wasserstein coordinate
\(\Hilbert=L^2(\mu_\ast)\) shows how to subtract two distribution-valued
responses and treat the difference as a vector.  The second is statistical.
Bickel's tangent space \(L_0^2(P)\) describes how the joint observation law
\(P=\Law(\mathbf X,Y)\) can be locally perturbed.  The influence function above is
therefore a map from one observation to a value in \(\Hilbert\): the argument
belongs to the observation-law world, while the output is a Wasserstein
tangent-coordinate residual.  Geometry supplies the coordinate in which the
target is linear; semiparametric tangent-space theory says which perturbations
of the data law determine the first-order uncertainty. \qedmark
\end{example}

\begin{example}[Cumulative hazard]
For right-censored event-time data, the Nelson--Aalen estimator can be written
as an integral of observed counting increments divided by the at-risk process.
Its first-order fluctuation is an integral with respect to a martingale:
\[
  \hat A(t)-A(t)
  \approx
  \int_0^t \frac{1}{Y(u)}\,M(du).
\]
This is an influence-function statement in continuous time.  The next
continuous-time chapters develop the martingale and compensator language that
makes this display precise.

\noindent\textit{Derivation.}
For subject \(i\), let \(N_i(t)=\ind{T_i\le t,\Delta_i=1}\) and
\(Y_i(t)=\ind{Y_i\ge t}\).  Under independent censoring and a cumulative
hazard \(A\), write
\[
  M_i(t)=N_i(t)-\int_0^t Y_i(u)\,A(du)
\]
for the counting-process martingale.  The Nelson--Aalen estimator is
\[
  \hat A(t)
  =
  \int_0^t \frac{\ind{Y_\cdot(u)>0}}{Y_\cdot(u)}\,N_\cdot(du),
  \qquad
  Y_\cdot=\sum_iY_i,\quad N_\cdot=\sum_iN_i.
\]
Substitute \(N_\cdot(du)=Y_\cdot(u)A(du)+M_\cdot(du)\).  On time intervals
where \(Y_\cdot(u)>0\),
\[
  \hat A(t)-A(t)
  =
  \int_0^t \frac{1}{Y_\cdot(u)}\,M_\cdot(du).
\]
If \(Y_\cdot(u)/n\to y(u)>0\) uniformly on \([0,\tau]\), then
\[
  \sqrt n\{\hat A(t)-A(t)\}
  =
  \frac1{\sqrt n}\sum_{i=1}^n
  \int_0^t \frac{1}{y(u)}\,M_i(du)
  +
  o_{\Prob}(1)
\]
uniformly on \([0,\tau]\).  The integral inside the sum is the subject-level
influence contribution. \qedmark
\end{example}

\begin{example}[Transition probabilities in a multi-state model]
The cumulative-hazard example has a matrix-valued extension that is central in
event-history analysis.  Suppose each subject moves among finite states
\(E=\{1,\ldots,K\}\).  In oncology these might be
\[
  \hbox{stable disease}\longrightarrow
  \hbox{progression}\longrightarrow
  \hbox{death};
\]
in credit risk they might be current, delinquent, default, and paid off; in
reliability they might be normal operation, degraded operation, failure, and
repair.  For subject \(i\), let \(X_i(t)\) be the state just after time \(t\).
For \(h\ne j\), define the transition counting process and risk indicator
\[
  N_{ihj}(t)
  =
  \sum_{0<u\le t}\ind{X_i(u-)=h,\ X_i(u)=j},
  \qquad
  Y_{ih}(t)=\ind{X_i(t-)=h}.
\]
Thus \(N_{ihj}\) counts jumps from \(h\) to \(j\), and \(Y_{ih}\) records
whether subject \(i\) is at risk for leaving state \(h\).

In a nonhomogeneous Markov multi-state model, the compensator of \(N_{ihj}\)
has the form
\[
  \Lambda_{ihj}(t)
  =
  \int_{(0,t]}Y_{ih}(u)\,A_{hj}(du),
  \qquad
  M_{ihj}=N_{ihj}-\Lambda_{ihj},
\]
where \(A_{hj}\) is the cumulative transition hazard from \(h\) to \(j\).
Collect the off-diagonal \(A_{hj}\)'s into a matrix-valued cumulative hazard
\(\mathbf A\) by setting
\[
  \mathbf A_{hj}=A_{hj}\quad(h\ne j),
  \qquad
  \mathbf A_{hh}=-\sum_{j\ne h}A_{hj}.
\]
The transition probability matrix is the product integral
\[
  \mathbf P(s,t)
  =
  \Prodi_{(s,t]}\{I+\mathbf A(du)\},
  \qquad
  p_{ab}(s,t)=\mathbf P_{ab}(s,t).
\]
It is the solution of the forward equation
\[
  \mathbf P(s,t)
  =
  I+\int_{(s,t]}\mathbf P(s,u-)\,\mathbf A(du),
\]
so \(p_{ab}(s,t)\) is the probability of being in state \(b\) at time \(t\)
given state \(a\) at time \(s\).

For \(n\) independent subjects, write
\[
  Y_{\cdot h}(u)=\sum_{i=1}^nY_{ih}(u),
  \qquad
  N_{\cdot hj}(u)=\sum_{i=1}^nN_{ihj}(u).
\]
The Nelson--Aalen transition-hazard estimator is
\[
  \widehat A_{hj}(t)
  =
  \int_{(0,t]}
  \frac{\ind{Y_{\cdot h}(u)>0}}{Y_{\cdot h}(u)}
  N_{\cdot hj}(du),
  \qquad h\ne j,
\]
with diagonal entry
\(\widehat{\mathbf A}_{hh}=-\sum_{j\ne h}\widehat A_{hj}\).  The plug-in
transition estimator
\[
  \widehat{\mathbf P}(s,t)
  =
  \Prodi_{(s,t]}\{I+\widehat{\mathbf A}(du)\}
\]
is the Aalen--Johansen estimator \citep{aalen1978empirical,andersen1993statistical}.

The functional delta method enters through the map
\[
  \Phi:\mathbf A\mapsto \mathbf P(s,t).
\]
If \(\mathbf H\) is a small matrix-valued bounded-variation perturbation, the
product-integral derivative, summarized in
\Appref{sec:appC-product-integrals}, is
\[
  \Phi_{\mathbf A}'(\mathbf H)(s,t)
  =
  \int_{(s,t]}
  \mathbf P(s,u-)\,\mathbf H(du)\,\mathbf P(u,t),
\]
with the usual product-integral convention around jumps.  For the scalar
target \(p_{ab}(s,t)\), perturbing only the off-diagonal transition \(h\to j\)
contributes
\[
  \int_{(s,t]}
  \mathbf P_{ah}(s,u-)
  \{\mathbf P_{jb}(u,t)-\mathbf P_{hb}(u,t)\}\,
  H_{hj}(du).
\]
The difference term appears because adding hazard from \(h\) to \(j\) also
removes probability mass from remaining in \(h\).

Now suppose \(Y_{\cdot h}(u)/n\to y_h(u)>0\) on \([s,t]\).  For each
off-diagonal entry,
\[
  \sqrt n\{\widehat A_{hj}(r)-A_{hj}(r)\}
  =
  \frac1{\sqrt n}\sum_{i=1}^n
  \int_{(0,r]}
  \frac{1}{y_h(u)}\,M_{ihj}(du)
  +o_{\Prob}(1).
\]
Substituting this martingale expansion into the derivative of \(\Phi\) gives
\[
  \sqrt n\{\widehat p_{ab}(s,t)-p_{ab}(s,t)\}
  =
  \frac1{\sqrt n}\sum_{i=1}^n\varphi_{i,ab}(s,t)
  +o_{\Prob}(1),
\]
where
\[
  \varphi_{i,ab}(s,t)
  =
  \sum_{h\ne j}
  \int_{(s,t]}
  \frac{
    \mathbf P_{ah}(s,u-)
    \{\mathbf P_{jb}(u,t)-\mathbf P_{hb}(u,t)\}
  }{y_h(u)}
  \,M_{ihj}(du).
\]
This is the subject-level influence contribution for the transition
probability.  The coefficient has a direct reading: at time \(u\), a surprise
jump \(h\to j\) matters for \(a\to b\) prediction in proportion to the chance
of having reached \(h\) from \(a\), multiplied by how much being sent to \(j\)
instead of staying in \(h\) changes the later chance of ending in \(b\).

The same formula explains why the application background is not decoration.
In a clinical illness-death model, \(p_{13}(0,24)\) may be a two-year death
probability under a treatment pathway.  In credit risk, \(p_{aD}(0,12)\) may
be a one-year default probability from rating \(a\).  In manufacturing,
\(p_{\mathrm{degraded},\mathrm{failed}}(s,t)\) may determine a preventive
maintenance rule.  The data vocabulary changes; the counting processes,
product integral, and delta-method derivative are the same.
\qedmark
\end{example}

\section{Bootstrap and Resampling Variability}
\label{sec:ch15-bootstrap}
\conceptindexes{bootstrap, resampling, empirical distribution, bootstrap consistency}

The bootstrap enters naturally after the derivative has identified the
first-order noise.  Its job is not to re-solve the whole asymptotic problem
from scratch, but to reproduce the same local linear fluctuation with the
empirical law standing in for the population law.

The nonparametric bootstrap replaces $P$ by $P_n$ and draws
$X_1^*,\ldots,X_n^*$ from $P_n$.  For smooth functionals, the bootstrap works
because it reproduces the same first-order empirical fluctuation:
\[
  \sqrt n\{T(P_n^*)-T(P_n)\}
\]
conditionally mimics
\[
  \sqrt n\{T(P_n)-T(P)\}.
\]
When $T$ is Hadamard differentiable, the functional delta method applies both
to the original empirical process and to the bootstrap empirical process.  The
bootstrap is therefore not magic resampling.  It is a way of re-reading the
linearized residual noise with $P_n$ standing in for $P$.

The same statement also explains bootstrap failures.  If the functional is not
sufficiently smooth, if the derivative is discontinuous, or if the estimator
converges at a non-$\sqrt n$ rate, the ordinary bootstrap may read the wrong
local geometry.

\begin{example}[Bootstrap bands for a censored survival surface]
Paired censoring gives a compact example of why influence functions need not be
one-dimensional.  The target is the surface
\[
  S(s,t)=\Prob(T_1>s,T_2>t),
\]
but the observation is only
\[
  O_i=(Y_{1i},Y_{2i},\Delta_{1i},\Delta_{2i}),
  \qquad
  Y_{ji}=T_{ji}\wedge C_{ji}.
\]
In the bivariate Kaplan--Meier setting, the estimator can be summarized at
this level as a product-limit functional of the observed law:
\[
  \widehat S_n=\Psi(P_n),\qquad S=\Psi(P),
\]
on regions where the bivariate risk set is not exhausted
\citep{dabrowska1988kaplan,dabrowskaStochasticProcessesCommunication}.  This notation deliberately
suppresses the estimator algebra.  The statistical point is that \(\Psi\)
translates the observable marked risk-set distribution into the unobserved
survival surface.

Under the smoothness and censoring conditions used for weak convergence,
\(\Psi\) has a first-order expansion in a two-parameter function space:
\[
  \sqrt n\{\widehat S_n-S\}
  =
  \frac1{\sqrt n}\sum_{i=1}^n \phi_P(O_i)
  +o_{\Prob}(1),
\]
where \(\phi_P(O_i)\) is a mean-zero random surface.  The bootstrap repeats the
same local reading with \(P_n\) replacing \(P\).  If
\(O_1^*,\ldots,O_n^*\) are drawn with replacement from the observed records and
\(\widehat S_n^*=\Psi(P_n^*)\), then
\[
  \sqrt n\{\widehat S_n^*-\widehat S_n\}
\]
conditionally mimics the Gaussian surface limit of
\(\sqrt n\{\widehat S_n-S\}\)
\citep{dabrowska1989kaplan,dabrowskaAdvancedProbabilityCommunication}.

A simultaneous band over a rectangle
\([0,\tau_1]\times[0,\tau_2]\) is therefore built from bootstrap replicates of
\[
  T_b^*
  =
  \sqrt n
  \sup_{0\le s\le\tau_1,\ 0\le t\le\tau_2}
  |\widehat S_{n,b}^*(s,t)-\widehat S_n(s,t)|.
\]
If \(c_\alpha^*\) is the empirical \((1-\alpha)\)-quantile of these statistics,
then
\[
  \widehat S_n(s,t)\pm c_\alpha^*/\sqrt n
\]
is the surface-valued analogue of a bootstrap confidence interval.  The
important lesson is not the special product-limit formula; it is the same
principle used throughout this chapter: identify the local linear surface
noise, then ask the bootstrap to reproduce that noise rather than the original
biological or clinical history. \qedmark
\end{example}

\subsection{Application Vignettes Across Domains}
\label{sec:ch15-application-vignettes}
\conceptindexes{application vignettes, clinical endpoints, genomics, functional data, deployed systems}

These vignettes are placed after the theory to return to the book's recurring
applications with a new question.  The earlier chapters asked what the data
object is, what target is scientifically meaningful, and how an estimator can
be selected.  Here the same examples ask how the last empirical wobble becomes
reported uncertainty.  A domain scientist names a target; the statistician
asks which functional of which law, hazard, curve, or stochastic process that
target is.  The delta method then says how uncertainty moves through the
translation, and the influence function says which observations, subjects,
events, or histories drive the first-order noise.

\begin{description}[leftmargin=2.15em,style=nextline]
\item[Biopharma and public health.]
A trial may report the same kinds of summaries that appeared in the opening
clinical examples: a risk ratio, log odds, restricted mean survival time,
progression-free survival probability, or multi-state transition probability.
The target is not the spreadsheet column itself; it is a functional of event
times, censoring, competing risks, and treatment assignment.  For example,
\(p_{13}(0,24)\) in an illness-death model can mean the probability of death by
24 months starting from stable disease.  The Aalen--Johansen expansion above
shows how patient-level transition martingales become the standard error of
that probability.  The log-risk and log-odds calculations at the start of the
chapter are the finite-dimensional version of the same translation.

\item[Ecology, language, and rare-type inventories.]
The missing-species and Shakespeare-vocabulary examples begin with a simple
question: how much probability mass sits in rare or unseen types?  For a fixed
rare class \(A\), the local target \(P(A)\) has influence function
\[
  \ind{X\in A}-P(A).
\]
This says exactly which observations move the estimate at first order.  In the
full missing-species problem the set of relevant rare classes may grow with
the sample size, so the challenge is no longer one Bernoulli fluctuation but a
field of rare-type fluctuations.  That is why the empirical-process language
from Chapter~11 is needed before this chapter can translate the uncertainty.

\item[Credit risk, insurance, and actuarial analytics.]
Rating migration is naturally a multi-state process.  If \(D\) is default and
\(a\) is an initial rating, a one-year default target is
\[
  \psi_a=p_{aD}(0,1).
\]
A portfolio loss functional may be
\[
  L=\sum_a w_a\,p_{aD}(0,1),
\]
where \(w_a\) is exposure in rating class \(a\).  Once each \(p_{aD}\) has an
influence expansion, \(L\) inherits one by linearity.  The same local
calculation turns transition-data uncertainty into uncertainty for capital,
pricing, or stress-test summaries.

\item[Reliability, manufacturing, and operations.]
A machine can move from normal operation to degraded operation, failure, and
repair.  A maintenance rule may depend on
\[
  p_{\mathrm{degraded},\mathrm{failed}}(s,t)
\]
or on the probability of avoiding failure until the next planned service
window.  Counting-process martingales separate expected wear-out from surprise
failures.  Influence functions then identify which machines, shifts, batches,
or usage regimes are responsible for the uncertainty in a reliability curve.

\item[Digital platforms and policy learning.]
In advertising, pricing, recommendation, and marketplace ranking, the observed
outcome is produced by a logging policy.  For a new policy \(\pi\), a common
value target is
\[
  \psi(\pi)=\Expect\left\{\sum_a \pi(a\mid X)m_a(X)\right\},
  \qquad
  m_a(X)=\Expect(Y\mid X,A=a).
\]
If \(e(a\mid X)\) is the logging propensity, the familiar orthogonal influence
function is
\[
  \varphi(O)
  =
  \sum_a \pi(a\mid X)m_a(X)-\psi(\pi)
  +
  \frac{\pi(A\mid X)}{e(A\mid X)}
  \{Y-m_A(X)\}.
\]
This is the missing-outcome influence function in platform form: observed
clicks, purchases, or retention outcomes must be reweighted because the
platform decided what each user saw.

\item[Climate, energy, and environmental risk.]
High quantiles and return levels are functionals of a distribution or a
smoothed proxy process.  This is the local-inference version of the
Zhu/Chu Ko-chen climate thread: after historical or proxy records have been
assembled into a distributional target, uncertainty in a tail summary is read
through a derivative.  The Ge--Hao--Zheng--Shao reconstruction above showed
the same idea for period means and warm-period contrasts; tail summaries use
the same template with a different functional.  If \(q_p=F^{-1}(p)\), the
quantile influence function is
\[
  \frac{p-\ind{X\le q_p}}{f(q_p)}.
\]
The denominator is the application warning.  When the density near the target
tail level is small, a modest amount of empirical noise can produce a large
change in the estimated return level.  This matters for heat-wave thresholds,
flood-risk design levels, energy-load peaks, and climate-proxy reconstructions.

\item[Imaging, omics, and quality systems.]
An ROC curve, an AUC, a cell-state proportion, a pathway-enrichment score, or a
defect-discovery rate is a functional of a high-dimensional empirical object.
The influence-function question is concrete: which images, cells, features,
labels, or batches move the target at first order?  That reading connects
uncertainty quantification with diagnostics, data cleaning, and design changes.
\end{description}

Across these settings, the mathematical habit is the same.  Name the target
functional, identify the empirical fluctuation that estimates its input, check
whether the map is differentiable in the relevant direction, and then read the
linear term.  Domain knowledge enters twice: it defines the target, and it says
which perturbations are plausible enough to deserve a standard error.
This is why the same chapter can contain clinical odds, rare vocabulary,
climate quantiles, Wasserstein coordinates, transition probabilities, and
platform policy values without becoming a catalogue: each is an instance of
the same uncertainty-translation template.

\begin{tcolorbox}[
  enhanced,
  breakable,
  colback=chaptercream,
  colframe=bookblue!88!black,
  boxrule=0.62pt,
  arc=4pt,
  boxsep=1pt,
  left=0.95em,
  right=0.9em,
  top=0.72em,
  bottom=0.72em,
  before skip=0.9\baselineskip,
  after skip=1.0\baselineskip
]
\noindent\textbf{Closing the loop.}
Chapter~13 ended with estimators represented by first-order empirical noise.
Chapter~14 placed that noise inside local testing experiments.  This chapter
has turned the representation into a portable uncertainty language.
The same local template now covers finite-dimensional smooth transforms,
distributional functionals, event-history product integrals, observation-level
influence diagnostics, bootstrap bands, and semiparametric efficiency.

This completes the estimation arc of the book.  Earlier chapters asked what
the data object is and when empirical summaries are stable.  Chapter~13 showed
how stable empirical objects choose estimates.  Chapter~14 showed how nearby
laws are separated and compared.  This chapter showed how the remaining local
noise becomes uncertainty for the target actually reported to a scientific,
industrial, or policy audience.  The recurring examples now have their
uncertainty layer: risk summaries get delta-method standard errors, rare-type
inventories get influence contributions, distributional regression gets
Hilbert-valued residuals, climate tails get quantile derivatives, and event
histories get martingale influence terms.  Chapter~16 then changes the clock:
it develops the continuous-time process language behind the martingale terms
that appeared here in survival and transition-probability examples.
\end{tcolorbox}

\section{Exercises}
\label{sec:ch15-exercises}
\conceptindexes{delta-method exercises, influence-function exercises, bootstrap exercises}

\begin{exercise}[Delta method for a ratio]
Let $(\bar X_n,\bar Y_n)$ satisfy a bivariate central limit theorem with mean
$(\mu_X,\mu_Y)$ and $\mu_Y\ne0$.  Derive the asymptotic distribution of
$\bar X_n/\bar Y_n$.
\end{exercise}

\begin{exercise}[Influence function for a covariance]
For $T(P)=\Cov_P(X,Y)$, derive $\IF((x,y);T,P)$.
\end{exercise}

\begin{exercise}[Quantile from the functional derivative]
Starting from $F(Q(p))=p$, perturb $F$ to $F+th$ and derive the quantile
derivative $-h(Q(p))/f(Q(p))$.
\end{exercise}

\begin{exercise}[A Z-estimator expansion]
Suppose $\Psi_n(\theta)=P_n\psi_\theta$, $\Psi(\theta_0)=0$, and
$\Psi$ is differentiable at $\theta_0$ with nonsingular derivative.  State
conditions under which
\[
  \sqrt n(\hat\theta_n-\theta_0)
  =
  -\dot\Psi_{\theta_0}^{-1}\mathbb G_n\psi_{\theta_0}
  +o_{\Prob}(1)
\]
holds.
\end{exercise}

\begin{exercise}[Bootstrap and nonsmoothness]
Investigate why the ordinary bootstrap for the sample maximum does not mimic
the limiting distribution of the centered maximum under a continuous
distribution with bounded support.
\end{exercise}

\begin{exercise}[Bootstrap band for a censored survival surface]
In the bivariate right-censoring example above, explain why the bootstrap must
resample the observed subject-level quadruples
$(Y_{1i},Y_{2i},\delta_{1i},\delta_{2i})$ rather than resampling the two
coordinates independently.  What dependence would be destroyed by the latter
procedure?
\end{exercise}

\begin{exercise}[Transition-probability influence term]
In the multi-state transition example, take three states
\(1=\) healthy, \(2=\) ill, and \(3=\) dead, with transitions
\(1\to2\), \(1\to3\), and \(2\to3\).  Write the coefficient multiplying
\(M_{i12}(du)\) in the influence function for \(p_{13}(0,t)\).  Explain in
words why the coefficient compares the later chance of death after entering
state \(2\) with the later chance of death after remaining in state \(1\).
\end{exercise}

\begin{exercise}[Calibration at a risk threshold]
Let \(R\) be a fitted risk score and \(Y\in\{0,1\}\).  For a fixed threshold
\(c\), define
\[
  T(P)=\Expect_P\{Y\mid R\in[c-\epsilon,c+\epsilon]\}.
\]
Treating \(R\) as fixed, derive the first-order influence form for this
localized calibration target.  What changes when \(R\) is itself estimated
from the same data?
\end{exercise}

\begin{exercise}[AUC as a pairwise functional]
For a binary outcome \(Y\), write the area under the ROC curve as
\[
  T(P)=\Prob\{R_1>R_0\}+\frac12\Prob\{R_1=R_0\},
\]
where \(R_1\) and \(R_0\) are independent scores from cases and controls.
Sketch the influence-function calculation by perturbing the case distribution
and the control distribution separately.
\end{exercise}

\begin{exercise}[Doubly robust policy score]
For logged data \(O=(H,A,b(A\mid H),Y)\), consider the value of a policy
\(\pi\).  Write the doubly robust score using an outcome regression
\(m(a,H)\) and the importance ratio \(\pi(A\mid H)/b(A\mid H)\).  Explain the
two ways in which first-order bias can disappear and why positivity remains a
separate condition.
\end{exercise}

\begin{exercise}[Net benefit as a functional]
For a binary clinical decision rule \(d_c(X)=\ind{\hat p(X)>c}\), define a net
benefit target that rewards true positives and penalizes false positives.
Write the target as a functional of the joint law of \((X,Y)\).  Which part of
the influence calculation is ordinary empirical averaging, and which part is
nonsmooth because the rule depends on a threshold?
\end{exercise}

\section*{Sources and Further Reading}
\addcontentsline{toc}{section}{Sources and Further Reading}

The finite-dimensional delta method used at the beginning of the chapter is
standard in mathematical statistics; see \citet{bickel1977mathematical} and
\citet{vaart1998asymptotic}.  The functional delta method, Hadamard
differentiability, and bootstrap versions are developed systematically in
\citet{vaart2023weak}.  The bootstrap connection used later in the chapter goes
back to \citet{efron1979bootstrap}.

The influence-curve terminology and its robust-statistics interpretation trace
to \citet{hampel1974influence}; the monograph treatment is
\citet{hampel1986robust}.  Huber's location estimator in
\citet{huber1964robust} is the canonical example linking bounded influence,
local robustness, and asymptotic normality.  The likelihood and information
viewpoint behind the parametric score calculations is rooted in
\citet{fisher1922foundations}.

The semiparametric tangent-space and projection formulation follows
\citet{bickel1993efficient}.  For complementary references, Newey's efficiency
bound paper \citep{newey1990semiparametric} gives a compact econometric route
to the same projection idea, \citet{tsiatis2006semiparametric} develops the
missing-data side in detail, and \citet{kosorok2008empirical} connects
semiparametric inference with empirical-process tools.  In this chapter the
link to Wasserstein regression is conceptual but mathematically specific:
Bickel tangent spaces live in \(L_0^2(P)\) for the observation law, while the
one-dimensional Wasserstein regression example uses
\(L^2(\mu_\ast)\) as the tangent coordinate for distribution-valued responses.
The latter geometry is developed in \citet{villani2009optimal},
\citet{panaretos2019statistical}, \citet{petersen2019frechet}, and
\citet{chen2023wasserstein}.

The transition-probability example follows the multi-state counting-process
and product-integral treatment of \citet{aalen1978empirical},
\citet{gill1990survey}, and \citet{andersen1993statistical}.  The bivariate
Kaplan--Meier example is a deliberately compressed paraphrase of
\citet{dabrowska1988kaplan}, with weak convergence, LIL, and bootstrap
confidence bands from \citet{dabrowska1989kaplan}.  The surrounding
delta-method, weak-convergence, martingale, and stochastic-process language is
guided by \citet{dabrowskaAdvancedProbabilityCommunication} and
\citet{dabrowskaStochasticProcessesCommunication}.
Counting-process influence representations for
survival analysis are developed in \citet{aalen1978nonparametric},
\citet{andersen1982cox}, \citet{andersen1993statistical}, and
\citet{fleming1991counting}.

%% file: chapters/ch16_processes_continuous_time.tex
\chapter{Continuous-Time Processes, Event Histories, and Martingales}
\label{chap:continuous-time-processes}
\conceptindexes{continuous-time processes, stochastic basis, stopping times, marked point processes, martingales, compensators, survival analysis, Markov models}

\providecommand{\sig}{\sigma}
\providecommand{\St}{\mathcal{S}}
\providecommand{\Ht}{\mathcal{H}}
\providecommand{\Gt}{\mathcal{G}}
\providecommand{\Pt}{\mathcal{P}}
\providecommand{\DeltaState}{\Delta}
\providecommand{\barF}{\overline{F}}
\providecommand{\matA}{\mathbf{A}}
\providecommand{\matI}{\mathbf{I}}
\providecommand{\matP}{\mathbf{P}}
\providecommand{\matQ}{\mathbf{Q}}
\let\cad\relax
\let\cag\relax
\let\cadlag\relax
\let\caglad\relax
\newcommand{\cad}{\ifmmode\mathrm{c\grave{a}d}\else c\`ad\fi}
\newcommand{\cag}{\ifmmode\mathrm{c\grave{a}g}\else c\`ag\fi}
\newcommand{\cadlag}{\ifmmode\mathrm{c\grave{a}dl\grave{a}g}\else c\`adl\`ag\fi}
\newcommand{\caglad}{\ifmmode\mathrm{c\grave{a}gl\grave{a}d}\else c\`agl\`ad\fi}

\begin{tcolorbox}[
  enhanced,
  breakable,
  colback=chaptercream,
  colframe=bookblue!88!black,
  boxrule=0.72pt,
  arc=5pt,
  boxsep=1pt,
  left=1.0em,
  right=0.95em,
  top=0.82em,
  bottom=0.82em,
  before skip=0.55\baselineskip,
  after skip=1.0\baselineskip
]
\noindent\textbf{Chapter overview.}
This chapter makes information through time explicit before the book turns to
use, risk, and feedback. Its rule is simple: use the past, not the future.
Filtrations record what has been seen; stopping times encode decisions made
without looking ahead; predictable processes formalize quantities fixed just
before the next instant; compensators turn observed jumps into martingales. The
statistical center is event-history data: survival, multi-state models,
intensities, product integrals, and continuous-time Markov chains.
\end{tcolorbox}

\noindent This chapter is not a full treatment of stochastic calculus, Markov
process theory, or martingale theory.  It uses continuous-time probability as
statistical grammar.  The route is deliberately layered.  Section~16.1 builds
the information structure.  Section~16.2 turns event logs into marked point
processes and predictable objects.  Section~16.3 explains why compensators
create martingales.  Section~16.4 stays with statistical event-history targets:
survival curves, panel counts, multi-state probabilities, and semi-Markov
regression.  Sections~16.5 and~16.6 then widen the same grammar to Poisson,
renewal, Markov jump, Feller, interacting-particle, diffusion, and
jump-diffusion systems.

\input{figures/ch15_event_history_clock}

Chapter~1 began with a claim about statistical work: before there is a model,
there is a data structure, and before there is a data structure, there is a
reason some part of the world was observed. Continuous time makes that claim
sharp. A patient's disease course, a machine's failure record, or a user's
stream of clicks does not arrive as a hazard function. It arrives as a growing
record: dates, visits, scans, state labels, interruptions, missing pieces, and
endpoints chosen by a protocol.

The central rule is no peeking. At calendar time \(t\), an analyst may use what
has been recorded by \(t\); a prediction for the next instant must be based on
what was known just before that instant. A rule such as ``stop when the first
relapse is observed'' is legitimate because it can be carried out from the
current record. A dose scheduled before the next visit is predictable. The
event itself, and especially its mark, is not predictable until it occurs.
These distinctions are the statistical face of the general theory of processes
\citep{doob1953stochastic,dellacherie1978probabilities,jacod1987limit}.

The martingale bookkeeping is the payoff.  A compensator is the predictable
clock that the history makes foreseeable; subtract that clock from the observed
jump count, and what remains is the surprise. This is the idea behind the
counting-process formulation of survival and event-history analysis
\citep{aalen1978nonparametric,andersen1982cox,andersen1993statistical,fleming1991counting}.

We now turn from discrete time to real time. Chapter 6 supplied the
product-space foundation: a stochastic process is a random object assembled
from coordinates. The new ingredient here is information ordered by time. A
stochastic process is a family
\[
  X=\{X(t):t\ge0\}
\]
of random variables defined on a probability space $(\Omega,\fieldF,P)$. The
state space is a measurable space $(S,\St)$, often a complete separable metric
space such as $\Real^d$ with its Borel $\sigma$-algebra. It is useful to attach
one extra point $\DeltaState\notin S$, called the cemetery state or empty mark,
and to write
\[
  S_{\Delta}=S\cup\{\DeltaState\},
  \qquad
  \St_{\Delta}=\sigma(\St,\{\DeltaState\}).
\]
Unless stated otherwise, processes are assumed measurable as maps
\[
  (\omega,t)\longmapsto X_t(\omega)
  \quad\text{from}\quad
  (\Omega\times\Rplus,\fieldF\otimes\Borel(\Rplus))
  \quad\text{to}\quad (S_{\Delta},\St_{\Delta}).
\]

The first running object is a marked point process.  It records event times and
their marks,
\[
  (T_n,X_n),\qquad T_0=0,\quad 0<T_1<T_2<\cdots,
\]
where each $X_n$ takes values in $(S,\St)$. The corresponding step process is
formally
\[
  X(t)=\sum_{n\ge0}X_n\,\ind{T_n\le t<T_{n+1}}.
\]
This resembles a renewal reward process, but the interarrival times need not be
independent or identically distributed, and the parameter is real time rather
than a discrete index.  For marked point processes as random measures, see
\citet{karr1986point} and \citet{jacod1987limit}.

We use the standard path acronyms:
\[
\begin{array}{@{}r@{\quad}l@{}}
\text{\normalfont\scshape cad} & \text{right-continuous},\\
\text{\normalfont\scshape cag} & \text{left-continuous},\\
\text{\normalfont\scshape cadlag} & \text{right-continuous with left limits},\\
\text{\normalfont\scshape caglad} & \text{left-continuous with right limits}.
\end{array}
\]
For a \cadlag\ path $Y$, write
\[
  Y_{t-}=\lim_{s\uparrow t}Y_s,\qquad
  \Delta Y_t=Y_t-Y_{t-}.
\]
For a marked point process, the random measure notation
\[
  N((0,t]\times B)=N_t(B)
\]
will be used interchangeably with the counting-process notation; the same
convention applies to compensators $\Lambda(dt,du)$.

A process valued in a complete separable metric space, whose paths are
right-continuous and whose coordinates are random variables, is jointly
measurable. The analogous statement holds for the left-continuous and
one-sided-limit variants.  The technical notation below exists to separate
what is known at a time from what is known immediately before that time. At a
jump time \(T\), \(\fieldF_T\) may contain the jump and its mark, while
\(\fieldF_{T-}\) contains only the pre-jump history. Counting-process
likelihoods, survival estimators, and transition martingales all rely on that
separation.

\section{Stochastic Bases and Stopping Times}
\conceptindexes{stochastic bases, filtrations, adapted processes, stopping times, stopped sigma-field}

\subsection{Stochastic Basis}
\label{sec:ch16-stochastic-basis}
\conceptindexes{stochastic basis, filtration, adapted process, information over time}

Let $(\Omega,\fieldF,P)$ be a probability space.

\begin{definition}[Filtration and stochastic basis]
A family of $\sigma$-fields
\[
  \mathbb{F}=\{\fieldF_t:t\ge0\},\qquad \fieldF_t\subseteq\fieldF,
\]
is called a filtration if it is increasing:
\[
  \fieldF_s\subseteq\fieldF_t,\qquad 0\le s\le t.
\]
The filtration is called standard if it is also right-continuous,
\[
  \fieldF_t=\bigcap_{s>t}\fieldF_s,
\]
and complete, meaning that every $\fieldF_t$ contains all $P$-null sets of
$\fieldF$. The quadruple
\[
  (\Omega,\fieldF,\mathbb{F},P)
\]
is called a stochastic basis.
\end{definition}

Chapter~3 defined \(\sigma(X)\) as the information carried by a random object.
A filtration is the time-indexed version of that idea.  If \(X=\{X_t:t\ge0\}\)
is an observed process, its natural filtration is
\[
  \fieldF_t^X=\sigma(X_s:0\le s\le t),
\]
usually after completion and right-continuous augmentation.  If \(N\) is a
marked random measure on \(\Rplus\times S\), the corresponding observed history
is
\[
  \fieldF_t^N
  =
  \sigma\{N((0,s]\times B):0\le s\le t,\ B\in\St\}.
\]
Thus \(\fieldF_t\) records exactly which events about the underlying outcome
have become observable by time \(t\).

We use the notation
\[
  \fieldF_{t+}=\bigcap_{s>t}\fieldF_s,\qquad
  \fieldF_{t-}=\sigma\!\left(\bigcup_{s<t}\fieldF_s\right),
  \qquad
  \fieldF_{0-}=\fieldF_0
\]
where the last convention says that, at time zero, the only available
``past'' information is the initial information.

\begin{definition}[Adapted process]
A process $X=\{X_t:t\ge0\}$ is adapted to the filtration
$\mathbb{F}=\{\fieldF_t:t\ge0\}$ if $X_t$ is $\fieldF_t$-measurable for every
$t\ge0$; equivalently,
\[
  \{\omega:X_t(\omega)\in B\}\in\fieldF_t,
  \qquad B\in\St_{\Delta}.
\]
\end{definition}

Adaptedness is a statement about what is known at each fixed time. It does not
by itself impose path regularity on $(\omega,t)\mapsto X_t(\omega)$. The
filtration is often interpreted as history: $\fieldF_t$ contains the
information collected over $[0,t]$.

Two navigation tables are useful before the terminology multiplies.
Table~\ref{tab:ch16-measurability-map} gives a compact map of the notions;
Table~\ref{tab:ch16-measurability-questions} reads the same ideas from the
question an analyst is trying to answer. They are guides to the definitions,
not substitutes for them.

\begingroup
\footnotesize
\setlength{\LTpre}{0.45\baselineskip}
\setlength{\LTpost}{0.45\baselineskip}
\setlength{\tabcolsep}{0.35em}
\renewcommand{\arraystretch}{1.12}
\begin{longtable}{@{}>{\raggedright\arraybackslash}p{0.22\linewidth}
                >{\raggedright\arraybackslash}p{0.37\linewidth}
                >{\raggedright\arraybackslash}p{0.32\linewidth}@{}}
\caption{A navigation map for process measurability}
\label{tab:ch16-measurability-map}\\
\toprule
\textbf{\textcolor{bookblue}{Notion}} &
\textbf{\textcolor{bookblue}{Short definition}} &
\textbf{\textcolor{bookblue}{Intuition and common use}} \\
\midrule
\endfirsthead
\caption[]{A navigation map for process measurability (continued)}\\
\toprule
\textbf{\textcolor{bookblue}{Notion}} &
\textbf{\textcolor{bookblue}{Short definition}} &
\textbf{\textcolor{bookblue}{Intuition and common use}} \\
\midrule
\endhead
\bottomrule
\endlastfoot
Random element / vector &
\(X:\Omega\to E\) is measurable; for \(E=\mathbb R^d\), this is equivalent to
measurability of all coordinates. &
The event ``\(X\) takes values in \(B\)'' has a probability; finite-dimensional
distributions. \\
\addlinespace[0.22em]
Pointwise process measurability &
For each fixed \(t\), \(X_t:\Omega\to E\) is measurable. &
The weakest process-level requirement: every time slice is a random variable. \\
\addlinespace[0.22em]
Cylinder/path-map measurability &
\(X:\Omega\to E^T\) is measurable for the coordinate-cylinder
\(\sigma\)-field. &
The law of a process viewed through finite-dimensional coordinates;
Kolmogorov extension. \\
\addlinespace[0.22em]
Joint/product measurability &
\((t,\omega)\mapsto X_t(\omega)\) is
\(\mathcal B(T)\otimes\fieldF\)-measurable. &
Time and randomness can be integrated together; deterministic time integrals
and random fields. \\
\addlinespace[0.22em]
Borel / Lebesgue / universal measurability &
Borel uses the topology; Lebesgue uses completed Lebesgue sets; universal means
measurable under every completed probability measure. &
Topological random objects, integration, projections, and optimization. \\
\addlinespace[0.22em]
Adaptedness &
\(X_t\) is \(\fieldF_t\)-measurable for every \(t\). &
No looking into the future; martingales and filtered stochastic processes. \\
\addlinespace[0.22em]
Progressive measurability &
For every \(T\), the restriction to \([0,T]\times\Omega\) is
\(\mathcal B([0,T])\otimes\fieldF_T\)-measurable. &
The whole past up to time \(T\) is observable; It\^o integrals and SDEs. \\
\addlinespace[0.22em]
Optional measurability &
Measurability with respect to the optional \(\sigma\)-field, generated by
adapted right-continuous paths. &
Processes that can be read at stopping times; optional sampling and
Doob--Meyer theory. \\
\addlinespace[0.22em]
Predictability &
Measurability with respect to the predictable \(\sigma\)-field, generated by
left-continuous adapted paths. &
Information known just before time \(t\); semimartingale integration and
trading strategies. \\
\addlinespace[0.22em]
\(C[0,T]\) / \(D[0,T]\) path measurability &
\(X:\Omega\to C[0,T]\) or \(D[0,T]\) is Borel measurable under the usual
uniform or Skorokhod topology. &
Continuous processes, jump processes, counting processes, and weak convergence
on path spaces. \\
\end{longtable}
\endgroup

The table deliberately uses shorthand. Basic random elements, product
measurability, and Borel/Lebesgue/universal measurability are reviewed in
Chapter~6 and in standard measure-theoretic sources such as
\citet{billingsley1995probability}, \citet{bogachev2007measure}, and
\citet{kallenberg2002foundations}. For cylinder and Skorokhod path-space
structures, see \citet{billingsley1999convergence} and
\citet{ethierKurtz1986markov}. For the optional, progressive, and predictable
\(\sigma\)-fields, see the general theory in \citet{doob1953stochastic},
\citet{dellacherie1978probabilities}, and \citet{jacod1987limit}.

\begingroup
\footnotesize
\setlength{\LTpre}{0.45\baselineskip}
\setlength{\LTpost}{0.45\baselineskip}
\setlength{\tabcolsep}{0.35em}
\renewcommand{\arraystretch}{1.12}
\begin{longtable}{@{}>{\raggedright\arraybackslash}p{0.33\linewidth}
                >{\raggedright\arraybackslash}p{0.24\linewidth}
                >{\raggedright\arraybackslash}p{0.35\linewidth}@{}}
\caption{Questions that determine the right measurability concept}
\label{tab:ch16-measurability-questions}\\
\toprule
\textbf{\textcolor{bookblue}{Question}} &
\textbf{\textcolor{bookblue}{Concept}} &
\textbf{\textcolor{bookblue}{Where it reappears}} \\
\midrule
\endfirsthead
\caption[]{Questions that determine the right measurability concept (continued)}\\
\toprule
\textbf{\textcolor{bookblue}{Question}} &
\textbf{\textcolor{bookblue}{Concept}} &
\textbf{\textcolor{bookblue}{Where it reappears}} \\
\midrule
\endhead
\bottomrule
\endlastfoot
What information is carried by one observed random object? &
Generated \(\sigma\)-field \(\sigma(X)\) &
Chapter~3; then reused here as process history. \\
\addlinespace[0.22em]
Is \(X_t\) a legitimate random variable at a fixed time? &
Pointwise measurability &
Chapter~6 and Table~\ref{tab:ch16-measurability-map}. \\
\addlinespace[0.22em]
Can the whole process be treated as one random object? &
Cylinder or path-space measurability &
Chapter~6 for construction; Chapter~10 for weak convergence. \\
\addlinespace[0.22em]
Can we integrate \(\int_0^T X_t\,dt\) over deterministic time? &
Joint/product measurability &
Chapter~6, through product \(\sigma\)-fields and Fubini. \\
\addlinespace[0.22em]
Can the model use only what is known by time \(t\)? &
Filtration and adaptedness &
This section, via \(\fieldF_t^X=\sigma(X_s:s\le t)\). \\
\addlinespace[0.22em]
Can the process be stopped and read at a stopping time? &
Progressive or optional measurability &
Stopping times, optional processes, and progressive processes in this chapter. \\
\addlinespace[0.22em]
Can the process serve as an integrand against a semimartingale or event clock? &
Predictability &
Predictable processes, compensators, and martingale integrals. \\
\addlinespace[0.22em]
Do projections, suprema, or optimizers escape Borel measurability? &
Universal measurability and measurable selection &
Chapter~6 and the measure-theoretic appendix. \\
\end{longtable}
\endgroup

A useful mnemonic, subject to the usual regularity assumptions, is
\[
  \text{predictable}
  \subset
  \text{optional}
  \subset
  \text{progressive}
  \subset
  \text{adapted and jointly measurable}.
\]
This display is not a replacement for the definitions. It is only a guide to
which level of measurability is normally demanded by which operation.

The right-continuity condition is best read as a convention about
infinitesimal observation. Passing from $[0,t]$ to $[0,t+h]$, with $h>0$
arbitrarily small, should not create a new event that was invisible at time
$t$ but visible at every later time. Many stopping-time identities fail without
this condition because one can otherwise hide information at a right limit.

Completeness plays a different role. It allows the analysis to ignore events that differ
only on null sets when defining stopped random variables such as $X_T$.
Starting from any filtration $\mathbb{F}$, one usually replaces it by the
completed filtration
\[
  \fieldF_t^{P}
  =
  \{A\cup N:A\in\fieldF_t,\ N\subseteq N_0
    \text{ for some }N_0\in\fieldF,\ P(N_0)=0\}.
\]
This operation is mathematically convenient and does not change the
observable model under $P$.

\subsection{Stopping Times}
\conceptindexes{stopping times, optional information, stopped process, stopping comparison}

\begin{definition}[Stopping time]
Let $(\Omega,\fieldF)$ be a measurable space and let
$\mathbb{F}=\{\fieldF_t:t\ge0\}$ be an increasing filtration. A random variable
$T$ taking values in $\Rplus\cup\{\infty\}$ is a stopping time if
\[
  \{T\le t\}\in\fieldF_t,\qquad t\ge0.
\]
\end{definition}

If $T$ is a stopping time, then
\[
  \{T<t\}=
  \bigcup_{\substack{q<t\\ q\in\Rat_{+}}}\{T\le q\}\in\fieldF_t,
  \qquad
  \{T=t\}\in\fieldF_t.
\]
Conversely, if the filtration is right-continuous and $\{T<t\}\in\fieldF_t$
for all $t\ge0$, then $T$ is a stopping time.

The stopped $\sigma$-field $\fieldF_T$ is the collection of events
$A\in\fieldF$ such that
\[
  A\cap\{T\le t\}\in\fieldF_t,\qquad t\ge0.
\]
The strict past $\fieldF_{T-}$ is generated by $\fieldF_0$ and all sets of the
form $A\cap\{T>t\}$ with $A\in\fieldF_t$.

The distinction between $\fieldF_T$ and $\fieldF_{T-}$ is small in notation but
large in meaning. At a jump time $T$, $\fieldF_T$ may contain the jump itself;
$\fieldF_{T-}$ contains only the information available immediately before the
jump. Predictability will always refer to this pre-jump information. This is
the convention used in the general theory of processes
\citep{doob1953stochastic,dellacherie1978probabilities,jacod1987limit}.

\begin{lemma}[Elementary operations]
\label{lem:ch16-elementary-stopping}
Let $S,T$ and $T_n$, $n\ge1$, be stopping times.
\begin{enumerate}
\item $S\wedge T$ and $S\vee T$ are stopping times. If $\mathbb F$ is
right-continuous, then $S+T$ is also a stopping time.
\item If $\mathbb{F}$ is right-continuous, then
$\inf_nT_n$, $\sup_nT_n$, $\liminf_nT_n$, and $\limsup_nT_n$ are stopping
times.
\end{enumerate}
\end{lemma}

\noindent\textit{Proof.}
The minimum and maximum follow from
\[
  \{S\wedge T\le t\}=\{S\le t\}\cup\{T\le t\},
  \qquad
  \{S\vee T\le t\}=\{S\le t\}\cap\{T\le t\}.
\]
For the sum, first show that
\[
  \{S+T<t\}
  =
  \bigcup_{\substack{q\in\Rat_+\\q<t}}
  \{S<q\}\cap\{T<t-q\}\in\fieldF_t .
\]
Right-continuity then turns the strict inequality into
\[
  \{S+T\le t\}=\bigcap_{m\ge1}\{S+T<t+1/m\}\in\fieldF_t .
\]
The proof for $\inf_nT_n$ is the same: strict inequalities are immediate, and
right-continuity gives the closed inequality. The supremum is easier because
\[
  \{\sup_nT_n\le t\}=\bigcap_{n\ge1}\{T_n\le t\}.
\]
Finally,
\[
  \liminf_nT_n=\sup_m\inf_{n\ge m}T_n,
  \qquad
  \limsup_nT_n=\inf_m\sup_{n\ge m}T_n.
\]
\qedmark

\begin{lemma}[Stopped information]
\label{lem:ch16-stopped-information}
Let $S$ and $T$ be stopping times.
\begin{enumerate}
\item $T$ is $\fieldF_{T-}$-measurable.
\item If $A\in\fieldF_S$, then
\[
  A\cap\{S<T\}\in\fieldF_{T-},\qquad
  A\cap\{S\le T\}\in\fieldF_T,\qquad
  A\cap\{S=T\}\in\fieldF_T.
\]
\item The events $\{T<S\}$, $\{T\le S\}$, and $\{T=S\}$ belong to
$\fieldF_T\cap\fieldF_S$.
\end{enumerate}
\end{lemma}

\noindent\textit{Proof.}
The event $\{T>t\}$ is one of the generators of $\fieldF_{T-}$, so
$\{T\le t\}\in\fieldF_{T-}$ for every $t$. Hence $T$ is
$\fieldF_{T-}$-measurable. The key strict-time identity is
\[
  A\cap\{S<T\}
  =
  \bigcup_{q\in\Rat_+}
  \bigl(A\cap\{S\le q\}\bigr)\cap\{T>q\},
\]
when $A\in\fieldF_S$. Since $A\cap\{S\le q\}\in\fieldF_q$, each term in the
union is a generator of $\fieldF_{T-}$.

For \(A\cap\{S\le T\}\), use
\[
  A\cap\{S\le T\}\cap\{T\le t\}
  =
  \bigcap_{m\ge1}
  \bigcup_{q\in\Rat_+\cap[0,t+1/m]}
  \bigl(A\cap\{S\le q\}\bigr)
  \cap
  \{q-1/m<T\le t\}.
\]
Here \(A\cap\{S\le q\}\in\fieldF_q\subseteq\fieldF_{t+1/m}\), and the event
\(\{q-1/m<T\le t\}\) is in \(\fieldF_t\subseteq\fieldF_{t+1/m}\).  The
right-continuity of the filtration then places the intersection over \(m\) in
\(\fieldF_t\).  Thus \(A\cap\{S\le T\}\in\fieldF_T\).  Since
\[
  A\cap\{S=T\}
  =
  A\cap\{S\le T\}\cap\{T\le S\},
\]
the equality event belongs to \(\fieldF_T\) once the same argument is applied
with \(A=\Omega\) and the roles of \(S\) and \(T\) interchanged.  Taking
\(A=\Omega\) in the strict and non-strict statements gives
\(\{T<S\}\), \(\{T\le S\}\), and \(\{T=S\}\) in both stopped
\(\sigma\)-fields.
\qedmark

\begin{lemma}[Comparison]
\label{lem:ch16-stopping-comparison}
Let $S,T$ and $T_n$, $n\ge1$, be stopping times.
\begin{enumerate}
\item If $\mathbb{F}$ is complete and $P(T\le S)=1$, then
$\fieldF_T\subseteq\fieldF_S$.
\item If $\mathbb{F}$ is complete and right-continuous and
$T_n\downarrow T$ almost surely, then
$\fieldF_T=\bigcap_n\fieldF_{T_n}$.
\end{enumerate}
\end{lemma}

\noindent\textit{Proof.}
If $T\le S$ surely and $A\in\fieldF_T$, then
\[
  A\cap\{S\le t\}
  =
  A\cap\{T\le t\}\cap\{S\le t\}\in\fieldF_t .
\]
Completeness makes it possible to ignore the null set on which the order may fail. This
proves $\fieldF_T\subseteq\fieldF_S$. If $T_n\downarrow T$, the same inclusion
gives $\fieldF_T\subseteq\fieldF_{T_n}$ for every $n$. Conversely, if
$A\in\bigcap_n\fieldF_{T_n}$, then
\[
  A\cap\{T\le t\}
  =
  \bigcap_{m\ge1}\bigcup_{n\ge1}
  \bigl(A\cap\{T_n\le t+1/m\}\bigr).
\]
For fixed $m$, every set in the union belongs to $\fieldF_{t+1/m}$. Hence the
intersection belongs to $\bigcap_m\fieldF_{t+1/m}=\fieldF_t$ by
right-continuity. Thus $A\in\fieldF_T$.
\qedmark

\begin{example}[Hitting a level]
Let $X_t$ be a real-valued adapted process with right-continuous increasing
paths. For $a\ge0$ set
\[
  T_a=\inf\{t:X_t\ge a\}.
\]
Then $T_a$ is a stopping time, since
\[
  \{T_a\le t\}=\{X_t\ge a\}\in\fieldF_t.
\]
The equality uses monotonicity. If the path has crossed level $a$ by time
$t$, then the value at $t$ is already at least $a$; conversely, if $X_t\ge a$,
then the first crossing time is no later than $t$. This example is deliberately
simple: it shows how path regularity and adaptedness cooperate. Adaptedness
alone shows $X_t$ is observable at time $t$; monotonicity turns the whole
past crossing event into the single event $\{X_t\ge a\}$.
\end{example}

\begin{example}[Hitting a closed set]
Suppose that $X=\{X(t):t\ge0\}$ is adapted and has continuous paths. If
$A\subseteq S$ is closed, then
\[
  T_A=\inf\{t:X(t)\in A\}
\]
is a stopping time. If the filtration is right-continuous, the same conclusion
also holds for open sets.

For closed $A$, put $B=A^c$. The event $\{T_A>t\}$ means that
$X_s\in B$ for every $0\le s\le t$. Since $B$ is open and the path is
continuous, it is enough to check rational times:
\[
  \{T_A>t\}
  =
  \bigcap_{\substack{q\in\Rat_{+}\\q\le t}}
  \{X_q\in B\}\in\fieldF_t.
\]
Taking complements gives $\{T_A\le t\}\in\fieldF_t$. For open $A$, write
$A=\bigcup_m A_m$ with $A_m$ closed and increasing. Then
\[
  \{T_A<t\}=\bigcup_m\{T_{A_m}<t\}\in\fieldF_t,
\]
and right-continuity of the filtration converts the strict inequality
criterion into the stopping-time criterion.
\end{example}

\begin{example}[A cadlag warning]
If $X$ is merely \cadlag\ and $A$ is closed, the first time
$X_t\in A$ need not behave as cleanly as in the continuous-path case, because
the path can jump across the boundary. What remains robust is the enlarged
hitting time
\[
  T_A^{*}
  =
  \inf\{t:X_t\in A\text{ or }X_{t-}\in A\}.
\]
For a right-continuous complete filtration, $T_A^{*}$ is a stopping time. This
minor-looking modification is one reason that left limits enter the theory so
early. In jump-process problems, the relevant question is often not only
whether the post-jump state lies in a set, but whether the pre-jump state did.
\end{example}

\section{Marked Point Processes and Information}
\label{sec:ch16-marked-point-processes}
\conceptindexes{marked point processes, information, counting process, marks, progressive measurability, predictable process}

\subsection{Marked Point Processes in Real Time}
\label{sec:ch16-marked-point-processes-real-time}
\conceptindexes{marked point processes, event times, marks, real-time data}

We first consider countable mark spaces.

Chapter 6 introduced point processes as random measures. Here that same object
is placed in real time, so the event time must be a stopping time and the mark
is information revealed at the jump. The formulation below is the standard
bridge between multivariate counting processes and marked point processes
\citep{jacod1975multivariate,karr1986point,jacod1987limit}.

The London bombing capsule gives the finite-grid shadow of this process
language.  In the grouped data there are 576 cells and 537 hits, so the fitted
homogeneous Poisson rate is \(\hat\lambda=0.932\) hits per cell; the grouped
Pearson diagnostic is \(X^2=1.169\).  Chapter~2 used the example to discipline
visual clustering.  Here the same result is read as a point-process lesson:
counts, marks, intensities, and compensators are the objects that turn a map
from a picture into a stochastic process.

\input{figures/marked_point_process_timeline}

\begin{definition}[Multivariate counting process]
Let $S$ be finite or countable and let $(\Omega,\fieldF,\mathbb{F},P)$ be a
standard stochastic basis. A vector
\[
  N(t)=\{N_j(t):j\in S\}
\]
is a multivariate counting process if:
\begin{enumerate}
\item $N$ is adapted to $\mathbb{F}$;
\item with probability one, each $N_j$ has $\cadlag$ step paths, $N_j(0)=0$,
all jumps have size $+1$, and no two components jump at the same time.
\end{enumerate}
\end{definition}

The aggregate process
\[
  N_{\cdot}(t)=\sum_{j\in S}N_j(t)
\]
counts the total number of events up to time $t$. Define
\[
  T_n=\inf\{t:N_{\cdot}(t)=n\}
      =\inf\{t:N_{\cdot}(t)\ge n\},
\]
and
\[
  X_n=
  \begin{cases}
  j, & T_n<\infty\ \text{and}\ \Delta N_j(T_n)=1,\\
  \DeltaState, & T_n=\infty.
  \end{cases}
\]
Then $(T_n,X_n)_{n\ge1}$ is the marked point process associated with $N$, and
almost surely
\[
  N_j(t)=\sum_{n\ge1}\ind{T_n\le t,\ X_n=j},\qquad j\in S.
\]

The definition says two things at once. The random time $T_n$ is the time of
the $n$th event, and the mark $X_n$ records which component jumped. The
assumption that no two components jump at the same time is not merely a
convenience of notation. It is what makes the mark $X_n$ unambiguous. If two
components could jump simultaneously, then a mark would have to be a subset of
$S$, or a vector of simultaneous event types, rather than a single element.

For each $n$, $T_n$ is a stopping time because
\[
  \{T_n\le t\}=\{N_{\cdot}(t)\ge n\}\in\fieldF_t.
\]
On $\{T_n<\infty\}$ the mark is $\fieldF_{T_n}$-measurable. Indeed, for a
fixed $j$,
\[
  \{X_n=j\}\cap\{T_n\le t\}
  =
  \{N_j(T_n)-N_j(T_n-)=1\}\cap\{T_n\le t\},
\]
and the right side is determined by the stopped path up to time $t$. This is
the first appearance of the slogan that event times are stopping times and
marks are revealed at those stopping times.

\begin{example}[General mark space]
Let $(S,\St)$ be a complete separable metric space. A marked point process is a
sequence $(T_n,X_n)_{n\ge0}$ such that:
\begin{enumerate}
\item $T_0=0$ almost surely and $T_1>0$ almost surely;
\item $T_n<T_{n+1}$ on $\{T_n<\infty\}$ and $T_n=T_{n+1}$ on $\{T_n=\infty\}$;
\item $X_n$ is $\fieldF_{T_n}$-measurable on $\{T_n<\infty\}$;
\item $X_n=\DeltaState$ on $\{T_n=\infty\}$.
\end{enumerate}
For $B\in\St$ define
\[
  N_t(B)=\sum_{n\ge1}\ind{T_n\le t,\ X_n\in B}.
\]
For disjoint $B_j$, the collection $\{N_t(B_j):j\ge1,t\ge0\}$ forms a
multivariate counting process.
\end{example}

The explosion time is
\[
  T_{\infty}=\lim_{n\to\infty}T_n.
\]
Usually one assumes $P(T_{\infty}=\infty)=1$. If this probability is less than
one, the marked point process is called explosive.

The non-explosion assumption guarantees that the random intervals
\[
  [T_0,T_1),\ [T_1,T_2),\ldots
\]
cover every finite time horizon. Without it, infinitely many events could
accumulate before a finite calendar time and the step process would require an
extra rule after the accumulation point. Continuous-time Markov chains on
countable state spaces encounter this issue directly; finite-state chains do
not.

For a general mark space it is often cleaner to treat $N$ as a random measure
on $\Rplus\times S$:
\[
  N((0,t]\times B)=N_t(B)
  =
  \sum_{n\ge1}\ind{T_n\le t,\ X_n\in B}.
\]
Then integration against $N$ is just summation over observed events:
\[
  \int_{(0,t]\times S}C(u,x)\,N(du,dx)
  =
  \sum_{n:T_n\le t}C(T_n,X_n).
\]
This notation is the natural one for likelihoods and compensator identities,
because the same integrand $C$ can depend on time, on the mark, and on the
history before the event.

\subsection{Progressive Measurability}
\label{sec:ch16-progressive-measurability}
\conceptindexes{progressive measurability, adapted process, stochastic integration}

\begin{definition}[Progressive measurability]
An adapted process $X=\{X_t:t\ge0\}$ is progressively measurable if, for every
$t\ge0$ and every Borel set $B\subseteq\Real^d$,
\[
  \{(\omega,s):X_s(\omega)\in B,\ 0\le s\le t\}
  \in \fieldF_t\otimes\Borel([0,t]).
\]
\end{definition}

Progressive measurability is stronger than adaptedness but weaker than asking
for every path to be continuous. It says that, if we stop the movie at time
$t$, the whole visible segment $\{X_s:0\le s\le t\}$ is measurable using only
the information in $\fieldF_t$. This is precisely the amount of measurability
needed when a deterministic time is replaced by a stopping time.

\begin{example}[One-sided continuous paths]
Every adapted $\cad$ or $\cag$ process is progressively measurable. We give the
argument because it is the prototype for many later stopping-time
constructions.

Fix $t>0$. Suppose first that $X$ has right-continuous paths. On $[0,t]$, set
\[
  X_s^{(n)}
  =
  X_0\,\ind{s=0}
  +
  \sum_{k=1}^{n}X_{kt/n}\,
  \ind{(k-1)t/n<s\le kt/n}.
\]
Each coefficient $X_{kt/n}$ is $\fieldF_{kt/n}$-measurable and hence
$\fieldF_t$-measurable. Therefore $(\omega,s)\mapsto X_s^{(n)}(\omega)$ is
$\fieldF_t\otimes\Borel([0,t])$-measurable. For fixed $s$, the grid point on
the right decreases to $s$, so right-continuity gives
\[
  X_s^{(n)}(\omega)\longrightarrow X_s(\omega).
\]
The pointwise limit is jointly measurable on $\Omega\times[0,t]$.

If $X$ has left-continuous paths, use instead the left-endpoint approximation
\[
  \widetilde X_s^{(n)}
  =
  X_0\,\ind{s=0}
  +
  \sum_{k=0}^{n-1}X_{kt/n}\,
  \ind{kt/n<s\le (k+1)t/n}.
\]
The grid point on the left increases to $s$, and the same measurability
argument applies. In particular, every adapted $\cadlag$ process is
progressively measurable.
\end{example}

\begin{lemma}[Stopping a progressive process]
Let $X$ be progressively measurable and let $T$ be a stopping time. If
$P(T<\infty)=1$, then $X_T$ is $\fieldF_T$-measurable and $X_{T\wedge t}$ is
$\fieldF_t$-measurable for each $t$.
\end{lemma}

\noindent\textit{Proof.}
Fix $t$ and put $S=T\wedge t$. Since $T$ is a stopping time, $S$ is
$\fieldF_t$-measurable. The mapping
\[
  \omega\longmapsto(\omega,S(\omega))
\]
from $(\Omega,\fieldF_t)$ into
$(\Omega\times[0,t],\fieldF_t\otimes\Borel([0,t]))$ is measurable. By
progressive measurability, the map
\[
  (\omega,s)\longmapsto X_s(\omega)
\]
is measurable on this same product space. Hence the composition
\[
  \omega\longmapsto X_{S(\omega)}(\omega)=X_{T\wedge t}(\omega)
\]
is $\fieldF_t$-measurable.

To prove $X_T\in\fieldF_T$, let $B$ be a Borel set. On $\{T\le t\}$ one has
$T=T\wedge t$, and therefore
\[
  \{X_T\in B\}\cap\{T\le t\}
  =
  \{X_{T\wedge t}\in B\}\cap\{T\le t\}\in\fieldF_t.
\]
This is exactly the defining condition for $\{X_T\in B\}\in\fieldF_T$. The
assumption $P(T<\infty)=1$ makes the value assigned to $X_\infty$ irrelevant.
\qedmark

\begin{definition}[Optional sigma-field]
The optional \(\sigma\)-field \(\mathcal O\) on \(\Omega\times\Rplus\) is the
\(\sigma\)-field generated by adapted right-continuous real-valued processes.
A process \(X=\{X_t:t\ge0\}\) is optional if
\((\omega,t)\mapsto X_t(\omega)\) is \(\mathcal O\)-measurable.
\end{definition}

Optional processes are the processes that can be evaluated at stopping times
under the usual hypotheses. Every adapted \(\cadlag\) process is optional, and
optional measurability sits between predictability and progressive
measurability in the standard hierarchy.

\subsection{Predictable Processes}
\label{sec:ch16-predictable-processes}
\conceptindexes{predictable processes, predictable sigma-field, left-continuity, pre-event information}

Intuitively, a predictable process has its value at time $t$ determined by
information accumulated strictly before $t$. In discrete time this is almost
invisible: ``predictable at time $n$'' just means measurable at time $n-1$.
In continuous time the phrase has a left-limit interpretation: a predictable
process may use the information available just before a possible jump, but it
cannot inspect the jump itself before it happens.
The joint-measurability and integration facts from Chapter 6 are the static
base; predictability is their real-time, no-looking-ahead refinement.

\begin{definition}[Predictable sigma-field]
The predictable $\sigma$-field $\Pt$ on $\Omega\times\Rplus$ is generated by
left-continuous adapted processes. A process $X=\{X_t:t\ge0\}$ is predictable
if it is measurable with respect to $\Pt$.
\end{definition}

Equivalently, $\Pt$ is generated by predictable rectangles
\[
  A\times\{0\},\quad A\in\fieldF_0,
  \qquad\text{and}\qquad
  A\times(s,t],\quad A\in\fieldF_s,\ s<t.
\]
Simple predictable processes have the form
\[
  X_t(\omega)
  =
  a_0\,\ind{t=0,\omega\in A}
  +
  \sum_{i=1}^{m}a_i\,\ind{s_i<t\le t_i,\omega\in A_i},
\]
where $A\in\fieldF_0$ and $A_i\in\fieldF_{s_i}$.

Here is the reason for the rectangle description. Let $\Pt_R$ be the
$\sigma$-field generated by the rectangles above. Every simple process of the
displayed form is $\Pt_R$-measurable. Conversely, every nonnegative
left-continuous adapted process can be approximated pointwise by simple
processes that only look at left endpoints. For instance, on $[0,\infty)$ set
\[
  X^{(n)}_t
  =
  X_0\,\ind{t=0}
  +
  \sum_{m\ge1}
  X_{(m-1)/n}\,
  \ind{(m-1)/n<t\le m/n}.
\]
The coefficient $X_{(m-1)/n}$ is $\fieldF_{(m-1)/n}$-measurable, so each
summand is predictable-rectangle measurable. Left-continuity gives
$X^{(n)}_t\to X_t$ at every $t$. Splitting a real-valued process into positive
and negative parts then gives the equivalence between the two definitions of
$\Pt$.

The phrase ``known just before time $t$'' is encoded in the interval
$(s,t]$ rather than $[s,t)$: the value on that interval is chosen using
$\fieldF_s$, information available strictly before all times in $(s,t]$.
This convention is what makes stochastic integrals against jump processes use
left-limit values of the integrand.

\begin{lemma}[Predictable stopping]
\label{lem:ch16-predictable-stopping}
If $X$ is predictable and $T$ is a stopping time, then
$X_T\ind{T<\infty}$ is $\fieldF_{T-}$-measurable and the stopped process
$X_{T\wedge t}$ is predictable.
\end{lemma}

\noindent\textit{Proof.}
It is enough to check the claim first for indicators of predictable rectangles.
For
\[
  X_u=\ind{A}\ind{s<u\le r},
  \qquad A\in\fieldF_s,
\]
one has
\[
  X_T\ind{T<\infty}
  =
  \ind{A}\ind{s<T\le r}.
\]
The condition $A\cap\{T>s\}\in\fieldF_{T-}$ comes directly from the definition
of $\fieldF_{T-}$, and $\{T\le r\}$ is
$\fieldF_{T-}$-measurable because $T$ is. Thus the stopped value is
$\fieldF_{T-}$-measurable for rectangle indicators. The same rectangle
calculation shows that $u\mapsto X_{T\wedge u}$ is predictable. A monotone
class argument extends the two conclusions to all bounded predictable
processes; positive and negative decompositions give the general case.
\qedmark

\begin{lemma}[Predictable integrals]
\label{lem:ch16-predictable-integrals}
Let $A=\{A_t:t\ge0\}$ be a $\cadlag$ adapted process with $A_0=0$ and paths
that are increasing or of bounded variation. If both $X$ and $A$ are
predictable, then
\[
  Y_t(\omega)=\int_{(0,t]}X(\omega,s)\,A(ds,\omega)
\]
is predictable whenever the pathwise integral is well defined.
\end{lemma}

\noindent\textit{Proof.}
The last lemma is the finite-variation version of stochastic integration. It
is enough for the present chapter because counting processes and their
compensators have finite variation on stopped intervals. For a simple
predictable integrand
\[
  X_u=\sum_{i=1}^m \xi_i\ind{s_i<u\le r_i},
  \qquad \xi_i\in\fieldF_{s_i},
\]
the integral has the form
\[
  Y_t
  =
  \sum_{i=1}^m
  \xi_i\{A_{t\wedge r_i}-A_{t\wedge s_i}\}.
\]
The predictable-stopping lemma says that each stopped process
$A_{t\wedge r_i}$ and $A_{t\wedge s_i}$ is predictable, so $Y$ is predictable.
For nonnegative predictable $X$, approximate by increasing simple predictable
processes and use monotone convergence. For signed integrands, apply the same
argument to the positive and negative parts, assuming the total variation
integral is finite on the interval under consideration.
\qedmark

\subsection{Internal History of a Marked Point Process}
\conceptindexes{internal history, counting-process history, marked-event history}

Let $(T_n,X_n)_{n\ge0}$ be a marked point process on a standard stochastic
basis. For $B\in\St$ define
\[
  N_t(B)=\sum_{n\ge1}\ind{T_n\le t,\ X_n\in B}.
\]
Let $\Ht_0\subseteq\fieldF_0$ contain the initial information, including
$\sigma(X_0)$. The internal history of the process is
\[
  \Ht_t=\sigma\{N_s(B):s\le t,\ B\in\St\}\vee \Ht_0.
\]
It is the smallest filtration to which the marked point process is adapted.

The internal history has the following standard properties. Each $T_n$ is a
stopping time with respect to $\Ht=\{\Ht_t:t\ge0\}$, and
\[
  \Ht_{T_n}
  =
  \sigma\bigl((T_1,X_1),\ldots,(T_n,X_n)\bigr)\vee\Ht_0,
\]
while
\[
  \Ht_{T_n-}
  =
  \sigma\bigl((T_1,X_1),\ldots,(T_{n-1},X_{n-1}),T_n\bigr)\vee\Ht_0.
\]
An adapted process $C_t$ is predictable relative to the internal history iff it
can be written in the form
\[
  C_t=C_0+\sum_{n\ge0}C_t^{(n)}\,\ind{T_n<t\le T_{n+1}},
\]
where $C_0$ is $\Ht_0$-measurable and $C^{(n)}$ is
$\Ht_{T_n}\otimes\Borel(\Rplus)$-measurable.

There are two companion facts that are often useful in proofs. First,
$A\in\Ht_t$ if and only if, for each $n\ge0$, there exists
$A_n\in\Ht_{T_n}$ such that
\[
  A\cap\{t<T_{n+1}\}
  =
  A_n\cap\{t<T_{n+1}\}.
\]
On the event that the process has not yet reached its $(n+1)$st event by time
$t$, the history up to $t$ is exactly the first $n$ event times and marks,
together with the information that no further event has occurred before $t$.

Second, if $T$ is a finite stopping time for the internal history, then on
the slice $\{T_n\le T<T_{n+1}\}$ the stopped information is the same as the
information available at $T_n$ plus the value of the waiting time after $T_n$.
Equivalently, for each $n$ there is a nonnegative random variable $U_n$ such
that
\[
  T\wedge T_{n+1}=(T_n+U_n)\wedge T_{n+1}
  \quad\text{on }\{T_n\le T\}.
\]
This is the real-time analogue of saying that, between two events, the only
new information is the passage of time without an event.

The representation of predictable processes follows the same logic. Between
$T_n$ and $T_{n+1}$ a predictable quantity may depend on the previous events
$(T_1,X_1),\ldots,(T_n,X_n)$ and on the current calendar time, but it cannot
depend on $X_{n+1}$, because that mark is not observed until the jump occurs.
This is exactly the form needed by intensities of marked point processes; this
representation is closely tied to the point-process compensator theory of
\citep{jacod1975multivariate,karr1986point}.

\subsection{Localization}
\conceptindexes{localization, local martingales, stopped processes, local property}

\begin{definition}[Local property]
A process $X=\{X_t:t\ge0\}$ has a property locally if there exists an
increasing sequence of stopping times $T_n\uparrow\infty$ almost surely such
that the stopped process $X(t\wedge T_n)$ has the property on $\{T_n>0\}$ for
every $n$. The sequence $\{T_n\}$ is called a localizing sequence.
\end{definition}

Localization is used when the right theorem is true under boundedness or
integrability assumptions, while the process seen in applications has those
properties only before it becomes too large, too old, or too close to an
explosion. We prove or define the object on the stopped intervals
$[0,T_n]$, where ordinary martingale arguments apply, and then let the
localizing times tend to infinity. Thus a local statement is not weaker in
practice; it is the way continuous-time theory records that the process can be
controlled on every finite piece selected without looking ahead.

The deterministic prototype is harmless but important. A continuous increasing
function $A(t)$ may grow from $0$ to infinity as $t\to\infty$, and the
difference $A_1(t)-A_2(t)$ may make sense on every finite interval even when
$A_1(\infty)-A_2(\infty)$ has the meaningless form $\infty-\infty$.
Localization is the stochastic version of working on finite intervals first.
The intervals are allowed to be random, but the endpoints must be stopping
times, so that we do not choose a convenient interval by looking into the
future.

For example, $X$ is locally bounded if there are constants $c_n<\infty$ and
stopping times $T_n\uparrow\infty$ such that
\[
  |X(t\wedge T_n)|\le c_n
  \quad\text{on}\quad \{T_n>0\}.
\]
Every adapted $\cadlag$ process with bounded jump sizes and $X_0=0$ is locally
bounded, with a possible localizing sequence
\[
  T_n=n\wedge\inf\{t\ge0:|X_t|\ge n\}.
\]
If the jumps are bounded by $K$, then the stopped paths are bounded by
$n+K$.

\begin{example}[Local boundedness of a multivariate counting process]
Let $N(t)=\{N_j(t):j\in S\}$ be a multivariate counting process and let
\[
  N_{\cdot}(t)=\sum_{j\in S}N_j(t),\qquad
  T_m=\inf\{t:N_{\cdot}(t)=m\}.
\]
If the process is non-explosive, then $T_m\uparrow\infty$ almost surely.
Since each event increases exactly one component by one,
\[
  N_{\cdot}(t\wedge T_m)\le m,\qquad t\ge0,
\]
on $\{T_m>0\}$. Hence each component $N_j$ is locally bounded, and so is the
aggregate counting process. This is the basic reason that compensator and
martingale statements for counting processes are often first formulated
locally: before the $m$th event occurs, the process has only accumulated a
bounded number of jumps.
\end{example}

Another useful characterization is domination by a left-continuous envelope.
Suppose $X_0=0$ and $X$ is locally bounded with localizing sequence $T_n$ and
bounds $c_n$. After replacing $T_n$ by an increasing sequence if necessary,
the process
\[
  Y_t=\sum_{n\ge1}c_n\,\ind{T_{n-1}<t\le T_n},
  \qquad T_0=0,
\]
is adapted and \caglad, and it dominates $|X_t|$ on each localized interval.
Conversely, if $|X_t|\le Y_t$ for an adapted \caglad\ process $Y$ that is
finite on compact time intervals, then
\[
  T_n=n\wedge\inf\{t:\sup_{s\le t}Y_{s+}\ge n\}
\]
localizes $X$. Thus local boundedness is not an exotic condition; it says that
the process admits an observable envelope on finite time horizons.

A \cadlag\ adapted process $M$ is a local martingale if $M_{t\wedge T_n}$ is a
martingale for some localizing sequence. Similarly, finite variation, square
integrability, and compensator identities are often stated locally. This is the
form needed for counting processes: a model may have unbounded calendar time or
unbounded state-dependent rates, but after stopping at a bounded time, before
explosion, or before the risk set degenerates, the familiar martingale
calculus is available.

The point is not to hide bad behavior. If the process explodes with positive
probability before a finite time, no localizing sequence tending to infinity
can remove that fact. Localization only separates behavior on each finite,
observable piece from behavior at the limiting boundary.

\section{Martingales, Compensators, and Variation}
\label{sec:ch16-martingales-compensators-variation}
\conceptindexes{martingales, compensators, variation, quadratic variation, predictable variation}

\subsection{Martingales, Submartingales, and Compensators}
\label{sec:ch16-martingales-compensators}
\conceptindexes{martingales, submartingales, compensators, Doob--Meyer decomposition}

Chapter 6 constructed random measures and their finite-dimensional laws. This
section asks what part of an observed random measure was predictable from the
history. The answer is the compensator; subtracting it leaves the martingale
fluctuation.

\begin{definition}[Continuous-time martingale]
A $\cadlag$ process $M=\{M_t:t\ge0\}$ on a stochastic basis is a martingale if:
\begin{enumerate}
\item $M$ is adapted;
\item $\Expect\!\left[|M_t|\right]<\infty$ for every $t\ge0$;
\item $\Expect\!\left[M_t\mid\fieldF_s\right]=M_s$ almost surely whenever $s<t$.
\end{enumerate}
It is a submartingale if the last equality is replaced by
\[
  \Expect\!\left[M_t\mid\fieldF_s\right]\ge M_s,
\]
and a supermartingale if the reverse inequality holds.
\end{definition}

Linear combinations of martingales on a common stochastic basis are again
martingales. Jensen's inequality gives a second useful rule: if $\varphi$ is
convex and $\Expect\!\left[|\varphi(M_t)|\right]<\infty$, then
$\varphi(M_t)$ is a submartingale.
In particular, a square-integrable martingale has $M_t^2$ as a submartingale.

\begin{example}[Increasing processes]
Let $N=\{N_t:t\ge0\}$ be an adapted increasing $\cadlag$ process with
$N_0=0$ and $\Expect\!\left[N_t\right]<\infty$ for all $t$. Then $N$ is a
submartingale because
\[
  \Expect\!\left[N_t\mid\fieldF_s\right]
  =
  \Expect\!\left[N_t-N_s\mid\fieldF_s\right]+N_s
  \ge N_s.
\]
\end{example}

If the integrability condition fails globally, the same example is usually
read locally. Stop the process at
\[
  T_m=m\wedge\inf\{t:N_t\ge m\}.
\]
Then $N_{t\wedge T_m}\le m+1$ for a unit-jump counting process, so the stopped
process is integrable and is a submartingale. This is why compensators of
counting processes are often introduced first as local objects and only later,
under additional assumptions, as integrable objects on the whole half-line.

The Doob--Meyer decomposition says, under suitable integrability conditions,
that a submartingale can be written as
\[
  X_t=M_t+\Lambda_t,
\]
where $M$ is a martingale and $\Lambda$ is a predictable increasing process
with $\Lambda_0=0$. The process $\Lambda$ is called the compensator.
Informally, the compensator is the predictable part of the infinitesimal
increment:
\[
  d\Lambda_t \approx \Expect\!\left[dX_t\mid\fieldF_{t-}\right].
\]
Along partitions
$0=t_0<t_1<\cdots<t_k=t$ whose mesh tends to zero, it is natural to think of
\[
  \Lambda_t
  \approx
  \sum_{i=1}^{k}
  \Expect\!\left(X_{t_i}-X_{t_{i-1}}\mid\fieldF_{t_{i-1}}\right),
\]
with convergence understood in the appropriate probabilistic sense. The
formula is heuristic, but it captures the essential idea: the compensator is
the part one could have forecast just before the increment occurs.

This viewpoint is less mysterious than the word ``martingale'' may suggest. In
applications one often starts with an observed count, subtracts the amount that
was predictable from the current history, and studies the remaining surprise.
For survival data the observed count is a death or failure count; the predictable
part is the accumulated hazard over the risk set; the martingale is the
fluctuation after this risk has been accounted for.

The statement used here is the finite-variation form of the compensator
decomposition. For the broader theorem and its role in the general theory, see
\citet{dellacherie1978probabilities} and \citet{jacod1987limit}.

\begin{example}[Extended Poisson process]
Let $N=\{N_t:t\ge0\}$ be a counting process adapted to $\mathbb{F}$ and suppose
$A_t=\Expect\!\left[N_t\right]<\infty$ for all $t$. If $N_t-N_s$ is
independent of $\fieldF_s$ for
$0<s<t$, then $N$ is called an extended Poisson process. In this case
\[
  M_t=N_t-A_t
\]
is a martingale. If $A$ is absolutely continuous,
\[
  A(t)=\int_{0}^{t}\alpha(u)\,du,
\]
then $\alpha$ is the intensity of the process.

Indeed, for $0\le s<t$,
\[
  \Expect[N_t-N_s\mid\fieldF_s]
  =
  \Expect[N_t-N_s]
  =
  A_t-A_s,
\]
so
\[
  \Expect[M_t-M_s\mid\fieldF_s]=0.
\]
The deterministic function $A$ is predictable. If $A$ is continuous and tends
to infinity, the process is the usual non-homogeneous Poisson process; if
$A(t)=t$, it reduces to the unit-rate homogeneous Poisson process.
\end{example}

\begin{theorem}[Doob--Meyer decomposition, finite-variation form; \citealp{doob1953stochastic}]
Let $X$ be an adapted \cadlag\ process of locally integrable variation on a
standard stochastic basis. There exists a predictable process $\Lambda$, called
the compensator of $X$, such that $\Lambda$ has locally integrable variation and
\[
  M=X-\Lambda
\]
is a local martingale. The decomposition is almost surely unique. Moreover:
\begin{enumerate}
\item if $H$ is predictable and
$Y(t)=\int_{(0,t]}H(s)\,dX(s)$ has locally integrable variation, then
$\int_{(0,t]}H(s)\,d\Lambda(s)$ is the compensator of $Y$;
\item for any bounded stopping time $T$,
$\Expect\!\left[X_T\right]=\Expect\!\left[\Lambda_T\right]$, whenever the
expectations are well defined.
\end{enumerate}
\end{theorem}

\subsection{Compensator of a Marked Point Process}
\conceptindexes{marked point process!compensator, compensator, intensity, martingale residual}

Let $(T_n,X_n)_{n\ge0}$ be a marked point process and define
\[
  N_t(B)=\sum_{n\ge1}\ind{T_n\le t,\ X_n\in B},\qquad B\in\St.
\]
Relative to the internal history $\Ht$, the process $N_t(B)$ is increasing and
adapted. Its compensator is a predictable process $\Lambda_t(B)$ such that
\[
  N_t(B)-\Lambda_t(B)
\]
is a martingale.

Let $G_n$ denote the conditional distribution of $(T_{n+1},X_{n+1})$ given
$\Ht_{T_n}$, and let $\barF_n$ be the corresponding conditional survival
function of $T_{n+1}$:
\[
  \barF_n(u)=P(T_{n+1}\ge u\mid\Ht_{T_n})
  =G_n([u,\infty]\times S).
\]
On $t\in(T_n,T_{n+1}]$, the compensator has the recurrent form
\[
  \Lambda_t(B)
  =
  \Lambda_{T_n}(B)
  +
  \int_{(T_n,t]\times B}
  \frac{G_n(du,dv)}{\barF_n(u-)}.
\]
The fraction is the conditional hazard measure of the next marked event. On
the event that the process has survived without an additional jump up to
$u-$, the conditional chance of the next event occurring at $u$ with mark in
$dv$ is $G_n(du,dv)/\barF_n(u-)$. The compensator simply accumulates these
conditional risks until the next observed jump.
This is the recurrent compensator calculation for marked point processes in
\citet{jacod1975multivariate}; \citet{karr1986point} gives a statistical
treatment of the same construction.

When $G_n$ has a conditional density of the form
\[
  G_n(du,dv)=g_n(u,v)\,du\,\nu(dv),
\]
the recurrent formula becomes
\[
  \Lambda_t(B)
  =
  \Lambda_{T_n}(B)
  +
  \int_{T_n}^{t\wedge T_{n+1}}
  \int_B
  \frac{g_n(u,v)}{\barF_n(u)}\,\nu(dv)\,du.
\]
Thus the intensity kernel on $(T_n,T_{n+1}]$ is
\[
  \lambda_n(u,dv)=\frac{G_n(du,dv)}{\barF_n(u-)}.
\]
Specifying a marked point process by the conditional laws $G_n$, or by the
predictable kernel $\lambda_n$, is therefore the same modeling act in two
different languages.
If $B_j$, $j\ge1$, are disjoint, then the compensators
$\Lambda_t(B_j)$ describe the multivariate counting process
$\{N_t(B_j):j\ge1,t\ge0\}$.

More generally, for nonnegative predictable $C$ on $\Rplus\times S$,
\[
  \Expect\!\left[\int_{\Rplus\times S}C(t,u)\,N(dt,du)\right]
  =
  \Expect\!\left[\int_{\Rplus\times S}C(t,u)\,\Lambda(dt,du)\right].
\]

This identity is the defining calculation behind compensators. It says that
for every nonnegative predictable way of sampling the future event stream, the
expected observed sum equals the expected compensated sum. In applications
$C(t,u)$ is often an at-risk indicator, a covariate process, a score function,
or a residual weight.

\subsection{Quadratic and Predictable Variation}
\conceptindexes{quadratic variation, predictable variation, predictable covariation, integration by parts}

Martingales become most useful when their accumulated variation is itself
measurable in a predictable way. The square-variation calculus below is the
continuous-time analogue of summing squared martingale increments in discrete
time. Standard bracket notation and predictable covariation are treated in
\citet{liptser1974statistics} and \citet{jacod1987limit}.

We use the bounded-variation integration-by-parts identity proved in
Chapter~6.  In interval notation, if \(f\) and \(g\) are \cadlag\ functions of
bounded variation on \(I=(a,b]\), then
\[
\begin{aligned}
f(b)g(b)-f(a)g(a)
&=\int_I f(u-)\,g(du)+\int_I g(u)\,f(du)\\
&=\int_I f(u)\,g(du)+\int_I g(u-)\,f(du)\\
&=\int_I f(u-)\,g(du)+\int_I g(u-)\,f(du)
  +\sum_{u\in I}\Delta f(u)\Delta g(u).
\end{aligned}
\]
The special case \(f=g\) and \(f(a)=0\) is the form needed below:
\[
  f(t)^2
  =
  2\int_{(a,t]}f(u-)\,f(du)
  +
  \sum_{a<u\le t}(\Delta f(u))^2.
\]

If $M$ is a pure-jump local martingale of finite variation, its quadratic
variation is
\[
  [M]_t=\sum_{u\le t}(\Delta M_u)^2.
\]
For a general \cadlag\ local martingale one adds the continuous martingale
part, so that
\[
  [M]_t=[M]^c_t+\sum_{u\le t}(\Delta M_u)^2.
\]
The counting-process martingales in this chapter are pure-jump, so the jump
sum is the operative formula.

The reason $[M]$ appears is visible from the preceding identity:
\[
  M_t^2
  =
  2\int_{(0,t]}M_{u-}\,M(du)+[M]_t.
\]
The stochastic integral term is a local martingale when $M$ is a local
martingale and the integrand is predictable. Therefore the finite-variation
part that compensates $M^2$ must come from the jump-sum process $[M]$.

The predictable variation $\langle M\rangle_t$ is the compensator of
$[M]_t$, so that
\[
  M_t^2-\langle M\rangle_t
\]
is a local martingale. For two local martingales,
\[
  [M,M']_t=\sum_{u\le t}\Delta M_u\,\Delta M'_u
\]
and the predictable covariation $\langle M,M'\rangle$ is defined as the
compensator of $[M,M']$.

\begin{theorem}[Predictable covariation]
If $M$ and $M'$ are locally square-integrable martingales, then there exists a
predictable locally integrable process $\langle M,M'\rangle$ such that
\[
  MM'-\langle M,M'\rangle
\]
is a local martingale. If $H$ and $H'$ are predictable and
\[
  Y(t)=\int_{(0,t]}H(s)\,dM(s),
  \qquad
  Y'(t)=\int_{(0,t]}H'(s)\,dM'(s),
\]
then
\[
  \langle Y,Y'\rangle_t
  =
  \int_{(0,t]}H(s)H'(s)\,\langle M,M'\rangle(ds).
\]
\end{theorem}

\begin{example}[Counting process variation]
Let $\{N_j:j\in S\}$ be a multivariate counting process with compensators
$\Lambda_j$ and martingales $M_j=N_j-\Lambda_j$. Since each $N_j$ jumps by
one, the predictable variation has the form
\[
  \langle M_j\rangle_t
  =
  \int_{(0,t]}(1-\Delta\Lambda_j(s))\,\Lambda_j(ds).
\]
For $j\ne k$,
\[
  \langle M_j,M_k\rangle_t
  =
  -\int_{(0,t]}\Delta\Lambda_k(s)\,\Lambda_j(ds).
\]
Thus distinct components of a multivariate counting process are typically
orthogonal only when their compensators do not share predictable jumps.
In the common continuous-intensity case, $\Delta\Lambda_j(s)=0$ for all $j$
and all $s$, so
\[
  \langle M_j\rangle_t=\Lambda_j(t),
  \qquad
  \langle M_j,M_k\rangle_t=0,\quad j\ne k.
\]
This is the formula most often used in large-sample theory for survival and
event-history models. Predictable jumps are the extra terms needed when there
are fixed appointment times, administrative visit times, or other atoms in the
event-time distribution.
\end{example}

\section{Event Histories and Survival Targets}
\conceptindexes{event-history models, survival analysis, cumulative hazard, multi-state models, semi-Markov transformation models, transition probabilities}

This section applies the preceding machinery to statistical event-history
targets. It starts with the one-jump survival clock, adds censoring and panel
observation, then moves to multi-state and semi-Markov models. The reader
should track the same ledger in each subsection: observed event history,
predictable exposure, target functional, estimator, uncertainty, and the
limitation introduced by the observation scheme.

\subsection{Cumulative Hazards}
\conceptindexes{cumulative hazard, hazard rate, survival function}

The cumulative hazard is the deterministic version of a compensator. It
measures accumulated conditional risk: not the probability of failure itself,
but the risk of failure at the next instant given survival up to the previous
instant.

Let $A$ be an increasing $\cadlag$ function on $[0,\varphi]$, with $A(0-)=0$
and $0\le\Delta A(u)\le1$. A cumulative hazard determines a survival function
through the product integral
\[
  S(t)=\Prodi_{(0,t]}(1-dA(u))
  =
  \prod_{u\le t}(1-\Delta A(u))\exp\{-A^c(t)\},
\]
where $A^c$ is the continuous part of $A$.
The exponential term records continuous accumulation of hazard; the product
records the survival losses at fixed atoms.

If $A$ is continuous, this reduces to the familiar expression
$S(t)=\exp\{-A(t)\}$. If $A$ has a jump at $u$, survival is multiplied by
$1-\Delta A(u)$. Thus a fixed-time failure atom is not represented by an
infinite density; it is represented by a discrete hazard jump. The condition
$0\le\Delta A(u)\le1$ simply says that an atom cannot remove more than all
remaining survival probability.

\begin{theorem}[Cumulative hazard representation; \citealp{aalen1978nonparametric}]
Let $A$ be increasing and $\cadlag$ on $[0,\varphi]$, with
$0\le\Delta A(u)<1$ for $u<\varphi$, and suppose either
\[
  A(\varphi-)<\infty,\quad \Delta A(\varphi)=1,
\]
or
\[
  A(\varphi-)=\infty,\quad \Delta A(\varphi)=0.
\]
Then the product integral above is the survival function of a nonnegative
random variable. Conversely, if $S$ is such a survival function, the
cumulative hazard is
\[
  A(t)=\int_{(0,t]}\frac{F(du)}{S(u-)}.
\]
\end{theorem}

This product-integral representation follows the cumulative-hazard
correspondence in \citet{jacod1975multivariate}. For a systematic treatment
of product integration, including the matrix-valued version used in multi-state
models, see \citet{gill1990survey}; numerical product-integral approximations
are discussed by \citet{moller1992numerical}.  A compact definition and the
properties used below are collected in
\Appref{sec:appC-product-integrals}.

The upper support point is
\[
  \varphi=\inf\{t:S(t)=0\}.
\]
If the distribution has a final atom at $\varphi$, then
$\Delta A(\varphi)=1$. If the distribution reaches zero survival continuously
at $\varphi$, then $A(\varphi-)=\infty$. These two cases are the two standard
ways a proper survival function exhausts its remaining mass at the right
endpoint.

\subsection{One-Jump Counting Process}
\conceptindexes{one-jump counting process, event time, at-risk process}

Let $T$ be a nonnegative random variable with distribution function $F$ and
survival function $S(t)=P(T>t)$. The cumulative hazard is
\[
  A(t)=\int_{(0,t]}\frac{F(du)}{S(u-)}.
\]
Define the elementary counting process
\[
  N(t)=\ind{T\le t},
\]
and let $\fieldF_t=\sigma(N(s):s\le t)$. Then
\[
  \Lambda(t)=A(t\wedge T)
  =
  \int_{(0,t]}\ind{T\ge u}\,A(du)
\]
is predictable and
\[
  M_t=N_t-\Lambda_t
\]
is a martingale.
Before $T$ occurs the process is still at risk, and after $T$ occurs there can
be no second jump. Thus the predictable accumulation is not $A(du)$ alone but
the at-risk version $\ind{T\ge u}A(du)$.

Let us verify the martingale identity, since this calculation is the template
for censored data. First,
\[
  \Expect[\Lambda(t)]
  =
  \int_{(0,t]}\Prob(T\ge u)\,\frac{F(du)}{S(u-)}
  =
  \int_{(0,t]}F(du)
  =
  \Expect[N(t)].
\]
Now fix $0\le s<t$. On the event $\{T\le s\}$ both $N$ and $\Lambda$ have no
further increment after $s$. On the event $\{T>s\}$,
\[
  \Expect[N_t-N_s\mid T>s]
  =
  \frac{F(t)-F(s)}{S(s)}
\]
and
\[
\begin{aligned}
  \Expect[\Lambda_t-\Lambda_s\mid T>s]
  &=
  \frac{1}{S(s)}
  \int_{(s,t]}\Prob(T\ge u)\frac{F(du)}{S(u-)}  \\
  &=
  \frac{F(t)-F(s)}{S(s)}.
\end{aligned}
\]
Since the sigma-field generated by $\{N(r):r\le s\}$ distinguishes exactly
whether $T\le s$ or $T>s$, this proves
\[
  \Expect[M_t-M_s\mid\fieldF_s]=0.
\]

\subsection{Survival Analysis as a Counting Process}
\conceptindexes{survival analysis, counting process, Kaplan--Meier estimator, Nelson--Aalen estimator, Cox model}

\begingroup
\setlength{\abovedisplayskip}{0.58\baselineskip}
\setlength{\belowdisplayskip}{0.58\baselineskip}
\setlength{\abovedisplayshortskip}{0.38\baselineskip}
\setlength{\belowdisplayshortskip}{0.48\baselineskip}

Survival analysis is the one-jump example with statistics attached, but in
practice the jump is rarely an abstract failure. In an oncology trial it may be
progression or death; in a subscription business it may be churn; in reliability
engineering it may be the first field failure of a device. The raw database is
not a hazard curve. It is an event log: who entered service, who was still
under observation yesterday, whose event appeared today, and whose record ended
because observation stopped.

For subject, customer, or device $i$, let $T_i$ be the event time and $C_i$ the
censoring time. The observed record and the two processes used by the analysis
are
\[
\begin{gathered}
  Z_i=T_i\wedge C_i,\qquad
  \delta_i=\ind{T_i\le C_i},\\
  N_i(t)=\ind{Z_i\le t,\ \delta_i=1},\qquad
  Y_i(t)=\ind{Z_i\ge t}.
\end{gathered}
\]
Here $N_i$ is the dashboard counter that jumps when the event is observed, and
$Y_i$ is the live risk-set indicator. A patient whose last scan was before
progression, a user lost to follow-up after uninstalling the app, or a machine
removed from service no longer contributes exposure after censoring.

If censoring does not reveal extra information about the future event time
after the observed history is taken into account, and the event time has
cumulative hazard $A$, then
\[
  \Lambda_i(t)=\int_{(0,t]}Y_i(u)\,A(du),
  \qquad
  M_i(t)=N_i(t)-\Lambda_i(t)
\]
with $M_i$ a martingale. This is the central industrial bookkeeping identity:
\[
  \text{observed events}
  =
  \text{predictable exposure to risk}
  +
  \text{unexplained shock}.
\]
It says exactly what an operations team wants to know. If more events occur
than the exposure clock predicted, something has changed: the treated arm is
doing worse than expected, a product cohort is churning unusually fast, or a
hardware batch is failing in the field.

For $n$ independent records, aggregate the event log by
\[
  N_{\cdot}(t)=\sum_{i=1}^{n}N_i(t),\qquad
  Y_{\cdot}(t)=\sum_{i=1}^{n}Y_i(t),\qquad
  J(t)=\ind{Y_{\cdot}(t)>0}.
\]
The Nelson--Aalen estimator divides each observed event by the number still at
risk just before it happens:
\[
\begin{aligned}
  \widehat A(t)
  &=
  \int_{(0,t]}\frac{J(u)}{Y_{\cdot}(u)}\,N_{\cdot}(du),\\
  \widehat A(t)-\int_{(0,t]}J(u)\,A(du)
  &=
  \int_{(0,t]}\frac{J(u)}{Y_{\cdot}(u)}\,M_{\cdot}(du),
\end{aligned}
\]
where $M_{\cdot}=N_{\cdot}-\int Y_{\cdot}\,dA$. Thus the estimator is the
true cumulative hazard, stopped when the risk set is empty, plus a martingale
integral. The formula is not merely elegant; it is the reason a clinical trial
or reliability platform can update a cumulative risk curve whenever a new
event arrives without refitting a parametric lifetime model.

On the survival scale the same update is multiplicative:
\[
  \widehat S(t)
  =
  \Prodi_{(0,t]}\{1-\widehat A(du)\}.
\]
Hazards accumulate additively, survival accumulates multiplicatively, and the
martingale term records the fluctuation around the predictable clock. This is
why the same counting-process engine can support Kaplan--Meier-style reporting
in a trial, churn monitoring in a product cohort, warranty forecasting for a
fleet of devices, and default-risk dashboards in finance.

A regulatory-grade version of this workflow appears in oncology drug
development at a multinational sponsor. The protocol and statistical analysis
plan define a time-to-event endpoint such as progression-free survival (PFS):
the clock often starts at randomization, the event is the first documented
progression or death, and censoring rules specify what happens at data cutoff,
loss to follow-up, missed tumor assessments, or use of new anti-cancer therapy.
Those rules are then implemented in a traceable analysis dataset, commonly an
ADaM time-to-event dataset, before the clinical study report displays
Kaplan--Meier curves, log-rank tests, Cox model summaries, and sensitivity
analyses \citep{fda2018oncologyendpoints,ich2021e9r1,cdisc2016adamtte}.

In counting-process notation the sponsor is defining, endpoint by endpoint,
\[
  N_i^{\mathrm{PFS}}(t)
  =
  \ind{Z_i^{\mathrm{PFS}}\le t,\ \delta_i^{\mathrm{PFS}}=1},
  \qquad
  Y_i^{\mathrm{PFS}}(t)
  =
  \ind{Z_i^{\mathrm{PFS}}\ge t}.
\]
The censoring indicator is therefore not a clerical afterthought. It encodes
which clinical histories are counted as PFS events, which histories remain at
risk, and which histories stop contributing exposure under the chosen
estimand. This is why the ICH estimand language matters for time-to-event
analysis: different choices for handling intercurrent events can change the
risk set, the censoring distribution, and even the interpretation of familiar
estimators such as log-rank and Cox summaries
\citep{rufibach2019tteestimands}. Counting processes are the compact
mathematical form of this industrial chain: protocol question, endpoint
definition, analysis dataset, estimator, and uncertainty statement.

The counting-process formulation follows the martingale approach of
\citet{aalen1978nonparametric}, \citet{andersen1982cox},
\citet{andersen1993statistical}, and \citet{fleming1991counting}.

\begin{example}[Panel counts as interval-censored event histories]
Right censoring hides the future after a last observation time.  Panel count
data hide something different: the process is inspected only at visits.  For
subject \(i\), suppose the recurrent-event counting process \(N_i(t)\) is seen
at random visit times
\[
  0<T_{i1}<\cdots<T_{iK_i},
\]
and the observed record is
\[
  O_i
  =
  \{K_i,T_{i1},\ldots,T_{iK_i},
    N_i(T_{i1}),\ldots,N_i(T_{iK_i})\}.
\]
The exact event times between visits are missing.  The increments
\[
  \Delta N_{ij}
  =
  N_i(T_{ij})-N_i(T_{i,j-1})
\]
show how many events occurred in \((T_{i,j-1},T_{ij}]\), not when they
occurred.  This is interval censoring for a counting process.

The natural target is the mean function
\[
  \Lambda_0(t)=\Expect\{N_i(t)\}.
\]
Under a non-homogeneous Poisson working model, the increment
\(\Delta N_{ij}\) has mean
\[
  \Lambda_0(T_{ij})-\Lambda_0(T_{i,j-1}),
\]
so likelihood calculations become monotone-function estimation rather than
ordinary survival estimation.  Wellner and Zhang studied two estimators of
\(\Lambda_0\) for this panel-count observation scheme, including the
nonparametric pseudo-likelihood estimator and the full nonparametric maximum
likelihood estimator \citep{wellnerZhang2000panel}.  A key point for this book
is their robustness message: consistency of these estimators can be obtained
even when the Poisson assumption is used as a working likelihood rather than
as the literal data-generating law.

The example is a useful bridge between Chapters~5 and~15.  Chapter~5 says that
observed data are functions of a larger world; here the larger world is the
unobserved path \(N_i(t)\), and the observed data are its values on a
subject-specific inspection grid.  Chapter~16 says that event histories are
counting processes with predictable clocks; panel counts remind the analyst that the
clock itself may be observed only through a visit schedule.
\end{example}

\begin{example}[Famine relief as an event history]
The Chinese famine-relief thread from Chapters~4 and~5 is a historical version
of the same information problem. Drought reports, harvest shortfalls, grain
prices, granary stocks, migration signals, memorials, transport orders, and
relief actions arrive as marked events under an administrative filtration
\(\fieldF_t\). A decision to open granaries, move grain, remit taxes, or send
relief should be representable as a stopping time with respect to that
filtration: it may depend on reports received by time \(t\), but not on later
mortality, later price data, or later political evaluation.

This does not turn famine history into a clean clinical trial. It clarifies the
statistical pressure. The analyst must distinguish the latent crisis process
from the reporting process, administrative delay, strategic understatement,
transport capacity, and relief response. A counting-process representation
therefore protects the central historical question from hindsight leakage:
what could the state have known, when could it have known it, and which action
was taken under that information set?
\qedmark
\end{example}

\endgroup

\subsection{Multi-State Survival Models}
\conceptindexes{multi-state models, transition intensity, event-history analysis}

The same bookkeeping becomes the standard language for multi-state event
history models. A clinical story is first discretized into states, visits,
transition times, and censoring indicators; only then do hazards and transition
matrices enter. Figure~\ref{fig:multi-state-translation} shows the
bookkeeping.

\begin{realdatacapsule}{Multi-state survival file}
\item[Data object.] Subject-level state histories with entry time, transition
times, censoring, treatment, covariates, and visit schedule.
\item[Observation mechanism.] Clinical states are observed at visits or
adjudicated event times; censoring, interval observation, and competing events
decide which transitions enter the file.
\item[Target.] Transition probabilities \(P_{jk}(s,t)\), state-occupation
probabilities, restricted time in state, or treatment contrasts on those
functionals.
\item[Model.] Counting-process intensities, Markov or semi-Markov transition
hazards, product integrals, and Aalen--Johansen estimators turn histories into
probability matrices.
\item[Uncertainty.] Martingale variance, influence functions, subject-level
bootstrap, and sensitivity to state definitions quantify the reported
transition curve.
\item[Limitation.] State labels, visit schedules, and censoring rules are part
of the estimand; a different state definition can change the target as much as
the estimator.
\end{realdatacapsule}

\input{figures/ch15_multi_state_model}

Let $E=\{1,\ldots,K\}$ be a finite state space and let
$X_t$ be a \cadlag\ state process. For $j\ne k$, define the transition count
and the at-risk indicator
\[
  N_{jk}(t)=\sum_{0<u\le t}\ind{X_{u-}=j,\ X_u=k},
  \qquad
  Y_j(t)=\ind{X_{t-}=j}.
\]
If the transition intensity from $j$ to $k$ is the predictable process
$\lambda_{jk}(t)$, then the compensator of $N_{jk}$ is
\[
  \Lambda_{jk}(t)=\int_{(0,t]}Y_j(u)\lambda_{jk}(u)\,du,
  \qquad
  M_{jk}(t)=N_{jk}(t)-\Lambda_{jk}(t)
\]
and $M_{jk}$ is the transition martingale. In matrix notation, the
off-diagonal cumulative transition hazards are collected in the matrix
$\matA$ by setting $(\matA)_{jk}=A_{jk}$ for $j\ne k$ and
$(\matA)_{jj}=-\sum_{k\ne j}A_{jk}$. The transition probability matrix is then
written as the product integral
\[
  \matP(s,t)=\Prodi_{(s,t]}\{\matI+\matA(du)\}.
\]
Replacing $\matA$ by the Nelson--Aalen transition estimator gives the
Aalen--Johansen estimator.
The estimator originates in \citet{aalen1978empirical}; the product-integral
calculus behind the transition matrix is developed in \citet{gill1990survey}.

This framework also makes clear where Markov and semi-Markov models differ. A
Markov multi-state model lets $\lambda_{jk}(t)$ depend on the calendar time and
current state. A semi-Markov model replaces calendar time by
the sojourn age
\[
  B_t=t-\sup\{u<t:X_u\ne X_{t-}\},
\]
with the supremum taken as $0$ before the first transition,
so that, for example,
\[
  \Lambda_{jk}(t)=\int_0^tY_j(u)h_{jk}(B_{u-})\,du.
\]
Thus the semi-Markov model keeps the same counting-process bookkeeping, but
changes the predictable clock: transition risk is indexed by time since entry
into the current state rather than by calendar time alone.

There is a corresponding bootstrap story for transition probabilities in
semi-Markov or Markov renewal models.  A Markov renewal model keeps the embedded
state chain and the sojourn-time distribution together: the next state and the
waiting time are governed by a transition kernel rather than by a calendar-time
intensity alone.  Nonparametric transition-probability estimators are therefore
functionals of the empirical transition histories and the product-integral or
renewal equations used to turn those histories into probabilities.  The
bootstrap example in \citet{dabrowska1995transition} resamples the observed
subject-level histories and recomputes the transition-probability estimator.
The point is the same as in the bivariate Kaplan--Meier surface: bootstrap the
empirical object at the level at which the dependence and censoring structure
are observed, then push the resampled object through the same product-integral
or renewal map.

\subsection{Semi-Markov Transformation Models}
\label{sec:ch16-semi-markov-transformation}
\conceptindexes{semi-Markov transformation models, Markov renewal models, sojourn times, transformation models}

The semi-Markov distinction becomes statistical once covariates enter the
holding-time law. Let \(S_m\) be transition times, \(J_m=X_{S_m}\) the embedded
state chain, and
\[
  Y_{m+1}=S_{m+1}-S_m
\]
the sojourn time after entering state \(J_m\). A Markov renewal model describes
the next state and the next waiting time through a kernel
\[
  Q_{ij}(t\mid z)
  =
  P\{J_{m+1}=j,\ Y_{m+1}\le t
    \mid J_m=i,\ Z_m=z\},
\]
where \(Z_m\) denotes covariates available when the subject enters state \(i\).
Equivalently, one separates a next-state probability from a transition-specific
sojourn distribution,
\[
  Q_{ij}(dt\mid z)
  =
  p_{ij}(z)\,H_{ij}(dt\mid z).
\]

Because Chapter~11 uses this same paper for estimation theory, the present
section records only the model skeleton. Dabrowska's notation is a little more
structural than the one-line survival regression shorthand.  The paper works
with transition times \(T_m\), states \(J_m\), sojourns
\(X_{m+1}=T_{m+1}-T_m\), and marks \(V_m=(J_m,\widetilde Z_m)\).  For the next
displays we switch to Dabrowska's one-index transition notation, where \(j\)
denotes the ordered pair \(j=(j_1,j_2)\).  The counting process is
\[
  \widetilde N_j(t)
  =
  \sum_{m\ge0}\ind{T_{m+1}\le t,\ J_m=j_1,\ J_{m+1}=j_2}.
\]
The model is specified through its compensator: for
\(t\in(T_m,T_{m+1}]\),
\[
  \Lambda_j(t)
  =
  \Lambda_j(T_m)
  +
  \int_0^{t-T_m}
    \ind{J_m=j_1}\,
    \alpha_j\{\Gamma_{(j_1,\cdot)}(u),\theta,Z_{j_1m}\}\,
    \Gamma_j(du).
\]
Here \(\Gamma_j\) is an unknown increasing transformation for transition
\(j\), \(\Gamma_{(j_1,\cdot)}\) collects the transformations for all
transitions that may leave state \(j_1\), \(\theta\) is Euclidean, and
\(\alpha_j\) is a parametric hazard-scale function.  Thus the transformation is
carried by the pair \((\alpha_j,\Gamma_j)\), not by a separate scalar operator
\(\mathcal T\).

The associated semi-Markov kernel is written in the paper as
\[
  F_j(x\mid Z)
  =
  P\{X_{m+1}\le x,\ J_{m+1}=j_2\mid J_m=j_1,\ Z\},
\]
and, in the latent competing-risk construction,
\[
  F_j(x\mid Z)
  =
  \int_0^x
    \bar F_{(j_1,\cdot)}(u\mid Z)\,
    \alpha_j\{\Gamma_{(j_1,\cdot)}(u),\theta,Z\}\,
    \Gamma_j(du),
\]
where \(\bar F_{(j_1,\cdot)}\) is the survival function of the sojourn time in
state \(j_1\).  Proportional-hazards and proportional-odds versions arise from
particular choices of this transformation architecture.  In the local notation
of this book, the earlier display \(Q_{ij}=p_{ij}H_{ij}\) is the compatible
Markov-renewal, one-step-kernel version of the same modeling architecture, not
a symbol-by-symbol translation of Dabrowska's display.

The counting-process view clarifies the observation mechanism. For each
transition \(i\to j\), the process
\[
  N_{ij}(t)=\sum_m\ind{S_m\le t,\ J_{m-1}=i,\ J_m=j}
\]
is observed only up to censoring or administrative cutoff, and the predictable
exposure depends on the current state, the sojourn age, and covariates. The
target is not just a regression coefficient. It may be a transition
probability, a state-occupation curve, a restricted time in state, or a
covariate contrast after the transformed sojourn laws have been pushed through
the Markov-renewal equations.

This is the continuous-time version of a recurring lesson in the book: a
statistical target is often a functional of a larger fitted law. Estimation
combines transition counts, censored sojourn-time likelihoods or estimating
equations, and the renewal/product-integral map that returns probabilities.
Inference uses the same empirical-process, martingale, and bootstrap logic as
the surrounding survival sections; Example~\ref{ex:ch11-dabrowska-semi-markov-application}
looks at the same Dabrowska paper from the U-process, Volterra-equation, and
multiplier-band side. Testing asks whether calendar time is enough or sojourn
age is needed, whether the chosen transformation architecture is adequate, and
whether covariates act on the next-state probabilities, the holding-time law, or
both. This is the role of the semi-Markov transformation model studied by
\citet{dabrowska2012semiMarkovTransformation}; it extends the
semiparametric Markov-renewal transition-probability work in
\citet{dabrowska1995transition}.

\section{Poisson, Renewal, and Markov Jump Systems}
\conceptindexes{Poisson processes, renewal processes, continuous-time Markov chains, Feller processes, interacting particle systems, generator}

The previous section stayed close to survival and multi-state statistical
targets. This section keeps the same compensator grammar but enlarges the
state space and the mechanism. Non-homogeneous Poisson processes show how a
deterministic clock changes time; renewal processes show how the current age
changes the next risk; continuous-time Markov chains show how transition
counts form a marked point process; Feller and interacting-particle systems
show how the same ideas survive when the state space is too large for a
transition matrix to be the main object.

\subsection{Non-Homogeneous Poisson Process}
\label{sec:ch16-nonhomogeneous-poisson}
\conceptindexes{non-homogeneous Poisson process, time change, arrival times}

Let $N$ be a non-homogeneous Poisson process with continuous mean function
$A(t)$ and $A(t)\to\infty$. Define
\[
  A^{-1}(x)=\inf\{s:A(s)>x\},\qquad x\ge0,
\]
and
\[
  \widetilde N(x)=N(A^{-1}(x)).
\]
The function $A$ is the clock seen by the process. In calendar time the jumps
may accelerate or slow down; in $A$-time the same process becomes a unit-rate
Poisson process.
Chapter 6 viewed the same object through independent increments and
finite-dimensional laws; here the mean function is read as a compensating
clock.

\begin{theorem}[Time change; \citealp{watanabe1964discontinuous}]
The process $\{\widetilde N(x):x\ge0\}$ is a homogeneous Poisson process with
rate one, adapted to $\Gt_x=\fieldF_{A^{-1}(x)}$.
\end{theorem}

\noindent\textit{Proof.}
The process is adapted by definition of $\Gt_x$. If $0\le x<x+y$, then
\[
  \widetilde N(x+y)-\widetilde N(x)
  =
  N(A^{-1}(x+y))-N(A^{-1}(x)).
\]
The increment on the right is independent of
$\fieldF_{A^{-1}(x)}=\Gt_x$ and has mean
\[
  A(A^{-1}(x+y))-A(A^{-1}(x))=y,
\]
using continuity of $A$. Thus the time-changed process is an extended Poisson
process with mean function $x$. The unit-rate homogeneous Poisson process is
the counting process with independent increments and mean increment equal to
elapsed time. \qedmark

It follows that the number of jumps in $(s,t]$ has a Poisson distribution with
mean $A(t)-A(s)$.
This time-change viewpoint is a classical route into non-homogeneous Poisson
processes; see \citet{karr1986point} and \citet{karlin1975first} for parallel
treatments.

Equivalently, the compensator of $N$ is $A$, and
\[
  N_t-A(t)
\]
is a martingale. If $A(t)=\int_0^t\alpha(u)\,du$, then $\alpha(t)$ is the
instantaneous intensity in calendar time.

\begin{theorem}[Arrival times]
Let $T_n$ be the arrival times of a non-homogeneous Poisson process with
continuous mean function $A$, and set $W_{n+1}=T_{n+1}-T_n$. Then
$A(T_n)$ are the arrival times of a homogeneous Poisson process, and
\[
  P(W_{n+1}>w\mid T_1,\ldots,T_n)
  =
  \begin{cases}
  1, & w<0,\\
  \exp\{-[A(T_n+w)-A(T_n)]\}, & w\ge0.
  \end{cases}
\]
\end{theorem}

\noindent\textit{Proof.}
By the time-change theorem, $A(T_n)$ are the jump times of a homogeneous
unit-rate Poisson process. Hence
\[
  A(T_{n+1})-A(T_n)
\]
is conditionally standard exponential given $T_1,\ldots,T_n$. For $w\ge0$,
monotonicity and continuity of $A$ give
\[
\begin{aligned}
  \Prob(W_{n+1}>w\mid T_1,\ldots,T_n)
  &=
  \Prob(T_{n+1}>T_n+w\mid T_1,\ldots,T_n)\\
  &=
  \Prob(A(T_{n+1})-A(T_n)>
        A(T_n+w)-A(T_n)\mid T_1,\ldots,T_n)\\
  &=
  \exp\{-[A(T_n+w)-A(T_n)]\}.
\end{aligned}
\]
For $w<0$ the event is certain. \qedmark

\subsection{Renewal Processes}
\conceptindexes{renewal process, renewal times, interarrival times}

Let $N(t)$ be a renewal process with iid interarrival distribution $F$ and
cumulative hazard $H$. If $T_n<t\le T_{n+1}$, then the compensator is
\[
  \Lambda(t)
  =
  \Lambda(T_n)
  +
  \int_{(0,t-T_n]}dH(u).
\]
The formula says that after the $n$th jump, the renewal process forgets the
calendar date but remembers the age since the last jump. The next conditional
risk is governed by the cumulative hazard of a fresh interarrival time.

Equivalently, writing $T_0=0$,
\[
  \Lambda(t)
  =
  \sum_{m=0}^{N(t-)-1} H(T_{m+1}-T_m)
  +
  H(t-T_{N(t-)})
\]
when $H$ is continuous; with jumps, the same expression is read as a
Stieltjes integral over the current age. This display makes the reset
structure explicit: each observed renewal closes one interarrival clock and
opens a fresh one.

For a delayed renewal process, the delay distribution has its own cumulative
hazard $H_d$ on the first interval $(0,T_1]$. If $H$ has hazard rate $h$, then
\[
  \Lambda(t)=\int_0^t h(\delta_{u-})\,du,
\]
where $\delta_u$ is the backward recurrence time.

Here
\[
  \delta_u=u-T_{N(u)}
\]
is the current age of the renewal process, with the left limit
$\delta_{u-}$ used at possible jump times. This is the continuous-time
counterpart of conditioning on the current state of an embedded chain: the
observable history is compressed into the age process.

For a delayed renewal process with first waiting-time hazard $h_d$, the same
formula is
\[
  \Lambda(t)
  =
  \int_0^{t\wedge T_1} h_d(u)\,du
  +
  \int_{(T_1,t]}h(\delta_{u-})\,du.
\]
Thus the first interval carries its own clock, while all later intervals share
the renewal clock. This is the precise compensator distinction between an
ordinary renewal process and a delayed one.

\subsection{Continuous-Time Markov Chains as Marked Point Processes}
\conceptindexes{continuous-time Markov chains, transition rates, jump chains}

The next natural special case is a continuous-time Markov chain. If the state
space is countable and the process jumps at times $T_n$ with marks
$X_n=X(T_n)$, then the marked point process records the successive states
visited by the chain. For a finite state space $E$, define
\[
  N_{jk}(t)=\sum_{0<u\le t}\ind{X_{u-}=j,\ X_u=k},
  \qquad j\ne k.
\]
The Markov-chain assumption is a restriction on the compensators of these
transition counts. In the time-inhomogeneous case one writes
\[
  \Lambda_{jk}(t)
  =
  \int_{(0,t]}\ind{X_{u-}=j}\,A_{jk}(du),
\]
where $A_{jk}$ is the cumulative transition hazard from $j$ to $k$. The
transition-hazard matrix $\matA$ is defined by
\[
  (\matA)_{jk}(t)=A_{jk}(t)\quad(j\ne k),
  \qquad
  (\matA)_{jj}(t)=-\sum_{k\ne j}A_{jk}(t).
\]
The transition probability matrix $\matP$ is then the matrix product integral
\[
  \matP(s,t)=\Prodi_{(s,t]}\{\matI+\matA(du)\}.
\]
For a homogeneous chain, $A_{jk}(t)=q_{jk}t$ for $j\ne k$; equivalently
$\matA(t)=t\matQ$, where $\matQ=(q_{jk})$ is the familiar generator matrix.
In the finite-state case this gives
$\matP(s,t)=\exp\{(t-s)\matQ\}$.

This bridge explains why the present chapter spends so much time on marked
point processes. A continuous-time Markov chain is not a separate species of
object; it is a marked point process whose compensator depends on the past only
through the present state, and sometimes through calendar time.

\subsection{Large State Spaces: Feller Processes and Local Jump Models}
\label{sec:ch16-feller-interacting-particles}
\conceptindexes{Feller processes, interacting particle systems, local jump models, Markov semigroup}

Finite-state Markov chains are only the first layer. A generator matrix works
well when the state space fits on the page; it becomes awkward when the state is
an opinion on every network node, an infection status at every contact, a
machine state at every sensor, or molecule counts across many compartments. The
Feller move is to stop printing the whole transition matrix and instead ask how
the process moves observable summaries, that is, functions of the state.

This subsection keeps the same clock as the preceding Markov-chain subsection
but replaces the small state space by a large or infinite configuration space.
Time is still \(t\ge0\), paths are usually right-continuous with jumps, and
local rates describe what may happen next given the present configuration. This
is the continuous-time version of several Chapter~2 examples: polls and votes,
contact networks, multi-omics dynamics, and autonomous wet-lab histories become
snapshot, log, or noisy-projection observations of a larger evolving state.

If \(X\) is a time-homogeneous Markov process on a state space \(E\), let
\(P_x\) denote its law when \(X_0=x\), and write
\[
  \mathsf P_t f(x)=\Expect_x f(X_t),\qquad t\ge0.
\]
Here \(P_x\) is a probability measure, while \(\mathsf P_t\) is an operator
that propagates observable summaries. Think of \(f\) as a readable quantity:
number infected in a neighborhood, fraction of cells with an open promoter,
local disagreement in an opinion network, or current yield in a robot lab.
Then
\[
  x
  \longmapsto
  \mathsf P_t f(x)
  =
  \Expect_x f(X_t)
\]
answers a concrete question: if the huge system starts in configuration \(x\),
what do we expect that summary to show after time \(t\)?

When \(E\) is locally compact, a Feller semigroup is a Markov semigroup
\((\mathsf P_t)_{t\ge0}\) on \(C_0(E)\), the continuous functions vanishing at
infinity, following the operator viewpoint in Feller's second volume
\citep{feller1971introductionVol2}, such that
\[
  \mathsf P_t C_0(E)\subseteq C_0(E),
  \qquad
  \|\mathsf P_t f-f\|_\infty\longrightarrow0
  \quad\text{as }t\downarrow0
\]
for every \(f\in C_0(E)\). A Markov process with such a semigroup is called a
Feller process, and its generator is
\[
  \mathcal Lf
  =
  \lim_{t\downarrow0}\frac{\mathsf P_t f-f}{t},
\]
defined on the functions for which the limit exists. This is the infinite-state
analogue of the finite generator matrix \(\matQ\).  The matrix \(\matQ\) tells
which state jumps to which other state.  The operator \(\mathcal L\) tells how
every nice summary begins to move:
\[
  \mathcal Lf(x)
  \approx
  \frac{\mathsf P_{\Delta t}f(x)-f(x)}{\Delta t}
  \qquad(\Delta t\downarrow0).
\]
That is why the generator is more than notation.  It is the local script from
which the whole random movie is assembled.

\begin{corollary}[Dynkin formula as the statistical bridge; \citealp{dynkin1965markov}]
\label{cor:ch16-feller-dynkin}
Let \(X\) be a Feller process with semigroup \((\mathsf P_t)_{t\ge0}\) and
generator \(\mathcal L\). If \(f\) belongs to the domain of \(\mathcal L\), then
\[
  M_t^f
  =
  f(X_t)-f(X_0)-\int_0^t \mathcal Lf(X_s)\,ds
\]
is a martingale under \(P_x\). Consequently,
\[
  \Expect_x f(X_t)
  =
  f(x)+
  \Expect_x\int_0^t\mathcal Lf(X_s)\,ds .
\]
\end{corollary}

\noindent\textit{Proof.}
For \(f\) in the generator domain of a strongly continuous Markov semigroup,
\[
  \mathsf P_t f-f
  =
  \int_0^t \mathsf P_s\mathcal Lf\,ds .
\]
Applying this identity after time \(u\) and using the Markov property gives, for
\(0\le u<t\),
\[
  \Expect\{f(X_t)-f(X_u)\mid\mathcal F_u\}
  =
  \Expect\left\{
    \int_u^t \mathcal Lf(X_s)\,ds
    \,\middle|\,\mathcal F_u
  \right\}.
\]
This is exactly the martingale increment condition for \(M^f\). Taking
expectations gives the displayed Dynkin formula. \qedmark

\noindent\textit{How this returns to estimation, inference, and testing.}
The corollary turns an infinite-state process into moment equations. If a model
specifies a parametric generator \(\mathcal L_\theta\), then the true parameter
should make
\[
  f(X_t)-f(X_0)-\int_0^t\mathcal L_\theta f(X_s)\,ds
\]
look like martingale noise for many readable functions \(f\). Estimation can
therefore be written as a \(Z\)- or \(M\)-estimation problem: choose
\(\theta\) so that these residuals have no systematic drift, or use the
complete jump-path likelihood when all relevant jumps and rates are observed.
Inference then uses the same martingale central-limit logic as the earlier
counting-process sections: predictable variation estimates standard errors,
and residual processes diagnose where the generator is missing structure.
Testing asks whether a proposed generator leaves drift behind. For interacting
particle systems, local functions \(f\) test local scientific claims: neighbor
imitation in a voter model, infection pressure in a contact process, exclusion
rules in a transport model, or reaction propensities in a laboratory process.
Thus the Feller language is not detached from statistics; it is the
large-state-space version of the book's route from model, to estimating
equation, to uncertainty, to lack-of-fit check.  Relative to Chapter~6, the
generator is a continuous-time way to specify the local kernels that assemble a
path law.  In the vocabulary of Chapter~8, the observed object may be a path, a
snapshot sequence, or a policy log; the target is a piece of the generator or a
functional of its path law.  In the vocabulary of Chapters~13 and~15, Dynkin
residuals are estimating equations whose predictable variation supplies
uncertainty.

An interacting particle system is a model family, often built on a
configuration space such as \(\{0,1\}^V\). For finite \(V\) it is a large
continuous-time Markov chain; for infinite \(V\), the Feller viewpoint is one
standard way to turn local rates into a well-defined process. This is the
pleasant part of the theory: local rules can be tiny, but the state space they
animate can be enormous. The application lesson is simple. When local units
update because neighboring units, local chemistry, or nearby environments
change, independence of units is no longer a reasonable first grammar.

\begin{example}[Interacting particle systems in voting and contact networks]
Let \(V\) be a finite or countable set of agents, sites, or network nodes, and let
\[
  E=\{0,1\}^{V}.
\]
A configuration \(\eta\in E\) records the current state of every site. The state
could be a vote, infection status, machine alarm, open/closed gene state, or
binary treatment adoption. For a local function \(f\), a simple spin-flip
generator has the form
\[
  \mathcal Lf(\eta)
  =
  \sum_{x\in V}c_x(\eta)\{f(\eta^x)-f(\eta)\},
\]
where \(\eta^x\) flips the state at site \(x\). The rate \(c_x(\eta)\) may
depend on nearby coordinates, so the observed states are not independent
samples. This is the continuous-time analogue of watching a configuration
change by locally triggered events rather than by resampling all nodes in
synchronized rounds.

In the voter model from Chapter~7, agent \(x\) chooses a neighbor \(y\)
according to a kernel \(p(x,y)\) and copies \(y\)'s opinion. For binary
opinions,
\[
  c_x(\eta)
  =
  \sum_{y\in V}p(x,y)\ind{\eta(y)\ne\eta(x)}.
\]
The data object is not a vector of independent votes. It is a path
\[
  \{\eta_t:t\ge0\}
  \quad\text{in}\quad
  \{0,1\}^V,
\]
shaped by network exposure and local imitation. The natural questions are
dynamic: probability of consensus, time to consensus, coexistence, sensitivity
to stubborn agents or shocks, and what can be learned from partial snapshots.

This connects directly to Chapter~2's polls-and-votes entry point. A poll at a
single date is a snapshot of \(\eta_t\), not a complete movie. If campaigns,
social exposure, media shocks, or peer contact influence transition rates, then
the target is not merely a population proportion at one time. It may be a
forecast of consensus, a comparison of intervention policies, or a diagnostic
for whether observed polarization is compatible with local interaction. The
same grammar applies to epidemic contact networks: infection and recovery are
jumps, contacts shape the rates, and a case report is a delayed partial
observation of the configuration path.
\qedmark
\end{example}

\begin{example}[Reaction networks, single-cell traces, and robot labs]
The Feller paragraph matters here because the transition matrix has
disappeared.  The model is specified by local propensities, the generator turns
those propensities into path laws, and the statistical object is a martingale
residual or likelihood built from observed reaction events.
Chemical reaction networks give a biochemical version of the same construction.
If \(X_t\) is the molecular-count vector and reaction channel \(r\) changes the
state by \(\nu_r\), then
\[
  X_t=X_0+\sum_r \nu_r N_r(t),
  \qquad
  \Lambda_r(t)=\int_0^t a_r(X_{u-})\,du
\]
is the counting-process form of the chemical master equation. Gillespie's
algorithm generates this jump process exactly \citep{gillespie1977exact}, while
Qian's nonequilibrium view reads flux, dissipation, and irreversibility as path
functionals of such processes
\citep{qian2002mesoscopic,qian2006open,qianKou2014single}.

This is the continuous-time version of the Chapter~2 autonomous-laboratory
story. A robot log records interventions, sensor traces, failures, and yields,
but the chemistry between two logged times is a path of reaction events. A
policy may choose the next condition from the observed history, while the
unobserved reaction network evolves through jumps. The filtration says what the
platform has seen; the intensities say what the chemistry is expected to do
next; the martingale residual says what surprised the model.

The same grammar appears in single-cell gene expression. Chapter~2's PBMC3k
matrix is a snapshot count table, useful for learning what a high-dimensional
biological record is. If instead we follow a cell lineage, a live-cell reporter,
or a perturbation experiment over time, the latent biological state becomes a
continuous-time process. A promoter state \(S_t\) may switch while a molecular
abundance or fluorescence signal \(Y_t\) changes through synthesis,
degradation, and feedback. For stochastic oscillations, a target may be not only
a mean curve but a second-order process law:
\[
  C(\tau)=\Cov(Y_{t+\tau},Y_t),
  \qquad
  S(\omega)=\int_{-\infty}^{\infty} e^{-i\omega\tau}C(\tau)\,d\tau .
\]
The power spectrum \(S(\omega)\) turns an irregular single-cell time trace into
evidence about oscillation, feedback, burstiness, and switching time scales
\citep{jiaQianZhang2024spectrum,golding2005realtime,cai2006stochastic}.
The statistical point is the same as in Chapters~1 and~2: the data object is
not the mechanism itself. Molecular events generate a Markov or hybrid
continuous-time process, while measurements see only noisy trajectories,
snapshot count matrices, or policy-selected experimental outcomes.
\qedmark
\end{example}

\section{Diffusions and Jump-Diffusions}
\label{sec:ch16-jump-diffusions}
\conceptindexes{diffusions, jump-diffusions, stochastic differential equations, jump measures, high-frequency data}

The last continuous-time model in this chapter is the place where the two
grammars meet. Counting processes make surprise arrive as a visible jump.
Diffusions make surprise accumulate continuously. A jump-diffusion allows both:
a system may drift and fluctuate most of the time, then move abruptly when a
shock, intervention, failure, trade, relapse, or reaction event occurs.

The connection to the preceding Feller discussion is the generator.  In a
large jump system, \(\mathcal L\) is built from local rates; in a diffusion,
\(\mathcal L\) is built from drift and quadratic variation; in a
jump-diffusion, both pieces appear together.  The statistical reading is the
same as before: specify the local law, derive martingale residuals, estimate
the generator pieces, and test whether the residual variation matches the
observation mechanism.  This is why the examples below should be read through
the Chapter~8 target ledger rather than as isolated stochastic-calculus
models.

In its simplest Markov form, a jump-diffusion may be written as
\[
  dX_t
  =
  b_\theta(X_{t-})\,dt
  +
  \sigma_\theta(X_{t-})\,dW_t
  +
  \int_{\mathcal Z}\gamma_\theta(X_{t-},z)\,\mu(dt,dz),
\]
where \(W\) is Brownian motion, \(\mu(dt,dz)\) is a random measure recording
jump times and marks, and \(\gamma_\theta(x,z)\) is the state change caused by
mark \(z\). The compensator of the jump measure is typically written
\[
  \nu_\theta(X_{t-},dz)\,dt .
\]
Thus the diffusion coefficient \(\sigma_\theta\) describes continuous
martingale variation, while \(\nu_\theta\) describes the predictable rate and
mark distribution of jumps. The notation deliberately echoes the earlier
compensator notation for marked point processes.

For smooth \(f\), the generator has the form
\[
\begin{aligned}
  \mathcal L_\theta f(x)
  &=
  b_\theta(x)^\top \nabla f(x)
  +
  \frac12
  \operatorname{tr}\!\left\{
    \sigma_\theta(x)\sigma_\theta(x)^\top \nabla^2 f(x)
  \right\}  \\
  &\qquad
  +
  \int_{\mathcal Z}
    \{f(x+\gamma_\theta(x,z))-f(x)\}\,
    \nu_\theta(x,dz),
\end{aligned}
\]
in the finite-activity case. For infinite-activity small jumps, the integral is
read with the usual local compensation term. The important statistical point is
not the extra technical term, but the decomposition: drift, continuous
variation, and jump mechanism all appear as separate pieces of the same
generator.

\begin{corollary}[It\^o--Dynkin formula for jump-diffusions; \citealp{ito1944stochastic,dynkin1965markov}]
\label{cor:ch16-jump-diffusion-dynkin}
For a jump-diffusion with generator \(\mathcal L_\theta\), and for \(f\) in a
suitable smooth domain,
\[
  f(X_t)-f(X_0)-\int_0^t\mathcal L_\theta f(X_s)\,ds
\]
is a martingale.
\end{corollary}

This is the same bridge as Corollary~\ref{cor:ch16-feller-dynkin}, now with
continuous martingale noise and jump noise in the same process. Estimation can
therefore proceed through martingale estimating equations, likelihoods, or
quasi-likelihoods. If the path is fully observed and the jump measure is
identified, the jump contribution has the familiar point-process likelihood
shape
\[
  \sum_{T_k\le T}\log \nu_\theta(X_{T_k-},Z_k)
  -
  \int_0^T \nu_\theta(X_{u-},\mathcal Z)\,du,
\]
while the continuous component is handled by diffusion likelihood or
Girsanov-type changes of drift. If the process is observed only at discrete
times, the theory becomes high-frequency inference: quadratic variation sees
both diffusion volatility and jumps, while bipower and threshold methods try to
separate the continuous variation from isolated discontinuities
\citep{barndorffNielsenShephard2004power,aitSahaliaJacod2009testing}.

The model is also a clean testing example. A pure diffusion null says that all
large short-time moves should be explainable by the continuous volatility
clock. A jump-diffusion alternative says that some moves come from a separate
event mechanism. Testing may ask whether jumps exist, whether the jump
intensity depends on covariates, whether jump sizes have the proposed law, or
whether an apparent jump is only a measurement artifact. This makes the model a
direct continuation of Chapter~14: hypotheses are sets of path laws, and the
test statistic must respect the observation mechanism.

\begin{example}[Three jump-diffusion data capsules]
In financial returns, the Black--Scholes diffusion
\[
  dS_t=\mu S_t\,dt+\sigma S_t\,dW_t
\]
captures continuous price fluctuation, while Merton's jump-diffusion adds
discontinuous moves for news, liquidity shocks, or market breaks
\citep{black1973pricing,merton1976option}. The data object may be tick-level
prices or daily returns; the target may be volatility, jump intensity, tail
risk, or option value; uncertainty comes from both Brownian variation and jump
arrival; and the limitation is microstructure noise, irregular trading, and the
fact that extreme events are rare precisely when they matter most.

In clinical or real-world evidence settings, a biomarker may drift under
ordinary disease progression but jump after treatment, relapse, infection, or a
recorded adverse event. The observation mechanism is visit-time sampling, often
with censoring and informative follow-up. The target may be treatment effect on
drift, shock size after an intervention, or risk of abrupt deterioration. The
same process can be read as a longitudinal model with event-history marks, so
the jump-diffusion connects the multi-state and biomarker parts of the book
instead of replacing them.

In an autonomous laboratory, a sensor trace may evolve smoothly between
interventions while reagent addition, clogging, phase transition, contamination,
or instrument reset creates a jump. The data object is a policy-selected log:
actions, times, sensor paths, failures, and yields. The target may be a
mechanism parameter, a control policy, or an anomaly detector. The limitation is
that the policy changes the observation law, so residual checks must condition
on what the robot knew when it acted.
\qedmark
\end{example}

\section{Chapter Summary: Event-Time Objects and Martingales}
\conceptindexes{event-time objects, event histories, martingales, dynamic inference}

Continuous-time process language adds one last constraint to the book's route:
an analysis must know what was available when. Filtrations, stopping times,
predictable processes, hazards, and compensators are not decorative
probability. They are the grammar that keeps time, information, and decision in
the right order.

The same discipline has appeared in different forms throughout the book. A
cluster of bomb impacts is not automatically a mechanism; a missing species is
not automatically absent; a historical climate proxy is not climate itself; a
textual feature is not authorship itself; a single-cell count matrix is not a
likelihood by itself; and a survival curve is not just a curve once censoring,
risk sets, and treatment timing enter. Each example asks the same question in a
different register: what is the observed object, what hidden object or target
does it speak about, and what mathematical language is needed to keep that
speech honest?

This is the reason the technical route has moved from measure to products,
from products to stability, from stability to indexed procedures, and from
indexed procedures to time. The point is not to make every statistical problem
look like a stochastic process. The point is to train a habit of reading: find
the data structure, identify the information set, state the target, choose the
law or criterion that connects them, and report the uncertainty that remains.

That habit is exactly the bridge to the final chapter.  Once the reported
uncertainty is used to rank an item, assign a treatment, stop a trial, trigger
an alarm, choose a molecule, or schedule the next measurement, the statistical
object is no longer only observed through time.  It is coupled to action.
Filtrations become the record of what a system knew before acting; stopping
times become legitimate intervention clocks; predictable rules become policies;
and future observations become partly consequences of earlier statistical
choices.  Continuous-time processes teach that information evolves.  The final
chapter asks what happens when statistical systems act on that evolving
information.

\section{Exercises}
\conceptindexes{continuous-time exercises, martingale exercises, survival-analysis exercises}

\begin{exercise}[Right-continuity and stopping times]
Let $\mathbb F=\{\fieldF_t:t\ge0\}$ be a filtration and let
$T:\Omega\to[0,\infty]$ be a random time.
\begin{enumerate}
\item Show that, if $T$ is a stopping time, then $\{T<t\}\in\fieldF_t$ for
      every $t\ge0$.
\item Assume now that $\mathbb F$ is right-continuous. Show that the condition
      $\{T<t\}\in\fieldF_t$ for every $t\ge0$ implies that $T$ is a stopping
      time.
\item Let $S$ and $T$ be stopping times. Show that $S\wedge T$ and $S\vee T$
      are stopping times. If $\mathbb F$ is right-continuous, show that
      $S+T$ is a stopping time.
\item Assume that $\mathbb F$ is right-continuous and that $T_n$, $n\ge1$, are
      stopping times. Prove that $\inf_nT_n$, $\sup_nT_n$, $\liminf_nT_n$, and
      $\limsup_nT_n$ are stopping times. Identify where right-continuity is
      used.
\end{enumerate}
\end{exercise}

\begin{exercise}[Stopped sigma-fields]
Let $S$ and $T$ be stopping times.
\begin{enumerate}
\item If $A\in\fieldF_S$, prove
\[
  A\cap\{S<T\}
  =
  \bigcup_{q\in\Rat_+}
  \bigl(A\cap\{S\le q\}\bigr)\cap\{T>q\},
\]
      and conclude that $A\cap\{S<T\}\in\fieldF_{T-}$.
\item Prove the remaining assertions in Lemma~\ref{lem:ch16-stopped-information}.
      In particular, show that $A\cap\{S\le T\}$ and $A\cap\{S=T\}$ belong to
      $\fieldF_T$ whenever $A\in\fieldF_S$.
\item Assume that the filtration is complete and that $S\le T$ almost surely.
      Show that $\fieldF_S\subseteq\fieldF_T$.
\item If $X$ is adapted and has right-continuous paths, show that $X_T$ is
      $\fieldF_T$-measurable on $\{T<\infty\}$.
\item Let $X$ be \cadlag. Show that $X_{T-}$ is $\fieldF_{T-}$-measurable when
      $T>0$.
\item For a one-jump counting process $N_t=\ind{T\le t}$, describe explicitly
      the difference between $\fieldF_T$ and $\fieldF_{T-}$ under the natural
      filtration of $N$.
\end{enumerate}
\end{exercise}

\begin{exercise}[Internal history of a marked point process]
Let $(T_n,X_n)$ be a marked point process with mark space $(S,\St)$ and natural
internal history $\mathbb H=\{\Ht_t:t\ge0\}$, where
\[
  \Ht_t
  =
  \sigma\bigl(
    \{T_n\le s\},\ \{T_n\le s,\ X_n\in B\}:
    0\le s\le t,\ B\in\St,\ n\ge1
  \bigr).
\]
\begin{enumerate}
\item Show that $T_n$ is a stopping time with respect to $\mathbb H$.
\item Show that $X_n$ is $\Ht_{T_n}$-measurable.
\item Decide whether $X_n$ must be $\Ht_{T_n-}$-measurable. Give a proof or a
      counterexample.
\item Let $C_t$ be constant between successive event times and suppose its
      value on $(T_n,T_{n+1}]$ is determined by
      $(T_1,X_1),\ldots,(T_n,X_n)$. Show that $C$ is predictable.
\end{enumerate}
\end{exercise}

\begin{exercise}[Random-measure notation]
For a marked point process define
\[
  N((0,t]\times B)=\sum_{n\ge1}\ind{T_n\le t,\ X_n\in B}.
\]
\begin{enumerate}
\item For fixed $B\in\St$, show that $N_t(B)$ is a counting process.
\item If $B_1,\ldots,B_m$ are disjoint, show that
      $(N_t(B_1),\ldots,N_t(B_m))$ is a multivariate counting process.
\item Express the jump $\Delta N_t(B)$ in terms of the mark observed at time
      $t$.
\item Let $g:S\to\Real$ be bounded and measurable. Show that
\[
  \int_{(0,t]\times S}g(x)\,N(du,dx)
  =
  \sum_{n:T_n\le t}g(X_n).
\]
\end{enumerate}
\end{exercise}

\begin{exercise}[Predictable stopping and finite-variation integrals]
Let $T$ be a stopping time.
\begin{enumerate}
\item For a predictable rectangle
      $X_u=\ind{A}\ind{s<u\le r}$ with $A\in\fieldF_s$, show directly that
      $X_T\ind{T<\infty}$ is $\fieldF_{T-}$-measurable.
\item Show directly that the stopped process $u\mapsto X_{T\wedge u}$ is
      predictable for the same rectangle indicator.
\item Use a monotone class argument to prove
      Lemma~\ref{lem:ch16-predictable-stopping} for bounded predictable $X$.
\item Let $A$ be predictable and of finite variation. Prove
      Lemma~\ref{lem:ch16-predictable-integrals} first for simple predictable
      $X$, then extend to nonnegative predictable $X$ by monotone
      approximation.
\end{enumerate}
\end{exercise}

\begin{exercise}[Predictable integrands and compensators]
Suppose the marked point process has compensator $\Lambda(du,dx)$ and let
$H(u,x)$ be bounded and predictable.
\begin{enumerate}
\item State the martingale obtained by integrating $H$ against
      $N-\Lambda$.
\item Specialize your expression to the case
      $H(u,x)=C_u\ind{x\in B}$, where $C$ is bounded and predictable.
\item Suppose
      $\Lambda(du,dx)=Y_u\nu_u(dx)\,du$, where $Y$ is predictable and
      $\nu_u$ is a predictable transition kernel. Write the compensator of
      $N_t(B)$.
\item Explain why predictability of $H$ is the mathematical version of
      ``choosing the stake before seeing the next jump.''
\end{enumerate}
\end{exercise}

\begin{exercise}[One-jump processes and cumulative hazards]
Let $T$ be a nonnegative random variable with survival function $S(t)=P(T>t)$
and cumulative hazard
\[
  A(t)=\int_{(0,t]}\frac{F(du)}{S(u-)}.
\]
Put $N_t=\ind{T\le t}$ and $Y_t=\ind{T\ge t}$.
\begin{enumerate}
\item Show that $\Lambda_t=\int_{(0,t]}Y_uA(du)$ is predictable under the
      natural filtration generated by $N$.
\item Verify directly that $M_t=N_t-\Lambda_t$ is a martingale.
\item If $P(T=t_0)>0$, compute $\Delta A(t_0)$ and interpret it as a discrete
      conditional failure probability.
\item Recover $S$ from $A$ using the product integral
      $S(t)=\Prodi_{(0,t]}(1-dA(u))$.
\end{enumerate}
\end{exercise}

\begin{exercise}[Non-homogeneous Poisson process]
Let $N$ be a counting process whose compensator is a deterministic continuous
increasing function $A(t)$ with $A(0)=0$.
\begin{enumerate}
\item Show that, for $0\le s<t$, the increment $N_t-N_s$ has mean
      $A(t)-A(s)$.
\item If $N$ has independent increments, derive the distribution of
      $N_t-N_s$.
\item Assume $A$ is strictly increasing and continuous. Show that the time
      changed process $\widetilde N_u=N_{A^{-1}(u)}$ is a unit-rate Poisson
      process.
\item If $A(t)=\int_0^t\alpha(u)\,du$, write the joint density of the first
      $m$ event times on $\{0<t_1<\cdots<t_m\le t\}$.
\end{enumerate}
\end{exercise}

\begin{exercise}[Renewal compensator]
Let $N$ be a renewal process with iid interarrival times having density $f$,
survival function $\barF$, and hazard rate $h=f/\barF$. Let
\[
  \delta_t=t-T_{N(t)}
\]
be the backward recurrence time.
\begin{enumerate}
\item Show that $\delta_{t-}$ is predictable.
\item Argue that the conditional chance of a renewal in $(t,t+dt]$, given the
      internal history just before $t$, is approximately $h(\delta_{t-})dt$.
\item Conclude that
\[
  M_t=N_t-\int_0^t h(\delta_{u-})\,du
\]
is the natural renewal martingale.
\item Redo the calculation when the first waiting time has hazard $h_d$ and all
      later waiting times have hazard $h$.
\end{enumerate}
\end{exercise}

\begin{exercise}[Nelson--Aalen estimator]
For independent right-censored observations define
\[
  N_i(t)=\ind{Z_i\le t,\delta_i=1},
  \qquad
  Y_i(t)=\ind{Z_i\ge t},
\]
and assume
\[
  N_i(t)-\int_{(0,t]}Y_i(u)A(du)
\]
is a martingale for each subject.
\begin{enumerate}
\item Derive the martingale representation
\[
  \widehat A(t)-\int_{(0,t]}J(u)A(du)
  =
  \int_{(0,t]}\frac{J(u)}{Y_{\cdot}(u)}M_{\cdot}(du).
\]
\item Compute the predictable variation of the right-hand side.
\item Give a plug-in estimator of this variation using observed increments
      $N_{\cdot}(du)$.
\item Explain why the estimator is stopped when $Y_{\cdot}(u)=0$.
\end{enumerate}
\end{exercise}

\begin{exercise}[Product-integral survival estimator]
Let the distinct observed failure times be $u_1<\cdots<u_m$. Write
$d_j=\Delta N_{\cdot}(u_j)$ and $r_j=Y_{\cdot}(u_j)$.
\begin{enumerate}
\item Show that the Nelson--Aalen estimator has jumps $d_j/r_j$.
\item Show that the product-integral survival estimator is
\[
  \widehat S(t)
  =
  \prod_{u_j\le t}\left(1-\frac{d_j}{r_j}\right).
\]
\item Compare this estimator with $\exp\{-\widehat A(t)\}$ when there are no
      tied failures.
\item Explain why the product-integral form is the natural one when the
      cumulative hazard has jumps.
\end{enumerate}
\end{exercise}

\begin{exercise}[Multiplicative intensity model]
Suppose that, for subject $i$,
\[
  \Lambda_i(t;\beta,A_0)
  =
  \int_{(0,t]}Y_i(u)\exp\{\beta^\top Z_i(u)\}\,A_0(du),
\]
where $Z_i$ is predictable.
\begin{enumerate}
\item Identify the martingale $M_i(t;\beta_0,A_0)$ under the true parameter
      $\beta_0$.
\item For a bounded predictable vector process $H_i$, write a martingale
      integral involving $\sum_i\int H_i\,dM_i$.
\item Choose $H_i(u)$ so that the resulting integrand resembles the Cox score
      contribution at time $u$.
\item Explain why the covariate process must be predictable in this model.
\end{enumerate}
\end{exercise}

\begin{exercise}[Competing risks as a multi-state model]
Consider the state space $\{0,1,2\}$, where state $0$ is alive and event-free
and states $1$ and $2$ are absorbing failure types. Let $A_{01}$ and $A_{02}$
be the cumulative transition hazards.
\begin{enumerate}
\item Write the transition-hazard matrix $\matA(t)$.
\item Show that the survival probability in state $0$ is
      $S_0(t)=\Prodi_{(0,t]}\{1-A_{01}(du)-A_{02}(du)\}$.
\item Derive the cumulative incidence formula
\[
  P_{0k}(0,t)=\int_{(0,t]}S_0(u-)\,A_{0k}(du),
  \qquad k=1,2.
\]
\item Describe the corresponding Aalen--Johansen estimators.
\end{enumerate}
\end{exercise}

\begin{exercise}[Aalen--Johansen product integral]
Let $X_t$ be a finite-state process with transition counts $N_{jk}$ and
at-risk indicators $Y_j(t)=\ind{X_{t-}=j}$. Define
\[
  \widehat A_{jk}(t)
  =
  \int_{(0,t]}\frac{\ind{Y_j(u)>0}}{Y_j(u)}\,N_{jk}(du),
  \qquad j\ne k,
\]
and set the diagonal entries so rows sum to zero.
\begin{enumerate}
\item Explain why $\widehat{\matA}$ is a matrix-valued Nelson--Aalen estimator.
\item Write the Aalen--Johansen estimator as a product integral.
\item For a time interval containing exactly one observed transition
      $j\to k$, compute the one-step factor
      $\matI+\widehat{\matA}(du)$.
\item In the two-state survival case, show that this reduces to the
      product-integral survival estimator.
\end{enumerate}
\end{exercise}

\begin{exercise}[Continuous-time Markov chains]
Let $X$ be a homogeneous finite-state Markov chain with generator
$\matQ=(q_{jk})$.
\begin{enumerate}
\item For $j\ne k$, define $N_{jk}(t)$ and show that its compensator is
      $\int_0^t\ind{X_{u-}=j}q_{jk}\,du$.
\item Show that the cumulative transition-hazard matrix is $\matA(t)=t\matQ$.
\item Use the product integral to recover
      $\matP(s,t)=\exp\{(t-s)\matQ\}$.
\item Work out the explicit transition matrix for a two-state chain with rates
      $q_{01}=\alpha$ and $q_{10}=\beta$.
\end{enumerate}
\end{exercise}

\begin{exercise}[Voter model as an interacting particle system]
Let \(V\) be a finite graph and let \(\eta_t\in\{0,1\}^{V}\) be the voter
model with copying kernel \(p(x,y)\).
\begin{enumerate}
\item Define the configuration \(\eta^{x}\) obtained by flipping the opinion at
      site \(x\), and verify that the flip rate is
      \(c_x(\eta)=\sum_y p(x,y)\ind{\eta(y)\ne\eta(x)}\).
\item Let \(N_x(t)\) count the number of flips at site \(x\).  Write its
      compensator in terms of \(c_x(\eta_{u-})\).
\item Show that the all-zero and all-one configurations are absorbing.
\item Explain why treating the final votes as iid Bernoulli observations would
      ignore the main statistical structure of the model.
\end{enumerate}
\end{exercise}

\begin{exercise}[Localization for counting-process martingales]
Let $N$ be a non-explosive counting process with compensator $\Lambda$, and let
$M=N-\Lambda$.
\begin{enumerate}
\item Define $T_m=\inf\{t:N_t\ge m\}$ and show that $T_m\uparrow\infty$ almost
      surely.
\item Show that $N_{t\wedge T_m}$ has bounded paths.
\item Give conditions under which $M^{T_m}=M_{t\wedge T_m}$ is a true
      martingale.
\item Explain how this proves that $M$ is a local martingale.
\end{enumerate}
\end{exercise}

\begin{exercise}[Designing a clock]
Suppose an event-history model has two possible event types, $a$ and $b$.
After an event of type $a$, the risk of another type-$a$ event depends on time
since the last type-$a$ event; after an event of type $b$, the risk of a
type-$a$ event depends on calendar time. Propose a marked point process model
for this situation.
\begin{enumerate}
\item Define the mark space and the transition counts.
\item Define predictable clock processes that encode the two kinds of elapsed
      time.
\item Write a compensator for the type-$a$ counting process.
\item State what must be checked to ensure that your clock processes are
      predictable.
\end{enumerate}
\end{exercise}

\newpage
\section*{Sources and Further Reading}
\addcontentsline{toc}{section}{Sources and Further Reading}

This chapter follows the standard route from filtrations and stopping times to
point processes, compensators, martingales, and quadratic variation. The
presentation is deliberately selective: it emphasizes formulations that are
stable across probability theory, mathematical statistics, and event-history
applications, while leaving more specialized extensions to the references below.
The exercises are original, but their emphasis follows the process-level
training style of \citet{cinlar2011probability} and the counting-process
survival-analysis viewpoint of \citet{fleming1991counting}.  The route through
filtrations, progressive and predictable processes, compensators, product
integrals, renewal processes, Markov jump processes, and Brownian examples is
also informed by \citet{dabrowskaStochasticProcessesCommunication}.

\begin{description}[leftmargin=0pt,labelsep=0.65em,style=unboxed,font=\normalfont,itemsep=0.45\baselineskip]
\item[\textsc{Filtrations, stopping times, and stochastic bases.}]
The measure-theoretic language of a stochastic process as a family of random
variables adapted to an increasing family of $\sigma$-fields is part of the
post-Kolmogorov development of probability. \citet{doob1953stochastic} is the
classical reference for the martingale and stopping-time viewpoint, while
\citet{dellacherie1978probabilities} and \citet{jacod1987limit} give the later
``general theory of processes'' formulation used in modern continuous-time
probability.

\item[\textsc{Optional, progressive, and predictable processes.}]
The distinction between being observable at time $t$, being observable along
the whole time interval, and being known just before time $t$ is one of the
technical achievements of the general theory. Predictability is indispensable
for compensators and stochastic integration because it separates what can be
forecast from the current history from the new jump or innovation. The present
chapter uses only the part of this theory needed for counting processes and
finite-variation compensators; more systematic treatments appear in
\citet{dellacherie1978probabilities} and \citet{jacod1987limit}.

\item[\textsc{Marked point processes and histories.}]
Encoding a sequence of event times together with marks is a natural way to pass
from renewal processes and Markov chains to more general event-history models.
The internal history generated by the marked process records exactly the
information available from past event times and marks. \citet{karr1986point}
gives a statistical account of point processes and their inference, while
\citet{jacod1987limit} and \citet{bremaud1981point} develop the random-measure
and martingale machinery behind the same objects.

\item[\textsc{Compensators and intensities.}]
The compensator is the continuous-time analogue of subtracting the predictable
part of a process. In the submartingale setting this idea is formalized by the
Doob--Meyer decomposition; for a counting process it becomes the representation
\[
  N_t=\Lambda_t+M_t,
\]
where $\Lambda$ is predictable and increasing and $M$ is a martingale.
\citet{watanabe1964discontinuous} is one classical source for the connection
between jumps, predictable characteristics, and martingales. The same language
underlies modern intensity models for point processes, queues, survival data,
and Markov jump processes.

\item[\textsc{Localization and local martingales.}]
Localization is the device that allows martingale arguments to be stated under
finite-time or boundedness assumptions and then transferred to processes that
are only locally integrable. This is especially useful for stochastic
integrals, likelihood calculations, and quadratic-variation identities, where
the natural hypotheses are often local rather than global. The treatment here
keeps only the stopping-time version needed for compensators and brackets.

\item[\textsc{Quadratic and predictable variation.}]
Quadratic variation records the realized second-order accumulation of a
martingale, while predictable variation records its compensating clock. For
counting processes this distinction is particularly transparent: jumps
contribute to $[M]$, and their predictable rates contribute to $\langle M\rangle$.
This is the route by which martingale central limit theorems enter
large-sample theory for point processes and survival models; see
\citet{liptser1974statistics}, \citet{jacod1987limit}, and
\citet{andersen1993statistical}.

\item[\textsc{Hazards, survival analysis, and semi-Markov models.}]
Cumulative hazards entered statistics through life-table, reliability, and
survival-analysis problems before being absorbed into the counting-process
martingale framework. \citet{nelson1972theory} on hazard plotting and
\citet{aalen1978nonparametric} on nonparametric counting-process inference are
the main sources behind the Nelson--Aalen estimator. \citet{cox1972regression}
made hazard-based modeling central in applied statistics, and
\citet{andersen1982cox} recast Cox regression in counting-process form. The
product-integral formulation for transition probabilities leads to the
Aalen--Johansen estimator for non-homogeneous multi-state models
\citep{aalen1978empirical,gill1990survey}.  Bootstrap inference for
transition probabilities in semiparametric Markov renewal models is developed
by \citet{dabrowska1995transition}, and semi-Markov transformation modeling for
covariate-dependent sojourn laws is treated in
\citet{dabrowska2012semiMarkovTransformation}. The local exposition here is
also informed by \citet{dabrowskaStochasticProcessesCommunication} for
cumulative hazards, counting-process compensators, and stochastic-process
grammar.

\item[\textsc{Poisson, renewal, and Markov jump processes.}]
The non-homogeneous Poisson process illustrates the compensator as a change of
clock: after time transformation by its mean measure, it becomes a unit-rate
Poisson process. Renewal processes show the next level of history dependence,
where the predictable clock is driven by the age since the last renewal.
Continuous-time Markov chains are the finite-state marked point processes whose
transition compensators depend on the past through the present state.
\citet{karlin1975first} gives a classical entry point for this Markov-chain
material.

\item[\textsc{Repulsive and determinantal point processes.}]
The Poisson process is the independent-count baseline for point patterns.
Determinantal point processes replace that baseline with kernel determinants
for joint intensities, thereby encoding negative dependence and repulsion.
\citet{macchi1975coincidence} is a classical source,
\citet{houghKrishnapurPeresVirag2009zeros} connects DPPs with zeros of random
functions and random-matrix phenomena, \citet{kuleszaTaskar2012dpp} develops
machine-learning uses such as diverse subset selection, and
\citet{lavancier2015dpp} treats spatial statistical modeling and inference.

\item[\textsc{Feller processes and interacting particle systems.}]
Feller's semigroup viewpoint connects Markov processes with operators on
function spaces \citep{feller1952parabolic,feller1971introductionVol2}; modern
treatments include \citet{ethierKurtz1986markov}. Interacting particle systems
turn local transition rules into Markov processes on configuration spaces such
as \(\{0,1\}^{V}\); standard references are
\citet{liggett1985interacting,liggett1999stochastic}, with the voter model
traced to \citet{holleyLiggett1975voter}.

\item[\textsc{Diffusions, jump-diffusions, and high-frequency inference.}]
It\^o's stochastic integral is the continuous-martingale counterpart to the
chapter's counting-process martingales \citep{ito1944stochastic}.
Black--Scholes gives the canonical diffusion example in finance, while
Merton's model
adds discontinuous returns through jumps
\citep{black1973pricing,merton1976option}. Modern high-frequency inference
uses quadratic, bipower, and threshold-type variation ideas to separate
continuous volatility from jumps; the local discussion points to
\citet{barndorffNielsenShephard2004power} and
\citet{aitSahaliaJacod2009testing}. Full stochastic calculus is outside this
chapter's scope; the local message is the shared decomposition into
predictable evolution, continuous martingale innovation, and jump innovation.
\end{description}

%% file: figures/ch15_event_history_clock.tex
\begin{center}
\includegraphics[width=0.98\linewidth]{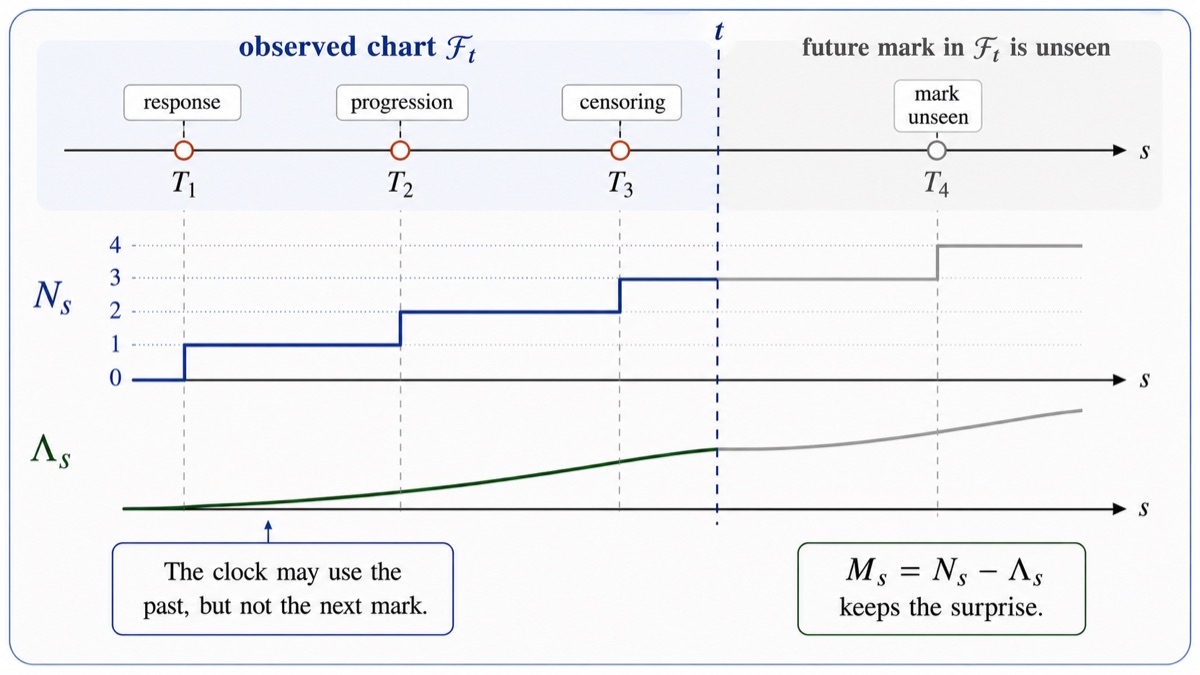}
\bookmanualfigure{fig:event-history-clock}{An event history as a growing chart}
\par\smallskip
\small Figure~\thefigure: An event history as a growing chart. The filtration
contains what has been recorded by time $t$; the compensator is the predictable
clock built from that history.
\end{center}

%% file: figures/marked_point_process_timeline.tex
\par\smallskip
\noindent\begin{minipage}{\linewidth}
\centering
\includegraphics[
  width=0.98\linewidth,
  trim=8bp 8bp 8bp 8bp,
  clip
]{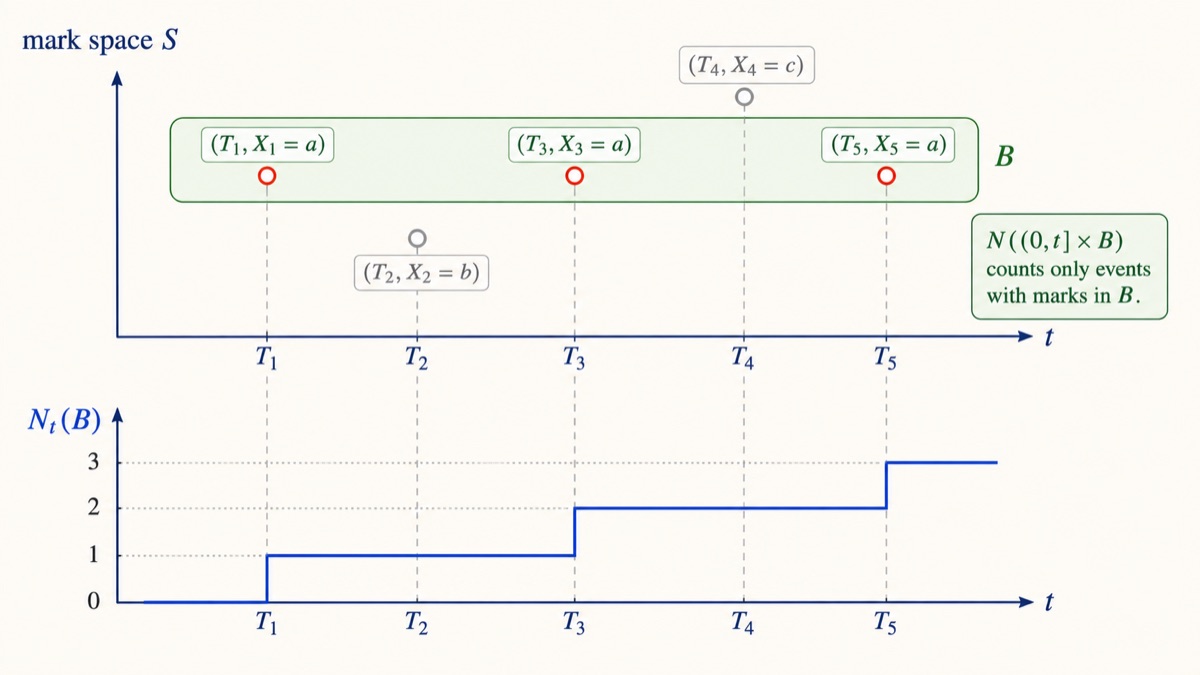}
\bookmanualfigure{fig:marked-point-process}{A marked point process as a random measure}
\vspace{0.10\baselineskip}

{\small Figure~\thefigure: A marked point process viewed as a random measure
on time and mark space. The counting process $N_t(B)$ records only those events
whose marks fall in $B$.}
\end{minipage}
\par\smallskip

%% file: figures/ch15_multi_state_model.tex
\par\smallskip
\noindent\begin{minipage}{\linewidth}
\centering
\includegraphics[
  width=0.98\linewidth,
  trim=8bp 8bp 8bp 8bp,
  clip
]{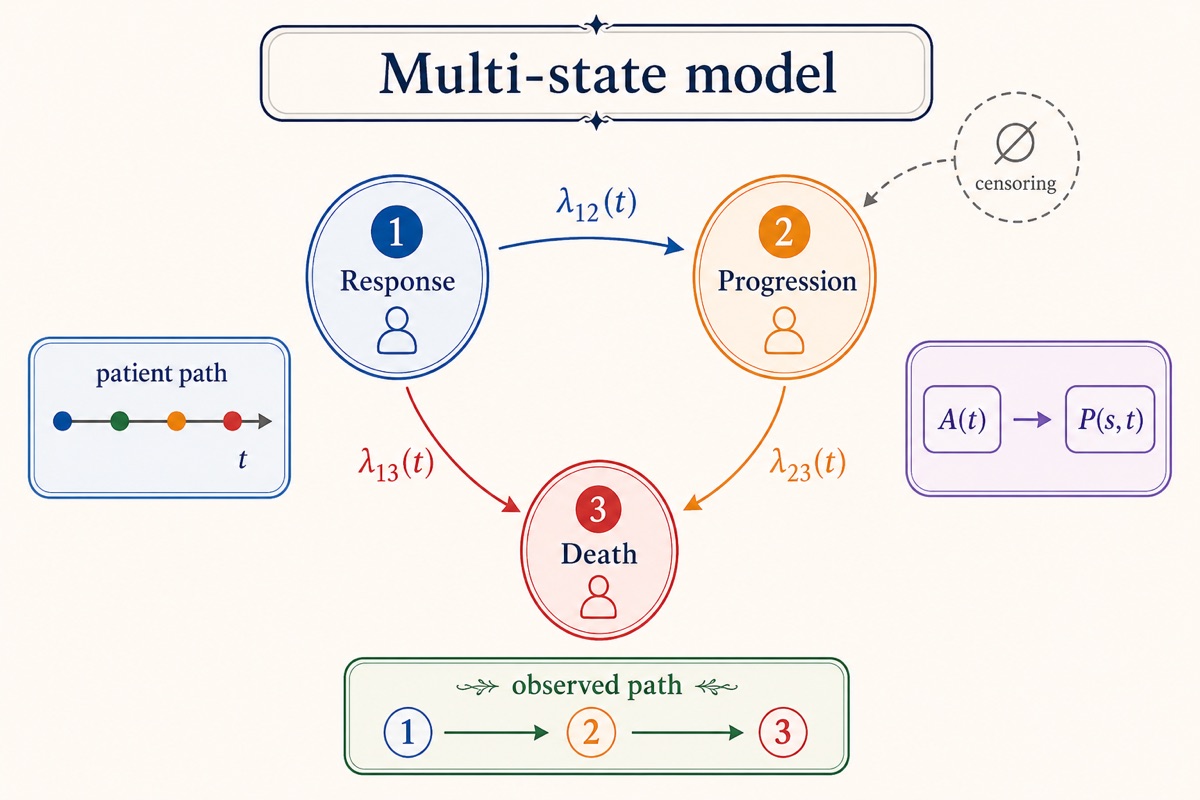}
\bookmanualfigure{fig:multi-state-translation}{A multi-state model as transition-specific hazards}
\vspace{0.12\baselineskip}

{\small Figure~\thefigure: A multi-state model turns an observed patient path
into transition-specific hazards.  The hazard graph is then collected in the
matrix cumulative hazard $\matA(t)$ and read as transition probabilities
$\matP(s,t)$ through a product integral.}
\end{minipage}
\par\smallskip

%% file: chapters/ch17_future_statistics.tex
\chapter{From Inference to Use: Prediction, Risk, Feedback, and Deployment}
\label{chap:future-statistics}
\conceptindexes{prediction, risk, feedback, deployment, statistical use, data science, automated experimentation}

\begin{tcolorbox}[
  enhanced,
  breakable,
  colback=chaptercream,
  colframe=bookblue!88!black,
  boxrule=0.72pt,
  arc=5pt,
  boxsep=1pt,
  left=1.0em,
  right=0.95em,
  top=0.82em,
  bottom=0.82em,
  before skip=0.55\baselineskip,
  after skip=1.0\baselineskip
]
\noindent\textbf{Chapter overview.}
This final chapter closes the loop from inference to use: prediction, risk,
feedback, and scientific or industrial deployment. Inference produces claims;
use requires loss, information flow, target, and feedback to be named.
Recommendation systems, automated experimentation, scientific AI, and deployed
uncertainty are not new territories added at the end. They are stress tests for
the same compass used throughout the book. Recorded logs define targets under
use, and use changes the next measurement. The final question is how
statistical discipline keeps that loop visible inside systems that learn from
their own consequences.
\end{tcolorbox}

\section{Statistics After Data Science}
\conceptindexes{statistics after data science, modern problems, data structures, statistical ledger}

The future of statistics is not a request for more techniques alone. It is a
request for a larger statistical imagination: data gathering, preparation,
exploration, modeling, computation, communication, reproducibility, and
prediction all belong to the same work
\citep{tukey1962future,tukey1977exploratory,donoho2017fifty}. The old tension
between statistical modeling and algorithmic prediction remains real
\citep{breiman2001two_cultures}, but modern practice needs a broader grammar
than either side by itself can provide.

The earlier chapters built a ledger for this broader grammar: measurement,
data structure, assumptions, target, procedure, uncertainty, and use.  The
order matters. Without measurement, prediction is just pattern. Without
probability, uncertainty becomes decoration. Without design, data collection is
treated as fate. Without a target, inference may be precise about the wrong
question. Without a use setting, inference may remain correct but unanchored.

Once statistical procedures act on the world, they become part of the
data-generating mechanism.  This is the closing pressure on the whole book.
The procedure is no longer merely a statistic computed after observation.  It
may change what is measured next, who appears in the record, which labels are
created, which experiments are run, and which errors become visible.

The statistical world of 2050 will not be smaller. It will contain clinical
trials and electronic health records, single-cell atlases and spatial omics,
robot laboratories and chemical synthesis platforms, recommender systems and
online experiments, climate archives and text corpora, wearable devices and
event streams. The unifying object will not be a table, a regression, or a
neural network. It will be the ledger that lets an analyst ask: what was
observed, by what mechanism, under what information set, for what target, with
what uncertainty, for whose use, and under what feedback?

\subsection{Modern Problems as Data Structures}
\conceptindexes{modern problems, data structures, statistical object, deployment data}

The modern-problem portfolio introduced in Chapter~1 has been distributed
through the technical chapters rather than expanded into a separate case-study
part. Missing species teach unseen structure; spatial point patterns teach
skepticism about visual clusters; genomics teaches multiplicity and selection;
single-cell and spatial omics teach observation mechanisms and simulation;
images, networks, text, and distributions teach representation; recommendation
and automated experimentation teach feedback. The examples are valuable because
each makes the data structure difficult in a different way.

The technical chapters enter at the record level.  A species list becomes an
empirical measure with missing mass.  A clinical event-history file becomes a
censored counting process.  A single-cell matrix becomes a noisy
high-dimensional random object.  A platform log becomes a sequence of
histories, actions, propensities, and outcomes.  A robot-laboratory campaign
becomes an adaptive design with a changing observation law.

The bridge is therefore a small ledger that can be used for any example:
observed record, observation mechanism, mathematical object, target, estimator
or decision rule, uncertainty, and feedback.  If any entry is missing, the
analysis may still be computationally impressive, but the statistical claim is
not yet anchored.  This chapter uses that ledger to return the book's abstract
tools to the kinds of records that created them.

Real-data analysis capsules make the ledger operational.  They record the
time origin, censoring rule, exposure definition, logging mechanism, metric
hierarchy, dependence structure, or policy loss that a theorem needs before it
can speak to a real claim.

The current capsules make that bridge concrete.  A climate reconstruction gives
a functional contrast and then a century-level Wasserstein regression; the
London bombing grid turns visual clustering into a Poisson point-process
diagnostic; the PBMC3k capsule turns a single-cell matrix into observation
summaries, a simulation check, and an explicit warning that no pseudotime target
is supported by the fallback data.  These are small computations, but they span
environmental history, spatial processes, genomics, simulation, and
quality-control thinking.  The shared ledger lets academic examples and
industrial systems use the same checks without pretending that their records
have the same shape.

\begin{realdatacapsule}{Online experimentation platform}
\item[Data object.] Assignment logs, exposure records, user histories, metric
events, guardrail outcomes, and marketplace context.
\item[Observation mechanism.] Users are randomized or targeted, exposed
through product surfaces, logged by instrumentation, and filtered by bot,
privacy, or quality rules.
\item[Target.] Short-run metric lift, long-run value, heterogeneous treatment
effect, guardrail violation, or policy value under a deployment population.
\item[Model.] Randomization-based estimators, regression adjustment,
sequential monitoring, interference checks, and off-policy evaluation connect
logs to targets.
\item[Uncertainty.] Clustered standard errors, variance reduction diagnostics,
multiple-metric adjustment, and sensitivity to logging defects report
uncertainty.
\item[Limitation.] A clean A/B contrast for one metric may not identify
network effects, long-run behavior, welfare, or the target after the product
changes user behavior.
\end{realdatacapsule}

\begin{realdatacapsule}{Robot lab and scientific AI loop}
\item[Data object.] Proposed compounds or conditions, robot actions, sensor
curves, yields, failures, model scores, and next-experiment decisions
\citep{burger2020mobile,abolhasani2023rise,xtalpi2025autonomousLab}.
\item[Observation mechanism.] The dataset is generated by the acquisition
policy: the robot records what it chose to try, plus failures and quality
filters.
\item[Target.] A scientific optimum, response surface, uncertainty-guided
acquisition rule, or laboratory decision that trades yield, cost, safety, and
exploration.
\item[Model.] Bayesian optimization, active learning, mechanistic simulators,
and foundation-model predictions supply candidate actions and uncertainty.
\item[Uncertainty.] Posterior predictive uncertainty, calibration checks,
replicate experiments, and policy regret summaries decide whether the loop is
learning.
\item[Limitation.] Policy-dependent data can create blind spots: an adaptive
system may become confident in regions it has barely explored.
\end{realdatacapsule}

\begin{realdatacapsule}{Deployed financial risk monitor}
\item[Data object.] Streaming returns, positions, covariances, liquidity
measures, breaches, and risk-limit decisions.
\item[Observation mechanism.] Market data arrive on trading calendars, are
cleaned and synchronized, and are shaped by portfolio decisions that may react
to previous risk reports.
\item[Target.] Forecasted loss distribution, value at risk, expected
shortfall, stress loss, or probability of limit breach under current exposure.
\item[Model.] Volatility filters, factor models, extreme-value diagnostics,
stress scenarios, and backtesting rules map histories to decisions.
\item[Uncertainty.] Exceedance tests, rolling recalibration, scenario
sensitivity, and block bootstrap quantify sampling and model risk.
\item[Limitation.] The risk report can change the portfolio and therefore the
future data-generating law; model validation must watch feedback and regime
change.
\end{realdatacapsule}

\subsection{Final Statistical Ledger}
\label{sec:ch17-closing-grammar}
\conceptindexes{statistical ledger, inference to use, prediction target, risk, feedback, uncertainty}

The final grammar has four pieces.  They are ordinary enough to be overlooked,
and that is why they deserve to be named explicitly.

\begin{description}[leftmargin=0pt,labelsep=0.55em,style=unboxed,font=\bookdescriptionlabelfont,itemsep=0.45\baselineskip]
\item[Prediction target.]
A prediction rule is not only a fitted function.  It is a map from an
information set to an output space.  In a time-indexed system one may write
\[
  a_t=\pi_t(H_t),
\]
where \(H_t\) is the history available just before the rule is used.  The
target must say whether the rule is being evaluated on the original sampling
law, a benchmark split, a future deployment population, a counterfactual
exposure regime, or a scientific simulator.

\item[Loss and risk.]
Prediction becomes a statistical target only after a loss and a law are named:
\[
  R_P(\pi)=\Expect_P\{L(\pi(H),Y)\}.
\]
Changing \(L\) changes what is good.  Changing \(P\) changes where it is good.
A ranking system, a diagnostic score, a laboratory acquisition rule, and a
weather forecast may all be called prediction, but their risks live on
different populations, time scales, and consequences.

\item[Feedback.]
When the rule changes exposure, treatment, attention, sampling, or future
experiments, the next observed law is no longer the law that would have been
seen without the rule.  It is useful to write \(P^\pi\) for a law generated
under a rule \(\pi\).  The notation is not decoration: it reminds the analyst that the
procedure has entered the data-generating process.  Recommendation logs,
adaptive trials, active-learning experiments, and autonomous laboratories are
all examples where this feedback is the main statistical fact.

\item[Deployed uncertainty.]
Uncertainty after deployment is wider than a confidence interval for a fixed
parameter.  It includes sampling error, model misspecification, label
construction, measurement drift, censoring, missingness, optimization error,
and the possibility that the target itself has moved.  The task is not to
attach a generic uncertainty band to a powerful model.  The task is to say
which uncertainty matters for the use at hand.
\end{description}

This grammar keeps the finale from becoming a slogan about artificial
intelligence.  A model may be large, adaptive, and multimodal, but the
statistical questions remain concrete: which record was observed, which law
generated it, which law defines the target, and which uncertainty should travel
with the result?

\section{Prediction, Risk, and Feedback Systems}
\conceptindexes{prediction, risk, feedback systems, deployed models, decision systems}

Chapter~8 separated targets from estimators and risks from algorithms.  That
distinction becomes urgent once a procedure is deployed.  A prediction rule is
not only a map \(x\mapsto \hat y\).  It is a rule used by someone, under an
information set, with a loss function, in a population that may change after
the rule is used.  A sequential rule is the same idea with time made explicit:
it maps the current history to the next action.

Recommendation systems show the point in concrete form.  A Netflix-style
rating matrix is not merely a sparse array of preferences
\citep{bennett2007netflix}.  It is a record of exposure, choice, missingness,
ranking, feedback, and later behavior.  The real data are not only ratings.
They are logs.

\begin{example}[A deployment log as a probability object]
Suppose an online service, adaptive trial, or robot laboratory records
\[
  O_t=(H_t,A_t,b_t(A_t\mid H_t),Y_t),\qquad t=1,\ldots,n.
\]
Here \(H_t\) is the history available just before time \(t\), \(A_t\) is the
action actually taken, \(b_t(A_t\mid H_t)\) is the logging probability under
the rule that generated the data, and \(Y_t\) is a later outcome.  The row may
look like engineering bookkeeping, but it is already a probability object.  The
observed law is \(P^b\); the law that would be generated by a candidate rule
\(\pi\) is \(P^\pi\).

One target is the value of \(\pi\),
\[
  V(\pi)=\Expect_{P^\pi}u(Y_t),
\]
for a utility or loss-derived outcome \(u\).  Under consistency, no unrecorded
confounding given \(H_t\), and positivity
\(\pi_t(a\mid H_t)>0\Rightarrow b_t(a\mid H_t)>0\), a one-step version of the
target can be represented under the observed law:
\[
  V(\pi)
  =
  \Expect_{P^b}\!\left[
    \frac{\pi_t(A_t\mid H_t)}{b_t(A_t\mid H_t)}\,u(Y_t)
  \right].
\]
Replacing \(P^b\) by the empirical measure \(P_n\) gives an estimator.  But the
formula is not only an estimator.  It is a map from real data to theory:
conditional laws and kernels describe the logging mechanism; the likelihood
ratio is a Radon--Nikodym derivative; empirical-process bounds control search
over many candidate policies; \(M\)- and \(Z\)-estimation organize
optimization and uncertainty; influence functions give standard errors; and
filtrations enforce the rule that \(A_t\) can depend on \(H_t\), not on
outcomes that arrive later.
\end{example}

This example is deliberately ordinary.  It is the same statistical shape as an
online experiment \citep{kohavi2020trustworthy}, a clinical alert system, an
adaptive dose-finding trial, or a robot acquisition campaign.  Prediction asks
for a function.  Explanation asks what mechanism should be believed.  Use asks
what action follows and what law that action creates.  These questions can
share data and algorithms, but they do not share the same target.

\begin{realdatacapsule}{AlphaGo as policy-generated data}
\item[Data object.] Board histories, legal moves, search statistics, self-play
games, expert games, policy probabilities, and game outcomes
\citep{silver2016mastering}.
\item[Observation mechanism.] The record is generated by a mixture of human
games, self-play, and tree-search-guided action selection.
\item[Target.] A policy value \(V^\pi(s)\), a move distribution
\(\pi(a\mid s)\), or a decision rule that maximizes win probability from the
current state.
\item[Model.] Policy networks, value networks, reinforcement learning, and
Monte Carlo tree search turn histories into actions.
\item[Uncertainty.] Simulation error, search depth, value calibration, and
self-play distribution shift determine how reliable the chosen move is.
\item[Limitation.] The benchmark target is clean because Go has fixed rules and
terminal outcomes; deployed social or clinical policies rarely have that
luxury.
\end{realdatacapsule}

\begin{realdatacapsule}{RFdiffusion as a scientific random object}
\item[Data object.] Protein backbones, residue coordinates, conditioning
motifs, sequence candidates, structure-prediction scores, and experimental
validation records \citep{watson2023denovo}.
\item[Observation mechanism.] Training structures come from curated structural
databases, while proposed designs are generated by a learned denoising process
and filtered before laboratory testing.
\item[Target.] A conditional distribution over structures, a functional design
objective, or the probability that a generated molecule works in the intended
assay.
\item[Model.] A diffusion generative model and structure-prediction network
turn noise and constraints into candidate protein structures.
\item[Uncertainty.] In-silico confidence, diversity checks, assay noise, and
wet-lab replication separate design plausibility from scientific evidence.
\item[Limitation.] A generated structure is not itself the scientific target;
the target is tied to function, mechanism, manufacturability, and validation.
\end{realdatacapsule}

\begin{realdatacapsule}{ChatGPT-style language model in deployment}
\item[Data object.] Prompts, retrieved context, tool calls, model responses,
ratings, edits, refusals, and downstream user actions.
\item[Observation mechanism.] Pretraining corpora, instruction data, human
preference labels, safety filters, product logging, and user selection all
shape the observed record
\citep{vaswani2017attention,brown2020language,ouyang2022training}.
\item[Target.] Benchmark risk, helpfulness, factuality, calibration, task
success, or policy-compliant behavior under a deployment population.
\item[Model.] Transformer language models, instruction tuning, preference
optimization, retrieval, and tool use turn text histories into sequential
actions.
\item[Uncertainty.] Evaluation variance, prompt sensitivity, distribution
shift, label disagreement, and hidden tool failures must be reported with the
score.
\item[Limitation.] A benchmark score is not a deployment estimand; once users
adapt to the system, the model becomes part of the data-generating mechanism.
\end{realdatacapsule}

Even the explanation layer has a target.  A Shapley or SHAP display allocates a
model score, loss, or utility value across features only after a background
distribution and coalition value have been chosen.  The display is therefore
part of the deployed statistical system, not a decorative caption attached to
it.

This is where Chapter~16 returns.  A deployed statistical system is a process:
a clock, an information set, actions, outcomes, and feedback.
In an online platform, the filtration records what the system has already
shown and observed. In a trial, it records enrolled patients, accumulated
outcomes, interim looks, and stopping rules. In a robot laboratory, it records
experiments already run, sensor readings, failures, and the next acquisition
choice. The no-looking-ahead principle is not just a theorem about stopping
times. It is a discipline for deployed analysis: do not evaluate a rule as if
it had access to information that arrived only after the rule was used.

\subsection{Scientific AI and Automated Experimentation}
\conceptindexes{scientific AI, automated experimentation, self-driving laboratories, active learning}

Scientific AI makes the design chapter more important, not less. In chemistry
and drug discovery, active learning and Bayesian optimization choose future
experiments from previous measurements
\citep{vamathevan2019applications,reker2017active,frazier2018tutorial,
shields2021bayesian}. Autonomous laboratories extend this idea into physical
systems: robots run experiments, sensors record results, models update, and an
acquisition rule chooses the next intervention
\citep{burger2020mobile,abolhasani2023rise,xtalpi2025autonomousLab}. The
result is not only a larger dataset. It is a closed-loop stochastic system.

The statistical question in such a system is not only whether the final model
predicts well. It is whether the observation process, action policy, and
scientific target were aligned. Did the system explore enough to learn, or did
it exploit early guesses too aggressively? Did a batch failure become a
scientific signal or a missingness mechanism? Did the robot choose experiments
that are convenient for the instrument rather than informative for the
question? Did uncertainty guide action, or was it computed after the action had
already been optimized away?

This is why the old distinction between method and design becomes porous.
Modern design is often algorithmic, adaptive, and sequential. Modern modeling
is often judged by the experiments it recommends. The statistician's task is to
make the loop legible: measurement, model, uncertainty, action, new
measurement. A closed-loop system without this legibility may be efficient and
still be scientifically confused.

A parallel future appears in physical and biological systems. In
single-molecule biophysics, stochastic thermodynamics, and gene regulation, the
data are trajectories, dwell times, reaction events, fluctuating molecular
states, and noisy cellular time traces. The mathematical language is Markov
jump processes, chemical master equations, nonequilibrium steady states, hybrid
dynamics, autocovariances, power spectra, and path functionals such as flux and
entropy production
\citep{qian2002mesoscopic,qian2006open,qianKou2014single,
jiaQianZhang2024spectrum,golding2005realtime,cai2006stochastic}. This line
belongs in a statistics book because it shows that stochastic processes are not
only probability examples. They are measurement languages for living and
chemical systems whose scientific meaning depends on time, noise,
irreversibility, feedback, and energy flow.

\subsection{Uncertainty in Deployed Systems}
\conceptindexes{uncertainty in deployment, monitoring, calibration, feedback loops}

Uncertainty in a deployed system is not only a standard error. It includes
sampling variation, model misspecification, measurement error, label error,
selection bias, censoring, feedback, distribution shift, optimization error,
and institutional incentives. Some of these uncertainties can be written as
probability. Some appear as sensitivity analyses. Some are design questions.
Some are warnings that the target has changed.

The recurrent examples in this book were chosen to make that point visible. A
cluster of bomb impacts can invite false mechanism; missing species show that
absence may be information; the Zhu/Chu Ko-chen curve shows how historical
climate is reconstructed from proxies; basic economic regions show that a
region can be a latent spatial-economic object; famine relief shows that
reports and actions are filtered by institutions; \emph{Dream of the Red
Chamber} shows that textual evidence becomes statistical only after features,
editions, and validation are specified. Single-cell genomics, clinical event
histories, recommender systems, and autonomous laboratories are modern forms of
the same problem.

More powerful models make the question harder, not easier.  When a model can
fit images, language, molecular assays, and time series at once, the
statistical object becomes easier to confuse with the world itself.  The old
warning remains: the map is not the territory.  The modern addition is that the
map may now move, recommend, intervene, and learn from its own consequences.

\section{Feedback Loops and Recurring Statistical Questions}
\conceptindexes{feedback loop, statistical loop, model use, data generation}

The future of statistics is not a catalog of new domains. It is the return of
the same problem under harder conditions. The unit of analysis may be an
episode, a path, a policy, a simulation run, a laboratory campaign, or a
human--machine interaction rather than a row in a table. Yet the first question
is still the one asked at the beginning of the book: what trace did the world
leave, and through what observation mechanism did that trace become data?

The second question is target and use. A curve, graph, image, event history,
platform log, or simulator output is not a statistical object merely because it
is complicated. It becomes one when a target, loss, comparison, prediction, or
decision is named. A benchmark score, a clinical estimand, a scientific
mechanism, and a deployed policy value are different targets even when the same
model class is used to fit them.

The third question is feedback. Adaptive trials, recommender systems,
autonomous laboratories, and learning platforms produce designed streams rather
than passive samples. The next observation may depend on the procedure
currently being used. The important question is therefore not only what was
observed, but what information was available, what action was chosen, what
could have been chosen, and how that choice changed the future record.
Filtrations, stopping rules, causal estimands, and loss under use are different
views of that single problem.

Simulation fits this loop rather than standing outside it. Digital twins,
mechanistic simulators, agent-based models, and large generative models may
help test procedures before deployment. But simulated evidence is not free
evidence. A simulator has parameters, calibration data, numerical error,
structural assumptions, and blind spots. The statistical question is when the
simulator is an estimator, when it is a prior, when it is a stress test, and
when it is only a story with numbers.

\subsection{Computational validity}
\conceptindexes{computational validity, algorithmic error, MCMC, variational inference, EM algorithm, bootstrap, Monte Carlo, stochastic approximation, particle swarm optimization}

Computation should enter the book, but not as a detached algorithms course.  It
enters whenever the reported statistical object is not the exact mathematical
estimator but an algorithmic approximation to it.  A useful mental
decomposition is
\[
  \text{reported error}
  =
  \text{statistical error}
  +
  \text{algorithmic error}
  +
  \text{model or approximation error}.
\]
The first term is the familiar sampling uncertainty.  The second asks whether
the algorithm actually reached, sampled, or approximated the intended object.
The third asks whether the object being computed is itself only a surrogate for
the scientific target.

\begin{center}
\small
\setlength{\tabcolsep}{0.48em}
\renewcommand{\arraystretch}{1.18}
\begin{longtable}{@{}>{\raggedright\arraybackslash}p{0.24\linewidth}
                  >{\raggedright\arraybackslash}p{0.32\linewidth}
                  >{\raggedright\arraybackslash}p{0.32\linewidth}@{}}
\caption{Computational questions that affect statistical claims\label{tab:ch17-computational-validity}}\\
\toprule
Computational route & What is computed & Statistical question \\
\midrule
\endfirsthead
\caption[]{Computational questions that affect statistical claims (continued)}\\
\toprule
Computational route & What is computed & Statistical question \\
\midrule
\endhead
MCMC &
Monte Carlo averages from a Markov chain
\citep{metropolis1953equation,hastings1970monte} &
Has the chain mixed for the target distribution, and is Monte Carlo error small
relative to posterior or sampling uncertainty? \\

Variational inference &
An optimized approximation inside a tractable family &
What projection is being optimized, and how large is the approximation gap
between the variational target and the posterior or likelihood target? \\

EM algorithm &
Iterates that increase an observed-data likelihood
\citep{dempsterLairdRubin1977em} &
Which local optimum was reached, and does monotonicity of the likelihood imply
anything about the scientific target? \\

Bootstrap and resampling &
A plug-in sampling law based on the observed empirical distribution
\citep{efron1979bootstrap} &
Is the resampling law valid for the statistic, especially under nonsmooth
functionals, dependence, selection, or boundary behavior? \\

Monte Carlo simulation &
Random approximation to an integral, risk, likelihood, or policy value &
How does simulation error combine with sampling error, and is the simulation
law calibrated to the target population? \\

Optimization and stochastic search &
An approximate maximizer, minimizer, saddle point, or best-so-far value from a
random search such as particle swarm optimization
\citep{kennedyEberhart1995pso} &
Are convexity, curvature, initialization, stopping tolerances, and seed
sensitivity strong enough for the statistical expansion to remain valid? \\

Stochastic approximation and SGD &
Noisy recursive updates \citep{robbinsMonro1951stochastic} &
Do step sizes, stability, averaging, and dependence control the gap between the
algorithmic iterate and the estimator or risk minimizer? \\
\bottomrule
\end{longtable}
\end{center}

For MCMC, the measure-theoretic object is the path law induced by an invariant
kernel.  Chapter~6 constructs a transition kernel \(K\) with target invariant
law \(\pi\), and Ionescu--Tulcea then turns the initial law and repeated use of
\(K\) into a probability measure on \(E^{\Nat}\).  The reported computation is
usually an empirical measure
\[
  \Pi_m=\frac1m\sum_{b=1}^m\delta_{X_b}
\]
or an average \(\Pi_m g=m^{-1}\sum_{b=1}^m g(X_b)\).  Under suitable
ergodicity conditions, \(\Pi_m g\) converges to \(\pi g\); under stronger
conditions a Markov-chain central limit theorem replaces the iid variance by a
long-run variance.  Thus an MCMC standard error is not the usual independent
Monte Carlo standard error unless the chain has effectively lost its
dependence.  Burn-in, mixing, thinning, multiple chains, and diagnostics are
practical attempts to control the gap between the path actually simulated and
the invariant law being invoked.

The same reporting issue becomes sharper for modern MCMC.  Hamiltonian and
manifold samplers add numerical integration error and tuning choices;
pseudo-marginal and particle MCMC add auxiliary Monte Carlo noise inside each
transition; stochastic-gradient Langevin methods add minibatch noise and
discretization bias unless corrected; nonreversible samplers change the
geometry of exploration; coupling methods turn convergence assessment into a
statement about a joint path law.  These are not merely software details.  They
decide whether the output is an exact invariant-kernel calculation, an
asymptotically exact approximation, or an approximate algorithm whose error
must be reported alongside sampling uncertainty.

This table should be read together with Chapter~13.  There, an approximate
\(M\)-estimator is allowed to miss the optimum by \(r_n\), and an approximate
\(Z\)-estimator is allowed to miss the root by \(r_n\).  Computational theory
asks whether the algorithmic tolerance, Monte Carlo noise, or approximation gap
is small on the scale of the inferential claim.  If it is not, the computation
has become part of the estimand rather than a harmless implementation detail
\citep{pakes1989simulation}.

Metaheuristics such as particle swarm optimization are best read in this
spirit.  They can often be represented as Markov chains after the algorithmic
state is augmented to include velocities, personal bests, and swarm bests, as
in Chapter~6.  But their output is a stochastic optimizer or best-so-far value,
not a posterior sample with an intended stationary law.  Statistical reporting
therefore has to treat initialization, repeated runs, premature convergence,
and seed sensitivity as part of the computational evidence.

The deepest risk is that prediction will look like understanding. A model may
forecast well because the world is stable, because the system has shaped the
world to match the model, or because the evaluation has forgotten people,
mechanisms, costs, or time scales outside the training record. Future
statistical work will need uncertainty for generated objects, uncertainty for
policies, uncertainty under distribution shift, and uncertainty about whether
the target itself has moved. These are not separate future topics. They are
ways in which the compass can fail if the loop is hidden.

\section{Practical Checklist for Deployment}
\conceptindexes{statistical compass, transparency, assumptions, use}

The statistician of 2050 will need mathematical literacy: probability,
geometry, asymptotics, stochastic processes, optimization, and the ability to
recognize when a theorem's assumptions are doing real work. She will need
computational literacy: simulation, reproducibility, numerical approximation,
software, and the ability to diagnose algorithms rather than merely run them.
She will need domain and institutional literacy: what generated the data, who
was measured, who was excluded, who benefits, who is harmed, and who must act
on the result.

The Chinese line named in Chapter~1 gives one concrete way to read this
combination. Hsu stands for probability and mathematical statistics as deep
structure; Chung for stochastic-process thinking and exposition; Chen for
rigorous asymptotic statistical training; Fang for design as a computational
and scientific act. Their work does not make the future of statistics less
international. It makes the international story less one-sided: theory,
computation, experiment, and scientific judgment have always needed one
another.

Two mathematical source lines matter here only because they keep the compass
usable for richer traces. Infinite-dimensional statistical models supply the
language for unknown functions, signals, densities, inverse problems, and
confidence sets in spaces where the parameter is not a finite vector
\citep{gineNickl2015foundations}. Nonparametric Bayesian inference supplies a
parallel language for priors on functions, distributions, and processes.  The
Dirichlet-process line begins with priors on probability measures
\citep{ferguson1973bayesian,ferguson1974prior}; in modern asymptotic theory,
posterior contraction replaces finite-dimensional normal approximation as the
main large-sample question
\citep{ghosalVdv2017nonparametricBayes,ghosalGhoshVdv2000rates,vaartZanten2008rates}.
Their role in this book is selective: whenever the data object is a function,
curve, density, field, or process, the reader should ask what prior, metric,
testing argument, minimax benchmark, or credible set would make uncertainty
honest for the target under use.

Most of all, she will need judgment. Judgment is not the opposite of
mathematics. It is the ability to know which mathematics a problem deserves.
Sometimes the right move is a randomized design. Sometimes it is a likelihood.
Sometimes it is a sensitivity analysis, a bootstrap, an estimating equation, a
Bayesian hierarchy, a graph, a filtration, or a sequential rule. Sometimes the
right move is to say that the data cannot answer the question as stated.

This is the final sense in which the book is a compass. A compass does not walk
for the traveler. It does not replace terrain, weather, memory, or judgment. It
keeps direction visible. For statistics, that direction is the disciplined
movement from world to measurement, from measurement to data structure, from
data structure to target and model, from model to uncertainty, and from
uncertainty back to action in the world. When that action changes the next
measurement, the route begins again. The future statistician's job is to keep
that loop visible.

\section{Exercises}
\label{sec:ch17-exercises}
\conceptindexes{deployment exercises, feedback exercises, capstone exercises}

\noindent\textbf{Topic families.}
The exercises deliberately span ten frontiers: language-model evaluation,
recommender logs, autonomous laboratories, privacy and federated learning,
medical AI, spatial omics, climate and energy risk, manufacturing monitoring,
financial feedback, and human-in-the-loop fairness.

\begin{exercise}[Deployment log and positivity]
A platform log records \(H_t,A_t,b_t(A_t\mid H_t),Y_t\), where \(b_t\) is the
logging policy.  For a candidate policy \(\pi\), identify the histories on
which positivity fails.  Explain why no amount of sample size can recover the
value of actions that were never tried under comparable histories.
\end{exercise}

\begin{exercise}[Benchmark drift]
For a language model, image model, or molecular-prediction model, describe how
a benchmark can become stale after repeated public use.  Name the original
target law, the post-publication law, and one diagnostic that would reveal
that the benchmark is now partly measuring adaptation to the benchmark rather
than general performance.
\end{exercise}

\begin{exercise}[Monitoring after launch]
Design a post-deployment monitoring plan for a clinical alert, fraud detector,
or risk model.  Include at least one calibration diagnostic, one drift
diagnostic, one harm or guardrail metric, and one rule for deciding when the
model must be retrained or withdrawn.
\end{exercise}

\begin{exercise}[Simulator stress test]
A team wants to use a simulator, digital twin, or agent-based model to test a
policy before deployment.  List the simulator's calibration data, structural
assumptions, numerical approximations, and outputs.  Which uncertainty should
be represented probabilistically, and which should be handled by sensitivity
analysis?
\end{exercise}

\begin{exercise}[Privacy noise and target drift]
In a federated or differentially private analysis, local devices or hospitals
contribute noisy summaries rather than raw data.  State the target population,
the observation law after privacy noise, and one way client heterogeneity can
make the private estimate converge to a target different from the scientific
target.
\end{exercise}

\begin{exercise}[Human-in-the-loop fairness]
In a hiring, lending, clinical-triage, or content-moderation system, a model
routes uncertain cases to human reviewers.  Draw the data-generating loop:
model score, routing decision, human label, action, and later outcome.  Which
fairness target is observable from the reviewed cases, and which target
requires assumptions about cases never reviewed?
\end{exercise}

\begin{exercise}[Autonomous-lab audit]
An autonomous laboratory proposes experiments using an acquisition rule based
on predicted yield, uncertainty, cost, and safety.  Write the history \(H_t\),
the action \(A_t\), the observed outcome \(Y_t\), and the rule \(\pi_t\).  What
would count as evidence of exploration, exploitation, calibration, and
scientific discovery?
\end{exercise}

\begin{exercise}[Final project: one loop, all the way through]
Write a short project proposal for one deployed or deployable statistical
system: a clinical AI tool, a recommender system, an autonomous laboratory, a
single-cell atlas, a financial risk monitor, a precision-agriculture platform,
or a manufacturing sensor network.  The proposal must fill in the full ledger
\[
  \begin{aligned}
  \text{observed record}
    &\to \text{observation mechanism}
    \to \text{probability object}
    \to \text{target}\\
    &\to \text{procedure}
    \to \text{uncertainty}
    \to \text{deployment feedback}.
  \end{aligned}
\]
For each arrow, name what can break.  Also include a validation plan and a
failure mode that would make the target unidentifiable from the available
data.  The point is not to choose the most advanced method, but to keep the
statistical loop visible.
\end{exercise}

\section*{Sources and Further Reading}
\addcontentsline{toc}{section}{Sources and Further Reading}

This closing chapter returns to the historical line introduced in Chapter~1:
data analysis, data science, prediction, and statistical modeling as overlapping
answers to the same problem
\citep{tukey1962future,tukey1977exploratory,donoho2017fifty,
breiman2001two_cultures}. The modern-problem portfolio points back to sources
used throughout the manuscript: Fisher--Corbet--Williams, Good--Turing, and
Efron--Thisted for unseen structure; Lander--Botstein and Benjamini--Hochberg
for high-dimensional genomics and multiplicity; ImageNet, Wasserstein
regression, networks, text, and single-cell examples for complex modern data
structures. The recommender-system and deployment-log discussion is anchored
by the Netflix Prize description of \citet{bennett2007netflix} and the online
experimentation perspective of \citet{kohavi2020trustworthy}. The automated
experimentation and scientific-AI discussion draws on active-learning and
drug-discovery sources
\citep{vamathevan2019applications,reker2017active,frazier2018tutorial,
shields2021bayesian} and on autonomous-laboratory discussions
\citep{burger2020mobile,abolhasani2023rise,xtalpi2025autonomousLab}. The
physical and biological process line is represented by mesoscopic
nonequilibrium thermodynamics, open-system nonequilibrium steady states,
single-molecule biophysics, and analytic stochastic-oscillation models for
single-cell gene expression
\citep{qian2002mesoscopic,qian2006open,qianKou2014single,jiaQianZhang2024spectrum}.
For the high-dimensional data-science side, \citet{fanLiZhangZou2021sfds}
is a useful modern anchor: dimensionality, sparsity, covariance and factor
learning, machine learning, and inference are treated as one statistical
foundation rather than as a loose list of algorithms.
The infinite-dimensional and Bayesian nonparametric source lines are anchored
by \citet{gineNickl2015foundations},
\citet{ferguson1973bayesian}, \citet{ferguson1974prior}, and
\citet{ghosalVdv2017nonparametricBayes}; representative posterior contraction
papers include \citet{ghosalGhoshVdv2000rates} and
\citet{vaartZanten2008rates}.

%% file: chapters/ch13_m_z_estimation.tex
\chapter{M- and Z-Estimation: Consistency and Asymptotic Normality}
\label{chap:m-z-estimation}
\conceptindexes{M-estimation, Z-estimation, consistency, asymptotic normality, argmax, estimating equations, sandwich covariance}

\begin{tcolorbox}[
  enhanced,
  breakable,
  colback=chaptercream,
  colframe=bookblue!88!black,
  boxrule=0.72pt,
  arc=5pt,
  boxsep=1pt,
  left=1.0em,
  right=0.95em,
  top=0.82em,
  bottom=0.82em,
  before skip=0.55\baselineskip,
  after skip=1.0\baselineskip
]
\noindent\textbf{Chapter overview.}
This chapter shows how stable empirical objects choose estimates: as peaks of
random criteria or roots of random equations. Examples include least squares,
maximum likelihood, quantiles, robust regression, clinical-trial estimating
equations, penalized prediction losses, single-cell likelihoods, Weibull
likelihoods, exponential families, James--Stein shrinkage, RKHS smoothing, and
median regression. The
proofs first separate true and false peaks or crossings, then zoom in locally
until random fluctuation becomes asymptotic normality.
\end{tcolorbox}

\begin{center}
\centering
\includegraphics[width=0.96\linewidth]{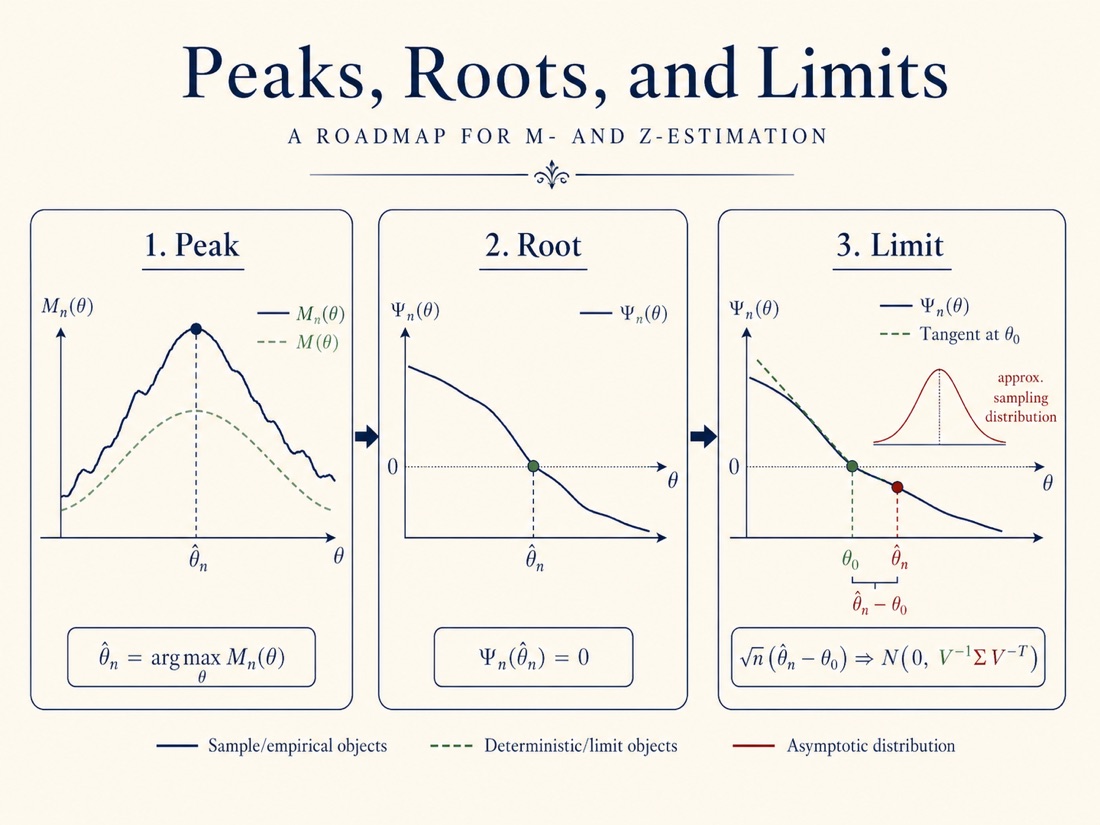}
\bookmanualfigure{fig:ch13-peaks-roots-limits-roadmap}{Peaks, roots, and local limits}
\par\smallskip
\parbox{0.92\linewidth}{\small\textbf{Figure~\thefigure.} A roadmap for the
chapter.  Global arguments first locate estimators as peaks or roots of random
objects; local arguments then turn the remaining displacement into an
asymptotic distribution.}
\end{center}

The preceding chapters studied stability before selection.  A fixed empirical
average can translate data into a population statement; an indexed empirical
field can remain stable even while a procedure scans it.  This chapter uses
that stability to choose.  An estimator is no longer just a number computed
from data.  It is a location extracted from a random landscape: the high point
of a criterion, the low point of a loss, or the crossing point of an empirical
equation.  The next chapter will then zoom in again and ask how the small
remaining displacement should be read.

This placement is deliberate.  Chapter~11 supplies stability for searched
fields; Chapter~8 says what the selected location is meant to estimate;
Chapter~12 shows that some fields are first represented by coordinates such as
functional principal component scores.  M- and Z-estimation now add the
selection step.  They turn a stable field or representation into a point, while
the proofs keep track of what the point is allowed to mean.

\begin{tcolorbox}[
  enhanced,
  breakable,
  colback=noteback,
  colframe=bookgold!75!black,
  boxrule=0.55pt,
  arc=4pt,
  boxsep=1pt,
  left=0.95em,
  right=0.9em,
  top=0.7em,
  bottom=0.7em,
  before skip=0.9\baselineskip,
  after skip=1.0\baselineskip
]
\noindent\textbf{Reading protocol for this chapter.}
For each estimator, ask six questions in order.  What data structure created
the empirical object?  What criterion \(M_n\) or equation \(\Psi_n\) translates
that object into a searchable field?  What population target \(M\) or
\(\Psi\) defines the scientific claim?  Which uniform law prevents a false
peak or false crossing?  Which local expansion converts residual fluctuation
into an error distribution?  Finally, what has been preserved or lost when the
answer is translated back to the original problem?
\end{tcolorbox}

\noindent This chapter is a bridge, not a replacement for a monograph on
M-estimation or Z-estimation.  Its purpose is to give the reader enough
grammar to recognize a statistical procedure as a selected point of a random
field, to see why uniform convergence is needed before optimization is safe,
and to understand why the local expansion in Chapter~14 is the next natural
move, followed by the uncertainty calculus of Chapter~15.  The proofs are
complete where they carry that route; the bibliographic remarks point to
sharper generality.

Across this chapter, data types enter by translation rather than by taxonomy.
The data-object vocabulary of Chapter~\ref{chap:grammar-modern-data-structures}
and the target vocabulary of Chapter~\ref{chap:data-structures-to-targets}
decide what has been observed and what claim is being made; the present
chapter asks what searchable field those choices create.  A table, count
matrix, curve, event history, graph-derived feature, or voxel field belongs
here once the analysis has become an optimization problem or an estimating
equation.  Its special assumptions still matter, but they matter through the
criterion, the law that controls it, and the local expansion that reads it.

\section{M- and Z-Estimators: Definitions and Examples}
\conceptindexes{M-estimators, Z-estimators, maximum likelihood, generalized linear models, robust estimation, cross-entropy, information geometry, KL projection, constrained likelihood}

\begin{definition}[M- and Z-estimators]
Let \(X_n\) be the observed data object and let \(\Theta\) be the space in
which the target value lives.

An \(M\)-estimator is an exact or approximate optimizer of a random criterion
\[
  M_n(X_n,\theta),\qquad \theta\in\Theta .
\]
For maximization, this means that \(\hat\theta_n\) satisfies
\[
  M_n(X_n,\hat\theta_n)
  \ge
  \sup_{\theta\in\Theta}M_n(X_n,\theta)-r_n,
\]
where the tolerance \(r_n\) is negligible on the scale of the statement being
proved.  Minimizers are treated by replacing \(M_n\) with \(-M_n\).  The
population target is usually the optimizer of a deterministic criterion
\[
  M(\theta)=P m_\theta,
  \qquad
  \theta_0\in\argmax_{\theta\in\Theta}M(\theta).
\]

A \(Z\)-estimator is an exact or approximate root of a random estimating
equation
\[
  \Psi_n(X_n,\theta)=0.
\]
In approximate form, \(\hat\theta_n\) satisfies
\[
  \|\Psi_n(X_n,\hat\theta_n)\|\le r_n,
\]
with \(r_n\) again negligible at the relevant scale.  The population target is
usually a root of \(\Psi(\theta)=P\psi_\theta\), so that
\[
  \Psi(\theta_0)=0.
\]
\end{definition}

Alternative names for \(M\)-estimators are minimum-distance, minimum-contrast,
or empirical-risk estimators, depending on the sign convention and field.
Equations arising in \(Z\)-estimation are called estimating equations.  In
smooth problems the two descriptions meet because the derivative of a criterion
is an estimating equation.

The definition is broad on purpose.  Least squares, likelihood, quantiles,
robust regression, penalized empirical risk, and median regression all fit into
the same grammar once their criterion or estimating equation has been named.
The examples below are meant to keep both pictures visible at once.

Every example should therefore be read twice.  The first reading is
substantive: what feature of the world is being estimated?  The second reading
is empirical-process: which indexed class is being scanned, and which
stability result from Chapter 11 allows the selected location to be trusted?
This double reading is what keeps the chapter aligned with the book's opening
rule that methods are translations, not free-floating formulas.

Let $X=(X_1,\ldots,X_n)$ be a random vector taking values in
$\mathcal X=\R^n$ or $(\R^d)^n$.  We assume that the distribution $P$ of $X$
belongs to a parametric model
\[
  \mathcal P=\{P_\theta:\theta\in\Theta\},
  \qquad \Theta\subseteq\R^p,
\]
and that the parametrization is identifiable:
\[
  P_\theta=P_{\theta'} \quad\text{iff}\quad \theta=\theta'.
\]
The hypothetical true value is denoted by $\theta_0$.  $M$- and
$Z$-estimators generalize least-squares and maximum-likelihood estimators.

\begin{example}[Fixed design regression]
The example to keep in mind is linear regression
\[
  \mathbf X_n=\mathbf Z_n\beta_0+\mathbf e_n,
\]
where $\mathbf X_n=(X_1,\ldots,X_n)^T$ is a vector of response variables,
$\mathbf e_n=(e_1,\ldots,e_n)^T$ is a vector of independent error terms, and
$\mathbf Z_n$ is an $n\times p$ fixed design matrix of full column rank.  The
regression coefficient $\beta_0\in\R^p$ is unknown.  For identifiability, assume
\[
  \Expect e_i=0,\qquad 0<\sigma^2=\Var(e_i)<\infty .
\]
The least-squares estimate is
\[
  \hat\beta=(\mathbf Z_n^T\mathbf Z_n)^{-1}\mathbf Z_n^T\mathbf X_n .
\]
If $e_i\sim \Normal(0,\sigma^2)$, then $\mathbf c^T\hat\beta$ is the UMVUE of
$\mathbf c^T\beta_0$ for any $\mathbf c\in\R^p$ and
\[
  \mathbf c^T(\hat\beta-\beta_0)
  \sim \Normal\{0,\sigma^2\mathbf c^T(\mathbf Z_n^T\mathbf Z_n)^{-1}\mathbf c\}.
\]
Least-squares regression does not require normal errors.  By the Gauss-Markov
theorem, $\mathbf c^T\hat\beta$ remains the best linear unbiased estimator of
$\mathbf c^T\beta_0$ under zero mean and the Gauss--Markov covariance condition
$\Var(\mathbf e_n)=\sigma^2I_n$.  The exact distribution is then no longer the
normal distribution displayed above, and the estimator is not necessarily UMVU.
With a known nonspherical covariance matrix, the corresponding BLUE is instead
the generalized least-squares estimator.
\qedmark
\end{example}

In the notation of the definition, least squares is an \(M\)-estimator because,
if \(\mathbf z_{ni}^T\) is the \(i\)th row of \(\mathbf Z_n\), the estimate
minimizes
\[
  \sum_{i=1}^n (x_i-\mathbf z_{ni}^T\beta)^2
\]
with respect to \(\beta\).  It is also a \(Z\)-estimator, since it solves the
normal equation
\[
  \sum_{i=1}^n \mathbf z_{ni}(x_i-\mathbf z_{ni}^T\beta)=0.
\]

\begin{example}[Maximum likelihood and KL projection]
$M$- and $Z$-estimators also generalize maximum likelihood estimators.  Suppose
that $X=(X_1,\ldots,X_n)$ has density $f_{n,\theta}(x)$ with respect to a
$\sigma$-finite dominating measure.  The maximum likelihood estimator maximizes
\[
  L_n(x,\theta)=\log\operatorname{lik}(\theta,x)=\log f_{n,\theta}(x)
\]
with respect to $\theta$.

This is a special case of $M$-estimation.  The population criterion is
connected to Kullback-Leibler divergence.  If two probability laws $P_1$ and
$P_2$ on $(\Omega,\mathcal F)$ satisfy $P_1\ll P_2$, define
\[
  K(P_1,P_2)
  =
  -\Expect_{P_2}\log\frac{dP_1}{dP_2},
\]
and set $K(P_1,P_2)=\infty$ if $P_1$ is not absolutely continuous with respect
to $P_2$.
\end{example}

\begin{theorem}[Kullback--Leibler nonnegativity]
By Jensen's inequality, $K(P_1,P_2)\ge0$, with equality iff
$dP_1/dP_2=1$ $P_2$-almost surely.
\end{theorem}

In particular, if $\mathcal P=\{P_{n,\theta}:\theta\in\Theta\}$ is identifiable
and each $P_{n,\theta}$ has density $f_{n,\theta}$, then the true value
$\theta_0$ is the unique minimizer of
\[
  K(\theta_0,\theta)
  =
  -\Expect_{\theta_0}
  \log\frac{f_{n,\theta}(X)}{f_{n,\theta_0}(X)} .
\]
Although $\theta_0$ is unknown, an empirical version can be minimized.  If
$X_1,\ldots,X_n$ are iid with density $f_\theta$, minimizing
\[
  -\frac1n\sum_{i=1}^n
  \log\frac{f_\theta(X_i)}{f_{\theta_0}(X_i)}
\]
is equivalent to maximizing
\[
  \frac1n\sum_{i=1}^n\log f_\theta(X_i),
\]
because the denominator does not depend on $\theta$.  When the log-likelihood
is Fréchet differentiable, the MLE is often found from the score equation
\[
  U_n(\theta)=0.
\]
Thus MLEs may also be viewed as $Z$-estimators.  Maximizing likelihood and
solving the score equation are not identical operations, since score roots may
be minima or saddle points.  As with least squares, maximum-likelihood
estimators derived from a model may estimate well-defined parameters even when
the model is not exactly true.

\begin{tcolorbox}[
  enhanced,
  breakable,
  colback=noteback,
  colframe=bookgold!75!black,
  boxrule=0.55pt,
  arc=4pt,
  boxsep=1pt,
  left=0.95em,
  right=0.9em,
  top=0.7em,
  bottom=0.7em,
  before skip=0.9\baselineskip,
  after skip=1.0\baselineskip
]
\noindent\textbf{Information-theory reading.}
If \(P\) is the true law and \(Q\) is a fitted law with densities \(p\) and
\(q\), the Shannon entropy of \(P\) is
\[
  H(P)=-P\log p(X),
\]
while the cross-entropy of using \(Q\) to code data generated from \(P\) is
\[
  H(P,Q)=-P\log q(X).
\]
Their difference is the Kullback--Leibler divergence:
\[
  D_{\mathrm{KL}}(P\|Q)=H(P,Q)-H(P).
\]
Since \(H(P)\) does not depend on \(Q\), maximum likelihood is also empirical
cross-entropy minimization.  A classifier trained by negative conditional
log-likelihood is fitting \(p_\theta(y\mid x)\) by the same logic: each
observation pays the code length \(-\log p_\theta(Y_i\mid X_i)\), and the
population target is the conditional law in the fitted family closest in
Kullback--Leibler divergence.  This information-theory language does not
change the estimation theorem; it explains why log-loss, likelihood, and
classification cross-entropy are the same mathematical object wearing
different applied clothing \citep{shannon1948mathematical}.
\end{tcolorbox}

\begin{example}[Generalized linear models as likelihood \(Z\)-estimators]
\label{ex:ch13-glm}
Generalized linear models are a good place to see why likelihood belongs in
this chapter rather than in a separate applied-regression detour
\citep{nelderWedderburn1972glm,mccullaghNelder1989glm}.  They take the
ordinary regression habit from Chapter~1--relating an outcome to covariates--
and adapt it to outcomes that are not well represented by additive Gaussian
noise: binary endpoints, grouped proportions, event counts, sequencing counts,
and rates measured with different exposures.

For one observation, the response distribution has exponential-family form
\[
  p(y;\theta,\phi)
  =
  \exp\left\{
    \frac{y\theta-b(\theta)}{a(\phi)}
    +c(y,\phi)
  \right\}.
\]
Then
\[
  \mu=\Expect(Y)=b'(\theta),
  \qquad
  \Var(Y)=a(\phi)b''(\theta).
\]
A generalized linear model connects the mean to covariates by
\[
  g(\mu_i)=\eta_i=x_i^T\beta ,
\]
where \(g\) is the link function.  For a canonical link, \(\theta_i=\eta_i\),
and the log-likelihood is
\[
  \ell_n(\beta)
  =
  \sum_{i=1}^n
  \left[
    \frac{Y_i x_i^T\beta-b(x_i^T\beta)}{a(\phi)}
    +c(Y_i,\phi)
  \right].
\]
The score equation is
\[
  U_n(\beta)
  =
  \frac1{a(\phi)}
  \sum_{i=1}^n x_i\{Y_i-\mu_i(\beta)\}
  =
  0,
  \qquad
  \mu_i(\beta)=b'(x_i^T\beta).
\]
Thus the GLM estimator is both an \(M\)-estimator, because it maximizes
\(\ell_n(\beta)\), and a \(Z\)-estimator, because it solves a score equation.
The observed curvature is
\[
  -\dot U_n(\beta)
  =
  \frac1{a(\phi)}
  \sum_{i=1}^n x_ix_i^T b''(x_i^T\beta),
\]
which is the Fisher information for \(\beta\) in the canonical-link model.

The examples below are deliberately the same kinds of examples that opened the
book.  GLM is not a new world; it is one way of turning those data structures
into likelihood equations.

\smallskip
\noindent\textit{Binary endpoints and public-risk examples.}
Suppose \(Y_i\in\{0,1\}\) records whether a patient has an adverse event, a
user converts after an online exposure, or a public-risk table records death
within a fixed interval for a person with covariates \(x_i\).  Logistic
regression uses
\[
  \Prob(Y_i=1\mid x_i)=p_i(\beta),
  \qquad
  \log\frac{p_i(\beta)}{1-p_i(\beta)}=x_i^T\beta .
\]
The log-likelihood is
\[
  \ell_n(\beta)
  =
  \sum_{i=1}^n
  \{Y_i x_i^T\beta-\log(1+e^{x_i^T\beta})\},
\]
and the score equation is
\[
  \sum_{i=1}^n x_i\{Y_i-p_i(\beta)\}=0 .
\]
If one component of \(x_i\) is a treatment indicator, its coefficient is a
log-odds ratio conditional on the other covariates.  The statistical target is
not ``the dataset has more ones than zeros.''  It is a conditional law for a
binary trace.  This is the Chapter~1 discipline in likelihood form.

\smallskip
\noindent\textit{Grouped proportions and clinical summaries.}
If \(Y_j\) successes are observed among \(m_j\) comparable units in group
\(j\), a binomial GLM uses
\[
  Y_j\sim\Binomial(m_j,p_j),
  \qquad
  \logit(p_j)=x_j^T\beta .
\]
The score becomes
\[
  \sum_j x_j\{Y_j-m_jp_j(\beta)\}=0 .
\]
This is the grouped version of the same logistic likelihood.  It is useful
when Chapter~1's risk summaries are recorded as counts out of denominators:
cases out of persons at risk, responses out of treated subjects, or failures
out of inspected devices.  The denominator \(m_j\) is not a nuisance detail;
it is part of the observation mechanism.

\smallskip
\noindent\textit{Counts, exposure, and absence as information.}
For event counts or vocabulary counts, a Poisson GLM writes
\[
  Y_i\sim\Poisson(\mu_i),
  \qquad
  \log\mu_i=\log e_i+x_i^T\beta ,
\]
where \(e_i>0\) is an exposure: person-time, sequencing depth, sampled area,
document length, or observation window.  The likelihood is
\[
  \ell_n(\beta)
  =
  \sum_i
  \{Y_i(\log e_i+x_i^T\beta)-e_i e^{x_i^T\beta}-\log(Y_i!)\},
\]
and the score equation is
\[
  \sum_i x_i\{Y_i-e_i e^{x_i^T\beta}\}=0 .
\]
This form is a natural cousin of Chapter~2's missing-species and Shakespeare
vocabulary examples.  A zero count is not merely a blank; it is a likelihood
contribution under a specified exposure.  If one chapter of text is twice as
long as another, or one biological sample has twice the sequencing depth, the
offset \(\log e_i\) prevents length from being mistaken for intensity.

\smallskip
\noindent\textit{Single-cell counts and overdispersion.}
For gene-expression counts, the Poisson GLM is often the first likelihood
grammar but not the final biological model.  Counts can be overdispersed,
zero-inflated, batched, and dependent through cell states.  A negative-binomial
GLM keeps the same mean equation
\[
  \log\mu_{ig}=\log s_i+x_i^T\beta_g
\]
with size factor \(s_i\), but enlarges the variance.  In the language of this
chapter, the score equation still targets a mean trend or perturbation effect;
the variance model controls the curvature and uncertainty.  This is why the
single-cell examples in Chapters~1 and~2 return here as likelihood and
\(Z\)-estimation problems rather than as ordinary least squares.

\smallskip
\noindent\textit{Online and robot-lab feedback.}
In an online experiment, \(Y_i\) may be a conversion indicator and \(x_i\) may
include assignment, user history, and exposure context.  In a robot laboratory,
\(Y_i\) may be a binary success/failure indicator or a count of events during a
run.  A GLM can describe the conditional law at one decision point, but the
feedback system itself is larger: later assignments may depend on earlier
outcomes.  The GLM score is then one local piece of a sequential data
structure.  Chapter~16 supplies the filtration language for that time order,
and Chapter~17 returns to deployment and feedback.

Under standard fixed-design regularity conditions,
\[
  A_n(\beta_0)
  =
  \frac1n\sum_{i=1}^n
  \frac{x_ix_i^T b''(x_i^T\beta_0)}{a(\phi)}
  \to A(\beta_0)
\]
with \(A(\beta_0)\) nonsingular, and a central limit theorem applies to
\[
  \frac1{\sqrt n}\sum_{i=1}^n
  \frac{x_i\{Y_i-\mu_i(\beta_0)\}}{a(\phi)}.
\]
The score expansion gives
\[
  \sqrt n(\hat\beta_n-\beta_0)
  =
  A(\beta_0)^{-1}
  \frac1{\sqrt n}\sum_{i=1}^n
  \frac{x_i\{Y_i-\mu_i(\beta_0)\}}{a(\phi)}
  +o_{\Prob}(1).
\]
This is the same linear representation that drives the local testing chapter
and the influence-function chapter.  A coefficient estimate, a fitted risk
\(g^{-1}(x^T\hat\beta)\), and a predicted rate \(e\exp(x^T\hat\beta)\) are
different reported functionals of the same likelihood root.
\qedmark
\end{example}

\begin{tcolorbox}[
  enhanced,
  breakable,
  colback=noteback,
  colframe=bookgold!75!black,
  boxrule=0.55pt,
  arc=4pt,
  boxsep=1pt,
  left=0.95em,
  right=0.9em,
  top=0.7em,
  bottom=0.7em,
  before skip=0.9\baselineskip,
  after skip=1.0\baselineskip
]
\noindent\textbf{Remark 13.1.1.}
Unless necessary, we suppress the dependence of $M_n$ and $\Psi_n$ on the
actual observation $x$.  If $X$ is defined on $(\Omega,\mathcal F,\Prob)$, then
$M_n(X(\omega),\theta)$ is a stochastic process indexed by $\theta\in\Theta$,
usually denoted simply by $M_n(\theta)$.

In many circumstances $M_n$ has no exact maximizer, or $\Psi_n(\theta)=0$ has
no exact root.  A near maximizer satisfies
\[
  M_n(\hat\theta_n)
  \ge
  \sup_{\theta\in\Theta}M_n(\theta)-o_{\Prob}(1),
\]
and a near root satisfies
\[
  \Psi_n(\hat\theta_n)=o_{\Prob}(1).
\]
\end{tcolorbox}

\begin{example}[Median as an \(M\)-estimator]
This is the robust-estimation reading of the median emphasized by
\citet{huber1964robust}.
Let $X$ have distribution function $F$.  A median $\theta$ of $F$ can be
defined as a minimizer of
\[
  \Expect_F\{|X-\theta|-|X|\}.
\]
For iid $X_1,\ldots,X_n$, the sample median is a minimizer of
\[
  M_n(\theta)
  =
  \frac1n\sum_{i=1}^n \{|X_i-\theta|-|X_i|\}.
\]
For fixed $x$, $m_\theta(x)=|x-\theta|-|x|$ is absolutely continuous in
$\theta$ with derivative $-\sign(x-\theta)$.  Thus the median can also be
viewed as a near solution of
\[
  \Psi_n(\theta)
  =
  -\frac1n\sum_{i=1}^n\sign(X_i-\theta)
  =
  o_{\Prob}(1).
\]

More generally, the $q$th quantile may be represented as a minimizer of
\[
  M_n(\theta)
  =
  \frac1n\sum_{i=1}^n \{m_\theta(X_i)-m_0(X_i)\},
\]
where
\[
  m_\theta(x)
  =
  -\min(x-\theta,0)+\frac{q}{1-q}\max(x-\theta,0).
\]
The derivative exists except at $x=\theta$ and is
\[
  \dot m_\theta(x)=\psi_\theta(x)
  =
  \ind{x<\theta}
  -
  \frac{q}{1-q}\ind{x>\theta}.
\]
The $Z$-equation may be written
\[
  \Psi_n(\theta)
  =
  F_n(\theta-)-\frac{q}{1-q}\{1-F_n(\theta)\}
  =
  o_{\Prob}(1),
\]
where
\[
  F_n(\theta)=\frac1n\sum_{i=1}^n\ind{X_i\le\theta}.
\]
Multiplying by $1-q$ gives
\[
  \widetilde\Psi_n(\theta)
  =
  (1-q)F_n(\theta-)-q\{1-F_n(\theta)\}
  =
  F_n(\theta-)+q\,F_n\{\Delta\theta\}-q,
\]
where
\[
  F_n\{\Delta\theta\}
  =
  \frac1n\sum_{i=1}^n\ind{X_i=\theta}
\]
is the jump of $F_n$ at $\theta$.  Under continuity, the empirical jumps have
size $1/n$ with probability one, so empirical quantiles are near roots.
\qedmark
\end{example}

\begin{example}[Information geometry: MLE as KL projection]
Suppose \(X_1,\ldots,X_n\) take values in a finite alphabet
\(\{1,\ldots,K\}\).  Write
\[
  \hat p_j=\frac1n\sum_{i=1}^n\ind{X_i=j},
  \qquad
  \hat p=(\hat p_1,\ldots,\hat p_K),
\]
and let a smooth model be a \(p\)-dimensional surface
\[
  \mathcal M=\{q_\theta:\theta\in\Theta\}
  \subset
  \Delta_{K-1},
  \qquad
  q_{\theta j}>0,\quad \sum_{j=1}^K q_{\theta j}=1 .
\]
The log-likelihood per observation is
\[
  M_n(\theta)
  =
  \sum_{j=1}^K \hat p_j\log q_{\theta j}.
\]
Since \(\sum_j\hat p_j\log\hat p_j\) does not depend on \(\theta\),
maximizing \(M_n\) is equivalent to minimizing
\[
  D_{\mathrm{KL}}(\hat p\|q_\theta)
  =
  \sum_{j=1}^K
  \hat p_j\log\frac{\hat p_j}{q_{\theta j}} .
\]
Thus the MLE is the KL projection of the empirical distribution onto the model
surface \(\mathcal M\).  This is the finite-alphabet version of the
information-geometric picture: an estimator is the point on a statistical
manifold closest to the empirical object in a directed divergence
\citep{csiszar1975idivergence,amari2016information}.

The same object is also a \(Z\)-estimator.  For
\[
  s_a(j;\theta)=\partial_a\log q_{\theta j},
  \qquad a=1,\ldots,p,
\]
the score equations are
\[
  0
  =
  \partial_a M_n(\theta)
  =
  \sum_{j=1}^K \hat p_j s_a(j;\theta),
  \qquad a=1,\ldots,p.
\]
Equivalently, if
\(\dot q_a(j;\theta)=\partial_a q_{\theta j}=q_{\theta j}s_a(j;\theta)\), then
at an interior solution \(\hat\theta\),
\[
  \sum_{j=1}^K
  \frac{\hat p_j-q_{\hat\theta j}}{q_{\hat\theta j}}
  \dot q_a(j;\hat\theta)
  =
  0,
  \qquad a=1,\ldots,p.
\]
The residual \(\hat p-q_{\hat\theta}\) is therefore orthogonal to every tangent
direction of the model surface under the Fisher inner product
\[
  \langle u,v\rangle_{q}
  =
  \sum_{j=1}^K\frac{u_jv_j}{q_j}.
\]
This is the same Fisher metric that, when integrated along paths on the
statistical manifold, gives Rao's intrinsic distance
\citep{rao1945information,amari2016information}.
For an exponential family
\[
  q_{\theta j}
  =
  \exp\{\theta^TT(j)-K(\theta)\}h_j,
\]
the same equations become the moment-matching condition
\[
  \sum_{j=1}^K \hat p_jT(j)
  =
  \sum_{j=1}^K q_{\hat\theta j}T(j).
\]
The geometry says what the algebra is doing: the fitted model keeps precisely
the features \(T\) that span the model's tangent directions and discards the
remaining empirical texture.
\qedmark
\end{example}

\begin{example}[Robust linear regression and nonlinear regression]
Suppose a real response satisfies
\[
  X=\mathbf Z^T\beta+\sigma e,
\]
where $e$ is independent of $\mathbf Z$ and has known distribution.  More generally,
\[
  X=g_\beta(\mathbf Z)+\sigma e.
\]
Given iid $(X_i,\mathbf Z_i)$, $M$-estimates minimize
\[
  \sum_{i=1}^n
  \rho\!\left(\frac{X_i-g_\beta(\mathbf Z_i)}{\sigma}\right),
\]
where $\rho\ge0$ and $\Expect\rho(e)<\infty$.  Alternatively, $\beta$ and
$\sigma$ may be obtained as near roots of
\[
  \frac1n\sum_{i=1}^n
  w_1\!\left(\frac{X_i-g_\beta(\mathbf Z_i)}{\sigma}\right)\dot g_\beta(\mathbf Z_i)
  =
  o_{\Prob}(1),
\]
and
\[
  \frac1n\sum_{i=1}^n
  w_2\!\left(\frac{X_i-g_\beta(\mathbf Z_i)}{\sigma}\right)
  =
  o_{\Prob}(1).
\]
These estimators can be studied under weaker assumptions than known error
distribution, but stronger identifiability assumptions are then needed.  For
linear LAD regression, $\hat\beta$ minimizes
\[
  \sum_{i=1}^n\{|X_i-\mathbf Z_i^T\beta|-|X_i|\}
\]
or solves the near equation
\[
  \sum_{i=1}^n \mathbf Z_i\sign(X_i-\mathbf Z_i^T\beta)=o_{\Prob}(1).
\]
If the error median is zero, the model may include an intercept.  A scale
parameter can be included when a scale normalization, such as LAD equal to one,
is imposed on the error distribution.
\qedmark
\end{example}

\begin{example}[Clinical trials: generalized estimating equations]
In a longitudinal clinical trial, subject $i$ may contribute a vector
$Y_i=(Y_{i1},\ldots,Y_{im_i})^T$ of repeated outcomes, together with treatment
assignment and baseline covariates.  A marginal model specifies
\[
  \mu_i(\beta)=\Expect(Y_i\mid A_i,X_i),
  \qquad
  g\{\mu_{ij}(\beta)\}=\eta_{ij}(\beta),
\]
where $\beta$ includes, for example, a treatment effect.  Generalized
estimating equations \citep{liang1986longitudinal} solve
\[
  \Psi_n(\beta)
  =
  \frac1n\sum_{i=1}^n
  \dot\mu_i(\beta)^T
  V_i(\beta,\alpha)^{-1}
  \{Y_i-\mu_i(\beta)\}
  =
  0,
\]
where $\dot\mu_i(\beta)$ is the derivative matrix and $V_i(\beta,\alpha)$ is a
working covariance.  This is a clean $Z$-estimation example: the estimating
equation can remain unbiased for the marginal mean parameter even when the
working correlation is imperfect, and the asymptotic covariance then takes the
sandwich form in \eqref{eq:z-sandwich}.
\qedmark
\end{example}

\begin{example}[Instrumental variables as moment equations]
\conceptindexes{instrumental variables, moment equations, generalized method of moments, weak instruments}
Chapter~8 named instrumental variables as an identification device.  Here they
return as \(Z\)-estimation.  In the simplest linear form, observe
\(O_i=(Y_i,A_i,Z_i)\), where \(A_i\) is an exposure or treatment and \(Z_i\) is
an instrument.  A structural slope \(\beta_0\) is encoded by the population
moment
\[
  \Psi(\beta_0)=\Expect\{Z(Y-A\beta_0)\}=0,
\]
with the understanding that the validity of this moment is a causal and
design assumption, not an algebraic fact.  The sample equation
\[
  \Psi_n(\beta)
  =
  \frac1n\sum_{i=1}^n Z_i(Y_i-A_i\beta)=0
\]
gives, when \(Z\) is scalar,
\[
  \hat\beta_{\mathrm{IV}}
  =
  \frac{\sum_i Z_iY_i}{\sum_i Z_iA_i},
\]
after centering or including an intercept as appropriate.  With covariates, the
instrument is usually residualized against them; with several instruments or
several parameters, the moment vector may be overidentified and the generalized
method of moments chooses \(\beta\) by minimizing
\[
  \Psi_n(\beta)^T W_n\Psi_n(\beta)
\]
for a positive semidefinite weight matrix \(W_n\)
\citep{hansen1982large}.

This is a clean example of the book's division of labor.  The target chapter
states relevance, exclusion, exchangeability, monotonicity, or sensitivity
conditions.  This chapter studies the root or minimum-distance point once those
conditions have produced a moment equation.  If the first-stage relation between
\(Z\) and \(A\) is weak, the Jacobian of \(\Psi\) is nearly singular, and the
ordinary local normal approximation can be misleading; the problem is then not
just small sample size, but weak curvature of the estimating equation.
\qedmark
\end{example}

\begin{example}[Language and vision: penalized empirical risk]
Many prediction systems are $M$-estimators once their training objective is
written down.  In sequence labeling for natural language processing,
conditional random fields \citep{lafferty2001conditional} maximize a regularized
conditional log-likelihood
\[
  M_n(\theta)
  =
  \frac1n\sum_{i=1}^n
  \left[
    \theta^TF(x_i,y_i)
    -
    \log\sum_y \exp\{\theta^TF(x_i,y)\}
  \right]
  -
  \frac{\lambda_n}{2}\|\theta\|^2 .
\]
The first-order condition is the $Z$-equation
\[
  \frac1n\sum_{i=1}^n
  \left[
    F(x_i,y_i)-\Expect_\theta\{F(x_i,Y)\mid x_i\}
  \right]
  -
  \lambda_n\theta
  =
  0.
\]
In computer vision, a neural image classifier trained by cross-entropy and
weight decay, as in the ImageNet setting of
\citet{krizhevsky2012imagenet}, minimizes
\[
  \frac1n\sum_{i=1}^n -\log p_\theta(Y_i\mid I_i)+\lambda_nJ(\theta).
\]
This is also penalized $M$-estimation, although the nonconvex and
high-dimensional geometry means that the classical fixed-$p$ theorems in this
chapter are only a starting grammar, not a complete theory.
\qedmark
\end{example}

\begin{example}[\emph{Dream of the Red Chamber}: authorship as criterion comparison]
The stylometric authorship problem from Chapters~2 and~8 also has an
$M$-estimation reading. Let \(X_c\) be a feature vector for chapter \(c\):
function-word frequencies, character or word \(n\)-grams, sentence patterns, or
verse/prose indicators. A one-style model estimates a common style parameter by
minimizing
\[
  \widehat M_1
  =
  \inf_{\theta}
  \sum_{c=1}^{120}\ell_{\theta}(X_c),
\]
whereas a two-style model estimates separate parameters for the first eighty
chapters and the last forty chapters by minimizing
\[
  \widehat M_2
  =
  \inf_{\theta_A,\theta_B}
  \left\{
  \sum_{c=1}^{80}\ell_{\theta_A}(X_c)
  +
  \sum_{c=81}^{120}\ell_{\theta_B}(X_c)
  \right\}.
\]
The contrast \(\widehat M_1-\widehat M_2\), with an appropriate complexity
penalty or resampling calibration, is a model-comparison statistic rather than
a literary verdict. It asks whether the chapter-level features look more stable
under one stylistic law or under a change in stylistic regime. This is the
statistical form of the analyses of \citet{hu2014redchamber},
\citet{tu2013redchamber}, and \citet{zhu2021redchamber}: feature design,
edition choice, segmentation, and validation matter as much as the final
classifier or contrast.
\qedmark
\end{example}

\begin{example}[Single-cell methods: observation, generation, and inference]
The single-cell thread from Chapter 11 can be read through the same $M/Z$ lens.
scImpute
\citep{li2018scimpute} begins at the observation layer: an observed zero in
single-cell RNA-seq is modeled as either biological absence or technical
dropout.  The fitted dropout and expression components can be viewed as
likelihood-based or classification-flavored $M$-estimation; the final imputed
matrix is a downstream functional of that fitted model.

scDesign3 \citep{song2024scdesign3} lives at the generation layer.  Its fitted
simulator estimates marginal expression laws and dependence structure so that
synthetic cells preserve biologically relevant summaries.  The fitting stage is
therefore most naturally a penalized or composite $M$-estimation problem, while
its score equations or penalized first-order conditions give the associated
$Z$-description.  The simulator itself is not a root; it is the generative law
obtained after the roots or optima have been found.

A vine dependence layer gives a concrete criterion
\citep{aas2009pair,dissmann2013selecting}.  After fitting marginal models,
form probability-scale pseudo-observations
\[
  \hat u_{ig}=F_g(x_{ig}\mid c_i;\hat\theta_g),
  \qquad i=1,\ldots,n,\quad g=1,\ldots,G.
\]
For a proposed vine structure \(\mathcal V\), pair-copula families
\(\mathcal B\), and pair parameters
\(\eta=(\eta_e:e\in\mathcal V)\), the dependence fit can be written as the
pseudo-likelihood $M$-estimator
\[
  \hat\eta
  \in
  \operatorname*{arg\,max}_{\eta}
  \sum_{i=1}^n
  \sum_{e\in\mathcal V}
  \log c_e
  \{\hat u_{i,j_e\mid D_e},
    \hat u_{i,k_e\mid D_e};\eta_e\}.
\]
Here \(j_e\) and \(k_e\) are the two conditioned variables on edge \(e\), and
\(D_e\) is the conditioning set determined by the vine.  If the conditional
copulas are allowed to depend on \(u_{D_e}\), that dependence is part of the
edge model; the simplified vine treats it as absent after transforming to
conditional probability scale.  Selecting the tree order, pair families, or a
penalty is then model selection around the same empirical landscape.  For
standard errors, the estimated margins are not free: their first-stage
uncertainty must either be included in the estimating equations or handled by
resampling.

In the PBMC3k capsule used by the book, a small scDesign3 diagnostic compares
reference and synthetic summaries for 30 genes and 500 cells.  The mean
absolute zero-rate difference is 0.00773, while the mean absolute gene-mean
difference is 0.7402.  Those two numbers are deliberately read as estimating
diagnostics, not as biological discoveries: they say which summaries the
fitted generative law currently preserves and which summaries still need
scrutiny before downstream inference.

scGTM \citep{cui2022scgtm} belongs at the inferential layer.  It estimates an
interpretable trend shape along pseudotime by constrained maximum likelihood.
This makes it a useful modern example for the local $M$-estimation results
below: after the trend has been located, Fisher information gives an
approximate covariance for parameters such as activation strength, repression
strength, and change time.
The same real-data audit explains when this example should not be run: the
PBMC3k fallback matrix has no real pseudotime coordinate, so a scGTM-style
trend would be a target invented by the analyst rather than supported by the
record.  In $M$-estimation language, the criterion can be written before the
estimand is justified; the book's discipline is to refuse that shortcut.
\qedmark
\end{example}

\section{Consistency of M and Z Estimators}
\conceptindexes{consistency, M-estimator consistency, Z-estimator consistency, separation, argmax theorem}

Consistency is the first promise an estimator has to keep.  The guiding
picture is simple: if the random criterion $M_n$ comes uniformly close to a
deterministic landscape $M$, and if $M$ has a well-separated summit at
$\theta_0$, then a near-summit of $M_n$ cannot wander far from $\theta_0$.
For estimating equations, the same idea reappears after turning a root into a
minimum of a norm.

This is the first payoff from the empirical-process chapter.  A pointwise law
of large numbers would only say that each fixed location in the landscape is
well estimated.  Selection needs more: the whole landscape must be controlled
well enough that random texture cannot create a convincing false summit or
false crossing after the procedure searches over \(\Theta\).  Thus the uniform
law is not a decorative condition in the theorem below.  It is exactly the
book's translation discipline applied to estimation: the selected point is
credible only if the searched field is stable.

\begin{center}
\centering
\includegraphics[width=0.96\linewidth]{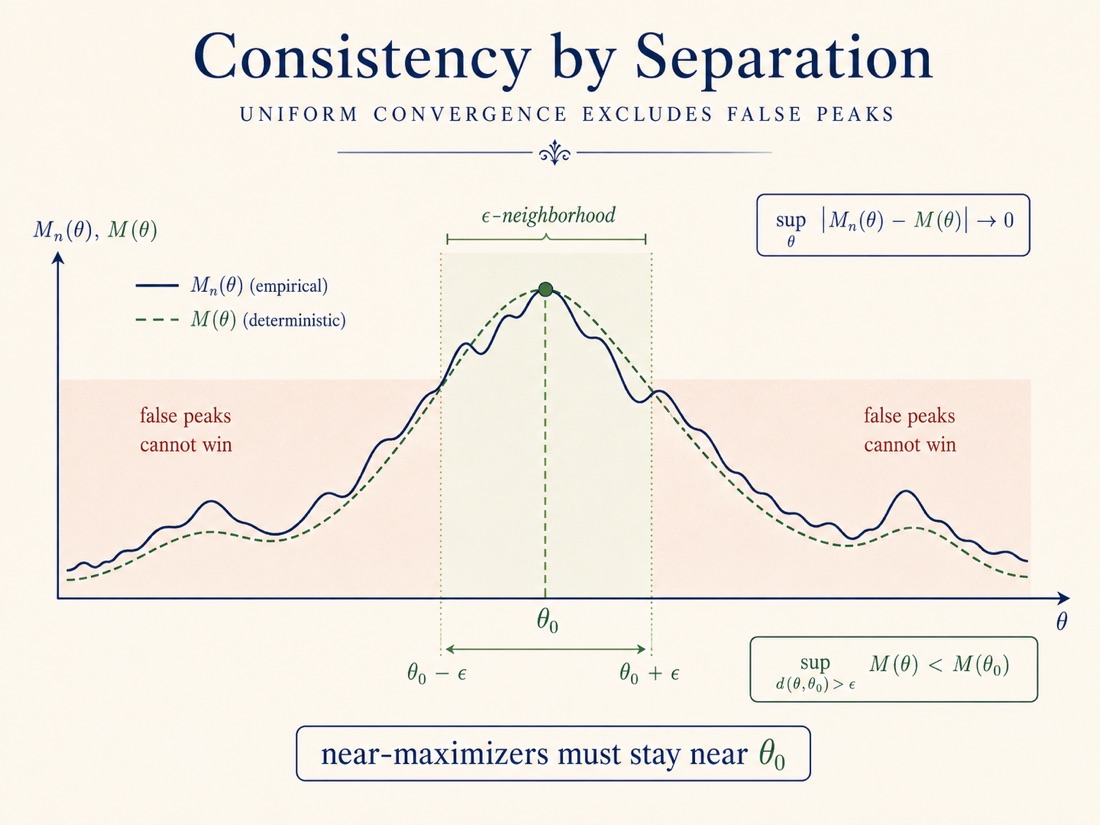}
\bookmanualfigure{fig:ch13-consistency-separation}{Consistency by uniform separation}
\par\smallskip
\parbox{0.92\linewidth}{\small\textbf{Figure~\thefigure.} The separation
picture behind consistency of maximizers.  Uniform convergence keeps the
empirical landscape close to its deterministic limit, while strict separation
away from $\theta_0$ prevents false peaks from winning.}
\end{center}

\subsection{Master Theorem for Global Maximizers}
\conceptindexes{argmax theorem, global maximizers, M-estimation master theorem}

The word ``master'' is used here in the informal sense of a reusable template.
The theorem is not a separate historical object with that exact name; it is the
one argument that will be invoked again and again.  Uniform convergence gives
control of the whole random landscape, while separation of the limiting
landscape keeps all false peaks below the true one.

Assume $\Theta\subseteq\R^d$ and let $d(\theta,\theta')$ be any metric
compatible with the Euclidean topology.

\begin{theorem}[Global argmax consistency]
Suppose that there exists a deterministic function
$M:\Theta\to\R$ such that
\[
  \sup_{\theta\in\Theta}|M_n(\theta)-M(\theta)|
  \toPstar 0.
\]
Assume also that for every $\varepsilon>0$,
\[
  \sup_{\theta:d(\theta,\theta_0)>\varepsilon}M(\theta)
  <
  M(\theta_0).
\]
Equivalently, for every $\varepsilon>0$ there is $\delta>0$ such that
\[
  \{\theta:d(\theta,\theta_0)>\varepsilon\}
  \subset
  \{\theta:M(\theta)<M(\theta_0)-\delta\}.
\]
Then any estimator satisfying
\[
  M_n(\hat\theta_n)\ge M_n(\theta_0)-\opstar(1)
\]
converges in outer probability to $\theta_0$.
\end{theorem}

\noindent\textit{Proof.}
Fix $\varepsilon>0$ and choose $\delta>0$ as above.  Put
\[
  A_n=\{M(\theta_0)-M(\hat\theta_n)>\delta\}.
\]
The event $\{d(\hat\theta_n,\theta_0)>\varepsilon\}$ is contained in $A_n$.
By near-maximality,
\[
  M_n(\hat\theta_n)\ge M_n(\theta_0)-\opstar(1),
\]
and the uniform convergence assumption gives \(M_n(\theta_0)\toPstar M(\theta_0)\).
Hence
\[
  M_n(\hat\theta_n)\ge M(\theta_0)-\opstar(1).
\]
It follows that
\[
  M(\theta_0)-M(\hat\theta_n)
  \le
  \{M_n(\hat\theta_n)-M(\hat\theta_n)\}+\opstar(1),
\]
which is bounded by
\[
  \sup_{\theta\in\Theta}|M_n(\theta)-M(\theta)|+\opstar(1)
  \toPstar 0.
\]
Thus \(\outerProb(A_n)\to0\), and the result follows.
\qedmark

\begin{example}
Suppose
\[
  M_n(\theta)=\frac1n\sum_{i=1}^n m_\theta(X_i),
\]
where $X_i$ are iid.  The uniform convergence condition holds if
\begin{enumerate}
\item $\Theta$ is a bounded Borel subset of $\R^p$;
\item $\Expect_{\theta_0}|m_{\theta_0}(X)|<\infty$;
\item there is an integrable random variable $V(X)$ such that
\[
  |m_\theta(X)-m_{\theta'}(X)|
  \le
  \|\theta-\theta'\|V(X).
\]
\end{enumerate}
The second condition gives pointwise convergence at $\theta_0$, while bounded
$\Theta$ and the Lipschitz envelope give the uniform law of large numbers.
\qedmark
\end{example}

The theorem also applies to $Z$-estimators, since solving
$\Psi_n(\theta)=o_{\Prob}(1)$ is equivalent to near-maximizing
$-\|\Psi_n(\theta)\|$ for any norm on $\R^d$.

\begin{example}[Quantiles]
The $q$th quantile may be defined as a near root of
\[
  \Psi_n(\theta)
  =
  \frac1n\sum_{i=1}^n \psi_\theta(X_i)
  =
  o_{\Prob}(1),
\]
where
\[
  \psi_\theta(x)
  =
  \begin{cases}
    1-q, & x<\theta,\\
    0, & x=\theta,\\
    -q, & x>\theta .
  \end{cases}
\]
Equivalently,
\[
  \widetilde\Psi_n(\theta)
  =
  (1-q)F_n(\theta-)-q\{1-F_n(\theta)\}.
\]
For fixed $\theta$,
\[
  \widetilde\Psi_n(\theta)
  \toP
  \Psi(\theta)
  =
  (1-q)\Prob(X<\theta)-q\Prob(X>\theta).
\]
By the Glivenko-Cantelli theorem,
\[
  \sup_\theta|F_n(\theta)-F(\theta)|\to0
  \quad\text{and}\quad
  \sup_\theta|F_n(\theta-)-F(\theta-)|\to0
  \quad\text{a.s.},
\]
and hence
\[
  \sup_{\theta\in\R}
  |\widetilde\Psi_n(\theta)-\Psi(\theta)|
  \to0
  \quad\text{a.s.}
\]
If $\theta_0$ is the unique $q$th quantile, then for every
$\varepsilon>0$,
\[
  \Psi(\theta_0-\varepsilon)<0<\Psi(\theta_0+\varepsilon),
\]
or equivalently
\[
  \Prob(X<\theta_0-\varepsilon)<q<\Prob(X<\theta_0+\varepsilon).
\]
The master theorem gives consistency of sample quantiles.  This separation
condition holds, for example, if $F$ is continuous and strictly increasing near
$\theta_0$.
\qedmark
\end{example}

\subsection{Huber's Theorem for Scalar Parameters}
\conceptindexes{Huber theorem, scalar parameters, estimating equations, crossing argument}

The global theorem is powerful, but sometimes it asks for more uniform control
than we want to prove.  In one dimension, monotonicity can replace that global
view.  Huber's theorem says that if the limiting equation crosses zero at
$\theta_0$, then any continuous or monotone empirical equation that crosses in
approximately the same way must put its root nearby.

\begin{center}
\centering
\includegraphics[width=0.96\linewidth]{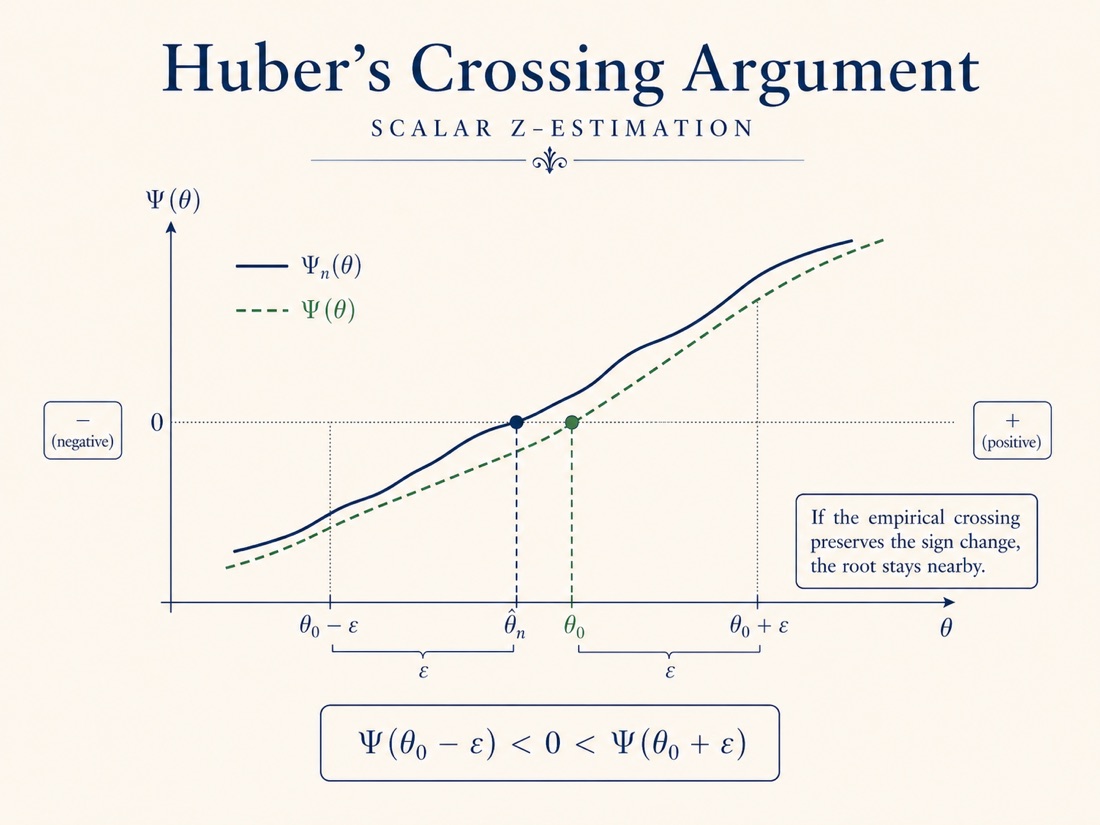}
\bookmanualfigure{fig:ch13-huber-crossing}{Huber's one-dimensional crossing argument}
\par\smallskip
\parbox{0.92\linewidth}{\small\textbf{Figure~\thefigure.} Huber's
one-dimensional crossing argument.  Once the limiting equation has opposite
signs at $\theta_0-\varepsilon$ and $\theta_0+\varepsilon$, pointwise control at
those two endpoints, together with continuity or monotonicity, traps the
empirical root nearby.}
\end{center}

\begin{theorem}[Huber's one-dimensional crossing theorem; \citealp{huber1967behavior}]
Suppose $\Theta$ is a Borel subset of the real line and
\[
  \Psi_n(\theta)\toP\Psi(\theta)
  \quad\text{pointwise in }\theta.
\]
Assume that for each $n$, either $\theta\mapsto\Psi_n(\theta)$ is continuous
and has exactly one zero $\hat\theta_n$, or it is monotone increasing and
$\Psi_n(\hat\theta_n)=o_{\Prob}(1)$.  If $\theta_0$ satisfies
\[
  \Psi(\theta_0-\varepsilon)<0<\Psi(\theta_0+\varepsilon)
  \qquad\text{for all }\varepsilon>0,
  \label{eq:huber-crossing}
\]
then $\hat\theta_n\toP\theta_0$.
\end{theorem}

\noindent\textit{Proof.}
First suppose that $\Psi_n$ is continuous and has exactly one root
$\hat\theta_n$.  Then
\[
  \Prob\{\Psi_n(\theta_0-\varepsilon)<0<
      \Psi_n(\theta_0+\varepsilon)\}
  \le
  \Prob\{\theta_0-\varepsilon<\hat\theta_n<\theta_0+\varepsilon\}.
\]
The left side tends to one by pointwise convergence and
\eqref{eq:huber-crossing}.

Now suppose $\hat\theta_n$ is a near root and $\Psi_n$ is monotone increasing.
Let
\[
  A_n=\{\hat\theta_n<\theta_0-\varepsilon\}.
\]
For any $\eta>0$,
\[
  A_n\subset
  \{|\Psi_n(\hat\theta_n)|>\eta\}
  \cup
  \{\Psi_n(\theta_0-\varepsilon)\ge-\eta\}.
\]
Choose $\eta>0$ so that
$\Psi(\theta_0-\varepsilon)<-2\eta$.  The first probability tends to zero by
near-rootness, and the second by pointwise convergence.  The event
$\{\hat\theta_n>\theta_0+\varepsilon\}$ is handled similarly.
\qedmark

\begin{example}[Quantiles as near roots]
This is the scalar crossing example used by \citet{huber1967behavior}.
For the quantile score in Example 13.2.2, the strong law for binomial random
variables gives
\[
  \Psi_n(\theta)\to\Psi(\theta)
  \quad\text{a.s.}
\]
for each fixed $\theta$.  Moreover, the empirical quantile is a near root with
error of order $O(1/n)$.
\qedmark
\end{example}

\begin{example}[Median survival under right censoring]
\conceptindexes{Kaplan--Meier estimator, median survival, right censoring}
Let \(T_i\) be a survival time and \(C_i\) an independent censoring time.  We
observe
\[
  Y_i=T_i\wedge C_i,
  \qquad
  \Delta_i=\ind{T_i\le C_i},
  \qquad i=1,\ldots,n.
\]
Write \(S(t)=\Prob(T>t)\) and let \(\hat S_n\) be the Kaplan--Meier estimator
\citep{kaplan1958nonparametric}.  Suppose that the target median survival time
\(\theta_0\) is separated in the sense that, for every \(\varepsilon>0\) small
enough,
\[
  S(\theta_0-\varepsilon)>1/2>S(\theta_0+\varepsilon),
\]
and suppose the censoring distribution leaves positive probability at risk in a
neighborhood of \(\theta_0\).  Define
\[
  \Psi_n(t)=\frac12-\hat S_n(t),
  \qquad
  \Psi(t)=\frac12-S(t).
\]
Then \(\Psi_n\) is monotone increasing and, at each continuity point of \(S\)
where the at-risk probability is positive, \(\Psi_n(t)\to\Psi(t)\) in
probability.  The empirical median
\[
  \hat\theta_n=\inf\{t:\hat S_n(t)\le1/2\}
\]
is a near root: its residual is at most the jump size of the Kaplan--Meier
curve near the crossing, which is \(o_{\Prob}(1)\) under the usual local
no-large-jump condition, for instance when the event-time law is continuous and
the censoring survival is positive near \(\theta_0\).  Huber's crossing theorem
therefore gives
\[
  \hat\theta_n\toP\theta_0 .
\]
This is the survival-analysis version of a sample quantile, but the estimator
is now a product-limit curve rather than an empirical cdf.  Confidence intervals
for median survival, such as those of \citet{brookmeyer1982confidence}, are
built on this same crossing picture.
\qedmark
\end{example}

\begin{example}[Profile likelihood estimation]
Let $X_1,\ldots,X_n$ be iid from the Weibull density
\[
  f_{\alpha,\lambda}(x)
  =
  \alpha\lambda x^{\alpha-1}\exp(-\lambda x^\alpha)\ind{x>0},
  \qquad \alpha>0,\ \lambda>0.
\]
The log-likelihood is
\[
  L_n(\alpha,\lambda)
  =
  \sum_{i=1}^n
  \ell_{\alpha,\lambda}(X_i),
\]
where
\[
  \ell_{\alpha,\lambda}(x)
  =
  \log\lambda+\log\alpha+(\alpha-1)\log x-\lambda x^\alpha .
\]
The score equations are
\[
  U_{1n}(\alpha,\lambda)
  =
  \sum_{i=1}^n
  \left\{\frac1\alpha+\log X_i-\lambda X_i^\alpha\log X_i\right\},
\]
and
\[
  U_{2n}(\alpha,\lambda)
  =
  \sum_{i=1}^n
  \left\{\frac1\lambda-X_i^\alpha\right\}.
\]
For fixed $\alpha$ the second equation gives
\[
  \hat\lambda(\alpha)
  =
  \frac{n}{\sum_{i=1}^n X_i^\alpha}.
\]
Substitution into $U_{1n}$ yields
\[
  U_{1n}\{\alpha,\hat\lambda(\alpha)\}
  =
  \frac n\alpha
  +
  \sum_{i=1}^n\log X_i
  -
  n\sum_{i=1}^n w_i(\alpha)\log X_i,
\]
where
\[
  w_i(\alpha)
  =
  \frac{X_i^\alpha}{\sum_{j=1}^n X_j^\alpha}
  =
  \frac{(X_i/X_{n:n})^\alpha}
       {\sum_{j=1}^n (X_j/X_{n:n})^\alpha}.
\]
The profile score is continuous and decreasing in $\alpha$; it tends to
$\infty$ as $\alpha\downarrow0$ and to
\[
  \sum_{i=1}^n\log X_i-n\log X_{n:n}<0
\]
as $\alpha\uparrow\infty$.  Hence it has a unique root $\hat\alpha$, and the
corresponding estimate of $\lambda$ is
\[
  \hat\lambda=\hat\lambda(\hat\alpha).
\]

To prove consistency, suppose the true parameter is
$\theta_0=(\alpha_0,\lambda_0)$.  Then $\lambda_0X^{\alpha_0}$ is standard
exponential.  Also
\[
  \Gamma(t+1)=t\Gamma(t),\qquad
  \Gamma'(t)=\int_0^\infty (\log u)u^{t-1}e^{-u}\,du.
\]
Let $\psi_0(t)=\Gamma'(t)/\Gamma(t)$ be the digamma function.  Then
\[
  \Expect_{\theta_0}X^\alpha
  =
  \Gamma\!\left(\frac{\alpha}{\alpha_0}+1\right)
  \lambda_0^{-\alpha/\alpha_0},
\]
\[
  \Expect_{\theta_0}\log X
  =
  \frac1{\alpha_0}\Gamma'(1)-\frac1{\alpha_0}\log\lambda_0,
\]
and
\[
  \Expect_{\theta_0}\{X^\alpha\log X\}
  =
  \frac1{\alpha_0}
  \left\{
  \psi_0\!\left(\frac{\alpha}{\alpha_0}\right)
  +
  \frac{\alpha_0}{\alpha}
  -
  \log\lambda_0
  \right\}
  \Expect_{\theta_0}X^\alpha .
\]
Therefore
\[
  \Psi_n(\alpha)
  =
  \frac1nU_{1n}\{\alpha,\hat\lambda(\alpha)\}
  \to
  \Psi(\alpha)
  \quad P_{\theta_0}\text{-a.s.},
\]
where
\[
  \Psi(\alpha)
  =
  \frac1\alpha+\Expect_{\theta_0}\log X
  -
  \frac{\Expect_{\theta_0}\{X^\alpha\log X\}}
       {\Expect_{\theta_0}X^\alpha}
  =
  \frac1{\alpha_0}
  \left\{\psi_0(1)-\psi_0\!\left(\frac{\alpha}{\alpha_0}\right)\right\}.
\]
Since the digamma function is strictly increasing and continuous on
$\Rplus$, $\Psi$ has a unique root at $\alpha_0$.  Huber's theorem
\citep{huber1967behavior} gives
\[
  \hat\alpha\xrightarrow{P_{\theta_0}}\alpha_0.
\]
The continuous mapping theorem then gives
\[
  \hat\lambda=\hat\lambda(\hat\alpha)\xrightarrow{P_{\theta_0}}\lambda_0.
\]
\end{example}

\begin{theorem}[Convexity converts pointwise convergence to consistency]
Let $\Theta$ be an open convex subset of $\R^p$ and let $M_n(\theta)$ be a
sequence of real random concave functions on $\Theta$ such that
\[
  M_n(\theta)\toP M(\theta)
\]
for some deterministic function $M$.  Then $M$ is concave and
\[
  \sup_{\theta\in K}|M_n(\theta)-M(\theta)|\toP 0
\]
for every compact $K\subset\Theta$.  In addition, if $\theta_0$ is a unique
maximum of $M$, then any maximizer $\hat\theta_n$ of $M_n$ satisfies
$\hat\theta_n\toP\theta_0$.
\end{theorem}

\noindent\textit{Proof.}
Fix $\theta_1,\theta_2\in\Theta$ and $0\le t\le1$.  Concavity gives
\[
  M_n\{(1-t)\theta_1+t\theta_2\}
  \ge
  (1-t)M_n(\theta_1)+tM_n(\theta_2).
\]
The three terms converge jointly in probability to their deterministic limits,
so passing to the limit gives the same inequality for $M$.  Hence $M$ is
concave.  Since finite concave functions on open convex sets are continuous,
$M$ is continuous on $\Theta$.

We next upgrade pointwise convergence to compact uniform convergence.  Let
$K\subset\Theta$ be compact.  It is enough to show that every subsequence has a
further subsequence along which
\[
  \sup_{\theta\in K}|M_n(\theta)-M(\theta)|\to0
  \quad\text{almost surely}.
\]
Choose a countable dense set $D\subset\Theta$.  From pointwise convergence in
probability and a diagonal argument, every subsequence has a further subsequence
for which $M_n(q)\to M(q)$ almost surely for every $q\in D$.  On this
probability-one event, the deterministic convex-analysis theorem for finite
concave functions says that pointwise convergence on a dense set, with a finite
concave limit on an open domain, implies uniform convergence on compact
subsets.  This proves
\[
  \sup_{\theta\in K}|M_n(\theta)-M(\theta)|\toP0.
\]

It remains to prove convergence of the maximizers.  Let $\varepsilon>0$ and
choose $0<a\le\varepsilon$ such that the closed ball
$\overline B(\theta_0,a)$ is contained in $\Theta$.  By continuity and the
unique-maximizer assumption,
\[
  \Delta
  =
  M(\theta_0)-
  \sup_{\|\theta-\theta_0\|=a}M(\theta)
  >0.
\]
If $\|\hat\theta_n-\theta_0\|>\varepsilon$, define
\[
  \theta_n^*
  =
  \theta_0+
  a\frac{\hat\theta_n-\theta_0}{\|\hat\theta_n-\theta_0\|}.
\]
Convexity of $\Theta$ puts $\theta_n^*$ on the segment between $\theta_0$ and
$\hat\theta_n$, and concavity plus maximality of $\hat\theta_n$ gives
\[
  M_n(\theta_n^*)\ge M_n(\theta_0).
\]
Therefore the event $\{\|\hat\theta_n-\theta_0\|>\varepsilon\}$ is contained in
\[
  \left\{
    \sup_{\|\theta-\theta_0\|=a} M_n(\theta)
    \ge M_n(\theta_0)
  \right\}.
\]
Uniform convergence on the compact sphere together with convergence at
$\theta_0$ makes the probability of this event tend to zero, because on the
event where both approximation errors are smaller than $\Delta/3$ the displayed
inequality is impossible.  Hence $\hat\theta_n\toP\theta_0$.
\qedmark

\begin{example}[Graphical lasso]
Let $\mathbf X_1,\ldots,\mathbf X_n$ be iid mean-zero $p$-vectors with finite
second moments and covariance $\Sigma_0\in\mathbb S^p_{++}$, and suppose $p$ is
fixed.  Write
\[
  S_n=\frac1n\sum_{i=1}^n \mathbf X_i\mathbf X_i^T .
\]
The graphical lasso \citep{yuan2007model,friedman2008sparse} estimates the
precision matrix $\Omega=\Sigma^{-1}$ by maximizing, over the open convex cone
$\mathbb S^p_{++}$,
\[
  M_n(\Omega)
  =
  \log\det\Omega-\tr(S_n\Omega)
  -
  \lambda_n\|\Omega\|_{1,\mathrm{off}},
  \qquad
  \|\Omega\|_{1,\mathrm{off}}
  =
  \sum_{i\ne j}|\Omega_{ij}|.
\]
This is a concave random criterion: $\log\det\Omega$ is strictly concave,
$-\tr(S_n\Omega)$ is linear, and
$-\lambda_n\|\Omega\|_{1,\mathrm{off}}$ is concave because the penalty is
convex.  In fact the criterion is strictly concave, so any maximizer is unique.
If \(S_n\toP\Sigma_0\) and \(\lambda_n\to\lambda\ge0\), then for every fixed
\(\Omega\in\mathbb S^p_{++}\),
\[
  M_n(\Omega)\toP
  M_\lambda(\Omega)
  =
  \log\det\Omega-\tr(\Sigma_0\Omega)
  -
  \lambda\|\Omega\|_{1,\mathrm{off}} .
\]
The theorem therefore upgrades pointwise convergence to compact uniform
convergence on every compact subset of the cone.

It remains only to check that the maximizers exist and do not escape to the
boundary.  The population criterion has a unique maximizer \(\Omega_\lambda\).
Indeed, as \(\lambda_{\min}(\Omega)\downarrow0\),
\(\log\det\Omega\to-\infty\); and as \(\|\Omega\|\to\infty\),
\[
  \tr(\Sigma_0\Omega)\ge\lambda_{\min}(\Sigma_0)\tr(\Omega)\to\infty .
\]
Thus the upper level sets of \(M_\lambda\) are compact subsets of
\(\mathbb S^p_{++}\).  The same localization argument holds for \(M_n\) on the
event
\[
  E_n=\{S_n\succeq cI_p,\ \lambda_n\le C\},
\]
for fixed constants \(0<c<\lambda_{\min}(\Sigma_0)\) and
\(C>\lambda+1\).  Since \(S_n\toP\Sigma_0\) in fixed dimension,
\(\Prob(E_n)\to1\).  Indeed, on \(E_n\),
\[
  M_n(\Omega)\le \log\det\Omega-c\tr(\Omega),
\]
and the right side tends to \(-\infty\) both at the boundary of
\(\mathbb S^p_{++}\) and as \(\|\Omega\|\to\infty\).  Since
\(M_n(\Omega_\lambda)\toP M_\lambda(\Omega_\lambda)\), the maximizer
\(\hat\Omega_n\), defined arbitrarily off \(E_n\), lies with probability tending
to one in a deterministic compact set \(K\subset\mathbb S^p_{++}\) containing
\(\Omega_\lambda\).  Compact uniform convergence on \(K\) and the separated
unique maximum of \(M_\lambda\) then give
\[
  \hat\Omega_n\toP\Omega_\lambda .
\]
When $\lambda=0$, the population maximizer satisfies
\[
  \Omega_0=\Sigma_0^{-1},
\]
so tuning with $\lambda_n\to0$ gives consistency for the true precision matrix.
For a fixed positive penalty, the limit is instead the sparse population target
$\Omega_\lambda$, not generally $\Sigma_0^{-1}$.  This is the important
statistical distinction: concavity gives a clean consistency theorem, but it
shows precisely which target the penalized estimator is consistent for.
\qedmark
\end{example}

\begin{example}[Median regression]
\label{ex:ch13-median-regression-consistency}
Consider
\[
  Y_i=\mathbf X_i^T\beta+\epsilon_i,
\]
where $\epsilon_i$ are iid and $\mathbf X_i$ is a vector of covariates.  Assume the
conditional distribution of $\epsilon_i$ given $\mathbf X_i$ is continuous and has
unique median zero.  The true coefficient $\beta_0$ minimizes
\[
  M(\beta)=\Expect\{|Y_i-\mathbf X_i^T\beta|-|Y_i|\},
\]
and the sample analogue minimizes
\[
  M_n(\beta)
  =
  \frac1n\sum_{i=1}^n
  \{|Y_i-\mathbf X_i^T\beta|-|Y_i|\}.
\]
Assume
\begin{enumerate}
\item $(Y_i,\mathbf X_i)$ are iid;
\item $\Expect(\mathbf X_i\mathbf X_i^T)$ is finite positive definite;
\item the conditional distribution of the errors is continuous and has
      unique median zero $P_{\mathbf X}$-a.s.
\end{enumerate}
Putting $h=\beta-\beta_0$,
\[
  M_n(\beta)-M_n(\beta_0)
  =
  \frac1n\sum_{i=1}^n
  \{|\epsilon_i-\mathbf X_i^T h|-|\epsilon_i|\}.
\]
By the strong law this converges to
\[
  M(\beta)-M(\beta_0)
  =
  \Expect\{|\epsilon_i-\mathbf X_i^T h|-|\epsilon_i|\}.
\]
The median-zero assumption makes this difference nonnegative, and it is
strictly positive for $\beta\ne\beta_0$ under the positive-definiteness
condition.  Since the criterion is convex, the convexity theorem of
\citet{rockafellar1970convex} and its stochastic form in
\citet{andersen1982cox} give consistency of any minimizer.
\qedmark
\end{example}

\begin{example}[Logistic likelihood and separation]
\label{ex:ch13-logistic-separation}
Let fixed covariates $\mathbf x_i\in\R^p$ have full column rank and let
$Y_i\in\{0,1\}$ be independent with
\[
  P_\beta(Y_i=1\mid \mathbf x_i)
  =
  \pi_i(\beta)
  =
  \frac{\exp(\mathbf x_i^T\beta)}
       {1+\exp(\mathbf x_i^T\beta)} .
\]
The normalized log-likelihood is
\[
  M_n(\beta)
  =
  \frac1n\sum_{i=1}^n
  \{Y_i\mathbf x_i^T\beta-\log(1+\exp(\mathbf x_i^T\beta))\}.
\]
It is a concave $M$-criterion, with score
\[
  \Psi_n(\beta)=\dot M_n(\beta)
  =
  \frac1n\sum_{i=1}^n\mathbf x_i\{Y_i-\pi_i(\beta)\}.
\]
Because the Hessian is
\[
  \ddot M_n(\beta)
  =
  -\frac1n\sum_{i=1}^n
  \pi_i(\beta)\{1-\pi_i(\beta)\}\mathbf x_i\mathbf x_i^T ,
\]
full column rank makes the criterion strictly concave.  Thus any finite
maximizer is unique and is also the unique finite root of the score equation.
\end{example}

\begin{theorem}[Logistic separation]
In the fixed-design logistic model of Example~\ref{ex:ch13-logistic-separation},
\citep{silvapulle1981existence,albert1984existence}, the following are
equivalent:
\begin{enumerate}
\item $M_n$ has a finite maximizer in $\R^p$;
\item the score equation $\Psi_n(\beta)=0$ has a solution in $\R^p$;
\item there is no nonzero vector $a\in\R^p$ such that
\[
  a^T\mathbf x_i\ge0\quad\text{whenever }Y_i=1,
  \qquad
  a^T\mathbf x_i\le0\quad\text{whenever }Y_i=0 .
\]
\end{enumerate}
\end{theorem}

The third condition is the familiar absence of complete or quasi-complete
separation.  If such a vector $a$ exists, then
$t\mapsto M_n(\beta+ta)$ is nondecreasing for every fixed $\beta$ and converges
to a finite limit as $t\to\infty$; the likelihood peak has moved to infinity.
Equivalently, the score is trying to send fitted probabilities to $1$ for one
side of the separating hyperplane and to $0$ for the other side, so no finite
root can solve the estimating equation.

This is the convex-support theorem for exponential families
\citep{bickel1977mathematical} in a form that practitioners actually meet:
separation appears in small clinical trials, rare-event safety endpoints, and
high-dimensional binary classifiers whenever a covariate pattern predicts the
response too perfectly.  Bias-reduced or penalized likelihoods, such as
Firth's correction \citep{firth1993bias,heinze2002solution}, deliberately
change the objective so that the optimizer remains finite.
\qedmark

\subsection{Penalized M-Estimation}
\conceptindexes{penalized M-estimation, shrinkage, James--Stein estimator, lasso, ridge regression, smoothing penalties, graphical lasso, RKHS regression, nonparametric regression, nonlocal priors}

Many modern criteria include a penalty.  With the normalization used in this
chapter, write
\[
  M_n^{\mathrm{pen}}(\theta)
  =
  \frac1n\sum_{i=1}^n m_\theta(X_i)-\rho_nJ(\theta),
  \qquad
  \hat\theta_n\in\argmax_\theta M_n^{\mathrm{pen}}(\theta).
\]
This is still $M$-estimation: the landscape has merely acquired an additional
deterministic ridge or barrier.  If $\rho_n\to\rho$, the limiting criterion is
\[
  M^{\mathrm{pen}}(\theta)=M(\theta)-\rho J(\theta).
\]
Thus consistency targets the maximizer of $M-\rho J$.  If the scientific target
is the unpenalized maximizer $\theta_0$, one usually needs $\rho_n\to0$,
together with enough control of $J$ on the region where the estimator lives, so
that the penalty vanishes at the consistency scale.

\begin{center}
\centering
\includegraphics[width=0.96\linewidth]{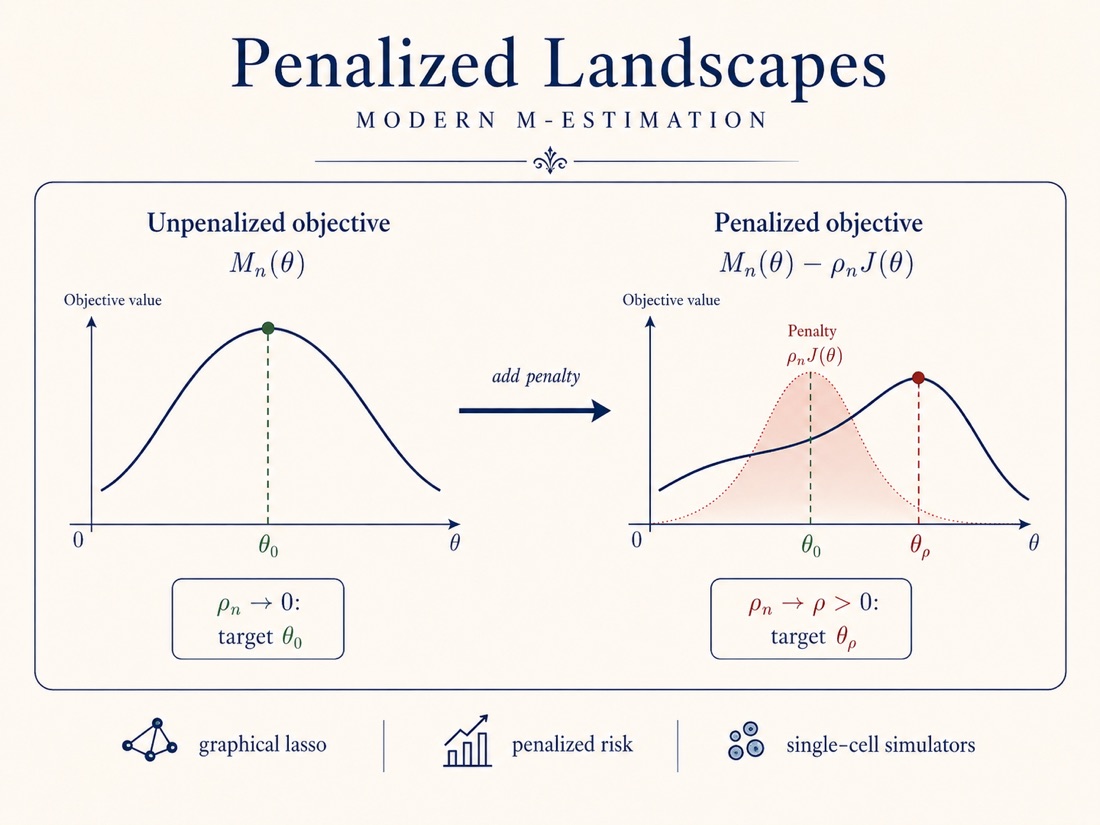}
\bookmanualfigure{fig:ch13-penalized-landscape}{Penalization changes the optimized landscape}
\par\smallskip
\parbox{0.92\linewidth}{\small\textbf{Figure~\thefigure.} Penalization changes
the landscape being optimized.  If the penalty weight vanishes, the
unpenalized target can remain the asymptotic target; if the weight converges to
a positive value, the limiting target is the maximizer of the penalized
criterion.}
\end{center}

At the local scale the size and smoothness of the penalty matter.  If
$J$ is differentiable at $\theta_0$ and $\sqrt n\,\rho_n\to\rho_\infty$, then
the first-order expansion becomes
\[
  \sqrt n(\hat\theta_n-\theta_0)
  =
  -V^{-1}
  \left\{
    \frac1{\sqrt n}\sum_{i=1}^n\dot m_{\theta_0}(X_i)
    -\rho_\infty\dot J(\theta_0)
  \right\}
  +o_{P_0}(1),
\]
with the sign convention for maximization.  If
$\sqrt n\,\rho_n\to0$, the penalty has no first-order effect; if
$\sqrt n\,\rho_n$ tends to a nonzero constant, it creates an asymptotic bias.
For nonsmooth penalties such as the lasso \citep{tibshirani1996regression},
the first-order condition is better written as a KKT inclusion,
\[
  0\in
  \frac1n\sum_{i=1}^n\dot m_{\hat\theta_n}(X_i)
  -
  \rho_n\,\partial J(\hat\theta_n),
\]
so the local limit may contain thresholding, model selection, or boundary
behavior rather than a single ordinary normal law.

The same idea has an older decision-theoretic face: shrinkage.  A shrinkage rule
deliberately moves an unstable empirical estimate toward a simpler point, model,
or subspace.  Ridge regression \citep{hoerlKennard1970ridge}, lasso
\citep{tibshirani1996regression}, empirical-Bayes posterior means, and smoothing
penalties are all different ways to trade bias for risk, stability, or
selection.  The point is not that bias is harmless.  The point is that the target
must say which risk is being optimized.

\begin{example}[James--Stein shrinkage as a risk warning]
\label{ex:ch13-james-stein}
Let \(Y\sim N_p(\theta,\sigma^2 I_p)\), with \(p\ge3\), and measure loss by
\[
  L(\theta,\delta)=\norm{\delta(Y)-\theta}^2,
  \qquad
  R(\theta,\delta)=\Expect_\theta L(\theta,\delta).
\]
The usual estimator \(\delta_0(Y)=Y\) is unbiased and has constant risk
\[
  R(\theta,\delta_0)=p\sigma^2.
\]
The James--Stein estimator \citep{stein1956inadmissibility,jamesStein1961estimation}
shrinks \(Y\) toward the origin:
\[
  \delta_{\mathrm{JS}}(Y)
  =
  \left(1-\frac{(p-2)\sigma^2}{\norm{Y}^2}\right)Y .
\]
For \(p\ge3\), this rule has no larger risk than \(\delta_0\) at any
\(\theta\), and strictly smaller risk at every \(\theta\) for which the
expectation below is finite in the usual way:
\[
  R(\theta,\delta_{\mathrm{JS}})
  =
  p\sigma^2
  -
  (p-2)^2\sigma^4\,
  \Expect_\theta\frac1{\norm{Y}^2}
  <
  p\sigma^2 .
\]
Thus the coordinatewise unbiased estimator is inadmissible under squared-error
risk.  The positive-part rule
\[
  \delta_{\mathrm{JS}}^+(Y)
  =
  \left(1-\frac{(p-2)\sigma^2}{\norm{Y}^2}\right)_+Y
\]
improves the ordinary James--Stein rule further
\citep{baranchik1970family}.
\end{example}

\paragraph{Risk calculation.}
For a rule of the form
\[
  \delta_a(Y)=\left(1-\frac{a}{\norm{Y}^2}\right)Y
  =
  Y+g(Y),
  \qquad
  g(y)=-a\,y/\norm y^2,
\]
Stein's identity gives
\[
  R(\theta,\delta_a)
  =
  p\sigma^2
  +
  \Expect_\theta\left\{
    \norm{g(Y)}^2+2\sigma^2\nabla\cdot g(Y)
  \right\}.
\]
Since
\[
  \norm{g(y)}^2=\frac{a^2}{\norm y^2},
  \qquad
  \nabla\cdot g(y)=-a\,\frac{p-2}{\norm y^2},
\]
we obtain
\[
  R(\theta,\delta_a)
  =
  p\sigma^2
  +
  \{a^2-2a\sigma^2(p-2)\}
  \Expect_\theta\frac1{\norm{Y}^2}.
\]
Taking \(a=(p-2)\sigma^2\) gives the displayed improvement.
\qedmark

This example is small but philosophically important for the whole chapter.
Penalization and shrinkage do not merely make an optimization problem easier.
They may change which estimator has better risk.  They also change the target:
with a fixed penalty, the limiting object is the penalized or shrunken target;
with a vanishing penalty, the unpenalized target may return but only after a
rate calculation.  Classical shrinkage is therefore the bridge from normal means
to modern regularization, empirical Bayes, and high-dimensional prediction.

This is the right way to place penalized generalized additive models and
single-cell simulators.  For example, the marginal-model fitting inside
scDesign3 can be described as penalized
$M$-estimation, with smoothing penalties of the kind studied for GAMs by
\citet{wood2011fast,wood2017generalized}.  Once smoothing parameters are fixed
or estimated as nuisance quantities, the penalized score or KKT equations give
the corresponding $Z$-view.  In the text, it is usually clearest to call
scDesign3 an $M$-estimation example at the model-fitting stage and a simulator
after the fitted generative law is used to create synthetic cells.

\begin{example}[RKHS and nonparametric regression]
\conceptindexes{RKHS, kernel ridge regression, representer theorem, sieve M-estimation}
Let \(\mathcal H\) be a reproducing-kernel Hilbert space on a covariate space
\(\mathcal X\), with kernel \(K\).  Given iid pairs
\((X_i,Y_i)\), kernel ridge regression, or spline smoothing in RKHS form
\citep{wahba1990spline}, estimates a regression function by
\[
  \hat f_{\lambda_n}
  \in
  \argmin_{f\in\mathcal H}
  \left\{
    \frac1n\sum_{i=1}^n\{Y_i-f(X_i)\}^2
    +
    \lambda_n\norm{f}_{\mathcal H}^2
  \right\}.
\]
Equivalently, it maximizes the negative of this criterion, so it is a
penalized \(M\)-estimator with an infinite-dimensional parameter \(f\).  If
\(\lambda_n\to\lambda>0\), the population target is
\[
  f_\lambda
  =
  \argmin_{f\in\mathcal H}
  \left\{
    \mathbb E\{Y-f(X)\}^2+\lambda\norm{f}_{\mathcal H}^2
  \right\},
\]
not necessarily the unpenalized conditional mean.  If the scientific target is
the unpenalized regression function, the tuning sequence must vanish and the
usual uniform-law or entropy conditions must control the growing effective
complexity of \(\mathcal H\).

The representer theorem \citep{kimeldorf1971some,scholkopf2001generalized}
explains why this infinite-dimensional \(M\)-problem is computationally finite.
Let
\[
  \mathcal S_n=\operatorname{span}\{K(X_1,\cdot),\ldots,K(X_n,\cdot)\}.
\]
Decompose \(f=f_\parallel+f_\perp\), with \(f_\parallel\in\mathcal S_n\) and
\(f_\perp\perp\mathcal S_n\).  By the reproducing property,
\[
  f_\perp(X_i)=\langle f_\perp,K(X_i,\cdot)\rangle_{\mathcal H}=0,
  \qquad i=1,\ldots,n.
\]
Thus the empirical loss depends only on \(f_\parallel\), while
\(\norm{f}_{\mathcal H}^2=\norm{f_\parallel}_{\mathcal H}^2+
\norm{f_\perp}_{\mathcal H}^2\).  Any minimizer therefore has
\(f_\perp=0\) and can be written as
\[
  \hat f_{\lambda_n}(x)=\sum_{i=1}^n a_i K(X_i,x).
\]
This is the same statistical geometry as finite-dimensional penalized
\(M\)-estimation, but now the penalty both regularizes the target and turns an
infinite-dimensional nonparametric regression problem into a finite
optimization problem.  More general nonparametric regression estimators can be
viewed similarly as sieve \(M\)-estimators, with \(\mathcal H\) replaced by a
sequence of finite-dimensional or increasingly rich function classes.
\qedmark
\end{example}

\section{Asymptotic Normality of M and Z Estimates}
\conceptindexes{asymptotic normality, local expansion, sandwich covariance, Hessian, score}

After consistency has located the estimator, asymptotic normality zooms in on
the local geometry around $\theta_0$.  At this scale the random equation is
replaced by its tangent line: stochastic fluctuation supplies the intercept,
and curvature or slope supplies the matrix that converts that fluctuation into
estimation error.  This is the source of the familiar sandwich covariance.
In the language of Chapter 11, the global field has already been tamed, and
only a shrinking local field remains.
Finite-dimensional central limit theorems may identify the intercept, but
stochastic equicontinuity is what lets the random plug-in location
\(\hat\theta_n\) be replaced by the fixed target \(\theta_0\) inside the
first-order expansion.
The kernel-density preview in Chapter~9 is a useful contrast: there the rate is
$\sqrt{nh_n}$ and the summands change with the bandwidth, so the fluctuation
comes from a triangular-array central limit theorem rather than an ordinary
root-$n$ fixed-parameter M/Z expansion.

\begin{center}
\centering
\includegraphics[width=0.96\linewidth]{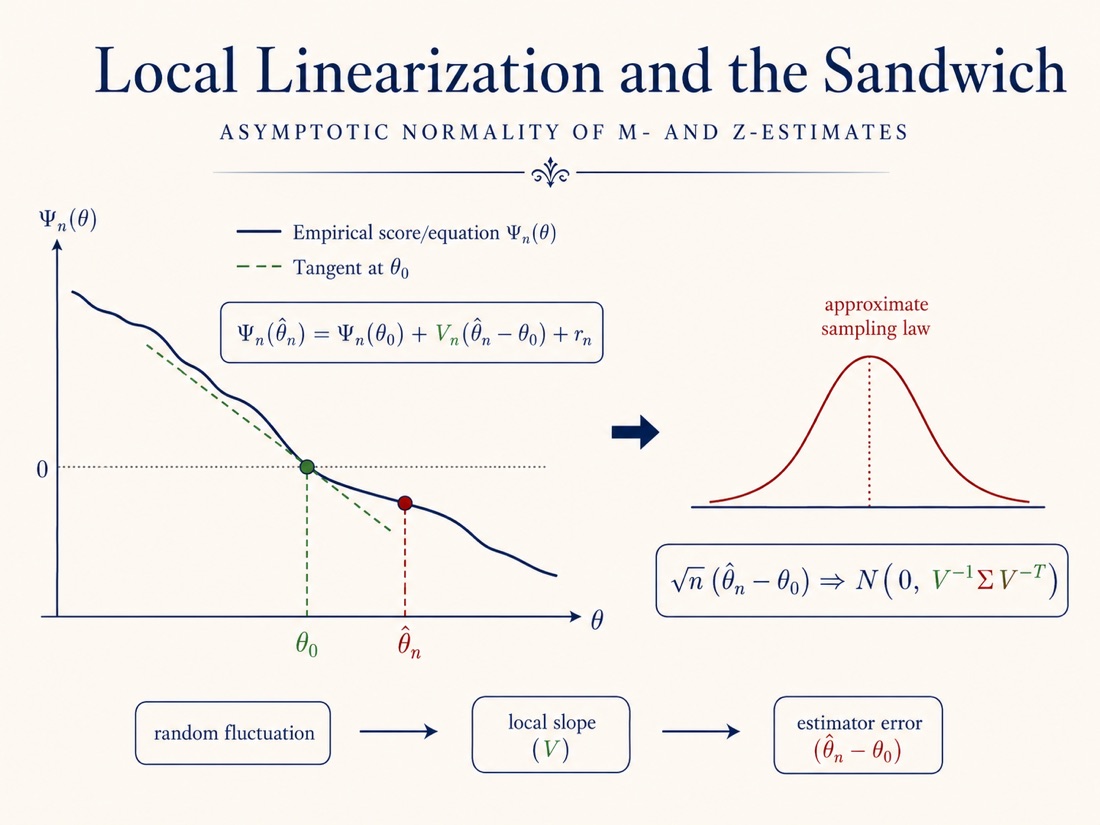}
\bookmanualfigure{fig:ch13-local-sandwich}{Local linearization and sandwich covariance}
\par\smallskip
\parbox{0.92\linewidth}{\small\textbf{Figure~\thefigure.} The local
linearization picture for asymptotic normality.  The empirical fluctuation at
$\theta_0$ is converted into estimator error by the local slope matrix, giving
the usual sandwich covariance in misspecified or estimating-equation problems.}
\end{center}

Let $P_0$ denote the true distribution of the observations.  If the model is
parametric and $\theta_0$ is the true parameter, then $P_0=P_{\theta_0}$.
For $Z$-estimators, the central idea is to approximate $\Psi_n$ near
$\theta_0$ by a linear map:
\[
  \Psi_n(\hat\theta_n)
  =
  \Psi_n(\theta_0)+V_n(\hat\theta_n-\theta_0)
  +r_n(\theta_0,\hat\theta_n).
  \label{eq:z-linearization}
\]
When $\Psi_n$ is differentiable, $V_n$ is typically $\dot\Psi_n(\theta_0)$.
Suppose $\Sigma$ and $V$ are nonsingular and
\[
  \sqrt n\,\Psi_n(\theta_0)
  \weakto
  \Normal(0,\Sigma),
  \qquad
  V_n\toP V.
\]
If $\Psi_n(\hat\theta_n)=o_{P_0}(n^{-1/2})$ and the remainder in
\eqref{eq:z-linearization} is also $o_{P_0}(n^{-1/2})$, then
\[
  \sqrt n(\hat\theta_n-\theta_0)
  =
  -V^{-1}\sqrt n\,\Psi_n(\theta_0)+o_{P_0}(1),
\]
and
\[
  \sqrt n(\hat\theta_n-\theta_0)
  \weakto
  \Normal(0,V^{-1}\Sigma V^{-T}).
  \label{eq:z-sandwich}
\]
For MLEs, $\Sigma$ and $V$ coincide up to sign conventions; for general
estimating equations they need not.

\subsection{Master Theorems for M- and Z-Asymptotic Normality}
\conceptindexes{master theorems, local asymptotic normality, stochastic equicontinuity, fixed points}

There are two recurring ways to manufacture a root with the right local
expansion.  The first uses monotonicity: a monotone field cannot oscillate
wildly, so a local linear approximation pins down the root.  The second uses a
fixed-point map: once the equation is rewritten as $\theta=\phi(\theta)$, a
contraction argument gives existence, uniqueness, and the expansion in one
stroke.

\begin{definition}
A function $f:\R^d\to\R^d$ is called a monotone field iff for every
$\theta_0$ and $s\in\R^d$, the function
\[
  \varepsilon\mapsto
  s^T\{f(\theta_0+\varepsilon s)-f(\theta_0)\}
\]
is monotone increasing in $\varepsilon$.  It is strictly monotone if the
displayed function is strictly increasing.
\end{definition}

Gradients of convex functions are monotone fields.  Monotone score processes
arise naturally in minimization of convex criteria and maximization of concave
criteria.

\begin{theorem}[Monotone Z-estimator limit theorem]
Let $\Psi_n(\theta)$, $\theta\in\R^d$, be a random function which, with
probability tending to one, is a strictly monotone field.  Consider the score
equation $\Psi_n(\theta)=0$.  Suppose
\begin{enumerate}
\item $\sqrt n\,\Psi_n(\theta_0)\weakto \Normal(0,\Sigma(\theta_0))$;
\item for every fixed $h\in\R^d$,
\[
  \sqrt n\,\Psi_n(\theta_0+n^{-1/2}h)
  =
  \sqrt n\,\Psi_n(\theta_0)+V_nh+o_{P_0}(1),
\]
where $V_n\toP V$ and $V$ is positive definite;
\item $\Psi_n$ is differentiable.
\end{enumerate}
Then, with probability tending to one, the estimating equation has a root
$\hat\theta_n$, and
\[
  \sqrt n(\hat\theta_n-\theta_0)
  \weakto
  \Normal(0,V^{-1}\Sigma(\theta_0)V^{-T}).
\]
\end{theorem}

\begin{example}
For the Weibull profile likelihood, the score equation for the shape parameter
is
\[
  \Psi_n(\alpha)
  =
  \frac1\alpha
  +
  \frac1n\sum_{i=1}^n\log X_i
  -
  \frac{A_n(\alpha)}{B_n(\alpha)},
\]
where
\[
  A_n(\alpha)=\frac1n\sum_{i=1}^nX_i^\alpha\log X_i,
  \qquad
  B_n(\alpha)=\frac1n\sum_{i=1}^nX_i^\alpha .
\]
This process is monotone decreasing and has a unique root.  It is not a simple
sum of iid variables, but at $\alpha_0$ it has the iid first-order
approximation
\[
  \sqrt n\,\Psi_n(\alpha_0)
  =
  \frac1{\sqrt n}\sum_{i=1}^n \psi_{\alpha_0}(X_i)+o_{P_0}(1),
\]
where
\[
  \psi_\alpha(X_i)
  =
  \frac1\alpha+\log X_i
  -
  \frac{X_i^\alpha\log X_i}{\Expect B_n(\alpha)}
  +
  \frac{\Expect A_n(\alpha)}{\{\Expect B_n(\alpha)\}^2}X_i^\alpha
  -
  \frac{\Expect A_n(\alpha)}{\Expect B_n(\alpha)}.
\]
At $\alpha_0$, $\Expect\psi_{\alpha_0}(X)=0$, and
\[
  \Sigma(\alpha_0)
  =
  \Var\{\psi_{\alpha_0}(X)\}
  =
  \frac{\pi^2}{6\alpha_0^2}.
\]
The derivative of the profile score converges to
$-\Sigma(\alpha_0)$, giving
\[
  \sqrt n(\hat\alpha-\alpha_0)
  \weakto
  \Normal\{0,\Sigma(\alpha_0)^{-1}\}.
\]
\qedmark
\end{example}

Solving a $Z$-equation can also be viewed as solving a fixed-point problem
$\theta=\phi(\theta)$.  Two standard deterministic results are useful.

\begin{definition}
Let $(\mathcal X,d)$ be a metric space.  A map
$\phi:\mathcal X\to\mathcal X$ is a contraction if there exists
$0\le c<1$ such that
\[
  d\{\phi(x),\phi(y)\}\le c\,d(x,y)
  \qquad x,y\in\mathcal X.
\]
The constant $c$ is the contraction coefficient.
\end{definition}

\begin{theorem}[Banach fixed point theorem]
If $(\mathcal X,d)$ is complete and $\phi$ is a contraction of
$\mathcal X$ into itself, then $\phi(x)=x$ has a unique solution.  Moreover,
starting from any $x_0$, the iterates
\[
  x_{n+1}=\phi(x_n),\qquad n=0,1,\ldots,
\]
converge to the solution.
\end{theorem}

\noindent\textit{Proof.}
For $n\ge1$,
\[
  d(x_{n+1},x_n)\le c\,d(x_n,x_{n-1}),
\]
and hence
\[
  d(x_{n+1},x_n)\le c^n d(x_1,x_0).
\]
If $n<m$,
\[
  d(x_n,x_m)
  \le
  \sum_{k=n+1}^m d(x_k,x_{k-1})
  \le
  \frac{c^n}{1-c}d(x_1,x_0).
\]
Thus $\{x_n\}$ is Cauchy and converges to some $x_\infty$.  Continuity of
$\phi$ gives $\phi(x_\infty)=x_\infty$.  Uniqueness follows from the
contraction inequality.
\qedmark

\begin{theorem}[Schauder fixed point theorem]
Let $C$ be a nonempty compact convex subset of $\R^d$ and let $\phi$ be a
continuous mapping of $C$ into itself.  Then $\phi$ has a fixed point.  If
$\phi$ is a contraction, the fixed point is unique.
\end{theorem}

For closed balls in Euclidean space this is Brouwer's fixed point theorem.  A
version of Schauder's theorem \citep{schauder1930fixpunktsatz} also holds in
Banach spaces.

\begin{theorem}[Local root theorem for Z-estimators]
Let $B(\theta_0,\varepsilon_n)=\{\theta:\|\theta-\theta_0\|\le\varepsilon_n\}$,
where $\varepsilon_n\downarrow0$ and $\sqrt n\,\varepsilon_n\to\infty$.
Suppose
\begin{enumerate}
\item $\sqrt n\,\Psi_n(\theta_0)\weakto \Normal(0,\Sigma(\theta_0))$;
\item $V_n\toP V$;
\item $\Sigma(\theta_0)$ and $V$ are nonsingular;
\item
\[
  \Psi_n(\theta)-\Psi_n(\theta_0)
  =
  V_n(\theta-\theta_0)+r_n(\theta)-r_n(\theta_0),
\]
where
\[
  b_n=
  \sup_{\substack{\theta\ne\theta'\\
       \theta,\theta'\in B(\theta_0,\varepsilon_n)}}
  \frac{\|r_n(\theta)-r_n(\theta')\|}
       {\|\theta-\theta'\|}
  =
  o_{P_0}(1).
  \label{eq:local-lipschitz-rem}
\]
\end{enumerate}
Then, with probability tending to one, $\Psi_n(\theta)=0$ has a unique root
$\hat\theta_n$ in $B(\theta_0,\varepsilon_n)$, and
\[
  \sqrt n(\hat\theta_n-\theta_0)
  \weakto
  \Normal(0,V^{-1}\Sigma(\theta_0)V^{-T}).
\]
\end{theorem}

\noindent\textit{Proof.}
Let
\[
  a_n=\|I-V^{-1}V_n\|+\|V^{-1}\|b_n=o_{P_0}(1),
  \qquad
  A_n=V^{-1}\Psi_n(\theta_0)=O_{P_0}(n^{-1/2}),
\]
and define
\[
  \phi_n(\theta)=\theta-V^{-1}\Psi_n(\theta).
\]
Starting at $\theta_0$, the iterates satisfy
\[
  \|\theta_n^{(m)}-\theta_0\|
  \le
  a_n\|\theta_n^{(m-1)}-\theta_0\|+\|A_n\|.
\]
Thus $\phi_n$ maps the random ball
\[
  B_n=
  \left\{\theta:\|\theta-\theta_0\|\le
  \frac{\|A_n\|}{1-a_n}\right\}
\]
into itself.  Since $\|A_n\|=O_{P_0}(n^{-1/2})$, $a_n=o_{P_0}(1)$, and
$\sqrt n\varepsilon_n\to\infty$, this ball lies inside
$B(\theta_0,\varepsilon_n)$ with probability tending to one.  Schauder's
theorem \citep{schauder1930fixpunktsatz} gives existence.  Moreover,
\[
  \|\phi_n(\theta)-\phi_n(\theta')\|
  \le
  a_n\|\theta-\theta'\|,
\]
so with probability tending to one $\phi_n$ is a contraction on $B_n$; Banach's
theorem \citep{banach1922operations} gives uniqueness.  The fixed-point
equation gives
\[
  \hat\theta_n-\theta_0
  =
  -V^{-1}\Psi_n(\theta_0)+o_{P_0}(n^{-1/2}),
\]
which implies the stated normal limit.
\qedmark

\begin{definition}
An estimator $\tilde\theta_n$ is root-$n$ consistent if
\[
  \sqrt n(\tilde\theta_n-\theta_0)=O_{P_0}(1).
\]
\end{definition}

\begin{theorem}[One-step Z-estimator expansion]
Suppose the assumptions of Theorem 13.3.4 hold and let $\tilde\theta_n$ be
root-$n$ consistent.  Let $\hat V_n\toP V$ and define
\[
  \hat\theta_n
  =
  \tilde\theta_n-\hat V_n^{-1}\Psi_n(\tilde\theta_n).
\]
Then
\[
  \hat\theta_n-\theta_0
  =
  -V^{-1}\Psi_n(\theta_0)+o_{P_0}(n^{-1/2}).
\]
The estimator constructed in this fashion is called a one-step
$M$-estimator.
\end{theorem}

\noindent\textit{Proof.}
By \eqref{eq:local-lipschitz-rem},
\[
  r_n(\tilde\theta_n)-r_n(\theta_0)
  =
  o_{P_0}(\|\tilde\theta_n-\theta_0\|)
  =
  o_{P_0}(n^{-1/2}).
\]
Therefore
\[
\begin{aligned}
  \hat\theta_n-\theta_0
  &=
  \tilde\theta_n-\theta_0
  -\hat V_n^{-1}\Psi_n(\tilde\theta_n)\\
  &=
  (I-\hat V_n^{-1}V_n)(\tilde\theta_n-\theta_0)
  -\hat V_n^{-1}\Psi_n(\theta_0)
  -\hat V_n^{-1}\{r_n(\tilde\theta_n)-r_n(\theta_0)\}\\
  &=
  -V^{-1}\Psi_n(\theta_0)+o_{P_0}(n^{-1/2}).
\end{aligned}
\]
\qedmark

\begin{example}
Let $X_1,\ldots,X_n$ be iid from the Cauchy location density
\[
  p_\theta(x)
  =
  \frac1{\pi\{1+(x-\theta)^2\}}.
\]
The likelihood score equation is
\[
  \Psi_n(\theta)
  =
  \frac1n\sum_{i=1}^n\psi_\theta(X_i),
  \qquad
  \psi_\theta(x)
  =
  \frac{\dot p_\theta(x)}{p_\theta(x)}
  =
  \frac{2(x-\theta)}{1+(x-\theta)^2}.
\]
The score equation can have many roots.  Let $\tilde\theta_n$ be the sample
median.  Since
\[
  \sqrt n(\tilde\theta_n-\theta_0)
  \weakto
  N\!\left(0,\frac1{4p_{\theta_0}(0)^2}\right),
\]
it is root-$n$ consistent.  Also
\[
  \sqrt n\,\Psi_n(\theta_0)
  \weakto
  \Normal\{0,\Expect_{\theta_0}\psi_{\theta_0}(X)^2\}
  =
  \Normal(0,1/2),
\]
and
\[
  \Expect_{\theta_0}\dot\psi_{\theta_0}(X)=-1/2.
\]
Thus a one-step likelihood estimator selects the local efficient root.
\qedmark
\end{example}

\subsection{Iid M- and Z-Estimators}
\conceptindexes{iid observations, M-estimator asymptotic normality, Z-estimator asymptotic normality, influence representation}

The previous theorems are geometric; this subsection translates that geometry
into the iid language used most often in applications.  Instead of differentiating
every sample path of the random score, we differentiate its expectation and use
empirical-process control to show that the random part is locally stable.  This
is the bridge from estimating equations and criteria to practical asymptotic
linear representations.

In practice one does not use all the theorems in this chapter for a single
problem.  The workflow is sequential.  First establish existence and
consistency, using the global argmax theorem, compactness, convexity, separation,
or a problem-specific identifiability argument.  Then choose the local theorem
by the form in which the estimator is naturally reported.  If it is a root of
an equation, use a \(Z\)-theorem; if it is a maximizer of an average criterion,
use an \(M\)-theorem; if the scalar equation is nonsmooth but monotone, use
Huber's crossing theorem; if a rough estimator is improved by one Newton step,
use the one-step theorem.  The output is always the same object: an influence
representation, followed by a covariance formula.

In the preceding results the score process was differentiable in $\theta$.
The following results replace differentiability of the random score by
differentiability of its expectation.  Let
\[
  \Psi_n(\theta)=\frac1n\sum_{i=1}^n\psi_\theta(X_i),
  \qquad
  \Psi(\theta)=\Expect_{P_0}\psi_\theta(X).
\]

\begin{theorem}[Monotone iid Z-estimator expansion; \citealp{huber1967behavior}]
Let $X_1,\ldots,X_n$ be iid with distribution $P_0$.  Let $\theta_0$ be an
isolated root of $\Psi(\theta)=0$, meaning
$|\Expect_{P_0}\psi_{\theta_0\pm\varepsilon}(X)|>0$ for all
$\varepsilon>0$.  Suppose
\begin{enumerate}
\item $\psi_\theta(X)$ is monotone in $\theta$;
\item $\Psi(\theta)$ is differentiable at $\theta_0$ and
      $\dot\Psi(\theta_0)\ne0$;
\item $\Expect_{P_0}\psi_\theta(X)^2<\infty$ and is continuous near
      $\theta_0$.
\end{enumerate}
Then every exact solution, and every near solution satisfying
$\Psi_n(\hat\theta_n)=o_{P_0}(n^{-1/2})$, satisfies
\[
  \hat\theta_n-\theta_0
  =
  -\dot\Psi(\theta_0)^{-1}
  \frac1n\sum_{i=1}^n\psi_{\theta_0}(X_i)
  +o_{P_0}(n^{-1/2}).
\]
In particular,
\[
  \sqrt n(\hat\theta_n-\theta_0)
  \weakto
  \Normal\{0,\sigma^2(\theta_0)\},
  \qquad
  \sigma^2(\theta_0)
  =
  \frac{\Expect_{P_0}\psi_{\theta_0}(X)^2}
       {\dot\Psi(\theta_0)^2}.
\]
\end{theorem}

\noindent\textit{Proof.}
Assume, for definiteness, that $\Psi_n$ is monotone decreasing.  Then for all
$t$,
\[
  P_0\{\Psi_n(t)<0\}
  \le
  P_0\{\hat\theta_n\le t\}
  \le
  P_0\{\Psi_n(t)\le0\}.
\]
Set
\[
  t_n=\theta_0+n^{-1/2}t\,\sigma(\theta_0).
\]
It is enough to show that
\[
  P_0\{\Psi_n(t_n)\le0\}\to\Phi(t)
\]
and similarly with $<$.  Let
\[
  Y_{ni}
  =
  \frac{\psi_{t_n}(X_i)-\Psi(t_n)}{v(t_n)},
  \qquad
  v(\theta)^2=\Var_{P_0}\{\psi_\theta(X)\}.
\]
Then $Y_{ni}$ are iid with mean zero and variance one.  By differentiability,
\[
  \sqrt n\,\Psi(t_n)
  \to
  \dot\Psi(\theta_0)t\sigma(\theta_0),
  \qquad
  v(t_n)\to v(\theta_0),
\]
and the standardized boundary tends to $t$.  Lindeberg's condition follows
from monotonicity: for small $\eta>0$,
$|\psi_{t_n}(X)|$ is bounded by
\[
  Z=\max\{|\psi_{\theta_0-\eta}(X)|,\ |\psi_{\theta_0+\eta}(X)|\},
\]
and $\Expect Z^2<\infty$.  Hence the triangular array CLT applies and proves
the result.
For a near solution with residual $o_{P_0}(n^{-1/2})$, the same bracketing
argument is applied to the events
$\{\Psi_n(t_n)<-\eta_n\}$ and $\{\Psi_n(t_n)\le \eta_n\}$ with
$\sqrt n\,\eta_n\to0$.
\qedmark

\begin{example}[Quantile asymptotic linearity]
This is the iid quantile version of Huber's scalar monotone argument
\citep{huber1967behavior}.
For the $q$th quantile, use
\[
  \psi_\theta(x)
  =
  \begin{cases}
    1-q, & x<\theta,\\
    0, & x=\theta,\\
    -q, & x>\theta .
  \end{cases}
\]
If $\theta_0$ is the $q$th quantile, $F$ is continuous at $\theta_0$, and
$F'(\theta_0)>0$, then
\[
  \Psi(\theta)
  =
  (1-q)F(\theta-)-q\{1-F(\theta)\},
\]
so $\Psi(\theta_0)=0$ and $\dot\Psi(\theta_0)=F'(\theta_0)$.  Also
\[
  \Expect\psi_{\theta_0}(X)^2=q(1-q).
\]
Therefore
\[
  \sqrt n(\hat\theta_n-\theta_0)
  \weakto
  N\!\left(0,\frac{q(1-q)}{F'(\theta_0)^2}\right).
\]
\qedmark
\end{example}

\begin{theorem}[Iid Z-estimator asymptotic linearity]
Let $\Theta_0$ be a bounded open neighborhood of $\theta_0$ such that
\begin{enumerate}
\item $\theta\mapsto\Expect_{P_0}\psi_\theta(X)$ is differentiable at
      $\theta_0$ with nonsingular finite derivative matrix $V$;
\item $\Expect_{P_0}\|\psi_{\theta_0}(X)\|^2<\infty$;
\item there is a random variable $W(X)$ with $\Expect W(X)^2<\infty$ such that
\[
  \|\psi_{\theta_1}(X)-\psi_{\theta_2}(X)\|
  \le
  W(X)\|\theta_1-\theta_2\|
\]
for all $\theta_1,\theta_2\in\Theta_0$.
\end{enumerate}
If $\hat\theta_n\xrightarrow{P_0}\theta_0$ and
$\Psi_n(\hat\theta_n)=o_{P_0}(n^{-1/2})$, then
\[
  \sqrt n(\hat\theta_n-\theta_0)
  =
  -V^{-1}\frac1{\sqrt n}\sum_{i=1}^n\psi_{\theta_0}(X_i)
  +o_{P_0}(1).
\]
Thus the estimator is asymptotically normal with covariance
\[
  V^{-1}\Sigma(\theta_0)V^{-T},
  \qquad
  \Sigma(\theta_0)
  =
  \Expect_{P_0}\{\psi_{\theta_0}(X)\psi_{\theta_0}(X)^T\}.
  \label{eq:vander-z-cov}
\]
\end{theorem}

\noindent\textit{Proof.}
The Lipschitz envelope implies the stochastic equicontinuity expansion
\[
  \frac1n\sum_{i=1}^n
  \{\psi_{\hat\theta_n}(X_i)
    -\Expect_{P_0}\psi_{\hat\theta_n}(X)\}
  -
  \frac1n\sum_{i=1}^n\psi_{\theta_0}(X_i)
  =
  o_{P_0}(n^{-1/2})
\]
for every consistent $\hat\theta_n$.  Since
$\Expect\psi_{\theta_0}(X)=0$ and
\[
  \Expect\psi_{\hat\theta_n}(X)
  =
  V(\hat\theta_n-\theta_0)+o_{P_0}(\|\hat\theta_n-\theta_0\|),
\]
near-rootness gives
\[
  o_{P_0}(n^{-1/2})
  =
  \frac1n\sum_{i=1}^n\psi_{\theta_0}(X_i)
  +
  V(\hat\theta_n-\theta_0)
  +
  o_{P_0}(\|\hat\theta_n-\theta_0\|).
\]
The first term is $O_{P_0}(n^{-1/2})$ by the CLT, implying root-$n$
consistency.  Rearranging yields the displayed linear representation.
\qedmark

\begin{theorem}[Iid M-estimator asymptotic linearity]
Let
\[
  M_n(\theta)=\frac1n\sum_{i=1}^n m_\theta(X_i),
\]
with iid $X_i$.  Let $\Theta_0$ be an open neighborhood of $\theta_0$.
Suppose
\begin{enumerate}
\item $\theta\mapsto m_\theta(x)$ is differentiable in quadratic mean at
      $\theta_0$ with derivative $\dot m_{\theta_0}(x)$:
\[
  \Expect_{P_0}
  \{m_\theta(X)-m_{\theta_0}(X)
    -(\theta-\theta_0)^T\dot m_{\theta_0}(X)\}^2
  =
  o(\|\theta-\theta_0\|^2);
\]
\item there is $W$ with $\Expect W^2<\infty$ such that
\[
  |m_\theta(X)-m_{\theta'}(X)|
  \le
  W(X)\|\theta-\theta'\|;
\]
\item $\theta\mapsto\Expect m_\theta(X)$ is twice Fréchet differentiable at
      its maximum $\theta_0$ with nonsingular symmetric second derivative
      matrix $V$.
\end{enumerate}
If $\hat\theta_n\xrightarrow{P_0}\theta_0$ and
\[
  M_n(\hat\theta_n)
  \ge
  \sup_{\theta\in\Theta_0}M_n(\theta)-o_{P_0}(n^{-1}),
\]
then
\[
  \sqrt n(\hat\theta_n-\theta_0)
  =
  -V^{-1}\frac1{\sqrt n}\sum_{i=1}^n\dot m_{\theta_0}(X_i)
  +o_{P_0}(1).
\]
Hence the limit covariance is
\[
  V^{-1}
  \Expect\{\dot m_{\theta_0}(X)\dot m_{\theta_0}(X)^T\}
  V^{-T}.
\]
\end{theorem}

\begin{example}[Bradley--Terry preference utilities]
The Bradley--Terry model is an \(M\)-estimator for pairwise comparisons.  There
are \(K\) items, item \(K\) is used as the baseline, and
\(\beta=(\beta_1,\ldots,\beta_{K-1})\in\R^{K-1}\) records relative utilities.
Let \(b_j=e_j\) for \(j<K\) and \(b_K=0\).  A comparison observes
\[
  O=(A,B,Y),\qquad A\ne B,\qquad Y\in\{0,1\},
\]
where \(Y=1\) means that item \(A\) beats item \(B\).  Put
\[
  X=b_A-b_B.
\]
The model says
\[
  P_{\beta}\{Y=1\mid A,B\}
  =
  p_{\beta}(X)
  =
  \frac{\exp(X^T\beta)}{1+\exp(X^T\beta)}.
\]
The average log-likelihood criterion is
\[
  M_n(\beta)=\frac1n\sum_{i=1}^n m_\beta(O_i),
  \qquad
  m_\beta(O)=Y X^T\beta-\log\{1+\exp(X^T\beta)\}.
\]
At the true value \(\beta_0\),
\[
  \dot m_{\beta_0}(O)=X\{Y-p_{\beta_0}(X)\}.
\]
The population second derivative is
\[
  V
  =
  -\Expect\!\left[
    p_{\beta_0}(X)\{1-p_{\beta_0}(X)\}XX^T
  \right]
  =
  -H.
\]
If the comparison design identifies all utility contrasts, equivalently
\(H\) is nonsingular, and the empirical maximizer is consistent, the iid
\(M\)-estimator theorem gives
\[
  \sqrt n(\hat\beta_n-\beta_0)
  =
  H^{-1}
  \frac1{\sqrt n}\sum_{i=1}^n
  X_i\{Y_i-p_{\beta_0}(X_i)\}
  +o_P(1).
\]
Thus the ranking estimator is asymptotically linear even though the parameter
is not a mean: it is a latent utility vector learned from pairwise wins and
losses.
\qedmark
\end{example}

\begin{example}[Nonlinear dose-response curves]
Let \(O=(T,Y)\), where \(T\) is a dose or time and
\[
  \Expect(Y\mid T)=f_{\theta_0}(T)
\]
for a smooth finite-dimensional response curve \(f_\theta\).  Examples include
Emax curves, growth curves, and pharmacokinetic concentration curves.  Estimate
\(\theta_0\) by nonlinear least squares, written as maximization of
\[
  m_\theta(O)
  =
  -\frac12\{Y-f_\theta(T)\}^2,
  \qquad
  M_n(\theta)=P_nm_\theta.
\]
Assume \(f_\theta(t)\) is continuously differentiable in a neighborhood of
\(\theta_0\), has a square-integrable local Lipschitz envelope, and
\[
  H=\Expect\{\dot f_{\theta_0}(T)\dot f_{\theta_0}(T)^T\}
\]
is nonsingular.  Here \(\dot f_{\theta_0}\) is the gradient with respect to
\(\theta\).  Then
\[
  \dot m_{\theta_0}(O)
  =
  \{Y-f_{\theta_0}(T)\}\dot f_{\theta_0}(T),
\]
and, because \(\Expect\{Y-f_{\theta_0}(T)\mid T\}=0\),
\[
  V
  =
  \left.\frac{\partial^2}{\partial\theta\partial\theta^T}
  \Expect m_\theta(O)\right|_{\theta=\theta_0}
  =
  -H.
\]
Therefore any consistent near-maximizer satisfies
\[
  \sqrt n(\hat\theta_n-\theta_0)
  =
  H^{-1}
  \frac1{\sqrt n}\sum_{i=1}^n
  \dot f_{\theta_0}(T_i)\{Y_i-f_{\theta_0}(T_i)\}
  +o_P(1).
\]
If \(\Var(Y\mid T)=\sigma^2(T)\), the limit covariance is the sandwich
\[
  H^{-1}
  \Expect\{\sigma^2(T)\dot f_{\theta_0}(T)\dot f_{\theta_0}(T)^T\}
  H^{-1}.
\]
The theorem is doing the familiar nonlinear-regression calculation without
requiring the fitted curve itself to be linear in the parameter.
\qedmark
\end{example}

\begin{example}[Finite-dimensional cross-entropy classifiers]
For a fixed-dimensional multiclass classifier, let \(O=(Z,Y)\), where
\(Z\in\R^p\) and \(Y\in\{1,\ldots,K\}\).  Use class \(K\) as the baseline and
write
\[
  \beta=(\beta_1,\ldots,\beta_{K-1})\in\R^{p(K-1)},
  \qquad
  \beta_K=0.
\]
For \(k<K\), define
\[
  p_{\beta k}(z)
  =
  \frac{\exp(z^T\beta_k)}
       {1+\sum_{\ell=1}^{K-1}\exp(z^T\beta_\ell)},
  \qquad
  p_{\beta K}(z)
  =
  \frac{1}
       {1+\sum_{\ell=1}^{K-1}\exp(z^T\beta_\ell)}.
\]
The cross-entropy criterion is the multinomial log-likelihood
\[
  m_\beta(O)
  =
  \sum_{k=1}^{K-1}\ind{Y=k}Z^T\beta_k
  -
  \log\left\{1+\sum_{\ell=1}^{K-1}\exp(Z^T\beta_\ell)\right\}.
\]
The target \(\beta_0\) is the maximizer of \(\Expect m_\beta(O)\).  Under
correct specification it is the true conditional logit parameter; under
misspecification it is the finite-dimensional KL projection.  The derivative
has \(k\)th block
\[
  \dot m_{\beta_0,k}(O)
  =
  Z\{\ind{Y=k}-p_{\beta_0 k}(Z)\},
  \qquad k=1,\ldots,K-1.
\]
Let \(p_0(Z)=(p_{\beta_0 1}(Z),\ldots,p_{\beta_0,K-1}(Z))^T\) and
\[
  \Omega_0(Z)=\diag\{p_0(Z)\}-p_0(Z)p_0(Z)^T .
\]
Then the population second derivative is
\[
  V
  =
  -\Expect\{\Omega_0(Z)\otimes ZZ^T\}.
\]
If \(\Expect\|Z\|^2<\infty\), the local Hessian is nonsingular, and the
cross-entropy maximizer is consistent, the theorem gives
\[
  \sqrt n(\hat\beta_n-\beta_0)
  =
  \left[\Expect\{\Omega_0(Z)\otimes ZZ^T\}\right]^{-1}
  \frac1{\sqrt n}\sum_{i=1}^n
  \dot m_{\beta_0}(O_i)
  +o_P(1).
\]
This is the fixed-dimensional statistical version of softmax training: logits
are estimated by maximizing an empirical cross-entropy, and the theorem turns
that optimizer into an influence-function average.
\qedmark
\end{example}

\begin{example}[Circular phase and mean direction]
Some data are angles rather than points on a line: cell-cycle phase, circadian
phase, wind direction, seasonal timing, or neural oscillation phase.  Let
\(X\in[0,2\pi)\) and choose a local branch around the mean direction
\(\theta_0\).  A natural criterion is
\[
  m_\theta(X)=\cos(X-\theta),
  \qquad
  M_n(\theta)=\frac1n\sum_{i=1}^n\cos(X_i-\theta).
\]
The maximizer is the sample mean direction, equivalently the argument of
\(\sum_i e^{iX_i}\), as long as the resultant is nonzero.  The derivative and
second derivative are
\[
  \dot m_{\theta_0}(X)=\sin(X-\theta_0),
  \qquad
  \ddot m_{\theta_0}(X)=-\cos(X-\theta_0).
\]
If
\[
  R=\Expect\{\cos(X-\theta_0)\}>0,
\]
then \(\theta_0\) is a local maximum of the population criterion and
\[
  V=-R.
\]
Since
\[
  |\cos(X-\theta)-\cos(X-\theta')|\le |\theta-\theta'|,
\]
the local Lipschitz condition holds with envelope \(W=1\).  The theorem gives
\[
  \sqrt n(\hat\theta_n-\theta_0)
  =
  \frac1R
  \frac1{\sqrt n}\sum_{i=1}^n\sin(X_i-\theta_0)
  +o_P(1),
\]
and therefore
\[
  \sqrt n(\hat\theta_n-\theta_0)
  \weakto
  \Normal\!\left(
    0,\,
    \frac{\Var\{\sin(X-\theta_0)\}}{R^2}
  \right).
\]
The same iid \(M\)-estimator theorem therefore covers a non-Euclidean data
type once the problem is written in a local coordinate.
\qedmark
\end{example}

\begin{example}[scGTM as a constrained likelihood $M$-estimator]
For one gene observed across cells with pseudotimes $t_1,\ldots,t_C$, a
simplified hill-shaped Poisson version of scGTM \citep{cui2022scgtm} writes
\[
  Y_c\sim\Poisson\{\tau_c(\vartheta)\},
  \qquad
  \vartheta=(\mu_{\mathrm{mag}},k_1,k_2,t_0)^T,
\]
and
\[
  \log\{\tau_c(\vartheta)+1\}
  =
  \begin{cases}
    \mu_{\mathrm{mag}}\exp\{-k_1(t_c-t_0)^2\}, & t_c\le t_0,\\
    \mu_{\mathrm{mag}}\exp\{-k_2(t_c-t_0)^2\}, & t_c>t_0 .
  \end{cases}
\]
Let $\hat\vartheta_C$ maximize the constrained log-likelihood criterion
\[
  M_C(\vartheta)
  =
  \frac1C\sum_{c=1}^C
  \left[
    Y_c\log \tau_c(\vartheta)
    -
    \tau_c(\vartheta)
    -
    \log(Y_c!)
  \right]
\]
over the parameter box used to keep the trend interpretable.  Thus scGTM is an
$M$-estimator.  At an interior point, differentiating the criterion gives the
score equation
\[
  \Psi_C(\vartheta)
  =
  \frac1C\sum_{c=1}^C
  \frac{Y_c-\tau_c(\vartheta)}{\tau_c(\vartheta)}
  \dot\tau_c(\vartheta)
  =
  0,
\]
so the same fitted object is also a $Z$-estimator when viewed through its score.

Here is a more explicit verification, written conditionally on the pseudotimes.
Let
\[
  \Theta
  =
  [\mu_-,\mu_+]\times[k_-,k_+]^2\times[\epsilon,1-\epsilon],
  \qquad
  0<\mu_-<\mu_+<\infty,\quad
  0<k_-<k_+<\infty ,
\]
and suppose the true value
$\vartheta_0=(\mu_0,k_{10},k_{20},t_{0,0})^T$ is in the interior of
$\Theta$.  Given $t_1,\ldots,t_C$, assume that the $Y_c$ are independent
$\Poisson\{\tau(t_c;\vartheta_0)\}$.  Write
$Q_C=C^{-1}\sum_{c=1}^C\delta_{t_c}$ for the empirical design measure.  Assume
$Q_C\weakto Q$, no design point lies exactly at the true switch
$t_{0,0}$, and the design is not allowed to pile up too fast at the switch:
for some $A<\infty$,
\[
  \sup_C Q_C([t_{0,0}-r,t_{0,0}+r])\le Ar
  \qquad\text{for all small }r>0 .
  \tag{D}
\]
This local condition is automatic for regular deterministic grids and holds
with high probability for random pseudotimes with a bounded density.

Set $\eta(t;\vartheta)=\log\{\tau(t;\vartheta)+1\}$.  On each side of
$t=t_0$, $\eta$ is $C^\infty$ in $\vartheta$.  At $t=t_0$, the two pieces have
the same value and the same first derivative.  Indeed, if
$\delta=t-t_0$, then on the left side
\[
  \dot\eta(t;\vartheta)
  =
  e^{-k_1\delta^2}
  \{1,\ -\mu_{\mathrm{mag}}\delta^2,\ 0,\ 2\mu_{\mathrm{mag}}k_1\delta\}^T,
\]
and the analogous right-hand derivative is obtained by replacing the second
component by $0$, the third by $-\mu_{\mathrm{mag}}\delta^2$, and $k_1$ by
$k_2$ in the last component.  Both limits at $\delta=0$ equal
$(1,0,0,0)^T$.  Hence $\vartheta\mapsto\tau(t;\vartheta)$ is continuously
differentiable for each fixed $t$, and $\dot\tau(t;\vartheta)$ is uniformly
Lipschitz on $\Theta$.  The model is generally not twice continuously
differentiable in $t_0$: for example,
$\partial^2\eta/\partial t_0^2$ has one-sided limits
$-2\mu_{\mathrm{mag}}k_1$ and $-2\mu_{\mathrm{mag}}k_2$ at the switch.  The
first-order likelihood theory below only needs the $C^1$ and local Lipschitz
properties; the possible second-derivative jump contributes no first-order
boundary term because the first derivatives agree at the switch.

The likelihood class is Glivenko--Cantelli under these conditions.  Since
$\tau(t;\vartheta)$ is bounded away from $0$ and $\infty$ on
$[0,1]\times\Theta$, and since $\|\dot\tau(t;\vartheta)\|$ is bounded, the
single-observation log-likelihood
\[
  \ell_\vartheta(y,t)
  =
  y\log\tau(t;\vartheta)-\tau(t;\vartheta)-\log(y!)
\]
satisfies, for a constant $B$ independent of $C$,
\[
  |\ell_\vartheta(Y_c,t_c)-\ell_{\vartheta'}(Y_c,t_c)|
  \le B(1+Y_c)\|\vartheta-\vartheta'\|.
\]
The Poisson means are uniformly bounded, so the triangular-array covering
argument gives
\[
  \sup_{\vartheta\in\Theta}
  |M_C(\vartheta)-\Expect_{\vartheta_0}M_C(\vartheta)|
  \toP0 .
\]
Moreover, by weak convergence of $Q_C$ and the equicontinuity just noted,
$\Expect_{\vartheta_0}M_C(\vartheta)$ converges uniformly on $\Theta$ to
\[
  M(\vartheta)
  =
  \int
  \{\tau(t;\vartheta_0)\log\tau(t;\vartheta)-\tau(t;\vartheta)\}\,dQ(t)
  +
  \text{constant}.
\]
For every $\vartheta$,
\[
  M(\vartheta)-M(\vartheta_0)
  =
  -
  \int
  \left[
    \tau_0(t)\log\frac{\tau_0(t)}{\tau(t;\vartheta)}
    +
    \tau(t;\vartheta)-\tau_0(t)
  \right]\,dQ(t)
  \le0,
\]
where $\tau_0(t)=\tau(t;\vartheta_0)$.  Equality holds only when
$\tau(t;\vartheta)=\tau_0(t)$ for $Q$-almost every $t$.  Thus, under the
identifiability condition
\[
  \tau(t;\vartheta)=\tau(t;\vartheta_0)\quad Q\text{-a.e.}
  \qquad\Longrightarrow\qquad
  \vartheta=\vartheta_0 ,
\]
$M$ has a unique maximum at $\vartheta_0$.  Compactness makes this maximum
separated, and the argmax theorem gives
$\hat\vartheta_C\toP\vartheta_0$.

For the limit distribution, define the score
\[
  s_\vartheta(y,t)
  =
  \dot\ell_\vartheta(y,t)
  =
  \frac{y-\tau(t;\vartheta)}{\tau(t;\vartheta)}
  \dot\tau(t;\vartheta).
\]
At $\vartheta_0$, $\Expect_{\vartheta_0}\{s_{\vartheta_0}(Y_c,t_c)\mid t_c\}=0$
and
\[
  \mathcal I_C(\vartheta_0)
  =
  \frac1C\sum_{c=1}^C
  \frac{1}{\tau_c(\vartheta_0)}
  \dot\tau_c(\vartheta_0)\dot\tau_c(\vartheta_0)^T .
\]
The uniformly bounded Poisson moments and derivatives imply Lindeberg's
condition, so if
$\mathcal I_C(\vartheta_0)\to\mathcal I(\vartheta_0)$ with
$\mathcal I(\vartheta_0)$ nonsingular, then
\[
  \frac1{\sqrt C}\sum_{c=1}^C s_{\vartheta_0}(Y_c,t_c)
  \weakto
  \Normal\{0,\mathcal I(\vartheta_0)\}.
\]
The same smoothness calculation, with condition (D) controlling observations
whose side of the switch changes under $C^{-1/2}$ perturbations of $t_0$, gives
the local score expansion
\[
  \sup_{\|h\|\le M}
  \left\|
    \sqrt C\{\Psi_C(\vartheta_0+C^{-1/2}h)-\Psi_C(\vartheta_0)\}
    +
    \mathcal I_C(\vartheta_0)h
  \right\|
  \toP0
\]
for each fixed $M<\infty$.  Since $\vartheta_0$ is an interior point and
$\hat\vartheta_C\toP\vartheta_0$, the constrained maximizer is an ordinary
interior likelihood root with probability tending to one.  Therefore
\[
  \sqrt C(\hat\vartheta_C-\vartheta_0)
  =
  \mathcal I(\vartheta_0)^{-1}
  \frac1{\sqrt C}\sum_{c=1}^C s_{\vartheta_0}(Y_c,t_c)
  +o_{\Prob}(1)
  \weakto
  \Normal\{0,\mathcal I(\vartheta_0)^{-1}\}.
\]
This is the theoretical form behind the plug-in Fisher-information confidence
intervals used in scGTM.  The applied paper also flags the right caveats:
choosing hill versus valley shape from the same data, boundary constraints, and
nuisance count-distribution parameters can make naive intervals too optimistic
unless those steps are included in the inferential calculation.

For the reader trying to connect this example back to computation, the workflow
is now clear.  A scDesign3-style analysis first fits a generative law and uses
simulation to ask whether candidate designs or downstream procedures preserve
the empirical summaries of interest.  A scGTM-style analysis then fixes an
ordered cellular coordinate, chooses an interpretable trend catalogue, optimizes
the constrained likelihood, and reads the score, information, and local
expansion as uncertainty statements for that trend.  The two tools therefore
occupy different slots in the same statistical translation: simulation checks
the law that could have produced the assay, while constrained likelihood names
and estimates the biological map to be reported.
\qedmark
\end{example}

\begin{example}[Median regression]
Consider the median regression model of
Example~\ref{ex:ch13-median-regression-consistency}; the differentiability
target is the median, \(q=1/2\), case of regression quantiles
\citep{koenker1978regression}, and the differentiability calculation follows
the treatment in
\citet{dabrowskaAdvancedProbabilityCommunication}:
\[
  Y_i=\mathbf X_i^T\beta_0+\epsilon_i,
\]
where the conditional distribution of $\epsilon_i$ given $\mathbf X_i$ is continuous
and has a unique median equal to zero.  Define $\beta$ as the minimizer of
\[
  M_n(\beta)=\frac1n\sum_{i=1}^n m_\beta(Y_i,\mathbf X_i),
  \qquad
  m_\beta(y,\mathbf x)=|y-\mathbf x^T\beta|-|y|.
\]
By the triangle inequality,
\[
  |m_\beta(Y,\mathbf X)-m_{\beta'}(Y,\mathbf X)|
  \le
  \|\beta-\beta'\|\,\|\mathbf X\|,
\]
so the Lipschitz condition reduces to square integrability of $\mathbf X$.

The function $t\mapsto |t|$ is differentiable except at $t=0$.  Therefore the
criterion is differentiable at $\beta=\beta_0$ provided
$P_0(Y_i=\beta_0^T\mathbf X_i)=P_0(\epsilon_i=0)=0$, which follows from continuity of
the conditional error distribution.  For any $\beta$, define
\[
  \dot m_\beta(Y,\mathbf X)
  =
  -\mathbf X\,\sign(Y-\mathbf X^T\beta)
\]
at points where $Y\ne \mathbf X^T\beta$.  Put $\beta'=\beta+h$.  Then
\[
  m_{\beta'}(Y,\mathbf X)-m_\beta(Y,\mathbf X)
  =
  h^T\dot m_\beta(Y,\mathbf X)+\operatorname{rem}(\beta,h,Y,\mathbf X),
\]
where
\[
\begin{aligned}
  \operatorname{rem}(\beta,h,Y,\mathbf X)
  &=
  -2(Y-\mathbf X^T\beta')
  \Bigl[
    \ind{0<Y-\mathbf X^T\beta<\mathbf X^T h} \\
  &\hspace{6.2em}
    -\ind{\mathbf X^T h<Y-\mathbf X^T\beta<0}
  \Bigr].
\end{aligned}
\]
Consequently
\[
  \operatorname{rem}(\beta,h,Y,\mathbf X)^2
  \le
  4(h^T\mathbf X\mathbf X^T h)\ind{|Y-\mathbf X^T\beta|\le |h^T\mathbf X|}.
\]
Thus
\[
  \Expect_{P_0}\operatorname{rem}(\beta,h,Y,\mathbf X)^2
  \le
  4\|h\|^2
  \Expect\!\left[
    \|\mathbf X\|^2
    P_0\{|Y-\mathbf X^T\beta|\le \|h\|\,\|\mathbf X\|\mid \mathbf X\}
  \right].
\]
Setting $\delta=\beta-\beta_0$, the conditional probability is
\[
  P_0\{|Y-\mathbf X^T\beta|\le \|h\|\,\|\mathbf X\|\mid \mathbf X=\mathbf x\}
  =
  F(\mathbf x^T\delta+\|\mathbf x\|\|h\|\mid \mathbf x)
  -
  F(\mathbf x^T\delta-\|\mathbf x\|\|h\|\mid \mathbf x),
\]
where $F(\cdot\mid \mathbf x)$ is the conditional distribution function of the error.
If $F(\cdot\mid \mathbf x)$ is continuous for almost all $\mathbf x$, the right hand side
tends to zero as $\|h\|\to0$ and is bounded by one.  Dominated convergence
then gives
\[
  \Expect_{P_0}\operatorname{rem}(\beta,h,Y,\mathbf X)^2=o(\|h\|^2),
\]
so the criterion is quadratic-mean differentiable.

The expectation has derivative
\[
  \dot M(\beta)
  =
  -\Expect\{\mathbf X\sign(Y-\mathbf X^T\beta)\},
\]
where the expectation is under $P_0$.  Assume that
$F(\cdot\mid \mathbf x)$ has density $f(\cdot\mid \mathbf x)$ with respect to Lebesgue measure.
For fixed $\beta$ and $h$, with $\beta'=\beta+h$ and
$\delta=\beta-\beta_0$,
\[
  \dot M(\beta')-\dot M(\beta)
  -
  2\Expect\{\mathbf X\mathbf X^T f(\mathbf X^T\delta\mid \mathbf X)\}h
  =
  r(h,\delta),
\]
where
\[
  r(h,\delta)
  =
  2\Expect\{\mathbf X\mathbf X^T h\,g(h,\delta,\mathbf X)\},
\]
and
\[
  g(h,\delta,\mathbf x)
  =
  \frac1{\mathbf x^T h}
  \int_{\min(\mathbf x^T h,0)}^{\max(\mathbf x^T h,0)}
  \{f(v+\mathbf x^T\delta\mid \mathbf x)-f(\mathbf x^T\delta\mid \mathbf x)\}\,dv
\]
when $\mathbf x^T h\ne0$, with $g(h,\delta,\mathbf x)=0$ when $\mathbf x^T h=0$.  Equivalently, if
$e=\sign(\mathbf x^T h)$,
\[
  g(h,\delta,\mathbf x)
  =
  \frac1{|\mathbf x^T h|}
  \int_0^{|\mathbf x^T h|}
  \{f(ev+\mathbf x^T\delta\mid \mathbf x)-f(\mathbf x^T\delta\mid \mathbf x)\}\,dv.
\]
Suppose that for almost all $\mathbf x$, the function
$u\mapsto f(u+\mathbf x^T\delta\mid \mathbf x)$ is uniformly continuous near zero.  Then
\[
  |g(h,\delta,\mathbf x)|
  \le
  \sup\{|f(v+\mathbf x^T\delta\mid \mathbf x)-f(\mathbf x^T\delta\mid \mathbf x)|:
       |v|\le \|h\|\,\|\mathbf x\|\}
  \to0
\]
as $\|h\|\to0$.  If, for almost all $\mathbf x$, the density
$f(\cdot+\mathbf x^T\delta\mid \mathbf x)$ is bounded near zero by a measurable function
$C(\mathbf x)$ satisfying $\Expect\{\|\mathbf X\mathbf X^T\|C(\mathbf X)\}<\infty$, dominated convergence
implies $r(h,\delta)=o(\|h\|)$.

At $\beta=\beta_0$, the linear part is governed by
\[
  2\Expect\{f(0\mid \mathbf X)\mathbf X\mathbf X^T\}h.
\]
Combining the preceding calculations gives the following proposition.
\end{example}

\begin{proposition}[Linear median-regression limit]
Let $(Y_i,\mathbf X_i)$ be independent random variables satisfying the linear median
regression model.  Suppose
\begin{enumerate}
\item $\Expect(\mathbf X_i\mathbf X_i^T)$ is finite positive definite;
\item the error has conditional density $f(\cdot\mid \mathbf x)$ such that, for
      $P_{\mathbf X}$-almost all $\mathbf x$, $f(\cdot\mid \mathbf x)$ is continuous on an interval
      $(-\|\mathbf x\|\varepsilon,\varepsilon\|\mathbf x\|)$ and
      $f(\cdot\mid \mathbf x)\le C(\mathbf x)$ with
      $\Expect\{\|\mathbf X_i\|^2C(\mathbf X_i)\}<\infty$;
\item the conditional median is zero and $f(0\mid \mathbf x)>0$ almost surely.
\end{enumerate}
Then
\[
  \sqrt n(\hat\beta-\beta_0)
  \weakto
  \Normal(0,V^{-1}\Expect[\mathbf X\mathbf X^T]V^{-T}),
\]
where
\[
  V=2\Expect\{f(0\mid \mathbf X)\mathbf X\mathbf X^T\}.
\]
\qedmark
\end{proposition}

\begin{tcolorbox}[
  enhanced,
  breakable,
  colback=noteback,
  colframe=bookgold!75!black,
  boxrule=0.55pt,
  arc=4pt,
  boxsep=1pt,
  left=0.95em,
  right=0.9em,
  top=0.7em,
  bottom=0.7em,
  before skip=0.9\baselineskip,
  after skip=1.0\baselineskip
]
\noindent\textbf{Hand-off to Chapters~14 and~15.}
This chapter has done two jobs.  The global job was to show that a random
criterion or estimating equation points to the right target.  The local job
was to show that, after the target has been found, the remaining displacement
has a first-order representation such as
\[
  \sqrt n(\hat\theta_n-\theta_0)
  =
  \mathbb G_n\varphi+o_{\Prob}(1).
\]
That display is the end of the estimation story, but only the beginning of
the inference story.  A scientific report rarely stops at
\(\hat\theta_n\).  It reports a transformed risk, a quantile, a survival
curve, a transition probability, a prediction contrast, a confidence interval,
a band, or an efficiency comparison.

Chapter~14 takes the local likelihood representation as its input and asks how
nearby laws are separated by tests.  Chapter~15 takes the linear
representation above as its input.  Its question is not how the estimator was
selected; that was the work of this chapter.  Its question is how the residual
empirical noise should be pushed through the reported functional, decomposed
into observation-level influence, compared across procedures, and re-read by
the bootstrap.  This is why the chapters are adjacent but distinct:
Chapter~13 finds the point, Chapter~14 compares nearby laws, and Chapter~15
reads the noise left around the point.
\end{tcolorbox}

\section{Exercises}
\conceptindexes{M-estimation exercises, Z-estimation exercises, asymptotic-normality exercises}

The exercises revisit the chapter's three motifs: peaks of criteria, roots of
estimating equations, and local limits.  Some problems ask for direct
calculations in familiar models; others ask for the structural checks that make
the master theorems usable.  The point is to practice recognizing which door,
$M$ or $Z$, opens the cleanest route.
For classroom use, Exercises~1--7 and 14--16 form the main route.  The
remaining distribution-specific problems are a model laboratory: they preserve
the book's technical depth but can be assigned selectively.

\begin{exercise}[Winsorized location as \(M\)- and \(Z\)-estimation]
Let $X_1,\ldots,X_n$ be iid from a distribution with density symmetric about
$\theta$.  Let $\theta_0$ be the true center of symmetry and define
$\Psi(\theta)=\Expect_{\theta_0}\psi(X-\theta)$, where
\[
  \psi(y)
  =
  \begin{cases}
    y, & |y|\le k,\\
    k\sign(y), & |y|>k.
  \end{cases}
\]
The sample winsorized mean $\hat\theta_n$ is a near solution of
\[
  \frac1n\sum_{i=1}^n\psi(X_i-\theta)=o_{\Prob}(1).
\]
It is also a near minimizer of
\[
  M_n(\theta)=\frac1n\sum_{i=1}^n m(X_i-\theta),
\]
where
\[
  m(y)
  =
  \begin{cases}
    \frac12 y^2, & |y|\le k,\\
    k|y|-\frac12k^2, & |y|>k.
  \end{cases}
\]
\begin{enumerate}
\item Show that $\Psi(\theta_0)=0$.
\item Show that $\hat\theta_n$ is consistent.
\item Show that $\sqrt n(\hat\theta_n-\theta_0)\weakto
      \Normal(0,\sigma^2(\theta_0))$ for some $\sigma^2(\theta_0)>0$ and determine
      its form.
\end{enumerate}
\end{exercise}

\begin{exercise}[Gamma shape from the arithmetic-geometric ratio]
Let $\mathcal P_0$ be the family of distributions of nonnegative random
variables such that $\Expect X<\infty$ and $\Expect|\log X|<\infty$.  Put
\[
  \mu=\Expect X,\qquad \rho=\Expect\log X.
\]
Let $X_1,\ldots,X_n$ be iid from $P\in\mathcal P_0$, and let $A_n$ and $G_n$
be the arithmetic and geometric means.
\begin{enumerate}
\item Show that $C_n=G_n/A_n$ converges almost surely to a constant $c(P)$ and
      express it as a function of $(\mu,\rho)$.
\item If $X$ and $\log X$ have finite variances, show that
      $\sqrt n(C_n-c)\weakto \Normal(0,\tau_1^2)$ and find $\tau_1^2$.
\item Let $\psi(\alpha)=\Gamma'(\alpha)/\Gamma(\alpha)$ be the digamma
      function.  Show that
\[
  \log c=\psi(\alpha)-\log\alpha,\qquad c\in(0,1],
\]
      has a unique positive root $\alpha$.
\item Let $\alpha(P)$ be this root with $c=c(P)$, and let $\hat\alpha_n$ be
      obtained by replacing $c$ with $C_n$.  Show that
      $\hat\alpha_n\to\alpha(P)$ almost surely.
\item Show that $\sqrt n(\hat\alpha_n-\alpha)\weakto
      \Normal(0,\tau_2^2)$ and find $\tau_2^2$.
\item For a gamma density
\[
  f_{\alpha,\sigma}(x)
  =
  \frac{1}{\sigma^\alpha\Gamma(\alpha)}
  x^{\alpha-1}e^{-x/\sigma}\ind{x>0},
\]
      show that the MLE is $(\hat\alpha,\hat\sigma)$, where
      $\hat\alpha$ is as above and $\hat\sigma=A_n/\hat\alpha$.
\item Under the gamma model, find the joint asymptotic distribution of
      $\sqrt n(\hat\alpha-\alpha_0,\hat\sigma-\sigma_0)$.
\end{enumerate}
\end{exercise}

\setcounter{exercise}{2}

\begin{exercise}[Location-scale information]
Let $X_1,\ldots,X_n$ be iid from the location-scale model of Example 13.1.4.
Suppose the density $f$ is supported on the whole line, differentiable, and
has tails satisfying $\lim_{x\to\pm\infty}x^kf(x)=0$ for $k=0,1$.
\begin{enumerate}
\item Derive the Fisher information.
\item Verify the information for normal, Cauchy, and type-I extreme value
      distributions.
\end{enumerate}
\end{exercise}

\begin{exercise}[Monotone location score]
Let $f_0$ be a continuous density on $\R$, positive everywhere, with
$f_0(x)\to0$ as $x\to\pm\infty$.
\begin{enumerate}
\item If $f_0$ is differentiable and
      $\psi(x)=f_0'(x)/f_0(x)$ is strictly decreasing, continuous, and square
      integrable, show that $\psi$ must change sign.
\item For the location family $f_\theta(x)=f_0(x-\theta)$, show that the score
      equation has a unique root and that the root is consistent.
\item Verify the conditions for normal and logistic distributions, and explain
      why they fail for the Cauchy distribution.
\end{enumerate}
\end{exercise}

\begin{exercise}[Monotone scale score]
Let $f_0$ be a continuous density on $\R$, symmetric about zero, with
$f_0(x)\to0$ and $xf_0(x)\to0$ as $x\to\infty$.
\begin{enumerate}
\item If $g(x)=xf_0'(x)/f_0(x)$ is continuous, square integrable, and
      decreasing for $x>0$, show that $\psi(x)=1+g(x)$ must change sign.
\item For the scale family $f_\theta(x)=\theta^{-1}f_0(x/\theta)$, show that
      the score equation has a unique root and that the root is consistent.
\item Verify the conditions for the Cauchy distribution.
\end{enumerate}
\end{exercise}

\begin{exercise}[Nonlinear least squares]
Let $(Y_i,X_i)$ be iid from
\[
  Y_i=g_\theta(X_i)+\epsilon_i,
\]
where $X_i\in\R^d$, $\Expect(\epsilon_i\mid X_i)=0$, and $g_\theta$ is a smooth
function of $\theta\in\Theta\subseteq\R^p$.  Propose regularity conditions on
$g_\theta$ under which the least-squares estimator minimizing
\[
  M_n(\theta)=\frac1n\sum_{i=1}^n\{Y_i-g_\theta(X_i)\}^2
\]
exists with probability tending to one, is consistent, and is asymptotically
normal.  Find the asymptotic covariance matrix.
\end{exercise}

\begin{exercise}[Shifted lognormal one-step inference]
Let $X_1,\ldots,X_n$ be iid from the shifted lognormal model discussed earlier.
\begin{enumerate}
\item Give the form of the log-likelihood expansion.
\item Propose preliminary root-$n$ consistent estimators of the three
      parameters based on quantiles or moments.
\item Construct an efficient one-step estimator.
\item Test $H_0:\gamma=0$ against local alternatives
      $\gamma_n=h/\sqrt n$ using a Fisher-Rao test; give the rejection region
      and asymptotic power.
\end{enumerate}
\end{exercise}

\begin{exercise}[Marshall--Olkin boundary irregularity]
Let $(X_i,Y_i)$ be iid from the Marshall-Olkin model with survival function
\[
  S_\lambda(x,y)
  =
  \exp\{-\lambda_1x-\lambda_2y-\lambda_3\max(x,y)\},
  \qquad x,y>0,
\]
and $S_\lambda(x,y)=1$ if $x<0$ or $y<0$.  The parameter space is
\[
  A=\{\lambda=(\lambda_1,\lambda_2,\lambda_3):
      \lambda_1>0,\lambda_2>0,\lambda_3\ge0\}.
\]
The distribution has density, with respect to the sum of two-dimensional
Lebesgue measure and Lebesgue measure on the diagonal,
\[
  f_\lambda(x,y)
  =
  \begin{cases}
    \lambda_2(\lambda_1+\lambda_3)e^{-(\lambda_1+\lambda_3)x-\lambda_2y},
      & x>y,\\
    \lambda_1(\lambda_2+\lambda_3)e^{-\lambda_1x-(\lambda_2+\lambda_3)y},
      & x<y,\\
    \lambda_3e^{-(\lambda_1+\lambda_2+\lambda_3)x},
      & x=y .
  \end{cases}
\]
Show that the model is not Hellinger differentiable at boundary points
$\lambda_3=0$.
\end{exercise}

\begin{exercise}[Constrained normal mean]
Let $X_1,\ldots,X_n$ be iid $N(\mu,1)$ with $\mu\ge0$.  The MLE is
\[
  \hat\mu=\max(\bar X_n,0).
\]
At $\mu=0$, show that $\sqrt n\,\hat\mu$ converges to a random variable with
distribution function
\[
  F(x)=0\ (x<0),\qquad F(0)=1/2,\qquad F(x)=\Phi(x)\ (x>0).
\]
Thus constraints can produce non-normal limits of MLEs in otherwise regular
models.
\end{exercise}

\begin{exercise}[Multinomial efficient information]
Let $X_1,\ldots,X_n$ be iid multinomial observations with one trial and
probabilities
\[
  \theta=(\theta_1,\ldots,\theta_{k+1}),
  \qquad
  \theta_{k+1}=1-\sum_{j=1}^k\theta_j,\quad 0<\theta_j<1.
\]
Find efficient scores, information, and MLEs for:
\begin{enumerate}
\item the full parameter $\theta_1,\ldots,\theta_k$;
\item a subvector $\xi=(\theta_1,\ldots,\theta_m)$ with the remaining
      coordinates nuisance;
\item the same subvector when the nuisance coordinates are known.
\end{enumerate}
\end{exercise}

\begin{exercise}[Extreme-value likelihood equations]
Let $X_1,\ldots,X_n$ be iid from the type-I extreme value distribution
\[
  F_{\mu,\sigma}(x)=F_0\!\left(\frac{x-\mu}{\sigma}\right),
  \qquad
  F_0(x)=\exp(-e^{-x}).
\]
\begin{enumerate}
\item Show that the likelihood equations reduce to
\[
  \hat\sigma
  =
  \bar X-
  \frac{\sum_i X_i e^{-X_i/\hat\sigma}}
       {\sum_i e^{-X_i/\hat\sigma}},
  \qquad
  \hat\mu
  =
  -\hat\sigma
  \log\left(\frac1n\sum_{i=1}^ne^{-X_i/\hat\sigma}\right).
\]
\item Show that the first equation has a unique root.
\item Show that $\hat\sigma$ can be written using the fitted distribution
      function, and compare it with the plug-in estimator using the empirical
      distribution function.
\end{enumerate}
\end{exercise}

\begin{exercise}[Beta method-of-moments test]
Let $X_1,\ldots,X_n$ be a sample from the beta distribution with parameters
$\alpha$ and $\beta$.
\begin{enumerate}
\item Show that
\[
  \hat\alpha_n=
  \frac{n^{-1}\sum_i X_i(1-X_i)}{S_n^2}\bar X_n,
  \qquad
  \hat\beta_n=
  \frac{n^{-1}\sum_i X_i(1-X_i)}{S_n^2}(1-\bar X_n)
\]
      are method-of-moments estimators.
\item Show that they are root-$n$ consistent.
\item Use these estimators to test $H_0:\alpha=\beta$.
\end{enumerate}
\end{exercise}

\begin{exercise}[Inverse Gaussian exponential-family inference]
Let $X_1,\ldots,X_n$ be iid inverse Gaussian with density
\[
  f_{\mu,\lambda}(x)
  =
  \left(\frac{\lambda}{2\pi}\right)^{1/2}
  x^{-3/2}
  \exp\!\left\{-\frac{\lambda}{2\mu^2x}(x-\mu)^2\right\}
  \ind{x>0}.
\]
\begin{enumerate}
\item Show that this is an exponential family and identify the canonical
      parametrization.
\item Find sufficient statistics, cumulant generating function, mean vector,
      and covariance matrix.
\item Show that the MLE exists with probability tending to one and find its
      asymptotic distribution.
\item Give likelihood-ratio, Wald, and Fisher-Rao tests for $H_0:\lambda=1$.
\end{enumerate}
\end{exercise}

\begin{exercise}[Probit model]
Let $Y_{ij}$, $i=1,\ldots,n_j$, $j=1,\ldots,k$, be independent binary variables
with
\[
  \Prob(Y_{ij}=1)=\Phi(\theta_0+\theta_1z_j),
\]
where $z_1<\cdots<z_k$ are distinct covariate levels.  Find the Cramer-Rao
information bound for unbiased estimators of:
\begin{enumerate}
\item $\theta=(\theta_0,\theta_1)^T$;
\item the vector $(p_1(\theta),\ldots,p_k(\theta))^T$.
\end{enumerate}
\end{exercise}

\begin{exercise}[Parity-observed Poisson model]
Let $X_1,\ldots,X_n$ be iid Poisson with mean $\lambda$, but observe only
the parity
\[
  Y_i=\ind{X_i\text{ is odd}}.
\]
\begin{enumerate}
\item Give the Cramer-Rao bound for unbiased estimation of $\lambda$ from the
      $Y_i$'s.
\item Derive the log-likelihood and score equation for $\lambda$; describe
      estimation of the asymptotic variance of the MLE.
\item Compare the MLE with the frequency method obtained by estimating
      $p=\Prob(X\text{ is odd})$ by $\bar Y_n$.
\end{enumerate}
\end{exercise}

\begin{exercise}[Pareto type-II one-step estimation]
The Pareto type-II density is
\[
  f_{\alpha,\beta}(x)
  =
  \frac{\alpha\beta}{(1+\beta x)^{\alpha+1}}\ind{x>0},
  \qquad \alpha,\beta>0.
\]
Its $p$th quantile satisfies
\[
  \log Q(p)
  =
  -\log\beta+\alpha^{-1}\log\frac1{1-p}.
\]
\begin{enumerate}
\item Show that this is a Hellinger differentiable model.
\item Propose one-step MLEs for the two parameters.
\item Find their asymptotic distribution.
\end{enumerate}
\end{exercise}

\begin{exercise}[Pareto boundary estimation]
Let $X_1,\ldots,X_n$ be iid from Pareto type I,
\[
  f_{\alpha,\theta}(x)
  =
  \frac{\alpha\theta^\alpha}{x^{\alpha+1}}\ind{x>\theta},
  \qquad \alpha,\theta>0.
\]
The MLEs are
\[
  \hat\theta=X_{1:n},
  \qquad
  \hat\alpha
  =
  \frac{n}{\sum_{i=1}^n\log(X_i/X_{1:n})}.
\]
\begin{enumerate}
\item Explain why Cramer's conditions do not apply.
\item Is the model Hellinger differentiable?
\item Show that the estimators are consistent.
\item Find the normal limit of $\sqrt n(\hat\alpha-\alpha)$ and the
      exponential limit of $n(\hat\theta-\theta)$; describe the joint limit.
\end{enumerate}
\end{exercise}

\begin{exercise}[Parametric proportional hazards]
Let $Y\ge0$ have survival function $S(y)=\Prob(Y>y)$ and density $f(y)$.  Its
hazard rate is
\[
  a(y)=\frac{f(y)}{S(y)}
  =
  \lim_{h\downarrow0}
  \frac1h \Prob(Y\in[y,y+h]\mid Y\ge y),
\]
and
\[
  A(y)=\int_0^y a(u)\,du
\]
is the cumulative hazard.  We have $-\log S(y)=A(y)$.
\begin{enumerate}
\item Show that $A(Y)$ has a unit exponential distribution.
\item In the proportional hazards regression model
\[
  \Prob(X\le x\mid \mathbf Z)
  =
  1-\exp\{-A(x,\theta)e^{\beta^T\mathbf Z}\},
\]
      derive likelihood, score, and information for $(\beta,\theta)$.
\item Specialize the information to $A(x,\theta)=e^\theta x$.
\end{enumerate}
\end{exercise}

\begin{exercise}[Index of dispersion]
Let $X_1,\ldots,X_n$ be iid from a discrete distribution on nonnegative
integers with mean $\mu>0$, variance $\sigma^2$, and finite fourth central
moment.  Put
\[
  \xi=\frac{\sigma^2}{\mu},
  \qquad
  \hat\xi_n=\frac{S_n^2}{\bar X_n}.
\]
\begin{enumerate}
\item Show that $\sqrt n(\hat\xi_n-\xi)\weakto \Normal(0,v^2)$ and find
      $v^2$.
\item The statistic $\hat\xi_n$ is often used to test for a Poisson
      distribution by rejecting when $|\hat\xi_n-1|$ is large.  If the $X_i$
      are Poisson with mean $\lambda$, show that the asymptotic variance in
      part (a) does not depend on $\lambda$, and describe the normal
      approximation critical values.
\end{enumerate}
\end{exercise}

\section*{Sources and Further Reading}
\addcontentsline{toc}{section}{Sources and Further Reading}

The notation $M$-estimator and the systematic treatment of estimating
equations are modern, but the two ideas behind them are old: optimize a
criterion, or solve an equation that expresses balance.  The references below
mark the main mathematical routes used in this chapter: uniform laws of large
numbers, convexity, monotonicity, local expansions, and empirical-process
arguments.  The local route through empirical-process consistency,
asymptotic linearity, and plug-in weak-convergence arguments is also informed
by \citet{dabrowskaAdvancedProbabilityCommunication} and
\citet{dabrowskaStochasticProcessesCommunication}.

\begin{description}[leftmargin=0pt,labelsep=0.65em,style=unboxed,font=\normalfont,itemsep=0.45\baselineskip]
\item[\textsc{Likelihood, contrast, and information.}]
The likelihood examples are classical, but the chapter uses them through the
more general language of contrasts.  The Kullback--Leibler calculation is tied
to Shannon's information-theoretic inequality \citep{shannon1948mathematical}:
the true distribution minimizes the population contrast when the model is
identifiable.  This is why maximum likelihood can be read as an $M$-estimator
even before differentiating the likelihood.

\item[\textsc{Huber's program.}]
\citet{huber1964robust} made robust location estimation a central example of
estimating by minimizing a criterion that is not necessarily quadratic.
\citet{huber1967behavior} then gave influential consistency results for
maximum likelihood and related estimators under nonstandard conditions.  The
scalar crossing theorem in this chapter reflects that point of view: roots can
be controlled by monotonicity and sign changes, not only by smooth likelihood
calculus.  The censored median example uses the product-limit estimator of
\citet{kaplan1958nonparametric}; \citet{brookmeyer1982confidence} develop the
corresponding confidence-interval construction for median survival.

\item[\textsc{Argmax theorems and empirical processes.}]
The global consistency theorem is a version of the argmax principle: uniform
convergence of $M_n$ to a deterministic criterion $M$ transfers maximizers of
$M_n$ to maximizers of $M$.  The formulation used here is close to
\citet{pakes1989simulation} and the empirical-process treatment of
\citet{vaart2023weak}.  It is called a master theorem only in the informal
sense that many later consistency proofs are substitutions into this one
template.

\item[\textsc{Convexity and concavity.}]
Convex analysis enters because convex and concave random criteria often need
less direct uniform control.  The deterministic background is
\citet{rockafellar1970convex}; the stochastic convexity result used for
likelihood and regression arguments is associated here with
\citet{andersen1982cox}.  These tools explain why maximum likelihood in
exponential families, graphical lasso, and least absolute deviation or median
regression can be handled with the same geometry.

\item[\textsc{Applied and penalized examples.}]
Generalized linear models enter through \citet{nelderWedderburn1972glm} and
\citet{mccullaghNelder1989glm}; they are used here as likelihood and
score-equation examples for binary endpoints, grouped risks, counts, and
rates.  Generalized estimating equations enter through the longitudinal-data
framework of \citet{liang1986longitudinal}, a natural clinical-trial example
of $Z$-estimation.  Conditional random fields
\citep{lafferty2001conditional} and cross-entropy image classifiers
\citep{krizhevsky2012imagenet} are included to make clear that modern
prediction training objectives are also empirical criteria.  Penalized
likelihood and empirical risk are represented by the lasso
\citep{tibshirani1996regression}, RKHS smoothing and kernel ridge regression
\citep{wahba1990spline,kimeldorf1971some,scholkopf2001generalized}, and
penalized GAMs \citep{wood2011fast,wood2017generalized}; they should be read
as $M$-estimation problems whose first-order or KKT equations provide the
corresponding $Z$-view.

\item[\textsc{Single-cell genomics.}]
The single-cell examples are organized as observation, generation, and
inference.  scImpute \citep{li2018scimpute} models the observation mechanism
behind zeros; scDesign3 estimates a generative law
for realistic synthetic single-cell and spatial-omics data; scGTM
\citep{cui2022scgtm} is used here as a constrained likelihood example whose
plug-in Fisher information gives approximate uncertainty for interpretable
pseudotime-trend parameters.

\item[\textsc{Monotone fields and fixed points.}]
The multiparameter $Z$-estimation results draw on monotonicity ideas from
\citet{brown1985multiparameter} and \citet{ritov1987tightness}.  The fixed-point
proofs also use deterministic theorems of \citet{banach1922operations} and
\citet{schauder1930fixpunktsatz}: existence, uniqueness, and local asymptotic
linearity are obtained by turning a score equation into a map that sends a
small ball into itself.

\item[\textsc{One-step and semiparametric viewpoints.}]
The one-step estimator and the iid asymptotic-normality theorems anticipate the
semiparametric efficiency language of \citet{bickel1993efficient}.  The key
object is the asymptotic linear representation: after consistency, the
estimator behaves like an average of influence-type functions plus a negligible
remainder.  The median-regression calculation at the end of the chapter is a
concrete example of this philosophy in the nonsmooth regression-quantile
tradition of \citet{koenker1978regression}.
\end{description}

%% file: chapters/ch03_probability_measure.tex
\chapter{Probability and Measure: Events, Laws, and Integration}
\label{chap:probability-measure}
\conceptindexes{probability measure, measure theory, observability, events, measurable spaces, random variables, pushforward laws, integration}

\begin{tcolorbox}[
  enhanced,
  breakable,
  colback=chaptercream,
  colframe=bookblue!88!black,
  boxrule=0.72pt,
  arc=5pt,
  boxsep=1pt,
  left=1.0em,
  right=0.95em,
  top=0.82em,
  bottom=0.82em,
  before skip=0.55\baselineskip,
  after skip=1.0\baselineskip
]
\noindent\textbf{Chapter overview.}
This chapter turns the opening examples into the formal language of
observability, random elements, laws, and integration:
\[
\begin{array}{@{}lll@{}}
\text{possible worlds} & \Omega & \text{what could have happened},\\
\text{observable distinctions} & \fieldF & \text{what counts as an event},\\
\text{ways of assigning size} & \mu,\ P & \text{what is large, rare, or likely},\\
\text{statistical summaries} & X & \text{measurable maps out of }\Omega.
\end{array}
\]
Every example exposes a mathematical phenomenon; every theorem explains why the
examples can be trusted.
\end{tcolorbox}

The examples from Chapter~2 should stay in the background as the formal
language begins.  A cluster on a map becomes a question about sets of point
patterns; a singleton species count becomes a function of a sample; a proxy
climate trace or a text feature becomes evidence only after we decide what the
record can distinguish.  The definitions below give names to these quiet
choices.  They do not replace the stories; they keep the stories from slipping
into impression.

The first surprise in measure-theoretic probability is that probability does
not begin with a number.  Before saying how likely an event is, we must say what
counts as an event.  This is not pedantry.  In the London bombing story, the raw
data are locations, but the questions are set-valued: did too many impacts fall
in a district, did the nearest-neighbor distances look unusual, did the pattern
fit a homogeneous point process?  In the missing-species story, the data are
labels, but the unseen part of the population is inferred through events and
summaries built from counts.  The unseen leaves evidence in the seen, but only
after we decide which features can be read.

There is also a social lesson hiding in the mathematics.  Borges's mock
taxonomy and Scott's discussion of legibility both remind the analyst that classification
is powerful because it decides what can be named and compared
\citep{borges1964johnwilkins,scott1998seeing}.  A \(\sigma\)-field is the
mathematical version of that decision.  It is a disciplined language of
observable distinctions.

\section{Events and Measurable Sets}
\label{sec:ch3-rich-events}
\conceptindexes{events, observability, event systems, sample space, sigma-field}

\subsection{Measurable spaces}
\conceptindexes{measurable spaces, sample space, sigma-field, measurable sets}

A sample space \(\Omega\) is a universe of possible outcomes.  It may be small,
as in a coin toss; geometric, as in a cloud of spatial points; functional, as in
a random curve; or combinatorial, as in a sequence of sampled species labels.
The event system is a family of subsets of \(\Omega\).  We do not usually allow
all subsets, because all subsets can be too large and unstable for the kind of
measure we want.

\begin{definition}[Measurable space]
A measurable space is a pair \((\Omega,\fieldF)\), where \(\Omega\) is a set and
\(\fieldF\) is a \(\sigma\)-field of subsets of \(\Omega\):
\[
  \Omega\in\fieldF,\qquad
  A\in\fieldF\Rightarrow A^c\in\fieldF,\qquad
  A_n\in\fieldF\Rightarrow \bigcup_{n=1}^{\infty}A_n\in\fieldF.
\]
The elements of \(\fieldF\) are called measurable sets or events.
\end{definition}

A useful halfway object is an algebra of sets: a family containing \(\Omega\)
that is closed under complements and finite unions.  A \(\sigma\)-field is an
algebra that is also closed under countable unions.  This distinction matters
because finite classification is often easy to describe, while limiting
questions require the countable closure.

The word countable is not a cosmetic restriction.  It is the compromise that
makes limits legal without forcing us to accept arbitrary uncountable unions.
Countable unions describe repeated trials, set limits, tail events, and
approximation by increasingly fine resolution.  Arbitrary uncountable unions can
break measurability and are not protected by countable additivity.

\begin{example}[Coarse observations and finite partitions]
Suppose a city is divided into districts
\[
  \Omega=A_1\biguplus\cdots\biguplus A_m .
\]
If we observe only the district in which an event occurs, then the natural event
system is not the whole power set of the city.  It is the finite algebra
\[
  \fieldF
  =
  \left\{\bigcup_{i\in I}A_i:I\subseteq\{1,\ldots,m\}\right\}.
\]
This algebra records every question answerable at district resolution and
forgets every question below that resolution.  This display is the simplest
model of statistical coarsening: the event system is not just a mathematical
container, but a record of what the measurement protocol can distinguish.
\end{example}

\begin{example}[The countable--cocountable field]
Let \(\Omega\) be uncountable and put
\[
  \fieldF
  =
  \{A\subseteq\Omega:A\text{ is countable or }A^c\text{ is countable}\}.
\]
Then \(\fieldF\) is a \(\sigma\)-field.  It can distinguish countable
exceptions from co-countable truths, but it cannot distinguish most intermediate
subsets.  This example is intentionally austere: it shows that a
\(\sigma\)-field can be mathematically valid while being scientifically too
coarse for many questions.
\end{example}

\begin{example}[Four event languages]
The same word event means different things in different sample spaces.
\begin{enumerate}
\item If \(\Omega=\{0,1\}^n\), an event might be ``at least \(k\) successes.''
\item If \(\Omega\) is the set of finite point configurations in a region
\(D\subseteq\Real^2\), an event might be ``at least ten points fall in
\(B\subseteq D\).''
\item If \(\Omega=S^n\) is a species-label sample, an event might be ``exactly
\(r\) species are observed once.''
\item If \(\Omega=C[0,1]\), an event might be ``the path crosses level \(a\)
before time \(t\).''
\end{enumerate}
The event is always a subset of \(\Omega\), but the scientific question shows
which subsets matter.
\end{example}

\subsection{Generated event systems}
\conceptindexes{generated sigma-field, event systems, generators, observable distinctions}

In practice, event systems are not listed one set at a time.  They are generated
from simpler observable questions.  If \(\classS\subseteq\Pow(\Omega)\), define
\[
  \sigma(\classS)
  =
  \bigcap\{\fieldF:\classS\subseteq\fieldF,\ 
  \fieldF\text{ is a }\sigma\text{-field on }\Omega\}.
\]
This is the smallest \(\sigma\)-field containing \(\classS\).
\Appref{app:set-theory-compass} reviews the set-theoretic convention behind
this intersection.

\begin{example}[Borel sets as observable geometry]
Let \(X\) be a topological space.  The Borel \(\sigma\)-field is
\[
  \Borel(X)=\sigma\{G:G\subseteq X\text{ is open}\}.
\]
Equivalently, it is generated by the closed sets.  On \(\Real^k\),
\[
  \Borel(\Real^k)
  =
  \sigma\{(a,b]:a,b\in\Rat^k,\ a_j<b_j,\ j=1,\ldots,k\}.
\]
Thus the continuous space can be read through a countable family of rational
rectangles.  The technical proof is in \Appref{sec:appB-borel-rectangles}; the main point is conceptual: a small
generating language can produce a large and stable event system.
\end{example}

\begin{tcolorbox}[
  enhanced,
  breakable,
  colback=noteback,
  colframe=bookgold!75!black,
  boxrule=0.55pt,
  arc=4pt,
  boxsep=1pt,
  left=0.95em,
  right=0.9em,
  top=0.7em,
  bottom=0.7em,
  before skip=0.85\baselineskip,
  after skip=0.95\baselineskip
]
\noindent\textbf{Borel operating rules.}
Open sets and closed sets are Borel.  An \(F_\sigma\)-set is a countable union
of closed sets, and a \(G_\delta\)-set is a countable intersection of open sets;
both are Borel.  Complements, countable unions, and countable intersections of
Borel sets are Borel.  Continuous preimages of Borel sets are Borel:
\[
  f:X\to Y\text{ continuous},\ B\in\Borel(Y)
  \quad\Longrightarrow\quad
  f^{-1}(B)\in\Borel(X).
\]
Images are subtler than preimages.  Homeomorphisms, translations, scalings, and
nonsingular linear maps on Euclidean space are safe cases; arbitrary images of
Borel sets require more structure.
\end{tcolorbox}

\begin{example}[Familiar sets in the Borel hierarchy]
The rationals are Borel because
\[
  \Rat=\bigcup_{q\in\Rat}\{q\},
\]
and each singleton is closed.  Hence \(\Rat\) is an \(F_\sigma\)-set.  Its
complement \(\Real-\Rat\) is therefore a \(G_\delta\)-set.  The Cantor set
is Borel because it is closed, but it is not small in the sense of cardinality:
it is uncountable.  These examples separate three notions that beginners often
blur: topological size, measure-theoretic size, and cardinality.  The full
Borel hierarchy continues these countable operations through the countable
ordinals; readers who want that descriptive-set-theoretic construction can
start with \citet{kechris1995classical}.
\end{example}

\begin{example}[A legal countable language can still be rich]
On \(\Real^k\), the Borel \(\sigma\)-field is generated by rational
half-open rectangles.  This means that the following event descriptions all
belong to the same countably generated language:
\[
  \{x:x_1\le c\},\qquad
  \{x:\|x\|<r\},\qquad
  \{x:x_1\in C,\ x_2\in[0,1]\}\quad(k\ge2),
\]
where \(c\in\Real\), \(r>0\), and \(C\) is the Cantor set.  Indeed, the first
set is \(\pi_1^{-1}((-\infty,c])\), where \(\pi_1(x)=x_1\) is continuous.  The
second is the open ball \(\|\,\cdot\,\|^{-1}((-\infty,r))\).  For \(k\ge2\), the
third is
\[
  \pi_1^{-1}(C)\cap\pi_2^{-1}([0,1])
  =
  C\times[0,1]\times\Real^{k-2},
\]
with the last factor omitted when \(k=2\).  Since \(C\) and \([0,1]\) are
closed and coordinate projections are continuous, it is Borel.  Countably
generated does not mean finite or simple; it means that every Borel question can
be built by countably many legal operations from a countable vocabulary.
\qedmark
\end{example}

\begin{example}[A map makes a question measurable]
Let \(\Omega_D\) be a space of point configurations in a city region \(D\).
An element \(\omega\in\Omega_D\) is one possible realized map of points, for
example \(\omega=\{x_1,\ldots,x_n\}\), not a single point.  For a district
\(B\subset D\), summarize this map by
\[
  X_B(\omega)=\#\{x\in\omega:x\in B\},
\]
the number of points falling in \(B\).  The event
\[
  \{\omega:X_B(\omega)\ge 10\}
\]
is \(X_B^{-1}(\{10,11,\ldots\})\), hence is measurable once the counting map
\(X_B\) is measurable.  A vague phrase such as ``the points look targeted''
becomes mathematical only after it is translated into a measurable rule, such as
a statistic of cell counts or nearest-neighbor distances.
\end{example}

\subsection{Rings, algebras, and semi-rings}
\conceptindexes{rings of sets, algebras of sets, semi-rings of sets, finite set systems}

The word \(\sigma\)-field is the final event language, but many constructions
begin with smaller set systems.  They are useful because simple sets are easier
to draw, count, partition, and assign mass to.

\begin{definition}[Finite set systems]
Let \(\mathcal A\) be a family of subsets of \(\Omega\).
\begin{enumerate}
\item \(\mathcal A\) is a \(\pi\)-system if it is closed under finite
intersections.
\item \(\mathcal A\) is a ring if it is closed under finite unions and relative
differences.
\item \(\mathcal A\) is an algebra if \(\Omega\in\mathcal A\), and whenever
\(A,B\in\mathcal A\),
\[
  A^c=\Omega-A\in\mathcal A,
  \qquad
  A\cup B\in\mathcal A.
\]
Equivalently, an algebra is closed under complements and finite unions.  It is
therefore also closed under finite intersections and relative differences.
\end{enumerate}
\end{definition}

\begin{center}
\small
\textbf{Closure properties of basic set systems.}\par\smallskip
\renewcommand{\arraystretch}{1.12}
\begin{tabular}{@{}lcccccc@{}}
\toprule
Family & \(A\cup B\) & \(A\cap B\) & \(A-B\) & \(A^c\) & \(\emptyset\) & \(\Omega\) \\
\midrule
\(\pi\)-system & & \(+\) & & & & \\
ring & \(+\) & \(+\) & \(+\) & & \(+\) & \\
algebra & \(+\) & \(+\) & \(+\) & \(+\) & \(+\) & \(+\) \\
\(\sigma\)-field & countable & countable & \(+\) & \(+\) & \(+\) & \(+\) \\
\bottomrule
\end{tabular}
\end{center}

The table is not meant to be memorized as taxonomy.  It shows which tool is
appropriate for which task.  A \(\pi\)-system is enough for uniqueness
arguments.  A ring is enough for finite disjoint decompositions.  A
\(\sigma\)-field is needed once limiting events enter.

\begin{definition}[Semi-ring]
A family \(\classS\) of subsets of \(\Omega\) is a semi-ring if
\[
  \emptyset\in\classS,\qquad
  A,B\in\classS\Longrightarrow A\cap B\in\classS,
\]
and for every \(A,B\in\classS\), the difference \(B-A\) can be written as a
finite disjoint union of sets from \(\classS\).
\end{definition}

If \(\classS\) is a semi-ring, then the family of finite disjoint unions
\[
  \classR(\classS)
  =
  \left\{\biguplus_{i=1}^m A_i:m<\infty,\ A_i\in\classS\right\}
\]
is a ring.  This is the route from simple geometry to measurable geometry:
start with intervals or rectangles, close under finite cutting and pasting, and
then pass to the generated \(\sigma\)-field.

\begin{example}[Half-open intervals and rectangles]
On the line, the class
\[
  \{\emptyset\}\cup\{(a,b]:-\infty<a<b<\infty\}
\]
is a semi-ring.  The intersection of two half-open intervals is empty or
half-open, and the difference of two half-open intervals is a finite disjoint
union of half-open intervals.  In \(\Real^k\), the same is true for half-open
rectangles
\[
  (a,b]=\prod_{j=1}^k(a_j,b_j].
\]
These rectangles are why interval probabilities and distribution functions can
eventually define full Borel probability laws.
\end{example}

\begin{example}[Dyadic graph paper]
For \(n\in\Nat\) and \(m=(m_1,\ldots,m_k)\in\Int^k\), put
\[
  I_{n,m}
  =
  \prod_{j=1}^k\left(\frac{m_j}{2^n},\frac{m_j+1}{2^n}\right].
\]
For fixed \(n\), the sets \(I_{n,m}\) form a countable partition of
\(\Real^k\).  As \(n\) increases, the grid becomes finer.  Every open set can
be approximated by countable unions of such cells.  This is the geometric
version of the chapter's recurring theme: continuous questions can often be
reached through countable discrete scaffolding.
\end{example}

\subsection{Products and cylinders}
\conceptindexes{products, cylinders, product event systems, coordinate maps}

Many statistical outcomes are not single numbers.  They are vectors, tables,
sequences, trajectories, or marked event histories.  Product \(\sigma\)-fields
turn coordinatewise observations into events on the whole object.

\begin{definition}[Product event system]
For measurable spaces \((S_1,\mathcal S_1),\ldots,(S_k,\mathcal S_k)\), define
\[
  \bigotimes_{j=1}^k\mathcal S_j
  =
  \sigma\{A_1\times\cdots\times A_k:A_j\in\mathcal S_j\}.
\]
The generating sets \(A_1\times\cdots\times A_k\) are measurable rectangles.
\end{definition}

If \(S_j=\Real\) with its usual Borel \(\sigma\)-field, then
\[
  \Borel(\Real^k)=\Borel(\Real)^{\otimes k}.
\]
Indeed, each coordinate projection \(\pi_j:\Real^k\to\Real\) is continuous, so
every Borel rectangle belongs to \(\Borel(\Real^k)\).  Hence
\(\Borel(\Real)^{\otimes k}\subset\Borel(\Real^k)\).  Conversely, rational open
rectangles form a countable basis for \(\Real^k\), and each such rectangle is in
\(\Borel(\Real)^{\otimes k}\).  Therefore every open set, and hence every Borel
set in \(\Real^k\), belongs to the product \(\sigma\)-field. \qedmark

\begin{definition}[Countable product event system]
For measurable spaces \((S_n,\mathcal S_n)_{n\ge1}\), let
\(S=\prod_{n\ge1}S_n\) and let \(\pi_i:S\to S_i\) be the \(i\)-th coordinate
projection.  The product \(\sigma\)-field is
\[
  \bigotimes_{n\ge1}\mathcal S_n
  =
  \sigma\{\pi_i^{-1}(A):i\ge1,\ A\in\mathcal S_i\}.
\]
Equivalently, it is generated by finite-dimensional cylinders
\[
  \{\omega:(\omega_{i_1},\ldots,\omega_{i_m})\in B\},
  \qquad
  B\in\bigotimes_{\ell=1}^m\mathcal S_{i_\ell}.
\]
It is the smallest event system that makes all coordinate projections
measurable.
\end{definition}

The same coordinate criterion works for arbitrary index sets and stochastic
processes; it is stated where process laws are built
in Chapter~\ref{chap:product-spaces-processes},
Proposition~\ref{prop:ch06-coordinate-process-measurability}.

\begin{example}[A sequence can be observed finitely at a time]
Let \(\Omega=\Real^{\Nat}\) and let \(X_n(\omega)=\omega_n\).  The event
\[
  \{\omega:X_1\le0,\ X_7\in(1,2],\ X_{20}>5\}
\]
is a cylinder event.  The event
\[
  \{\omega:X_n\to0\}
  =
  \bigcap_{r=1}^{\infty}\bigcup_{m=1}^{\infty}
  \bigcap_{n\ge m}\{\omega:|X_n(\omega)|<1/r\}
\]
is also measurable, because it is built from countably many cylinder events.
This is why product event systems are large enough for limits, not just for
finite-dimensional questions.
\end{example}

\section{Measures and Size}
\label{sec:ch3-rich-measures}
\conceptindexes{measures, probability measures, size, countable additivity}

\subsection{Measures and probability measures}
\conceptindexes{measure, probability measure, countable additivity, null sets}

Once the event system has been chosen, a measure assigns size to events.
Probability is the special case in which the whole sample space has size one.

\begin{definition}[Measure]
Let \((\Omega,\fieldF)\) be a measurable space.  A measure is a function
\(\mu:\fieldF\to[0,\infty]\) such that \(\mu(\emptyset)=0\) and, for disjoint
events \(A_1,A_2,\ldots\),
\[
  \mu\left(\bigcup_{n=1}^{\infty}A_n\right)
  =
  \sum_{n=1}^{\infty}\mu(A_n).
\]
If \(\mu(\Omega)=1\), then \(\mu\) is a probability measure.  In this book we
usually denote such a measure by \(P\), reserving \(\mu,\nu\) for general
measures.
\end{definition}

\begin{proposition}[Basic consequences]
Let \(\mu\) be a measure on \((\Omega,\fieldF)\).  If \(A\subseteq B\), then
\[
  \mu(A)\le \mu(B).
\]
If \(\mu(A)<\infty\), then
\[
  \mu(B-A)=\mu(B)-\mu(A)
  \qquad (A\subseteq B).
\]
For arbitrary events \(A_n\),
\[
  \mu\left(\bigcup_{n=1}^{\infty}A_n\right)
  \le
  \sum_{n=1}^{\infty}\mu(A_n).
\]
\end{proposition}

\noindent\textit{Proof.}
If \(A\subseteq B\), then \(B=A\biguplus(B-A)\), so
\(\mu(B)=\mu(A)+\mu(B-A)\ge\mu(A)\).  If \(\mu(A)<\infty\), the same identity
gives \(\mu(B-A)=\mu(B)-\mu(A)\).  For subadditivity, disjointify the union by
setting
\[
  B_1=A_1,\qquad
  B_n=A_n-\bigcup_{i<n}A_i,\quad n\ge2.
\]
Then the \(B_n\)'s are disjoint, \(B_n\subseteq A_n\), and
\(\bigcup_nA_n=\biguplus_nB_n\).  Countable additivity and monotonicity give
\[
  \mu\left(\bigcup_nA_n\right)
  =
  \sum_n\mu(B_n)
  \le
  \sum_n\mu(A_n).
\]
\qedmark

\begin{proposition}[Boole's inequality]
For any events \(A_1,A_2,\ldots\),
\[
  P\left(\bigcup_{n=1}^{\infty}A_n\right)
  \le
  \sum_{n=1}^{\infty}P(A_n).
\]
In particular, for finitely many events,
\[
  P(A_1\cup\cdots\cup A_m)
  \le
  P(A_1)+\cdots+P(A_m).
\]
\end{proposition}

Boole's inequality is often the first probability bound one uses before any
model-specific structure is available.  It is crude because it ignores overlap,
but it is universal.  In high-dimensional statistics, multiple testing, and
concentration arguments, this simple inequality is the reason a large number of
small error probabilities can still be controlled.

\begin{example}[The union bound as bookkeeping]
Suppose \(m\) diagnostic checks are run on the same dataset, and the probability
that check \(j\) falsely raises an alarm is at most \(\alpha_j\).  Without any
independence assumption,
\[
  P(\text{at least one false alarm})
  \le
  \sum_{j=1}^m\alpha_j.
\]
If the right-hand side is small, the family of checks is jointly reliable.  If
it is large, the bound may be uninformative, but it still correctly names the
problem: many small chances can accumulate.
\end{example}

\begin{definition}[Finite, \(\sigma\)-finite, and probability measures]
A measure \(\mu\) is finite if \(\mu(\Omega)<\infty\).  It is
\(\sigma\)-finite if
\[
  \Omega=\bigcup_{n=1}^{\infty}\Omega_n,
  \qquad
  \Omega_n\in\fieldF,\quad \mu(\Omega_n)<\infty.
\]
A probability measure is a finite measure normalized to total mass one.
\end{definition}

\(\sigma\)-finiteness is the usual compromise between finite probability and
infinite geometric size.  Lebesgue measure on \(\Real^k\) is not finite, but it
is \(\sigma\)-finite because \(\Real^k\) is the countable union of bounded
cubes.  This condition is exactly what allows many uniqueness and extension
arguments to reduce infinite spaces to finite pieces.

\subsection{Common Measures}
\conceptindexes{counting measure, Lebesgue measure, Dirac measure, Hausdorff measure, probability distributions}

The definition of a measure is deliberately spare.  It becomes interesting only
because it can hold many kinds of size: point mass, cardinality, area, curve
length, fractal size, random counts, empirical distributions, and probability
laws in infinite-dimensional spaces.

\begin{center}
\small
\textbf{Examples of measures and their statistical roles.}\par\smallskip
\setlength{\tabcolsep}{0.42em}
\renewcommand{\arraystretch}{1.12}
\begin{tabular}{@{}p{0.27\linewidth}p{0.27\linewidth}p{0.36\linewidth}@{}}
\toprule
Measure & What it sees & Typical statistical role \\
\midrule
\(\delta_x\) & one fixed point & deterministic observation or atom \\
counting measure & number of elements & discrete likelihoods and event counts \\
Lebesgue measure & length, area, volume & densities and continuous models \\
Stieltjes measure & jumps and continuous mass & distribution functions and mixed laws \\
Hausdorff measure & scale-sensitive geometry & curves, surfaces, and fractals \\
empirical measure \(P_n\) & observed data as a law & nonparametric statistics \\
counting-process measure \(N\) & events in time and marks & survival and event-history data \\
Poisson random measure & random scattered mass & spatial processes and jumps \\
Gaussian measure on \(\Hilbert\) & random functions or signals & infinite-dimensional models \\
\bottomrule
\end{tabular}
\end{center}

\begin{example}[Atoms, counts, and empirical laws]
Let \((\Omega,\fieldF)\) be a measurable space.
\begin{enumerate}
\item For \(x\in\Omega\), the Dirac measure at \(x\) is
\[
  \delta_x(A)=\ind{x\in A}.
\]
It turns a deterministic outcome into a probability measure.
\item The counting measure is
\[
  \#(A)=
  \begin{cases}
  \text{the number of elements of }A, & A\text{ finite},\\
  \infty, & A\text{ infinite}.
  \end{cases}
\]
On \(\Pow(\Omega)\), counting measure is \(\sigma\)-finite exactly when
\(\Omega\) is countable.
\item Given observations \(X_1,\ldots,X_n\) in \((S,\mathcal S)\), the empirical
measure is
\[
  P_n=\frac1n\sum_{i=1}^{n}\delta_{X_i},
  \qquad
  P_n(B)=\frac1n\sum_{i=1}^{n}\ind{X_i\in B}.
\]
This is one of the most important translations in statistics: a dataset becomes
a probability measure.
\end{enumerate}
\end{example}

\begin{example}[Lebesgue and Stieltjes measures]
Lebesgue measure \(\lambda^k\) assigns length, area, or volume in \(\Real^k\).
It begins as a Borel measure and becomes the usual Lebesgue measure after
completion by subsets of null sets.

A one-dimensional Lebesgue--Stieltjes measure is generated by a nondecreasing
right-continuous function \(F:\Real\to\Real\):
\[
  \mu_F((a,b])=F(b)-F(a).
\]
When \(F\) is a distribution function, \(\mu_F\) is the probability law with
\[
  \mu_F((-\infty,x])=F(x).
\]
Thus distribution functions are not just plotting devices.  They are compressed
instructions for building probability measures.  Later we use the same
construction for bounded-variation functions; then the resulting
Lebesgue--Stieltjes measure is generally signed, obtained by subtracting the
two increasing parts in a Jordan decomposition.
\end{example}

\begin{example}[Densities, atoms, and mixed distributions]
If \(f\ge0\) and \(\int_\Real f\,d\lambda=1\), then
\[
  P(A)=\int_A f(x)\,dx
\]
defines a probability measure absolutely continuous with respect to Lebesgue
measure.  But many statistical laws are not purely continuous.  A measurement
with a detection limit may have an atom at zero and a density above zero:
\[
  P=p\,\delta_0+(1-p)P_{\mathrm{cont}},
  \qquad 0\le p\le1.
\]
The corresponding distribution function has a jump at zero and a smooth part
elsewhere.  The Stieltjes viewpoint handles both pieces at once; the density
viewpoint alone does not.
\end{example}

\begin{example}[Hausdorff measure and geometric scale]
Lebesgue measure can say that a curve in the plane has area zero.  That does
not mean the curve has no size.  On a metric space \((S,d)\), define
\[
\begin{aligned}
\mathcal{H}^s_\varepsilon(A)
&=
\inf\Bigl\{\sum_i(\diam U_i)^s:
A\subseteq\bigcup_i U_i,\ 
\diam U_i\le\varepsilon\Bigr\},\\
\mathcal{H}^s(A)
&=
\lim_{\varepsilon\downarrow0}\mathcal{H}^s_\varepsilon(A).
\end{aligned}
\]
Up to conventional normalizing constants, \(\mathcal{H}^0\) is counting
measure, \(\mathcal{H}^1\) measures curve length, and noninteger \(s\)'s make
fractal size visible.

The middle-third Cantor set is the example to keep in mind.  Lebesgue measure
sees it as null.  Hausdorff measure asks at what scale it becomes visible.  The
critical exponent is
\[
  s=\frac{\log 2}{\log 3},
\]
because at construction level \(n\) there are \(2^n\) intervals of length
\(3^{-n}\), and
\[
  2^n(3^{-n})^s=(2\,3^{-s})^n.
\]
The equation \(2\,3^{-s}=1\) marks the dimension at which the cover neither
explodes nor vanishes.  This is a perfect example of measure as a choice of
resolution.  \Appref{sec:appB-hausdorff} records the matching lower-bound
idea behind the Hausdorff dimension calculation.
\end{example}

\begin{example}[Gaussian measures in Hilbert space]
Let \(\Hilbert\) be a separable Hilbert space with orthonormal basis
\(e_1,e_2,\ldots\).  If \(Z_i\) are independent standard normal variables and
\(\lambda_i\ge0\) with \(\sum_i\lambda_i<\infty\), then
\[
  X=\sum_{i=1}^{\infty}\sqrt{\lambda_i}\,Z_i e_i
\]
converges in \(\Hilbert\) almost surely.  The law of \(X\) is a Gaussian Borel
probability measure on \(\Hilbert\).

Here is the short measure-theoretic check.  Since
\[
  \Expect\sum_{i=1}^{\infty}\lambda_i Z_i^2
  =
  \sum_{i=1}^{\infty}\lambda_i<\infty,
\]
one has \(\sum_i\lambda_i Z_i^2<\infty\) almost surely.  Thus the partial sums
are Cauchy in \(\Hilbert\) almost surely and converge to an
\(\Hilbert\)-valued random element.  For any \(h\in\Hilbert\),
\[
  \langle X,h\rangle
  =
  \sum_{i=1}^{\infty}\sqrt{\lambda_i}\,Z_i\langle e_i,h\rangle
\]
is a real Gaussian random variable with variance
\[
  \sum_{i=1}^{\infty}\lambda_i\langle e_i,h\rangle^2.
\]
Equivalently, the covariance operator is \(Qe_i=\lambda_i e_i\).

The trace condition \(\sum_i\lambda_i<\infty\) is not a decorative assumption.
It says that the random series lands in the Hilbert space.  A formal Gaussian
with identity covariance would require \(\sum_i1<\infty\), which fails.  This is
one of the first places where infinite-dimensional probability refuses to be
Euclidean probability with more coordinates.

This measure is the natural language for random curves, spectra, images, and
fields after expanding them in a basis.  In manufacturing, the coordinates might
be principal components of sensor traces from a production line.  In biotech,
they might be coefficients of smooth gene-expression trends or spatial
intensity fields.  In physics and chemistry, they may describe noisy signals,
energy spectra, or random fields over space.  The important point is that a
probability law need not have a density on \(\Real^k\) to be useful; it may be a
Borel probability measure on a function space.
\end{example}

\begin{example}[Counting-process measures]
If event times and marks are \((T_i,X_i)\), with marks in a measurable space
\((S,\mathcal A)\), then each outcome \(\omega\) defines a counting measure
\[
  N_\omega((0,t]\times B)
  =
  \sum_{i\ge1}\ind{T_i(\omega)\le t,\ X_i(\omega)\in B}.
\]
For fixed \((0,t]\times B\), the value \(N((0,t]\times B)\) is a random
variable.  As \(t\) varies, the object becomes a counting process.  This example
is the bridge from measure theory to the later chapters on product spaces,
stochastic processes, and event-history data.
\end{example}

\begin{example}[Poisson random measures]
A Poisson random measure is the random version of counting measure guided by a
deterministic intensity measure.  Let \((S,\mathcal S)\) be a measurable space
and let \(\Lambda\) be a \(\sigma\)-finite measure on it.  A random measure \(N\)
is called Poisson with intensity \(\Lambda\) if
\[
  N(A)\sim \Poisson(\Lambda(A))
  \qquad\text{whenever }\Lambda(A)<\infty,
\]
and if \(N(A_1),\ldots,N(A_m)\) are independent whenever
\(A_1,\ldots,A_m\) are disjoint.

The definition is compact but expressive.  In a spatial point pattern,
\(\Lambda(A)\) is the expected number of points in region \(A\).  In a jump
process, \(\Lambda\) says how often jumps of different sizes should appear.  In
survival analysis, it is the ancestor of the cumulative hazard: deterministic
mass controls random event counts.

Return now to the London bombing example from Chapter~2.  If \(S\) is the map
region and \(\lambda^2\) is area measure, the homogeneous Poisson model uses
\[
  \Lambda(A)=\rho\,\lambda^2(A),
\]
so the observed impact pattern is the random counting measure
\[
  N=\sum_i\delta_{X_i},\qquad N(A)=\#\{i:X_i\in A\}.
\]
Choosing grid cells \(A_1,\ldots,A_m\) turns the map into data
\((N(A_1),\ldots,N(A_m))\).  The model says that disjoint cells have
independent Poisson counts, with means proportional to their areas.  The
unknown rate \(\rho\) is estimated from the total count divided by total area,
and inference asks whether the observed clustering summary, such as the
numbers of empty, one-hit, two-hit, and many-hit cells, is rare under that
fitted null law \citep{clarke1946poisson,feller1968introduction}.

Thus a Poisson random measure is not merely a decorative abstract object.  It
records the whole statistical motion:
\[
  \begin{aligned}
  &\text{question}
  \longrightarrow
  \text{locations}
  \longrightarrow
  \text{random counting measure} \\
  &\hspace{2.6em}\longrightarrow
  \text{intensity model}
  \longrightarrow
  \text{comparison of observed and expected patterns}.
  \end{aligned}
\]
It is the same example, now written in the language that lets the idea travel
from maps to jumps, event histories, spatial biology, and point processes.
Later, the same object will be constructed as a random measure; still later,
the mean measure will reappear dynamically through intensities and
compensators.
\end{example}

\begin{example}[Determinantal point processes]
A Poisson random measure makes disjoint counts independent.  A determinantal
point process (DPP) keeps the same counting-measure object
\[
  N=\sum_i\delta_{X_i},
\]
but changes the law so that points tend to repel.  Relative to a reference
measure, a DPP with kernel \(K\) has product densities of the form
\[
  \rho_m(x_1,\ldots,x_m)
  =
  \det\{K(x_a,x_b)\}_{a,b=1}^m,
\]
under the usual positivity and contraction conditions on the kernel.  If two
candidate locations are nearly identical under \(K\), two rows of this matrix
nearly agree, the determinant is small, and observing both points together is
unlikely.

This determinant is the mathematical reason DPPs are useful for modeling
random-matrix-style eigenvalue repulsion, inhibited spatial patterns, and
diversity-promoting subset selection in machine learning, recommendations,
and summarization
\citep{macchi1975coincidence,houghKrishnapurPeresVirag2009zeros,
kuleszaTaskar2012dpp,lavancier2015dpp}.  The lesson for this chapter is
structural: once a point pattern is written as a random counting measure,
changing the law from independent Poisson counts to a repulsive DPP changes
the dependence, not the measurable object.
\end{example}

\subsection{Integration as weighted measurement}
\conceptindexes{integration, expectation, weighted measurement, simple functions}

Measures assign size to sets.  Integrals assign weighted size to functions.
For a nonnegative simple function
\[
  s=\sum_{j=1}^m a_j\indset{A_j},
  \qquad a_j\ge0,\ A_j\in\fieldF,
\]
define
\[
  \int s\,d\mu=\sum_{j=1}^m a_j\mu(A_j).
\]
For a nonnegative measurable function \(f\), define
\[
  \int f\,d\mu
  =
  \sup\left\{\int s\,d\mu:0\le s\le f,\ s\text{ simple}\right\}.
\]
For a probability measure \(P\), expectation is integration:
\[
  \Expect X=\int_\Omega X\,dP
\]
whenever the positive and negative parts are not both infinite.
The convergence rules that make this definition useful in limiting arguments
are collected in \Appref{sec:appB-integration-engine}.

\begin{example}[The same integral, three readings]
The expression \(\int f\,d\mu\) changes its meaning with the measure.
\begin{enumerate}
\item If \(\mu=\lambda\), it is area under a curve or volume under a surface.
\item If \(\mu=\#\) is counting measure on a countable set, it is a sum:
\[
  \int f\,d\#=\sum_x f(x).
\]
\item If \(\mu=P_n=n^{-1}\sum_i\delta_{X_i}\), it is an empirical average:
\[
  \int f\,dP_n=\frac1n\sum_{i=1}^n f(X_i).
\]
\end{enumerate}
This is why the empirical measure is more than notation.  It turns sample
averages into ordinary integrals.
\end{example}

\subsection{Comparing measures}
\label{subsec:ch3-rich-comparing-measures}
\conceptindexes{absolute continuity, singularity, Radon--Nikodym derivative, comparing measures}

A measure is never just a pile of numbers.  It is a way of seeing a space.  Two
measures on the same measurable space may see the same negligible sets, or they
may live on nearly disjoint parts of the space.  This distinction is not
technical decoration; it is the language behind densities, likelihood ratios,
atoms, singular distributions, and many statistical pathologies.

\begin{definition}[Absolute continuity and singularity]
Let \(\mu\) and \(\nu\) be measures on \((\Omega,\fieldF)\).
\begin{enumerate}
\item We say \(\mu\) is absolutely continuous with respect to \(\nu\), written
\[
  \mu\ll\nu,
\]
if \(\nu(A)=0\) implies \(\mu(A)=0\).
\item We say \(\mu\) and \(\nu\) are singular, written
\[
  \mu\perp\nu,
\]
if there is a measurable set \(B\) such that \(\mu(B^c)=0\) and
\(\nu(B)=0\).
\end{enumerate}
\end{definition}

Absolute continuity says that \(\mu\) does not assign mass to places invisible
to \(\nu\).  Singularity says that the two measures can be separated onto
disjoint measurable worlds.  A density is a way of translating one measure into
another.

\begin{theorem}[Radon--Nikodym theorem, reading version; \citealp{radon1913theorie,nikodym1930generalisation}]
If \(\mu\) and \(\nu\) are \(\sigma\)-finite measures and \(\mu\ll\nu\), then
there exists a nonnegative measurable function \(f\) such that
\[
  \mu(A)=\int_A f\,d\nu
  \qquad(A\in\fieldF).
\]
The function \(f\), unique up to \(\nu\)-null sets, is written
\[
  f=\frac{d\mu}{d\nu}.
\]
\end{theorem}

\Appref{sec:appB-radon-nikodym} gives the technical formulation and the
Lebesgue decomposition theorem.  For now the message is enough: densities are
not intrinsic functions floating above the space.  They are derivatives of one
measure relative to another measure.

\begin{example}[Densities depend on a reference measure]
The standard normal law has density
\[
  \phi(x)=\frac{1}{\sqrt{2\pi}}e^{-x^2/2}
\]
with respect to Lebesgue measure.  A Bernoulli law has mass function
\[
  p^x(1-p)^{1-x},\qquad x\in\{0,1\},
\]
with respect to counting measure on \(\{0,1\}\).  These are the same idea:
\[
  P(A)=\int_A f\,d\nu.
\]
The reference measure \(\nu\) names what kind of support the model is using.
Continuous models usually use Lebesgue measure; discrete models usually use
counting measure; mixed models need a reference measure that can see both atoms
and continuous pieces.
\end{example}

\begin{example}[Singular but not atomic: the Cantor distribution]
Let \(B_1,B_2,\ldots\) be independent Bernoulli variables with
\[
  P(B_n=0)=P(B_n=1)=\frac12
\]
and define
\[
  X=\sum_{n=1}^{\infty}\frac{2B_n}{3^n}.
\]
Then \(X\in C\) almost surely, where \(C\) is the middle-third Cantor set.  The
law \(P_X\) therefore satisfies
\[
  P_X(C)=1,\qquad \lambda(C)=0,
\]
so \(P_X\perp\lambda\).  But \(P_X\) has no atoms: fixing one point of \(C\)
fixes an infinite sequence of binary choices, whose probability is
\[
  \prod_{n=1}^{\infty}\frac12=0.
\]
Thus \(P_X\) is singular continuous.  It is not discrete, not Lebesgue-density
continuous, and not exotic for the sake of being exotic.  It shows that
``continuous distribution'' and ``has a density'' are different statements.
\end{example}

\begin{example}[Likelihood ratios as measure ratios]
Suppose \(P_\theta\ll\nu\) for every \(\theta\) in a model, and write
\[
  f_\theta=\frac{dP_\theta}{d\nu}.
\]
For observed data \(x\), likelihood compares the densities \(f_\theta(x)\).
Changing the common reference measure changes all densities in a coordinated
way, but the likelihood ratio
\[
  \frac{f_{\theta_1}(x)}{f_{\theta_0}(x)}
\]
is the derivative \(dP_{\theta_1}/dP_{\theta_0}\) at points where the two laws
are mutually absolutely continuous.  This is one reason absolute continuity is
not a side condition in statistics: it is the condition under which models can
be compared by ratios rather than by incompatible supports.
\end{example}

\begin{example}[Empirical measures are close, yet singular]
Let \(X_1,X_2,\ldots\) be iid from a continuous distribution \(P\) on
\(\Real\), and let
\[
  P_n=\frac1n\sum_{i=1}^n\delta_{X_i}.
\]
For each fixed \(n\), \(P_n\) is concentrated on finitely many observed points,
while \(P\) gives those points probability zero.  Thus \(P_n\perp P\) almost
surely when \(P\) is non-atomic.  Nevertheless, for many classes of functions,
\[
  \int f\,dP_n \to \int f\,dP
\]
as \(n\to\infty\).  Singularity is a statement about exact support at a fixed
sample size; statistical convergence is a statement about how integrals behave
as information accumulates.  Confusing these two ideas is a common source of
unnecessary mystery.
\end{example}

\section{Random Variables and Pushforward Laws}
\label{sec:ch3-rich-random-variables}
\conceptindexes{random variables, random elements, measurable maps, pushforward laws, distributions}

A random variable is not a variable that behaves magically.  It is a measurable
map from an outcome space into a value space.

\begin{definition}[Measurable map]
Let \((\Omega,\fieldF)\) and \((S,\mathcal S)\) be measurable spaces.  A map
\(X:\Omega\to S\) is measurable if
\[
  X^{-1}(B)=\{\omega:X(\omega)\in B\}\in\fieldF
  \qquad\text{for every }B\in\mathcal S.
\]
When \(S=\Real\) and \(\mathcal S=\Borel(\Real)\), \(X\) is a real-valued random
variable.
\end{definition}

Preimages behave better than images:
\[
  X^{-1}\left(\bigcup_nB_n\right)=\bigcup_nX^{-1}(B_n),
  \qquad
  X^{-1}(B^c)=X^{-1}(B)^c.
\]
This is why measurability is formulated with preimages.

\begin{definition}[Distribution or pushforward law]
If \(X:(\Omega,\fieldF,P)\to(S,\mathcal S)\) is measurable, its distribution is
the probability measure \(P_X\) on \((S,\mathcal S)\) defined by
\[
  P_X(B)=P(X\in B)=P(X^{-1}(B)).
\]
Equivalently, \(P_X=P\circ X^{-1}\).
\end{definition}

\subsection{Information Carried by a Random Variable}
\conceptindexes{information, generated sigma-field, sigma-field generated by a random variable, random vector, stochastic process, random field, random measure, observation map, distribution}

\begin{definition}[Generated \(\sigma\)-field]
Let \(X:(\Omega,\fieldF)\to(E,\mathcal E)\) be a random element.  The
\(\sigma\)-field generated by \(X\) is
\[
  \sigma(X)=X^{-1}(\mathcal E)=\{X^{-1}(B):B\in\mathcal E\}.
\]
Equivalently, \(\sigma(X)\) is the smallest sub-\(\sigma\)-field of
\(\fieldF\) that makes \(X\) measurable.  It is the information about
\(\omega\) available after observing \(X(\omega)\).
\end{definition}

Preimage rules show that \(X^{-1}(\mathcal E)\) is a \(\sigma\)-field.  The
minimality is built into the definition: any \(\sigma\)-field that makes \(X\)
measurable must contain all events \(X^{-1}(B)\).

The same notation applies to larger random objects:
\[
\begin{array}{rcl}
X=(X_1,\ldots,X_k) &\Longrightarrow&
  \sigma(X)=\sigma(X_1,\ldots,X_k),\\[0.35em]
X=\{X_t:t\in T\} &\Longrightarrow&
  \sigma(X)=\sigma(X_t:t\in T),\\[0.35em]
M\text{ a random measure on }(S,\mathcal S) &\Longrightarrow&
  \sigma(M)=\sigma\{M(A):A\in\mathcal S\}.
\end{array}
\]
For a process or random field, this means the information carried by all
coordinate variables.  For a random measure, it means the information carried by
all evaluation counts or masses; often a smaller determining class, such as
intervals, rectangles, or grid cells, gives the same \(\sigma\)-field.
The product-space and random-measure constructions are developed
in Chapter~\ref{chap:product-spaces-processes}, especially
Proposition~\ref{prop:ch06-coordinate-process-measurability} and
Section~\ref{sec:ch06-random-objects}.
The time-indexed version of the same idea is the natural filtration:
\[
  \mathcal F_t^X=\sigma(X_s:s\le t)
\]
for a process, and similarly the history generated by a random measure is built
from its evaluation variables up to time \(t\).
\ifdefined\CHAPTERTHREE
\else
Chapter~\ref{chap:continuous-time-processes} uses this language for event
histories, stopping times, predictable processes, and compensators.
\fi

If \(Y=g(X)\) for a measurable map \(g\), then
\[
  \sigma(Y)\subseteq\sigma(X).
\]
The inclusion says that applying a summary cannot create information about the
original outcome.  It can only discard information or re-express it.

\begin{example}[Statistics as measurable compression]
Let \(X_1,\ldots,X_n\) be real observations.  The sample mean, empirical
distribution function, maximum, and number of observations in a set \(B\) are
all measurable functions of the data vector:
\[
  \bar X,\qquad
  F_n(t)=\frac1n\sum_{i=1}^n\ind{X_i\le t},\qquad
  M_n=\max_iX_i,\qquad
  \sum_{i=1}^n\ind{X_i\in B}.
\]
They are different compressions of the same outcome.  A statistic is useful
when the compression keeps the feature needed for the question and discards
what is irrelevant.
\end{example}

\begin{example}[Rare species as a measurable summary]
In the missing-species problem, an outcome may be a sequence of sampled labels.
Let
\[
  F_r(\omega)=\#\{j:\text{species }j\text{ appears exactly }r
  \text{ times in }\omega\}.
\]
The count \(F_1\) of singletons is a random variable.  It carries information
about what has not been observed because rare observed labels are evidence of
unseen labels.  The statistical idea is vivid, but measure theory supplies the
permission: once \(F_1\) is measurable, it has a law, an expectation, and a
limiting behavior \citep{fisher1943species,good1953population,efronThisted1976unseen}.
\end{example}

\begin{proposition}[Change of variables through a pushforward law]
\label{prop:ch3-pushforward-integration}
If \(X:(\Omega,\fieldF,P)\to(S,\mathcal S)\) is measurable and \(g:S\to\Real\)
is nonnegative or integrable, then
\[
  \int_\Omega g(X(\omega))\,P(d\omega)
  =
  \int_S g(x)\,P_X(dx).
\]
\end{proposition}

\noindent\textit{Proof.}
Check the identity first for indicators \(g=\indset{B}\), where it is exactly
the definition of \(P_X\):
\[
  \int_\Omega \indset B(X(\omega))\,P(d\omega)
  =
  P(X\in B)
  =
  P_X(B)
  =
  \int_S\indset B(x)\,P_X(dx).
\]
Linearity extends the identity to nonnegative simple functions.  For
nonnegative measurable \(g\), choose simple \(g_m\uparrow g\) and apply monotone
convergence on both sides.  If \(g\) is integrable, apply the nonnegative result
to \(g^+\) and \(g^-\) and subtract the two finite values. \qedmark

\section{Limits: Why Countability Matters}
\label{sec:ch3-rich-limits}
\conceptindexes{limits, countability, convergence modes, complete convergence, continuity of measures}

Probability becomes useful because it survives limiting operations.  Larger
samples, longer follow-up, repeated opportunities, and finer partitions all
force finite statements to pass to limits.

\begin{theorem}[Continuity of measures]
Let \(\mu\) be a measure on \((\Omega,\fieldF)\).
If \(A_n\uparrow A\), then
\[
  \mu(A)=\lim_{n\to\infty}\mu(A_n).
\]
If \(A_n\downarrow A\) and \(\mu(A_1)<\infty\), then
\[
  \mu(A)=\lim_{n\to\infty}\mu(A_n).
\]
\end{theorem}

\noindent\textit{Proof.}
For the increasing case, write
\[
  A=A_1\biguplus\bigcup_{n\ge2}(A_n-A_{n-1})
\]
as a disjoint union.  Countable additivity gives
\[
  \mu(A)=\mu(A_1)+\sum_{n\ge2}\mu(A_n-A_{n-1})
  =
  \lim_{m\to\infty}\mu(A_m).
\]
For the decreasing case, the sets \(A_1-A_n\) increase to \(A_1-A\).  By the
increasing case,
\[
  \mu(A_1-A_n)\uparrow\mu(A_1-A).
\]
Since \(\mu(A_1)<\infty\), subtracting from \(\mu(A_1)\) gives
\(\mu(A_n)\downarrow\mu(A)\). \qedmark

\begin{theorem}[Fatou's lemma for sets]
For events \(A_n\),
\[
  \mu(\liminf_nA_n)\le \liminf_n\mu(A_n).
\]
\end{theorem}

\noindent\textit{Proof.}
Write
\[
  \liminf_n A_n
  =
  \bigcup_{m=1}^{\infty}\bigcap_{n\ge m}A_n.
\]
The sets \(C_m=\bigcap_{n\ge m}A_n\) increase with \(m\), so continuity from
below gives
\[
  \mu(\liminf_nA_n)=\lim_{m\to\infty}\mu(C_m).
\]
For every \(n\ge m\), \(C_m\subseteq A_n\), hence
\[
  \mu(C_m)\le \inf_{n\ge m}\mu(A_n).
\]
Letting \(m\to\infty\) gives the inequality. \qedmark

The liminf is the event that all but finitely many \(A_n\)'s occur.  The
limsup is the event that infinitely many occur:
\[
  \limsup_nA_n
  =
  \bigcap_{m=1}^{\infty}\bigcup_{n\ge m}A_n.
\]
\Appref{sec:app-set-limits} gives the set-theoretic identities.

\begin{tcolorbox}[
  enhanced,
  breakable,
  colback=noteback,
  colframe=bookgold!75!black,
  boxrule=0.55pt,
  arc=4pt,
  boxsep=1pt,
  left=0.95em,
  right=0.9em,
  top=0.7em,
  bottom=0.7em,
  before skip=0.85\baselineskip,
  after skip=0.95\baselineskip
]
\noindent\textbf{Set limits as statistical limits.}
\[
\begin{aligned}
  \limsup_nA_n
  &=
  \{\omega:\omega\in A_n\text{ for infinitely many }n\},\\
  \liminf_nA_n
  &=
  \{\omega:\omega\in A_n\text{ for all but finitely many }n\}.
\end{aligned}
\]
Thus \(\limsup\) describes repeated occurrence and \(\liminf\) describes
eventual permanence.  The indicators obey
\[
  \indset{\limsup_nA_n}=\limsup_n\indset{A_n},
  \qquad
  \indset{\liminf_nA_n}=\liminf_n\indset{A_n}.
\]
These identities are often the bridge from statements about events to
statements about random variables.
\end{tcolorbox}

\begin{example}[Convergence expressed as an event]
For random variables \(X_n\) and \(X\),
\[
  \{X_n\to X\}
  =
  \bigcap_{r=1}^{\infty}\bigcup_{m=1}^{\infty}
  \bigcap_{n\ge m}\{|X_n-X|<1/r\}.
\]
This formula explains why almost sure convergence is an event in the underlying
\(\sigma\)-field.  It also shows why countable operations are enough: the real
quantifier ``for every \(\varepsilon>0\)'' can be replaced by the countable
family \(\varepsilon=1/r\).
\end{example}

\begin{definition}[Four common modes of convergence]
For random variables \(X_n\) and \(X\), we say:
\begin{enumerate}
\item \(X_n\to X\) almost surely if \(P(X_n\to X)=1\).
\item \(X_n\to X\) in probability if, for every \(\varepsilon>0\),
\[
  P(|X_n-X|>\varepsilon)\to0.
\]
\item \(X_n\to X\) in \(L^p\), \(p\ge1\), if
\[
  E|X_n-X|^p\to0.
\]
\item \(X_n\to X\) in distribution if the laws \(P_{X_n}\) converge weakly to
the law \(P_X\); on \(\Real\), this means
\[
  P(X_n\le x)\to P(X\le x)
\]
at every continuity point \(x\) of the distribution function of \(X\).
\end{enumerate}
Almost sure convergence is the one that is literally an event in \(\Omega\);
the other three are convergence statements about probabilities, expectations,
or probability laws.
\end{definition}

\noindent\textit{Remark.}
These four names are not the full convergence vocabulary.  On a general finite measure
space, ``convergence in probability'' is usually called convergence in measure.
For measures on a locally compact space, vague convergence means
\[
  \int f\,d\mu_n\to\int f\,d\mu
  \qquad\text{for every }f\in C_c(S),
\]
and is the natural language for random measures and point processes.  Complete
convergence, introduced by \citet{hsuRobbins1947complete}, strengthens almost
sure convergence by requiring
\[
  \sum_{n=1}^{\infty}P(|X_n-X|>\varepsilon)<\infty
  \qquad(\varepsilon>0).
\]
Almost uniform convergence asks for uniform convergence outside a set of
arbitrarily small measure.  Egorov's theorem turns almost sure convergence into
almost uniform convergence on finite measure spaces; Riesz-type subsequence
principles relate convergence in measure to almost sure subsequential
convergence.  Lusin's theorem is a nearby but different small-exception-set
result: it makes a measurable function continuous outside a set of small
measure.

\begin{theorem}[First Borel--Cantelli lemma]
Let \(A_n\in\fieldF\).  If
\[
  \sum_{n=1}^{\infty}P(A_n)<\infty,
\]
then
\[
  P(A_n\ \text{infinitely often})=0.
\]
\end{theorem}

\noindent\textit{Proof.}
The event that the \(A_n\)'s occur infinitely often is
\[
  \limsup_n A_n
  =
  \bigcap_{m=1}^{\infty}\bigcup_{n\ge m}A_n.
\]
The sets \(B_m=\bigcup_{n\ge m}A_n\) decrease as \(m\) increases, so continuity
from above gives
\[
  P(\limsup_n A_n)=\lim_{m\to\infty}P(B_m).
\]
For each \(m\), countable subadditivity gives
\[
  P(B_m)
  \le
  \sum_{n\ge m}P(A_n).
\]
Since \(\sum_nP(A_n)<\infty\), the tail sums on the right converge to \(0\).
Hence \(P(\limsup_nA_n)=0\), which is exactly
\(P(A_n\ \text{infinitely often})=0\). \qedmark

The first Borel--Cantelli lemma turns summable rarity into eventual absence.
This is a statistical statement in measure-theoretic clothing: infinitely many
chances do not guarantee infinitely many surprises if the total probability
budget is finite.

\begin{example}[Almost sure language]
An event \(A\) happens almost surely if \(P(A)=1\).  This does not mean the
exceptional set is empty; it means the exceptional set is null.  For example,
if \(U\) is uniform on \([0,1]\), then
\[
  P(U\notin\Rat)=1
\]
because the rationals are countable and have Lebesgue measure zero.  Almost
sure statements are therefore stronger than high-probability statements, but
weaker than pointwise statements.  They are the natural language of limits in
probability.
\end{example}

\begin{example}[Rare alarms over time]
Suppose a monitoring system raises alarm \(A_n\) on day \(n\), and suppose a
calibration argument gives \(P(A_n)\le n^{-2}\).  Then
\[
  \sum_{n=1}^{\infty}P(A_n)<\infty,
\]
so the first Borel--Cantelli lemma says that only finitely many alarms occur
almost surely.  The theorem does not say when the last alarm occurs.  It says
that an infinite stream of sufficiently rare alarms eventually quiets down with
probability one.
\end{example}

\section{Completion and Null Sets}
\label{sec:ch3-rich-completion}
\conceptindexes{completion, null sets, outer measure, inner measure, measurable envelope}

Null sets are easy to underestimate.  They often behave like negligible dust,
but they also reveal why the distinction between countable and uncountable
operations matters.

\begin{definition}[Completion]
Let \((\Omega,\fieldF,\mu)\) be a measure space.  A subset \(N\subseteq\Omega\)
is \(\mu\)-null if \(N\subseteq D\) for some \(D\in\fieldF\) with
\(\mu(D)=0\).  The completion of \(\fieldF\) is
\[
  \fieldF^\mu
  =
  \{A=C\cup N:C\in\fieldF,\ N\text{ is }\mu\text{-null}\}.
\]
\end{definition}

The completed measure gives \(A=C\cup N\) the same mass as \(C\).  The detailed
well-definedness proof belongs in \Appref{sec:appB-completion}; the
reason is simple: changing a set by a subset of a null set should not change
its size.

\subsection{Outer and Inner Approximation}
\conceptindexes{outer measure, inner measure, nonmeasurable sets, measurability repairs}

Completion is one way to enlarge a measurable world.  Outer and inner
approximation are another.  Given a measure first defined on a rich
\(\sigma\)-field, the outer size of a set \(E\) asks how cheaply measurable
sets can cover it:
\[
  \mu^*(E)=\inf\{\mu(A):E\subseteq A,\ A\in\fieldF\}.
\]
The inner size asks how much measurable mass can be placed inside it:
\[
  \mu_*(E)=\sup\{\mu(A):A\subseteq E,\ A\in\fieldF\}.
\]
For finite measures, a set is measurable precisely when these two sizes agree.
The slogan is simple: a set is measurable when its size can be pinned down from
outside and inside by legal sets.

\begin{example}[Lebesgue regularity]
Lebesgue measurable sets in \(\Real^k\) can be approximated from outside by
open sets and from inside by compact sets.  If \(A\) is Lebesgue measurable and
\(\lambda^k(A)<\infty\), then for every \(\varepsilon>0\) there exist a compact
set \(K\) and an open set \(G\) such that
\[
  K\subseteq A\subseteq G,
  \qquad
  \lambda^k(G-K)<\varepsilon.
\]
Thus Lebesgue measurability is not merely closure under abstract operations; it
also has geometric content.
\end{example}

\begin{example}[The Cantor set]
The middle-third Cantor set is
\[
  C=\left\{\sum_{n=1}^{\infty}\frac{a_n}{3^n}:a_n\in\{0,2\}\right\}.
\]
It is closed, hence Borel.  It is obtained by removing intervals whose total
length is
\[
  \frac13+\frac{2}{9}+\frac{4}{27}+\cdots
  =
  \sum_{n=0}^{\infty}\frac{2^n}{3^{n+1}}
  =
  1,
\]
so \(C\) has Lebesgue measure zero.

Since Lebesgue measure is complete, every subset of \(C\) is Lebesgue
measurable with measure zero.  But \(C\) has cardinality \(\mathfrak c\), so it
has \(2^{\mathfrak c}\) subsets.  There are only \(\mathfrak c\) Borel subsets
of \(\Real\).  Thus the Lebesgue measurable sets are strictly larger than the
Borel sets.
\end{example}

\begin{example}[A null set whose sum is large]
Let
\[
  C+C=\{x+y:x,y\in C\}.
\]
Since \(C\subseteq[0,1]\), \(C+C\subseteq[0,2]\).  We show the reverse
inclusion.  Fix \(t\in[0,2]\) and put \(u=t/2\).  Choose a ternary expansion
\[
  u=\sum_{n=1}^{\infty}\frac{a_n}{3^n},\qquad a_n\in\{0,1,2\},
\]
using any such expansion; for the endpoint \(u=1\), use
\[
  1=\sum_{n=1}^{\infty}\frac{2}{3^n}=0.222\ldots_3.
\]
Define
\[
  b_n=c_n=a_n\quad(a_n=0\text{ or }2),\qquad
  b_n=0,\ c_n=2\quad(a_n=1).
\]
Then \(b_n,c_n\in\{0,2\}\), so
\[
  x=\sum_{n=1}^{\infty}\frac{b_n}{3^n},
  \qquad
  y=\sum_{n=1}^{\infty}\frac{c_n}{3^n}
\]
belong to \(C\).  Also \(b_n+c_n=2a_n\), hence \(x+y=2u=t\).  Therefore
\[
  C+C=[0,2].
\]

This does not contradict translation invariance.  A single translate \(x+C\)
has measure zero, but
\[
  C+C=\bigcup_{x\in C}(x+C)
\]
is an uncountable union.  Countable additivity controls countable unions; it
does not control arbitrary uncountable unions.
\end{example}

\section{Extension Theorems: Infrastructure for Laws}
\label{sec:ch3-rich-extension}
\conceptindexes{extension theorems, uniqueness, existence, generated laws, Caratheodory extension}

Statistical models often specify probabilities on simple events: intervals,
rectangles, cylinder sets, or finite-dimensional projections.  Extension
theorems explain why those assignments determine a full probability law.

\subsection{Uniqueness before existence}
\conceptindexes{uniqueness, pi-lambda theorem, generating class, Dynkin system}

There are two separate questions.  Uniqueness asks whether two measures that
agree on simple events must agree everywhere.  Existence asks whether an
assignment on simple events actually comes from a measure.  It is usually
easier to understand uniqueness first.

\begin{theorem}[Uniqueness from a generating class]
Let \(\classS\) be a \(\pi\)-system and let \(\fieldF=\sigma(\classS)\).  If
\(\mu_1\) and \(\mu_2\) are finite measures on \(\fieldF\), agree on
\(\classS\), and satisfy \(\mu_1(\Omega)=\mu_2(\Omega)\), then
\[
  \mu_1=\mu_2\quad\text{on }\fieldF.
\]
\end{theorem}

This is the reason distribution functions, moment-generating functions when
valid, finite-dimensional distributions, and rectangle probabilities can be so
powerful: agreement on a generating class can force agreement everywhere.
\Appref{sec:appB-dynkin} gives the \(\pi\)-\(\lambda\) proof.

\begin{example}[A distribution is determined by its intervals]
If \(P\) and \(Q\) are probability measures on \(\Borel(\Real)\) and
\[
  P((-\infty,x])=Q((-\infty,x])
  \qquad\text{for every }x\in\Real,
\]
then \(P=Q\).  The intervals \((-\infty,x]\) form a \(\pi\)-system that
generates \(\Borel(\Real)\), and both measures have total mass one.  This is
the formal reason a cumulative distribution function determines a
one-dimensional probability law.
\end{example}

\subsection{Existence from simple assignments}
\conceptindexes{existence, premeasure, extension from semi-ring, simple assignments}

Existence begins with a class \(\classS\) of simple sets and a set function
\(\gamma:\classS\to[0,\infty]\).  The assignment must be internally consistent:
whenever a simple set is decomposed into disjoint simple pieces, the assigned
mass must add correctly.  Semi-rings are designed so that this finite
bookkeeping can be done.

\begin{theorem}[Extension from simple building blocks]
Let \(\classS\) be a semi-ring and let \(\classR(\classS)\) be the ring of
finite disjoint unions of sets from \(\classS\).  Suppose
\(\gamma_R:\classR(\classS)\to[0,\infty]\) is a \(\sigma\)-finite premeasure:
\(\gamma_R(\emptyset)=0\), and whenever \(E=\biguplus_{n\ge1}E_n\) with
\(E,E_n\in\classR(\classS)\), one has
\[
  \gamma_R(E)=\sum_{n=1}^{\infty}\gamma_R(E_n).
\]
Then \(\gamma_R\) extends uniquely to a measure on \(\sigma(\classS)\).
\end{theorem}

The full Carath\'eodory construction is in
\Appref{sec:appB-outer-extension}.  The theorem buys a practical freedom: specify
the mass of intervals, rectangles, or cylinders, then let the theorem create
the measure on the generated event system.

\begin{example}[From length on intervals to Lebesgue measure]
Start with half-open intervals and define
\[
  \gamma((a,b])=b-a.
\]
Finite additivity is just telescoping along adjacent intervals.  Countable
additivity is subtler; it uses compactness of closed bounded intervals and
approximation by slightly enlarged covers.  Once this is checked on the
semi-ring, the extension theorem creates Lebesgue measure on
\(\Borel(\Real)\), and completion gives the usual Lebesgue measurable sets.
\end{example}

\begin{example}[Distribution functions as extension data]
A right-continuous nondecreasing distribution function \(F\) specifies
\[
  \gamma((a,b])=F(b)-F(a).
\]
The extension theorem turns this interval assignment into a probability measure
on \(\Borel(\Real)\).  In higher dimensions, the same idea uses alternating
rectangle increments.  This is why a cumulative distribution function is not
merely descriptive; it is a construction manual for a law.
\end{example}

\begin{example}[Multivariate rectangle increments]
For a distribution function \(F\) on \(\Real^2\), the mass assigned to a
half-open rectangle should be
\[
\begin{aligned}
  \gamma_F((a_1,b_1]\times(a_2,b_2])
  &=
  F(b_1,b_2)-F(a_1,b_2)\\
  &\quad{}-F(b_1,a_2)+F(a_1,a_2).
\end{aligned}
\]
The alternating signs are not decorative; they remove the two side rectangles
that were counted too many times and restore the lower-left corner.  In
\(\Real^k\), the same inclusion-exclusion pattern runs over the \(2^k\)
vertices of the rectangle.
\end{example}

\begin{example}[Cylinder probabilities]
For a stochastic process \(X_t\), finite-dimensional statements have the form
\[
  P\{(X_{t_1},\ldots,X_{t_k})\in B\}.
\]
These are probabilities of cylinder events.  Kolmogorov's extension theorem,
introduced later with product spaces, says that a compatible family of
finite-dimensional distributions determines a probability law on path space.
The philosophy is the same: simple observable questions generate the event
system, and consistency lets them define a law.
\end{example}

\begin{example}[Why consistency is not optional]
Suppose someone specifies a process by saying
\[
  P(X_1=1)=0.9,\qquad
  P(X_2=1)=0.9,\qquad
  P(X_1=1,X_2=1)=0.1.
\]
These numbers cannot define a probability law, because
\[
  P(X_1=1,X_2=1)\le P(X_1=1)
\]
is true but not enough; the marginal constraint also forces
\[
  P(X_1=1,X_2=1)\ge P(X_1=1)+P(X_2=1)-1=0.8.
\]
Finite-dimensional specifications must agree with their lower-dimensional
marginals and with the elementary inequalities of probability.  Extension
theorems are powerful only after these consistency tests have been passed.
\end{example}

\section{Independence and Tail Events}
\label{sec:ch3-rich-independence}
\conceptindexes{independence, tail events, zero-one laws, exchangeability, de Finetti theorem}

Independence is not just a computational shortcut.  It is a structural claim
about how event systems factor.

\begin{definition}[Independent events and classes]
Events \(A_1,\ldots,A_k\) are independent if, for every nonempty
\(I\subseteq\{1,\ldots,k\}\),
\[
  P\left(\bigcap_{i\in I}A_i\right)
  =
  \prod_{i\in I}P(A_i).
\]
Families of sets \(\mathcal A_1,\ldots,\mathcal A_k\) are independent if every
finite choice \(A_i\in\mathcal A_i\) is independent.  Random variables are
independent when their generated \(\sigma\)-fields are independent.
\end{definition}

The generated-\(\sigma\)-field formulation matters.  It makes it possible to prove
independence on convenient generating classes, then use Dynkin's theorem to
carry it to the full event systems.

\begin{proposition}[Checking independence on generators]
Let \(\mathcal A\) and \(\mathcal B\) be \(\pi\)-systems.  If
\[
  P(A\cap B)=P(A)P(B)
  \qquad(A\in\mathcal A,\ B\in\mathcal B),
\]
then \(\sigma(\mathcal A)\) and \(\sigma(\mathcal B)\) are independent.
\end{proposition}

\noindent\textit{Proof.}
Fix \(A\in\mathcal A\) and let
\[
  \mathcal D_A=\{B:P(A\cap B)=P(A)P(B)\}.
\]
This is a \(\lambda\)-system: it contains \(\Omega\), is closed under relative
differences of nested sets by finite additivity, and is closed under increasing
unions by continuity from below.  By assumption it contains \(\mathcal B\), so
Dynkin's theorem gives \(\sigma(\mathcal B)\subseteq\mathcal D_A\).  Thus, for
every \(B\in\sigma(\mathcal B)\),
\[
  P(A\cap B)=P(A)P(B)
  \qquad(A\in\mathcal A).
\]
Now fix such a \(B\) and define
\[
  \mathcal E_B=\{A\in\sigma(\mathcal A):P(A\cap B)=P(A)P(B)\}.
\]
The same \(\lambda\)-system argument shows that \(\mathcal E_B\) contains
\(\sigma(\mathcal A)\).  Hence every event in \(\sigma(\mathcal A)\) is
independent of every event in \(\sigma(\mathcal B)\). \qedmark

\begin{theorem}[Second Borel--Cantelli lemma]
If \(A_1,A_2,\ldots\) are independent events and
\[
  \sum_{n=1}^{\infty}P(A_n)=\infty,
\]
then
\[
  P(A_n\ \text{infinitely often})=1.
\]
\end{theorem}

\noindent\textit{Proof.}
For fixed \(m\),
\[
  P\left(\bigcap_{n=m}^{N}A_n^c\right)
  =
  \prod_{n=m}^{N}(1-P(A_n))
  \le
  \exp\left\{-\sum_{n=m}^{N}P(A_n)\right\}.
\]
Because the series \(\sum_nP(A_n)\) diverges, letting \(N\to\infty\) gives
\[
  P\left(\bigcap_{n\ge m}A_n^c\right)=0.
\]
Thus
\[
  P\left(\bigcup_{n\ge m}A_n\right)=1
  \qquad(m\ge1).
\]
Intersecting these probability-one events over \(m\) gives
\[
  P\left(\bigcap_{m=1}^{\infty}\bigcup_{n\ge m}A_n\right)=1.
\]
The event in parentheses is exactly \(A_n\) occurring infinitely often.
\qedmark

Together with the first Borel--Cantelli lemma, this theorem says that
independent repeated opportunities obey a sharp zero-one pattern: summable
probabilities give only finitely many occurrences; divergent probabilities give
infinitely many occurrences.

\begin{definition}[Tail \(\sigma\)-field]
\conceptindexes{tail sigma-field, tail events, Kolmogorov zero-one law}
For a sequence of random variables \(X_1,X_2,\ldots\), the tail \(\sigma\)-field
is
\[
  \mathcal T
  =
  \bigcap_{m=1}^{\infty}\sigma(X_m,X_{m+1},\ldots).
\]
Its events do not depend on any finite initial segment of the sequence.
\end{definition}

\begin{theorem}[Kolmogorov zero-one law; \citealp{kolmogorov1933grundbegriffe}]
If \(X_1,X_2,\ldots\) are independent, then every event \(A\in\mathcal T\)
satisfies
\[
  P(A)\in\{0,1\}.
\]
\end{theorem}

\noindent\textit{Proof.}
If \(A\in\mathcal T\), then for each \(m\) the event \(A\) belongs to
\(\sigma(X_{m+1},X_{m+2},\ldots)\), so independence of the variables gives
independence of \(A\) and \(\sigma(X_1,\ldots,X_m)\).  To pass from finite
initial blocks to the whole sequence, define
\[
  \mathcal D
  =
  \{B\in\sigma(X_1,X_2,\ldots):P(A\cap B)=P(A)P(B)\}.
\]
The class \(\mathcal D\) is a \(\lambda\)-system: it contains \(\Omega\), is
closed under complements inside \(\sigma(X_1,X_2,\ldots)\), and is closed under
countable disjoint unions by countable additivity.  Also
\(\sigma(X_1,\ldots,X_m)\subset\mathcal D\) for every \(m\).  The increasing
union
\[
  \mathcal F_0=\bigcup_{m=1}^{\infty}\sigma(X_1,\ldots,X_m)
\]
is an algebra, hence a \(\pi\)-system, and
\(\sigma(\mathcal F_0)=\sigma(X_1,X_2,\ldots)\).  Dynkin's \(\pi\)-\(\lambda\)
theorem gives \(\mathcal D=\sigma(X_1,X_2,\ldots)\).  But
\(A\in\mathcal T\subseteq\sigma(X_1,X_2,\ldots)\), so \(A\in\mathcal D\); in
other words, \(A\) is independent of itself.  Therefore
\[
  P(A)=P(A\cap A)=P(A)^2,
\]
and \(P(A)\) is either \(0\) or \(1\). \qedmark

\begin{example}[Limsup, liminf, and convergence are tail objects]
For any sequence \(X_1,X_2,\ldots\), independence is not needed to see that
\[
  \{\limsup_nX_n>r\}\in\mathcal T,
  \qquad
  \{\liminf_nX_n<r\}\in\mathcal T
  \qquad(r\in\Real).
\]
Indeed, deleting finitely many terms does not change \(\limsup\) or
\(\liminf\), so each displayed event belongs to
\(\sigma(X_m,X_{m+1},\ldots)\) for every \(m\).  The event that the sequence has
an extended-real limit is also tail-measurable, since
\[
  \{\lim_nX_n\text{ exists in }\overline{\Real}\}^c
  =
  \bigcup_{r\in\Rat}
  \{\liminf_nX_n<r\}\cap\{\limsup_nX_n>r\}.
\]
If the \(X_n\)'s are independent, Kolmogorov's zero-one law makes this
convergence event have probability \(0\) or \(1\).
\end{example}

There is a neighboring symmetry principle that is often more statistical than
the tail formulation.  A tail event ignores the first finitely many observations;
a permutation-invariant event ignores their labels.

\begin{definition}[Exchangeability and finite-permutation events]
\conceptindexes{exchangeability, finite permutations, Hewitt--Savage zero-one law, de Finetti theorem}
A sequence \(X_1,X_2,\ldots\) is exchangeable if for every \(n\) and every
permutation \(\sigma\) of \(\{1,\ldots,n\}\),
\[
  (X_1,\ldots,X_n)
  \quad\text{and}\quad
  (X_{\sigma(1)},\ldots,X_{\sigma(n)})
\]
have the same distribution.  An event \(A\in\sigma(X_1,X_2,\ldots)\) is
invariant under finite permutations if \(T_\pi^{-1}A=A\) for every permutation
\(\pi\) of the positive integers that moves only finitely many indices, where
\[
  T_\pi(x_1,x_2,\ldots)=(x_{\pi(1)},x_{\pi(2)},\ldots).
\]
\end{definition}

\begin{theorem}[Hewitt--Savage zero-one law; \citealp{hewitt1955symmetric}]
If \(X_1,X_2,\ldots\) are iid real-valued random variables, then every event
\(A\in\sigma(X_1,X_2,\ldots)\) that is invariant under finite permutations
satisfies
\[
  P(A)\in\{0,1\}.
\]
\end{theorem}

\noindent\textit{Proof.}
The proof has the same final shape as Kolmogorov's law, but the approximation
uses symmetry rather than tails.  Let
\(\mathcal F_0=\bigcup_n\sigma(X_1,\ldots,X_n)\).  This is an algebra and
\(\sigma(\mathcal F_0)=\sigma(X_1,X_2,\ldots)\).  The class of events that can
be approximated in probability by sets in \(\mathcal F_0\) is a \(\sigma\)-field,
so every event in \(\sigma(X_1,X_2,\ldots)\) has such approximations.  Choose
\(B\in\sigma(X_1,\ldots,X_n)\) with \(P(A\triangle B)<\varepsilon\).  Pick a
finite permutation \(\pi\) that moves the first \(n\) coordinates to a disjoint
block, say \(m+1,\ldots,m+n\), and put \(B^\pi=T_\pi^{-1}B\).  The iid
assumption makes the sequence law invariant under \(T_\pi\), while
\(T_\pi^{-1}A=A\); hence
\[
  P(A\triangle B^\pi)=P(A\triangle B)<\varepsilon.
\]
The events \(B\) and \(B^\pi\) depend on disjoint coordinate blocks, so they are
independent.  Therefore
\[
  |P(A)-P(A)^2|
  \le
  |P(A)-P(B\cap B^\pi)|+|P(B)^2-P(A)^2|
  \le 4\varepsilon.
\]
Letting \(\varepsilon\downarrow0\) gives \(P(A)=P(A)^2\). \qedmark

Kolmogorov's law says that remote properties of independent variables are
decisive.  Hewitt--Savage says something subtler: under iid sampling, any
property that depends only on the unordered infinite sample is also decisive.
This makes exchangeability a natural next step.  It weakens independence while
retaining the idea that labels carry no information.

\begin{example}[Zero-one laws in monitoring systems]
Zero-one laws are useful as model diagnostics in industrial monitoring.  In a
semiconductor fab, let \(X_i=1\) mean that wafer \(i\) triggers a critical defect
alarm.  In an online service, let \(X_i=1\) mean that deployment \(i\) produces a
major incident.  In a repeated toxicity screen, let \(X_i=1\) mean that compound
\(i\) crosses a safety threshold.  The event
\[
  \{X_i=1\ \text{infinitely often}\}
\]
is a tail event: changing any finite initial batch, deployment, or assay cannot
change whether the event occurs forever often.  Under an independent repeated
model, Kolmogorov's law forces its probability to be zero or one.  If the model
instead treats such a long-run event as having an irreducible middle probability,
that middle probability is usually not a property of the tail event itself; it
is a sign that there is an unmodeled source of heterogeneity, drift, or
adaptation.

Hewitt--Savage gives a related check for label-free claims.  Suppose a quality
engineer says that only the unordered collection of infinitely many test results
matters, not which serial number received which result.  For iid data, every
finite-permutation-invariant event is again zero or one.  This does not say that
industrial systems are iid.  It says what the iid assumption would imply, and
therefore where it is too rigid for production lines, platform experiments, and
biological screens with batch effects. \qedmark
\end{example}

\begin{theorem}[Bernoulli de Finetti theorem; \citealp{definetti1937prevision}]
Let \(X_1,X_2,\ldots\) be an exchangeable sequence taking values in
\(\{0,1\}\).  Then there is a unique probability measure \(\nu\) on \([0,1]\)
such that, for every \(n\) and every \(x_1,\ldots,x_n\in\{0,1\}\),
\[
  P(X_1=x_1,\ldots,X_n=x_n)
  =
  \int_0^1 z^k(1-z)^{n-k}\,\nu(dz),
  \qquad
  k=\sum_{i=1}^n x_i.
\]
Equivalently, there is a random variable \(Z\in[0,1]\) with law \(\nu\) such
that, conditionally on \(Z\), the variables \(X_1,X_2,\ldots\) are iid
Bernoulli\((Z)\).
\end{theorem}

The finite-dimensional identity is the part needed here.  The product-space
machinery behind the conditional-iid wording comes later, when kernels, product
measures, and extension from consistent finite-dimensional laws are developed
as tools for building full sequence and process laws.

\noindent\textit{Proof.}
Write \(S_N=X_1+\cdots+X_N\).  For a fixed pattern
\((x_1,\ldots,x_n)\) with \(k=\sum_{i=1}^n x_i\), exchangeability implies that,
conditional on \(S_N=s\), all binary strings of length \(N\) with \(s\) ones are
equally likely.  Hence, for \(N\ge n\),
\[
  P(X_1=x_1,\ldots,X_n=x_n)
  =
  \Expect
  \frac{(S_N)_k(N-S_N)_{n-k}}{(N)_n},
\]
where \((y)_m=y(y-1)\cdots(y-m+1)\) and \((y)_0=1\).

Let \(\nu_N\) be the law of \(S_N/N\) on \([0,1]\).  Since \([0,1]\) is compact,
some subsequence \(\nu_{N_j}\) converges weakly to a probability measure
\(\nu\).  The functions
\[
  g_N(z)=\frac{(Nz)_k(N(1-z))_{n-k}}{(N)_n}
\]
converge uniformly on \([0,1]\) to \(z^k(1-z)^{n-k}\).  Letting
\(N_j\to\infty\) gives
\[
  P(X_1=x_1,\ldots,X_n=x_n)
  =
  \int_0^1 z^k(1-z)^{n-k}\,\nu(dz).
\]

The mixing measure is unique.  Indeed, taking \(x_1=\cdots=x_m=1\) gives the
moments
\[
  \int_0^1z^m\,\nu(dz)=P(X_1=\cdots=X_m=1),\qquad m\ge0.
\]
Moments determine a probability measure on the compact interval \([0,1]\),
because polynomials are dense in \(C[0,1]\).  Finally, the mixture law
\[
  \int_0^1 \Bernoulli(z)^{\otimes\infty}\,\nu(dz)
\]
has exactly the displayed cylinder probabilities.  Cylinder events generate the
sequence \(\sigma\)-field, so uniqueness from the generating class identifies it
with the law of \((X_i)\).  This is precisely the conditional iid
representation. \qedmark

Exchangeability is therefore not independence, but for binary infinite sequences
it is exactly a mixture of iid laws.  The randomness that remains after
forgetting labels can be represented by the latent success probability \(Z\).

\begin{example}[P\'olya urn and beta mixing]
An urn starts with \(b>0\) black balls and \(r>0\) red balls.  After each draw,
the selected ball is returned together with \(c>0\) additional balls of the same
color.  Let \(X_i=1\) if draw \(i\) is black and \(X_i=0\) otherwise.  For the
event that the first \(m\) draws are all black,
\[
  P(X_1=\cdots=X_m=1)
  =
  \prod_{i=1}^{m}
  \frac{b+(i-1)c}{b+r+(i-1)c}.
\]
Put \(a=b/c\) and \(d=r/c\).  Using
\[
  \prod_{i=1}^{m}(\alpha+i-1)=\frac{\Gamma(\alpha+m)}{\Gamma(\alpha)}
  \qquad(\alpha>0),
\]
one obtains
\[
  P(X_1=\cdots=X_m=1)
  =
  \frac{\Gamma(a+m)\Gamma(a+d)}{\Gamma(a)\Gamma(a+d+m)}
  =
  \frac{\Beta(a+m,d)}{\Beta(a,d)}.
\]
If \(Z\sim\Beta(a,d)\), then this last quantity is
\[
  \Expect Z^m.
\]
More generally, for any particular binary sequence \(x_1,\ldots,x_n\) with
\(k=\sum_i x_i\) black draws,
\[
  P(X_1=x_1,\ldots,X_n=x_n)
  =
  \frac{\prod_{j=0}^{k-1}(b+jc)\prod_{\ell=0}^{n-k-1}(r+\ell c)}
       {\prod_{i=0}^{n-1}(b+r+ic)}
  =
  \frac{\Beta(a+k,d+n-k)}{\Beta(a,d)}.
\]
The probability depends on the sequence only through \(k\), so the draws are
exchangeable.  The final expression also equals
\[
  \Expect\{Z^k(1-Z)^{n-k}\},
  \qquad Z\sim\Beta(a,d).
\]
Thus the P\'olya urn is a concrete de Finetti mixture: conditionally on the
latent \(Z\), the draws behave like iid Bernoulli\((Z)\), while marginally the
reinforcement creates dependence. \qedmark
\end{example}

\begin{example}[Beta-binomial thinking across industries]
The urn notation is old, but the modeling pattern is contemporary.  Start with
\(Z\sim\Beta(a,d)\), and conditionally on \(Z\), let
\[
  X_i\mid Z\sim\Bernoulli(Z)
  \qquad\text{independently}.
\]
After \(n\) observations with \(k\) ones, the predictive probability for the next
one is
\[
  P(X_{n+1}=1\mid X_1,\ldots,X_n)
  =
  \frac{a+k}{a+d+n}.
\]
This is the same predictive rule as the P\'olya urn.  The interpretation changes
with the field.
\begin{enumerate}
\item In manufacturing, \(X_i=1\) may mean that unit \(i\) from a line, shift, or
supplier lot is defective.  The latent \(Z\) is the defect probability under the
current calibration, temperature, operator mix, and material batch.  Marginal
dependence between units is not mysterious; it comes from sharing the same
unknown production condition.
\item In IT and digital products, \(X_i=1\) may mean conversion, churn, fraud, or
an incident in a nominally comparable stream of users, requests, or releases.
The latent \(Z\) represents the feature, traffic mix, or reliability state that
is not fully observed.
\item In pharma and biotech, \(X_i=1\) may mean response, dose-limiting toxicity,
gene-edit success, contamination, or assay hit.  Exchangeability within a dose
cohort, batch, donor stratum, or plate is often a reasonable first-order
approximation, while the beta mixing distribution records uncertainty about the
underlying response or failure rate.
\item In biology, chemistry, and physics, the same model appears as colony
growth versus no growth, reaction success versus failure, binding versus
non-binding, or detector hit versus no hit under a shared experimental
condition.  The mathematics is not about balls in an urn; it is about repeated
binary outcomes whose dependence is explained by a common latent environment.
\end{enumerate}
The warning is as important as the convenience.  Time trends, learning effects,
depletion, interference, and changing protocols can break exchangeability.  In
those cases the beta-binomial model is a useful baseline, not a license to
ignore process knowledge.  For the beta-binomial conjugate model, see
\citet{gelman2013bayesian}; representative applied entry points include
statistical quality control \citep{montgomery2020quality}, online experiments
\citep{kohavi2020trustworthy}, and Bayesian clinical trials
\citep{berry2010bayesian}. \qedmark
\end{example}

\section{Exercises}
\conceptindexes{exercises, measure-theory exercises, probability exercises}

\noindent\textbf{Suggested route.}
For a first course, the core path is Exercises~1--13, 16--20, 24--25, and
30.  The remaining problems can be used as technical enrichment or as a
second-pass route through singular measures, fractal support, Gaussian series,
and tail events.

\begin{exercise}[Event systems]
For each object below, propose a sample space and three measurable events: a
spatial point pattern, a species sample, a survival dataset, and a random curve
on \([0,1]\).
\end{exercise}

\begin{exercise}[From stories to measurable objects]
Return to one story from Chapter~2 and translate it into the language of this
chapter.  Fill in the four columns below for that story, then write two
sentences explaining which part of the translation is most fragile.
\begin{center}
\small
\textbf{Translation worksheet.}\par\smallskip
\setlength{\tabcolsep}{0.35em}
\renewcommand{\arraystretch}{1.12}
\begin{tabular}{@{}>{\raggedright\arraybackslash}p{0.19\linewidth}
                >{\raggedright\arraybackslash}p{0.23\linewidth}
                >{\raggedright\arraybackslash}p{0.25\linewidth}
                >{\raggedright\arraybackslash}p{0.23\linewidth}@{}}
\toprule
Story pressure & Possible worlds \(\Omega\) & Observable events \(\fieldF\)
& Random summaries \(X\) \\
\midrule
Bomb impacts look clustered & & & \\
Species or words are unseen & & & \\
Proxy records hint at climate & & & \\
Text hints at authorship & & & \\
\bottomrule
\end{tabular}
\end{center}
For example, a cluster on a map should become a set of point patterns or
cell-count arrays, not merely a visual impression.  A singleton species count
should become a measurable function of a sample, not merely a phrase about
rarity.
\end{exercise}

\begin{exercise}[Borel discipline]
Show that every countable subset of \(\Real\) is Borel.  Explain why this does
not imply that every subset of \(\Real\) is Borel.
\end{exercise}

\begin{exercise}[Generated algebra from a partition]
Let \(\Omega=A_1\biguplus\cdots\biguplus A_m\).  Prove that
\[
  \left\{\bigcup_{i\in I}A_i:I\subseteq\{1,\ldots,m\}\right\}
\]
is the smallest algebra containing \(A_1,\ldots,A_m\).  How many sets does it
contain if all \(A_i\)'s are nonempty?
\end{exercise}

\begin{exercise}[Countable partition]
Let \(A_1,A_2,\ldots\) be a countable partition of \(\Omega\).  Describe
\(\sigma(A_1,A_2,\ldots)\).  Show that every element of this
\(\sigma\)-field is a union of partition cells.
\end{exercise}

\begin{exercise}[Countable-cocountable field]
Let \(\Omega\) be uncountable and let
\[
  \fieldF=\{A\subseteq\Omega:A\text{ is countable or }A^c\text{ is countable}\}.
\]
Prove that \(\fieldF\) is a \(\sigma\)-field.  Define a probability measure on
\(\fieldF\) that gives countable sets mass zero and co-countable sets mass one.
\end{exercise}

\begin{exercise}[Half-open intervals]
Show that the family \(\{\emptyset\}\cup\{(a,b]:a<b\}\) is a semi-ring on
\(\Real\).  Give all possible forms of the difference \((a,b]-(c,d]\).
\end{exercise}

\begin{exercise}[Product measurability]
Let \(X:(\Omega,\fieldF)\to(S,\mathcal S)\) and
\(Y:(\Omega,\fieldF)\to(T,\mathcal T)\) be measurable.  Prove that
\[
  \omega\mapsto (X(\omega),Y(\omega))
\]
is measurable from \((\Omega,\fieldF)\) to
\((S\times T,\mathcal S\otimes\mathcal T)\).
\end{exercise}

\begin{exercise}[Cylinder event]
Let \(\Omega=\Real^\Nat\) with the product \(\sigma\)-field.  Prove that
\[
  \{\omega:\limsup_n \omega_n\le a\}
\]
is measurable.
\end{exercise}

\begin{exercise}[Empirical measure]
Let \(X_1,\ldots,X_n\) be observations in \((S,\mathcal S)\).  Prove that the
empirical measure \(P_n\) is a probability measure.
\end{exercise}

\begin{exercise}[Expectation as integration]
Let \(P_n=n^{-1}\sum_{i=1}^n\delta_{X_i}\).  Prove that
\[
  \int f\,dP_n=\frac1n\sum_{i=1}^n f(X_i)
\]
for every nonnegative measurable \(f\).
\end{exercise}

\begin{exercise}[Mixed distribution]
Let \(P=p\delta_0+(1-p)P_{\mathrm{cont}}\), where \(P_{\mathrm{cont}}\) has
density \(f\) on \(\Real\).  Write the distribution function of \(P\).  Where
does it jump?
\end{exercise}

\begin{exercise}[Absolute continuity]
Let \(\nu\) be counting measure on a countable set \(S\).  Show that every
measure \(\mu\) on \((S,\Pow(S))\) is absolutely continuous with respect to
\(\nu\).  What is \(d\mu/d\nu\)?
\end{exercise}

\begin{exercise}[Singular measures]
Show that \(\delta_0\perp\lambda\) on \(\Real\).  More generally, if
\(P=p\delta_0+(1-p)P_{\mathrm{cont}}\) with \(0<p<1\) and
\(P_{\mathrm{cont}}\ll\lambda\), identify the part of \(P\) that is singular
with respect to \(\lambda\).
\end{exercise}

\begin{exercise}[Cantor distribution]
Let \(X=\sum_{n\ge1}2B_n/3^n\), where the \(B_n\)'s are independent
Bernoulli\((1/2)\) variables.  Prove that \(P(X\in C)=1\), where \(C\) is the
middle-third Cantor set.  Prove also that \(P(X=x)=0\) for every fixed
\(x\in\Real\).
\end{exercise}

\begin{exercise}[Likelihood ratios]
Let \(P_0\) and \(P_1\) have densities \(f_0\) and \(f_1\) with respect to a
common \(\sigma\)-finite measure \(\nu\).  Assuming \(P_1\ll P_0\), prove that
\[
  \frac{dP_1}{dP_0}=\frac{f_1}{f_0}
\]
\(P_0\)-almost surely, with the usual convention on the set \(\{f_0=0\}\).
\end{exercise}

\begin{exercise}[Poisson random measure]
Let \(N\) be a Poisson random measure with intensity \(\Lambda\).  If
\(A_1,\ldots,A_m\) are disjoint sets with finite \(\Lambda\)-mass, compute the
joint probability
\[
  P\{N(A_1)=n_1,\ldots,N(A_m)=n_m\}.
\]
Then compute the distribution of \(N(A_1\cup\cdots\cup A_m)\).
\end{exercise}

\begin{exercise}[Boole's inequality]
Prove Boole's inequality by disjointifying \(\bigcup_nA_n\).  Then give an
example with two events where the inequality is strict.
\end{exercise}

\begin{exercise}[Sigma-finiteness]
Show that Lebesgue measure on \(\Real^k\) is \(\sigma\)-finite.  Show that
counting measure on an uncountable set is not \(\sigma\)-finite.
\end{exercise}

\begin{exercise}[Distribution determines law]
Let \(P\) and \(Q\) be probability measures on \(\Borel(\Real)\).  Suppose
\[
  P((-\infty,x])=Q((-\infty,x])
  \qquad(x\in\Real).
\]
Use the \(\pi\)-\(\lambda\) theorem to prove \(P=Q\).
\end{exercise}

\begin{exercise}[Rectangle increment]
Write the alternating rectangle increment associated with a function
\(F:\Real^3\to\Real\).  Explain why every vertex of the rectangle appears.
\end{exercise}

\begin{exercise}[Hausdorff scaling]
For the middle-third Cantor set, use the cover by \(2^n\) intervals of length
\(3^{-n}\) to explain why \(\log 2/\log 3\) is the critical exponent for
Hausdorff dimension.
\end{exercise}

\begin{exercise}[Gaussian series]
Let \(e_i\) be an orthonormal basis of a Hilbert space and let
\[
  X=\sum_i\sqrt{\lambda_i}Z_ie_i,
\]
where \(Z_i\) are independent standard normal variables.  Show that
\(\sum_i\lambda_i<\infty\) implies convergence in \(L^2\).  What goes wrong
when \(\lambda_i=1\) for all \(i\)?
\end{exercise}

\begin{exercise}[Pushforward law]
Let \(X\) be a real-valued random variable and let \(Y=aX+b\) with \(a\ne0\).
Express the law of \(Y\) as a pushforward of the law of \(X\).  If \(X\) has
density \(f\), derive the density of \(Y\).
\end{exercise}

\begin{exercise}[Borel--Cantelli]
Let \(A_n\) be events with \(P(A_n)\le n^{-2}\).  Prove that
\[
  P(A_n\text{ infinitely often})=0.
\]
Give a short interpretation in the language of repeated rare alarms.
\end{exercise}

\begin{exercise}[Normal extremes]
Let \(X_1,X_2,\ldots\) be iid standard normal random variables, and put
\[
  M_n^+=\max_{1\le i\le n}X_i,
  \qquad
  M_n^-=\min_{1\le i\le n}X_i.
\]
Use Gaussian tail bounds and Borel--Cantelli to prove
\[
  P\left(\limsup_{n\to\infty}\frac{M_n^+}{\sqrt{2\log n}}=1\right)=1,
  \qquad
  P\left(\liminf_{n\to\infty}\frac{M_n^-}{\sqrt{2\log n}}=-1\right)=1.
\]
Explain why these are tail statements.
\end{exercise}

\begin{exercise}[Tail events]
Let \(X_1,X_2,\ldots\) be independent Bernoulli variables.  Show that the event
\(\{X_n=1\text{ infinitely often}\}\) belongs to the tail \(\sigma\)-field.
\end{exercise}

\begin{exercise}[Zero-one law]
Let \(X_1,X_2,\ldots\) be independent and let \(A\) be a tail event.  Fill in
the proof that \(A\) is independent of \(\sigma(X_1,X_2,\ldots)\), and conclude
that \(P(A)\in\{0,1\}\).
\end{exercise}

\begin{exercise}[Series and products as tail events]
Let \(X_1,X_2,\ldots\) be independent real-valued random variables.  Show that
\[
  A=\left\{\omega:\sum_{n=1}^{\infty}X_n(\omega)
  \text{ converges in }\Real\right\}
\]
has probability \(0\) or \(1\).  If \(P(X_n=0)=0\) for every \(n\), show that
\[
  B=\left\{\omega:\prod_{n=1}^{\infty}X_n(\omega)
  \text{ converges in }\Real\right\}
\]
also has probability \(0\) or \(1\).  Why is a condition excluding finite zero
factors needed?  Give a counterexample when \(X_1\) can be zero and the later
variables are deterministic.
\end{exercise}

\begin{exercise}[Modeling choice]
For each of the following, say whether the natural measure is closer to
Lebesgue measure, counting measure, a Stieltjes measure, an empirical measure,
a Hausdorff measure, or a counting-process measure: RNA read counts, a random
curve observed on a grid, event times with causes, a fractal boundary, a
detection-limited biomarker, and an iid sample.
\end{exercise}

\section*{Sources and Further Reading}
\addcontentsline{toc}{section}{Sources and Further Reading}

This chapter keeps the conventional measure-theoretic route while making the
examples do visible work.  Standard references for the formal development
include \citet{billingsley1995probability},
\citet{durrett2019probability}, \citet{folland1999real}, and
\citet{bogachev2007measure}; \citet{kallenberg2002foundations} is a useful
modern probability reference for the same measure-theoretic backbone.  The
route through set systems, probability measures, measurable maps,
integration, Radon--Nikodym derivatives, and finite products is also informed
by \citet{dabrowskaAdvancedProbabilityCommunication}.
For weak and vague convergence, see \citet{billingsley1999convergence},
\citet{vaart1998asymptotic}, and the random-measure treatment in
\citet{kallenberg2002foundations}.  Complete convergence was introduced by
\citet{hsuRobbins1947complete}.
Carath\'eodory's extension viewpoint traces back to
\citet{caratheodory1914lineare}, while the countable-probability and zero-one
themes are historically tied to \citet{borel1909probabilites},
\citet{cantelli1917probabilita}, and \citet{kolmogorov1933grundbegriffe}.
The P\'olya urn, exchangeability, de Finetti representation, and the
Hewitt--Savage law enter through \citet{polya1930sur},
\citet{definetti1937prevision}, and \citet{hewitt1955symmetric}; modern
treatments can be found in \citet{kallenberg2002foundations}.

The London bombing point-pattern example follows \citet{clarke1946poisson} and
\citet{feller1968introduction}.  The missing-species thread begins with
\citet{fisher1943species}, continues through \citet{good1953population}, and
connects to the Shakespeare vocabulary example of
\citet{efronThisted1976unseen}.  For Hausdorff measure and fractal geometry,
see \citet{falconer2014fractal}.  For Poisson processes, random counting
measures, and point-process inference, see \citet{kingman1993poisson},
\citet{karr1986point}, and \citet{bremaud1981point}.  The counting-process
view of event-history data is developed in \citet{aalen1978nonparametric},
\citet{andersen1993statistical}, and \citet{fleming1991counting}.  The
classification and legibility language in the opening paragraphs echoes
\citet{borges1964johnwilkins} and \citet{scott1998seeing}.

%% file: chapters/ch06_product_spaces_processes.tex
\chapter{How Data Objects Are Built: Products, Kernels, and Processes}
\label{chap:product-spaces-processes}
\conceptindexes{product spaces, product measures, transition kernels, Fubini theorem, stochastic processes, process laws, random objects}

\begin{tcolorbox}[
  enhanced,
  breakable,
  colback=chaptercream,
  colframe=bookblue!88!black,
  boxrule=0.72pt,
  arc=5pt,
  boxsep=1pt,
  left=1.0em,
  right=0.95em,
  top=0.82em,
  bottom=0.82em,
  before skip=0.55\baselineskip,
  after skip=1.0\baselineskip
]
\noindent\textbf{Chapter overview.}
As the last chapter of Part~II, this chapter answers a practical question:
how do tables, longitudinal records, event histories, random fields, adaptive
experiments, and process-valued observations become probability objects?
Product spaces explain how many-coordinate records are built; kernels explain
sequential generation; Fubini explains conditioning and iterated averaging;
Tulcea and Kolmogorov explain how local specifications become process laws;
coupling and optimal transport explain how dependence can be engineered once
marginals are fixed; decoupling explains when dependence can be compared with
independent copies;
and separability, analytic sets, and outer probability explain why
stochastic-process suprema can be treated as legitimate random objects. The
parallel repair for argmax notation, measurable selection, is kept in
\Appref{sec:set-valued-measurability-optimization}.
\end{tcolorbox}

A single random variable is rarely the unit of modern statistics. We observe
vectors, panels, event histories, sample paths, random functions, empirical
processes, and objective functions indexed by parameters. This chapter is not
included to display measure theory for its own sake.  It is included because
the examples from Chapter~2 eventually ask: what kind of probability object
was observed, and how is that object assembled from simpler pieces?  The
mathematical question is always the same: which subsets of the resulting
product space are observable, and which operations preserve measurability? The
answer begins with finite product measures and ends with a warning.
Finite-dimensional distributions are essential, but they do not by themselves
control path properties such as continuity, hitting times, or the random
variable
\[
  \sup_{t\in T} X_t .
\]
The historical development of stochastic-process theory can be read as a series
of repairs to this problem: build versions with separable or regular paths,
work in Polish path spaces, use analytic sets for projections, and in empirical
process theory use outer probability until measurability has been verified.

\input{figures/ch06_concept_map}

\noindent
Figure~\ref{fig:ch06-concept-map} is meant to be read as a route through the
chapter, not merely as a summary diagram. The core path runs from product event
systems to product measures and kernels, then to Fubini/Tonelli,
finite-dimensional laws, and finally processes and random objects.
The advanced path begins when uncountable operations enter the statistical
argument: suprema, hitting times, path regularity, analytic sets, outer
probability, and separability. The analogous selection machinery for
optimization is recorded in \Appref{sec:set-valued-measurability-optimization}.
A first reading can follow the core path and treat the advanced path as
infrastructure to return to when empirical processes, M-estimation, and
continuous-time models require it.

\begin{realdatacapsule}{Contact or protein-interaction network}
\item[Data object.] A graph: people and contacts, proteins and inferred
interactions, or documents and citations \citep{wasserman1994social,newman2010networks}.
\item[Observation mechanism.] Nodes and edges are sampled, reported, inferred,
thresholded, or crawled by an instrument or platform; missing edges are rarely
simple zeros.
\item[Target.] Degree distribution, community structure, contagion risk,
latent positions, or the effect of a network intervention.
\item[Model.] Product spaces hold the adjacency array; exchangeable graph
models, stochastic blocks, preferential attachment, or local dependence models
encode different structural assumptions.
\item[Uncertainty.] Resampling by nodes or egos, posterior graph uncertainty,
sensitivity to edge thresholds, and null-model comparison show which features
survive observation choices.
\item[Limitation.] A graph observed through one crawl, assay, or platform API
is not the full relation network; unobserved nodes and false edges can dominate
the scientific conclusion.
\end{realdatacapsule}

\section{Product Measures, Kernels, and Fubini}
\label{sec:ch06-products-kernels-fubini}
\conceptindexes{product measures, product sigma-field, transition kernels, kernel product measure, Tonelli theorem, Fubini theorem}

\subsection{Transition Kernels and Product Measures}
\conceptindexes{transition kernels, product measures, kernel product measure, Markov kernels}

\noindent\textbf{Statistical thread.}
A product measure is the mathematical version of a data-generating story. A
kernel says: first draw one part of the data, then draw the next part
conditionally on what has already been seen. This is the language of regression,
Markov chains, missing-data models, Bayesian hierarchies, and sequential
experiments.

\begin{definition}[Product sigma-field]
For measurable spaces \((\Omega_i,\mathcal F_i)\), \(i=1,\ldots,k\), the
product sigma-field on \(\Omega_1\times\cdots\times\Omega_k\) is
\[
  \bigotimes_{i=1}^k \mathcal F_i
  =
  \sigma\{A_1\times\cdots\times A_k:A_i\in\mathcal F_i\}.
\]
For a countable product \(\prod_{i\ge1}\Omega_i\), the product sigma-field is
generated by finite-dimensional cylinders,
\[
  \{\omega:(\omega_{i_1},\ldots,\omega_{i_m})\in B\},
  \qquad
  B\in\bigotimes_{\ell=1}^m \mathcal F_{i_\ell}.
\]
Equivalently, it is the smallest sigma-field that makes every coordinate
projection measurable.
\end{definition}

This definition is the backbone of the chapter.  Chapter~3 introduced product
event systems as part of the language of measurable observation; here they
become the machinery for building joint laws, kernels, iterated integrals, and
process distributions.

\begin{tcolorbox}[
  enhanced,
  breakable,
  colback=noteback,
  colframe=bookgold!75!black,
  boxrule=0.55pt,
  arc=4pt,
  boxsep=1pt,
  left=0.95em,
  right=0.9em,
  top=0.7em,
  bottom=0.7em,
  before skip=0.85\baselineskip,
  after skip=0.95\baselineskip
]
\noindent\textbf{Product, Borel, and ball sigma-fields.}
These names describe different generators.
The product sigma-field
\(\bigotimes_i\mathcal F_i\) is generated by coordinate events and finite
cylinders; it does not require a topology.  A Borel sigma-field
\(\Borel(S)\) is generated by the open sets of a topological space \(S\).
If \(S\) is a metric space with distance \(d\), the ball sigma-field is
\[
  \sigma\{B_d(x,r):x\in S,\ r>0\},
  \qquad
  B_d(x,r)=\{y:d(x,y)<r\}.
\]
Every open ball is open, so the ball sigma-field is contained in
\(\Borel(S)\).  They are equal when the metric space is separable, equivalently
second countable for metric spaces, because every open set is then a countable
union of balls from a countable base.  Without separability the inclusion can
be strict: in an uncountable set with the discrete metric, open balls generate
the countable--cocountable sigma-field, while the Borel sigma-field is the
whole power set.

For finite or countable products of second-countable spaces \(S_i\), with
\(\mathcal F_i=\Borel(S_i)\), the coordinate product sigma-field agrees with
the Borel sigma-field of the product topology; for example
\[
  \Borel(\R^k)=\Borel(\R)^{\otimes k}.
\]
For arbitrary products or nonseparable spaces, one should not assume equality.
This chapter builds process laws on the coordinate product sigma-field; later,
when paths are viewed as elements of a topological function space, we explicitly
check measurability into that path-space Borel sigma-field.
\end{tcolorbox}

\begin{definition}[Transition kernel]
Let $(\Omega_1,\mathcal F_1)$ and $(\Omega_2,\mathcal F_2)$ be measurable
spaces. A transition kernel from $\Omega_1$ to $\Omega_2$ is a function
\[
  K:\Omega_1\times\mathcal F_2\longrightarrow [0,\infty]
\]
such that
\begin{enumerate}
\item for each $\omega_1\in\Omega_1$, $K(\omega_1,\cdot)$ is a measure on
      $\mathcal F_2$;
\item for each $B\in\mathcal F_2$, $K(\cdot,B)$ is
      $\mathcal F_1/\Borel([0,\infty])$ measurable.
\end{enumerate}
If $K(\omega_1,\Omega_2)=1$ for all $\omega_1$, then $K$ is called a
probability kernel or Markov kernel.
\end{definition}

\begin{theorem}[Kernel product measure]
Let $\mu$ be a measure on $(\Omega_1,\mathcal F_1)$ and let $K$ be a transition
kernel from $\Omega_1$ to $\Omega_2$. There is a unique measure, denoted
$\mu\otimes K$, on
$\mathcal F_1\otimes\mathcal F_2$ satisfying
\[
  (\mu\otimes K)(A_1\times A_2)
  =
  \int_{A_1} K(\omega_1,A_2)\,\mu(d\omega_1),
  \qquad A_i\in\mathcal F_i .
\]
If $\mu$ is sigma-finite and $K$ is sigma-finite in the usual local sense, then
$\mu\otimes K$ is sigma-finite.
\end{theorem}

\noindent\textit{Proof.}
Let
\[
  \mathcal S=\{A_1\times A_2:A_i\in\mathcal F_i\}
\]
be the semiring of measurable rectangles. Define $Q_0$ on rectangles by the
formula in the theorem. If
$A\times B$ is decomposed into finitely many disjoint rectangles
$A_m\times B_m$, then
\[
  \indset{A}(\omega_1)\indset{B}(\omega_2)
  =
  \sum_m \indset{A_m}(\omega_1)\indset{B_m}(\omega_2).
\]
Integrating the identity first in $\omega_2$ against
$K(\omega_1,d\omega_2)$ gives
\[
  \indset{A}(\omega_1)K(\omega_1,B)
  =
  \sum_m \indset{A_m}(\omega_1)K(\omega_1,B_m).
\]
Integrating the last display in $\omega_1$ gives finite additivity of $Q_0$ on
$\mathcal S$. To check countable additivity on $\mathcal S$, suppose that
$A\times B=\bigcup_{n\ge1}(A_n\times B_n)$ is a disjoint union of rectangles.
Then for each $(\omega_1,\omega_2)$,
\[
  \indset{A}(\omega_1)\indset{B}(\omega_2)
  =
  \sum_{n\ge1}\indset{A_n}(\omega_1)\indset{B_n}(\omega_2).
\]
Integrating first with respect to $K(\omega_1,d\omega_2)$ and then with respect
to $\mu(d\omega_1)$ gives, by monotone convergence,
\[
  Q_0(A\times B)=\sum_{n\ge1}Q_0(A_n\times B_n).
\]
Thus $Q_0$ is a premeasure on the rectangle semiring. Caratheodory extension
therefore gives a measure, denoted $\mu\otimes K$, on
$\sigma(\mathcal S)=\mathcal F_1\otimes\mathcal F_2$. If $Q_1$ and $Q_2$ are
two such extensions, the class
\[
  \mathcal D=\{C\in\mathcal F_1\otimes\mathcal F_2:Q_1(C)=Q_2(C)\}
\]
is a Dynkin class containing all rectangles. Since rectangles are a pi-system
that generates the product sigma-field, the pi-lambda theorem gives
$\mathcal D=\mathcal F_1\otimes\mathcal F_2$. This proves uniqueness.
\qedmark

\begin{example}[Ordinary product measure]
If $K(\omega_1,B)=\nu(B)$ does not depend on $\omega_1$, then
$\mu\otimes K=\mu\otimes\nu$.
In statistics this is the independence construction: a pair $(X,Y)$ has joint
law $\mu\otimes\nu$ exactly when $X$ and $Y$ are independent with marginal laws
$\mu$ and $\nu$.
\qedmark
\end{example}

\begin{tcolorbox}[
  enhanced,
  breakable,
  colback=noteback,
  colframe=bookgold!75!black,
  boxrule=0.55pt,
  arc=4pt,
  boxsep=1pt,
  left=0.95em,
  right=0.9em,
  top=0.7em,
  bottom=0.7em,
  before skip=0.85\baselineskip,
  after skip=0.95\baselineskip
]
\noindent\textbf{Warning: complete after the joint law is built.}
Let $\lambda$ be Lebesgue measure on $\R$. Then
\[
  \Borel(\R^2)=\Borel(\R)\otimes\Borel(\R)
  \subset
  \mathcal L(\R)\otimes\mathcal L(\R)
  \subset
  \mathcal L(\R^2),
\]
and both inclusions may be strict. If $V\subset[0,1]$ is a Vitali set, then
$V\times\{0\}$ is contained in the two-dimensional null set
$[0,1]\times\{0\}$, hence belongs to the completed plane sigma-field
$\mathcal L(\R^2)$.  But $V\times\{0\}$ does not belong to
$\mathcal L(\R)\otimes\mathcal L(\R)$: if it did, its section at second
coordinate \(0\),
\[
  (V\times\{0\})^{0}
  =
  \{x\in\R:(x,0)\in V\times\{0\}\}
  =
  V,
\]
would be Lebesgue measurable, a contradiction.  The practical message is not
to chase Vitali sets.  It is to build products on the intended measurable
spaces, construct the joint law, and complete only after the joint law has
been fixed.
\end{tcolorbox}

\begin{definition}[Sigma-finite kernel]
Let $\mu$ be sigma-finite on $(\Omega_1,\mathcal F_1)$. A kernel $K$ is
sigma-finite $\mu$-a.e. if there are increasing measurable sets
$A_n\uparrow\Omega_1$, $B_n\uparrow\Omega_2$, and a set
$\Omega_0\in\mathcal F_1$ with $\mu(\Omega_0^c)=0$ such that
\[
  \mu(A_n)<\infty,\qquad
  \sup_{\omega_1\in\Omega_0}K(\omega_1,B_n)<\infty .
\]
Under this condition the kernel product measure $\mu\otimes K$ is sigma-finite
on the sets $(A_n\cap\Omega_0)\times B_n$.
\end{definition}

\subsection{Sections, Tonelli, and Fubini}
\conceptindexes{sections, Tonelli theorem, Fubini theorem, iterated integrals}

\noindent\textbf{Statistical thread.}
Fubini is the engine behind conditioning, marginalization, and simulation. It
shows when a two-coordinate experiment can be read row by row or column by
column without changing the expected answer.

For $C\subseteq\Omega_1\times\Omega_2$, define the sections
\[
  C_{\omega_1}=\{\omega_2:(\omega_1,\omega_2)\in C\},
  \qquad
  C^{\omega_2}=\{\omega_1:(\omega_1,\omega_2)\in C\}.
\]
If $C\in\mathcal F_1\otimes\mathcal F_2$, then
$C_{\omega_1}\in\mathcal F_2$ and $C^{\omega_2}\in\mathcal F_1$ for all fixed
coordinates.

\begin{theorem}[Fubini for a kernel product]
Let $\mu\otimes K$ be the measure induced by a sigma-finite measure $\mu$ and
a sigma-finite kernel $K$. If
$f:(\Omega_1\times\Omega_2,\mathcal F_1\otimes\mathcal F_2)\to[0,\infty]$ is
measurable, then
\[
  g(\omega_1)
  =
  \int_{\Omega_2}f(\omega_1,\omega_2)K(\omega_1,d\omega_2)
\]
is measurable and
\begin{equation}
  \int_{\Omega_1\times\Omega_2}f\,d(\mu\otimes K)
  =
  \int_{\Omega_1}g(\omega_1)\,\mu(d\omega_1).
  \label{eq:kernel-fubini}
\end{equation}
If $f$ is signed and $\int |f|\,d(\mu\otimes K)<\infty$, the same identity holds
for $f$.
\end{theorem}

\noindent\textit{Proof.}
For $f=\indset{A\times B}$,
\[
  g(\omega_1)=\indset{A}(\omega_1)K(\omega_1,B)
\]
is measurable by the kernel property, and \eqref{eq:kernel-fubini} is exactly
the definition of $(\mu\otimes K)(A\times B)$. Let $\mathcal C$ be the
collection of nonnegative measurable functions for which both measurability of
$g$ and the identity \eqref{eq:kernel-fubini} hold. The preceding paragraph puts all
rectangle indicators in $\mathcal C$. By nonnegative linearity, all nonnegative
simple functions of the form
\[
  s=\sum_{m=1}^M a_m\indset{A_m\times B_m},\qquad a_m\ge0,
\]
also belong to $\mathcal C$. If $0\le f_n\uparrow f$ and each $f_n\in\mathcal C$,
then
\[
  \int f_n(\omega_1,\omega_2)K(\omega_1,d\omega_2)
  \uparrow
  \int f(\omega_1,\omega_2)K(\omega_1,d\omega_2)
\]
for every $\omega_1$ by monotone convergence. The pointwise limit of measurable
functions is measurable, and applying monotone convergence once more to the
outer integral proves \eqref{eq:kernel-fubini} for $f$. The monotone class
theorem now gives all nonnegative measurable $f$. If $f$ is signed and
$\int |f|\,d(\mu\otimes K)<\infty$, apply the nonnegative result to $f^+$ and
$f^-$; both transformed integrals are finite because their sum is the
transformed integral of $|f|$.
\qedmark

\begin{corollary}[Tonelli and Fubini for products]
If $(\Omega_i,\mathcal F_i,\mu_i)$ are sigma-finite, then for every
nonnegative measurable $f$,
\begin{align*}
  \int f\,d(\mu_1\otimes\mu_2)
  &=
  \int_{\Omega_1}\int_{\Omega_2} f(\omega_1,\omega_2)
    \,\mu_2(d\omega_2)\,\mu_1(d\omega_1)\\
  &=
  \int_{\Omega_2}\int_{\Omega_1} f(\omega_1,\omega_2)
    \,\mu_1(d\omega_1)\,\mu_2(d\omega_2).
\end{align*}
The same equalities hold for integrable signed $f$.
\end{corollary}

\noindent\textit{Proof.}
Apply the kernel Fubini theorem twice. For the first displayed equality, take
$K(\omega_1,B)=\mu_2(B)$; then the kernel product measure is
$\mu_1\otimes\mu_2$. For the second equality, interchange the two coordinates
and take the constant kernel $L(\omega_2,A)=\mu_1(A)$; uniqueness of product
measure gives the same measure $\mu_1\otimes\mu_2$ on
$\mathcal F_1\otimes\mathcal F_2$. The nonnegative case is therefore Tonelli's
identity in both orders. If $f$ is integrable and signed, apply the
nonnegative identity to $f^+$ and $f^-$ separately. Since
\[
  \int f^+\,d(\mu_1\otimes\mu_2)<\infty,\qquad
  \int f^-\,d(\mu_1\otimes\mu_2)<\infty,
\]
the two equalities can be subtracted without an undefined
$\infty-\infty$ expression.
\qedmark

\begin{example}[Double series]
Let $a_{ij}\ge0$. Counting measure on $\Nat^2$ gives
\[
  \sum_{i=1}^{\infty}\sum_{j=1}^{\infty}a_{ij}
  =
  \sum_{j=1}^{\infty}\sum_{i=1}^{\infty}a_{ij}.
\]
Indeed, put $f(i,j)=a_{ij}$ and let $\#\otimes\#$ be product counting measure.
Tonelli gives
\[
  \int_{\Nat^2} f\,d(\#\otimes\#)
  =
  \int_{\Nat}\left\{\int_{\Nat}f(i,j)\,\#(dj)\right\}\#(di)
  =
  \sum_i\sum_j a_{ij},
\]
and the same product integral can be read in the other order. The theorem is
not saying that a rearrangement is harmless by magic; it is saying that
nonnegativity makes the partial sums increase monotonically toward one common
extended value.
If the terms have signs, the equality of iterated sums is guaranteed when
$\sum_{i,j}|a_{ij}|<\infty$.
\qedmark
\end{example}

\begin{example}[Expectations live on the pushforward law]
If $X:(\Omega,\mathcal F,\Prob)\to(S,\mathcal S)$ has law
$\mu=\Prob X^{-1}$, then
\[
  \Expect h(X)=\int_S h(x)\,\mu(dx)
\]
for every nonnegative or integrable $h$.  For $h=\indset{B}$ this is just
\[
  \Expect\indset{B}(X)=\Prob(X\in B)=\mu(B),
\]
and the general statement follows from simple functions and monotone
convergence.  When $S=\R$ and $X$ has distribution function $F$, the same
identity says
\[
  \Expect h(X)=\int_{\R}h(x)\,\mu_F(dx),
  \qquad
  \mu_F((a,b])=F(b)-F(a).
\]
This is the one-coordinate shadow of the product-space theorem: an integral
may be read either on the original sample space or on the distributional image.
\qedmark
\end{example}

\section{Transformations and Distributional Calculus}
\conceptindexes{transformations, image measures, change of variables, convolution, tail identities, distributional calculus}

\subsection{Image Measures and Change of Variables}
\conceptindexes{image measure, change of variables, pushforward law, Jacobian}

\noindent\textbf{Statistical thread.}
Change of variables, integration by parts, convolutions, rank symmetries,
quantile maps, and beta-gamma identities are not decorative calculus. They are
the ways a statistical distribution changes when the question changes.

\begin{theorem}[Image measure]
\label{thm:ch06-image-measure}
Let $\Psi:(\Omega,\mathcal F,\mu)\to(S,\mathcal S)$ be measurable. The set
function
\[
  \mu_{\Psi}(B)=\mu(\Psi^{-1}B),\qquad B\in\mathcal S,
\]
is a measure on $(S,\mathcal S)$. For every nonnegative or integrable $h$,
\[
  \int_S h(s)\,\mu_{\Psi}(ds)
  =
  \int_{\Omega}h(\Psi(\omega))\,\mu(d\omega).
\]
\end{theorem}

\noindent\textit{Proof.}
First, \(\mu_\Psi(\emptyset)=\mu(\emptyset)=0\).  If
\((B_k)_{k\ge1}\) are disjoint sets in \(\mathcal S\), then the inverse
images \(\Psi^{-1}B_k\) are disjoint sets in \(\mathcal F\), and
\[
  \Psi^{-1}\Bigl(\bigcup_{k\ge1}B_k\Bigr)
  =
  \bigcup_{k\ge1}\Psi^{-1}B_k .
\]
Countable additivity of \(\mu\) gives
\[
  \mu_\Psi\Bigl(\bigcup_{k\ge1}B_k\Bigr)
  =
  \sum_{k\ge1}\mu_\Psi(B_k),
\]
so \(\mu_\Psi\) is a measure.  For the integral identity, begin with
\(h=\ind_B\).  Then
\[
  \int_S\ind_B(s)\,\mu_\Psi(ds)
  =
  \mu_\Psi(B)
  =
  \mu(\Psi^{-1}B)
  =
  \int_\Omega \ind_B(\Psi(\omega))\,\mu(d\omega).
\]
Linearity proves the identity for nonnegative simple functions.  If
\(h\ge0\) is measurable, take simple \(h_m\uparrow h\) and apply monotone
convergence on both sides.  If \(h\) is integrable, apply the nonnegative
case to \(h^+\) and \(h^-\) and subtract the two finite quantities.
\qedmark

\begin{theorem}[Lebesgue change of variables]
Let $G,H\subseteq\R^d$ be open and let $\Psi:G\to H$ be a one-to-one $C^1$ map
whose inverse is $C^1$. If $J_{\Psi}(x)=|\det D\Psi(x)|$, then for every
nonnegative or integrable $h$,
\[
  \int_H h(y)\,dy
  =
  \int_G h(\Psi(x))J_{\Psi}(x)\,dx .
\]
\end{theorem}

\noindent\textit{Proof.}
Write $m$ for Lebesgue measure. The geometric part of the theorem is the set
identity
\begin{equation}
  m(\Psi A)=\int_A J_{\Psi}(x)\,dx,\qquad A\in\Borel(G).
  \label{eq:change-of-variables-set-identity}
\end{equation}
For affine maps this is exactly the determinant formula for the volume of a
parallelepiped. For a $C^1$ diffeomorphism one proves
\eqref{eq:change-of-variables-set-identity} by covering compact subsets of $G$
with small cubes on which $D\Psi$ changes little, comparing $\Psi$ on each cube
with its affine approximation, and then using inner and outer regularity of
Lebesgue measure to pass from compact sets and open sets to Borel sets. We now
show carefully how the integral formula follows from this set identity.

First take $h=\indset{B}$, where $B\in\Borel(H)$. Since $\Psi$ is a bijection from
$G$ onto $H$ and $\Psi^{-1}(B)\in\Borel(G)$,
\[
  \int_H \indset{B}(y)\,dy
  =m(B)
  =m\{\Psi(\Psi^{-1}B)\}
  =\int_{\Psi^{-1}B}J_{\Psi}(x)\,dx
  =\int_G \indset{B}(\Psi(x))J_{\Psi}(x)\,dx .
\]
Thus the formula is true for indicator functions. If
\[
  s(y)=\sum_{k=1}^n a_k \indset{B_k}(y),
  \qquad a_k\ge 0,\quad B_k\in\Borel(H),
\]
is a nonnegative simple function, then linearity gives
\[
  \int_H s(y)\,dy
  =\sum_{k=1}^n a_k m(B_k)
  =\sum_{k=1}^n a_k
    \int_G \indset{B_k}(\Psi(x))J_{\Psi}(x)\,dx
  =\int_G s(\Psi(x))J_{\Psi}(x)\,dx .
\]
Now let $h\ge0$ be measurable. Choose nonnegative simple functions
$s_n\uparrow h$. Then $s_n\circ\Psi\uparrow h\circ\Psi$, and the monotone
convergence theorem gives
\[
  \int_H h(y)\,dy
  =\lim_{n\to\infty}\int_H s_n(y)\,dy
  =\lim_{n\to\infty}\int_G s_n(\Psi(x))J_{\Psi}(x)\,dx
  =\int_G h(\Psi(x))J_{\Psi}(x)\,dx .
\]
Finally, if $h$ is integrable, apply the nonnegative case to $h^+$ and $h^-$.
It also gives
\[
  \int_G |h(\Psi(x))|J_{\Psi}(x)\,dx=\int_H |h(y)|\,dy<\infty,
\]
so both transformed positive and negative parts are finite and the two
identities may be subtracted.
\qedmark

\begin{example}[Translations and linear transformations]
If $Y=X+a$ in $\R^d$, then the density of $Y$ is $f_Y(y)=f_X(y-a)$. If
$Y=AX+b$ with $A$ nonsingular, then
\[
  f_Y(y)=f_X(A^{-1}(y-b))\,|\det A|^{-1}.
\]
The Jacobian is not a technical ornament; it is the price paid for changing the
coordinate system in which probability mass is measured.
\qedmark
\end{example}

\subsection{Integration by Parts and Tail Identities}
\conceptindexes{integration by parts, tail identities, survival function, covariance identity, stochastic integral, quadratic covariation}

Let $f$ and $g$ be right-continuous functions of bounded variation on $[0,t]$,
and write $d f$ and $d g$ for the signed Lebesgue--Stieltjes measures generated
by their increments:
\[
  d f((a,b])=f(b)-f(a),\qquad
  d g((a,b])=g(b)-g(a).
\]
This is the signed version of the Stieltjes construction from Chapter~3:
a bounded-variation function is a difference of two increasing functions, so
the corresponding increment measure is a difference of two positive
Stieltjes measures.  Here \(d f\) is measure notation, not a derivative.  Write
\(\Delta f(u)=f(u)-f(u-)\), and write \(f(u-)\) for the left limit of \(f\) at
\(u\).

\begin{theorem}[Integration by parts]
For bounded-variation functions as above,
\begin{equation}
  f(t)g(t)-f(0)g(0)
  =
  \int_{(0,t]} f(u-)\,d g(u)
  +
  \int_{(0,t]} g(u)\,d f(u).
  \label{eq:bv-integration-by-parts}
\end{equation}
Equivalently,
\[
  f(t)g(t)-f(0)g(0)
  =
  \int_{(0,t]} g(u-)\,d f(u)
  +
  \int_{(0,t]} f(u)\,d g(u).
\]
A fully left-limit version makes the jump correction explicit:
\[
\begin{aligned}
  f(t)g(t)-f(0)g(0)
  &=
  \int_{(0,t]} f(u-)\,d g(u) \\
  &\quad+
  \int_{(0,t]} g(u-)\,d f(u) \\
  &\quad+
  \sum_{0<u\le t}\Delta f(u)\Delta g(u).
\end{aligned}
\]
\end{theorem}

\noindent\textit{Proof.}
First assume that \(f\) and \(g\) are increasing, and put
\(\mu=d f\), \(\nu=d g\), and \(I=(0,t]\). Then
\[
  f(t)-f(0)=\mu(I),\qquad
  g(t)-g(0)=\nu(I).
\]
Expanding
\[
  f(t)g(t)-f(0)g(0)
  =
  \mu(I)\nu(I)+f(0)\nu(I)+g(0)\mu(I).
\]
It remains to split the product term.  The square \(I\times I\), with first
coordinate charged by \(\mu\) and second coordinate charged by \(\nu\), is the
disjoint union of
\[
  \{(r,s): r<s\}\quad\text{and}\quad \{(r,s): s\le r\}.
\]
By Tonelli--Fubini,
\[
\begin{aligned}
  \mu(I)\nu(I)
  &=
  \int_I \mu((0,s))\,\nu(ds)
  +\int_I \nu((0,r])\,\mu(dr) \\
  &=
  \int_I \{f(s-)-f(0)\}\,d g(s)
  +\int_I \{g(r)-g(0)\}\,d f(r).
\end{aligned}
\]
Substitution gives \eqref{eq:bv-integration-by-parts} for increasing \(f\)
and \(g\).  If \(f\) and \(g\) have bounded variation, write their signed
increment measures as differences of positive Stieltjes measures, or
equivalently write each path as a constant plus a difference of two increasing
right-continuous functions.  Applying the increasing case to the four
increasing pairs and adding the terms with signs proves
\eqref{eq:bv-integration-by-parts} in general.

Interchanging the roles of \(f\) and \(g\) gives the displayed equivalent
form.  Finally,
\[
  \int_{(0,t]} f(u)\,d g(u)
  =
  \int_{(0,t]} f(u-)\,d g(u)
  +\sum_{0<u\le t}\Delta f(u)\Delta g(u),
\]
because \(f(u)-f(u-)\) is nonzero only at jumps and
\(d g(\{u\})=\Delta g(u)\).  Combining this identity with the equivalent form
gives the fully left-limit version.
\qedmark

\noindent\textit{Stochastic-integral preview.}
The notation above is chosen to match the notation used later for random
integrals. If \(A=(A_u)_{0\le u\le t}\) has right-continuous bounded-variation
sample paths, then each fixed \(\omega\) gives a Lebesgue--Stieltjes measure
\(dA_\omega\), and a pathwise random Stieltjes integral is
\[
  (H\cdot A)_t(\omega)
  =
  \int_{(0,t]} H_u(\omega)\,dA_u(\omega),
\]
whenever the sample-path integral is defined.  For a counting process \(N\),
this reduces to the jump sum
\[
  \int_{(0,t]}H_u\,dN_u
  =
  \sum_{0<u\le t} H_u\,\Delta N_u .
\]
This is the part that returns later for marked point processes and counting
process compensators.  If \(N\) has predictable compensator \(\Lambda\), then
the random fluctuation is \(M=N-\Lambda\), and predictable integrands satisfy
\[
  \int_{(0,t]} H_u\,dN_u
  =
  \int_{(0,t]} H_u\,d\Lambda_u
  +
  \int_{(0,t]} H_u\,dM_u .
\]
The last term is the martingale noise used in event-history likelihoods,
Nelson--Aalen estimators, and marked point-process residuals.  Thus the
counting-process theory uses the finite-variation and pure-jump part of the
stochastic-integral language.  The full semimartingale integral extends this
picture to integrators such as continuous martingales.  Its product rule is
\[
  X_tY_t-X_0Y_0
  =
  \int_{(0,t]}X_{u-}\,dY_u
  +
  \int_{(0,t]}Y_{u-}\,dX_u
  +
  [X,Y]_t ,
\]
where \([X,Y]\) is quadratic covariation. For bounded-variation paths,
\([X,Y]_t=\sum_{0<u\le t}\Delta X_u\Delta Y_u\), so the formula above is exactly
the jump-corrected Stieltjes identity. Brownian It\^o calculus is the
continuous-martingale counterpart: for Brownian motion \([B,B]_t=t\), so
stochastic integration by parts is no longer ordinary calculus with random
functions \citep{ito1944stochastic,jacod1987limit}.

\begin{example}[Tail integration]
If $X\ge0$ and $S(x)=\Prob(X>x)$, then
\[
  \Expect X=\int_0^\infty S(x)\,dx
\]
whenever either side is finite. More generally, for $\alpha>0$,
\[
  \Expect X^{\alpha}
  =
  \alpha\int_0^\infty x^{\alpha-1}\Prob(X>x)\,dx .
\]
The first identity is Tonelli applied to the pointwise representation
\[
  X(\omega)=\int_0^\infty \ind{x<X(\omega)}\,dx.
\]
Taking expectations and interchanging the two nonnegative integrals gives
\[
  \Expect X
  =
  \int_0^\infty \Prob(X>x)\,dx.
\]
For the power identity, use
\[
  X^\alpha=\int_0^X \alpha x^{\alpha-1}\,dx
  =
  \int_0^\infty \alpha x^{\alpha-1}\ind{X>x}\,dx
\]
and apply Tonelli again.
\qedmark
\end{example}

\begin{example}[A covariance identity]
Let $X,Y\ge0$ have joint survival function
$S(x,y)=\Prob(X>x,Y>y)$ and marginal survivals $S_X,S_Y$. If the integrals are
finite, then
\[
  \Expect(XY)=\int_0^\infty\int_0^\infty S(x,y)\,dx\,dy
\]
and
\begin{equation}
  \Cov(X,Y)
  =
  \int_0^\infty\int_0^\infty
  \{S(x,y)-S_X(x)S_Y(y)\}\,dx\,dy .
  \label{eq:survival-covariance}
\end{equation}
The product has the nonnegative integral representation
\[
  XY
  =
  \int_0^\infty\int_0^\infty
  \ind{x<X,\ y<Y}\,dx\,dy.
\]
Tonelli gives the formula for $\Expect(XY)$. Applying the one-dimensional tail
identity to each margin gives
\[
  \Expect X\,\Expect Y
  =
  \left(\int_0^\infty S_X(x)\,dx\right)
  \left(\int_0^\infty S_Y(y)\,dy\right)
  =
  \int_0^\infty\int_0^\infty S_X(x)S_Y(y)\,dx\,dy.
\]
Subtracting the two displays yields \eqref{eq:survival-covariance}. Thus
positive quadrant dependence is visible directly as positive covariance.
\qedmark
\end{example}

\begin{example}[Rank regions in a product space]
Let \(X_1,\ldots,X_n\) be iid with continuous law \(P\). For each permutation
\(\pi\) of \(\{1,\ldots,n\}\), define the strict ordering region
\[
  C_\pi=\{x_{\pi(1)}<x_{\pi(2)}<\cdots<x_{\pi(n)}\}\subseteq\R^n .
\]
The sets \(C_\pi\) are product-measurable and partition \(\R^n\) up to the
diagonals \(x_i=x_j\), which have \(P^{\otimes n}\)-probability zero. Since
the product law is invariant under coordinate permutations, every strict
ordering region has probability \(1/n!\). Thus the rank vector is uniform on
the \(n!\) permutations, and the rank
\[
  R_i=1+\sum_{j\ne i}\ind{X_j\le X_i}
\]
is uniform on \(\{1,\ldots,n\}\).

If \(P\) has density \(f\) and distribution function \(F\), the usual density
of the \(k\)th order statistic is just the one-coordinate version of the same
partition:
\[
  f_{k:n}(x)
  =
  \frac{n!}{(k-1)!(n-k)!}
  F(x)^{k-1}\{1-F(x)\}^{n-k}f(x).
\]
The important object is the measurable finite partition of the product space;
the density formula is only one shadow of that symmetry.
\qedmark
\end{example}

\begin{definition}[Convolution]
For finite measures $\mu$ and $\nu$ on $\R^d$, their convolution is the image of
$\mu\otimes\nu$ under addition:
\[
  (\mu*\nu)(B)
  =
  \int_{\R^d}\int_{\R^d}\ind{x+y\in B}\,\mu(dx)\nu(dy).
\]
\end{definition}

\begin{example}[Sums of independent random variables]
If $X$ and $Y$ are independent with laws $\mu$ and $\nu$, then $X+Y$ has law
$\mu*\nu$. If both variables have densities $f$ and $g$, then
\[
  (f*g)(z)=\int_{\R^d}f(z-y)g(y)\,dy
\]
is the density of $X+Y$.
Indeed,
\[
  \Prob(X+Y\in B)
  =
  \int_{\R^d}\int_{\R^d}\ind{x+y\in B}f(x)g(y)\,dx\,dy.
\]
For fixed $y$, substitute $z=x+y$ to get
\[
  \Prob(X+Y\in B)
  =
  \int_B\left\{\int_{\R^d}f(z-y)g(y)\,dy\right\}dz,
\]
which identifies the displayed convolution density.
\qedmark
\end{example}

\begin{example}[Convolution kernel density estimates]
Convolution is also a statistical smoothing device. Let $P$ be a probability
law on $\R$, let $P_n=n^{-1}\sum_{i=1}^n\delta_{X_i}$ be the empirical law, and
let $K$ be an integrable kernel with $\int K(u)\,du=1$. For $h>0$ set
\[
  K_h(u)=h^{-1}K(u/h).
\]
The smoothed population target is
\[
  f_h(x)
  =
  (K_h*P)(x)
  :=
  \int K_h(x-z)\,P(dz).
\]
If $P$ has density $f$, this is the ordinary convolution
\[
  f_h(x)=\int K_h(x-z)f(z)\,dz=(K_h*f)(x).
\]
Replacing $P$ by $P_n$ gives the plug-in estimator
\[
  \hat f_h(x)
  =
  (K_h*P_n)(x)
  =
  \frac1n\sum_{i=1}^n K_h(x-X_i)
  =
  \frac1{nh}\sum_{i=1}^n K\left(\frac{x-X_i}{h}\right).
\]
Thus a kernel density estimator is literally a convolution of the empirical
measure with a smoothing kernel. When $K$ is a density and $V$ has density $K$,
independent of $X\sim P$, $f_h$ is also the density of $X+hV$. Indeed, for a
Borel set $B$,
\[
  \Prob(X+hV\in B)
  =
  \int\int \ind{x+hv\in B}K(v)\,dv\,P(dx).
\]
For fixed $x$, substitute $y=x+hv$ to obtain
\[
  \Prob(X+hV\in B)
  =
  \int_B\left\{\int K_h(y-x)\,P(dx)\right\}dy,
\]
which identifies the convolution density. This is the product-measure view of
the convolution kernel estimates discussed in \citet{bickel1977mathematical}.

Local-polynomial smoothing uses the same localizing kernel idea, but in general
it is not a fixed convolution operator: the equivalent weights depend on the
local design through a weighted least-squares moment matrix
\citep{fanGijbels1996local}.
\qedmark
\end{example}

\begin{example}[Discrete convolution in convolutional neural networks]
The word convolution in a convolutional neural network has the same algebraic
spine, now on a discrete product space. Let
$x_b:\Int^d\to\R$, $b=1,\ldots,c_{\mathrm{in}}$, be input channels, let
$\Lambda\subset\Int^d$ be a finite filter window, and let
$w_{ab}:\Lambda\to\R$ be learned weights from input channel $b$ to output
channel $a$. A convolution layer forms
\[
  z_a(r)
  =
  \beta_a+\sum_{b=1}^{c_{\mathrm{in}}}\sum_{u\in\Lambda}
  w_{ab}(u)x_b(r-u),
  \qquad r\in\Int^d,
\]
followed by a pointwise nonlinearity such as
$x^{\mathrm{new}}_a(r)=\sigma\{z_a(r)\}$. Many software libraries implement the
closely related cross-correlation with $r+u$ instead of $r-u$; the learned
filter can absorb this reversal, so the statistical idea is the same.

The key property is translation equivariance. If $(T_sx)_b(r)=x_b(r-s)$ is a
shift of the input, then, before boundary corrections,
\[
\begin{aligned}
  (w*T_sx)_a(r)
  &=
  \sum_{b=1}^{c_{\mathrm{in}}}\sum_{u\in\Lambda}
  w_{ab}(u)x_b(r-u-s)  \\
  &=
  (w*x)_a(r-s)
  =
  T_s(w*x)_a(r).
\end{aligned}
\]
Thus the same local detector is reused at every location. Statistically, a
convolutional layer is a parameter-sharing assumption: the feature ``seen'' by
one translated patch should be meaningful at another translated patch. This is
one reason convolutional networks became central in image recognition
\citep{lecun1998gradient,krizhevsky2012imagenet}. As a map from the finite
product space of pixel arrays to another finite product space, each layer is
measurable whenever its activation function is Borel measurable.
\qedmark
\end{example}

\subsection{Gamma, Beta, and Stein Calculus}
\conceptindexes{gamma distribution, beta distribution, normal distribution, Stein operator, distributional calculus}

\begin{example}[Independent gammas produce a beta law]
Let $G_1$ and $G_2$ be independent gamma variables with common scale one and
shape parameters $\alpha,\beta>0$. Their product density is
\[
  \frac{1}{\Gamma(\alpha)\Gamma(\beta)}
  g_1^{\alpha-1}g_2^{\beta-1}e^{-(g_1+g_2)},
  \qquad g_1,g_2>0,
\]
where
\[
  \Gamma(a)=\int_0^\infty x^{a-1}e^{-x}\,dx .
\]
Set
\[
  S=G_1+G_2,\qquad U=\frac{G_1}{G_1+G_2}.
\]
The inverse map is $g_1=su$, $g_2=s(1-u)$, with Jacobian $s$. Hence the joint
density of $(S,U)$ is
\[
  \frac{1}{\Gamma(\alpha)\Gamma(\beta)}
  s^{\alpha+\beta-1}e^{-s}u^{\alpha-1}(1-u)^{\beta-1},
  \qquad s>0,\ 0<u<1.
\]
Since this factors as
\[
  \left\{
  \frac{1}{\Gamma(\alpha+\beta)}
  s^{\alpha+\beta-1}e^{-s}
  \right\}
  \left\{
  \frac{\Gamma(\alpha+\beta)}{\Gamma(\alpha)\Gamma(\beta)}
  u^{\alpha-1}(1-u)^{\beta-1}
  \right\},
\]
$S$ and $U$ are independent,
$S\sim\operatorname{Gamma}(\alpha+\beta,1)$, and
$U\sim\Beta(\alpha,\beta)$. This is more than a beta-gamma identity: it is a
product law, a change of variables, and a marginalization in one calculation.
The recursion $\Gamma(a+1)=a\Gamma(a)$ is the one-dimensional integration by
parts shadow of the same calculus.
\qedmark
\end{example}

\begin{example}[Gaussian integration by parts and Stein's operator]
Let $Z\sim\Normal(0,1)$ with density $\phi$. If $f$ is continuously
differentiable, $f(x)\phi(x)\to0$ as $|x|\to\infty$, and the following
expectations are finite, then
\[
  \Expect f'(Z)=\Expect\{Zf(Z)\}.
\]
Indeed, $\phi'(x)=-x\phi(x)$, so ordinary integration by parts gives
\[
  \int f'(x)\phi(x)\,dx
  =
  -\int f(x)\phi'(x)\,dx
  =
  \int x f(x)\phi(x)\,dx .
\]
Equivalently,
\[
  \Expect\{(\mathcal A f)(Z)\}=0,
  \qquad
  (\mathcal A f)(x)=f'(x)-xf(x).
\]
The differential operator $\mathcal A$ is the normal Stein operator
\citep{stein1972bound}. Taking
$f(x)=x^m$ gives the moment recursion
\[
  \Expect Z^{m+1}=m\Expect Z^{m-1},\qquad m\ge1,
\]
so the usual normal moments are a corollary rather than the main event.
\qedmark
\end{example}

\begin{example}[A Stein--Chen bound from a product space]
Let $X_1,\ldots,X_n$ be independent with
$X_i\sim\Bernoulli(p_i)$, and put
\[
  W=\sum_{i=1}^n X_i,\qquad \lambda=\sum_{i=1}^n p_i.
\]
For each $i$, write $W_i=W-X_i$. The product construction makes $W_i$
independent of $X_i$. Hence for any bounded
$f:\{0,1,2,\ldots\}\to\R$,
\[
\begin{aligned}
  \Expect\{Wf(W)\}
  &=
  \sum_i \Expect\{X_i f(W)\}
   =
  \sum_i p_i\Expect f(W_i+1),
\end{aligned}
\]
and therefore
\[
  \Expect\{\lambda f(W+1)-Wf(W)\}
  =
  \sum_i p_i\,\Expect\{f(W+1)-f(W_i+1)\}.
\]
Because $W=W_i+X_i$, the difference inside the expectation is nonzero only
when $X_i=1$. Thus
\[
  \left|\Expect\{\lambda f(W+1)-Wf(W)\}\right|
  \le
  \|\Delta f\|_\infty\sum_i p_i^2,
  \qquad
  \Delta f(k)=f(k+1)-f(k).
\]
The classical Poisson Stein equation for $Z_\lambda\sim\Poisson(\lambda)$ is
\[
  \lambda f_A(w+1)-w f_A(w)
  =
  \indset{A}(w)-\Prob(Z_\lambda\in A).
\]
Its standard bounded solution satisfies
$\|\Delta f_A\|_\infty\le 1\wedge\lambda^{-1}$
\citep{chen1975poisson,barbourHolstJanson1992poisson}. Taking the supremum over
$A\subseteq\{0,1,2,\ldots\}$ yields the classical product-space bound
\[
  d_{\mathrm{TV}}\{\mathcal L(W),\Poisson(\lambda)\}
  \le
  (1\wedge\lambda^{-1})\sum_i p_i^2.
\]
The calculation is the Stein--Chen method in miniature: independence supplies
the leave-one-out identity, and the Stein operator turns that identity into a
distributional approximation.
\qedmark
\end{example}

\begin{example}[Normal tail bounds]
Let $Z\sim\Normal(0,1)$ and let $\bar\Phi(t)=\Prob(Z>t)$ for $t>0$.
Integration by parts gives Mills-type bounds
\[
  \frac{t}{1+t^2}\phi(t)
  \le
  \bar\Phi(t)
  \le
  \frac{\phi(t)}{t},
\]
where $\phi$ is the standard normal density.
The upper bound uses $x/t\ge1$ for $x\ge t$:
\[
  \bar\Phi(t)
  =
  \int_t^\infty \phi(x)\,dx
  \le
  \frac1t\int_t^\infty x\phi(x)\,dx
  =
  \frac{\phi(t)}{t}.
\]
For the lower bound, integrate by parts using
$d\{-\phi(x)\}=x\phi(x)\,dx$:
\[
\begin{aligned}
  \bar\Phi(t)
  &=
  \int_t^\infty \frac1x\,x\phi(x)\,dx\\
  &=
  \frac{\phi(t)}{t}
  -
  \int_t^\infty \frac{\phi(x)}{x^2}\,dx\\
  &\ge
  \frac{\phi(t)}{t}
  -
  \frac1{t^2}\bar\Phi(t).
\end{aligned}
\]
Moving the last term to the left gives
\[
  \bar\Phi(t)\ge \frac{t}{1+t^2}\phi(t).
\]
These two inequalities are rough but extremely useful: the normal tail is of
order $\phi(t)/t$.
\qedmark
\end{example}

\subsection{Product Calculus: Dependence, Tails, and Geometry}
\conceptindexes{dependence, tails, geometry, product calculus, rank symmetry, coupling, optimal transport, quantile coupling, maximal coupling, decoupling}

\begin{example}[Marshall--Olkin survival copula and common shocks]
Let \(E_1,E_2,E_{12}\) be independent exponential variables with rates
\(\lambda_1,\lambda_2,\lambda_{12}\), and set
\[
  T_1=E_1\wedge E_{12},\qquad
  T_2=E_2\wedge E_{12}.
\]
The shared shock \(E_{12}\) can kill both components at the same time. The
joint survival function of \((T_1,T_2)\) is
\[
  \Prob(T_1>t_1,T_2>t_2)
  =
  \exp\{-\lambda_1t_1-\lambda_2t_2-\lambda_{12}(t_1\vee t_2)\},
  \qquad t_1,t_2\ge0.
\]
The marginal survivals are
\[
  S_1(t)=e^{-(\lambda_1+\lambda_{12})t},
  \qquad
  S_2(t)=e^{-(\lambda_2+\lambda_{12})t}.
\]
Let
\[
  \alpha_1=\frac{\lambda_{12}}{\lambda_1+\lambda_{12}},
  \qquad
  \alpha_2=\frac{\lambda_{12}}{\lambda_2+\lambda_{12}}.
\]
Writing \(u=S_1(t_1)\) and \(v=S_2(t_2)\), the survival copula is
\[
  \bar C(u,v)
  =
  u^{1-\alpha_1}v^{1-\alpha_2}
  \min\{u^{\alpha_1},v^{\alpha_2}\},
  \qquad 0<u,v\le1.
\]
Indeed, \(t_1=-\log u/(\lambda_1+\lambda_{12})\) and
\(t_2=-\log v/(\lambda_2+\lambda_{12})\), so
\[
  e^{-\lambda_{12}(t_1\vee t_2)}
  =
  \min\{e^{-\lambda_{12}t_1},e^{-\lambda_{12}t_2}\}
  =
  \min\{u^{\alpha_1},v^{\alpha_2}\}.
\]
This copula has an absolutely continuous part and a singular part. The
singular mass comes from the event that the common shock arrives first:
\[
  \Prob(E_{12}<E_1\wedge E_2)
  =
  \frac{\lambda_{12}}{\lambda_1+\lambda_2+\lambda_{12}}.
\]
On this event \(T_1=T_2\), which becomes the curve
\[
  v=u^{(\lambda_2+\lambda_{12})/(\lambda_1+\lambda_{12})}
\]
in survival-copula coordinates. Thus the Marshall--Olkin law
\citep{marshallOlkin1967multivariate} is a useful warning against thinking
``copula'' means ``copula density'': a realistic common-shock mechanism can
put positive probability on lower-dimensional geometry.
\qedmark
\end{example}

\begin{example}[Singular laws live on lower-dimensional geometry]
Let \(\Theta\sim\Unif(0,2\pi)\) and put
\[
  Z=(X,Y)=(\cos\Theta,\sin\Theta).
\]
Then \(Z\) is an ordinary random vector in \(\R^2\), but its law is supported
on the unit circle
\[
  S^1=\{(x,y):x^2+y^2=1\}.
\]
Since two-dimensional Lebesgue measure satisfies \(m_2(S^1)=0\), the law of
\(Z\) is singular with respect to \(m_2\); there is no density \(f(x,y)\) such
that \(\Prob(Z\in A)=\int_A f(x,y)\,dx\,dy\).  The problem is not the
distribution.  It is the reference measure.  If \(s\) denotes arc-length
measure on \(S^1\), then
\[
  \Prob(Z\in A)=\frac{1}{2\pi}s(A\cap S^1),
  \qquad A\in\Borel(\R^2).
\]
Thus the same law has density \(1/(2\pi)\) with respect to arc length on its
natural support.

The same phenomenon appears whenever one coordinate is a function of another.
If \(X\sim P\) on \(\R\), \(Y=T(X)\), and \(T:\R\to\R\) is Borel measurable,
then the joint law is the image
\[
  \gamma=(\mathrm{id},T)_\#P.
\]
It is concentrated on the graph
\[
  G_T=\{(x,T(x)):x\in\R\}.
\]
The graph is Borel and has two-dimensional Lebesgue measure zero, because each
vertical section contains at most one point.  Hence \(\gamma\) is singular
with respect to \(dx\,dy\).  Perfect dependence is therefore not a failure of
probability calculus; it is a reminder that joint laws need not be described
by a density relative to planar volume.
\qedmark
\end{example}

\begin{example}[Exchangeable ranks and predictive \(p\)-values]
Let \(S_1,\ldots,S_{n+1}\) be exchangeable real-valued scores. Think of
\(S_1,\ldots,S_n\) as calibration scores and \(S_{n+1}\) as the score of a new
case. If there may be ties, attach iid \(V_1,\ldots,V_{n+1}\sim\Unif(0,1)\),
independent of the scores, and order the pairs \((S_i,V_i)\) lexicographically.
Define
\[
  R
  =
  1+\sum_{i=1}^n
  \ind{(S_i,V_i)\le_{\mathrm{lex}}(S_{n+1},V_{n+1})}.
\]
Exchangeability says that the index \(n+1\) is equally likely to occupy any of
the \(n+1\) rank positions, so
\[
  \Prob(R=r)=\frac{1}{n+1},\qquad r=1,\ldots,n+1.
\]
For a score where large values are unusual, the rank \(p\)-value
\[
  p_{n+1}
  =
  \frac{1+\sum_{i=1}^n
  \ind{(S_i,V_i)\ge_{\mathrm{lex}}(S_{n+1},V_{n+1})}}{n+1}
  =
  \frac{n+2-R}{n+1}
\]
satisfies
\[
  \Prob(p_{n+1}\le \alpha)
  =
  \frac{\lfloor (n+1)\alpha\rfloor}{n+1}
  \le \alpha,\qquad 0\le\alpha\le1.
\]
This is the finite-sample product-space symmetry behind conformal predictive
\(p\)-values \citep{vovk2005algorithmic}: no density estimation is needed, only
exchangeability and a measurable rank map.
\qedmark
\end{example}

\begin{example}[Probability integral transform]
If $F$ is continuous and $X\sim F$, then $U=F(X)$ is uniform on $[0,1]$ when
$F$ is strictly increasing. In general, the randomized transform
\[
  U=F(X-)+V\{F(X)-F(X-)\},\qquad V\sim\Unif(0,1),
\]
with $V$ independent of $X$, is uniform. The extra coordinate $V$ is a product
space device: it spreads each atom of $X$ uniformly across the jump of $F$.
To verify this, fix $u\in[0,1]$. On an atom $x$ with jump
$\Delta F(x)=F(x)-F(x-)$, the conditional value of $U$ is uniform on
$(F(x-),F(x)]$. On the nonatomic part, $U=F(X)$. Therefore the preimage
$\{U\le u\}$ collects exactly total $F$-mass $u$: whole jumps before $u$, the
proper fraction of the jump that contains $u$, and the continuous mass before
that point. Hence $\Prob(U\le u)=u$.
\qedmark
\end{example}

\begin{example}[Dirichlet law from gamma variables]
Let $G_1,\ldots,G_m$ be independent gamma variables with common scale one and
shape parameters $\alpha_1,\ldots,\alpha_m$. Put
\[
  S=\sum_{i=1}^mG_i,\qquad Y_i=G_i/S,\quad i=1,\ldots,m-1.
\]
The change of variables
\[
  g_i=sy_i,\quad i<m,\qquad
  g_m=s\left(1-\sum_{i=1}^{m-1}y_i\right)
\]
has Jacobian $s^{m-1}$. The joint density factors into a gamma density for
$S$ and a Dirichlet density for $Y$:
\[
  c\,
  s^{\alpha_1+\cdots+\alpha_m-1}e^{-s}
  \prod_{i=1}^{m}y_i^{\alpha_i-1}.
\]
Starting from the independent gamma density
\[
  \prod_{i=1}^m
  \frac{1}{\Gamma(\alpha_i)}
  g_i^{\alpha_i-1}e^{-g_i},
\]
substitution gives $\sum_i g_i=s$ and
\[
  \prod_{i=1}^m g_i^{\alpha_i-1}
  =
  s^{\alpha_1+\cdots+\alpha_m-m}
  \prod_{i=1}^m y_i^{\alpha_i-1}.
\]
Multiplying by the Jacobian $s^{m-1}$ yields
$s^{\alpha_1+\cdots+\alpha_m-1}e^{-s}\prod_i y_i^{\alpha_i-1}$, which factors
into a function of $s$ times a function of $y$ on the simplex. Normalizing the
$y$ part gives the Dirichlet constant
\[
  \frac{\Gamma(\alpha_1+\cdots+\alpha_m)}
       {\prod_{i=1}^m\Gamma(\alpha_i)}.
\]
Thus $S$ and $Y$ are independent. This calculation is a compact example of
everything in the chapter so far: product density, change of variables,
Jacobian, and marginalization.
\qedmark
\end{example}

\begin{example}[Couplings as product-space laws]
Let \(\mu\) and \(\nu\) be probability measures on measurable spaces
\((S,\mathcal S)\) and \((T,\mathcal T)\). A \emph{coupling} of \(\mu\) and
\(\nu\) is a probability measure \(\gamma\) on
\((S\times T,\mathcal S\otimes\mathcal T)\) whose marginals are \(\mu\) and
\(\nu\):
\[
  \gamma(A\times T)=\mu(A),\qquad
  \gamma(S\times B)=\nu(B).
\]
Write the set of all such couplings as \(\Pi(\mu,\nu)\). The product measure
\(\mu\otimes\nu\) is the independent coupling, but it is only one point in the
larger set \(\Pi(\mu,\nu)\). If \(T_0:S\to T\) is measurable and
\((T_0)_\#\mu=\nu\), then
\[
  \gamma=(\mathrm{id},T_0)_\#\mu
\]
is another coupling, concentrated on the graph of \(T_0\). Thus independence,
perfect dependence, common-shock singular laws, and transport maps are all
different choices of a product-space law with prescribed marginals.

Optimal transport asks which coupling is best for a given cost
\(c:S\times T\to[0,\infty]\):
\[
  \inf_{\gamma\in\Pi(\mu,\nu)}
  \int_{S\times T} c(x,y)\,\gamma(dx,dy).
\]
For \(S=T=\R^d\) and \(c(x,y)=\|x-y\|^p\), the \(p\)th root of this value is
the Wasserstein distance \(W_p(\mu,\nu)\). This is not a departure from the
chapter's product-space theme. It is exactly the same theme with the product
law no longer fixed in advance: hold the two marginals fixed and search over
all admissible joint laws. See \citet{villani2009optimal}; for statistical
uses of Wasserstein distances, see \citet{panaretos2019statistical}.
\qedmark
\end{example}

\begin{example}[Maximal coupling and the price of disagreement]
Coupling can turn an abstract distance between laws into an event.  Suppose
\(P\) and \(Q\) have densities \(p\) and \(q\) with respect to a common
measure \(\mu\), and define
\[
  a=\int \min(p,q)\,d\mu
  =
  1-\norm{P-Q}_{\mathrm{TV}}.
\]
Construct \((X,Y)\) as follows.  With probability \(a\), draw
\[
  Z\sim \frac{\min(p,q)}{a}\,d\mu
\]
and set \(X=Y=Z\).  With the remaining probability \(1-a\), draw \(X\) from
the residual density \(\{p-\min(p,q)\}/(1-a)\) and \(Y\) from
\(\{q-\min(p,q)\}/(1-a)\).  These residual densities live on disjoint parts of
the space, so \(X\ne Y\) on the residual event.  The resulting marginals are
\(X\sim P\) and \(Y\sim Q\), while
\[
  \Prob(X=Y)=a,
  \qquad
  \Prob(X\ne Y)=\norm{P-Q}_{\mathrm{TV}}.
\]
No coupling can do better.  Indeed, for every coupling and every measurable
set \(A\),
\[
  P(A)-Q(A)
  =
  \Prob(X\in A,Y\notin A)-\Prob(X\notin A,Y\in A)
  \le \Prob(X\ne Y).
\]
Taking the supremum over \(A\) gives
\[
  \norm{P-Q}_{\mathrm{TV}}\le \Prob(X\ne Y).
\]
Thus total variation is the smallest possible probability that two random
objects with the prescribed marginals fail to be equal.  The same formula will
reappear in Chapter~14 as a testing distance; here it is a statement about how
well two laws can be placed on one probability space.
\qedmark
\end{example}

\begin{example}[Quantile coupling as a product-space map]
Let \(F\) and \(G\) be distribution functions on \(\R\), with quantile
functions
\[
  Q_F(u)=\inf\{x:F(x)\ge u\},
  \qquad
  Q_G(u)=\inf\{y:G(y)\ge u\}.
\]
If \(U\sim\Unif(0,1)\), then \(Q_F(U)\sim F\) and \(Q_G(U)\sim G\). Hence
\[
  \gamma=(Q_F,Q_G)_\#\lambda_{[0,1]}
\]
is a coupling of the two laws. It is not an arbitrary coupling: it pairs equal
quantile levels. In one dimension this monotone coupling is optimal for the
quadratic transport cost, so for square-integrable laws
\[
  W_2^2(F,G)
  =
  \int_0^1\{Q_F(u)-Q_G(u)\}^2\,du;
\]
see, for example, \citet{villani2009optimal}.

The empirical law is the finite-sample version of the same construction. If
\[
  P_n=\frac1n\sum_{i=1}^n\delta_{X_i},
  \qquad
  Q_n(u)=\inf\{x:P_n(-\infty,x]\ge u\},
\]
then, after ordering the sample,
\[
  Q_n(u)=X_{k:n},
  \qquad \frac{k-1}{n}<u\le \frac{k}{n}.
\]
Thus \(Q_n(U)\), conditionally on the data, has law \(P_n\), and
\[
  \int_0^1 h(Q_n(u))\,du=\frac1n\sum_{i=1}^n h(X_i)
\]
for every measurable \(h\) for which either side is defined. Order statistics
are therefore best viewed here as coordinates of an image measure, not as a
catalogue of special densities.
\qedmark
\end{example}

\begin{example}[Decoupling: from shared indices to independent copies]
Coupling builds a useful dependence structure.  Decoupling goes in the other
direction: it asks when a statistic that reuses the same coordinates can be
compared with a statistic built on a larger product space with fresh copies.
Let \(X_1,\ldots,X_n\) be iid observations with common law \(P\) on \(S\).
Here the word \emph{kernel} is not the transition kernel of
Section~\ref{sec:ch06-products-kernels-fubini}.  It is the standard
U-statistic word for a measurable score function
\[
  h:S\times S\to\R.
\]
Given two observations \(x\) and \(y\), \(h(x,y)\) records the pairwise
feature one wants to aggregate: a similarity score, an interaction contrast, a
distance-based discrepancy, or a centered product such as
\(h(x,y)=g(x)g(y)\) with \(\int g\,dP=0\).  The quadratic statistic
\[
  U=\sum_{1\le i<j\le n} h(X_i,X_j)
\]
is a measurable map on the product space \(S^n\), but it is not a sum of
independent terms: \(h(X_i,X_j)\) and \(h(X_i,X_k)\) share the coordinate
\(X_i\).

The product-space repair is to add a second iid sample
\[
  X'_1,\ldots,X'_n
  \quad\text{independent of}\quad
  X_1,\ldots,X_n,
\]
that lives on the enlarged space \(S^n\times S^n\) with law
\(P^n\otimes P^n\).  The decoupled object is
\[
  U^{\mathrm{dec}}
  =
  \sum_{1\le i<j\le n} h(X_i,X'_j).
\]
This is the same operation as replacing a reused coordinate by an independent
coordinate drawn from the constant transition kernel \(K(x,\cdot)=P(\cdot)\).
For a single pair, both \((X_i,X_j)\) and \((X_i,X'_j)\) have law
\(P\otimes P\).  The difference is global: in \(U\), many summands reuse the
same \(X_i\)'s, while in \(U^{\mathrm{dec}}\) the two argument slots of \(h\)
are supplied by two independent coordinate arrays.

The cleanest inequalities are stated for \emph{canonical} kernels.  For order
two this means
\[
  \int h(x,y)\,P(dy)=0\quad\text{for \(P\)-a.e. }x,
  \qquad
  \int h(x,y)\,P(dx)=0\quad\text{for \(P\)-a.e. }y.
\]
Canonicality removes the one-observation main effects and leaves a genuinely
pairwise fluctuation.  In that case, and in much greater generality for
Banach-space-valued sums, decoupling inequalities control tails and moments of
\(U\) by those of \(U^{\mathrm{dec}}\), up to constants depending only on the
order of the statistic \citep{delapenaGine1999decoupling}.  A typical schematic
statement is
\[
  \norm{U}_{L^r}
  \le C_r\,\norm{U^{\mathrm{dec}}}_{L^r},
  \qquad r\ge1,
\]
with the exact hypotheses and constants supplied by the decoupling theorem
being used.

This is a product-space idea with teeth.  The dependent statistic lives on
one diagonal product space, where coordinates are reused.  The decoupled
statistic lives on a larger product space with fresh copies, where conditional
arguments, symmetrization, and concentration inequalities become available.
The later empirical-process chapter uses this as proof technology for
U-processes.  Here the lesson is more basic: changing the product space changes
the dependence pattern even when each individual pair still has the same
two-coordinate law.
\qedmark
\end{example}

\begin{example}[Gumbel noise turns utilities into a softmax draw]
This example belongs in the transformation section because the categorical
choice below is the image of a finite product law under an argmax map.  The
numbers \(\eta_1,\ldots,\eta_m\) are deterministic utilities, also called
logits: larger \(\eta_k\) means that option \(k\) is preferred before random
noise is added.  They are not statistical score functions.
Let \(G_1,\ldots,G_m\) be iid standard Gumbel variables, with
\[
  F(g)=\exp\{-e^{-g}\},\qquad f(g)=e^{-g}\exp\{-e^{-g}\}.
\]
Given these utilities, define the product-space map
\[
  A=\argmax_{1\le k\le m}\{\eta_k+G_k\}.
\]
Ties have probability zero.  Conditioning on \(G_j=g\),
\[
\begin{aligned}
  \Prob(A=j)
  &=
  \int_{-\infty}^{\infty}
    f(g)\prod_{k\ne j}F(g+\eta_j-\eta_k)\,dg \\
  &=
  \int_{-\infty}^{\infty}
    e^{-g}
    \exp\left\{
      -e^{-g}\sum_{k=1}^{m}e^{\eta_k-\eta_j}
    \right\}\,dg .
\end{aligned}
\]
With \(u=e^{-g}\), this becomes
\[
  \Prob(A=j)
  =
  \int_0^\infty
  \exp\left\{-u\sum_{k=1}^m e^{\eta_k-\eta_j}\right\}\,du
  =
  \frac{e^{\eta_j}}{\sum_{k=1}^m e^{\eta_k}}.
\]
Thus an argmax of independent noisy utilities is a categorical draw with softmax
probabilities.
\qedmark
\end{example}

\section{Building Process Laws}
\label{sec:ch06-building-process-laws}
\conceptindexes{process laws, countable products, uncountable products, finite-dimensional distributions, Kolmogorov extension theorem, Ionescu--Tulcea theorem}

\subsection{Countable Products and Ionescu--Tulcea}
\conceptindexes{countable products, Ionescu--Tulcea theorem, sequential laws, stochastic recursion}

\noindent\textbf{Statistical thread.}
Tulcea's theorem turns a sequential recipe into a probability law on an entire
infinite history. This is the construction behind Markov chains, repeated
measurements, longitudinal data, and every model where tomorrow is specified
conditionally on today's record.

Let $(\Omega_n,\mathcal F_n)$, $n\ge1$, be measurable spaces and let
\[
  \Omega^\infty=\prod_{n=1}^{\infty}\Omega_n,
  \qquad
  \mathcal F^\infty=\bigotimes_{n=1}^{\infty}\mathcal F_n.
\]
For $\alpha_n=\{1,\ldots,n\}$ write
\[
  \Omega_{\alpha_n}=\prod_{i=1}^n\Omega_i,
  \qquad
  \mathcal F_{\alpha_n}=\bigotimes_{i=1}^n\mathcal F_i,
\]
and let $\pi_{\alpha_n}$ be the projection onto the first $n$ coordinates.

\begin{theorem}[Ionescu--Tulcea; \citealp{ionescuTulcea1949mesures}]
Let $\nu_1$ be an initial probability measure on
$(\Omega_1,\mathcal F_1)$. For
$n\ge2$, let
\[
  K_n:\Omega_{\alpha_{n-1}}\times\mathcal F_n\to[0,1]
\]
be a probability kernel from
\((\Omega_{\alpha_{n-1}},\mathcal F_{\alpha_{n-1}})\) to
\((\Omega_n,\mathcal F_n)\). Then there exists a unique probability measure
$P^\infty$ on \((\Omega^\infty,\mathcal F^\infty)\) such that for rectangles
$A_1\times\cdots\times A_n$,
\[
\begin{aligned}
  &P^\infty\{\omega_1\in A_1,\ldots,\omega_n\in A_n\} \\
  &\quad =
  \int_{A_1}\nu_1(d\omega_1)
  \int_{A_2}K_2(\omega_1,d\omega_2)
  \cdots
  \int_{A_n}K_n(\omega_1,\ldots,\omega_{n-1},d\omega_n).
\end{aligned}
\]
\end{theorem}

\noindent\textit{Proof.}
See \Appref{app:measure-theoretic-toolkit}, Theorem~\ref{thm:appB-ionescu-tulcea}.

\begin{example}[Independent coordinates]
If $K_n(\omega_1,\ldots,\omega_{n-1},A)=\mu_n(A)$ does not depend on the past,
Tulcea's theorem gives the infinite product measure $\bigotimes_{n\ge1}\mu_n$.
For a cylinder depending on the first $n$ coordinates,
\[
  \left(\bigotimes_{i\ge1}\mu_i\right)
    (A_1\times\cdots\times A_n\times\Omega_{n+1}\times\cdots)
  =
  \prod_{i=1}^n \mu_i(A_i).
\]
Thus independence is not an extra property added after construction; it is
encoded by the fact that each kernel ignores the past.
\qedmark
\end{example}

\begin{example}[Markov chains]
Let $E$ be a measurable state space, let $\mu$ be the initial law, and let
$K(x,A)$ be a Markov transition kernel. Taking
\[
  K_n(x_1,\ldots,x_{n-1},A)=K(x_{n-1},A)
\]
constructs a path law $P^\infty$ for a time-homogeneous Markov chain on
$E^{\Nat}$.
For example,
\[
  P^\infty\{X_1\in A_1,\ldots,X_n\in A_n\}
  =
  \int_{A_1}\mu(dx_1)\int_{A_2}K(x_1,dx_2)\cdots
  \int_{A_n}K(x_{n-1},dx_n).
\]
The only memory carried into the next step is the current state $x_{n-1}$,
which is exactly the Markov property written as a product-space construction.
\qedmark
\end{example}

\begin{example}[Markov chain Monte Carlo (MCMC) as an invariant-kernel construction]
Markov chain Monte Carlo belongs naturally in the present chapter because its
first object is not an estimator but a transition kernel.  Let
\((E,\mathcal E)\) be the state space and let \(\pi\) be a target probability
measure on it, such as a posterior law or a difficult normalizing-constant
law.  A Markov chain Monte Carlo algorithm constructs a kernel \(K\) for which
\[
  \pi K(A):=\int_E K(x,A)\,\pi(dx)=\pi(A),
  \qquad A\in\mathcal E .
\]
Thus \(\pi\) is invariant for the chain.  If the chain is started from
\(\pi\), every coordinate has law \(\pi\); if it is started elsewhere,
ergodicity is what justifies treating a long run as approximately drawn from
\(\pi\).

The Metropolis--Hastings construction gives a clean measure-theoretic example
\citep{metropolis1953equation,hastings1970monte}.  Let \(Q(x,dy)\) be a
proposal kernel.  On \(E\times E\), define the forward proposal measure
\[
  M(dx,dy)=\pi(dx)Q(x,dy).
\]
Let \(s:E\times E\to E\times E\) be the coordinate-swap map
\(s(x,y)=(y,x)\).  The reversed proposal measure is the pushforward
\(M^\leftarrow=M\circ s^{-1}\), so
\[
  M^\leftarrow(A\times B)=M(B\times A),
  \qquad A,B\in\mathcal E .
\]
In the formal differential notation this is \(M^\leftarrow(dx,dy)=M(dy,dx)\):
the same proposal-pair law, but with current and proposed states interchanged.
On the part where the two measures are mutually comparable, set
\[
  \alpha(x,y)
  =
  1\wedge \frac{dM^\leftarrow}{dM}(x,y),
\]
with the usual convention on null sets.  The transition kernel is
\[
  K(x,A)
  =
  \int_A \alpha(x,y)\,Q(x,dy)
  +
  \ind{x\in A}
  \left\{1-\int_E\alpha(x,y)\,Q(x,dy)\right\}.
\]
The second term is the holding probability: rejected proposals leave the chain
at its current state.  This kernel satisfies detailed balance,
\[
  \pi(dx)K(x,dy)=\pi(dy)K(y,dx),
\]
and hence leaves \(\pi\) invariant.  In the common density notation this reads
\[
  \alpha(x,y)
  =
  1\wedge
  \frac{\pi(y)q(y,x)}{\pi(x)q(x,y)}.
\]
The density formula is only a convenient special case; the kernel-and-measure
form is the actual object.

Gibbs sampling is another invariant-kernel construction.  If
\(E=E_1\times\cdots\times E_d\) and regular conditional distributions
\(\pi_j(dy_j\mid x_{-j})\) are available, the \(j\)th coordinate-update kernel
is
\[
  K_j(x,dy)
  =
  \delta_{x_{-j}}(dy_{-j})\,\pi_j(dy_j\mid x_{-j}).
\]
Each \(K_j\) leaves \(\pi\) invariant, and a systematic-scan Gibbs sampler
uses the product kernel \(K=K_1K_2\cdots K_d\).  Chapter~17 later asks the
statistical question that this construction does not answer by itself: whether
the realized path has mixed enough, and whether Monte Carlo error is small on
the scale of the reported uncertainty.

Modern MCMC methods are best read as variations on the same kernel idea, often
after enlarging the state space.  Hamiltonian Monte Carlo augments \(x\) by a
momentum variable \(p\) and constructs a kernel on \(E\times P\) whose invariant
law has \(x\)-marginal \(\pi\); the No-U-Turn sampler adapts the simulated
Hamiltonian path within a transition while preserving the same invariant target
after tuning \citep{girolamiCalderhead2011riemann,hoffmanGelman2014nuts}.
Pseudo-marginal MCMC works on an extended space even more explicitly.  If
\(\widehat L(\theta,U)\) is a nonnegative unbiased estimate of an intractable
likelihood \(L(\theta)\), with auxiliary law \(m_\theta(du)\), then the extended
target
\[
  \widetilde\pi(d\theta,du)
  \propto
  p(\theta)\widehat L(\theta,u)m_\theta(du)
\]
has \(\theta\)-marginal proportional to \(p(\theta)L(\theta)\).  A
Metropolis--Hastings kernel on \((\theta,u)\) can therefore be exact for the
desired marginal even though the likelihood is never evaluated exactly; particle
MCMC is the sequential version of this idea, with a particle filter supplying
the auxiliary random likelihood estimate
\citep{andrieuRoberts2009pseudo,andrieuDoucetHolenstein2010particle}.

Other modern samplers relax a different part of the recipe.  Stochastic-gradient
Langevin methods replace full-data gradients by noisy minibatch gradients and
are therefore best viewed as approximate kernels unless a correction restores
the intended invariant law \citep{wellingTeh2011sgld}.  Nonreversible samplers
remind us that detailed balance is sufficient, not necessary: the real
requirement is invariance, \(\pi K=\pi\), while breaking reversibility may
improve exploration \citep{bierkensFearnheadRoberts2019zigzag}.  Coupling-based
diagnostics and unbiased MCMC estimators go one level higher still: they build
a joint kernel \(\bar K\) whose marginals are both \(K\), then use meeting times
of two chains to study convergence and Monte Carlo bias
\citep{jacobOLearyAtchade2020unbiased}.  The common grammar is unchanged:
specify the measurable state, the target law or its marginal, the transition
kernel, and the path law induced by repeated application of that kernel.
\qedmark
\end{example}

\begin{example}[Particle swarm optimization as an augmented Markov chain]
Stochastic algorithms often become Markov chains after the state is enlarged
enough.  Particle swarm optimization (PSO) \citep{kennedyEberhart1995pso}
is a useful example because the visible particle positions alone are not
Markov.  Suppose an objective $h:E\to\mathbb R$ is being minimized by
particles with positions $X_{i,t}$, velocities $V_{i,t}$, personal bests
$B_{i,t}$, and current swarm best $G_t$.  A typical update has the form
\[
  V_{i,t+1}
  =
  \omega V_{i,t}
  + c_1 U_{i,t}(B_{i,t}-X_{i,t})
  + c_2 W_{i,t}(G_t-X_{i,t}),
  \qquad
  X_{i,t+1}=X_{i,t}+V_{i,t+1},
\]
where the random multipliers $U_{i,t}$ and $W_{i,t}$ are newly drawn at time
$t$.  The personal and swarm bests are then updated by comparing objective
values.  If
\[
  S_t=(X_{1:N,t},V_{1:N,t},B_{1:N,t},G_t),
\]
then the conditional law of $S_{t+1}$ depends on the past only through $S_t$.
Thus PSO defines a transition kernel
\[
  K(S_t,\cdot)=\mathcal L(S_{t+1}\mid S_t).
\]
The point is not that PSO is MCMC.  It is usually an optimization heuristic,
not a chain designed to have a prescribed stationary distribution.  The
Markov-kernel viewpoint instead tells us exactly where the algorithmic
randomness lives, and why convergence, stagnation, and sensitivity to random
seeds are questions about the stochastic recursion itself.
\qedmark
\end{example}

\begin{example}[Hidden Markov observations]
Let $(E,\mathcal E)$ be a hidden-state space and $(F,\mathcal F)$ an
observation space. Suppose the latent chain has initial law $\mu$ and transition
kernel $K$, and suppose observations are emitted from the kernel $G(x,\cdot)$.
Chapter~6's construction is applied not to $X_n$ alone but to the product
coordinate
\[
  Z_n=(X_n,Y_n)\in E\times F .
\]
The first-coordinate law and the later kernels on $E\times F$ are
\[
  \nu_1(dz_1)=\mu(dx_1)G(x_1,dy_1),
  \qquad z_1=(x_1,y_1),
\]
and, for $n\ge2$,
\[
  L_n(z_1,\ldots,z_{n-1},d z_n)
  =
  K(x_{n-1},dx_n)G(x_n,dy_n),
  \qquad z_n=(x_n,y_n).
\]
Ionescu--Tulcea now gives a unique probability law on
$(E\times F)^{\Nat}$ with these finite-prefix kernels.  For a finite
observed-hidden history, the corresponding cylinder probability factors as
\[
  \mu(dx_1)G(x_1,dy_1)
  \prod_{k=2}^n K(x_{k-1},dx_k)G(x_k,dy_k).
\]
Thus the usual hidden-Markov graphical model is exactly a product-space
construction: the path object is a coordinate process, the model is a sequence
of kernels, and the theorem turns those local kernels into one probability
measure. The conditional-independence reading is then visible in the formula:
the hidden state propagates through $K$, and $Y_n$ depends on the past only
through the current hidden state $X_n$ \citep{rabiner1989tutorial,cappeMoulinesRyden2005hmm}.
\qedmark
\end{example}

\subsection{Uncountable Products and Kolmogorov Extension}
\label{sec:ch06-kolmogorov-extension}
\conceptindexes{uncountable products, Kolmogorov extension theorem, finite-dimensional distributions, consistency}

\noindent\textbf{Statistical thread.}
Kolmogorov's theorem begins with finite-dimensional shadows and asks whether
they can belong to one full random object. The statistical lesson is sharp:
finite-dimensional distributions describe what every finite analysis can see,
but path questions require more measurable structure.

Let $T$ be an arbitrary index set. For a finite set
\(\alpha=\{t_1,\ldots,t_k\}\subset T\), write
\[
  \Omega_\alpha=\prod_{t\in\alpha}\Omega_t,
  \qquad
  \mathcal F_\alpha=\bigotimes_{t\in\alpha}\mathcal F_t,
\]
and let \(\pi_\alpha:\prod_{t\in T}\Omega_t\to\Omega_\alpha\) be the coordinate
projection. The coordinate product sigma-field is
\[
  \Omega^T=\prod_{t\in T}\Omega_t,
  \qquad
  \mathcal F^T=\bigotimes_{t\in T}\mathcal F_t
  =\sigma\{\pi_\alpha^{-1}B:
    \alpha\subset T \text{ finite},\ B\in\mathcal F_\alpha\}.
\]
Every set in $\mathcal F^T$ depends on at most countably many coordinates. This
simple fact is the source of many stochastic-process measurability surprises.
The coordinate criterion below follows the standard advanced-probability
treatment; see \citet{dabrowskaAdvancedProbabilityCommunication}.

\begin{proposition}[Coordinate criterion for process measurability]
\label{prop:ch06-coordinate-process-measurability}
Let \((S_t,\mathcal S_t)_{t\in T}\) be measurable spaces and let
\[
  X:\Omega\longrightarrow \prod_{t\in T}S_t,
  \qquad
  X(\omega)=(X_t(\omega):t\in T).
\]
For a finite \(\alpha\subset T\), put \(X_\alpha=\pi_\alpha\circ X\).  The
following are equivalent:
\[
\begin{array}{ll}
\mathrm{(i)} & X \text{ is } \mathcal F/\bigotimes_{t\in T}\mathcal S_t
  \text{ measurable},\\[0.25em]
\mathrm{(ii)} & X_t \text{ is } \mathcal F/\mathcal S_t
  \text{ measurable for every }t\in T,\\[0.25em]
\mathrm{(iii)} & X_\alpha \text{ is }
  \mathcal F/\bigotimes_{t\in\alpha}\mathcal S_t
  \text{ measurable for every finite }\alpha\subset T.
\end{array}
\]
When these conditions hold,
\[
  \sigma(X)
  =
  \sigma(X_t:t\in T)
  =
  \sigma\bigl(\,\bigcup_{\alpha\subset T\ \mathrm{finite}}\sigma(X_\alpha)\bigr).
\]
\end{proposition}

\noindent\textit{Proof.}
If \(X\) is measurable, then \(X_\alpha=\pi_\alpha\circ X\) is measurable for
every finite \(\alpha\), because coordinate projections are measurable by the
definition of the product sigma-field. Thus (i) implies (iii), and (iii)
implies (ii) by taking one-point sets.

Conversely, suppose each coordinate \(X_t\) is measurable. For finite
\(\alpha=\{t_1,\ldots,t_k\}\), rectangles in
\(\bigotimes_{t\in\alpha}\mathcal S_t\) pull back to finite intersections
\[
  X_\alpha^{-1}(A_{t_1}\times\cdots\times A_{t_k})
  =
  \bigcap_{j=1}^k X_{t_j}^{-1}(A_{t_j})\in\mathcal F.
\]
Since such rectangles generate the finite product sigma-field, \(X_\alpha\) is
measurable. Hence every finite-coordinate cylinder satisfies
\[
  X^{-1}\!\left(\pi_\alpha^{-1}(B)\right)=X_\alpha^{-1}(B)\in\mathcal F.
\]
The cylinders generate \(\bigotimes_{t\in T}\mathcal S_t\), so \(X\) is
measurable. Finally,
\[
  X^{-1}\!\left(\bigotimes_{t\in T}\mathcal S_t\right)
  =
  \sigma\{X_t^{-1}(A):t\in T,\ A\in\mathcal S_t\},
\]
which is exactly the displayed identity for the generated sigma-field.
\qedmark

\begin{definition}[Finite-dimensional distributions]
For a process $X=\{X_t:t\in T\}$ with state spaces $(S_t,\mathcal S_t)$, the
finite-dimensional distributions are the laws
\[
  \mu_{t_1,\ldots,t_k}
  =
  \Law(X_{t_1},\ldots,X_{t_k}),
  \qquad t_1,\ldots,t_k\in T.
\]
They are consistent if deleting or permuting coordinates gives the
corresponding lower-dimensional law.
\end{definition}

\begin{theorem}[Kolmogorov extension theorem; \citealp{kolmogorov1933grundbegriffe}]
Suppose each $(S_t,\mathcal S_t)$ is a standard Borel space. Every consistent
family of finite-dimensional probability measures
\[
  \{\mu_{t_1,\ldots,t_k}:k\ge1,\ t_i\in T\}
\]
defines a unique probability measure $\mu$ on
$(\prod_{t\in T}S_t,\bigotimes_{t\in T}\mathcal S_t)$ whose finite-dimensional
projections have the prescribed laws.
\end{theorem}

\noindent\textit{Proof.}
See \Appref{app:measure-theoretic-toolkit}, Theorem~\ref{thm:appB-kolmogorov-extension}.

\begin{definition}[Equivalence and indistinguishability]
Two processes $X$ and $Y$ indexed by $T$ are stochastically equivalent if
\[
  \Prob(X_t=Y_t)=1,\qquad t\in T.
\]
They are indistinguishable if
\[
  \Prob(X_t=Y_t\text{ for all }t\in T)=1.
\]
For countable $T$ these notions coincide; for uncountable $T$ they do not.
\end{definition}

\begin{example}[Finite-dimensional laws do not see path regularity]
Let $(\Omega,\mathcal F,\Prob)=([0,1],\Borel[0,1],\lambda)$ and define
\[
  X_t(\omega)=0,\qquad
  Y_t(\omega)=\ind{\omega=t},\qquad t\in[0,1].
\]
For each fixed $t$, $Y_t=0$ almost surely, so $X$ and $Y$ have the same
one-dimensional distributions. More generally, for finitely many times
$t_1,\ldots,t_k$,
\[
  \Prob\{Y_{t_1}=\cdots=Y_{t_k}=0\}
  =
  1-\lambda(\{t_1,\ldots,t_k\})
  =
  1,
\]
so every finite vector of $Y$ equals the zero vector almost surely, just like
the corresponding vector of $X$. But for a fixed sample point $\omega$, the path
$t\mapsto Y_t(\omega)$ has a single spike at $t=\omega$, while the path of $X$
is identically zero. Thus the set $C[0,1]$ of continuous paths cannot be a
cylinder-measurable subset of $\R^{[0,1]}$: otherwise the two induced
product-space laws would assign it the same probability.
\qedmark
\end{example}

\begin{example}[Hitting times are not automatically measurable]
Let $T=\Rplus$ and let $X_t$ be a coordinate process on a product space
$S^{\Rplus}$. For a Borel set $B\subseteq S$ the hitting time
\[
  \tau_B=\inf\{t\ge0:X_t\in B\}
\]
is defined path by path. But measurability of $\tau_B$ does not follow from
measurability of each coordinate $X_t$. Indeed,
\[
  \{\tau_B<t\}
  =
  \bigcup_{0\le s<t}\{X_s\in B\}
\]
is an uncountable union of measurable cylinder sets. A sigma-field is closed
under countable unions, not arbitrary unions. If the path is right-continuous
and $B$ is open, then rational times repair the problem:
\[
  \{\tau_B<t\}
  =
  \bigcup_{\substack{q<t\\q\in\Rat_+}}\{X_q\in B\}.
\]
This is the same separability idea in its simplest hitting-time form.
\qedmark
\end{example}

\begin{example}[The continuity event depends on uncountably many coordinates]
The event
\[
  \{x\in\R^{[0,1]}:x\text{ is continuous}\}
\]
is not measurable in the raw product sigma-field. On the smaller path space
$C[0,1]$ it is the whole space and is Borel. The difference is conceptual:
the product space records only finite-dimensional observations, while the path
space builds continuity into the state space itself.
\qedmark
\end{example}

\begin{example}[Brownian motion as two different objects]
Kolmogorov's theorem constructs a probability law on $\R^{[0,\infty)}$ with
the Brownian finite-dimensional distributions. Kolmogorov's continuity theorem
then gives a continuous modification, which can be viewed as a random element
of $C[0,\infty)$. Statistical functionals such as
\[
  \sup_{0\le t\le1}W_t,\qquad
  \inf\{t:W_t=a\}
\]
are naturally measurable on the continuous path space. They are not justified
by finite-dimensional distributions alone.
\qedmark
\end{example}

\section{Basic Concepts of Stochastic Processes}
\label{sec:ch06-basic-process-concepts}
\conceptindexes{stochastic processes, random fields, random measures, random elements, process equivalence, Dirichlet process, Poisson random measure}

\noindent\textbf{Statistical thread.}
A stochastic process is a random variable whose value is itself a function-like
object. Once data become curves, event histories, point patterns, or empirical
processes, the question ``what is the distribution?'' becomes ``which path
features are measurable?''

\begin{definition}[Stochastic process]
Let $(\Omega,\mathcal F,\Prob)$ be a probability space and let
$(S,\mathcal S)$ be a measurable state space. A stochastic process indexed by
$T$ is a family
\[
  X=\{X_t:t\in T\}
\]
of $\mathcal F/\mathcal S$ measurable maps. Equivalently, it is a measurable
map from $\Omega$ into $S^T$ equipped with the product sigma-field, by
Proposition~\ref{prop:ch06-coordinate-process-measurability}.
\end{definition}

\subsection{Processes, Fields, Random Measures, and Random Elements}
\label{sec:ch06-random-objects}
\conceptindexes{random objects, stochastic process, random field, random measure, random element}

The term stochastic process is flexible. In its narrowest use it means a family
indexed by time. In modern statistics it is better read as a first example of a
random object:
\[
\begin{gathered}
\text{random variable}
\subset
\text{random vector}
\subset
\text{stochastic process}
\\
\subset
\text{random field / random measure / random function}.
\end{gathered}
\]
The inclusions are not formal set inclusions; they are a hierarchy of modeling
languages. Each step adds coordinates, and each added coordinate brings the same
question back: which maps from the object to the real line are measurable?

\input{figures/ch06_random_object_taxonomy}

\begin{definition}[Random objects]
A \emph{random field} is a stochastic process whose index set has spatial or
multiparameter structure. Thus, instead of $X_t$ for time $t$, one may have
\[
  X_s,\qquad s\in D\subseteq\R^d,
\]
where typically $D\subseteq\R^d$ or $D$ is another multiparameter index set.

Let $(S,\mathcal S)$ be a measurable space and let $\mathcal M(S)$ be a class of
measures on $(S,\mathcal S)$, equipped with the evaluation sigma-field
\[
  \sigma\{m\mapsto m(A):A\in\mathcal S\}.
\]
A \emph{random measure} is a measurable map
$M:(\Omega,\mathcal F)\to \mathcal M(S)$. Equivalently, $M(\omega,A)$ is a
kernel in the following sense: for each $A\in\mathcal S$, $M(A)$ is a random
variable, and for each $\omega$, $A\mapsto M(\omega,A)$ is a measure.

More generally, if $(E,\mathcal E)$ is a measurable space, a \emph{random
element} of $E$ is a measurable map
\[
  X:(\Omega,\mathcal F)\longrightarrow (E,\mathcal E).
\]
Random variables, random vectors, stochastic processes, random functions, and
random probability measures are all random elements with different choices of
state space $E$ and sigma-field $\mathcal E$.
\end{definition}

This vocabulary is not decorative. Random fields appear in spatial statistics,
Gaussian random fields, longitudinal images, and environmental surfaces. Their
finite-dimensional laws still describe $(X_{s_1},\ldots,X_{s_k})$, but
scientifically important features are often geometric: maxima over regions,
level sets, excursion sets, smoothness, and spatial correlation.

\begin{example}[Brain voxels as a three-dimensional random field]
\label{ex:ch06-brain-voxels-random-field}
A voxel is the three-dimensional analogue of a pixel.  After a brain image has
been registered to a common coordinate system, a structural MRI volume can be
read as a random field
\[
  X(s),\qquad s=(x,y,z)\in D\subseteq\R^3,
\]
where \(s\) indexes a brain location and \(X(s)\) might be an intensity,
gray-matter summary, or other local measurement.  For subject-level data one
would write \(X_i(s)\), \(i=1,\ldots,n\).  In an fMRI study the index set is
larger:
\[
  X_i(s,t),\qquad (s,t)\in D\times T,
\]
because the object has both spatial and temporal coordinates.

This notation is not only a storage convention.  If diseased and control
subjects are compared, the target might be the whole contrast field
\[
  \Delta(s)=
  \Expect\{X_i(s)\mid \text{disease}\}
  -
  \Expect\{X_i(s)\mid \text{control}\}.
\]
The question is no longer the value of one mean, but which regions of the
three-dimensional field show a stable difference.  Nearby voxels are usually
correlated because acquisition, anatomy, registration, and smoothing all carry
local information across neighboring locations.  Treating voxel values as iid
coordinates would therefore erase the object being studied.  Voxelwise
multiple testing, region-of-interest summaries, smoothing, cluster-level
inference, and spatial random-field approximations are different ways of
responding to the same fact: the index set is spatial, and the geometry of that
index set matters.

The same data can also be re-indexed.  An atlas replaces voxels by brain
regions \(a\in\mathcal A\); a connectivity analysis replaces locations by
edges \(e\in E\) of a brain network; a learned embedding replaces the image by
a feature vector.  The random object changes when the index set changes, even
when the original scanner record is the same.
\qedmark
\end{example}

Two central statistical random measures are
\[
  P_n(A)=\frac1n\sum_{i=1}^n\indset{A}(X_i),
  \qquad
  N((0,t]\times B)=\sum_{n\ge1}\ind{T_n\le t,\ Z_n\in B}.
\]
The first is the empirical measure, a random probability measure. The second is
a point process written as an integer-valued random measure. In Bayesian
nonparametrics, the Dirichlet process is another canonical random probability
measure. Chapter 16 returns to the same random-measure notation after adding a
filtration: event counts then acquire intensities, predictable compensators, and
martingale residuals.

\begin{example}[Voting configurations as random elements]
Let \(V\) be a finite or countable set of agents and let each agent hold a
binary state.  At a single time, a voting profile is a configuration
\[
  \eta\in\{0,1\}^{V},
\]
where \(\eta(x)=1\) may mean that agent \(x\) supports a proposal.  A dynamic
voting system is therefore not merely a random vector of final ballots.  It is
a random element such as
\[
  \eta=\{\eta_t(x):t\in T,\ x\in V\}
  \in \{0,1\}^{T\times V},
\]
or, after imposing path regularity, a random path in
\(D(T,\{0,1\}^{V})\).  Its coordinates ask for the state of selected agents at
selected times, while its scientifically interesting summaries ask for events
such as consensus, persistence of a minority, or the first time a coalition
crosses a threshold.

This is the product-space face of interacting particle systems.  In the voter
model of \citet{holleyLiggett1975voter}, an agent updates by copying a neighbor
chosen from a network transition rule.  Thus the observed votes are coupled
through exposure and imitation; they are not iid reports from a fixed
population distribution.  Chapter 7 uses this as a grammar for multi-agent
network data, and Chapter 16 returns to the same object with generators,
Feller semigroups, and continuous-time transition rates
\citep{liggett1985interacting}.
\qedmark
\end{example}

\begin{example}[Basic economic regions as spatial-economic objects]
Historical China's basic economic regions are a useful example of why random
objects need not be simple vectors.  A region can be encoded as a partition of
counties, a graph of market connections, a spatial field of grain prices and
transport costs, or a latent macroregional system.  Skinner's work on rural
marketing and macroregions treated region formation as a structured
spatial-economic object rather than as a fixed administrative label
\citep{skinner1964marketing,skinner1977regional}.

One possible statistical encoding is
\[
  X
  =
  \{(G_t,Z_t,\Pi_t):t\in T\},
\]
where \(G_t\) is a market or transport graph, \(Z_t(s)\) is a spatial field of
prices, harvests, population, or reports over locations \(s\), and \(\Pi_t\)
is a partition into economic regions.  The measurable space is no longer just
\(\R^p\).  It contains graphs, fields, and partitions.  This example is the
historical counterpart of modern spatial omics and network data: the first
question is not which regression to run, but what kind of random element a
``region'' has become.
\qedmark
\end{example}

\begin{definition}[Dirichlet process]
Let $(S,\mathcal S)$ be a measurable space and let $\eta$ be a finite nonzero
measure on it. A random probability measure $G$ on $(S,\mathcal S)$ is a
\emph{Dirichlet process} with parameter measure $\eta$, written
$G\sim\DP(\eta)$, if for every finite measurable partition
$(A_1,\ldots,A_k)$ of $S$,
\[
  (G(A_1),\ldots,G(A_k))
  \sim
  \Dirichlet\{\eta(A_1),\ldots,\eta(A_k)\}.
\]
If $\eta=aH$, where $a>0$ and $H$ is a probability measure, also write
$G\sim\DP(aH)$. Cells with $\eta(A_i)=0$ are understood to
receive zero mass almost surely; equivalently, they may be deleted before
writing the ordinary Dirichlet density.
\end{definition}

\begin{theorem}[Existence of Dirichlet processes; \citealp{ferguson1973bayesian}]
Let $(S,\mathcal S)$ be a standard Borel space and let $\eta$ be a finite
nonzero measure on $(S,\mathcal S)$. There exists a probability measure
$\Pi_\eta$ on
\[
  \left([0,1]^{\mathcal S},
  \bigotimes_{A\in\mathcal S}\Borel[0,1]\right)
\]
such that the coordinate process $G(A)(x)=x_A$ has the Dirichlet finite-partition
laws in the preceding definition. Thus the Dirichlet process exists as a
random-measure coordinate process. On standard Borel spaces this coordinate
law is the usual law of a random probability measure with evaluation
sigma-field, as in Ferguson's construction \citep{ferguson1973bayesian}.
\end{theorem}

\noindent\textit{Proof.}
The proof follows the explicit disjointification argument in
\citet{cui2024nature}. We construct the finite-dimensional laws for the
coordinates
\[
  \{G(A):A\in\mathcal S\}
\]
and then apply Kolmogorov's extension theorem. Fix a finite list
$A_1,\ldots,A_m\in\mathcal S$. For $v=(v_1,\ldots,v_m)\in\{0,1\}^m$, define
\[
  A_j^1=A_j,\qquad A_j^0=A_j^c,\qquad
  B_v=\bigcap_{j=1}^m A_j^{v_j}.
\]
The sets $B_v$ form a finite measurable partition of $S$, after deleting empty
cells. Let
\[
  (Y_v:v\in\{0,1\}^m)
  \sim
  \Dirichlet\{\eta(B_v):v\in\{0,1\}^m\},
\]
with $Y_v=0$ for cells satisfying $\eta(B_v)=0$. Define
\[
  Z_i=\sum_{\substack{v\in\{0,1\}^m\\ v_i=1}}Y_v,\qquad i=1,\ldots,m.
\]
Let $\mu_{A_1,\ldots,A_m}$ be the law of
$Z=(Z_1,\ldots,Z_m)$ on $[0,1]^m$. This is the proposed law of
$(G(A_1),\ldots,G(A_m))$.

We must check consistency. Permuting $A_1,\ldots,A_m$ merely permutes the
coordinates of $v$ and relabels the same partition cells $B_v$, so the law
$\mu_{A_1,\ldots,A_m}$ is consistent under coordinate permutations.

Now delete the last set from the list. For $u=(u_1,\ldots,u_{m-1})$, set
\[
  C_u=\bigcap_{j=1}^{m-1}A_j^{u_j}.
\]
Then
\[
  C_u=B_{(u,0)}\cup B_{(u,1)}
\]
as a disjoint union, and hence
\[
  \eta(C_u)=\eta(B_{(u,0)})+\eta(B_{(u,1)}).
\]
We use the aggregation property of the Dirichlet distribution. To see it
directly, write the Dirichlet vector through independent gamma variables:
take independent $E_v$ with common rate one and shape $\eta(B_v)$, putting
$E_v=0$ when $\eta(B_v)=0$, and set
\[
  Y_v=\frac{E_v}{\sum_wE_w}.
\]
For each $u$, the sum
\[
  E_u^-:=E_{(u,0)}+E_{(u,1)}
\]
is gamma with shape $\eta(C_u)$ and rate one, and the variables
$E_u^-$ are independent over different $u$. Therefore
\[
  Y_u^-:=Y_{(u,0)}+Y_{(u,1)}
  =
  \frac{E_u^-}{\sum_{u'}E_{u'}^-}
\]
has Dirichlet distribution with parameters $\{\eta(C_u):u\in\{0,1\}^{m-1}\}$.
For $i<m$,
\[
\begin{aligned}
  Z_i
  &=
  \sum_{\substack{v\in\{0,1\}^m\\ v_i=1}}Y_v  \\
  &=
  \sum_{\substack{u\in\{0,1\}^{m-1}\\ u_i=1}}
  \{Y_{(u,0)}+Y_{(u,1)}\}
  =
  \sum_{\substack{u\in\{0,1\}^{m-1}\\ u_i=1}}Y_u^- .
\end{aligned}
\]
This is exactly the construction of
$\mu_{A_1,\ldots,A_{m-1}}$. Hence the marginal law of
$(Z_1,\ldots,Z_{m-1})$ under $\mu_{A_1,\ldots,A_m}$ equals
$\mu_{A_1,\ldots,A_{m-1}}$. Repeated deletion gives consistency for every
sublist.

Kolmogorov's extension theorem, applied with index set $\mathcal S$ and common
state space $[0,1]$, now gives a probability measure $\Pi_\eta$ on
$[0,1]^{\mathcal S}$ whose finite-dimensional coordinate laws are the
$\mu_{A_1,\ldots,A_m}$ just constructed.

It remains only to check that the finite-partition distributions are the
Dirichlet ones. If $(A_1,\ldots,A_k)$ is a measurable partition of $S$, then
the only nonempty cells in the disjointification are $A_1,\ldots,A_k$
themselves, together with possibly empty cells. Thus the coordinate vector
$(G(A_1),\ldots,G(A_k))$ has law
\[
  \Dirichlet\{\eta(A_1),\ldots,\eta(A_k)\},
\]
which is the required Dirichlet-process specification.
\qedmark

This finite-partition specification is exactly a product-space idea: local
finite-dimensional laws are required to cohere into a law on random probability
measures. Ferguson's construction of the Dirichlet process is one of the
statistical descendants of this viewpoint
\citep{ferguson1973bayesian,ferguson1974prior}.  The conceptual move is simple
but large: the unknown object is no longer a finite vector $\theta$, but a law
$G\in\mathcal P(S)$, and the prior is a probability law on that space of laws.

\begin{example}[A prior on a probability measure]
Let \(G\sim\DP(\alpha H)\), where \(\alpha>0\) and \(H\) is a probability
measure on \(S\).  Conditionally on \(G\), let \(X_1,\ldots,X_n\) be
independent observations with common law \(G\).  The posterior random
probability measure is again a Dirichlet process:
\[
  G\mid X_{1:n}
  \sim
  \DP\left(\alpha H+\sum_{i=1}^n\delta_{X_i}\right).
\]
Consequently, for every measurable \(A\subseteq S\),
\[
  \Prob(X_{n+1}\in A\mid X_{1:n})
  =
  \Expect\{G(A)\mid X_{1:n}\}
  =
  \frac{\alpha H(A)+\sum_{i=1}^n\indset{A}(X_i)}
       {\alpha+n}.
\]
This formula is the statistical content of the phrase ``prior on a space of
probability measures.''  The update does not choose one parametric density.
It updates the whole random law \(G\), and prediction averages over the
posterior uncertainty in that law.

The same display also explains the species-sampling reading.  If \(H\) has no
atoms and the observed distinct values are \(x_1^\ast,\ldots,x_K^\ast\), with
counts \(m_1,\ldots,m_K\), then
\[
  \Prob(X_{n+1}\in dx\mid X_{1:n})
  =
  \frac{\alpha}{\alpha+n}H(dx)
  +
  \sum_{k=1}^K\frac{m_k}{\alpha+n}\delta_{x_k^\ast}(dx).
\]
Thus the next observation is new with probability \(\alpha/(\alpha+n)\), and
otherwise repeats a previously seen type with probability proportional to its
current count.  This is the Polya-urn predictive scheme of
\citet{blackwellMacQueen1973ferguson}; in later language it is the
Chinese-restaurant picture behind the Dirichlet process.  The missing-species
example in Chapter~2 asks a frequentist version of the same question:
what do the rare observed types say about the chance that the next type has not
yet appeared?  Indian-buffet processes belong to a neighboring but different
story: feature allocation rather than probability measures, so they are better
left outside the main line here.
\qedmark
\end{example}

The random-element language is useful because modern data often are not scalars
or vectors. The state space $E$ may itself be a function space, a path space, a
metric space of probability measures, or a manifold. The mathematical task
remains the same as in ordinary probability: choose the state space and
sigma-field so that the statistics of interest are measurable.

\begin{example}[Distribution-valued data as random elements]
Let $D\subseteq\R$ be an interval and let $\mathcal W_2(D)$ be the space of
probability measures on $D$ with finite second moment, equipped with the
$L^2$-Wasserstein metric. If $\mu,\nu\in\mathcal W_2(D)$ have distribution
functions $F_\mu,F_\nu$ and quantile functions
$Q_\mu=F_\mu^{-1}$, $Q_\nu=F_\nu^{-1}$, then in one dimension
\[
  d_W^2(\mu,\nu)
  =
  \int_0^1\{Q_\mu(u)-Q_\nu(u)\}^2\,du .
\]
A distribution-valued observation is therefore a random element
\[
  Y:(\Omega,\mathcal F)\longrightarrow
  \bigl(\mathcal W_2(D),\Borel(\mathcal W_2(D))\bigr),
\]
and its quantile process
\[
  Q_Y(u)=F_Y^{-1}(u),\qquad 0<u<1,
\]
is a stochastic process indexed by $u$. This is the same chapter in another
costume: a random object is controlled through coordinates, and useful
statistics require measurability of maps such as
\[
  Y\longmapsto d_W^2(Y,\nu),
  \qquad
  Y\longmapsto \int_0^1 Q_Y(u)h(u)\,du .
\]
For example, the unconditional Fréchet mean is any minimizer
\[
  \mu_\oplus\in
  \argmin_{\nu\in\mathcal W_2(D)}
  \Expect d_W^2(Y,\nu),
\]
and, when the quantile process is square-integrable, its quantile function is
\[
  Q_{\mu_\oplus}(u)=\Expect Q_Y(u).
\]
Wasserstein regression builds on this random-element viewpoint. For an
atomless reference distribution $\mu_\ast$ with cdf $F_\ast$, the logarithmic
coordinate used by \citet{chen2023wasserstein} is
\[
  \operatorname{Log}_{\mu_\ast}(\mu)
  =
  Q_\mu\circ F_\ast-\operatorname{id}
  \in L^2(\mu_\ast),
\]
and the corresponding exponential map sends a tangent coordinate back to a
distribution by
\[
  \operatorname{Exp}_{\mu_\ast}(g)=(g+\operatorname{id})_\#\mu_\ast .
\]
Thus a nonlinear space of probability measures is studied through measurable
Hilbert-space coordinates, while the final object remains a random probability
measure. This is the same product-space theme as before, now wearing modern
data clothes; see also the random-object framework of
\citet{petersen2019frechet}.
\qedmark
\end{example}

\begin{example}[Gaussian processes]
A real-valued process $X=\{X_t:t\in T\}$ is Gaussian if every finite vector
$(X_{t_1},\ldots,X_{t_k})$ is multivariate normal. Its law is determined by
the mean function $m(t)=\Expect X_t$ and covariance function
\[
  K(s,t)=\Cov(X_s,X_t),
\]
where $K$ is symmetric and nonnegative definite:
\[
  \sum_{i,j=1}^k a_i a_j K(t_i,t_j)\ge0.
\]
Kolmogorov's theorem constructs a Gaussian process from any such $m$ and $K$.
For any finite list $t_1,\ldots,t_k$, the vector is assigned mean
\[
  (m(t_1),\ldots,m(t_k))
\]
and covariance matrix $\{K(t_i,t_j)\}_{i,j\le k}$. Nonnegative definiteness is
exactly the condition that this matrix be a legitimate covariance matrix. The
finite-dimensional normal laws are automatically consistent because deleting a
coordinate from a multivariate normal vector deletes the corresponding row and
column of the covariance matrix. Kolmogorov then turns this finite-dimensional
recipe into a process law.
\qedmark
\end{example}

\begin{example}[Brownian finite-dimensional distributions]
Standard Brownian motion has $W_0=0$, independent increments, and
\[
  W_t-W_s\sim\Normal(0,t-s),\qquad 0\le s<t.
\]
Its covariance function is
\[
  \Cov(W_s,W_t)=s\wedge t.
\]
For $0<t_1<\cdots<t_k$, write the vector as cumulative sums of independent
increments:
\[
  (W_{t_1},\ldots,W_{t_k})
  =
  (\Delta_1,\Delta_1+\Delta_2,\ldots,\Delta_1+\cdots+\Delta_k),
\]
where $\Delta_i=W_{t_i}-W_{t_{i-1}}$ and $t_0=0$. The increments are independent
normals with variances $t_i-t_{i-1}$. From this representation,
\[
  \Cov(W_s,W_t)=\Var(W_{s\wedge t})=s\wedge t.
\]
Kolmogorov's theorem gives a process with these finite-dimensional
distributions; the continuity theorem below gives a continuous modification.
\qedmark
\end{example}

\begin{example}[Poisson process]
\label{ex:ch06-poisson-process-fdds}
A counting process $N=\{N_t:t\ge0\}$ with $N_0=0$ is a non-homogeneous Poisson
process with mean measure $A$ if it has independent increments and
\[
  N_t-N_s\sim\Poisson\{A(t)-A(s)\},
  \qquad 0\le s<t.
\]
If $A(t)=\alpha t$, it is homogeneous with rate $\alpha$.
For times $0<t_1<\cdots<t_k$, the finite-dimensional law is obtained by writing
\[
  N_{t_j}=\sum_{i=1}^j \Delta_i,\qquad
  \Delta_i=N_{t_i}-N_{t_{i-1}},
\]
with independent
\[
  \Delta_i\sim\Poisson\{A(t_i)-A(t_{i-1})\}.
\]
Thus the path is built from product-measure increments, but the cumulative
coordinates enforce monotonicity.
\qedmark
\end{example}

\begin{theorem}[Existence of Poisson random measures]
\label{thm:ch06-poisson-random-measures}
Let $(S,\mathcal S)$ be a measurable space and let $\nu$ be a sigma-finite
measure on it. There exists a counting random measure $N$ such that, for every
finite list of disjoint sets $A_1,\ldots,A_m\in\mathcal S$ satisfying
$\nu(A_i)<\infty$,
\[
  N(A_1),\ldots,N(A_m)
  \quad\text{are independent,}
  \qquad
  N(A_i)\sim\Poisson\{\nu(A_i)\}.
\]
If $S$ is locally compact and second countable and $\nu$ is finite on compact
sets, then $N$ is locally finite almost surely. This is the Poisson random
measure with mean measure $\nu$.
\end{theorem}

\noindent\textit{Proof.}
First suppose $\nu(S)=M<\infty$. If $M=0$, set $N(A)=0$ for all $A$. If
$M>0$, let
\[
  K\sim\Poisson(M)
\]
and, independently of $K$, let $\xi_1,\xi_2,\ldots$ be iid points in $S$ with
common law $\nu/M$. Define, path by path,
\[
  N(A)=\sum_{r=1}^{K}\indset{A}(\xi_r),\qquad A\in\mathcal S,
\]
with the empty sum interpreted as zero. For each outcome, $A\mapsto N(A)$ is a
finite counting measure because it is the sum of $K$ point masses.

Let $A_1,\ldots,A_m$ be disjoint measurable sets. Put
\[
  p_i=\frac{\nu(A_i)}{M},\qquad
  p_0=1-\sum_{i=1}^{m}p_i .
\]
For $0\le z_1,\ldots,z_m\le1$, conditioning on $K$ gives
\[
\begin{aligned}
  \Expect\!\left[
    \prod_{i=1}^{m}z_i^{N(A_i)}
    \,\middle|\, K
  \right]
  &=
  \left(p_0+\sum_{i=1}^{m}p_i z_i\right)^K .
\end{aligned}
\]
Taking expectation over the Poisson variable $K$,
\[
\begin{aligned}
  \Expect\prod_{i=1}^{m}z_i^{N(A_i)}
  &=
  \exp\left\{
    M\left(p_0+\sum_{i=1}^{m}p_i z_i-1\right)
  \right\}  \\
  &=
  \exp\left\{\sum_{i=1}^{m}\nu(A_i)(z_i-1)\right\}  \\
  &=
  \prod_{i=1}^{m}\exp\{\nu(A_i)(z_i-1)\}.
\end{aligned}
\]
This is the product of the probability generating functions of independent
Poisson variables with means $\nu(A_i)$. Thus the finite-measure construction
has the required count laws.

Now let $\nu$ be sigma-finite. Choose measurable sets
$S_1,S_2,\ldots$ such that
\[
  S=\bigcup_{j\ge1}S_j,\qquad
  S_i\cap S_j=\emptyset\ (i\ne j),\qquad
  \nu(S_j)<\infty .
\]
Such a partition is obtained by disjointifying any countable finite-measure
cover. For each $j$, apply the finite construction to the measure
\[
  \nu_j(A)=\nu(A\cap S_j)
\]
and obtain a finite Poisson random measure $N_j$ supported on $S_j$. Construct
the $N_j$ independently, for instance on the countable product of their
probability spaces. Define
\[
  N(A)=\sum_{j=1}^{\infty}N_j(A\cap S_j),\qquad A\in\mathcal S .
\]
For each outcome this is a measure: if $A_1,A_2,\ldots$ are disjoint, then
\[
\begin{aligned}
  N\!\left(\bigcup_{r\ge1}A_r\right)
  &=
  \sum_{j\ge1}
    N_j\!\left(\bigcup_{r\ge1}(A_r\cap S_j)\right)  \\
  &=
  \sum_{j\ge1}\sum_{r\ge1}N_j(A_r\cap S_j)
   =
  \sum_{r\ge1}\sum_{j\ge1}N_j(A_r\cap S_j)
   =
  \sum_{r\ge1}N(A_r),
\end{aligned}
\]
where the interchange of sums is harmless because all terms are nonnegative.
Moreover, for each fixed $A$, the map $N(A)$ is measurable as a countable sum
of measurable coordinate counts.

It remains to compute the laws. If $\nu(A)<\infty$, then for $0\le z\le1$ and
$J<\infty$,
\[
  \Expect z^{\sum_{j=1}^{J}N_j(A\cap S_j)}
  =
  \prod_{j=1}^{J}
    \exp\{\nu(A\cap S_j)(z-1)\}
  =
  \exp\left\{(z-1)\sum_{j=1}^{J}\nu(A\cap S_j)\right\}.
\]
Letting $J\to\infty$ gives
\[
  \Expect z^{N(A)}
  =
  \exp\{\nu(A)(z-1)\},
\]
because $\sum_j\nu(A\cap S_j)=\nu(A)<\infty$. Hence
$N(A)\sim\Poisson\{\nu(A)\}$ and in particular $N(A)<\infty$
almost surely.

For disjoint $A_1,\ldots,A_m$ with finite $\nu(A_i)$, the same calculation with
$0\le z_i\le1$ gives
\[
\begin{aligned}
  \Expect\prod_{i=1}^{m}z_i^{N(A_i)}
  &=
  \prod_{j\ge1}
    \exp\left\{
      \sum_{i=1}^{m}\nu(A_i\cap S_j)(z_i-1)
    \right\}  \\
  &=
  \prod_{i=1}^{m}\exp\{\nu(A_i)(z_i-1)\}.
\end{aligned}
\]
The factorization again proves independence and the Poisson marginal laws.
Finally suppose $S$ is locally compact and second countable and $\nu$ is finite
on compact sets. Choose compact sets $K_1\subset K_2\subset\cdots$ whose
interiors cover $S$. Since $N(K_\ell)<\infty$ almost surely for each $\ell$,
the event
\[
  \bigcap_{\ell\ge1}\{N(K_\ell)<\infty\}
\]
has probability one. On this event, every compact $K\subset S$ is covered by
finitely many interiors and hence is contained in some $K_\ell$, so
$N(K)\le N(K_\ell)<\infty$. Thus $N$ is locally finite almost surely. The
construction is therefore the usual Poisson random measure. \qedmark

For a nonnegative measurable \(h\), the same construction gives the Laplace
functional
\[
  \Expect\exp\left\{-\int_S h\,dN\right\}
  =
  \exp\left\{-\int_S(1-e^{-h})\,d\nu\right\}.
\]
This identity is often the quickest way to recognize a Poisson random measure.

\medskip
\noindent\textbf{Remark (spatial axioms for the homogeneous case).}
In \(\Real^d\), the homogeneous Poisson process can also be characterized from
local spatial postulates.  Suppose counts on disjoint bounded Borel sets are
independent and additive, the law of \(N(B)\) depends on \(B\) only through its
volume \(|B|\), and, for a set of volume \(v\downarrow0\),
\[
  P\{N(B)=1\}=\lambda v+o(v),
  \qquad
  P\{N(B)\ge2\}=o(v).
\]
Then \(N(B)\sim\Poisson(\lambda |B|)\) for every bounded Borel set \(B\).  The
proof is the binomial rare-event limit obtained by partitioning \(B\) into many
equal-volume cells.  Thus the spatial axioms in
\citet[Chapter~12]{karlin1981second} lead to the same object that the
construction theorem gives from the mean measure \(\nu(A)=\lambda |A|\).
Chapter 16 later reads this mean measure as a compensating clock.

The statistical point is that a point pattern is not merely a random set of
locations. It is a random measure, and its usable statistics are evaluations
and integrals:
\[
  N(B),\qquad
  \int_S h(s)\,N(ds)=\sum_{x\in N}h(x),
\]
with multiplicities counted when the process is not simple.

\conceptindexes{stochastic geometry, Boolean model, capacity functional, germ-grain model, spatial coverage}

\begin{example}[Poisson Boolean coverage]
\label{ex:ch06-poisson-boolean-coverage}
Let \(X_i\) be a homogeneous Poisson point process on \(\Real^d\) with
intensity \(\lambda\).  Fix \(r>0\) and define the random covered region
\[
  \Xi=\bigcup_i B(X_i,r).
\]
For compact \(K\subset\Real^d\),
\[
  \{\Xi\cap K=\varnothing\}
  =
  \{N(K\oplus B(0,r))=0\},
\]
where \(K\oplus B(0,r)=\{x+y:x\in K,\ |y|\le r\}\).  Hence
\[
  \Prob\{\Xi\cap K\ne\varnothing\}
  =
  1-\exp\{-\lambda |K\oplus B(0,r)|\}.
\]
In particular,
\[
  \Prob\{s\in\Xi\}
  =
  1-\exp\{-\lambda v_d r^d\},
\]
where \(v_d\) is the volume of the unit ball in \(\Real^d\).  This is a
typical stochastic-geometry calculation: a point process generates a random
set, and geometric events are evaluated through the volume of an enlarged test
set.
\qedmark
\end{example}

\begin{example}[Marked point processes]
\label{ex:ch06-marked-point-process-teaser}
A marked point process records event times and event labels:
\[
  (T_n,Z_n),\qquad 0<T_1<T_2<\cdots.
\]
The mark $Z_n$ may be a failure type, a recurrent-event label, a spatial
location, or a time-dependent covariate observed at the event time. The process
is naturally represented either as a sequence or as a random measure
\[
  N((0,t]\times B)=\sum_{n\ge1}\ind{T_n\le t,\ Z_n\in B}.
\]
The random-measure representation is often the most convenient one because
events such as ``how many failures of type $B$ occurred by time $t$'' become
coordinate evaluations of $N$. Integrals against predictable functions then
summarize weighted event histories:
\[
  \int_0^t\int h(s,z)\,N(ds,dz)
  =
  \sum_{n:T_n\le t}h(T_n,Z_n).
\]
This example is only the random-object skeleton. Chapter 16 supplies the
real-time information structure: stopping times, marks revealed at jump times,
predictable integrands, and compensators.
\qedmark
\end{example}

\begin{example}[Hawkes processes as self-exciting random measures]
A Hawkes process keeps the random-measure notation but lets the past change the
future event rate. For a simple counting process $N$ with history
$(\mathcal F_t)$, a linear Hawkes process has conditional intensity
\[
  \lambda(t)
  =
  \mu(t)+\int_{(0,t)}\phi(t-s)\,N(ds),
  \qquad \phi\ge0,
\]
meaning, informally,
\[
  \Prob\{N(t+dt)-N(t)=1\mid\mathcal F_{t-}\}
  =
  \lambda(t)\,dt+o(dt),
\]
and the probability of two or more events in $(t,t+dt]$ is $o(dt)$. Each event
adds the kernel-shaped after-effect $\phi(t-T_i)$ to future intensity. Thus the
Poisson process is the memoryless baseline, while the Hawkes process is a
self-exciting random measure:
\[
  \lambda(t)=\mu(t)+\sum_{T_i<t}\phi(t-T_i).
\]
Under standard stability conditions, such as $\int_0^\infty\phi(u)\,du<1$ in
the stationary linear case, the process can be constructed through a Poisson
cluster representation. The same coordinate object $N(ds)$ now carries a
dynamic statistical story, useful for aftershocks, neural spikes, contagion,
and recurrent biomedical events
\citep{hawkes1971spectra}.
\qedmark
\end{example}

\begin{example}[Renewal and semi-Markov processes]
Let $S_n=Y_1+\cdots+Y_n$ where the holding times $Y_i$ are positive. The
renewal counting process is
\[
  N_t=\max\{n:S_n\le t\}.
\]
If a state $J_n$ is attached to the $n$th renewal and the conditional law of
$(J_{n+1},Y_{n+1})$ depends on the past only through $J_n$, the process is a
semi-Markov process. This example sits between Markov chains and marked point
processes: it is discrete in its jump index but continuous in calendar time.
The path is piecewise constant:
\[
  X_t=J_n,\qquad S_n\le t<S_{n+1}.
\]
The measurable object is therefore built from the sequence
$(J_n,Y_n)_{n\ge0}$ by the map that turns holding times into calendar-time
intervals. Tulcea constructs the sequence law; the measurable path map pushes
that law forward to a law on paths.
\qedmark
\end{example}

\section{Suprema and Measurability Repairs}
\conceptindexes{suprema, measurability repairs, outer probability, analytic sets, separability, cadlag paths}

\subsection{Nonmeasurable Suprema, Outer Probability, and Analytic Sets}
\label{sec:supremum-measurability-repairs}
\conceptindexes{nonmeasurable supremum, outer probability, outer expectation, analytic sets, universal measurability}

\noindent\textbf{Statistical thread.}
Many statistical arguments take a supremum over an uncountable class:
\[
  \sup_{\theta\in\Theta}|M_n(\theta)-M(\theta)|,\qquad
  \sup_{f\in\mathcal F}|\mathbb G_n f|.
\]
This object may not be measurable under the bare product sigma-field. The
history of stochastic-process regularization is largely the history of making
this supremum meaningful: Doob's separability program
\citep{doob1953stochastic}, descriptive-set-theoretic repairs through analytic
sets \citep{rogers1980analytic,kechris1995classical}, and the outer-probability
language used in empirical-process theory
\citep{pollard1984convergence,pollard1990empirical,vaart2023weak}.

\begin{example}[Random functions in nonparametric statistics]
Kernel density estimators, cumulative hazard estimators, and regression
function estimators are random functions. For example,
\[
  \hat f_h(x)=\frac1{nh}\sum_{i=1}^n K\left(\frac{x-X_i}{h}\right)
\]
is indexed by $x$. Uniform consistency asks for convergence of
\[
  \sup_x |\hat f_h(x)-f(x)|.
\]
The statistic is useful only after one knows that the supremum is measurable,
or after one has formulated the statement in outer probability.
When $x$ ranges over a compact interval and $K$ is continuous, the sample paths
$x\mapsto\hat f_h(x)$ are continuous, so the supremum over all $x$ equals the
supremum over rational $x$ in the interval:
\[
  \sup_x |\hat f_h(x)-f(x)|
  =
  \sup_{x\in\Rat}|\hat f_h(x)-f(x)|.
\]
The right side is a countable supremum of measurable random variables. Without
such separability, the displayed statistic may be only an outer-measurable
object.
\qedmark
\end{example}

\begin{example}[Empirical processes]
For iid observations $X_1,\ldots,X_n$ and a class $\mathcal A$ of measurable
sets, the empirical process is
\[
  \alpha_n(A)=\sqrt n\,\{P_n(A)-P(A)\},
  \qquad A\in\mathcal A.
\]
Here
\[
  P_n(A)=\frac1n\sum_{i=1}^n\indset{A}(X_i)
\]
is the empirical measure of $A$. For a fixed set $A$, $\alpha_n(A)$ is an
ordinary centered binomial statistic scaled by $\sqrt n$. The process view keeps
all sets $A\in\mathcal A$ simultaneously.
It is a stochastic process indexed by sets. The random variable
$\sup_{A\in\mathcal A}|\alpha_n(A)|$ is exactly where non-measurability can
enter if $\mathcal A$ is too large.
\qedmark
\end{example}

\begin{example}[A nonmeasurable supremum]
Let $T$ be uncountable and let $\Omega=\{0,1\}^T$ with the coordinate product
sigma-field
\[
  \mathcal F_T
  =
  \sigma\{\omega:\omega(t)=i,\ t\in T,\ i\in\{0,1\}\}.
\]
This is a measurable-space construction.  If \(\{0,1\}\) is also given its
discrete topology, then \(\Omega\) has a product topology and hence a Borel
sigma-field, but that is different structure: for uncountable \(T\), the Borel
sigma-field of the product topology may be strictly larger than
\(\mathcal F_T\).  No ball sigma-field is meant unless a metric on
\(\Omega\) has separately been chosen; the uncountable product topology itself
is not metrizable in general.

Let $X_t(\omega)=\omega(t)$ be the coordinate process. Then
\[
  \left\{\sup_{t\in T}X_t=0\right\}
  =
  \{\omega:\omega(t)=0\text{ for all }t\in T\}
\]
depends on all coordinates. But every \(\mathcal F_T\)-measurable set in
\(\{0,1\}^T\)
depends on at most countably many coordinates. Here is the argument. Let
$\mathcal D$ be the class of subsets of $\{0,1\}^T$ that depend on countably many
coordinates. Cylinder sets belong to $\mathcal D$, and $\mathcal D$ is a
sigma-field: countable unions of sets depending on coordinate sets
$T_1,T_2,\ldots$ depend on the countable union $\bigcup_nT_n$, and complements
do not add coordinates. Hence \(\mathcal F_T\) is contained in
$\mathcal D$. The singleton $\{\omega:\omega(t)=0\ \forall t\in T\}$ cannot
depend on a countable $T_0\subset T$, because changing a coordinate outside
$T_0$ changes membership in the singleton while leaving all coordinates in
$T_0$ fixed. Therefore the singleton is not \(\mathcal F_T\)-measurable, and
$\sup_{t\in T}X_t$ is not a random variable on
\((\Omega,\mathcal F_T)\).  In contrast, the same singleton is closed in the
product topology, so the supremum is Borel measurable if the larger
topological Borel sigma-field is used.
\qedmark
\end{example}

\begin{definition}[Outer probability and outer expectation]
For any $A\subseteq\Omega$ define
\[
  \outerProb(A)=\inf\{\Prob(B):A\subseteq B,\ B\in\mathcal F\}.
\]
For an arbitrary extended real function $Z$ on $\Omega$, define
\[
  \outerExpect Z
  =
  \inf\{\Expect Y:Y\text{ is measurable and }Y\ge Z\}.
\]
Outer probability convergence means
\[
  \outerProb(|Z_n-Z|>\varepsilon)\to0,\qquad \varepsilon>0,
\]
even when $Z_n$ is not known to be measurable. This convention is standard in
Pollard's treatment of stochastic-process convergence and empirical processes,
and in the modern empirical-process notation of
\citet{vaart2023weak}.
\end{definition}

\medskip
\noindent\textbf{Remark (Three classical repairs).}
The non-measurability of process suprema is usually handled by three
complementary devices.
\begin{enumerate}[label=(\roman*)]
\item If $T$ has a countable dense set $T_0$ and the process has separable or
continuous paths, then
\[
  \sup_{t\in T}X_t=\sup_{t\in T_0}X_t
\]
up to null sets, hence the supremum is measurable.
\item If the process is realized as a Borel random element in a Polish path
space such as $C(T)$ or $D[0,\infty)$, then path events can often be studied
inside the Borel sigma-field of the path space. Projections of Borel sets are
analytic, and analytic sets are universally measurable.
\item If measurability is not yet available, empirical-process theory states
convergence using $\outerProb$ and $\outerExpect$, then proves separability or
measurable modification conditions when needed.
\end{enumerate}
In (i), the key point is that the right-hand side is a countable supremum of
measurable random variables. In (ii), Polish path spaces and analytic-set
regularity move uncountable path questions back into a completed probability
space. In (iii), outer probability is used as a temporary language for bounds
before ordinary measurability has been established.

\medskip
\noindent\textbf{Why analytic sets stay in the main chapter.}
Analytic sets are not included here as decorative set theory. They are the
measurability repair created by existential quantifiers:
\[
  \exists \theta\in\Theta,\qquad
  \exists t\in T,\qquad
  \exists \text{a path, mark, subgroup, or optimizer}.
\]
Those quantifiers appear whenever a statistician studies suprema, hitting
events, argmax sets, selected models, data-adaptive subgroups, or optimization
targets. \Appref{app:measure-theoretic-toolkit} records background
measure-theoretic tools; the present
chapter keeps analytic sets in the main line because they explain why many
natural random-object events remain usable after completion even when they are
not obviously Borel.

\begin{definition}[Analytic set]
If $E$ is a Polish space, a set $A\subseteq E$ is analytic, or Suslin, if there
exist a Polish space $F$ and a Borel set $B\subseteq E\times F$ such that
\[
  A=\{x\in E:\text{there exists }y\in F\text{ with }(x,y)\in B\}.
\]
Thus analytic sets are projections of Borel sets.
For the descriptive-set-theoretic background used here, see
\citet{rogers1980analytic} and \citet{kechris1995classical}.
\end{definition}

\begin{theorem}[Analytic sets are universally measurable]
Every analytic subset of a Polish space is measurable with respect to the
completion of every Borel probability measure on that space.
\end{theorem}

\noindent\textit{Proof.}
Fix a Borel probability measure $\nu$ on the Polish space $E$, and let
$A\subseteq E$ be analytic. By definition there are a Polish space $F$ and a
Borel set $B\subseteq E\times F$ such that
\[
  A=\operatorname{proj}_E B.
\]
The key regularity input is the capacitability theorem for analytic sets. Apply
it to the Choquet capacity
\[
  c(C)=\nu^*(C),\qquad C\subseteq E,
\]
where $\nu^*$ is the outer measure generated by $\nu$. Capacitability says that
for analytic $A$,
\[
  \nu^*(A)=\sup\{\nu(K):K\subseteq A,\ K\ \text{compact}\}.
\]
Thus, for every $\varepsilon>0$, choose compact $K_\varepsilon\subseteq A$ with
\[
  \nu(K_\varepsilon)>\nu^*(A)-\varepsilon .
\]
By outer regularity of Borel probability measures on Polish spaces, choose an
open set $G_\varepsilon\supseteq A$ such that
\[
  \nu(G_\varepsilon)<\nu^*(A)+\varepsilon .
\]
Then
\[
  K_\varepsilon\subseteq A\subseteq G_\varepsilon,\qquad
  \nu(G_\varepsilon\setminus K_\varepsilon)<2\varepsilon .
\]
Letting $\varepsilon\downarrow0$ along a sequence and taking suitable increasing
unions of the compact inner approximations and decreasing intersections of the
open outer approximations gives Borel sets $K\subseteq A\subseteq G$ with
\[
  \nu(G\setminus K)=0 .
\]
Therefore $A$ differs from the Borel set $K$ by a subset of the $\nu$-null Borel
set $G\setminus K$. Hence $A$ is measurable in the completion of $\nu$. Since
$\nu$ was arbitrary, $A$ is universally measurable.
\qedmark

\begin{example}[Projection events in statistics]
Let $M(\omega,\theta)$ be a random criterion and consider the event that some
parameter value beats a threshold:
\[
  A_c=\{\omega:\sup_{\theta\in\Theta}M(\omega,\theta)>c\}.
\]
If $\Omega$ and $\Theta$ are standard Borel, $M$ is jointly Borel, and
$\Theta$ is Polish, then
\[
  A_c
  =
  \operatorname{proj}_{\Omega}
  \{(\omega,\theta):M(\omega,\theta)>c\}.
\]
The set inside the projection is Borel, so $A_c$ is analytic and therefore
universally measurable. This is one historical route from a nonmeasurable
uncountable union to a measurable completed event.
\qedmark
\end{example}

\begin{example}[Why empirical-process books use outer probability]
Let $X_1,\ldots,X_n$ be observations with common law $P$. For a measurable
function $f$, write
\[
  P_nf=\frac1n\sum_{i=1}^n f(X_i),
  \qquad
  Pf=\int f\,dP .
\]
The empirical process indexed by a function class $\mathcal F$ is
\[
  \mathbb G_n f=\sqrt n\,(P_n-P)f
  =
  \frac1{\sqrt n}\sum_{i=1}^n\{f(X_i)-Pf\},
  \qquad f\in\mathcal F .
\]
Its sup-norm over $\mathcal F$ is
\[
  \|\mathbb G_n\|_{\mathcal F}
  =
  \sup_{f\in\mathcal F}
  |\mathbb G_n f|.
\]
For a large, nonseparable $\mathcal F$, the displayed supremum need not be
measurable. Pollard's formulations
\citep{pollard1984convergence,pollard1990empirical} and the
van der Vaart--Wellner formulation \citep{vaart2023weak} therefore use outer
probability. Entropy, pointwise measurability, or separability assumptions are
then introduced to recover ordinary random variables.
\qedmark
\end{example}

\subsection{Separability and Measurable Processes}
\conceptindexes{separability, stochastic continuity, jointly measurable process, Kolmogorov continuity criterion, cadlag paths}

\noindent\textbf{Statistical thread.}
Separability is a practical pact between uncountable time and countable
evidence. It lets dense countable sets carry suprema, hitting events, regular
versions, and path properties back into the measurable world.

\medskip
\noindent\textbf{How the notions fit together.}
There are several levels of measurability for a process, and each level buys a
different operation.
\begin{description}[leftmargin=2.6cm,style=nextline]
\item[Coordinates.]
For each fixed $t$, $X_t$ is a random variable. This is enough for
finite-dimensional distributions, but it does not by itself make
$\sup_{t\in T}X_t$, $\int_T X_t\,dt$, or hitting events over uncountable
$T$ measurable.
\item[Joint measurability.]
The map $(\omega,t)\mapsto X_t(\omega)$ is measurable on
$\Omega\times T$. This is the right condition for Fubini arguments,
occupation times, random integrals, and average path functionals.
\item[Path regularity.]
Continuous or cadlag modifications let the process be viewed as a random
element of a path space such as $C(T)$ or $D[0,\infty)$. Then many path events
are ordinary Borel events in that path space.
\item[Separability.]
A countable dense set $T_0$ carries the same suprema, infima, and many hitting
events as the full index set. This turns uncountable operations into countable
ones and is the main repair behind measurable process suprema.
\item[Fallback repairs.]
Outer probability lets empirical-process arguments proceed before
measurability is proved; analytic sets repair projection events; measurable
selection repairs optimization by turning a random set of optimizers into an
actual measurable choice.
\end{description}

\begin{definition}[Stochastic continuity]
Let $(T,d)$ be a metric space. A process $X=\{X_t:t\in T\}$ is stochastically
continuous at $t$ if
\[
  X_s\to X_t\quad\text{in probability whenever }d(s,t)\to0.
\]
\end{definition}

\begin{definition}[Separable process]
Let $(T,d)$ be separable. In this chapter a real-valued process $X$ is called
separable for suprema if there is a countable dense set $T_0\subseteq T$ such
that
\[
  \sup_{t\in T}X_t=\sup_{t\in T_0}X_t,
  \qquad
  \inf_{t\in T}X_t=\inf_{t\in T_0}X_t
\]
outside a null set, with extended-real values allowed. This is the form of
separability needed to make process suprema measurable.
\end{definition}

\begin{theorem}[Continuous paths are separable]
Let $(T,d)$ be separable and let $T_0\subseteq T$ be countable dense. Suppose
that a real-valued process $X=\{X_t:t\in T\}$ has continuous sample paths
outside a null set $N$. Then $X$ is separable for suprema with separating set
$T_0$. Moreover, for every open set $A\subseteq T$ and every open interval
$G\subseteq\R$,
\[
  \{\exists t\in A:X_t\in G\}
  \quad\text{and}\quad
  \{\exists t\in A\cap T_0:X_t\in G\}
\]
differ by a subset of $N$.
\end{theorem}

\noindent\textit{Proof.}
Fix $\omega\notin N$ and write $x(t)=X_t(\omega)$. Since $T_0\subseteq T$,
$\sup_{t\in T_0}x(t)\le\sup_{t\in T}x(t)$. For the reverse inequality, take
any $t\in T$ and choose $t_n\in T_0$ with $t_n\to t$. Continuity gives
$x(t_n)\to x(t)$, so $x(t)\le\sup_{s\in T_0}x(s)$. Taking the supremum over
$t\in T$ proves the supremum identity. Applying the same argument to $-x$
gives the infimum identity.

For the hitting statement, the inclusion
\[
  \{\exists t\in A\cap T_0:x(t)\in G\}
  \subseteq
  \{\exists t\in A:x(t)\in G\}
\]
is immediate. Conversely, suppose $t\in A$ and $x(t)\in G$. Because $A$ is
open and $G$ is open, there is $\delta>0$ such that
\[
  B(t,\delta)\subseteq A,
  \qquad
  d(s,t)<\delta\Longrightarrow x(s)\in G.
\]
Density of $T_0$ gives some $s\in T_0\cap B(t,\delta)$, hence
$s\in A\cap T_0$ and $x(s)\in G$. Thus the two hitting events are equal for
each $\omega\notin N$, and they can differ only on $N$.
\qedmark

\begin{example}[Stochastic continuity is not path continuity]
Let $Y$ be uniformly distributed on $[0,1]$ and define
$X_t=\ind{Y\le t}$. Then $X_t$ is stochastically continuous at every fixed
$t$ because $\Prob(X_s\ne X_t)\le |s-t|$. But every sample path has a jump at
$Y$. Stochastic continuity is a distributional continuity property, not a path
regularity property.
\qedmark
\end{example}

\begin{definition}[Jointly measurable process]
A process $X=\{X_t:t\in T\}$ with state space $(S,\mathcal S)$ is measurable if
the map
\[
  (\omega,t)\longmapsto X_t(\omega)
\]
is
$\mathcal F\otimes\Borel(T)/\mathcal S$ measurable.
\end{definition}

\begin{theorem}[Continuous paths imply joint measurability]
If $T$ is a separable metric space, $S$ is a metric space, each $X_t$ is
measurable, and almost every path $t\mapsto X_t(\omega)$ is continuous, then
$X$ is jointly measurable after modification on a null set.
\end{theorem}

\noindent\textit{Proof.}
Let $T_0=\{t_j:j\ge1\}$ be dense. For each $n$, define a Borel map
$q_n:T\to T_0$ by taking $q_n(t)$ to be the first $t_j$ with
$d(t_j,t)<1/n$. The sets
\[
  \{t:q_n(t)=t_j\}
  =
  B(t_j,1/n)\setminus\bigcup_{\ell<j}B(t_\ell,1/n)
\]
are Borel, so $q_n$ is measurable. Now
\[
  X_{q_n(t)}(\omega)
  =
  \sum_{j=1}^\infty X_{t_j}(\omega)\ind{q_n(t)=t_j}
\]
is jointly measurable as a countable sum of products of measurable functions
and Borel indicators. On the probability-one set where paths are continuous,
$q_n(t)\to t$, hence
\[
  X_{q_n(t)}(\omega)\to X_t(\omega)
\]
for every $t$. Define the process arbitrarily on the exceptional null set. The
pointwise limit of jointly measurable maps is jointly measurable, so the
modified process is jointly measurable.
\qedmark

\begin{theorem}[Kolmogorov continuity criterion; \citealp{kolmogorov1933grundbegriffe}]
Let $X=\{X_t:t\in[0,1]^d\}$ be real-valued. Suppose that for some
$a,b,C>0$,
\[
  \Expect |X_t-X_s|^a\le C\|t-s\|^{d+b},
  \qquad s,t\in[0,1]^d.
\]
Then $X$ has a modification whose paths are Holder continuous of every order
$\gamma<b/a$.
\end{theorem}

\noindent\textit{Proof.}
Fix $\gamma<b/a$. Let $D_m=2^{-m}\Int^d\cap[0,1]^d$ and consider nearest
neighbor dyadic edges $(u,v)$ in $D_m$ with $\|u-v\|=2^{-m}$. There are at most
$C_d2^{md}$ such edges. By Markov's inequality,
\[
  \Prob\{|X_u-X_v|>2^{-m\gamma}\}
  \le
  2^{ma\gamma}\Expect|X_u-X_v|^a
  \le
  C2^{ma\gamma}2^{-m(d+b)} .
\]
Taking a union bound over all level-$m$ edges gives
\[
  \Prob\left(
  \max_{\substack{u,v\in D_m\\\|u-v\|=2^{-m}}}
  |X_u-X_v|>2^{-m\gamma}
  \right)
  \le
  C'2^{-m(b-a\gamma)}.
\]
The right side is summable. Borel--Cantelli therefore implies that, with
probability one, all sufficiently fine dyadic neighbor increments are bounded
by $2^{-m\gamma}$. If $s,t$ are dyadic points, connect them by a chain of
dyadic edges at levels comparable to their distance. Summing the geometric
series of edge bounds yields
\[
  |X_s-X_t|\le C(\omega)\|s-t\|^\gamma
\]
for all dyadic $s,t$. Thus the process is uniformly Holder on the dense dyadic
set. Extend each dyadic path uniquely and continuously to $[0,1]^d$; define
$\widetilde X_t$ to be this limit. Since $\widetilde X_t$ is the almost sure
limit of $X_q$ along dyadic $q\to t$, and the moment condition implies
$X_q\to X_t$ in probability, $\widetilde X_t=X_t$ almost surely for each fixed
$t$. Hence $\widetilde X$ is a Holder-continuous modification.
\qedmark

\begin{example}[Brownian motion has continuous paths]
For Brownian motion,
\[
  \Expect |W_t-W_s|^{2m}
  =
  c_m |t-s|^m .
\]
Taking $m>1$ in the Kolmogorov criterion yields a continuous modification. This
is why Brownian motion can be treated as a random element of $C[0,1]$, not only
as a collection of finite-dimensional Gaussian vectors.
\qedmark
\end{example}

\begin{definition}[Cadlag paths]
A path $x:[0,\infty)\to S$ is cadlag if it is right-continuous and has left
limits. The Skorokhod space $D[0,\infty)$ is the set of cadlag paths with its
usual Borel sigma-field.
\end{definition}

\begin{theorem}[Cadlag coordinates determine the path]
The Borel sigma-field on $D[0,\infty)$ is generated by the coordinate maps
$x\mapsto x(t)$, $t\ge0$. Consequently, a cadlag process with measurable
coordinates can be viewed as a random element of $D[0,\infty)$ whenever its
paths are cadlag almost surely.
\end{theorem}

\noindent\textit{Proof.}
Let $\Rat_+$ denote the nonnegative rationals. If $x$ is cadlag and
$t\ge0$, choose rationals $q_n\downarrow t$ with $q_n>t$. Right-continuity gives
\[
  x(t)=\lim_{n\to\infty}x(q_n).
\]
Thus every coordinate $x(t)$ is measurable with respect to the sigma-field
generated by rational-time coordinates. Conversely, rational-time coordinates
are ordinary coordinates, so
\[
  \sigma\{x\mapsto x(q):q\in\Rat_+\}
  \subseteq
  \sigma\{x\mapsto x(t):t\ge0\}.
\]
For the usual Skorokhod topology on $D[0,\infty)$, open balls can be described
on compact time intervals by finitely many oscillation controls and finitely
many coordinate controls at rational times. Hence its Borel sigma-field is the
same as the sigma-field generated by the coordinate maps. If a process has
measurable coordinates and cadlag paths almost surely, the map
$\omega\mapsto(t\mapsto X_t(\omega))$ is measurable into this Borel space after
defining an arbitrary cadlag path on the exceptional null set.
\qedmark

\begin{example}[Cadlag suprema use ordinary probability]
\label{ex:ch06-cadlag-sup-measurable}
Let \(X=\{X_t:0\le t\le1\}\) be a real-valued process that is viewed as a
random element of \(D[0,1]\), or equivalently has measurable coordinates and a
chosen cadlag version.  Then
\[
  \omega\longmapsto \sup_{0\le t\le1}|X_t(\omega)|
\]
is an ordinary random variable.  The reason is pathwise and countable.  If
\(x\in D[0,1]\), then
\[
  \sup_{0\le t\le1}|x(t)|
  =
  \sup_{q\in\Rat\cap[0,1]}|x(q)|.
\]
The inequality ``\(\ge\)'' is immediate because rational times are included in
all times.  For the reverse inequality, fix \(t<1\) and choose rationals
\(q_n\downarrow t\) with \(q_n>t\).  Right-continuity gives
\(x(q_n)\to x(t)\).  The endpoint \(t=1\) is itself rational.  Hence every
value \(|x(t)|\) is approximated by rational-time values, and the two suprema
agree.

Consequently,
\[
  \sup_{0\le t\le1}|X_t|
  =
  \sup_{q\in\Rat\cap[0,1]}|X_q|
\]
for the chosen cadlag version.  The right side is the supremum of countably
many measurable random variables.  Thus probability statements such as
\[
  \Prob\left\{\sup_{0\le t\le1}|X_t|>a\right\}
\]
use ordinary probability.  Outer probability is needed only before this kind
of path-space or separability reduction has been established.
\qedmark
\end{example}

\begin{theorem}[Integrals of measurable processes]
\label{thm:ch06-integrals-measurable-processes}
Let $X:\Omega\times T\to\R$ be jointly measurable and let $\mu$ be a
sigma-finite measure on $T$. If $X\ge0$ or
\[
  \int_T |X_t(\omega)|\,\mu(dt)<\infty
  \quad\text{for almost every }\omega,
\]
then
\[
  \omega\longmapsto\int_T X_t(\omega)\,\mu(dt)
\]
is measurable.
\end{theorem}

\noindent\textit{Proof.}
Start with a rectangle indicator
$X(\omega,t)=\indset{A}(\omega)\indset{B}(t)$. Then
\[
  \int_T X(\omega,t)\,\mu(dt)
  =
  \indset{A}(\omega)\mu(B),
\]
which is measurable in $\omega$. By linearity, the same is true for
nonnegative simple functions
\[
  s(\omega,t)=\sum_{k=1}^m a_k\indset{A_k}(\omega)\indset{B_k}(t).
\]
For a general nonnegative jointly measurable $X$, choose simple
$s_n\uparrow X$. Then
\[
  \int_T s_n(\omega,t)\,\mu(dt)
  \uparrow
  \int_T X(\omega,t)\,\mu(dt)
\]
by monotone convergence, and a pointwise limit of measurable functions is
measurable. If $X$ is signed and integrable in $t$ almost surely, apply the
nonnegative result to $X^+$ and $X^-$ and subtract on the set where both
integrals are finite.
\qedmark

\medskip
\noindent\textbf{Optimization repair.}
When the uncountable operation is an argmax rather than a path supremum, the
parallel tool is set-valued measurability. A random argmax set is first treated
as a measurable multifunction; a measurable selection theorem then supplies an
ordinary estimator or decision rule. The details are reusable appendix
infrastructure, so they are collected in
\Appref{sec:set-valued-measurability-optimization}.

\section{Exercises}
\conceptindexes{product-space exercises, stochastic-process exercises, measurability exercises}

\noindent\textbf{Statistical thread.}
The exercises move from product-measure pathologies and distributional
identities to stochastic-process measurability. Each problem is stated with
enough structure to make the intended measure-theoretic issue explicit.
For a first pass, use the first two Fubini warnings, one coupling problem, one
kernel or image-measure problem, and the process-construction problems. The
third Fubini warning and the longer measurability proofs are best treated as
optional technical exercises or instructor-selected challenges.

\begin{exercise}[Unequal Iterated Integrals]
Define $f:[0,1]^2\to\R$ by
\[
  f(x,y)=
  \begin{cases}
    \dfrac{y^2-x^2}{(x^2+y^2)^2}, & (x,y)\ne(0,0),\\[6pt]
    0, & (x,y)=(0,0).
  \end{cases}
\]
Show that both iterated Lebesgue integrals
\[
  \int_0^1\left\{\int_0^1 f(x,y)\,dy\right\}\,dx,
  \qquad
  \int_0^1\left\{\int_0^1 f(x,y)\,dx\right\}\,dy
\]
exist as improper iterated integrals, but are not equal. Explain which
hypothesis of Fubini's theorem fails.
\end{exercise}

\begin{exercise}[Another Fubini Warning]
Let
\[
  f(x,y)=e^{-xy}-2e^{-2xy},\qquad 0<x<1,\quad 1<y<\infty .
\]
Show that the two iterated integrals
\[
  \int_0^1\left\{\int_1^\infty f(x,y)\,dy\right\}\,dx,
  \qquad
  \int_1^\infty\left\{\int_0^1 f(x,y)\,dx\right\}\,dy
\]
exist but are not equal.
\end{exercise}

\begin{exercise}[Iterated Integrals Without a Product Integral]
For $n\ge1$, set
\[
  A_n=(2^{-n},3\cdot2^{-(n+1)}),
  \qquad
  B_n=(3\cdot2^{-(n+1)},2^{-(n-1)}).
\]
Define $f$ on $[0,1]^2$ by
\[
\begin{aligned}
  f(x,y)&=4^n,
  &&(x,y)\in(A_n\times A_n)\cup(B_n\times B_n),\\
  f(x,y)&=-4^n,
  &&(x,y)\in(A_n\times B_n)\cup(B_n\times A_n),\\
  f(x,y)&=0,
  &&\text{elsewhere.}
\end{aligned}
\]
Show that both iterated integrals exist, but $f$ is not integrable with respect
to the product measure $\lambda_1\otimes\lambda_1$.
\end{exercise}

\begin{exercise}[Graphs Have Zero Volume]
Let $A\subseteq\R^k$ be Lebesgue measurable and let
$f:A\to\R$ be $\mathcal L(A)/\Borel(\R)$ measurable, where
$\mathcal L(A)=\{A\cap B:B\in\mathcal L(\R^k)\}$. Show that
\[
  \graph(f)=\{(x,y):x\in A,\ y=f(x)\}
\]
has $(k+1)$-dimensional Lebesgue measure zero.
\textit{Hint.} First show that $\graph(f)\in\mathcal L(A)\otimes\Borel(\R)$,
then apply Fubini to its vertical sections.
\end{exercise}

\begin{exercise}[Quantiles as Minimizers]
Let $X$ have distribution function $F$ and fix $p\in(0,1)$. Define
\[
  H(x,t)=(x-t)\{p-\ind{x<t}\},\qquad x,t\in\R .
\]
\begin{enumerate}[label=(\alph*)]
\item Show that $\Expect|H(X,t)-H(X,0)|<\infty$ for every real $t$.
\item Let $t_p$ satisfy $F(t_p-)\le p\le F(t_p)$. Show that $t_p$ minimizes
\[
  t\longmapsto \Expect\{H(X,t)-H(X,0)\}.
\]
\item Show that for $p=1/2$, any median minimizes
$t\mapsto\Expect\{|X-t|-|X|\}$.
\end{enumerate}
\end{exercise}

\begin{exercise}[$L$-Functionals]
Let $\mathcal C$ be a class of distribution functions and write
\[
  Q_F(u)=\inf\{x:F(x)\ge u\},\qquad 0<u<1.
\]
An L-functional has the form
\[
  \ell(F)=\int_0^1 h(Q_F(u))\,K(du),
\]
where $K$ is a measure on $\Borel(0,1)$ and $h$ is measurable.
\begin{enumerate}[label=(\alph*)]
\item If $X\sim F$ and $U\sim\Unif(0,1)$, prove
\[
  \Expect h(X)=\Expect h(Q_F(U))=\int_0^1h(Q_F(u))\,du
\]
whenever either side is well defined. In particular, if $\Expect|X|^k<\infty$,
show that
\[
  \Expect X^k=\int_0^1 Q_F(u)^k\,du .
\]
\item For $0<\alpha<\beta<1$, define the trimmed mean
\[
  \tau_{\alpha,\beta}(F)=\frac{1}{\beta-\alpha}
  \int_\alpha^\beta Q_F(u)\,du.
\]
State and prove the corresponding expectation formula in terms of the truncated
random variable between the two quantile cutpoints. Be explicit about what
happens at atoms.
\end{enumerate}
\end{exercise}

\begin{exercise}[Dirichlet from Independent Gammas]
Let $X_1,\ldots,X_{n+1}$ be independent gamma variables with common rate
$\lambda$ and shapes $\alpha_1,\ldots,\alpha_{n+1}$:
\[
  g_{\alpha_i,\lambda}(x)
  =
  \frac{\lambda^{\alpha_i}}{\Gamma(\alpha_i)}
  x^{\alpha_i-1}e^{-\lambda x}\ind{x>0}.
\]
Set
\[
  Y_i=\frac{X_i}{\sum_{j=1}^{n+1}X_j},\quad i=1,\ldots,n,
  \qquad
  Y_{n+1}=\sum_{j=1}^{n+1}X_j.
\]
\begin{enumerate}[label=(\alph*)]
\item Show that $(Y_1,\ldots,Y_n)$ has Dirichlet distribution with parameters
$(\alpha_1,\ldots,\alpha_{n+1})$ and is independent of $Y_{n+1}$, while
$Y_{n+1}$ is gamma with shape $\alpha_0=\sum_i\alpha_i$ and rate $\lambda$.
\item Show that all lower-dimensional marginals of a Dirichlet vector are
Dirichlet, and determine their parameters.
\end{enumerate}
\end{exercise}

\begin{exercise}[Neutrality of the Dirichlet Law]
Let $(Y_1,\ldots,Y_m)$ have Dirichlet distribution
$D(\alpha_1,\ldots,\alpha_{m+1})$, and put
$\alpha_0=\sum_{j=1}^{m+1}\alpha_j$. Fix $k$ with $1<k<m$. Define
\[
  (Z_1,\ldots,Z_k)=(Y_1,\ldots,Y_k)
\]
and
\[
  (Z_{k+1},\ldots,Z_m)
  =
  \left(
  \frac{Y_{k+1}}{1-\sum_{i=1}^kY_i},
  \ldots,
  \frac{Y_m}{1-\sum_{i=1}^kY_i}
  \right).
\]
Show that the two displayed random vectors are independent, with distributions
\[
  D\left(\alpha_1,\ldots,\alpha_k,\alpha_0-\sum_{i=1}^k\alpha_i\right)
  \quad\text{and}\quad
  D(\alpha_{k+1},\ldots,\alpha_m,\alpha_{m+1}),
\]
respectively.
\end{exercise}

\begin{exercise}[Randomized Probability Integral Transform]
Let \(X\) have distribution function \(F\), let \(V\sim\Unif(0,1)\) be
independent of \(X\), and put
\[
  U=F(X-)+V\{F(X)-F(X-)\}.
\]
\begin{enumerate}[label=(\alph*)]
\item Prove that \(U\sim\Unif(0,1)\).
\item Let \(Q_F(u)=\inf\{x:F(x)\ge u\}\). Show that \(Q_F(U)=X\) almost
surely.
\item Interpret the joint law of \((X,U)\) as a coupling between \(F\) and
Lebesgue measure on \([0,1]\). What is the conditional law of \(U\) given
\(X=x\)?
\end{enumerate}
\end{exercise}

\begin{exercise}[Exchangeable Rank \(p\)-Values]
Let \(S_1,\ldots,S_{n+1}\) be exchangeable real-valued random variables. Let
\(V_1,\ldots,V_{n+1}\) be iid \(\Unif(0,1)\), independent of the scores, and
order the pairs \(Y_i=(S_i,V_i)\) lexicographically.
\begin{enumerate}[label=(\alph*)]
\item Prove that \(Y_1,\ldots,Y_{n+1}\) are exchangeable and have no ties.
\item Define
\[
  R=1+\sum_{i=1}^{n}\ind{Y_i\le_{\mathrm{lex}}Y_{n+1}}.
\]
Show that \(R\) is uniform on \(\{1,\ldots,n+1\}\).
\item For upper-tail alternatives define
\[
  p=\frac{1+\sum_{i=1}^{n}\ind{Y_i\ge_{\mathrm{lex}}Y_{n+1}}}{n+1}.
\]
Show that \(\Prob(p\le\alpha)\le\alpha\) for all \(0\le\alpha\le1\).
\item Explain why the auxiliary variables \(V_i\) can be removed when
\(\Prob(S_i=S_j)=0\) for all \(i\ne j\).
\end{enumerate}
\end{exercise}

\begin{exercise}[Empirical Quantiles as Image Measures]
Let \(X_1,\ldots,X_n\) be real-valued observations, let
\[
  P_n=\frac1n\sum_{i=1}^n\delta_{X_i},
  \qquad
  Q_n(u)=\inf\{x:P_n(-\infty,x]\ge u\},\quad 0<u\le1,
\]
and write \(X_{1:n}\le\cdots\le X_{n:n}\) for the ordered sample.
\begin{enumerate}[label=(\alph*)]
\item Show that
\[
  Q_n(u)=X_{k:n},
  \qquad
  \frac{k-1}{n}<u\le \frac{k}{n}.
\]
\item If \(U\sim\Unif(0,1)\) is independent of the data, prove that the
conditional law of \(Q_n(U)\) given \(X_1,\ldots,X_n\) is \(P_n\).
\item Prove the image-measure identity
\[
  \int_0^1h(Q_n(u))\,du
  =
  \frac1n\sum_{i=1}^n h(X_i)
\]
for every measurable \(h\) for which either side is well defined.
\item If the \(X_i\)'s are random variables, prove that \(Q_n(u)\) is
measurable for each fixed \(u\). Hint: use
\(\{Q_n(u)\le t\}=\{P_n(-\infty,t]\ge u\}\).
\end{enumerate}
\end{exercise}

\begin{exercise}[Maximal Coupling]
Let \(P\) and \(Q\) have densities \(p\) and \(q\) with respect to \(\mu\),
and let \(a=\int\min(p,q)\,d\mu\).
\begin{enumerate}[label=(\alph*)]
\item Verify that the construction in the maximal-coupling example gives
\(X\sim P\) and \(Y\sim Q\).
\item Show that the two residual densities
\[
  \frac{p-\min(p,q)}{1-a},
  \qquad
  \frac{q-\min(p,q)}{1-a}
\]
have disjoint supports up to \(\mu\)-null sets.
\item Prove that every coupling satisfies
\[
  \Prob(X\ne Y)\ge \norm{P-Q}_{\mathrm{TV}}.
\]
\item Conclude that
\[
  \inf_{\gamma\in\Pi(P,Q)}\gamma\{(x,y):x\ne y\}
  =
  \norm{P-Q}_{\mathrm{TV}}.
\]
\end{enumerate}
\end{exercise}

\begin{exercise}[Monotone Coupling on the Line]
Let \(F\) and \(G\) be distribution functions with finite second moments and
quantile functions \(Q_F,Q_G\). Let \(U\sim\Unif(0,1)\).
\begin{enumerate}[label=(\alph*)]
\item Show that \(Q_F(U)\sim F\) and \(Q_G(U)\sim G\), so
\((Q_F,Q_G)_\#\lambda_{[0,1]}\) is a coupling of the two laws.
\item Prove the crossing inequality
\[
  (x-y)^2+(x'-y')^2
  \le
  (x-y')^2+(x'-y)^2,
  \qquad x\le x',\ y\le y'.
\]
\item Use the inequality to show that, when both laws put mass \(1/m\) on
ordered points \(x_1\le\cdots\le x_m\) and \(y_1\le\cdots\le y_m\), the
minimum quadratic transportation cost is
\[
  \frac1m\sum_{i=1}^m(x_i-y_i)^2.
\]
\item By approximation with empirical quantile grids, justify the
one-dimensional formula
\[
  W_2^2(F,G)=\int_0^1\{Q_F(u)-Q_G(u)\}^2\,du .
\]
\end{enumerate}
\end{exercise}

\begin{exercise}[A Decoupled Quadratic Statistic]
Let \(X_1,\ldots,X_n\) be iid, let \(X'_1,\ldots,X'_n\) be an independent
copy, and let \(h\) be a measurable kernel with \(\Expect h(X_1,X_2)=0\) and
\(\Expect h^2(X_1,X_2)<\infty\). Define
\[
  U=\sum_{1\le i<j\le n}h(X_i,X_j),
  \qquad
  U^{\mathrm{dec}}=\sum_{1\le i<j\le n}h(X_i,X'_j).
\]
\begin{enumerate}[label=(\alph*)]
\item Show that the summands in \(U\) are generally dependent by identifying
two summands that share a coordinate.
\item Condition on \(X_1,\ldots,X_n\).  Which summands in
\(U^{\mathrm{dec}}\) are independent under this conditional law, and which
still share a copied coordinate?
\item Replace \(X'_j\) by independent copies \(X'_{ij}\), one for each pair
\((i,j)\), and define
\[
  \widetilde U=\sum_{1\le i<j\le n}h(X_i,X'_{ij}).
\]
Explain why \(\widetilde U\) is easier to analyze conditionally on
\(X_1,\ldots,X_n\).
\item In words, state what a decoupling inequality adds beyond this
conditioning calculation.
\end{enumerate}
\end{exercise}

\begin{exercise}[Image Measures]
Prove the image-measure theorem: if
$\Psi:(\Omega,\mathcal F,\mu)\to(S,\mathcal S)$ is
measurable and $\mu_\Psi(B)=\mu(\Psi^{-1}B)$, then $\mu_\Psi$ is a measure and
\[
  \int_S h\,d\mu_\Psi=\int_\Omega h(\Psi(\omega))\,\mu(d\omega)
\]
for every nonnegative or integrable measurable $h$. Your proof should proceed
from indicators to nonnegative simple functions, then to nonnegative measurable
functions by monotone convergence, and finally to integrable signed functions by
positive and negative parts.
\end{exercise}

\begin{exercise}[Sections and monotone classes]
Let $C\in\mathcal F_1\otimes\mathcal F_2$. Prove that every section
$C_{\omega_1}$ belongs to $\mathcal F_2$ and every section
$C^{\omega_2}$ belongs to $\mathcal F_1$. Then use a monotone-class argument to
show that, for every nonnegative measurable $f$ on
$\Omega_1\times\Omega_2$, the section $f(\omega_1,\cdot)$ is
$\mathcal F_2$-measurable for each fixed $\omega_1$.
\end{exercise}

\begin{exercise}[Tulcea consistency]
Let $\nu_1$ be an initial law and let $K_n$ be probability kernels as in
Ionescu--Tulcea's theorem. Write the finite-dimensional law
$\mu_{1:n}$ explicitly on rectangles
$A_1\times\cdots\times A_n$. Prove the marginal consistency identity
\[
  \mu_{1:n+1}(B\times\Omega_{n+1})=\mu_{1:n}(B),
  \qquad B\in\mathcal F_1\otimes\cdots\otimes\mathcal F_n.
\]
\textit{Hint.} First prove the identity for rectangles, then extend by a
pi-lambda argument.
\end{exercise}

\begin{exercise}[Path equivalence]
On $([0,1],\Borel[0,1],\lambda)$ define
\[
  X_t(\omega)=0,\qquad Y_t(\omega)=\ind{\omega=t},\qquad t\in[0,1].
\]
Show that $X$ and $Y$ are stochastically equivalent but not indistinguishable.
Identify exactly where uncountability enters the proof.
\end{exercise}

\begin{exercise}[Nonmeasurable supremum]
Let $T$ be uncountable and let $\Omega=\{0,1\}^T$ with the product sigma-field.
For $t\in T$, set $X_t(\omega)=\omega(t)$. Prove that every product-measurable
set depends on at most countably many coordinates. Use this to show that
\[
  \left\{\sup_{t\in T}X_t=0\right\}
\]
is not product-measurable.
\end{exercise}

\begin{exercise}[Outer probability as a repair]
Let
\[
  \outerProb(A)=\inf\{\Prob(B):A\subseteq B,\ B\in\mathcal F\}.
\]
Show that $\outerProb$ is an outer measure. Show also that if
$A\in\mathcal F$, then $\outerProb(A)=\Prob(A)$. Finally, explain why this
definition permits probability bounds for nonmeasurable suprema.
\end{exercise}

\begin{exercise}[Separable suprema]
Let $T$ be compact metric, let $T_0\subseteq T$ be countable dense, and let
$X=\{X_t:t\in T\}$ have continuous sample paths. Prove
\[
  \sup_{t\in T}X_t=\sup_{t\in T_0}X_t
\]
path by path. Conclude that the supremum is a measurable random variable.
\end{exercise}

\begin{exercise}[Kolmogorov continuity, dyadic version]
Let $X=\{X_t:0\le t\le1\}$ satisfy
\[
  \Expect|X_t-X_s|^a\le C|t-s|^{1+b}.
\]
Fix $\gamma<b/a$. Work through the dyadic proof: apply Markov's inequality to
dyadic increments, sum over all dyadic edges at level $m$, apply
Borel--Cantelli, and then use chaining to construct a Holder-continuous
modification.
\end{exercise}

\begin{exercise}[Measurable argmax]
Let $\Theta$ be compact metric and let $M:\Omega\times\Theta\to\R$ be
measurable in $\omega$ and continuous in $\theta$. Prove that
\[
  V(\omega)=\sup_{\theta\in\Theta}M(\omega,\theta)
\]
is measurable. Then show that
\[
  A(\omega)=\{\theta\in\Theta:M(\omega,\theta)=V(\omega)\}
\]
is a nonempty compact-valued weakly measurable multifunction.
\end{exercise}

\begin{exercise}[Measurable selector for estimators]
Under the assumptions of the previous exercise, assume $\Theta$ is complete and
separable. Use a countable dense subset of $\Theta$ to construct a measurable
sequence of approximate maximizers. Show how a measurable selection theorem
turns the compact-valued argmax set into an ordinary estimator
$\hat\theta(\omega)\in A(\omega)$.
\end{exercise}

\begin{exercise}[Gaussian process consistency]
Let $T$ be an arbitrary index set, let $m:T\to\R$, and let
$K:T\times T\to\R$ be symmetric and nonnegative definite. For every finite
ordered list $(t_1,\ldots,t_k)$ of distinct points of $T$, let
\[
  \mu_{t_1,\ldots,t_k}
  =
  \mathcal N_k(m_{t_1,\ldots,t_k},K_{t_1,\ldots,t_k}),
\]
where $m_{t_1,\ldots,t_k}=(m(t_1),\ldots,m(t_k))$ and
$K_{t_1,\ldots,t_k}=(K(t_i,t_j))_{i,j}$.
\begin{enumerate}[label=(\alph*)]
\item Prove that these finite-dimensional laws are consistent under
permutations of the coordinates.
\item Prove consistency under deletion of one coordinate. That is, show that
the marginal law of $(X_{t_1},\ldots,X_{t_{k-1}})$ under
$\mu_{t_1,\ldots,t_k}$ is $\mu_{t_1,\ldots,t_{k-1}}$.
\item Use Kolmogorov's extension theorem to construct a process
$X=\{X_t:t\in T\}$ with these finite-dimensional laws.
\item Show that every finite linear combination $\sum_{i=1}^k a_iX_{t_i}$ is
normal with mean $\sum_i a_im(t_i)$ and variance
$\sum_{i,j}a_ia_jK(t_i,t_j)$. Explain why this confirms that the constructed
process is Gaussian.
\end{enumerate}
\end{exercise}

\begin{exercise}[Brownian finite-dimensional laws]
Fix $0=t_0<t_1<\cdots<t_k$. Let
$\Delta_i$ be independent normal random variables with
$\Delta_i\sim\Normal(0,t_i-t_{i-1})$, and set
$W_{t_j}=\sum_{i=1}^j\Delta_i$.
\begin{enumerate}[label=(\alph*)]
\item Show that $(W_{t_1},\ldots,W_{t_k})$ is multivariate normal with mean
zero and covariance matrix
\[
  \Cov(W_{t_i},W_{t_j})=t_i\wedge t_j .
\]
\item If one time point, say $t_r$, is deleted, prove directly from the
increment representation that the marginal law is the Brownian
finite-dimensional law on the remaining time points.
\item Conclude that Kolmogorov's theorem constructs a process on
$\R_+^{[0,\infty)}$ whose finite-dimensional distributions are the Brownian
ones.
\item Explain why this construction alone does not prove that the sample paths
are continuous. Which additional result in this chapter supplies a continuous
modification?
\end{enumerate}
\end{exercise}

\begin{exercise}[Stopping times from right-continuous paths]
Let $X=\{X_t:t\ge0\}$ be adapted to a filtration
$(\mathcal F_t)_{t\ge0}$ and suppose that every sample path is
right-continuous. Let $G$ be an open subset of the state space and define the
hitting time
\[
  \tau_G=\inf\{t\ge0:X_t\in G\}.
\]
\begin{enumerate}[label=(\alph*)]
\item Prove that, for every $t>0$,
\[
  \{\tau_G<t\}
  =
  \bigcup_{\substack{q<t\\ q\in\Rat_+}}\{X_q\in G\}.
\]
Be explicit about the direction that uses openness of $G$ and
right-continuity of the path.
\item Deduce that $\{\tau_G<t\}\in\mathcal F_t$ for every $t>0$.
\item If the filtration is right-continuous, prove that
$\{\tau_G\le t\}\in\mathcal F_t$ and hence $\tau_G$ is a stopping time.
\end{enumerate}
\end{exercise}

\begin{exercise}[Poisson process as a random measure]
Let $N=\{N_t:t\ge0\}$ be a Poisson process with rate $\lambda$. Define, first
on half-open intervals,
\[
  M((s,t])=N_t-N_s,\qquad 0\le s<t<\infty .
\]
\begin{enumerate}[label=(\alph*)]
\item Show that $M((s,t])$ is integer-valued and has
$\Poisson\{\lambda(t-s)\}$ distribution.
\item For disjoint half-open intervals $(s_1,t_1],\ldots,(s_k,t_k]$, prove
finite additivity:
\[
  M\left(\bigcup_{j=1}^k(s_j,t_j]\right)
  =
  \sum_{j=1}^k M((s_j,t_j]).
\]
Begin with adjacent intervals and then reduce the general finite-union case by
ordering endpoints.
\item State the monotone-class or extension step that turns this premeasure on
finite unions of half-open intervals into a Borel random measure on
$[0,\infty)$. Identify the mean measure $\Expect M(B)$ for bounded Borel sets
$B$.
\item Explain why the increments of $N$ are the coordinate evaluations of this
integer-valued random measure.
\end{enumerate}
\end{exercise}

\begin{exercise}[Empirical process indexed by sets]
Let $X_1,\ldots,X_n$ be iid with law $P$, and let $\mathcal A$ be a countable
class of measurable sets. Define
\[
  P_n(A)=\frac1n\sum_{i=1}^n\indset{A}(X_i),
  \qquad
  \alpha_n(A)=\sqrt n\{P_n(A)-P(A)\},
  \qquad A\in\mathcal A .
\]
\begin{enumerate}[label=(\alph*)]
\item Show that $\alpha_n=\{\alpha_n(A):A\in\mathcal A\}$ is a stochastic
process indexed by $\mathcal A$.
\item Prove that
\[
  \|\alpha_n\|_{\mathcal A}
  =
  \sup_{A\in\mathcal A}|\alpha_n(A)|
\]
is measurable.
\item Explain precisely where countability was used. What can fail if
$\mathcal A$ is uncountable and no separability assumption is imposed?
\end{enumerate}
\end{exercise}

\begin{exercise}[Random field rectangles]
Let $X=\{X_s:s\in[0,1]^d\}$ be a real-valued random field with continuous
sample paths. Let
$R=\prod_{j=1}^d[a_j,b_j]\subseteq[0,1]^d$ be a compact rectangle and set
$D_R=R\cap\Rat^d$.
\begin{enumerate}[label=(\alph*)]
\item Prove path by path that
\[
  \sup_{s\in R}X_s=\sup_{s\in D_R}X_s .
\]
Conclude that $\sup_{s\in R}X_s$ is measurable.
\item Show that the random set
\[
  A(\omega)=\argmax_{s\in R}X_s(\omega)
\]
is nonempty and compact for every $\omega$.
\item Suppose the maximizer is unique for every $\omega$, and write it as
$\hat s(\omega)$. Prove that $\hat s$ is measurable as follows. First show that
for every compact $F\subseteq R$,
\[
  \sup_{s\in F}X_s
  =
  \inf_{\ell\ge1}
  \sup\{X_r:r\in D_R,\ d(r,F)<1/\ell\},
\]
so the supremum over $F$ is measurable. Then prove, for every open
$O\subseteq R$, that
\[
  \{\hat s\in O\}
  =
  \bigcup_{m\ge1}
  \left\{
  \sup_{s\in F_m}X_s>\sup_{s\in R\setminus O}X_s
  \right\},
  \qquad
  F_m=\{s\in R:d(s,R\setminus O)\ge1/m\}.
\]
Conclude that $\hat s$ is Borel measurable.
\end{enumerate}
\end{exercise}

\begin{exercise}[Dirichlet-process finite-partition consistency]
Let $G\sim\DP(\alpha H)$, so that for every finite measurable
partition $(A_1,\ldots,A_k)$,
\[
  (G(A_1),\ldots,G(A_k))
  \sim
  \Dirichlet\{\alpha H(A_1),\ldots,\alpha H(A_k)\}.
\]
Suppose $A_k=B_1\cup B_2$ with $B_1,B_2$ disjoint and measurable.
\begin{enumerate}[label=(\alph*)]
\item Starting from the $(k+1)$-cell partition
$(A_1,\ldots,A_{k-1},B_1,B_2)$, prove that
\[
  (G(A_1),\ldots,G(A_{k-1}),G(B_1)+G(B_2))
\]
has the $k$-cell Dirichlet law assigned to $(A_1,\ldots,A_k)$.
\item Give the proof using the gamma representation of the Dirichlet law.
\item Explain why this calculation is the finite-partition analogue of the
consistency condition in Kolmogorov's extension theorem.
\end{enumerate}
\end{exercise}

\newpage
\section*{Sources and Further Reading}
\addcontentsline{toc}{section}{Sources and Further Reading}

This chapter follows the measure-theoretic route from product measures to
stochastic processes, but the historical emphasis is on measurability. The
central lesson is that finite-dimensional probability laws are not enough to
control uncountable operations such as suprema, hitting times, and path
regularity.  The local route through products, kernels, process construction,
separability, measurable processes, and path regularity is also informed by
\citet{dabrowskaAdvancedProbabilityCommunication} and
\citet{dabrowskaStochasticProcessesCommunication}.

\begin{description}[leftmargin=0pt,labelsep=0.65em,style=unboxed,font=\normalfont,itemsep=0.45\baselineskip]
\item[\textsc{Product measures and Fubini.}]
The extension approach through rectangles, kernels, and monotone-class
arguments is standard in modern measure-theoretic probability; see
\citet{billingsley1995probability} and \citet{kallenberg2002foundations}. The
kernel formulation is the finite-step ancestor of Tulcea's theorem and the
conditional-law notation used throughout stochastic-process theory.

\item[\textsc{Coupling, transport, and decoupling.}]
Coupling is the product-space view of dependence with fixed marginals.  The
optimal-transport examples follow the standard modern reference
\citet{villani2009optimal}; statistical uses of Wasserstein geometry are
surveyed by \citet{panaretos2019statistical}.  The maximal-coupling example is
the event-level form of total variation distance.  Decoupling inequalities,
especially for dependent sums, U-statistics, and chaos, are developed
systematically in \citet{delapenaGine1999decoupling}.  Conceptually, coupling
moves from marginals toward a joint law; decoupling moves from a difficult
joint law toward independent copies without losing the probabilistic scale of
the problem.

\item[\textsc{Tulcea and Kolmogorov extension.}]
Kolmogorov's extension theorem is one of the structural achievements of
\citet{kolmogorov1933grundbegriffe}. Tulcea's theorem supplies the sequential
version needed for Markov chains and general stochastic recursions. The
presentation here uses the standard modern formulation rather than the original
historical notation. Hidden Markov models give a statistical instance of this
kernel-to-path-law construction: the latent transition and emission kernels
specify finite prefixes, and the extension theorem supplies the joint law on
the whole observed-hidden sequence. For the classical HMM formulation and its
statistical inference theory, see \citet{rabiner1989tutorial} and
\citet{cappeMoulinesRyden2005hmm}.

\item[\textsc{Dirichlet processes.}]
The Dirichlet process was introduced by \citet{ferguson1973bayesian};
\citet{ferguson1974prior} framed the broader idea of priors on spaces of
probability measures, and \citet{blackwellMacQueen1973ferguson} gave the
Polya-urn predictive representation. The existence proof in this chapter
follows the explicit disjointification proof in \citet{cui2024nature}, writing
Ferguson's finite-partition construction as a Kolmogorov-consistency argument:
arbitrary measurable sets are first disjointified into Boolean atoms,
Dirichlet aggregation supplies compatible finite-dimensional laws, and
Kolmogorov extension places the resulting law on the coordinate space of
random probabilities.

\item[\textsc{Point processes.}]
The spatial Poisson axioms and their finite-dimensional construction are
classical; see \citet[Chapter~12]{karlin1981second}. The Hawkes process was
introduced by \citet{hawkes1971spectra} as a self-exciting and mutually
exciting point-process model. In this chapter it appears only to emphasize the
random-measure viewpoint: the same coordinate object $N(ds)$ can be either a
memoryless Poisson count or a history-dependent event process through its
conditional intensity.

\item[\textsc{Stochastic geometry.}]
For the Poisson Boolean model and related spatial-coverage calculations, see
\citet{chiuStoyanKendallMecke2013stochastic}.

\item[\textsc{Interacting particle systems.}]
The voting-configuration example treats a multi-agent system as a random
element in a configuration or path space.  The classical voter model was
introduced in the interacting-particle-systems literature by
\citet{holleyLiggett1975voter}; the broader framework is developed in
\citet{liggett1985interacting,liggett1999stochastic}.  Feller semigroups and
their generators enter formally in Chapter 16, following the Markov-process
tradition of \citet{feller1952parabolic} and
\citet{ethierKurtz1986markov}.

\item[\textsc{Historical regions.}]
The basic-economic-region example uses Skinner's work on rural marketing and
macroregional structure in China \citep{skinner1964marketing,skinner1977regional}.
It is included to emphasize that partitions, graphs, and spatial fields are
also statistical random objects once historical records are translated into
data.

\item[\textsc{The nonmeasurable supremum problem.}]
In an uncountable product space the product sigma-field sees only countably
many coordinates at a time. As a result, path events and process suprema may
fail to be measurable. This is not a minor technicality: it is exactly the
issue that appears in empirical processes when one writes
$\sup_{f\in\mathcal F}|\mathbb G_n f|$ for a large class $\mathcal F$.

\item[\textsc{Separable modifications and path regularity.}]
One classical repair is to replace the process by a separable or path-regular
modification. \citet{doob1953stochastic} is the classical source for this
viewpoint. Once continuous or cadlag paths are obtained, suprema over
uncountable time sets can often be reduced to suprema over countable dense
sets.

\item[\textsc{Analytic sets and projections.}]
Another repair comes from descriptive set theory. Projections of Borel sets in
Polish spaces are analytic sets, and analytic sets are universally measurable.
This is why many hitting and projection events become measurable after
completion, even when they are not obviously Borel. Standard references include
\citet{rogers1980analytic} and \citet{kechris1995classical}.

\item[\textsc{Outer probability in empirical process theory.}]
David Pollard's books made outer probability and outer expectation a practical
language for asymptotic statistics when measurability of suprema has not yet
been established; see \citet{pollard1984convergence} and
\citet{pollard1990empirical}. The same convention is central in
\citet{vaart2023weak}. It lets one state convergence theorems first and then
verify separability, pointwise measurability, or entropy conditions that turn
outer statements into ordinary probabilistic statements.

\item[\textsc{Set-valued measurability and optimization.}]
Measurable selection theorems are the optimization analogue of separability:
they turn random closed sets of maximizers into measurable estimators. Because
the proof is reusable infrastructure, the theorem statements and examples are
collected in \Appref{sec:set-valued-measurability-optimization}. The selection
viewpoint is aligned with \citet{himmelberg1975measurable} and
\citet{aubin1990set}; it also explains why compactness and continuity
assumptions are so common in maximum-likelihood and $M$-estimation theory.
\end{description}

%% file: figures/ch06_concept_map.tex
\begin{center}
\includegraphics[width=0.92\linewidth]{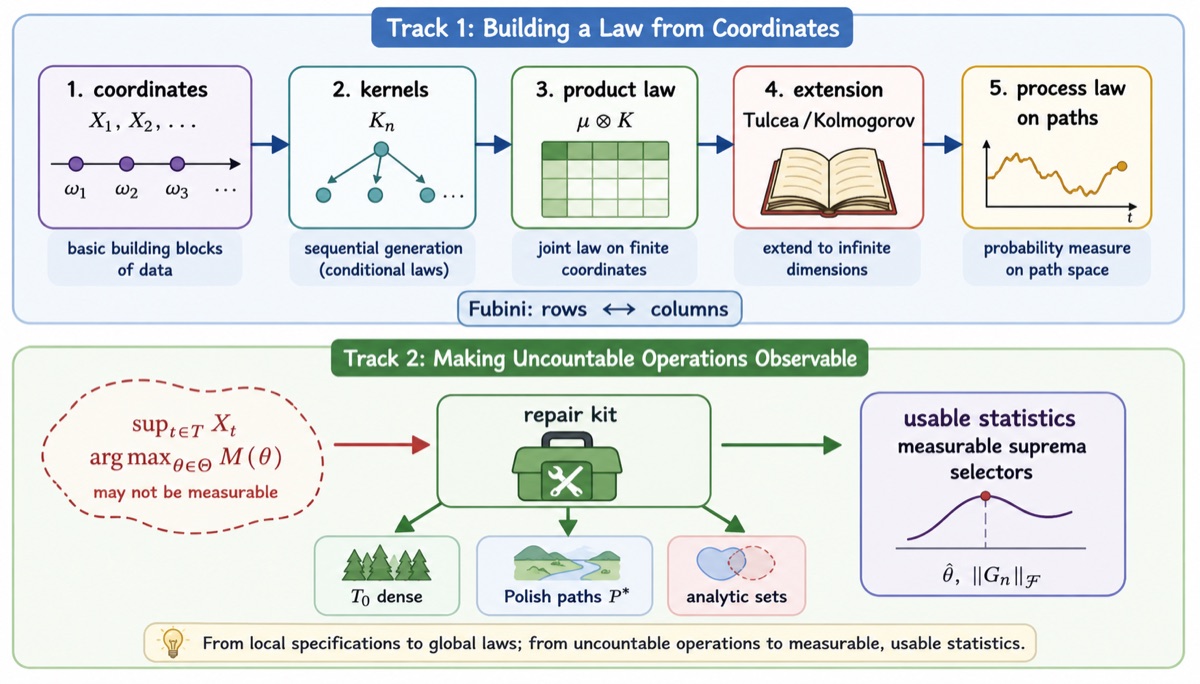}
\bookmanualfigure{fig:ch06-concept-map}{The architecture of randomness}
\par\smallskip
\small Figure~\thefigure: The architecture of randomness.
\end{center}

%% file: figures/ch06_random_object_taxonomy.tex
\begin{center}
\includegraphics[width=0.92\linewidth,trim=0 135bp 0 0,clip]{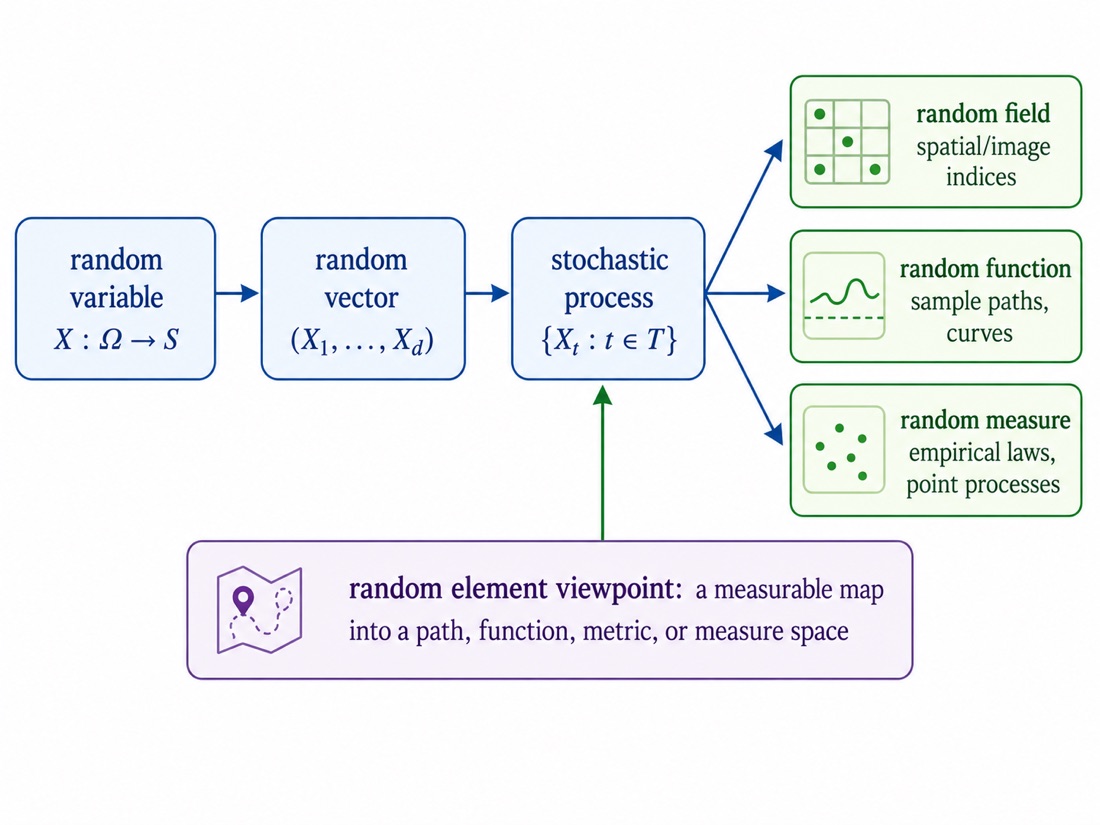}
\bookmanualfigure{fig:ch06-random-object-taxonomy}{From single coordinates to random objects}
\par\smallskip
\small Figure~\thefigure: From single coordinates to random objects.
\end{center}

%% file: chapters/ch08_common_grammar_modern_data_structures.tex
\chapter{Modern Data Objects: What Kind of Object Was Observed?}
\label{chap:grammar-modern-data-structures}
\conceptindexes{data structures, data-object schema, statistical language, counting processes, random graphs, empirical objects}

\begin{tcolorbox}[
  enhanced,
  breakable,
  colback=chaptercream,
  colframe=bookblue!88!black,
  boxrule=0.72pt,
  arc=5pt,
  boxsep=1pt,
  left=1.0em,
  right=0.95em,
  top=0.82em,
  bottom=0.82em,
  before skip=0.55\baselineskip,
  after skip=1.0\baselineskip
]
\noindent\textbf{Chapter overview.}
This is the central data-object chapter. Together with Chapter~8, it keeps the
foundations from becoming detached from inference: first name the empirical
object, then name the target. Modern statistics rarely begins with a clean
rectangular table and one chosen summary. It begins with labels, locations,
event histories, images, networks, sparse matrices, spatial assays, text,
rankings, and feedback. The common pattern is
\[
  \begin{aligned}
  \hbox{object}
  &\longrightarrow
  \hbox{observation mechanism}
  \longrightarrow
  \hbox{empirical object} \\
  &\longrightarrow
  \hbox{structure}
  \longrightarrow
  \hbox{target functional}
  \longrightarrow
  \hbox{uncertainty or decision}.
  \end{aligned}
\]
The purpose is not to solve every listed problem here. It is to teach the
reader to recognize the statistical object before choosing the model.
\end{tcolorbox}

One reason statistics can feel fragmented is that different fields name their
objects differently.  A biologist may provide a cell-by-gene count matrix.  A
clinician may provide visit times, censoring indicators, and endpoints.  A
spatial scientist may provide a map of cases and controls.  A recommender
system may provide sparse ratings, exposures, and later choices.  The surface
forms differ, but the statistical work begins with the same questions:

\begin{enumerate}
\item What is the mathematical object that represents the world or population?
\item Through what mechanism was that object observed?
\item What empirical object does the observed data naturally form?
\item What structure makes inference possible?
\item What target functional is scientifically or operationally meaningful?
\item What uncertainty statement or decision rule will be justified?
\end{enumerate}

This chapter calls that checklist the grammar of modern data structures.  It is
not a new model.  It is a way to avoid choosing a model before knowing what
kind of sentence the data can speak.

\section{The Six-Slot Data-Object Schema}
\label{sec:ch08-six-slot-grammar}
\conceptindexes{six-slot data-object schema, units, observation space, record map, target object, procedure, use}

Let \(W\) denote the object we wish we could see, \(O\) the object we actually
observe, and \(P\) the law, mechanism, or population distribution that governs
them.  The simplest iid textbook problem hides this distinction because
\[
  O_i=W_i=X_i,\qquad X_i\sim P,
\]
and the empirical measure
\[
  P_n=\frac1n\sum_{i=1}^n \delta_{X_i}
\]
is the obvious summary.  Once data have censoring, spatial coordinates,
adaptive collection, missing labels, image annotations, or sequencing
protocols, the equality \(O_i=W_i\) is no longer credible.  A useful analysis
must keep the slots separate.

\begin{definition}[Statistical grammar of a data structure]
A statistical grammar for a problem consists of six connected choices:
\begin{enumerate}
\item an object space \(\mathcal W\) for the target or latent object \(W\);
\item an observation space \(\mathcal O\) and observation mechanism
\(K(d o\mid w)\) or \(P_{O\mid W}\);
\item an empirical object \(E_n=E(O_1,\ldots,O_n)\);
\item a structural assumption that makes the problem learnable;
\item a target functional \(T(P)\) or \(T(W,P)\);
\item an uncertainty statement, comparison, prediction rule, or decision.
\end{enumerate}
\end{definition}

The observation mechanism may be explicit, as in a censoring model, a
randomized trial, or a sequencing simulator.  It may also be implicit, as in
a benchmark dataset whose labels were created by a human annotation pipeline.
Either way, it shapes what the empirical object is allowed to mean.

\begin{example}[The ordinary mean as a degenerate grammar]
For iid real-valued data with finite mean, the object is a distribution
\(P\) on \(\R\), the observation mechanism is direct sampling, the empirical
object is \(P_n\), the structural assumption is integrability, the target is
\[
  T(P)=\int x\,dP(x),
\]
and the uncertainty statement is a law of large numbers, a concentration
bound, or a central limit approximation.  The grammar looks trivial only
because every slot is aligned.
\end{example}

\begin{example}[Why modern data need the slots]
In a single-cell RNA-seq experiment, the biological state of a cell is not the
same as the observed UMI count vector.  Dissociation, capture efficiency,
library size, batch, sequencing depth, ambient RNA, and computational
preprocessing are part of the observation mechanism.  A method that treats the
count matrix as if it were the biological object itself is making a strong
shortcut.  Sometimes that shortcut is useful.  Sometimes it erases the very
structure the study was designed to find.
\end{example}

\begin{tcolorbox}[
  enhanced,
  breakable,
  colback=noteback,
  colframe=bookgold!75!black,
  boxrule=0.55pt,
  arc=4pt,
  boxsep=1pt,
  left=0.95em,
  right=0.9em,
  top=0.7em,
  bottom=0.7em,
  before skip=0.9\baselineskip,
  after skip=1.0\baselineskip
]
\noindent\textbf{Compass rule.}
Before asking whether an estimator is unbiased, efficient, predictive, or
interpretable, ask what object it estimates and which observation mechanism
made that object visible.
\end{tcolorbox}

\noindent\textbf{Application spine.}
The same six slots organize the real-data and industrial examples that return
later.  The table is deliberately compact: its purpose is to show where the
statistical grammar lives, not to turn this chapter into a case-study catalog.
The industrial anchors below are documented through platform experimentation
papers and engineering reports
\citep{kohavi2020trustworthy,xu2015infrastructure,kaufman2017democratizing,
uber2018experimentation}, oncology real-world-evidence and estimand sources
\citep{griffith2019tumorBurden,fda2018rweframework,ich2021e9r1}, AlphaFold
reports \citep{jumper2021alphafold,abramson2024alphafold3}, and John Deere's
See \& Spray documentation \citep{johnDeere2026seeSpray}.

\begin{center}
\small
\setlength{\tabcolsep}{0.34em}
\renewcommand{\arraystretch}{1.12}
\begin{longtable}{@{}>{\raggedright\arraybackslash}p{0.20\linewidth}
                  >{\raggedright\arraybackslash}p{0.25\linewidth}
                  >{\raggedright\arraybackslash}p{0.25\linewidth}
                  >{\raggedright\arraybackslash}p{0.22\linewidth}@{}}
\caption{Application spine for modern data structures}\\
\toprule
System & Object \(W\) & Observed record \(O\) & Target pressure \\
\midrule
\endfirsthead
\caption[]{Application spine for modern data structures (continued)}\\
\toprule
System & Object \(W\) & Observed record \(O\) & Target pressure \\
\midrule
\endhead
Online experimentation &
User, session, network, or marketplace response under possible product states &
Assignment, exposure, metrics, histories, guardrails, and logging metadata &
Metric lift is only one target; long-run value, interference, and safety
guardrails may define different functionals. \\

Real-world oncology evidence &
Disease course, treatment history, progression, survival, and care pathway &
EHR visits, imaging reports, chart abstractions, treatment lines, and death
links &
Routine-care traces must be translated into real-world endpoints or
trial-like estimands before survival machinery is meaningful. \\

Molecular prediction &
Biomolecular structure, interaction, conformation, and physical mechanism &
Sequence, templates, alignments, structural database entries, ligands, and
benchmark splits &
Benchmark structure risk, confidence, binding claims, and laboratory decisions
are related but distinct targets. \\

Precision agriculture &
Crop state, weed state, field conditions, and operational constraints &
Images, GPS, detections, nozzle actions, machine logs, and as-applied maps &
The policy target depends on missed weeds, crop injury, chemical cost, drift,
and timing, not image accuracy alone. \\
\bottomrule
\end{longtable}
\end{center}

These examples are industrial in scale, but the mathematical problem is the
same one visible in smaller scientific studies.  The analyst must not let the
record \(O\) impersonate the object \(W\), and must not let an available score
impersonate the target \(T(P)\).  Chapter~8 takes the next step by naming the
targets, losses, and estimands that these records can support.

\section{Counting Processes as Data Objects}
\label{sec:ch08-counting-process-template}
\conceptindexes{counting processes, event-history data objects, event histories, stochastic clock}

Survival analysis is a clean example because its grammar is not hidden.  For
subject \(i\), let \(T_i\) be an event time and \(C_i\) a censoring time.  We
observe
\[
  \tilde T_i=T_i\wedge C_i,\qquad
  \Delta_i=\ind{T_i\le C_i},
\]
possibly with covariate history \(\mathbf Z_i(t)\).  The event history can be
written as a counting process
\[
  N_i(t)=\ind{\tilde T_i\le t,\Delta_i=1},
\]
and the risk process is
\[
  Y_i(t)=\ind{\tilde T_i\ge t}.
\]
The modeling step is to describe the next possible event conditional on the
information available just before time \(t\).  A subject can contribute event
risk only while still under observation, which is the role of \(Y_i(t)\).
Among those at risk, the baseline clock \(\lambda_0(t)\) is multiplied by the
covariate history through \(\exp\{\beta^T\mathbf Z_i(t)\}\).  Thus, under an
intensity model,
\[
  \lambda_i(t\mid \mathcal F_{t-})
  =
  Y_i(t)\lambda_0(t)\exp\{\beta^T\mathbf Z_i(t)\},
\]
the compensator is
\[
  \Lambda_i(t)=\int_0^t
  Y_i(s)\lambda_0(s)\exp\{\beta^T\mathbf Z_i(s)\}\,ds,
\]
and
\[
  M_i(t)=N_i(t)-\Lambda_i(t)
\]
is the martingale noise.  Here \(\Lambda_i\) is the predictable event clock
accumulated by subject \(i\), and \(M_i\) is what remains after that clock has
been subtracted from the observed event history.  This is the
counting-process grammar behind modern large-sample survival analysis
\citep{aalen1978nonparametric,andersen1982cox,andersen1993statistical,fleming1991counting}.

\begin{center}
\small
\textbf{Counting-process grammar for survival analysis.}\par\smallskip
\setlength{\tabcolsep}{0.46em}
\renewcommand{\arraystretch}{1.16}
\begin{tabular}{@{}>{\raggedright\arraybackslash}p{0.25\linewidth}
                >{\raggedright\arraybackslash}p{0.65\linewidth}@{}}
\toprule
Grammar slot & Survival-analysis version \\
\midrule
Object & Event-time process, disease history, treatment/covariate history \\
Observation mechanism & Censoring, delayed entry, visit schedule, endpoint definition \\
Empirical object & \(N_i(t)\), \(Y_i(t)\), covariate history, aggregated risk sets \\
Structure & Intensity, compensator, proportional hazards, independent censoring \\
Target functional & Hazard, survival curve, cumulative incidence, regression effect \\
Uncertainty/decision & Martingale variance, confidence bands, tests, treatment comparison \\
\bottomrule
\end{tabular}
\end{center}

The strength of this framework is that it refuses to treat censoring as an
annoyance added after the fact.  Censoring is part of the observed-data
mechanism.  The risk set \(Y_i(t)\) is not bookkeeping; it says who is still
observable at time \(t\).  The compensator is not a formal decoration; it is
the predictable part of the observed event process.  The martingale is not
generic noise; it is the residual fluctuation after the history has been
accounted for.

The chapter's larger claim is that this is not special to survival analysis.
Many modern data structures become tractable when we find their analogue of
\[
  (N,Y,\Lambda,M,T).
\]
That analogue may be an empirical distribution over labels, a point process in
space, a field of \(p\)-values, a count matrix with latent biological states,
a graph adjacency measure, a Wasserstein barycenter, or a policy-value process.
The names change.  The grammar remains.

\section{Selected Modern Data Structures}
\label{sec:ch08-short-atlas}
\conceptindexes{data-structure atlas, tables, images, text, graphs, curves, point patterns}

The table below is a deliberately compressed atlas.  Each row names the
canonical empirical object and the kind of structure that turns observation
into inference.  Later chapters and later parts of the book can expand these
rows into full stories.  Here they serve as translation practice.

\begin{center}
\small
\setlength{\tabcolsep}{0.34em}
\renewcommand{\arraystretch}{1.15}
\begin{longtable}{@{}>{\raggedright\arraybackslash}p{0.18\linewidth}
                  >{\raggedright\arraybackslash}p{0.24\linewidth}
                  >{\raggedright\arraybackslash}p{0.25\linewidth}
                  >{\raggedright\arraybackslash}p{0.22\linewidth}@{}}
\caption{Modern data structures as empirical objects and targets}\\
\toprule
Problem family & Empirical object & Structure & Target functional \\
\midrule
\endfirsthead
\caption[]{Modern data structures as empirical objects and targets (continued)}\\
\toprule
Problem family & Empirical object & Structure & Target functional \\
\midrule
\endhead
Missing species &
Occupancy counts \(f_1,f_2,\ldots\), empirical label distribution &
Exchangeable sampling from a label law; rare-type regularity &
Unseen mass, support size, number of undiscovered types
\citep{fisher1943species,good1953population,efronThisted1976unseen} \\

Spatial clustering &
Point pattern or counting measure \(N(A)\) over regions \(A\) &
Poisson baseline, DPP repulsion, intensity, marks &
Clustering or inhibition, disease risk surface, environmental association
\citep{clarke1946poisson,kingman1993poisson,karr1986point} \\

Genomics and GWAS &
Vector or field of test statistics, \(p\)-values, genotypes, phenotypes &
Multiplicity, dependence, sparsity, linkage, signal strength &
Discoveries, FDR, selected loci, interpretable biological signal
\citep{landerBotstein1989mapping,benjamini1995fdr} \\

Single-cell and spatial omics &
Cell-by-feature count matrix, modality layers, spatial marks &
Latent cell state, measurement noise, batch, zero inflation, smooth trends,
copula or vine dependence &
Cell states, trajectories, imputation, simulation, differential expression
\citep{li2018scimpute,cui2022scgtm} \\

Images, text, and networks &
Functions, tensors, token sequences, graphs, embeddings &
Geometry, invariance, exchangeability, representation, metric structure &
Risk, classification, similarity, communities, interpretable features
\citep{wasserman1994social,breiman2001two_cultures} \\

Functional trajectories &
Curves \(Y_i(t)\), warped curves \(Y_i\{h_i(t)\}\), irregular time grids &
Smooth covariance, amplitude variation, phase variation, common shape &
Functional principal component scores, registered mean curve, timing
differences, functional regression signal
\citep{ramsaySilverman2005fda,telescaInoue2008curve} \\

Distributional and geometric data &
Random probability measures, shapes, covariance objects, manifolds &
Metric geometry, transport, Fr\'echet means, principal curves, tangent
approximations, nearest-neighbor geometry &
Mean object, regression functional, principal-curve target, variation around a
template, intrinsic dimension
\citep{petersen2019frechet,chen2023wasserstein,villani2009optimal,
levinaBickel2004intrinsic,hastieStuetzle1989principal,kegl2000principal,
cuiShao2024metricPrincipalCurve} \\

Recommendation and feedback &
Sparse user-item-event history, exposure log, policy trace &
Missing-not-at-random observation, adaptivity, exploration, feedback loops &
Prediction risk, calibration, ranking error, feedback bias
\citep{bennett2007netflix} \\
\bottomrule
\end{longtable}
\end{center}

Two warnings keep the atlas honest.  First, the same raw file can support
different empirical objects.  A spatial assay can be read as a count matrix, a
marked point pattern, a spatial field, a graph, or a discretized image.
Second, the same empirical object can serve different targets.  A graph may be
used for community detection, prediction, causal exposure mapping, privacy
auditing, or uncertainty propagation.  The object does not determine the task
by itself.

The rows should therefore be read as real-record templates, not as a taxonomy
of toy examples.  London bombing counts, Shakespeare's frequency table,
interval-censored clinical visits, benchmark image labels, and recommendation
logs will reappear in Chapter~8 as target audits.  The audit asks what claim a
particular record can support, and which claim would require a different
observation law or additional assumptions.

Two neighboring languages sit just outside the book's main route.  Algebraic
statistics studies statistical models whose constraints are polynomial or
combinatorial, as in contingency tables, conditional-independence models, and
phylogenetic or graphical model varieties.  Its lesson for this chapter is that
``model structure'' may be an algebraic constraint on possible laws, not only a
smoothness or independence assumption.  Quantum statistics changes the sample
space more radically: states are density operators, observations are
measurements, and probability enters through the measurement map rather than
through a single classical random variable.  Both subjects fit the six-slot
grammar, but they require specialized machinery beyond the recurring examples
of this book.

\section{Random Graphs and Complex Networks}
\label{sec:ch08-random-graphs-complex-networks}
\conceptindexes{random graphs, complex networks, graph models, network dynamics, message passing}

Networks deserve a little more than a row in the atlas because they are a
useful test of the chapter's grammar.  Let
\[
  V_n=\{1,\ldots,n\}
\]
be the node set and let \(A=(A_{ij})_{1\leq i,j\leq n}\) be an adjacency
matrix.  For a simple undirected network,
\[
  A_{ij}=A_{ji}\in\{0,1\},\qquad A_{ii}=0.
\]
Here \(A_{ij}=1\) means that nodes \(i\) and \(j\) are linked.  Directed,
weighted, temporal, spatial, and bipartite networks change the state space but
not the basic idea: the empirical object is no longer a list of independent
rows, but a relational object whose entries share nodes.

Some summaries have immediate interpretations:
\[
  d_i=\sum_{j\ne i}A_{ij},\qquad
  \hat p=\frac{2}{n(n-1)}\sum_{i<j}A_{ij},\qquad
  T(A)=\sum_{i<j<k}A_{ij}A_{ik}A_{jk}.
\]
The degree \(d_i\) measures how connected node \(i\) is, \(\hat p\) is the
observed edge density, and \(T(A)\) counts triangles.  Triangles matter because
many real networks have transitivity: two people with a common friend, two
proteins in the same pathway, or two hospitals sharing many referrals are more
likely to be connected than two arbitrary nodes.

The six slots keep the modeling problem honest:
\[
\begin{array}{rcl}
\hbox{object} &:& \hbox{population network, latent relation, or evolving graph},\\
\hbox{observation} &:& \hbox{survey, API log, sensor threshold, sequencing pipeline},\\
\hbox{empirical object} &:& \hbox{adjacency matrix, edge list, weighted graph, snapshots},\\
\hbox{structure} &:& \hbox{exchangeability, sparsity, communities, geometry, closure},\\
\hbox{target} &:& \hbox{communities, links, diffusion risk, hubs, intervention value},\\
\hbox{uncertainty} &:& \hbox{model checks, resampling, subsampling, sensitivity analysis}.
\end{array}
\]
A contact network in an epidemic, for example, is not observed by magic.  It
may be reconstructed from diaries, Bluetooth proximity, clinic reports, or
partial tracing.  The target may be outbreak size, not the exact graph itself.
A protein interaction network may have edges inferred by noisy experiments;
then edge uncertainty is part of the empirical object, not an annoyance to be
ignored after drawing the graph.

\begin{realdatacapsule}{Network data target audit}
\item[Data object.] An observed adjacency matrix, edge list, or sequence of
network snapshots for contacts, citations, referrals, or protein interactions.
\item[Observation mechanism.] Edges arise through surveys, sensors, platform
logs, assays, thresholds, or crawlers; node inclusion and edge detection are
part of the sampling design.
\item[Target.] Community labels, diffusion risk, hub ranking, link prediction,
or the effect of an intervention on a network-mediated outcome.
\item[Model.] Erd\H{o}s--R\'enyi baselines, stochastic block models,
exchangeable graphons, preferential attachment, or local-dependence models
encode different structural questions.
\item[Uncertainty.] Edge-error sensitivity, node subsampling, model-based
posterior variation, and null comparisons decide which graph features are
stable.
\item[Limitation.] Network dependence makes row-wise iid reasoning misleading;
unobserved nodes and measurement thresholds may change both structure and
target.
\end{realdatacapsule}

\subsection{Null, Community, and Exchangeable Graph Models}
\label{subsec:ch08-network-null-community-exchangeable}
\conceptindexes{null graph model, community detection, exchangeable graph model, stochastic block model}

The first family of network models asks how much structure is already visible
in the static relational object.  A baseline model names what would happen
without extra organization; community and exchangeable models then describe
the larger patterns that remain.

\paragraph{Erd\H{o}s--R\'enyi baseline.}
The simplest random graph is \(G(n,p)\): for \(1\leq i<j\leq n\)
\citep{erdosRenyi1959random},
\[
  A_{ij}\ \hbox{are mutually independent},\qquad
  A_{ij}\sim \Bernoulli(p).
\]
Then
\[
  \Expect d_i=(n-1)p,\qquad
  \Expect T(A)=\binom{n}{3}p^3.
\]
This model is rarely a final scientific model, but it is an excellent null
language.  If the observed network has far more triangles, hubs, or disconnected
components than \(G(n,p)\) predicts at the same density, the graph is telling us
which structure is missing.

\paragraph{Communities and block models.}
A stochastic block model adds latent node types
\citep{holland1983stochastic}
\[
  Z_i\in\{1,\ldots,K\},\qquad \Prob(Z_i=a)=\pi_a,
\]
and a matrix \(B=(B_{ab})\) of link probabilities:
\[
  \Prob(A_{ij}=1\mid Z_i=a,Z_j=b)=B_{ab},\qquad i<j.
\]
If \(B_{aa}\) is larger than \(B_{ab}\) for \(a\ne b\), nodes connect more
often within their own communities.  In applications, \(Z_i\) might represent
social groups, cell-state modules, hospital referral regions, or user segments.
A degree-corrected version replaces \(B_{Z_iZ_j}\) by a term proportional to
\(\theta_i\theta_jB_{Z_iZ_j}\), separating community structure from node-level
popularity.  That distinction matters in complex networks, where hubs can look
like communities if degree variation is not modeled.

\paragraph{Graphons and exchangeability.}
For dense networks, exchangeability leads to a nonparametric model
\citep{lovasz2012large}.  Let
\[
  U_i\stackrel{\mathrm{iid}}{\sim}\Unif(0,1),\qquad
  A_{ij}\mid U_i,U_j \sim
  \Bernoulli\{W(U_i,U_j)\},
\]
where \(W:[0,1]^2\to[0,1]\) is symmetric.  The function \(W\) is called a
graphon.  It plays the role of a regression surface for edge probabilities:
nearby latent positions have similar connection behavior.  The catch is that
node labels have no intrinsic meaning.  If \(\phi\) is a measure-preserving
relabeling of \([0,1]\), then \(W(u,v)\) and
\(W\{\phi(u),\phi(v)\}\) describe the same unlabeled random-network law.
Thus the parameter is geometric and relational, not an ordinary Euclidean
vector.

\subsection{Local Dependence, Growth, and Network Dynamics}
\label{subsec:ch08-network-local-growth-dynamics}
\conceptindexes{local dependence, graph growth, network dynamics, preferential attachment}

The second family asks where relational dependence comes from.  Sometimes the
model is built from local features such as triangles and homophily; sometimes
it is built from a growth rule or a stochastic update rule over time.

\paragraph{Exponential random graph models.}
When the scientific language is built from local network features, an
exponential random graph model writes \citep{frank1986markov}
\[
  P_\theta(A=a)
  =\frac{\exp\{\theta^\top g(a)\}}
  {\sum_{b}\exp\{\theta^\top g(b)\}},
\]
where \(g(a)\) might contain the number of edges, triangles, \(k\)-stars, or
same-attribute links.  Positive triangle coefficients encode triadic closure.
Positive homophily coefficients encode a tendency to connect to similar nodes.
The advantage is interpretability.  The statistical warning is that the
normalizing constant sums over all graphs on \(n\) nodes and some parameter
values generate nearly empty or nearly complete graphs.  A clear application
story and model diagnostics are not optional here.

\paragraph{Small worlds and preferential growth.}
Complex-network models often explain shapes that are not well captured by
static independent-edge models.  A small-world model starts from a local
lattice and rewires edges with probability \(\rho\), producing high clustering
together with short path lengths \citep{watts1998collective}.  A
preferential-attachment model grows the network over time
\citep{barabasi1999emergence,newman2010networks}:
\[
  \Prob(\hbox{new node attaches to }i\mid \mathcal F_t)
  \propto d_i(t)+\alpha,\qquad \alpha\geq 0.
\]
The phrase ``the rich get richer'' is the modeling idea: nodes that already
have many links attract more links.  This mechanism can produce hubs and heavy
degree tails, which are common in citation networks, web graphs, social media
networks, and some biological interaction maps.

\paragraph{Interacting agents and voting.}
A multi-agent voting system is not automatically an iid sample of opinions.  If
agents influence one another over a network, the object may be a dynamic
configuration
\[
  \eta_t\in\{0,1\}^{V},
\]
where \(V\) is the set of agents and \(\eta_t(x)\) is the opinion or action of
agent \(x\) at time \(t\).  In the voter model, an agent randomly copies a
neighbor's opinion, so the observed votes are states in an interacting particle
system rather than independent reports \citep{holleyLiggett1975voter,
liggett1985interacting}.  The target is then dynamic: consensus probability,
time to consensus, polarization, coexistence, or the effect of changing the
interaction graph.  Chapter~16 returns to this example through Feller
processes, local generators, and counting-process compensators.

\subsection{Learned Message Passing and Statistical Targets}
\label{subsec:ch08-network-message-passing}
\conceptindexes{message passing, graph neural networks, statistical targets, learned representations}

The third family is computational rather than purely probabilistic.  It keeps
the relational object but uses learned local updates to build predictions,
embeddings, classifications, or decisions.

\paragraph{Graph neural networks and learned message passing.}
A graph neural network is a computational model for graph-structured data, not
a new probability model by itself.  It takes a graph \(G=(V,E)\), node features
\(x_i\), edge features \(e_{ij}\), and repeatedly updates node representations
by aggregating information from neighbors \citep{scarselli2009graph,
gilmer2017neural,kipfWelling2017semi}.  A generic layer has the form
\[
  m_i^{(\ell)}
  =
  \operatorname{AGG}_{j\in\mathcal N(i)}
  \phi_\ell\{h_i^{(\ell)},h_j^{(\ell)},e_{ij}\},
  \qquad
  h_i^{(\ell+1)}
  =
  \psi_\ell\{h_i^{(\ell)},m_i^{(\ell)}\}.
\]
This is message passing in the same relational sense as a voter model, but the
meaning is different.  In a voter model, local updates are stochastic events in
time.  In a graph neural network, local updates are usually learned
deterministic transformations used for prediction, embedding, classification,
or control.  The shared grammar is locality: the state of a node is interpreted
through its neighborhood.  The statistical warning is that the graph, the node
features, and the labels are all observed through a sampling or platform
mechanism; a powerful neural architecture does not erase the observation
problem.  The broader graph-network view treats relational structure as an
inductive bias for learning on structured data
\citep{battaglia2018relational}.

\begin{example}[Contact network and epidemic risk]
Suppose \(A_{ij}=1\) means that people \(i\) and \(j\) had enough contact for
possible transmission during a fixed week.  A purely descriptive target is the
degree distribution.  A public-health target might be the expected outbreak
size under a transmission model on the graph.  A decision target might be the
risk reduction from vaccinating the highest-degree \(5\%\) of nodes.  These are
different functionals of the same empirical object.  Sampling also matters:
diary data miss brief contacts, phone proximity data miss context, and contact
tracing oversamples edges near detected cases.
\end{example}

\begin{example}[Networks in biology and platforms]
In a gene or protein interaction network, an edge may be the output of a
measurement pipeline rather than a directly observed biological bond.  Modules
can suggest pathways, but uncertainty in edge construction should travel into
uncertainty about modules.  In a recommender system, a user-item bipartite
graph records only exposed and acted-upon items.  Missing edges are therefore
not ordinary zeros: the observation policy helped create the empirical graph.
\end{example}

The practical lesson is not that one model wins.  It is that networks force the analysis
to say which dependence is scientific signal and which dependence is an
artifact of observation.  Erd\H{o}s--R\'enyi graphs give a baseline, block
models give communities, graphons give exchangeable nonparametric structure,
exponential random graph models give local feature interpretation, and
small-world or preferential-growth models give mechanisms for paths, clustering,
and hubs.  The right choice depends on the target functional and on how the
graph entered the dataset.

Functional trajectories are another useful preview of Chapter~12.  There the
curve is kept as a curve long enough for its covariance operator to reveal
dominant modes of variation; only then is it reduced to scores, regressors, or
diagnostic summaries.

\section{What Structure Provides}
\label{sec:ch08-what-structure-buys}
\conceptindexes{structure, invariance, dependence, compression, prediction, interpretation}

The word ``structure'' is easy to overuse.  In this book it has a concrete
meaning: a structural assumption is a restriction that makes a target
functional learnable from the observed object.  The restriction may be
probabilistic, geometric, algebraic, causal, or computational.

\begin{description}[leftmargin=2.15em,style=nextline]
\item[Smoothness.]
Nearby inputs have nearby outputs.  This is the structure behind kernels,
nonparametric regression, spatial smoothing, and many trajectory models.

\item[Sparsity.]
Only a small part of a large ambient object matters.  This is the structure
behind rare-signal genomics, high-dimensional regression, and some
interpretable representation problems.

\item[Low rank or latent factors.]
Observed variation lives near a lower-dimensional structure.  This is the
structure behind matrix completion, recommendation, factor analysis, and many
multi-omics integration methods.

\item[Modular dependence.]
Margins and dependence can be separated, then the dependence can be assembled
from smaller blocks.  This is the structure behind copulas, pair-copula
constructions, and vine models for high-dimensional joint laws.

\item[Exchangeability.]
Labels may be permuted without changing the law.  This is the structure behind
species sampling, random partitions, network models, and some Bayesian
nonparametric constructions.

\item[Geometry.]
The object space has meaningful distances, geodesics, or transport maps.  This
is the structure behind Wasserstein data analysis, shape analysis, and
distributional regression.

\item[Adapted time order.]
Only the past may inform the next action.  This is the structure behind
filtrations, stopping times, martingales, adaptive trials, online learning,
and self-driving laboratories.
\end{description}

Structure buys three things.  It buys \emph{identifiability}: the target has a
meaning that can be separated from nuisance variation.  It buys
\emph{stability}: the empirical object does not change the conclusion wildly
under small perturbations or additional sampling.  It buys
\emph{transport}: a result learned in one sample can speak about another
population, time, condition, or decision.

Without structure, more data may only produce a more detailed record of the
observation mechanism.  With the wrong structure, an analysis may be precise
about the wrong object.  The discipline is to choose the weakest structure
that makes the target estimable and the strongest structure that the science
can defend.

\begin{example}[Missing species]
Let \(p_j\) be the probability of type \(j\), and let \(f_r\) be the number of
types observed exactly \(r\) times in a sample.  The target may be the missing
mass
\[
  U_n=\sum_j p_j\ind{N_j=0},
\]
or the number of new types expected in a future sample.  The empirical object
is not the ordered list of labels; it is the frequency-of-frequency profile.
The Good--Turing idea estimates missing mass from singletons because the
sampling structure connects what has appeared once with what has not appeared
yet \citep{good1953population}.
\end{example}

\begin{example}[Spatial clustering]
A map of disease cases becomes statistical only after a reference mechanism is
named.  Under a homogeneous Poisson baseline, counts in disjoint equal-area
regions are independent with equal expected counts.  Under an inhomogeneous
baseline, the intensity varies with population density or environmental
covariates.  The same visual cluster can mean different things under those two
structures.
\end{example}

\begin{example}[Multiple testing]
In a high-throughput experiment, the empirical object is often a vector of
test statistics or \(p\)-values.  The target is not one parameter but a
selection property: how many reported discoveries are false, or whether a set
of selected hypotheses is stable enough to interpret.  The Benjamini--Hochberg
procedure made the false discovery rate a central functional of the selected
set \citep{benjamini1995fdr}.  The grammar changes the question from
``which test is significant?'' to ``what property of the selection process is
controlled?''
\end{example}

\begin{example}[Vine copulas: pairwise dependence as architecture]
A copula model separates marginal behavior from dependence
\citep{sklar1959fonctions}.  For continuous \((X_1,\ldots,X_d)\) with marginal
distribution functions \(F_j\) and densities \(f_j\), define
\[
  U_j=F_j(X_j),\qquad j=1,\ldots,d.
\]
Then the joint density can be written as
\[
  f(x_1,\ldots,x_d)
  =
  c\{F_1(x_1),\ldots,F_d(x_d)\}
  \prod_{j=1}^d f_j(x_j),
\]
where \(c\) is the density of the uniform vector
\((U_1,\ldots,U_d)\).  The margins describe one-feature behavior; the copula
describes how ranks move together.

A vine copula makes the dependence density \(c\) modular
\citep{bedfordCooke2002vines,aas2009pair}.  In three dimensions, choose
variable \(2\) as the middle node and write
\[
  u_j=F_j(x_j),\qquad
  u_{1\mid 2}=F_{1\mid 2}(x_1\mid x_2),\qquad
  u_{3\mid 2}=F_{3\mid 2}(x_3\mid x_2).
\]
The chain rule and the conditional-copula identity give
\[
\begin{aligned}
  f(x_1,x_2,x_3)
  &=
  f_1(x_1)f_2(x_2)f_3(x_3)\,
  c_{12}(u_1,u_2)\,c_{23}(u_2,u_3) \\
  &\quad{}\times
  c_{13\mid 2}\{u_{1\mid 2},u_{3\mid 2}\mid x_2\}.
\end{aligned}
\]
The last factor is a copula density for the conditional law of
\((X_1,X_3)\) given \(X_2=x_2\).  In a general conditional copula it may depend
on the conditioning value \(x_2\).  The common \emph{simplified vine} model
assumes that this factor depends on \(x_2\) only through
\((u_{1\mid 2},u_{3\mid 2})\), so the final term is written
\[
  c_{13\mid 2}(u_{1\mid 2},u_{3\mid 2}).
\]
This is not a theorem forced by Sklar's theorem; it is an additional
structural assumption that buys tractability.

The same idea extends to higher dimensions by multiplying pair-copula factors
over a sequence of trees,
\[
  c(u_1,\ldots,u_d)
  =
  \prod_{m=1}^{d-1}\prod_{e\in E_m}
  c_{j_e k_e\mid D_e}
  \{u_{j_e\mid D_e},u_{k_e\mid D_e}\mid u_{D_e}\},
\]
with the conditioning sets \(D_e\) determined by the vine structure.  A regular
vine is therefore a grammar for dependence: choose the margins, choose a tree
architecture, choose pair-copula families on the edges, and then estimate the
parameters \citep{dissmann2013selecting}.  The grammar is especially useful
when a full multivariate density is too rigid, but scientific knowledge still
suggests which low-dimensional dependences matter.
\end{example}

\begin{example}[Singular copulas: dependence without a density]
A copula is a probability law on \([0,1]^d\) with uniform margins.  It need not
have a Lebesgue density.  Let \(U\sim\Unif(0,1)\).  The comonotone pair
\[
  (U,V)=(U,U)
\]
has copula
\[
  C_+(u,v)=\Prob(U\le u,\ U\le v)=\min(u,v),
\]
and all its mass lies on the diagonal \(v=u\).  The countermonotone pair
\[
  (U,V)=(U,1-U)
\]
has copula
\[
  C_-(u,v)=\Prob(U\le u,\ 1-U\le v)=\max(u+v-1,0),
\]
and all its mass lies on the anti-diagonal \(v=1-u\).  Both copulas have
uniform margins, but both are singular with respect to two-dimensional
Lebesgue measure.

Thus the density factorization in the vine example is an absolute-continuity
model, not the definition of a copula.  In applications this distinction
matters: rank dependence can be smooth, discrete, singular, or a mixture of
these, depending on the observation mechanism and scientific constraints.
\end{example}

\begin{example}[Curve registration]
Functional data are not only vertical measurements over time.  Many biological
and industrial curves have both amplitude variation and phase variation.  A
gene-expression profile may rise earlier in one condition than another; a
wearable-sensor trajectory may show the same daily pattern shifted by sleep
schedule; a manufacturing signal may contain the same thermal shape but with a
delayed peak.  A simple model is
\[
  Y_i(t)
  =
  a_i\,m\{h_i(t)\}+b_i+\varepsilon_i(t),
\]
where \(m\) is a common shape, \(a_i\) and \(b_i\) describe amplitude and
baseline variation, and \(h_i\) is a monotone time-warping function.

The empirical object is therefore not just the cloud of pairs
\((t,Y_i(t))\).  It is a collection of curves together with a registration
problem: which variation is vertical, which variation is timing, and which
variation is scientifically meaningful?  The Bayesian hierarchical curve
registration model of \citet{telescaInoue2008curve} is a compact example of
the grammar.  It treats the common shape and the time transformations as
unknown objects, borrows strength across curves, and was motivated in part by
time-course microarray applications.  In the book's language, registration is
the step that decides what it means for two paths to be comparable before a
later regression, clustering, or network analysis reads them.
\end{example}

\section{Empirical Objects, Not Just Data Tables}
\label{sec:ch08-empirical-objects}
\conceptindexes{empirical objects, empirical measures, empirical fields, object-valued data}

A table is one possible empirical object, but it is not the default form of
modern data.  This book repeatedly uses five empirical objects.

\begin{enumerate}
\item \textbf{Empirical measures.}
For iid observations,
\[
  P_n=\frac1n\sum_{i=1}^n\delta_{X_i}
\]
turns data into a random probability measure.  Averages, risks, empirical
cdf's, and many estimating equations are functionals of \(P_n\).

\item \textbf{Counting measures and point processes.}
A point pattern or event history is naturally summarized by
\[
  N(A)=\sum_i \ind{X_i\in A}.
\]
When \(A\) ranges over space-time regions, this object remembers geometry,
marks, and time order.  A Poisson process is the independent-count baseline;
a determinantal point process uses a kernel determinant to encode repulsion,
so it becomes a natural model when the target is regular spacing,
competition, or diversity rather than clustering
\citep{macchi1975coincidence,kuleszaTaskar2012dpp,lavancier2015dpp}.

\item \textbf{Random fields.}
An indexed collection
\[
  \{Z(t):t\in T\}
\]
represents curves, maps, image intensities, empirical processes, and scan
statistics.  The index set \(T\) is part of the data structure.

\item \textbf{Matrices and tensors.}
Count matrices, genotype matrices, user-item arrays, image tensors, and
multi-omics blocks carry row, column, layer, and batch structure.  Treating
them as one long vector usually loses information about the mechanism.

\item \textbf{Graphs and metric objects.}
Networks, trees, embeddings, distributions, shapes, and manifolds require
distances, adjacency, transport, or combinatorial structure.  Their summaries
are often functionals on non-Euclidean spaces.  A principal curve is one such
target: a one-dimensional curve chosen to summarize a distribution or metric
cloud, not just a plotted embedding.
\end{enumerate}

The empirical object decides what later probability theory must prove.
Chapter~8 will pause at the target-functional slot: what claim, risk,
estimand, or decision is the empirical object supposed to support?
Chapter~10 will ask when empirical objects converge weakly as random elements.
Chapter~11 will ask when a whole indexed family \(P_nf\) is uniformly close to
\(Pf\).  Chapter~13 will turn empirical landscapes into estimates.  Chapter~14
will ask how nearby laws are tested.  Chapter~15 will ask how small
perturbations propagate through a functional.  Chapter~16
will use filtrations, compensators, and martingales for event-time objects.

\clearpage

\section{From Data-Object Schema to Inference}
\label{sec:ch08-from-grammar-to-inference}
\conceptindexes{data-object schema to inference, target definition, inferential machinery, data translation}

The grammar becomes useful when it prevents category mistakes.

One compact way to say this is to separate the scientific object from the
recording mechanism.  Let \(W\) denote the object in the world, let
\(O\sim K(\cdot\mid W)\) be the observed record produced by an observation
kernel \(K\), and let \(E_n=\Phi(O_1,\ldots,O_n)\) be the empirical object
that an algorithm or estimator actually sees.  A prediction score is often a
risk under the observed mechanism,
\(R_K(f)=\Expect_{P,K}\{\ell(f,O)\}\), whereas the scientific or policy
question may be a different functional \(\psi(P)\), a counterfactual risk, or
an estimand defined after intervention on the mechanism.  The examples below
are all instances of this same separation.

\begin{example}[Prediction is not the same target as explanation]
In a recommender system, the observed ratings are not a random sample of all
user-item preferences.  They are conditioned on exposure, interface design,
previous recommendations, and user choice.  If \(A_{ui}\) records whether item
\(i\) was visible to user \(u\), then logged ratings describe a distribution
conditional on \(A_{ui}=1\), not the full matrix of possible preferences.  A
predictor can be excellent for the platform's observed rating task while
answering a narrower question than ``what would this user have liked if all
items had been equally visible?''  The grammar separates prediction risk under
the logging mechanism from a counterfactual preference functional
\citep{bennett2007netflix,schnabel2016recommendations}.
\end{example}

\begin{example}[A benchmark is an observation mechanism]
An image benchmark is not just a set of pictures.  It includes label
definitions, annotator behavior, inclusion rules, preprocessing, train-test
splits, and evaluation metrics.  The reported score
\(\frac1n\sum_i\ell\{f(X_i),\tilde Y_i\}\) is therefore a functional of the
benchmark mechanism as well as a statement about visual recognition.  Changing
the image source, label ontology, or annotation pipeline changes the target
even when the task name stays fixed.  This does not make benchmarks useless.
It makes their object explicit
\citep{deng2009imagenet,torralba2011unbiased,recht2019imagenet,
gebru2021datasheets}.
\end{example}

\begin{example}[An EHR cohort is not a trial]
An electronic-health-record cohort is produced by care delivery, billing,
testing, documentation, and follow-up processes.  The record \(O\) may contain
diagnosis codes, laboratory values, medication orders, narrative entries, visits, and
death links, but the target trial has to define eligibility, time zero,
treatment strategies, outcomes, censoring, and the contrast of interest.  A
model can predict an observed code or utilization event without estimating the
causal or clinical estimand the study is meant to support.  The grammar forces
the analyst to name the bridge from routine-care traces to a target-trial or
real-world-evidence functional
\citep{johnson2016mimic,hernan2016targettrial,fda2018rweframework,ich2021e9r1}.
\end{example}

\begin{example}[Simulation as statistical modeling]
A single-cell simulator is not merely a convenience for software comparison.
It proposes a generative law for count matrices, cell states, batches,
modalities, and spatial locations.  When the simulator is used to benchmark an
imputation, integration, or clustering method, it defines both the object and
the target by construction: performance is measured under the simulator's
chosen law for biological variation and measurement noise.  This is why data
generation can be part of statistical modeling, not an afterthought
\citep{zappia2017splatter,luecken2019singlecell,sun2021scdesign2}.
\end{example}

The rest of the book can now be read through a stable lens.  A probability
space is not enough; we also need the object space.  An observation is not
enough; we need the mechanism that made it observable.  A statistic is not
enough; we need the empirical object it compresses.  A model is not enough; we
need the structural assumption it imposes.  An estimate is not enough; we need
the target functional and the decision it supports.  The next chapter is the
hinge: it asks how a grammar of data structures becomes a grammar of
estimands, risks, policies, and scientific claims.  In this sense, the chapter
does not add a competing structure to the book.  It makes Chapter~1's compass
usable for the modern data forms previewed in Chapter~2.

\section{Exercises}
\label{sec:ch08-exercises}
\conceptindexes{data-structure exercises, graph exercises, grammar exercises}

\begin{exercise}[Write the grammar]
Choose one of the following data structures: a spatial point pattern, a
cell-by-gene count matrix, a network, a collection of images, or a recommender
system log.  Fill in the six grammar slots: object, observation mechanism,
empirical object, structure, target functional, and uncertainty or decision.
\end{exercise}

\begin{exercise}[Two targets, one object]
For a vector of \(p\)-values from a high-throughput experiment, describe two
different target functionals.  One should be a testing or selection target;
the other should be a biological interpretation target.  Explain why the same
empirical object does not determine the target by itself.
\end{exercise}

\begin{exercise}[Counting-process analogy]
For spatial disease mapping, propose analogues of the survival-analysis
objects \(N_i(t)\), \(Y_i(t)\), \(\Lambda_i(t)\), and \(M_i(t)\).  Which parts
of the analogy are clean, and which parts break down because space has no
natural time order?
\end{exercise}

\begin{exercise}[Observation mechanism]
In a single-cell experiment, list three steps between the biological state of a
cell and the observed count vector.  For each step, say whether it mainly
affects bias, variance, missingness, dependence, or interpretation.
\end{exercise}

\begin{exercise}[Structure and falsifiability]
Choose one structural assumption from Section~\ref{sec:ch08-what-structure-buys}.
Give an example where the assumption is plausible and an example where it is
dangerous.  What diagnostic, sensitivity analysis, or design change would make
the assumption more defensible?
\end{exercise}

\begin{exercise}[Multimodal patient record]
An oncology system stores pathology images, radiology images, genomic panels,
clinician notes, treatment lines, adverse events, and survival outcomes.  Write
the grammar for this record: object, observation mechanism, empirical object,
structure, target functional, and uncertainty or decision.  Which part of the
record is naturally a table, which part is a path, which part is an image or
text object, and which part is an event history?
\end{exercise}

\begin{exercise}[Foundation-model evaluation log]
Suppose a language model is evaluated by prompts, model completions, automatic
scores, human preference labels, and later deployment incidents.  Treat the
evaluation log as a data structure rather than as a single benchmark score.
Name the unit, the observed object, the missing or censored pieces, and one
dependence created by reusing prompts, raters, or model versions.
\end{exercise}

\begin{exercise}[Spatial omics neighborhood object]
In a spatial transcriptomics experiment, each spot or cell has a location,
gene-expression vector, segmentation quality score, and local neighborhood.
Write two possible empirical objects: one matrix-based and one neighborhood or
graph-based.  For each object, state one target that becomes natural and one
target that would be misleading if spatial dependence were ignored.
\end{exercise}

\begin{exercise}[Digital twin as a statistical object]
A manufacturing team uses a simulator as a digital twin of a production line.
The simulator records inputs, machine states, sensor traces, quality outcomes,
and simulated counterfactual settings.  Which parts of this object are
measured data, which are model outputs, and which are design choices?  State
one assumption under which simulator output could support a real-world target,
and one diagnostic that would challenge that assumption.
\end{exercise}

\begin{exercise}[Human review as data structure]
In a content-safety or clinical-triage system, model scores are sent to human
reviewers only when they cross a threshold.  Describe the observed data
structure, including unreviewed cases.  What target is identified from the
reviewed cases alone, and what target would require assumptions about the
unreviewed cases?
\end{exercise}

\section*{Sources and Further Reading}
\addcontentsline{toc}{section}{Sources and Further Reading}

The chapter's organizing idea is deliberately synthetic.  It combines the
observed-data viewpoint developed earlier in the book with the random-element
language of Chapter~6 and the stability language of Chapter~9.  The
counting-process example follows the survival-analysis tradition of
\citet{aalen1978nonparametric}, \citet{andersen1982cox},
\citet{andersen1993statistical}, and \citet{fleming1991counting}.  Missing
species and unseen mass trace through \citet{fisher1943species},
\citet{good1953population}, and \citet{efronThisted1976unseen}.  The
point-pattern entry point uses \citet{clarke1946poisson} and the broader
Poisson-process references \citet{kingman1993poisson} and
\citet{karr1986point}.  Determinantal point processes are included as the
repulsive counterpart to the Poisson baseline: \citet{macchi1975coincidence}
is the classical entry point, \citet{kuleszaTaskar2012dpp} develops the
machine-learning applications, and \citet{lavancier2015dpp} gives statistical
modeling and inference for spatial DPPs.  The high-dimensional testing row points to
\citet{landerBotstein1989mapping} and \citet{benjamini1995fdr}; later
chapters can decide how much of modern selective inference and empirical Bayes
belongs in the main text.  The network section uses \citet{wasserman1994social}
for social-network background, \citet{erdosRenyi1959random} for the classical
random-graph baseline, \citet{holland1983stochastic} for stochastic block
models, \citet{frank1986markov} for exponential random graph models,
\citet{lovasz2012large} for graphons, and
\citet{watts1998collective}, \citet{barabasi1999emergence}, and
\citet{newman2010networks} for complex-network mechanisms and applications.
The random-object and Wasserstein rows connect with
\citet{petersen2019frechet}, \citet{chen2023wasserstein}, and
\citet{villani2009optimal}; the intrinsic-dimension entry uses
\citet{levinaBickel2004intrinsic} as a compact bridge from manifold learning
intuition to likelihood-based statistical estimation.  Principal curves enter
only as geometric statistical targets: \citet{hastieStuetzle1989principal}
introduced the classical self-consistency formulation, \citet{kegl2000principal}
studied learning and design, and
\citet{cuiShao2024metricPrincipalCurve} gives a recent metric-based
one-dimensional-manifold version.  The vine-copula example uses Sklar's copula
representation \citep{sklar1959fonctions}, regular vines
\citep{bedfordCooke2002vines}, pair-copula constructions
\citep{aas2009pair}, and regular-vine selection and estimation
\citep{dissmann2013selecting}.  The single-cell row uses \citet{li2018scimpute},
\citet{sun2021scdesign2}, scDesign3, and \citet{cui2022scgtm} as recurring
examples of observation, simulation, and
structured biological trend estimation.
Algebraic statistics is represented only as a neighboring language; compact
entry points are \citet{pachterSturmfels2005algebraic} and
\citet{drtonSturmfelsSullivant2009lectures}.  Quantum statistics is likewise
only signposted; useful anchors are \citet{holevo1982probabilistic},
\citet{barndorffNielsenGillJupp2003quantum}, and
\citet{petz2008quantum}.

The closing examples connect the same grammar to applied machine learning,
biomedicine, and single-cell analysis.  Recommendation logs are treated as
policy-dependent observations in the spirit of
\citet{schnabel2016recommendations}.  Image benchmarks and dataset
documentation draw on ImageNet \citep{deng2009imagenet}, dataset-bias
diagnostics \citep{torralba2011unbiased,recht2019imagenet}, and datasheets
for datasets \citep{gebru2021datasheets}.  The electronic-health-record
example uses the MIMIC-III data resource \citep{johnson2016mimic} as a
representative clinical database and the target-trial/RWE/estimand literature
\citep{hernan2016targettrial,fda2018rweframework,ich2021e9r1} as the
statistical bridge from routine-care records to scientific claims.
Single-cell simulation is anchored by Splatter \citep{zappia2017splatter},
best-practice discussions of single-cell preprocessing and analysis
\citep{luecken2019singlecell}, and the scDesign family
\citep{sun2021scdesign2,song2024scdesign3}.

%% file: chapters/ch09_data_structures_to_targets.tex
\chapter{Statistical Targets: What Claim Is Being Made?}
\label{chap:data-structures-to-targets}
\conceptindexes{statistical targets, target functionals, data structures, observation law, target law, loss, risk, prediction}

\begin{tcolorbox}[
  enhanced,
  breakable,
  colback=chaptercream,
  colframe=bookblue!88!black,
  boxrule=0.72pt,
  arc=5pt,
  boxsep=1pt,
  left=1.0em,
  right=0.95em,
  top=0.82em,
  bottom=0.82em,
  before skip=0.55\baselineskip,
  after skip=1.0\baselineskip
]
\noindent\textbf{Chapter overview.}
This chapter is the book's hinge from data structure to estimand, loss, risk,
scientific claim, and interpretation. Chapter~7 asks what kind of empirical
object the data have become; this chapter asks what claim the object is
supposed to support. A dataset does not determine its own estimand: the same
observed object can support a descriptive summary, prediction risk, causal
contrast, scientific mechanism, or regulatory estimand. Later inference
chapters are easier to read once this target layer has been named.  The rule is
simple: a model is not chosen for data alone; it is chosen for a target under
an observation law.  To keep the chapter from simply retelling the opening
examples, most of the new anchors here come from industrial and
interdisciplinary systems where companies have had to make the target layer
explicit.
\end{tcolorbox}

Chapter~7 ended with a warning: a model is not enough until the target
functional and the interpretation it supports have been named.  This chapter
makes that warning operational.  No model before target: the statistical target
decides which summaries matter, which assumptions are relevant, and which
inference machinery is worth building.  In a classical parametric problem, the
target may look obvious because the notation has already chosen it.  If
\(X_1,\ldots,X_n\) are iid from \(P_\theta\), then one often says that the
target is \(\theta\).  But even there, the statement hides choices.  Is
\(\theta\) a scientific mechanism, a convenient index, a best approximation
inside a misspecified family, a prediction device, or a parameter required by a
regulatory question?

Modern data make the ambiguity impossible to ignore.  A cell-by-gene matrix
can be used to estimate a mean expression contrast, cluster cells, infer a
trajectory, simulate a perturbation, predict a phenotype, or plan a follow-up
experiment.  A product experiment at a platform company can target a short-run
lift in a metric, a long-run effect on retention, a guardrail for safety or
fairness, or a marketplace equilibrium.  An oncology real-world-evidence
dataset can target survival, real-world progression, treatment duration, or an
external-control comparison.  A molecular-prediction system can target a
benchmark structure loss, a binding pose, an uncertainty score, or a laboratory
decision.  Those are not different wordings of one target.  They are different
targets.

\section{Real-Data Anchors and Analysis Capsules}
\label{sec:ch09-real-data-anchors}
\conceptindexes{real-data anchors, datasets, recurring examples, analysis capsules}

The book uses real data in a controlled way.  A dataset should not appear only
because it is famous, available, or visually appealing.  It should do
mathematical work.  A real-data anchor earns its place when it clarifies at
least one of the book's recurring translations:
\[
\begin{aligned}
  \text{world}
  &\longrightarrow
  \text{observed record}
  \longrightarrow
  \text{data structure}
  \longrightarrow
  \text{target} \\
  &\longrightarrow
  \text{model or procedure}
  \longrightarrow
  \text{claim or interpretation}.
\end{aligned}
\]

There are three levels of real-data use.
\begin{description}[leftmargin=0pt,labelsep=0.55em,style=unboxed,font=\bookdescriptionlabelfont,itemsep=0.35\baselineskip]
\item[Opening-chapter bridges.]
Some classic examples are allowed to return, but only briefly.  London bombing
counts, missing vocabulary, historical climate proxies, and stylometry have
already done their main work in Chapters~1 and~2.  In this chapter they serve
as small reminders that a target must be named; they should not occupy the main
example budget again.

\item[Reproducible mini-analyses.]
Some examples should be rerun in companion code because the computation is
small, public, and pedagogically useful: an online-experiment metric
calculation, a financial-return volatility plot, a functional principal
component analysis (FPCA) of public curve data, a small survival curve, a compact
benchmark-prediction example, or an empirical-distribution example.

\item[Restricted or institutional anchors.]
Some data regimes are central even when raw records are unavailable or
inappropriate to distribute.  Platform experimentation logs, oncology
real-world-evidence records, ADNI, A4, regulated clinical-study analysis
datasets, proprietary agricultural-machine telemetry, and industrial
laboratory records can be discussed through protocols, standards, product
descriptions, published summaries, or benchmark descriptions.  The
observed-data mechanism is still real; the book simply does not claim to
reproduce restricted patient-level, user-level, or proprietary records.
\end{description}

The following table is the working policy.  It is meant to prevent two failure
modes at once: a purely theoretical book with no empirical traction, and a
data-analysis cookbook in which every chapter imports a new dataset without a
structural reason.

\begin{center}
\footnotesize
\setlength{\tabcolsep}{0.30em}
\renewcommand{\arraystretch}{1.15}
\begin{longtable}{@{}>{\raggedright\arraybackslash}p{0.18\linewidth}
                  >{\raggedright\arraybackslash}p{0.25\linewidth}
                  >{\raggedright\arraybackslash}p{0.25\linewidth}
                  >{\raggedright\arraybackslash}p{0.22\linewidth}@{}}
\caption{Empirical anchors and their roles in the statistical compass}\\
\toprule
Anchor & Real record & Result used in the book & Role in the compass \\
\midrule
\endfirsthead
\caption[]{Empirical anchors and their roles in the statistical compass (continued)}\\
\toprule
Anchor & Real record & Result used in the book & Role in the compass \\
\midrule
\endhead
Opening examples from Chapters~1--2 &
London cell counts, unseen vocabulary, historical proxies, and stylometric
features &
Used only as brief callbacks to published analyses &
Prevents example repetition: the chapter now asks what target those records
would support, not whether the stories are interesting. \\

Online experimentation platforms &
Randomized assignments, exposure logs, user or marketplace outcomes, guardrail
metrics, and experiment metadata &
Published industrial accounts from Microsoft-style experimentation, LinkedIn,
Booking.com, and Uber &
Average treatment effects, heterogeneous effects, interference, guardrail
metrics, and the distinction between metric lift and product value. \\

Real-world oncology evidence &
Electronic health records, abstracted progression assessments, treatment
lines, mortality links, and oncology chart review &
Flatiron Health work on EHR-derived real-world tumor-burden/progression
endpoints, alongside FDA and ICH estimand guidance &
Observed-data law versus target endpoint; how routine-care traces become
real-world survival or progression estimands. \\

Molecular structure prediction &
Protein sequences, templates, multiple-sequence alignments, structural
databases, ligand or nucleic-acid context, and benchmark splits &
DeepMind/Isomorphic Labs AlphaFold and AlphaFold~3 reports &
Prediction risk under a benchmark law; structure, confidence, and downstream
experimental decision are distinct targets. \\

Precision agriculture and robotics &
Field images, GPS locations, crop and weed detections, nozzle actions, machine
logs, and as-applied maps &
John Deere See \& Spray product documentation &
Action-policy targets: a weed classifier matters only through a loss involving
missed weeds, crop injury, chemical use, drift, cost, and timing. \\

Functional curves &
Growth, weather, gait, glucose, or pharmacokinetic curves observed on time or
dose grids &
Reproducible mini-analysis planned for a public curve dataset; published FDA
examples guide the mathematical reading
\citep{ramsaySilverman2005fda,yaoMullerWang2005pace} &
Random curves, covariance operators, FPCA scores, registration, and functional
targets. \\

Financial returns &
Daily prices, returns, rolling covariances, and portfolio losses &
Reproducible mini-analysis planned for a frozen public return series; published
RiskMetrics and stock-return results supply context
\citep{jpmorgan1996riskmetrics,fama1965behavior} &
Time order, volatility, martingale differences, model idealization, risk, and
backtesting. \\

Single-cell and spatial omics &
Count matrices, batches, donors, modalities, pseudotime, and spatial positions &
Published scImpute/scDesign3/scGTM examples now; a small public matrix can be
added as companion computation later
\citep{li2018scimpute,cui2022scgtm} &
Observation mechanism, generative simulation, imputation, empirical fields, and
structured biological targets. \\

Clinical event histories &
Trial visits, endpoints, censoring indicators, and analysis-ready time-to-event
records &
Endpoint definitions, ADaM time-to-event conventions, Kaplan--Meier/Cox-style
published reporting, and panel-count examples
\citep{cdisc2016adamtte,ich2021e9r1,wellnerZhang2000panel} &
Risk sets, censoring, compensators, martingale residuals, and the separation
between an observed event log and the clinical estimand. \\
\bottomrule
\end{longtable}
\end{center}

This policy also decides where code belongs.  The main text should show the
record, the fitted object, the result, and the interpretation.  Companion
scripts should carry data ingestion, cleaning, and plotting.  When a result is
taken from the literature, the text should say so openly.  Reproducibility is
not the same as rerunning every historical analysis; it is the discipline of
being clear about which claims were computed here, which were inherited from a
published source, and which depend on restricted or institutional records.

\subsection{Real-Data Analysis Capsules}
\label{sec:ch09-real-data-analysis-capsules}
\conceptindexes{real-data analysis capsules, capsule registry, analysis manifest, reproducible analysis}

An example becomes more than a motivating story when it is analyzed.  The book
should therefore use a small number of real-data analysis capsules.  A capsule
has four parts:
\begin{enumerate}
\item the observed record, or a published summary when the raw record is
restricted;
\item the computation: an estimate, curve, diagnostic, resampling result,
model check, or sensitivity analysis;
\item the mathematical object that makes the computation interpretable;
\item the feedback from the data back to the theory: the assumption, target,
loss, or information set that had to be sharpened.
\end{enumerate}
The last part prevents the analysis from becoming decoration.  A plot or table
should end by changing the statistical question, not merely by illustrating a
formula that had already been chosen.

Several public sources can support these capsules without making the manuscript
depend on proprietary records: Hillstrom's randomized e-mail campaign data for
an online-experiment style analysis \citep{hillstrom2008email}; Kenneth
French's data library for return and factor series \citep{frenchDataLibrary};
R package survival example data for event-history calculations
\citep{therneau2024survival}; the NOAA/WDS Ge et al. climate reconstruction for
a proxy-curve and Wasserstein-regression capsule
\citep{ge2013temperature}; and the 10x Genomics PBMC data for a compact
single-cell count-matrix analysis \citep{tenx2016pbmc3k}.  Other capsules can
use published summaries or standards rather than raw records, as with oncology
real-world evidence and analysis-ready clinical event-time data
\citep{griffith2019tumorBurden,cdisc2016adamtte,ich2021e9r1}.

\begin{center}
\footnotesize
\setlength{\tabcolsep}{0.26em}
\renewcommand{\arraystretch}{1.16}
\begin{longtable}{@{}>{\raggedright\arraybackslash}p{0.18\linewidth}
                  >{\raggedright\arraybackslash}p{0.24\linewidth}
                  >{\raggedright\arraybackslash}p{0.25\linewidth}
                  >{\raggedright\arraybackslash}p{0.23\linewidth}@{}}
\caption{Real-data analysis capsules and their feedback to theory}\\
\toprule
Capsule & Real record & Computation & Feedback to theory \\
\midrule
\endfirsthead
\caption[]{Real-data analysis capsules and their feedback to theory (continued)}\\
\toprule
Capsule & Real record & Computation & Feedback to theory \\
\midrule
\endhead
Randomized customer campaign &
Assignment, segment, visit, conversion, and spend fields from a public campaign
file &
Estimate intent-to-treat effects for revenue and conversion; compare subgroup
summaries and guardrail metrics &
Randomization identifies a contrast, but the target changes with the metric;
heterogeneity and multiple metrics move the problem toward Chapter~11. \\

Market-return series &
Daily returns or factor returns from a frozen public financial series &
Plot returns, rolling volatility, tail events, and simple risk exceedances &
LLN and CLT intuition must be checked against time dependence, changing
volatility, and tails; Chapter~16's process language becomes natural. \\

Quant-trading rule audit &
Returns, characteristics, quotes, order-book states, fills, costs, positions,
and risk limits from a reproducible market record or published backtest
protocol &
Define a signal-to-weight map, estimate net out-of-sample value, and record the
searched signal/asset/horizon/threshold class &
The target is not the winning backtest number.  It is the value of a selected
procedure under transaction costs and constraints; Chapter~11 treats the
searched class as an empirical field. \\

Event-history file &
Entry times, event indicators, censoring, covariates, and follow-up from a
public survival dataset or published trial summary &
Build risk-set tables, Kaplan--Meier curves, Cox-style contrasts, and a
censoring sensitivity check &
The target is not a column called time.  Time zero, risk set membership,
censoring, and intercurrent events define the estimand before asymptotics. \\

Functional-curve record &
Growth, weather, gait, glucose, or pharmacokinetic curves observed on grids &
Smooth curves, estimate a mean and covariance surface, and extract FPCA scores &
The random object lives in a function space; registration, sparse sampling, and
measurement noise decide whether covariance modes are interpretable. \\

Century-level climate distributions &
Ten decadal temperature anomalies inside each century from the Ge et al.
reconstruction &
Treat each century as an empirical distribution; regress quantile functions on
century midpoint in one-dimensional \(W_2\) geometry &
The target is a covariate-indexed distribution, not a mean curve alone.
Small within-century sample size makes the limitation part of the target
audit. \\

Single-cell count matrix &
UMI counts, genes, cells, batches, library sizes, and cell annotations from a
small public matrix &
Inspect library-size variation and zero rates; run normalization, low-rank
summaries, or a simple simulation diagnostic &
The observed matrix is not the biological state.  Capture, sequencing depth,
batch, and preprocessing are part of the observation mechanism. \\

Restricted industrial or regulatory record &
Published platform, oncology, agricultural, or clinical-study summaries when raw
records cannot be shared &
Reconstruct the target audit: unit, exposure, endpoint, loss, estimand,
uncertainty, and decision &
Restricted access does not make the example less statistical.  It makes the
target ledger explicit: which claim is computed here, inherited from a source,
or impossible without the raw record. \\
\bottomrule
\end{longtable}
\end{center}

The current companion analyses already give this idea numerical content.  The
results below are deliberately small: they are meant to teach how a statistical
object is named, checked, and then carried forward, not to replace a full
domain analysis.

\begin{center}
\scriptsize
\setlength{\tabcolsep}{0.22em}
\renewcommand{\arraystretch}{1.14}
\begin{longtable}{@{}>{\raggedright\arraybackslash}p{0.18\linewidth}
                  >{\raggedright\arraybackslash}p{0.29\linewidth}
                  >{\raggedright\arraybackslash}p{0.25\linewidth}
                  >{\raggedright\arraybackslash}p{0.20\linewidth}@{}}
\caption{Companion capsule results and their mathematical roles}\\
\toprule
Capsule result & Computed result & Bridge to theory & Cross-domain use \\
\midrule
\endfirsthead
\caption[]{Companion capsule results and their mathematical roles (continued)}\\
\toprule
Capsule result & Computed result & Bridge to theory & Cross-domain use \\
\midrule
\endhead
Ge et al. climate curve &
The cleaned reconstruction has 200 decadal points from years 5--1995.  The PLS
and PC reconstructions have correlation 0.860 and RMSE 0.121 deg C.  The
ensemble period contrast between 1921--2000 and 1321--1920 is 0.424 deg C. &
A historical proxy record becomes a functional target.  The period contrast
depends on the covariance of the reconstructed curve, not only on pointwise
error bars. &
Environmental history, functional data analysis, and uncertainty accounting. \\

Climate Wasserstein regression &
The same reconstruction gives 20 century-level empirical distributions, each
made from ten decadal anomalies.  On a 19-point quantile grid, the fitted
one-dimensional Wasserstein regression has integrated tangent-coordinate
\(R^2=0.029\). &
The Chapter~2 climate curve becomes a distribution-valued response.  The target
is a quantile-indexed slope \(\beta(u)\), not one scalar trend. &
Optimal transport geometry as a practical distributional-regression language. \\

Feller London bombing counts &
The grouped grid has 576 cells and 537 hits.  The fitted Poisson rate is
\(\hat\lambda=0.932\), and the grouped Pearson diagnostic is \(X^2=1.169\). &
Visible clustering is not enough.  The example turns a map into a point-process
null and a stability check against that null. &
Spatial process thinking for historical, safety, operations, and reliability
records. \\

PBMC3k single-cell capsule &
The fallback public matrix has 2,700 cells and 100 retained genes.  Median
cell-level zero rate is 0.03.  The small scDesign3 diagnostic gives mean
absolute zero-rate difference 0.00773 between reference and synthetic genes;
scGTM is skipped because PBMC3k has no real pseudotime field. &
The count matrix is not the biological state.  Sequencing depth, zeros, and
missing trajectory information are part of the observation mechanism and the
target audit. &
Genomics, simulation-based checking, and industrial-style data-quality
protocols. \\
\bottomrule
\end{longtable}
\end{center}

\begin{realdatacapsule}{Quant-trading rule audit}
\item[Data object.] A frozen market record: time-stamped returns
\((R_{t+1,j})\), signals or characteristics \(X_{t,j}\), tradability and cost
fields \(C_{t,j}\), and, for market making, quotes, fills, cancellations,
inventory, and cash.  The record should be public, synthetic, or described by a
published backtest protocol; famous firms are context, not evidence.
\item[Observation mechanism.] At decision time \(t\), the rule may use only the
information set \(\mathcal F_t\).  Listing rules, liquidity filters,
survivorship, corporate actions, venue access, latency, fill simulation, and
transaction-cost assumptions decide which market state is actually observed.
\item[Target.] For a fixed rule, the target is a net-performance functional
\[
  \psi(\theta)=E_P\{G_{t+1}(\theta)\},
\]
where \(G_{t+1}\) subtracts transaction costs and risk penalties from portfolio
return.  For a research pipeline, the target is the value of the selected
procedure produced by a documented walk-forward or train/test protocol, not
the largest in-sample backtest number.
\item[Model.] A score map \(f_\theta(X_{t,j})\), a constrained portfolio map
\(w_{\theta,t}=\Pi_{\mathcal C_t}\{A_t f_\theta(X_t)\}\), a pairs-trading
threshold rule, or a limit-order-book intensity model supplies the operational
rule class.  Chapter~11 writes these examples as an empirical field indexed by
\(\theta\).
\item[Uncertainty.] Report calendar-split performance, block or
dependence-aware resampling, and selection uncertainty for
\(\sup_{\theta\in\Theta}\hat\psi_n(\theta)\), rather than a standard error for
the winning rule as if it had been prespecified.
\item[Limitation.] The capsule does not infer proprietary systems used by
Citadel, Morgan Stanley, Renaissance Technologies, or Jane Street.  It studies
the public statistical problem: nonstationary markets, searched signal classes,
market impact, capacity limits, and feedback from trading decisions to future
observations.
\end{realdatacapsule}

\begin{realdatacapsule}{PBMC3k target audit}
\item[Data object.] A cell-by-gene UMI count matrix with cell barcodes, gene
labels, library sizes, and derived quality-control summaries
\citep{tenx2016pbmc3k}.
\item[Observation mechanism.] Capture efficiency, sequencing depth, alignment,
filtering, normalization, and batch handling determine which biological signal
survives into the matrix.
\item[Target.] A count-matrix target such as stable gene/cell summaries,
synthetic-data fidelity, or a downstream cluster/trajectory claim if the
required design variables exist.
\item[Model.] Negative-binomial or related count models, low-rank
representations, and scDesign3-style simulation checks connect the matrix to a
generative law.
\item[Uncertainty.] Bootstrap or simulation variation, zero-rate diagnostics,
library-size sensitivity, and gene-set perturbations report which conclusions
are stable.
\item[Limitation.] PBMC3k lacks a real pseudotime coordinate in the fallback
capsule, so scGTM-style trajectory inference is not a supported target here.
\end{realdatacapsule}

\section{Targets Are Named Maps}
\label{sec:ch09-targets-named-maps}
\conceptindexes{targets, named maps, statistical target functional, estimand, parameter}

The simplest formal language is functional language.  Let \(\mathcal P\) be a
class of probability laws for the object we want to read.  A statistical target
is a map
\[
  T:\mathcal P\longrightarrow \mathcal T,
\]
where \(\mathcal T\) is the space in which the answer lives.  The target might
be a number, vector, function, distribution, graph feature, survival curve,
treatment contrast, prediction target, or scientific summary.

\begin{definition}[Statistical target functional]
A statistical target functional is a specified map \(T\) from a law, model, or
structured population object to an answer space.  It becomes an estimand when
it is tied to a substantive question, an observation regime, and an
interpretation.
\end{definition}

This definition is intentionally plain.  It says that the target is not the
estimator and not the algorithm.  It is the object the estimator is trying to
recover, approximate, compare, or optimize.  The difference matters because
the same empirical object can be routed toward different targets.

\begin{example}[One dataset, several targets]
Suppose a trial records baseline covariates \(X\), treatment \(A\), event time
\(T\), censoring time \(C\), and the observed pair
\[
  O=(X,A,\min(T,C),\ind{T\le C}).
\]
Possible targets include a survival probability \(S_a(t)\), a restricted mean
survival time contrast, a cause-specific hazard, a marginal causal contrast, a
covariate-specific survival summary, or a prediction risk for an individual
risk score.  The observed object is shared, but the target changes the role of
censoring, covariates, time, and interpretation.
\end{example}

A target functional is therefore part of the statistical object, not a
decorative label attached after the analysis.  It decides which nuisance
features matter, which assumptions are identifying assumptions, which
convergence theorem is relevant, and which uncertainty statement should be
reported.

\begin{example}[Intrinsic dimension as a target audit]
\label{ex:ch09-intrinsic-dimension-target}
High-dimensional data often arrive in an ambient space \(\R^p\), but the
scientific or algorithmic claim may be that the data lie near a lower-dimensional
structure.  Intrinsic dimension is not automatically a property of the file.  It
becomes a target only after a metric, a local scale, and a sampling law are
named.

\citet{levinaBickel2004intrinsic} make this translation explicit.  Suppose
\(X_1,\ldots,X_n\in\R^p\) are embedded observations, locally generated from an
\(m\)-dimensional smooth distribution.  Fix a point \(x\), and let
\[
  N_x(t)=\sum_{i=1}^n \ind{\|X_i-x\|\le t}
\]
count neighbors within radius \(t\).  If the density is nearly constant near
\(x\), then for small \(t\)
\[
  \Expect\{N_x(t)\}\approx c_x t^m,
\]
where \(c_x\) absorbs the local density and unit-ball volume.  Equivalently,
the neighbor-counting process is locally approximated by a Poisson process with
rate
\[
  \lambda_x(t)=c_x m t^{m-1}.
\]
Let \(T_j(x)\) be the distance from \(x\) to its \(j\)th nearest neighbor.
Using the likelihood of this local point process and conditioning on the
\(k\)th neighbor radius gives the local estimator
\[
  \widehat m_k(x)
  =
  \left\{
    \frac{1}{k-1}\sum_{j=1}^{k-1}
    \log\frac{T_k(x)}{T_j(x)}
  \right\}^{-1}.
\]
A global estimate averages these local estimates over points, and often over a
small range of \(k\)'s.

The statistical kernel is the target audit.  The estimand is a local geometric
exponent under a chosen metric and scale, not an intrinsic truth detached from
sampling.  Small \(k\) reduces curvature and density bias but increases
variance; larger \(k\) stabilizes the estimate but may mix scales, densities,
or manifold pieces.  Thus the method is useful here not because it adds a new
machine-learning topic, but because it turns a geometric slogan into an
estimable target with assumptions.
\end{example}

\subsection{Target Audits in Industrial and Scientific Systems}
\label{sec:ch09-target-audits}
\conceptindexes{target audits, industrial systems, scientific systems, metric design}

Before choosing a theorem, it helps to perform a target audit.  The audit is
small: write the observed record, name the law the record appears to sample,
state the target as a functional of a law or structured object, and then say
one claim the record does \emph{not} support without extra assumptions.  The
exercise sounds almost bureaucratic.  In practice it is where many attractive
analyses either become statistical or quietly fall apart.  Chapters~1 and~2
already used historical and benchmark examples to train the eye.  The audits
below use industry and cross-disciplinary settings to show the same discipline
at work in systems that companies and scientific teams actually operate.

\begin{example}[Online experiments: a metric is not the whole target]
A product experiment records more than a treatment indicator and one outcome.
An observed unit may look like
\[
  O_i=(Z_i,E_i,Y_i,H_i,G_i),
\]
where \(Z_i\) is assignment, \(E_i\) is actual exposure, \(Y_i\) contains
clicks, sessions, revenue, or retention, \(H_i\) is pre-experiment history, and
\(G_i\) describes social, marketplace, or device context.  For a metric \(m\),
one target might be
\[
  \Delta_m
  =
  \Expect\{m(Y_i^{(1)},H_i,G_i)-m(Y_i^{(0)},H_i,G_i)\},
\]
under the randomization and use law of the experiment.  Industrial accounts of
experimentation at Microsoft-style platforms, LinkedIn, Booking.com, and Uber
make this target discipline explicit: assignment, exposure, logging,
guardrail metrics, and interference are part of the statistical object
\citep{kohavi2020trustworthy,xu2015infrastructure,
kaufman2017democratizing,uber2018experimentation}.

A short-run lift in a click metric is therefore not automatically the target
``product value,'' ``user welfare,'' or ``marketplace health.''  Those are
different targets, often with different time scales and losses.  A platform
can estimate \(\Delta_m\) cleanly while still needing extra assumptions to
translate that metric effect into a business, ethical, or long-run behavioral
claim.
\end{example}

\begin{example}[Real-world oncology endpoints: chart traces are not trial endpoints]
In oncology real-world-evidence studies, the observed record is built from
routine care rather than from a trial visit calendar.  A compressed notation is
\[
  O_i=(X_i,A_i,V_i,R_i,D_i),
\]
where \(X_i\) are baseline features, \(A_i\) is a treatment or treatment
sequence, \(V_i\) contains visits, imaging dates, clinician entries, and
abstracted chart information, \(R_i\) records real-world progression
abstractions, and \(D_i\) records death information when available.  A target
may be real-world overall survival, real-world progression, real-world
progression-free survival, treatment duration, or an external-control
contrast.  Flatiron Health's work on EHR-derived tumor-burden and progression
endpoints illustrates the industrial version of this target-audit problem
\citep{griffith2019tumorBurden}; FDA real-world-evidence and ICH estimand
guidance give the regulatory vocabulary \citep{fda2018rweframework,ich2021e9r1}.

The audit matters because a radiology report or chart abstraction is not the
same object as a protocol-scheduled RECIST assessment.  The target may be a
useful real-world endpoint, but it is not automatically the clinical-trial
endpoint, and it is not automatically a causal treatment effect.  The target
law, the observation law, and the abstraction process all have to be named
before survival or progression machinery can be trusted.
\end{example}

\begin{example}[Molecular structure prediction: benchmark accuracy is a target]
A molecular-prediction system turns biological and chemical objects into a
structured prediction problem.  For a protein or biomolecular complex, the
record may contain a sequence, templates, a multiple-sequence alignment,
ligand or nucleic-acid context, and a reference structure when one is available.
AlphaFold made the statistical target visible at scale: predict a structure and
associated confidence under a benchmark law built from available structural
biology data \citep{jumper2021alphafold}.  AlphaFold~3 extends the setting to
biomolecular interactions, where proteins, nucleic acids, small molecules, and
ions can all be part of the structured output \citep{abramson2024alphafold3}.

One target is expected structure loss under the benchmark population,
\[
  R_P(a)=\Expect_P\,d\{a(X),S\},
\]
where \(X\) is the available molecular information, \(S\) is the reference
structure, and \(d\) is a structural discrepancy.  That is not the same target
as predicting all conformational states in a cell, proving a biochemical
mechanism, estimating binding free energy, or guaranteeing that a designed
molecule will succeed in the laboratory.  The scientific excitement is real,
but the target audit keeps benchmark prediction, physical mechanism, and
experimental decision from collapsing into one sentence.
\end{example}

\begin{example}[Precision agriculture: the classifier is only part of the policy]
Computer-vision agriculture systems are a compact example of prediction
becoming action.  John Deere's See \& Spray documentation describes sprayers
that use camera and machine-learning systems to distinguish crops, weeds, and
bare ground and then target nozzle-level spraying decisions
\citep{johnDeere2026seeSpray}.  The observed record is not just an image.  It
includes location, crop stage, weed detections, machine speed, nozzle actions,
field conditions, and an as-applied map.

The statistical target is therefore closer to a policy
\[
  \pi^\star
  \in
  \argmin_{\pi\in\Pi}
  \Expect\,L\{\pi(X),W,C\},
\]
where \(X\) is the sensor record, \(W\) is the latent agronomic state, and
\(C\) includes chemical cost, drift, timing, crop injury, and operational
constraints.  Pixel-level weed accuracy may be an important subtarget, but it
is not the whole target.  The industrial question is whether the action rule
reduces the relevant loss in the field where it is used.
\end{example}

\section{Observation Law and Target Law}
\label{sec:ch09-observation-law-target-law}
\conceptindexes{observation law, target law, identification, transport, measurement map}

Chapter~5 separated the scientific world, designed full data, and observed
data.  The target often belongs to one layer, while estimation must operate on
another.  A useful shorthand is
\[
  W \longrightarrow O,
  \qquad
  P_W \longrightarrow P_O,
  \qquad
  T(P_W)\ \hbox{estimated through}\ P_O .
\]
The arrow from \(W\) to \(O\) may represent censoring, missingness,
coarsening, label creation, preprocessing, exposure, platform logging, or
selection.  If the target is a feature of \(P_W\) but the analyst observes
\(O\), then identification is the bridge from \(P_O\) back to \(T(P_W)\).

\begin{definition}[Identification]
A target \(T(P_W)\) is identified from an observed-data law \(P_O\) over a
model class if any two full-data laws in the class that induce the same
observed-data law also have the same target value.  In that case the target can
be written as an observed-data functional \(\tau(P_O)\).
\end{definition}

Identification is not estimation.  It says what would be knowable with
infinite observed data under the stated assumptions.  Estimation asks how well
finite data can approximate the identified functional.  The distinction is a
major source of clarity in missing-data, causal, survival, and platform-data
problems.

\begin{example}[Missingness changes the target unless it is named]
Let \(Y\) be a response and \(R\) indicate whether \(Y\) is observed.  The
scientific target might be \(E(Y)\), but the complete-case average estimates
\(E(Y\mid R=1)\) unless assumptions connect the two.  The missingness
mechanism is not a nuisance detail after the estimator has been chosen.  It is
part of the route from the observed law to the target law
\citep{rubin1976inference}.
\end{example}

\subsection{Loss, Risk, and Prediction Targets}
\label{sec:ch09-loss-risk-prediction}
\conceptindexes{loss, risk, prediction targets, decision rules, expected risk}

Prediction targets are easiest to state through loss.  Let \(a\) be a
prediction rule and let \(L(a,O)\) be the loss incurred when prediction \(a\)
meets observation \(O\).  The risk is
\[
  R_P(a)=\Expect_P L(a,O).
\]
The target may be the risk of a fixed rule, the risk difference between two
rules, or the optimal rule
\[
  a^\star\in\argmin_{a\in\mathcal A} R_P(a).
\]
This is the statistical core of prediction as a target: the number reported by
a model is meaningful only relative to a loss, population, information set, and
use.

\begin{example}[Prediction risk is mechanism-specific]
A classifier, structure predictor, or field-action rule estimates risk under
the law that created its validation record.  AlphaFold-style structure accuracy
is a target under a structural-biology benchmark law; a weed/crop vision model
inside a precision sprayer is useful only under the field, crop, weather, and
machine-action law where spraying decisions are made
\citep{jumper2021alphafold,johnDeere2026seeSpray}.  Benchmark risk can be a
useful target.  It is not automatically the risk under a clinic, laboratory,
farm, market, or platform whose population and measurement process differ.
\end{example}

The book uses this loss language to name prediction targets.  When testing or
decision language appears, it is only a local grammar for naming a reference
law, a discrepancy, or a target under a specified information set.
Modern machine learning often uses the same mathematics with vocabulary such as
validation risk, calibration, and generalization error.  The statistical
question is stable: what law is the risk taken under, what information is
available to the rule, and what discrepancy does the loss notice?

\begin{example}[Shapley attribution makes explanation a target]
A fitted predictor \(\hat f\) does not by itself determine an explanation.  To
allocate a prediction or loss value across features, one must also
choose a background law and a value function.  Let
\([p]=\{1,\ldots,p\}\), and let \(v(S)\) be the value assigned to the coalition
of features \(S\subseteq[p]\).  The Shapley attribution for feature \(j\) is
\[
  \phi_j(v)
  =
  \sum_{S\subseteq [p]\setminus\{j\}}
  \frac{|S|!(p-|S|-1)!}{p!}
  \{v(S\cup\{j\})-v(S)\}.
\]
The formula is an allocation rule, not a complete statistical interpretation.
The statistical choice is hidden in \(v\).  If
\(v(S)=\Expect\{\hat f(X)\mid X_S=x_S\}\), the attribution depends on the
background population and on how conditional laws are estimated.  If
\(v(S)=\Expect\{\hat f(x_S,X_{-S})\}\), the attribution uses a marginal
replacement rule that may combine feature values that rarely occur together.
If \(v(S)\) is a loss reduction, the same Shapley formula is explaining an
error target rather than a fitted score.  Thus feature
importance is not a property of a model alone.  It is a functional of a model,
a population law, a coalition rule, and a value scale
\citep{shapley1953value,strumbeljKononenko2014feature,
lundbergLee2017unified,aasJullumLoland2021dependent,
janzingMinoricsBlobaum2020feature}.
\end{example}

\subsection{Causal, Clinical, and Scientific Targets}
\label{sec:ch09-causal-clinical-scientific-targets}
\conceptindexes{causal targets, clinical targets, scientific targets, treatment effects, estimands}

Not every target is a prediction target.  Some targets describe a scientific
population; some compare possible interventions; some define a regulatory
question.  It is useful to separate three
families.

\begin{description}[leftmargin=0pt,labelsep=0.55em,style=unboxed,font=\bookdescriptionlabelfont,itemsep=0.35\baselineskip]
\item[Descriptive targets.]
These summarize a law or object: a mean, quantile, distribution function,
survival curve, covariance operator, missing mass, graph density, clustering
coefficient, or functional principal component.

\item[Causal and clinical targets.]
These compare worlds under different interventions or treatment strategies.
Potential-outcome, causal-diagram, target-trial, and clinical-estimand
language all force the same discipline: name the population, intervention,
endpoint, intercurrent-event strategy, summary measure, and identifying
assumptions before choosing the estimator
\citep{pearl1995causal,hernan2016targettrial,ich2021e9r1}.

\item[Prediction and risk targets.]
These evaluate forecasts, scores, classifiers, or rankings
under a specified distribution and loss.  The target is not merely a fitted
function; it is the performance of that function under a stated use.
\end{description}

These families can overlap.  A clinical prediction model may support a
treatment comparison.  A causal effect may be reported as a risk difference.  A
scientific target may be chosen because it informs a future experiment.  The
point of the taxonomy is not to sort every analysis into a pure category.  The
point is to force the analyst to name the role the target is playing.

\begin{example}[Instrumental variables as target discipline]
\conceptindexes{instrumental variables, Wald estimand, local average treatment effect, identification}
Instrumental variables belong here because they are not just another regression
column.  Suppose \(Z\in\{0,1\}\) is an encouragement, assignment, distance,
policy rule, or other source of variation; \(A\) is an exposure or treatment;
and \(Y\) is the outcome.  If \(A\) is confounded, the coefficient of \(A\) in an
ordinary regression of \(Y\) on \(A\) need not target a causal effect.  An
instrumental-variable target instead asks whether variation in \(Z\) changes
\(A\), affects \(Y\) only through \(A\), and is as good as random for the
potential outcomes under the target population.

In the simplest binary case, the Wald ratio
\[
  \psi_{\mathrm{IV}}
  =
  \frac{\Expect(Y\mid Z=1)-\Expect(Y\mid Z=0)}
       {\Expect(A\mid Z=1)-\Expect(A\mid Z=0)}
\]
is not a descriptive slope.  Under relevance, independence, exclusion, and a
monotonicity condition, it is read as a local average treatment effect for
compliers \citep{angristImbensRubin1996iv}.  Without those assumptions it is
only a ratio of two observational contrasts.  This is why the target chapter
must come before the estimating-equation chapter: Chapter~13 can solve the
moment equation, but this chapter decides what the moment identifies.
\end{example}

\section{From Targets to Mathematical Machinery}
\label{sec:ch09-targets-to-machinery}
\conceptindexes{mathematical machinery, target-to-method translation, asymptotic tools, likelihood tools}

Once the target is named, the later mathematical machinery has a job.  The
following table is a local map for the rest of the book.

\begin{center}
\small
\setlength{\tabcolsep}{0.36em}
\renewcommand{\arraystretch}{1.16}
\begin{longtable}{@{}>{\raggedright\arraybackslash}p{0.23\linewidth}
                  >{\raggedright\arraybackslash}p{0.28\linewidth}
                  >{\raggedright\arraybackslash}p{0.35\linewidth}@{}}
\caption{Target pressures and the later mathematical machinery}\\
\toprule
Target pressure & Typical empirical object & Later machinery \\
\midrule
\endfirsthead
\caption[]{Target pressures and the later mathematical machinery (continued)}\\
\toprule
Target pressure & Typical empirical object & Later machinery \\
\midrule
\endhead
Population mean, proportion, fixed risk &
Empirical average \(P_n f\) &
Laws of large numbers and concentration stabilize the fixed translation. \\

Distribution, curve, path, random measure &
Empirical distribution function, process, point measure, survival curve &
Weak convergence names the state space and topology for object-level limits. \\

Searched criterion or flexible model class &
\(\{P_n f:f\in\mathcal F\}\), likelihood surface, loss landscape &
Empirical-process theory controls searching, selection, and uniform error. \\

Parameter, estimating equation, optimality condition &
Argmax, argmin, score, moment equation, calibration equation &
M- and Z-estimation convert stable empirical objects into selected locations. \\

Reported functional of an estimator &
Quantile, odds ratio, survival contrast, FPCA score, Wasserstein contrast &
Delta methods and influence functions propagate local uncertainty. \\

Deployed action or decision policy &
Experiment assignment, recommender exposure, field spray decision, laboratory
next-action log &
Loss, causal identification, empirical-process control, and stochastic-process
language separate prediction accuracy from intervention value. \\

Time-ordered event-history target &
Counting process, filtration, risk set, treatment history, observation log &
Stochastic-process language keeps information, events, censoring, and updating
in temporal order. \\
\bottomrule
\end{longtable}
\end{center}

The table explains why target language belongs before asymptotics.  A theorem
does not know which scientific question we meant to ask.  It stabilizes,
transports, or approximates a mathematical object after that object has been
chosen.  Naming the target makes the theorem useful rather than impressive.

\subsection{A Reading Protocol for the Technical Chapters}
\label{sec:ch09-reading-protocol}
\conceptindexes{reading protocol, technical chapters, target-first reading}

The rest of the book can be read with a short protocol:
\[
  \begin{aligned}
  \hbox{data structure}
  &\longrightarrow
  \hbox{target}
  \longrightarrow
  \hbox{empirical procedure} \\
  &\longrightarrow
  \hbox{uncertainty}
  \longrightarrow
  \hbox{scientific interpretation}.
  \end{aligned}
\]
For each technical chapter, ask which arrow is being formalized.  If a proof is
long, ask what target pressure made the proof necessary.  If an example is
memorable, ask which target it clarifies.  If an estimator looks natural, ask
what it estimates under the observation law actually available.

This protocol also sets boundaries.  The weak-convergence chapter is not a
full treatise on probability in metric spaces; it supplies the object-level
limit language needed once the target is a curve, path, measure, or field.  The
empirical-process chapter is not a monograph; it supplies the uniform language
needed once a target is found by searching.  The influence-function chapter is
not all of semiparametric theory; it supplies the local language needed once a
reported target is a functional of a law.  The continuous-time chapter is not
all of stochastic calculus; it supplies the information-through-time language
needed once targets depend on event histories and observation schedules.

\section{Exercises}
\label{sec:ch09-exercises}
\conceptindexes{target exercises, risk exercises, identification exercises}

\begin{exercise}[One object, three targets]
Choose one observed data object from an online experiment, an EHR-derived
oncology study, a molecular-prediction benchmark, or a precision-agriculture
system.  Write three distinct targets: one descriptive, one predictive or
risk-based, and one causal, clinical, operational, or scientific-mechanism
target.  For each target, state what law or object the target is a functional
of.
\end{exercise}

\begin{exercise}[Observation law versus target law]
Give an example in which the observed-data law \(P_O\) differs from the
scientific law \(P_W\).  State a target \(T(P_W)\), then write one assumption
under which it could be represented as a functional of \(P_O\).  You may use a
routine-care oncology endpoint, a platform exposure log, a field-spraying
telemetry stream, or another system you know well.
\end{exercise}

\begin{exercise}[Loss makes prediction specific]
For a binary classifier in medicine, agriculture, or platform safety, compare
squared-error loss, misclassification loss, and a loss that makes false
negatives five times as costly as false positives.  Explain how the target
rule changes even when the data structure is unchanged.
\end{exercise}

\begin{exercise}[Target-to-machinery map]
Pick a target from an online experiment, a clinical event-history problem, a
single-cell experiment, a molecular-prediction system, or a precision
agriculture policy.  Decide which later chapter supplies the first
mathematical tool you would need: concentration, weak convergence, empirical
processes, M/Z-estimation, influence functions, or filtrations.  Justify the
choice in terms of the target, not the method name.
\end{exercise}

\begin{exercise}[Benchmark target versus deployment target]
For a medical-image model or language-model assistant, compare a benchmark
accuracy target with a deployment target.  Specify the law under which the
benchmark target is defined, the law under which deployment performance should
be judged, and one mechanism by which the two laws can differ.
\end{exercise}

\begin{exercise}[Off-policy value target]
A recommender system log contains histories \(H_t\), actions \(A_t\), logging
probabilities \(b_t(A_t\mid H_t)\), and outcomes \(Y_t\).  Define the value of
a new policy \(\pi\) as a target under \(P^\pi\).  Then state the assumptions
needed to represent that value as a functional of the observed logging law
\(P^b\).
\end{exercise}

\begin{exercise}[Fairness as a target functional]
For a risk score used in lending, hiring, or hospital triage, write two
possible fairness targets: one based on error rates conditional on a group
variable and one based on calibration conditional on a group variable.  Explain
why both are target functionals, not merely software diagnostics.
\end{exercise}

\begin{exercise}[Net benefit for clinical AI]
A diagnostic rule sends a patient to treatment when \(\hat p(X)>c\).  Define a
loss or utility that gives different weights to false positives and false
negatives, and write the corresponding risk target.  How does the target
change when the clinical threshold \(c\), treatment harm, or disease
prevalence changes?
\end{exercise}

\begin{exercise}[When the target moves]
Give an example in which deploying a model changes the future target law: a
fraud detector, a recommendation policy, a sepsis alert, an adaptive trial, or
an autonomous laboratory.  Write \(P\) for the pre-deployment law and
\(P^\pi\) for the law under the deployed rule.  Which target was estimated,
and which target is needed after deployment?
\end{exercise}

\section*{Sources and Further Reading}
\addcontentsline{toc}{section}{Sources and Further Reading}

The chapter uses ``target'' broadly, covering estimands, functionals, risks,
and scientific summaries.  The classical examples from Chapters~1 and~2 remain
available as callbacks, but the main audits here are deliberately industrial
and interdisciplinary.  The online-experiment discussion draws on
\citet{kohavi2020trustworthy}, LinkedIn's large-scale experimentation account
\citep{xu2015infrastructure}, Booking.com's experimentation paper
\citep{kaufman2017democratizing}, and Uber's engineering description
\citep{uber2018experimentation}.  The real-world oncology example uses
Flatiron Health's EHR-derived progression endpoint work
\citep{griffith2019tumorBurden}.  The molecular-prediction discussion points to
AlphaFold and AlphaFold~3 \citep{jumper2021alphafold,abramson2024alphafold3}.
The precision-agriculture example cites John Deere product documentation to
document the deployed system, not as an independent performance validation
\citep{johnDeere2026seeSpray}.  The missing-data distinction between
observed and full data follows the line of \citet{rubin1976inference}.  The
clinical estimand language is aligned with ICH E9(R1)
\citep{ich2021e9r1}, while target-trial language is represented by
\citet{hernan2016targettrial}.  The causal target discussion points to
\citet{pearl1995causal}.  The loss and risk vocabulary is used only to name
prediction targets.  The Shapley attribution
example starts from the cooperative-game allocation rule of
\citet{shapley1953value}; modern prediction-attribution versions include
\citet{strumbeljKononenko2014feature} and \citet{lundbergLee2017unified}, with
dependent-feature and causal cautions emphasized by
\citet{aasJullumLoland2021dependent} and
\citet{janzingMinoricsBlobaum2020feature}.  The chapter is meant as a hinge.
Hypothesis testing, decision theory, prediction, explainable machine learning,
and regulatory clinical-trial methodology remain background dialects rather
than separate destinations.  The intrinsic-dimension example follows
\citet{levinaBickel2004intrinsic}, whose nearest-neighbor likelihood turns
low-dimensional manifold intuition into a statistical target involving metric,
scale, local density, and a point-process approximation.

%% file: chapters/ch07_laws_concentration.tex
\chapter{Laws of Large Numbers and Concentration: Stability of Empirical Summaries}
\label{chap:laws-concentration}
\conceptindexes{stability, laws of large numbers, concentration inequalities, tail inequalities, empirical averages, heavy tails, martingales}

\begin{tcolorbox}[
  enhanced,
  breakable,
  colback=chaptercream,
  colframe=bookblue!88!black,
  boxrule=0.72pt,
  arc=5pt,
  boxsep=1pt,
  left=1.0em,
  right=0.95em,
  top=0.82em,
  bottom=0.82em,
  before skip=0.55\baselineskip,
  after skip=1.0\baselineskip
]
\noindent\textbf{Chapter overview.}
This chapter asks when an empirical translation is stable enough to support a
target, model, or decision. Markov and Chebyshev inequalities give the first
tail grammar; laws of large numbers explain why averages settle down;
concentration inequalities explain how large a finite-sample departure must be
before it should surprise us. Hoeffding, Azuma, Bernstein, random projections,
quantiles, kernels, Kaplan--Meier survival surfaces, and empirical distribution
functions are treated as versions of one question: when may a noisy trace be
trusted as a population signal?
\end{tcolorbox}

The eye notices accident before it notices stability.  A few observations can
make a treatment arm look unusually successful, a hospital look unusually
dangerous, a phrase look unusually common, or a cell type look newly discovered.
Chapter 2 warned us that visible patterns are not automatically mechanisms:
random points cluster, rare words disappear, and the unseen leaves marks in the
seen.  The present chapter adds the complementary lesson.  Some random
summaries are disciplined by repetition.  They fluctuate, but their fluctuation
can be bounded, compared, and eventually trusted.

Think of a running clinical dashboard.  On Monday the response rate is
impressive, on Tuesday a few late outcomes make it ordinary, and by the next
month the same number begins to look less like a mood swing and more like a
feature of the treatment population.  The mathematical question is not whether
the first display was noisy.  It was.  The question is what kind of repetition
turns that noise into a stable translation.

The basic object is almost embarrassingly simple.  If
$X_1,\ldots,X_n$ are observations and $f$ is a measurable function, define
\[
  P_n f=\frac1n\sum_{i=1}^n f(X_i),
  \qquad
  P f=\Expect f(X).
\]
The notation is deliberately suggestive.  $P$ is the population law, while
$P_n$ is the empirical law that places mass $1/n$ at each observed point.  The
quantity $P_n f$ is a statistic, but it is also a sentence: ``the data say that
the population average of $f$ is approximately this.''  Large-sample theory
begins by asking when that sentence becomes reliable.

The surrounding chapters keep returning to this sentence.  Chapter~7
organizes modern data structures by object, observation mechanism, empirical
object, structure, functional, and decision.  Chapter~8 names the target layer:
estimands, risks, policies, and scientific claims.  Chapter~10 asks how whole
random objects, such as paths and empirical curves, converge.  Chapter~11 asks
what happens when $f$ ranges over a whole class.  Chapter~13 uses such uniform
stability to locate peaks and roots.  Chapter~15 asks how the remaining error
is read once the estimator has found its target.  The present chapter is the
fixed-$f$ foundation underneath that later movement.

\begin{realdatacapsule}{Market-return stability check}
\item[Data object.] Daily returns or factor returns from a frozen financial
series, with calendar dates, missing trading days, and portfolio losses.
\item[Observation mechanism.] Prices are observed only when markets trade,
then cleaned, adjusted, and differenced into returns; the sampling clock is
part of the record.
\item[Target.] A stable mean return, volatility, tail probability, or risk
exceedance rate over a stated period.
\item[Model.] Empirical averages, Bernstein-style bounds for bounded or
truncated losses, block resampling, and conditional-volatility models represent
different stability claims.
\item[Uncertainty.] Concentration diagnostics, rolling-window variation,
exceedance counts, and block bootstrap intervals show whether finite-sample
fluctuation is compatible with the reported target.
\item[Limitation.] Heavy tails, volatility clustering, and regime change make
iid concentration a baseline check rather than a complete risk analysis.
\end{realdatacapsule}

\section{Tail Inequalities and Stability}
\label{sec:ch07-tail-grammar}
\conceptindexes{tail inequalities, Markov's inequality, Chebyshev's inequality, stability bounds}

Before averages can be trusted, deviations need a language.  The most basic
sentences in that language convert an amount of integrability into a bound on
the probability of being far away.  They are deliberately crude.  Their value
is that they require little structure and reveal exactly what kind of moment
control is being spent.

\begin{theorem}[Markov's inequality]
If $Y\ge0$ and $\Expect Y<\infty$, then for every $t>0$,
\[
  \Prob(Y\ge t)\le \frac{\Expect Y}{t}.
\]
\end{theorem}

\noindent\textit{Proof.}
Since $Y\ge t\ind{Y\ge t}$, taking expectations gives
\[
  \Expect Y\ge t\Prob(Y\ge t).
\]
Dividing by $t$ proves the claim.
\qedmark

Markov's inequality is weak only if it is asked to do more than the available
information permits.  A first moment cannot show the exact shape of the
tail; it can only say that a large tail event must be paid for by enough mass
in the expectation.

\begin{theorem}[Chebyshev's inequality]
If $X$ has mean $\mu$ and variance $\sigma^2<\infty$, then for every
$\varepsilon>0$,
\[
  \Prob(|X-\mu|\ge\varepsilon)
  \le
  \frac{\sigma^2}{\varepsilon^2}.
\]
\end{theorem}

\noindent\textit{Proof.}
Apply Markov's inequality to the nonnegative variable $(X-\mu)^2$:
\[
  \Prob(|X-\mu|\ge\varepsilon)
  =
  \Prob\{(X-\mu)^2\ge\varepsilon^2\}
  \le
  \frac{\Expect(X-\mu)^2}{\varepsilon^2}.
\]
\qedmark

Chebyshev is the first stability inequality for averages because variances add
under orthogonality.  It does not need a normal approximation, a smooth
density, or a tail model.  It only needs second-order size.

The same Markov idea also has an exponential form.  If $Z$ is any random
variable and $\lambda>0$, then
\[
  \Prob(Z\ge t)
  =
  \Prob(e^{\lambda Z}\ge e^{\lambda t})
  \le
  e^{-\lambda t}\Expect e^{\lambda Z}.
\]
Equivalently, if
\[
  \log\Expect e^{\lambda Z}\le \phi(\lambda)
\]
on a useful range of $\lambda>0$, then
\[
  \Prob(Z\ge t)
  \le
  \inf_{\lambda>0}
  \exp\{-\lambda t+\phi(\lambda)\}.
\]
This is the engine behind Hoeffding, Bernstein, Chernoff, and many martingale
concentration bounds: prove a cumulant-generating-function inequality, then
optimize the exponential Markov bound.

\begin{tcolorbox}[
  enhanced,
  breakable,
  colback=noteback,
  colframe=bookgold!75!black,
  boxrule=0.55pt,
  arc=4pt,
  boxsep=1pt,
  left=0.95em,
  right=0.9em,
  top=0.7em,
  bottom=0.7em,
  before skip=0.9\baselineskip,
  after skip=1.0\baselineskip
]
\noindent\textbf{Toolbox boundary.}
This chapter uses only the inequalities needed to read LLNs and finite-sample
deviation bounds.  Berry--Esseen bounds answer a different question: how fast
a standardized sum is approximated by a normal law.  Minkowski's inequality is
an $L^p$ norm tool, useful when moment bounds are built in normed spaces.
Burkholder--Davis--Gundy inequalities belong to the continuous-time martingale
toolkit, where the size of a martingale is compared with its quadratic
variation.  Those ideas are important later, but the present chapter needs
only their discrete-time shadow: conditional centering plus predictable
variation.
\end{tcolorbox}

\section{Empirical Averages and Laws of Large Numbers}
\label{sec:ch07-empirical-averages}
\conceptindexes{empirical averages, weak law of large numbers, strong law of large numbers, stability}

A data analysis often begins with a chosen feature.  In a clinical study,
$f(X)$ may be a response indicator, a transformed biomarker, a loss under a
prediction rule, or a score contribution.  In a spatial problem, it may be the
indicator that a point falls in a particular region.  In a missing-species
problem, it may record whether a type appears once.  The statistical act is to
turn many local observations of $f(X_i)$ into a population statement $P f$.

\begin{definition}[Weak and strong stability]
Let $X_1,X_2,\ldots$ be random variables and let $a_n>0$ be a normalizing
sequence with $a_n\to\infty$.  A normalized sum
\[
  \frac{1}{a_n}\sum_{i=1}^n (X_i-\mu_i)
\]
is said to satisfy a weak law if it converges to $0$ in probability, and a
strong law if it converges to $0$ almost surely.  In the iid average case,
$a_n=n$ and $\mu_i=\Expect X_i=\mu$.
\end{definition}

The distinction is practical.  Convergence in probability says that for a large
fixed experiment the chance of a substantial error is small.  Almost sure
convergence says that along the whole infinite sequence of experiments the
errors eventually settle down, except on a null set.  A statistician usually
uses a finite $n$, but the stronger statement often gives the cleaner story:
the empirical translation does not merely work often; it eventually stays
working along almost every data stream.

\begin{theorem}[Chebyshev weak law]
Let $X_1,\ldots,X_n$ be uncorrelated random variables with common mean $\mu$
and variances bounded by $C<\infty$.  Then
\[
  \frac1n\sum_{i=1}^n X_i \toP \mu .
\]
\end{theorem}

\noindent\textit{Proof.}
Let $\bar X_n=n^{-1}\sum_i X_i$.  Then
\[
  \Var(\bar X_n)
  =
  \frac1{n^2}\sum_{i=1}^n\Var(X_i)
  \le \frac{C}{n}.
\]
Chebyshev's inequality gives
\[
  \Prob\{|\bar X_n-\mu|>\varepsilon\}
  \le
  \frac{C}{n\varepsilon^2}\to0 .
\]
\qedmark

This proof is almost too clean, and that is why it is useful.  It shows what
the simplest law of large numbers is buying: averaging reduces variance.  The
theorem does not care whether the variables are normal, whether their
distribution is symmetric, or whether the statistic has a pleasing formula.  It
only needs enough second-moment control to make the random errors cancel.

\begin{example}[Sample proportions]
If $Z_i$ are iid observations and $A$ is a measurable event with
$P\ind{A}=p$, then
\[
  P_n\ind{A}=\frac1n\sum_{i=1}^n\ind{A}(Z_i)
  \toP p.
\]
\noindent\textit{Proof.}
Put $X_i=\ind{A}(Z_i)$.  Then $\Expect X_i=p$ and
$\Var(X_i)=p(1-p)\le1/4$.  The Chebyshev weak law gives, for every
$\varepsilon>0$,
\[
  \Prob\{|P_n\ind{A}-p|>\varepsilon\}
  \le
  \frac{p(1-p)}{n\varepsilon^2}
  \to0.
\]
The empirical proportion is therefore a stable translation of the event
probability.  In Chapter 2, counts in boxes looked
irregular; here the average count over repeated independent trials becomes
regular.
\qedmark
\end{example}

\begin{example}[Empirical risk]
Let $\ell_\theta(X)$ be a loss function for a prediction rule indexed by
$\theta$.  For fixed $\theta$, the empirical risk
\[
  P_n\ell_\theta
  =
  \frac1n\sum_{i=1}^n\ell_\theta(X_i)
\]
estimates the population risk $P\ell_\theta$.  The law of large numbers tells
us that each fixed rule can be evaluated by data.
\noindent\textit{Proof.}
If $P|\ell_\theta|<\infty$, Khintchine's weak law applied to
$U_i=\ell_\theta(X_i)$ gives
\[
  P_n\ell_\theta-P\ell_\theta\toP0.
\]
If $P\ell_\theta^2<\infty$, the simpler Chebyshev proof gives the explicit
bound
\[
  \Prob\{|P_n\ell_\theta-P\ell_\theta|>\varepsilon\}
  \le
  \frac{\Var\{\ell_\theta(X)\}}{n\varepsilon^2}.
\]
This does not yet show that the rule chosen after scanning many
$\theta$'s is reliable.  That uniform question is the next chapter's subject.
\qedmark
\end{example}

\subsection{Truncation and the Cost of Heavy Tails}
\label{sec:ch07-truncation}
\conceptindexes{truncation, heavy tails, Khintchine weak law, Kolmogorov strong law}

The Chebyshev proof is a small machine with a visible fuel source: finite
variance.  But statistics often meets variables whose variance is large,
unstable, or infinite.  A few patients may have extremely long hospital stays,
a few claims may dominate a cost distribution, a few expression measurements
may be very large, and a few games may have outrageous payoffs.  The right
question is not whether averaging always works.  It is which part of the tail
must be controlled for the average to become a population statement.

The standard repair is truncation.  Write
\[
  Y_{n,i}=X_i\ind{|X_i|\le n}.
\]
The sum of the $Y_{n,i}$ has bounded pieces, while the discarded tail is
controlled separately.  This is not a technical trick alone.  It is a way of
making the story explicit: the average is stable if the extreme observations
are rare enough not to carry macroscopic mass.

\begin{theorem}[Khintchine's weak law]
Let $X_1,X_2,\ldots$ be iid with $\Expect |X_1|<\infty$ and
$\mu=\Expect X_1$.  Then
\[
  \frac1n\sum_{i=1}^n X_i \toP \mu .
\]
\end{theorem}

\noindent\textit{Proof.}
It is enough to prove the result for $\mu=0$, after replacing $X_i$ by
$X_i-\mu$.  Define the triangular truncation
\[
  Y_{n,i}=X_i\ind{|X_i|\le n},
  \qquad
  R_{n,i}=X_i-Y_{n,i}=X_i\ind{|X_i|>n}.
\]
First control the truncated centered average.  Since $Y_{n,1},\ldots,Y_{n,n}$
are iid for each fixed $n$,
\[
  \Var\left(\frac1n\sum_{i=1}^nY_{n,i}\right)
  =
  \frac1n\Var(Y_{n,1})
  \le
  \frac1n\Expect Y_{n,1}^2.
\]
Let $Z=|X_1|$.  The tail integral identity gives
\[
  \Expect\{Z^2\ind{Z\le n}\}
  \le
  \int_0^n 2t\,\Prob(Z>t)\,dt .
\]
We claim that
\[
  \frac1n\int_0^n 2t\,\Prob(Z>t)\,dt\to0.
\]
Indeed, fix $M<n$.  Then
\[
  \frac1n\int_0^n2t\,\Prob(Z>t)\,dt
  \le
  \frac{M^2}{n}
  +
  2\int_M^\infty\Prob(Z>t)\,dt.
\]
First let $n\to\infty$ and then $M\to\infty$.  Since
$\Expect Z=\int_0^\infty\Prob(Z>t)\,dt<\infty$, the right side tends to zero.
Thus
\[
  \Var\left(\frac1n\sum_{i=1}^nY_{n,i}\right)\to0.
\]
Chebyshev's inequality implies
\[
  \frac1n\sum_{i=1}^n\{Y_{n,i}-\Expect Y_{n,1}\}\toP0.
\]
The truncation bias vanishes by dominated convergence:
\[
  \Expect Y_{n,1}\to\Expect X_1=0.
\]
Finally, the discarded tail is negligible in probability:
\[
  \Prob\left\{\frac1n\sum_{i=1}^n|R_{n,i}|>\varepsilon\right\}
  \le
  \frac1\varepsilon\Expect\{|X_1|\ind{|X_1|>n}\}
  \to0.
\]
Combining the centered truncated part, the bias, and the tail gives
$n^{-1}\sum_{i=1}^nX_i\toP0$.
\qedmark

The theorem says that finite mean is enough for weak stability of iid averages.
It also reveals where the theorem can fail.  If the tail is so heavy that the
sample is repeatedly dominated by rare gigantic observations, the empirical
average may refuse to translate the population in the ordinary $n^{-1}$ scale.

\begin{example}[The St. Petersburg Paradox]
In the St. Petersburg game, a payoff of $2^k$ occurs with probability
$2^{-k}$, $k=1,2,\ldots$.  Let $X$ denote one payoff and let
$S_n=X_1+\cdots+X_n$ for iid plays.

\noindent\textit{Feller's normalization.}
A single play has infinite mean, because every dyadic level contributes exactly
one unit:
\[
  \Expect X
  =
  \sum_{k=1}^\infty 2^k\Prob(X=2^k)
  =
  \sum_{k=1}^\infty 1
  =
  \infty .
\]
But a block of $n$ plays has a stable typical scale.  Put
$m_n=\lfloor\log_2 n\rfloor$ and
\[
  b_n=\lfloor \log_2(nm_n)\rfloor .
\]
Truncate only beyond the level \(2^{b_n}\), which is of order \(n\log_2 n\):
\[
  Z_{n,i}=X_i\ind{X_i\le 2^{b_n}}.
\]
Then
\[
  \Expect Z_{n,i}=b_n,
  \qquad
  \Expect Z_{n,i}^2
  =
  \sum_{k=1}^{b_n}4^k2^{-k}
  =
  2^{b_n+1}-2
  \le 2nm_n.
\]
Since \(b_n/m_n\to1\),
\[
  \Expect\left\{\frac{1}{nm_n}\sum_{i=1}^n Z_{n,i}\right\}
  =
  \frac{b_n}{m_n}
  \to1,
\]
and
\[
  \Var\left(\frac{1}{nm_n}\sum_{i=1}^n Z_{n,i}\right)
  \le
  \frac{\Expect Z_{n,i}^2}{nm_n^2}
  \le
  \frac{2}{m_n}
  \to0.
\]
Thus
\[
  \frac{1}{nm_n}\sum_{i=1}^n Z_{n,i}
  \to
  1
  \qquad\text{in probability.}
\]
The truncation almost never changes the sample sum, because
\[
  \Prob\left\{\max_{1\le i\le n}X_i>2^{b_n}\right\}
  \le
  n2^{-b_n}
  \le
  \frac{2}{m_n}
  \to0.
\]
Therefore \(S_n=\sum_i Z_{n,i}\) with probability tending to one.  Since
\(m_n\sim\log_2 n\),
\[
  \frac{S_n}{n\log_2 n}
  \to
  1
  \qquad\text{in probability.}
\]
So the fair price is not a fixed price per play.  For a contract of \(n\)
plays, the natural total entrance fee is about \(n\log_2 n\), or about
\(\log_2 n\) per play.  The longer the contract, the more tail levels the
buyer is likely to see.
\qedmark
\end{example}

Strong laws sharpen the same story.  For iid variables, the classical result is
as clean as one could hope.

\begin{theorem}[Kolmogorov strong law for iid averages; \citealp{kolmogorov1933grundbegriffe}]
Let $X_1,X_2,\ldots$ be iid.  Then
\[
  \frac1n\sum_{i=1}^n X_i \to \mu\quad\text{almost surely}
\]
for a finite constant $\mu$ if and only if $\Expect |X_1|<\infty$, in which case
$\mu=\Expect X_1$.
\end{theorem}

\noindent\textit{Proof.}
We prove the standard iid version because it also explains why the integrability
condition is sharp.

First suppose $\Expect|X_1|<\infty$.  It is enough to consider
$\mu=0$.  Define
\[
  Y_i=X_i\ind{|X_i|\le i}.
\]
Since
\[
  \sum_{i=1}^\infty \Prob(X_i\ne Y_i)
  =
  \sum_{i=1}^\infty \Prob(|X_1|>i)
  \le
  \Expect |X_1|
  <\infty,
\]
Borel--Cantelli implies $X_i=Y_i$ for all sufficiently large $i$, almost
surely.  Thus it is enough to prove
\[
  \frac1n\sum_{i=1}^n\{Y_i-\Expect Y_i\}\to0
  \quad\text{a.s.}
\]
For the centered independent variables $Z_i=Y_i-\Expect Y_i$,
\[
  \sum_{i=1}^\infty \frac{\Var(Z_i)}{i^2}
  \le
  \sum_{i=1}^\infty
  \frac{\Expect\{X_1^2\ind{|X_1|\le i}\}}{i^2}
  =
  \Expect\left[
    X_1^2\sum_{i\ge |X_1|}\frac1{i^2}
  \right]
  \le
  2\Expect |X_1|
  <\infty.
\]
Kolmogorov's convergence criterion therefore gives convergence of
$\sum_i Z_i/i$ almost surely; Kronecker's lemma then gives
$n^{-1}\sum_{i=1}^nZ_i\to0$ almost surely.  For completeness, here is the
short argument behind this step.  If $W_i$ are independent, mean zero, and
$\sum_i\Var(W_i)<\infty$, Kolmogorov's maximal inequality applied to the tails
gives
\[
  \Prob\left\{
    \max_{m<k\le N}\left|\sum_{i=m+1}^k W_i\right|>\varepsilon
  \right\}
  \le
  \varepsilon^{-2}\sum_{i=m+1}^N\Var(W_i).
\]
The maximal inequality follows by splitting according to the first time the
partial sum crosses $\varepsilon$; on each first-crossing event the future
increments have conditional mean zero, so the cross term with the remaining
tail has expectation zero, and the total second moment
$\sum_{i=m+1}^N\Var(W_i)$ dominates $\varepsilon^2$ times the crossing
probability.
Letting $N\to\infty$ and then $m\to\infty$ shows that the tail sums are
Cauchy in probability uniformly over the end point.  To get almost sure
convergence, choose $m_j$ so large that
$\sum_{i>m_j}\Var(W_i)\le \varepsilon^2 2^{-j}$.  The displayed bound and
Borel--Cantelli imply that, almost surely, the suprema of the tails after
$m_j$ are eventually at most $\varepsilon$.  Taking
$\varepsilon=1,1/2,1/3,\ldots$ shows that the partial sums are Cauchy almost
surely, so $\sum_iW_i$ converges almost surely.  Apply this with $W_i=Z_i/i$.
If $T_n=\sum_{i=1}^nZ_i/i\to T$, summation by parts gives
\[
  \frac1n\sum_{i=1}^n Z_i
  =
  T_n-\frac1n\sum_{i=1}^{n-1}T_i
  \to
  T-T=0,
\]
which is Kronecker's lemma in this case.  Finally,
$\Expect Y_i\to\Expect X_1=0$ by dominated convergence, so Cesaro's theorem
gives $n^{-1}\sum_{i=1}^n\Expect Y_i\to0$.

Conversely, suppose $n^{-1}\sum_{i=1}^nX_i\to\mu$ almost surely for a finite
$\mu$.  Then
\[
  \frac{X_n}{n}
  =
  \frac{S_n}{n}
  -
  \frac{n-1}{n}\frac{S_{n-1}}{n-1}
  \to
  \mu-\mu=0
  \quad\text{a.s.}
\]
Thus $\Prob(|X_n|>n\ \text{i.o.})=0$.  The events $\{|X_n|>n\}$ are
independent, so the converse Borel--Cantelli lemma yields
\[
  \sum_{n=1}^\infty \Prob(|X_1|>n)<\infty.
\]
Since
\[
  \Expect|X_1|
  =
  \int_0^\infty \Prob(|X_1|>t)\,dt
  \le
  1+\sum_{n=1}^\infty\Prob(|X_1|>n),
\]
one has $\Expect|X_1|<\infty$.  The finite limit must then equal
$\Expect X_1$ by the already proved sufficiency applied to the integrable
variable.
\qedmark

The ``only if'' direction matters statistically.  A sample average that seems
reasonable in a finite display may be trying to estimate an object that does
not exist.  Before asking for a better estimator, one should ask whether the
target has been named on a scale where stability is possible.

\section{Martingale Differences and Conditional Stability}
\label{sec:ch07-martingale-lln}
\conceptindexes{martingale differences, conditional stability, martingale weak law}

Independence is the easiest route to cancellation, but it is not the only one.
Sequential data often arrive with a past: adaptive trials, autoregressive
models, reinforcement-learning logs, clinical monitoring, and event histories.
In such problems the right object is not unconditional independence but a
filtration
\[
  \mathcal F_0\subseteq \mathcal F_1\subseteq\cdots,
\]
where $\mathcal F_i$ is the information available after the $i$th step.

A martingale difference sequence satisfies
\[
  \Expect(X_i\mid \mathcal F_{i-1})=0 .
\]
The increment may depend on the past, but after the past is known it has no
predictable drift.  This is the probabilistic grammar behind residuals in
sequential models and behind counting-process martingales later in the book.

\begin{theorem}[A martingale weak law]
Let $X_{n,1},\ldots,X_{n,r_n}$ be a triangular array of square-integrable
martingale differences with respect to filtrations
$\mathcal F_{n,j}$.  Put $S_n=\sum_{j=1}^{r_n}X_{n,j}$ and let $v_n\to\infty$.
If
\[
  \frac1{v_n^2}\sum_{j=1}^{r_n}\Expect X_{n,j}^2\to0,
\]
then $S_n/v_n\toP0$.
\end{theorem}

\noindent\textit{Proof.}
Conditionally on the past, the centered increments are orthogonal.  The
variance of the sum is therefore governed by the predictable variation
\[
  \sum_{j=1}^{r_n}\Expect(X_{n,j}^2\mid\mathcal F_{n,j-1}).
\]
Taking expectations gives
\[
  \Expect S_n^2=\sum_{j=1}^{r_n}\Expect X_{n,j}^2.
\]
Chebyshev's inequality then yields
\[
  \Prob(|S_n|>\varepsilon v_n)
  \le
  \frac{1}{\varepsilon^2v_n^2}
  \sum_{j=1}^{r_n}\Expect X_{n,j}^2\to0.
  \]
\qedmark

\begin{example}[Adaptive residual averages]
Suppose a binary outcome is observed sequentially.  Before patient $i$ enters,
the design may use the past to choose a treatment rule, a dose, or an
enrollment stratum.  Let
\[
  p_i=\Prob(Y_i=1\mid\mathcal F_{i-1})
\]
be the conditional success probability under the actual adaptive design, and
put $X_i=Y_i-p_i$.  Then
\[
  \Expect(X_i\mid\mathcal F_{i-1})=0,
  \qquad |X_i|\le1.
\]
The martingale weak law gives
\[
  \frac1n\sum_{i=1}^n\{Y_i-p_i\}\toP0.
\]
Thus the observed number of successes is close to the predictable number of
successes, even though the conditional probabilities may change with the
history.  Independence is not the right promise in such a design; conditional
centering is.
\end{example}

This result is a useful bridge.  It shows that large-sample stability can be
proved by controlling predictable variation, not only by proving independence.
The same logic will reappear when continuous-time counting processes are
written as
\[
  N(t)-\Lambda(t),
\]
an observed count minus its compensator.

\begin{tcolorbox}[
  enhanced,
  breakable,
  colback=noteback,
  colframe=bookgold!75!black,
  boxrule=0.55pt,
  arc=4pt,
  boxsep=1pt,
  left=0.95em,
  right=0.9em,
  top=0.7em,
  bottom=0.7em,
  before skip=0.9\baselineskip,
  after skip=1.0\baselineskip
]
\noindent\textbf{Optional convergence principle.}
The martingale analogue of Kolmogorov's three-series theorem gives a precise
answer to when an adapted series $\sum_nX_n$ converges almost surely.  For a
truncation level $c>0$, write
$X_n^{(c)}=X_n\ind{|X_n|\le c}$.  One standard form asks for three predictable
objects: the large-jump series
\[
  \sum_n\Prob(|X_n|>c\mid\mathcal F_{n-1}),
\]
the truncated drift series
\[
  \sum_n\Expect(X_n^{(c)}\mid\mathcal F_{n-1}),
\]
and the truncated conditional variance series
\[
  \sum_n\Var(X_n^{(c)}\mid\mathcal F_{n-1}).
\]
Their convergence is the conditional version of controlling rare jumps,
compensated drift, and residual variation.  In a martingale difference
sequence the untruncated drift is zero, but the truncated drift need not be;
large jumps leave a compensation term behind.
\end{tcolorbox}

\section{From Convergence to Concentration}
\label{sec:ch07-concentration}
\conceptindexes{concentration inequalities, Hoeffding inequality, Azuma--Hoeffding inequality, Bernstein inequality, random projections}

A law of large numbers is asymptotic.  It says that the empirical object will
eventually settle.  A scientist with $n=80$ patients or $n=5000$ cells needs a
more local question: how surprising is this observed deviation now?

Concentration inequalities answer that question.  The engine was already
visible in Section~\ref{sec:ch07-tail-grammar}: exponential Markov converts a
cumulant-generating-function bound into a tail bound.  The proof is usually a
small piece of calculus, but the interpretation is large: concentration tells
us how much random texture a stable mechanism can still generate.

\begin{theorem}[Hoeffding's inequality; \citealp{hoeffding1963probability}]
Let $X_1,\ldots,X_n$ be independent random variables with
$a_i\le X_i\le b_i$ and $\Expect X_i=0$.  Then for all $t>0$,
\[
  \Prob\left\{\sum_{i=1}^n X_i>t\right\}
  \le
  \exp\left\{-\frac{2t^2}{\sum_{i=1}^n(b_i-a_i)^2}\right\}.
\]
Consequently,
\[
  \Prob\left\{\left|\sum_{i=1}^n X_i\right|>t\right\}
  \le
  2\exp\left\{-\frac{2t^2}{\sum_{i=1}^n(b_i-a_i)^2}\right\}.
\]
\end{theorem}

\noindent\textit{Proof.}
We first prove Hoeffding's lemma.  If $a\le X\le b$ and $\Expect X=0$, then
for every $\lambda\in\R$,
\[
  \Expect e^{\lambda X}
  \le
  \exp\left\{\frac{\lambda^2(b-a)^2}{8}\right\}.
\]
To see this, use convexity of $x\mapsto e^{\lambda x}$ on $[a,b]$:
\[
  e^{\lambda x}
  \le
  \frac{b-x}{b-a}e^{\lambda a}
  +
  \frac{x-a}{b-a}e^{\lambda b}.
\]
Taking expectations and using $\Expect X=0$ gives
\[
  \Expect e^{\lambda X}
  \le
  \frac{b}{b-a}e^{\lambda a}
  -
  \frac{a}{b-a}e^{\lambda b}.
\]
Writing $p=-a/(b-a)$ and $u=\lambda(b-a)$, the right side becomes
(after shifting the interval)
\[
  (1-p)e^{-pu}+pe^{(1-p)u}.
\]
The logarithm of this expression has second derivative bounded by $1/4$ and
has value and first derivative equal to zero at $u=0$, hence it is at most
$u^2/8$.  This proves the lemma.

For $S_n=\sum_iX_i$, independence gives
\[
  \Expect e^{\lambda S_n}
  =
  \prod_{i=1}^n\Expect e^{\lambda X_i}
  \le
  \exp\left\{
    \frac{\lambda^2}{8}\sum_{i=1}^n(b_i-a_i)^2
  \right\}.
\]
Markov's inequality yields, for $\lambda>0$,
\[
  \Prob(S_n>t)
  \le
  \exp\left\{
    -\lambda t+
    \frac{\lambda^2}{8}\sum_{i=1}^n(b_i-a_i)^2
  \right\}.
\]
The minimizing value is
\[
  \lambda=\frac{4t}{\sum_i(b_i-a_i)^2},
\]
which gives the one-sided bound.  Applying the same argument to $-S_n$ and
using the union bound gives the two-sided inequality.
\qedmark

For Bernoulli averages, this gives
\[
  \Prob(|P_n\ind{A}-P\ind{A}|>\varepsilon)
  \le 2e^{-2n\varepsilon^2}.
\]
Compare this with Chebyshev's inequality.  Chebyshev gives a polynomial
$1/(n\varepsilon^2)$ bound; Hoeffding gives an exponential bound.  The same
stability is now measured with a sharper ruler.

\begin{example}[A finite list of prediction rules]
Let $\ell_\theta(X)$ be a loss bounded between $0$ and $1$, and suppose a
model builder wants to compare a finite list
$\theta_1,\ldots,\theta_M$.  For each fixed rule, Hoeffding gives
\[
  \Prob\{|P_n\ell_{\theta_m}-P\ell_{\theta_m}|>\varepsilon\}
  \le
  2e^{-2n\varepsilon^2}.
\]
A union bound then yields
\[
  \Prob\left\{
    \max_{1\le m\le M}
    |P_n\ell_{\theta_m}-P\ell_{\theta_m}|>\varepsilon
  \right\}
  \le
  2M e^{-2n\varepsilon^2}.
\]
Finite search costs only a logarithm of the number of rules.  This is the
simplest bridge from one empirical average to the uniform questions of the
next chapter.
\end{example}

\begin{theorem}[Azuma--Hoeffding inequality; \citealp{azuma1967weighted,hoeffding1963probability}]
Let $M_n=\sum_{i=1}^n X_i$ be a martingale with differences satisfying
$a_i\le X_i\le b_i$ almost surely.  Then
\[
  \Prob\{M_n-M_0>t\}
  \le
  \exp\left\{-\frac{2t^2}{\sum_{i=1}^n(b_i-a_i)^2}\right\}.
\]
\end{theorem}

\noindent\textit{Proof.}
Let $\mathcal F_i$ be the filtration and write $X_i=M_i-M_{i-1}$.  Conditional
on $\mathcal F_{i-1}$, the variable $X_i$ has mean zero and lies in
$[a_i,b_i]$.  Hoeffding's lemma applied conditionally gives
\[
  \Expect\{e^{\lambda X_i}\mid\mathcal F_{i-1}\}
  \le
  \exp\left\{\frac{\lambda^2(b_i-a_i)^2}{8}\right\}.
\]
Iterating conditional expectations,
\[
  \Expect e^{\lambda(M_n-M_0)}
  \le
  \exp\left\{
    \frac{\lambda^2}{8}\sum_{i=1}^n(b_i-a_i)^2
  \right\}.
\]
The same Markov optimization used in Hoeffding's inequality gives the stated
bound.
\qedmark

The martingale version is often the one that statistics needs.  It allows the
conditional success probability of a binary outcome to depend on the past, as
long as the centered increment remains bounded.  In an adaptive experiment,
the data-generating mechanism may learn; the deviation inequality still sees a
controlled martingale.

\begin{example}[Sequential binary monitoring]
Return to the adaptive binary sequence in the previous section, and let
$p_i=\Prob(Y_i=1\mid\mathcal F_{i-1})$.  The centered process
\[
  M_n=\sum_{i=1}^n(Y_i-p_i)
\]
is a martingale with increments in $[-1,1]$.  Azuma--Hoeffding gives
\[
  \Prob\left\{
    \left|\sum_{i=1}^nY_i-\sum_{i=1}^np_i\right|>t
  \right\}
  \le
  2\exp\left\{-\frac{t^2}{2n}\right\}.
\]
The bound compares observed successes with their conditional expectation
under the actual adaptive rule.  This is the finite-sample form of a residual
diagnostic: if the observed total is far from the compensator, the design's
conditional model is being challenged.
\end{example}

\subsection{Bernstein Bounds and Regression Examples}
\label{sec:ch07-bernstein}
\conceptindexes{Bernstein inequality, regression examples, sub-exponential tails}

Hoeffding uses boundedness but ignores variance.  Bernstein inequalities add
variance back into the story.  They say, roughly, that moderate deviations are
governed by variance and very large deviations are governed by the largest
increment.

\begin{theorem}[A Bernstein inequality; \citealp{bernstein1924theory}]
Let $X_1,\ldots,X_n$ be independent mean-zero variables with
$|X_i|\le C$ almost surely and let
\[
  v=\sum_{i=1}^n \Var(X_i).
\]
Then for all $t>0$,
\[
  \Prob\left\{\sum_{i=1}^n X_i>t\right\}
  \le
  \exp\left\{
    -\frac{t^2}{2(v+Ct/3)}
  \right\}.
\]
\end{theorem}

\noindent\textit{Proof.}
If $v=0$, then $\Var(X_i)=0$ for every $i$.  Since also $\Expect X_i=0$, each
$X_i=0$ almost surely, and therefore
\[
  \Prob\left\{\sum_iX_i>t\right\}=0
\]
for every $t>0$.  The bound is then trivial.  Hence assume $v>0$.  In this
nontrivial case $C>0$ as well.

For $|u|<3$,
\[
  e^u
  \le
  1+u+\frac{u^2}{2(1-|u|/3)}.
\]
This follows by comparing the power series of $e^u$ with the geometric series
that bounds $\sum_{k\ge2}|u|^k/k!$.  If $|X_i|\le C$ and $\Expect X_i=0$, then
for $0<\lambda<3/C$,
\[
  \Expect e^{\lambda X_i}
  \le
  1+
  \frac{\lambda^2\Var(X_i)}{2(1-\lambda C/3)}
  \le
  \exp\left\{
    \frac{\lambda^2\Var(X_i)}{2(1-\lambda C/3)}
  \right\}.
\]
By independence,
\[
  \Expect e^{\lambda\sum_iX_i}
  \le
  \exp\left\{
    \frac{\lambda^2 v}{2(1-\lambda C/3)}
  \right\}.
\]
Markov's inequality gives
\[
  \Prob\left\{\sum_iX_i>t\right\}
  \le
  \exp\left\{
    -\lambda t+
    \frac{\lambda^2v}{2(1-\lambda C/3)}
  \right\}.
\]
Choose
\[
  \lambda=\frac{t}{v+Ct/3}.
\]
Then $0<\lambda<3/C$, and substitution gives
\[
  -\lambda t+
  \frac{\lambda^2v}{2(1-\lambda C/3)}
  =
  -\frac{t^2}{2(v+Ct/3)}.
\]
This proves the inequality.
\qedmark

The denominator has two regimes.  If $t$ is small relative to $v/C$, the bound
looks Gaussian: $\exp\{-t^2/(2v)\}$.  If $t$ is large, the linear $Ct$ term
prevents the inequality from pretending that bounded variables have perfectly
Gaussian tails.  This is why Bernstein-type bounds are so useful in modern
statistics: they remember both local variance and worst-case size.

\begin{example}[Fixed-design regression]
In the fixed-design linear model
\[
  Y_i=\mathbf z_i^T\beta_0+e_i,
\]
least squares solves
\[
  \sum_{i=1}^n \mathbf z_i(Y_i-\mathbf z_i^T\beta)=0.
\]
For a fixed contrast $\mathbf a^T(\hat\beta-\beta_0)$, the stochastic part is a
weighted sum of errors.  If the errors satisfy a Bernstein moment condition,
then Bernstein's inequality controls the finite-sample probability that the
estimated contrast deviates from its target.  The normal approximation used in
standard regression is therefore not the only route to uncertainty; a
nonasymptotic tail bound can be read directly from the score sum.

\noindent\textit{Proof.}
Let $\mathbf Z$ be the design matrix with rows $\mathbf z_i^T$ and assume
$\mathbf Z^T\mathbf Z$ is nonsingular.
The least-squares estimator satisfies
\[
  \hat\beta-\beta_0
  =
  (\mathbf Z^T\mathbf Z)^{-1}\mathbf Z^T\mathbf e.
\]
For a fixed vector $\mathbf a$, define
\[
  w_i=\mathbf a^T(\mathbf Z^T\mathbf Z)^{-1}\mathbf z_i.
\]
Then
\[
  \mathbf a^T(\hat\beta-\beta_0)
  =
  \sum_{i=1}^nw_ie_i.
\]
If the errors are independent, mean zero, and satisfy $|e_i|\le C_e$ almost
surely, then $X_i=w_ie_i$ are independent mean-zero variables with
\[
  |X_i|\le C_e|w_i|,
  \qquad
  \sum_{i=1}^n\Var(X_i)=\sum_{i=1}^nw_i^2\Var(e_i).
\]
The bounded-variable Bernstein inequality, with
\[
  C=C_e\max_i|w_i|,
  \qquad
  v=\sum_{i=1}^nw_i^2\Var(e_i),
\]
gives
\[
  \Prob\{\mathbf a^T(\hat\beta-\beta_0)>t\}
  \le
  \exp\left\{
    -\frac{t^2}{2(v+Ct/3)}
  \right\}.
\]
Applying the same bound to $-\mathbf a^T(\hat\beta-\beta_0)$ gives the two-sided
version.  Moment-condition Bernstein inequalities replace the bounded
$C_e$ assumption by factorial moment bounds and lead to the same variance plus
size structure.
\qedmark
\end{example}

\begin{example}[Autoregression]
For the autoregressive model
\[
  Y_i=\beta_0Y_{i-1}+e_i,
\]
the least-squares estimating equation is
\[
  \sum_{i=1}^n Y_{i-1}(Y_i-\beta Y_{i-1})=0.
\]
The terms $Y_{i-1}e_i$ are martingale differences with respect to the past.
Their conditional variance is proportional to $\sum_i Y_{i-1}^2$.  Bernstein
or Freedman-type martingale inequalities therefore condition the deviation of
$\hat\beta$ on the observed amount of information in the time series.

\noindent\textit{Proof.}
Solving the least-squares equation gives
\[
  \hat\beta-\beta_0
  =
  \frac{\sum_{i=1}^nY_{i-1}e_i}
       {\sum_{i=1}^nY_{i-1}^2},
\]
whenever $\sum_iY_{i-1}^2>0$.  Let
\[
  \mathcal F_i=\sigma(Y_0,e_1,\ldots,e_i).
\]
If $\Expect(e_i\mid\mathcal F_{i-1})=0$, then
$Y_{i-1}e_i$ is a martingale difference.  Suppose additionally that
$|e_i|\le C_e$ and
$\Expect(e_i^2\mid\mathcal F_{i-1})=\sigma^2$.  Put
\[
  T_n=\sum_{i=1}^nY_{i-1}^2,
  \qquad
  M_n=\max_{1\le i\le n}|Y_{i-1}|.
\]
For deterministic numbers $0<s\le\tau$ and $m>0$, consider the event
\[
  A_{s,\tau,m}=\{s\le T_n\le\tau,\ M_n\le m\}.
\]
On this event the denominator is at least $s$, the predictable variance of the
numerator is at most $\sigma^2\tau$, and each martingale increment is bounded
by $C_em$.

Here is the martingale Bernstein calculation.  For
$0<\lambda<3/(C_em)$, the conditional version of the Bernstein mgf bound gives,
as long as $M_n\le m$ is enforced by stopping,
\[
  \Expect\left\{
    \exp\left[
      \lambda Y_{i-1}e_i
      -
      \frac{\lambda^2\sigma^2Y_{i-1}^2}
           {2(1-\lambda C_em/3)}
    \right]
    \middle|\mathcal F_{i-1}
  \right\}
  \le
  1.
\]
Iterating conditional expectations and applying exponential Markov therefore
yields
\[
  \Prob\left\{\sum_{i=1}^nY_{i-1}e_i>t,\ A_{s,\tau,m}\right\}
  \le
  \exp\left\{
    -\frac{t^2}{2(\sigma^2\tau+C_emt/3)}
  \right\}.
\]
Apply the same bound to the negative sum and take $t=us$.  Since
$|\hat\beta-\beta_0|>u$ and $T_n\ge s$ imply
$|\sum_iY_{i-1}e_i|>us$, one obtains
\[
  \Prob\{|\hat\beta-\beta_0|>u,\ A_{s,\tau,m}\}
  \le
  2\exp\left\{
    -\frac{u^2s^2}{2(\sigma^2\tau+C_emus/3)}
  \right\}.
\]
The denominator $T_n$ is the observed information in this simple time-series
problem.  More information makes the contrast more stable; a large maximum
lagged value $M_n$ records the price paid for a large single increment.
\qedmark
\end{example}

These examples show why concentration is not merely a probability topic.  It
is a way of attaching finite-sample control to the same empirical objects that
later become estimating equations.

\subsection{Random Projections and Johnson--Lindenstrauss Embeddings}
\label{sec:ch07-random-projections}
\conceptindexes{random projections, Johnson--Lindenstrauss lemma, embeddings, Gaussian projection}

Concentration also explains a phenomenon that at first sounds almost
unreasonable.  A high-dimensional data cloud can often be pushed into many
fewer dimensions without seriously changing the distances among its points.
This is the content of the Johnson--Lindenstrauss embedding lemma
\citep{johnson1984extensions}.  For statistics and machine learning, the
lesson is geometric: random compression can be disciplined when it preserves
the features on which a procedure depends.

Suppose that $\mathbf x_1,\ldots,\mathbf x_N\in\R^d$ are fixed points.  A
random projection chooses a matrix $\mathbf A\in\R^{k\times d}$ and replaces
each $\mathbf x_i$ by $\mathbf A\mathbf x_i$.  The ambient dimension $d$ may be
enormous.  The target dimension $k$ will depend only on the number of points
and the allowed distortion.  The goal is
\[
  (1-\varepsilon)\|\mathbf x_i-\mathbf x_j\|^2
  \le
  \|\mathbf A\mathbf x_i-\mathbf A\mathbf x_j\|^2
  \le
  (1+\varepsilon)\|\mathbf x_i-\mathbf x_j\|^2
\]
simultaneously for all pairs $i,j$.

\input{figures/ch10_johnson_lindenstrauss_projection}

The proof is a perfect example of the chapter's finite-sample logic.  First
prove concentration for one fixed direction.  Then use a union bound over the
finite list of pairwise differences.

\begin{lemma}[One-vector norm preservation]
Let $\mathbf A$ be a $k\times d$ random matrix with independent
$\Normal(0,1/k)$ entries.  For any fixed $\mathbf u\in\R^d$ and
$0<\varepsilon<1$,
\[
  \Prob\left\{
    \left|\|\mathbf A\mathbf u\|^2-\|\mathbf u\|^2\right|>
    \varepsilon\|\mathbf u\|^2
  \right\}
  \le
  2\exp\left\{-\frac{k\varepsilon^2}{8}\right\}.
\]
\end{lemma}

\noindent\textit{Proof.}
If $\mathbf u=0$, the claim is trivial.  Otherwise the coordinates of
$\mathbf A\mathbf u$ are independent $\Normal(0,\|\mathbf u\|^2/k)$ variables, so
\[
  \frac{k\|\mathbf A\mathbf u\|^2}{\|\mathbf u\|^2}\sim \chi_k^2.
\]
Let $Z\sim\chi_k^2$.  The Chernoff method applied to the moment generating
function of $Z$ gives
\[
  \Prob\{Z\ge k(1+\varepsilon)\}
  \le
  \exp\left\{
    -\frac{k}{2}\{\varepsilon-\log(1+\varepsilon)\}
  \right\}
  \le
  \exp\left\{-\frac{k\varepsilon^2}{8}\right\}
\]
for $0<\varepsilon<1$.  The lower tail is similar:
\[
  \Prob\{Z\le k(1-\varepsilon)\}
  \le
  \exp\left\{
    -\frac{k}{2}\{-\varepsilon-\log(1-\varepsilon)\}
  \right\}
  \le
  \exp\left\{-\frac{k\varepsilon^2}{4}\right\}.
\]
Adding the two tails proves the displayed bound.
\qedmark

\begin{theorem}[Johnson--Lindenstrauss lemma, Gaussian form; \citealp{johnson1984extensions}]
Let $\mathbf x_1,\ldots,\mathbf x_N\in\R^d$, let $0<\varepsilon<1$, and let
$0<\delta<1$.  Let $\mathbf A$ be a $k\times d$ random matrix with independent
$\Normal(0,1/k)$ entries.  If
\[
  k\ge
  \frac{8}{\varepsilon^2}
  \log\frac{N(N-1)}{\delta},
\]
then with probability at least $1-\delta$,
\[
  (1-\varepsilon)\|\mathbf x_i-\mathbf x_j\|^2
  \le
  \|\mathbf A\mathbf x_i-\mathbf A\mathbf x_j\|^2
  \le
  (1+\varepsilon)\|\mathbf x_i-\mathbf x_j\|^2
\]
for every pair $1\le i<j\le N$.
\end{theorem}

\noindent\textit{Proof.}
For each pair, apply the preceding lemma to
$\mathbf u_{ij}=\mathbf x_i-\mathbf x_j$.  If $\mathbf u_{ij}=0$, the pair is
automatically preserved.  If $\mathbf u_{ij}\ne0$, the probability that its
squared length is distorted by more than $\varepsilon$ is at most
$2\exp\{-k\varepsilon^2/8\}$.  There are $N(N-1)/2$ pairs.  By the union bound,
the probability that at least one pair fails is no more than
\[
  \frac{N(N-1)}{2}\,
  2\exp\left\{-\frac{k\varepsilon^2}{8}\right\}
  \le
  \delta .
\]
On the complementary event, all pairwise squared distances are preserved.
\qedmark

The remarkable feature is the absence of $d$ from the target dimension.  The
ambient space may have thousands or millions of coordinates, but a finite
cloud of $N$ points can be embedded into $k$ dimensions with
$k$ proportional to $\varepsilon^{-2}\log N$.  Randomness is not merely noise
here; it is a construction device whose failure probability can be budgeted.

The Gaussian assumption is not the heart of the result.  It is used above
because the length of one projected vector has an exact chi-square
distribution.  Rademacher matrices, with independent entries
$\pm1/\sqrt k$, and sparse Achlioptas-type matrices also give
Johnson--Lindenstrauss embeddings, with different constants.  What matters is
that the projection is isotropic and that each fixed direction satisfies a
one-vector concentration bound.  Once that bound is available, the union-bound
argument over the $N(N-1)/2$ pairwise differences is unchanged.

\begin{tcolorbox}[
  enhanced,
  breakable,
  colback=noteback,
  colframe=bookgold!75!black,
  boxrule=0.55pt,
  arc=4pt,
  boxsep=1pt,
  left=0.95em,
  right=0.9em,
  top=0.7em,
  bottom=0.7em,
  before skip=0.9\baselineskip,
  after skip=1.0\baselineskip
]
\noindent\textbf{Finite complexity before empirical processes.}
The Johnson--Lindenstrauss proof controls a finite family of directions, so
the price of simultaneous control is the logarithm of the number of pairs.
The next chapter asks what replaces this finite union bound when the
statistician scans an infinite function class.  Entropy, brackets, and
stochastic equicontinuity are the empirical-process versions of the same
stabilizing idea.
\end{tcolorbox}

\subsection{Quantiles and Kernel Smoothing}
\label{sec:ch07-quantiles-kernels}
\conceptindexes{quantiles, kernels, empirical distribution function, sample quantiles}

Means are not the only translations of data into population information.
Quantiles and kernel estimates make the same stability question more geometric.

\begin{example}[Quantile concentration]
Let $F$ be a distribution function and let
\[
  Q(p)=\inf\{x:F(x)\ge p\},\qquad 0<p<1.
\]
The empirical quantile is
\[
  Q_n(p)=\inf\{x:F_n(x)\ge p\}.
\]
If $Q(p)$ is unique, then the event $\{Q_n(p)>Q(p)+\varepsilon\}$ is the same
kind of event as
\[
  F_n(Q(p)+\varepsilon)<p.
\]
Thus a quantile deviation is converted into a Bernoulli-average deviation.

\noindent\textit{Proof.}
Let $q=Q(p)$ and fix $\varepsilon>0$.  If $Q_n(p)>q+\varepsilon$, then the
empirical distribution has not yet reached level $p$ at $q+\varepsilon$:
\[
  F_n(q+\varepsilon)<p.
\]
Hence
\[
  F(q+\varepsilon)-F_n(q+\varepsilon)
  >
  F(q+\varepsilon)-p.
\]
Similarly, if $Q_n(p)<q-\varepsilon$, then
\[
  F_n(q-\varepsilon)\ge p,
\]
so
\[
  F_n(q-\varepsilon)-F(q-\varepsilon)
  \ge
  p-F(q-\varepsilon).
\]
Assume the two separation quantities are positive and put
\[
  \eta(\varepsilon)
  =
  \min\{F(q+\varepsilon)-p,\,
        p-F(q-\varepsilon)\}.
\]
Both $F_n(q+\varepsilon)$ and $F_n(q-\varepsilon)$ are Bernoulli averages.
Hoeffding's inequality and the union bound give
\[
  \Prob\{|Q_n(p)-Q(p)|>\varepsilon\}
  \le
  2\exp\{-2n\eta(\varepsilon)^2\},
\]
with the displayed $\eta(\varepsilon)$.  If $F$ has a positive density $f$ at
$q$, then
\[
  F(q+\varepsilon)-p=f(q)\varepsilon+o(\varepsilon),
  \qquad
  p-F(q-\varepsilon)=f(q)\varepsilon+o(\varepsilon).
\]
Thus the concentration of the empirical CDF becomes concentration of the
empirical quantile.
\qedmark
\end{example}

\begin{example}[Kernel density estimate]
Kernel density estimation tells a parallel story.  Let $X_1,\ldots,X_n$ be iid
with density $f$, let $h_n\downarrow0$, and define
\[
  \hat f_n(x)=\frac1{nh_n}\sum_{i=1}^n
  K\left(\frac{x-X_i}{h_n}\right),
\]
where, for a clean proof, $K$ is a bounded compactly supported density satisfying
\[
  \int rK(r)\,dr=0,\qquad
  \int r^2K(r)\,dr<\infty,\qquad
  \int K^2(r)\,dr<\infty .
\]
Assume $f(x)>0$ and that $f''$ is continuous at $x$ and bounded in a neighborhood
of $x$.  The same calculation extends to non-compact kernels by truncating the
kernel tails under the displayed moment conditions.

\noindent\textit{Bias calculation.}
Set
\[
  Y_{n,i}=\frac1{h_n}K\left(\frac{x-X_i}{h_n}\right).
\]
Then $\hat f_n(x)=n^{-1}\sum_iY_{n,i}$ and, changing variables
$r=(x-u)/h_n$,
\[
  \Expect Y_{n,i}
  =
  \int \frac1{h_n}K\left(\frac{x-u}{h_n}\right)f(u)\,du
  =
  \int K(r)f(x-rh_n)\,dr.
\]
Taylor expansion gives
\[
  f(x-rh_n)
  =
  f(x)-rh_nf'(x)+\frac12r^2h_n^2f''(x)+o(h_n^2r^2)
\]
uniformly over the compact support of $K$.  Since $\int K=1$ and
$\int rK(r)\,dr=0$,
\[
  \Expect\hat f_n(x)-f(x)
  =
  \frac12h_n^2f''(x)\int r^2K(r)\,dr+o(h_n^2).
\]

\noindent\textit{Variance calculation.}
Again by the same change of variables,
\[
  \Expect Y_{n,i}^2
  =
  \frac1{h_n}
  \int K^2(r)f(x-rh_n)\,dr
  =
  \frac{f(x)}{h_n}\int K^2(r)\,dr+o(h_n^{-1}).
\]
Therefore
\[
  \Var\{\hat f_n(x)\}
  =
  \frac1n\Var(Y_{n,1})
  =
  \frac{f(x)}{nh_n}\int K^2(r)\,dr+o\{(nh_n)^{-1}\}.
\]
If $nh_n\to\infty$, the variance tends to zero, while the bias tends to zero
when $h_n\to0$.

\noindent\textit{Bernstein concentration.}
Suppose, for simplicity, that $|K|\le K_\infty$.  Apply Bernstein's inequality
to the centered variables $Y_{n,i}-\Expect Y_{n,i}$.  They are bounded in
absolute value by $2K_\infty/h_n$ and have total variance
\[
  \sum_{i=1}^n\Var(Y_{n,i})
  =
  \frac{n f(x)}{h_n}\int K^2(r)\,dr+o(nh_n^{-1}).
\]
For $\varepsilon>0$,
\[
  \Prob\{|\hat f_n(x)-\Expect\hat f_n(x)|>\varepsilon\}
  \le
  2\exp\left[
    -\frac{nh_n\varepsilon^2}
    {2\{f(x)v_K+o(1)+2K_\infty\varepsilon/3\}}
  \right],
\]
where $v_K=\int K^2(r)\,dr$.  In particular, for sufficiently large constants
$a$, if $nh_n/\log n\to\infty$,
\[
  |\hat f_n(x)-\Expect\hat f_n(x)|
  =
  O_{\text{a.s.}}\left(\sqrt{\frac{\log n}{nh_n}}\right)
\]
because the probability of the event
\[
  |\hat f_n(x)-\Expect\hat f_n(x)|
  >
  a\sqrt{\frac{\log n}{nh_n}}
\]
is bounded by $2n^{-c a^2}$ for some $c>0$ and all large $n$, and
Borel--Cantelli applies once $a$ is chosen large enough.  Combining this
stochastic term with the bias gives
\[
  |\hat f_n(x)-f(x)|
  =
  O_{\text{a.s.}}\left(\sqrt{\frac{\log n}{nh_n}}+h_n^2\right).
\]

\noindent\textit{Normal approximation preview.}
The same triangular-array viewpoint gives the pointwise normal approximation
that later reappears in asymptotic inference.  In this paragraph only,
$Z_n\weakto Z$ means ordinary one-dimensional convergence in distribution.
Since
\[
  \Var(Y_{n,1})
  =
  \frac{f(x)}{h_n}v_K+o(h_n^{-1}),
  \qquad
  v_K=\int K^2(r)\,dr,
\]
and $|Y_{n,1}-\Expect Y_{n,1}|\le 2K_\infty/h_n$, the largest summand is
negligible after normalization whenever $nh_n\to\infty$:
\[
  \frac{h_n^{-1}}{\sqrt{n\Var(Y_{n,1})}}
  =
  O\!\left((nh_n)^{-1/2}\right)
  \to0.
\]
Thus Lindeberg's condition holds and
\[
  \sqrt{nh_n}\{\hat f_n(x)-\Expect\hat f_n(x)\}
  \weakto
  \Normal\{0,f(x)v_K\}.
\]
The bias calculation above gives, with
$\mu_{2,K}=\int r^2K(r)\,dr$,
\[
  \sqrt{nh_n}\{\hat f_n(x)-f(x)\}
  =
  \sqrt{nh_n}\{\hat f_n(x)-\Expect\hat f_n(x)\}
  +
  \frac12\sqrt{nh_n^5}\,f''(x)\mu_{2,K}
  +
  o(\sqrt{nh_n^5}).
\]
If $nh_n^5\to0$, the bias is invisible at the stochastic scale and
\[
  \sqrt{nh_n}\{\hat f_n(x)-f(x)\}
  \weakto
  \Normal\{0,f(x)v_K\}.
\]
If instead $nh_n^5\to\gamma\in(0,\infty)$, the same approximation has the
nonzero mean shift
$\frac12\sqrt\gamma\,f''(x)\mu_{2,K}$.

This kernel example illustrates the language of the present chapter: the
bandwidth is a contract between random fluctuation and smoothing bias.
\qedmark
\end{example}

\begin{example}[A proxy climate curve]
The Zhu/Chu Ko-chen climate example from Chapter~2 can be read through the
same smoothing lens.  Suppose proxy observations are recorded at historical
times \(t_i\):
\[
  Y_i=m(t_i)+b(t_i)+\varepsilon_i,
\]
where \(m\) is a latent climate curve, \(b\) is a systematic recording or proxy
bias, and \(\varepsilon_i\) is random error.  A kernel smoother
\[
  \hat m_h(t)
  =
  \frac{\sum_iK\{(t-t_i)/h\}Y_i}
       {\sum_iK\{(t-t_i)/h\}}
\]
has the same two terms as the density example: stochastic fluctuation decreases
when many nearby proxy records support the estimate, while smoothing bias
increases when the window is too wide for a changing climate signal.  The new
historical term is \(b(t)\).  More records reduce random error, but they do not
automatically remove a dynasty-specific reporting habit, a regional archival
imbalance, or a proxy calibration error.  This is why concentration and
smoothing are necessary but not sufficient: they control random texture after
the observation system has been named.
\qedmark
\end{example}

\subsubsection{The Empirical Distribution Function and Uniform Convergence}
\label{sec:ch07-edf-bridge}
\conceptindexes{empirical distribution function, Glivenko--Cantelli theorem, Dvoretzky--Kiefer--Wolfowitz inequality}

The empirical distribution function
\[
  F_n(t)=P_n\ind{X\le t}
\]
is the simplest object that refuses to remain one-dimensional.  For each fixed
$t$, the strong law gives $F_n(t)\to F(t)$ almost surely.  But statistical
procedures often scan over all $t$: quantiles, Kolmogorov--Smirnov tests,
confidence bands, and residual diagnostics all look at the whole curve.

\begin{theorem}[Glivenko--Cantelli theorem; \citealp{glivenko1933sulla,cantelli1933sulla}]
If $X_1,X_2,\ldots$ are iid with distribution function $F$, then
\[
  \sup_{t\in\R}|F_n(t)-F(t)|\to0
  \quad\text{almost surely}.
\]
\end{theorem}

\noindent\textit{Proof.}
Fix $\varepsilon>0$.  First suppose that $F$ is continuous.  Choose points
\[
  -\infty=t_0<t_1<\cdots<t_m=\infty
\]
so that
\[
  F(t_j)-F(t_{j-1})\le \varepsilon
  \qquad j=1,\ldots,m.
\]
For every finite endpoint $t_j$, the ordinary strong law gives
$F_n(t_j)\to F(t_j)$ almost surely.  On the probability-one event where this
holds for all finitely many endpoints, define
\[
  D_n=\max_{1\le j\le m-1}|F_n(t_j)-F(t_j)|.
\]
Then $D_n\to0$.  If $t_{j-1}\le t<t_j$, monotonicity gives
\[
  F_n(t)-F(t)
  \le
  F_n(t_j)-F(t_{j-1})
  =
  \{F_n(t_j)-F(t_j)\}
  +
  \{F(t_j)-F(t_{j-1})\}
  \le D_n+\varepsilon,
\]
and similarly
\[
  F(t)-F_n(t)
  \le
  F(t_j)-F_n(t_{j-1})
  \le
  \varepsilon+D_n.
\]
Thus $\sup_t|F_n(t)-F(t)|\le D_n+\varepsilon$ eventually, and then
$\limsup_n\sup_t|F_n(t)-F(t)|\le\varepsilon$.  Since $\varepsilon$ is
arbitrary, the supremum converges to zero almost surely.

For a general distribution, there are only finitely many atoms with mass larger
than $\varepsilon$.  Include those atoms as special cutpoints, and partition
the remaining intervals so that their $F$-mass is at most $\varepsilon$.  At a
large atom $a$, the jump size is estimated by the fixed empirical average
\[
  P_n\ind{X=a}\to P\{X=a\}
  \quad\text{a.s.}
\]
The same monotonicity argument applies on the intervals between the selected
atoms, while the selected jumps are controlled by these finitely many strong
laws.  Sending $\varepsilon\downarrow0$ proves the general case.
\qedmark

The finite-sample companion is the Dvoretzky--Kiefer--Wolfowitz inequality:
\[
  \Prob\left\{
  \sup_{t\in\R}|F_n(t)-F(t)|>\varepsilon
  \right\}
  \le
  2e^{-2n\varepsilon^2}.
\]
The remarkable feature is that the bound does not depend on $F$.  A single
inequality controls every distribution function on the line.

\begin{example}[A censored survival surface as a uniform-law stress test]
Let \(T=(T_1,T_2)\) be paired event times and let \(C=(C_1,C_2)\) be paired
censoring times.  The observed record is
\[
  O=(Y_1,Y_2,\Delta_1,\Delta_2),
  \qquad
  Y_j=T_j\wedge C_j,\quad
  \Delta_j=\ind{T_j\le C_j}.
\]
The target is the bivariate survival surface
\[
  S(s,t)=\Prob(T_1>s,T_2>t),
\]
but the data do not reveal the indicator \(\ind{T_1>s,T_2>t}\) directly.  They
reveal a marked two-dimensional risk-set process: for each \((s,t)\), the
observation says whether the pair is still observable beyond \((s,t)\), and
which coordinates have failed rather than been censored.  In the bivariate
Kaplan--Meier construction, these marked risk-set surfaces are assembled into
a product-limit estimate of \(S\)
\citep{dabrowska1988kaplan,dabrowskaAdvancedProbabilityCommunication,dabrowskaStochasticProcessesCommunication}.

For this chapter, the estimator's algebra is less important than the empirical
process lesson.  On a rectangle
\([0,\tau_1]\times[0,\tau_2]\) where the observable risk set remains bounded
away from zero, the empirical risk and failure surfaces satisfy a uniform law
of large numbers.  The bivariate Kaplan--Meier map is a continuous
transformation of those surfaces at such well-behaved laws.  Therefore the
estimated survival surface is uniformly consistent on the rectangle:
\[
  \sup_{0\le s\le\tau_1,\ 0\le t\le\tau_2}
  |\widehat S_n(s,t)-S(s,t)|
  \to0
  \qquad\text{a.s.}
\]
This is the part worth carrying forward.  The univariate Kaplan--Meier curve is
not merely copied twice; dependence between \(T_1\) and \(T_2\) lives in the
joint risk-set geometry.  The uniform law applies to observable marked
rectangles, and the survival estimator is a stable functional of those
observable summaries.
\end{example}

\begin{example}[Fluctuation scale for the same censored surface]
The same survival surface also illustrates why a law of large numbers is not
the end of the story.  Once the surface has settled toward its target, the next
question is the scale and shape of the remaining error.  The weak-convergence,
LIL, and bootstrap analysis in \citet{dabrowska1989kaplan} shows that, under
regularity conditions, the centered bivariate Kaplan--Meier surface has a
first-order empirical-process representation
\citep{dabrowskaAdvancedProbabilityCommunication}.  Schematically,
\[
  \widehat S_n-S
  =
  \frac1n\sum_{i=1}^n \phi_{P}(O_i)
  +\hbox{smaller remainder},
\]
where \(\phi_P(O_i)\) is a mean-zero random surface determined by the censored
observation law.  Thus the residual error is governed by the same grammar as
the rest of this chapter: a uniform law identifies the target, while a
functional central limit theorem and law-of-the-iterated-logarithm scale
describe the remaining surface noise.
\end{example}

This is the doorway to the next chapter.
The object
\[
  \sup_{t\in\R}|P_n\ind{X\le t}-P\ind{X\le t}|
\]
is no longer a fixed average.  It is the supremum of a whole indexed field of
empirical averages.  The statistician is no longer checking one fingerprint;
she is reading a whole surface of them at once.  Once we write the problem that
way, the next question becomes inevitable:
\[
  \sup_{f\in\mathcal F}|P_n f-P f|.
\]
The law of large numbers has become an empirical-process problem.

\section{Exercises}
\label{sec:ch07-exercises}
\conceptindexes{stability exercises, concentration exercises, law-of-large-numbers exercises}

\begin{exercise}[A weighted weak law]
Let $X_1,\ldots,X_n$ be independent with common mean $\mu$ and variances
$\sigma_i^2$.  Define
\[
  T_n=\sum_{i=1}^n w_iX_i,\qquad
  w_i=\frac{\sigma_i^{-2}}{\sum_{j=1}^n\sigma_j^{-2}}.
\]
Find a condition on $\sum_i\sigma_i^{-2}$ that guarantees $T_n\toP\mu$.
\end{exercise}

\begin{exercise}[Weak law with fading dependence]
Let $X_1,X_2,\ldots$ be mean-zero variables with
$\sup_i\Var(X_i)<\infty$.  Suppose there is a deterministic sequence
$r_h\to0$ such that
\[
  |\Cov(X_i,X_j)|\le r_{|i-j|}
  \qquad (i\ne j).
\]
Show that $n^{-1}\sum_{i=1}^nX_i\toP0$.  The point is to split the covariance
sum into nearby lags and far lags rather than to assume independence.
\end{exercise}

\begin{exercise}[Spreading fixed designs]
Consider the fixed-design regression model
\[
  Y_i=\alpha+\beta z_i+e_i,\qquad i=1,\ldots,n,
\]
where $z_i$ are nonrandom, $\Expect e_i=0$, and $\Var(e_i)=\sigma^2<\infty$.
Let $\hat\alpha_n,\hat\beta_n$ be the least-squares estimators.  Find a simple
condition on
\[
  S_{zz,n}=\sum_{i=1}^n(z_i-\bar z_n)^2
\]
that guarantees $\hat\beta_n\toP\beta$ and
$\hat\alpha_n-\alpha-(\beta-\hat\beta_n)\bar z_n\toP0$.  What additional
condition makes $\hat\alpha_n\toP\alpha$?
\end{exercise}

\begin{exercise}[Pareto averages on different scales]
Let $X_1,X_2,\ldots$ be iid nonnegative variables with
\[
  \Prob(X_1>x)=x^{-\alpha},\qquad x\ge1,
\]
where $\alpha>0$.

\begin{enumerate}
\item For $\alpha=1$, show that $S_n/(n\log n)\toP1$.
\item For $0<\alpha<1$, find a deterministic scale $a_n$ larger than
$n^{1/\alpha}$ by a slowly varying factor such that $S_n/a_n\toP0$.
\item For $\alpha>1$, explain why the ordinary average converges to the
finite mean, and identify where the proof of the first two parts breaks.
\end{enumerate}
\end{exercise}

\begin{exercise}[A martingale average]
Let $X_i$ be martingale differences with
$\Expect(X_i^2\mid\mathcal F_{i-1})\le C$ almost surely.  Show that
$n^{-1}\sum_{i=1}^nX_i\toP0$.  Under bounded increments, use
Azuma--Hoeffding to show almost sure convergence.
\end{exercise}

\begin{exercise}[Predictable weights]
Let $X_i$ be square-integrable martingale differences and let $V_i$ be
$\mathcal F_{i-1}$-measurable weights.  Define
\[
  Z_n=\sum_{i=1}^n V_iX_i.
\]
Show that $Z_n$ is a martingale when the summands are integrable.  If
\[
  \sum_{i=1}^\infty
  \Expect(V_i^2X_i^2\mid\mathcal F_{i-1})<\infty
  \quad\text{a.s.},
\]
explain why the martingale convergence theorem suggests almost sure
convergence of $Z_n$ after a suitable localization argument.
\end{exercise}

\begin{exercise}[Sampling without replacement]
A finite population contains numbers $a_1,\ldots,a_N$ with mean $\mu_N$.
Draw $X_1,\ldots,X_n$ successively without replacement and let
$S_m=\sum_{i=1}^mX_i$.  Show that
\[
  Z_m=\frac{S_m-m\mu_N}{N-m},\qquad 0\le m<N,
\]
is a martingale with respect to the sampling history.  Use this martingale to
explain why sampling without replacement has smaller variance than sampling
with replacement at the same sample size.
\end{exercise}

\begin{exercise}[Conditional Borel--Cantelli]
Let $A_i\in\mathcal F_i$ be events and put
\[
  p_i=\Prob(A_i\mid\mathcal F_{i-1}).
\]
Show that
\[
  M_n=\sum_{i=1}^n\{\ind{A_i}-p_i\}
\]
is a martingale.  If $\sum_i p_i<\infty$ almost surely, use martingale
convergence or a maximal inequality to show that only finitely many $A_i$
occur almost surely.
\end{exercise}

\begin{exercise}[Quantile concentration]
Assume that $F$ has density $f$ satisfying $f(x)\ge c>0$ on
$[Q(p)-\delta,Q(p)+\delta]$.  Use the quantile bound above to derive an
exponential inequality for $|Q_n(p)-Q(p)|$ when the error is less than
$\delta$.
\end{exercise}

\begin{exercise}[Kernel bias and variance]
For the kernel estimator in Subsection~\ref{sec:ch07-quantiles-kernels}, assume
that $K$ is bounded, integrates to one, has zero first moment, and has finite
second moment.  Derive the leading bias term when $f$ is twice continuously
differentiable near $x$, and compute the order of the variance.
\end{exercise}

\begin{exercise}[A random-projection dimension budget]
Let $\mathbf A$ be a Gaussian random projection as in
Subsection~\ref{sec:ch07-random-projections}.  For a fixed set of
$N$ points, find a value of $k$ that makes the probability of any pairwise
distance distortion larger than $\varepsilon$ at most $\delta$.  Compare the
answer with a naive coordinate projection that keeps $k$ coordinates.
\end{exercise}

\begin{exercise}[Rademacher random projections]
Let $\mathbf A$ have independent entries
$A_{rj}=\varepsilon_{rj}/\sqrt k$, where
$\Prob(\varepsilon_{rj}=1)=\Prob(\varepsilon_{rj}=-1)=1/2$.

\begin{enumerate}
\item Show that for every fixed $\mathbf u\in\R^d$,
\[
  \Expect\|\mathbf A\mathbf u\|^2=\|\mathbf u\|^2.
\]
\item Suppose that the following one-vector bound is available for some
universal constant $c>0$:
\[
  \Prob\left\{
    \left|\|\mathbf A\mathbf u\|^2-\|\mathbf u\|^2\right|>
    \varepsilon\|\mathbf u\|^2
  \right\}
  \le
  2\exp\{-ck\varepsilon^2\},
  \qquad 0<\varepsilon<1.
\]
Use the same union-bound argument as in the Gaussian proof to obtain a
Johnson--Lindenstrauss embedding dimension for $N$ fixed points.
\item Identify which step in the Gaussian proof used rotational invariance and
the chi-square distribution, and explain why the Rademacher proof needs a
different concentration input.
\end{enumerate}
\end{exercise}

\begin{exercise}[Sparse random projections]
Let $\eta$ take values $\sqrt3$, $0$, and $-\sqrt3$ with probabilities
$1/6$, $2/3$, and $1/6$, respectively, and let
$A_{rj}=\eta_{rj}/\sqrt k$ with independent copies $\eta_{rj}$.

\begin{enumerate}
\item Check that $\Expect\eta=0$ and $\Expect\eta^2=1$, and conclude that
$\Expect\|\mathbf A\mathbf u\|^2=\|\mathbf u\|^2$ for every fixed
$\mathbf u$.
\item Assume a one-vector concentration bound of the same form as in the
Rademacher exercise, possibly with a different constant.  Deduce the
Johnson--Lindenstrauss dimension $k\asymp\varepsilon^{-2}\log N$.
\item Explain why the zeros make this projection more attractive
computationally when the original dimension $d$ is large.
\end{enumerate}
\end{exercise}

\begin{exercise}[Records grow logarithmically]
Let $X_1,X_2,\ldots$ be iid with a continuous distribution.  Say that time
$j$ is a record time if $X_j>\max_{i<j}X_i$, with $j=1$ counted as a record.
Let $R_n$ be the number of records up to time $n$.

\begin{enumerate}
\item Show that $\Prob(j\text{ is a record})=1/j$.
\item Show that the record indicators are independent.
\item Prove that $R_n/\log n\to1$ almost surely.
\end{enumerate}
\end{exercise}

\begin{exercise}[From DKW to a confidence band]
Use the Dvoretzky--Kiefer--Wolfowitz inequality to construct a simultaneous
$1-\alpha$ confidence band for $F$ of the form
$F_n(t)\pm c_{n,\alpha}$.
\end{exercise}

\section*{Sources and Further Reading}
\addcontentsline{toc}{section}{Sources and Further Reading}

The chapter follows a standard large-sample route: weak laws through
truncation and triangular arrays, strong laws through Borel--Cantelli and
martingale inequalities, and concentration through exponential bounds.
Classical probability references include
\citet{feller1968introduction}, \citet{chung1974course},
\citet{billingsley1995probability}, and \citet{durrett2019probability}.
The chapter's path through conditional expectation, martingale differences,
concentration, sample quantiles, empirical distribution functions, and
Glivenko--Cantelli arguments is also informed by
\citet{dabrowskaAdvancedProbabilityCommunication}.
The opening Markov, Chebyshev, and exponential Markov inequalities are included
as a minimal toolbox rather than as a separate theory of inequalities.  Normal
approximation rates such as Berry--Esseen bounds, norm inequalities such as
Minkowski's inequality, and continuous-time martingale inequalities such as
Burkholder--Davis--Gundy are deliberately kept as boundary markers here; they
belong more naturally with weak convergence, functional approximation, and the
continuous-time martingale chapter.
Complete convergence, a strengthening of the strong-law viewpoint that asks
for summability of tail probabilities, enters the modern literature through
the Hsu--Robbins theorem \citep{hsuRobbins1947complete}.  This is one place
where Pao-Lu Hsu's work belongs directly inside the technical spine of
probability theory, not only in historical background.
Hoeffding's bounded-difference inequality is due to
\citet{hoeffding1963probability}; the martingale extension is due to
\citet{azuma1967weighted}.  Bernstein inequalities go back to
\citet{bernstein1924theory} and now appear in many martingale and empirical
process forms.  The Johnson--Lindenstrauss embedding lemma is due to
\citet{johnson1984extensions}; the random-projection proof used here follows
the elementary route of \citet{dasgupta2003elementary}.  Computationally
friendly variants include \citet{achlioptas2003database}.  The
Dvoretzky--Kiefer--Wolfowitz inequality was proved by
\citet{dvoretzky1956asymptotic}; \citet{massart1990tight} identified the
sharp constant.  The proxy climate curve example points back to the Zhu/Chu
Ko-chen reconstruction and modern climate reconstructions
\citep{chu1973climatic,ge2013temperature,ge2016recent}; it is used here only
to illustrate the bias--variance logic of smoothing historical traces.  The
bivariate Kaplan--Meier examples are paraphrases of the statistical idea in
\citet{dabrowska1988kaplan} and \citet{dabrowska1989kaplan}, with the surrounding
uniform-law and weak-convergence grammar aligned with
\citet{dabrowskaAdvancedProbabilityCommunication}.  The counting-process and
cumulative-hazard background is also consistent with
\citet{dabrowskaStochasticProcessesCommunication}.  The empirical
CDF material points forward to
\citet{pollard1984convergence}, \citet{pollard1990empirical}, and
\citet{vaart2023weak}.

%% file: figures/ch10_johnson_lindenstrauss_projection.tex
\begin{center}
\includegraphics[width=0.98\linewidth]{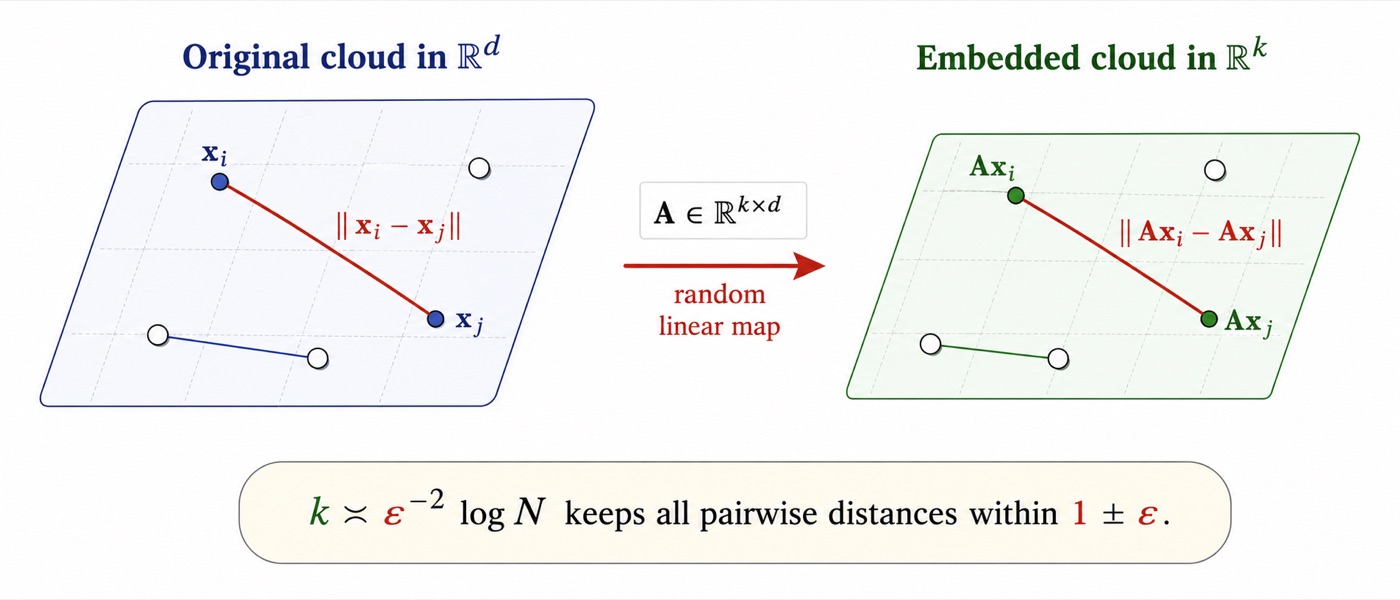}
\bookmanualfigure{fig:ch09-johnson-lindenstrauss}{A Johnson--Lindenstrauss embedding}
\par\smallskip
\small Figure~\thefigure: A Johnson--Lindenstrauss embedding.  A random linear
map compresses an $N$-point cloud from $\R^d$ to $\R^k$ while preserving all
pairwise distances up to a controlled multiplicative error.
\end{center}

%% file: appendices/set_theory_compass.tex
\chapter{Set-Theoretic Language for Measurability}
\label{app:set-theory-compass}
\conceptindexes{set theory, sets, functions, products, generated families, countability, set limits}

Before probability can measure uncertainty, set theory has to say what kind of
uncertainty can be named. This appendix is not a substitute for a set-theory text. It is a
small working glossary for \ifdefined\APPENDICES Chapter~3\else
Chapter~\ref{chap:probability-measure}\fi: enough language to read
$\sigma$-fields, generated classes, measurable maps, and set limits without
losing the main thread.

\section{Basic Set Language and Algebra}
\conceptindexes{set language, set algebra, complements, unions, intersections, disjointification}
\label{sec:app-basic-set-language}

\subsection{Naive Language, Fixed Universes}
\conceptindexes{fixed universe, naive set language, ambient set}
\label{sec:app-naive-language}

Throughout the probability chapters, ordinary set language is used, but always
inside a fixed universe. A sample space \(\Omega\) is the universe of possible
outcomes, and an event is a subset of \(\Omega\). This convention matters because
a complement is not absolute: \(A^c\) means \(\Omega-A\), not ``everything that is
not \(A\)'' in some universe of all objects.

The power set \(\Pow(\Omega)\) is the family of all subsets of \(\Omega\). It is
the largest possible event family. Measure theory usually works with a smaller
family \(\fieldF\subseteq\Pow(\Omega)\), because \(\Pow(\Omega)\) can be too large
to carry the required kind of measure.

\subsection{Sets and Operations}
\conceptindexes{sets, set operations, complement, relative difference, symmetric difference}
\label{sec:app-sets-operations}

We write \(A\subseteq B\) when every element of \(A\) also belongs to \(B\). The
complement of \(A\subseteq\Omega\) is
\[
  A^c=\Omega-A=\{\omega\in\Omega:\omega\notin A\}.
\]
The union \(A\cup B\) means ``\(A\) or \(B\),'' and the intersection
\(A\cap B\) means ``\(A\) and \(B\).'' The relative difference
\[
  A-B=A\cap B^c
\]
keeps the part of \(A\) outside \(B\). The symmetric difference
\[
  A\triangle B=(A-B)\cup(B-A)
\]
records where two sets disagree.

A few identities are used constantly:
\[
  A\cap B=A-(A-B),\qquad
  A-B=A\cap B^c,\qquad
  A\triangle B=(A\cup B)-(A\cap B).
\]
De Morgan's laws are
\[
  (A\cup B)^c=A^c\cap B^c,\qquad
  (A\cap B)^c=A^c\cup B^c.
\]
More generally, if \(\{A_t:t\in T\}\) is an indexed family, then
\[
  \left(\bigcup_{t\in T}A_t\right)^c=\bigcap_{t\in T}A_t^c,
  \qquad
  \left(\bigcap_{t\in T}A_t\right)^c=\bigcup_{t\in T}A_t^c.
\]

Indicator functions turn set algebra into ordinary algebra. For \(A\subseteq
\Omega\),
\[
  \ind{A}(\omega)=
  \begin{cases}
  1, & \omega\in A,\\
  0, & \omega\notin A.
  \end{cases}
\]
Then
\begin{align*}
  \ind{A^c}&=1-\ind{A},&
  \ind{A\cap B}&=\ind{A}\ind{B},\\
  \ind{A\cup B}&=\ind{A}\vee\ind{B},&
  \ind{A\triangle B}&=|\ind{A}-\ind{B}|.
\end{align*}
This is often the fastest way to check a set identity.

\subsection{Countable Operations}
\conceptindexes{countable operations, countable union, countable intersection}
\label{sec:app-countable-operations}

If \(\{A_t:t\in T\}\) is an indexed family, then
\begin{align*}
  \bigcup_{t\in T}A_t
  &=\{\omega:\omega\in A_t\text{ for at least one }t\in T\},\\
  \bigcap_{t\in T}A_t
  &=\{\omega:\omega\in A_t\text{ for every }t\in T\}.
\end{align*}
For a sequence of sets \(A_1,A_2,\ldots\), this becomes
\[
  \bigcup_{n=1}^{\infty}A_n
  =\{\omega:\omega\in A_n\text{ for at least one }n\},
  \qquad
  \bigcap_{n=1}^{\infty}A_n
  =\{\omega:\omega\in A_n\text{ for every }n\}.
\]
Measure theory is built around countable operations. Finite closure is too
weak for limits and approximation; closure under arbitrary uncountable unions
is usually too strong.

A sequence \(A_n\) is increasing, written \(A_n\uparrow A\), if
\[
  A_1\subseteq A_2\subseteq\cdots,
  \qquad
  A=\bigcup_{n=1}^{\infty}A_n.
\]
It is decreasing, written \(A_n\downarrow A\), if
\[
  A_1\supseteq A_2\supseteq\cdots,
  \qquad
  A=\bigcap_{n=1}^{\infty}A_n.
\]
These two notations are the set-theoretic version of monotone convergence.

\subsection{Disjoint Families and the Disjointification Trick}
\conceptindexes{disjoint families, disjointification trick}
\label{sec:app-partitions}

Sets \(A_i\) are disjoint if \(A_i\cap A_j=\emptyset\) whenever \(i\ne j\).
A partition of \(\Omega\) is a family of disjoint sets whose union is
\(\Omega\). Partitions are important because they represent finite or countable
levels of resolution. If the cells of a map partition London into squares,
then the corresponding event language can initially distinguish only unions of
those squares.

When a union is disjoint, it is sometimes written
\[
  \bigsqcup_i A_i
\]
to remind the reader that no point is counted twice.

Every countable union can be turned into a disjoint union. Given arbitrary sets
\(A_1,A_2,\ldots\), define
\[
  B_1=A_1,\qquad
  B_n=A_n-\bigcup_{i=1}^{n-1}A_i,\quad n\ge2.
\]
Then the \(B_n\)'s are disjoint and
\[
  \bigcup_{n=1}^{\infty}A_n=\biguplus_{n=1}^{\infty}B_n.
\]
This small device is what lets countable additivity imply subadditivity and
continuity from below.

\section{Products, Maps, and Generated Families}
\conceptindexes{Cartesian products, projections, cylinders, functions, images, preimages, generated families}
\label{sec:app-products-maps-generated}

\subsection{Cartesian Products, Projections, and Cylinders}
\conceptindexes{Cartesian product, projections, cylinders}
\label{sec:app-products-cylinders}

For sets \(X\) and \(Y\), the Cartesian product is
\[
  X\times Y=\{(x,y):x\in X,\ y\in Y\}.
\]
If \(A\subseteq X\) and \(B\subseteq Y\), then \(A\times B\) is called a
rectangle. It need not look like a geometric rectangle; it is a product-shaped
set.

The basic identities are
\[
  (A_1\times B_1)\cap(A_2\times B_2)
  =(A_1\cap A_2)\times(B_1\cap B_2),
\]
and, relative to \(X\times Y\),
\[
  (A\times B)^c=(A^c\times Y)\cup(X\times B^c).
\]
These identities explain why rectangles generate product \(\sigma\)-fields.

For an infinite product \(X_T=\prod_{t\in T}X_t\), the coordinate projection
\(\pi_t:X_T\to X_t\) is \(\pi_t(x)=x_t\). If
\(\alpha=(t_1,\ldots,t_k)\) is finite, then
\[
  \pi_\alpha(x)=(x_{t_1},\ldots,x_{t_k}).
\]
A cylinder set is a preimage of a finite-dimensional set:
\[
  \pi_\alpha^{-1}(B)
  =\{x\in X_T:(x_{t_1},\ldots,x_{t_k})\in B\}.
\]
Cylinder sets are how infinite product spaces remain readable: they ask only
finitely many coordinate questions at a time.

\subsection{Functions, Images, and Preimages}
\conceptindexes{functions, images, preimages, inverse images}
\label{sec:app-preimages}

For a function \(X:\Omega\to S\) and a set \(B\subseteq S\), the preimage of
\(B\) is
\[
  X^{-1}(B)=\{\omega\in\Omega:X(\omega)\in B\}.
\]
This is the event that the value of \(X\) falls in \(B\). A map
\(X:(\Omega,\fieldF)\to(S,\mathcal S)\) is measurable if
\[
  X^{-1}(B)\in\fieldF
  \qquad\text{for every }B\in\mathcal S.
\]
Thus measurability means that every question about the reported value of \(X\)
corresponds to an observable event in the original outcome space.

Preimages preserve the set operations that matter for measure theory:
\[
  X^{-1}(B^c)=X^{-1}(B)^c,\qquad
  X^{-1}\left(\bigcup_tB_t\right)=\bigcup_tX^{-1}(B_t),
\]
and
\[
  X^{-1}\left(\bigcap_tB_t\right)=\bigcap_tX^{-1}(B_t).
\]
Images do not behave as well: \(X(A\cap C)\subseteq X(A)\cap X(C)\) can be
strict. This is why measurability is stated using preimages, not images.

\subsection{Generated Families}
\conceptindexes{generated families, generated sigma-field, closure operation}
\label{sec:app-generated-families}

If \(\classS\) is a family of subsets of \(\Omega\), then
\(\sigma(\classS)\) denotes the smallest \(\sigma\)-field containing
\(\classS\). The word ``smallest'' means intersection of all \(\sigma\)-fields
that contain \(\classS\):
\[
  \sigma(\classS)
  =
  \bigcap\{\fieldF:\classS\subseteq\fieldF,\ \fieldF
  \text{ is a }\sigma\text{-field on }\Omega\}.
\]
Generated families allow simple observable events to serve as starting points, then close them
under the operations needed for probability. For example, the Borel
\(\sigma\)-field on \(\Real\) is generated by open intervals. On \(\Real^d\), it
is often generated by half-open rectangles with rational endpoints.

\section{Size, Extended Values, and Limits}
\conceptindexes{countability, uncountability, extended nonnegative reals, limits of sets}
\label{sec:app-size-values-limits}

\subsection{Countable Versus Uncountable}
\conceptindexes{countable sets, uncountable sets, sequences}
\label{sec:app-countable-uncountable}

A set is countable if it is finite or can be listed as a sequence. The basic
rules are:
\[
  \Nat\times\Nat\text{ is countable},\qquad
  \bigcup_{n=1}^{\infty}C_n\text{ is countable if each }C_n\text{ is countable}.
\]
The real line is uncountable. The power set \(\Pow(\Nat)\) has the same
cardinality as \(\Real\), often denoted by \(\mathfrak c\).

This distinction is not bookkeeping. A \(\sigma\)-field is closed under countable
unions and intersections, but not under arbitrary uncountable ones. This is why a
set constructed by taking uncountably many individually measurable pieces may
fail to be measurable, and why product spaces and path spaces require care.

For most of this book, the practical rule is simple: finite and countable set
operations are safe inside a \(\sigma\)-field; uncountable operations must be
justified.

\subsection{Extended Nonnegative Reals}
\conceptindexes{extended nonnegative reals, infinity, monotone limits}
\label{sec:app-extended-nonnegative}

Measures take values in \([0,\infty]\), not just in finite real numbers. The
symbol \(\infty\) is therefore part of the arithmetic. For \(a\in[0,\infty]\),
\[
  a+\infty=\infty,\qquad
  a\cdot\infty=
  \begin{cases}
  0, & a=0,\\
  \infty, & a>0,
  \end{cases}
\]
with the usual convention that \(\infty-\infty\) is undefined. This is why
statements such as
\[
  \mu(B-A)=\mu(B)-\mu(A)
\]
need hypotheses like \(\mu(A)<\infty\). Without a finite anchor, subtraction
from infinity is not a legal operation.

\subsection{Limits of Sets}
\conceptindexes{set limits, limsup of sets, liminf of sets, monotone set sequences}
\label{sec:app-set-limits}

For events \(A_n\), the limsup and liminf are
\[
  \limsup_{n\to\infty} A_n
  =
  \bigcap_{m=1}^{\infty}\bigcup_{n\ge m} A_n,
  \qquad
  \liminf_{n\to\infty} A_n
  =
  \bigcup_{m=1}^{\infty}\bigcap_{n\ge m} A_n.
\]
The event \(\limsup_n A_n\) means that infinitely many of the events \(A_n\)
occur. The event \(\liminf_n A_n\) means that all but finitely many occur.
These two constructions are why countable unions and intersections appear so
early in probability.

The indicator-function version is often useful:
\[
  \ind{\limsup_nA_n}=\limsup_n\ind{A_n},
  \qquad
  \ind{\liminf_nA_n}=\liminf_n\ind{A_n}.
\]
Set convergence \(A_n\to A\) means
\[
  \limsup_nA_n=\liminf_nA_n=A.
\]
In particular, \(A_n\uparrow A\) and \(A_n\downarrow A\) are special cases of set
convergence.

\section{How This Set Language Serves the Book}
\conceptindexes{set-theoretic compass, appendix navigation, mathematical infrastructure}
\label{sec:appA-serving-book}

The set language above has one job: it keeps the book's questions well formed
before probability assigns them numbers.  A set is the mathematical form of a
yes-or-no question; a family of sets says which questions are being allowed; a
generated family says how simple observable questions are closed under the
operations needed for probability.

This is why the same small vocabulary keeps reappearing.  Products and
projections describe records with several coordinates.  Preimages translate
questions about a recorded value back into events on the underlying sample
space.  Countable unions and intersections make room for repeated sampling,
stopping events, tail events, and limiting statements.  Indicators turn event
logic into algebra, so that a probability of a set can become an expectation of
a function.

In the main chapters, this compass should be read as grammar rather than as
extra ornament.  When a later argument asks whether a feature is measurable,
whether a cylinder event is enough to describe a process, or whether an event
occurs infinitely often, it is asking a set-theoretic question first.  The
measure-theoretic tools of \Appref{app:measure-theoretic-toolkit} then
put sizes, integrals, and conditional laws on top of that grammar.

\section*{Sources and Further Reading}
\addcontentsline{toc}{section}{Sources and Further Reading}

This appendix uses standard set-theoretic and measure-theoretic grammar in the
form needed for probability.  The local route through axioms, functions,
preimages, generated families, countability, products, indicators, and set
limits is informed by \citet{dabrowskaAdvancedProbabilityCommunication}.  For
full background on the measure-theoretic uses of this language, see
\citet{billingsley1995probability}, \citet{folland1999real}, and
\citet{bogachev2007measure}.

%% file: appendices/measure_theoretic_toolkit.tex
\chapter{Measure-Theoretic Toolkit}
\label{app:measure-theoretic-toolkit}
\conceptindexes{measure-theoretic toolkit, generators, Borel spaces, extension theorem, integration, reference measures, conditioning, product laws}
\providecommand{\classD}{\mathcal{D}}
\providecommand{\classH}{\mathcal{H}}

This appendix is a reusable toolkit for the measure-theoretic arguments that
appear throughout the book.  \Appref{app:set-theory-compass} supplies the
basic set language; this appendix supplies the construction and proof tools:
Dynkin systems, semi-rings, outer measures, extension arguments behind product
constructions, integration, Radon--Nikodym derivatives, conditional
expectations, regular conditional laws, Hausdorff measures, and the zero-one
arguments behind repeated random phenomena.  The local organization is informed by
\citet{dabrowskaAdvancedProbabilityCommunication}, alongside the standard
references cited below.

\noindent\textbf{Reading note.}
This appendix is organized by measure-theoretic dependency, not by a first
course's order of exposition.  The order is now arranged so that the compact
integration review comes before Radon--Nikodym derivatives and density
translations.  This keeps the dependency visible: a density is a way to express
one measure through integrals with respect to another measure.

\section{Generators, Borel Structure, and Uniqueness}
\label{sec:appB-generators-borel-uniqueness}
\conceptindexes{generators, Borel structure, uniqueness, Dynkin theorem, monotone class theorem}

\subsection{Set Systems Used in Extension Arguments}
\label{sec:appB-set-systems}
\conceptindexes{set systems, extension arguments, pi-system, lambda-system}

A \(\pi\)-system is a nonempty family \(\classS\subseteq\Pow(\Omega)\) such that
\[
  A,B\in\classS \quad\Longrightarrow\quad A\cap B\in\classS.
\]
It is the right level of structure for uniqueness: if two finite measures agree
on a \(\pi\)-system, then the \(\pi\)-\(\lambda\) argument can often carry the
agreement to the generated \(\sigma\)-field.

A \(\lambda\)-system, also called a Dynkin system, is a family
\(\classD\subseteq\Pow(\Omega)\) such that
\[
  \Omega\in\classD,\qquad
  A\subseteq B,\ A,B\in\classD \Longrightarrow B-A\in\classD,
\]
and, whenever \(A_1\subseteq A_2\subseteq\cdots\) with \(A_n\in\classD\),
\[
  \bigcup_{n=1}^{\infty}A_n\in\classD.
\]
Equivalently, a \(\lambda\)-system contains \(\Omega\), is closed under
complements, and is closed under countable disjoint unions.

A ring of sets is a nonempty family \(\classR\) closed under finite unions and
relative differences.  An algebra is a ring that contains \(\Omega\).  A
semi-ring is a family \(\classS\) with \(\emptyset\in\classS\) such that
\[
  A,B\in\classS \Longrightarrow A\cap B\in\classS,
\]
and for every \(A,B\in\classS\) the difference \(A-B\) can be written as a
finite disjoint union of sets in \(\classS\).  A semi-algebra is a semi-ring
that contains \(\Omega\).

The important hierarchy is
\[
  \text{semi-ring}
  \longrightarrow
  \text{ring of finite disjoint unions}
  \longrightarrow
  \sigma\text{-field}.
\]
This is the route by which a set function specified on rectangles or intervals
becomes a measure on the full event system.

\subsection{Dynkin's Theorem and Uniqueness}
\label{sec:appB-dynkin}
\conceptindexes{Dynkin theorem, uniqueness, pi-lambda theorem}

Let \(\lambda(\classS)\) denote the smallest \(\lambda\)-system containing
\(\classS\), obtained by intersecting all such \(\lambda\)-systems.

\begin{theorem}[Dynkin's theorem]
If \(\classS\) is a \(\pi\)-system, then
\[
  \lambda(\classS)=\sigma(\classS).
\]
\end{theorem}

\noindent\textit{Proof.}
Since every \(\sigma\)-field is a \(\lambda\)-system, one always has
\(\lambda(\classS)\subseteq\sigma(\classS)\).  It remains to show that
\(\lambda(\classS)\) is itself a \(\sigma\)-field.

Put \(\classD=\lambda(\classS)\).  For fixed \(A\in\classS\), define
\[
  \classD_A=\{B\in\classD:A\cap B\in\classD\}.
\]
Because \(\classS\) is a \(\pi\)-system, \(\classS\subseteq\classD_A\).  The
closure properties of a \(\lambda\)-system show that \(\classD_A\) is also a
\(\lambda\)-system.  Hence \(\classD\subseteq\classD_A\), so
\[
  A\in\classS,\ B\in\classD \Longrightarrow A\cap B\in\classD.
\]
Now fix \(B\in\classD\) and define
\[
  \classD^B=\{A\in\classD:A\cap B\in\classD\}.
\]
The previous paragraph gives \(\classS\subseteq\classD^B\), and the same closure
check shows that \(\classD^B\) is a \(\lambda\)-system.  Thus
\(\classD\subseteq\classD^B\).  Therefore \(\classD\) is closed under finite
intersections.  A \(\lambda\)-system that is also a \(\pi\)-system is a
\(\sigma\)-field, so \(\sigma(\classS)\subseteq\classD\). \qedmark

\begin{theorem}[Uniqueness from a generating class]
Let \(\classS\) be a \(\pi\)-system and let
\(\fieldF=\sigma(\classS)\).  If \(\mu_1\) and \(\mu_2\) are finite measures on
\((\Omega,\fieldF)\), agree on \(\classS\), and satisfy
\(\mu_1(\Omega)=\mu_2(\Omega)\), then \(\mu_1=\mu_2\) on \(\fieldF\).
\end{theorem}

\noindent\textit{Proof.}
Let
\[
  \classD=\{A\in\fieldF:\mu_1(A)=\mu_2(A)\}.
\]
The set \(\Omega\) belongs to \(\classD\).  If \(A\subseteq B\) and both sets
belong to \(\classD\), then finiteness gives
\[
  \mu_1(B-A)=\mu_1(B)-\mu_1(A)
  =\mu_2(B)-\mu_2(A)=\mu_2(B-A).
\]
If \(A_n\uparrow A\) and \(A_n\in\classD\), continuity from below gives
\[
  \mu_1(A)=\lim_n\mu_1(A_n)=\lim_n\mu_2(A_n)=\mu_2(A).
\]
Thus \(\classD\) is a \(\lambda\)-system containing \(\classS\).  Dynkin's
theorem gives \(\fieldF=\sigma(\classS)\subseteq\classD\). \qedmark

\begin{theorem}[Monotone class theorem for functions]
Let \(\classH\) be a vector space of bounded real functions on \(\Omega\) that
contains the constants and is closed under bounded increasing pointwise limits.
If \(\classH\) contains the indicators of a \(\pi\)-system \(\classS\), then
\(\classH\) contains every bounded \(\sigma(\classS)\)-measurable function.
\end{theorem}

\noindent\textit{Proof.}
First define
\[
  \classD=\{A\subseteq\Omega:\indset A\in\classH\}.
\]
The set \(\Omega\) belongs to \(\classD\), because constants belong to
\(\classH\).  If \(A\subseteq B\) and \(A,B\in\classD\), then
\(\indset{B-A}=\indset B-\indset A\in\classH\), so \(B-A\in\classD\).  If
\((A_n)\) are disjoint members of \(\classD\), then the finite partial sums
\(\sum_{k=1}^n\indset{A_k}\) belong to \(\classH\), increase pointwise to
\(\indset{\cup_k A_k}\), and are bounded by \(1\).  Hence
\(\cup_kA_k\in\classD\).  Thus \(\classD\) is a \(\lambda\)-system.  Since
\(\classS\subseteq\classD\), Dynkin's theorem gives
\(\sigma(\classS)\subseteq\classD\).

Therefore indicators of \(\sigma(\classS)\)-sets belong to \(\classH\).  By
linearity, all bounded simple \(\sigma(\classS)\)-measurable functions belong to
\(\classH\).  If \(f\) is a bounded nonnegative
\(\sigma(\classS)\)-measurable function, choose simple functions
\(s_n\uparrow f\) with \(0\le s_n\le \|f\|_\infty\); the closure assumption gives
\(f\in\classH\).  For a general bounded \(f\), choose \(M\) such that
\(f+M\ge0\), apply the previous step to \(f+M\), and subtract the constant
\(M\).  Hence every bounded \(\sigma(\classS)\)-measurable function lies in
\(\classH\).
\qedmark

This theorem is the workhorse behind many ``check it on rectangles'' arguments.
For instance, once an integral identity is verified for indicators of a
generating class, the monotone class theorem extends it to bounded measurable
functions, and then positive and integrable functions follow by standard
approximation.

\subsection{Half-Open Rectangles and Borel Sets}
\label{sec:appB-borel-rectangles}
\conceptindexes{half-open rectangles, Borel sets, Euclidean Borel sigma-field}

For \(a,b\in\Real^k\) with \(a_j<b_j\), write
\[
  (a,b]=\prod_{j=1}^k(a_j,b_j].
\]
The class of bounded half-open rectangles in \(\Real^k\), together with
\(\emptyset\), is a semi-ring.  Intersections of two such rectangles are again
empty or half-open rectangles.  Differences split into finitely many disjoint
half-open rectangles by cutting along the finitely many coordinate faces that
appear in the two rectangles.

If rational endpoints are allowed, then
\[
  \Borel(\Real^k)
  =
  \sigma\{(a,b]:a,b\in\Rat^k,\ a_j<b_j,\ j=1,\ldots,k\}.
\]
Indeed, every rational half-open rectangle is Borel, so the generated
\(\sigma\)-field is contained in \(\Borel(\Real^k)\).  Conversely, every open
rectangle can be written as a countable union of rational half-open rectangles,
and every open subset of \(\Real^k\) is a countable union of rational open
rectangles.  Since open sets generate \(\Borel(\Real^k)\), the reverse inclusion
follows.

This is the technical reason probability laws can often be specified on
intervals, rectangles, or cylinders and then extended to Borel events.

\subsection{Polish and Standard Borel Spaces}
\label{sec:appB-polish-standard-borel}
\conceptindexes{Polish spaces, standard Borel spaces, Borel isomorphism, non-Polish spaces, non-standard Borel spaces}

\begin{definition}[Polish space and Borel \(\sigma\)-field]
A metric space \((S,d)\) is \emph{Polish} if it is separable and complete: it
has a countable dense subset, and every Cauchy sequence converges in \(S\).
More generally, a topological space \(S\) is called Polish if its topology is
induced by some complete separable metric.  Its Borel \(\sigma\)-field, denoted
\(\Borel(S)\), is the \(\sigma\)-field generated by the open sets.
\end{definition}

\begin{definition}[Standard Borel space]
A measurable space \((S,\mathcal S)\) is \emph{standard Borel} if it is
measurably isomorphic to a Borel subset of a Polish space.  Equivalently for
the purposes of this book, it is a measurable space that can be given a Polish
topology whose Borel sets are exactly \(\mathcal S\).  Standard Borel structure
remembers the measurable sets, not the particular metric that produced them.
\end{definition}

Euclidean spaces, countable discrete spaces, separable Banach spaces with their
Borel \(\sigma\)-fields, \(C[0,1]\) with the uniform topology, and many
Skorokhod path spaces are the standard examples.

The two adjectives should not be conflated.  A topology may fail to be Polish
while the measurable space it generates is still standard Borel.  For example,
\(\Rat\) with the usual subspace topology is not Polish: it is a countable
metric space with no isolated points, so the Baire category theorem rules out
complete metrizability for that topology.  But \((\Rat,\Borel(\Rat))\) is
standard Borel, because it is a countable Borel subset of \(\Real\), and
\(\Borel(\Rat)=\Pow(\Rat)\).

There are also genuinely non-standard Borel spaces.  If \(S\) is uncountable
and carries the discrete \(\sigma\)-field \(\Pow(S)\), then
\((S,\Pow(S))\) is not standard Borel.  Likewise, the Lebesgue-completed
measurable space \(([0,1],\mathcal L)\) is not standard Borel, although its
Borel core \(([0,1],\Borel([0,1]))\) is.  These examples are why the book
states standard-Borel or Polish hypotheses before using regular conditional
laws, disintegration, measurable selection, or weak-convergence representation
theorems.  One useful consequence, proved using Borel codes or the transfinite
Borel hierarchy, is that a second-countable topology generates at most
continuum many Borel sets; see \citet{kechris1995classical} for that fuller
descriptive-set-theoretic viewpoint.

The reason Polish and standard Borel spaces recur in the book is not
topological decoration.  They are the regular state spaces on which probability
behaves like statistics needs it to behave: Borel probability measures are
tight and inner regular, projections lead to analytic sets rather than
pathological arbitrary images, regular conditional laws and disintegrations
exist, and weak-convergence tools such as Prokhorov and Skorokhod--Dudley
representations have their cleanest form.  In practice, the phrase
``assume the state space is Polish'' is a compact way of saying that the random
object lives in a space where kernels, limits, path laws, and measurable
optimization can be handled without exceptional-set bookkeeping; see
\citet{billingsley1999convergence}, \citet{kallenberg2002foundations}, and
\citet{kechris1995classical} for full treatments.

\subsection{Borel Maps and Safe Operations}
\label{sec:appB-borel-maps}
\conceptindexes{Borel maps, safe operations, measurable operations}

Let \(X\) and \(Y\) be topological spaces.  If \(f:X\to Y\) is continuous, then
\[
  f^{-1}(B)\in\Borel(X)
  \qquad(B\in\Borel(Y)).
\]
Indeed, the class
\[
  \classD=\{B\subseteq Y:f^{-1}(B)\in\Borel(X)\}
\]
is a \(\sigma\)-field on \(Y\), and it contains every open set by continuity.
Therefore it contains \(\Borel(Y)\).

In Euclidean space, this immediately gives the following safe operations:
\[
  A\in\Borel(\Real^k),\ h\in\Real^k
  \quad\Longrightarrow\quad
  A+h\in\Borel(\Real^k),
\]
because translation is a homeomorphism.  Similarly, nonsingular linear maps,
scalings by nonzero constants, rotations, coordinate projections, and coordinate
embeddings preserve Borel measurability in the expected preimage direction.

The warning is that images are not controlled by the definition of measurable
maps.  A continuous image of a Borel set in a Polish space is analytic, and
analytic sets are universally measurable, but they need not be Borel without
extra assumptions.  The safe cases used throughout this book are preimage-based
arguments, translations, homeomorphisms, and nonsingular linear maps.

\subsection{Set-Valued Measurability and Optimization}
\label{sec:set-valued-measurability-optimization}
\conceptindexes{set-valued measurability, measurable multifunctions, measurable maximum theorem, measurable selection, random sets}

\noindent\textbf{Statistical thread.}
Statistical procedures often end as optimization problems: choose a minimizer,
a maximizer, or a decision from a random feasible set. Selection theorems show
when such choices are honest random objects rather than informal argmin
notation.

\begin{definition}[Measurable multifunction]
Let $(\Omega,\mathcal F)$ be a measurable space and let $S$ be a metric space.
A multifunction $\Gamma:\Omega\rightrightarrows S$ is weakly measurable if
\[
  \{\omega:\Gamma(\omega)\cap G\ne\emptyset\}\in\mathcal F
\]
for every open $G\subseteq S$. Its graph is
\[
  \graph(\Gamma)=\{(\omega,s):s\in\Gamma(\omega)\}.
\]
\end{definition}

\begin{theorem}[Measurable maximum theorem; \citealp{himmelberg1975measurable}]
Let $S$ be a compact metric space, let $\Gamma(\omega)$ be nonempty compact
valued and weakly measurable, and let
$M:\Omega\times S\to\R$ be jointly measurable with $M(\omega,\cdot)$ continuous.
Then
\[
  V(\omega)=\sup_{s\in\Gamma(\omega)}M(\omega,s)
\]
is measurable, and the argmax multifunction
\[
  A(\omega)=\argmax_{s\in\Gamma(\omega)}M(\omega,s)
\]
is nonempty compact valued and weakly measurable.
\end{theorem}

\noindent\textit{Proof.}
Choose a countable dense set $\{s_j:j\ge1\}$ in compact $S$. For fixed $s_j$,
the distance
\[
  d(s_j,\Gamma(\omega))=\inf_{u\in\Gamma(\omega)}d(s_j,u)
\]
is measurable because
\[
  \{d(s_j,\Gamma(\omega))<r\}
  =
  \{\Gamma(\omega)\cap B(s_j,r)\ne\emptyset\}.
\]
For $n\ge1$, define the countable approximation
\[
  V_n(\omega)=
  \sup_{j:\ d(s_j,\Gamma(\omega))<1/n} M(\omega,s_j),
\]
with the convention that the supremum over an empty set is $-\infty$. Each
$V_n$ is measurable as a countable supremum of measurable terms restricted by
measurable events. Compactness of $\Gamma(\omega)$ and continuity of
$M(\omega,\cdot)$ imply
\[
  V(\omega)=\lim_{n\to\infty}V_n(\omega),
\]
so $V$ is measurable.

The argmax set is nonempty and compact because a continuous function attains
its maximum on the nonempty compact set $\Gamma(\omega)$. To prove weak
measurability, let $G\subseteq S$ be open and define
\[
  V_G(\omega)=\sup_{s\in\Gamma(\omega)\cap G}M(\omega,s).
\]
The same countable approximation, using only dense points eventually lying in
$G$, shows that $V_G$ is measurable. Then
\[
  \{A(\omega)\cap G\ne\emptyset\}
  =
  \bigcap_{m=1}^\infty
  \{V_G(\omega)>V(\omega)-1/m\},
\]
because a point in $G$ can attain the maximum exactly when the supremum over
$\Gamma(\omega)\cap G$ can be made arbitrarily close to $V(\omega)$. Hence
$A$ is weakly measurable.
\qedmark

\begin{theorem}[Measurable selection; \citealp{himmelberg1975measurable}]
If $S$ is a complete separable metric space and $\Gamma$ is weakly measurable
with nonempty closed values, then there exists a measurable selector
$\xi:\Omega\to S$ such that $\xi(\omega)\in\Gamma(\omega)$ for all $\omega$.
\end{theorem}

\noindent\textit{Proof.}
Let $\{s_j:j\ge1\}$ be dense in $S$. For each $n$, cover $S$ by balls
$B(s_j,2^{-n})$. Define $j_1(\omega)$ to be the first index $j$ such that
\[
  \Gamma(\omega)\cap B(s_j,2^{-1})\ne\emptyset.
\]
For example,
\[
  \{j_1=j\}
  =
  \{\omega:\Gamma(\omega)\cap B(s_j,2^{-1})\ne\emptyset\}
  \cap
  \bigcap_{\ell<j}
  \{\omega:\Gamma(\omega)\cap B(s_\ell,2^{-1})=\emptyset\},
\]
so $\{j_1=j\}\in\mathcal F$ by weak measurability and complementation.
At stage $n+1$, define $j_{n+1}(\omega)$ to be the first index $j$ such that
\[
  \Gamma(\omega)\cap B(s_j,2^{-(n+1)})
  \cap B(s_{j_n(\omega)},2^{-n+2})\ne\emptyset.
\]
If $j_n$ is measurable, then the candidate event
\[
  H_{n+1,j}
  =
  \{\omega:\Gamma(\omega)\cap B(s_j,2^{-(n+1)})
  \cap B(s_{j_n(\omega)},2^{-n+2})\ne\emptyset\}
\]
is measurable because
\[
  H_{n+1,j}
  =
  \bigcup_{k=1}^\infty
  \left[
    \{j_n=k\}
    \cap
    \{\omega:\Gamma(\omega)\cap B(s_j,2^{-(n+1)})
    \cap B(s_k,2^{-n+2})\ne\emptyset\}
  \right],
\]
and the second event in brackets is another open-ball hitting event for
$\Gamma$. Hence
\[
  \{j_{n+1}=j\}
  =
  H_{n+1,j}\cap\bigcap_{\ell<j}H_{n+1,\ell}^{c}
\]
is measurable. Thus every $j_n$ is measurable by induction. Put
$\xi_n(\omega)=s_{j_n(\omega)}$. The consistency condition forces
$\{\xi_n(\omega)\}$ to be Cauchy: consecutive centers are within a constant
multiple of $2^{-n}$ of a common point of $\Gamma(\omega)$. Completeness gives a
limit $\xi(\omega)=\lim_n\xi_n(\omega)$, and measurability of $\xi$ follows
because it is a pointwise limit of measurable simple maps. Since
$d(\xi_n(\omega),\Gamma(\omega))\le2^{-n}$ and $\Gamma(\omega)$ is closed, the
limit belongs to $\Gamma(\omega)$.
\qedmark

\begin{example}[Measurable estimators]
Let $\Theta$ be compact and let $M_n(\omega,\theta)$ be measurable in $\omega$
and continuous in $\theta$. Then
\[
  \hat\Theta_n(\omega)=\argmax_{\theta\in\Theta}M_n(\omega,\theta)
\]
is a measurable compact-valued multifunction. A measurable selection theorem
then produces an estimator $\hat\theta_n\in\hat\Theta_n$.

The logic is two-step. The measurable maximum theorem first says that the
random set of maximizers $\hat\Theta_n$ is a weakly measurable closed-valued
multifunction. This still gives a set, not a single estimator. The measurable
selection theorem then supplies a measurable tie-breaking rule
$\omega\mapsto\hat\theta_n(\omega)$ with
$\hat\theta_n(\omega)\in\hat\Theta_n(\omega)$. If the maximizer is unique,
this selector is the usual argmax. If there are ties, the theorem says that
one can choose among the tied maximizers without leaving the category of
random variables.
\qedmark
\end{example}

\begin{example}[K-means centers as measurable selections]
Let $X_1,\ldots,X_n$ be observations taking values in a compact set
$K\subset\R^d$, and fix the number of clusters $k$. A codebook is
$c=(c_1,\ldots,c_k)\in K^k$, and the empirical $k$-means loss is
\[
  L_n(\omega,c)
  =
  \sum_{i=1}^n
  \min_{1\le a\le k}\|X_i(\omega)-c_a\|^2 .
\]
For fixed $c$, this is measurable in $\omega$; for fixed $\omega$, it is
continuous in $c$. Since $K^k$ is compact, the set
\[
  A_n(\omega)=\argmin_{c\in K^k}L_n(\omega,c)
\]
is nonempty and compact. Applying the measurable maximum theorem to
$M_n=-L_n$ shows that $A_n$ is a measurable compact-valued multifunction.
The measurable selection theorem then gives a measurable codebook
$\hat c_n(\omega)\in A_n(\omega)$.

This example is useful because nonuniqueness is not an edge case: permuting
cluster labels gives another optimizer, and ties can create further
minimizers. The theorem is not finding the clusters; it is certifying that the
phrase ``choose a global $k$-means solution'' can be implemented as a genuine
random object.
\qedmark
\end{example}

\section{Constructing Measures from Simple Data}
\label{sec:appB-measure-construction}
\conceptindexes{measure construction, premeasure, outer measure, Caratheodory extension, sigma-finite uniqueness}

\subsection{From Semi-Rings to Rings}
\label{sec:appB-semirings-rings}
\conceptindexes{semi-rings, rings, premeasure, finite additivity}

Let \(\classS\) be a semi-ring.  Define
\[
  \classR(\classS)
  =
  \left\{\biguplus_{i=1}^m A_i:
  m<\infty,\ A_i\in\classS\right\}.
\]
Then \(\classR(\classS)\) is the smallest ring containing \(\classS\).
Indeed, finite disjoint unions of semi-ring sets are closed under finite unions
after refining the pieces to be disjoint.  They are also closed under
differences: subtract one semi-ring piece at a time and use the semi-ring
difference property to split the remainder into finitely many semi-ring pieces.
Any ring containing \(\classS\) must contain all such finite disjoint unions, so
\(\classR(\classS)\) is minimal.

Suppose \(\gamma:\classS\to[0,\infty]\) is finitely additive in the following
semi-ring sense: whenever \(A,A_1,\ldots,A_m\in\classS\) and
\[
  A=\biguplus_{i=1}^m A_i,
\]
one has
\[
  \gamma(A)=\sum_{i=1}^m\gamma(A_i).
\]
Then \(\gamma\) extends to \(\classR(\classS)\) by
\[
  \gamma_R\left(\biguplus_{i=1}^m A_i\right)
  =
  \sum_{i=1}^m\gamma(A_i).
\]
This definition is well posed: if the same set has two finite disjoint
representations, refine both by all intersections \(A_i\cap B_j\), then use
the semi-ring difference property to decompose the remaining pieces.  The two
sums agree after passing to the common refinement.

If \(\gamma\) is countably additive on \(\classS\), then \(\gamma_R\) is
countably additive on \(\classR(\classS)\).  In applications, this is how an
assignment on half-open rectangles becomes a premeasure on the ring of finite
unions of such rectangles.

\begin{proposition}[Countable additivity via subadditivity]
Let \(\classS\) be a semi-ring and let \(\classR=\classR(\classS)\).  Suppose
\(\gamma:\classS\to[0,\infty]\) is finitely additive and let \(\gamma_R\) be
its extension to \(\classR\).
\begin{enumerate}
\item If \(\gamma\) is countably subadditive on \(\classS\), then it is
countably additive on \(\classS\).
\item If \(\gamma_R\) is countably subadditive on \(\classR\), then it is
countably additive on \(\classR\).
\item If \(\gamma\) is countably additive on \(\classS\), then
\(\gamma_R\) is countably additive on \(\classR\).
\end{enumerate}
\end{proposition}

\noindent\textit{Proof.}
For the first statement, let
\[
  A=\biguplus_{n=1}^{\infty}A_n
\]
with \(A,A_n\in\classS\).  For each \(k\),
\[
  B_k=\biguplus_{n=1}^k A_n,\qquad C_k=A-B_k
\]
are finite disjoint unions of semi-ring sets.  Finite additivity of \(\gamma_R\)
on the ring gives
\[
  \gamma(A)=\gamma_R(B_k)+\gamma_R(C_k)\ge \sum_{n=1}^k\gamma(A_n).
\]
Letting \(k\to\infty\) gives
\[
  \gamma(A)\ge\sum_{n=1}^{\infty}\gamma(A_n).
\]
The reverse inequality is exactly countable subadditivity.  The same argument
works on the ring \(\classR\), with \(\gamma_R\) in place of \(\gamma\), proving
the second statement.

For the third statement, write
\[
  B=\biguplus_{n=1}^{\infty}B_n,
  \qquad B,B_n\in\classR.
\]
Choose finite disjoint decompositions
\[
  B=\biguplus_{\ell=1}^mD_\ell,\qquad
  B_n=\biguplus_{j=1}^{r(n)}C_{nj},
  \qquad D_\ell,C_{nj}\in\classS.
\]
Then each \(D_\ell\) is decomposed by the intersections
\[
  D_\ell\cap C_{nj}\in\classS,
\]
and countable additivity of \(\gamma\) on \(\classS\) yields
\[
\begin{aligned}
  \gamma_R(B)
  &=\sum_{\ell=1}^m\gamma(D_\ell)\\
  &=\sum_{\ell=1}^m\sum_{n=1}^{\infty}
    \sum_{j=1}^{r(n)}\gamma(D_\ell\cap C_{nj})\\
  &=\sum_{n=1}^{\infty}\sum_{j=1}^{r(n)}
    \sum_{\ell=1}^m\gamma(D_\ell\cap C_{nj})
   =\sum_{n=1}^{\infty}\gamma_R(B_n).
\end{aligned}
\]
All sums contain nonnegative terms, so the interchange of summation is legal.
\qedmark

\begin{example}[Lebesgue--Stieltjes assignment on intervals]
Let \(F:\Real\to\Real\) be nondecreasing and right-continuous, and define
\[
  \gamma_F((a,b])=F(b)-F(a),\qquad \gamma_F(\emptyset)=0.
\]
Finite additivity on half-open intervals follows by telescoping.  If
\[
  (a,b]=\biguplus_{n=1}^{\infty}(a_n,b_n],
\]
then countable additivity is proved by the usual compactness argument: shrink
\((a,b]\) slightly from the left, enlarge each \((a_n,b_n]\) slightly from the
right, extract a finite subcover, and let the two errors tend to zero.  This is
the one-dimensional prototype for the rectangle construction below.
\end{example}

\subsection{Rectangle Functions and Multivariate Distribution Data}
\label{sec:appB-rectangle-functions}
\conceptindexes{rectangle functions, multivariate distribution functions, distribution data}

Let \(F:\Real^k\to\Real\).  For a half-open rectangle
\[
  (a,b]=\prod_{j=1}^k(a_j,b_j],
\]
define the alternating rectangle increment
\[
  \Delta_F(a,b)
  =
  \sum_{\varepsilon\in\{0,1\}^k}
  (-1)^{k-\sum_j\varepsilon_j}
  F(x_\varepsilon),
\]
where
\[
  x_{\varepsilon,j}=\varepsilon_j b_j+(1-\varepsilon_j)a_j.
\]
For \(k=2\), this reads
\[
  \Delta_F(a,b)
  =
  F(b_1,b_2)-F(a_1,b_2)-F(b_1,a_2)+F(a_1,a_2).
\]
For \(k=3\), it is
\[
\begin{aligned}
  \Delta_F(a,b)={}&F(b_1,b_2,b_3)\\
  &-\{F(a_1,b_2,b_3)+F(b_1,a_2,b_3)+F(b_1,b_2,a_3)\}\\
  &+\{F(a_1,a_2,b_3)+F(a_1,b_2,a_3)+F(b_1,a_2,a_3)\}\\
  &-F(a_1,a_2,a_3).
\end{aligned}
\]

A function \(F\) is rectangle-increasing if \(\Delta_F(a,b)\ge0\) for every
half-open rectangle.  If \(F\) is also right-continuous in the coordinatewise
sense and has the correct limits at infinity, then
\[
  \gamma_F((a,b])=\Delta_F(a,b)
\]
is the rectangle assignment of a Borel measure on \(\Real^k\).

The finite additivity proof is a repeated telescoping argument.  In two
dimensions, fix \(I_2=(a_2,b_2]\) and write
\[
  \phi_1(x,I_2)=F(x,b_2)-F(x,a_2).
\]
Then
\[
  \Delta_F((a_1,b_1]\times I_2)
  =
  \phi_1(b_1,I_2)-\phi_1(a_1,I_2).
\]
Adjacent rectangles in the first coordinate therefore add by the
one-dimensional telescoping identity.  The same argument works in the second
coordinate, and a finite grid refinement reduces arbitrary finite unions of
adjacent rectangles to these one-coordinate telescopes.

Countable additivity uses compactness.  Suppose
\[
  A=\biguplus_{n=1}^{\infty}A_n
\]
where \(A,A_n\) are bounded half-open rectangles.  Shrink \(A\) slightly to a
closed rectangle \(K\subset A\) with \(\gamma_F(A)-\varepsilon<\gamma_F(K)\).
Enlarge \(A_n\) slightly to open rectangles \(G_n\supset A_n\) with
\[
  \gamma_F(G_n)<\gamma_F(A_n)+\varepsilon2^{-n}.
\]
The sets \(G_n\) cover \(K\), so compactness gives a finite subcover.  Finite
subadditivity then yields
\[
  \gamma_F(A)-\varepsilon
  <
  \gamma_F(K)
  \le
  \sum_{n=1}^{N}\gamma_F(G_n)
  \le
  \sum_{n=1}^{\infty}\gamma_F(A_n)+\varepsilon.
\]
Letting \(\varepsilon\downarrow0\) gives countable subadditivity, hence
countable additivity by the preceding proposition.

\subsection{Outer Measures and the Extension Theorem}
\label{sec:appB-outer-extension}
\conceptindexes{outer measures, Caratheodory measurability, extension theorem}

Let \(\gamma_R\) be a premeasure on a ring \(\classR\).  For an arbitrary set
\(E\subseteq\Omega\), define the outer measure
\[
  \mu^*(E)
  =
  \inf\left\{
  \sum_{n=1}^{\infty}\gamma_R(R_n):
  E\subseteq\bigcup_{n=1}^{\infty}R_n,\ R_n\in\classR
  \right\}.
\]
The infimum over covers makes \(\mu^*\) defined on every subset of \(\Omega\).
It satisfies
\[
  \mu^*(\emptyset)=0,\qquad
  A\subseteq B\Longrightarrow\mu^*(A)\le\mu^*(B),
\]
and countable subadditivity:
\[
  \mu^*\left(\bigcup_{n=1}^{\infty}E_n\right)
  \le
  \sum_{n=1}^{\infty}\mu^*(E_n).
\]

A set \(A\subseteq\Omega\) is Carath\'eodory measurable if
\[
  \mu^*(E)=\mu^*(E\cap A)+\mu^*(E\cap A^c)
  \qquad\text{for every }E\subseteq\Omega.
\]

\begin{theorem}[Carath\'eodory extension theorem; \citealp{caratheodory1914lineare}]
The Carath\'eodory measurable sets form a \(\sigma\)-field
\(\fieldF^*\).  The restriction \(\mu=\mu^*|_{\fieldF^*}\) is a measure, and
\(\classR\subseteq\fieldF^*\).  Moreover \(\mu\) agrees with \(\gamma_R\) on
\(\classR\).  If \(\gamma_R\) is \(\sigma\)-finite, this extension is unique on
\(\sigma(\classR)\).
\end{theorem}

\noindent\textit{Proof.}
The subadditivity of \(\mu^*\) always gives
\[
  \mu^*(E)\le\mu^*(E\cap A)+\mu^*(E\cap A^c).
\]
Measurability asks for the reverse inequality.

First take \(A\in\classR\).  If \(E\subseteq\bigcup_nR_n\) with
\(R_n\in\classR\), then
\[
  E\cap A\subseteq\bigcup_n(R_n\cap A),
  \qquad
  E\cap A^c\subseteq\bigcup_n(R_n-A).
\]
Since \(\classR\) is a ring and \(\gamma_R\) is finitely additive,
\[
  \gamma_R(R_n)=\gamma_R(R_n\cap A)+\gamma_R(R_n-A).
\]
Taking infima over all covers of \(E\) gives
\[
  \mu^*(E\cap A)+\mu^*(E\cap A^c)\le\mu^*(E),
\]
so every set in \(\classR\) is Carath\'eodory measurable.

The measurable sets are closed under complements by symmetry of the defining
equality.  If \(A\) and \(B\) are measurable, splitting first by \(A\) and then
by \(B\) shows that \(A\cap B\), \(A-B\), and \(A\cup B\) are measurable.  For
disjoint measurable \(A_1,A_2,\ldots\), repeated splitting gives
\[
  \mu^*\!\left(E\cap\bigcup_{n=1}^N A_n\right)
  =
  \sum_{n=1}^N\mu^*(E\cap A_n).
\]
Letting \(N\to\infty\) and using countable subadditivity gives closure under
countable disjoint unions and countable additivity on measurable disjoint
unions.  Together with complements, this proves that the measurable sets form a
\(\sigma\)-field and that \(\mu^*\) restricted to them is a measure.

For \(R\in\classR\), the one-set cover gives \(\mu^*(R)\le\gamma_R(R)\).  For the
reverse inequality, take any cover \(R\subseteq\bigcup_nR_n\) by ring sets and
define
\[
  D_n=R\cap R_n-\bigcup_{j<n}R_j .
\]
Then \(D_n\in\classR\), the \(D_n\)'s are disjoint, \(R=\biguplus_nD_n\), and
\(D_n\subseteq R_n\).  Hence
\[
  \gamma_R(R)=\sum_n\gamma_R(D_n)\le\sum_n\gamma_R(R_n).
\]
Taking the infimum over covers gives \(\gamma_R(R)\le\mu^*(R)\), so the
extension agrees with \(\gamma_R\) on \(\classR\).  If \(\gamma_R\) is
\(\sigma\)-finite, uniqueness on \(\sigma(\classR)\) follows from the
sigma-finite uniqueness theorem applied on finite-measure pieces. \qedmark

\subsection{Sigma-Finite Uniqueness and Approximation}
\label{sec:appB-sigma-finite-approximation}
\conceptindexes{sigma-finite uniqueness, approximation, finite-measure exhaustion}

The finite uniqueness theorem extends to \(\sigma\)-finite measures by cutting
the space into finite pieces.

\begin{theorem}[Sigma-finite uniqueness]
Let \(\classS\) be a \(\pi\)-system and let
\(\fieldF=\sigma(\classS)\).  Let \(\mu_1\) and \(\mu_2\) be
\(\sigma\)-finite measures on \((\Omega,\fieldF)\).  Suppose
\[
  \mu_1(A)=\mu_2(A)\qquad(A\in\classS),
\]
and suppose there are sets \(C_n\in\classS\) such that
\[
  C_n\uparrow\Omega,\qquad \mu_1(C_n)=\mu_2(C_n)<\infty.
\]
Then \(\mu_1=\mu_2\) on \(\fieldF\).
\end{theorem}

\noindent\textit{Proof.}
For each \(n\), define finite measures
\[
  \mu_i^{(n)}(A)=\mu_i(A\cap C_n),\qquad A\in\fieldF,\quad i=1,2.
\]
They agree on \(\classS\), because \(\classS\) is closed under finite
intersections.  They also have the same total mass \( \mu_i(C_n) \).  The
finite uniqueness theorem gives \(\mu_1^{(n)}=\mu_2^{(n)}\) on \(\fieldF\).
Finally,
\[
  \mu_i(A)=\lim_{n\to\infty}\mu_i(A\cap C_n)
\]
by continuity from below. \qedmark

\begin{proposition}[Approximation by finite-measure pieces]
Let \(\mu\) be \(\sigma\)-finite on \((\Omega,\fieldF)\).  There exist
disjoint sets \(D_1,D_2,\ldots\in\fieldF\) such that
\[
  \Omega=\biguplus_{n=1}^{\infty}D_n,
  \qquad
  \mu(D_n)<\infty.
\]
There also exist increasing sets \(C_n\uparrow\Omega\) with
\(\mu(C_n)<\infty\).
\end{proposition}

\noindent\textit{Proof.}
Choose \(E_n\in\fieldF\) with \(\mu(E_n)<\infty\) and \(\Omega=\bigcup_nE_n\).
Set
\[
  D_1=E_1,\qquad
  D_n=E_n-\bigcup_{j<n}E_j,\quad n\ge2.
\]
Then the \(D_n\)'s are disjoint, cover \(\Omega\), and satisfy
\(\mu(D_n)\le\mu(E_n)<\infty\).  Put \(C_n=\bigcup_{j=1}^nD_j\). \qedmark

\begin{example}[Why sigma-finiteness matters]
Counting measure on an uncountable set \(\Omega\) is not \(\sigma\)-finite,
because finite-measure sets are finite and a countable union of finite sets is
countable.  Many uniqueness theorems fail without some condition of this kind:
one cannot reduce an uncountable counting space to countably many finite-mass
pieces.
\end{example}

\section{Integration and Averages}
\label{sec:appB-integration-engine}
\conceptindexes{integration, averages, convergence theorems, uniform integrability, expectation}

An expectation is an integral, and an empirical average is an integral with
respect to an empirical measure:
\[
  P_nf=\int f\,dP_n=\frac1n\sum_{i=1}^n f(X_i).
\]
This is why integration is more than a way to compute means.  It is the
language that decides when a limiting model can replace a finite sample, when a
likelihood expansion may pass a limit through an integral, and when rare large
observations are too dangerous to ignore.  It also prepares the next section:
a density \(d\mu/d\nu\) is precisely the function that makes
\(\mu(A)=\int_A(d\mu/d\nu)\,d\nu\).

\begin{theorem}[Three convergence permissions]
\label{thm:appB-convergence-permissions}
Let \((\Omega,\fieldF,\mu)\) be a measure space.
\begin{enumerate}
\item If \(0\le f_n\uparrow f\), then
\[
  \int f_n\,d\mu\uparrow \int f\,d\mu .
\]
\item If \(f_n\ge0\), then
\[
  \int \liminf_n f_n\,d\mu
  \le
  \liminf_n\int f_n\,d\mu .
\]
\item If \(f_n\to f\) almost everywhere and \(|f_n|\le g\) for an integrable
function \(g\), then \(f\) is integrable and
\[
  \int |f_n-f|\,d\mu\to0,
  \qquad
  \int f_n\,d\mu\to\int f\,d\mu .
\]
\end{enumerate}
\end{theorem}

\noindent\textit{Proof.}
The first statement is the monotone convergence theorem, the construction
principle behind the Lebesgue integral.  Standard proofs build the integral
from simple functions and then use the increasing approximation
\(f_n\uparrow f\); see \citet[Ch.~2]{folland1999real} or
\citet[Sec.~16]{billingsley1995probability}.

For the second statement, put
\[
  h_m=\inf_{n\ge m}f_n.
\]
Then \(h_m\uparrow \liminf_n f_n\).  By monotone convergence,
\[
  \int\liminf_n f_n\,d\mu
  =
  \lim_m\int h_m\,d\mu.
\]
Since \(h_m\le f_n\) for every \(n\ge m\),
\[
  \int h_m\,d\mu
  \le
  \inf_{n\ge m}\int f_n\,d\mu.
\]
Letting \(m\to\infty\) gives Fatou's lemma.

For the third statement, \(g+f_n\ge0\) and \(g-f_n\ge0\).  Fatou's lemma gives
\[
  \int(g+f)\,d\mu
  \le
  \liminf_n\int(g+f_n)\,d\mu
  =
  \int g\,d\mu+\liminf_n\int f_n\,d\mu
\]
and
\[
  \int(g-f)\,d\mu
  \le
  \liminf_n\int(g-f_n)\,d\mu
  =
  \int g\,d\mu-\limsup_n\int f_n\,d\mu .
\]
Therefore
\[
  \limsup_n\int f_n\,d\mu
  \le
  \int f\,d\mu
  \le
  \liminf_n\int f_n\,d\mu,
\]
so the integrals converge.  Applying the same conclusion to
\(|f_n-f|\), which converges to \(0\) and is dominated by \(2g\), gives
\(\int |f_n-f|\,d\mu\to0\). \qedmark

\begin{example}[A rare enormous value defeats naive convergence]
Let
\[
  X_n=
  \begin{cases}
  n, & \text{with probability }1/n,\\
  0, & \text{with probability }1-1/n.
  \end{cases}
\]
Then \(X_n\to0\) in probability, because
\[
  \Prob(|X_n|>\varepsilon)=1/n\to0
\]
for every \(\varepsilon>0\).  But
\[
  \Expect X_n=n\cdot(1/n)=1.
\]
The limit in probability does not control the mean.  A clinical version is a
lab value that is usually quiet but occasionally enormous; a sequencing version
is a count that is usually absent but sometimes explodes because of depth,
batch, or amplification.  Moving expectations through limits requires a
dominating guard or a uniform-integrability guard.
\end{example}

\begin{definition}[Uniform integrability]
A family \(\mathcal X\) of integrable random variables is uniformly integrable
if
\[
  \lim_{K\to\infty}
  \sup_{X\in\mathcal X}
  \Expect\{|X|\ind{|X|>K}\}=0.
\]
\end{definition}

\begin{proposition}[Uniform integrability closes the expectation gap]
\label{prop:appB-ui-vitali}
If \(X_n\to X\) in probability and \(\{X_n:n\ge1\}\) is uniformly integrable,
then \(X\) is integrable and
\[
  \Expect|X_n-X|\to0.
\]
In particular, \(\Expect X_n\to\Expect X\).
\end{proposition}

\noindent\textit{Proof.}
Uniform integrability implies bounded first moments: choose \(K\) so that
\(\sup_n\Expect\{|X_n|\ind{|X_n|>K}\}\le1\), and then
\[
  \sup_n\Expect|X_n|\le K+1.
\]
By Fatou's lemma applied along a subsequence that converges almost surely to
\(X\), \(\Expect|X|\le K+1\), so \(X\) is integrable.

Let \(D_n=X_n-X\).  The family \(\{D_n:n\ge1\}\) is uniformly integrable:
\[
  |D_n|\ind{|D_n|>K}
  \le
  |X_n|\ind{|X_n|>K/2}
  +
  |X|\ind{|X|>K/2}.
\]
The first term is controlled uniformly by the uniform integrability of
\(\{X_n\}\), and the second tends to zero because \(X\in L^1\).

Fix \(\eta>0\).  Choose \(K\) so large that
\[
  \sup_n\Expect\{|D_n|\ind{|D_n|>K}\}<\eta.
\]
For the bounded part,
\[
  \Expect(|D_n|\wedge K)
  \le
  \varepsilon
  +
  K\,\Prob(|D_n|>\varepsilon).
\]
Since \(D_n\to0\) in probability, the right side has limit at most
\(\varepsilon\).  Let \(\varepsilon\downarrow0\), and then use the tail bound
\(\eta\).  Thus \(\Expect|D_n|\to0\). \qedmark

\section{Reference Measures, Geometry, and Densities}
\label{sec:appB-reference-measures}
\conceptindexes{reference measures, densities, Hausdorff measure, Lebesgue measure, Radon--Nikodym derivative, Lebesgue decomposition}

\subsection{Hausdorff Measure and Dimension}
\label{sec:appB-hausdorff}
\conceptindexes{Hausdorff measure, Hausdorff dimension, fractal dimension}

Let \((S,d)\) be a metric space and let \(s\ge0\).  For \(\varepsilon>0\), set
\[
  \mathcal H^s_\varepsilon(A)
  =
  \inf\left\{
  \sum_i(\diam U_i)^s:
  A\subseteq\bigcup_i U_i,\ \diam U_i\le\varepsilon
  \right\}.
\]
As \(\varepsilon\downarrow0\), the allowed covers become more restrictive, so
\(\mathcal H^s_\varepsilon(A)\) increases.  Hence the limit
\[
  \mathcal H^s(A)=\lim_{\varepsilon\downarrow0}
  \mathcal H^s_\varepsilon(A)
\]
exists in \([0,\infty]\).  The resulting set function is an outer measure.  Its
restriction to the Borel sets is a measure; this follows from the metric outer
measure criterion, because sets at positive distance split additively.

The Hausdorff dimension of \(A\) is
\[
  \dim_H(A)
  =
  \inf\{s:\mathcal H^s(A)=0\}
  =
  \sup\{s:\mathcal H^s(A)=\infty\}.
\]
The equality follows from the monotonicity in \(s\): if \(t>s\) and
\(\diam U_i\le\varepsilon\), then
\[
  \sum_i(\diam U_i)^t
  \le
  \varepsilon^{t-s}\sum_i(\diam U_i)^s.
\]
Thus finiteness of \(\mathcal H^s(A)\) forces \(\mathcal H^t(A)=0\) for
\(t>s\).  Hausdorff dimension is the critical exponent where the measurement
changes from infinite to zero.

For the middle-third Cantor set \(C\), the elementary cover by \(2^n\) intervals
of length \(3^{-n}\) gives
\[
  \mathcal H^s(C)
  \le
  \liminf_{n\to\infty}2^n3^{-ns}.
\]
This upper bound vanishes when \(s>\log2/\log3\).  A matching lower bound can be
proved using the natural Cantor probability measure \(\nu\), which assigns mass
\(2^{-n}\) to each level-\(n\) interval.  If an interval \(I\) has length
\(|I|\), then \(\nu(I)\le C_s|I|^s\) for \(s=\log2/\log3\).  Any cover
\(\{U_i\}\) of \(C\) therefore satisfies
\[
  1=\nu(C)\le\sum_i\nu(U_i)\le C_s\sum_i(\diam U_i)^s.
\]
Taking infima shows \(\mathcal H^s(C)>0\), and hence
\[
  \dim_H(C)=\frac{\log2}{\log3}.
\]

\subsection{Lebesgue and Lebesgue--Stieltjes Measures}
\label{sec:appB-stieltjes}
\conceptindexes{Lebesgue measure, Lebesgue--Stieltjes measure, distribution function}

Lebesgue measure on \(\Real\) starts from the semi-ring of half-open intervals
\((a,b]\), with
\[
  \gamma((a,b])=b-a.
\]
The extension theorem produces a measure on \(\Borel(\Real)\).  Completing that
measure gives the usual Lebesgue \(\sigma\)-field and Lebesgue measure.

More generally, let \(F:\Real\to\Real\) be nondecreasing and right-continuous.
Define
\[
  \gamma_F((a,b])=F(b)-F(a).
\]
Then \(\gamma_F\) is countably additive on the interval semi-ring, and the
extension theorem gives the Lebesgue--Stieltjes measure \(\mu_F\) satisfying
\[
  \mu_F((a,b])=F(b)-F(a).
\]
If \(F\) is a distribution function, \(\mu_F\) is the probability law with
distribution function \(F\).

For \(\Real^k\), a multivariate distribution function \(F\) assigns mass to a
half-open rectangle through alternating increments.  In two dimensions,
\[
  \gamma_F((a_1,b_1]\times(a_2,b_2])
  =
  F(b_1,b_2)-F(a_1,b_2)-F(b_1,a_2)+F(a_1,a_2).
\]
Higher dimensions use the corresponding alternating sum over vertices.  The
extension theorem then turns these rectangle probabilities into a probability
measure on \(\Borel(\Real^k)\).

\subsection{Absolute Continuity and Radon--Nikodym Derivatives}
\label{sec:appB-radon-nikodym}
\conceptindexes{absolute continuity, Radon--Nikodym derivative, density}

Let \(\mu\) and \(\nu\) be measures on \((\Omega,\fieldF)\).  The notation
\[
  \mu\ll\nu
\]
means that every \(\nu\)-null set is also \(\mu\)-null.  The notation
\[
  \mu\perp\nu
\]
means that there is \(B\in\fieldF\) such that
\[
  \mu(B^c)=0,\qquad \nu(B)=0.
\]
Thus \(\mu\) lives on \(B\), while \(\nu\) lives on \(B^c\), up to null sets.

\begin{theorem}[Radon--Nikodym; \citealp{radon1913theorie,nikodym1930generalisation}]
Let \(\mu\) and \(\nu\) be \(\sigma\)-finite measures on
\((\Omega,\fieldF)\).  If \(\mu\ll\nu\), then there exists a nonnegative
measurable function \(f\) such that
\[
  \mu(A)=\int_A f\,d\nu
  \qquad(A\in\fieldF).
\]
If \(g\) is another such function, then \(f=g\) \(\nu\)-almost everywhere.
\end{theorem}

\noindent\textit{Proof.}
We first prove the result when both \(\mu\) and \(\nu\) are finite.  Put
\(\rho=\mu+\nu\).  The map
\[
  L(g)=\int g\,d\mu
\]
is a bounded linear functional on \(L^2(\rho)\), since
\[
  |L(g)|
  \le
  \mu(\Omega)^{1/2}\left(\int g^2\,d\mu\right)^{1/2}
  \le
  \mu(\Omega)^{1/2}\|g\|_{L^2(\rho)} .
\]
By the Hilbert-space representation theorem, there is
\(h\in L^2(\rho)\) such that
\[
  \int g\,d\mu=\int gh\,d\rho
  \qquad(g\in L^2(\rho)).
\]
Taking \(g=\indset A\) gives \(\mu(A)=\int_A h\,d\rho\).  Since the left side
is nonnegative for every \(A\), \(h\ge0\) \(\rho\)-almost everywhere.  Also
\[
  \nu(A)=\rho(A)-\mu(A)=\int_A(1-h)\,d\rho,
\]
so \(1-h\ge0\) \(\rho\)-almost everywhere.  Thus \(0\le h\le1\).

On the set \(\{h=1\}\), the measure \(\nu\) is zero.  Since
\(\mu\ll\nu\), also \(\mu\{h=1\}=0\).  But
\(\mu\{h=1\}=\int_{\{h=1\}}h\,d\rho=\rho\{h=1\}\), so
\(\rho\{h=1\}=0\).  Hence \(h<1\) \(\rho\)-almost everywhere.  Define
\[
  f=\frac{h}{1-h}.
\]
For every \(A\in\fieldF\),
\[
  \int_A f\,d\nu
  =
  \int_A \frac{h}{1-h}(1-h)\,d\rho
  =
  \int_A h\,d\rho
  =
  \mu(A).
\]
This proves existence in the finite case.

For the \(\sigma\)-finite case, choose a disjoint measurable partition
\(\Omega=\biguplus_{n\ge1}E_n\) such that
\(\mu(E_n)<\infty\) and \(\nu(E_n)<\infty\) for every \(n\).  Apply the finite
case to the restricted measures on \(E_n\), obtaining \(f_n\) with
\[
  \mu(A\cap E_n)=\int_{A\cap E_n}f_n\,d\nu .
\]
Set \(f=f_n\) on \(E_n\).  Summing over \(n\) gives
\[
  \mu(A)
  =
  \sum_n \mu(A\cap E_n)
  =
  \sum_n \int_{A\cap E_n}f\,d\nu
  =
  \int_A f\,d\nu .
\]

For uniqueness, suppose \(f\) and \(g\) both represent \(\mu\).  Then
\(\int_A(f-g)\,d\nu=0\) for every \(A\).  Taking
\(A=\{f>g\}\) and \(A=\{g>f\}\) gives \(f=g\) \(\nu\)-almost everywhere.
\qedmark

The function \(f\) is denoted by \(d\mu/d\nu\).  It should be read literally as
a derivative of one measure with respect to another.  If \(\nu=\lambda\) is
Lebesgue measure, then \(f\) is an ordinary density.  If \(\nu\) is counting
measure on a countable set, then \(f(x)=\mu(\{x\})\) is a mass function.  The
same theorem covers both cases.

\begin{proposition}[Chain rule]
If \(\mu\ll\nu\ll\rho\), then
\[
  \frac{d\mu}{d\rho}
  =
  \frac{d\mu}{d\nu}\frac{d\nu}{d\rho}
  \qquad \rho\text{-almost everywhere}.
\]
If \(\mu\) and \(\nu\) are mutually absolutely continuous, then
\[
  \frac{d\nu}{d\mu}
  =
  \left(\frac{d\mu}{d\nu}\right)^{-1}
  \qquad \mu\text{-almost everywhere}.
\]
\end{proposition}

\noindent\textit{Proof.}
For any \(A\in\fieldF\),
\[
  \mu(A)
  =
  \int_A \frac{d\mu}{d\nu}\,d\nu
  =
  \int_A \frac{d\mu}{d\nu}\frac{d\nu}{d\rho}\,d\rho.
\]
Uniqueness in the Radon--Nikodym theorem identifies the derivative.  The inverse
formula follows by applying the first identity to
\(\nu\ll\mu\ll\nu\):
\[
  1
  =
  \frac{d\nu}{d\nu}
  =
  \frac{d\nu}{d\mu}\frac{d\mu}{d\nu}
  \qquad \nu\text{-almost everywhere}.
\]
Because \(\mu\) and \(\nu\) have the same null sets, the same identity holds
\(\mu\)-almost everywhere, and the displayed inverse formula follows. \qedmark

\begin{theorem}[Lebesgue decomposition; \citealp{lebesgue1910discontinues,radon1913theorie}]
Let \(\mu\) and \(\nu\) be \(\sigma\)-finite measures on
\((\Omega,\fieldF)\).  Then \(\mu\) decomposes uniquely as
\[
  \mu=\mu_{\mathrm{ac}}+\mu_{\mathrm{s}},
\]
where
\[
  \mu_{\mathrm{ac}}\ll\nu,\qquad \mu_{\mathrm{s}}\perp\nu.
\]
Consequently,
\[
  \mu(A)=\int_A f\,d\nu+\mu_{\mathrm{s}}(A)
  \qquad(A\in\fieldF)
\]
for a nonnegative measurable \(f\), unique \(\nu\)-almost everywhere.
\end{theorem}

This theorem is the mathematical version of a familiar modeling split.  A
distribution can have a density relative to the chosen background measure, and
also a part that lives where the background measure sees nothing.  For
\(\nu=\lambda\) on \(\Real\), atoms are singular, but not all singular measures
are atomic: the Cantor distribution is singular and continuous.

\begin{example}[Dominating a mixed model]
Let
\[
  P=p\delta_0+(1-p)P_{\mathrm{cont}},
  \qquad P_{\mathrm{cont}}\ll\lambda.
\]
Then \(P\) is not absolutely continuous with respect to \(\lambda\) when
\(p>0\).  But it is absolutely continuous with respect to
\[
  \nu=\delta_0+\lambda.
\]
If \(P_{\mathrm{cont}}\) has Lebesgue density \(f\), then
\[
  \frac{dP}{d\nu}(x)
  =
  p\,\ind{x=0}+(1-p)f(x)\ind{x\ne0}
\]
up to \(\nu\)-null sets.  Choosing a reference measure is therefore part of the
modeling act: it determines what counts as a density.
\end{example}

\begin{example}[Empirical versus continuous laws]
If \(P\) is non-atomic on \(\Real\) and
\[
  P_n=\frac1n\sum_{i=1}^n\delta_{X_i},
\]
then \(P_n\perp P\) almost surely for each fixed \(n\).  Indeed,
\[
  P_n(\{X_1,\ldots,X_n\})=1,
  \qquad
  P(\{X_1,\ldots,X_n\})=0
\]
almost surely.  Weak convergence of \(P_n\) to \(P\), when it holds, is
therefore not absolute continuity gradually appearing.  It is convergence of
integrals against test functions.
\end{example}

\section{Conditioning Toolkit}
\label{sec:appB-conditioning-toolkit}
\conceptindexes{conditioning toolkit, conditional expectation, regular conditional law, Markov kernel, conditional independence}

The observation and likelihood chapter uses conditioning as a language for
information: what has been observed, what is being held fixed, and what
randomness remains.  This section records the technical facts behind that
language.  The main chapter only needs the reading; this appendix gives the
measure-theoretic support.

\subsection{Conditional Expectation from Radon--Nikodym}
\label{sec:appB-conditional-expectation-rn}
\conceptindexes{conditional expectation, Radon--Nikodym theorem, projection}

Let \((\Omega,\fieldF,\Prob)\) be a probability space and let
\(\mathcal H\subseteq\fieldF\) be a sub-\(\sigma\)-field.  If \(Y\) is
integrable, define a signed measure on \(\mathcal H\) by
\[
  \nu(H)=\int_H Y\,d\Prob,\qquad H\in\mathcal H.
\]
The measure \(\nu\) is absolutely continuous with respect to
\(\Prob|_{\mathcal H}\).  The Radon--Nikodym theorem therefore gives an
\(\mathcal H\)-measurable random variable \(Z\) such that
\[
  \int_H Z\,d\Prob
  =
  \int_H Y\,d\Prob,
  \qquad H\in\mathcal H.
\]
This \(Z\) is a version of \(\Expect(Y\mid\mathcal H)\).

\begin{theorem}[Existence and uniqueness of conditional expectation]
If \(Y\in L^1(\Prob)\) and \(\mathcal H\subseteq\fieldF\), then
\(\Expect(Y\mid\mathcal H)\) exists and is unique up to almost sure equality.
\end{theorem}

\noindent\textit{Proof.}
Existence follows from the Radon--Nikodym argument above.  If \(Z_1\) and
\(Z_2\) both satisfy the defining identity, then for every
\(H\in\mathcal H\),
\[
  \int_H (Z_1-Z_2)\,d\Prob=0.
\]
Taking \(H=\{Z_1>Z_2\}\) and \(H=\{Z_2>Z_1\}\), both of which belong to
\(\mathcal H\), gives \(Z_1=Z_2\) almost surely.
\qedmark

Chapter~5 gives the \(L^2\) projection interpretation of conditional
expectation.  The appendix keeps only the construction and rule-checking facts
needed when that geometric picture is not enough.

\subsection{Rules for Conditional Expectation}
\label{sec:appB-conditioning-rules}
\conceptindexes{conditional expectation rules, tower property, pull-out property}

Let \(Y,Y_n\) be integrable whenever the expressions below are used.
Conditional expectation obeys the following rules:
\[
  \Expect\{\Expect(Y\mid\mathcal H)\}=\Expect(Y),
\]
\[
  Y\ \mathcal H\text{-measurable}
  \quad\Longrightarrow\quad
  \Expect(Y\mid\mathcal H)=Y\quad\text{a.s.},
\]
\[
  Y_1\le Y_2\quad\Longrightarrow\quad
  \Expect(Y_1\mid\mathcal H)\le
  \Expect(Y_2\mid\mathcal H)\quad\text{a.s.},
\]
and, for constants \(a,b\),
\[
  \Expect(aY_1+bY_2\mid\mathcal H)
  =
  a\Expect(Y_1\mid\mathcal H)+b\Expect(Y_2\mid\mathcal H).
\]

\begin{proposition}[Tower and pull-out rules]
If \(\mathcal H_0\subseteq\mathcal H_1\subseteq\fieldF\), then
\[
  \Expect\{\Expect(Y\mid\mathcal H_1)\mid\mathcal H_0\}
  =
  \Expect(Y\mid\mathcal H_0)
  \quad\text{a.s.}
\]
If \(Z\) is bounded and \(\mathcal H\)-measurable, then
\[
  \Expect(ZY\mid\mathcal H)
  =
  Z\,\Expect(Y\mid\mathcal H)
  \quad\text{a.s.}
\]
\end{proposition}

\noindent\textit{Proof.}
For the tower rule, integrate both sides over \(H\in\mathcal H_0\) and use
\(H\in\mathcal H_1\).  For the pull-out rule, first prove it for
\(Z=\indset A\), \(A\in\mathcal H\), then extend by linearity to bounded
simple functions and by bounded approximation to bounded
\(\mathcal H\)-measurable functions.
\qedmark

\begin{proposition}[Conditional Jensen inequality]
Let \(g:\Real\to\Real\) be convex.  If \(Y\) and \(g(Y)\) are integrable, then
\[
  g\{\Expect(Y\mid\mathcal H)\}
  \le
  \Expect\{g(Y)\mid\mathcal H\}
  \quad\text{a.s.}
\]
\end{proposition}

\noindent\textit{Proof.}
For a convex \(g\) on an interval, choose a countable family of affine
minorants \(a_m y+b_m\), with rational slopes and intercepts, whose supremum
equals \(g(y)\) at every point of the interval.  This countable reduction is
obtained by taking supporting lines with rational coefficients and then
approximating the supporting constants from below.  For each \(m\),
\[
  a_m\Expect(Y\mid\mathcal H)+b_m
  =
  \Expect(a_mY+b_m\mid\mathcal H)
  \le
  \Expect\{g(Y)\mid\mathcal H\},
\]
because \(a_my+b_m\le g(y)\) pointwise and conditional expectation preserves
order.  Taking the supremum over \(m\) gives
\[
  g\{\Expect(Y\mid\mathcal H)\}
  =
  \sup_m\{a_m\Expect(Y\mid\mathcal H)+b_m\}
  \le
  \Expect\{g(Y)\mid\mathcal H\}
\]
almost surely.
\qedmark

Conditional convergence theorems mirror the ordinary convergence theorems:
monotone convergence, Fatou, and dominated convergence remain valid after
conditioning, provided the same integrability hypotheses are imposed.  These
facts justify passing limits through conditional expectations in martingale
arguments, sequential likelihood calculations, and posterior predictive
approximations.

\subsection{Conditional Probabilities and Kernels}
\label{sec:appB-regular-conditional-laws}
\conceptindexes{conditional probability, regular conditional law, probability kernel}

For an event \(A\in\fieldF\),
\[
  \Prob(A\mid\mathcal H)
  =
  \Expect(\indset A\mid\mathcal H).
\]
For each fixed \(A\), this is an \(\mathcal H\)-measurable random variable
defined up to almost sure equality.  The subtlety is that the exceptional null
set can depend on \(A\).  A regular conditional probability chooses versions
simultaneously so that, for almost every \(\omega\), the map
\[
  A\mapsto K(\omega,A)
\]
is an honest probability measure.

\begin{definition}[Regular conditional law]
Let \(X:(\Omega,\fieldF)\to(S,\mathcal S)\) be measurable and let
\(\mathcal H\subseteq\fieldF\).  A regular conditional law of \(X\) given
\(\mathcal H\) is a function
\[
  K:\Omega\times\mathcal S\to[0,1]
\]
such that \(K(\omega,\cdot)\) is a probability measure for each \(\omega\),
\(K(\cdot,B)\) is \(\mathcal H\)-measurable for each \(B\in\mathcal S\), and
\[
  K(\cdot,B)
  =
  \Prob(X\in B\mid\mathcal H)
  \quad\text{a.s.}
\]
\end{definition}

\begin{theorem}[Existence on standard Borel spaces; \citealp{rokhlin1949fundamental}]
\label{thm:appB-rcp-standard-borel}
Let \(X:(\Omega,\fieldF,\Prob)\to(S,\mathcal S)\) be a random element, where
\((S,\mathcal S)\) is a standard Borel space, and let
\(\mathcal H\subseteq\fieldF\).  Then a regular conditional law of \(X\) given
\(\mathcal H\) exists.
\end{theorem}

\noindent\textit{Proof.}
First suppose \(S=\Real\) and \(\mathcal S=\Borel(\Real)\).  For each rational
number \(q\), choose a version
\[
  F_q(\omega)=\Prob(X\le q\mid\mathcal H)(\omega).
\]
Because the rationals are countable, the versions can be modified on one
\(\mathcal H\)-measurable null set so that, for every \(\omega\),
\[
  0\le F_q(\omega)\le F_r(\omega)\le 1
  \qquad(q\le r,\ q,r\in\Rat).
\]
For \(r\in\Rat\), put
\[
  Z_r(\omega)
  =
  \inf_{\substack{q\in\Rat\\q>r}}
  \{F_q(\omega)-F_r(\omega)\}.
\]
Then \(Z_r\) is nonnegative and \(\mathcal H\)-measurable.  For
\(H\in\mathcal H\),
\[
  0
  \le
  \int_H Z_r\,d\Prob
  \le
  \inf_{\substack{q\in\Rat\\q>r}}
  \left[
    \Prob\{H\cap(X\le q)\}
    -
    \Prob\{H\cap(X\le r)\}
  \right]
  =
  0,
\]
where the last equality follows by continuity from above along rational
\(q\downarrow r\).  Hence \(Z_r=0\) almost surely.  Another countable
modification gives right-continuity on the rational grid for every \(\omega\).
The same argument applied to
\[
  \inf_{q\in\Rat}F_q(\omega),
  \qquad
  \inf_{q\in\Rat}\{1-F_q(\omega)\}
\]
and the events \(\{X\le q\}\downarrow\emptyset\) as \(q\downarrow-\infty\) and
\(\{X\le q\}\uparrow\Omega\) as \(q\uparrow\infty\) gives the endpoint limits
0 and 1, again after one countable modification.  On the exceptional null set,
replace the rational grid by any fixed distribution function \(F_0\).

For \(x\in\Real\), define
\[
  F(\omega,x)
  =
  \inf_{\substack{q\in\Rat\\q>x}} F_q(\omega).
\]
For each \(\omega\), \(x\mapsto F(\omega,x)\) is a distribution function.  For
each fixed \(x\), the map \(\omega\mapsto F(\omega,x)\) is
\(\mathcal H\)-measurable.  If \(q_n\downarrow x\) through rationals, then
\(\indset{\{X\le q_n\}}\downarrow\indset{\{X\le x\}}\); conditional monotone
convergence gives
\[
  F(\cdot,x)
  =
  \Prob(X\le x\mid\mathcal H)
  \quad\text{a.s.}
\]

For each \(\omega\), let \(K(\omega,\cdot)\) be the probability measure on
\(\Borel(\Real)\) determined by the distribution function \(F(\omega,\cdot)\).
The class of Borel sets \(B\) for which \(K(\cdot,B)\) is
\(\mathcal H\)-measurable is a Dynkin system containing the intervals
\((-\infty,x]\), hence all Borel sets.  The class of Borel sets \(B\) for which
\[
  K(\cdot,B)
  =
  \Prob(X\in B\mid\mathcal H)
  \quad\text{a.s.}
\]
is also a Dynkin system, by countable additivity of \(K(\omega,\cdot)\) and
conditional monotone convergence.  It contains the same generating intervals,
so it contains \(\Borel(\Real)\).  Thus \(K\) is a regular conditional law in
the real-valued case.

Now let \((S,\mathcal S)\) be standard Borel.  We use the standard
representation fact that such a space is measurably isomorphic to a Borel
subset \(A\) of \(\Real\).  Let \(\psi:S\to A\) be a measurable bijection with
measurable inverse, and set \(Y=\psi(X)\).  By the real-valued case, \(Y\) has a
regular conditional law \(L(\omega,\cdot)\) on \(\Borel(\Real)\).  Since
\(Y\in A\) almost surely, \(L(\omega,A)=1\) outside an
\(\mathcal H\)-measurable null set.  On that null set replace \(L\) by a fixed
point mass \(\delta_{a_0}\) with \(a_0\in A\).  The modified kernel, still
called \(L\), is supported on \(A\) for every \(\omega\).  Define
\[
  K(\omega,B)=L(\omega,\psi(B)),
  \qquad B\in\mathcal S.
\]
Then \(K(\omega,\cdot)\) is a probability measure on \(\mathcal S\),
\(K(\cdot,B)\) is \(\mathcal H\)-measurable, and
\[
  K(\cdot,B)
  =
  \Prob(Y\in\psi(B)\mid\mathcal H)
  =
  \Prob(X\in B\mid\mathcal H)
  \quad\text{a.s.}
\]
This is the desired regular conditional law of \(X\) given \(\mathcal H\).
\qedmark

Standard Borel spaces include the spaces used in ordinary statistics: Euclidean
spaces, countable spaces, Polish spaces with their Borel \(\sigma\)-fields, and
products of such spaces.  This is the technical reason one can usually write
\[
  \Prob(X\in dx\mid Z=z)
\]
as a probability kernel without apology.

\begin{corollary}[Disintegration over a random variable]
\label{cor:appB-disintegration-random-variable}
Let \((S,\mathcal S)\) and \((T,\mathcal T)\) be standard Borel spaces, and let
\(X:\Omega\to S\) and \(Y:\Omega\to T\) be random elements on the same
probability space \((\Omega,\fieldF,\Prob)\).  The pair map
\[
  (X,Y):\Omega\to S\times T,\qquad
  \omega\mapsto (X(\omega),Y(\omega)),
\]
is measurable into the product \(\sigma\)-field
\(\mathcal S\otimes\mathcal T\).  Its pushforward law
\[
  M=\Prob\circ(X,Y)^{-1}
\]
is called the \emph{joint law} of \((X,Y)\); on rectangles,
\[
  M(B\times C)=\Prob(X\in B,\ Y\in C).
\]
The marginal law of \(Y\) is
\(\mu=P_Y=\Prob\circ Y^{-1}\).  Thus the joint law is not an extra object
assumed into existence; it is exactly the pushforward law construction from
Chapter~3, applied to the pair map.  Then there is a probability kernel
\[
  K:T\times\mathcal S\to[0,1]
\]
such that
\[
  M(B\times C)
  =
  \int_C K(y,B)\,\mu(dy),
  \qquad B\in\mathcal S,\ C\in\mathcal T.
\]
Equivalently, \(K(Y,\cdot)\) is a regular conditional law of \(X\) given
\(\sigma(Y)\).  For every integrable measurable \(h:S\to\Real\),
\[
  \Expect\{h(X)\mid\sigma(Y)\}
  =
  \int_S h(x)\,K(Y,dx)
  \quad\text{a.s.}
\]
\end{corollary}

\noindent\textit{Proof.}
Apply Theorem~\ref{thm:appB-rcp-standard-borel} to the first coordinate under
the joint law \(M\), conditioning on the \(\sigma\)-field generated by the
second coordinate.  The resulting conditional probabilities are measurable with
respect to the second coordinate, so the Doob--Dynkin factorization gives
functions of \(y\).  Choosing these functions on a countable generating class
and extending by a monotone-class argument gives a probability kernel \(K\).
Integrating the defining conditional probabilities over the set \(\{Y\in C\}\)
gives the displayed disintegration identity.  The final conditional-expectation
formula follows first for indicators \(h=\indset B\), then for simple functions,
and finally by the usual positive and integrable decompositions.
\qedmark

\noindent\textit{Density translation.}
The disintegration statement above is measure-theoretic: it only uses the
joint law \(M\), the marginal law \(\mu=P_Y\), and a probability kernel
\(K\).  When a common dominating measure exists, the same identity can be
written in the calculus language of densities.  The next example is this
translation; the integration facts behind it are summarized in
\Appref{sec:appB-integration-engine}.

\begin{example}[Conditional density]
Suppose \((X,Y)\) has joint density \(f_{X,Y}(x,y)\) with respect to a product
dominating measure, and let
\[
  f_Y(y)=\int f_{X,Y}(x,y)\,dx.
\]
For \(f_Y(y)>0\),
\[
  f_{X\mid Y}(x\mid y)
  =
  \frac{f_{X,Y}(x,y)}{f_Y(y)}
\]
defines a version of the conditional density.  This familiar formula is a
special case of regular conditional probability.
\end{example}

\begin{proposition}[Bayesian updating as disintegration]
Let \((\Theta,\mathcal A)\) be a standard Borel parameter space and let
\((\mathcal O,\mathcal G)\) be a standard Borel observation space.  Let
\(\Pi\) be a prior probability measure on \((\Theta,\mathcal A)\), and let
\[
  \theta\mapsto P_\theta(G),
  \qquad G\in\mathcal G,
\]
be an observed-data probability kernel from parameters to observations.  Define
the joint law
\[
  M(B\times G)
  =
  \int_B P_\theta(G)\,\Pi(d\theta),
  \qquad B\in\mathcal A,\ G\in\mathcal G,
\]
and the prior predictive, or evidence, measure
\[
  m(G)=M(\Theta\times G)
  =
  \int_{\Theta}P_\theta(G)\,\Pi(d\theta).
\]
Then there is a posterior probability kernel
\[
  o\mapsto \Pi(\cdot\mid o)
\]
from \((\mathcal O,\mathcal G)\) to \((\Theta,\mathcal A)\) such that
\[
  M(B\times G)
  =
  \int_G \Pi(B\mid o)\,m(do),
  \qquad B\in\mathcal A,\ G\in\mathcal G.
\]
Equivalently, under the joint law \(M\), \(\Pi(\cdot\mid O)\) is a regular
conditional law of the parameter coordinate given the observed coordinate
\(O\).
\end{proposition}

\noindent\textit{Proof.}
Apply the existence theorem for regular conditional laws to the parameter
coordinate on the product space \(\Theta\times\mathcal O\), conditioning on
the \(\sigma\)-field generated by the second coordinate \(O\).  The defining
property of conditional probability gives, for rectangles \(B\times G\),
\[
  M(B\times G)
  =
  \int_{\{O\in G\}}\Pi(B\mid O)\,dM
  =
  \int_G\Pi(B\mid o)\,m(do),
\]
because \(m\) is the marginal law of \(O\).  Rectangles determine the product
law, so the displayed identity is the desired disintegration.
\qedmark

\noindent\textit{Dominated-model translation.}
The posterior kernel above exists before any density is chosen.  If the
observed-data laws are dominated by a common measure \(\nu\), write
\[
  p_\theta(o)=\frac{dP_\theta}{d\nu}(o).
\]
Tonelli's theorem gives the prior predictive density
\[
  \bar p(o)
  =
  \int_{\Theta}p_\theta(o)\,\Pi(d\theta),
  \qquad
  m(do)=\bar p(o)\,\nu(do).
\]
For \(B\in\mathcal A\), the finite measure
\[
  G\mapsto M(B\times G)
\]
has \(\nu\)-density \(\int_B p_\theta(o)\,\Pi(d\theta)\).  Comparing
Radon--Nikodym derivatives in the disintegration identity yields
\[
  \Pi(B\mid o)
  =
  \frac{\int_B p_\theta(o)\,\Pi(d\theta)}
       {\int_{\Theta}p_\theta(o)\,\Pi(d\theta)}
\]
for \(m\)-almost every \(o\) with positive denominator.  This is Bayes'
formula as a Radon--Nikodym identity: the likelihood is a density of the
observed-data law, and the posterior is a conditional law under the joint
prior-predictive construction.  If a fixed design \(d\) is part of the model,
one simply restores the subscript \(d\) on \(P_\theta\), \(M\), \(m\), and
\(\Pi(\cdot\mid o)\); the construction is unchanged.

\subsection{Conditional Independence and Markov Kernels}
\label{sec:appB-conditional-independence}
\conceptindexes{conditional independence, Markov kernels, conditional law}

Sub-\(\sigma\)-fields \(\mathcal F_1\) and \(\mathcal F_2\) are conditionally
independent given \(\mathcal H\) if
\[
  \Prob(A\cap B\mid\mathcal H)
  =
  \Prob(A\mid\mathcal H)\Prob(B\mid\mathcal H)
  \quad\text{a.s.}
\]
for all \(A\in\mathcal F_1\), \(B\in\mathcal F_2\).  Equivalently, for bounded
\(\mathcal F_1\)-measurable \(U\) and bounded \(\mathcal F_2\)-measurable
\(V\),
\[
  \Expect(UV\mid\mathcal H)
  =
  \Expect(U\mid\mathcal H)\Expect(V\mid\mathcal H).
\]
Another equivalent form is
\[
  \Expect(U\mid\mathcal H\vee\mathcal F_2)
  =
  \Expect(U\mid\mathcal H)
  \quad\text{a.s.}
\]
for every bounded \(\mathcal F_1\)-measurable \(U\).

If \(\mathcal F_{\mathrm{past}}\), \(\mathcal F_{\mathrm{present}}\), and
\(\mathcal F_{\mathrm{future}}\) satisfy
\[
  \mathcal F_{\mathrm{past}}
  \perp\!\!\!\perp
  \mathcal F_{\mathrm{future}}
  \mid
  \mathcal F_{\mathrm{present}},
\]
then the future depends on the past only through the present.  This is the
\(\sigma\)-field form of the Markov property.  In later chapters, the same
idea becomes transition kernels, filtrations, intensities, and martingale
residuals.

\section{Product Laws and Extension Details}
\label{sec:appB-product-extension-details}
\conceptindexes{product laws, cylinder consistency, Kolmogorov extension theorem}

\subsection{Cylinder Consistency and Extension Details}
\label{sec:appB-products-cylinders}
\conceptindexes{cylinder consistency, finite-dimensional distributions, extension details}

The product-spaces chapter owns the main definition and use of product
\(\sigma\)-fields.  This appendix only records the technical facts that are
useful when one wants to check generated event systems or invoke an extension
theorem.

For measurable spaces \((S_i,\mathcal S_i)\), \(i=1,\ldots,k\), one starts from
measurable rectangles and generates
\[
  \bigotimes_{i=1}^k\mathcal S_i
  =
  \sigma\left\{
  A_1\times\cdots\times A_k:A_i\in\mathcal S_i
  \right\}.
\]
If \(S_i=\Real\) with the usual Borel \(\sigma\)-field, then
\[
  \Borel(\Real^k)=\Borel(\Real)^{\otimes k}.
\]
The equality follows because rational half-open rectangles generate both
sides.  This identity is used constantly, often silently, when a random vector
is treated as a measurable map into \(\Real^k\).

For a sequence of spaces \((S_i,\mathcal S_i)\), the countable product event
system on
\[
  S^\infty=\prod_{i=1}^{\infty}S_i
\]
is generated by cylinder sets:
\[
  \{\omega:(\omega_{i_1},\ldots,\omega_{i_k})\in B\},
  \qquad
  B\in\bigotimes_{j=1}^k\mathcal S_{i_j}.
\]
Equivalently, it is the smallest \(\sigma\)-field that makes every coordinate
projection measurable.  Cylinder sets are the finite questions about an
infinite object.

\begin{theorem}[Ionescu--Tulcea extension theorem; \citealp{ionescuTulcea1949mesures}]
\label{thm:appB-ionescu-tulcea}
Let \((S_n,\mathcal S_n)\), \(n\ge1\), be measurable spaces.  Let
\(\nu_1\) be a probability measure on \((S_1,\mathcal S_1)\).  For
\(n\ge2\), let
\[
  K_n:S_{1:n-1}\times\mathcal S_n\to[0,1],
  \qquad
  S_{1:n-1}=\prod_{i=1}^{n-1}S_i,
\]
be a probability kernel from
\((S_{1:n-1},\bigotimes_{i=1}^{n-1}\mathcal S_i)\) to
\((S_n,\mathcal S_n)\).  Then there exists a unique probability measure
\(\mu\) on
\[
  \left(\prod_{n=1}^{\infty}S_n,\bigotimes_{n=1}^{\infty}\mathcal S_n\right)
\]
whose finite-prefix probabilities are the iterated kernel probabilities.
That is, for \(A_i\in\mathcal S_i\),
\[
\begin{aligned}
  &\mu\{\omega_1\in A_1,\ldots,\omega_n\in A_n\} \\
  &\quad =
  \int_{A_1}\nu_1(d\omega_1)
  \int_{A_2}K_2(\omega_1,d\omega_2)
  \cdots
  \int_{A_n}K_n(\omega_1,\ldots,\omega_{n-1},d\omega_n).
\end{aligned}
\]
\end{theorem}

\noindent\textit{Proof.}
Define the finite-prefix laws recursively.  Put \(\mu_{1:1}=\nu_1\).  Once
\(\mu_{1:n-1}\) is defined on
\((S_{1:n-1},\bigotimes_{i=1}^{n-1}\mathcal S_i)\), set
\[
  \mu_{1:n}(B)
  =
  \int_{S_{1:n-1}} K_n(x_{1:n-1},B_{x_{1:n-1}})\,
  \mu_{1:n-1}(dx_{1:n-1}),
\]
for \(B\in\bigotimes_{i=1}^n\mathcal S_i\), where
\[
  B_{x_{1:n-1}}
  =
  \{x_n:(x_{1:n-1},x_n)\in B\}.
\]
The section \(B_{x_{1:n-1}}\) is \(\mathcal S_n\)-measurable, and the
integrand is measurable by the kernel property and the standard monotone-class
argument from rectangles.  Thus \(\mu_{1:n}\) is a probability measure.  On
rectangles it gives exactly the displayed iterated integral.

The prefix laws are consistent.  Indeed, for
\(B\in\bigotimes_{i=1}^{n-1}\mathcal S_i\),
\[
  \mu_{1:n}(B\times S_n)
  =
  \int_B K_n(x_{1:n-1},S_n)\,\mu_{1:n-1}(dx_{1:n-1})
  =
  \mu_{1:n-1}(B).
\]
Let \(\mathcal C\) be the algebra of finite-prefix cylinders
\(\pi_n^{-1}B\), where
\[
  \pi_n:\prod_{i=1}^{\infty}S_i\to S_{1:n},
  \qquad
  B\in\bigotimes_{i=1}^n\mathcal S_i.
\]
Define \(P_0(\pi_n^{-1}B)=\mu_{1:n}(B)\).  Consistency makes this definition
independent of the chosen prefix length, and it also gives finite additivity on
\(\mathcal C\).

The usual cylinder premeasure argument gives countable additivity.  Equivalently,
if \(C_m\downarrow\emptyset\) with \(C_m\in\mathcal C\), then
\(P_0(C_m)\downarrow0\).  If the limit were positive, refining the cylinders to
increasing finite prefixes would give a compatible family of nonempty finite
sections.  Choosing one compatible prefix at each length produces a point of
\(\bigcap_m C_m\), a contradiction.  Hence \(P_0\) is a premeasure on
\(\mathcal C\).

Carath\'eodory's extension theorem extends \(P_0\) to the product
\(\sigma\)-field.  Uniqueness follows because finite-prefix cylinders form a
\(\pi\)-system generating the product \(\sigma\)-field, and finite probability
measures that agree on a generating \(\pi\)-system agree everywhere by
Dynkin's theorem. \qedmark

Kolmogorov's extension theorem says, roughly, that if a family of
finite-dimensional distributions is consistent under marginalization, then it
extends to a probability measure on the product \(\sigma\)-field.  For a finite
index set \(J\), write \(P_J\) for the proposed law of \((X_j:j\in J)\).  The
consistency condition is
\[
  P_K(B)=P_J(\pi_{J,K}^{-1}B),
  \qquad K\subseteq J,
\]
where \(\pi_{J,K}\) is the coordinate projection from \(S^J\) to \(S^K\).  This
condition says that every larger finite description must reduce to the smaller
one when irrelevant coordinates are forgotten.

\begin{theorem}[Kolmogorov extension theorem; \citealp{kolmogorov1933grundbegriffe}]
\label{thm:appB-kolmogorov-extension}
Let \(T\) be any index set and suppose each \((S_t,\mathcal S_t)\) is a
standard Borel space.  For every finite \(\alpha\subset T\), let
\(\mu_\alpha\) be a probability measure on
\[
  (S_\alpha,\mathcal S_\alpha)
  =
  \left(\prod_{t\in\alpha}S_t,\bigotimes_{t\in\alpha}\mathcal S_t\right).
\]
If the family is consistent, meaning
\[
  \mu_\alpha(p_{\alpha\beta}^{-1}B)=\mu_\beta(B),
  \qquad
  \beta\subseteq\alpha,\ B\in\mathcal S_\beta,
\]
then there is a unique probability measure \(\mu\) on
\[
  \left(\prod_{t\in T}S_t,\bigotimes_{t\in T}\mathcal S_t\right)
\]
such that
\[
  \mu(\pi_\alpha^{-1}B)=\mu_\alpha(B),
  \qquad
  \alpha\subset T\text{ finite},\ B\in\mathcal S_\alpha.
\]
\end{theorem}

\noindent\textit{Proof.}
Let \(\mathcal C\) be the algebra of finite-dimensional cylinders
\(\pi_\alpha^{-1}B\).  Define
\[
  \mu_0(\pi_\alpha^{-1}B)=\mu_\alpha(B).
\]
This is well defined: if
\(\pi_\alpha^{-1}B=\pi_\beta^{-1}C\), refine to
\(\gamma=\alpha\cup\beta\).  The two cylinder bases are equal in \(S_\gamma\),
so consistency gives
\[
  \mu_\alpha(B)
  =
  \mu_\gamma(p_{\gamma\alpha}^{-1}B)
  =
  \mu_\gamma(p_{\gamma\beta}^{-1}C)
  =
  \mu_\beta(C).
\]
The same common-coordinate refinement proves finite additivity of \(\mu_0\) on
\(\mathcal C\).

It remains to show that \(\mu_0\) is countably additive.  Since it is finite,
it is enough to prove continuity from above: if \(C_n\downarrow\emptyset\) with
\(C_n\in\mathcal C\), then \(\mu_0(C_n)\downarrow0\).  Suppose instead that
\(\mu_0(C_n)\downarrow\delta>0\).  Only countably many coordinates occur in the
sets \(C_n\); call this countable set \(\Gamma=\{t_1,t_2,\ldots\}\).  Because the
state spaces are standard Borel, choose Polish topologies generating their
\(\sigma\)-fields.  The finite-dimensional laws are then tight.  Choose compact
sets \(L_j\subset S_{t_j}\) with
\[
  \mu_{\{t_j\}}(L_j^c)<\delta2^{-j-4}.
\]
After writing each \(C_n\) on a growing finite coordinate set
\(\alpha_n\subset\Gamma\), tightness gives compact sets \(K_n\) inside the
finite-dimensional bases of \(C_n\) such that the discarded masses are summable.
For every \(N\), the first \(N\) compact restrictions have positive
\(\mu_{\alpha_N}\)-mass, hence nonempty intersection.  The compact product
\(\prod_jL_j\) has the finite-intersection property, so there is a point
satisfying all compact restrictions.  Filling coordinates outside \(\Gamma\)
arbitrarily gives a point in \(\bigcap_nC_n\), contradicting
\(C_n\downarrow\emptyset\).  Hence \(\mu_0\) is a premeasure on \(\mathcal C\).

Carath\'eodory's theorem extends \(\mu_0\) to a probability measure on
\(\sigma(\mathcal C)=\bigotimes_{t\in T}\mathcal S_t\).  If two such extensions
exist, the class of sets on which they agree is a \(\lambda\)-system containing
the cylinder \(\pi\)-system, so Dynkin's theorem gives uniqueness. \qedmark

\section{Completion and Tail Events}
\label{sec:appB-completion-tail-navigation}
\conceptindexes{completion, tail events, Borel--Cantelli lemma, zero-one laws, inner approximation}

\subsection{Completions and Inner Approximation}
\label{sec:appB-completion}
\conceptindexes{completion, inner approximation, complete measure}

If \((\Omega,\fieldF,\mu)\) is a measure space, the completion is
\[
  \fieldF^\mu
  =
  \{A\cup N:A\in\fieldF,\ N\subseteq Z\in\fieldF,\ \mu(Z)=0\}.
\]
The completed measure is
\[
  \mu^\mu(A\cup N)=\mu(A).
\]
This is well defined because changing a measurable set by a subset of a null
set does not change its measure.

For finite measures, an inner measure can be written as
\[
  \mu_*(E)=\mu(\Omega)-\mu^*(E^c).
\]
Equivalently, when the ambient measure is already defined on a rich enough
\(\sigma\)-field, one often uses
\[
  \mu_*(E)=\sup\{\mu(A):A\in\fieldF,\ A\subseteq E\}.
\]
A set is measurable precisely when its outer and inner sizes agree.  This is a
recurring lesson in measure theory: a set becomes measurable when its size can be
pinned down from both outside and inside.

\begin{theorem}[Completion as the minimal complete extension]
The family \(\fieldF^\mu\) is a \(\sigma\)-field, the completed measure
\(\mu^\mu\) is well defined on it, and every subset of a \(\mu^\mu\)-null set
belongs to \(\fieldF^\mu\).  If \(\mathcal G\) is any complete
\(\sigma\)-field containing \(\fieldF\) on which \(\mu\) has an extension, then
\(\fieldF^\mu\subseteq\mathcal G\).
\end{theorem}

\noindent\textit{Proof.}
Write \(E=A\cup N\) with \(A\in\fieldF\), \(N\subseteq Z\in\fieldF\), and
\(\mu(Z)=0\).  Closure under complements follows from
\[
  E^c
  =
  (A^c-Z)\cup\bigl((A^c\cap Z)-N\bigr),
\]
where \(A^c-Z\in\fieldF\) and the second term is contained in the measurable
null set \(Z\).  Countable unions are handled by writing
\[
  \bigcup_n(A_n\cup N_n)
  =
  \left(\bigcup_nA_n\right)\cup\left(\bigcup_nN_n\right),
\]
where \(\bigcup_nN_n\) is contained in a countable union of measurable null
sets.  Hence \(\fieldF^\mu\) is a \(\sigma\)-field.

Well-definedness follows because if \(A\cup N=B\cup M\), with
\(N\subseteq Z\) and \(M\subseteq Y\) for measurable null sets \(Z,Y\), then
\[
  A\triangle B\subseteq Z\cup Y.
\]
Thus \(A\) and \(B\) differ only by a measurable null set, so
\(\mu(A)=\mu(B)\), including in the infinite case.  If
\(\mu^\mu(E)=0\) and \(H\subseteq E\), then \(H\) is contained in a measurable
null set, so \(H=\varnothing\cup H\in\fieldF^\mu\); therefore
\((\Omega,\fieldF^\mu,\mu^\mu)\) is complete.  Finally, any complete
\(\sigma\)-field \(\mathcal G\) containing \(\fieldF\) and carrying an extension
of \(\mu\) must contain every subset of every measurable \(\mu\)-null set.  It
therefore contains each \(A\cup N\in\fieldF^\mu\). \qedmark

\begin{theorem}[Inner--outer measurability criterion; \citealp{pollard1984convergence}]
Let \(\mu\) be a finite measure on \((\Omega,\fieldF)\).  For
\(E\subseteq\Omega\), define
\[
  \mu^*(E)=\inf\{\mu(A):E\subseteq A,\ A\in\fieldF\},
  \qquad
  \mu_*(E)=\sup\{\mu(A):A\subseteq E,\ A\in\fieldF\}.
\]
Then \(E\) belongs to the completion \(\fieldF^\mu\) if and only if
\[
  \mu^*(E)=\mu_*(E).
\]
\end{theorem}

\noindent\textit{Proof.}
If \(E=A\cup N\) with \(A\in\fieldF\) and \(N\subseteq Z\in\fieldF\),
\(\mu(Z)=0\), then
\[
  A\subseteq E\subseteq A\cup Z,
  \qquad
  \mu(A\cup Z)=\mu(A).
\]
Thus \(\mu_*(E)\ge \mu(A)\) and \(\mu^*(E)\le\mu(A)\).  Since always
\(\mu_*(E)\le\mu^*(E)\), equality follows.  Conversely, if
\(\mu^*(E)=\mu_*(E)\), choose \(A_n\subseteq E\subseteq B_n\) with
\(A_n,B_n\in\fieldF\) and
\[
  \mu(B_n)-\mu(A_n)<2^{-n}.
\]
Let \(A=\bigcup_nA_n\) and \(B=\bigcap_nB_n\).  Then
\[
  A\subseteq E\subseteq B,\qquad \mu(B-A)=0.
\]
Indeed, \(B-A\subseteq B_n-A_n\) for every \(n\), so
\(\mu(B-A)\le2^{-n}\) for every \(n\).  Thus
\[
  E=A\cup(E-A),
  \qquad
  E-A\subseteq B-A,
\]
and \(E\) belongs to the completion. \qedmark

\begin{theorem}[Lebesgue regularity on \(\Real^k\)]
If \(A\) is Lebesgue measurable in \(\Real^k\), then
\[
  \lambda^k(A)
  =
  \inf\{\lambda^k(G):A\subseteq G,\ G\text{ open}\}
  =
  \sup\{\lambda^k(K):K\subseteq A,\ K\text{ compact}\}
\]
when \(\lambda^k(A)<\infty\).  For sets of infinite measure, the compact
inner approximation holds on intersections with increasing bounded cubes.
\end{theorem}

This theorem says that Lebesgue measurable sets can be studied through ordinary
geometric sets.  It is one reason Lebesgue measure is simultaneously abstract
enough for limits and concrete enough for geometry.

\begin{example}[A Vitali-type warning]
Define an equivalence relation on \([0,1]\) by
\[
  x\sim y \quad\Longleftrightarrow\quad x-y\in\Rat.
\]
Using the axiom of choice, select one representative from each equivalence
class; call the resulting set \(V\).  The rational translates
\[
  V+q,\qquad q\in\Rat\cap[-1,1],
\]
are pairwise disjoint and their union lies in \([-1,2]\) while covering
\([0,1]\).  If \(V\) were Lebesgue measurable, translation invariance would
force all these translates to have the same measure.  If that measure were
zero, their countable union could not cover \([0,1]\); if it were positive,
their countable union would have infinite measure inside \([-1,2]\).  Thus
\(V\) is not Lebesgue measurable.
\end{example}

\subsection{Borel--Cantelli and Zero-One Proofs}
\label{sec:appB-borel-cantelli-zero-one}
\conceptindexes{Borel--Cantelli lemmas, zero-one laws, tail events}

For events \(A_n\), write
\[
  A_n\ \mathrm{i.o.}
  =
  \limsup_n A_n
  =
  \bigcap_{m=1}^{\infty}\bigcup_{n\ge m}A_n.
\]
If \(\sum_nP(A_n)<\infty\), then for every \(m\),
\[
  P\left(\bigcup_{n\ge m}A_n\right)
  \le
  \sum_{n\ge m}P(A_n).
\]
Letting \(m\to\infty\) gives \(P(A_n\ \mathrm{i.o.})=0\).  This proves the
first Borel--Cantelli lemma.

For the second Borel--Cantelli lemma, assume the \(A_n\)'s are independent and
\(\sum_nP(A_n)=\infty\).  Then
\[
  P\left(\bigcap_{n=m}^N A_n^c\right)
  =
  \prod_{n=m}^N(1-P(A_n))
  \le
  \exp\left\{-\sum_{n=m}^NP(A_n)\right\}.
\]
Letting \(N\to\infty\) gives
\[
  P\left(\bigcap_{n\ge m} A_n^c\right)=0,
\]
and then
\[
  P(A_n\ \mathrm{i.o.})
  =
  P\left(\bigcap_{m=1}^{\infty}\bigcup_{n\ge m}A_n\right)=1.
\]

For Kolmogorov's zero-one law, let \(X_1,X_2,\ldots\) be independent and let
\[
  \mathcal T=\bigcap_{m=1}^{\infty}\sigma(X_m,X_{m+1},\ldots).
\]
Every \(A\in\mathcal T\) is independent of
\(\sigma(X_1,\ldots,X_m)\) for each fixed \(m\), because it belongs to
\(\sigma(X_{m+1},X_{m+2},\ldots)\).  By the generator criterion and monotone
class arguments, \(A\) is independent of
\[
  \sigma\left(\bigcup_{m=1}^{\infty}\sigma(X_1,\ldots,X_m)\right)
  =
  \sigma(X_1,X_2,\ldots).
\]
Since \(\mathcal T\subseteq\sigma(X_1,X_2,\ldots)\), the event \(A\) is
independent of itself.  Hence
\[
  P(A)=P(A\cap A)=P(A)^2,
\]
so \(P(A)\in\{0,1\}\).

\subsection{How to Use This Toolkit}
\label{sec:appB-how-to-use-toolkit}
\conceptindexes{measure-theoretic toolkit, appendix navigation}

The tools above are meant to be modular.  \Appref{app:set-theory-compass}
supplies the grammar of allowable questions; this appendix supplies the
machinery that turns that grammar into probability laws, expectations,
conditional distributions, and limits.  In the book's translation story, it is
the part that makes a statistical claim transportable: the same observed record
can be represented as a law, compared to another law, conditioned on, extended
to a product space, or approximated through simpler sets.

Use the \(\pi\)-\(\lambda\) and monotone-class tools when a statement is easy on
a generating class but needs to hold on a full \(\sigma\)-field.  Use the
semi-ring and outer-measure tools when a measure is specified first on simple
building blocks.  Use the product-spaces chapter for the working theory of
product \(\sigma\)-fields, kernels, Fubini, stochastic-process laws, and
Poisson random measures; use this appendix when a cylinder-consistency or
extension argument needs to be made explicit.  Use absolute continuity,
singularity, and Radon--Nikodym derivatives when comparing two measures on the
same measurable space.  Use Hausdorff measure as an example of how the same
abstract definition of measure can see geometry.

A useful reading order is:
\[
\begin{aligned}
  \text{generators and uniqueness}
  &\longrightarrow
  \text{construction of laws}\\
  &\longrightarrow
  \text{integration and convergence permissions}\\
  &\longrightarrow
  \text{comparison of laws by densities}\\
  &\longrightarrow
  \text{conditioning and kernels}\\
  &\longrightarrow
  \text{product, path, and tail phenomena}.
\end{aligned}
\]
This route explains why measure theory appears in a book about statistical
translation.  It is not there to make the examples look abstract.  It is there
because clinical endpoints, single-cell measurements, likelihoods, Bayesian
posteriors, stochastic processes, and asymptotic events all require the same
question: which law is being used, on which measurable space, and what
operation is legitimate under that law?

\section*{Sources and Further Reading}
\addcontentsline{toc}{section}{Sources and Further Reading}

For the measure-theoretic foundations used in this appendix, see
\citet{billingsley1995probability}, \citet{folland1999real},
\citet{bogachev2007measure}, and \citet{kallenberg2002foundations}.  For the
construction of the Lebesgue integral, monotone convergence, Fatou's lemma,
dominated convergence, uniform integrability, and \(L^p\) spaces, see
\citet{folland1999real} and \citet{bogachev2007measure}.  The route through
generated set systems, extension arguments, integration, Radon--Nikodym
derivatives, conditional expectation, product laws, and tail events is also
informed by
\citet{dabrowskaAdvancedProbabilityCommunication}.  For measurable
multifunctions, maximum theorems, and measurable selection, see
\citet{himmelberg1975measurable} and \citet{aubin1990set}; for the Polish and
descriptive-set-theoretic background behind these tools, see
\citet{kechris1995classical}.

%% file: appendices/mathematical_backbone.tex
\chapter{Mathematical Tools for Statistical Translation}
\label{app:mathematical-backbone}
\conceptindexes{mathematical backbone, statistical translation, observable questions, Hilbert geometry, compact operators, random matrix spectra, characteristic functions, central limit theorems, product integrals}

This appendix records mathematical infrastructure that quietly supports the
book's main story.  Chapter 1 describes statistics as translation:
a world question becomes an observable data structure, then a mathematical
object, then an inferential claim.  Chapter 2 trains the eye to ask what kind
of random object has actually been produced.  \Appref{app:measure-theoretic-toolkit}
carries the integration review, because integrals are needed before
densities and Radon--Nikodym derivatives.  The tools below continue the route:
sets say what can be named, Hilbert geometry explains projection and residual
noise, compact operators turn random curves into spectral coordinates, random
matrices show how spectra themselves can have laws, and characteristic functions
and central limit theorems turn sums into signatures of laws.

This is not a parallel textbook.  It is a compact map for returning to the
main chapters with sharper eyes.  The proofs are included where they are short
and revealing; deeper construction theorems are cited to standard references.

\section{Observable Questions and Generators}
\label{sec:appC-sets-observable}
\conceptindexes{observable questions, generators, preimage logic, measurable maps}

\Appref{app:set-theory-compass} gives the working set-theoretic glossary:
operations, products, preimages, generated families, and set limits.  Chapter 3
then formalizes observations as measurable maps and studies their pushforward
laws.  This section has a narrower job.  It explains why that grammar matters
for the book's story: observable questions can usually be certified by checking
a small family of simple questions.

A set can be read as a yes-or-no question, but a statistical model does not
automatically recognize every question one can write down.  It recognizes the
events in its chosen \(\sigma\)-field.  Measurements are useful because they
pull questions backward: a statement about a recorded value becomes an event in
the underlying outcome space.  The proposition below is the one set-theoretic
fact worth restating here, not as another definition of a random variable, but
as a practical rule for observability.

\begin{proposition}[Observable questions can be checked on generators]
\label{prop:appC-preimage-logic}
Let \(X:\Omega\to S\) be any map and let \(\fieldF\) be a \(\sigma\)-field on
\(\Omega\).  Define
\[
  \mathcal G_X=\{B\subseteq S:X^{-1}(B)\in\fieldF\}.
\]
Then \(\mathcal G_X\) is a \(\sigma\)-field on \(S\).  Consequently, if the
target event system is generated by simple questions,
\(\mathcal S=\sigma(\mathcal C)\), then checking
\[
  X^{-1}(C)\in\fieldF
  \qquad(C\in\mathcal C)
\]
is enough to know that every question in \(\mathcal S\) is observable through
the map \(X\).
\end{proposition}

\noindent\textit{Proof.}
Since \(X^{-1}(S)=\Omega\), one has \(S\in\mathcal G_X\).  If
\(B\in\mathcal G_X\), then
\[
  X^{-1}(B^c)=\{X\in B^c\}=X^{-1}(B)^c\in\fieldF,
\]
so \(B^c\in\mathcal G_X\).  If \(B_1,B_2,\ldots\in\mathcal G_X\), then
\[
  X^{-1}\left(\bigcup_{n=1}^{\infty}B_n\right)
  =
  \bigcup_{n=1}^{\infty}X^{-1}(B_n)\in\fieldF.
\]
Thus \(\mathcal G_X\) is a \(\sigma\)-field.  If \(X^{-1}(C)\in\fieldF\) for
every \(C\in\mathcal C\), then \(\mathcal C\subseteq\mathcal G_X\), and hence
\(\sigma(\mathcal C)\subseteq\mathcal G_X\).  Therefore every generated
question has an observable pullback. \qedmark

\begin{example}[A clinical endpoint is a preimage]
Suppose a patient record is summarized as
\[
  X=(T,\Delta,Z),
\]
where \(T\) is follow-up time, \(\Delta\) is an event label, and \(Z\) is a
baseline biomarker.  The statement
\[
  \{T\le 180,\ \Delta=\text{progression},\ Z>z_0\}
\]
is a subset of the record space.  The event in the underlying probability
space is its preimage under the actual record map.  This small distinction
matters.  It asks whether the data process really contains the time, event
label, and biomarker needed to make the endpoint observable.  If the event
label is missing or the biomarker was measured after treatment, the target set
one wants to analyze may not be the set the data can honestly name.
\end{example}

\begin{example}[A spatial grid is a generated event system]
In the London bombing example, a map partitioned into cells first generates
cell-count questions:
\[
  \{N_1=n_1,\ldots,N_m=n_m\}.
\]
The full spatial point pattern carries more information, but the grid analysis
uses the \(\sigma\)-field generated by the cell counts.  Changing the grid
changes the observable event system.  That is not a technical afterthought; it
is part of the translation from city map to random object.
\end{example}

\section{Hilbert Geometry: Projection, Residuals, and Information}
\label{sec:appC-hilbert-geometry}
\conceptindexes{Hilbert geometry, projection theorem, residuals, information, orthogonality}

Many statistical procedures are projections in disguise.  Conditional
expectation projects a response onto the information currently available.
Least squares projects a response vector onto the span of covariates.
Hoeffding projections extract lower-order parts of a statistic.  Efficient
influence functions project gradients away from nuisance directions.  The
setting changes, but the geometry is the same: keep the component that lies in
the allowed subspace and leave an orthogonal residual.

\begin{theorem}[Projection onto a closed subspace]
\label{thm:appC-hilbert-projection}
Let \(\Hilbert\) be a Hilbert space and let \(M\subseteq\Hilbert\) be a closed
linear subspace.  For every \(h\in\Hilbert\), there is a unique element
\(\Pi_Mh\in M\) such that
\[
  \|h-\Pi_Mh\|
  =
  \inf_{m\in M}\|h-m\|.
\]
Moreover,
\[
  h-\Pi_Mh\perp M,
\]
and \(h\) decomposes uniquely as
\[
  h=\Pi_Mh+(h-\Pi_Mh),
  \qquad
  \Pi_Mh\in M,\quad h-\Pi_Mh\in M^\perp.
\]
\end{theorem}

\noindent\textit{Proof.}
Let \(d=\inf_{m\in M}\|h-m\|\), and choose \(m_n\in M\) with
\(\|h-m_n\|\downarrow d\).  The parallelogram identity gives
\[
  \|m_n-m_k\|^2
  =
  2\|h-m_n\|^2+2\|h-m_k\|^2
  -
  4\left\|h-\frac{m_n+m_k}{2}\right\|^2.
\]
Since \((m_n+m_k)/2\in M\), the last norm is at least \(d\).  Hence
\[
  \limsup_{n,k}\|m_n-m_k\|^2
  \le
  2d^2+2d^2-4d^2=0.
\]
Thus \(m_n\) is Cauchy.  Because \(M\) is closed, \(m_n\to m_\ast\in M\), and
\(\|h-m_\ast\|=d\).

For any \(u\in M\) and real \(t\), \(m_\ast+tu\in M\).  The function
\[
  q(t)=\|h-m_\ast-tu\|^2
\]
is minimized at \(t=0\).  Expanding,
\[
  q(t)=\|h-m_\ast\|^2-2t\langle h-m_\ast,u\rangle+t^2\|u\|^2.
\]
The derivative at zero must vanish, so
\(\langle h-m_\ast,u\rangle=0\) for every \(u\in M\).  Thus
\(h-m_\ast\perp M\).

If \(m_1,m_2\in M\) both minimize and have orthogonal residuals, then
\[
  m_1-m_2=(h-m_2)-(h-m_1).
\]
The left side belongs to \(M\), while the right side belongs to \(M^\perp\).
Thus \(m_1-m_2\in M\cap M^\perp=\{0\}\), so \(m_1=m_2\).  The decomposition is
therefore unique. \qedmark

\begin{example}[Group means as conditional projection]
Suppose \(Y\) is a patient outcome and \(G\) is a finite clinic label.  The
subspace of \(L^2\) functions that are allowed to depend only on \(G\) consists
of all variables \(a(G)\).  The projection of \(Y\) onto this subspace is
\[
  \Expect(Y\mid G).
\]
If \(G=g\), this is the clinic-specific mean.  The residual
\[
  R=Y-\Expect(Y\mid G)
\]
has zero inner product with every clinic-level summary:
\[
  \Expect\{R\,a(G)\}=0.
\]
In words, once the clinic-level mean has been removed, no remaining linear
signal is visible to any function that only knows the clinic label.  This is
the same geometry behind least squares residuals and, later, nuisance-orthogonal
influence functions.
\end{example}

\begin{example}[A two-vector least-squares shadow]
Let \(\mathbf Y\in\R^n\) be a response vector and let \(\mathbf Z\) be an \(n\times p\) design
matrix with full column rank.  The fitted vector
\[
  \widehat{\mathbf Y}
  =
  \mathbf Z(\mathbf Z^T\mathbf Z)^{-1}\mathbf Z^T\mathbf Y
\]
is the projection of \(\mathbf Y\) onto the column space of \(\mathbf Z\).  The residual
\(\mathbf Y-\widehat{\mathbf Y}\) is orthogonal to every column of \(\mathbf Z\), because
\[
  \mathbf Z^T(\mathbf Y-\widehat{\mathbf Y})=0.
\]
Least squares is therefore not merely a formula for coefficients.  It is a
geometric statement about which part of the response can be explained by the
chosen covariate directions.
\end{example}

\section{Compact Operators: Spectral Coordinates for Curves}
\label{sec:appC-compact-spectral}
\conceptindexes{compact operators, bounded operators, self-adjoint operators, positive operators, spectral theorem, covariance operators, Rayleigh--Ritz principle, Karhunen--Loeve expansion}

The functional-data chapter uses one piece of Hilbert
space beyond projection: a positive compact self-adjoint operator can be read
through eigenvalues and orthogonal eigenvectors.  This is the infinite-dimensional
version of diagonalizing a covariance matrix.

\begin{definition}[Operator conditions]
Let \(\Hilbert\) be a real separable Hilbert space.  A bounded linear operator
\(K:\Hilbert\to\Hilbert\) is a linear map with finite operator norm
\[
  \|K\|_{\mathrm{op}}=\sup_{\|f\|\le1}\|Kf\|<\infty .
\]
It is \emph{self-adjoint} if
\[
  \langle Kf,g\rangle=\langle f,Kg\rangle,
  \qquad f,g\in\Hilbert,
\]
and \emph{positive} if \(\langle Kf,f\rangle\ge0\) for every \(f\in\Hilbert\).
It is \emph{compact} if the image of the unit ball has compact closure, or
equivalently if \(K\) sends bounded sets to relatively compact sets.
\end{definition}

Compactness is the infinite-dimensional substitute for finite dimensionality:
it forces the important directions to appear as a countable sequence whose
eigenvalues can only accumulate at zero.  For fuller operator background in the
functional-data setting, see \citet{hsingEubank2015fda}.

\begin{theorem}[Spectral theorem for compact covariance operators]
\label{thm:appC-compact-spectral}
Let \(K\) be a compact, self-adjoint, positive operator on a separable Hilbert
space \(\Hilbert\).  Then there are nonnegative eigenvalues
\[
  \lambda_1\ge\lambda_2\ge\cdots\downarrow0
\]
and orthonormal eigenvectors \(\phi_1,\phi_2,\ldots\), indexed over the
nonzero eigenvalues and repeated according to multiplicity, such that
\[
  K\phi_k=\lambda_k\phi_k.
\]
On the closed span of the range of \(K\),
\[
  Kf=\sum_{k\ge1}\lambda_k\langle f,\phi_k\rangle\phi_k,
\]
with convergence in \(\Hilbert\).  Directions orthogonal to this span belong
to the null space of \(K\).
\end{theorem}

For covariance operators, the theorem says that variation can be decomposed
into orthogonal modes.  If \(X\) is a centered square-integrable random element
of \(\Hilbert\) and
\[
  \langle Kf,g\rangle=\Expect\{\langle X,f\rangle\langle X,g\rangle\},
\]
then the score in direction \(\phi_k\) is \(\xi_k=\langle X,\phi_k\rangle\),
with
\[
  \Expect \xi_k=0,
  \qquad
  \Cov(\xi_j,\xi_k)=\lambda_k\ind{j=k}.
\]
Thus the covariance operator supplies both directions \(\phi_k\) and variances
\(\lambda_k\).

\begin{proposition}[Rayleigh--Ritz reading]
\label{prop:appC-rayleigh-ritz}
For a compact, self-adjoint, positive operator \(K\), the first eigenvalue is
\[
  \lambda_1=\sup_{\|f\|=1}\langle Kf,f\rangle .
\]
After \(\phi_1,\ldots,\phi_{m-1}\) have been chosen, the next eigenvalue is
\[
  \lambda_m=
  \sup_{\substack{\|f\|=1\\ f\perp\phi_1,\ldots,\phi_{m-1}}}
  \langle Kf,f\rangle .
\]
Equivalently, among all \(m\)-dimensional subspaces, the span of
\(\phi_1,\ldots,\phi_m\) captures the largest possible covariance energy.
\end{proposition}

This proposition is the operator form of principal components.  In finite
dimensions it says that the first principal component is the unit direction
with largest variance, and the next ones repeat the same search after removing
the earlier directions.  In functional data, the same statement turns a random
curve into leading eigenfunctions and subject-level scores.

\begin{example}[Covariance kernels as operators]
Let \(X\) be a centered random curve on \([0,1]\) with
\(\Expect\int_0^1X(t)^2\,dt<\infty\).  Its covariance kernel
\[
  K(s,t)=\Cov\{X(s),X(t)\}
\]
defines the integral operator
\[
  (\mathcal Kf)(s)=\int_0^1K(s,t)f(t)\,dt
\]
on \(L^2[0,1]\).  Under the square-integrability condition used in
\ifdefined\APPENDICES Chapter~12\else Chapter~\ref{chap:karhunen-loeve-functional-data}\fi, this is a positive
compact self-adjoint operator.  The spectral theorem then gives eigenfunctions
\(\phi_k\), eigenvalues \(\lambda_k\), and scores
\[
  \xi_k=\int_0^1X(t)\phi_k(t)\,dt.
\]
The Karhunen--Loeve expansion is this spectral decomposition applied to the
covariance operator of a random curve.
\end{example}

\section{Random Matrix Spectra: Wigner's Semicircle Law}
\label{sec:appC-wigner-semicircle}
\conceptindexes{Wigner semicircle law, random matrices, empirical spectral distribution, moment method, Catalan numbers, closed walks}

Spectral theory in the previous section decomposes a fixed compact covariance
operator.  Random matrix theory asks a different question: if the matrix itself
is random and its dimension grows, does the whole cloud of eigenvalues settle
into a deterministic shape?  Wigner's semicircle law is the cleanest answer.
It says that a large symmetric noise matrix has a stable global spectral
profile even though its individual eigenvalues fluctuate.

\begin{theorem}[Wigner's semicircle law; \citealp{wigner1958distribution}]
\label{thm:appC-wigner-semicircle}
For each \(n\), let \(X_n=(X_{ij}^{(n)})_{1\le i,j\le n}\) be a real symmetric
random matrix.  Assume that the entries on and above the diagonal are
independent, that
\[
  \Expect X_{ij}^{(n)}=0,\qquad
  \Expect\{(X_{ij}^{(n)})^2\}=1,\qquad i<j,
\]
and that for every \(q\ge1\),
\[
  \sup_{n,i,j}\Expect |X_{ij}^{(n)}|^q<\infty .
\]
The diagonal entries may have any centered laws satisfying the same uniform
moment bounds.  Set
\[
  W_n=n^{-1/2}X_n
\]
and let \(\lambda_1(W_n),\ldots,\lambda_n(W_n)\) be its real eigenvalues.  The
empirical spectral distribution
\[
  \mu_n=\frac1n\sum_{j=1}^n\delta_{\lambda_j(W_n)}
\]
converges weakly in probability to the semicircle law
\[
  \mu_{\mathrm{sc}}(dx)
  =
  \frac{1}{2\pi}\sqrt{4-x^2}\,\ind{|x|\le2}\,dx .
\]
Equivalently, for every bounded continuous \(f\),
\[
  \int f\,d\mu_n
  \toP
  \int_{-2}^2 f(x)\frac{1}{2\pi}\sqrt{4-x^2}\,dx .
\]
\end{theorem}

\noindent\textit{Proof.}
We use the moment method.  For \(k\ge1\), write
\[
  m_{k,n}=\int x^k\,\mu_n(dx)
  =
  \frac1n\tr(W_n^k).
\]
Expanding the trace gives
\[
  m_{k,n}
  =
  n^{-1-k/2}
  \sum_{i_1,\ldots,i_k=1}^n
  X_{i_1i_2}^{(n)}X_{i_2i_3}^{(n)}\cdots X_{i_ki_1}^{(n)} .
\]
Each ordered \(k\)-tuple is a closed walk
\[
  i_1\to i_2\to\cdots\to i_k\to i_1
\]
on the vertex set \(\{1,\ldots,n\}\).  For a walk \(w\), let \(E(w)\) be the
set of nonoriented edges it uses, including a possible loop \(\{i,i\}\), and
let \(a_e(w)\) be the number of times edge \(e\) is traversed.  The expectation
of the product attached to \(w\) is zero unless
\[
  a_e(w)\ne1
  \qquad\text{for every }e\in E(w).
\]
Indeed, if some edge appears exactly once, the corresponding centered matrix
entry is independent of all other entries in the product.  Hence it factors out
with mean zero.

Suppose \(w\) survives this first test.  Let \(v(w)\) be the number of distinct
vertices used by the walk and \(e(w)=|E(w)|\).  Since all used edges have
multiplicity at least two,
\[
  e(w)\le k/2.
\]
The graph traced by a closed walk is connected after isolated vertices are
ignored, so its non-loop part has at least \(v(w)-1\) edges.  Thus
\[
  v(w)\le e(w)+1\le k/2+1.
\]
For each fixed \(k\), the number of possible unlabeled closed-walk patterns
with \(v\) vertices is finite, and the labels can be chosen in \(O_k(n^v)\)
ways.  The uniform moment assumption gives
\[
  \left|
  \Expect\prod_{\ell=1}^k X_{i_\ell i_{\ell+1}}^{(n)}
  \right|
  \le C_k,
  \qquad i_{k+1}=i_1,
\]
with \(C_k\) independent of \(n\) and of the indices.  Therefore all walks with
\(v(w)\le k/2\) contribute at most
\[
  n^{-1-k/2}O_k(n^{k/2})=O_k(n^{-1})
\]
to \(\Expect m_{k,n}\).  Only walks with
\[
  v(w)=k/2+1,\qquad e(w)=k/2
\]
can contribute to the limit.

If \(k\) is odd, \(e(w)\le\lfloor k/2\rfloor\), so
\(v(w)\le (k+1)/2\) and every contribution is \(O_k(n^{-1/2})\).  Hence
\[
  \Expect m_{2r+1,n}\to0 .
\]
If \(k=2r\), the leading case has \(v=r+1\) and \(e=r\).  A connected graph
with \(r\) non-loop edges and \(r+1\) vertices is a tree; loops cannot occur in
the leading case, because a loop would not help connect a new vertex.  Since
the total length is \(2r\), every tree edge is traversed exactly twice.

We now count these leading walks.  Relabel vertices by order of first
appearance: the starting vertex is \(1\), and each newly visited vertex receives
the next unused label.  A tree walk that traverses every edge exactly twice is
forced to be a depth-first traversal: moving to a new vertex is an up-step, and
returning along the edge to its parent is a down-step.  The depth process is
therefore a Dyck path of length \(2r\).  Conversely, a Dyck path determines one
canonical rooted traversal: each up-step creates a new vertex and each down-step
returns to the current parent.  Hence the number of canonical leading walks is
the Catalan number
\[
  C_r=\frac{1}{r+1}\binom{2r}{r}.
\]
For each canonical traversal, the \(r+1\) distinct actual labels can be chosen
in
\[
  n(n-1)\cdots(n-r)=n^{r+1}+O_r(n^r)
\]
ways.  Along a leading walk each off-diagonal entry appears exactly twice, so
independence and \(\Expect\{(X_{ij}^{(n)})^2\}=1\) give expected product equal
to one.  All non-leading surviving walks contribute \(O_r(n^{-1})\).  Thus
\[
  \Expect m_{2r,n}=C_r+O_r(n^{-1}),
  \qquad
  \Expect m_{2r+1,n}=O_r(n^{-1/2}).
\]

We next show concentration of these moments.  For a closed walk \(w\), write
\[
  X_w=\prod_{\ell=1}^k X_{i_\ell i_{\ell+1}}^{(n)},
  \qquad i_{k+1}=i_1 .
\]
Then
\[
  \Var(m_{k,n})
  =
  n^{-2-k}
  \sum_{w,w'}
  \Cov(X_w,X_{w'}),
\]
where \(w,w'\) range over all length-\(k\) closed walks.  If the two walks use
disjoint edge sets, then \(X_w\) and \(X_{w'}\) are independent and the
covariance is zero.  If some edge appears exactly once in the combined product
\(X_wX_{w'}\), then
\[
  \Expect(X_wX_{w'})=0
\]
by the same centering argument; moreover that edge appears in only one of the
two walks, so the corresponding single-walk expectation is also zero.  Hence
the covariance is zero.  Therefore a nonzero covariance can occur only when
the two walks share at least one edge and every edge in the union has combined
multiplicity at least two.

For such a surviving pair, the union graph is connected, because the graph of
each closed walk is connected and the two graphs share an edge.  If the union
has \(e\) distinct edges and \(v\) distinct vertices, then \(e\le k\), since the
combined product has length \(2k\) and every union edge appears at least twice.
Connectivity gives
\[
  v\le e+1\le k+1.
\]
For fixed \(k\), there are only finitely many unlabeled paired-walk patterns,
and their labels can be chosen in \(O_k(n^v)\le O_k(n^{k+1})\) ways.  Uniform
moment bounds give \(|\Cov(X_w,X_{w'})|\le C_k\).  Hence
\[
  \Var(m_{k,n})
  \le
  n^{-2-k}O_k(n^{k+1})
  =
  O_k(n^{-1})
  \to0 .
\]
Combining the expectation and variance calculations, for every fixed \(r\),
\[
  m_{2r,n}\toP C_r,
  \qquad
  m_{2r+1,n}\toP0 .
\]

It remains to justify why convergence of moments gives weak convergence.  The
limiting moment sequence is the moment sequence of the semicircle law.  Indeed,
by symmetry its odd moments are zero, and with \(x=2\cos\theta\),
\[
\begin{aligned}
  \int_{-2}^2 x^{2r}\frac{1}{2\pi}\sqrt{4-x^2}\,dx
  &=
  \frac{2^{2r+1}}{\pi}
  \int_0^\pi \cos^{2r}\theta\,\sin^2\theta\,d\theta  \\
  &=
  \frac{(2r)!}{r!(r+1)!}
  =
  C_r .
\end{aligned}
\]
This is a Hamburger moment problem on \(\R\).  The Catalan moments satisfy
Carleman's condition \citep{shohatTamarkin1943problem}:
\[
  \sum_{r=1}^{\infty} C_r^{-1/(2r)}=\infty,
\]
because \(C_r\le4^r\), so \(C_r^{-1/(2r)}\ge1/2\).  Therefore there is at most
one probability law on \(\R\) with these moments.  Since the semicircle law has
them, it is the unique law with this moment sequence.

Finally, \(m_{2,n}\toP1\), so \(\mu_n\) is tight in probability by Markov's
inequality:
\[
  \mu_n(|x|>R)\le R^{-2}m_{2,n}.
\]
To make the last step explicit, take any subsequence.  Since every moment
converges in probability, a further subsequence can be chosen along which all
moments converge almost surely:
\[
  \int x^k\,\mu_{n_j}(dx)\to s_k,
  \qquad
  s_{2r}=C_r,\quad s_{2r+1}=0.
\]
For almost every sample point on this event, tightness gives weakly convergent
subsubsequences.  If \(\mu_{n_{j_\ell}}\weakto\nu\), then the convergence of
higher even moments gives uniform integrability for every fixed power: for a
given \(k\), choose \(p\) with \(2p>k\) and use the boundedness of
\(\int |x|^{2p}\,\mu_{n_{j_\ell}}(dx)\).  Using a continuous cutoff and then
letting the cutoff grow shows
\[
  \int x^k\,\nu(dx)=s_k,\qquad k=0,1,2,\ldots .
\]
By Carleman's determinacy criterion, \(\nu=\mu_{\mathrm{sc}}\).  Thus every
subsequential weak limit is the semicircle law, and the original random
measures converge weakly in probability to \(\mu_{\mathrm{sc}}\).  This is
equivalent to the bounded-continuous test-function convergence stated in the
theorem. \qedmark

\begin{example}[What the theorem says statistically]
A Wigner matrix is not a covariance matrix estimated from observations; it is a
model for pure symmetric noise.  The theorem says that even pure noise has a
predictable global spectral shape after the \(n^{-1/2}\) scaling.  In high
dimensional statistics, this is the baseline warning behind many spectral
methods: an eigenvalue or eigenvector is meaningful only after it is compared
with the random spectral bulk that noise alone would create.
\end{example}

\section{Characteristic Functions and Fourier Transforms of Laws}
\label{sec:appC-characteristic-functions}
\conceptindexes{characteristic functions, Fourier signatures, convolution, central limit theorem}

A characteristic function records a probability law through oscillating test
functions:
\[
  \varphi_X(t)=\Expect e^{\mathrm{i}tX},
  \qquad t\in\Real.
\]
It may look less tangible than a density or distribution function, but it has a
decisive advantage: sums of independent variables become products of
characteristic functions.  This turns additive noise into multiplicative
algebra.

\begin{proposition}[Independent sums multiply]
\label{prop:appC-cf-products}
If \(X\) and \(Y\) are independent real-valued random variables, then
\[
  \varphi_{X+Y}(t)=\varphi_X(t)\varphi_Y(t),
  \qquad t\in\Real.
\]
Consequently, if \(X_1,\ldots,X_n\) are independent, then
\[
  \varphi_{X_1+\cdots+X_n}(t)
  =
  \prod_{j=1}^n\varphi_{X_j}(t).
\]
\end{proposition}

\noindent\textit{Proof.}
Since \(e^{\mathrm{i}t(X+Y)}=e^{\mathrm{i}tX}e^{\mathrm{i}tY}\) and the two
factors are bounded functions of independent random variables,
\[
  \Expect e^{\mathrm{i}t(X+Y)}
  =
  \Expect\{e^{\mathrm{i}tX}e^{\mathrm{i}tY}\}
  =
  \Expect e^{\mathrm{i}tX}\,\Expect e^{\mathrm{i}tY}.
\]
Iterating gives the finite-product formula. \qedmark

\begin{example}[Poisson counts add by multiplying signatures]
If \(N\sim\Poisson(\lambda)\), then
\[
  \varphi_N(t)
  =
  \sum_{k=0}^{\infty}e^{\mathrm{i}tk}e^{-\lambda}\frac{\lambda^k}{k!}
  =
  \exp\{\lambda(e^{\mathrm{i}t}-1)\}.
\]
If \(N_1\) and \(N_2\) are independent Poisson counts with means
\(\lambda_1\) and \(\lambda_2\), then
\[
  \varphi_{N_1+N_2}(t)
  =
  \exp\{(\lambda_1+\lambda_2)(e^{\mathrm{i}t}-1)\}.
\]
Thus \(N_1+N_2\sim\Poisson(\lambda_1+\lambda_2)\).  This is the
Fourier version of a common modeling move: independent spatial cell counts,
read counts, or event counts combine by adding intensities.
\end{example}

\begin{theorem}[Central limit theorem through characteristic functions]
\label{thm:appC-cf-clt}
Let \(X_1,X_2,\ldots\) be iid with mean \(\mu\) and variance
\(0<\sigma^2<\infty\).  Then
\[
  \frac{\sum_{j=1}^n(X_j-\mu)}{\sigma\sqrt n}
  \weakto \Normal(0,1).
\]
\end{theorem}

\noindent\textit{Proof.}
Put \(Y=(X_1-\mu)/\sigma\), so \(\Expect Y=0\) and \(\Expect Y^2=1\).  The
Taylor expansion of the exponential, together with dominated convergence using
\(\Expect Y^2<\infty\), gives
\[
  \varphi_Y(u)=1-\frac{u^2}{2}+o(u^2),
  \qquad u\to0.
\]
If
\[
  S_n=\frac{Y_1+\cdots+Y_n}{\sqrt n},
\]
then independence gives
\[
  \varphi_{S_n}(t)
  =
  \left\{\varphi_Y(t/\sqrt n)\right\}^n
  =
  \left\{1-\frac{t^2}{2n}+o(n^{-1})\right\}^n
  \to
  e^{-t^2/2}.
\]
The function \(e^{-t^2/2}\) is the characteristic function of \(\Normal(0,1)\).
Levy's continuity theorem identifies pointwise convergence of characteristic
functions, with a continuous limit at zero, as weak convergence of the
corresponding laws.  Therefore \(S_n\weakto \Normal(0,1)\). \qedmark

\begin{example}[Why the proof matches the story]
The CLT proof says that many small independent contributions do not have to
look normal individually.  Their Fourier signatures multiply.  After
centering, scaling, and multiplying \(n\) nearby factors, the quadratic term in
the local expansion survives while higher-order terms vanish.  This is why
normal approximations appear in sample means, score equations, smoothed
estimators, and many instrument-noise calculations: the limit is produced by
local second-order behavior plus multiplication of independent signatures.
\end{example}

\subsection{Selected Central Limit Theorems}
\label{sec:appC-clt-atlas}
\conceptindexes{central limit theorem, Lindeberg condition, functional central limit theorem, Banach-valued central limit theorem, martingale central limit theorem, Markov chain central limit theorem, Markov process central limit theorem, diffusion central limit theorem, renewal process central limit theorem, Hawkes process central limit theorem, random measure central limit theorem, random field central limit theorem, Bernstein--von Mises theorem, U-statistic central limit theorem, self-normalized central limit theorem, stable law}

The iid theorem above is the cleanest entrance, but modern statistics rarely
sees only iid scalar sums.  The same normal limit reappears after the random
object, the clock, and the dependence structure are changed.  The statements
below are meant as a comparison chart, not as a replacement for the specialized
limit theory cited near each statement and at the end of the appendix.

\begin{center}
\small
\setlength{\tabcolsep}{0.35em}
\renewcommand{\arraystretch}{1.16}
\begin{longtable}{@{}>{\raggedright\arraybackslash}p{0.23\linewidth}
                  >{\raggedright\arraybackslash}p{0.34\linewidth}
                  >{\raggedright\arraybackslash}p{0.33\linewidth}@{}}
\caption{Central-limit-theorem variants and their random structures}\\
\toprule
Version & Random structure & What changes from iid \\
\midrule
\endfirsthead
\caption[]{Central-limit-theorem variants and their random structures (continued)}\\
\toprule
Version & Random structure & What changes from iid \\
\midrule
\endhead
Lindeberg--Levy &
Iid scalar summands with finite variance. &
The clock is \(n\); the variance is one number. \\
Lindeberg--Feller &
Independent triangular arrays. &
No summand may carry a visible fraction of the final variance. \\
Functional CLT &
Partial-sum paths in \(D[0,1]\). &
The limit is a Brownian path, so tightness joins finite-dimensional CLT. \\
Banach-valued CLT &
Sums taking values in a function space. &
Geometry and tightness of the ambient space become part of the theorem. \\
Martingale/Rebolledo &
Vector martingales with predictable covariation matrices. &
The deterministic clock is replaced by a predictable covariance clock. \\
Markov-chain CLT &
Additive functionals of a stationary ergodic chain. &
Dependence changes the variance through autocovariances or a Poisson equation. \\
Markov/Feller-process CLT &
Time integrals along a continuous-time Markov process or diffusion. &
The Poisson equation turns the integral into a martingale plus a small boundary term. \\
Renewal-process CLT &
Counting process generated by iid waiting times. &
The random count is the inverse of a partial-sum clock. \\
Hawkes-process CLT &
Self-exciting point process with stable branching ratio. &
Past events inflate the effective variance. \\
Random-measure CLT &
Centered measures tested on sets or functions. &
The limit is a Gaussian random measure or Gaussian process. \\
Random-field CLT &
Spatially indexed sums over growing boxes. &
Dependence is controlled by distance, and the variance sums covariances over lags. \\
Bernstein--von Mises &
Posterior laws in regular parametric models. &
The posterior itself becomes Gaussian after local \(n^{1/2}\) rescaling. \\
\bottomrule
\end{longtable}
\end{center}

Classical accounts of triangular-array CLTs include
\citet{feller1971introductionVol2}, \citet{billingsley1995probability}, and
\citet{kallenberg2002foundations}.

\begin{theorem}[Lindeberg--Feller triangular-array CLT]
\label{thm:appC-lindeberg-feller}
For each \(n\), let \(X_{n1},\ldots,X_{nk_n}\) be independent mean-zero
variables.  Suppose
\[
  \sum_{j=1}^{k_n}\Var(X_{nj})\to\sigma^2\in(0,\infty)
\]
and, for every \(\varepsilon>0\),
\[
  \sum_{j=1}^{k_n}
  \Expect\{X_{nj}^2\ind{|X_{nj}|>\varepsilon}\}\to0.
\]
Then
\[
  \sum_{j=1}^{k_n}X_{nj}\weakto \Normal(0,\sigma^2).
\]
\end{theorem}

The condition says that the limiting Gaussian is not produced by one rare
dominant observation.  It is the triangular-array form behind many regression,
stratified-sampling, and changing-design normal approximations.

The path version is Donsker's invariance principle
\citep{donsker1952justification}; modern treatments place it in \(D[0,1]\)
and related function spaces \citep{billingsley1999convergence,vaart2023weak}.

\begin{theorem}[Functional CLT, or Donsker invariance principle; \citealp{donsker1952justification}]
\label{thm:appC-functional-clt}
Let \(X_1,X_2,\ldots\) be iid with mean \(0\) and variance \(\sigma^2<\infty\).
Define
\[
  W_n(t)=\frac{1}{\sqrt n}\sum_{j=1}^{\lfloor nt\rfloor}X_j,
  \qquad 0\le t\le1.
\]
Then
\[
  W_n\weakto \sigma W
  \qquad\text{in }D[0,1],
\]
where \(W\) is standard Brownian motion.
\end{theorem}

This is the path version of the CLT.  Finite-dimensional convergence gives the
normal limits at finitely many times; tightness says the whole path has no
uncontrolled oscillation between those times.  This is exactly the language
used in Chapter~\ref{chap:weak-convergence-random-objects}, especially
Sections~\ref{sec:ch10-fidi-tightness} and
\ref{sec:ch10-continuous-mapping}, and it becomes the empirical-process
grammar of Chapter~\ref{chap:uniform-laws-empirical-processes}.

\begin{example}[Banach-valued CLT]
\label{ex:appC-banach-valued-clt}
Let \(B\) be a separable Banach space, and let \(X_1,X_2,\ldots\) be iid
centered \(B\)-valued random elements.  A Banach-valued CLT asks when
\[
  \frac{1}{\sqrt n}\sum_{i=1}^n X_i
  \weakto
  G
  \qquad\text{in }B,
\]
where \(G\) is a centered Gaussian random element satisfying
\[
  \ell(G)\sim \Normal\{0,\Var(\ell(X_1))\}
  \qquad\text{for every }\ell\in B^*.
\]
The Hilbert-space case behaves much like finite-dimensional Euclidean space
under a finite second-moment assumption.  In a general Banach space, however,
one must also control whether the candidate Gaussian law exists and whether
the normalized sums are tight.  This is why Banach-valued CLTs are statements
about both probability and the geometry of the normed space; see
\citet{araujoGine1980cltBanach} and
\citet{ledouxTalagrand1991probabilityBanach}.
\end{example}

The following statement is a compact multivariate version of Rebolledo's
martingale CLT; see \citet{rebolledo1980central} for the local-martingale
theorem, \citet{jacod1987limit} for stochastic-process limit theory, and
\citet{andersen1993statistical} for the counting-process statistical form.

\begin{theorem}[Multivariate martingale CLT, Rebolledo form; \citealp{rebolledo1980central}]
\label{thm:appC-rebolledo-clt}
Let \(\mathbf M_n(t)=(M_{n1}(t),\ldots,M_{nd}(t))^\top\), \(0\le t\le T\), be
square-integrable \(\mathbb R^d\)-valued martingales with \(\mathbf M_n(0)=0\).  Let
\(\langle \mathbf M_n\rangle(t)\) denote the predictable quadratic covariation matrix,
whose \((a,b)\) entry is \(\langle M_{na},M_{nb}\rangle(t)\).  Suppose that
\[
  \sup_{t\le T}\|\langle \mathbf M_n\rangle(t)-\mathbf V(t)\|\to0
  \quad\text{in probability},
\]
where \(\mathbf V(0)=0\) is a continuous deterministic matrix-valued function and
\(\mathbf V(t)-\mathbf V(s)\) is nonnegative definite for \(0\le s\le t\le T\).  If the
martingale jumps are asymptotically negligible, for example
\[
  \sup_{t\le T}\|\Delta \mathbf M_n(t)\|\to0
  \quad\text{in probability},
\]
then
\[
  \mathbf M_n \weakto \mathbf G
  \qquad\text{in }D[0,T],
\]
where \(\mathbf G\) is a continuous centered Gaussian martingale with covariance
\[
  \Expect\{\mathbf G(s)\mathbf G(t)^\top\}=\mathbf V(s\wedge t).
\]
In particular,
\[
  \mathbf M_n(T)\weakto \Normal\{0,\mathbf V(T)\}.
\]
\end{theorem}

The martingale theorem is the form that survival analysis, counting processes,
adaptive designs, and online experiments keep using
\citep{rebolledo1980central,jacod1987limit,andersen1993statistical}.  The
variance is not a fixed sample size times a variance; it is the limiting
predictable quadratic covariation.  The scalar version is the one-dimensional
shadow; the multivariate statement is what is usually needed when several
hazards, scores, or treatment contrasts are followed at once.

For Markov chains, additive-function CLTs can be read either through
autocovariances or through a Poisson equation for the transition operator; see
\citet{karlin1981second}, \citet{kallenberg2002foundations}, and
\citet{ethierKurtz1986markov}.

\begin{theorem}[Markov-chain additive-functional CLT; \citealp{kipnisVaradhan1986central}]
\label{thm:appC-markov-chain-clt}
Let \(X_0,X_1,\ldots\) be a stationary irreducible aperiodic Markov chain on a
finite state space with invariant law \(\pi\), and let \(f\) satisfy
\(\pi f=0\).  Then
\[
  \sigma_f^2
  =
  \Var_\pi\{f(X_0)\}
  +
  2\sum_{k=1}^{\infty}\Cov_\pi\{f(X_0),f(X_k)\}
\]
is finite and nonnegative, and
\[
  \frac{1}{\sqrt n}\sum_{k=1}^n f(X_k)
  \weakto
  \Normal(0,\sigma_f^2).
\]
\end{theorem}

For Markov chains the variance remembers dependence.  The iid variance is
replaced by the long-run variance; equivalently, in many treatments it is read
through the Poisson equation for the transition operator.

Continuous-time Markov processes and diffusions use the same idea in generator
language.  The following schematic form is the one used behind many Feller and
diffusion limits; see \citet{feller1952parabolic},
\citet{ethierKurtz1986markov}, and \citet{kipnisVaradhan1986central}.

\begin{theorem}[Markov-process additive-functional CLT, Poisson-equation form; \citealp{kipnisVaradhan1986central}]
\label{thm:appC-markov-process-clt}
Let \(X=(X_t)_{t\ge0}\) be a stationary ergodic Markov process with invariant
law \(\pi\) and generator \(\mathcal L\).  Let \(f\) be centered,
\(\pi f=0\), and suppose the Poisson equation
\[
  -\mathcal L g=f
\]
has a sufficiently regular solution \(g\).  If the martingale
\[
  M_t^g
  =
  g(X_t)-g(X_0)+\int_0^t f(X_s)\,ds
\]
satisfies
\[
  \frac{\langle M^g\rangle_t}{t}\to\sigma_f^2
  \qquad\text{in probability}
\]
and the boundary term \(g(X_t)-g(X_0)\) is \(o_{\Prob}(\sqrt t)\), then
\[
  \frac{1}{\sqrt t}\int_0^t f(X_s)\,ds
  \weakto
  \Normal(0,\sigma_f^2).
\]
\end{theorem}

For a diffusion, \(\mathcal L\) is the differential generator and the
martingale part comes from It\^o's formula.  For a Feller process, the same
statement is read through the semigroup and generator on a function space.

Classical renewal-process CLTs are treated in \citet{feller1971introductionVol2},
\citet{karlin1975first}, and \citet{karlin1981second}.

\begin{theorem}[Renewal counting-process CLT]
\label{thm:appC-renewal-clt}
Let \(\xi_1,\xi_2,\ldots\) be iid positive interarrival times with
\[
  m=\Expect \xi_1\in(0,\infty),
  \qquad
  \tau^2=\Var(\xi_1)<\infty,
\]
and let
\[
  N(t)=\max\{n:\xi_1+\cdots+\xi_n\le t\}.
\]
Then
\[
  \frac{N(t)-t/m}{\sqrt{\tau^2t/m^3}}
  \weakto
  \Normal(0,1),
  \qquad t\to\infty.
\]
\end{theorem}

The renewal theorem is a CLT for an inverse clock.  The partial sums
\(\xi_1+\cdots+\xi_n\) satisfy an ordinary CLT; the counting process asks how
many arrivals fit before time \(t\), so the variance is transformed by the
inverse map.

For Hawkes processes, the original self-exciting construction is
\citet{hawkes1971spectra}; the point-process and martingale machinery is
developed in \citet{bremaud1981point}, \citet{jacod1987limit}, and
\citet{andersen1993statistical}.

\begin{theorem}[Linear Hawkes-process CLT, one-dimensional form; \citealp{hawkes1971spectra}]
\label{thm:appC-hawkes-clt}
Let \(N(t)\) be a stationary linear Hawkes process with baseline rate
\(\nu>0\), excitation kernel \(h\ge0\), and branching ratio
\[
  \eta=\int_0^\infty h(s)\,ds<1.
\]
Its stationary mean intensity is
\[
  \bar\lambda=\frac{\nu}{1-\eta}.
\]
Under standard moment and stability conditions,
\[
  \frac{N(t)-\bar\lambda t}{\sqrt t}
  \weakto
  \Normal\!\left(0,\frac{\bar\lambda}{(1-\eta)^2}\right)
  =
  \Normal\!\left(0,\frac{\nu}{(1-\eta)^3}\right).
\]
\end{theorem}

The Hawkes limit shows how self-excitation changes uncertainty.  The mean rate
is inflated by \(1/(1-\eta)\), and the long-run variance is inflated again
because events arrive in clusters rather than as independent Poisson points.

Random-measure limits are a natural language for Poisson processes and
empirical processes; see \citet{kingman1993poisson},
\citet{kallenberg2002foundations}, and \citet{jacod1987limit}.

\begin{theorem}[Poisson random-measure CLT]
\label{thm:appC-random-measure-clt}
Let \(N_n\) be a Poisson random measure on \((S,\mathcal S)\) with intensity
measure \(n\Lambda\).  For sets \(A_1,\ldots,A_k\) with
\(\Lambda(A_j)<\infty\), define
\[
  Z_n(A_j)=\frac{N_n(A_j)-n\Lambda(A_j)}{\sqrt n}.
\]
Then
\[
  \{Z_n(A_1),\ldots,Z_n(A_k)\}
  \weakto
  \{G(A_1),\ldots,G(A_k)\},
\]
where \(G\) is a centered Gaussian random measure with covariance
\[
  \Cov\{G(A),G(B)\}=\Lambda(A\cap B).
\]
Equivalently, for square-integrable test functions,
\[
  \frac{\int f\,dN_n-n\int f\,d\Lambda}{\sqrt n}
  \weakto
  \Normal\!\left(0,\int f^2\,d\Lambda\right).
\]
\end{theorem}

This random-measure form is the common ancestor of grid-cell count limits,
Poisson-process approximations, and empirical-process CLTs.  Once the random
object is a measure, a CLT is read by testing it on sets or functions.

Random-field CLTs extend time-series normal approximation to spatially indexed
objects.  The mixing version below is represented by
\citet{bolthausen1982central}.

\begin{theorem}[Stationary mixing random-field CLT; \citealp{bolthausen1982central}]
\label{thm:appC-random-field-clt}
Let \(\{X_z:z\in\mathbb Z^d\}\) be a stationary mean-zero random field with
\(\Expect |X_0|^{2+\delta}<\infty\) for some \(\delta>0\).  Assume a standard
strong-mixing condition fast enough for the Bolthausen random-field CLT, and
assume that
\[
  \sigma^2
  =
  \sum_{h\in\mathbb Z^d}\Cov(X_0,X_h)
  \in(0,\infty).
\]
For rectangular boxes
\[
  B_n=\{1,\ldots,n_1\}\times\cdots\times\{1,\ldots,n_d\},
  \qquad |B_n|\to\infty,
\]
one has
\[
  \frac{1}{\sqrt{|B_n|}}\sum_{z\in B_n}X_z
  \weakto
  \Normal(0,\sigma^2).
\]
Under stronger mixing and tightness assumptions, the rectangular partial-sum
process converges to a Brownian sheet with variance coefficient \(\sigma^2\).
\end{theorem}

The random-field theorem is the spatial analogue of the time-series CLT.
Nearby lattice sites may be dependent, but sufficiently distant regions behave
almost independently, and the limiting variance is the sum of all covariance
lags.

The Bernstein--von Mises theorem is the Bayesian version of local asymptotic
normality: the random posterior law itself becomes Gaussian after local
rescaling.  Standard routes go through \citet{leCam1986asymptotic},
\citet{vaart1998asymptotic}, and \citet{vaart2023weak}.

\begin{theorem}[Bernstein--von Mises theorem, regular parametric form]
\label{thm:appC-bernstein-von-mises}
Let \(X_1,\ldots,X_n\) be iid from a regular \(p\)-dimensional parametric model
\(\{P_\theta:\theta\in\Theta\}\), with true value \(\theta_0\) in the interior
of \(\Theta\), nonsingular Fisher information \(I_{\theta_0}\), and local
asymptotic normality at \(\theta_0\).  Let the prior have a density that is
positive and continuous at \(\theta_0\), and assume posterior consistency.  If
\(\hat\theta_n\) is an efficient estimator such as the maximum likelihood
estimator, then, in \(P_{\theta_0}\)-probability,
\[
  \sup_{A\in\mathcal B(\mathbb R^p)}
  \left|
    \Pi\!\left\{\sqrt n(\theta-\hat\theta_n)\in A\mid X_1,\ldots,X_n\right\}
    -
    \Normal(0,I_{\theta_0}^{-1})(A)
  \right|
  \to0.
\]
\end{theorem}

This is a CLT for a random probability measure.  The likelihood supplies the
local quadratic shape, the prior is locally flat at the \(n^{-1/2}\) scale, and
the posterior mass concentrates on the same Gaussian neighborhood that governs
efficient frequentist estimation.

\begin{example}[Three nearby limit statements]
The atlas can be extended in several directions without changing the central
message; these point to \citet{hoeffding1948class},
\citet{student1908probable}, and \citet{feller1971introductionVol2}.

\begin{itemize}[leftmargin=*]
\item \emph{U-statistics.}  If \(h\) is a symmetric square-integrable kernel of
order \(m\), \(U_n\) is the associated U-statistic, and the first Hoeffding
projection
\[
  h_1(x)=\Expect\{h(x,X_2,\ldots,X_m)\}-\theta
\]
has positive variance, then
\[
  \sqrt n\,(U_n-\theta)
  \weakto
  \Normal\{0,m^2\Var(h_1(X))\}.
\]
The Gaussian part comes from the first projection; degenerate kernels lead to
different, often non-Gaussian, chaos limits \citep{hoeffding1948class}.

\item \emph{Self-normalized sums.}  If \(X_i\) are iid with mean \(\mu\) and
finite nonzero variance, then the Studentized mean
\[
  \frac{\sqrt n(\bar X_n-\mu)}{s_n},
  \qquad
  s_n^2=\frac{1}{n-1}\sum_{i=1}^n(X_i-\bar X_n)^2,
\]
converges to \(\Normal(0,1)\).  The interesting point is that the scale is
estimated from the same data, so the statement already mixes a CLT with a law
of large numbers \citep{student1908probable}.

\item \emph{Stable-law boundary.}  If the summands have tails in the domain of
attraction of an \(\alpha\)-stable law with \(0<\alpha<2\), then suitable
centering and scaling give
\[
  \frac{\sum_{i=1}^n X_i-b_n}{a_n}\weakto S_\alpha
\]
for a non-Gaussian stable law \(S_\alpha\).  This is not a CLT in the Gaussian
sense, but it explains why the finite-variance assumption is not cosmetic
\citep{feller1971introductionVol2}.
\end{itemize}
\end{example}

\section{Product Integrals for Event-History Maps}
\label{sec:appC-product-integrals}
\conceptindexes{product integrals, cumulative hazards, transition probabilities, Aalen--Johansen estimator, matrix product integral}

Product integrals are the event-time analogue of exponentials.  An ordinary
integral accumulates additive increments.  A product integral accumulates
multiplicative increments in their time order.  That order is irrelevant for
scalar hazards but essential for matrices, because matrix increments at
different times need not commute.

Let \(A\) be a \(d\times d\) matrix-valued function of bounded variation on
\([s,t]\).  For a partition
\[
  \pi=\{s=t_0<t_1<\cdots<t_m=t\},
  \qquad
  \Delta_\ell A=A(t_\ell)-A(t_{\ell-1}),
\]
define the ordered product
\[
  \Pi_\pi(s,t)
  =
  \{\matI+\Delta_1A\}\{\matI+\Delta_2A\}\cdots\{\matI+\Delta_mA\}.
\]
When these ordered products converge as the mesh of \(\pi\) tends to zero, the
limit is written
\[
  \Prodi_{(s,t]}\{\matI+A(du)\}.
\]
The factor for earlier time stands to the left.  This is the row-vector
convention used in the event-history chapters: if \(p(s)\) is a row vector,
then \(p(t)=p(s)\matP(s,t)\).

\begin{proposition}[Product-integral calculus]
\label{prop:appC-product-integral-calculus}
Let
\[
  \matP_A(s,t)=\Prodi_{(s,t]}\{\matI+A(du)\}
\]
exist for a bounded-variation matrix \(A\).  Then the following identities hold
whenever the displayed quantities are defined.

\begin{enumerate}[leftmargin=*]
\item \emph{Identity and composition.}
\[
  \matP_A(s,s)=\matI,
  \qquad
  \matP_A(s,t)=\matP_A(s,u)\matP_A(u,t),
  \quad s\le u\le t.
\]

\item \emph{Integral equation.}
\[
  \matP_A(s,t)
  =
  \matI+\int_{(s,t]}\matP_A(s,u-)\,A(du).
\]
If \(A(t)=\int_s^t Q(r)\,dr\), then
\[
  \frac{\partial}{\partial t}\matP_A(s,t)
  =
  \matP_A(s,t)Q(t),
  \qquad
  \matP_A(s,s)=\matI.
\]
For constant \(Q\), this gives \(\matP_A(s,t)=\exp\{(t-s)Q\}\).

\item \emph{Jump update.}
At a jump time \(u\),
\[
  \matP_A(s,u)
  =
  \matP_A(s,u-)\{\matI+\Delta A(u)\}.
\]

\item \emph{Scalar reduction.}
For a scalar bounded-variation function \(B\),
\[
  \Prodi_{(s,t]}(1+dB(u))
  =
  \exp\{B^c(t)-B^c(s)\}
  \prod_{s<u\le t}\{1+\Delta B(u)\}.
\]
Thus the survival formula in Chapter~16 is the special case \(B=-A\):
\[
  \Prodi_{(0,t]}(1-dA(u))
  =
  \exp\{-A^c(t)\}
  \prod_{0<u\le t}\{1-\Delta A(u)\}.
\]

\item \emph{Transition matrices.}
Suppose \(A\) is a transition-hazard matrix: the off-diagonal entries
\(A_{jk}\) are increasing, and
\[
  A_{jj}=-\sum_{k\ne j}A_{jk}.
\]
If every jump matrix \(\matI+\Delta A(u)\) has nonnegative entries, then
\(\matP_A(s,t)\) is row-stochastic.  Hence product integration turns local
transition hazards into global transition probabilities.

\item \emph{First variation.}
For a bounded-variation perturbation \(H\), the first-order derivative of
\(\Phi(A)=\matP_A(s,t)\) is
\[
  \Phi_A'(H)(s,t)
  =
  \int_{(s,t]}
    \matP_A(s,u-)\,H(du)\,\matP_A(u,t),
\]
with the usual product-integral convention around jumps.
\end{enumerate}
\end{proposition}

For survival analysis, the scalar product integral converts a cumulative
hazard into a survival function.  For multi-state analysis, the matrix product
integral converts transition hazards into a transition probability matrix.
Replacing \(A\) by a Nelson--Aalen transition-hazard estimator gives the
Aalen--Johansen estimator:
\[
  \widehat{\matP}(s,t)
  =
  \Prodi_{(s,t]}\{\matI+\widehat{\matA}(du)\}.
\]
The derivative above is the delta-method bridge used when transition
probabilities are targets rather than nuisance summaries.

\section{How These Mathematical Tools Serve the Book}
\label{sec:appC-serving-book}
\conceptindexes{mathematical tools, appendix navigation, statistical translation}

The tools above support one route:
\[
\begin{aligned}
  \text{observable questions}
  &\longrightarrow
  \text{measure and integration tools}\\
  &\longrightarrow
  \text{stable averages and limits}\\
  &\longrightarrow
  \text{projected residual noise}\\
  &\longrightarrow
  \text{spectral coordinates and laws}\\
  &\longrightarrow
  \text{event-time product maps}\\
  &\longrightarrow
  \text{inferential claims}.
\end{aligned}
\]
Set theory keeps the word ``observable'' honest.  The measure and integration
tools in \Appref{app:measure-theoretic-toolkit} explain why averages,
likelihoods, risks, and empirical measures can be compared to their population
targets.  Hilbert geometry explains why conditional means, least-squares fits,
influence functions, and efficient scores are all projection problems.  Compact
spectral theory explains why random curves can be represented by covariance
modes before later inference reads those modes.  Wigner's law explains how the
global spectrum of a large random noise matrix can itself stabilize.  Product
integrals explain how local hazards assemble into survival and transition
probabilities.  Characteristic functions give one clean route from sums to
distributional limits.

The point is not that every analysis must explicitly cite these tools.  The
point is that when a proof later asks for measurability, dominated convergence,
uniform integrability, orthogonality, or a weak limit, it is not leaving the
book's opening story.  It is protecting the translation from the world to a
mathematical claim and back again.

\section*{Sources and Further Reading}
\addcontentsline{toc}{section}{Sources and Further Reading}

The set and measure language behind this appendix follows the standard
measure-theoretic probability route in \citet{billingsley1995probability},
\citet{durrett2019probability}, and \citet{kallenberg2002foundations}, with
additional influence from \citet{dabrowskaAdvancedProbabilityCommunication}.
The projection theorem is a basic Hilbert-space fact; its statistical uses
appear throughout least squares, conditional expectation, and semiparametric
efficiency theory, with the latter developed systematically by
\citet{bickel1993efficient}.  The compact-operator spectral
facts used for covariance operators and functional principal components are
standard Hilbert-space material; for their statistical use in functional data
analysis, see \citet{hsingEubank2015fda}.  The semicircle law for random
matrix spectra is due to \citet{wigner1958distribution}; the moment-determinacy
step uses Carleman's criterion for the Hamburger moment problem, as treated in
\citet{shohatTamarkin1943problem}.  Characteristic functions, Levy's
continuity theorem, and the characteristic-function proof of the CLT are
classical; see \citet{feller1968introduction}, \citet{chung1974course}, and
\citet{billingsley1995probability}.  Functional and empirical-process CLTs are
treated in \citet{donsker1952justification}, \citet{billingsley1999convergence},
and \citet{vaart2023weak}.  Martingale and counting-process versions are
developed in \citet{rebolledo1980central}, \citet{jacod1987limit}, and
\citet{andersen1993statistical}.  Markov-process and renewal-process limit
theory is part of the broader framework of \citet{ethierKurtz1986markov},
\citet{karlin1981second}, \citet{kallenberg2002foundations}, and
\citet{kipnisVaradhan1986central}.  Banach-valued CLTs are treated in
\citet{araujoGine1980cltBanach} and
\citet{ledouxTalagrand1991probabilityBanach}.  For Poisson and related random
measures, see \citet{kingman1993poisson}; for the Hawkes
process, see \citet{hawkes1971spectra}.  Random-field CLTs under mixing
conditions are represented by \citet{bolthausen1982central}.  The
Bernstein--von Mises theorem and its local-asymptotic-normality background are
part of the asymptotic-statistics route in \citet{leCam1986asymptotic},
\citet{vaart1998asymptotic}, and \citet{vaart2023weak}.  U-statistic normal
limits go back to \citet{hoeffding1948class}; Studentized and
self-normalized normal approximations trace to \citet{student1908probable};
stable-law alternatives are treated classically in
\citet{feller1971introductionVol2}. Product integration for survival and
multi-state models is surveyed by \citet{gill1990survey}; the
Aalen--Johansen estimator is due to \citet{aalen1978empirical}.

%% file: appendices/computational_biomedical_translation.tex
\chapter{Computational and Biomedical Translation}
\label{app:computational-biomedical-translation}
\conceptindexes{computational translation, biomedical translation, algorithmic primitives, bioinformatics objects, clinical vocabulary, stochastic dynamics}

This appendix is a short field guide rather than a separate treatment of
computer science, bioinformatics, biochemistry, or medicine.  Its purpose is to help a
reader translate common modern terms back into the probability language used in
the main chapters.  It is also a guardrail against a common modern mistake:
letting a software object, assay name, benchmark, or clinical label stand in
for the statistical object.  The recurring question is always the same: what is
the random object, what generated the observation, what task is being
performed, what target or loss is being named, and what uncertainty should
travel back to the scientific, clinical, or deployed world?

\section{Algorithmic Primitives}
\label{sec:appD-algorithmic-primitives}
\conceptindexes{Monte Carlo, bootstrap, cross-validation, optimization, stochastic gradient, conformal prediction, Bayesian optimization}

\begin{center}
\small
\setlength{\tabcolsep}{0.35em}
\renewcommand{\arraystretch}{1.18}
\begin{longtable}{@{}>{\raggedright\arraybackslash}p{0.21\linewidth}
                  >{\raggedright\arraybackslash}p{0.34\linewidth}
                  >{\raggedright\arraybackslash}p{0.35\linewidth}@{}}
\caption{Algorithmic primitives as probability objects}\\
\toprule
Computational term & Probability reading & Statistical warning \\
\midrule
\endfirsthead
\caption[]{Algorithmic primitives as probability objects (continued)}\\
\toprule
Computational term & Probability reading & Statistical warning \\
\midrule
\endhead
Monte Carlo simulation &
Approximate an integral, risk, power, posterior expectation, or operating
characteristic by random draws. &
The simulation law is part of the model; changing it changes the claim
\citep{metropolis1953equation,hastings1970monte}. \\

Bootstrap &
Replace the unknown population law by an empirical or fitted law, then
resample the statistic as a random object. &
The bootstrap copies the smoothness and support of the resampling law; it can
fail when the target functional is nonsmooth or boundary-sensitive
\citep{efron1979bootstrap}. \\

Cross-validation &
Estimate prediction risk by repeatedly splitting the observed data into
training and evaluation roles. &
The split must respect dependence, time order, clustering, leakage, and the
future deployment population \citep{stone1974cross}. \\

Empirical risk minimization &
Choose a function or parameter by minimizing an observed average loss. &
The loss defines the target; changing the loss changes the scientific meaning
of the estimator \citep{vapnik1998statistical}. \\

Stochastic gradient methods &
Use noisy local estimates of a gradient or score to move through a criterion. &
Optimization error and statistical error are different; an algorithm can
converge computationally to the wrong statistical target
\citep{robbinsMonro1951stochastic}. \\

Monotone operator splitting &
Rewrite an optimality condition as a fixed-point problem or a monotone
inclusion, then use resolvents, proximal maps, or forward steps. &
The fixed point solves the computational problem; the sampling law, loss, and
observation mechanism still determine the statistical target. \\

Calibration and conformal checking &
Compare nominal probabilities, ranks, scores, or prediction sets with their
empirical behavior under a data-generating regime. &
Coverage and calibration are conditional on the exchangeability, sampling, and
deployment assumptions used to justify the check \citep{vovk2005algorithmic}. \\

Adaptive experimentation and active learning &
Use earlier observations to choose the next design point, treatment, molecule,
or measurement. &
The policy becomes part of the data-generating mechanism; future records are
not neutral iid rows from a fixed design \citep{frazier2018tutorial}. \\
\bottomrule
\end{longtable}
\end{center}

The rows in this table are deliberately written in the language of the main
chapters.  Computation does not merely ``implement'' a statistical method.  It
often supplies the randomization, the approximation, the search path, or the
future sampling policy.  The statistical reading is therefore always two-layer:
what mathematical object does the algorithm approximate, and what scientific
target made that object worth approximating?

Monte Carlo is the computational face of expectation.  If a quantity can be
written as \(\Expect_P g(X)\), then simulation replaces the integral by
\[
  m^{-1}\sum_{b=1}^m g(X_b), \qquad X_b\sim P.
\]
When \(\Var_P\{g(X)\}<\infty\), the computational uncertainty has the familiar
form
\[
  \sqrt m\{\bar g_m-\Expect_P g(X)\}
  \weakto
  \Normal\!\left(0,\Var_P\{g(X)\}\right),
  \qquad
  \bar g_m=m^{-1}\sum_{b=1}^m g(X_b).
\]
The usual simulation standard error is
\[
  \widehat{\mathrm{se}}(\bar g_m)
  =
  \frac{\hat\sigma_g}{\sqrt m},
  \qquad
  \hat\sigma_g^2
  =
  \frac{1}{m-1}\sum_{b=1}^m\{g(X_b)-\bar g_m\}^2.
\]
This standard error measures Monte Carlo noise under the simulated law; it does
not measure whether the simulated law is scientifically right.
The notation hides the most important modeling choice: which \(P\) is being
simulated.  In a clinical trial, \(P\) may encode dropout, toxicity, enrollment
rates, and stopping rules.  In single-cell simulation, \(P\) may encode
library size, gene correlation, batches, dropout, and spatial structure.  A
simulation study is therefore a designed experiment on models.

The bootstrap is a simulation study in which the simulation law is learned from
the data.  In the nonparametric bootstrap, \(P\) is replaced by \(P_n\).  In a
model-based bootstrap, \(P\) is replaced by \(P_{\hat\theta}\).  In a cluster,
time-series, or event-history problem, the resampling unit must match the
dependence structure; resampling rows independently is a claim of iid
generation, not a neutral default.

Cross-validation is a risk-estimation device.  It is most honest when the
training/evaluation split imitates the future use case.  Random row splits may
be appropriate for iid prediction, but longitudinal, spatial, family,
hospital, donor, batch, and time-indexed data often require grouped or
forward-in-time splits.  The probability question is not merely whether the
average validation loss is small; it is whether the validation experiment
samples from the same target task that the deployed method will face.

Calibration and conformal prediction are useful precisely because they turn a
forecasting claim back into an observable frequency question.  But the
frequency statement is never floating in space.  Marginal conformal coverage,
for example, is tied to exchangeability or to whatever weaker sampling
structure replaces it.  When patients, images, hospitals, donors, batches, or
time periods shift, the check must be re-read as a claim about that deployment
law, not as a universal certificate.

Empirical risk, likelihood, penalized objectives, and neural-network losses all
fit the same \(M\)-estimation template:
\[
  \hat\theta\in\argmin_{\theta\in\Theta} P_n\ell_\theta .
\]
The computational form may be modern, but the statistical question is old:
what population criterion \(P\ell_\theta\) is being approximated, and what
scientific target does its minimizer represent?
For a target minimizer \(\theta_0\in\argmin_{\theta}P\ell_\theta\), the basic
comparison is
\[
  P\ell_{\hat\theta}-P\ell_{\theta_0},
\]
which separates the scientific loss scale from the optimization path used to
find \(\hat\theta\).

Stochastic gradient methods add another stochastic layer.  The iterates form an
adapted sequence: at step \(k\), the update is measurable with respect to the
history of previous mini-batches, gradients, and parameter values.  This is why
the method belongs to the same family of ideas as martingales, stochastic
approximation, and adaptive designs.  A small training loss is not, by itself,
a statistical guarantee; one must still ask which population criterion is being
optimized and how optimization error, sampling error, and approximation error
are being separated.

Monotone-operator methods give a compact language for many algorithms that
appear in modern statistical computation.  In the formulation emphasized by
Wotao Yin and collaborators, a convex or constrained problem is often rewritten
as
\[
  0\in A(u)+B(u)+C(u),
\]
where the operators encode smooth loss, nonsmooth penalty, constraints, or dual
variables.  Iterations built from resolvents
\(J_{\gamma A}=(I+\gamma A)^{-1}\), gradients, and projections then recover
proximal gradient, Douglas--Rachford splitting, ADMM, and three-operator
splitting as variations on the same theme
\citep{davisYin2017threeOperator,ryuYin2022largeScale}.  This is related to
the book because Chapters~6, 13, and 14 repeatedly turn statistical questions
into minimizers, roots, and gradients.  It is not a replacement for the
probabilistic analysis: algorithmic convergence says where the computation
goes, while consistency, limiting distribution, and sensitivity still depend on
the data-generating law.

\section{Bioinformatics Data Objects}
\label{sec:appD-bioinformatics-objects}
\conceptindexes{bioinformatics data objects, count matrix, single-cell RNA sequencing, pseudotime, gene regulatory network, spatial transcriptomics}

\begin{center}
\small
\setlength{\tabcolsep}{0.35em}
\renewcommand{\arraystretch}{1.18}
\begin{longtable}{@{}>{\raggedright\arraybackslash}p{0.23\linewidth}
                  >{\raggedright\arraybackslash}p{0.33\linewidth}
                  >{\raggedright\arraybackslash}p{0.34\linewidth}@{}}
\caption{Bioinformatics terms as statistical objects}\\
\toprule
Bioinformatics term & Data-structure reading & Probability question \\
\midrule
\endfirsthead
\caption[]{Bioinformatics terms as statistical objects (continued)}\\
\toprule
Bioinformatics term & Data-structure reading & Probability question \\
\midrule
\endhead
Count matrix &
Cells, samples, or spots by genes, transcripts, peaks, proteins, or features. &
What is the unit: molecule, cell, donor, tissue, library, or processed
feature? \\

Library size and normalization &
Observed totals and transformations used to make samples comparable. &
Which variation is biological signal, and which is measurement or exposure? \\

Dropout and zero inflation &
Zeros created by biological absence, technical failure, limited capture, or
thresholding. &
Is zero an event in the scientific process, the observation process, or both? \\

Batch and donor effects &
Dependence and shifts induced by preparation, instrument, lab, time, donor, or
site. &
Which factors are nuisance variation, and which define the target population? \\

Pseudotime and trajectory &
A learned ordering or curve used to represent progression. &
Is the trajectory observed time, latent state, algorithmic embedding, or
scientific hypothesis? \\

Multi-omics and spatial omics &
Matched or partially matched molecular layers plus coordinates or images. &
What is the joint random object, and what missingness is induced by modality
alignment? \\
\bottomrule
\end{longtable}
\end{center}

A bioinformatics matrix is rarely just a matrix.  It is the visible part of an
assay pipeline.  In RNA-seq or single-cell RNA-seq, the entries have passed
through capture chemistry, sequencing, alignment or quantification,
normalization, filtering, and often dimension reduction.  The statistical
object is therefore not only \(Y_{ij}\), a count for feature \(j\) in unit
\(i\).  It is the pair consisting of a biological state and an observation
mechanism that turns that state into \(Y_{ij}\).

One minimal count model writes
\[
  Y_{ij}\mid \mu_{ij},s_i \sim \Poisson(s_i\mu_{ij}),
  \qquad
  \text{or}\qquad
  Y_{ij}\mid \mu_{ij},s_i,\phi_j \sim
  \NegBin(s_i\mu_{ij},\phi_j),
\]
where \(s_i\) is an exposure or library-size factor and \(\mu_{ij}\) is the
biological abundance on the chosen scale.  A dropout layer, if used, changes the
observation map rather than the biological state itself:
\[
  Y_{ij}^{\mathrm{obs}}=D_{ij}Y_{ij},\qquad
  D_{ij}\mid Y_{ij},X_i \sim \Bernoulli\{p_{ij}(Y_{ij},X_i)\}.
\]
The practical question is whether zeros, size factors, and overdispersion are
features of biology, measurement, or both.

This is why the single-cell examples in the main chapters are organized as
observation, generation, and inference.  scImpute begins with the question of
what a zero means; scDesign3 asks for a realistic generative law for synthetic
single-cell and spatial-omics data; scGTM treats a structured biological map
as an inferential object rather than as a mere visualization
\citep{li2018scimpute,cui2022scgtm}.  The common thread is
that computation does not arrive after statistics.  It is part of how the
random object is named.

Normalization should be read as a change of scale and sometimes as a change of
target.  Dividing by library size, using size factors, transforming counts,
selecting highly variable genes, and correcting batches can be necessary.  But
each step decides which variation will be treated as technical, which variation
will be allowed to remain biological, and which downstream uncertainty will be
ignored unless it is explicitly propagated.

Pseudotime is especially easy to overread.  It may be a useful coordinate for a
biological process, but it is usually learned from the same data that are later
analyzed along the coordinate.  In probability language, pseudotime is a random
function of the data.  Treating it as fixed can be useful as a first
approximation, but the approximation should be recognized.

\section{Clinical and Medical Vocabulary}
\label{sec:appD-clinical-vocabulary}
\conceptindexes{clinical vocabulary, estimand, endpoint, censoring, target trial, real-world evidence, electronic health records}

\begin{center}
\small
\setlength{\tabcolsep}{0.35em}
\renewcommand{\arraystretch}{1.18}
\begin{longtable}{@{}>{\raggedright\arraybackslash}p{0.22\linewidth}
                  >{\raggedright\arraybackslash}p{0.34\linewidth}
                  >{\raggedright\arraybackslash}p{0.34\linewidth}@{}}
\caption{Clinical vocabulary as estimand and observation language}\\
\toprule
Clinical term & Statistical translation & Typical pressure \\
\midrule
\endfirsthead
\caption[]{Clinical vocabulary as estimand and observation language (continued)}\\
\toprule
Clinical term & Statistical translation & Typical pressure \\
\midrule
\endhead
Estimand &
The target quantity: population, treatment condition, endpoint, summary, and
intercurrent-event handling. &
The estimator is secondary; first state what comparison is meant. \\

Endpoint &
The rule that turns clinical experience into an analyzable variable or event. &
Endpoint definitions create data; they do not merely label data. \\

Censoring &
Partial observation of an event time or longitudinal history. &
Independent censoring is an assumption about the observation mechanism. \\

Hazard and risk set &
Instantaneous event tendency among subjects still under observation and at
risk. &
The risk set is a changing denominator produced by entry, follow-up, events,
and censoring. \\

Biomarker &
Measured biological feature used for diagnosis, prognosis, enrichment, safety,
or mechanism. &
The assay, timing, cutoff, and missingness are part of the statistical object. \\

Adherence and treatment switching &
Post-assignment behavior that affects exposure and interpretation. &
The trial can estimate different effects depending on how this behavior enters
the estimand. \\

Target trial and real-world evidence &
A hypothetical or explicit trial protocol used to organize routine-care data. &
Design thinking must precede estimator choice
\citep{hernan2016targettrial,fda2018rweframework}. \\
\bottomrule
\end{longtable}
\end{center}

Clinical language often looks administrative, but it is mathematical once it
defines what can be observed and compared.  An endpoint is a function from a
clinical history to an analysis object.  Progression-free survival, for
example, requires rules for imaging schedule, progression criteria, death,
missed visits, new therapy, and censoring.  Before a Kaplan--Meier curve, Cox
model, or counting-process martingale appears, the clinical history has already
been discretized into an event-time record.

In notation, a patient history \(H_i\) is mapped to an analysis variable by an
endpoint rule
\[
  (Z_i,\delta_i)=\Phi(H_i),
  \qquad
  N_i(t)=\ind{Z_i\le t,\ \delta_i=1},
  \qquad
  Y_i(t)=\ind{Z_i\ge t}.
\]
The function \(\Phi\) contains the clinical rules: what counts as an event,
when follow-up stops, and how intercurrent events are handled.  The counting
processes \(N_i\) and \(Y_i\) are therefore not raw biology; they are biology
after endpoint construction.

An estimand is the clinical counterpart of the book's translation principle.
It asks: in which population, under which treatment strategies, with which
endpoint, under which handling of intercurrent events, and summarized by which
quantity?  This framing is formalized in ICH E9(R1)
\citep{ich2021e9r1}.  Once this target is fixed, likelihoods, estimating equations,
inverse-probability weights, \(g\)-methods, or sensitivity analyses become
candidate routes rather than substitutes for the target.

Real-world data make the same issue sharper.  An EHR record is not a direct
copy of disease.  It is a care trace: visits, orders, codes, lab results,
medications, referrals, insurance, and institution-specific recording habits.
Target-trial language is useful because it forces a routine-care archive to be
read as an observed-data map from a clinical question, not as a convenient
table waiting for adjustment
\citep{weiskopf2013methods,goldstein2017opportunities,hernan2016targettrial}.

\section{Biochemistry and Stochastic Dynamics}
\label{sec:appD-biochemistry-dynamics}
\conceptindexes{biochemistry, stochastic dynamics, reaction networks, Gillespie algorithm, self-driving laboratories}

Biochemistry enters the book only where it naturally touches stochastic
processes.  A reaction network gives species, reactions, propensities, and
state changes.  In the simplest stochastic reading, the molecular count vector
\(X(t)\) is a continuous-time Markov chain, and each reaction channel has an
intensity depending on the current state.  The chemical master equation
describes the law of \(X(t)\); Gillespie's algorithm simulates exact reaction
events under the model \citep{gillespie1977exact}.

If reaction \(r\) has state-change vector \(\nu_r\) and propensity
\(a_r(x)\), the random-time-change representation is
\[
  X(t)
  =
  X(0)
  +
  \sum_r \nu_r\,
  Y_r\left(\int_0^t a_r\{X(s)\}\,ds\right),
\]
where the \(Y_r\)'s are independent unit-rate Poisson processes.  This formula
is the bridge between biochemical language and the point-process language of
intensities and compensators.

The statistical point is not that every biochemical system should be reduced
to a Markov jump process.  The point is that the same objects from the main
text reappear with scientific names: states, transitions, intensities,
filtrations, stopping times, path likelihoods, and observation error.  In
single-molecule experiments or gene-expression time traces, the raw scientific
record may be a noisy path, a sequence of dwell times, a burst process, or a
partially observed reaction history.  The stochastic-process language says
what information is present before asking which estimator or classifier to use.

Autonomous laboratories add one more layer.  The experiment is no longer a
fixed list of reactions or compounds.  It is an adapted process: a model reads
previous results, proposes a new experiment, receives a new measurement, and
updates the next proposal.  This is the same statistical loop as an adaptive
trial or recommender system, with chemistry or biology as the domain that
generates the traces
\citep{frazier2018tutorial,shields2021bayesian,abolhasani2023rise}.

\section{How to Use This Appendix}
\label{sec:appD-how-to-use}
\conceptindexes{translation appendix, biomedical glossary, computational glossary}

When a modern term appears, translate it in five steps.
\begin{enumerate}
\item Name the world question: clinical comparison, biological mechanism,
prediction task, design decision, policy value, or scientific explanation.
\item Name the random object: vector, matrix, graph, path, event time,
function, distribution, image, text feature class, or adaptive log.
\item Name the observation mechanism: assay, instrument, missingness,
censoring, filtering, labeling, batching, exposure, or decision policy.
\item Name the target and loss: estimand, functional, likelihood, risk,
utility, calibration claim, or operating characteristic.
\item Name the uncertainty that must be returned: sampling variation,
measurement error, model misspecification, optimization error, selection,
feedback, domain shift, or sensitivity to untestable assumptions.
\end{enumerate}
This checklist is the whole book in miniature.  Measure theory says which
questions are observable; product spaces and processes say what kind of random
object is present; design and observation say how the record was generated;
targets and losses say what the analysis is trying to learn; asymptotics,
influence functions, bootstrap, and calibration say which uncertainty can be
carried back to the user.  The appendix keeps the book from needing miniature
textbooks in every neighboring field while still making clear that
statistics becomes modern only when it can translate across them without
forgetting its own grammar.

\section*{Sources and Further Reading}
\addcontentsline{toc}{section}{Sources and Further Reading}

This appendix is a translator's guide rather than a separate technical
treatment.  The algorithmic rows draw on classical Monte Carlo
\citep{metropolis1953equation,hastings1970monte}, stochastic approximation
\citep{robbinsMonro1951stochastic}, cross-validation
\citep{stone1974cross}, the bootstrap \citep{efron1979bootstrap}, empirical
risk and statistical learning \citep{vapnik1998statistical}, conformal
prediction \citep{vovk2005algorithmic}, Bayesian optimization
\citep{frazier2018tutorial}, and operator-splitting perspectives on modern
optimization \citep{davisYin2017threeOperator,ryuYin2022largeScale}.  The
bioinformatics discussion is anchored by the single-cell examples used in the
book, especially \citet{li2018scimpute} and \citet{cui2022scgtm}.  For the
clinical language of estimands, endpoint construction, target-trial emulation,
and real-world evidence, see
\citet{ich2021e9r1}, \citet{hernan2016targettrial},
\citet{weiskopf2013methods}, \citet{goldstein2017opportunities}, and
\citet{fda2018rweframework}.  The stochastic-dynamics reading of biochemical
reaction systems follows the chemical master-equation and exact-simulation
route represented by \citet{gillespie1977exact}; autonomous experimentation
and self-driving laboratories are represented by
\citet{shields2021bayesian} and \citet{abolhasani2023rise}.

%% file: frontmatter/preface.tex
\cleardoublepage
\phantomsection
\chapter*{Preface and User's Guide}
\addcontentsline{toc}{chapter}{Preface and User's Guide}
\markboth{Preface and User's Guide}{Preface and User's Guide}
\conceptindexes{preface, user guide, measure-theoretic language, reading strategy, real-data examples}

Statistics does not begin with a formula.  It begins when some piece of the
world is allowed to become a record.  Someone chooses a unit, a clock, an
instrument, an inclusion rule, a denominator, a coding convention, a cleaning
rule, a linkage procedure, or a stopping time.  By the time a dataset reaches
an analyst as a matrix, curve, image, graph, event history, spatial pattern, or
sequence of decisions, it already has a biography.  A clean formula can hide
that biography.  This book tries to keep it visible.

The route followed throughout the book is simple enough to remember as one
chain:
\[
  \begin{aligned}
  \hbox{world and question}
    &\longrightarrow \hbox{observation mechanism}
    \longrightarrow \hbox{data object}
    \longrightarrow \hbox{probability law} \\
    &\longrightarrow \hbox{target}
    \longrightarrow \hbox{procedure}
    \longrightarrow \hbox{uncertainty}
    \longrightarrow \hbox{use}.
  \end{aligned}
\]
It is also demanding enough to occupy a book.  Each arrow can break.  The
observed record may be a coarsened version of a fuller world.  The sampling
clock may matter.  Missingness may be informative.  A target may be easier to
name than to estimate.  A procedure may optimize the wrong loss.  A theorem
may stabilize the wrong object.  A deployment may change the system it was
meant to measure.  The point is not to make statistics gloomy.  The point is
to make it honest and therefore more useful.

The book's recurring ledger is practical rather than decorative: observation
mechanism, data structure, probability language, assumptions, target,
procedure, uncertainty, and use.  Whenever an argument moves from a real
question to a theorem and back to a claim, those entries should remain on the
page.  If one of them disappears, the reader should feel the draft under the
door.

\section*{Why Measure-Theoretic Language?}
\conceptindexes{measure-theoretic language, measurable spaces, random elements, stochastic processes}

Measure theory is used here because the book needs one grammar for many kinds
of records.  In elementary examples, probability can look like arithmetic on
events or densities on familiar Euclidean spaces.  Modern statistical problems
are less uniform.  A dataset may be a table, a curve, an image, a graph, a
point pattern, a censored event history, a marked process, an electronic health
record, or a many-layer object produced by preprocessing and linkage.  A design
may randomize treatments, adapt doses, choose spatial locations, stop early, or
decide which measurement is taken next.  A model may describe full data that
are never completely observed, while the analyst sees a selected, transformed,
coarsened, or censored version of that world.

The measure-theoretic language lets these cases share a spine.  A
\(\sigma\)-field records what distinctions are observable.  A random element is
a measurable map from the world into the space where the data live.  A design
can be represented by an index, a law, a kernel, or an observation map.  A
statistical model is a family of probability measures on a stated measurable
space.  Likelihoods are Radon--Nikodym derivatives when a common reference
measure is available.  Conditioning, missingness, censoring, empirical
measures, weak convergence, filtrations, stopping times, compensators, and
martingales all use the same basic grammar.

The abstraction is not included to make simple things look impressive.  It is
included to keep different applications from being mistaken for unrelated
tricks.  The same structural questions return across the book: What is
observable?  What is the random object?  Which law generates it?  Which map
turns the fuller world into the recorded data?  Which assumptions make
conditioning, estimation, prediction, testing, or decision possible?  For this
book, measure theory is not ornament.  It is the infrastructure that keeps the
translation from data to probability to inference from becoming a magic act.

\section*{Who the Book Is For}
\conceptindexes{audience, targeted audience, graduate students, practitioners, biomedical statistics}

The targeted audience has three overlapping groups.  The first is graduate
students and early researchers in statistics, biostatistics, data science,
machine learning, bioinformatics, clinical research, and adjacent quantitative
fields.  These readers need a map that is mathematically honest without
turning every chapter into a specialist monograph.  The second is the applied
scientist, clinical investigator, data scientist, or industry researcher who
works with biomedical, genomic, imaging, environmental, platform, or deployed
AI data and wants to know what statistical reasoning tries to preserve when a
substantive question becomes a model.  The third is the expert reader or course
designer who wants a recognizable probability-and-inference spine, but also
wants that spine connected to modern data objects and targets.

Readers should be willing to work with mathematical notation, but they need
not already be specialists in measure theory, empirical processes, or
stochastic-process asymptotics.  The book is meant to build a map: design,
likelihood, data-object grammar, target definition, laws of large numbers, weak
convergence, empirical processes, event histories, testing, efficiency, and
decision under feedback are not separate territories.  They are different
views of the same movement from observed structure to uncertain claim.
Advanced undergraduates with strong preparation can read the book for
orientation, postponing the more technical proofs to a second pass.

\section*{How to Read}
\conceptindexes{reading strategy, real-data capsules, mathematical mechanism, statistical question}

The book can be read at two speeds.  At the first speed, read for the
statistical question: What is being observed?  What is the unit?  What is the
task?  What comparison, prediction, explanation, test, or decision is wanted?
What uncertainty must be reported?  At the second speed, read for the
mathematical mechanism: Which measurable space holds the object?  Which law,
kernel, filtration, or product construction generates it?  Which convergence
theorem, likelihood factorization, estimating equation, testing geometry, or
process representation supports the claim?

A useful first reading keeps five questions nearby:
\begin{enumerate}
\item What is the data structure?
\item What observation mechanism produced it?
\item What target or task is being named?
\item What assumptions make that target or task mathematically possible?
\item What is gained, lost, or hidden by those assumptions?
\end{enumerate}
These questions are simple enough to carry through every chapter, but they are
not elementary.  They are the questions that keep a statistical analysis from
becoming a ritual application of formulas.

Real-data examples are used as anchors, not as a separate cookbook track.  A
classic example may train the eye; a small public dataset may support a
companion calculation; an industrial or institutional system may appear through
published documentation, standards, or peer-reviewed summaries when raw data
cannot be distributed.  In all three cases the example earns space only if it
does mathematical work: it must clarify an observation mechanism, a data
structure, a target, a loss, an uncertainty statement, or a feedback loop.
This is why the same example may return in different chapters with a different
job.  A platform experiment may first illustrate design, later define a target
effect, and finally become a feedback system.  An oncology real-world-evidence
record may first illustrate observation, later become an estimand problem, and
then reappear as an event history.  A molecular-prediction benchmark or
precision-agriculture system may begin as a data structure and later become a
loss, policy, or deployment question.

Some of these threads should become real-data analysis capsules.  A capsule is
more than a named example.  It includes a record or published summary, a small
computation, the mathematical object that computation estimates, and the way the
result pushes back on the theory.  The pushback is often the most important
part: a survival curve forces the analysis to name censoring and time zero; a financial
return plot makes iid intuition fragile; a single-cell count matrix makes the
observation mechanism visible; a platform experiment separates metric lift from
the value or policy target.  In this sense the book does not only apply theory
to data.  It also lets data teach the reader which theory is needed.

\section*{Example Threads}
\conceptindexes{example threads, real-data capsules, case studies, applied threads, reader navigation}

The book does not include a full frontmatter list of examples.  Many examples
are local teaching devices: a two-line distributional calculation, a
notation-checking toy model, or a citation-only bridge.  Listing every one of
them would make the frontmatter heavier without giving the reader a better
route.  The navigation below instead records the recurring threads that carry
substantive data objects or case-study logic across chapters; the separate
Real-Data Capsules and Case Studies page names the compact worked anchors.

\begin{description}[
  leftmargin=0pt,
  labelsep=0.65em,
  style=unboxed,
  font=\normalfont\scshape\color{bookteal!72!black},
  itemsep=0.20\baselineskip,
  topsep=0.26\baselineskip
]
\item[Clinical and event history.]
Trial design, real-world oncology endpoints, clinical-trial paths, survival
records, multi-state files, at-risk processes, cumulative hazards, and
event-time martingales connect Chapters~4--6, 10--11, and 16--17.

\item[Single-cell.]
PBMC count matrices, dropout, generative simulation, pseudotime, feature
fields, and constrained likelihood link the observation chapters to empirical
processes, M/Z-estimation, and local uncertainty.

\item[Spatial.]
Bombing grids, spatial point patterns, spatial omics summaries, climate
curves, and Wasserstein responses show how location and geometry change the
observed object before they change the theorem.

\item[Finance risk.]
RiskMetrics returns, market-stability checks, value-at-risk fields,
quant-trading audits, and deployed risk monitors keep iid intuition, searched
targets, and feedback under pressure.

\item[Robot lab.]
Autonomous laboratory logs and scientific-AI loops turn design into a
sequential policy problem: the next measurement depends on what the current
record has already made knowable.

\item[Online experiments.]
Platform experiments, metric lift, sequential monitoring, interference,
off-policy evaluation, and deployment risk return the book's inference
language to products that keep changing while they are measured.

\item[Text, network, and image.]
Text features, contact or protein-interaction networks, image benchmarks,
policy-generated game records, protein-generation systems, and deployed
language models show why modern data objects need probability spaces larger
than tables.
\end{description}

Chapter~1 fixes the route, and Chapter~2 fixes the habit of mind.  Part~I asks
the reader to see data as observed structure.  Part~II, Chapters~3--6, gives
the grammar of observation: probability spaces, measurable objects, design,
conditioning, likelihood, products, kernels, and process laws.  Part~III,
Chapters~7--8, is the book's hinge: first name the modern data object, then
name the target or claim it is meant to support.  Part~IV, Chapters~9--12,
builds stability, limit, and representation tools: laws of large numbers,
concentration, weak convergence, empirical processes, and functional
coordinates.  Part~V, Chapters~13--16, builds inference through local
geometry, influence functions, and event time: estimation, testing,
linearization, filtrations, stopping, compensators, and event histories.
Part~VI, Chapter~17, is the closing return from inference to use, where
prediction, intervention, policy, deployment, and feedback make the statistical
object part of a larger system.

The technical chapters should therefore be read as grammar and bridges, not as
mini-monographs.  Readers who want the mathematical spine can follow
Chapters~3, 6, 9, 10, 11, 13, 14, 15, and 16 in sequence.  Readers working with
biomedical, clinical, or industrial data can instead follow the example threads
on design, observation, missingness, censoring, event histories, feedback, and
deployment.  The two readings are meant to meet: theory should sharpen the
examples, and examples should keep theory from floating away.

\section*{What the Book Does Not Replace}
\conceptindexes{scope, prerequisites, mathematical statistics, machine learning}

The book assumes that readers are willing to meet mathematical notation and to
slow down when definitions matter. At the same time, it does not replace full
courses in measure-theoretic probability, mathematical statistics, convex
optimization, causal inference, Bayesian computation, machine learning,
survival analysis, or stochastic processes. Those subjects each deserve their
own sustained treatment. This book instead tries to explain why they belong to
one statistical project.

The guiding promise is simple: the reader should come away with a clearer
account of how data become random objects, how assumptions shape models, how
procedures turn observations into claims, and how uncertainty supports
interpretation and use.  If the book succeeds, a theorem will feel less like a
closed box and more like a carefully labeled passage from a question to a
claim.

\cleardoublepage

%% file: frontmatter/notation.tex
\cleardoublepage
\phantomsection
\chapter*{Notation and Conventions}
\addcontentsline{toc}{chapter}{Notation and Conventions}
\markboth{Notation and Conventions}{Notation and Conventions}
\conceptindexes{notation, conventions, overloaded symbols, probability notation, stochastic-process notation}

This table is a reading key for notation used repeatedly across the book, not
an exhaustive glossary.  A definition inside a theorem, proof, example, or
chapter always takes precedence.  When no local definition is given, the
conventions below apply.

The order follows the statistical compass in Chapter~\ref{chap:introduction}.
Spaces and measures describe where observations live; random variables turn
that world into data; empirical measures translate data back into
distributions; inferential notation records the model map used to answer a
question.  Typographically, scalar random quantities are usually italic,
finite-dimensional linear-algebra objects are bold roman, and named operators
are upright.  The main probability convention is this: use blackboard-bold
\(\Prob\) and \(\Expect\) for the ambient probability and expectation operators
when the law is not itself being compared, varied, or estimated; use ordinary
italic \(P,Q,P_\theta,P_0\), and \(P_n\) for named laws, model laws, true laws,
empirical measures, and integration operators.

\begin{center}
\small
\setlength{\tabcolsep}{0.55em}
\renewcommand{\arraystretch}{1.18}
\begin{longtable}{@{}>{\raggedright\arraybackslash}p{0.33\textwidth}>{\raggedright\arraybackslash}p{0.59\textwidth}@{}}
\multicolumn{2}{@{}l}{\textbf{Notation and core symbols}}\\[0.25em]
\toprule
Symbol & Meaning \\
\midrule
\endfirsthead
\multicolumn{2}{@{}l}{\textbf{Notation and core symbols (continued)}}\\[0.25em]
\toprule
Symbol & Meaning \\
\midrule
\endhead
\bottomrule
\endfoot
\multicolumn{2}{@{}l}{\textbf{Sets, Spaces, and Measures}}\\[0.2em]
$\Omega$ & underlying set or sample space. \\
$\emptyset$, $A^c$, $A-B$, $A\triangle B$ & empty set, complement, relative difference, and symmetric difference. \\
$\Pow(\Omega)$ & power set of $\Omega$. \\
$\Nat$, $\Int$, $\Rat$, $\Real$, $\Real^d$, $\Rplus$ & positive integers, integers, rational numbers, real line, Euclidean space, and $[0,\infty)$. \\
$\Hilbert$ & a Hilbert space, usually separable when random elements or Gaussian measures are involved. \\
$(\Omega,\fieldF)$ & measurable space; $\fieldF$ is a $\sigma$-algebra. \\
$(\Omega,\fieldF,\mu)$ & measure space with measure $\mu$. \\
$(\Omega,\fieldF,\Prob)$ or $(\Omega,\fieldF,P)$ & probability space; $\Prob$ marks the ambient operator on a background sample space, while $P$ is used once the law is named, compared with other laws, indexed by a parameter, or passed into a functional. \\
$\classS,\classR,\classG,\classC$ & generic classes of sets, often a semi-ring, ring, agreement class, or auxiliary class. \\
$\sigma(\classS)$ & $\sigma$-algebra generated by $\classS$. \\
$\lambda(\classS)$ & Dynkin, or $\lambda$-, system generated by $\classS$; this is not Lebesgue measure. \\
$\Borel(S)$, $\mathcal B(S)$ & Borel $\sigma$-algebra on a topological space $S$. \\
$\mathcal L(\R^d)$ & Lebesgue measurable sets in $\R^d$. \\
$\lambda^d$ & $d$-dimensional Lebesgue measure; $\lambda=\lambda^1$ on $\R$ unless context says otherwise. \\
$\mathcal E/\mathcal S$ measurable & shorthand for a map $f:(E,\mathcal E)\to(S,\mathcal S)$ satisfying $f^{-1}(B)\in\mathcal E$ for every $B\in\mathcal S$. \\
$\mu,\nu$ & generic measures. \\
$\mu\perp\nu$ & mutually singular measures; this is a measure-theoretic or geometric orthogonality symbol, not the stochastic-independence symbol below. \\
$\mu^*$, $\mu_*$ & outer and inner measures induced by $\mu$, when these are defined. \\
$M_{\mu^*}$ & $\mu^*$-measurable sets in the sense of Carath\'eodory. \\
$\delta_x$ & point mass at $x$. \\
$\mu\otimes\nu$ & product measure. \\
$K(\omega,B)$ & transition kernel: a measure in $B$ and a measurable function in $\omega$; Chapter~6 gives the formal definition. \\
\addlinespace
\multicolumn{2}{@{}l}{\textbf{Random Variables and Distributions}}\\[0.2em]
$X,Y,Z$ & random variables or random elements.  When a finite-dimensional vector or matrix is used as a linear-algebra object, it is written in bold roman, for example $\mathbf X$, $\mathbf y$, or $\mathbf Z$; scalar coordinates such as $X_i$ and process values such as $X(t)$ remain ordinary italic.  Greek parameter vectors such as $\beta$ are kept in Greek form unless a local definition says otherwise. \\
$X:(\Omega,\fieldF)\to(S,\mathcal S)$ & random element, equivalently an $\fieldF/\mathcal S$ measurable map from the sample space to the state space. \\
$\Law(X)$ & distribution, or law, of $X$.  We use $\Law$ rather than $\mathcal L(X)$ to avoid conflict with $\mathcal L(\R^d)$. \\
$\Prob(A)$, $P(A)$, $\Expect X$ & probability of $A$ under the ambient operator $\Prob$, probability of $A$ under a named law $P$, and expectation under the ambient operator. \\
$P,Q,P_\theta,P_0$ & named probability laws: two laws being compared, a model law indexed by $\theta$, and the true law.  We write $P_\theta(A)$ and $P_\theta f$, not $\Prob_\theta(A)$ or $\Prob_\theta f$, when the law is named. \\
$\Expect_P$, $\Expect_{P_0}$ & expectation under a named law $P$ and under the true law $P_0$. \\
$\Var(X)$, $\Cov(X,Y)$ & variance and covariance. \\
$\Corr(X,Y)$ & correlation. \\
$X\indep Y$, $X\indep Y\mid Z$ & stochastic independence and conditional independence.  The double symbol \(\indep\) is used to distinguish independence of random objects from orthogonality or mutual singularity, written with \(\perp\). \\
$\ind{A}$, $\indset{A}$ & indicators: $\mathbf 1\{A\}$ and $\mathbf 1_A$. \\
$\Normal(m,\Sigma)$ & normal distribution with mean $m$ and covariance $\Sigma$. \\
$\Poisson(\lambda)$, $\Bernoulli(p)$, $\Binomial(n,p)$, $\NegBin(\mu,\phi)$ & common discrete laws.  The parameterization is stated locally when it matters; in negative-binomial examples, $\phi$ is an overdispersion parameter, not the dose-finding target below. \\
$\Beta(a,b)$, $\Unif(0,1)$, $\Dirichlet(\alpha)$, $\DP(\eta)$ & other named laws, or in the case of $\Beta$, the beta function when the local formula defines it.  Named objects are written in upright roman through macros. \\
$\chi^2_\nu$ & chi-square distribution with $\nu$ degrees of freedom. \\
$X_n\toP X$ & convergence in probability. \\
$X_n\weakto X$ & weak convergence, or convergence in distribution, used consistently throughout the book. \\
$\outerProb$, $\innerProb$ & outer and inner probability. \\
$\outerExpect$, $\innerExpect$ & outer and inner expectation. \\
$Z_n\toPstar 0$, $R_n=\opstar(a_n)$ & convergence and small-order notation in outer probability; the star disappears when the relevant random quantities are measurable. \\
\addlinespace
\multicolumn{2}{@{}l}{\textbf{Design and Data Collection}}\\[0.2em]
$d\in\mathcal D$ & design and design space. \\
$P_{\theta,d}$ & design-indexed law under parameter $\theta$ and design $d$, usually for the full designed data before an observation map is applied. \\
$O$ & observed data object received by the analyst. \\
$\mathcal O,\mathcal G$ & observed-data space and its $\sigma$-algebra. \\
$\Phi_d$ & observed-data map, possibly design-dependent. \\
$P^O_{\theta,d}$ & observed-data law, often $P_{\theta,d}\circ\Phi_d^{-1}$. \\
$Z_i$ & treatment-assignment indicator in randomized-design examples. \\
$Y_i(1),Y_i(0)$ & potential outcomes under treatment and control, when causal design notation is used. \\
$\xi=\sum_j w_j\delta_{x_j}$ & approximate design measure with support points $x_j$ and weights $w_j$. \\
$M(\xi,\theta)$, $M(\xi)$ & information matrix under design $\xi$, with $\theta$ included when the model is nonlinear or locally parameter-dependent. \\
$\Phi\{M(\xi,\theta)\}$ & optimal-design criterion applied to the information matrix. \\
$\phi$ & target toxicity probability in dose-finding sections.  Elsewhere $\phi$ may be a scalar map, density, kernel, or eigenfunction, always fixed by the local definition. \\
\addlinespace
\multicolumn{2}{@{}l}{\textbf{Empirical Measures and Function Classes}}\\[0.2em]
$X_1,\ldots,X_n$ & observed sample, usually iid with common law $P$ when no dependence structure is specified. \\
$\mathcal X,\mathcal A$ & observation space and its $\sigma$-algebra. \\
$Pf$ & integral $\int f\,dP$; for a random variable $X\sim P$, this is $\Expect_P f(X)$. \\
$P_n$ & empirical measure $n^{-1}\sum_{i=1}^n\delta_{X_i}$. \\
$P_nf$ & empirical average $n^{-1}\sum_{i=1}^n f(X_i)$. \\
$\mathcal P(S)$ & probability measures on a state space $S$, especially in weak-convergence and empirical-measure discussions. \\
$\mathcal F$ & function class in empirical-process chapters; this is distinct from the sample-space $\sigma$-algebra $\fieldF$ by context. \\
$\|P_n-P\|_{\mathcal F}$ & uniform empirical deviation $\sup_{f\in\mathcal F}|P_nf-Pf|$. \\
$\mathbb G_n f$ & empirical process $\sqrt n(P_n-P)f$. \\
$\ell^\infty(\mathcal F)$ & bounded real functions on $\mathcal F$ with the supremum norm. \\
\addlinespace
\multicolumn{2}{@{}l}{\textbf{Inference and Asymptotics}}\\[0.2em]
$\theta$, $\theta_0$, $\Theta$ & parameter, true parameter, and parameter space. \\
$\mathcal P=\{P_\theta:\theta\in\Theta\}$ & statistical model as a family of laws. \\
$p_{\theta,d}$ & Radon--Nikodym derivative $dP^O_{\theta,d}/d\nu$, read as a density or mass function for the observed-data law with respect to a stated dominating measure. \\
$L(\theta;o,d)$ & likelihood after observing $O=o$ under design $d$, defined up to factors not depending on $\theta$. \\
$\Lambda(\theta_1,\theta_0;O)$ & likelihood ratio comparing $\theta_1$ with $\theta_0$ using the observed data. \\
$\ell_\theta$, $\dot\ell_\theta$ & log-likelihood contribution and score, when a likelihood model is in use. \\
$\logit(p)$, $\logit^{-1}(x)$ & log-odds transform and inverse-logit link. \\
$\Phi$, $\phi$ & generic maps or functionals when locally introduced; $\Phi$ is also used for the standard normal cdf in a few normal-approximation exercises. \\
$M_n(\theta)$, $M(\theta)$ & sample and population objective functions in $M$-estimation. \\
$m_\theta$ & observation-level criterion, so that often $M_n(\theta)=P_nm_\theta$ and $M(\theta)=Pm_\theta$. \\
$\Psi_n(\theta)$, $\Psi(\theta)$ & sample and population estimating equations or score-like maps in $Z$-estimation. \\
$\psi_\theta$ & observation-level estimating function, so that often $\Psi_n(\theta)=P_n\psi_\theta$. \\
$\hat\theta_n$, $\tilde\theta_n$ & estimator or approximate estimator. \\
$T(P)$, $T(P_n)$ & statistical functional and its plug-in estimator. \\
$\IF(x;T,P)$ & influence function of the functional $T$ at distribution $P$, evaluated at point $x$. \\
$I(\theta_0)$ & Fisher information at the true parameter in regular likelihood models. \\
$o_{\Prob}(a_n)$, $O_{\Prob}(a_n)$ & stochastic little-oh and big-oh under the ambient probability operator. \\
$o_P(a_n)$, $O_P(a_n)$, $o_{P_0}(a_n)$, $O_{P_0}(a_n)$ & stochastic little-oh and big-oh under a named law $P$ and under the true law $P_0$; indexed versions such as $o_{P_n}(1)$ name a sequence of laws. \\
$\argmax$, $\argmin$ & sets of maximizers and minimizers. \\
$\dot m_\theta$, $\dot\Psi_\theta$ & derivatives with respect to the parameter, when differentiability is assumed. \\
$\tr(A)$, $\diag(a_1,\ldots,a_p)$ & trace and diagonal matrix/operator notation. \\
\addlinespace
\multicolumn{2}{@{}l}{\textbf{Stochastic Processes and Event-Time Data}}\\[0.2em]
$\mathbb F=\{\fieldF_t:t\ge0\}$ & filtration, interpreted as information available up to time $t$. \\
$(\Omega,\fieldF,\mathbb F,P)$ & stochastic basis; here \(P\) is the named probability law supporting the filtered process. \\
$\fieldF\otimes\Borel(T)/\mathcal S$ measurable & jointly measurable map $X:\Omega\times T\to S$ with domain $\sigma$-field $\fieldF\otimes\Borel(T)$. \\
$\fieldF_{t-}$, $\fieldF_T$, $\fieldF_{T-}$ & information just before $t$, at a stopping time $T$, and just before $T$. \\
$T$, $T_n$ & stopping time, event time, or jump time, depending on context. \\
$X_t$, $X_{t-}$, $\Delta X_t$ & process value, left limit, and jump $X_t-X_{t-}$. \\
$S_\Delta=S\cup\{\Delta\}$ & state space with cemetery or empty mark attached. \\
$N_t$, $N(t)$ & counting process. \\
$N((0,t]\times B)$, $N_t(B)$ & marked counting process counting events up to $t$ with mark in $B$. \\
$Y_t$, $Y_i(t)$ & at-risk or exposure process. \\
$\Lambda_t$, $\Lambda_t(B)$ & compensator of a counting process or marked counting process. \\
$M_t=N_t-\Lambda_t$ & martingale residual associated with a counting process. \\
$A(t)$ & cumulative hazard, cumulative rate, or mean function; its meaning is specified locally. \\
$\lambda(t)$, $\lambda_{jk}(t)$ & intensity, hazard rate, transition rate, or model parameter, as specified locally. \\
$[M]_t$, $\langle M\rangle_t$ & quadratic variation and predictable quadratic variation. \\
$F(t)$ & distribution function. \\
$S(t)$, $\bar F(t)$ & survival or tail function.  Typically \(S(t)=\Prob(T>t)\) under the ambient law, or \(S_P(t)=P(T>t)\) when a named law \(P\) is being compared.  \(\bar F\) is used for the same survival/tail idea, sometimes with the at-risk convention \(T\ge t\) and left limits written explicitly. \\
$P_x$, $\mathsf P_t$, $\matP(s,t)$ & process notation: \(P_x\) is the law of a Markov process started at \(x\), \(\mathsf P_t f(x)=\Expect_x f(X_t)\) is a transition semigroup acting on functions, and \(\matP(s,t)\) is a transition matrix.  These are locally introduced process objects, not the ambient operator \(\Prob\). \\
$\mathcal P$ & predictable $\sigma$-field in continuous-time process sections; statistical model in inference sections. \\
\end{longtable}
\end{center}

\noindent\textbf{Overloaded symbols.}
Some symbols deliberately carry standard meanings in different parts of the book.
The symbol $\lambda^d$ always denotes Lebesgue measure on $\R^d$; the expression
$\lambda(\classS)$ denotes the $\lambda$-system generated by $\classS$; and
$\lambda(t)$ or $\lambda_{jk}(t)$ denotes a rate, intensity, hazard, or transition
rate only when a process or statistical model has introduced it.  Similarly,
$\mathcal L(\R^d)$ is reserved for the Lebesgue $\sigma$-algebra, while
$\Law(X)$ denotes the distribution of a random element.  The symbol $\Delta$ may
mean symmetric difference, a jump operator, or a cemetery state; the surrounding
definition determines which meaning is active.
The ordinary letter \(P\) may also appear in process notation as a named law
\(P_x\), a transition matrix entry \(P_{ij}(t)\), or a semigroup written
\(\mathsf P_t\); local typography distinguishes these from the ambient
probability operator \(\Prob\).
The calligraphic letter $\mathcal F$ also has two roles: $\fieldF$ denotes a
$\sigma$-algebra or filtration component, while $\mathcal F$ denotes a function
class in empirical-process arguments.  The symbol $\mathcal P$ may denote a
statistical model, a predictable $\sigma$-field, or the space $\mathcal P(S)$ of
probability measures on \(S\); the argument or chapter fixes the meaning.
Similarly, $M_n(\theta)$ is an objective function in $M$-estimation, whereas
$M_t$ is a martingale residual in process chapters; $T$ may be a stopping time,
event time, or statistical functional, and is always fixed by the local
definition.  The letters $\Phi$, $\phi$, $\Psi$, and $\psi$ are reserved for
locally defined maps, criteria, functionals, or estimating functions, except
for standard normal cdf/density notation where that use is explicitly stated.

\cleardoublepage